\documentclass[corporate, monograph, usegrafix]{adqoas}
\usepackage[mathscr]{eucal}
\usepackage{makeidx}
\makeindex

\newcommand{\unab}{\nabla_u}
\newcommand{\ucurl}[1]{{\let\nab\unab\curl{#1}}}
\newcommand{\udivg}[1]{{\let\nab\unab\dive{#1}}}
\newcommand{\ugrad}[1]{{\let\nab\unab\grad{#1}}}
\newcommand{\ulapl}[1]{{\let\nab\unab\lapl{#1}}}
\newcommand{\ugdiv}[2]{{\let\nab\unab\gdiv{#1}{#2}}}

\newcommand{\scalt}{\scal{t}}
\newcommand{\scalb}{\scal{b}}
\newcommand{\scalh}{\scal{h}}
\newcommand{\scalz}{\scal{z}}
\newcommand{\scalf}{\scal{f}}
\newcommand{\scals}{\scal{s}}
\newcommand{\scalq}{\scal{q}}
\newcommand{\scalr}{\scal{r}}
\newcommand{\scalu}{\scal{u}}
\newcommand{\scalc}{\scal{c}}
\newcommand{\scalw}{\scal{w}}
\newcommand{\scala}{\scal{a}}
\newcommand{\scalm}{\scal{m}}
\newcommand{\scaln}{\scal{n}}

\newcommand{\vectr}{\vect{r}}
\newcommand{\vecta}{\vect{a}}
\newcommand{\vectu}{\vect{u}}

\newcommand{\vectz}{\vect{z}}
\newcommand{\vectw}{\vect{w}}
\newcommand{\vectc}{\vect{c}}
\newcommand{\vectg}{\vect{g}}
\newcommand{\vecth}{\vect{h}}
\newcommand{\vecte}{\vect{e}}

\newcommand{\xirep}{\xi}
\newcommand{\murep}{\mu}
\newcommand{\Pirep}{\Pi}
\newcommand{\pirep}{\pi}
\newcommand{\nurep}{\nu}

\newcommand{\etarep}{\eta}
\newcommand{\Phirep}{\Phi}
\newcommand{\Psirep}{\Psi}
\newcommand{\phirep}{\phi}
\newcommand{\psirep}{\psi}
\newcommand{\taurep}{\tau}
\newcommand{\rhorep}{\rho}
\newcommand{\betrep}{\beta}
\newcommand{\zetarep}{\zeta}
\newcommand{\Gamrep}{\Gamma}
\newcommand{\Delrep}{\Delta}
\newcommand{\delrep}{\delta}
\newcommand{\thtrep}{\theta}
\newcommand{\Thtrep}{\Theta}
\newcommand{\sigrep}{\sigma}

\newcommand{\alprep}{\alpha}
\newcommand{\kaprep}{\kappa}
\newcommand{\Omerep}{\Omega}
\newcommand{\omerep}{\omega}
\newcommand{\gamrep}{\gamma}
\newcommand{\vpirep}{\varpi}
\newcommand{\Lamrep}{\Lambda}
\newcommand{\lamrep}{\lambda}
\newcommand{\upsrep}{\upsilon}

\newcommand{\Upsrep}{\Upsilon}
\newcommand{\vphirep}{\varphi}
\newcommand{\vrhorep}{\varrho}
\newcommand{\vkaprep}{\varkappa}
\newcommand{\vsigrep}{\varsigma}
\newcommand{\vthtrep}{\vartheta}
\newcommand{\vepsrep}{\varepsilon}

\newcommand{\vectups}{\boldsymbol{\upsrep}}
\newcommand{\vectkap}{\boldsymbol{\kaprep}}
\newcommand{\vectOme}{\boldsymbol{\Omerep}}
\newcommand{\vectLam}{\boldsymbol{\Lamrep}}
\newcommand{\vectbet}{\boldsymbol{\betrep}}
\newcommand{\vectsig}{\boldsymbol{\sigrep}}

\newcommand{\vg}{\vectg}
\newcommand{\vga}{\vg_1}
\newcommand{\vgb}{\vg_2}
\newcommand{\vgc}{\vg_3}
\newcommand{\vgd}{\vg_4}
\newcommand{\vge}{\vg_5}
\newcommand{\vgf}{\vg_6}
\newcommand{\vgg}{\vg_7}
\newcommand{\vgh}{\vg_8}
\newcommand{\vgi}{\vg_9}

\newcommand{\vj}{\boldsymbol{\jmath}}
\newcommand{\vja}{\vj_1}
\newcommand{\vjb}{\vj_2}
\newcommand{\vjc}{\vj_3}
\newcommand{\vjd}{\vj_4}
\newcommand{\vje}{\vj_5}
\newcommand{\vjf}{\vj_6}
\newcommand{\vjg}{\vj_7}
\newcommand{\vjh}{\vj_8}
\newcommand{\vji}{\vj_9}
\newcommand{\vjj}{\vj_0}

\newcommand{\vscr}{\boldsymbol{\mathscr J}}
\newcommand{\vscra}{\vscr_a}
\newcommand{\vscrb}{\vscr_b}
\newcommand{\vscrc}{\vscr_c}
\newcommand{\vscrd}{\vscr_d}
\newcommand{\vscre}{\vscr_e}
\newcommand{\vscrf}{\vscr_f}
\newcommand{\vscrg}{\vscr_g}
\newcommand{\vscrh}{\vscr_h}
\newcommand{\vscri}{\vscr_i}
\newcommand{\vscrj}{\vscr_j}
\newcommand{\vscrk}{\vscr_k}
\newcommand{\vscrl}{\vscr_l}
\newcommand{\vscrn}{\vscr_n}
\newcommand{\vscro}{\vscr_o}
\newcommand{\vscrp}{\vscr_p}
\newcommand{\vscrq}{\vscr_q}
\newcommand{\vscrr}{\vscr_r}

\newcommand{\veus}{\boldsymbol{\mathscr S}}
\newcommand{\veusa}{\veus_a}
\newcommand{\veusb}{\veus_b}
\newcommand{\veusc}{\veus_c}
\newcommand{\veusd}{\veus_d}
\newcommand{\veuse}{\veus_e}
\newcommand{\veusf}{\veus_f}
\newcommand{\veusg}{\veus_g}
\newcommand{\veush}{\veus_h}
\newcommand{\veusi}{\veus_i}
\newcommand{\veusj}{\veus_j}
\newcommand{\veusk}{\veus_k}
\newcommand{\veusl}{\veus_l}
\newcommand{\veusm}{\veus_m}
\newcommand{\veusn}{\veus_n}
\newcommand{\veuso}{\veus_o}
\newcommand{\veusp}{\veus_p}
\newcommand{\veusq}{\veus_q}
\newcommand{\veusr}{\veus_r}
\newcommand{\veuss}{\veus_s}
\newcommand{\veust}{\veus_t}
\newcommand{\veusu}{\veus_u}

\newcommand{\epsv}{\vepsrep}
\newcommand{\epsva}{\epsv_a}
\newcommand{\epsvb}{\epsv_b}
\newcommand{\epsvc}{\epsv_c}
\newcommand{\epsvd}{\epsv_d}
\newcommand{\epsve}{\epsv_e}
\newcommand{\epsvf}{\epsv_f}
\newcommand{\epsvg}{\epsv_g}
\newcommand{\epsvh}{\epsv_h}
\newcommand{\epsvi}{\epsv_i}
\newcommand{\epsvj}{\epsv_j}
\newcommand{\epsvk}{\epsv_k}
\newcommand{\epsvl}{\epsv_l}
\newcommand{\epsvm}{\epsv_m}
\newcommand{\epsvn}{\epsv_n}
\newcommand{\epsvo}{\epsv_o}

\newcommand{\vphix}{\vphirep}
\newcommand{\vphia}{\vphix_a}
\newcommand{\vphib}{\vphix_b}
\newcommand{\vphic}{\vphix_c}
\newcommand{\vphid}{\vphix_d}
\newcommand{\vphie}{\vphix_e}
\newcommand{\vphif}{\vphix_f}
\newcommand{\vphig}{\vphix_g}
\newcommand{\vphih}{\vphix_h}
\newcommand{\vphii}{\vphix_i}
\newcommand{\vphij}{\vphix_j}
\newcommand{\vphik}{\vphix_k}
\newcommand{\vphil}{\vphix_l}
\newcommand{\vphim}{\vphix_m}
\newcommand{\vphin}{\vphix_n}
\newcommand{\vphio}{\vphix_o}
\newcommand{\vphip}{\vphix_p}
\newcommand{\vphiq}{\vphix_q}
\newcommand{\vphir}{\vphix_r}
\newcommand{\vphis}{\vphix_s}
\newcommand{\vphit}{\vphix_t}
\newcommand{\vphiu}{\vphix_u}
\newcommand{\vphiv}{\vphix_v}

\newcommand{\dltv}{\delrep}
\newcommand{\dltva}{\dltv_a}
\newcommand{\dltvb}{\dltv_b}
\newcommand{\dltvc}{\dltv_c}
\newcommand{\dltvd}{\dltv_d}
\newcommand{\dltve}{\dltv_e}
\newcommand{\dltvf}{\dltv_f}
\newcommand{\dltvg}{\dltv_g}
\newcommand{\dltvh}{\dltv_h}
\newcommand{\dltvi}{\dltv_i}
\newcommand{\dltvj}{\dltv_j}
\newcommand{\dltvo}{\dltv_o}
\newcommand{\dltvk}{\dltv_k}
\newcommand{\dltvl}{\dltv_l}
\newcommand{\dltvm}{\dltv_m}
\newcommand{\dltvn}{\dltv_n}
\newcommand{\dltvp}{\dltv_p}
\newcommand{\dltvq}{\dltv_q}
\newcommand{\dltvr}{\dltv_r}
\newcommand{\dltvs}{\dltv_s}
\newcommand{\dltvt}{\dltv_t}
\newcommand{\dltvu}{\dltv_u}

\newcommand{\ethv}{\mathfrak x}
\newcommand{\ethva}{\ethv_a}
\newcommand{\ethvb}{\ethv_b}
\newcommand{\ethvc}{\ethv_c}
\newcommand{\ethvd}{\ethv_d}
\newcommand{\ethve}{\ethv_e}
\newcommand{\ethvf}{\ethv_f}
\newcommand{\ethvg}{\ethv_g}
\newcommand{\ethvh}{\ethv_h}
\newcommand{\ethvi}{\ethv_i}
\newcommand{\ethvj}{\ethv_j}
\newcommand{\ethvk}{\ethv_k}
\newcommand{\ethvl}{\ethv_l}
\newcommand{\ethvm}{\ethv_m}
\newcommand{\ethvn}{\ethv_n}
\newcommand{\ethvo}{\ethv_o}
\newcommand{\ethvp}{\ethv_p}
\newcommand{\ethvq}{\ethv_q}
\newcommand{\ethvr}{\ethv_r}
\newcommand{\ethvs}{\ethv_s}
\newcommand{\ethvt}{\ethv_t}
\newcommand{\ethvu}{\ethv_u}
\newcommand{\ethvv}{\ethv_v}
\newcommand{\ethvw}{\ethv_w}
\newcommand{\ethvx}{\ethv_x}
\newcommand{\ethvy}{\ethv_y}
\newcommand{\ethvz}{\ethv_z}

\newcommand{\vch}{\nurep}
\newcommand{\vcha}{\vch_1}
\newcommand{\vchb}{\vch_2}
\newcommand{\vchc}{\vch_3}
\newcommand{\vchd}{\vch_4}
\newcommand{\vche}{\vch_5}
\newcommand{\vchf}{\vch_6}
\newcommand{\vchg}{\vch_7}
\newcommand{\vchh}{\vch_8}

\newcommand{\efkt}{\hslash}
\newcommand{\efkta}{\efkt_a}
\newcommand{\efktb}{\efkt_b}
\newcommand{\efktc}{\efkt_c}
\newcommand{\efktd}{\efkt_d}
\newcommand{\efkte}{\efkt_e}
\newcommand{\efktf}{\efkt_f}
\newcommand{\efktg}{\efkt_g}
\newcommand{\efkth}{\efkt_h}
\newcommand{\efkti}{\efkt_i}
\newcommand{\efktj}{\efkt_j}
\newcommand{\efktk}{\efkt_k}
\newcommand{\efktl}{\efkt_l}
\newcommand{\efktm}{\efkt_m}
\newcommand{\efktn}{\efkt_n}
\newcommand{\efkto}{\efkt_o}
\newcommand{\efktp}{\efkt_p}
\newcommand{\efktq}{\efkt_q}
\newcommand{\efktr}{\efkt_r}
\newcommand{\efkts}{\efkt_s}
\newcommand{\efktt}{\efkt_t}
\newcommand{\efktu}{\efkt_u}
\newcommand{\efktv}{\efkt_v}
\newcommand{\efktw}{\efkt_w}
\newcommand{\efktx}{\efkt_x}

\newcommand{\vakp}{{\vkaprep}}
\newcommand{\vakpa}{\vakp_a}
\newcommand{\vakpb}{\vakp_b}
\newcommand{\vakpc}{\vakp_c}
\newcommand{\vakpd}{\vakp_d}
\newcommand{\vakpe}{\vakp_e}
\newcommand{\vakpf}{\vakp_f}
\newcommand{\vakpg}{\vakp_g}
\newcommand{\vakph}{\vakp_h}
\newcommand{\vakpi}{\vakp_i}
\newcommand{\vakpj}{\vakp_j}
\newcommand{\vakpk}{\vakp_k}
\newcommand{\vakpl}{\vakp_l}
\newcommand{\vakpm}{\vakp_m}
\newcommand{\vakpn}{\vakp_n}
\newcommand{\vakpo}{\vakp_o}
\newcommand{\vakpp}{\vakp_p}
\newcommand{\vakpq}{\vakp_q}
\newcommand{\vakpr}{\vakp_r}
\newcommand{\vakps}{\vakp_s}
\newcommand{\vakpt}{\vakp_t}
\newcommand{\vakpu}{\vakp_u}
\newcommand{\vakpv}{\vakp_v}
\newcommand{\vakpw}{\vakp_w}
\newcommand{\vakpx}{\vakp_x}
\newcommand{\vakpy}{\vakp_y}
\newcommand{\vakpz}{\vakp_z}

\newcommand{\efkc}{{\mathfrak f}}
\newcommand{\efkca}{\efkc_a}
\newcommand{\efkcb}{\efkc_b}
\newcommand{\efkcc}{\efkc_c}
\newcommand{\efkcd}{\efkc_d}
\newcommand{\efkce}{\efkc_e}
\newcommand{\efkcf}{\efkc_f}

\newcommand{\efko}{{\mathfrak o}}
\newcommand{\efkoa}{\efko_a}
\newcommand{\efkob}{\efko_b}
\newcommand{\efkoc}{\efko_c}
\newcommand{\efkod}{\efko_d}
\newcommand{\efkoe}{\efko_e}
\newcommand{\efkof}{\efko_f}
\newcommand{\efkog}{\efko_g}
\newcommand{\efkoh}{\efko_h}
\newcommand{\efkoi}{\efko_i}
\newcommand{\efkoj}{\efko_j}
\newcommand{\efkok}{\efko_k}
\newcommand{\efkol}{\efko_l}
\newcommand{\efkom}{\efko_m}
\newcommand{\efkon}{\efko_n}
\newcommand{\efkoo}{\efko_o}
\newcommand{\efkop}{\efko_p}
\newcommand{\efkoq}{\efko_q}
\newcommand{\efkor}{\efko_r}
\newcommand{\efkos}{\efko_s}
\newcommand{\efkot}{\efko_t}
\newcommand{\efkou}{\efko_u}
\newcommand{\efkov}{\efko_v}
\newcommand{\efkow}{\efko_w}
\newcommand{\efkox}{\efko_x}
\newcommand{\efkoy}{\efko_y}
\newcommand{\efkoz}{\efko_z}

\newcommand{\parv}{{\mathfrak v}}
\newcommand{\parva}{\parv_a}
\newcommand{\parvb}{\parv_b}
\newcommand{\parvc}{\parv_c}
\newcommand{\parvd}{\parv_d}
\newcommand{\parve}{\parv_e}
\newcommand{\parvf}{\parv_f}
\newcommand{\parvg}{\parv_g}
\newcommand{\parvh}{\parv_h}
\newcommand{\parvi}{\parv_i}
\newcommand{\parvj}{\parv_j}
\newcommand{\parvk}{\parv_k}
\newcommand{\parvl}{\parv_l}
\newcommand{\parvm}{\parv_m}
\newcommand{\parvn}{\parv_n}
\newcommand{\parvo}{\parv_o}
\newcommand{\parvp}{\parv_p}
\newcommand{\parvq}{\parv_q}
\newcommand{\parvr}{\parv_r}
\newcommand{\parvs}{\parv_s}
\newcommand{\parvt}{\parv_t}
\newcommand{\parvu}{\parv_u}
\newcommand{\parvv}{\parv_v}
\newcommand{\parvw}{\parv_w}
\newcommand{\parvx}{\parv_x}
\newcommand{\parvy}{\parv_y}
\newcommand{\parvz}{\parv_z}

\newcommand{\frkt}{{\mathfrak d}}
\newcommand{\frkta}{\frkt_a}
\newcommand{\frktb}{\frkt_b}
\newcommand{\frktc}{\frkt_c}
\newcommand{\frktd}{\frkt_d}
\newcommand{\frkte}{\frkt_e}
\newcommand{\frktf}{\frkt_f}
\newcommand{\frktg}{\frkt_g}
\newcommand{\frkth}{\frkt_h}
\newcommand{\frkti}{\frkt_i}
\newcommand{\frktj}{\frkt_j}
\newcommand{\frktk}{\frkt_k}
\newcommand{\frktl}{\frkt_l}
\newcommand{\frktm}{\frkt_m}
\newcommand{\frktn}{\frkt_n}
\newcommand{\frkto}{\frkt_o}
\newcommand{\frktp}{\frkt_p}
\newcommand{\frktq}{\frkt_q}
\newcommand{\frktr}{\frkt_r}
\newcommand{\frkts}{\frkt_s}
\newcommand{\frktt}{\frkt_t}
\newcommand{\frktu}{\frkt_u}
\newcommand{\frktv}{\frkt_v}
\newcommand{\frktw}{\frkt_w}
\newcommand{\frktx}{\frkt_x}
\newcommand{\frkty}{\frkt_y}
\newcommand{\frktz}{\frkt_z}
\newcommand{\frkx}{{\mathfrak p}}
\newcommand{\frkxa}{\frkx_a}
\newcommand{\frkxb}{\frkx_b}
\newcommand{\frkxc}{\frkx_c}
\newcommand{\frkxd}{\frkx_d}
\newcommand{\frkxe}{\frkx_e}
\newcommand{\frkxf}{\frkx_f}
\newcommand{\frkxg}{\frkx_g}
\newcommand{\frkxh}{\frkx_h}
\newcommand{\frkxi}{\frkx_i}

\newcommand{\frky}{{\mathfrak y}}
\newcommand{\frkya}{\frky_a}
\newcommand{\frkyb}{\frky_b}
\newcommand{\frkyc}{\frky_c}
\newcommand{\frkyd}{\frky_d}
\newcommand{\frkye}{\frky_e}
\newcommand{\frkyf}{\frky_f}
\newcommand{\frkyg}{\frky_g}
\newcommand{\frkyh}{\frky_h}
\newcommand{\frkyi}{\frky_i}
\newcommand{\frkyj}{\frky_j}
\newcommand{\frkyk}{\frky_k}
\newcommand{\frkyl}{\frky_l}
\newcommand{\frkym}{\frky_m}
\newcommand{\frkyn}{\frky_n}
\newcommand{\frkyo}{\frky_o}
\newcommand{\frkyp}{\frky_p}
\newcommand{\frkyq}{\frky_q}
\newcommand{\frkyr}{\frky_r}
\newcommand{\frkys}{\frky_s}
\newcommand{\frkyt}{\frky_t}
\newcommand{\frkyu}{\frky_u}
\newcommand{\frkyv}{\frky_v}
\newcommand{\frkyw}{\frky_w}
\newcommand{\frkyx}{\frky_x}
\newcommand{\frkyy}{\frky_y}
\newcommand{\frkyz}{\frky_z}

\newcommand{\vek}{{\mathfrak K}}
\newcommand{\veka}{\vek_a}
\newcommand{\vekb}{\vek_b}
\newcommand{\vekc}{\vek_c}
\newcommand{\vekd}{\vek_d}
\newcommand{\veke}{\vek_e}
\newcommand{\vekf}{\vek_f}
\newcommand{\vekg}{\vek_g}
\newcommand{\vekh}{\vek_h}
\newcommand{\veki}{\vek_i}
\newcommand{\vekj}{\vek_j}
\newcommand{\vekk}{\vek_k}
\newcommand{\vekl}{\vek_l}
\newcommand{\vekm}{\vek_m}
\newcommand{\vekn}{\vek_n}
\newcommand{\veko}{\vek_o}
\newcommand{\vekp}{\vek_p}
\newcommand{\vekq}{\vek_q}
\newcommand{\vekr}{\vek_r}
\newcommand{\veks}{\vek_s}
\newcommand{\vekt}{\vek_t}
\newcommand{\veku}{\vek_u}
\newcommand{\vekv}{\vek_v}
\newcommand{\vekw}{\vek_w}
\newcommand{\vekx}{\vek_x}
\newcommand{\veky}{\vek_y}
\newcommand{\vekz}{\vek_z}

\newcommand{\vps}{{\mathfrak H}}
\newcommand{\vpsa}{\vps_a}
\newcommand{\vpsb}{\vps_b}
\newcommand{\vpsc}{\vps_c}
\newcommand{\vpsd}{\vps_d}
\newcommand{\vpse}{\vps_e}
\newcommand{\vpsf}{\vps_f}
\newcommand{\vpsg}{\vps_g}
\newcommand{\vpsh}{\vps_h}
\newcommand{\vpsi}{\vps_i}
\newcommand{\vpsj}{\vps_j}
\newcommand{\vpsk}{\vps_k}
\newcommand{\vpsl}{\vps_l}

\newcommand{\alep}{\aleph}
\newcommand{\alepha}{\alep_1}
\newcommand{\alephb}{\alep_2}
\newcommand{\alephc}{\alep_3}
\newcommand{\alephd}{\alep_4}
\newcommand{\alephe}{\alep_5}
\newcommand{\alephf}{\alep_6}

\newcommand{\ima}{\Re}
\newcommand{\imaa}{\ima_1}
\newcommand{\imab}{\ima_2}
\newcommand{\imac}{\ima_3}
\newcommand{\imad}{\ima_4}
\newcommand{\imae}{\ima_5}

\newcommand{\vrho}{\vrhorep}
\newcommand{\vrhoa}{\vrho_a}
\newcommand{\vrhob}{\vrho_b}
\newcommand{\vrhoc}{\vrho_c}
\newcommand{\vrhod}{\vrho_d}
\newcommand{\vrhoe}{\vrho_e}
\newcommand{\vrhof}{\vrho_f}
\newcommand{\vrhog}{\vrho_g}
\newcommand{\vrhoh}{\vrho_h}
\newcommand{\vrhoi}{\vrho_i}
\newcommand{\vrhoj}{\vrho_j}
\newcommand{\vrhok}{\vrho_k}
\newcommand{\vrhol}{\vrho_l}
\newcommand{\vrhom}{\vrho_m}
\newcommand{\vrhon}{\vrho_n}
\newcommand{\vrhoo}{\vrho_o}
\newcommand{\vrhop}{\vrho_p}
\newcommand{\vrhoq}{\vrho_q}
\newcommand{\vrhor}{\vrho_r}
\newcommand{\vrhos}{\vrho_s}
\newcommand{\vrhot}{\vrho_t}
\newcommand{\vrhou}{\vrho_u}
\newcommand{\vrhov}{\vrho_v}
\newcommand{\vrhow}{\vrho_w}
\newcommand{\vrhox}{\vrho_x}
\newcommand{\vrhoy}{\vrho_y}
\newcommand{\vrhoz}{\vrho_z}

\newcommand{\vsig}{\vsigrep}
\newcommand{\vsiga}{\vsig_a}
\newcommand{\vsigb}{\vsig_b}
\newcommand{\vsigc}{\vsig_c}
\newcommand{\vsigd}{\vsig_d}
\newcommand{\vsige}{\vsig_e}
\newcommand{\vsigf}{\vsig_f}
\newcommand{\vsigg}{\vsig_g}
\newcommand{\vsigh}{\vsig_h}
\newcommand{\vsigi}{\vsig_i}
\newcommand{\vsigj}{\vsig_j}
\newcommand{\vsigk}{\vsig_k}
\newcommand{\vsigl}{\vsig_l}
\newcommand{\vsigm}{\vsig_m}
\newcommand{\vsign}{\vsig_n}
\newcommand{\vsigo}{\vsig_o}
\newcommand{\vsigp}{\vsig_p}
\newcommand{\vsigq}{\vsig_q}
\newcommand{\vsigr}{\vsig_r}
\newcommand{\vsigs}{\vsig_s}
\newcommand{\vsigt}{\vsig_t}
\newcommand{\vsigu}{\vsig_u}
\newcommand{\vsigv}{\vsig_v}

\newcommand{\vtt}{\imath}
\newcommand{\vtta}{\vtt_0}
\newcommand{\vttb}{\vtt_1}
\newcommand{\vttc}{\vtt_2}
\newcommand{\vttd}{\vtt_3}
\newcommand{\vtte}{\vtt_4}
\newcommand{\vttf}{\vtt_5}
\newcommand{\vttg}{\vtt_6}
\newcommand{\vtth}{\vtt_7}
\newcommand{\vtti}{\vtt_8}
\newcommand{\vttj}{\vtt_9}

\newcommand{\vbb}{\scalb}
\newcommand{\vbba}{\vbb_1}
\newcommand{\vbbb}{\vbb_2}
\newcommand{\vbbc}{\vbb_3}
\newcommand{\vbbd}{\vbb_4}
\newcommand{\vbbe}{\vbb_5}
\newcommand{\vbbf}{\vbb_6}

\newcommand{\szer}{\scals_0}
\newcommand{\sone}{\scals_1}
\newcommand{\stwo}{\scals_2}
\newcommand{\sthr}{\scals_3}
\newcommand{\sfou}{\scals_4}

\newcommand{\fzer}{\scalf_0}
\newcommand{\fone}{\scalf_1}
\newcommand{\ftwo}{\scalf_2}

\newcommand{\bcal}{\mathcal{B}}
\newcommand{\dcal}{\mathcal{D}}
\newcommand{\acal}{\mathcal{A}}
\newcommand{\hcal}{\mathcal{H}}
\newcommand{\ecal}{\mathcal{E}}
\newcommand{\lcal}{\mathcal{L}}
\newcommand{\ncal}{\mathcal{N}}
\newcommand{\fcal}{\mathcal{F}}
\newcommand{\gcal}{\mathcal{G}}
\newcommand{\pcal}{\mathcal{P}}
\newcommand{\rcal}{\mathcal{R}}
\newcommand{\jcal}{\mathcal{X}}
\newcommand{\xcala}{\jcal_1}
\newcommand{\xcalb}{\jcal_2}
\newcommand{\xcalc}{\jcal_3}
\newcommand{\xcald}{\jcal_4}

\newcommand{\toppo}{\top}
\newcommand{\botto}{\bot}
\newcommand{\absv}[1]{{\left|#1\right|}}

\newcommand{\bbk}{\mathbb{K}}
\newcommand{\bbt}{\mathbb{T}}
\newcommand{\bbr}{\mathbb{R}}

\newcommand{\bbkbar}{\overline{\bbk}}
\newcommand{\bbtbar}{\overline{\bbt}}
\newcommand{\bbrbar}{\overline{\bbr}}

\newcommand{\frt}{\boldsymbol{\ell}_t}
\newcommand{\frb}{\boldsymbol{\ell}_b}
\newcommand{\frn}{\boldsymbol{\ell}_n}
\newcommand{\frc}{\boldsymbol{\ell}_c}

\newcommand{\oell}{\boldsymbol{\overline{\ell}}}
\newcommand{\frtbar}{\oell_t}
\newcommand{\frbbar}{\oell_b}
\newcommand{\frnbar}{\oell_n}
\newcommand{\frcbar}{\oell_c}

\newcommand{\pointPM}{PM}
\newcommand{\pointQN}{QN}
\newcommand{\pointPR}{PR}
\newcommand{\pointQS}{QS}

\newcommand{\primedW}{X}
\newcommand{\primedE}{Y}
\newcommand{\primedSW}{SX}
\newcommand{\primedSE}{SY}

\newcommand{\dif}[1]{{#1}^{\prime}}
\newcommand{\fdota}{\fdot{\vecta}}
\newcommand{\ffdota}{\ffdot{\vecta}}
\newcommand{\fffdota}{\fffdot{\vecta}}
\newcommand{\fdote}{\fdot{\vecte}}
\newcommand{\ffdote}{\ffdot{\vecte}}
\newcommand{\fffdote}{\fffdot{\vecte}}
\newcommand{\fdotg}{\fdot{\gamrep}}
\newcommand{\ffdotg}{\ffdot{\gamrep}}
\newcommand{\fffdotg}{\fffdot{\gamrep}}
\newcommand{\fdott}{\fdot{\taurep}}
\newcommand{\ffdott}{\ffdot{\taurep}}
\newcommand{\fffdott}{\fffdot{\taurep}}

\newcommand{\vectro}{\vectr_o}
\newcommand{\vectuo}{\vectu_o}
\newcommand{\scalro}{\scalr_o}
\newcommand{\scaluo}{\scalu_o}
\newcommand{\phio}{\phirep_o}
\newcommand{\betao}{\betrep_o}
\newcommand{\Omegao}{\Omerep_o}
\newcommand{\thetao}{\thtrep_o}
\newcommand{\lambdao}{\lamrep_o}

\newcommand{\auxva}{x}
\newcommand{\auxvb}{y}
\newcommand{\auxvao}{x_o}
\newcommand{\auxvbo}{y_o}
\newcommand{\retronp}{\vepsrep_1}
\newcommand{\retrofq}{\vepsrep_2}
\newcommand{\retrofp}{\vepsrep_3}
\newcommand{\retrolq}{\vepsrep_4}
\newcommand{\retronpx}{\vepsrep_5}
\newcommand{\retrolqx}{\vepsrep_6}
\newcommand{\azimuth}{\nurep}
\newcommand{\munorm}{\murep_\bot}
\newcommand{\muparr}{\murep_\parallel}
\newcommand{\accnorm}{\scala_\bot}
\newcommand{\accparr}{\scala_\parallel}
\newcommand{\signorm}{\sigrep_\bot}
\newcommand{\sigparr}{\sigrep_\parallel}

\newcommand{\anglea}{A}
\newcommand{\angleb}{B}
\newcommand{\precvw}{\hslash}
\newcommand{\precvv}{{\mathscr S}}
\newcommand{\precvu}{{\mathfrak d}}
\newcommand{\precrate}{\Gamrep}
\newcommand{\precvect}{\boldsymbol{\Gamrep}}
\newcommand{\precdir}{\widehat{\precvect}}
\newcommand{\latitm}{\vepsrep}
\newcommand{\latitr}{\nurep}
\newcommand{\longtm}{\vpirep}
\newcommand{\longtr}{\vphirep}
\newcommand{\rotacc}{\Pirep}
\newcommand{\GamOme}{\Upsrep}
\newcommand{\redOme}{\tilde{\Omerep}}
\newcommand{\redGam}{\tilde{\precrate}}
\newcommand{\precsx}{\mathcal{X}}
\newcommand{\precsy}{\mathcal{Y}}
\newcommand{\precsz}{\mathcal{Z}}

\newcommand{\anodif}{\vpirep}
\newcommand{\potent}{\Phirep}
\newcommand{\hoverc}{\scal{b}}
\newcommand{\tanomaly}{\nurep}
\newcommand{\panomaly}{\nurep_o}
\newcommand{\brquot}{{\mathfrak K}}
\newcommand{\dilfact}{\Gamrep}
\newcommand{\orbspd}{\tilde{\vepsrep}}
\newcommand{\atilde}{\tilde{\scal{a}}}
\newcommand{\zhratio}{\tilde{\scal{z}}}
\newcommand{\qhratio}{\tilde{\scal{q}}}
\newcommand{\semilat}{\tilde{\scal{p}}}
\newcommand{\eccentty}{\tilde{\scal{e}}}

\newcommand{\dragf}{d}

\newcommand{\kcons}{\mho}
\newcommand{\plusmin}{}
\newcommand{\absign}{\pm}
\newcommand{\xcons}{\pirep}
\newcommand{\aspua}{\Thtrep}
\newcommand{\rdshft}{{\mathfrak{z}}}
\newcommand{\zeropi}{0^\circ}
\newcommand{\halfpi}{90^\circ}
\newcommand{\fullpi}{180^\circ}
\newcommand{\threeqpi}{270^\circ}
\newcommand{\doublepi}{360^\circ}
\newcommand{\pfreq}{\omerep_o}
\newcommand{\afreq}{{\omerep^\prime}}
\newcommand{\cdkt}{\mathscr{Y}}
\newcommand{\angus}{\mathfrak{X}}

\newcommand{\unitvx}{\widehat{\vect{x}}}
\newcommand{\unitvy}{\widehat{\vect{y}}}
\newcommand{\unitvz}{\widehat{\vect{z}}}
\newcommand{\unitpos}{\widehat{\vectr}}
\newcommand{\unitkap}{\widehat{\vectkap}}
\newcommand{\unitplz}{\widehat{\vect{p}}}
\newcommand{\unitjap}{\widehat{\boldsymbol{\jmath}}}
\newcommand{\unitiap}{\widehat{\boldsymbol{\imath}}}

\begin{document}

\workno{2001}{03A}
\worker{A. I. A. Adewole}
\monthdate{March 2001}
\mailto{aiaa@adequest.ca}
\work{Classical Aberration And Obliquation}
\shortwork{Classical Aberration And Obliquation}
\makefront

\setcounter{tocdepth}{2}
\tableofcontents
\newpage
\numberwithin{equation}{section}

\begin{abstract}{A Tribute To Faraday}
Paradise is a world without interpretations, a world where every mind agrees with every other
mind about everything. Our world, however, is a world full of interpretations, where many
minds disagree with many other minds about so many things. One of the peculiar aspects of our
world is that, when properly exercised, our intuitive faculties permit us to catch a
glimpse of paradise now and then, but such glimpses are beyond the reach of any mind for which
intuition is a lost art. In the entire history of the subject matter of this monograph,
perhaps no one has demonstrated the effectiveness of intuition as a living art in a more
fruitful and a more engaging way than
M.~Faraday. His thoughts, work and interpretation of electromagnetic phenomena occupied
the most capable thinkers in and beyond his generation, and to this day, they continue to
inspire us as they have inspired J.~C.~Maxwell, H.~Hertz and many others. We shall take it
upon ourselves in this series of monographs to study certain problems raised by the work of
H.~Hertz regarding the optics of accelerated systems (with the understanding that Hertz's
theory is Hertz's system of equations), because solutions to these problems appear to
have many unexplored applications relevant to science and technology, and at the same
time represent an unfinished chapter in the history of the investigations begun by Faraday.

Our first problem arises from the fact that a ray of light propagating relative
to a moving observer can be associated
with two directions in space. One is the true direction the ray would have if
the observer were stationary, the other is the apparent direction the ray is
actually observed to have. If the apparent direction is the
same for two observers moving relatively to each other, then the true direction
is likely to be different for the observers, and we may refer to this effect as
aberration. Likewise, if the true direction is the same for the observers, then
the apparent direction is likely to be different for the observers, and we may refer to
this effect as obliquation. We shall show that while the wavefront of light is tilted in
aberration, there is no such tilt in obliquation, and for an observer in accelerated
translational, rotational or gravitational motion, we shall give a rigorous treatment
of the apparent angular displacement of a light source due to obliquation, the rate of change of
this displacement with the observer's velocity, the apparent path traced by the light source, the
apparent geometry of a light ray from the source to the observer, and the apparent frequency
of the ray. Readers with only a casual interest in the subject are advised that \partref{parttwo}
may be omitted without an appreciable loss of continuity.

My thanks and gratitude go to Dr. T. E. Phipps Jr. for bringing the aberration problem in Hertz's
electrodynamics to my attention many years ago. It gives me great pleasure to dedicate this series to
all those whose minds are hardly ever at rest, and for whom, therefore, $\mathcal{P}-\mathcal{C}
\triangleq\mathcal{N}$.
\end{abstract}

\part{Foundations}\label{partone}
\begin{wisdom}{Ren\'e Descartes (1596-1650)}
It is true, however, that it is not customary to pull down all the houses
of a town with the single design of rebuilding them differently, and thereby
rendering the streets more handsome; but it often happens that a private
individual takes down his own with the view of erecting it anew, and that
people are even sometimes constrained to this when their houses are in danger
of falling from age, or when the foundations are insecure.
\end{wisdom}

\section{Introduction}\label{INTRO}
\art{Review of previous results}
We have studied elsewhere~\cite{Adewole01b} the problem of light propagation for accelerated
observers in a stationary, homogeneous and isotropic medium using Hertz's version of Maxwell's
theory. We showed in that work that a linearly polarized regular plane ray propagates relative
to a translating, rotating or gravitating observer with a velocity $\vectups$ given by
\begin{equation}\label{main1}
\vectups=\vectc\dragf+\vectw-\vectu,\quad
\vectc = \scalc\unitkap
\end{equation}
where $\scalc$ is determined by the permittivity and the permeability of the medium in the
usual way, $\vectu(\vectr, \scalt)$ is the velocity of the observer at position \vectr\
and time \scalt, $\unitkap$ is the unit or normalized wave vector, and the quantities
$\dragf, \vectw$ depend on the wave vector, the wave polarization, and the observer's
acceleration $\vecta=\fdot{\vectu}$ in a fairly complicated way. It was shown,
more precisely, that if $\unitplz$ is a unit vector in the polarization direction and
$\vectkap$ is the wave vector, then
\begin{subequations}\label{main2}
\begin{equation}\label{main2a}
\dragf=\gamrep\left\{\frac{1+\sqrt{1+\vthtrep^2}}{2}\right\}^{1/2},\quad
\vectw=\rhorep\vecta-\taurep\vectkap+\vecte
\end{equation}
where (all square roots being nonnegative),
\begin{equation}\label{main2b}
\rhorep=\xcons/(4\dragf\pfreq),\quad
\xcons=\vthtrep/(1+\vthtrep^2)^{1/2},\quad
\vthtrep=\alprep/(\gamrep\pfreq)^{2},\quad
\alprep=\dprod{\vectkap}{\vecta},\quad
\pfreq=\scalc\kaprep
\end{equation}
\end{subequations}
and the quantities $\gamrep$, \vecta, $\taurep$ and \vecte\ are given for each type of motion
as follow. For an observer translating with acceleration \vecta,
\begin{subequations}\label{main3}
\begin{equation}\label{main3a}
\gamrep=1,\quad
\vecta=\vecta(\scalt),\quad
\taurep=2\rhorep\alprep\kaprep^{-2},\quad
\vecte=\zvect
\end{equation}
where
\begin{equation}\label{main3b}
\vectu=\vectu(\scalt),\quad
\dprod{\unitkap}{\unitplz}=0.
\end{equation}
\end{subequations}
For an observer rotating with angular velocity $\vectOme(\scalt)$,
\begin{subequations}\label{main4}
\begin{equation}\label{main4a}
\gamrep=\left|1+\frac{\szer}{\etarep\pfreq^2}\right|^{1/2}\ne0,\quad
\vecta=\cprod{\vectOme}{\vectu}+\cprod{\vectLam}{\vectr}
=(\dprod{\vectOme}{\vectr})\vectOme-\Omerep^2\vectr+\cprod{\vectLam}{\vectr}
\end{equation}
\begin{equation}\label{main4b}
\taurep=2(\rhorep\alprep-\etarep\stwo)\kaprep^{-2},\quad
\vecte=\stwo(\cprod{\unitplz}{\vectOme})+\sthr\vectOme-\sfou\unitplz
\end{equation}
where
\begin{equation}\label{main4c}
\begin{split}
&\qquad\qquad\vectu=\cprod{\vectOme}{\vectr},\quad
\etarep=\dprod{\vectkap}{(\cprod{\unitplz}{\vectOme})}\ne0,\quad
\vectLam=\fdot{\vectOme}\\
&\xirep=2(\dprod{\vectOme}{\unitplz})(\dprod{\vectOme}{\vectkap})
  -\Omerep^2(\dprod{\vectkap}{\unitplz}),\quad
\zetarep=(\dprod{\vectkap}{\unitplz})(\dprod{\vectOme}{\vectLam})
  -(\dprod{\vectLam}{\unitplz})(\dprod{\vectOme}{\vectkap})
\end{split}
\end{equation}
and
\begin{equation}\label{main4d}
\begin{split}
\szer&=\zetarep-(\xirep^2/\etarep),\quad
\sone=\frac{2\rhorep\alprep-\dragf\pfreq}
  {2\etarep(\szer+\etarep\pfreq^2)},\quad
\stwo=\szer\sone\\
\sthr&=\bigl\{\etarep(\dprod{\vectLam}{\unitplz})
  +4\xirep(\dprod{\vectOme}{\unitplz})\bigr\}\sone,\quad
\sfou=\bigl\{2\xirep\Omerep^2+\etarep(\dprod{\vectOme}{\vectLam})\bigr\}\sone.
\end{split}
\end{equation}
\end{subequations}
For an observer gravitating with acceleration $\vecta$ due to gravity,
\begin{subequations}\label{main5}
\begin{equation}\label{main5a}
\gamrep=\left|1-\frac{\kcons^2}{\kaprep^2}\right|^{1/2},\quad
\vecta=\vecta(\vectr)=-\scalq\vectr/\scalr^3
\end{equation}
\begin{equation}\label{main5b}
\taurep=(2\rhorep\alprep-\scaln\kcons^2\fzer)\kaprep^{-2},\quad
\vecte=\fone(\cprod{\vectr}{\vecth})+\ftwo\unitplz
\end{equation}
where, if $\unitpos$ be a unit vector in the direction of $\vectr$,
\begin{equation}\label{main5c}
\begin{split}
&\quad\vectu=\scalh^{-2}(\cprod{\vectz}{\vecth}+\cprod{\scalq\unitpos}{\vecth}),\quad
\vecth=\cprod{\vectu}{\vectr},\quad
\vectz=\cprod{\vecth}{\vectu}-\scalq\unitpos\\
&\scalm=\dprod{\vectkap}{(\cprod{\vectr}{\vecth})},\quad
\scaln=\dprod{\unitplz}{(\cprod{\vectr}{\vecth})}\ne0,\quad
\kcons^2=(\scalm/\scaln)(\dprod{\vectkap}{\unitplz})\ne\kaprep^2
\end{split}
\end{equation}
and (\vecth, \vectz, \scalq\ being arbitrary constant quantities which
determine the orbit of the observer according to Newton's theory),
\begin{equation}\label{main5d}
\fzer=\frac{2\rhorep\alprep-\dragf\pfreq}{\scaln(\kaprep^2-\kcons^2)},\quad
\fone=\frac{(\dprod{\vectkap}{\unitplz})\fzer}{2},\quad
\ftwo=\frac{\scalm\fzer}{2}.
\end{equation}
\end{subequations}
For the three types of motion to be studied in this work, it is convenient to
introduce the angles shown in \figref{FIG1} as well as the quantities
\begin{equation}\label{main6}
\vectbet=\vectu/\scalc\ne\zvect,\quad
\vectsig=\vecta/\scalc,\quad
\murep=\sigrep/\pfreq,\quad
\cdkt=\scalc\dragf-\kaprep\taurep.
\end{equation}
We recall that the conditions for a linearly polarized plane ray to be regular
(i.e., to propagate with a velocity that depends on $c$) are $\dprod{\vectkap}{\unitplz}=0$
for a translating observer, $\etarep\ne0\text{ and }\gamrep\ne0$ for a rotating observer,
and $\kcons\ne\kaprep$ for a gravitating observer~\cite{Adewole01b}. When these conditions
are not satisfied, the ray may either propagate with a velocity that is independent of $c$
or cease to propagate altogether. Furthermore, for a gravitating observer, the ray propagates only
if $\scaln\ne0$; otherwise, it degenerates into a nonpropagating mode\footnote{This implies
for example that a gravitating observer cannot perceive the ray
if the ray is linearly polarized along the position vector $\vectr$ or along
a normal vector $\vecth$ to the plane of the observer's orbit. We note here that our
vector $\vecth$ points to the south ecliptic pole instead of pointing to the more customary
north ecliptic pole.}.

\begin{figure}[hbt]
\centering\input{FIG1.LP}
\caption{\footnotesize Common angular parameters. Rectilinear motion is defined to be such that
the velocity $\vectu$ and the acceleration $\vecta$ are parallel ($\thtrep=0, \phirep=\lamrep$),
radial motion to be such that the velocity $\vectu$ and the wave vector $\vectkap$ are parallel
($\phirep=0, \thtrep=\lamrep$), and coradial motion to be such that the acceleration $\vecta$
and the wave vector $\vectkap$ are parallel ($\lamrep=0, \thtrep=\phirep$). Note that all
three angles may change with time.}\label{FIG1}
\end{figure}
\art{Scope of this work}
Our purpose in this work is to apply the above results to the phenomenon of light
aberration. This work is motivated in part by the need to correct the widespread
misconception that light aberration is inconsistent with Hertz's electrodynamics~\cite{Phipps91,
Phipps93}, with Fresnel's wave theory of light~\cite{Liebscher98},
or more generally, with classical physics~\cite{Bohm96, Atwater74}. The practical motivations are
to develop new imaging and visualization techniques for optical systems in accelerated
motion~\cite{Howard95, Rau98, Weiskopf99, Savage99},
to study the possibility of designing integrated accelerometers
based on the effects of acceleration on obliquation or
aberration~\cite{Sardin01, Kassner02, Collins61}, and to understand the corrections that must be applied
in software for high-precision astrometric measurements due, for example, to the rotational
and the orbital motion of the earth or a satellite~\cite{Tagaki56, Klioner01},
if possible without using solar system barycentric
velocity~\cite{Porter50, Stumpff79, Ron86} or elliptic e-terms~\cite{Scott64}.
We illustrate the difference between aberration and obliquation in the next section by
showing that what Bradley reported in 1729 was not aberration
but obliquation. A rigorous treatment of obliquation for an observer in accelerated translational,
rotational or gravitational motion is given in subsequent sections.

The treatment begins with a
general theory of obliquation in \secref{S_KIOB}. We apply the theory to a translating observer
in \secref{S_TRAOB}, to a rotating observer
in \secref{S_ROTOB}, and to a gravitating observer in \secref{S_GRAOB}. In each case, we study
the apparent angular displacement of a light source due to obliquation, the rate of change of
this angular displacement with the observer's velocity, the apparent path traced by the
light source, the apparent geometry of a light ray from the source to the observer,
and the apparent frequency of the ray. The remaining sections in the third part of the work
illustrate the types of problems that can be solved by our methods and calculations with a
number of interesting results that may be of practical importance.

\section{Bradley obliquation}\label{S_ANG}
\begin{figure}[th]
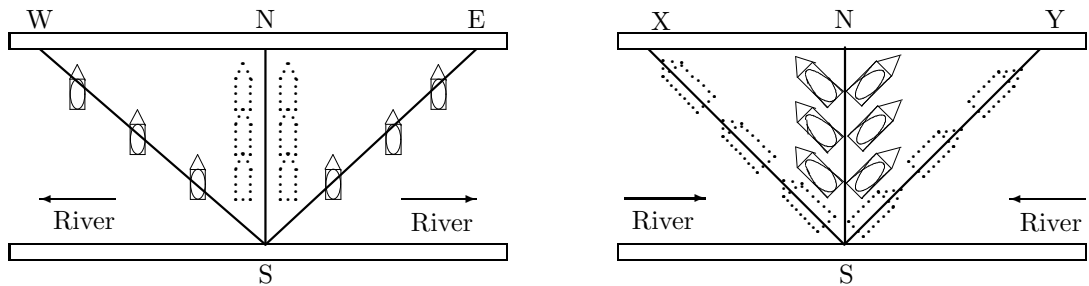

\subfigure[Obliquation of a ferry. The ferry is aimed along SN (true direction)
but drifts along SW or SE (apparent directions) due to the river. It is
parallel or untilted to SN on arrival at W or E.%
]{\input{FIG2A.LP}\label{FIG2A}}
\subfigure[Aberration of a ferry. The ferry is aimed along
\primedSW\ or \primedSE\ (true directions)
but drifts along SN (apparent direction) due to the river. It is
inclined or tilted to SN on arrival at N.%
]{\input{FIG2B.LP}\label{FIG2B}}
\caption{Aberration and obliquation of a ferry.} \label{FIG2}
\end{figure}

\art{The ferryman's problem}
Consider a ferry travelling from south S to north N at right
angles across a river as shown in \figref{FIG2A}. If the river flows due east, the ferry
will drift away from SN and will reach the other side of the river at some point E
east of N; whereas, if the river flows due west, the ferry will reach the other side of the
river at some point W west of N. But as the ferry drifts along SW or SE, it will
be pointing in a direction parallel to SN. Similarly, as two light rays
aimed in the same direction drift in different directions relative to two relatively moving
observers, the normal to their wavefronts will be pointing in the same
direction for both observers. We refer to this phenomenon as obliquation.
Suppose now that the ferry is to arrive at N in spite of the river, \figref{FIG2B}.
If the river flows due
west, then one must aim the ferry at some point \primedE\ east of N so that the river
will compel it to drift along SN, while if the river flows due east, one must aim
the ferry at some point \primedW\ west of N to compel it to drift along SN. In both cases,
as the ferry drifts along SN, it will be pointing in a direction inclined to SN.
Similarly, as two light rays aimed in different directions drift
in the same direction relative to two relatively moving observers, the normal to their
wavefronts will be pointing in different directions for both observers. We refer to
this phenomenon as aberration.

\art{Aberration versus obliquation}
Several things follow from the above illustration. We see clearly that obliquation
describes a situation in which one assumes the true direction of a light ray to be
the same for two observers in relative motion, and consequently infers that the
apparent direction of the ray must be different for the observers. Aberration on
the other hand describes a situation in which one assumes the apparent direction
of a light ray to be the same for two observers in relative motion, and
consequently infers that the true direction of the ray
must be different for the observers. We see further that while the wavefront of a
light ray is not tilted in obliquation, a telescope that is intended to perceive
the ray must be tilted from one direction (SW) to another (SE) when the motion of
the observer is reversed. For aberration, on the other hand, we see that although
the wavefront of a light ray is tilted, a telescope that is intended to perceive
the ray must be pointed in the same direction (SN) when the motion of the
observer is reversed.
\begin{figure}[th]
\centering\input{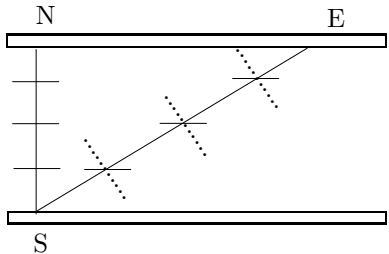}
\caption{\footnotesize Tilted and untilted wavefronts. When a ray of light aimed along SN
drifts relative to a moving observer along SE, its wavefronts remain perpendicular to SN.
If the wavefronts were tilted, they would be perpendicular to SE as shown by the dotted
lines.
}\label{FIG3}
\end{figure}
It follows from these considerations that Bradley's
observations~\cite{Bradley29} correspond to obliquation instead of aberration. This
conclusion is historically evident because in Bradley's time, there were no practical
means of observing a tilt (illustrated in \figref{FIG3}) in the wavefront of light.
Hence, as Fresnel~\cite{Liebscher98, Fresnel18} pointed out, what Bradley observed
was not aberration but obliquation~\footnote{What we call obliquation may also be
called ``ray aberration'' while what we call aberration may also be called ``wave
aberration''. In this terminology, Bradley's observation as well as all modern
devices such as those used in adaptive optics measure ray aberration rather than
wave aberration. The distinction is vital for any optical system in which the
direction of a ray does not coincide with that of a wave normal, although both
terms are often used interchangeably or even in different senses by others.
We shall continue to use our preferred terminology for clarity.}.

\begin{figure}[htb]
\centering\input{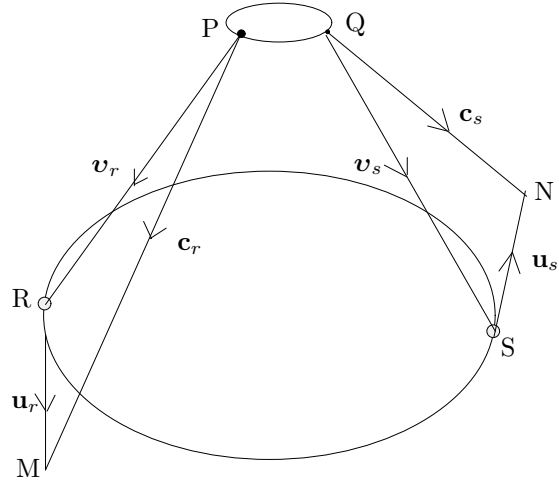}
\caption{\footnotesize Aberration and obliquation of light. Light rays from a star at P
that would have propagated along PM relative to an observer at R propagate
instead along PR due to the observer's motion.
Rays from the star at Q that would have propagated along QN if the observer %
were stationary at S propagate instead along QS due to the observer's motion.
The angle between PR and QS describes obliquation while the angle between
PM and QN describes aberration.
}\label{FIG4}
\end{figure}

\art{Formulae for classical aberration and obliquation}
Let us investigate the above conclusion with some rigour. We consider linearly
polarized regular plane light rays from a star at points P and Q relative to
an observer at points S and R as shown in \figref{FIG4}.
If $\vectu_s,\vectu_r$ are respectively the velocity of the observer at S and R,
and if we assume the observer to be translating with
negligible acceleration, then by \eqnref{main1}, \eqnref{main2} and \eqnref{main3},
the velocity of the rays relative to the observer at these points will be
\begin{subequations}\label{abob1}
\begin{align}
\vectups_s=\vectc_s-\vectu_s,\quad\vectc_s=\scalc\unitkap_s\label{abob1a}\\
\vectups_r=\vectc_r-\vectu_r,\quad\vectc_r=\scalc\unitkap_r\label{abob1b}
\end{align}
\end{subequations}
where $\vectc_s, \vectc_r$ are the velocities the rays would have if the observer
were stationary at S and R respectively. The obliquation angle $\Phirep$ for the star
is the angle between $\vectups_s$ and $\vectups_r$, and is therefore given by
\begin{equation}\label{abob2}
\tan\Phirep=\frac{\left|\cprod{\vectups_s}{\vectups_r}\right|}
  {\dprod{\vectups_s}{\vectups_r}}.
\end{equation}
If we refer to the directions of $\vectc_s, \vectc_r$ as true and to the directions
of $\vectups_s, \vectups_r$ as apparent, then the angle defined by \eqnref{abob2}
represents a change in the apparent direction of the star. We can also define
an aberration angle $\Psirep$
\begin{equation}\label{abob3}
\tan\Psirep=\frac{\left|\cprod{\vectc_s}{\vectc_r}\right|}
  {\dprod{\vectc_s}{\vectc_r}}
\end{equation}
which represents a change in the true direction of the star.

\art{Obliquation of starlight viewed from the earth}
As a concrete example, suppose that S and R correspond to the positions of the earth
at a six months interval, when the earth must have reversed its direction of motion,
so that
\begin{equation}\label{abob4}
\vectu_s = -\vectu_r = -\vectu\;(\text{say}).
\end{equation}
Suppose also that the true direction of the star remains unchanged, so that
\begin{equation}\label{abob5}
\vectc_s = \vectc_r = \vectc=\scalc\unitkap\;(\text{say}).
\end{equation}
Then by \eqnref{abob1} one has $\cprod{\vectups_s}{\vectups_r} = \cprod{(\vectc
+ \vectu)}{(\vectc-\vectu)} = 2\cprod{\vectu}{\vectc}$ and
$\dprod{\vectups_s}{\vectups_r} = \dprod{(\vectc + \vectu)}{(\vectc-\vectu)} = c^2 - u^2$.
By \eqnref{abob2}, the obliquation angle for the star is
\begin{subequations}\label{abob6}
\begin{equation}\label{abob6a}
\tan\Phirep=\frac{2\betrep\sin\phirep}{1-\betrep^2}
\end{equation}
where $\phirep$ is the angle between \vectc\ and \vectu\ (see \figref{FIG1}).
In particular if the true direction
of the star is at right angles to the earth's motion, so that $\phirep=\halfpi$, then
\begin{equation}\label{abob6b}
\tan\Phirep=\frac{2\betrep}{1-\betrep^2}
\end{equation}
which agrees with Bradley's observations.
\begin{figure}[th]
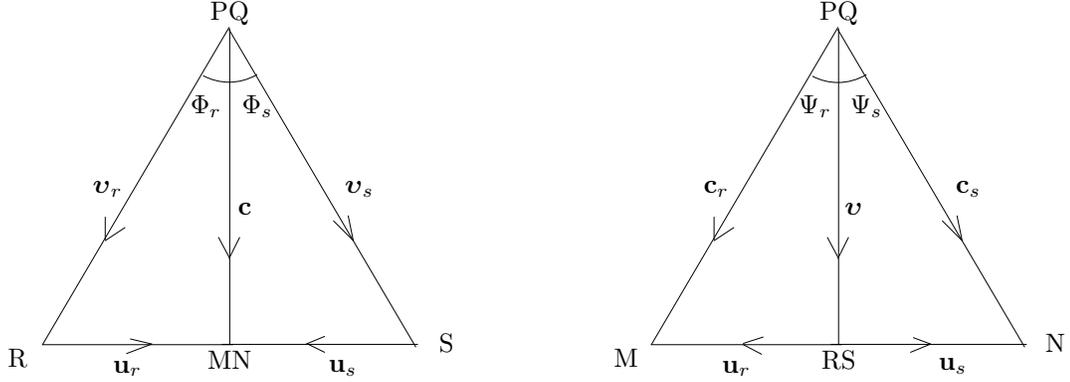

\subfigure[Transverse obliquation. The true direction of the star is assumed to
be unchanged (\pointPM$\,\parallel\,$\pointQN) as the earth reverses its
direction of motion. As a result, the apparent direction of the
star is observed to have changed (\pointPR$\,\angle\,$\pointQS).
]{\input{FIG5A.LP}\label{FIG5A}}
\subfigure[Transverse aberration. The apparent direction of the star is observed to
be unchanged (\pointPR$\,\parallel\,$\pointQS) as the earth reverses its
direction of motion. As a result, the true direction of the
star is assumed to have changed (\pointPM$\,\angle\,$\pointQN).
]{\input{FIG5B.LP}\label{FIG5B}}
\caption{Transverse aberration and obliquation.} \label{FIG5}
\end{figure}
The origin of the second order term in this equation is easy to explain~\cite{Rothman95}.
As shown in \figref{FIG5A}, when the observer is at positions
S and R, the star is displaced from its true direction by angles $\Phirep_s$ and $\Phirep_r$
such that
\begin{equation}\label{abob6c}
\tan\Phirep_s=\tan\Phirep_r=\betrep.
\end{equation}
\end{subequations}
Equation \eqnref{abob6b} results from applying the trigonometric identity
\begin{equation}\label{abob8}
\tan(A + B) =\frac{\tan A + \tan B}{1 - \tan A\tan B}
\end{equation}
with $A = \Phirep_s, B = \Phirep_r$.

\art{Aberration of starlight viewed from the earth}
Let us calculate the angle of aberration for the
star. For this purpose the apparent direction of the star is kept unchanged, as described
earlier, so that
\begin{equation}\label{abob9}
\vectups_s = \vectups_r = \vectups\;(\text{say}).
\end{equation}
From \eqnref{abob1}, \eqnref{abob4} and \eqnref{abob9}, one has
$\cprod{\vectc_s}{\vectc_r} = \cprod{(\vectups
- \vectu)}{(\vectups + \vectu)} = 2\cprod{\vectups}{\vectu}$ and
$\dprod{\vectc_s}{\vectc_r} = \dprod{(\vectups-\vectu)}{(\vectups+\vectu)} = \upsrep^2 - u^2$.
Substituting into \eqnref{abob3} gives
\begin{subequations}\label{abob10}
\begin{equation}\label{abob10a}
\tan\Psirep=\frac{2(\scalu/\upsrep)\sin\psirep}{1-(\scalu/\upsrep)^2}
\end{equation}
where $\psirep$ is the angle between $\vectups$ and \vectu. Writing \eqnref{abob1b} as
$\vectups+\vectu=\vectc_r$ and squaring gives, after solving the resulting quadratic
equation for $\upsrep$ and using the fact that $|\vectc_s|=|\vectc_r|=c$ by \eqnref{abob1},
\begin{equation}\label{abob10b}
\scalu/\upsrep=\betrep\left(-\betrep\cos\psirep+\sqrt{1-\betrep^2\sin^2\psirep}\right)^{-1}.
\end{equation}
\end{subequations}
From \eqnref{abob10a} and \eqnref{abob10b}, we finally get
\begin{subequations}\label{abob11}
\begin{equation}\label{abob11a}
\tan\Psirep=\frac{2\betrep\sin\psirep\left(-\betrep\cos\psirep+\sqrt{1-\betrep^2\sin^2\psirep}\right)}
{1+\betrep^2(1-2\sin^2\psirep)-2\betrep\cos\psirep\sqrt{1-\betrep^2\sin^2\psirep}}
\end{equation}
as the aberration angle for the star. In particular
if the apparent direction of the star is at right angles to the earth's motion, so that
$\psirep=\halfpi$, then
\begin{equation}\label{abob11b}
\tan\Psirep=\frac{2\betrep}{\sqrt{1-\betrep^2}}.
\end{equation}
As shown in \figref{FIG5B}, this corresponds to a displacement of the star from its
true directions at S and R by angles $\Psirep_s$ and $\Psirep_r$ such that~\cite{Einstein05, Brown01}
\begin{equation}\label{abob11c}
\tan\Psirep_s=\tan\Psirep_r=\frac{\betrep}{\sqrt{1-\betrep^2}}.
\end{equation}
\end{subequations}

\art{Bradley observed obliquation, not aberration}
Equations \eqnref{abob6} and \eqnref{abob11} show that obliquation and aberration are
not only distinct phenomena but also, generally speaking, are of different magnitudes.
Moreover, while both \eqnref{abob6c} and \eqnref{abob11c} agree with Bradley's observation,
\eqnref{abob6c} describes a change in the apparent direction of a star while \eqnref{abob11c}
describes a change in the true direction of the star. Therefore, since Bradley's observations
demonstrate a displacement in the apparent direction of a star, they correspond to \eqnref{abob6c}
rather than \eqnref{abob11c}. We shall confine our attention to phenomena related to obliquation
in the remainder of this work.

\art{Obliquation for accelerated observers}
Equation \eqnref{abob6c} is somewhat inaccurate because it requires the
observer to be translating with a negligible acceleration.
To obtain a more accurate equation that does not neglect the observer's acceleration,
let us generalize \eqnref{abob1} by using \eqnref{main1} to get
\begin{equation}\label{rad1}
\vectups_s=\vectc_s\dragf_s+\vectw_s-\vectu_s,\quad
\vectups_r=\vectc_r\dragf_r+\vectw_r-\vectu_r
\end{equation}
and consider the situation when the motion of the observer at S and R is
such that
\begin{equation}\label{rad2}
\vectkap_s=\vectkap_r=\vectkap\;(\text{say}),\quad
\vectu_s=-\vectu_r=-\vectu\;(\text{say}),\quad
\vecta_s=-\vecta_r=-\vecta\;(\text{say}).
\end{equation}
Equations \eqnref{main2} and \eqnref{main3} show that for a translating observer
in this situation,
\begin{equation}\label{rad3}
\vectc_s=\vectc_r=\vectc\;(\text{say}),\quad
\dragf_s=\dragf_r=\dragf\;(\text{say}),\quad
\vectw_s=\vectw_r=\vectw\;(\text{say}).
\end{equation}
Substituting \eqnref{rad1}, \eqnref{rad2} and \eqnref{rad3} into \eqnref{abob2} gives
\begin{equation}\label{rad4}
\tan\Phirep=\frac{\left|\cprod{(\vectc\dragf+\vectw+\vectu)}{(\vectc\dragf+\vectw-\vectu)}
  \right|}{\dprod{(\vectc\dragf+\vectw+\vectu)}{(\vectc\dragf+\vectw-\vectu)}}
\end{equation}
which can be simplified as follows.
Let $\phirep, \thtrep \text { and } \lamrep$ be the angles shown in \figref{FIG1}.
Then, from \eqnref{main2} and \eqnref{main3},
\begin{equation}\label{rad5}
\alprep=\scala\kaprep\cos\lamrep,\quad
\vectw=\rhorep(\vecta-2\unitkap\scala\cos\lamrep)
\end{equation}
and from these equations, we derive
\begin{subequations}\label{rad6}
\begin{align}\label{rad6a}
\dprod{\vectw}{\vectw}=\scalw^2
&=\rhorep^2\dprod{(\vecta-2\unitkap\scala\cos\lamrep)}{(\vecta-2\unitkap\scala\cos\lamrep)}\nonumber\\
&=\rhorep^2(\scala^2-4(\dprod{\unitkap}{\vecta})\scala\cos\lamrep+4\scala^2\cos^2\lamrep)\nonumber\\
&=\rhorep^2(\scala^2-4\scala^2\cos^2\lamrep+4\scala^2\cos^2\lamrep)
=\rhorep^2\scala^2
\end{align}
\begin{align}\label{rad6b}
\dprod{\vectw}{\vectc}
&=\rhorep\dprod{(\vecta-2\unitkap\scala\cos\lamrep)}{\vectc}\nonumber\\
&=\rhorep(\scala\scalc\cos\lamrep-2\scala\scalc\cos\lamrep)
=-\rhorep\scala\scalc\cos\lamrep
\end{align}
\begin{align}\label{rad6c}
\dprod{\vectw}{\vectu}
&=\rhorep\dprod{(\vecta-2\unitkap\scala\cos\lamrep)}{\vectu}\nonumber\\
&=\rhorep(\scala\scalu\cos\thtrep-2(\dprod{\unitkap}{\vectu})\scala\cos\lamrep)\nonumber\\
&=\rhorep(\scala\scalu\cos\thtrep-2\scala\scalu\cos\lamrep\cos\phirep)\nonumber\\
&=\rhorep\scala\scalu(\cos\thtrep-2\cos\lamrep\cos\phirep).
\end{align}
\end{subequations}
Using these equations, the denominator of \eqnref{rad4} becomes
\begin{align}\label{rad7}
\dprod{(\vectc\dragf+\vectw+\vectu)}{(\vectc\dragf+\vectw-\vectu)}
&=\scalc^2\dragf^2+2\dragf\dprod{\vectw}{\vectc}+\dprod{\vectw}{\vectw}-\scalu^2\nonumber\\
&=\scalc^2\dragf^2+2\dragf(-\rhorep\scala\scalc\cos\lamrep)+\rhorep^2\scala^2-\scalu^2\nonumber\\
&=\scalc^2(\dragf^2-\betrep^2+\rhorep^2\sigrep^2-2\dragf\rhorep\sigrep\cos\lamrep).
\end{align}
Expanding the numerator of \eqnref{rad4} directly, we have
\begin{subequations}\label{rad8}
\begin{equation}\label{rad8a}
\cprod{(\vectc\dragf+\vectw+\vectu)}{(\vectc\dragf+\vectw-\vectu)}
= 2\dragf\cprod{\vectu}{\vectc}+2\cprod{\vectu}{\vectw}.
\end{equation}
We square this equation to get, in view of \eqnref{rad6} and \eqnref{alg2},
\begin{align}\label{rad8b}
(2\dragf\cprod{\vectu}{\vectc}+2\cprod{\vectu}{\vectw})^2
&=4\dprod{(\cprod{\vectu}{\vectw}+\dragf\cprod{\vectu}{\vectc})}
  {(\cprod{\vectu}{\vectw}+\dragf\cprod{\vectu}{\vectc})}\nonumber\\
&=4[\dprod{(\cprod{\vectu}{\vectw})}{(\cprod{\vectu}{\vectw})}
  +2\dragf\dprod{(\cprod{\vectu}{\vectw})}{(\cprod{\vectu}{\vectc})}
  +\dragf^2\dprod{(\cprod{\vectu}{\vectc})}{(\cprod{\vectu}{\vectc})}
  ]\nonumber\\
&=4[\{\scalu^2\scalw^2-(\dprod{\vectu}{\vectw})^2\}
  +2\dragf\{\scalu^2(\dprod{\vectw}{\vectc})-(\dprod{\vectu}{\vectc})(\dprod{\vectu}{\vectw})\}
  +\dragf^2\{\scalu^2\scalc^2-(\dprod{\vectu}{\vectc})^2\}
  ]\nonumber\\
&=4[\scalu^2\scalw^2-(\dprod{\vectu}{\vectw})^2
  +2\dragf\scalu^2(\dprod{\vectw}{\vectc})-2\dragf\scalu\scalc(\dprod{\vectu}{\vectw})\cos\phirep
  +\dragf^2\scalu^2\scalc^2\sin^2\phirep
  ]\nonumber\\
\begin{split}
&=4[\rhorep^2\scala^2\scalu^2-\rhorep^2\scala^2\scalu^2(\cos\thtrep-2\cos\lamrep\cos\phirep)^2
  +2\dragf\scalu^2(-\rhorep\scala\scalc\cos\lamrep)\\
  &\qquad-2\dragf\rhorep\scala\scalc\scalu^2(\cos\thtrep-2\cos\lamrep\cos\phirep)\cos\phirep
  +\dragf^2\scalu^2\scalc^2\sin^2\phirep]
\end{split}
\nonumber\\
\begin{split}
&=4\betrep^2\scalc^4[\rhorep^2\sigrep^2-2\dragf\rhorep\sigrep\cos\lamrep
  -\rhorep^2\sigrep^2(\cos\thtrep-2\cos\lamrep\cos\phirep)^2\\
  &\qquad-2\dragf\rhorep\sigrep(\cos\thtrep-2\cos\lamrep\cos\phirep)\cos\phirep+\dragf^2\sin^2\phirep].
\end{split}
\end{align}
\end{subequations}
Substituting \eqnref{rad7} and the square root of \eqnref{rad8b} into \eqnref{rad4}
finally gives
\begin{subequations}\label{rad9}
\begin{equation}\label{rad9a}
\tan\Phirep=\frac{2\betrep\bigl(\epsva+\dragf^2\sin^2\phirep-\epsvb^2-2\dragf\epsvb\cos\phirep\bigr)^{1/2}}
{\epsva+\dragf^2-\betrep^2}
\end{equation}
where
\begin{equation}\label{rad9b}
\epsva=\rhorep\sigrep(\rhorep\sigrep-2\dragf\cos\lamrep),\quad
\epsvb=\rhorep\sigrep(\cos\thtrep-2\cos\lamrep\cos\phirep)
\end{equation}
and
\begin{equation}\label{rad9c}
\rhorep=\frac{\vthtrep(1+\vthtrep^2)^{-1/2}}{4\dragf\pfreq},\quad
\dragf=\left\{\frac{1+\sqrt{1+\vthtrep^2}}{2}\right\}^{1/2},\quad
\vthtrep=\murep\cos\lamrep
\end{equation}
\end{subequations}
as the complete set of equations describing obliquation for an observer in accelerated
translational motion.

\art{Effects of acceleration on obliquation}
We note that \eqnref{rad9a} reduces to \eqnref{abob6a} for a nonaccelerated observer
($\scala=0$) or for an observer whose acceleration is perpendicular to the true direction
of light ($\cos\lamrep=0$). We note also that the effect of acceleration on the obliquation
angle is extremely small and is determined to a large extent by the quantity $\murep^2$.
For an observer moving with $\scala\approx10\text{ms}^{-2}$ and for visible light with
$\kaprep\approx10^7\text{m}^{-1}$ and $\scalc\approx3\times10^{8}\text{ms}^{-1}$,
we get $\murep^2\approx10^{-46}$. This shows that for all practical purposes, translational
acceleration has no effect on the obliquation of visible light. Despite the smallness of
$\murep$, there are three respects in which \eqnref{rad9a}
differs significantly from \eqnref{abob6a} from a theoretical viewpoint. First, according to
\eqnref{rad9a}, the obliquation angle is a function of wavelength, which means that
the apparent direction of a star depends on the wavelength of light emitted
by the star, or loosely speaking, on the spectral type of the star~\cite{Thuring78}.
This dispersive effect does not exist for a nonaccelerated observer according to
\eqnref{abob6a}. Second, \eqnref{rad9a} implies that light from a star may be obliquated even
when the true direction of the star is parallel to the observer's velocity
($\phirep=0$). A look at \eqnref{abob6a} shows that this radial effect does not
exist for a nonaccelerated observer. Third, \eqnref{rad9a} shows that if the true
direction of a star can be such that~\footnote{We shall sometimes refer to the true direction of
light as the line of incidence or {\em loi}, and to the apparent direction of light as the line of
sight or {\em los}. We emphasize that for an accelerated observer, the direction
in which a light ray is incident differs from the direction in which the ray is sighted. Failure
to grasp the full implications of this distinction has resulted in many errors being
committed by those who are eager to find inadequacies in the classical treatment of the subject.}
\begin{equation}\label{rad10}
\epsva+\dragf^2\sin^2\phirep-\epsvb^2-2\dragf\epsvb\cos\phirep=0,\quad
\epsva+\dragf^2-\betrep^2\ne0,
\end{equation}
then light from the star will not be obliquated. This means that the apparent direction of
the star as observed on the earth will not change when the earth reverses its direction of
motion. All these effects may be observed with reasonable certainty
if an astronomical object with a measurable $\murep$ can be found.
Such will be the case, for example, if the object emits waves with
$\kaprep\approx\scala/\scalc^2$ which, for small values of \scala, implies that
the waves must be infraradio waves. At present, the implied frequency of these
waves is a great many orders of magnitude below the range of existing and proposed low
frequency radio arrays such as LOFAR~\cite{Hewitt00}. It is also well below
the ionospheric cutoff frequency and the range of detectable LF radio waves.

\section{Obliquation theory}\label{S_KIOB}
\art{Apparent direction to a light source}
The method of quantifying obliquation described in the previous section is
noninstantaneous because it depends on the velocity of light relative to an observer at two
separate instants. To develop a method that depends only on the instantaneous velocity $\vectups$
of light relative to an observer, we note that one only needs to have a standard direction with
respect to which the direction of $\vectups$ can be compared. We note further that for practical
reasons, it is essential that this standard direction be observable while the observer is moving.
This practical requirement rules out the possibility of using the true direction of
light as a standard since this direction is not always observable by a moving observer.
If we take the direction of the observer's velocity \vectu\ as standard under the assumption
that the observer can (in principle) determine this velocity by methods that assume nothing
regarding the optical properties of light, then we may compare
this direction with that of $\vectups$ at any instant by calculating the angle $\psirep$ defined by
\begin{equation}\label{grad1}
\tan\psirep=\frac{\left|\cprod{\vectu}{\vectups}\right|}
  {\dprod{\vectu}{\vectups}}.
\end{equation}
This definition implies that at any instant, the apparent direction of light is inclined
at angle $\psirep$ to the observer's velocity \vectu\ at that instant. For this reason
we shall call $\psirep$ the instantaneous angle of obliquation.

For the three types of motion that are of interest to us in this work, we can develop
\eqnref{grad1} by using \eqnref{main1} and \eqnref{main2} as follow. Let
$\phirep, \thtrep, \lamrep$ be the angles shown in \figref{FIG1}. Let also
\newcommand{\btcf}{\betrep^2\scalc^4}
\begin{subequations}\label{grad2}
\begin{align}
\bcal&=\dprod{\vecte}{\vectc},\quad
\dcal=\dprod{\vecte}{\vectu},\quad
\acal=\dprod{\vecte}{\vecta},\quad
\hcal=\dprod{\vecte}{\vectkap},\quad
\ecal=\dprod{\vecte}{\vecte}
\label{grad2a}\\
\lcal_0&=\dcal-\scalu\kaprep\taurep\cos\phirep,\quad
\lcal_1=\rhorep\scala\cos\thtrep-\kaprep\taurep\cos\phirep,\quad
\lcal_2=\lcal_1+\rhorep\scala\cos\thtrep
\label{grad2b}\\
\ncal_1&=2\dragf\scalu^2(\bcal-\taurep\pfreq),\quad
\ncal_2=2\dragf\scalu\scalc\lcal_0\cos\phirep,\quad
\ncal_3=\scalu^2[\ecal+2(\rhorep\acal-\taurep\hcal)]
\label{grad2c}\\
\ncal_4&=\scalu^2\kaprep\taurep(\kaprep\taurep-2\rhorep\scala\cos\lamrep),\quad
\ncal_5=\scalu^2\kaprep\taurep\lcal_2\cos\phirep,\quad
\ncal_6=\dcal(\dcal+2\scalu\lcal_1)
\label{grad2d}
\end{align}
\end{subequations}
\begin{subequations}\label{grad2w}
\begin{align}
\lcal&=\lcal_0/(\betrep\scalc^2)
\label{grad2wa}\\
\pcal&=[\ncal_1+\ncal_3+\ncal_4-2\scalu^2(\dcal+\scalu\lcal_1)]/(\btcf)
\label{grad2wb}\\
\ncal&=(\ncal_1-\ncal_2+\ncal_3+\ncal_4+\ncal_5-\ncal_6)/(\btcf)
\label{grad2wc}
\end{align}
\begin{align}
\gcal&=\lcal-\betrep+\dragf\cos\phirep+\rhorep\sigrep\cos\thtrep
\label{grad2xa}\\
\rcal&=[\pcal+\dragf^2+\betrep^2+\rhorep^2\sigrep^2+2\dragf(\rhorep\sigrep\cos\lamrep
  -\betrep\cos\phirep)]^{1/2}
\label{grad2xb}\\
\fcal&=[\ncal+\dragf^2\sin^2\phirep+\rhorep^2\sigrep^2\sin^2\thtrep
  +2\dragf\rhorep\sigrep(\cos\lamrep-\cos\phirep\cos\thtrep)]^{1/2}.
\label{grad2xc}
\end{align}
\end{subequations}
Using \eqnref{main2a}, we derive
\begin{subequations}\label{grad3}
\begin{align}\label{grad3a}
\dprod{\vectc}{\vectw}
&=\dprod{\vectc}{[\rhorep\vecta - \taurep\vectkap + \vecte]}\nonumber\\
&=\scalc(\rhorep\scala\cos\lamrep-\kaprep\taurep)+\bcal
\end{align}
\begin{align}\label{grad3b}
\dprod{\vectu}{\vectw}
&=\dprod{\vectu}{[\rhorep\vecta - \taurep\vectkap + \vecte]}\nonumber\\
&=\scalu(\rhorep\scala\cos\thtrep-\kaprep\taurep\cos\phirep)+\dcal
\end{align}
\begin{align}\label{grad3c}
\dprod{\vectw}{\vectw}
&=\dprod{[\rhorep\vecta - \taurep\vectkap + \vecte]}{[\rhorep\vecta
  - \taurep\vectkap + \vecte]}\nonumber\\
&=(\rhorep\vecta-\taurep\vectkap)^2+\dprod{2\vecte}
  {(\rhorep\vecta-\taurep\vectkap)}+\ecal\nonumber\\
&=\rhorep^2\scala^2-2\rhorep\scala\kaprep\taurep\cos\lamrep+\kaprep^2\taurep^2
  +2(\rhorep\acal-\taurep\hcal)+\ecal.
\end{align}
\end{subequations}
Similarly, using \eqnref{main1}, we obtain\footnote{Equation \eqnref{grad5a}
shows that one may regard $1/\rcal$ as an effective refractive index for the accelerating
observer. This suggests an obvious and reasonably comprehensive formalism for solving some problems
in classical kineoptics, including in particular, problems of kineoptic refraction
or acceleration-induced ``light bending''. We shall suppose throughout this work that the case
$\rcal=0$ is to be excluded from general consideration.}
\begin{subequations}\label{grad5}
\begin{align}\label{grad5a}
\upsrep^2
&=\dprod{(\vectc\dragf+\vectw-\vectu)}{(\vectc\dragf+\vectw-\vectu)}\nonumber\\
&=\scalc^2\dragf^2+\scalu^2-2\dragf(\dprod{\vectu}{\vectc})
  +2\dragf(\dprod{\vectw}{\vectc})-2\dprod{\vectw}{\vectu}+\scalw^2\nonumber\\
\begin{split}
&=\scalc^2\dragf^2+\scalu^2-2\dragf\scalu\scalc\cos\phirep
  +2\dragf[\bcal+\scalc(\rhorep\scala\cos\lamrep-\kaprep\taurep)]
  -2[\dcal+\scalu(\rhorep\scala\cos\thtrep-\kaprep\taurep\cos\phirep)]\\
  &\qquad+\rhorep^2\scala^2-2\rhorep\scala\kaprep\taurep\cos\lamrep+\kaprep^2\taurep^2
    +2(\rhorep\acal-\taurep\hcal)+\ecal\beqref{grad3}
\end{split}
\nonumber\\
\begin{split}
&=\scalc^2\dragf^2+\scalu^2+\rhorep^2\scala^2+2\dragf\scalc(\rhorep\scala\cos\lamrep-\scalu\cos\phirep)
  +2\dragf(\bcal-\scalc\kaprep\taurep)-2(\dcal+\scalu\lcal_1)\\
  &\qquad+\kaprep\taurep(\kaprep\taurep-2\rhorep\scala\cos\lamrep)
    +2(\rhorep\acal-\taurep\hcal)+\ecal\beqref{grad2b}
\end{split}
\nonumber\\
\begin{split}
&=\scalc^2\dragf^2+\scalu^2+\rhorep^2\scala^2+2\dragf\scalc(\rhorep\scala\cos\lamrep-\scalu\cos\phirep)
  +\scalu^{-2}\ncal_1-2(\dcal+\scalu\lcal_1)\\
  &\qquad+\scalu^{-2}\ncal_4+\scalu^{-2}\ncal_3
  \beqref{grad2c}\text{ \& }\eqnref{grad2d}
\end{split}
\nonumber\\
&=\scalc^2[\dragf^2+\betrep^2+\rhorep^2\sigrep^2+2\dragf(\rhorep\sigrep\cos\lamrep-\betrep\cos\phirep)
  +\pcal]\beqref{grad2wb}\nonumber\\
\therefore\upsrep
&=\scalc\rcal\beqref{grad2xb}.
\end{align}
The denominator of \eqnref{grad1} becomes
\begin{align}\label{grad5b}
\dprod{\vectu}{\vectups}
&=\dprod{\vectu}{(\vectc\dragf-\vectu+\vectw)}\beqref{main1}\nonumber\\
&=\dragf\scalu\scalc\cos\phirep-\scalu^2+\scalu(\rhorep\scala\cos\thtrep
  -\kaprep\taurep\cos\phirep)+\dcal\beqref{grad3b}\nonumber\\
&=\dragf\scalu\scalc\cos\phirep-\scalu^2+\rhorep\scala\scalu\cos\thtrep
  +(\dcal-\scalu\kaprep\taurep\cos\phirep)\nonumber\\
&=\dragf\scalu\scalc\cos\phirep-\scalu^2+\rhorep\scala\scalu\cos\thtrep
  +\lcal_0 \beqref{grad2b}\nonumber\\
&=\betrep\scalc^2(\dragf\cos\phirep-\betrep+\rhorep\sigrep\cos\thtrep+\lcal)
  \beqref{grad2wa}\nonumber\\
&=\betrep\scalc^2\gcal\beqref{grad2xa}.
\end{align}
Squaring the numerator of \eqnref{grad1} using \eqnref{main1}, we have
\begin{align*}
(\cprod{\vectu}{\vectups})^2
&=\dprod{[\cprod{\vectu}{(\vectc\dragf+\vectw-\vectu)}]}
  {[\cprod{\vectu}{(\vectc\dragf+\vectw-\vectu)}]}\nonumber\\
&=\dragf^2\dprod{(\cprod{\vectu}{\vectc})}{(\cprod{\vectu}{\vectc})}
  +2\dragf\dprod{(\cprod{\vectu}{\vectc})}{(\cprod{\vectu}{\vectw})}
  +\dprod{(\cprod{\vectu}{\vectw})}{(\cprod{\vectu}{\vectw})}\nonumber\\
\begin{split}
&=\dragf^2[\scalu^2\scalc^2-(\dprod{\vectu}{\vectc})^2]
  +2\dragf[\scalu^2(\dprod{\vectc}{\vectw})-(\dprod{\vectu}{\vectw})(\dprod{\vectu}{\vectc})]
  +[\scalu^2\scalw^2-(\dprod{\vectu}{\vectw})^2]\beqref{alg2}
\end{split}
\nonumber\\
&=\dragf^2\scalu^2\scalc^2\sin^2\phirep+2\dragf\scalu^2(\dprod{\vectc}{\vectw})
  -2\dragf\scalu\scalc(\dprod{\vectu}{\vectw})\cos\phirep
  +\scalu^2\scalw^2-(\dprod{\vectu}{\vectw})^2
\end{align*}
\begin{align*}
\begin{split}
&=\dragf^2\scalu^2\scalc^2\sin^2\phirep+2\dragf\scalu^2[\scalc(\rhorep\scala\cos\lamrep-\kaprep\taurep)+\bcal]
  -2\dragf\scalu\scalc[\scalu(\rhorep\scala\cos\thtrep-\kaprep\taurep\cos\phirep)+\dcal]\cos\phirep\\
  &\quad+\scalu^2[\rhorep^2\scala^2-2\rhorep\scala\kaprep\taurep\cos\lamrep+\kaprep^2\taurep^2
    +2(\rhorep\acal-\taurep\hcal)+\ecal]\\
  &\quad-[\scalu(\rhorep\scala\cos\thtrep-\kaprep\taurep\cos\phirep)+\dcal]^2\beqref{grad3}
\end{split}
\nonumber\\
\begin{split}
&=\dragf^2\scalu^2\scalc^2\sin^2\phirep+2\dragf\rhorep\scala\scalc\scalu^2\cos\lamrep
  +2\dragf\scalu^2(\bcal-\taurep\pfreq)-2\dragf\rhorep\scala\scalc\scalu^2\cos\phirep\cos\thtrep\\
  &\quad-2\dragf\scalu\scalc(\dcal-\scalu\kaprep\taurep\cos\phirep)\cos\phirep
   +\rhorep^2\scalu^2\scala^2+\scalu^2\kaprep\taurep(\kaprep\taurep-2\rhorep\scala\cos\lamrep)
   +\scalu^2[\ecal+2(\rhorep\acal-\taurep\hcal)]\\
  &\quad-\rhorep^2\scalu^2\scala^2\cos^2\thtrep
   +\scalu^2\kaprep\taurep(2\rhorep\scala\cos\thtrep-\kaprep\taurep\cos\phirep)\cos\phirep
    -\dcal[\dcal+2\scalu(\rhorep\scala\cos\thtrep-\kaprep\taurep\cos\phirep)]
\end{split}
\end{align*}
\begin{align}\label{grad5c}
\begin{split}
&=\dragf^2\scalu^2\scalc^2\sin^2\phirep+2\dragf\rhorep\scala\scalc\scalu^2\cos\lamrep+\ncal_1
  -2\dragf\rhorep\scala\scalc\scalu^2\cos\phirep\cos\thtrep-\ncal_2\\
  &\quad+\rhorep^2\scalu^2\scala^2+\ncal_4+\ncal_3
  -\rhorep^2\scalu^2\scala^2\cos^2\thtrep+\ncal_5-\ncal_6
  \text{\beqref{main2b}, \eqnref{grad2c} \& \eqnref{grad2d}}
\end{split}
\nonumber\\
\begin{split}
&=\dragf^2\scalu^2\scalc^2\sin^2\phirep+2\dragf\rhorep\scala\scalc\scalu^2(\cos\lamrep-\cos\phirep\cos\thtrep)
  +\rhorep^2\scalu^2\scala^2\sin^2\thtrep+\btcf\ncal\beqref{grad2w}
\end{split}
\nonumber\\
&=\btcf[\dragf^2\sin^2\phirep+2\dragf\rhorep\sigrep(\cos\lamrep-\cos\phirep\cos\thtrep)
  +\rhorep^2\sigrep^2\sin^2\thtrep+\ncal]\nonumber\\
\therefore|\cprod{\vectu}{\vectups}|
&=\betrep\scalc^2\fcal\beqref{grad2xc}.
\end{align}
We also have
\begin{align}\label{grad5d}
\scalu^2\upsrep^2
&=(\dprod{\vectu}{\vectups})^2+(\cprod{\vectu}{\vectups})^2\beqref{alg3}\nonumber\\
\scalu^2\scalc^2\rcal^2&=\btcf\gcal^2+\btcf\fcal^2
  \beqref{grad5a},\eqnref{grad5b},\text{ \& }\eqnref{grad5c}\nonumber\\
\therefore
\rcal^2&=\fcal^2+\gcal^2.
\end{align}
\end{subequations}
Substituting \eqnref{grad5b} and \eqnref{grad5c} into \eqnref{grad1} leads to
\begin{equation}\label{grad6}
\tan\psirep=\fcal/\gcal
=\frac{[\ncal+\dragf^2\sin^2\phirep+\rhorep^2\sigrep^2\sin^2\thtrep
  +2\dragf\rhorep\sigrep(\cos\lamrep-\cos\phirep\cos\thtrep)]^{1/2}}
  {\lcal-\betrep+\dragf\cos\phirep+\rhorep\sigrep\cos\thtrep}
\end{equation}
which, together with \eqnref{grad2w}, allows $\psirep$ to be calculated for an
observer in accelerated translational, rotational or gravitational motion.

\art{Apparent drift of a light source}
In general, we are interested not only in $\psirep$ but also in the rate at
which $\psirep$ is changing with \vectu. To quantify this rate, we introduce a
quantity $\angus$ defined by
\begin{subequations}\label{kas1}
\begin{equation}\label{kas1a}
\angus=\ugrad{\psirep}
\end{equation}
where the subscript on $\grad{}$ signifies that the gradient operation
is to be performed with respect to
\vectu. This definition implies that when the observer's velocity
changes from $\vectu_1$ to $\vectu_2$, the corresponding change in $\psirep$
can be calculated from \eqnref{kas1a} as
\begin{equation}\label{kas1b}
\psirep=\int_{\vectu_1}^{\vectu_2}\dprod{\angus}{\text{d\vectu}}.
\end{equation}
Moreover, the apparent motion or drift (defined as the total time derivative
of $\psirep$) per unit acceleration of the observer is given by
\begin{equation}\label{kas1c}
\aspua=\fdot{\psirep}/\scala,\quad\fdot{\psirep}=\dprod{\angus}{\vecta}.
\end{equation}
\end{subequations}
We shall refer to $\angus$ as the slope or gradient of obliquation, and to $\aspua$ as
the variation of obliquation.
To develop \eqnref{kas1} for the three kinds of motion we are investigating, let
\begin{subequations}\label{kas2}
\begin{align}\label{kas2a}
\begin{split}
&\vcha=\rhorep/\pfreq,\quad
\vchb=(\pfreq\gamrep)^{-2},\quad
\vchc=2\alprep\vchb\gamrep^{-1},\quad
\vchd=(1+\vthtrep^2)^{1/2},\quad
\vche=(4\dragf\pfreq\vchd)^{-1}\\
&\qquad\qquad\vchf=(2\alprep\vcha-\dragf)/\gamrep,\quad
\vchg=\vche\left[\frac{\vchb}{\vchd^2}-\frac{\vthtrep\vcha}{\dragf}\right],\quad
\vchh=\vche\left[\frac{\vchc}{\vchd^2}-\frac{\vthtrep\vchf}{\dragf}\right]
\end{split}
\end{align}
\begin{align}\label{kas2b}
\begin{split}
\vga&=\ugdiv{\vectkap}{\vecta},\quad
\vgb=\ucurl{\vecta},\quad
\vgc=\ugdiv{\vectu}{\vecta},\quad
\vgd=\ugdiv{\vectups}{\vecta},\quad
\vge=\ugrad{\gamrep}\\
\vgf&=\vga+\cprod{\vectkap}{\vgb},\quad
\vgg=\vchb\vgf-\vchc\vge,\quad
\vgh=\vcha\vgf-\vchf\vge,\quad
\vgi=\vchg\vgf-\vchh\vge\\
\end{split}
\end{align}
\begin{align}\label{kas2c}
\begin{split}
\vja&=\ucurl{\vecte},\quad
\vjb=\ugdiv{\vectu}{\vecte},\quad
\vjc=\ugdiv{\vectups}{\vecte},\quad
\vjd=\vja+\rhorep\vgb,\quad
\vje=\vjb+\rhorep\vgc\\
\vjf&=\vjc+\rhorep\vgd,\quad
\vjg=\ugrad{\taurep},\quad
\vjh=(\cprod{\vectu}{\vectups})/|\cprod{\vectu}{\vectups}|,\quad
\vji=\cprod{\vjh}{\vectu},\quad
\vjj=\gcal\vji-\fcal\vectu.
\end{split}
\end{align}
\end{subequations}
We derive
\begin{subequations}\label{kas3}
\begin{align}\label{kas3a}
\ugrad{\alprep}
&=\ugrad{(\dprod{\vectkap}{\vecta})}\beqref{main2b}\nonumber\\
&=\cprod{\vectkap}{(\ucurl{\vecta})}+\ugdiv{\vectkap}{\vecta}
  +\ugdiv{\vecta}{\vectkap}+\cprod{\vecta}{(\ucurl{\vectkap})}
\beqref{clc36}\nonumber\\
&=\cprod{\vectkap}{(\ucurl{\vecta})}+\ugdiv{\vectkap}{\vecta}
=\vga+\cprod{\vectkap}{\vgb}\beqref{kas2b}
\nonumber\\
&=\vgf\beqref{kas2b}
\end{align}
\begin{align}\label{kas3b}
\ugrad{\vthtrep}
&=\ugrad{(\alprep\gamrep^{-2}\pfreq^{-2})}\beqref{main2b}\nonumber\\
&=\pfreq^{-2}\ugrad{(\alprep\gamrep^{-2})}\nonumber\\
&=\pfreq^{-2}(\alprep\ugrad{\gamrep^{-2}}+\gamrep^{-2}\ugrad{\alprep})
  \beqref{clc33}\nonumber\\
&=\pfreq^{-2}(-2\alprep\gamrep^{-3}\ugrad{\gamrep}
  +\gamrep^{-2}\ugrad{\alprep})\beqref{clc32}\nonumber\\
&=\pfreq^{-2}\gamrep^{-2}(\ugrad{\alprep}-2\alprep\gamrep^{-1}\ugrad{\gamrep})\nonumber\\
&=\pfreq^{-2}\gamrep^{-2}(\vgf-2\alprep\gamrep^{-1}\vge)
  \beqref{kas3a}\text{ \& }\eqnref{kas2b}\nonumber\\
&=\vgg\beqref{kas2b}
\end{align}
\begin{align}\label{kas3c}
\ugrad{\dragf}
&=\ugrad{\left[\gamrep\left\{\frac{1+\sqrt{1+\vthtrep^2}}{2}\right\}^{1/2}\right]}
  \beqref{main2a}\nonumber\\
&=\dragf\gamrep^{-1}\ugrad{\gamrep}
  +\gamrep\ugrad{\left\{\frac{1+\sqrt{1+\vthtrep^2}}{2}\right\}^{1/2}}
  \beqref{clc33}\text{ \& }\eqnref{main2a}\nonumber\\
&=\dragf\gamrep^{-1}\ugrad{\gamrep}
  +\frac{\gamrep^2\vthtrep(1+\vthtrep^2)^{-1/2}}{4\dragf}\ugrad{\vthtrep}
  \beqref{clc32}\nonumber\\
&=\dragf\gamrep^{-1}\ugrad{\gamrep}
  +\rhorep\pfreq\gamrep^2\ugrad{\vthtrep}
  \beqref{main2b}\nonumber\\
&=\dragf\gamrep^{-1}\vge+\rhorep\pfreq\gamrep^2[\pfreq^{-2}\gamrep^{-2}(\vgf
  -2\alprep\gamrep^{-1}\vge)]\beqref{kas2}\text{ \& }\eqnref{kas3b}\nonumber\\
&=\pfreq^{-1}[\rhorep\vgf-\gamrep^{-1}(2\rhorep\alprep-\dragf\pfreq)\vge]\nonumber\\
&=\vgh\beqref{kas2b}
\end{align}
\begin{align}\label{kas3d}
\ugrad{\rhorep}
&=(4\pfreq)^{-1}\ugrad{[\dragf^{-1}\vthtrep(1+\vthtrep^2)^{-1/2}]}
  \beqref{main2b}\nonumber\\
&=(4\pfreq)^{-1}\{\vthtrep(1+\vthtrep^2)^{-1/2}\ugrad{\dragf^{-1}}
  +\dragf^{-1}\ugrad{[\vthtrep(1+\vthtrep^2)^{-1/2}]}\}
  \beqref{clc33}\nonumber\\
&=(4\pfreq)^{-1}[-\dragf^{-2}\vthtrep(1+\vthtrep^2)^{-1/2}\ugrad{\dragf}
  +\dragf^{-1}(1+\vthtrep^2)^{-3/2}\ugrad{\vthtrep}]
  \beqref{clc32}\nonumber\\
&=\left[\frac{(1+\vthtrep^2)^{-1/2}}{4\dragf\pfreq}\right]\left[
  -\frac{\vthtrep\ugrad{\dragf}}{\dragf}+\frac{\ugrad{\vthtrep}}{1+\vthtrep^2}\right]
=\vche\left[
  -\frac{\vthtrep\ugrad{\dragf}}{\dragf}+\frac{\ugrad{\vthtrep}}{\vchd^2}\right]
  \beqref{main2b}\text{ \& }\eqnref{kas2a}\nonumber\\
&=\vche[-(\vthtrep/\dragf)\vgh+(1/\vchd^2)\vgg]
  \beqref{kas3b}\text{ \& }\eqnref{kas3c}\nonumber\\
&=\vche[(\vthtrep/\dragf)(-\vcha\vgf+\vchf\vge)
  +(1/\vchd^2)(\vchb\vgf-\vchc\vge)]\beqref{kas2b}\nonumber\\
&=\vche[(\vchb/\vchd^2)-(\vthtrep\vcha/\dragf)]\vgf
  -\vche[(\vchc/\vchd^2)-(\vthtrep\vchf/\dragf)]\vge\nonumber\\
&=\vgi\beqref{kas2a}\text{ \& }\eqnref{kas2b}
\end{align}
\end{subequations}
\begin{subequations}\label{kas4}
\begin{align}\label{kas4a}
\ucurl{\vectw}
&=\ucurl{(\rhorep\vecta-\taurep\vectkap+\vecte)}\beqref{main2a}\nonumber\\
&=\ucurl{\vecte}+\ucurl{(\rhorep\vecta)}-\ucurl{(\taurep\vectkap)}\nonumber\\
&=\ucurl{\vecte}+\rhorep(\ucurl{\vecta})-\cprod{\vecta}{\ugrad{\rhorep}}
  -\taurep(\ucurl{\vectkap})+\cprod{\vectkap}{\ugrad{\taurep}}
  \beqref{clc24}\nonumber\\
&=\vja+\rhorep\vgb-\cprod{\vecta}{\vgi}+\cprod{\vectkap}{\vjg}
  \beqref{kas2b}\text{ \& }\eqnref{kas2c}\nonumber\\
&=\vjd+\cprod{\vectkap}{\vjg}-\cprod{\vecta}{\vgi}
  \beqref{kas2c}
\end{align}
\begin{align}
\ugdiv{\vectu}{\vectw}
&=\ugdiv{\vectu}{(\rhorep\vecta-\taurep\vectkap+\vecte)}\beqref{main2a}\nonumber\\
&=\ugdiv{\vectu}{\vecte}+\ugdiv{\vectu}{(\rhorep\vecta)}-\ugdiv{\vectu}{(\taurep\vectkap)}
 \nonumber\\
&=\ugdiv{\vectu}{\vecte}+\vecta(\dprod{\vectu}{\ugrad{\rhorep}})+\rhorep\ugdiv{\vectu}{\vecta}
  -\vectkap(\dprod{\vectu}{\ugrad{\taurep}})-\taurep\ugdiv{\vectu}{\vectkap}
  \beqref{clc42}\nonumber\\
&=\vjb+\vecta(\dprod{\vectu}{\vgi})+\rhorep\vgc-\kaprep(\dprod{\vectu}{\vjg})
  \beqref{kas2b}, \eqnref{kas2c}\text{ \& }\eqnref{kas3d}\nonumber\\
&=\vje+\vecta(\dprod{\vectu}{\vgi})-\vectkap(\dprod{\vectu}{\vjg})
  \beqref{kas2c}
\label{kas4b}\\
\thicksim\ugdiv{\vectups}{\vectw}
&=\vjf+\vecta(\dprod{\vectups}{\vgi})-\vectkap(\dprod{\vectups}{\vjg})
\label{kas4c}
\end{align}
\end{subequations}
\begin{subequations}\label{kas5}
\begin{align}\label{kas5a}
\ucurl{\vectups}
&=\ucurl{(\vectc\dragf+\vectw-\vectu)}\beqref{main1}\nonumber\\
&=\ucurl{(\vectc\dragf)}+\ucurl{\vectw}-\ucurl{\vectu}\nonumber\\
&=\dragf(\ucurl{\vectc})-\cprod{\vectc}{(\ugrad{\dragf})}+\ucurl{\vectw}
  \beqref{clc21}\text{ \& }\eqnref{clc24}\nonumber\\
&=\ucurl{\vectw}-\cprod{\vectc}{(\ugrad{\dragf})}\nonumber\\
&=\vjd+\cprod{\vectkap}{\vjg}-\cprod{\vecta}{\vgi}-\cprod{\vectc}{\vgh}
  \beqref{kas3c}\text{ \& }\eqnref{kas4a}
\end{align}
\begin{align}
\ugdiv{\vectu}{\vectups}
&=\ugdiv{\vectu}{(\vectc\dragf+\vectw-\vectu)}\beqref{main1}\nonumber\\
&=\ugdiv{\vectu}{(\vectc\dragf)}+\ugdiv{\vectu}{\vectw}-\ugdiv{\vectu}{\vectu}\nonumber\\
&=\vectc(\dprod{\vectu}{\ugrad{\dragf}})+\dragf\ugdiv{\vectu}{\vectc}+\ugdiv{\vectu}{\vectw}-\vectu
  \beqref{clc41}\text{ \& }\eqnref{clc42}\nonumber\\
&=-\vectu+\ugdiv{\vectu}{\vectw}+\vectc(\dprod{\vectu}{\vgh})
  \beqref{kas3c}\nonumber\\
&=-\vectu+\vje+\vecta(\dprod{\vectu}{\vgi})-\vectkap(\dprod{\vectu}{\vjg})
  +\vectc(\dprod{\vectu}{\vgh})\beqref{kas4b}
\label{kas5b}\\
\thicksim\ugdiv{\vectups}{\vectups}
&=-\vectups+\vjf+\vecta(\dprod{\vectups}{\vgi})-\vectkap(\dprod{\vectups}{\vjg})
  +\vectc(\dprod{\vectups}{\vgh})
\label{kas5c}
\end{align}
\end{subequations}
\begin{subequations}\label{kas6}
\begin{align}\label{kas6a}
\ugrad{(\dprod{\vectups}{\vectu})}
&=\vectups+\ugdiv{\vectu}{\vectups}+\cprod{\vectu}{(\ucurl{\vectups})}
  \beqref{clc34}\nonumber\\
\begin{split}
&=\vectups-\vectu+\vje+\vecta(\dprod{\vectu}{\vgi})
  -\vectkap(\dprod{\vectu}{\vjg})+\vectc(\dprod{\vectu}{\vgh})\\
 &\quad+\cprod{\vectu}{(\vjd+\cprod{\vectkap}{\vjg}-\cprod{\vecta}{\vgi}
   -\cprod{\vectc}{\vgh})}\beqref{kas5}
\end{split}
\nonumber\\
&=\vectups-\vectu+\vje+\vgi(\dprod{\vectu}{\vecta})-\vjg(\dprod{\vectu}{\vectkap})
  +\vgh(\dprod{\vectu}{\vectc})+\cprod{\vectu}{\vjd}\beqref{alg1}
\end{align}
\begin{align*}
|\cprod{\vectu}{\vectups}|(\ugrad{|\cprod{\vectu}{\vectups}|})
&=\scalu^2\upsrep(\ugrad{\upsrep})+\upsrep^2\vectu-(\dprod{\vectups}{\vectu})
  \ugrad{(\dprod{\vectups}{\vectu})}\beqref{clc37}\nonumber\\
&=\upsrep^2\vectu+\scalu^2[\cprod{\vectups}{(\ucurl{\vectups})}+\ugdiv{\vectups}{\vectups}]
  -(\dprod{\vectups}{\vectu})\ugrad{(\dprod{\vectups}{\vectu})}
  \beqref{clc35}\nonumber\\
\begin{split}
&=\upsrep^2\vectu+\scalu^2\cprod{\vectups}{[\vjd+\cprod{\vectkap}{\vjg}
  -\cprod{\vecta}{\vgi}-\cprod{\vectc}{\vgh}]}\\
 &\quad+\scalu^2[-\vectups+\vjf+\vecta(\dprod{\vectups}{\vgi})-\vectkap(\dprod{\vectups}{\vjg})
  +\vectc(\dprod{\vectups}{\vgh})]\\
 &\quad-(\dprod{\vectups}{\vectu})[\vectups-\vectu+\vje+\vgi(\dprod{\vectu}{\vecta})
  -\vjg(\dprod{\vectu}{\vectkap})\\
 &\qquad+\vgh(\dprod{\vectu}{\vectc})+\cprod{\vectu}{\vjd}]
  \beqref{kas5}\text{ \& }\eqnref{kas6a}
\end{split}
\nonumber\\
\begin{split}
&=\upsrep^2\vectu+\scalu^2(\cprod{\vectups}{\vjd})+\scalu^2[-\vectups+\vjf
  +\vgi(\dprod{\vectups}{\vecta})-\vjg(\dprod{\vectups}{\vectkap})
  +\vgh(\dprod{\vectups}{\vectc})]\\
 &\quad-(\dprod{\vectups}{\vectu})[\vectups-\vectu+\vje+\vgi(\dprod{\vectu}{\vecta})
  -\vjg(\dprod{\vectu}{\vectkap})+\vgh(\dprod{\vectu}{\vectc})+\cprod{\vectu}{\vjd}]
  \beqref{alg1}
\end{split}
\end{align*}
\begin{align}\label{kas6ax}
\begin{split}
&=\scalu^2\vjf-(\dprod{\vectups}{\vectu})\vje+(\upsrep^2+\dprod{\vectups}{\vectu})\vectu
  -(\scalu^2+\dprod{\vectups}{\vectu})\vectups\\
&\quad+\cprod{[\scalu^2\vectups-(\dprod{\vectups}{\vectu})\vectu]}{\vjd}
  +[\scalu^2(\dprod{\vectups}{\vecta})-(\dprod{\vectups}{\vectu})(\dprod{\vectu}{\vecta})]\vgi\\
&\quad-[\scalu^2(\dprod{\vectups}{\vectkap})-(\dprod{\vectups}{\vectu})(\dprod{\vectu}{\vectkap})]\vjg
  +[\scalu^2(\dprod{\vectups}{\vectc})-(\dprod{\vectups}{\vectu})(\dprod{\vectu}{\vectc})]\vgh
\end{split}
\nonumber\\
\begin{split}
&=\scalu^2\vjf-(\dprod{\vectups}{\vectu})\vje+(\upsrep^2+\dprod{\vectups}{\vectu})\vectu
  -(\scalu^2+\dprod{\vectups}{\vectu})\vectups\\
  &\quad-\cprod{[\cprod{\vectu}{(\cprod{\vectu}{\vectups})}]}{\vjd}
    +[\dprod{(\cprod{\vectu}{\vectups})}{(\cprod{\vectu}{\vecta})}]\vgi\\
  &\quad-[\dprod{(\cprod{\vectu}{\vectups})}{(\cprod{\vectu}{\vectkap})}]\vjg
    +[\dprod{(\cprod{\vectu}{\vectups})}{(\cprod{\vectu}{\vectc})}]\vgh
    \beqref{alg1}\text{ \& }\eqnref{alg2}
\end{split}
\end{align}
\begin{align}\label{kas6b}
\begin{split}
\ugrad{|\cprod{\vectu}{\vectups}|}
&=(\betrep/\fcal)\vjf-(\gcal/\fcal)\vje+[(\rcal^2+\betrep\gcal)/(\betrep\fcal)]\vectu
  -[(\betrep+\gcal)/\fcal]\vectups+\cprod{\vjd}{(\cprod{\vectu}{\vjh})}\\
  &\quad+[\dprod{\vjh}{(\cprod{\vectu}{\vecta})}]\vgi
  -[\dprod{\vjh}{(\cprod{\vectu}{\vectkap})}]\vjg
  +[\dprod{\vjh}{(\cprod{\vectu}{\vectc})}]\vgh
  \beqref{kas6ax},\eqnref{grad5}\text{ \& }\eqnref{kas2c}
\end{split}
\nonumber\\
\begin{split}
&=(\betrep/\fcal)\vjf-(\gcal/\fcal)\vje+[(\rcal^2+\betrep\gcal)/(\betrep\fcal)]\vectu
  -[(\betrep+\gcal)/\fcal]\vectups\\
 &\quad-\cprod{\vjd}{\vji}+(\dprod{\vecta}{\vji})\vgi-(\dprod{\vectkap}{\vji})\vjg
  +(\dprod{\vectc}{\vji})\vgh\beqref{alg4}\text{ \& }\eqnref{kas2c}
\end{split}
\end{align}
\begin{align}\label{kas6c}
\ugrad\psirep
&=\ugrad{\left[\arctan\left(\frac{|\cprod{\vectu}{\vectups}|}
  {\dprod{\vectu}{\vectups}}\right)\right]}\beqref{grad1}\nonumber\\
&=\left[1+\left(\frac{|\cprod{\vectu}{\vectups}|}{\dprod{\vectu}{\vectups}}\right)^2\right]^{-1}
  \ugrad{\left[\frac{|\cprod{\vectu}{\vectups}|}{\dprod{\vectu}{\vectups}}\right]}
  \beqref{clc32}\nonumber\\
&=\left[\frac{\dprod{\vectu}{\vectups}}{\scalu\upsrep}\right]^2
   \ugrad{\left[\frac{|\cprod{\vectu}{\vectups}|}{\dprod{\vectu}{\vectups}}\right]}
   \beqref{alg3}\nonumber\\
&=\left[\frac{\dprod{\vectu}{\vectups}}{\scalu\upsrep}\right]^2
  \left[\frac{\ugrad{|\cprod{\vectu}{\vectups}|}}{\dprod{\vectu}{\vectups}}
  +|\cprod{\vectu}{\vectups}|\ugrad{(\dprod{\vectu}{\vectups})^{-1}}\right]
  \beqref{clc33}\nonumber\\
&=\left[\frac{\dprod{\vectu}{\vectups}}{\scalu\upsrep}\right]^2
  \left[\frac{\ugrad{|\cprod{\vectu}{\vectups}|}}{\dprod{\vectu}{\vectups}}
  -\frac{|\cprod{\vectu}{\vectups}|\ugrad{(\dprod{\vectu}{\vectups})}}{(\dprod{\vectu}{\vectups})^2}\right]
  \beqref{clc32}\nonumber\\
&=(\scalu\upsrep)^{-2}[(\dprod{\vectu}{\vectups})\ugrad{|\cprod{\vectu}{\vectups}|}
  -|\cprod{\vectu}{\vectups}|\ugrad{(\dprod{\vectu}{\vectups})}]\nonumber\\
&=\betrep(\scalu\rcal)^{-2}[\gcal\ugrad{|\cprod{\vectu}{\vectups}|}
  -\fcal\ugrad{(\dprod{\vectu}{\vectups})}]\beqref{grad5}.
\end{align}
\end{subequations}
From \eqnref{kas1a}, \eqnref{kas6} and \eqnref{grad5d}, we obtain
\begin{subequations}\label{kas7}
\begin{equation}\label{kas7a}
\begin{split}
\angus
&=\frac{\betrep}{\scalu^2\rcal^2}\biggl[\xcala\vectu-\xcalb\vectups+\xcalc\vjf
  -\xcald\vje+(\dprod{\vectc}{\vjj})\vgh-(\dprod{\vectkap}{\vjj})\vjg
  +(\dprod{\vecta}{\vjj})\vgi+\cprod{\vjj}{\vjd}\biggr]
\end{split}
\end{equation}
\begin{align}\label{kas7b}
\xcala&=\frac{\rcal^2(\betrep+\gcal)}{\betrep\fcal},\quad
\xcalb=\frac{\rcal^2+\betrep\gcal}{\fcal},\quad
\xcalc=\frac{\betrep\gcal}{\fcal},\quad
\xcald=\frac{\rcal^2}{\fcal},\quad
\fcal\ne0
\end{align}
\end{subequations}
which, together with \eqnref{grad2w} and \eqnref{kas2}, allows $\angus$ (and
hence $\aspua$ as well as $\fdot{\psirep}$) to be calculated for an observer in accelerated
translational, rotational or gravitational motion.

\art{Apparent path of a light source}
If we draw several arrows from a fixed point in space to represent
the ray velocity $\vectups$ in magnitude
and direction at various instants, it is clear that the locus of the endpoints of these
arrows will represent the apparent trajectory or path of the light source as viewed by
the observer. By treating this locus or hodograph~\cite{Hamilton47} as a curve
in velocity space (which amounts to treating $\vectups$ as the position vector of the observer from the
light source), the curvature $\bbk$ and torsion $\bbt$ of the
apparent path of the light source can be calculated from the formulae
\begin{subequations}\label{kpath1}
\begin{equation}\label{kpath1a}
\bbk=\plusmin\frac{|\cprod{\fdot{\vectups}}{\ffdot{\vectups}}|}{|\fdot{\vectups}|^3},\qquad
\bbt=\frac{\dprod{\fffdot{\vectups}}{(\cprod{\fdot{\vectups}}{\ffdot{\vectups}})}}
  {|\cprod{\fdot{\vectups}}{\ffdot{\vectups}}|^2},\quad
\bbr=\plusmin1/\bbk
\end{equation}
where $\bbr$ is the radius of curvature of the path for $\bbk\ne0$.
If the apparent path is smooth or regular, so that $\fdot{\vectups}$ and
$\ffdot{\vectups}$ are nonzero everywhere on the path, then the curvature and torsion determine
the path up to geometric congruence, or up to a rigid translation and rotation.
It is easy to determine the path uniquely by evaluating $\vectups$ at an initial instant
and by computing the Frenet frame at that instant~\footnote{One can also determine the apparent
path of the light source by deriving an equation for the hodograph of the ray velocity $\vectups$.
This approach is not taken here because it is less formal and also because it requires the
introduction of a coordinate system. Interested readers will find examples of this approach
in~\protect\cite{Rothman95, Stumpff80}.}. The tangent vector $\frt$, the normal
vector $\frn$, and the binormal vector $\frb$ of the instantaneous Frenet frame at any point on the
apparent path of the light source can be computed from the formulae
\begin{equation}\label{kpath1b}
\frt=\frac{\fdot{\vectups}}{|\fdot{\vectups}|},\quad
\frb=\frac{\cprod{\fdot{\vectups}}{\ffdot{\vectups}}}
  {|\cprod{\fdot{\vectups}}{\ffdot{\vectups}}|},\quad
\frn=\cprod{\frb}{\frt},\quad
\frc=\bbr\,\frn
\end{equation}
in which $\frc$ gives the instantaneous center of curvature of the path,
while according to \eqnref{main1}, \eqnref{main2a} and \eqnref{main6}, $\vectups$ can be
evaluated at any instant from
\begin{equation}\label{kpath1c}
\vectups=\cdkt\unitkap+\rhorep\vecta-\vectu+\vecte.
\end{equation}
\end{subequations}
To develop the above equations for the three kinds of motion we are investigating, it
is convenient to introduce the quantities
\begin{subequations}\label{kpath2}
\begin{align}\label{kpath2a}
\begin{split}
&\qquad\vbba=\fdot{\rhorep}-1,\quad
\vbbb=2\fdot{\rhorep}-1,\quad
\vbbc=3\fdot{\rhorep}-1\\
&\quad\fdot{\cdkt}=\scalc\fdot{\dragf}-\kaprep\fdott,\quad
\ffdot{\cdkt}=\scalc\ffdot{\dragf}-\kaprep\ffdott,\quad
\fffdot{\cdkt}=\scalc\fffdot{\dragf}-\kaprep\fffdott\\
&\vbbd=\ffdot{\rhorep}\fdot{\cdkt}-\ffdot{\cdkt}\vbba,\quad
\vbbe=\fdot{\cdkt}\vbbb-\rhorep\ffdot{\cdkt},\quad
\vbbf=\vbba\vbbb-\rhorep\ffdot{\rhorep}
\end{split}
\end{align}
\begin{align}\label{kpath2b}
\begin{split}
&\vscra=\cprod{\unitkap}{\vecta},\quad
\vscrb=\cprod{\unitkap}{\fdota},\quad
\vscrc=\cprod{\unitkap}{\ffdota},\quad
\vscrd=\cprod{\unitkap}{\fdote},\quad
\vscre=\cprod{\unitkap}{\ffdote}\\
&\qquad\quad\vscrf=\cprod{\vecta}{\fdota},\quad
\vscrg=\cprod{\vecta}{\ffdota},\quad
\vscrh=\cprod{\vecta}{\fdote},\quad
\vscri=\cprod{\vecta}{\ffdote}\\
&\quad\vscrj=\cprod{\fdota}{\ffdota},\quad
\vscrk=\cprod{\fdota}{\fdote},\quad
\vscrl=\cprod{\fdota}{\ffdote},\quad
\vscrn=\cprod{\fdote}{\ffdota},\quad
\vscro=\cprod{\fdote}{\ffdote}
\end{split}
\end{align}
\begin{align}\label{kpath2c}
\begin{split}
&\qquad\vscrp=\vbbd\vscra+\vbbe\vscrb+\rhorep\fdot{\cdkt}\vscrc-\ffdot{\cdkt}\vscrd+\fdot{\cdkt}\vscre
  +\vbbf\vscrf+\rhorep\vbba\vscrg\\
&\qquad\vscrq=-\ffdot{\rhorep}\vscrh+\vbba\vscri+\rhorep^2\vscrj-\vbbb\vscrk+\rhorep\vscrl
  +\rhorep\vscrn+\vscro\\
&\qquad\vscrr=\fdot{\cdkt}\unitkap+\vbba\vecta+\rhorep\fdota+\fdote
\end{split}
\end{align}
\begin{align}\label{kpath2d}
\begin{split}
\alepha&=\dprod{\unitkap}{(\vscrp+\vscrq)},\quad
\alephb=\dprod{\vecta}{(\vscrp+\vscrq)},\quad
\alephc=\dprod{\fdota}{(\vscrp+\vscrq)}\\
\alephd&=\dprod{\ffdota}{(\vscrp+\vscrq)},\quad
\alephe=\dprod{\fffdota}{(\vscrp+\vscrq)},\quad
\alephf=\dprod{\fffdote}{(\vscrp+\vscrq)}.
\end{split}
\end{align}
\end{subequations}
Using primes to indicate the total time derivatives of complicated
expressions for convenience, we derive
\begin{subequations}\label{kpath4}
\begin{align}\label{kpath4a}
\fdot{\vectups}
&=\dif{(\cdkt\unitkap+\rhorep\vecta-\vectu+\vecte)}
  \beqref{kpath1c}\nonumber\\
&=\fdot{\cdkt}\unitkap+\fdot{\rhorep}\vecta+\rhorep\fdota-\vecta+\fdote
  \nonumber\\
&=\fdot{\cdkt}\unitkap+\vbba\vecta+\rhorep\fdota+\fdote\beqref{kpath2a}\nonumber\\
&=\vscrr\beqref{kpath2c}
\end{align}
\begin{align}\label{kpath4b}
\ffdot{\vectups}
&=\dif{[\fdot{\cdkt}\unitkap+(\fdot{\rhorep}-1)\vecta+\rhorep\fdota+\fdote]}
  \beqref{kpath4a}, \eqnref{kpath2a}\text{ \& }\eqnref{kpath2c}\nonumber\\
&=\ffdot{\cdkt}\unitkap+\ffdot{\rhorep}\vecta+(\fdot{\rhorep}-1)\fdota+\fdot{\rhorep}\fdota
  +\rhorep\ffdota+\ffdote\nonumber\\
&=\ffdot{\cdkt}\unitkap+\ffdot{\rhorep}\vecta+\vbbb\fdota+\rhorep\ffdota+\ffdote\beqref{kpath2a}
\end{align}
\begin{align}\label{kpath4c}
\fffdot{\vectups}
&=\dif{[\ffdot{\cdkt}\unitkap+\ffdot{\rhorep}\vecta+(2\fdot{\rhorep}-1)\fdota+\rhorep\ffdota+\ffdote]}
  \beqref{kpath4b}\text{ \& }\eqnref{kpath2a}\nonumber\\
&=\fffdot{\cdkt}\unitkap+\fffdot{\rhorep}\vecta+\ffdot{\rhorep}\fdota+2\ffdot{\rhorep}\fdota
  +(2\fdot{\rhorep}-1)\ffdota+\fdot{\rhorep}\ffdota+\rhorep\fffdota+\fffdote\nonumber\\
&=\fffdot{\cdkt}\unitkap+\fffdot{\rhorep}\vecta+3\ffdot{\rhorep}\fdota
  +\vbbc\ffdota+\rhorep\fffdota+\fffdote\beqref{kpath2a}
\end{align}
\begin{align*}
\begin{split}
\cprod{\fdot{\vectups}}{\ffdot{\vectups}}
&=\cprod{[\fdot{\cdkt}\unitkap+\vbba\vecta+\rhorep\fdota+\fdote]}
  {[\ffdot{\cdkt}\unitkap+\ffdot{\rhorep}\vecta+\vbbb\fdota+\rhorep\ffdota+\ffdote]}
  \beqref{kpath4a}, \eqnref{kpath4b}\text{ \& }\eqnref{kpath2c}
\end{split}
\nonumber\\
\begin{split}
&=\cprod{\fdot{\cdkt}\unitkap}
   {[\ffdot{\cdkt}\unitkap+\ffdot{\rhorep}\vecta+\vbbb\fdota+\rhorep\ffdota+\ffdote]}
  +\cprod{\vbba\vecta}
   {[\ffdot{\cdkt}\unitkap+\ffdot{\rhorep}\vecta+\vbbb\fdota+\rhorep\ffdota+\ffdote]}\\
  &\qquad+\cprod{\rhorep\fdota}
   {[\ffdot{\cdkt}\unitkap+\ffdot{\rhorep}\vecta+\vbbb\fdota+\rhorep\ffdota+\ffdote]}
  +\cprod{\fdote}
   {[\ffdot{\cdkt}\unitkap+\ffdot{\rhorep}\vecta+\vbbb\fdota+\rhorep\ffdota+\ffdote]}
\end{split}
\nonumber\\
\begin{split}
&=\ffdot{\rhorep}\fdot{\cdkt}(\cprod{\unitkap}{\vecta})
   +\fdot{\cdkt}\vbbb(\cprod{\unitkap}{\fdota})
   +\rhorep\fdot{\cdkt}(\cprod{\unitkap}{\ffdota})
   +\fdot{\cdkt}(\cprod{\unitkap}{\ffdote})
   +\ffdot{\cdkt}\vbba(\cprod{\vecta}{\unitkap})
   +\vbba\vbbb(\cprod{\vecta}{\fdota})\\
   &\qquad+\rhorep\vbba(\cprod{\vecta}{\ffdota})
   +\vbba(\cprod{\vecta}{\ffdote})
   +\rhorep\ffdot{\cdkt}(\cprod{\fdota}{\unitkap})
   +\rhorep\ffdot{\rhorep}(\cprod{\fdota}{\vecta})
   +\rhorep^2(\cprod{\fdota}{\ffdota})
   +\rhorep(\cprod{\fdota}{\ffdote})\\
  &\qquad+\ffdot{\cdkt}(\cprod{\fdote}{\unitkap})
   +\ffdot{\rhorep}(\cprod{\fdote}{\vecta})
   +\vbbb(\cprod{\fdote}{\fdota})
   +\rhorep(\cprod{\fdote}{\ffdota})
   +(\cprod{\fdote}{\ffdote})
\end{split}
\end{align*}
\begin{align}\label{kpath4d}
\begin{split}
&=(\ffdot{\rhorep}\fdot{\cdkt}-\ffdot{\cdkt}\vbba)(\cprod{\unitkap}{\vecta})
   +(\fdot{\cdkt}\vbbb-\rhorep\ffdot{\cdkt})(\cprod{\unitkap}{\fdota})
   +\rhorep\fdot{\cdkt}(\cprod{\unitkap}{\ffdota})
   -\ffdot{\cdkt}(\cprod{\unitkap}{\fdote})
   +\fdot{\cdkt}(\cprod{\unitkap}{\ffdote})\\
  &\qquad+(\vbba\vbbb-\rhorep\ffdot{\rhorep})(\cprod{\vecta}{\fdota})
   +\rhorep\vbba(\cprod{\vecta}{\ffdota})
   -\ffdot{\rhorep}(\cprod{\vecta}{\fdote})
   +\vbba(\cprod{\vecta}{\ffdote})
   +\rhorep^2(\cprod{\fdota}{\ffdota})\\
  &\qquad-\vbbb(\cprod{\fdota}{\fdote})
   +\rhorep(\cprod{\fdota}{\ffdote})
   +\rhorep(\cprod{\fdote}{\ffdota})
   +(\cprod{\fdote}{\ffdote})
\end{split}
\nonumber\\
\begin{split}
&=\vbbd\vscra+\vbbe\vscrb+\rhorep\fdot{\cdkt}\vscrc-\ffdot{\cdkt}\vscrd+\fdot{\cdkt}\vscre
  +\vbbf\vscrf+\rhorep\vbba\vscrg-\ffdot{\rhorep}\vscrh+\vbba\vscri\\
  &\qquad+\rhorep^2\vscrj-\vbbb\vscrk+\rhorep\vscrl+\rhorep\vscrn+\vscro
  \beqref{kpath2a}\text{ \& }\eqnref{kpath2b}
\end{split}
\nonumber\\
&=\vscrp+\vscrq\beqref{kpath2c}
\end{align}
\begin{align}\label{kpath4e}
\dprod{\fffdot{\vectups}}{(\cprod{\fdot{\vectups}}{\ffdot{\vectups}})}
&=\dprod{(\fffdot{\cdkt}\unitkap+\fffdot{\rhorep}\vecta+3\ffdot{\rhorep}\fdota
  +\vbbc\ffdota+\rhorep\fffdota+\fffdote)}{(\vscrp+\vscrq)}
  \beqref{kpath4c}\text{ \& }\eqnref{kpath4d}\nonumber\\
\begin{split}
&=\fffdot{\cdkt}[\dprod{\unitkap}{(\vscrp+\vscrq)}]+\fffdot{\rhorep}[\dprod{\vecta}{(\vscrp+\vscrq)}]
  +3\ffdot{\rhorep}[\dprod{\fdota}{(\vscrp+\vscrq)}]\\
  &\qquad+\vbbc[\dprod{\ffdota}{(\vscrp+\vscrq)}]+\rhorep[\dprod{\fffdota}{(\vscrp+\vscrq)}]
  +[\dprod{\fffdote}{(\vscrp+\vscrq)}]
\end{split}
\nonumber\\
&=\fffdot{\cdkt}\alepha+\fffdot{\rhorep}\alephb+3\ffdot{\rhorep}\alephc
  +\vbbc\alephd+\rhorep\alephe+\alephf\beqref{kpath2d}.
\end{align}
\end{subequations}
From \eqnref{main2b}, we have
\begin{align}\label{kpath5}
\fdot{\alprep}=\dprod{\vectkap}{\fdota},\quad
\ffdot{\alprep}=\dprod{\vectkap}{\ffdota},\quad
\fffdot{\alprep}=\dprod{\vectkap}{\fffdota}
\end{align}
so that
\begin{subequations}\label{kpath6}
\begin{align}\label{kpath6a}
\fdot{\vthtrep}
&=\frac{1}{\pfreq^2}\dif{\left[\frac{\alprep}{\gamrep^2}\right]}
=\frac{1}{\pfreq^2}\left[\frac{\fdot{\alprep}}{\gamrep^2}-\frac{2\alprep\fdot{\gamrep}}{\gamrep^3}\right]
=\frac{1}{\gamrep^2\pfreq^2}\left[\fdot{\alprep}-\frac{2\alprep\fdot{\gamrep}}{\gamrep}\right]
\end{align}
\begin{align}\label{kpath6b}
\ffdot{\vthtrep}
&=\dif{\left\{\frac{1}{\gamrep^2\pfreq^2}\left[\fdot{\alprep}-\frac{2\alprep\fdotg}
  {\gamrep}\right]\right\}}\beqref{kpath6a}\nonumber\\
&=\dif{\left[\frac{1}{\gamrep^2\pfreq^2}\right]}\left[\fdot{\alprep}
   -\frac{2\alprep\fdotg}{\gamrep}\right]
   +\frac{1}{\gamrep^2\pfreq^2}\dif{\left[\fdot{\alprep}-\frac{2\alprep\fdotg}{\gamrep}\right]}
  \nonumber\\
&=\frac{-2\fdotg}{\gamrep^2\pfreq^2}\left[\frac{\fdot{\alprep}}{\gamrep}-\frac{2\alprep\fdotg}{\gamrep^2}\right]
   +\frac{1}{\gamrep^2\pfreq^2}\left[\ffdot{\alprep}-\frac{2\fdot{\alprep}\fdotg}{\gamrep}
   -\frac{2\alprep\ffdotg}{\gamrep}+\frac{2\alprep\fdotg^2}{\gamrep^2}\right]\nonumber\\
&=\frac{1}{\gamrep^2\pfreq^2}\left[\ffdot{\alprep}-\frac{4\fdot{\alprep}\fdotg}{\gamrep}
   -\frac{2\alprep\ffdotg}{\gamrep}+\frac{6\alprep\fdotg^2}{\gamrep^2}\right]
\end{align}
\begin{align}\label{kpath6c}
\fffdot{\vthtrep}
&=\dif{\left\{\frac{1}{\gamrep^2\pfreq^2}\left[\ffdot{\alprep}-\frac{4\fdot{\alprep}\fdotg}{\gamrep}
   -\frac{2\alprep\ffdotg}{\gamrep}+\frac{6\alprep\fdotg^2}{\gamrep^2}\right]\right\}}
  \beqref{kpath6b}\nonumber\\
&=\frac{1}{\gamrep^2\pfreq^2}\dif{\left[\ffdot{\alprep}-\frac{4\fdot{\alprep}\fdotg}{\gamrep}
    -\frac{2\alprep\ffdotg}{\gamrep}+\frac{6\alprep\fdotg^2}{\gamrep^2}\right]}
    +\dif{\left[\frac{1}{\gamrep^2\pfreq^2}\right]}\left[\ffdot{\alprep}
    -\frac{4\fdot{\alprep}\fdotg}{\gamrep}-\frac{2\alprep\ffdotg}{\gamrep}
    +\frac{6\alprep\fdotg^2}{\gamrep^2}\right]
\nonumber\\
\begin{split}
&=\frac{1}{\gamrep^2\pfreq^2}\left[\fffdot{\alprep}-\frac{4\ffdot{\alprep}\fdotg}{\gamrep}
   -\frac{4\fdot{\alprep}\ffdotg}{\gamrep}+\frac{4\fdot{\alprep}\fdotg^2}{\gamrep^2}
   -\frac{2\fdot{\alprep}\ffdotg}{\gamrep}-\frac{2\alprep\fffdotg}{\gamrep}
   +\frac{2\alprep\fdotg\ffdotg}{\gamrep^2}
   +\frac{6\fdot{\alprep}\fdotg^2}{\gamrep^2}+\frac{12\alprep\fdotg\ffdotg}{\gamrep^2}
   -\frac{12\alprep\fdotg^3}{\gamrep^3}
     \right]\\
  &\qquad+\frac{-2\fdotg}{\gamrep^2\pfreq^2}\left[\frac{\ffdot{\alprep}}{\gamrep}
    -\frac{4\fdot{\alprep}\fdotg}{\gamrep^2}-\frac{2\alprep\ffdotg}{\gamrep^2}
    +\frac{6\alprep\fdotg^2}{\gamrep^3}\right]
\end{split}
\nonumber\\
\begin{split}
&=\frac{1}{\gamrep^2\pfreq^2}\left[\fffdot{\alprep}-\frac{6\ffdot{\alprep}\fdotg}{\gamrep}
   -\frac{6\fdot{\alprep}\ffdotg}{\gamrep}+\frac{18\fdot{\alprep}\fdotg^2}{\gamrep^2}
   -\frac{2\alprep\fffdotg}{\gamrep}+\frac{18\alprep\fdotg\ffdotg}{\gamrep^2}
   -\frac{24\alprep\fdotg^3}{\gamrep^3}\right]
\end{split}
\end{align}
\end{subequations}
and
\begin{subequations}\label{kpath7}
\begin{align}\label{kpath7a}
\fdot{\xcons}
&=\dif{[\vthtrep(1+\vthtrep^2)^{-1/2}]}\beqref{main2b}\nonumber\\
&=\fdot{\vthtrep}(1+\vthtrep^2)^{-1/2}+\vthtrep\dif{[(1+\vthtrep^2)^{-1/2}]}
  \nonumber\\
&=\fdot{\vthtrep}(1+\vthtrep^2)^{-1/2}-\vthtrep^2\fdot{\vthtrep}(1+\vthtrep^2)^{-3/2}
  \nonumber\\
&=\fdot{\vthtrep}(1+\vthtrep^2)^{-3/2}
\end{align}
\begin{align}\label{kpath7b}
\ffdot{\xcons}
&=\dif{[\fdot{\vthtrep}(1+\vthtrep^2)^{-3/2}]}\beqref{kpath7a}\nonumber\\
&=\ffdot{\vthtrep}(1+\vthtrep^2)^{-3/2}+\fdot{\vthtrep}\dif{[(1+\vthtrep^2)^{-3/2}]}\nonumber\\
&=\ffdot{\vthtrep}(1+\vthtrep^2)^{-3/2}-3\vthtrep\fdot{\vthtrep}^2(1+\vthtrep^2)^{-5/2}\nonumber\\
&=(1+\vthtrep^2)^{-5/2}[\ffdot{\vthtrep}(1+\vthtrep^2)-3\vthtrep\fdot{\vthtrep}^2]
\end{align}
\begin{align}\label{kpath7c}
\fffdot{\xcons}
&=\dif{\{(1+\vthtrep^2)^{-5/2}[\ffdot{\vthtrep}(1+\vthtrep^2)-3\vthtrep\fdot{\vthtrep}^2]\}}
  \beqref{kpath7b}\nonumber\\
&=\dif{[(1+\vthtrep^2)^{-5/2}]}[\ffdot{\vthtrep}(1+\vthtrep^2)-3\vthtrep\fdot{\vthtrep}^2]
  +(1+\vthtrep^2)^{-5/2}\dif{[\ffdot{\vthtrep}(1+\vthtrep^2)-3\vthtrep\fdot{\vthtrep}^2]}
  \nonumber\\
&=-5\vthtrep\fdot{\vthtrep}(1+\vthtrep^2)^{-7/2}[\ffdot{\vthtrep}(1+\vthtrep^2)
   -3\vthtrep\fdot{\vthtrep}^2]
  +(1+\vthtrep^2)^{-5/2}[\fffdot{\vthtrep}(1+\vthtrep^2)
   -4\vthtrep\fdot{\vthtrep}\ffdot{\vthtrep}-3\fdot{\vthtrep}^3]
\nonumber\\
&=(1+\vthtrep^2)^{-7/2}\{-5\vthtrep\fdot{\vthtrep}[\ffdot{\vthtrep}(1+\vthtrep^2)
   -3\vthtrep\fdot{\vthtrep}^2]+(1+\vthtrep^2)[\fffdot{\vthtrep}(1+\vthtrep^2)
   -3\fdot{\vthtrep}^3-4\vthtrep\fdot{\vthtrep}\ffdot{\vthtrep}]\}
\nonumber\\
\begin{split}
&=(1+\vthtrep^2)^{-7/2}\{-5\vthtrep\fdot{\vthtrep}\ffdot{\vthtrep}-5\vthtrep^3\fdot{\vthtrep}\ffdot{\vthtrep}
    +15\vthtrep^2\fdot{\vthtrep}^3\\
  &\qquad+\fffdot{\vthtrep}(1+\vthtrep^2)^2-3\fdot{\vthtrep}^3-4\vthtrep\fdot{\vthtrep}\ffdot{\vthtrep}
    -3\vthtrep^2\fdot{\vthtrep}^3-4\vthtrep^3\fdot{\vthtrep}\ffdot{\vthtrep}\}
\end{split}
\nonumber\\
&=(1+\vthtrep^2)^{-7/2}[\fffdot{\vthtrep}(1+\vthtrep^2)^2-3\fdot{\vthtrep}^3-9\vthtrep\fdot{\vthtrep}\ffdot{\vthtrep}
  +12\vthtrep^2\fdot{\vthtrep}^3-9\vthtrep^3\fdot{\vthtrep}\ffdot{\vthtrep}]
\nonumber\\
&=(1+\vthtrep^2)^{-7/2}[\fffdot{\vthtrep}(1+\vthtrep^2)^2-9\vthtrep\fdot{\vthtrep}\ffdot{\vthtrep}(1+\vthtrep^2)
  -3\fdot{\vthtrep}^3(1-4\vthtrep^2)].
\end{align}
\end{subequations}
We have also that
\begin{subequations}\label{kpath8}
\begin{align}\label{kpath8a}
\fdot{\dragf}
&=\dif{\left[\gamrep\left\{\frac{1+\sqrt{1+\vthtrep^2}}{2}\right\}^{1/2}\right]}
  \beqref{main2a}\nonumber\\
&=\fdotg\left[\frac{1+\sqrt{1+\vthtrep^2}}{2}\right]^{1/2}
  +\gamrep\left[\frac{\vthtrep\fdot{\vthtrep}(1+\vthtrep^2)^{-1/2}}{4}\right]
     \left[\frac{1+\sqrt{1+\vthtrep^2}}{2}\right]^{-1/2}\nonumber\\
&=\fdotg(\dragf/\gamrep)
  +\gamrep(\gamrep/\dragf)\left[\frac{\vthtrep\fdot{\vthtrep}(1+\vthtrep^2)^{-1/2}}{4}\right]
     \beqref{main2a}\nonumber\\
&=\frac{\fdotg\dragf}{\gamrep}+\frac{\xcons\gamrep^2\fdot{\vthtrep}}{4\dragf}\beqref{main2b}
\end{align}
\begin{align}\label{kpath8b}
\ffdot{\dragf}
&=\dif{\left[\frac{\fdotg\dragf}{\gamrep}+\frac{\xcons\gamrep^2\fdot{\vthtrep}}{4\dragf}\right]}
  \beqref{kpath8a}\nonumber\\
&=\frac{\ffdotg\dragf}{\gamrep}+\frac{\fdotg\fdot{\dragf}}{\gamrep}-\frac{\fdotg^2\dragf}{\gamrep^2}
  +\frac{1}{4}\left[\frac{\fdot{\xcons}\gamrep^2\fdot{\vthtrep}}{\dragf}
    +\frac{2\xcons\gamrep\fdotg\fdot{\vthtrep}}{\dragf}+\frac{\xcons\gamrep^2\ffdot{\vthtrep}}{\dragf}
    -\frac{\xcons\gamrep^2\fdot{\dragf}\fdot{\vthtrep}}{\dragf^2}
   \right]\nonumber\\
&=\dragf\biggl[\frac{\ffdotg}{\gamrep}-\frac{\fdotg^2}{\gamrep^2}\biggr]
   +\fdot{\dragf}\biggl[\frac{\fdotg}{\gamrep}-\frac{\xcons\gamrep^2\fdot{\vthtrep}}{4\dragf^2}\biggr]
   +\frac{\gamrep}{4\dragf}\biggl[\fdot{\xcons}\gamrep\fdot{\vthtrep}
    +2\xcons\fdotg\fdot{\vthtrep}+\xcons\gamrep\ffdot{\vthtrep}\biggr]
\end{align}
\begin{align*}
\fffdot{\dragf}
&=\dif{\left\{\dragf\biggl[\frac{\ffdotg}{\gamrep}-\frac{\fdotg^2}{\gamrep^2}\biggr]
   +\fdot{\dragf}\biggl[\frac{\fdotg}{\gamrep}-\frac{\xcons\gamrep^2\fdot{\vthtrep}}{4\dragf^2}\biggr]
   +\frac{\gamrep}{4\dragf}\biggl[\fdot{\xcons}\gamrep\fdot{\vthtrep}
    +2\xcons\fdotg\fdot{\vthtrep}+\xcons\gamrep\ffdot{\vthtrep}\biggr]
   \right\}}\beqref{kpath8b}\nonumber\\
\begin{split}
&=\fdot{\dragf}\biggl[\frac{\ffdotg}{\gamrep}-\frac{\fdotg^2}{\gamrep^2}\biggr]
  +\dragf\dif{\biggl[\frac{\ffdotg}{\gamrep}-\frac{\fdotg^2}{\gamrep^2}\biggr]}
  +\ffdot{\dragf}\biggl[\frac{\fdotg}{\gamrep}-\frac{\xcons\gamrep^2\fdot{\vthtrep}}{4\dragf^2}\biggr]
  +\fdot{\dragf}\dif{\biggl[\frac{\fdotg}{\gamrep}-\frac{\xcons\gamrep^2\fdot{\vthtrep}}{4\dragf^2}\biggr]}\\
  &\qquad+\dif{\biggl[\frac{\gamrep}{4\dragf}\biggr]}\biggl[\fdot{\xcons}\gamrep\fdot{\vthtrep}
    +2\xcons\fdotg\fdot{\vthtrep}+\xcons\gamrep\ffdot{\vthtrep}\biggr]
  +\frac{\gamrep}{4\dragf}\dif{\biggl[\fdot{\xcons}\gamrep\fdot{\vthtrep}
    +2\xcons\fdotg\fdot{\vthtrep}+\xcons\gamrep\ffdot{\vthtrep}\biggr]}
\end{split}
\end{align*}
\begin{align}\label{kpath8c}
\begin{split}
&=\dragf\biggl[\frac{\fffdotg}{\gamrep}-\frac{\fdotg\ffdotg}{\gamrep^2}-\frac{2\fdotg\ffdotg}{\gamrep^2}
     +\frac{2\fdotg^3}{\gamrep^3}\biggr]
  +\fdot{\dragf}\biggl[\frac{\ffdotg}{\gamrep}-\frac{\fdotg^2}{\gamrep^2}\biggr]
  +\fdot{\dragf}\biggl[\frac{\ffdotg}{\gamrep}-\frac{\fdotg^2}{\gamrep^2}
  -\frac{\fdot{\xcons}\gamrep^2\fdot{\vthtrep}}{4\dragf^2}-\frac{2\xcons\gamrep\fdotg\fdot{\vthtrep}}{4\dragf^2}
  -\frac{\xcons\gamrep^2\ffdot{\vthtrep}}{4\dragf^2}
  +\frac{2\xcons\gamrep^2\fdot{\dragf}\fdot{\vthtrep}}{4\dragf^3}\biggr]\\
  &\qquad+\ffdot{\dragf}\biggl[\frac{\fdotg}{\gamrep}-\frac{\xcons\gamrep^2\fdot{\vthtrep}}{4\dragf^2}\biggr]
   +\biggl[\frac{\fdotg}{4\dragf}-\frac{\gamrep\fdot{\dragf}}{4\dragf^2}\biggr]
    \biggl[\fdot{\xcons}\gamrep\fdot{\vthtrep}+2\xcons\fdotg\fdot{\vthtrep}+\xcons\gamrep\ffdot{\vthtrep}\biggr]\\
   &\qquad+\frac{\gamrep}{4\dragf}\biggl[\ffdot{\xcons}\gamrep\fdot{\vthtrep}+\fdot{\xcons}\fdotg\fdot{\vthtrep}
    +\fdot{\xcons}\gamrep\ffdot{\vthtrep}+2\fdot{\xcons}\fdotg\fdot{\vthtrep}+2\xcons\ffdotg\fdot{\vthtrep}
    +2\xcons\fdotg\ffdot{\vthtrep}+\fdot{\xcons}\gamrep\ffdot{\vthtrep}+\xcons\fdotg\ffdot{\vthtrep}
    +\xcons\gamrep\fffdot{\vthtrep}\biggr]
\end{split}
\nonumber\\
\begin{split}
&=\dragf\biggl[\frac{\fffdotg}{\gamrep}-\frac{3\fdotg\ffdotg}{\gamrep^2}+\frac{2\fdotg^3}{\gamrep^3}\biggr]
  +2\fdot{\dragf}\biggl[\frac{\ffdotg}{\gamrep}-\frac{\fdotg^2}{\gamrep^2}
    -\frac{\fdot{\xcons}\gamrep^2\fdot{\vthtrep}}{4\dragf^2}-\frac{\xcons\gamrep\fdotg\fdot{\vthtrep}}{2\dragf^2}
    -\frac{\xcons\gamrep^2\ffdot{\vthtrep}}{4\dragf^2}
    +\frac{\xcons\gamrep^2\fdot{\dragf}\fdot{\vthtrep}}{4\dragf^3}\biggr]
   +\ffdot{\dragf}\biggl[\frac{\fdotg}{\gamrep}-\frac{\xcons\gamrep^2\fdot{\vthtrep}}{4\dragf^2}\biggr]\\
   &\quad+\frac{\fdotg}{4\dragf}\biggl[\fdot{\xcons}\gamrep\fdot{\vthtrep}+2\xcons\fdotg\fdot{\vthtrep}
      +\xcons\gamrep\ffdot{\vthtrep}\biggr]
  +\frac{\gamrep}{4\dragf}\biggl[\ffdot{\xcons}\gamrep\fdot{\vthtrep}+3\fdot{\xcons}\fdotg\fdot{\vthtrep}
    +2\fdot{\xcons}\gamrep\ffdot{\vthtrep}+2\xcons\ffdotg\fdot{\vthtrep}
    +3\xcons\fdotg\ffdot{\vthtrep}+\xcons\gamrep\fffdot{\vthtrep}\biggr]
\end{split}
\end{align}
\end{subequations}
\begin{subequations}\label{kpath9}
\begin{align}
\fdot{\rhorep}
&=\dif{\biggl[\frac{\xcons}{4\dragf\pfreq}\biggr]}\beqref{main2b}\nonumber\\
&=\frac{\fdot{\xcons}}{4\dragf\pfreq}-\frac{\xcons\fdot{\dragf}}{4\dragf^2\pfreq}
=\frac{\dragf\fdot{\xcons}-\xcons\fdot{\dragf}}{4\pfreq\dragf^2}
\label{kpath9a}\\
\ffdot{\rhorep}
&=\dif{\left\{\frac{1}{4\pfreq\dragf^2}\biggl[\dragf\fdot{\xcons}-\xcons\fdot{\dragf}\biggr]
   \right\}}\beqref{kpath9a}\nonumber\\
&=\dif{\biggl[\frac{1}{4\pfreq\dragf^2}\biggr]}\biggl[\dragf\fdot{\xcons}-\xcons\fdot{\dragf}\biggr]
   +\frac{1}{4\pfreq\dragf^2}\dif{\biggl[\dragf\fdot{\xcons}-\xcons\fdot{\dragf}\biggr]}
=-\frac{\fdot{\dragf}(\dragf\fdot{\xcons}-\xcons\fdot{\dragf})}{2\pfreq\dragf^3}
   +\frac{\dragf\ffdot{\xcons}-\xcons\ffdot{\dragf}}{4\pfreq\dragf^2}
\label{kpath9b}
\end{align}
\begin{align*}
\fffdot{\rhorep}
&=\dif{\left\{-\frac{\fdot{\dragf}}{2\pfreq\dragf^3}\biggl[\dragf\fdot{\xcons}-\xcons\fdot{\dragf}\biggr]
   +\frac{1}{4\pfreq\dragf^2}\biggl[\dragf\ffdot{\xcons}-\xcons\ffdot{\dragf}\biggr]
  \right\}}\beqref{kpath9b}\nonumber\\
&=\dif{\biggl[-\frac{\fdot{\dragf}}{2\pfreq\dragf^3}\biggr]}\biggl[\dragf\fdot{\xcons}-\xcons\fdot{\dragf}\biggr]
  -\frac{\fdot{\dragf}}{2\pfreq\dragf^3}\dif{\biggl[\dragf\fdot{\xcons}-\xcons\fdot{\dragf}\biggr]}
  +\dif{\biggl[\frac{1}{4\pfreq\dragf^2}\biggr]}\biggl[\dragf\ffdot{\xcons}-\xcons\ffdot{\dragf}\biggr]
  +\frac{1}{4\pfreq\dragf^2}\dif{\biggl[\dragf\ffdot{\xcons}-\xcons\ffdot{\dragf}\biggr]}
\end{align*}
\begin{align}\label{kpath9c}
\begin{split}
&=\biggl[-\frac{\ffdot{\dragf}}{2\pfreq\dragf^3}+\frac{3\fdot{\dragf}^2}{2\pfreq\dragf^4}\biggr]
   \biggl[\dragf\fdot{\xcons}-\xcons\fdot{\dragf}\biggr]
   -\frac{\fdot{\dragf}}{2\pfreq\dragf^3}\biggl[\dragf\ffdot{\xcons}-\xcons\ffdot{\dragf}\biggr]
   -\frac{\fdot{\dragf}}{2\pfreq\dragf^3}\biggl[\dragf\ffdot{\xcons}-\xcons\ffdot{\dragf}\biggr]\\
  &\qquad+\frac{1}{4\pfreq\dragf^2}\biggl[\fdot{\dragf}\ffdot{\xcons}+\dragf\fffdot{\xcons}
     -\fdot{\xcons}\ffdot{\dragf}-\xcons\fffdot{\dragf}\biggr]
\end{split}
\nonumber\\
&=\frac{(3\fdot{\dragf}^2-\dragf\ffdot{\dragf})(\dragf\fdot{\xcons}-\xcons\fdot{\dragf})}{2\pfreq\dragf^4}
   -\frac{\fdot{\dragf}(\dragf\ffdot{\xcons}-\xcons\ffdot{\dragf})}{\pfreq\dragf^3}
  +\frac{\fdot{\dragf}\ffdot{\xcons}+\dragf\fffdot{\xcons}
     -\fdot{\xcons}\ffdot{\dragf}-\xcons\fffdot{\dragf}}{4\pfreq\dragf^2}.
\end{align}
\end{subequations}
From \eqnref{kpath1} and \eqnref{kpath4}, we get
\begin{subequations}\label{kpath11}
\begin{align}\label{kpath11a}
\bbk=\plusmin\frac{|\vscrp+\vscrq|}{|\vscrr|^3},\quad
\bbt=\frac{\fffdot{\cdkt}\alepha+\fffdot{\rhorep}\alephb+3\ffdot{\rhorep}\alephc
  +\vbbc\alephd+\rhorep\alephe+\alephf}{|\vscrp+\vscrq|^2}
\end{align}
\begin{align}\label{kpath11b}
\frt=\frac{\vscrr}{|\vscrr|},\quad
\frb=\frac{\vscrp+\vscrq}{|\vscrp+\vscrq|},\quad
\frn=\cprod{\frb}{\frt}
\end{align}
\end{subequations}
which, together with \eqnref{kpath2}, is the complete set of equations describing the apparent
path of the light source. To use these equations, one should first compute the quantities
$\fdota, \ffdota, \fffdota$ and $\fdotg, \ffdotg, \fffdotg$ for the type of observer's motion
in question, followed by a computation of the quantities given by \eqnref{kpath5} through
\eqnref{kpath9}. Then $\fdott,\ffdott, \fffdott$ and $\fdote, \ffdote, \fffdote$ as well as
the quantities defined in \eqnref{kpath2} can be computed, after which \eqnref{kpath11} can
finally be evaluated.

\art{Apparent geometry of obliquated rays}
Regarding a light ray as the curve described by a point moving with the ray velocity
$\vectups$ in configuration space, we see that the curvature $\bbkbar$, the torsion $\bbtbar$ and
the radius of curvature $\bbrbar$ of the ray can be computed from the formulae
\begin{subequations}\label{kray1}
\begin{equation}\label{kray1a}
\bbkbar=\plusmin\frac{|\cprod{\vectups}{\fdot{\vectups}}|}{|\vectups|^3},\qquad
\bbtbar=\frac{\dprod{\ffdot{\vectups}}{(\cprod{\vectups}{\fdot{\vectups}})}}
  {|\cprod{\vectups}{\fdot{\vectups}}|^2},\quad
\bbrbar=\plusmin1/\bbkbar
\end{equation}
while the tangent vector $\frtbar$, the normal vector $\frnbar$, and the binormal
vector $\frbbar$ of the instantaneous Frenet frame at any point on the ray can be
computed from the formulae
\begin{equation}\label{kray1b}
\frtbar=\frac{\vectups}{|\vectups|},\quad
\frbbar=\frac{\cprod{\vectups}{\fdot{\vectups}}}{|\cprod{\vectups}{\fdot{\vectups}}|},\quad
\frnbar=\cprod{\frbbar}{\frtbar},\quad
\frcbar=\bbrbar\,\frnbar
\end{equation}
\end{subequations}
where $\frcbar$ gives the instantaneous center of curvature of the ray~\footnote{Let $\fdot{\vect{x}}
=\vectups$. Then in order to determine the ray uniquely, one needs to evaluate $\vect{x}$ at an initial
instant and to compute the Frenet frame at that instant. When this is not done, the ray is determined up
to geometric congruence, which is good enough for our purposes here. We note that due diligence requires
the position vector $\vect{x}$ of a point on the ray to be distinguished from the position vector $\vectr$
of the point of observation.}. To develop these equations, we introduce
\begin{subequations}\label{kray2}
\begin{align}\label{kray2a}
\begin{split}
\veusa&=\cprod{\unitkap}{\vectu},\quad
\veusb=\cprod{\unitkap}{\vecta},\quad
\veusc=\cprod{\unitkap}{\vecte},\quad
\veusd=\cprod{\unitkap}{\fdota},\quad
\veuse=\cprod{\unitkap}{\fdote}\\
&\quad\veusf=\cprod{\vecta}{\vectu},\quad
\veusg=\cprod{\vecta}{\vecte},\quad
\veush=\cprod{\vecta}{\fdota},\quad
\veusi=\cprod{\vecta}{\fdote}\\
&\quad\veusj=\cprod{\vectu}{\fdota},\quad
\veusk=\cprod{\vectu}{\fdote},\quad
\veusl=\cprod{\vecte}{\fdota},\quad
\veusm=\cprod{\vecte}{\fdote}
\end{split}
\end{align}
\begin{align}\label{kray2b}
\begin{split}
&\qquad\veusn=\rhorep\veusd+\veuse,\quad
\veuso=\rhorep\veush+\veusi,\quad
\veusp=\veusa-\veusc\\
&\qquad\veusq=\veusf-\veusg,\quad
\veusr=\veusl-\veusj,\quad
\veuss=\veusm-\veusk\\
&\veust=(\cdkt\vbba-\rhorep\fdot{\cdkt})\veusb+\cdkt\veusn+\rhorep\veuso,\quad
\veusu=\fdot{\cdkt}\veusp+\vbba\veusq+\rhorep\veusr+\veuss
\end{split}
\end{align}
\begin{align}\label{kray2c}
\begin{split}
&\qquad\qquad\imaa=\dprod{\unitkap}{(\veust+\veusu)},\quad
\imab=\dprod{\vecta}{(\veust+\veusu)}\\
&\imac=\dprod{\fdota}{(\veust+\veusu)},\quad
\imad=\dprod{\ffdota}{(\veust+\veusu)},\quad
\imae=\dprod{\ffdote}{(\veust+\veusu)}
\end{split}
\end{align}
\end{subequations}
where $\vbba$ is given by \eqnref{kpath2a}, and we derive
\begin{subequations}\label{kray3}
\begin{align}\label{kray3a}
\cprod{\vectups}{\fdot{\vectups}}
&=\cprod{(\cdkt\unitkap+\rhorep\vecta-\vectu+\vecte)}{(\fdot{\cdkt}\unitkap+\vbba\vecta
  +\rhorep\fdota+\fdote)}\beqref{kpath1c}\text{ \& }\eqnref{kpath4a}\nonumber\\
\begin{split}
&=(\cdkt\vbba-\rhorep\fdot{\cdkt})(\cprod{\unitkap}{\vecta})
    +\rhorep\cdkt(\cprod{\unitkap}{\fdota})+\cdkt(\cprod{\unitkap}{\fdote})
    +\rhorep^2(\cprod{\vecta}{\fdota})+\rhorep(\cprod{\vecta}{\fdote})\\
  &\quad-\fdot{\cdkt}(\cprod{\vectu}{\unitkap})-\vbba(\cprod{\vectu}{\vecta})
    -\rhorep(\cprod{\vectu}{\fdota})-(\cprod{\vectu}{\fdote})
  +\fdot{\cdkt}(\cprod{\vecte}{\unitkap})+\vbba(\cprod{\vecte}{\vecta})\\
    &\quad+\rhorep(\cprod{\vecte}{\fdota})+(\cprod{\vecte}{\fdote})
\end{split}
\nonumber\\
\begin{split}
&=(\cdkt\vbba-\rhorep\fdot{\cdkt})\veusb+\cdkt(\rhorep\veusd+\veuse)
    +\rhorep(\rhorep\veush+\veusi)+\fdot{\cdkt}(\veusa-\veusc)+\vbba(\veusf-\veusg)\\
   &\qquad+\rhorep(\veusl-\veusj)+\veusm-\veusk\beqref{kray2a}
\end{split}
\nonumber\\
&=\veust+\veusu\beqref{kray2b}
\end{align}
\begin{align}\label{kray3b}
\dprod{\ffdot{\vectups}}{(\cprod{\vectups}{\fdot{\vectups}})}
&=\dprod{(\ffdot{\cdkt}\unitkap+\ffdot{\rhorep}\vecta+\vbbb\fdota+\rhorep\ffdota+\ffdote)}
  {(\veust+\veusu)}\beqref{kpath4b}\text{ \& }\eqnref{kray3a}\nonumber\\
\begin{split}
&=\ffdot{\cdkt}[\dprod{\unitkap}{(\veust+\veusu)}]
  +\ffdot{\rhorep}[\dprod{\vecta}{(\veust+\veusu)}]
  +\vbbb[\dprod{\fdota}{(\veust+\veusu)}]\\
  &\qquad+\rhorep[\dprod{\ffdota}{(\veust+\veusu)}]
  +\dprod{\ffdote}{(\veust+\veusu)}
\end{split}
\nonumber\\
&=\ffdot{\cdkt}\imaa+\ffdot{\rhorep}\imab+\vbbb\imac+\rhorep\imad+\imae
  \beqref{kray2c}.
\end{align}
\end{subequations}
Substituting \eqnref{kray3}, \eqnref{kpath1c} and \eqnref{grad5a} into
\eqnref{kray1} yields
\begin{subequations}\label{kray4}
\begin{equation}\label{kray4a}
\bbkbar=\plusmin\frac{|\veust+\veusu|}{\scalc^3\rcal^3},\qquad
\bbtbar=\frac{\ffdot{\cdkt}\imaa+\ffdot{\rhorep}\imab+\vbbb\imac+\rhorep\imad+\imae}
  {|\veust+\veusu|^2}
\end{equation}
\begin{equation}\label{kray4b}
\frtbar=\frac{\cdkt\unitkap+\rhorep\vecta-\vectu+\vecte}{\scalc\rcal},\quad
\frbbar=\frac{\veust+\veusu}{|\veust+\veusu|},\quad
\frnbar=\cprod{\frbbar}{\frtbar}
\end{equation}
\end{subequations}
as the complete set of equations describing the apparent geometry of obliquated rays.

\art{Apparent frequency of obliquated rays}
A change in the velocity of a light ray can be interpreted as a change in the
frequency of the ray, and this change in {\em ray frequency} can be
calculated as follows. From \eqnref{main2b} and \eqnref{grad5a}, we have
\begin{equation}\label{kfreq1}
\scalc=\frac{\pfreq}{\kaprep}=\frac{\upsrep}{\rcal}
\end{equation}
which leads to
\begin{equation}\label{kfreq2}
\afreq=\rcal\pfreq,\quad\afreq=\kaprep\upsrep
\end{equation}
where the quantity $\afreq$ represents the apparent frequency of the ray. If we introduce
the redshift
\begin{equation}\label{kfreq3}
\rdshft=\frac{\pfreq-\afreq}{\afreq}
\end{equation}
in order to facilitate comparison with other treatments, then from the foregoing equations,
we shall obtain
\begin{equation}\label{kfreq4}
\frac{1}{1+\rdshft}=\frac{\afreq}{\pfreq}=\frac{\upsrep}{\scalc}=\rcal
\end{equation}
which, with the help of \eqnref{grad2w}, allows $\rdshft$ and $\afreq$ to be calculated
for an observer in accelerated translational, rotational or gravitational motion~\footnote{
It may be worthwhile to emphasize here that all the quantities calculated in this work
are due to the motion of the observer and not to any kind of motion of the light source.
Kineoptic effects of moving light sources will be treated when we study the nature of the
intrinsic redshifts advocated by our very own Galileo, the most eminent Dr. Halton C. Arp,
and his coworkers.}.

\part{Calculations}\label{parttwo}
\begin{wisdom}{Immanuel Kant (1724-1804)}
The investigations and calculations of astronomers have taught us much that is wonderful; but the
most important lesson we have received from them is the discovery of the abyss of our ignorance in
relation to the universe --- an ignorance the magnitude of which reason, without the information thus
derived, could never have conceived.
\end{wisdom}

\section{Translational obliquation}\label{S_TRAOB}
\art{Apparent direction to a light source}
To evaluate \eqnref{grad6} for an observer in accelerated translation, we have from \eqnref{main2b}
that $\alprep=\kaprep\scala\cos\lamrep$, whence $\kaprep\taurep=2\rhorep\scala\cos\lamrep$ by
\eqnref{main3a}. Moreover, since $\vecte=\zvect$ by \eqnref{main3a}, we have
in \eqnref{grad2} that
\begin{equation}\label{trob1}
\begin{split}
\bcal&=0,\quad\dcal=0,\quad\acal=0,\quad\hcal=0,\quad\ecal=0\\
\lcal_0&=-2\rhorep\scala\scalu\cos\lamrep\cos\phirep,\quad
\lcal_1=\rhorep\scala(\cos\thtrep-2\cos\lamrep\cos\phirep),\quad
\lcal_2=2\rhorep\scala(\cos\thtrep-\cos\lamrep\cos\phirep)\\
\ncal_1&=-4\dragf\rhorep\scala\scalc\scalu^2\cos\lamrep,\quad
\ncal_2=-4\dragf\rhorep\scala\scalc\scalu^2\cos\lamrep\cos^2\phirep,\quad
\ncal_3=0,\quad\ncal_4=0\\
\ncal_5&=4\rhorep^2\scala^2\scalu^2\cos\lamrep\cos\phirep(\cos\thtrep-\cos\lamrep\cos\phirep),\quad
\ncal_6=0
\end{split}
\end{equation}
from which we obtain, by \eqnref{grad2w},
\begin{subequations}\label{trob2}
\begin{align}\label{trob2a}
\begin{split}
\lcal&=-2\rhorep\sigrep\cos\lamrep\cos\phirep\\
\pcal&=-2\rhorep\sigrep[2\dragf\cos\lamrep+\betrep(\cos\thtrep-2\cos\lamrep\cos\phirep)]\\
\ncal&=4\rhorep^2\sigrep^2\cos\lamrep\cos\phirep(\cos\thtrep-\cos\lamrep\cos\phirep)
  -4\dragf\rhorep\sigrep\cos\lamrep\sin^2\phirep
\end{split}
\end{align}
\begin{align}\label{trob2b}
\begin{split}
\gcal&=-\betrep+\dragf\cos\phirep+\rhorep\sigrep(\cos\thtrep-2\cos\lamrep\cos\phirep)\\
\rcal^2&=\dragf^2+\betrep^2-2\dragf\betrep\cos\phirep+\rhorep\sigrep[\rhorep\sigrep
  -2\dragf\cos\lamrep-2\betrep(\cos\thtrep-2\cos\lamrep\cos\phirep)]\\
\fcal^2&=\dragf^2\sin^2\phirep+\rhorep^2\sigrep^2[\sin^2\thtrep+4\cos\lamrep\cos\phirep
  (\cos\thtrep-\cos\lamrep\cos\phirep)]\\
 &\qquad+2\dragf\rhorep\sigrep(\cos\lamrep\cos2\phirep-\cos\phirep\cos\thtrep)
\end{split}
\end{align}
and by \eqnref{grad6},
\begin{align}\label{trob2b2}
\tan\psirep=\fcal/\gcal
\end{align}
while by \eqnref{main2}, \eqnref{main3} and \eqnref{main6},
\begin{align}\label{trob2c}
\begin{split}
&\rhorep=\frac{\xcons}{4\dragf\pfreq},\quad
\xcons=\frac{\vthtrep}{\sqrt{1+\vthtrep^2}},\quad
\dragf=\left\{\frac{1+\sqrt{1+\vthtrep^2}}{2}\right\}^{1/2}\\
&\vthtrep=\frac{\alprep}{\pfreq^2}=\murep\cos\lamrep,\quad
\cdkt=\scalc(\dragf-2\rhorep\sigrep\cos\lamrep),\quad
\dprod{\vectkap}{\unitplz}=0.
\end{split}
\end{align}
\end{subequations}
Equation \eqnref{trob2} gives a complete prescription for calculating $\psirep$ for a translating observer.
For an observer in radial motion ($\phirep=0,\thtrep=\lamrep$), we have from \eqnref{trob2b},
\begin{subequations}\label{trob3}
\begin{align}\label{trob3a}
\begin{split}
&\fcal=\absv{\rhorep}\sigrep\sin\lamrep,\quad
\gcal=\dragf-\betrep-\rhorep\sigrep\cos\lamrep\\
&\rcal^2=\rhorep^2\sigrep^2+(\dragf-\betrep)(\dragf-\betrep-2\rhorep\sigrep\cos\lamrep)
\end{split}
\end{align}
whence \eqnref{trob2b2} becomes
\begin{equation}\label{trob3b}
\tan\psirep=\frac{\absv{\rhorep}\sigrep\sin\lamrep}{\dragf-\betrep-\rhorep\sigrep\cos\lamrep}
\qquad\qquad(\phirep=0, \thtrep=\lamrep).
\end{equation}
\end{subequations}
Similarly, for an observer in rectilinear motion ($\thtrep=0,\phirep=\lamrep$), \eqnref{trob2b} yields
\begin{subequations}\label{trob4}
\begin{align}\label{trob4a}
\begin{split}
&\fcal=\absv{\dragf-2\rhorep\sigrep\cos\lamrep}\sin\lamrep,\quad
\gcal=\dragf\cos\lamrep-\betrep-\rhorep\sigrep\cos2\lamrep\\
&\rcal^2=\dragf^2+\betrep^2-2\dragf\betrep\cos\lamrep
    +\rhorep\sigrep(\rhorep\sigrep-2\dragf\cos\lamrep+2\betrep\cos2\lamrep)
\end{split}
\end{align}
which, when substituted into \eqnref{trob2b2}, yields
\begin{equation}\label{trob4b}
\tan\psirep=\frac{\absv{\dragf-2\rhorep\sigrep\cos\lamrep}\sin\lamrep}
  {\dragf\cos\lamrep-\betrep-\rhorep\sigrep\cos2\lamrep}
\qquad\qquad(\thtrep=0, \phirep=\lamrep).
\end{equation}
\end{subequations}
Finally, for an observer in coradial motion ($\lamrep=0,\thtrep=\phirep$), \eqnref{trob2b} gives
\begin{subequations}\label{trob5}
\begin{align}\label{trob5a}
\begin{split}
&\fcal=\absv{\dragf-\rhorep\sigrep}\sin\phirep,\quad
\gcal=-\betrep+(\dragf-\rhorep\sigrep)\cos\phirep\\
&\rcal^2=(\dragf-\rhorep\sigrep)^2+\betrep^2-(2\dragf+\rhorep\sigrep)\betrep\cos\phirep
\end{split}
\end{align}
so that, from \eqnref{trob2b2}, we get
\begin{equation}\label{trob5b}
\tan\psirep=\frac{\absv{\dragf-\rhorep\sigrep}\sin\phirep}
  {-\betrep+(\dragf-\rhorep\sigrep)\cos\phirep}
\qquad\qquad(\lamrep=0, \thtrep=\phirep).
\end{equation}
\end{subequations}
These results establish once again that the effect of translational acceleration on
obliquation is of second and higher orders in $\murep$, which implies that it is of practical
significance only in situations involving infraradio waves or ultrahigh accelerations or both.

\art{Apparent drift of a light source}\label{a_txpeed}
To evaluate \eqnref{kas7} for a translating observer, it is necessary to know the acceleration
$\vecta$ as a function of the velocity $\vectu$. In the simplest case when $\vecta$ is independent
of $\vectu$, we have from \eqnref{main3a} and \eqnref{kas2} that
\begin{subequations}\label{tspeed1}
\begin{equation}\label{tspeed1a}
\begin{split}
\vga&=\zvect,\quad
\vgb=\zvect,\quad
\vgc=\zvect,\quad
\vgd=\zvect,\quad
\vge=\zvect,\quad
\vgf=\zvect,\quad
\vgg=\zvect,\quad
\vgh=\zvect,\quad
\vgi=\zvect\\
&\qquad\qquad\qquad
\vja=\zvect,\quad
\vjb=\zvect,\quad
\vjc=\zvect,\quad
\vjd=\zvect,\quad
\vje=\zvect,\quad
\vjf=\zvect
\end{split}
\end{equation}
\begin{align}\label{tspeed1b}
\vjg&=\ugrad{\taurep}\beqref{kas2b}\nonumber\\
&=\ugrad{(2\rhorep\alprep\kaprep^{-2})}\beqref{main3a}\nonumber\\
&=2\kaprep^{-2}(\rhorep\ugrad{\alprep}+\alprep\ugrad{\rhorep})\beqref{clc33}\nonumber\\
&=2\kaprep^{-2}(\rhorep\vgf+\alprep\vgi)\beqref{kas3a}\text{ \& }\eqnref{kas3d}\nonumber\\
&=\zvect\beqref{tspeed1a}.
\end{align}
\end{subequations}
Substituting \eqnref{tspeed1} into \eqnref{kas7} yields
\begin{align}\label{tspeed2}
\angus
&=\betrep(\scalu\rcal)^{-2}(\xcala\vectu-\xcalb\vectups)\nonumber\\
&=\betrep(\scalu\rcal)^{-2}[\xcala\vectu-\xcalb(\vectc\dragf+\rhorep\vecta-\taurep\vectkap-\vectu)]
  \beqref{main1},\eqnref{main2a}\text{ \& }\eqnref{main3a}\nonumber\\
&=\betrep(\scalu\rcal)^{-2}[(\xcala+\xcalb)\vectu-\xcalb\rhorep\vecta
  -\xcalb(\scalc\dragf-\kaprep\taurep)\unitkap]\nonumber\\
&=\betrep(\scalu\rcal)^{-2}[(\xcala+\xcalb)\vectu-\xcalb\rhorep\vecta-\xcalb(\scalc\dragf
  -2\rhorep\scala\cos\lamrep)\unitkap]\beqref{main3a}
\end{align}
which, in view of the values of $\xcala$ and $\xcalb$ from \eqnref{kas7b}, may be rewritten as
\begin{subequations}\label{tspeed3}
\begin{equation}\label{tspeed3a}
\angus=\vtta\unitkap+\vttb\vectu-\vttc\vecta
\end{equation}
where, with $\fcal, \gcal, \rcal$ given by \eqnref{trob2b} and $\cdkt, \rhorep,\dragf$ given
by \eqnref{trob2c},
\begin{equation}\label{tspeed3b}
\vtta=-\vtte\cdkt,\quad
\vttb=\vttd+\vtte,\quad
\vttc=\vtte\rhorep,\quad
\vttd=\frac{\betrep+\gcal}{\fcal\scalu^2},\quad
\vtte=\frac{\betrep(\rcal^2+\betrep\gcal)}{\fcal\rcal^2\scalu^2}.
\end{equation}
\end{subequations}
From \eqnref{kas1c} and \eqnref{tspeed3a}, we get the variation of obliquation (or the apparent drift
of the light source per unit acceleration of the observer) as
\begin{equation}\label{tspeed4}
\aspua=\vtta\cos\lamrep+\vttb\scalu\cos\thtrep-\vttc\scala.
\end{equation}

\art{Apparent path of a light source}
Equation \eqnref{kpath11} can be evaluated for a translating observer as follows. With
$\rhorep, \xcons, \dragf, \vthtrep$ and $\alprep$ given by \eqnref{trob2c}, we introduce
\begin{subequations}\label{tpath1}
\begin{align}\label{tpath1xa}
\begin{split}
&\vsiga=\dprod{\vectkap}{\fdota},\quad
\vsigb=\dprod{\vectkap}{\ffdota},\quad
\vsigc=\dprod{\vectkap}{\fffdota},\quad
\vsigd=\dprod{\vecta}{\fdota},\quad
\vsige=\dprod{\vecta}{\ffdota},\quad
\vsigf=\dprod{\vecta}{\fffdota}\\
&\vsigg=\dprod{\fdota}{\fdota},\quad
\vsigh=\dprod{\fdota}{\ffdota},\quad
\vsigi=\dprod{\fdota}{\fffdota},\quad
\vsigj=\dprod{\ffdota}{\ffdota},\quad
\vsigk=\dprod{\ffdota}{\fffdota},\quad
\vsigl=\dprod{\fffdota}{\fffdota}\\
&\vsigm=\dprod{\unitkap}{(\cprod{\vecta}{\fdota})},\quad
\vsign=\dprod{\unitkap}{(\cprod{\vecta}{\ffdota})},\quad
\vsigo=\dprod{\unitkap}{(\cprod{\fdota}{\ffdota})},\quad
\vsigp=\dprod{\vecta}{(\cprod{\fdota}{\ffdota})}\\
&\qquad\vsigq=\dprod{\fffdota}{(\cprod{\unitkap}{\vecta})},\quad
\vsigr=\dprod{\fffdota}{(\cprod{\unitkap}{\fdota})},\quad
\vsigs=\dprod{\fffdota}{(\cprod{\unitkap}{\ffdota})}\\
&\qquad\vsigt=\dprod{\fffdota}{(\cprod{\vecta}{\fdota})},\quad
\vsigu=\dprod{\fffdota}{(\cprod{\vecta}{\ffdota})},\quad
\vsigv=\dprod{\fffdota}{(\cprod{\fdota}{\ffdota})}
\end{split}
\end{align}
\begin{align}\label{tpath1a}
\begin{split}
\vrhoa&=(1+\vthtrep^2)^{1/2},\quad
\vrhob=1-4\vthtrep^2,\quad
\vrhoc=\frac{1}{\vrhoa^3}-\frac{\xcons^2}{4\dragf^2},\quad
\vrhod=-1+\frac{\vsiga\vrhoc}{4\dragf\pfreq^3}\\
\vrhoe&=\frac{1}{\vrhoa^5\pfreq^2}\biggl[\vsigb\vrhoa^2-\frac{3\vthtrep\vsiga^2}{\pfreq^2}\biggr],\quad
\vrhof=\frac{1}{\vrhoa^7\pfreq^2}\biggl[\vsigc\vrhoa^4-\frac{3\vsiga}{\pfreq^2}
  \biggl(3\vthtrep\vsigb\vrhoa^2+\frac{\vrhob\vsiga^2}{\pfreq^2}\biggr)\biggr]
\end{split}
\end{align}
\begin{align}\label{tpath1b}
\begin{split}
\vrhog&=\frac{1}{4\dragf\pfreq^2}\biggl[\xcons\vsigb+\frac{\vrhoc\vsiga^2}{\pfreq^2}\biggr]\\
\vrhoh&=\frac{1}{4\dragf\pfreq^2}\biggl[\xcons\biggl\{\vsigc-\frac{\vrhog\vsiga}{\dragf}\biggr\}
    +\vsiga\biggl\{\vrhoe+\frac{2\vsigb}{\pfreq^2\vrhoa^3}\biggr\}
    -\frac{\xcons\vsiga}{2\dragf^2\pfreq^2}\biggl\{\xcons\vsigb+\frac{\vrhoc\vsiga^2}{\pfreq^2}\biggr\}\biggr]\\
\vrhoi&=\frac{1}{4\dragf\pfreq^2}\biggl[\pfreq\biggl\{\vrhoe-\frac{\xcons\vrhog}{\dragf}\biggr\}
    -\frac{\xcons\vrhoc\vsiga^2}{2\dragf^2\pfreq^3}\biggr]\\
\vrhoj&=\frac{1}{4\dragf\pfreq^2}\biggl[\pfreq\biggl\{\vrhof-\frac{\xcons\vrhoh}{\dragf}\biggr\}
   +\frac{3\vsiga}{\dragf\pfreq}\biggl\{\frac{\xcons}{4\dragf}\biggl(
   \frac{\xcons\vsiga^2\vrhoc}{2\dragf^2\pfreq^4}-\vrhoe\biggr)
   -\vrhog\biggl(\frac{1}{\vrhoa^3}-\frac{\xcons^2}{2\dragf^2}\biggr)\biggr\}\biggr]
\end{split}
\end{align}
\begin{align}\label{tpath1c}
\begin{split}
\vrhok&=\frac{\vsiga}{\kaprep}\biggl[-2\rhorep+\frac{1}{4\dragf\pfreq}\biggl\{\xcons
  -\frac{2\alprep\vrhoc}{\pfreq^2}\biggr\}\biggr],\quad
\vrhol=\frac{1}{\kaprep}\biggl[\pfreq\vrhog-2\biggl\{\rhorep\vsigb+\alprep\vrhoi
  +\frac{\vsiga^2\vrhoc}{2\dragf\pfreq^3}\biggr\}\biggr]\\
\vrhom&=\frac{1}{\kaprep}\biggl[\pfreq\vrhoh-2\biggl\{\rhorep\vsigc
  +\alprep\vrhoj+3\vsiga\biggl(\vrhoi+\frac{\vsigb\vrhoc}{4\dragf\pfreq^3}\biggr)\biggr\}\biggr]\\
\vrhon&=\vrhoi\vrhok-\vrhol\vrhod,\quad
\vrhoo=\vrhok(1+2\vrhod)-\rhorep\vrhol,\quad
\vrhop=\vrhod(1+2\vrhod)-\rhorep\vrhoi\\
\vrhoq&=\biggl[\scala^2\vrhod^2+\rhorep(\rhorep\vsigg+2\vrhod\vsigd)+
  \vrhok\biggl\{\vrhok+\frac{2(\alprep\vrhod+\rhorep\vsiga)}{\kaprep}\biggr\}\biggr]^{1/2}
\end{split}
\end{align}
\begin{align}\label{tpath1d}
\begin{split}
&\vrhor=\vrhon(\scala^2-\kaprep^{-2}\alprep^2)+\vrhoo(\vsigd-\kaprep^{-2}\alprep\vsiga)
  +\rhorep\vrhok(\vsige-\kaprep^{-2}\alprep\vsigb)+\vrhop\kaprep^{-1}(\alprep\vsigd-\vsiga\scala^2)\\
  &\qquad+\rhorep\vrhod\kaprep^{-1}(\alprep\vsige-\vsigb\scala^2)
  +\rhorep^2\kaprep^{-1}(\vsiga\vsige-\vsigb\vsigd)\\
&\vrhos=\vrhon(\vsigd-\kaprep^{-2}\alprep\vsiga)+\vrhoo(\vsigg-\kaprep^{-2}\vsiga^2)
  +\rhorep\vrhok(\vsigh-\kaprep^{-2}\vsigb\vsiga)+\vrhop\kaprep^{-1}(\alprep\vsigg-\vsiga\vsigd)\\
  &\qquad+\rhorep\vrhod\kaprep^{-1}(\alprep\vsigh-\vsigb\vsigd)
  +\rhorep^2\kaprep^{-1}(\vsiga\vsigh-\vsigb\vsigg)\\
&\vrhot=\vrhon(\vsige-\kaprep^{-2}\alprep\vsigb)+\vrhoo(\vsigh-\kaprep^{-2}\vsiga\vsigb)
  +\rhorep\vrhok(\vsigj-\kaprep^{-2}\vsigb^2)+\vrhop\kaprep^{-1}(\alprep\vsigh-\vsiga\vsige)\\
  &\qquad+\rhorep\vrhod\kaprep^{-1}(\alprep\vsigj-\vsigb\vsige)
  +\rhorep^2\kaprep^{-1}(\vsiga\vsigj-\vsigb\vsigh)
\end{split}
\end{align}
\begin{align}\label{tpath1e}
\begin{split}
&\vrhou=\vrhon\kaprep^{-1}(\alprep\vsigd-\scala^2\vsiga)+\vrhoo\kaprep^{-1}(\alprep\vsigg-\vsigd\vsiga)
  +\rhorep\vrhok\kaprep^{-1}(\alprep\vsigh-\vsige\vsiga)+\vrhop(\scala^2\vsigg-\vsigd^2)\\
  &\qquad+\rhorep\vrhod(\scala^2\vsigh-\vsige\vsigd)
  +\rhorep^2(\vsigd\vsigh-\vsige\vsigg)\\
&\vrhov=\vrhon\kaprep^{-1}(\alprep\vsige-\scala^2\vsigb)+\vrhoo\kaprep^{-1}(\alprep\vsigh-\vsigd\vsigb)
  +\rhorep\vrhok\kaprep^{-1}(\alprep\vsigj-\vsige\vsigb)+\vrhop(\scala^2\vsigh-\vsigd\vsige)\\
  &\qquad+\rhorep\vrhod(\scala^2\vsigj-\vsige^2)
  +\rhorep^2(\vsigd\vsigj-\vsige\vsigh)\\
&\vrhow=\vrhon\kaprep^{-1}(\vsiga\vsige-\vsigd\vsigb)+\vrhoo\kaprep^{-1}(\vsiga\vsigh-\vsigg\vsigb)
  +\rhorep\vrhok\kaprep^{-1}(\vsiga\vsigj-\vsigh\vsigb)+\vrhop(\vsigd\vsigh-\vsigg\vsige)\\
  &\qquad+\rhorep\vrhod(\vsigd\vsigj-\vsigh\vsige)
  +\rhorep^2(\vsigg\vsigj-\vsigh^2)
\end{split}
\end{align}
\begin{align}\label{tpath1f}
\begin{split}
&\vrhox=[\vrhon\vrhor+\vrhoo\vrhos+\rhorep\vrhok\vrhot+\vrhop\vrhou+\rhorep\vrhod\vrhov+\rhorep^2\vrhow]^{1/2}\\
&\vrhoy=\vrhom(\vrhop\vsigm+\rhorep\vrhod\vsign+\rhorep^2\vsigo)+\vrhoj(-\vrhoo\vsigm-\rhorep\vrhok\vsign+\rhorep^2\vsigp)
  +3\vrhoi(\vrhon\vsigm-\rhorep\vrhok\vsigo-\rhorep\vrhod\vsigp)\\
  &\qquad+(2+3\vrhod)(\vrhon\vsign+\vrhoo\vsigo+\vrhop\vsigp)+\rhorep(\vrhon\vsigq+\vrhoo\vsigr+\vrhop\vsigt)
  +\rhorep^2(\vrhok\vsigs+\vrhod\vsigu+\rhorep\vsigv).
\end{split}
\end{align}
\end{subequations}

\subart{Development of equations \eqnref{kpath5} through \eqnref{kpath9}}
From \eqnref{main3a}, \eqnref{kpath5}, \eqnref{kpath6} and \eqnref{tpath1xa}, we have
\begin{align}\label{tpath2}
\begin{split}
&\fdotg=0,\quad\ffdotg=0,\quad\fffdotg=0,\quad
\fdot{\alprep}=\vsiga,\quad\ffdot{\alprep}=\vsigb,\quad\fffdot{\alprep}=\vsigc\\
&\qquad\fdot{\vthtrep}=\vsiga/\pfreq^2,\quad\ffdot{\vthtrep}=\vsigb/\pfreq^2,\quad
\fffdot{\vthtrep}=\vsigc/\pfreq^2.
\end{split}
\end{align}
It is easy to see that
\begin{subequations}\label{tpath4}
\begin{align}\label{tpath4a}
\fdot{\xcons}
&=\fdot{\vthtrep}(1+\vthtrep^2)^{-3/2}\beqref{kpath7a}\nonumber\\
&=\vsiga/(\pfreq^2\vrhoa^3)\beqref{tpath1a}\text{ \& }\eqnref{tpath2}
\end{align}
\begin{align}\label{tpath4b}
\ffdot{\xcons}
&=(1+\vthtrep^2)^{-5/2}[\ffdot{\vthtrep}(1+\vthtrep^2)-3\vthtrep\fdot{\vthtrep}^2]
  \beqref{kpath7b}\nonumber\\
&=\vrhoa^{-5}[\vsigb\pfreq^{-2}\vrhoa^2-3\vthtrep(\vsiga\pfreq^{-2})^2]
  \beqref{tpath1a}\text{ \& }\eqnref{tpath2}\nonumber\\
&=\vrhoe\beqref{tpath1a}
\end{align}
\begin{align}\label{tpath4c}
\fffdot{\xcons}
&=(1+\vthtrep^2)^{-7/2}[\fffdot{\vthtrep}(1+\vthtrep^2)^2-9\vthtrep\fdot{\vthtrep}\ffdot{\vthtrep}(1+\vthtrep^2)
  -3\fdot{\vthtrep}^3(1-4\vthtrep^2)]\beqref{kpath7c}\nonumber\\
&=\vrhoa^{-7}[(\vsigc\pfreq^{-2})\vrhoa^4-9\vthtrep(\vsiga\pfreq^{-2})(\vsigb\pfreq^{-2})\vrhoa^2
  -3(\vsiga\pfreq^{-2})^3\vrhob]\beqref{tpath1a}\text{ \& }\eqnref{tpath2}\nonumber\\
&=\vrhof\beqref{tpath1a}
\end{align}
\end{subequations}
\begin{subequations}\label{tpath5}
\begin{align}\label{tpath5a}
\fdot{\dragf}
&=\frac{\fdotg\dragf}{\gamrep}+\frac{\xcons\gamrep^2\fdot{\vthtrep}}{4\dragf}
  \beqref{kpath8a}\nonumber\\
&=\frac{\xcons\vsiga}{4\dragf\pfreq^2}
  \beqref{tpath2}\text{ \& }\eqnref{main3a}
\end{align}
\begin{align}\label{tpath5b}
\ffdot{\dragf}
&=\dragf\biggl[\frac{\ffdotg}{\gamrep}-\frac{\fdotg^2}{\gamrep^2}\biggr]
   +\fdot{\dragf}\biggl[\frac{\fdotg}{\gamrep}-\frac{\xcons\gamrep^2\fdot{\vthtrep}}{4\dragf^2}\biggr]
   +\frac{\gamrep}{4\dragf}\biggl[\fdot{\xcons}\gamrep\fdot{\vthtrep}
    +2\xcons\fdotg\fdot{\vthtrep}+\xcons\gamrep\ffdot{\vthtrep}\biggr]
  \beqref{kpath8b}\nonumber\\
&=\fdot{\dragf}\biggl[-\frac{\xcons\fdot{\vthtrep}}{4\dragf^2}\biggr]
   +\frac{1}{4\dragf}\biggl[\fdot{\xcons}\fdot{\vthtrep}+\xcons\ffdot{\vthtrep}\biggr]
   \beqref{tpath2}\text{ \& }\eqnref{main3a}\nonumber\\
\begin{split}
&=\biggl[\frac{\xcons\vsiga}{4\dragf\pfreq^2}\biggr]\biggl[-\frac{\xcons}{4\dragf^2}\biggr]
    \biggl[\frac{\vsiga}{\pfreq^2}\biggr]
  +\frac{1}{4\dragf}\biggl[\biggl\{\frac{\vsiga}{\pfreq^2\vrhoa^3}\biggr\}\biggl\{\frac{\vsiga}{\pfreq^2}\biggr\}
    +\xcons\biggl\{\frac{\vsigb}{\pfreq^2}\biggr\}\biggr]
  \beqref{tpath2},\eqnref{tpath4a}\text{ \& }\eqnref{tpath5a}
\end{split}
\nonumber\\
&=-\frac{\xcons^2\vsiga^2}{16\dragf^3\pfreq^4}
  +\frac{1}{4\dragf}\biggl[\frac{\vsiga^2}{\pfreq^4\vrhoa^3}
  +\frac{\xcons\vsigb}{\pfreq^2}\biggr]
=\frac{1}{4\dragf\pfreq^2}\biggl[\xcons\vsigb+\frac{\vsiga^2}{\pfreq^2}
  \biggl\{\frac{1}{\vrhoa^3}-\frac{\xcons^2}{4\dragf^2}\biggr\}\biggr]\nonumber\\
&=\frac{1}{4\dragf\pfreq^2}\biggl[\xcons\vsigb+\frac{\vrhoc\vsiga^2}{\pfreq^2}\biggr]
  \beqref{tpath1a}\nonumber\\
&=\vrhog\beqref{tpath1b}
\end{align}
\begin{align*}
\begin{split}
\fffdot{\dragf}
&=\dragf\biggl[\frac{\fffdotg}{\gamrep}-\frac{3\fdotg\ffdotg}{\gamrep^2}+\frac{2\fdotg^3}{\gamrep^3}\biggr]
  +2\fdot{\dragf}\biggl[\frac{\ffdotg}{\gamrep}-\frac{\fdotg^2}{\gamrep^2}
    -\frac{\fdot{\xcons}\gamrep^2\fdot{\vthtrep}}{4\dragf^2}-\frac{\xcons\gamrep\fdotg\fdot{\vthtrep}}{2\dragf^2}
    -\frac{\xcons\gamrep^2\ffdot{\vthtrep}}{4\dragf^2}
    +\frac{\xcons\gamrep^2\fdot{\dragf}\fdot{\vthtrep}}{4\dragf^3}\biggr]
   +\ffdot{\dragf}\biggl[\frac{\fdotg}{\gamrep}-\frac{\xcons\gamrep^2\fdot{\vthtrep}}{4\dragf^2}\biggr]\\
   &\quad+\frac{\fdotg}{4\dragf}\biggl[\fdot{\xcons}\gamrep\fdot{\vthtrep}+2\xcons\fdotg\fdot{\vthtrep}
      +\xcons\gamrep\ffdot{\vthtrep}\biggr]
  +\frac{\gamrep}{4\dragf}\biggl[\ffdot{\xcons}\gamrep\fdot{\vthtrep}+3\fdot{\xcons}\fdotg\fdot{\vthtrep}
    +2\fdot{\xcons}\gamrep\ffdot{\vthtrep}+2\xcons\ffdotg\fdot{\vthtrep}
    +3\xcons\fdotg\ffdot{\vthtrep}+\xcons\gamrep\fffdot{\vthtrep}\biggr]
  \beqref{kpath8c}
\end{split}
\nonumber\\
&=2\fdot{\dragf}\biggl[-\frac{\fdot{\xcons}\fdot{\vthtrep}}{4\dragf^2}
    -\frac{\xcons\ffdot{\vthtrep}}{4\dragf^2}+\frac{\xcons\fdot{\dragf}\fdot{\vthtrep}}{4\dragf^3}\biggr]
   +\ffdot{\dragf}\biggl[-\frac{\xcons\fdot{\vthtrep}}{4\dragf^2}\biggr]
   +\frac{1}{4\dragf}\biggl[\ffdot{\xcons}\fdot{\vthtrep}
    +2\fdot{\xcons}\ffdot{\vthtrep}+\xcons\fffdot{\vthtrep}\biggr]
  \beqref{main3a}\text{ \& }\eqnref{tpath2}
\nonumber\\
&=2\fdot{\dragf}\biggl[-\frac{\fdot{\xcons}\vsiga}{4\dragf^2\pfreq^2}
    -\frac{\xcons\vsigb}{4\dragf^2\pfreq^2}+\frac{\xcons\fdot{\dragf}\vsiga}{4\dragf^3\pfreq^2}\biggr]
   +\ffdot{\dragf}\biggl[-\frac{\xcons\vsiga}{4\dragf^2\pfreq^2}\biggr]
   +\frac{1}{4\dragf}\biggl[\frac{\ffdot{\xcons}\vsiga}{\pfreq^2}
    +\frac{2\fdot{\xcons}\vsigb}{\pfreq^2}+\frac{\xcons\vsigc}{\pfreq^2}\biggr]
  \beqref{tpath2}
\end{align*}
\begin{align}\label{tpath5c}
\begin{split}
&=2\fdot{\dragf}\biggl[-\biggl\{\frac{\vsiga}{\pfreq^2\vrhoa^3}\biggr\}\biggl\{\frac{\vsiga}{4\dragf^2\pfreq^2}\biggr\}
    -\frac{\xcons\vsigb}{4\dragf^2\pfreq^2}+\frac{\xcons\fdot{\dragf}\vsiga}{4\dragf^3\pfreq^2}\biggr]
   +\ffdot{\dragf}\biggl[-\frac{\xcons\vsiga}{4\dragf^2\pfreq^2}\biggr]\\
   &\qquad+\frac{1}{4\dragf}\biggl[\frac{\vrhoe\vsiga}{\pfreq^2}
    +\biggl\{\frac{\vsiga}{\pfreq^2\vrhoa^3}\biggr\}\biggl\{\frac{2\vsigb}{\pfreq^2}\biggr\}
    +\frac{\xcons\vsigc}{\pfreq^2}\biggr]
   \beqref{tpath4}
\end{split}
\nonumber\\
\begin{split}
&=\frac{\xcons\vsiga}{2\dragf\pfreq^2}\biggl[-\frac{\vsiga^2}{4\dragf^2\pfreq^4\vrhoa^3}
    -\frac{\xcons\vsigb}{4\dragf^2\pfreq^2}
    +\biggl\{\frac{\xcons\vsiga}{4\dragf\pfreq^2}\biggr\}\biggl\{\frac{\xcons\vsiga}{4\dragf^3\pfreq^2}\biggr\}\biggr]
   -\frac{\xcons\vrhog\vsiga}{4\dragf^2\pfreq^2}
   +\frac{1}{4\dragf}\biggl[\frac{\vrhoe\vsiga}{\pfreq^2}
    +\frac{2\vsiga\vsigb}{\pfreq^4\vrhoa^3}+\frac{\xcons\vsigc}{\pfreq^2}\biggr]\\
    &\qquad\beqref{tpath5a}\text{ \& }\eqnref{tpath5b}
\end{split}
\nonumber\\
&=-\frac{\xcons\vsiga}{8\dragf^3\pfreq^4}\biggl[\xcons\vsigb+\frac{\vsiga^2}{\pfreq^2\vrhoa^3}
    -\frac{\xcons^2\vsiga^2}{4\dragf^2\pfreq^2}\biggr]
   +\frac{1}{4\dragf\pfreq^2}\biggl[\vrhoe\vsiga+\xcons\vsigc
    +\frac{2\vsiga\vsigb}{\pfreq^2\vrhoa^3}-\frac{\xcons\vrhog\vsiga}{\dragf}\biggr]
   \nonumber\\
&=\frac{1}{4\dragf\pfreq^2}\biggl[\xcons\biggl\{\vsigc-\frac{\vrhog\vsiga}{\dragf}\biggr\}
    +\vsiga\biggl\{\vrhoe+\frac{2\vsigb}{\pfreq^2\vrhoa^3}\biggr\}
    -\frac{\xcons\vsiga}{2\dragf^2\pfreq^2}\biggl\{\xcons\vsigb+\frac{\vsiga^2}{\pfreq^2}\biggl(\frac{1}{\vrhoa^3}
    -\frac{\xcons^2}{4\dragf^2}\biggr)\biggr\}\biggr]\nonumber\\
&=\vrhoh\beqref{tpath1a}\text{ \& }\eqnref{tpath1b}
\end{align}
\end{subequations}
\begin{subequations}\label{tpath6}
\begin{align}\label{tpath6a}
\fdot{\rhorep}
&=\frac{\dragf\fdot{\xcons}-\xcons\fdot{\dragf}}{4\pfreq\dragf^2}
  \beqref{kpath9a}\nonumber\\
&=\frac{1}{4\pfreq\dragf^2}\biggl[\dragf\biggl\{\frac{\vsiga}{\pfreq^2\vrhoa^3}\biggr\}
  -\xcons\biggl\{\frac{\xcons\vsiga}{4\dragf\pfreq^2}\biggr\}\biggr]
  \beqref{tpath4a}\text{ \& }\eqnref{tpath5a}\nonumber\\
&=\frac{\vsiga}{4\dragf\pfreq^3}\biggl[\frac{1}{\vrhoa^3}
  -\frac{\xcons^2}{4\dragf^2}\biggr]=\frac{\vsiga\vrhoc}{4\dragf\pfreq^3}\beqref{tpath1a}
\end{align}
\begin{align}\label{tpath6b}
\ffdot{\rhorep}
&=-\frac{\fdot{\dragf}(\dragf\fdot{\xcons}-\xcons\fdot{\dragf})}{2\pfreq\dragf^3}
   +\frac{\dragf\ffdot{\xcons}-\xcons\ffdot{\dragf}}{4\pfreq\dragf^2}
   \beqref{kpath9b}\nonumber\\
&=-\frac{1}{2\pfreq\dragf^3}\biggl[\frac{\xcons\vsiga}{4\dragf\pfreq^2}\biggr]
     \biggl[\dragf\biggl\{\frac{\vsiga}{\pfreq^2\vrhoa^3}\biggr\}
     -\xcons\biggl\{\frac{\xcons\vsiga}{4\dragf\pfreq^2}\biggr\}\biggr]
   +\frac{\dragf\vrhoe-\xcons\vrhog}{4\pfreq\dragf^2}
   \beqref{tpath4}\text{ \& }\eqnref{tpath5}\nonumber\\
&=-\frac{\xcons\vsiga^2}{8\dragf^3\pfreq^5}
     \biggl[\frac{1}{\vrhoa^3}-\frac{\xcons^2}{4\dragf^2}\biggr]
   +\frac{1}{4\dragf\pfreq}\biggl[\vrhoe-\frac{\xcons\vrhog}{\dragf}\biggr]\nonumber\\
&=\frac{1}{4\dragf\pfreq^2}\biggl[\pfreq\biggl\{\vrhoe-\frac{\xcons\vrhog}{\dragf}\biggr\}
    -\frac{\xcons\vsiga^2}{2\dragf^2\pfreq^3}\biggl\{\frac{1}{\vrhoa^3}-\frac{\xcons^2}{4\dragf^2}\biggr\}
   \biggr]\nonumber\\
&=\vrhoi\beqref{tpath1a}\text{ \& }\eqnref{tpath1b}
\end{align}
\begin{align*}
\fffdot{\rhorep}
&=\frac{(3\fdot{\dragf}^2-\dragf\ffdot{\dragf})(\dragf\fdot{\xcons}-\xcons\fdot{\dragf})}{2\pfreq\dragf^4}
   -\frac{\fdot{\dragf}(\dragf\ffdot{\xcons}-\xcons\ffdot{\dragf})}{\pfreq\dragf^3}
  +\frac{\fdot{\dragf}\ffdot{\xcons}+\dragf\fffdot{\xcons}
     -\fdot{\xcons}\ffdot{\dragf}-\xcons\fffdot{\dragf}}{4\pfreq\dragf^2}
  \beqref{kpath9c}\nonumber\\
\begin{split}
&=\frac{1}{2\dragf^4\pfreq}\biggl[3\biggl\{\frac{\xcons\vsiga}{4\dragf\pfreq^2}\biggr\}^2-\dragf\vrhog\biggr]
   \biggl[\dragf\biggl\{\frac{\vsiga}{\pfreq^2\vrhoa^3}\biggr\}
       -\xcons\biggl\{\frac{\xcons\vsiga}{4\dragf\pfreq^2}\biggr\}\biggr]
   -\frac{\dragf\vrhoe-\xcons\vrhog}{\dragf^3\pfreq}
       \biggl\{\frac{\xcons\vsiga}{4\dragf\pfreq^2}\biggr\}\\
  &\qquad+\frac{1}{4\dragf^2\pfreq}\biggl[\vrhoe\biggl\{\frac{\xcons\vsiga}{4\dragf\pfreq^2}\biggr\}
     +\dragf\vrhof-\vrhog\biggl\{\frac{\vsiga}{\pfreq^2\vrhoa^3}\biggr\}
     -\xcons\vrhoh\biggr]\beqref{tpath4}\text{ \& }\eqnref{tpath5}
\end{split}
\nonumber\\
\begin{split}
&=\frac{\vsiga}{2\dragf^2\pfreq^3}\biggl[\frac{3\xcons^2\vsiga^2}{16\dragf^3\pfreq^4}-\vrhog\biggr]
   \biggl[\frac{1}{\vrhoa^3}-\frac{\xcons^2}{4\dragf^2}\biggr]
   -\frac{\xcons\vsiga}{4\dragf^3\pfreq^3}\biggl[\vrhoe-\frac{\xcons\vrhog}{\dragf}\biggr]\\
  &\qquad+\frac{1}{4\dragf^2\pfreq}\biggl[\frac{\vsiga}{\pfreq^2}\biggl\{\frac{\xcons\vrhoe}{4\dragf}
     -\frac{\vrhog}{\vrhoa^3}\biggr\}+\dragf\vrhof-\xcons\vrhoh\biggr]
\end{split}
\end{align*}
\begin{align}\label{tpath6c}
\begin{split}
&=\frac{\vsiga}{4\dragf^2\pfreq^3}\biggl[2\vrhoc\biggl\{
   \frac{3\xcons^2\vsiga^2}{16\dragf^3\pfreq^4}-\vrhog\biggr\}
   -\frac{\xcons}{\dragf}\biggl\{\vrhoe-\frac{\xcons\vrhog}{\dragf}\biggr\}
   +\biggl\{\frac{\xcons\vrhoe}{4\dragf}-\frac{\vrhog}{\vrhoa^3}\biggr\}\biggr]\\
   &\qquad+\frac{1}{4\dragf\pfreq}\biggl\{\vrhof-\frac{\xcons\vrhoh}{\dragf}\biggr\}
     \beqref{tpath1a}
\end{split}
\nonumber\\
&=\frac{\vsiga}{4\dragf^2\pfreq^3}\biggl[\frac{3\xcons}{4\dragf}\biggl\{
   \frac{\xcons\vsiga^2\vrhoc}{2\dragf^2\pfreq^4}-\vrhoe\biggr\}
   -\vrhog\biggl\{2\vrhoc+\frac{1}{\vrhoa^3}-\frac{\xcons^2}{\dragf^2}\biggr\}\biggr]
   +\frac{1}{4\dragf\pfreq}\biggl\{\vrhof-\frac{\xcons\vrhoh}{\dragf}\biggr\}
  \nonumber\\
&=\frac{1}{4\dragf\pfreq^2}\biggl[\pfreq\biggl\{\vrhof-\frac{\xcons\vrhoh}{\dragf}\biggr\}
   +\frac{\vsiga}{\dragf\pfreq}\biggl\{\frac{3\xcons}{4\dragf}\biggl(
   \frac{\xcons\vsiga^2\vrhoc}{2\dragf^2\pfreq^4}-\vrhoe\biggr)
   -\vrhog\biggl(2\vrhoc+\frac{1}{\vrhoa^3}-\frac{\xcons^2}{\dragf^2}\biggr)\biggr\}\biggr]
  \nonumber\\
&=\frac{1}{4\dragf\pfreq^2}\biggl[\pfreq\biggl\{\vrhof-\frac{\xcons\vrhoh}{\dragf}\biggr\}
   +\frac{3\vsiga}{\dragf\pfreq}\biggl\{\frac{\xcons}{4\dragf}\biggl(
   \frac{\xcons\vsiga^2\vrhoc}{2\dragf^2\pfreq^4}-\vrhoe\biggr)
   -\vrhog\biggl(\frac{1}{\vrhoa^3}-\frac{\xcons^2}{2\dragf^2}\biggr)\biggr\}\biggr]
  \beqref{tpath1a}\nonumber\\
&=\vrhoj\beqref{tpath1b}.
\end{align}
\end{subequations}

\subart{Derivatives of $\taurep$ and $\vecte$}
From \eqnref{main3a}, we derive
\begin{subequations}\label{tpath7}
\begin{align}\label{tpath7a}
\begin{split}
\fdote&=\zvect,\quad\ffdote=\zvect,\quad\fffdote=\zvect\\
\fdott&=\dif{(2\rhorep\alprep\kaprep^{-2})}
  =2\kaprep^{-2}(\rhorep\fdot{\alprep}+\fdot{\rhorep}\alprep)
\end{split}
\end{align}
\begin{align}\label{tpath7b}
\ffdott
&=2\kaprep^{-2}\dif{(\rhorep\fdot{\alprep}+\fdot{\rhorep}\alprep)}
  \beqref{tpath7a}\nonumber\\
&=2\kaprep^{-2}(\fdot{\rhorep}\fdot{\alprep}+\rhorep\ffdot{\alprep}+\ffdot{\rhorep}\alprep
  +\fdot{\rhorep}\fdot{\alprep})\nonumber\\
&=2\kaprep^{-2}(\rhorep\ffdot{\alprep}+2\fdot{\rhorep}\fdot{\alprep}+\ffdot{\rhorep}\alprep)
\end{align}
\begin{align}\label{tpath7c}
\fffdott
&=2\kaprep^{-2}\dif{(\rhorep\ffdot{\alprep}+2\fdot{\rhorep}\fdot{\alprep}+\ffdot{\rhorep}\alprep)}
  \beqref{tpath7b}\nonumber\\
&=2\kaprep^{-2}(\fdot{\rhorep}\ffdot{\alprep}+\rhorep\fffdot{\alprep}+2\ffdot{\rhorep}\fdot{\alprep}
  +2\fdot{\rhorep}\ffdot{\alprep}+\fffdot{\rhorep}\alprep+\ffdot{\rhorep}\fdot{\alprep})\nonumber\\
&=2\kaprep^{-2}(\rhorep\fffdot{\alprep}+3\ffdot{\rhorep}\fdot{\alprep}
  +3\fdot{\rhorep}\ffdot{\alprep}+\fffdot{\rhorep}\alprep)
\end{align}
\end{subequations}
which leads to
\begin{subequations}\label{tpath8}
\begin{align}\label{tpath8a}
\fdott
&=2\kaprep^{-2}(\rhorep\fdot{\alprep}+\fdot{\rhorep}\alprep)
  \beqref{tpath7a}\nonumber\\
&=\frac{2}{\kaprep^2}\biggl[\rhorep\vsiga+\alprep
    \biggl\{\frac{\vsiga\vrhoc}{4\dragf\pfreq^3}\biggr\}\biggr]
  \beqref{tpath2}\text{ \& }\eqnref{tpath6a}\nonumber\\
&=\frac{2\vsiga}{\kaprep^2}\biggl[\rhorep+\frac{\alprep\vrhoc}{4\dragf\pfreq^3}\biggr]
\end{align}
\begin{align}\label{tpath8b}
\ffdott
&=2\kaprep^{-2}(\rhorep\ffdot{\alprep}+2\fdot{\rhorep}\fdot{\alprep}+\ffdot{\rhorep}\alprep)
  \beqref{tpath7b}\nonumber\\
&=\frac{2}{\kaprep^2}\biggl[\rhorep\vsigb+2\vsiga\biggl\{\frac{\vsiga\vrhoc}{4\dragf\pfreq^3}\biggr\}
  +\alprep\vrhoi\biggr]\beqref{tpath2}\text{ \& }\eqnref{tpath6}\nonumber\\
&=\frac{2}{\kaprep^2}\biggl[\rhorep\vsigb+\alprep\vrhoi
  +\frac{\vsiga^2\vrhoc}{2\dragf\pfreq^3}\biggr]
\end{align}
\begin{align}\label{tpath8c}
\fffdott
&=2\kaprep^{-2}(\rhorep\fffdot{\alprep}+3\ffdot{\rhorep}\fdot{\alprep}
  +3\fdot{\rhorep}\ffdot{\alprep}+\fffdot{\rhorep}\alprep)
  \beqref{tpath7c}\nonumber\\
&=\frac{2}{\kaprep^2}\biggl[\rhorep\vsigc+3\vrhoi\vsiga
  +3\vsigb\biggl\{\frac{\vsiga\vrhoc}{4\dragf\pfreq^3}\biggr\}+\alprep\vrhoj\biggr]
  \beqref{tpath2}\text{ \& }\eqnref{tpath6}\nonumber\\
&=\frac{2}{\kaprep^2}\biggl[\rhorep\vsigc+\alprep\vrhoj+3\vsiga\biggl\{\vrhoi
  +\frac{\vsigb\vrhoc}{4\dragf\pfreq^3}\biggr\}\biggr].
\end{align}
\end{subequations}

\subart{Development of equation \eqnref{kpath2a}}
We are now ready to evaluate the quantities defined by \eqnref{kpath2}. We start with
\begin{subequations}\label{tpath9}
\begin{align}\label{tpath9a}
\fdot{\cdkt}
&=\scalc\fdot{\dragf}-\kaprep\fdott\beqref{kpath2a}\nonumber\\
&=\scalc\biggl\{\frac{\xcons\vsiga}{4\dragf\pfreq^2}\biggr\}
  -\kaprep\biggl\{\frac{2\vsiga}{\kaprep^2}\biggr\}\biggl\{\rhorep
   +\frac{\alprep\vrhoc}{4\dragf\pfreq^3}\biggr\}
  \beqref{tpath5a}\text{ \& }\eqnref{tpath8a}\nonumber\\
&=-\frac{2\rhorep\vsiga}{\kaprep}+\frac{\scalc\vsiga}{4\dragf\pfreq^2}\biggl\{\xcons
  -\frac{2\alprep\vrhoc}{\kaprep\scalc\pfreq}\biggr\}
	\nonumber\\
&=\frac{\vsiga}{\kaprep}\biggl[-2\rhorep+\frac{1}{4\dragf\pfreq}\biggl\{\xcons
  -\frac{2\alprep\vrhoc}{\pfreq^2}\biggr\}\biggr]\nonumber\\
&=\vrhok\beqref{tpath1c}
\end{align}
\begin{align}\label{tpath9b}
\ffdot{\cdkt}
&=\scalc\ffdot{\dragf}-\kaprep\ffdott\beqref{kpath2a}\nonumber\\
&=\scalc\vrhog-\kaprep\biggl\{\frac{2}{\kaprep^2}\biggr\}\biggl\{\rhorep\vsigb+\alprep\vrhoi
  +\frac{\vsiga^2\vrhoc}{2\dragf\pfreq^3}\biggr\}
 \beqref{tpath5b}\text{ \& }\eqnref{tpath8b}\nonumber\\
&=\frac{1}{\kaprep}\biggl[\pfreq\vrhog-2\biggl\{\rhorep\vsigb+\alprep\vrhoi
  +\frac{\vsiga^2\vrhoc}{2\dragf\pfreq^3}\biggr\}\biggr]\nonumber\\
&=\vrhol\beqref{tpath1c}
\end{align}
\begin{align}\label{tpath9c}
\fffdot{\cdkt}
&=\scalc\fffdot{\dragf}-\kaprep\fffdott\beqref{kpath2a}\nonumber\\
&=\scalc\vrhoh-\kaprep\biggl\{\frac{2}{\kaprep^2}\biggr\}\biggl\{\rhorep\vsigc
  +\alprep\vrhoj+3\vsiga\biggl\{\vrhoi+\frac{\vsigb\vrhoc}{4\dragf\pfreq^3}\biggr\}\biggr\}
  \beqref{tpath5c}\text{ \& }\eqnref{tpath8c}\nonumber\\
&=\frac{1}{\kaprep}\biggl[\pfreq\vrhoh-2\biggl\{\rhorep\vsigc
  +\alprep\vrhoj+3\vsiga\biggl(\vrhoi+\frac{\vsigb\vrhoc}{4\dragf\pfreq^3}\biggr)\biggr\}\biggr]\nonumber\\
&=\vrhom\beqref{tpath1c}
\end{align}
\end{subequations}
\begin{subequations}\label{tpath10}
\begin{align}\label{tpath10a}
\vbba
&=\fdot{\rhorep}-1\beqref{kpath2a}\nonumber\\
&=\frac{\vsiga\vrhoc}{4\dragf\pfreq^3}-1\beqref{tpath6a}\nonumber\\
&=\vrhod\beqref{tpath1a}
\end{align}
\begin{align}\label{tpath10b}
\vbbb
&=2\fdot{\rhorep}-1=1+2\vbba\beqref{kpath2a}\nonumber\\
&=1+2\vrhod\beqref{tpath10a}
\end{align}
\begin{align}\label{tpath10c}
\vbbc
&=3\fdot{\rhorep}-1=2+3\vbba\beqref{kpath2a}\nonumber\\
&=2+3\vrhod\beqref{tpath10a}
\end{align}
\begin{align}\label{tpath10d}
\vbbd
&=\ffdot{\rhorep}\fdot{\cdkt}-\ffdot{\cdkt}\vbba\beqref{kpath2a}\nonumber\\
&=\vrhoi\vrhok-\vrhol\vrhod\beqref{tpath6b},\eqnref{tpath9}\text{ \& }\eqnref{tpath10a}\nonumber\\
&=\vrhon\beqref{tpath1c}
\end{align}
\begin{align}\label{tpath10e}
\vbbe
&=\fdot{\cdkt}\vbbb-\rhorep\ffdot{\cdkt}\beqref{kpath2a}\nonumber\\
&=\vrhok(1+2\vrhod)-\rhorep\vrhol\beqref{tpath9}\text{ \& }\eqnref{tpath10b}\nonumber\\
&=\vrhoo\beqref{tpath1c}
\end{align}
\begin{align}\label{tpath10f}
\vbbf
&=\vbba\vbbb-\rhorep\ffdot{\rhorep}\beqref{kpath2a}\nonumber\\
&=\vrhod(1+2\vrhod)-\rhorep\vrhoi\beqref{tpath10a},\eqnref{tpath10b}\text{ \& }\eqnref{tpath6b}\nonumber\\
&=\vrhop\beqref{tpath1c}.
\end{align}
\end{subequations}

\subart{Development of equations \eqnref{kpath2b} and \eqnref{kpath2c}}
Substituting the derivatives of $\vecte$ from \eqnref{tpath7a} into \eqnref{kpath2b}, we get
\begin{subequations}\label{tpath11}
\begin{align}\label{tpath11a}
\begin{split}
&\vscra=\cprod{\unitkap}{\vecta},\quad
\vscrb=\cprod{\unitkap}{\fdota},\quad
\vscrc=\cprod{\unitkap}{\ffdota},\quad
\vscrd=\zvect,\quad
\vscre=\zvect\\
&\qquad\quad\vscrf=\cprod{\vecta}{\fdota},\quad
\vscrg=\cprod{\vecta}{\ffdota},\quad
\vscrh=\zvect,\quad
\vscri=\zvect\\
&\quad\vscrj=\cprod{\fdota}{\ffdota},\quad
\vscrk=\zvect,\quad
\vscrl=\zvect,\quad
\vscrn=\zvect,\quad
\vscro=\zvect
\end{split}
\end{align}
from which we obtain, by \eqnref{kpath2c},
\begin{align}\label{tpath11b}
\begin{split}
\vscrp&=\vbbd\vscra+\vbbe\vscrb+\rhorep\fdot{\cdkt}\vscrc+\vbbf\vscrf+\rhorep\vbba\vscrg,\quad
\vscrq=\rhorep^2\vscrj,\quad
\vscrr=\fdot{\cdkt}\unitkap+\vbba\vecta+\rhorep\fdota.
\end{split}
\end{align}
\end{subequations}
Consequently, we have
\begin{subequations}\label{tpath14}
\begin{align}\label{tpath14a}
\vscrr
&=\fdot{\cdkt}\unitkap+\vbba\vecta+\rhorep\fdota\beqref{tpath11b}\nonumber\\
&=\vrhok\unitkap+\vrhod\vecta+\rhorep\fdota\beqref{tpath9a}\text{ \& }\eqnref{tpath10a}
\end{align}
\begin{align}\label{tpath14b}
\vscrp+\vscrq
&=\vbbd\vscra+\vbbe\vscrb+\rhorep\fdot{\cdkt}\vscrc
  +\vbbf\vscrf+\rhorep\vbba\vscrg+\rhorep^2\vscrj\beqref{tpath11b}\nonumber\\
&=\vbbd\vscra+\vbbe\vscrb+\rhorep\vrhok\vscrc
  +\vbbf\vscrf+\rhorep\vbba\vscrg+\rhorep^2\vscrj\beqref{tpath9a}\nonumber\\
\begin{split}
&=\vrhon(\cprod{\unitkap}{\vecta})+\vrhoo(\cprod{\unitkap}{\fdota})+\rhorep\vrhok(\cprod{\unitkap}{\ffdota})
  +\vrhop(\cprod{\vecta}{\fdota})+\rhorep\vrhod(\cprod{\vecta}{\ffdota})\\
  &\quad+\rhorep^2(\cprod{\fdota}{\ffdota})\beqref{tpath10}\text{ \& }\eqnref{tpath11a}
\end{split}
\end{align}
\end{subequations}
which in turn leads to
\begin{subequations}\label{tpath12}
\begin{align}\label{tpath12a}
&\dprod{(\cprod{\unitkap}{\vecta})}{(\vscrp+\vscrq)}
\nonumber\\
&=\dprod{(\cprod{\unitkap}{\vecta})}{[}\vrhon(\cprod{\unitkap}{\vecta})+\vrhoo(\cprod{\unitkap}{\fdota})
  +\rhorep\vrhok(\cprod{\unitkap}{\ffdota})+\vrhop(\cprod{\vecta}{\fdota})+\rhorep\vrhod(\cprod{\vecta}{\ffdota})
  +\rhorep^2(\cprod{\fdota}{\ffdota})]
\nonumber\\
\begin{split}
&=\vrhon[\dprod{(\cprod{\unitkap}{\vecta})}{(\cprod{\unitkap}{\vecta})}]
  +\vrhoo[\dprod{(\cprod{\unitkap}{\vecta})}{(\cprod{\unitkap}{\fdota})}]
  +\rhorep\vrhok[\dprod{(\cprod{\unitkap}{\vecta})}{(\cprod{\unitkap}{\ffdota})}]\\
  &\quad+\vrhop[\dprod{(\cprod{\unitkap}{\vecta})}{(\cprod{\vecta}{\fdota})}]
  +\rhorep\vrhod[\dprod{(\cprod{\unitkap}{\vecta})}{(\cprod{\vecta}{\ffdota})}]
  +\rhorep^2[\dprod{(\cprod{\unitkap}{\vecta})}{(\cprod{\fdota}{\ffdota})}]
\end{split}
\nonumber\\
\begin{split}
&=\vrhon[\scala^2-(\dprod{\unitkap}{\vecta})^2]
  +\vrhoo[(\dprod{\vecta}{\fdota})-(\dprod{\unitkap}{\fdota})(\dprod{\vecta}{\unitkap})]
  +\rhorep\vrhok[(\dprod{\vecta}{\ffdota})-(\dprod{\unitkap}{\ffdota})(\dprod{\vecta}{\unitkap})]\\
  &\quad+\vrhop[(\dprod{\unitkap}{\vecta})(\dprod{\vecta}{\fdota})-(\dprod{\unitkap}{\fdota})\scala^2]
  +\rhorep\vrhod[(\dprod{\unitkap}{\vecta})(\dprod{\vecta}{\ffdota})-(\dprod{\unitkap}{\ffdota})\scala^2]\\
  &\quad+\rhorep^2[(\dprod{\unitkap}{\fdota})(\dprod{\vecta}{\ffdota})-(\dprod{\unitkap}{\ffdota})(\dprod{\vecta}{\fdota})]
  \beqref{alg2}
\end{split}
\nonumber\\
\begin{split}
&=\vrhon(\scala^2-\kaprep^{-2}\alprep^2)
  +\vrhoo(\vsigd-\kaprep^{-2}\alprep\vsiga)
  +\rhorep\vrhok(\vsige-\kaprep^{-2}\alprep\vsigb)
  +\vrhop\kaprep^{-1}(\alprep\vsigd-\vsiga\scala^2)\\
  &\quad+\rhorep\vrhod\kaprep^{-1}(\alprep\vsige-\vsigb\scala^2)
  +\rhorep^2\kaprep^{-1}(\vsiga\vsige-\vsigb\vsigd)
  \beqref{tpath1xa}\text{ \& }\eqnref{main2b}
\end{split}
\nonumber\\
&=\vrhor\beqref{tpath1d}
\end{align}
\begin{align}\label{tpath12b}
&\dprod{(\cprod{\unitkap}{\fdota})}{(\vscrp+\vscrq)}\nonumber\\
&=\dprod{(\cprod{\unitkap}{\fdota})}{[}\vrhon(\cprod{\unitkap}{\vecta})+\vrhoo(\cprod{\unitkap}{\fdota})
  +\rhorep\vrhok(\cprod{\unitkap}{\ffdota})+\vrhop(\cprod{\vecta}{\fdota})+\rhorep\vrhod(\cprod{\vecta}{\ffdota})
  +\rhorep^2(\cprod{\fdota}{\ffdota})]
\nonumber\\
\begin{split}
&=\vrhon[\dprod{(\cprod{\unitkap}{\fdota})}{(\cprod{\unitkap}{\vecta})}]
  +\vrhoo[\dprod{(\cprod{\unitkap}{\fdota})}{(\cprod{\unitkap}{\fdota})}]
  +\rhorep\vrhok[\dprod{(\cprod{\unitkap}{\fdota})}{(\cprod{\unitkap}{\ffdota})}]\\
  &\quad+\vrhop[\dprod{(\cprod{\unitkap}{\fdota})}{(\cprod{\vecta}{\fdota})}]
  +\rhorep\vrhod[\dprod{(\cprod{\unitkap}{\fdota})}{(\cprod{\vecta}{\ffdota})}]
  +\rhorep^2[\dprod{(\cprod{\unitkap}{\fdota})}{(\cprod{\fdota}{\ffdota})}]
\end{split}
\nonumber\\
\begin{split}
&=\vrhon[(\dprod{\fdota}{\vecta})-(\dprod{\unitkap}{\vecta})(\dprod{\fdota}{\unitkap})]
  +\vrhoo[\fdota^2-(\dprod{\unitkap}{\fdota})^2]
  +\rhorep\vrhok[(\dprod{\fdota}{\ffdota})-(\dprod{\unitkap}{\ffdota})(\dprod{\fdota}{\unitkap})]\\
  &\quad+\vrhop[(\dprod{\unitkap}{\vecta})\fdota^2-(\dprod{\unitkap}{\fdota})(\dprod{\fdota}{\vecta})]
  +\rhorep\vrhod[(\dprod{\unitkap}{\vecta})(\dprod{\fdota}{\ffdota})-(\dprod{\unitkap}{\ffdota})(\dprod{\fdota}{\vecta})]\\
  &\quad+\rhorep^2[(\dprod{\unitkap}{\fdota})(\dprod{\fdota}{\ffdota})-(\dprod{\unitkap}{\ffdota})\fdota^2]
  \beqref{alg2}
\end{split}
\nonumber\\
\begin{split}
&=\vrhon(\vsigd-\kaprep^{-2}\alprep\vsiga)
  +\vrhoo(\vsigg-\kaprep^{-2}\vsiga^2)
  +\rhorep\vrhok(\vsigh-\kaprep^{-2}\vsigb\vsiga)
  +\vrhop\kaprep^{-1}(\alprep\vsigg-\vsiga\vsigd)\\
  &\quad+\rhorep\vrhod\kaprep^{-1}(\alprep\vsigh-\vsigb\vsigd)
  +\rhorep^2\kaprep^{-1}(\vsiga\vsigh-\vsigb\vsigg)
  \beqref{tpath1xa}\text{ \& }\eqnref{main2b}
\end{split}
\nonumber\\
&=\vrhos\beqref{tpath1d}
\end{align}
\begin{align}\label{tpath12c}
&\dprod{(\cprod{\unitkap}{\ffdota})}{(\vscrp+\vscrq)}\nonumber\\
&=\dprod{(\cprod{\unitkap}{\ffdota})}{[}\vrhon(\cprod{\unitkap}{\vecta})+\vrhoo(\cprod{\unitkap}{\fdota})
  +\rhorep\vrhok(\cprod{\unitkap}{\ffdota})+\vrhop(\cprod{\vecta}{\fdota})+\rhorep\vrhod(\cprod{\vecta}{\ffdota})
  +\rhorep^2(\cprod{\fdota}{\ffdota})]
\nonumber\\
\begin{split}
&=\vrhon[\dprod{(\cprod{\unitkap}{\ffdota})}{(\cprod{\unitkap}{\vecta})}]
  +\vrhoo[\dprod{(\cprod{\unitkap}{\ffdota})}{(\cprod{\unitkap}{\fdota})}]
  +\rhorep\vrhok[\dprod{(\cprod{\unitkap}{\ffdota})}{(\cprod{\unitkap}{\ffdota})}]\\
  &\quad+\vrhop[\dprod{(\cprod{\unitkap}{\ffdota})}{(\cprod{\vecta}{\fdota})}]
  +\rhorep\vrhod[\dprod{(\cprod{\unitkap}{\ffdota})}{(\cprod{\vecta}{\ffdota})}]
  +\rhorep^2[\dprod{(\cprod{\unitkap}{\ffdota})}{(\cprod{\fdota}{\ffdota})}]
\end{split}
\nonumber\\
\begin{split}
&=\vrhon[(\dprod{\ffdota}{\vecta})-(\dprod{\unitkap}{\vecta})(\dprod{\ffdota}{\unitkap})]
  +\vrhoo[(\dprod{\ffdota}{\fdota})-(\dprod{\unitkap}{\fdota})(\dprod{\ffdota}{\unitkap})]
  +\rhorep\vrhok[\ffdota^2-(\dprod{\unitkap}{\ffdota})^2]\\
  &\quad+\vrhop[(\dprod{\unitkap}{\vecta})(\dprod{\ffdota}{\fdota})-(\dprod{\unitkap}{\fdota})(\dprod{\ffdota}{\vecta})]
  +\rhorep\vrhod[(\dprod{\unitkap}{\vecta})\ffdota^2-(\dprod{\unitkap}{\ffdota})(\dprod{\ffdota}{\vecta})]\\
  &\quad+\rhorep^2[(\dprod{\unitkap}{\fdota})\ffdota^2-(\dprod{\unitkap}{\ffdota})(\dprod{\ffdota}{\fdota})]
  \beqref{alg2}
\end{split}
\nonumber\\
\begin{split}
&=\vrhon(\vsige-\kaprep^{-2}\alprep\vsigb)
  +\vrhoo(\vsigh-\kaprep^{-2}\vsiga\vsigb)
  +\rhorep\vrhok(\vsigj-\kaprep^{-2}\vsigb^2)
  +\vrhop\kaprep^{-1}(\alprep\vsigh-\vsiga\vsige)\\
  &\quad+\rhorep\vrhod\kaprep^{-1}(\alprep\vsigj-\vsigb\vsige)
  +\rhorep^2\kaprep^{-1}(\vsiga\vsigj-\vsigb\vsigh)
  \beqref{tpath1xa}\text{ \& }\eqnref{main2b}
\end{split}
\nonumber\\
&=\vrhot\beqref{tpath1d}
\end{align}
\begin{align}\label{tpath12d}
&\dprod{(\cprod{\vecta}{\fdota})}{(\vscrp+\vscrq)}\nonumber\\
&=\dprod{(\cprod{\vecta}{\fdota})}{[}\vrhon(\cprod{\unitkap}{\vecta})+\vrhoo(\cprod{\unitkap}{\fdota})
  +\rhorep\vrhok(\cprod{\unitkap}{\ffdota})+\vrhop(\cprod{\vecta}{\fdota})+\rhorep\vrhod(\cprod{\vecta}{\ffdota})
  +\rhorep^2(\cprod{\fdota}{\ffdota})]
\nonumber\\
\begin{split}
&=\vrhon[\dprod{(\cprod{\vecta}{\fdota})}{(\cprod{\unitkap}{\vecta})}]
  +\vrhoo[\dprod{(\cprod{\vecta}{\fdota})}{(\cprod{\unitkap}{\fdota})}]
  +\rhorep\vrhok[\dprod{(\cprod{\vecta}{\fdota})}{(\cprod{\unitkap}{\ffdota})}]\\
  &\quad+\vrhop[\dprod{(\cprod{\vecta}{\fdota})}{(\cprod{\vecta}{\fdota})}]
  +\rhorep\vrhod[\dprod{(\cprod{\vecta}{\fdota})}{(\cprod{\vecta}{\ffdota})}]
  +\rhorep^2[\dprod{(\cprod{\vecta}{\fdota})}{(\cprod{\fdota}{\ffdota})}]
\end{split}
\nonumber\\
\begin{split}
&=\vrhon[(\dprod{\vecta}{\unitkap})(\dprod{\fdota}{\vecta})-\scala^2(\dprod{\fdota}{\unitkap})]
  +\vrhoo[(\dprod{\vecta}{\unitkap})\fdota^2-(\dprod{\vecta}{\fdota})(\dprod{\fdota}{\unitkap})]
  +\rhorep\vrhok[(\dprod{\vecta}{\unitkap})(\dprod{\fdota}{\ffdota})-(\dprod{\vecta}{\ffdota})(\dprod{\fdota}{\unitkap})]\\
  &\quad+\vrhop[\scala^2\fdota^2-(\dprod{\vecta}{\fdota})^2]
  +\rhorep\vrhod[\scala^2(\dprod{\fdota}{\ffdota})-(\dprod{\vecta}{\ffdota})(\dprod{\fdota}{\vecta})]
  +\rhorep^2[(\dprod{\vecta}{\fdota})(\dprod{\fdota}{\ffdota})-(\dprod{\vecta}{\ffdota})\fdota^2]
  \beqref{alg2}
\end{split}
\nonumber\\
\begin{split}
&=\vrhon\kaprep^{-1}(\alprep\vsigd-\scala^2\vsiga)
  +\vrhoo\kaprep^{-1}(\alprep\vsigg-\vsigd\vsiga)
  +\rhorep\vrhok\kaprep^{-1}(\alprep\vsigh-\vsige\vsiga)
  +\vrhop(\scala^2\vsigg-\vsigd^2)\\
  &\quad+\rhorep\vrhod(\scala^2\vsigh-\vsige\vsigd)
  +\rhorep^2(\vsigd\vsigh-\vsige\vsigg)
  \beqref{tpath1xa}\text{ \& }\eqnref{main2b}
\end{split}
\nonumber\\
&=\vrhou\beqref{tpath1e}
\end{align}
\begin{align}\label{tpath12e}
&\dprod{(\cprod{\vecta}{\ffdota})}{(\vscrp+\vscrq)}\nonumber\\
&=\dprod{(\cprod{\vecta}{\ffdota})}{[}\vrhon(\cprod{\unitkap}{\vecta})+\vrhoo(\cprod{\unitkap}{\fdota})
  +\rhorep\vrhok(\cprod{\unitkap}{\ffdota})+\vrhop(\cprod{\vecta}{\fdota})+\rhorep\vrhod(\cprod{\vecta}{\ffdota})
  +\rhorep^2(\cprod{\fdota}{\ffdota})]
\nonumber\\
\begin{split}
&=\vrhon[\dprod{(\cprod{\vecta}{\ffdota})}{(\cprod{\unitkap}{\vecta})}]
  +\vrhoo[\dprod{(\cprod{\vecta}{\ffdota})}{(\cprod{\unitkap}{\fdota})}]
  +\rhorep\vrhok[\dprod{(\cprod{\vecta}{\ffdota})}{(\cprod{\unitkap}{\ffdota})}]\\
  &\quad+\vrhop[\dprod{(\cprod{\vecta}{\ffdota})}{(\cprod{\vecta}{\fdota})}]
  +\rhorep\vrhod[\dprod{(\cprod{\vecta}{\ffdota})}{(\cprod{\vecta}{\ffdota})}]
  +\rhorep^2[\dprod{(\cprod{\vecta}{\ffdota})}{(\cprod{\fdota}{\ffdota})}]
\end{split}
\nonumber\\
\begin{split}
&=\vrhon[(\dprod{\vecta}{\unitkap})(\dprod{\ffdota}{\vecta})-\scala^2(\dprod{\ffdota}{\unitkap})]
  +\vrhoo[(\dprod{\vecta}{\unitkap})(\dprod{\ffdota}{\fdota})-(\dprod{\vecta}{\fdota})(\dprod{\ffdota}{\unitkap})]
  +\rhorep\vrhok[(\dprod{\vecta}{\unitkap})\ffdota^2-(\dprod{\vecta}{\ffdota})(\dprod{\ffdota}{\unitkap})]\\
  &\quad+\vrhop[\scala^2(\dprod{\ffdota}{\fdota})-(\dprod{\vecta}{\fdota})(\dprod{\ffdota}{\vecta})]
  +\rhorep\vrhod[\scala^2\ffdota^2-(\dprod{\vecta}{\ffdota})^2]
  +\rhorep^2[(\dprod{\vecta}{\fdota})\ffdota^2-(\dprod{\vecta}{\ffdota})(\dprod{\ffdota}{\fdota})]
  \beqref{alg2}
\end{split}
\nonumber\\
\begin{split}
&=\vrhon\kaprep^{-1}(\alprep\vsige-\scala^2\vsigb)
  +\vrhoo\kaprep^{-1}(\alprep\vsigh-\vsigd\vsigb)
  +\rhorep\vrhok\kaprep^{-1}(\alprep\vsigj-\vsige\vsigb)
  +\vrhop(\scala^2\vsigh-\vsigd\vsige)\\
  &\quad+\rhorep\vrhod(\scala^2\vsigj-\vsige^2)
  +\rhorep^2(\vsigd\vsigj-\vsige\vsigh)
  \beqref{tpath1xa}\text{ \& }\eqnref{main2b}
\end{split}
\nonumber\\
&=\vrhov\beqref{tpath1e}
\end{align}
\begin{align}\label{tpath12f}
&\dprod{(\cprod{\fdota}{\ffdota})}{(\vscrp+\vscrq)}\nonumber\\
&=\dprod{(\cprod{\fdota}{\ffdota})}{[}\vrhon(\cprod{\unitkap}{\vecta})+\vrhoo(\cprod{\unitkap}{\fdota})
  +\rhorep\vrhok(\cprod{\unitkap}{\ffdota})+\vrhop(\cprod{\vecta}{\fdota})+\rhorep\vrhod(\cprod{\vecta}{\ffdota})
  +\rhorep^2(\cprod{\fdota}{\ffdota})]
\nonumber\\
\begin{split}
&=\vrhon[\dprod{(\cprod{\fdota}{\ffdota})}{(\cprod{\unitkap}{\vecta})}]
  +\vrhoo[\dprod{(\cprod{\fdota}{\ffdota})}{(\cprod{\unitkap}{\fdota})}]
  +\rhorep\vrhok[\dprod{(\cprod{\fdota}{\ffdota})}{(\cprod{\unitkap}{\ffdota})}]\\
  &\quad+\vrhop[\dprod{(\cprod{\fdota}{\ffdota})}{(\cprod{\vecta}{\fdota})}]
  +\rhorep\vrhod[\dprod{(\cprod{\fdota}{\ffdota})}{(\cprod{\vecta}{\ffdota})}]
  +\rhorep^2[\dprod{(\cprod{\fdota}{\ffdota})}{(\cprod{\fdota}{\ffdota})}]
\end{split}
\nonumber\\
\begin{split}
&=\vrhon[(\dprod{\fdota}{\unitkap})(\dprod{\ffdota}{\vecta})-(\dprod{\fdota}{\vecta})(\dprod{\ffdota}{\unitkap})]
  +\vrhoo[(\dprod{\fdota}{\unitkap})(\dprod{\ffdota}{\fdota})-\fdota^2(\dprod{\ffdota}{\unitkap})]
  +\rhorep\vrhok[(\dprod{\fdota}{\unitkap})\ffdota^2-(\dprod{\fdota}{\ffdota})(\dprod{\ffdota}{\unitkap})]\\
  &\quad+\vrhop[(\dprod{\fdota}{\vecta})(\dprod{\ffdota}{\fdota})-\fdota^2(\dprod{\ffdota}{\vecta})]
  +\rhorep\vrhod[(\dprod{\fdota}{\vecta})\ffdota^2-(\dprod{\fdota}{\ffdota})(\dprod{\ffdota}{\vecta})]
  +\rhorep^2[\fdota^2\ffdota^2-(\dprod{\fdota}{\ffdota})^2]
  \beqref{alg2}
\end{split}
\nonumber\\
\begin{split}
&=\vrhon\kaprep^{-1}(\vsiga\vsige-\vsigd\vsigb)
  +\vrhoo\kaprep^{-1}(\vsiga\vsigh-\vsigg\vsigb)
  +\rhorep\vrhok\kaprep^{-1}(\vsiga\vsigj-\vsigh\vsigb)
  +\vrhop(\vsigd\vsigh-\vsigg\vsige)\\
  &\quad+\rhorep\vrhod(\vsigd\vsigj-\vsigh\vsige)
  +\rhorep^2(\vsigg\vsigj-\vsigh^2)
  \beqref{tpath1xa}\text{ \& }\eqnref{main2b}
\end{split}
\nonumber\\
&=\vrhow\beqref{tpath1e}.
\end{align}
\end{subequations}
Using \eqnref{tpath14} and \eqnref{tpath12}, we derive
\begin{subequations}\label{tpath13}
\begin{align}\label{tpath13a}
|\vscrr|^2
&=\dprod{(\vrhok\unitkap+\vrhod\vecta+\rhorep\fdota)}
  {(\vrhok\unitkap+\vrhod\vecta+\rhorep\fdota)}\beqref{tpath14a}\nonumber\\
&=\vrhok^2+2\vrhod\vrhok(\dprod{\unitkap}{\vecta})+2\rhorep\vrhok(\dprod{\unitkap}{\fdota})
  +\vrhod^2(\dprod{\vecta}{\vecta})+2\rhorep\vrhod(\dprod{\vecta}{\fdota})
  +\rhorep^2(\dprod{\fdota}{\fdota})\nonumber\\
&=\vrhok^2+2\vrhod\vrhok\kaprep^{-1}(\dprod{\vectkap}{\vecta})
  +2\rhorep\vrhok\kaprep^{-1}(\dprod{\vectkap}{\fdota})
  +2\rhorep\vrhod(\dprod{\vecta}{\fdota})+\vrhod^2\scala^2
  +\rhorep^2\fdota^2\nonumber\\
&=\vrhok^2+2\alprep\vrhod\vrhok\kaprep^{-1}+2\rhorep\vrhok\vsiga\kaprep^{-1}
  +2\rhorep\vrhod\vsigd+\vrhod^2\scala^2+\rhorep^2\vsigg
\nonumber\\
&=\scala^2\vrhod^2+\rhorep(\rhorep\vsigg+2\vrhod\vsigd)+\vrhok[\vrhok+2\kaprep^{-1}(\alprep\vrhod+\rhorep\vsiga)]\nonumber\\
\therefore|\vscrr|
&=\vrhoq\beqref{tpath1c}
\end{align}
\begin{align}\label{tpath13b}
|\vscrp+\vscrq|^2
&=\dprod{(\vscrp+\vscrq)}{(\vscrp+\vscrq)}\nonumber\\
\begin{split}
&=\dprod{(\vscrp+\vscrq)}{[}\vrhon(\cprod{\unitkap}{\vecta})+\vrhoo(\cprod{\unitkap}{\fdota})
  +\rhorep\vrhok(\cprod{\unitkap}{\ffdota})
  +\vrhop(\cprod{\vecta}{\fdota})+\rhorep\vrhod(\cprod{\vecta}{\ffdota})\\
  &\quad+\rhorep^2(\cprod{\fdota}{\ffdota})]\beqref{tpath14b}
\end{split}
\nonumber\\
\begin{split}
&=\vrhon[\dprod{(\vscrp+\vscrq)}{(\cprod{\unitkap}{\vecta})}]
  +\vrhoo[\dprod{(\vscrp+\vscrq)}{(\cprod{\unitkap}{\fdota})}]
  +\rhorep\vrhok[\dprod{(\vscrp+\vscrq)}{(\cprod{\unitkap}{\ffdota})}]\\
  &\quad+\vrhop[\dprod{(\vscrp+\vscrq)}{(\cprod{\vecta}{\fdota})}]
  +\rhorep\vrhod[\dprod{(\vscrp+\vscrq)}{(\cprod{\vecta}{\ffdota})}]
  +\rhorep^2[\dprod{(\vscrp+\vscrq)}{(\cprod{\fdota}{\ffdota})}]
\end{split}
\nonumber\\
&=\vrhon\vrhor+\vrhoo\vrhos+\rhorep\vrhok\vrhot+\vrhop\vrhou+\rhorep\vrhod\vrhov+\rhorep^2\vrhow
\nonumber\\
\therefore|\vscrp+\vscrq|
&=\vrhox\beqref{tpath1f}.
\end{align}
\end{subequations}

\subart{Development of equation \eqnref{kpath2d}}
For the quantities defined in \eqnref{kpath2d}, we derive
\begin{subequations}\label{tpath15}
\begin{align}\label{tpath15a}
\alepha
&=\dprod{\unitkap}{(\vscrp+\vscrq)}\beqref{kpath2d}\nonumber\\
\begin{split}
&=\dprod{\unitkap}{}[\vrhon(\cprod{\unitkap}{\vecta})+\vrhoo(\cprod{\unitkap}{\fdota})
  +\rhorep\vrhok(\cprod{\unitkap}{\ffdota})
  +\vrhop(\cprod{\vecta}{\fdota})+\rhorep\vrhod(\cprod{\vecta}{\ffdota})\\
  &\quad+\rhorep^2(\cprod{\fdota}{\ffdota})]\beqref{tpath14b}
\end{split}
\nonumber\\
&=\dprod{\unitkap}{[\vrhop(\cprod{\vecta}{\fdota})+\rhorep\vrhod(\cprod{\vecta}{\ffdota})
  +\rhorep^2(\cprod{\fdota}{\ffdota})]}\nonumber\\
&=\vrhop\vsigm+\rhorep\vrhod\vsign+\rhorep^2\vsigo\beqref{tpath1xa}
\end{align}
\begin{align}\label{tpath15b}
\alephb
&=\dprod{\vecta}{(\vscrp+\vscrq)}\beqref{kpath2d}\nonumber\\
\begin{split}
&=\dprod{\vecta}{}[\vrhon(\cprod{\unitkap}{\vecta})+\vrhoo(\cprod{\unitkap}{\fdota})
  +\rhorep\vrhok(\cprod{\unitkap}{\ffdota})
  +\vrhop(\cprod{\vecta}{\fdota})+\rhorep\vrhod(\cprod{\vecta}{\ffdota})\\
  &\quad+\rhorep^2(\cprod{\fdota}{\ffdota})]\beqref{tpath14b}
\end{split}
\nonumber\\
&=-\vrhoo[\dprod{\unitkap}(\cprod{\vecta}{\fdota})]-\rhorep\vrhok[\dprod{\unitkap}{(\cprod{\vecta}{\ffdota})}]
  +\rhorep^2[\dprod{\vecta}{(\cprod{\fdota}{\ffdota})}]\beqref{alg4}\nonumber\\
&=-\vbbe\vsigm-\rhorep\vrhok\vsign+\rhorep^2\vsigp\beqref{tpath1xa}
\end{align}
\begin{align}\label{tpath15c}
\alephc
&=\dprod{\fdota}{(\vscrp+\vscrq)}\beqref{kpath2d}\nonumber\\
\begin{split}
&=\dprod{\fdota}{}[\vrhon(\cprod{\unitkap}{\vecta})+\vrhoo(\cprod{\unitkap}{\fdota})
  +\rhorep\vrhok(\cprod{\unitkap}{\ffdota})
  +\vrhop(\cprod{\vecta}{\fdota})+\rhorep\vrhod(\cprod{\vecta}{\ffdota})\\
  &\quad+\rhorep^2(\cprod{\fdota}{\ffdota})]\beqref{tpath14b}
\end{split}
\nonumber\\
&=\vrhon[\dprod{\unitkap}{(\cprod{\vecta}{\fdota})}]-\rhorep\vrhok[\dprod{\unitkap}{(\cprod{\fdota}{\ffdota})}]
  -\rhorep\vrhod[\dprod{\vecta}{(\cprod{\fdota}{\ffdota})}]\beqref{alg4}\nonumber\\
&=\vrhon\vsigm-\rhorep\vrhok\vsigo-\rhorep\vrhod\vsigp\beqref{tpath1xa}
\end{align}
\begin{align}\label{tpath15d}
\alephd
&=\dprod{\ffdota}{(\vscrp+\vscrq)}\beqref{kpath2d}\nonumber\\
\begin{split}
&=\dprod{\ffdota}{}[\vrhon(\cprod{\unitkap}{\vecta})+\vrhoo(\cprod{\unitkap}{\fdota})
  +\rhorep\vrhok(\cprod{\unitkap}{\ffdota})
  +\vrhop(\cprod{\vecta}{\fdota})+\rhorep\vrhod(\cprod{\vecta}{\ffdota})\\
  &\quad+\rhorep^2(\cprod{\fdota}{\ffdota})]\beqref{tpath14b}
\end{split}
\nonumber\\
&=\vrhon[\dprod{\unitkap}{(\cprod{\vecta}{\ffdota})}]+\vrhoo[\dprod{\unitkap}{(\cprod{\fdota}{\ffdota})}]
  +\vrhop[\dprod{\vecta}{(\cprod{\fdota}{\ffdota})}]\beqref{alg4}\nonumber\\
&=\vrhon\vsign+\vrhoo\vsigo+\vrhop\vsigp\beqref{tpath1xa}
\end{align}
\begin{align}
\alephe
&=\dprod{\fffdota}{(\vscrp+\vscrq)}\beqref{kpath2d}\nonumber\\
\begin{split}
&=\dprod{\fffdota}{}[\vrhon(\cprod{\unitkap}{\vecta})+\vrhoo(\cprod{\unitkap}{\fdota})
  +\rhorep\vrhok(\cprod{\unitkap}{\ffdota})
  +\vrhop(\cprod{\vecta}{\fdota})+\rhorep\vrhod(\cprod{\vecta}{\ffdota})\\
  &\quad+\rhorep^2(\cprod{\fdota}{\ffdota})]\beqref{tpath14b}
\end{split}
\nonumber\\
&=\vrhon\vsigq+\vrhoo\vsigr+\vrhop\vsigt+\rhorep(\vrhok\vsigs
  +\vrhod\vsigu+\rhorep\vsigv)\beqref{tpath1xa}
\label{tpath15e}\\
\alephf
&=\dprod{\fffdote}{(\vscrp+\vscrq)}\beqref{kpath2d}\nonumber\\
&=0\beqref{tpath7a}
\label{tpath15f}
\end{align}
\end{subequations}
\begin{align}\label{tpath16}
\begin{split}
&\fffdot{\cdkt}\alepha+\fffdot{\rhorep}\alephb+3\ffdot{\rhorep}\alephc+\vbbc\alephd+\rhorep\alephe+\alephf\\
&=\vrhom\alepha+\vrhoj\alephb+3\vrhoi\alephc+(2+3\vrhod)\alephd+\rhorep\alephe+\alephf
  \beqref{tpath9c},\eqnref{tpath6}\text{ \& }\eqnref{tpath10c}
\end{split}
\nonumber\\
\begin{split}
&=\vrhom(\vrhop\vsigm+\rhorep\vrhod\vsign+\rhorep^2\vsigo)
  +\vrhoj(-\vrhoo\vsigm-\rhorep\vrhok\vsign+\rhorep^2\vsigp)
  +3\vrhoi(\vrhon\vsigm-\rhorep\vrhok\vsigo-\rhorep\vrhod\vsigp)\\
  &\quad+(2+3\vrhod)(\vrhon\vsign+\vrhoo\vsigo+\vrhop\vsigp)
  +\rhorep(\vrhon\vsigq+\vrhoo\vsigr+\vrhop\vsigt)\\
  &\quad+\rhorep^2(\vrhok\vsigs+\vrhod\vsigu+\rhorep\vsigv)\beqref{tpath15}
\end{split}
\nonumber\\
&=\vrhoy\beqref{tpath1f}.
\end{align}

\subart{Results of the computations}
Substituting \eqnref{tpath14}, \eqnref{tpath13} and \eqnref{tpath16} into \eqnref{kpath11}, we finally get
\begin{align}\label{tpath17}
\begin{split}
&\qquad\qquad\qquad\qquad\bbk=\plusmin\frac{\vrhox}{(\vrhoq)^3},\quad
\bbt=\frac{\vrhoy}{(\vrhox)^2},\quad
\frt=\frac{1}{\vrhoq}\biggl[\vrhok\unitkap+\vrhod\vecta+\rhorep\fdota\biggr]\\
&\frb=\frac{1}{\vrhox}\biggl[\vrhon(\cprod{\unitkap}{\vecta})+\vrhoo(\cprod{\unitkap}{\fdota})+\vrhop(\cprod{\vecta}{\fdota})
  +\rhorep\vrhok(\cprod{\unitkap}{\ffdota})+\rhorep\vrhod(\cprod{\vecta}{\ffdota})
  +\rhorep^2(\cprod{\fdota}{\ffdota})\biggr]
\end{split}
\end{align}
as the set of equations that, together with \eqnref{tpath1}, completely describes the apparent
path of the light source.

\art{Apparent geometry of obliquated rays}
To evaluate \eqnref{kray4} for a translating observer, we introduce, in addition to \eqnref{tpath1},
\begin{subequations}\label{tray1}
\begin{align}\label{tray1a}
\begin{split}
&\vpsa=\dprod{\unitkap}{(\cprod{\vecta}{\vectu})},\quad
\vpsb=\dprod{\unitkap}{(\cprod{\fdota}{\vectu})},\quad
\vpsc=\dprod{\vecta}{(\cprod{\fdota}{\vectu})},\quad
\vpsd=\dprod{\ffdota}{(\cprod{\unitkap}{\vectu})}\\
&\qquad\vpse=\dprod{\ffdota}{(\cprod{\vecta}{\vectu})},\quad
\vpsf=\dprod{\ffdota}{(\cprod{\fdota}{\vectu})},\quad
\vpsg=\dprod{\unitkap}{\fdota},\quad
\vpsh=\dprod{\fdota}{\vectu}
\end{split}
\end{align}
\begin{align}\label{tray1b}
\begin{split}
\veka&=\cdkt\vrhod-\rhorep\vrhok,\quad
\vekb=\vrhod\vpsa+\rhorep(\vpsb+\rhorep\vsigm),\quad
\vekc=\rhorep(\vpsc-\cdkt\vsigm)-\vrhok\vpsa\\
\vekd&=\veka\vsigm-\vrhok\vpsb-\vrhod\vpsc,\quad
\veke=\veka\vsign+\vrhok\vpsd+\vrhod\vpse+\rhorep(\vpsf+\cdkt\vsigo+\rhorep\vsigp)
\end{split}
\end{align}
\begin{align}\label{tray1c}
\begin{split}
\vekf&=\scala^2\veka^2\sin^2\lamrep
   +2\rhorep\cdkt\veka(\vsigd-\scala\vpsg\cos\lamrep)
   +2\scala\rhorep^2\veka(\vsigd\cos\lamrep-\scala\vpsg)\\
   &\quad+\rhorep^2\cdkt^2(\vsigg-\vpsg^2)
   +2\rhorep^3\cdkt(\scala\vsigg\cos\lamrep-\vpsg\vsigd)
   +\rhorep^4(\scala^2\vsigg-\vsigd^2)\\
\vekg&=\rhorep^2(\scalu^2\vsigg-\vpsh^2)
   +\scalu^2[\vrhok^2\sin^2\phirep+\scala^2\vrhod^2\sin^2\thtrep
   +2\scala\vrhod\vrhok(\cos\lamrep-\cos\phirep\cos\thtrep)]\\
   &\quad+2\scalu\rhorep[\vrhok(\scalu\vpsg-\vpsh\cos\phirep)
   +\vrhod(\scalu\vsigd-\scala\vpsh\cos\thtrep)]
\end{split}
\end{align}
\begin{align}\label{tray1d}
\begin{split}
\vekh&=\scalu\veka[\rhorep(\scala\vpsg\cos\thtrep-\vsigd\cos\phirep)
     +\scala\vrhok(\cos\thtrep-\cos\lamrep\cos\phirep)
     +\scala^2\vrhod(\cos\lamrep\cos\thtrep-\cos\phirep)]\\
   &\quad+\rhorep\cdkt[\vrhok(\vpsh-\scalu\vpsg\cos\phirep)
     +\vrhod(\scala\vpsh\cos\lamrep-\scalu\vsigd\cos\phirep)]\\
   &\quad+\scala\rhorep^2[\vrhok(\vpsh\cos\lamrep-\scalu\vpsg\cos\thtrep)
     +\vrhod(\scala\vpsh-\scalu\vsigd\cos\thtrep)]\\
   &\quad+\rhorep^2[\cdkt(\vpsg\vpsh-\scalu\vsigg\cos\phirep)
     +\rhorep(\vsigd\vpsh-\scalu\scala\vsigg\cos\thtrep)].
\end{split}
\end{align}
\end{subequations}

\subart{Development of equation \eqnref{kray2}}
Using \eqnref{main3a} and \eqnref{tpath7a} in \eqnref{kray2a}, we have
\begin{subequations}\label{tray2}
\begin{align}\label{tray2a}
\begin{split}
\veusa&=\cprod{\unitkap}{\vectu},\quad
\veusb=\cprod{\unitkap}{\vecta},\quad
\veusc=\zvect,\quad
\veusd=\cprod{\unitkap}{\fdota},\quad
\veuse=\zvect\\
&\quad\veusf=\cprod{\vecta}{\vectu},\quad
\veusg=\zvect,\quad
\veush=\cprod{\vecta}{\fdota},\quad
\veusi=\zvect\\
&\quad\veusj=\cprod{\vectu}{\fdota},\quad
\veusk=\zvect,\quad
\veusl=\zvect,\quad
\veusm=\zvect
\end{split}
\end{align}
so that by \eqnref{kray2b}, \eqnref{tray2a}, \eqnref{tpath9a}, \eqnref{tpath10a}
and \eqnref{tray1b},
\begin{align}\label{tray2b}
\begin{split}
&\veusn=\rhorep(\cprod{\unitkap}{\fdota}),\quad
\veuso=\rhorep(\cprod{\vecta}{\fdota}),\quad
\veusp=\cprod{\unitkap}{\vectu},\quad
\veusq=\cprod{\vecta}{\vectu},\quad
\veusr=\cprod{\fdota}{\vectu},\quad
\veuss=\zvect\\
&\veust=\veka(\cprod{\unitkap}{\vecta})
   +\cdkt\rhorep(\cprod{\unitkap}{\fdota})+\rhorep^2(\cprod{\vecta}{\fdota}),\quad
\veusu=\vrhok(\cprod{\unitkap}{\vectu})
  +\vrhod(\cprod{\vecta}{\vectu})+\rhorep(\cprod{\fdota}{\vectu})
\end{split}
\end{align}
from which we obtain
\begin{align}\label{tray2c}
\veust+\veusu
=\vrhok(\cprod{\unitkap}{\vectu})+\veka(\cprod{\unitkap}{\vecta})+\cdkt\rhorep(\cprod{\unitkap}{\fdota})
   +\rhorep^2(\cprod{\vecta}{\fdota})+\vrhod(\cprod{\vecta}{\vectu})+\rhorep(\cprod{\fdota}{\vectu}).
\end{align}
\end{subequations}
We then derive
\begin{subequations}\label{tray3}
\begin{align}\label{tray3a}
\imaa
&=\dprod{\unitkap}{(\veust+\veusu)}\beqref{kray2c}\nonumber\\
&=\dprod{\unitkap}{[\vrhok(\cprod{\unitkap}{\vectu})+\veka(\cprod{\unitkap}{\vecta})
   +\cdkt\rhorep(\cprod{\unitkap}{\fdota})+\rhorep^2(\cprod{\vecta}{\fdota})
   +\vrhod(\cprod{\vecta}{\vectu})+\rhorep(\cprod{\fdota}{\vectu})]}
   \beqref{tray2c}\nonumber\\
&=\rhorep^2[\dprod{\unitkap}{(\cprod{\vecta}{\fdota})}]
   +\vrhod[\dprod{\unitkap}{(\cprod{\vecta}{\vectu})}]
   +\rhorep[\dprod{\unitkap}{(\cprod{\fdota}{\vectu})}]
   \nonumber\\
&=\rhorep^2\vsigm+\vrhod\vpsa+\rhorep\vpsb
   \beqref{tpath1xa}\text{ \& }\eqnref{tray1a}\nonumber\\
&=\vekb\beqref{tray1b}
\end{align}
\begin{align}\label{tray3b}
\imab
&=\dprod{\vecta}{(\veust+\veusu)}\beqref{kray2c}\nonumber\\
&=\dprod{\vecta}{[\vrhok(\cprod{\unitkap}{\vectu})+\veka(\cprod{\unitkap}{\vecta})
   +\cdkt\rhorep(\cprod{\unitkap}{\fdota})+\rhorep^2(\cprod{\vecta}{\fdota})
   +\vrhod(\cprod{\vecta}{\vectu})+\rhorep(\cprod{\fdota}{\vectu})]}
   \beqref{tray2c}\nonumber\\
&=-\vrhok[\dprod{\unitkap}{(\cprod{\vecta}{\vectu})}]
   -\cdkt\rhorep[\dprod{\unitkap}{(\cprod{\vecta}{\fdota})}]
   +\rhorep[\dprod{\vecta}{(\cprod{\fdota}{\vectu})}]
   \beqref{alg4}\nonumber\\
&=-\vrhok\vpsa-\rhorep\cdkt\vsigm+\rhorep\vpsc
   \beqref{tpath1xa}\text{ \& }\eqnref{tray1a}\nonumber\\
&=\vekc\beqref{tray1b}
\end{align}
\begin{align}\label{tray3c}
\imac
&=\dprod{\fdota}{(\veust+\veusu)}\beqref{kray2c}\nonumber\\
&=\dprod{\fdota}{[\vrhok(\cprod{\unitkap}{\vectu})+\veka(\cprod{\unitkap}{\vecta})
   +\cdkt\rhorep(\cprod{\unitkap}{\fdota})+\rhorep^2(\cprod{\vecta}{\fdota})
   +\vrhod(\cprod{\vecta}{\vectu})+\rhorep(\cprod{\fdota}{\vectu})]}
   \beqref{tray2c}\nonumber\\
&=-\vrhok[\dprod{\unitkap}{(\cprod{\fdota}{\vectu})}]
   +\veka[\dprod{\unitkap}{(\cprod{\vecta}{\fdota})}]
   -\vrhod[\dprod{\vecta}{(\cprod{\fdota}{\vectu})}]
   \beqref{alg4}\nonumber\\
&=-\vrhok\vpsb+\veka\vsigm-\vrhod\vpsc
   \beqref{tpath1xa}\text{ \& }\eqnref{tray1a}\nonumber\\
&=\vekd\beqref{tray1b}
\end{align}
\begin{align}\label{tray3d}
\imad
&=\dprod{\ffdota}{(\veust+\veusu)}\beqref{kray2c}\nonumber\\
&=\dprod{\ffdota}{[\vrhok(\cprod{\unitkap}{\vectu})+\veka(\cprod{\unitkap}{\vecta})
   +\cdkt\rhorep(\cprod{\unitkap}{\fdota})+\rhorep^2(\cprod{\vecta}{\fdota})
   +\vrhod(\cprod{\vecta}{\vectu})+\rhorep(\cprod{\fdota}{\vectu})]}
   \beqref{tray2c}\nonumber\\
\begin{split}
&=\vrhok[\dprod{\ffdota}{(\cprod{\unitkap}{\vectu})}]
   +\veka[\dprod{\unitkap}{(\cprod{\vecta}{\ffdota})}]
   +\cdkt\rhorep[\dprod{\unitkap}{(\cprod{\fdota}{\ffdota})}]
   +\rhorep^2[\dprod{\vecta}{(\cprod{\fdota}{\ffdota})}]\\
   &\qquad+\vrhod[\dprod{\ffdota}{(\cprod{\vecta}{\vectu})}]
   +\rhorep[\dprod{\ffdota}{(\cprod{\fdota}{\vectu})}]
   \beqref{alg4}
\end{split}
\nonumber\\
&=\vrhok\vpsd+\veka\vsign+\rhorep\cdkt\vsigo+\rhorep^2\vsigp+\vrhod\vpse+\rhorep\vpsf
   \beqref{tpath1xa}\text{ \& }\eqnref{tray1a}\nonumber\\
&=\veke\beqref{tray1b}
\end{align}
\begin{align}\label{tray3e}
\imae
&=\dprod{\ffdote}{(\veust+\veusu)}\beqref{kray2c}\nonumber\\
&=0\beqref{tpath7a}.
\end{align}
\end{subequations}

\subart{Magnitude of the vector $\veust+\veusu$}
We also have
\begin{subequations}\label{tray4}
\begin{align}\label{tray4a}
|\veust|^2
&=\dprod{[\veka(\cprod{\unitkap}{\vecta})+\cdkt\rhorep(\cprod{\unitkap}{\fdota})
   +\rhorep^2(\cprod{\vecta}{\fdota})]}{[\veka(\cprod{\unitkap}{\vecta})
   +\cdkt\rhorep(\cprod{\unitkap}{\fdota})+\rhorep^2(\cprod{\vecta}{\fdota})]}
   \beqref{tray2b}\nonumber\\
\begin{split}
&=\veka^2[\dprod{(\cprod{\unitkap}{\vecta})}{(\cprod{\unitkap}{\vecta})}]
   +2\rhorep\cdkt\veka[\dprod{(\cprod{\unitkap}{\vecta})}{(\cprod{\unitkap}{\fdota})}]
   +2\rhorep^2\veka[\dprod{(\cprod{\unitkap}{\vecta})}{(\cprod{\vecta}{\fdota})}]\\
   &\quad+\rhorep^2\cdkt^2[\dprod{(\cprod{\unitkap}{\fdota})}{(\cprod{\unitkap}{\fdota})}]
   +2\rhorep^3\cdkt[\dprod{(\cprod{\unitkap}{\fdota})}{(\cprod{\vecta}{\fdota})}]
   +\rhorep^4[\dprod{(\cprod{\vecta}{\fdota})}{(\cprod{\vecta}{\fdota})}]
\end{split}
\nonumber\\
\begin{split}
&=\veka^2[\unitkap^2\vecta^2-(\dprod{\unitkap}{\vecta})^2]
   +2\rhorep\cdkt\veka[\unitkap^2(\dprod{\vecta}{\fdota})-(\dprod{\unitkap}{\fdota})(\dprod{\unitkap}{\vecta})]
   +2\rhorep^2\veka[(\dprod{\unitkap}{\vecta})(\dprod{\vecta}{\fdota})-(\dprod{\unitkap}{\fdota})\vecta^2]\\
   &\quad+\rhorep^2\cdkt^2[\unitkap^2\fdota^2-(\dprod{\unitkap}{\fdota})^2]
   +2\rhorep^3\cdkt[(\dprod{\unitkap}{\vecta})\fdota^2-(\dprod{\unitkap}{\fdota})(\dprod{\vecta}{\fdota})]
   +\rhorep^4[\vecta^2\fdota^2-(\dprod{\vecta}{\fdota})^2]\beqref{alg2}
\end{split}
\nonumber\\
\begin{split}
&=\scala^2\veka^2\sin^2\lamrep
   +2\rhorep\cdkt\veka(\vsigd-\scala\vpsg\cos\lamrep)
   +2\rhorep^2\veka(\scala\vsigd\cos\lamrep-\vpsg\scala^2)
   +\rhorep^2\cdkt^2(\vsigg-\vpsg^2)\\
   &\quad+2\rhorep^3\cdkt(\scala\vsigg\cos\lamrep-\vpsg\vsigd)
   +\rhorep^4(\scala^2\vsigg-\vsigd^2)
   \beqref{tpath1xa}\text{ \& }\eqnref{tray1a}
\end{split}
\nonumber\\
&=\vekf\beqref{tray1c}
\end{align}
\begin{align}\label{tray4b}
|\veusu|^2
&=\dprod{[\vrhok(\cprod{\unitkap}{\vectu})+\vrhod(\cprod{\vecta}{\vectu})+\rhorep(\cprod{\fdota}{\vectu})]}
  {[\vrhok(\cprod{\unitkap}{\vectu})+\vrhod(\cprod{\vecta}{\vectu})+\rhorep(\cprod{\fdota}{\vectu})]}
  \beqref{tray2b}\nonumber\\
\begin{split}
&=\vrhok^2[\dprod{(\cprod{\unitkap}{\vectu})}{(\cprod{\unitkap}{\vectu})}]
   +2\vrhod\vrhok[\dprod{(\cprod{\unitkap}{\vectu})}{(\cprod{\vecta}{\vectu})}]
   +2\rhorep\vrhok[\dprod{(\cprod{\unitkap}{\vectu})}{(\cprod{\fdota}{\vectu})}]\\
   &\quad+\vrhod^2[\dprod{(\cprod{\vecta}{\vectu})}{(\cprod{\vecta}{\vectu})}]
   +2\rhorep\vrhod[\dprod{(\cprod{\vecta}{\vectu})}{(\cprod{\fdota}{\vectu})}]
   +\rhorep^2[\dprod{(\cprod{\fdota}{\vectu})}{(\cprod{\fdota}{\vectu})}]
\end{split}
\nonumber\\
\begin{split}
&=\vrhok^2[\unitkap^2\vectu^2-(\dprod{\unitkap}{\vectu})^2]
   +2\vrhod\vrhok[(\dprod{\unitkap}{\vecta})\vectu^2-(\dprod{\unitkap}{\vectu})(\dprod{\vecta}{\vectu})]
   +2\rhorep\vrhok[(\dprod{\unitkap}{\fdota})\vectu^2-(\dprod{\unitkap}{\vectu})(\dprod{\fdota}{\vectu})]\\
   &\quad+\vrhod^2[\vecta^2\vectu^2-(\dprod{\vecta}{\vectu})^2]
   +2\rhorep\vrhod[(\dprod{\vecta}{\fdota})\vectu^2-(\dprod{\vecta}{\vectu})(\dprod{\fdota}{\vectu})]
   +\rhorep^2[\fdota^2\vectu^2-(\dprod{\fdota}{\vectu})^2]\beqref{alg2}
\end{split}
\nonumber\\
\begin{split}
&=\scalu^2\vrhok^2\sin^2\phirep
   +2\vrhod\vrhok(\scalu^2\scala\cos\lamrep-\scalu^2\scala\cos\phirep\cos\thtrep)
   +2\rhorep\vrhok(\scalu^2\vpsg-\scalu\vpsh\cos\phirep)\\
   &\quad+\scala^2\scalu^2\vrhod^2\sin^2\thtrep
   +2\rhorep\vrhod(\vsigd\scalu^2-\scala\scalu\vpsh\cos\thtrep)
   +\rhorep^2(\scalu^2\vsigg-\vpsh^2)
   \beqref{tpath1xa}\text{ \& }\eqnref{tray1a}
\end{split}
\nonumber\\
\begin{split}
&=\rhorep^2(\scalu^2\vsigg-\vpsh^2)
   +\scalu^2[\vrhok^2\sin^2\phirep+\scala^2\vrhod^2\sin^2\thtrep
   +2\scala\vrhod\vrhok(\cos\lamrep-\cos\phirep\cos\thtrep)]\\
   &\quad+2\scalu\rhorep[\vrhok(\scalu\vpsg-\vpsh\cos\phirep)
   +\vrhod(\scalu\vsigd-\scala\vpsh\cos\thtrep)]
\end{split}
\nonumber\\
&=\vekg\beqref{tray1c}
\end{align}
\begin{align*}
\dprod{\veust}{\veusu}
&=\dprod{[\veka(\cprod{\unitkap}{\vecta})+\cdkt\rhorep(\cprod{\unitkap}{\fdota})+\rhorep^2(\cprod{\vecta}{\fdota})]}
   {[\vrhok(\cprod{\unitkap}{\vectu})+\vrhod(\cprod{\vecta}{\vectu})+\rhorep(\cprod{\fdota}{\vectu})]}
   \beqref{tray2b}\nonumber\\
\begin{split}
&=\veka\vrhok[\dprod{(\cprod{\unitkap}{\vectu})}{(\cprod{\unitkap}{\vecta})}]
   +\rhorep\cdkt\vrhok[\dprod{(\cprod{\unitkap}{\vectu})}{(\cprod{\unitkap}{\fdota})}]
   +\rhorep^2\vrhok[\dprod{(\cprod{\unitkap}{\vectu})}{(\cprod{\vecta}{\fdota})}]\\
   &\quad+\veka\vrhod[\dprod{(\cprod{\vecta}{\vectu})}{(\cprod{\unitkap}{\vecta})}]
   +\rhorep\cdkt\vrhod[\dprod{(\cprod{\vecta}{\vectu})}{(\cprod{\unitkap}{\fdota})}]
   +\rhorep^2\vrhod[\dprod{(\cprod{\vecta}{\vectu})}{(\cprod{\vecta}{\fdota})}]\\
   &\quad+\veka\rhorep[\dprod{(\cprod{\fdota}{\vectu})}{(\cprod{\unitkap}{\vecta})}]
   +\rhorep^2\cdkt[\dprod{(\cprod{\fdota}{\vectu})}{(\cprod{\unitkap}{\fdota})}]
   +\rhorep^3[\dprod{(\cprod{\fdota}{\vectu})}{(\cprod{\vecta}{\fdota})}]
\end{split}
\nonumber\\
\begin{split}
&=\veka\vrhok[\unitkap^2(\dprod{\vectu}{\vecta})-(\dprod{\unitkap}{\vecta})(\dprod{\unitkap}{\vectu})]
   +\rhorep\cdkt\vrhok[\unitkap^2(\dprod{\vectu}{\fdota})-(\dprod{\unitkap}{\fdota})(\dprod{\unitkap}{\vectu})]\\
   &\quad+\rhorep^2\vrhok[(\dprod{\unitkap}{\vecta})(\dprod{\vectu}{\fdota})-(\dprod{\unitkap}{\fdota})
     (\dprod{\vecta}{\vectu})]
   +\veka\vrhod[(\dprod{\vecta}{\unitkap})(\dprod{\vectu}{\vecta})-\vecta^2(\dprod{\unitkap}{\vectu})]\\
   &\quad+\rhorep\cdkt\vrhod[(\dprod{\vecta}{\unitkap})(\dprod{\vectu}{\fdota})-(\dprod{\vecta}{\fdota})
     (\dprod{\vectu}{\unitkap})]
   +\rhorep^2\vrhod[\vecta^2(\dprod{\vectu}{\fdota})-(\dprod{\vecta}{\fdota})(\dprod{\vecta}{\vectu})]\\
   &\quad+\veka\rhorep[(\dprod{\fdota}{\unitkap})(\dprod{\vectu}{\vecta})-(\dprod{\fdota}{\vecta})
     (\dprod{\unitkap}{\vectu})]
   +\rhorep^2\cdkt[(\dprod{\fdota}{\unitkap})(\dprod{\vectu}{\fdota})-\fdota^2(\dprod{\unitkap}{\vectu})]\\
   &\quad+\rhorep^3[(\dprod{\fdota}{\vecta})(\dprod{\vectu}{\fdota})-\fdota^2(\dprod{\vecta}{\vectu})]
   \beqref{alg2}
\end{split}
\end{align*}
\begin{align}\label{tray4c}
\begin{split}
&=\scalu\scala\veka\vrhok(\cos\thtrep-\cos\lamrep\cos\phirep)
   +\rhorep\cdkt\vrhok(\vpsh-\scalu\vpsg\cos\phirep)
   +\scala\rhorep^2\vrhok(\vpsh\cos\lamrep-\scalu\vpsg\cos\thtrep)\\
   &\quad+\scalu\scala^2\veka\vrhod(\cos\lamrep\cos\thtrep-\cos\phirep)
   +\rhorep\cdkt\vrhod(\scala\vpsh\cos\lamrep-\scalu\vsigd\cos\phirep)
   +\scala\rhorep^2\vrhod(\scala\vpsh-\scalu\vsigd\cos\thtrep)\\
   &\quad+\scalu\rhorep\veka(\scala\vpsg\cos\thtrep-\vsigd\cos\phirep)
   +\rhorep^2\cdkt(\vpsg\vpsh-\scalu\vsigg\cos\phirep)\\
   &\quad+\rhorep^3(\vsigd\vpsh-\scalu\scala\vsigg\cos\thtrep)
   \beqref{tpath1xa}\text{ \& }\eqnref{tray1a}
\end{split}
\nonumber\\
\begin{split}
&=\scalu\veka[\rhorep(\scala\vpsg\cos\thtrep-\vsigd\cos\phirep)
     +\scala\vrhok(\cos\thtrep-\cos\lamrep\cos\phirep)
     +\scala^2\vrhod(\cos\lamrep\cos\thtrep-\cos\phirep)]\\
   &\quad+\rhorep\cdkt[\vrhok(\vpsh-\scalu\vpsg\cos\phirep)
     +\vrhod(\scala\vpsh\cos\lamrep-\scalu\vsigd\cos\phirep)]\\
   &\quad+\scala\rhorep^2[\vrhok(\vpsh\cos\lamrep-\scalu\vpsg\cos\thtrep)
     +\vrhod(\scala\vpsh-\scalu\vsigd\cos\thtrep)]\\
   &\quad+\rhorep^2[\cdkt(\vpsg\vpsh-\scalu\vsigg\cos\phirep)
     +\rhorep(\vsigd\vpsh-\scalu\scala\vsigg\cos\thtrep)]
\end{split}
\nonumber\\
&=\vekh\beqref{tray1d}
\end{align}
\end{subequations}
\begin{subequations}\label{tray5}
\begin{align}\label{tray5a}
|\veust+\veusu|^2
&=|\veust|^2+2(\dprod{\veust}{\veusu})+|\veusu|^2\nonumber\\
&=\vekf+2\vekh+\vekg\beqref{tray4}
\end{align}
\begin{align}\label{tray5b}
&\ffdot{\cdkt}\imaa+\ffdot{\rhorep}\imab+\vbbb\imac+\rhorep\imad+\imae
  \nonumber\\
&=\ffdot{\cdkt}\vekb+\ffdot{\rhorep}\vekc+\vbbb\vekd+\rhorep\veke
  \beqref{tray3}\nonumber\\
&=\vrhol\vekb+\vrhoi\vekc+\vekd(1+2\vrhod)+\rhorep\veke
  \beqref{tpath6b},\eqnref{tpath9b}\text{ \& }\eqnref{tpath10b}.
\end{align}
\end{subequations}

\subart{Results of the computations}
Substituting \eqnref{tray2c}, \eqnref{tray5} and \eqnref{main3a} into \eqnref{kray4},
we obtain the equations describing the apparent geometry of the rays as
\begin{equation}\label{tray6}
\begin{split}
&\bbkbar=\plusmin\frac{(\vekf+2\vekh+\vekg)^{1/2}}{\scalc^3\rcal^3},\quad
\bbtbar=\frac{\vrhol\vekb+\vrhoi\vekc+\vekd(1+2\vrhod)+\rhorep\veke}{\vekf+2\vekh+\vekg},\quad
\frtbar=\frac{\cdkt\unitkap+\rhorep\vecta-\vectu}{\scalc\rcal}\\
&\qquad\frbbar=\frac{\vrhok(\cprod{\unitkap}{\vectu})+\veka(\cprod{\unitkap}{\vecta})
   +\vrhod(\cprod{\vecta}{\vectu})+\rhorep[(\cprod{\fdota}{\vectu})+\cdkt(\cprod{\unitkap}{\fdota})
   +\rhorep(\cprod{\vecta}{\fdota})]}{(\vekf+2\vekh+\vekg)^{1/2}}
\end{split}
\end{equation}
where $\rcal$ is given by \eqnref{trob2b}, $\cdkt$ is given by \eqnref{trob2c},
and all other quantities are given by \eqnref{tray1} or \eqnref{tpath1}.

\section{Rotational obliquation}\label{S_ROTOB}
\art{Apparent direction to a light source}
To evaluate \eqnref{grad6} for a rotating observer, it is convenient to introduce the quantities
\begin{subequations}\label{rot1}
\begin{align}\label{rot1a}
\begin{split}
&\epsva=\dprod{\vectOme}{\unitkap},\quad
\epsvb=\dprod{\vectOme}{\vectr},\quad
\epsvc=\dprod{\vectOme}{\unitplz},\quad
\epsvd=\dprod{\unitkap}{\vectr},\quad
\epsve=\dprod{\unitkap}{\unitplz},\quad
\epsvf=\dprod{\vectr}{\unitplz}\\
&\epsvg=\dprod{\vectOme}{\vectLam},\quad
\epsvh=\dprod{\vectr}{\vectLam},\quad
\epsvi=\dprod{\unitplz}{\vectLam},\quad
\epsvj=\dprod{\unitkap}{(\cprod{\unitplz}{\vectOme})}\ne0,\quad
\epsvk=\dprod{\unitkap}{(\cprod{\vectOme}{\vectr})}\\
&\epsvl=\dprod{\unitplz}{(\cprod{\vectOme}{\vectr})},\quad
\epsvm=\dprod{\vectOme}{(\cprod{\vectLam}{\vectr})},\quad
\epsvn=\dprod{\unitkap}{(\cprod{\vectLam}{\vectr})},\quad
\epsvo=\dprod{\unitplz}{(\cprod{\vectLam}{\vectr})}
\end{split}
\end{align}
\begin{align}\label{rot1b}
\begin{split}
&\vphia=(\Omerep^2\scalr^2-\epsvb^2)^{1/2}\ne0,\quad
\vphib=(\Omerep^2\vphia^2+2\epsvb\epsvm+\Lamrep^2\scalr^2-\epsvh^2)^{1/2},\quad
\vphic=\scalr^2\epsvg-\epsvb\epsvh\\
&\vphid=\epsva\epsvb-\Omerep^2\epsvd+\epsvn,\quad
\vphie=\epsvc\epsva-\Omerep^2\epsve,\quad
\vphif=\epsve\epsvg-\epsvi\epsva,\quad
\vphig=\vphif-4(\vphie^2/\epsvj)\\
&\vphih=(2\rhorep\vphid-\scalc\dragf)/[2\epsvj(\vphig+\epsvj\pfreq^2)],\quad
\vphii=\vphih(\epsvi\epsvj+8\vphie\epsvc),\quad
\vphij=\vphih(4\vphie\Omerep^2+\epsvj\epsvg)\\
&\vphik=2(\rhorep\vphid-\epsvj\vphig\vphih),\quad
\vphil=\left|1+[\vphig/(\epsvj\pfreq^2)]\right|^{1/2}\ne0,\quad
\vphim=\vphig\vphih\epsvj+\vphii\epsva-\vphij\epsve
\end{split}
\end{align}
\begin{align}\label{rot1c}
\begin{split}
&\vphin=\vphig\vphih(\epsvc\epsvb-\Omerep^2\epsvf)-\vphij\epsvl,\quad
\vphio=\vphig\vphih\epsvj+\vphii\epsva-\vphij\epsve\\
&\vphip=\vphig\vphih(-\epsvl\Omerep^2+\epsvi\epsvb-\epsvf\epsvg)+\vphii\epsvm
   -\vphij(\epsvc\epsvb-\epsvf\Omerep^2+\epsvo)\\
&\qquad\vphiq=[\vphig^2\vphih^2(\Omerep^2-\epsvc^2)+\vphii^2\Omerep^2-2\vphii\vphij\epsvc+\vphij^2]^{1/2}
\end{split}
\end{align}
\begin{align}\label{rot1d}
\begin{split}
\vphir=[\vphin+2(\rhorep\vphic&-\vphik\epsvk)]/\vphia^2,\quad
\vphis=[\vphik(2\rhorep\vphic-\vphik\epsvk)-2\scalc\dragf(\vphin-\vphik\epsvk)]/\vphia^2\\
&\vphit=\epsvk/\vphia,\quad\vphiu=\vphid/\vphib,\quad\vphiv=\vphic/(\vphia\vphib).
\end{split}
\end{align}
\end{subequations}

\subart{Development of equation \eqnref{main4}}
With the above quantities in view, we derive
\begin{subequations}\label{rot2}
\begin{align}\label{rot2a}
\scalu^2
&=\dprod{(\cprod{\vectOme}{\vectr})}{(\cprod{\vectOme}{\vectr})}
  \beqref{main4c}\nonumber\\
&=\Omerep^2\scalr^2-(\dprod{\vectOme}{\vectr})^2
  \beqref{alg2}\nonumber\\
&=\Omerep^2\scalr^2-\epsvb^2\beqref{rot1a}\nonumber\\
\therefore\scalu
&=\vphia\beqref{rot1b}
\end{align}
\begin{align}\label{rot2b}
\scala^2
&=\dprod{(\cprod{\vectOme}{\vectu}+\cprod{\vectLam}{\vectr})}
  {(\cprod{\vectOme}{\vectu}+\cprod{\vectLam}{\vectr})}
  \beqref{main4a}\nonumber\\
&=\dprod{(\cprod{\vectOme}{\vectu})}{(\cprod{\vectOme}{\vectu})}
  +2\dprod{(\cprod{\vectOme}{\vectu})}{(\cprod{\vectLam}{\vectr})}
  +\dprod{(\cprod{\vectLam}{\vectr})}{(\cprod{\vectLam}{\vectr})}
  \nonumber\\
&=[\Omerep^2\scalu^2-(\dprod{\vectOme}{\vectu})^2]
  +2[(\dprod{\vectOme}{\vectLam})(\dprod{\vectu}{\vectr})
     -(\dprod{\vectOme}{\vectr})(\dprod{\vectLam}{\vectu})]
  +[\Lamrep^2\scalr^2-(\dprod{\vectLam}{\vectr})^2]
  \beqref{alg2}\nonumber\\
&=\Omerep^2\scalu^2+2(\dprod{\vectOme}{\vectr})[\dprod{\vectOme}{(\cprod{\vectLam}{\vectr})}]
  +[\Lamrep^2\scalr^2-(\dprod{\vectLam}{\vectr})^2]
  \beqref{main4c}\text{ \& }\eqnref{alg4}\nonumber\\
&=\Omerep^2\vphia^2+2\epsvb\epsvm+\Lamrep^2\scalr^2-\epsvh^2
  \beqref{rot1a}\text{ \& }\eqnref{rot2a}\nonumber\\
\therefore\scala
&=\vphib\beqref{rot1b}
\end{align}
\end{subequations}
\begin{subequations}\label{rot3}
\begin{align}\label{rot3a}
\dprod{\vectu}{\vecta}
&=\dprod{\vectu}{(\cprod{\vectOme}{\vectu}+\cprod{\vectLam}{\vectr})}
  \beqref{main4a}\nonumber\\
&=\dprod{(\cprod{\vectOme}{\vectr})}{(\cprod{\vectLam}{\vectr})}
  \beqref{main4c}\nonumber\\
&=\scalr^2(\dprod{\vectOme}{\vectLam})-(\dprod{\vectOme}{\vectr})(\dprod{\vectLam}{\vectr})
  \beqref{alg2}\nonumber\\
&=\scalr^2\epsvg-\epsvb\epsvh\beqref{rot1a}\nonumber\\
&=\vphic\beqref{rot1b}
\end{align}
\begin{align}\label{rot3b}
\dprod{\vectkap}{\vecta}
&=\dprod{\vectkap}{(\cprod{\vectOme}{\vectu}+\cprod{\vectLam}{\vectr})}
  \beqref{main4a}\nonumber\\
&=\dprod{\vectkap}{[\cprod{\vectOme}{(\cprod{\vectOme}{\vectr})}]}
  +\dprod{\vectkap}{(\cprod{\vectLam}{\vectr})}\beqref{main4c}\nonumber\\
&=\dprod{\vectkap}{[\vectOme(\dprod{\vectOme}{\vectr})-\Omerep^2\vectr]}
  +\dprod{\vectkap}{(\cprod{\vectLam}{\vectr})}\beqref{alg1}\nonumber\\
&=(\dprod{\vectkap}{\vectOme})(\dprod{\vectOme}{\vectr})-\Omerep^2(\dprod{\vectkap}{\vectr})
  +\dprod{\vectkap}{(\cprod{\vectLam}{\vectr})}\nonumber\\
&=\kaprep(\epsva\epsvb-\Omerep^2\epsvd+\epsvn)\beqref{rot1a}\nonumber\\
&=\kaprep\vphid\beqref{rot1b}
\end{align}
\end{subequations}
\begin{subequations}\label{rot4}
\begin{align}
\alprep
&=\dprod{\vectkap}{\vecta}\beqref{main2b}\nonumber\\
&=\kaprep\vphid\beqref{rot3b}
\label{rot4a}\\
\etarep
&=\dprod{\kaprep}{(\cprod{\unitplz}{\vectOme})}\beqref{main4c}\nonumber\\
&=\kaprep\epsvj\beqref{rot1a}
\label{rot4b}
\end{align}
\begin{align}\label{rot4c}
\xirep
&=2(\dprod{\vectOme}{\unitplz})(\dprod{\vectOme}{\vectkap})
  -\Omerep^2(\dprod{\vectkap}{\unitplz})\beqref{main4c}\nonumber\\
&=2\kaprep(\epsvc\epsva-\Omerep^2\epsve)\beqref{rot1a}\nonumber\\
&=2\kaprep\vphie\beqref{rot1b}
\end{align}
\begin{align}\label{rot4d}
\zetarep
&=(\dprod{\vectkap}{\unitplz})(\dprod{\vectOme}{\vectLam})
  -(\dprod{\vectLam}{\unitplz})(\dprod{\vectOme}{\vectkap})\beqref{main4c}\nonumber\\
&=\kaprep(\epsve\epsvg-\epsvi\epsva)\beqref{rot1a}\nonumber\\
&=\kaprep\vphif\beqref{rot1b}
\end{align}
\end{subequations}
\begin{subequations}\label{rot5}
\begin{align}\label{rot5a}
\szer
&=\zetarep-(\xirep^2/\etarep)\beqref{main4d}\nonumber\\
&=\kaprep\vphif-[(2\kaprep\vphie)^2/(\kaprep\epsvj)]\beqref{rot4}\nonumber\\
&=\kaprep[\vphif-4(\vphie^2/\epsvj)]\nonumber\\
&=\kaprep\vphig\beqref{rot1b}
\end{align}
\begin{align}\label{rot5b}
\sone
&=(2\rhorep\alprep-\dragf\pfreq)/[2\etarep(\szer+\etarep\pfreq^2)]
  \beqref{main4d}\nonumber\\
&=(2\rhorep\kaprep\vphid-\dragf\pfreq)/[2(\kaprep\epsvj)(\kaprep\vphig+\kaprep\epsvj\pfreq^2)]
  \beqref{rot4}\text{ \& }\eqnref{rot5a}\nonumber\\
&=(2\rhorep\vphid-\scalc\dragf)/[2\kaprep\epsvj(\vphig+\epsvj\pfreq^2)]\nonumber\\
&=\vphih/\kaprep\beqref{rot1b}
\end{align}
\begin{align}\label{rot5c}
\stwo
&=\szer\sone\beqref{main4d}\nonumber\\
&=\vphig\vphih\beqref{rot5a}\text{ \& }\eqnref{rot5b}
\end{align}
\begin{align}\label{rot5d}
\sthr
&=[\etarep(\dprod{\vectLam}{\unitplz})+4\xirep(\dprod{\vectOme}{\unitplz})]\sone
   \beqref{main4d}\nonumber\\
&=[\kaprep\epsvj\epsvi+8\kaprep\vphie\epsvc](\vphih/\kaprep)
   \beqref{rot4}, \eqnref{rot1a}\text{ \& }\eqnref{rot5b}\nonumber\\
&=\vphih(\epsvi\epsvj+8\vphie\epsvc)\nonumber\\
&=\vphii\beqref{rot1b}
\end{align}
\begin{align}\label{rot5e}
\sfou
&=[2\xirep\Omerep^2+\etarep(\dprod{\vectOme}{\vectLam})]\sone
   \beqref{main4d}\nonumber\\
&=[2(2\kaprep\vphie)\Omerep^2+\kaprep\epsvj\epsvg](\vphih/\kaprep)
   \beqref{rot4}, \eqnref{rot1a}\text{ \& }\eqnref{rot5b}\nonumber\\
&=\vphih(4\vphie\Omerep^2+\epsvj\epsvg)\nonumber\\
&=\vphij\beqref{rot1b}
\end{align}
\end{subequations}
\begin{subequations}\label{rot6}
\begin{align}\label{rot6a}
\kaprep\taurep
&=2(\rhorep\alprep-\etarep\stwo)/\kaprep\beqref{main4b}\nonumber\\
&=2(\rhorep\kaprep\vphid-\kaprep\epsvj\vphig\vphih)/\kaprep
   \beqref{rot4}\text{ \& }\eqnref{rot5c}\nonumber\\
&=2(\rhorep\vphid-\epsvj\vphig\vphih)\nonumber\\
&=\vphik\beqref{rot1b}
\end{align}
\begin{align}\label{rot6b}
\gamrep^2
&=\left|1+[\szer/(\etarep\pfreq^2)]\right|\beqref{main4a}\nonumber\\
&=\left|1+[(\kaprep\vphig)/(\kaprep\epsvj\pfreq^2)]\right|
   \beqref{rot5a}\text{ \& }\eqnref{rot4b}\nonumber\\
&=\left|1+[\vphig/(\epsvj\pfreq^2)]\right|\nonumber\\
\therefore\gamrep
&=\vphil\beqref{rot1b}.
\end{align}
\end{subequations}

\subart{Development of equation \eqnref{grad2}}
The various quantities defined by \eqnref{grad2} become
\begin{subequations}\label{rot7}
\begin{align}\label{rot7a}
\bcal
&=\dprod{\vectc}{\vecte}\beqref{grad2a}\nonumber\\
&=\dprod{\scalc\unitkap}{[\stwo(\cprod{\unitplz}{\vectOme})+\sthr\vectOme-\sfou\unitplz]}
  \beqref{main1}\text{ \& }\eqnref{main4b}\nonumber\\
&=\scalc\stwo[\dprod{\unitkap}{(\cprod{\unitplz}{\vectOme})}]
  +\scalc\sthr(\dprod{\unitkap}{\vectOme})
  -\scalc\sfou(\dprod{\unitkap}{\unitplz})\nonumber\\
&=\scalc(\vphig\vphih\epsvj+\vphii\epsva-\vphij\epsve)
  \beqref{rot1a}\text{ \& }\eqnref{rot5}\nonumber\\
&=\scalc\vphim\beqref{rot1b}
\end{align}
\begin{align}\label{rot7b}
\dcal
&=\dprod{\vectu}{\vecte}\beqref{grad2a}\nonumber\\
&=\dprod{\vectu}{[\stwo(\cprod{\unitplz}{\vectOme})+\sthr\vectOme-\sfou\unitplz]}
   \beqref{main4b}\nonumber\\
&=\stwo\dprod{(\cprod{\vectOme}{\vectr})}{(\cprod{\unitplz}{\vectOme})}
   +\sthr\dprod{(\cprod{\vectOme}{\vectr})}{\vectOme}
   -\sfou\dprod{(\cprod{\vectOme}{\vectr})}{\unitplz}
   \beqref{main4c}\nonumber\\
&=\stwo[(\dprod{\unitplz}{\vectOme})(\dprod{\vectOme}{\vectr})-\Omerep^2(\dprod{\unitplz}{\vectr})]
   -\sfou[\dprod{\unitplz}{(\cprod{\vectOme}{\vectr})}]\beqref{alg2}\nonumber\\
&=\vphig\vphih(\epsvc\epsvb-\Omerep^2\epsvf)-\vphij\epsvl
  \beqref{rot1a}\text{ \& }\eqnref{rot5}\nonumber\\
&=\vphin\beqref{rot1c}
\end{align}
\begin{align*}
\acal
&=\dprod{\vecta}{\vecte}\beqref{grad2a}\nonumber\\
&=\dprod{\vecta}{[\stwo(\cprod{\unitplz}{\vectOme})+\sthr\vectOme-\sfou\unitplz]}
   \beqref{main4b}\nonumber\\
\begin{split}
&=\stwo[\dprod{(\cprod{\unitplz}{\vectOme})}{(\cprod{\vectOme}{\vectu}+\cprod{\vectLam}{\vectr})}]
   +\sthr[\dprod{\vectOme}{(\cprod{\vectOme}{\vectu}+\cprod{\vectLam}{\vectr})}]
   -\sfou[\dprod{\unitplz}{(\cprod{\vectOme}{\vectu}+\cprod{\vectLam}{\vectr})}]\\
   &\qquad\beqref{main4a}
\end{split}
\end{align*}
\begin{align}\label{rot7c}
\begin{split}
&=\stwo[\dprod{(\cprod{\unitplz}{\vectOme})}{(\cprod{\vectOme}{\vectu})}
      +\dprod{(\cprod{\unitplz}{\vectOme})}{(\cprod{\vectLam}{\vectr})}]
   +\sthr[\dprod{\vectOme}{(\cprod{\vectLam}{\vectr})}]\\
   &\qquad-\sfou[\dprod{(\cprod{\unitplz}{\vectOme})}{(\cprod{\vectOme}{\vectr})}
      +\dprod{\unitplz}{(\cprod{\vectLam}{\vectr})}]\beqref{alg4}\text{ \& }\eqnref{main4c}
\end{split}
\nonumber\\
\begin{split}
&=\stwo[(\dprod{\unitplz}{\vectOme})(\dprod{\vectOme}{\vectu})-(\dprod{\unitplz}{\vectu})\Omerep^2
      +(\dprod{\unitplz}{\vectLam})(\dprod{\vectOme}{\vectr})-(\dprod{\unitplz}{\vectr})(\dprod{\vectLam}{\vectOme})]
   +\sthr[\dprod{\vectOme}{(\cprod{\vectLam}{\vectr})}]\\
   &\qquad-\sfou[(\dprod{\unitplz}{\vectOme})(\dprod{\vectOme}{\vectr})-(\dprod{\unitplz}{\vectr})\Omerep^2
      +\dprod{\unitplz}{(\cprod{\vectLam}{\vectr})}]\beqref{alg2}
\end{split}
\nonumber\\
&=\vphig\vphih(-\epsvl\Omerep^2+\epsvi\epsvb-\epsvf\epsvg)+\vphii\epsvm
   -\vphij(\epsvc\epsvb-\epsvf\Omerep^2+\epsvo)
   \beqref{main4c}, \eqnref{rot1a}\text{ \& }\eqnref{rot5}
\nonumber\\
&=\vphip\beqref{rot1c}
\end{align}
\begin{align}\label{rot7d}
\hcal
&=\dprod{\vectkap}{\vecte}\beqref{grad2a}\nonumber\\
&=\dprod{\vectkap}{[\stwo(\cprod{\unitplz}{\vectOme})+\sthr\vectOme-\sfou\unitplz]}
   \beqref{main4b}\nonumber\\
&=\stwo[\dprod{\vectkap}{(\cprod{\unitplz}{\vectOme})}]
   +\sthr(\dprod{\vectkap}{\vectOme})-\sfou(\dprod{\vectkap}{\unitplz})\nonumber\\
&=\kaprep(\vphig\vphih\epsvj+\vphii\epsva-\vphij\epsve)
   \beqref{rot1a}\text{ \& }\eqnref{rot5}\nonumber\\
&=\kaprep\vphio\beqref{rot1c}
\end{align}
\begin{align}\label{rot7e}
\ecal
&=\dprod{\vecte}{\vecte}\beqref{grad2a}\nonumber\\
&=\dprod{[\stwo(\cprod{\unitplz}{\vectOme})+\sthr\vectOme-\sfou\unitplz]}
   {[\stwo(\cprod{\unitplz}{\vectOme})+\sthr\vectOme-\sfou\unitplz]}
   \beqref{main4b}\nonumber\\
\begin{split}
&=\stwo^2[\dprod{(\cprod{\unitplz}{\vectOme})}{(\cprod{\unitplz}{\vectOme})}]
   +2\stwo\sthr[\dprod{\vectOme}{(\cprod{\unitplz}{\vectOme})}]
   -2\stwo\sfou[\dprod{\unitplz}{(\cprod{\unitplz}{\vectOme})}]\\
   &\qquad+\sthr^2(\dprod{\vectOme}{\vectOme})
   -2\sthr\sfou(\dprod{\unitplz}{\vectOme})
   +\sfou^2(\dprod{\unitplz}{\unitplz})
\end{split}
\nonumber\\
&=\stwo^2[\Omerep^2-(\dprod{\unitplz}{\vectOme})^2]
   +\sthr^2\Omerep^2-2\sthr\sfou(\dprod{\unitplz}{\vectOme})+\sfou^2
   \beqref{alg2}\nonumber\\
&=\vphig^2\vphih^2(\Omerep^2-\epsvc^2)+\vphii^2\Omerep^2-2\vphii\vphij\epsvc+\vphij^2
   \beqref{rot1a}\text{ \& }\eqnref{rot5}\nonumber\\
&=\vphiq^2\beqref{rot1c}.
\end{align}
\end{subequations}
Using \eqnref{rot7} in \eqnref{grad2}, we have
\begin{subequations}\label{rot8}
\begin{align}\label{rot8a}
\lcal_0
&=\dcal-\scalu\kaprep\taurep\cos\phirep\beqref{grad2b}\nonumber\\
&=\vphin-\vphik(\dprod{\unitkap}{\vectu})
  \beqref{rot7b}\text{ \& }\eqnref{rot6a}\nonumber\\
&=\vphin-\vphik\epsvk
  \beqref{main4c}\text{ \& }\eqnref{rot1a}
\end{align}
\begin{align}
\lcal_1
&=\rhorep\scala\cos\thtrep-\kaprep\taurep\cos\phirep\beqref{grad2b}
\label{rot8b}\\
\lcal_2
&=\lcal_1+\rhorep\scala\cos\thtrep\beqref{grad2b}\nonumber\\
&=2\rhorep\scala\cos\thtrep-\kaprep\taurep\cos\phirep\beqref{rot8b}
\label{rot8c}
\end{align}
\end{subequations}
\begin{subequations}\label{rot9}
\begin{align}\label{rot9a}
\ncal_1
&=2\dragf\scalu^2(\bcal-\taurep\pfreq)\beqref{grad2c}\nonumber\\
&=2\scalc\dragf\scalu^2(\vphim-\kaprep\taurep)
   \beqref{rot7a}\text{ \& }\eqnref{main2b}\nonumber\\
&=2\scalc\dragf\scalu^2(\vphim-\vphik)\beqref{rot6a}
\end{align}
\begin{align}\label{rot9b}
\ncal_2
&=2\dragf\scalu\scalc\lcal_0\cos\phirep\beqref{grad2c}\nonumber\\
&=2\dragf\scalu\scalc(\vphin-\vphik\epsvk)\cos\phirep
   \beqref{rot8a}\nonumber\\
&=2\dragf\scalc(\dprod{\unitkap}{\vectu})(\vphin-\vphik\epsvk)\nonumber\\
&=2\scalc\dragf\epsvk(\vphin-\vphik\epsvk)
   \beqref{main4c}\text{ \& }\eqnref{rot1a}
\end{align}
\begin{align}\label{rot9c}
\ncal_3
&=\scalu^2[\ecal+2(\rhorep\acal-\taurep\hcal)]\beqref{grad2c}\nonumber\\
&=\scalu^2(\vphiq^2+2\rhorep\vphip-2\kaprep\taurep\vphio)\beqref{rot7}\nonumber\\
&=\scalu^2(\vphiq^2+2\rhorep\vphip-2\vphik\vphio)\beqref{rot6a}
\end{align}
\begin{align}\label{rot9d}
\ncal_4
&=\scalu^2\kaprep\taurep(\kaprep\taurep-2\rhorep\scala\cos\lamrep)
  \beqref{grad2d}\nonumber\\
&=\scalu^2\vphik[\vphik-2\rhorep(\dprod{\unitkap}{\vecta})]
  \beqref{rot6a}\nonumber\\
&=\scalu^2\vphik(\vphik-2\rhorep\vphid)
  \beqref{rot3b}
\end{align}
\begin{align}\label{rot9e}
\ncal_5
&=\scalu^2\kaprep\taurep\lcal_2\cos\phirep\beqref{grad2d}\nonumber\\
&=\scalu^2\vphik(2\rhorep\scala\cos\thtrep-\vphik\cos\phirep)\cos\phirep
   \beqref{rot8c}\text{ \& }\eqnref{rot6a}\nonumber\\
&=\vphik(\dprod{\unitkap}{\vectu})[2\rhorep(\dprod{\vectu}{\vecta})
   -\vphik(\dprod{\unitkap}{\vectu})]\nonumber\\
&=\vphik\epsvk(2\rhorep\vphic-\vphik\epsvk)
   \beqref{main4c}, \eqnref{rot1a}\text{ \& }\eqnref{rot3a}
\end{align}
\begin{align}\label{rot9f}
\ncal_6
&=\dcal(\dcal+2\scalu\lcal_1)\beqref{grad2d}\nonumber\\
&=\vphin[\vphin+2\scalu(\rhorep\scala\cos\thtrep-\kaprep\taurep\cos\phirep)]
   \beqref{rot7b}\text{ \& }\eqnref{rot8b}\nonumber\\
&=\vphin[\vphin+2\rhorep(\dprod{\vectu}{\vecta})-2\vphik(\dprod{\unitkap}{\vectu})]
  \beqref{rot6a}\nonumber\\
&=\vphin(\vphin+2\rhorep\vphic-2\vphik\epsvk)
   \beqref{main4c}, \eqnref{rot1a}\text{ \& }\eqnref{rot3a}
\end{align}
\end{subequations}
\begin{subequations}\label{rot10}
\begin{align}\label{rot10a}
&\ncal_1+\ncal_3+\ncal_4-2\scalu^2(\dcal+\scalu\lcal_1)\nonumber\\
\begin{split}
&=2\scalc\dragf\scalu^2(\vphim-\vphik)+\scalu^2(\vphiq^2+2\rhorep\vphip-2\vphik\vphio)
  +\scalu^2\vphik(\vphik-2\rhorep\vphid)\\
  &\qquad-2\scalu^2[\vphin+\scalu(\rhorep\scala\cos\thtrep-\kaprep\taurep\cos\phirep)]
  \beqref{rot9}, \eqnref{rot7b}\text{ \& }\eqnref{rot8b}
\end{split}
\nonumber\\
\begin{split}
&=\scalu^2[\vphiq^2+2\rhorep\vphip-2\vphik\vphio+\vphik(\vphik-2\rhorep\vphid)+2\scalc\dragf(\vphim-\vphik)\\
  &\qquad-2\vphin-2\rhorep(\dprod{\vectu}{\vecta})+2\vphik(\dprod{\unitkap}{\vectu})]\beqref{rot6a}
\end{split}
\nonumber\\
\begin{split}
&=\scalu^2[\vphiq^2+2\rhorep\vphip-2\vphik\vphio+\vphik(\vphik-2\rhorep\vphid)+2\scalc\dragf(\vphim-\vphik)\\
  &\qquad-2\vphin-2\rhorep\vphic+2\vphik\epsvk]
  \beqref{rot3a}, \eqnref{main4c}\text{ \& }\eqnref{rot1a}
\end{split}
\nonumber\\
\begin{split}
&=\scalu^2(\vphiq^2+\vphik^2)-2\scalu^2[\vphin-\rhorep(\vphip-\vphic-\vphik\vphid)
   +\vphik(\vphio-\epsvk)-\scalc\dragf(\vphim-\vphik)]
\end{split}
\end{align}
\begin{align}\label{rot10b}
&\ncal_1-\ncal_2+\ncal_3+\ncal_4+\ncal_5-\ncal_6\nonumber\\
\begin{split}
&=2\scalc\dragf\scalu^2(\vphim-\vphik)-2\scalc\dragf\epsvk(\vphin-\vphik\epsvk)
  +\scalu^2(\vphiq^2+2\rhorep\vphip-2\vphik\vphio)\\
  &\quad+\scalu^2\vphik(\vphik-2\rhorep\vphid)
  +\vphik\epsvk(2\rhorep\vphic-\vphik\epsvk)-\vphin(\vphin+2\rhorep\vphic-2\vphik\epsvk)
  \beqref{rot9}
\end{split}
\nonumber\\
\begin{split}
&=\scalu^2(\vphiq^2+\vphik^2)
  -2\scalu^2[\vphik\vphio-\rhorep(\vphip-\vphik\vphid)-\scalc\dragf(\vphim-\vphik)]\\
  &\quad+\epsvk[\vphik(2\rhorep\vphic-\vphik\epsvk)-2\scalc\dragf(\vphin-\vphik\epsvk)]
  -\vphin[\vphin+2(\rhorep\vphic-\vphik\epsvk)]
\end{split}
\nonumber\\
\begin{split}
&=\scalu^2(\vphiq^2+\vphik^2)
  -2\scalu^2[\vphik\vphio-\rhorep(\vphip-\vphik\vphid)-\scalc\dragf(\vphim-\vphik)]
  +\scalu^2(\epsvk\vphis-\vphin\vphir)\\
  &\quad\beqref{rot1d}\text{ \& }\eqnref{rot2a}.
\end{split}
\end{align}
\end{subequations}
Using \eqnref{rot1d} and the definition of the angles shown in \figref{FIG1}, we have
\begin{align}\label{rot11}
\begin{split}
&\cos\phirep=(\dprod{\unitkap}{\vectu})/\scalu=\epsvk/\vphia=\vphit,\quad
\cos\lamrep=(\dprod{\unitkap}{\vecta})/\scala=\vphid/\vphib=\vphiu\\
&\qquad\qquad\cos\thtrep=(\dprod{\vectu}{\vecta})/(\scalu\scala)=\vphic/(\vphia\vphib)=\vphiv.
\end{split}
\end{align}

\subart{Results of the computations}
Substituting \eqnref{rot2a}, \eqnref{rot8a}, \eqnref{rot10} and \eqnref{rot11} into \eqnref{grad2w}
yields
\begin{subequations}\label{rot12}
\begin{equation}\label{rot12a}
\begin{split}
\lcal\scalc&=(\vphin-\vphik\epsvk)/\vphia\\
\pcal\scalc^2&=\vphiq^2+\vphik^2-2[\vphin-\rhorep(\vphip-\vphic-\vphik\vphid)
   +\vphik(\vphio-\epsvk)-\scalc\dragf(\vphim-\vphik)]\\
\ncal\scalc^2&=\vphiq^2+\vphik^2+\epsvk\vphis-\vphin\vphir
  -2[\vphik\vphio-\rhorep(\vphip-\vphik\vphid)-\scalc\dragf(\vphim-\vphik)]
\end{split}
\end{equation}
\begin{equation}\label{rot12b}
\begin{split}
\gcal&=\lcal-\betrep+\dragf\vphit+\rhorep\sigrep\vphiv\\
\rcal^2&=\pcal+\dragf^2+\betrep^2+\rhorep^2\sigrep^2+2\dragf(\rhorep\sigrep\vphiu-\betrep\vphit)\\
\fcal^2&=\ncal+\dragf^2(1-\vphit^2)+\rhorep^2\sigrep^2(1-\vphiv^2)+2\dragf\rhorep\sigrep(\vphiu-\vphit\vphiv)
\end{split}
\end{equation}
while from \eqnref{rot12b} and \eqnref{grad6}, we get
\begin{equation}\label{rot12c}
\tan\psirep=
\frac{[\ncal+\dragf^2(1-\vphit^2)+\rhorep^2\sigrep^2(1-\vphiv^2)+2\dragf\rhorep\sigrep(\vphiu-\vphit\vphiv)]^{1/2}}
   {\lcal-\betrep+\dragf\vphit+\rhorep\sigrep\vphiv}
\end{equation}
and by \eqnref{main2}, \eqnref{main6}, \eqnref{rot2}, \eqnref{rot6b} and \eqnref{rot4a},
\begin{align}\label{rot12d}
\begin{split}
&\qquad\qquad\qquad\rhorep=\frac{\xcons}{4\dragf\pfreq},\quad
\xcons=\frac{\vthtrep}{\sqrt{1+\vthtrep^2}},\quad
\dragf=\gamrep\left\{\frac{1+\sqrt{1+\vthtrep^2}}{2}\right\}^{1/2}\\
&\vthtrep=\alprep/(\gamrep\pfreq)^{2},\quad
\alprep=\kaprep\vphid,\quad
\gamrep=\vphil,\quad
\betrep=\vphia/\scalc,\quad
\sigrep=\vphib/\scalc,\quad
\cdkt=\scalc\dragf-\vphik.
\end{split}
\end{align}
\end{subequations}
Equations \eqnref{rot1} and \eqnref{rot12} give a complete prescription for
calculating $\psirep$ for a rotating observer.

\art{Apparent drift of a light source}
To evaluate \eqnref{kas7} for a rotating observer, let us introduce the following quantities
in addition to those given by \eqnref{rot1} and \eqnref{rot12},
\begin{subequations}\label{rxpeed1}
\begin{align}\label{rxpeed1a}
\begin{split}
&\dltva=\dprod{\unitkap}{\vectLam},\quad
\dltvb=\dprod{\unitkap}{(\cprod{\unitplz}{\vectLam})},\quad
\dltvc=-\cdkt\epsvd+\rhorep\vphia^2-\vphig\vphih\epsvl-\vphii\epsvb+\vphij\epsvf\\
&\dltvd=\cdkt\epsva+\rhorep\epsvm+\vphii\Omerep^2-\vphij\epsvc,\quad
\dltve=\vphij\epsvj+\Omerep^2(\epsvd-\rhorep\epsvk+\vphig\vphih\epsve)
  +\rhorep(\epsvb\dltva-\epsvd\epsvg)\\
&\dltvf=\dltve-\epsva(\epsvb+\vphig\vphih\epsvc),\quad
\dltvg=(1+\vthtrep^2)^{1/2},\quad
\dltvh=(\dragf\gamrep^{-2}\pfreq^{-1}-\rhorep\vthtrep\dltvg^2)/(4\dragf^2\pfreq^2\dltvg^3)\\
&\dltvi=(\rhorep+2\kaprep\vphid\dltvh)/[2\kaprep\epsvj(\vphig+\epsvj\pfreq^2)],\quad
\dltvj=\epsvj\epsvi+8\vphie\epsvc,\quad
\dltvk=4\vphie\Omerep^2+\epsvj\epsvg
\end{split}
\end{align}
\begin{align}\label{rxpeed1b}
\begin{split}
&\dltvl=\kaprep\dltvi(\vphig\epsvj+\dltvj\epsva-\dltvk\epsve),\quad
\dltvm=\kaprep\dltvi(\dltvk\epsvc-\dltvj\Omerep^2),\quad
\dltvn=\kaprep\dltvi(\epsva\epsvb-\Omerep^2\epsvd)\\
&\dltvo=\dltvl+2\rhorep,\quad
\dltvp=\dltvn\dltvj+\rhorep\epsvb,\quad
\dltvq=\rhorep(\Omerep^2-\rhorep\epsvg),\quad
\dltvr=\kaprep\dltvi\dltvf\dltvj-\rhorep(\epsvb+\vphig\vphih\epsvc)\\
&\dltvs=\rhorep\vphig\vphih\Omerep^2-\kaprep\dltvi\dltvf\dltvk,\quad
\dltvt=\kaprep\dltvi\dltvf\vphig+\rhorep\vphij,\quad
\dltvu=2[\vphid\dltvh-\epsvj\vphig\dltvi+(\rhorep/\kaprep)]
\end{split}
\end{align}
\begin{align}\label{rxpeed1c}
\begin{split}
&\efkta=(\epsvb\dltvc+\scalr^2\dltvd)/(\betrep\scalc^2\fcal),\quad
\efktb=(\Omerep^2\dltvc+\epsvb\dltvd)/(\betrep\scalc^2\fcal),\quad
\efktc=\gcal(\efkta\epsva-\efktb\epsvd)-\fcal\epsvk\\
&\efktd=\gcal(\efkta\epsvm+\efktb\vphia^2)-\fcal\vphic,\quad
\efkte=\fcal(\dltvo\epsvb+\dltvm\epsvd),\quad
\efktf=\fcal(\dltvo\Omerep^2+\epsva\dltvm).
\end{split}
\end{align}
\end{subequations}

\subart{Development of equation \eqnref{kas2b}}
We derive, with a view to the above quantities,
\begin{subequations}\label{rxpeed2}
\begin{align*}
\cprod{\vectOme}{\vectups}
&=\cprod{\vectOme}{(\cdkt\unitkap+\rhorep\vecta-\vectu+\vecte)}
 \beqref{kpath1c}\nonumber\\
\begin{split}
&=\cdkt(\cprod{\vectOme}{\unitkap})
  +\rhorep\cprod{\vectOme}{(\cprod{\vectOme}{\vectu}+\cprod{\vectLam}{\vectr})}
  -\cprod{\vectOme}{(\cprod{\vectOme}{\vectr})}\\
  &\qquad+\cprod{\vectOme}{[\stwo(\cprod{\unitplz}{\vectOme})+\sthr\vectOme-\sfou\unitplz]}
  \beqref{main4}
\end{split}
\nonumber\\
\begin{split}
&=\cdkt(\cprod{\vectOme}{\unitkap})
  +\rhorep\cprod{\vectOme}{(\cprod{\vectOme}{\vectu})}
  +\rhorep\cprod{\vectOme}{(\cprod{\vectLam}{\vectr})}
  -\cprod{\vectOme}{(\cprod{\vectOme}{\vectr})}\\
  &\qquad+\vphig\vphih\cprod{\vectOme}{(\cprod{\unitplz}{\vectOme})}
  +\vphii(\cprod{\vectOme}{\vectOme})
  -\vphij(\cprod{\vectOme}{\unitplz})\beqref{rot5}
\end{split}
\end{align*}
\begin{align}\label{rxpeed2a}
\begin{split}
&=\cdkt(\cprod{\vectOme}{\unitkap})
  +\rhorep[(\dprod{\vectOme}{\vectu})\vectOme-\Omerep^2\vectu]
  +\rhorep[(\dprod{\vectOme}{\vectr})\vectLam-(\dprod{\vectOme}{\vectLam})\vectr]
  -[(\dprod{\vectOme}{\vectr})\vectOme-\Omerep^2\vectr]\\
  &\qquad+\vphig\vphih[\Omerep^2\unitplz-(\dprod{\vectOme}{\unitplz})\vectOme]
  -\vphij(\cprod{\vectOme}{\unitplz})\beqref{alg1}
\end{split}
\nonumber\\
\begin{split}
&=\cdkt(\cprod{\vectOme}{\unitkap})-\rhorep\Omerep^2(\cprod{\vectOme}{\vectr})
  +\rhorep(\epsvb\vectLam-\epsvg\vectr)-\epsvb\vectOme+\Omerep^2\vectr\\
  &\qquad+\vphig\vphih(\Omerep^2\unitplz-\epsvc\vectOme)
  -\vphij(\cprod{\vectOme}{\unitplz})\beqref{main4c}\text{ \& }\eqnref{rot1a}
\end{split}
\nonumber\\
\begin{split}
&=\cdkt(\cprod{\vectOme}{\unitkap})-\rhorep\Omerep^2(\cprod{\vectOme}{\vectr})
  +\vphij(\cprod{\unitplz}{\vectOme})+\rhorep\epsvb\vectLam+\vphig\vphih\Omerep^2\unitplz\\
  &\qquad+(\Omerep^2-\rhorep\epsvg)\vectr-(\epsvb+\vphig\vphih\epsvc)\vectOme
\end{split}
\end{align}
\begin{align}\label{rxpeed2b}
\cprod{\vectu}{\vectups}
&=\cprod{\vectu}{(\cdkt\unitkap+\rhorep\vecta-\vectu+\vecte)}
 \beqref{kpath1c}\nonumber\\
\begin{split}
&=\cdkt(\cprod{\vectu}{\unitkap})
  +\rhorep\cprod{\vectu}{(\cprod{\vectOme}{\vectu}+\cprod{\vectLam}{\vectr})}\\
  &\qquad+\cprod{\vectu}{[\vphig\vphih(\cprod{\unitplz}{\vectOme})+\vphii\vectOme-\vphij\unitplz]}
 \beqref{main4}\text{ \& }\eqnref{rot5}
\end{split}
\nonumber\\
\begin{split}
&=-\cdkt[\cprod{\unitkap}{(\cprod{\vectOme}{\vectr})}]
  +\rhorep[\cprod{\vectu}{(\cprod{\vectOme}{\vectu})}]
  +\rhorep[\cprod{\vectu}{(\cprod{\vectLam}{\vectr})}]\\
  &\qquad+\vphig\vphih[\cprod{\vectu}{(\cprod{\unitplz}{\vectOme})}]
  -\vphii[\cprod{\vectOme}{(\cprod{\vectOme}{\vectr})}]
  +\vphij[\cprod{\unitplz}{(\cprod{\vectOme}{\vectr})}]\beqref{main4c}
\end{split}
\nonumber\\
\begin{split}
&=-\cdkt[(\dprod{\unitkap}{\vectr})\vectOme-(\dprod{\unitkap}{\vectOme})\vectr]
  +\rhorep[\scalu^2\vectOme-(\dprod{\vectOme}{\vectu})\vectu]
  +\rhorep[(\dprod{\vectr}{\vectu})\vectLam-(\dprod{\vectLam}{\vectu})\vectr]\\
  &\qquad+\vphig\vphih[(\dprod{\vectOme}{\vectu})\unitplz-(\dprod{\unitplz}{\vectu})\vectOme]
  -\vphii[(\dprod{\vectr}{\vectOme})\vectOme-\Omerep^2\vectr]
  +\vphij[(\dprod{\unitplz}{\vectr})\vectOme-(\dprod{\vectOme}{\unitplz})\vectr]
  \beqref{alg1}
\end{split}
\nonumber\\
\begin{split}
&=-\cdkt[(\dprod{\unitkap}{\vectr})\vectOme-(\dprod{\unitkap}{\vectOme})\vectr]
  +\rhorep\scalu^2\vectOme+\rhorep[\dprod{\vectOme}{(\cprod{\vectLam}{\vectr})}]\vectr
  -\vphig\vphih[\dprod{\unitplz}{(\cprod{\vectOme}{\vectr})}]\vectOme\\
  &\qquad-\vphii[(\dprod{\vectr}{\vectOme})\vectOme-\Omerep^2\vectr]
  +\vphij[(\dprod{\unitplz}{\vectr})\vectOme-(\dprod{\vectOme}{\unitplz})\vectr]
  \beqref{main4c}\text{ \& }\eqnref{alg4}
\end{split}
\nonumber\\
&=(-\cdkt\epsvd+\rhorep\scalu^2-\vphig\vphih\epsvl-\vphii\epsvb+\vphij\epsvf)\vectOme
  +(\cdkt\epsva+\rhorep\epsvm+\vphii\Omerep^2-\vphij\epsvc)\vectr\beqref{rot1a}
\nonumber\\
&=\dltvc\vectOme+\dltvd\vectr\beqref{rxpeed1a}\text{ \& }\eqnref{rot2a}
\end{align}
\begin{align}\label{rxpeed2c}
\dprod{\vectups}{(\cprod{\vectkap}{\vectOme})}
&=\dprod{\vectkap}{(\cprod{\vectOme}{\vectups})}\beqref{alg4}\nonumber\\
\begin{split}
&=\dprod{\vectkap}{}[\cdkt(\cprod{\vectOme}{\unitkap})-\rhorep\Omerep^2(\cprod{\vectOme}{\vectr})
  +\vphij(\cprod{\unitplz}{\vectOme})+\rhorep\epsvb\vectLam+\vphig\vphih\Omerep^2\unitplz\\
  &\qquad+(\Omerep^2-\rhorep\epsvg)\vectr-(\epsvb+\vphig\vphih\epsvc)\vectOme]\beqref{rxpeed2a}
\end{split}
\nonumber\\
\begin{split}
&=\kaprep[-\rhorep\Omerep^2\epsvk+\vphij\epsvj+\rhorep\epsvb\dltva+\vphig\vphih\Omerep^2\epsve
  +\epsvd(\Omerep^2-\rhorep\epsvg)-\epsva(\epsvb+\vphig\vphih\epsvc)]\\
  &\qquad\beqref{rot1a}\text{ \& }\eqnref{rxpeed1a}
\end{split}
\nonumber\\
&=\kaprep[\vphij\epsvj+\Omerep^2(\epsvd-\rhorep\epsvk+\vphig\vphih\epsve)
  +\rhorep(\epsvb\dltva-\epsvd\epsvg)-\epsva(\epsvb+\vphig\vphih\epsvc)]
\nonumber\\
&=\kaprep\dltvf\beqref{rxpeed1a}
\end{align}
\end{subequations}
\begin{subequations}\label{rxpeed3}
\begin{align}\label{rxpeed3a}
\vga
&=\ugdiv{\vectkap}{\vecta}\beqref{kas2b}\nonumber\\
&=\ugdiv{\vectkap}{(\cprod{\vectOme}{\vectu}+\cprod{\vectLam}{\vectr})}
  \beqref{main4a}\nonumber\\
&=\ugdiv{\vectkap}{(\cprod{\vectOme}{\vectu})}\nonumber\\
&=\cprod{\vectOme}{[\ugdiv{\vectkap}{\vectu}]}
  -\cprod{\vectu}{[\ugdiv{\vectkap}{\vectOme}]}
  \beqref{clc43}\nonumber\\
&=\cprod{\vectOme}{\vectkap}\beqref{clc41}
\end{align}
\begin{align}\label{rxpeed3b}
\vgb
&=\ucurl{\vecta}\beqref{kas2b}\nonumber\\
&=\ucurl{(\cprod{\vectOme}{\vectu}+\cprod{\vectLam}{\vectr})}
  \beqref{main4a}\nonumber\\
&=\ucurl{(\cprod{\vectOme}{\vectu})}\nonumber\\
&=\vectOme(\udivg{\vectu})-\ugdiv{\vectOme}{\vectu}+\ugdiv{\vectu}{\vectOme}
  -\vectu(\udivg{\vectOme})\beqref{clc25}\nonumber\\
&=\vectOme(\udivg{\vectu})-\ugdiv{\vectOme}{\vectu}\nonumber\\
&=2\vectOme\beqref{clc11}\text{ \& }\eqnref{clc41}
\end{align}
\begin{align}
\vgc
&=\ugdiv{\vectu}{\vecta}\beqref{kas2b}\nonumber\\
&=\ugdiv{\vectu}{(\cprod{\vectOme}{\vectu}+\cprod{\vectLam}{\vectr})}
  \beqref{main4a}\nonumber\\
&=\ugdiv{\vectu}{(\cprod{\vectOme}{\vectu})}\nonumber\\
&=\cprod{\vectOme}{[\ugdiv{\vectu}{\vectu}]}
  -\cprod{\vectu}{[\ugdiv{\vectu}{\vectOme}]}
  \beqref{clc43}\nonumber\\
&=\cprod{\vectOme}{\vectu}\beqref{clc41}
\label{rxpeed3c}\\
\thicksim\vgd
&=\ugdiv{\vectups}{\vecta}
=\cprod{\vectOme}{\vectups}
\label{rxpeed3d}
\end{align}
\begin{align}\label{rxpeed3e}
\vge
&=\ugrad{\gamrep}\beqref{kas2b}\nonumber\\
&=\ugrad{\left|1+\frac{\szer}{\etarep\pfreq^2}\right|^{1/2}}
  \beqref{main4a}\nonumber\\
&=0\beqref{main4c}\text{ \& }\eqnref{main4d}
\end{align}
\begin{align}\label{rxpeed3f}
\vgf
&=\vga+\cprod{\vectkap}{\vgb}\beqref{kas2b}\nonumber\\
&=\cprod{\vectOme}{\vectkap}+2\cprod{\vectkap}{\vectOme}
  \beqref{rxpeed3a}\text{ \& }\eqnref{rxpeed3b}\nonumber\\
&=\cprod{\vectkap}{\vectOme}
\end{align}
\begin{align}\label{rxpeed3g}
\vgg
&=\vchb\vgf-\vchc\vge\beqref{kas2b}\nonumber\\
&=\vchb(\cprod{\vectkap}{\vectOme})
  \beqref{rxpeed3f}\text{ \& }\eqnref{rxpeed3e}
\end{align}
\begin{align}\label{rxpeed3h}
\vgh
&=\vcha\vgf-\vchf\vge\beqref{kas2b}\nonumber\\
&=\vcha(\cprod{\vectkap}{\vectOme})
  \beqref{rxpeed3f}\text{ \& }\eqnref{rxpeed3e}
\end{align}
\begin{align}\label{rxpeed3i}
\vgi
&=\vchg\vgf-\vchh\vge\beqref{kas2b}\nonumber\\
&=\vchg(\cprod{\vectkap}{\vectOme})
  \beqref{rxpeed3f}\text{ \& }\eqnref{rxpeed3e}
\end{align}
\end{subequations}
\begin{align}\label{rxpeed4}
\vchg
&=\vche\left[\frac{\vchb}{\vchd^2}-\frac{\vthtrep\vcha}{\dragf}\right]
  \beqref{kas2a}\nonumber\\
&=\frac{(1+\vthtrep^2)^{-1/2}}{4\dragf\pfreq}\left[
  \frac{1}{(1+\vthtrep^2)\gamrep^2\pfreq^2}-\frac{\vthtrep\rhorep}{\dragf\pfreq}\right]
  \beqref{kas2a}\nonumber\\
&=\frac{(1+\vthtrep^2)^{-3/2}}{4\dragf\pfreq}\left[
  \frac{1}{\gamrep^2\pfreq^2}-\frac{\rhorep\vthtrep(1+\vthtrep^2)}{\dragf\pfreq}\right]
=\frac{1}{4\dragf^2\pfreq^2\dltvg^3}\left[\frac{\dragf}{\gamrep^2\pfreq}-\rhorep\vthtrep\dltvg^2\right]
  \beqref{rxpeed1a}\nonumber\\
&=\dltvh\beqref{rxpeed1a}.
\end{align}

\subart{Development of equation \eqnref{kas2c}}
The foregoing expressions lead to
\begin{subequations}\label{rxpeed5}
\begin{align}\label{rxpeed5a}
\ugrad{\szer}&=0
  \beqref{main4d}\text{ \& }\eqnref{main4c}
\end{align}
\begin{align}
\ugrad{\sone}\label{rxpeed5b}
&=\ugrad{\left[\frac{2\rhorep\alprep-\dragf\pfreq}{2\etarep(\szer+\etarep\pfreq^2)}\right]}
  \beqref{main4d}\nonumber\\
&=\frac{2\etarep(\szer+\etarep\pfreq^2)\ugrad{(2\rhorep\alprep-\dragf\pfreq)}
  -(2\rhorep\alprep-\dragf\pfreq)\ugrad{[2\etarep(\szer+\etarep\pfreq^2)]}}
  {[2\etarep(\szer+\etarep\pfreq^2)]^2}
  \beqref{clc32}\text{ \& }\eqnref{clc33}\nonumber\\
&=[\ugrad{(2\rhorep\alprep-\dragf\pfreq)}]/[2\etarep(\szer+\etarep\pfreq^2)]
  \beqref{main4d}\text{ \& }\eqnref{main4c}\nonumber\\
&=[2\rhorep\ugrad{\alprep}+2\alprep\ugrad{\rhorep}-\pfreq\ugrad{\dragf}]/[2\etarep(\szer+\etarep\pfreq^2)]
  \beqref{clc33}\nonumber\\
&=[2\rhorep\vgf+2\alprep\vgi-\pfreq\vgh]/[2\etarep(\szer+\etarep\pfreq^2)]
  \beqref{kas3}\nonumber\\
&=[(2\rhorep+2\alprep\vchg-\pfreq\vcha)(\cprod{\vectkap}{\vectOme})]/[2\etarep(\szer+\etarep\pfreq^2)]
  \beqref{rxpeed3}\nonumber\\
&=[(\rhorep+2\kaprep\vphid\dltvh)(\cprod{\vectkap}{\vectOme})]/
  [2\kaprep\epsvj(\kaprep\vphig+\kaprep\epsvj\pfreq^2)]
  \beqref{kas2a}, \eqnref{rxpeed4}, \eqnref{rot4}\text{ \& }\eqnref{rot5a}\nonumber\\
&=[(\rhorep+2\kaprep\vphid\dltvh)(\cprod{\unitkap}{\vectOme})]/[2\kaprep\epsvj(\vphig+\epsvj\pfreq^2)]
  \nonumber\\
&=\dltvi(\cprod{\unitkap}{\vectOme})\beqref{rxpeed1a}
\end{align}
\begin{align}\label{rxpeed5c}
\ugrad{\stwo}
&=\ugrad{(\szer\sone)}\beqref{main4d}\nonumber\\
&=\szer\ugrad{\sone}\beqref{rxpeed5a}\nonumber\\
&=\kaprep\vphig\dltvi(\cprod{\unitkap}{\vectOme})
  \beqref{rot5a}\text{ \& }\eqnref{rxpeed5b}
\end{align}
\begin{align}\label{rxpeed5d}
\ugrad{\sthr}
&=\ugrad{[\etarep\sone(\dprod{\vectLam}{\unitplz})
  +4\xirep\sone(\dprod{\vectOme}{\unitplz})]}\beqref{main4d}\nonumber\\
&=[\etarep(\dprod{\vectLam}{\unitplz})+4\xirep(\dprod{\vectOme}{\unitplz})]\ugrad{\sone}
  \beqref{main4c}\nonumber\\
&=\kaprep\dltvi(\epsvj\epsvi+8\vphie\epsvc)(\cprod{\unitkap}{\vectOme})
  \beqref{rot1a}, \eqnref{rot4}\text{ \& }\eqnref{rxpeed5b}\nonumber\\
&=\kaprep\dltvj\dltvi(\cprod{\unitkap}{\vectOme})\beqref{rxpeed1a}
\end{align}
\begin{align}\label{rxpeed5e}
\ugrad{\sfou}
&=\ugrad{[2\xirep\sone\Omerep^2+\etarep\sone(\dprod{\vectOme}{\vectLam})]}
  \beqref{main4d}\nonumber\\
&=[2\xirep\Omerep^2+\etarep(\dprod{\vectOme}{\vectLam})]\ugrad{\sone}
  \beqref{main4c}\nonumber\\
&=\kaprep\dltvi(4\vphie\Omerep^2+\epsvj\epsvg)(\cprod{\unitkap}{\vectOme})
  \beqref{rot1a}, \eqnref{rot4}\text{ \& }\eqnref{rxpeed5b}\nonumber\\
&=\kaprep\dltvk\dltvi(\cprod{\unitkap}{\vectOme})\beqref{rxpeed1a}
\end{align}
\end{subequations}
in consequence of which we have
\begin{subequations}\label{rxpeed6}
\begin{align}\label{rxpeed6a}
\vja
&=\ucurl{\vecte}\beqref{kas2c}\nonumber\\
&=\ucurl{[\stwo(\cprod{\unitplz}{\vectOme})+\sthr\vectOme-\sfou\unitplz]}
  \beqref{main4b}\nonumber\\
\begin{split}
&=\stwo(\ucurl{(\cprod{\unitplz}{\vectOme})})-\cprod{(\cprod{\unitplz}{\vectOme})}{(\ugrad{\stwo})}
  +\sthr(\ucurl{\vectOme})-\cprod{\vectOme}{(\ugrad{\sthr})}\\
  &\qquad-\sfou(\ucurl{\unitplz})+\cprod{\unitplz}{(\ugrad{\sfou})}
    \beqref{clc24}
\end{split}
\nonumber\\
&=-\cprod{(\cprod{\unitplz}{\vectOme})}{(\ugrad{\stwo})}
  -\cprod{\vectOme}{(\ugrad{\sthr})}+\cprod{\unitplz}{(\ugrad{\sfou})}
\nonumber\\
&=-\kaprep\vphig\dltvi\cprod{(\cprod{\unitplz}{\vectOme})}{(\cprod{\unitkap}{\vectOme})}
  -\kaprep\dltvj\dltvi\cprod{\vectOme}{(\cprod{\unitkap}{\vectOme})}
  +\kaprep\dltvk\dltvi\cprod{\unitplz}{(\cprod{\unitkap}{\vectOme})}
  \beqref{rxpeed5}\nonumber\\
\begin{split}
&=-\kaprep\vphig\dltvi[\unitkap\{\dprod{\vectOme}{(\cprod{\unitplz}{\vectOme})}\}
    -\vectOme\{\dprod{\unitkap}{(\cprod{\unitplz}{\vectOme})}\}]
  -\kaprep\dltvj\dltvi[\unitkap\Omerep^2-\vectOme(\dprod{\unitkap}{\vectOme})]\\
  &\qquad+\kaprep\dltvk\dltvi[\unitkap(\dprod{\vectOme}{\unitplz})-\vectOme(\dprod{\unitkap}{\unitplz})]
  \beqref{alg1}
\end{split}
\nonumber\\
&=\kaprep\vphig\dltvi\epsvj\vectOme
  -\kaprep\dltvj\dltvi(\Omerep^2\unitkap-\epsva\vectOme)
  +\kaprep\dltvk\dltvi(\epsvc\unitkap-\epsve\vectOme)
  \beqref{rot1a}
\nonumber\\
&=\kaprep\dltvi(\vphig\epsvj+\dltvj\epsva-\dltvk\epsve)\vectOme
  +\kaprep\dltvi(\dltvk\epsvc-\dltvj\Omerep^2)\unitkap\nonumber\\
&=\dltvl\vectOme+\dltvm\unitkap\beqref{rxpeed1b}
\end{align}
\begin{align*}
\vjb
&=\ugdiv{\vectu}{\vecte}\beqref{kas2c}\nonumber\\
&=\ugdiv{\vectu}{[\stwo(\cprod{\unitplz}{\vectOme})+\sthr\vectOme-\sfou\unitplz]}
  \beqref{main4b}\nonumber\\
\begin{split}
&=(\cprod{\unitplz}{\vectOme})(\dprod{\vectu}{\ugrad{\stwo}})
   +\stwo\ugdiv{\vectu}{(\cprod{\unitplz}{\vectOme})}
   +\vectOme(\dprod{\vectu}{\ugrad{\sthr}})
   +\sthr\ugdiv{\vectu}{\vectOme}
   -\unitplz(\dprod{\vectu}{\ugrad{\sfou}})\\
   &\qquad-\sfou\ugdiv{\vectu}{\unitplz}\beqref{clc42}
\end{split}
\nonumber\\
&=(\cprod{\unitplz}{\vectOme})(\dprod{\vectu}{\ugrad{\stwo}})
  +\vectOme(\dprod{\vectu}{\ugrad{\sthr}})
  -\unitplz(\dprod{\vectu}{\ugrad{\sfou}})
\end{align*}
\begin{align}\label{rxpeed6b}
&=\kaprep\vphig\dltvi(\cprod{\unitplz}{\vectOme})[\dprod{\vectu}{(\cprod{\unitkap}{\vectOme})}]
  +\kaprep\dltvj\dltvi\vectOme[\dprod{\vectu}{(\cprod{\unitkap}{\vectOme})}]
  -\kaprep\dltvk\dltvi\unitplz[\dprod{\vectu}{(\cprod{\unitkap}{\vectOme})}]
  \beqref{rxpeed5}
\nonumber\\
&=\kaprep\dltvi[\vphig(\cprod{\unitplz}{\vectOme})+\dltvj\vectOme-\dltvk\unitplz]
  [\dprod{(\cprod{\vectOme}{\vectr})}{(\cprod{\unitkap}{\vectOme})}]\beqref{main4c}
\nonumber\\
&=\kaprep\dltvi[\vphig(\cprod{\unitplz}{\vectOme})+\dltvj\vectOme-\dltvk\unitplz]
  [(\dprod{\vectOme}{\unitkap})(\dprod{\vectr}{\vectOme})-\Omerep^2(\dprod{\unitkap}{\vectr})]\beqref{alg2}
\nonumber\\
&=\kaprep\dltvi(\epsva\epsvb-\Omerep^2\epsvd)[\vphig(\cprod{\unitplz}{\vectOme})+\dltvj\vectOme-\dltvk\unitplz]
  \beqref{rot1a}\nonumber\\
&=\dltvn[\dltvj\vectOme-\dltvk\unitplz+\vphig(\cprod{\unitplz}{\vectOme})]
  \beqref{rxpeed1b}
\end{align}
\begin{align}\label{rxpeed6c}
\vjc
&=\ugdiv{\vectups}{\vecte}\beqref{kas2c}\nonumber\\
&=\ugdiv{\vectups}{[\stwo(\cprod{\unitplz}{\vectOme})+\sthr\vectOme-\sfou\unitplz]}
  \beqref{main4b}\nonumber\\
\begin{split}
&=(\cprod{\unitplz}{\vectOme})(\dprod{\vectups}{\ugrad{\stwo}})
   +\stwo\ugdiv{\vectups}{(\cprod{\unitplz}{\vectOme})}
   +\vectOme(\dprod{\vectups}{\ugrad{\sthr}})
   +\sthr\ugdiv{\vectups}{\vectOme}\\
  &\qquad-\unitplz(\dprod{\vectups}{\ugrad{\sfou}})
   -\sfou\ugdiv{\vectups}{\unitplz}\beqref{clc42}
\end{split}
\nonumber\\
&=(\cprod{\unitplz}{\vectOme})(\dprod{\vectups}{\ugrad{\stwo}})
  +\vectOme(\dprod{\vectups}{\ugrad{\sthr}})
  -\unitplz(\dprod{\vectups}{\ugrad{\sfou}})\nonumber\\
&=\kaprep\dltvi[\vphig(\cprod{\unitplz}{\vectOme})+\dltvj\vectOme-\dltvk\unitplz]
  [\dprod{\vectups}{(\cprod{\unitkap}{\vectOme})}]\beqref{rxpeed5}
\nonumber\\
&=\kaprep\dltvi\dltvf[\dltvj\vectOme-\dltvk\unitplz+\vphig(\cprod{\unitplz}{\vectOme})]
  \beqref{rxpeed2c}
\end{align}
\begin{align}\label{rxpeed6d}
\vjd
&=\vja+\rhorep\vgb\beqref{kas2c}\nonumber\\
&=\dltvl\vectOme+\dltvm\unitkap+2\rhorep\vectOme
   \beqref{rxpeed6a}\text{ \& }\eqnref{rxpeed3b}\nonumber\\
&=(2\rhorep+\dltvl)\vectOme+\dltvm\unitkap\nonumber\\
&=\dltvo\vectOme+\dltvm\unitkap\beqref{rxpeed1b}
\end{align}
\begin{align}\label{rxpeed6e}
\vje
&=\vjb+\rhorep\vgc\beqref{kas2c}\nonumber\\
&=\dltvn[\dltvj\vectOme-\dltvk\unitplz+\vphig(\cprod{\unitplz}{\vectOme})]
  +\rhorep\cprod{\vectOme}{(\cprod{\vectOme}{\vectr})}
  \beqref{rxpeed6b},\eqnref{rxpeed3c}\text{ \& }\eqnref{main4c}\nonumber\\
&=\dltvn[\dltvj\vectOme-\dltvk\unitplz+\vphig(\cprod{\unitplz}{\vectOme})]
  +\rhorep[\vectOme(\dprod{\vectOme}{\vectr})-\Omerep^2\vectr]
  \beqref{alg1}\nonumber\\
&=\dltvn[\dltvj\vectOme-\dltvk\unitplz+\vphig(\cprod{\unitplz}{\vectOme})]
  +\rhorep(\epsvb\vectOme-\Omerep^2\vectr)
  \beqref{rot1a}\nonumber\\
&=-\rhorep\Omerep^2\vectr+(\dltvn\dltvj+\rhorep\epsvb)\vectOme
  -\dltvn\dltvk\unitplz+\dltvn\vphig(\cprod{\unitplz}{\vectOme})
  \nonumber\\
&=-\rhorep\Omerep^2\vectr+\dltvp\vectOme-\dltvn\dltvk\unitplz
  +\dltvn\vphig(\cprod{\unitplz}{\vectOme})\beqref{rxpeed1b}
\end{align}
\begin{align}\label{rxpeed6f}
\vjf
&=\vjc+\rhorep\vgd\beqref{kas2c}\nonumber\\
&=\kaprep\dltvi\dltvf\dltvj\vectOme-\kaprep\dltvi\dltvf\dltvk\unitplz
  +\kaprep\dltvi\dltvf\vphig(\cprod{\unitplz}{\vectOme})+\rhorep(\cprod{\vectOme}{\vectups})
  \beqref{rxpeed6c}\text{ \& }\eqnref{rxpeed3d}\nonumber\\
\begin{split}
&=\kaprep\dltvi\dltvf\dltvj\vectOme-\kaprep\dltvi\dltvf\dltvk\unitplz
  +\kaprep\dltvi\dltvf\vphig(\cprod{\unitplz}{\vectOme})+\rhorep[\cdkt(\cprod{\vectOme}{\unitkap})
  -\rhorep\Omerep^2(\cprod{\vectOme}{\vectr})+\vphij(\cprod{\unitplz}{\vectOme})\\
  &\quad+\rhorep\epsvb\vectLam+\vphig\vphih\Omerep^2\unitplz+(\Omerep^2-\rhorep\epsvg)\vectr
  -(\epsvb+\vphig\vphih\epsvc)\vectOme]\beqref{rxpeed2a}
\end{split}
\nonumber\\
\begin{split}
&=\rhorep^2\epsvb\vectLam+\rhorep(\Omerep^2-\rhorep\epsvg)\vectr
  +[\kaprep\dltvi\dltvf\dltvj-\rhorep(\epsvb+\vphig\vphih\epsvc)]\vectOme
  +(\rhorep\vphig\vphih\Omerep^2-\kaprep\dltvi\dltvf\dltvk)\unitplz\\
  &\quad+(\kaprep\dltvi\dltvf\vphig+\rhorep\vphij)(\cprod{\unitplz}{\vectOme})
  +\rhorep\cdkt(\cprod{\vectOme}{\unitkap})-\rhorep^2\Omerep^2(\cprod{\vectOme}{\vectr})
\end{split}
\nonumber\\
\begin{split}
&=\rhorep^2\epsvb\vectLam+\dltvq\vectr+\dltvr\vectOme+\dltvs\unitplz+\dltvt(\cprod{\unitplz}{\vectOme})
  +\rhorep\cdkt(\cprod{\vectOme}{\unitkap})-\rhorep^2\Omerep^2(\cprod{\vectOme}{\vectr})
  \beqref{rxpeed1b}
\end{split}
\end{align}
\begin{align}\label{rxpeed6g}
\vjg
&=\ugrad{\taurep}\beqref{kas2c}\nonumber\\
&=\ugrad{[2\kaprep^{-2}(\rhorep\alprep-\etarep\stwo)]}\beqref{main4b}\nonumber\\
&=2\kaprep^{-2}[\rhorep\ugrad{\alprep}+\alprep\ugrad{\rhorep}-\etarep\ugrad{\stwo}]
  \beqref{clc33}\text{ \& }\eqnref{main4c}\nonumber\\
&=2\kaprep^{-2}[\rhorep\vgf+\kaprep\vphid\vgi-\kaprep^2\epsvj\vphig\dltvi(\cprod{\unitkap}{\vectOme})]
  \beqref{kas3}, \eqnref{rot4}\text{ \& }\eqnref{rxpeed5c}\nonumber\\
&=2\kaprep^{-2}(\rhorep\kaprep+\kaprep^2\vphid\dltvh-\kaprep^2\epsvj\vphig\dltvi)(\cprod{\unitkap}{\vectOme})
  \beqref{rxpeed3f}, \eqnref{rxpeed3i}\text{ \& }\eqnref{rxpeed4}\nonumber\\
&=\dltvu(\cprod{\unitkap}{\vectOme})\beqref{rxpeed1b}
\end{align}
\begin{align}\label{rxpeed6h}
\vjh
&=(\cprod{\vectu}{\vectups})/|\cprod{\vectu}{\vectups}|\beqref{kas2c}\nonumber\\
&=(\betrep\scalc^2\fcal)^{-1}(\cprod{\vectu}{\vectups})\beqref{grad5c}\nonumber\\
&=(\betrep\scalc^2\fcal)^{-1}(\dltvc\vectOme+\dltvd\vectr)\beqref{rxpeed2b}
\end{align}
\begin{align}\label{rxpeed6i}
\vji
&=\cprod{\vjh}{\vectu}\beqref{kas2c}\nonumber\\
&=(\betrep\scalc^2\fcal)^{-1}\cprod{(\dltvc\vectOme+\dltvd\vectr)}{(\cprod{\vectOme}{\vectr})}
  \beqref{rxpeed6h}\text{ \& }\eqnref{main4c}\nonumber\\
&=(\betrep\scalc^2\fcal)^{-1}[\dltvc(\dprod{\vectOme}{\vectr})\vectOme-\dltvc\Omerep^2\vectr
  +\dltvd\scalr^2\vectOme-\dltvd(\dprod{\vectOme}{\vectr})\vectr]
  \beqref{alg1}\nonumber\\
&=(\betrep\scalc^2\fcal)^{-1}[(\dltvc\epsvb+\scalr^2\dltvd)\vectOme-(\Omerep^2\dltvc+\dltvd\epsvb)\vectr]
  \beqref{rot1a}\nonumber\\
&=\efkta\vectOme-\efktb\vectr\beqref{rxpeed1c}
\end{align}
\begin{align}\label{rxpeed6j}
\vjj
&=\gcal\vji-\fcal\vectu\beqref{kas2c}\nonumber\\
&=\gcal(\efkta\vectOme-\efktb\vectr)-\fcal\vectu\beqref{rxpeed6i}.
\end{align}
\end{subequations}
Furthermore, we derive
\begin{subequations}\label{rxpeed7}
\begin{align}\label{rxpeed7a}
\dprod{\unitkap}{\vjj}
&=\dprod{\unitkap}{[\gcal(\efkta\vectOme-\efktb\vectr)-\fcal\vectu]}
  \beqref{rxpeed6j}\nonumber\\
&=\gcal\efkta(\dprod{\unitkap}{\vectOme})-\gcal\efktb(\dprod{\unitkap}{\vectr})
  -\fcal(\dprod{\unitkap}{\vectu})\nonumber\\
&=\gcal(\efkta\epsva-\efktb\epsvd)-\fcal\epsvk
  \beqref{rot1a}\text{ \& }\eqnref{main4c}\nonumber\\
&=\efktc\beqref{rxpeed1c}
\end{align}
\begin{align}\label{rxpeed7b}
\dprod{\vectc}{\vjj}=\scalc(\dprod{\unitkap}{\vjj})=\scalc\efktc,\quad
\dprod{\vectkap}{\vjj}=\kaprep(\dprod{\unitkap}{\vjj})=\kaprep\efktc\beqref{rxpeed7a}
\end{align}
\begin{align}\label{rxpeed7c}
\dprod{\vecta}{\vjj}
&=\dprod{(\cprod{\vectOme}{\vectu}+\cprod{\vectLam}{\vectr})}
  {(\gcal\efkta\vectOme-\gcal\efktb\vectr-\fcal\vectu)}
  \beqref{main4a}\text{ \& }\eqnref{rxpeed6j}\nonumber\\
\begin{split}
&=\gcal\efkta[\dprod{\vectOme}{(\cprod{\vectOme}{\vectu})}+\dprod{\vectOme}{(\cprod{\vectLam}{\vectr})}]
  -\gcal\efktb[\dprod{\vectr}{(\cprod{\vectOme}{\vectu})}+\dprod{\vectr}{(\cprod{\vectLam}{\vectr})}]\\
  &\qquad-\fcal[\dprod{\vectu}{(\cprod{\vectOme}{\vectu})}+\dprod{\vectu}{(\cprod{\vectLam}{\vectr})}]
\end{split}
\nonumber\\
&=\gcal\efkta\dprod{\vectOme}{(\cprod{\vectLam}{\vectr})}
  -\gcal\efktb\dprod{\vectr}{[\cprod{\vectOme}{(\cprod{\vectOme}{\vectr})}]}
  -\fcal[\dprod{(\cprod{\vectOme}{\vectr})}{(\cprod{\vectLam}{\vectr})}]
  \beqref{main4c}\nonumber\\
&=\gcal\efkta\dprod{\vectOme}{(\cprod{\vectLam}{\vectr})}
  -\gcal\efktb\dprod{\vectr}{[(\dprod{\vectOme}{\vectr})\vectOme-\Omerep^2\vectr]}
  -\fcal[(\dprod{\vectOme}{\vectLam})\scalr^2-(\dprod{\vectOme}{\vectr})(\dprod{\vectLam}{\vectr})]
  \beqref{alg1}\text{ \& }\eqnref{alg2}\nonumber\\
&=\gcal[\efkta\epsvm+\efktb(\Omerep^2\scalr^2-\epsvb^2)]
  -\fcal(\scalr^2\epsvg-\epsvb\epsvh)\beqref{rot1a}\nonumber\\
&=\gcal(\efkta\epsvm+\efktb\vphia^2)-\fcal\vphic\beqref{rot1b}\nonumber\\
&=\efktd\beqref{rxpeed1c}
\end{align}
\end{subequations}
\begin{align*}
\cprod{\vjj}{\vjd}
&=\cprod{(\gcal\efkta\vectOme-\gcal\efktb\vectr-\fcal\vectu)}{(\dltvo\vectOme+\dltvm\unitkap)}
  \beqref{rxpeed6j}\text{ \& }\eqnref{rxpeed6d}\nonumber\\
\begin{split}
&=\gcal\efkta\dltvo(\cprod{\vectOme}{\vectOme})+\gcal\efkta\dltvm(\cprod{\vectOme}{\unitkap})
  -\gcal\efktb\dltvo(\cprod{\vectr}{\vectOme})-\gcal\efktb\dltvm(\cprod{\vectr}{\unitkap})\\
  &\qquad-\fcal\dltvo(\cprod{\vectu}{\vectOme})-\fcal\dltvm(\cprod{\vectu}{\unitkap})
\end{split}
\nonumber\\
\begin{split}
&=-\gcal\efkta\dltvm(\cprod{\unitkap}{\vectOme})
  -\gcal\efktb\dltvo(\cprod{\vectr}{\vectOme})+\gcal\efktb\dltvm(\cprod{\unitkap}{\vectr})\\
  &\qquad+\fcal\dltvo[\cprod{\vectOme}{(\cprod{\vectOme}{\vectr})}]
  +\fcal\dltvm[\cprod{\unitkap}{(\cprod{\vectOme}{\vectr})}]\beqref{main4c}
\end{split}
\end{align*}
\begin{align}\label{rxpeed8}
\begin{split}
&=-\gcal\efkta\dltvm(\cprod{\unitkap}{\vectOme})
  -\gcal\efktb\dltvo(\cprod{\vectr}{\vectOme})+\gcal\efktb\dltvm(\cprod{\unitkap}{\vectr})\\
  &\qquad+\fcal\dltvo[(\dprod{\vectOme}{\vectr})\vectOme-\Omerep^2\vectr]
  +\fcal\dltvm[(\dprod{\unitkap}{\vectr})\vectOme-(\dprod{\unitkap}{\vectOme})\vectr]\beqref{alg1}
\end{split}
\nonumber\\
\begin{split}
&=-\gcal\efkta\dltvm(\cprod{\unitkap}{\vectOme})
  -\gcal\efktb\dltvo(\cprod{\vectr}{\vectOme})+\gcal\efktb\dltvm(\cprod{\unitkap}{\vectr})\\
  &\qquad+\fcal(\dltvo\epsvb+\dltvm\epsvd)\vectOme
  -\fcal(\dltvo\Omerep^2+\epsva\dltvm)\vectr\beqref{rot1a}
\end{split}
\nonumber\\
&=\efkte\vectOme-\efktf\vectr
  -\gcal\efkta\dltvm(\cprod{\unitkap}{\vectOme})
  -\gcal\efktb\dltvo(\cprod{\vectr}{\vectOme})
  +\gcal\efktb\dltvm(\cprod{\unitkap}{\vectr})\beqref{rxpeed1c}
\end{align}
\begin{align*}
&\xcala\vectu-\xcalb\vectups+\xcalc\vjf-\xcald\vje+(\dprod{\vectc}{\vjj})\vgh
  -(\dprod{\vectkap}{\vjj})\vjg+(\dprod{\vecta}{\vjj})\vgi+\cprod{\vjj}{\vjd}
  \nonumber\\
&=\xcala\vectu-\xcalb\vectups+\xcalc\vjf-\xcald\vje+\scalc\efktc\vgh
  -\kaprep\efktc\vjg+\efktd\vgi+\cprod{\vjj}{\vjd}\beqref{rxpeed7}\nonumber\\
\begin{split}
&=\xcala(\cprod{\vectOme}{\vectr})-\xcalb(\cdkt\unitkap+\rhorep\vecta-\vectu+\vecte)
  +\xcalc[\rhorep^2\epsvb\vectLam+\dltvq\vectr+\dltvr\vectOme+\dltvs\unitplz+\dltvt(\cprod{\unitplz}{\vectOme})\\
    &\quad+\rhorep\cdkt(\cprod{\vectOme}{\unitkap})-\rhorep^2\Omerep^2(\cprod{\vectOme}{\vectr})]
  -\xcald[-\rhorep\Omerep^2\vectr+\dltvp\vectOme-\dltvn\dltvk\unitplz+\dltvn\vphig(\cprod{\unitplz}{\vectOme})]\\
  &\quad+\scalc\efktc[\vcha(\cprod{\vectkap}{\vectOme})]-\kaprep\efktc[\dltvu(\cprod{\unitkap}{\vectOme})]
  +\efktd[\vchg(\cprod{\vectkap}{\vectOme})]+\efkte\vectOme-\efktf\vectr-\gcal\efkta\dltvm(\cprod{\unitkap}{\vectOme})\\
  &\quad-\gcal\efktb\dltvo(\cprod{\vectr}{\vectOme})+\gcal\efktb\dltvm(\cprod{\unitkap}{\vectr})
  \beqref{main4c}, \eqnref{kpath1c}, \eqnref{rxpeed6}, \eqnref{rxpeed3}\text{ \& }\eqnref{rxpeed8}
\end{split}
\nonumber\\
\begin{split}
&=-\cdkt\xcalb\unitkap+\rhorep^2\epsvb\xcalc\vectLam+(\dltvs\xcalc+\dltvn\dltvk\xcald)\unitplz
  +(\dltvq\xcalc+\rhorep\Omerep^2\xcald-\efktf)\vectr+(\dltvr\xcalc-\dltvp\xcald+\efkte)\vectOme\\
  &\quad+(\scalc\kaprep\efktc\vcha-\kaprep\efktc\dltvu+\efktd\vchg\kaprep-\gcal\efkta\dltvm
    -\rhorep\cdkt\xcalc)(\cprod{\unitkap}{\vectOme})+\gcal\efktb\dltvm(\cprod{\unitkap}{\vectr})\\
  &\quad+(\dltvt\xcalc-\dltvn\vphig\xcald)(\cprod{\unitplz}{\vectOme})
  +(\xcala-\rhorep^2\Omerep^2\xcalc+\gcal\efktb\dltvo)(\cprod{\vectOme}{\vectr})\\
  &\quad-\xcalb[\rhorep(\cprod{\vectOme}{\vectu}+\cprod{\vectLam}{\vectr})-(\cprod{\vectOme}{\vectr})
  +\vphig\vphih(\cprod{\unitplz}{\vectOme})+\vphii\vectOme-\vphij\unitplz]
  \beqref{main4}\text{ \& }\eqnref{rot5}
\end{split}
\end{align*}
\begin{align}\label{rxpeed9}
\begin{split}
&=-\cdkt\xcalb\unitkap+\rhorep^2\epsvb\xcalc\vectLam+(\dltvs\xcalc+\dltvn\dltvk\xcald+\vphij\xcalb)\unitplz
  +(\dltvq\xcalc+\rhorep\Omerep^2\xcald-\efktf)\vectr\\
  &\quad+(\dltvr\xcalc-\dltvp\xcald+\efkte-\vphii\xcalb)\vectOme
  +(\scalc\kaprep\efktc\vcha-\kaprep\efktc\dltvu+\efktd\vchg\kaprep-\gcal\efkta\dltvm
  -\rhorep\cdkt\xcalc)(\cprod{\unitkap}{\vectOme})\\
  &\quad+\gcal\efktb\dltvm(\cprod{\unitkap}{\vectr})
  -\rhorep\xcalb(\cprod{\vectLam}{\vectr})
  +(\dltvt\xcalc-\dltvn\vphig\xcald-\vphig\vphih\xcalb)(\cprod{\unitplz}{\vectOme})\\
  &\quad+(\xcala-\rhorep^2\Omerep^2\xcalc+\gcal\efktb\dltvo+\xcalb)(\cprod{\vectOme}{\vectr})
  -\rhorep\xcalb[(\dprod{\vectOme}{\vectr})\vectOme-\Omerep^2\vectr]
  \beqref{main4c}\text{ \& }\eqnref{alg1}
\end{split}
\nonumber\\
\begin{split}
&=-\cdkt\xcalb\unitkap+\rhorep^2\epsvb\xcalc\vectLam+(\vphij\xcalb+\dltvs\xcalc+\dltvn\dltvk\xcald)\unitplz
  +[\rhorep\Omerep^2(\xcalb+\xcald)+\dltvq\xcalc-\efktf]\vectr\\
  &\quad+[\efkte-(\vphii+\rhorep\epsvb)\xcalb+\dltvr\xcalc-\dltvp\xcald]\vectOme
  +[(\rhorep-\kaprep\dltvu)\efktc+\kaprep\efktd\dltvh-\gcal\efkta\dltvm
    -\rhorep\cdkt\xcalc](\cprod{\unitkap}{\vectOme})\\
  &\quad+\gcal\efktb\dltvm(\cprod{\unitkap}{\vectr})
  -\rhorep\xcalb(\cprod{\vectLam}{\vectr})
  +[\dltvt\xcalc-(\vphih\xcalb+\dltvn\xcald)\vphig](\cprod{\unitplz}{\vectOme})\\
  &\quad+(\gcal\efktb\dltvo+\xcala+\xcalb-\rhorep^2\Omerep^2\xcalc)(\cprod{\vectOme}{\vectr})
  \beqref{rot1a}, \eqnref{rxpeed4}\text{ \& }\eqnref{kas2a}.
\end{split}
\end{align}

\subart{Results of the computations}
It follows from \eqnref{rxpeed9} and \eqnref{kas7a} that
\begin{subequations}\label{rxpeed10}
\begin{equation}\label{rxpeed10a}
\begin{split}
\angus&=\frac{1}{\betrep\scalc^2\rcal^2}\biggl[\vtta\unitkap+\vttb\vectLam+\vttc\unitplz
  +\vttd\vectr+\vtte\vectOme+\vttf(\cprod{\unitkap}{\vectOme})+\vttg(\cprod{\unitkap}{\vectr})\\
  &\qquad-\vtth(\cprod{\vectLam}{\vectr})+\vtti(\cprod{\unitplz}{\vectOme})+\vttj(\cprod{\vectOme}{\vectr})
  \biggr]
\end{split}
\end{equation}
where, with $\xcala, \xcalb, \xcalc, \xcald$ defined by \eqnref{kas7b}
wherein $\rcal, \fcal, \gcal, \betrep$ are given by \eqnref{rot12b} and \eqnref{rot12d},
\begin{align}\label{rxpeed10b}
\begin{split}
\vtta&=-\cdkt\xcalb,\quad
\vttb=\rhorep^2\epsvb\xcalc,\quad
\vttc=\vphij\xcalb+\dltvs\xcalc+\dltvn\dltvk\xcald\\
\vttd&=\rhorep\Omerep^2(\xcalb+\xcald)+\dltvq\xcalc-\efktf,\quad
\vtte=\efkte-(\vphii+\rhorep\epsvb)\xcalb+\dltvr\xcalc-\dltvp\xcald\\
\vttf&=(\rhorep-\kaprep\dltvu)\efktc+\kaprep\efktd\dltvh-\gcal\efkta\dltvm-\rhorep\cdkt\xcalc,\quad
\vttg=\gcal\efktb\dltvm,\quad
\vtth=\rhorep\xcalb\\
\vtti&=\dltvt\xcalc-(\vphih\xcalb+\dltvn\xcald)\vphig,\quad
\vttj=\gcal\efktb\dltvo+\xcala+\xcalb-\rhorep^2\Omerep^2\xcalc.
\end{split}
\end{align}
Also, from \eqnref{kas1c}, we obtain
\begin{align*}
\begin{split}
\aspua
&=(\dprod{\angus}{\vecta})/\scala
=\frac{1}{\betrep\scalc^2\rcal^2\vphib}
 \dprod{\biggl[\epsvb\vectOme-\Omerep^2\vectr+\cprod{\vectLam}{\vectr}\biggr]}{}
  \biggl[\vtta\unitkap+\vttb\vectLam+\vttc\unitplz+\vttd\vectr+\vtte\vectOme+\vttf(\cprod{\unitkap}{\vectOme})\\
  &\quad+\vttg(\cprod{\unitkap}{\vectr})-\vtth(\cprod{\vectLam}{\vectr})+\vtti(\cprod{\unitplz}{\vectOme})
  +\vttj(\cprod{\vectOme}{\vectr})\biggr]
  \beqref{main4a}, \eqnref{rot2b}, \eqnref{rot1a}\text{ \& }\eqnref{rxpeed10a}
\end{split}
\end{align*}
\begin{align*}
\begin{split}
&=\frac{1}{\betrep\scalc^2\rcal^2\vphib}\biggl\{
  \vtta\epsvb(\dprod{\vectOme}{\unitkap})
  +\vttb\epsvb(\dprod{\vectOme}{\vectLam})
  +\vttc\epsvb(\dprod{\vectOme}{\unitplz})
  +\vttd\epsvb(\dprod{\vectOme}{\vectr})
  +\vtte\epsvb\Omerep^2\\
  &\quad+\vttg\epsvb[\dprod{\vectOme}{(\cprod{\unitkap}{\vectr})}]
  -\vtth\epsvb[\dprod{\vectOme}{(\cprod{\vectLam}{\vectr})}]
  -\vtta\Omerep^2(\dprod{\vectr}{\unitkap})
  -\vttb\Omerep^2(\dprod{\vectr}{\vectLam})
  -\vttc\Omerep^2(\dprod{\vectr}{\unitplz})
  -\vttd\Omerep^2\scalr^2\\
  &\quad-\vtte\Omerep^2(\dprod{\vectr}{\vectOme})
  -\vttf\Omerep^2[\dprod{\vectr}{(\cprod{\unitkap}{\vectOme})}]
  -\vtti\Omerep^2[\dprod{\vectr}{(\cprod{\unitplz}{\vectOme})}]
  +\vtta[\dprod{\unitkap}{(\cprod{\vectLam}{\vectr})}]
  +\vttc[\dprod{\unitplz}{(\cprod{\vectLam}{\vectr})}]\\
  &\quad+\vtte[\dprod{\vectOme}{(\cprod{\vectLam}{\vectr})}]
  +\vttf[\dprod{(\cprod{\vectLam}{\vectr})}{(\cprod{\unitkap}{\vectOme})}]
  +\vttg[\dprod{(\cprod{\vectLam}{\vectr})}{(\cprod{\unitkap}{\vectr})}]
  -\vtth[\dprod{(\cprod{\vectLam}{\vectr})}{(\cprod{\vectLam}{\vectr})}]\\
  &\quad+\vtti[\dprod{(\cprod{\vectLam}{\vectr})}{(\cprod{\unitplz}{\vectOme})}]
  +\vttj[\dprod{(\cprod{\vectLam}{\vectr})}{(\cprod{\vectOme}{\vectr})}]
  \biggr\}
\end{split}
\nonumber\\
\begin{split}
&=\frac{1}{\betrep\scalc^2\rcal^2\vphib}\biggl\{
  \vtta\epsvb\epsva+\vttb\epsvb\epsvg+\vttc\epsvb\epsvc+\vttd\epsvb^2+\vtte\epsvb\Omerep^2
  -\vttg\epsvb\epsvk-\vtth\epsvb\epsvm-\vtta\Omerep^2\epsvd-\vttb\Omerep^2\epsvh\\
  &\quad-\vttc\Omerep^2\epsvf-\vttd\Omerep^2\scalr^2-\vtte\Omerep^2\epsvb-\vttf\Omerep^2\epsvk-\vtti\Omerep^2\epsvl
  +\vtta\epsvn+\vttc\epsvo+\vtte\epsvm\\
  &\quad+\vttf[(\dprod{\vectLam}{\unitkap})(\dprod{\vectr}{\vectOme})
     -(\dprod{\vectLam}{\vectOme})(\dprod{\unitkap}{\vectr})]
  +\vttg[\scalr^2(\dprod{\vectLam}{\unitkap})-(\dprod{\vectLam}{\vectr})(\dprod{\vectr}{\unitkap})]
  -\vtth[\Lamrep^2\scalr^2-(\dprod{\vectLam}{\vectr})^2]\\
  &\quad+\vtti[(\dprod{\vectLam}{\unitplz})(\dprod{\vectr}{\vectOme})
     -(\dprod{\vectLam}{\vectOme})(\dprod{\unitplz}{\vectr})]
  +\vttj[\scalr^2(\dprod{\vectLam}{\vectOme})-(\dprod{\vectLam}{\vectr})(\dprod{\vectr}{\vectOme})]
  \biggr\}\beqref{rot1a}, \eqnref{alg4}\text{ \& }\eqnref{alg5}
\end{split}
\end{align*}
\begin{align}\label{rxpeed10c}
\begin{split}
&=\frac{1}{\betrep\scalc^2\rcal^2\vphib}\biggl\{
  \vtta\epsvb\epsva+\vttb\epsvb\epsvg+\vttc\epsvb\epsvc+\vttd\epsvb^2+\vtte\epsvb\Omerep^2
  -\vttg\epsvb\epsvk-\vtth\epsvb\epsvm-\vtta\Omerep^2\epsvd-\vttb\Omerep^2\epsvh\\
  &\quad-\vttc\Omerep^2\epsvf-\vttd\Omerep^2\scalr^2-\vtte\Omerep^2\epsvb-\vttf\Omerep^2\epsvk-\vtti\Omerep^2\epsvl
  +\vtta\epsvn+\vttc\epsvo+\vtte\epsvm+\vttf(\dltva\epsvb-\epsvg\epsvd)\\
  &\quad+\vttg(\scalr^2\dltva-\epsvh\epsvd)-\vtth(\Lamrep^2\scalr^2-\epsvh^2)
  +\vtti(\epsvi\epsvb-\epsvg\epsvf)+\vttj(\scalr^2\epsvg-\epsvh\epsvb)
  \biggr\}\beqref{rot1a}\text{ \& }\eqnref{rxpeed1a}
\end{split}
\nonumber\\
\begin{split}
&=\frac{1}{\betrep\scalc^2\rcal^2\vphib}\biggl\{
  \vtta(\epsvn+\epsvb\epsva-\Omerep^2\epsvd)
  +\vttb(\epsvb\epsvg-\Omerep^2\epsvh)
  +\vttc(\epsvo+\epsvb\epsvc-\Omerep^2\epsvf)\\
  &\quad+\vttd(\epsvb^2-\Omerep^2\scalr^2)+\vtte\epsvm
  -\vttf(\dltva\epsvb-\epsvg\epsvd+\Omerep^2\epsvk)
  -\vttg(\epsvb\epsvk-\epsvh\epsvd+\scalr^2\dltva)\\
  &\quad-\vtth(\epsvb\epsvm-\Lamrep^2\scalr^2-\epsvh^2)
  -\vtti(\Omerep^2\epsvl+\epsvi\epsvb-\epsvg\epsvf)
  +\vttj(\scalr^2\epsvg-\epsvh\epsvb)\biggr\}.
\end{split}
\end{align}
\end{subequations}
Equations \eqnref{rxpeed10} and \eqnref{rxpeed1} completely determine $\angus$ and $\aspua$
(as well as the drift $\fdot{\psirep}$) for a rotating observer.

\art{Apparent path of a light source}
Equation \eqnref{kpath11} can be evaluated for a rotating observer as follows.
Let us introduce the following quantities in addition to those defined by \eqnref{rot1},
\eqnref{rot12} and \eqnref{rxpeed1},
\begin{subequations}\label{rpath1}
\begin{align}\label{rpath1a}
\begin{split}
&\vsiga=\dprod{\fdot{\vectLam}}{\unitplz},\quad
\vsigb=\dprod{\fdot{\vectLam}}{\unitkap},\quad
\vsigc=\dprod{\fdot{\vectLam}}{\vectOme},\quad
\vsigd=\dprod{\fdot{\vectLam}}{\vectLam},\quad
\vsige=\dprod{\fdot{\vectLam}}{\fdot{\vectLam}},\quad
\vsigf=\dprod{\ffdot{\vectLam}}{\unitplz}\\
&\vsigg=\dprod{\ffdot{\vectLam}}{\unitkap},\quad
\vsigh=\dprod{\ffdot{\vectLam}}{\vectOme},\quad
\vsigi=\dprod{\ffdot{\vectLam}}{\vectLam},\quad
\vsigj=\dprod{\ffdot{\vectLam}}{\fdot{\vectLam}},\quad
\vsigk=\dprod{\ffdot{\vectLam}}{\ffdot{\vectLam}},\quad
\vsigl=\dprod{\fffdot{\vectLam}}{\unitplz}\\
&\qquad\qquad\qquad\vsigm=\dprod{\fffdot{\vectLam}}{\vectOme},\quad
\vsign=\dprod{\vectr}{\fdot{\vectLam}},\quad
\vsigo=\dprod{\vectr}{\ffdot{\vectLam}},\quad
\vsigp=\dprod{\fffdot{\vectLam}}{\unitkap}\\
&\qquad\qquad\qquad\qquad\vsigq=\dprod{\fffdot{\vectLam}}{\vectr},\quad
\vsigr=\dprod{\fffdot{\vectLam}}{\vectLam},\quad
\vsigs=\dprod{\fffdot{\vectLam}}{\fdot{\vectLam}}
\end{split}
\end{align}
\begin{align}\label{rpath1b}
\begin{split}
&\frkta=\dprod{\unitkap}{(\cprod{\unitplz}{\fdot{\vectLam}})},\quad
\frktb=\dprod{\unitkap}{(\cprod{\unitplz}{\ffdot{\vectLam}})},\quad
\frktc=\dprod{\unitkap}{(\cprod{\fdot{\vectLam}}{\vectr})},\quad
\frktd=\dprod{\unitkap}{(\cprod{\ffdot{\vectLam}}{\vectr})}\\
&\frkte=\dprod{\unitkap}{(\cprod{\vectLam}{\vectOme})},\quad
\frktf=\dprod{\unitkap}{(\cprod{\vectOme}{\fdot{\vectLam}})},\quad
\frktg=\dprod{\unitkap}{(\cprod{\fffdot{\vectLam}}{\vectr})},\quad
\frkth=\dprod{\fdot{\vectLam}}{(\cprod{\vectLam}{\vectr})}\\
&\frkti=\dprod{\ffdot{\vectLam}}{(\cprod{\vectLam}{\vectr})},\quad
\frktj=\dprod{\ffdot{\vectLam}}{(\cprod{\fdot{\vectLam}}{\vectr})},\quad
\frktk=\dprod{\vectLam}{(\cprod{\vectOme}{\vectr})},\quad
\frktl=\dprod{\fdot{\vectLam}}{(\cprod{\vectOme}{\vectr})}\\
&\frktm=\dprod{\ffdot{\vectLam}}{(\cprod{\vectOme}{\vectr})},\quad
\frktn=\dprod{\unitplz}{(\cprod{\fdot{\vectLam}}{\vectr})},\quad
\frkto=\dprod{\unitplz}{(\cprod{\vectOme}{\vectLam})},\quad
\frktp=\dprod{\unitplz}{(\cprod{\vectOme}{\fdot{\vectLam}})}\\
&\frktq=\dprod{\unitplz}{(\cprod{\vectOme}{\ffdot{\vectLam}})},\quad
\frktr=\dprod{\unitplz}{(\cprod{\vectLam}{\fdot{\vectLam}})},\quad
\frkts=\dprod{\unitplz}{(\cprod{\vectLam}{\ffdot{\vectLam}})},\quad
\frktt=\dprod{\unitkap}{(\cprod{\vectr}{\unitplz})}
\end{split}
\end{align}
\begin{align}\label{rpath1b2}
\begin{split}
&\frktu=\dprod{\unitkap}{(\cprod{\vectLam}{\fdot{\vectLam}})},\quad
\frktv=\dprod{\vectOme}{(\cprod{\vectLam}{\fdot{\vectLam}})},\quad
\frktw=\dprod{\ffdot{\vectLam}}{(\cprod{\unitkap}{\vectOme})},\quad
\frktx=\dprod{\ffdot{\vectLam}}{(\cprod{\unitkap}{\vectLam})}\\
&\frkty=\dprod{\ffdot{\vectLam}}{(\cprod{\unitkap}{\fdot{\vectLam}})},\quad
\frktz=\dprod{\ffdot{\vectLam}}{(\cprod{\vectOme}{\vectLam})},\quad
\frkxa=\dprod{\ffdot{\vectLam}}{(\cprod{\vectOme}{\fdot{\vectLam}})},\quad
\frkxb=\dprod{\ffdot{\vectLam}}{(\cprod{\vectr}{\unitplz})}\\
&\frkxc=\dprod{\ffdot{\vectLam}}{(\cprod{\unitplz}{\fdot{\vectLam}})},\quad
\frkxd=\dprod{\ffdot{\vectLam}}{(\cprod{\vectLam}{\fdot{\vectLam}})},\quad
\frkxe=\dprod{\vectLam}{(\cprod{\fffdot{\vectLam}}{\vectr})},\quad
\frkxf=\dprod{\unitplz}{(\cprod{\fffdot{\vectLam}}{\vectr})}\\
&\qquad\qquad\frkxg=\dprod{\vectOme}{(\cprod{\fffdot{\vectLam}}{\vectr})},\quad
\frkxh=\dprod{\fdot{\vectLam}}{(\cprod{\fffdot{\vectLam}}{\vectr})},\quad
\frkxi=\dprod{\ffdot{\vectLam}}{(\cprod{\fffdot{\vectLam}}{\vectr})}
\end{split}
\end{align}
\begin{align}\label{rpath1c}
\begin{split}
&\vrhoa=\epsvi\epsva+\epsvc\dltva-\epsvg\epsve,\quad
\vrhob=\vsiga\epsva+2\epsvi\dltva+\epsvc\vsigb-\epsve(\Lamrep^2+\vsigc),\quad
\vrhoc=\epsve(3\vsigd+\vsigh)\\
&\vrhod=\vsigf\epsva+3\vsiga\dltva+3\epsvi\vsigb+\epsvc\vsigg-\vrhoc,\quad
\vrhoe=\epsve(\Lamrep^2+\vsigc)-\vsiga\epsva-\epsvi\dltva\\
&\vrhof=\vrhoc-\vsigf\epsva-2\vsiga\dltva-\epsvi\vsigb,\quad
\vrhog=\epsve(3\vsige+4\vsigi+\vsigm)-\vsigl\epsva-3\vsigf\dltva-3\vsiga\vsigb-\epsvi\vsigg\\
&\vrhoh=2\vrhoa-\dltvb(\vphie/\epsvj),\quad
\vrhoi=\vrhoe-4\vrhoh(\vphie/\epsvj),\quad
\vrhoj=\vrhoa^2+\vphie\vrhob,\quad
\vrhok=\vphie\frkta+\dltvb\vrhoh
\end{split}
\end{align}
\begin{align}\label{rpath1d}
\begin{split}
&\vrhol=2\gamrep\pfreq^2\epsvj,\quad
\vrhom=\vrhoi-\dltvb(\vphig/\epsvj),\quad
\vrhon=\vrhof-\frac{8}{\epsvj}\biggl[\vrhoj-\frac{\vphie}{2\epsvj}\biggl\{\vrhok+\dltvb\vrhoh\biggr\}\biggr]\\
\begin{split}
&\vrhoo=\vrhog-\frac{8}{\epsvj}\biggl[\vphie\vrhod+3\vrhoa\vrhob
  -\frac{1}{2\epsvj}\biggl\{\vphie^2\frktb+6\dltvb\vrhoj+6\vphie\vrhoa\frkta\biggr\}
  +\frac{3\vrhok\vphie\dltvb}{\epsvj^2}\biggr]
\end{split}
\\
&\vrhop=\absign\vrhom/\vrhol,\quad
\vrhoq=\frac{\vrhop}{\gamrep}+\frac{2\dltvb}{\epsvj},\quad
\vrhor=\absign\frac{1}{\vrhol}\biggl\{\vrhon-\vrhom\vrhoq-\frac{\vphig\frkta}{\epsvj}\biggr\}\\
\begin{split}
&\vrhos=\absign\frac{1}{\vrhol}\biggl[\vrhoo
  -\frac{1}{\epsvj}\biggl\{\vphig\frktb+3\vrhon\dltvb+3\vrhoi\frkta\biggr\}
  +\frac{6\dltvb}{\epsvj^2}\biggl\{\vrhoi\dltvb+\vphig\biggl(\frkta-\frac{\dltvb^2}{\epsvj}\biggr)\biggr\}\\
  &\qquad-\frac{1}{\gamrep}\biggl\{\vrhoi\vrhor+2\vrhon\vrhop\biggr\}
  +\frac{\vphig}{\gamrep\epsvj}\biggl\{2\frkta\vrhop+\dltvb\vrhor\biggr\}
  +\frac{2\vrhom\vrhop\vrhoq}{\gamrep}\biggr]
\end{split}
\end{split}
\end{align}
\begin{align}\label{rpath1e}
\begin{split}
&\vrhot=3\vsign-\Omerep^2\epsvb,\quad
\vrhou=\Omerep^4-4\vsigc-3\Lamrep^2,\quad
\vrhov=4\vsigo-4\Omerep^2\epsvh-5\epsvb\epsvg\\
&\vrhow=2\Omerep^2\epsvg-\vsigh-2\vsigd,\quad
\vrhox=\frktc+\epsvb\dltva+2\epsva\epsvh-3\epsvd\epsvg-\Omerep^2\epsvk\\
&\vrhoy=\frktd+\epsva\vrhot+\epsvd\vrhou+\epsvb(\vsigb+2\frkte)
  +3(\epsvh\dltva-\Omerep^2\epsvn-\epsvg\epsvk)\\
&\vrhoz=\frktg+\epsva\vrhov+\epsvk\vrhou+5\epsvd\vrhow+\epsvb(\vsigg-5\frktf)
  +\dltva(\vrhot+3\vsign)+\epsvh(4\vsigb+5\frkte)\\
  &\qquad-6(\Omerep^2\frktc+2\epsvg\epsvn)
\end{split}
\end{align}
\begin{align}\label{rpath1f}
\begin{split}
&\ethva=\frac{\kaprep}{\gamrep^2\pfreq^2}\biggl\{\vrhox-\frac{2\vphid\vrhop}{\gamrep}\biggr\},\quad
\ethvb=\frac{\kaprep}{\gamrep^2\pfreq^2}\biggl\{\vrhoy-\frac{4\vrhox\vrhop}{\gamrep}
   -\frac{2\vphid\vrhor}{\gamrep}+\frac{6\vphid\vrhop^2}{\gamrep^2}\biggr\}\\
&\ethvc=\frac{\kaprep}{\gamrep^2\pfreq^2}\biggl\{\vrhoz-\frac{6\vrhoy\vrhop}{\gamrep}
   -\frac{6\vrhox\vrhor}{\gamrep}+\frac{18\vrhox\vrhop^2}{\gamrep^2}
   -\frac{2\vphid\vrhos}{\gamrep}+\frac{18\vphid\vrhop\vrhor}{\gamrep^2}
   -\frac{24\vphid\vrhop^3}{\gamrep^3}\biggr\}\\
&\ethvd=\dltvg^{-5}(\ethvb\dltvg^2-3\vthtrep\ethva^2),\quad
\ethve=\dltvg^{-7}[\ethvc\dltvg^4-9\vthtrep\ethva\ethvb\dltvg^2-3\ethva^3(1-4\vthtrep^2)]\\
\end{split}
\end{align}
\begin{align}\label{rpath1g}
\begin{split}
&\ethvf=\frac{\dragf\vrhop}{\gamrep}+\frac{\xcons\gamrep^2\ethva}{4\dragf}\\
&\ethvg=\dragf\biggl\{\frac{\vrhor}{\gamrep}-\frac{\vrhop^2}{\gamrep^2}\biggr\}
   +\ethvf\biggl\{\frac{\vrhop}{\gamrep}-\frac{\xcons\gamrep^2\ethva}{4\dragf^2}\biggr\}
   +\frac{\gamrep}{4\dragf}\biggl\{\frac{\gamrep\ethva^2}{\dltvg^3}+2\xcons\vrhop\ethva
     +\xcons\gamrep\ethvb\biggr\}\\
&\ethvh=\dragf\biggl\{\frac{\vrhos}{\gamrep}-\frac{3\vrhop\vrhor}{\gamrep^2}
    +\frac{2\vrhop^3}{\gamrep^3}\biggr\}
  +\ethvg\biggl\{\frac{\vrhop}{\gamrep}-\frac{\xcons\gamrep^2\ethva}{4\dragf^2}\biggr\}
  +\frac{\vrhop}{4\dragf}\biggl\{\frac{\gamrep\ethva^2}{\dltvg^3}+2\xcons\vrhop\ethva
      +\xcons\gamrep\ethvb\biggr\}\\
  &\qquad+\frac{\gamrep}{4\dragf}\biggl\{\gamrep\ethva\ethvd+\frac{3\ethva^2\vrhop}{\dltvg^3}
    +\frac{2\gamrep\ethva\ethvb}{\dltvg^3}+2\xcons\vrhor\ethva+3\xcons\vrhop\ethvb+\xcons\gamrep\ethvc\biggr\}\\
  &\qquad+2\ethvf\biggl\{\frac{\vrhor}{\gamrep}-\frac{\vrhop^2}{\gamrep^2}
    -\frac{\gamrep^2\ethva^2}{4\dragf^2\dltvg^3}-\frac{\xcons\gamrep\vrhop\ethva}{2\dragf^2}
    -\frac{\xcons\gamrep^2\ethvb}{4\dragf^2}+\frac{\xcons\gamrep^2\ethvf\ethva}{4\dragf^3}\biggr\}
\end{split}
\end{align}
\begin{align}\label{rpath1h}
\begin{split}
&\ethvi=\frac{1}{4\pfreq\dragf^2}\biggl\{\frac{\dragf\ethva}{\dltvg^3}-\xcons\ethvf\biggr\},\quad
\ethvj=\frac{1}{4\pfreq\dragf^2}\biggl[\dragf\ethvd-\xcons\ethvg
   -\frac{2\ethvf}{\dragf}\biggl\{\frac{\dragf\ethva}{\dltvg^3}-\xcons\ethvf\biggr\}\biggr]\\
&\ethvk=\frac{1}{4\pfreq\dragf^2}\biggl[\ethvf\ethvd+\dragf\ethve-\xcons\ethvh-\frac{\ethva\ethvg}{\dltvg^3}
  +\frac{2(3\ethvf^2-\dragf\ethvg)}{\dragf^2}\biggl\{\frac{\dragf\ethva}{\dltvg^3}-\xcons\ethvf\biggr\}
  -\frac{4\ethvf}{\dragf}\biggl\{\dragf\ethvd-\xcons\ethvg\biggr\}\biggr]
\end{split}
\end{align}
\begin{align}\label{rpath1i}
\begin{split}
&\ethvl=\frac{2\vphid\ethvi+2\rhorep\vrhox-\scalc\ethvf
  -2\vrhoh(\dltvb\vphig+\epsvj\vrhoi+2\epsvj\dltvb\pfreq^2)}{2\epsvj(\vphig+\epsvj\pfreq^2)}\\
&\ethvm=\frac{2\vphid\ethvj+4\vrhox\ethvi+2\rhorep\vrhoy-\scalc\ethvg}
   {2\epsvj(\vphig+\epsvj\pfreq^2)}
  -\frac{4\ethvl(\dltvb\vphig+\epsvj\vrhoi+2\epsvj\dltvb\pfreq^2)}
   {2\epsvj(\vphig+\epsvj\pfreq^2)}\\
  &\qquad-\frac{2\vphih(\vphig\frkta+\epsvj\vrhon+2\dltvb\vrhoi
    +2\dltvb^2\pfreq^2+2\epsvj\frkta\pfreq^2)}
   {2\epsvj(\vphig+\epsvj\pfreq^2)}\\
&\ethvn=\frac{2\vphid\ethvk+2\rhorep\vrhoz+6\vrhox\ethvj+6\vrhoy\ethvi-\scalc\ethvh}
    {2\epsvj(\vphig+\epsvj\pfreq^2)}
  -\frac{6\ethvm(\dltvb\vphig+\epsvj\vrhoi+2\dltvb\epsvj\pfreq^2)}
    {2\epsvj(\vphig+\epsvj\pfreq^2)}\\
  &\quad-\frac{2\vphih(\vphig\frktb+\epsvj\vrhoo+3\vrhoi\frkta+3\dltvb\vrhon
    +6\dltvb\frkta\pfreq^2+2\epsvj\frktb\pfreq^2)}
    {2\epsvj(\vphig+\epsvj\pfreq^2)}\\
  &\quad-\frac{6\ethvl(\vphig\frkta+\epsvj\vrhon+2\dltvb\vrhoi
    +2\dltvb^2\pfreq^2+2\epsvj\frkta\pfreq^2)}
    {2\epsvj(\vphig+\epsvj\pfreq^2)}
\end{split}
\end{align}
\begin{align}\label{rpath1j}
\begin{split}
&\ethvo=\vphig\ethvl+\vrhoi\vphih,\quad
\ethvp=2\vrhoi\ethvl+\vphig\ethvm+\vrhon\vphih,\quad
\ethvq=3(\vrhon\ethvl+\vrhoi\ethvm)+\vphig\ethvn+\vrhoo\vphih\\
&\ethvr=\epsvj(\epsvi\ethvl+\vphih\vsiga)+\epsvi\vphih(\dltvb+8\vphie)
  +8\epsvc(\vphie\ethvl+\vrhoa\vphih)\\
&\ethvs=\epsvj(\vphih\vsigf+\epsvi\ethvm)+\epsvi(\frkta\vphih+2\dltvb\ethvl+16\vrhoa\vphih+16\vphie\ethvl)\\
  &\qquad+2\vsiga(\epsvj\ethvl+\dltvb\vphih+4\vphie\vphih)+8\epsvc(\vphie\ethvm+\vrhob\vphih+2\vrhoa\ethvl)\\
&\ethvt=\epsvj(\vphih\vsigl+\epsvi\ethvn)+\epsvi(\frktb\vphih+3\frkta\ethvl+3\dltvb\ethvm+24\vrhob\vphih
    +48\vrhoa\ethvl+24\vphie\ethvm)\\
  &\qquad+3\vsiga(\epsvj\ethvm+\frkta\vphih+2\dltvb\ethvl+8\vrhoa\vphih+8\vphie\ethvl)
  +\vsigf(3\epsvj\ethvl+3\dltvb\vphih+8\vphie\vphih)\\
  &\qquad+8\epsvc(\vphie\ethvn+\vrhod\vphih+3\vrhob\ethvl+3\vrhoa\ethvm)
\end{split}
\end{align}
\begin{align}\label{rpath1j2}
\begin{split}
&\ethvu=4\Omerep^2(\vrhoa\vphih+\vphie\ethvl)+\epsvg(\epsvj\ethvl+\dltvb\vphih+8\vphie\vphih)
    +\epsvj\vphih(\Lamrep^2+\vsigc)\\
&\ethvv=4\Omerep^2(\vrhob\vphih+\vphie\ethvm+2\vrhoa\ethvl)+\epsvg(\epsvj\ethvm+\frkta\vphih+2\dltvb\ethvl
    +16\vrhoa\vphih+16\vphie\ethvl)\\
  &\qquad+2(\Lamrep^2+\vsigc)(\epsvj\ethvl+\dltvb\vphih+4\vphie\vphih)+\epsvj\vphih(3\vsigd+\vsigh)\\
&\ethvw=4\Omerep^2(\vrhod\vphih+\vphie\ethvn+3\vrhob\ethvl+3\vrhoa\ethvm)+\epsvj\vphih(4\vsigi+3\vsige+\vsigm)\\
  &\qquad+\epsvg(\epsvj\ethvn+\frktb\vphih+3\frkta\ethvl+3\dltvb\ethvm+24\vrhob\vphih+48\vrhoa\ethvl
     +24\vphie\ethvm)\\
  &\qquad+3(\Lamrep^2+\vsigc)(\epsvj\ethvm+\frkta\vphih+2\dltvb\ethvl+8\vrhoa\vphih+8\vphie\ethvl)\\
  &\qquad+(\vsigh+3\vsigd)(3\epsvj\ethvl+3\dltvb\vphih+8\vphie\vphih)
\end{split}
\end{align}
\begin{align}\label{rpath1k}
\begin{split}
&\ethvx=\vrhod\ethvi+\rhorep\vrhox-\dltvb\vphig\vphih-\epsvj\ethvo\\
&\ethvy=\ethvj\vrhod+2\vrhox\ethvi+\rhorep\vrhoy-\frkta\vphig\vphih-2\dltvb\ethvo-\epsvj\ethvp\\
&\ethvz=\ethvk\vphid+3\ethvj\vrhox+3\ethvi\vrhoy+\rhorep\vrhoz-\frktb\vphig\vphih-3\frkta\ethvo
   -3\dltvb\ethvp-\epsvj\ethvq
\end{split}
\end{align}
\begin{align}\label{rpath1l}
\begin{split}
&\frkya=\Omerep^2\epsva-\vsigb,\quad
\frkyb=2\epsva\epsvb-3\Omerep^2\epsvd,\quad
\frkyc=2\epsvb\dltva+3\epsvd\epsvg,\quad
\frkyd=3\Omerep^2\dltva+3\epsva\epsvg-\vsigg\\
&\frkye=\epsva\ethvo+\vphig\vphih\dltva,\quad
\frkyf=\ethvp\epsva+2\ethvo\dltva+\vphig\vphih\vsigb,\quad
\frkyg=2\epsvh\Omerep^2-3\epsvg\epsvb\\
&\frkyh=\Omerep^2(\vsign+\epsvm)+\epsvb(\Lamrep^2-\vsigc)-\epsvg\epsvh-\frkth,\quad
\frkyi=\epsvb\vrhou+\Omerep^2\vrhot\\
&\frkyj=3\Omerep^2\vphia^2-3\epsvh^2-\epsvb\vsign-\vrhot\epsvb-\vrhou\scalr^2+2\epsvb\epsvm,\quad
\frkyk=3\scalr^2\Omerep^2\epsvg+2\Omerep^2\epsvb\epsvh-5\epsvg\epsvb^2\\
&\frkyl=3\Omerep^2\epsvb\epsvg-\epsvb\vsigh-3\Omerep^4\epsvh+\Omerep^2\vsigo+3\epsvh\Lamrep^2+\epsvb\vsigd
     +\vrhot\epsvg+\vrhou\epsvh+3\epsvg\epsvm-\frkti\\
&\frkym=\Omerep^2\epsvf-\epsvb\epsvc-\epsvo,\quad
\frkyn=\ethvr\epsvg+\vphii\Lamrep^2-\ethvu\epsvi,\quad
\frkyo=\vphig\vphih(\epsvb\epsvg-\epsvh\Omerep^2)+\ethvo\epsvm\\
&\frkyp=\ethvu\epsvf-\ethvr\epsvb-\vphii\epsvh+\vphig\vphih\frkym,\quad
\frkyq=\ethvs\epsvg+2\ethvr\Lamrep^2+\vphii\vsigd-\ethvv\epsvi
\end{split}
\end{align}
\begin{align}\label{rpath1m}
\begin{split}
&\frkyr=\ethvp\epsvm+2\ethvo(\epsvb\epsvg-\Omerep^2\epsvh)+\vphig\vphih(\epsvb\vsigc-\Omerep^2\vsign+\frkth)\\
&\frkys=\ethvv\epsvf-\ethvs\epsvb-2\ethvr\epsvh-\vphii\vsign+2\ethvo\frkym,\quad
\frkyt=6\epsvh^2-\vrhot\epsvb,\quad
\frkyu=2\vrhou\epsvh+3\vrhot\epsvg\\
&\frkyv=9\epsvh\epsvg+\vrhou\epsvb,\quad
\frkyw=5\Omerep^2\vphic+4\epsvg\vphia^2-2\epsvb\frktl,\quad
\frkyx=3\epsvh^2+\epsvb\vsign+\vrhot\epsvb+\vrhou\scalr^2\\
&\frkyy=7\epsvg\vphic+\Omerep^2\epsvb(\vrhot+2\frktk+\vsign)
    +\Omerep^2(\scalr^2\vrhou+3\epsvh^2)+2\epsvb(\frkth-\Lamrep^2\epsvb)+2\scalr^2\epsvg^2\\
&\frkyz=\Omerep^4(3\epsvm-\vrhot)+\Omerep^2(\frktm-3\frkth)+\Omerep^2\epsvb(3\Lamrep^2-\vrhou-\vsigc)
    +\epsvb(\fdot{\vectLam}^2-6\epsvg^2-\vsigi)\\
    &\qquad+\epsvh(3\vsigd-2\vsigh)+3\epsvg(\vsigo-\frktl)+\vrhot\vsigc+\vrhou\vsign-\frktj
\end{split}
\end{align}
\begin{align}\label{rpath1n}
\begin{split}
&\vakpa=2\vphii\epsvh-\ethvr\epsvb,\quad
\vakpb=\ethvr(\vsigc-\Omerep^4)+\vphii(\vsigd-\Omerep^2\epsvg)+\ethvu(\Omerep^2\epsvc-\vsiga)\\
&\vakpc=\ethvu\epsvf-\ethvr\epsvb-\vphii\epsvh,\quad
\vakpd=-2\epsvc\epsvh+3\epsvf\epsvg-\epsvb\epsvi-\frktn+\Omerep^2\epsvl\\
&\vakpe=\Omerep^2(2\epsvh\ethvo+\epsvm\vphig\vphih)-\ethvo(\frktl+2\epsvb\epsvg)
    +\vphig\vphih(\Lamrep^2\epsvb-\frkth-\epsvg\epsvh)\\
&\vakpf=\Omerep^2(\epsvl\ethvo-\epsvf\ethvu+\epsvb\ethvr+\epsvh\vphii)
    +\ethvo(3\epsvf\epsvg-2\epsvc\epsvh-\epsvb\epsvi-\frktn)\\
&\vakpg=4\ethvr\epsvh-\ethvs\epsvb,\quad
\vakph=\Omerep^2\epsvl-2\epsvh\epsvc+3\epsvg\epsvf-\epsvb\epsvi-\frktn\\
&\vakpi=2\ethvo(\epsvb\Lamrep^2-\epsvg\epsvh-\frkth)
    +\vphig\vphih(-\Omerep^2\frktl+2\epsvh\vsigc-3\epsvg\vsign+\epsvb\vsigd)\\
    &\qquad+2\Omerep^2(\ethvp\epsvh+\ethvo\epsvm)-\ethvp(2\epsvg\epsvb+\frktl)\\
&\vakpj=\Omerep^2(\ethvs\epsvb+2\ethvr\epsvh+\vphii\vsign-\ethvv\epsvf+\ethvp\epsvl)
    +\ethvp(3\epsvg\epsvf-2\epsvh\epsvc-\epsvb\epsvi-\frktn)\\
&\vakpk=\vphig\vphih(3\epsvg\epsvf-2\epsvh\epsvc-\epsvb\epsvi-\frktn+\Omerep^2\epsvl)
    -\ethvs\epsvb-2\ethvr\epsvh-\vphii\vsign+\ethvv\epsvf\\
&\vakpl=\ethvs(\vsigc-\Omerep^4)+2\ethvr(\vsigd-\Omerep^2\epsvg)
    +\vphii(\fdot{\vectLam}^2-\Omerep^2\vsigc)+\ethvv(\Omerep^2\epsvc-\vsiga)
\end{split}
\end{align}
\begin{align}\label{rpath1o}
\begin{split}
&\vakpm=3\epsvh\ethvr-\vrhot\vphii,\quad
\vakpn=\ethvr\epsvb+\vphii\epsvh-\ethvu\epsvf+\ethvo\epsvl+\vphig\vphih\epsvo\\
&\vakpo=-\ethvo(3\epsvh\epsvg+\epsvb\vsigc+\vrhot\Omerep^2+\vrhou\epsvb)
    +\vphig\vphih(-3\epsvh\Lamrep^2-\epsvb\vsigd-\vrhot\epsvg-\vrhou\epsvh)\\
&\vakpp=3\Omerep^2(2\ethvr\epsvg+\vphii\Lamrep^2-\ethvu\epsvi+\ethvo\frkto)
    +\vphii(3\epsvg^2-\vsigi)+\ethvu(\vsigf-3\epsvg\epsvc)\\
    &\qquad-\vphig\vphih(3\epsvg\frkto+\frkts)-\ethvo\frktq-\ethvr\vsigh\\
&\vakpq=\ethvo(3\epsvh\epsvi+\epsvb\vsiga+\vrhot\epsvc+\vrhou\epsvf)
    +2\epsvb(\ethvu\epsvi-\ethvo\frkto-\ethvr\epsvg-\vphii\Lamrep^2)\\
    &\qquad+3\epsvg(\ethvu\epsvf-\ethvo\epsvl-\vphig\vphih\epsvo-\epsvb\ethvr-\vphii\epsvh)\\
&\vakpr=\vphig\vphih(3\epsvh\epsvi+\epsvb\vsiga+\vrhot\epsvc+\vrhou\epsvf)
    -3\Omerep^2(\ethvr\epsvb+\vphii\epsvh-\ethvu\epsvf+\ethvo\epsvl+\vphig\vphih\epsvo)\\
    &\qquad+2\epsvb(\vphii\epsvg+\ethvr\Omerep^2-\ethvu\epsvc-\vphig\vphih\frkto)\\
&\vakps=2\ethvr^2-\ethvs\vphii,\quad
\vakpt=\ethvs\ethvu-\ethvv\ethvr,\quad
\vakpu=\ethvv\vphii-2\ethvr\ethvu,\quad
\vakpv=\ethvu-\epsvc\ethvr-\epsvi\vphii\\
&\vakpw=\Omerep^2(\ethvp\ethvr-\ethvs\ethvo)+\epsvg(\ethvp\vphii-2\ethvr\ethvo)
    +2\ethvo^2\frkto-\vsigc\vphii\ethvo+\epsvc(\ethvv\ethvo-\ethvp\ethvu)\\
    &\qquad+\epsvg(2\ethvo\ethvr-\ethvs\vphig\vphih)
    +2\Lamrep^2(\ethvo\vphii-\ethvr\vphig\vphih)+\epsvi(\ethvv\vphig\vphih-2\ethvo\ethvu)\\
    &\qquad+\vphig\vphih(-\vsiga\ethvu+\vsigc\ethvr+\ethvo\frktp-\ethvp\frkto+\vphig\vphih\frktr)\\
&\vakpx=\vphig\vphih(\vsiga\vphii-\ethvv)-\epsvc(2\ethvo\ethvr-\ethvs\vphig\vphih)
    -2\epsvi(\ethvo\vphii-\ethvr\vphig\vphih)+2\ethvo\ethvu)\\
&\vakpy=\ethvo(\vsiga\vphii-\ethvv)-\epsvc(\ethvp\ethvr-\ethvs\ethvo)-\epsvi(\ethvp\vphii-2\ethvr\ethvo)
    +\ethvp\ethvu)
\end{split}
\end{align}
\begin{align}\label{rpath1p}
\begin{split}
&\parva=\ethvi-1,\quad
\parvb=\scalc\ethvf-2\ethvx,\quad
\parvc=\epsvb\parva+2\rhorep\epsvh+\ethvr,\quad
\parvd=\Omerep^2\parva+3\rhorep\epsvg\\
&\parve=\rhorep\epsvb+\vphii,\quad
\parvf=\scalc\ethvg-2\ethvy,\quad
\parvg=\scalc\ethvh-2\ethvz,\quad
\parvh=2\ethvi-1,\quad
\parvi=3\ethvi-1\\
&\parvj=\ethvj\parvb-\parvf\parva,\quad
\parvk=\parvb\parvh-\rhorep\parvf,\quad
\parvl=\parva\parvh-\rhorep\ethvj
\end{split}
\end{align}
\begin{align}\label{rpath1q}
\begin{split}
&\parvm=\parvj\epsvb+2\parvk\epsvh\vrhot-\parvf\ethvr+\parvb(\rhorep+\ethvs),\quad
\parvn=(\rhorep\parvb\vrhou-\parvj\Omerep^2-3\parvk\epsvg)\\
&\parvo=\parvk\epsvb-\parvf\vphii+\parvb(3\rhorep\epsvh+2\ethvr),\quad
\parvp=\parvf\ethvu-\parvb\ethvv\\
&\parvq=\parvl\frkyg+\parva(\rhorep\frkyi+\ethvs\Omerep^2)+\rhorep(\rhorep\frkyu+3\ethvs\epsvg)
     +\ethvr(\rhorep\vrhou-\ethvj\Omerep^2-3\parvh\epsvg)\\
&\parvr=\vakps-\parvh\vakpa+\epsvb(\parvl\epsvb-\ethvj\vphii+\parva(3\rhorep\epsvh+2\ethvr))
     +\rhorep(\vakpg+\vakpm+\rhorep\frkyt)\\
&\parvs=\rhorep\epsvb(\ethvr+\parva\epsvb+2\rhorep\epsvh)+\vphii(\ethvr+\parva\epsvb+2\rhorep\epsvh)\\
&\parvt=\vakpt+\rhorep\vrhot\ethvu+\epsvb(\ethvj\ethvu-\parva\ethvv)+2\epsvh(\parvh\ethvu-\rhorep\ethvv)
\end{split}
\end{align}
\begin{align}\label{rpath1r}
\begin{split}
&\parvu=\vphii(3\parvh\epsvg+\ethvj\Omerep^2)-\Omerep^2(\parvl\epsvb+\parva(3\rhorep\epsvh+2\ethvr))
     -\rhorep(\rhorep\frkyv+6\ethvr\epsvg+\vrhou\vphii)\\
&\parvv=\rhorep\vrhou\ethvu+\Omerep^2(\parva\ethvv-\ethvj\ethvu)+3\epsvg(\rhorep\ethvv-\parvh\ethvu),\quad
\parvw=\vakpu-\parvh\ethvu\epsvb+\rhorep(\ethvv\epsvb-3\epsvh\ethvu)\\
&\parvx=\vakpx-\ethvj\frkyp+3\parvl\vphic+\parvj\epsvd+\parvb(\rhorep\frkyb-2\ethvo\epsve)+\parva(\frkys+\rhorep\frkyj)\\
     &\qquad+\rhorep(\vakpr+\rhorep\frkyw+2\ethvo\vakph)+\vphig\vphih(\parvf\epsve-\parvh\vakpd)\\
&\parvy=\parvk\frkya-\parvj\dltva+\parvl\frkyh-\ethvj\frkyn+\parva\frkyq-\parvh\vakpb
     +\rhorep(\vakpl+\vakpp+\rhorep\frkyz+\parvb\frkyd+\parva\frkyl)\\
&\parvz=\vakpw-\ethvj\frkyo+\parva\frkyr+\parvb\frkyf-\parvf\frkye-\parvh\vakpe+\rhorep(\vakpi+\vakpo)
\end{split}
\end{align}
\end{subequations}
\begin{subequations}\label{rpathx1}
\begin{align}\label{rpathx1a}
\begin{split}
&\efkoa=\vakpy-\parvh\vakpf+\Omerep^2(\parvl\vphia^2-\parvk\epsvd)+\ethvo(\parvf\epsve-\ethvj\frkym)
     +\rhorep(\parva\frkyk-\parvb\frkyc+\vakpj+\vakpq+\rhorep\frkyy)\\
     &\qquad+\ethvp(\parva\frkym-\parvb\epsve)\\
&\efkob=\parvk\epsvd-\parvl\vphia^2-\parvh\vakpc+\vphig\vphih(\vakpv-\parvb\epsve+\parva\frkym)+\rhorep(\vakpk-\rhorep\frkyx)\\
&\efkoc=\rhorep(\parvb\epsvd-\parva\vphia^2-3\rhorep\vphic+\vakpn)\\
&\efkod=\parvx\Lamrep^2+\parvy\epsvh+\parvz\epsvi+\efkoa\epsvg+\efkob\vsigd+\efkoc\vsigi-\parvm\frkte
     -\parvn\epsvn-\parvb\parve\frktu+\parvp\dltvb-\parvq\epsvm\\
     &\qquad-\parvs\frktv-\parvt\frkto-\parve\parvd\frkth+\parvv\epsvo+\ethvu\parve\frktr\\
&\efkoe=\parvx\epsvh+\parvy\scalr^2+\parvz\epsvf+\efkoa\epsvb+\efkob\vsign+\efkoc\vsigo+\parvm\epsvk
     +\parvo\epsvn+\parvb\parve\frktc-\parvp\frktt+\parvr\epsvm\\
     &\qquad-\parvs\frktl-\parvt\epsvl-\ethvu\parve\frktn+\parvw\epsvo-\parve^2\frkth
\end{split}
\end{align}
\begin{align}\label{rpathx1b}
\begin{split}
&\efkof=\parvx\epsvi+\parvy\epsvf+\parvz+\efkoa\epsvc+\efkob\vsiga+\efkoc\vsigf-\parvm\epsvj
     +\parvn\frktt-\parvo\dltvb-\parvb\parve\frkta+\parvq\epsvl\\
     &\qquad+\parvr\frkto+\parvs\frktp+\parve\parvd\frktn-\parvu\epsvo+\parve^2\frktr\\
&\efkog=\parvx\epsvg+\parvy\epsvb+\parvz\epsvc+\efkoa\Omerep^2+\efkob\vsigc+\efkoc\vsigh-\parvn\epsvk
     +\parvo\frkte-\parvb\parve\frktf+\parvp\epsvj-\parve\parvd\frktl\\
     &\qquad-\parvu\epsvm+\parvv\epsvl+\ethvu\parve\frktp-\parvw\frkto+\parve^2\frktv\\
&\efkoh=\parvx\vsigd+\parvy\vsign+\parvz\vsiga+\efkoa\vsigc+\efkob\vsige+\efkoc\vsigj+\parvm\frktf
     -\parvn\frktc+\parvo\frktu+\parvp\frkta+\parvq\frktl\\
     &\qquad+\parvr\frktv-\parvt\frktp-\parvu\frkth+\parvv\frktn+\parvw\frktr\\
&\efkoi=\parvx\vsigi+\parvy\vsigo+\parvz\vsigf+\efkoa\vsigh+\efkob\vsigj+\efkoc\vsigk+\parvm\frktw
     -\parvn\frktd+\parvo\frktx+\parvb\parve\frkty+\parvp\frktb\\
     &\qquad+\parvq\frktm+\parvr\frktz+\parvs\frkxa-\parvt\frktq+\parve\parvd\frktj-\parvu\frkti
     +\parvv\frkxb-\ethvu\parve\frkxc+\parvw\frkts+\parve^2\frkxd
\end{split}
\end{align}
\begin{align}\label{rpathx1c}
\begin{split}
&\efkoj=-\parvx\frkte+\parvy\epsvk-\parvz\epsvj+\efkob\frktf+\efkoc\frktw
  +\parvm(\Omerep^2-\epsva^2)+\parvn(\epsvb-\epsvd\epsva)+\parvo(\epsvg-\dltva\epsva)\\
  &\qquad+\parvb\parve(\vsigc-\vsigb\epsva)+\parvp(\epsvc-\epsve\epsva)
  +\parvq(\epsva\epsvb-\epsvd\Omerep^2)+\parvr(\epsva\epsvg-\dltva\Omerep^2)\\
  &\qquad+\parvs(\epsva\vsigc-\vsigb\Omerep^2)+\parvt(\epsva\epsvc-\epsve\Omerep^2)
  -\parve\parvd(\epsvd\vsigc-\vsigb\epsvb)+\parvu(\epsvd\epsvg-\dltva\epsvb)\\
  &\qquad+\parvv(\epsvd\epsvc-\epsve\epsvb)-\ethvu\parve(\epsve\vsigc-\vsigb\epsvc)
  +\parvw(\epsve\epsvg-\dltva\epsvc)+\parve^2(\dltva\vsigc-\vsigb\epsvg)\\
&\efkok=-\parvx\epsvn+\parvz\frktt-\efkoa\epsvk-\efkob\frktc-\efkoc\frktd
  +\parvm(\epsvb-\epsva\epsvd)+\parvn(\scalr^2-\epsvd^2)+\parvo(\epsvh-\dltva\epsvd)\\
  &\qquad+\parvb\parve(\vsign-\vsigb\epsvd)+\parvp(\epsvf-\epsve\epsvd)
  +\parvq(\epsva\scalr^2-\epsvd\epsvb)+\parvr(\epsva\epsvh-\dltva\epsvb)\\
  &\qquad+\parvs(\epsva\vsign-\vsigb\epsvb)+\parvt(\epsva\epsvf-\epsve\epsvb)
  -\parve\parvd(\epsvd\vsign-\vsigb\scalr^2)+\parvu(\epsvd\epsvh-\dltva\scalr^2)\\
  &\qquad+\parvv(\epsvd\epsvf-\epsve\scalr^2)-\ethvu\parve(\epsve\vsign-\vsigb\epsvf)
  +\parvw(\epsve\epsvh-\dltva\epsvf)+\parve^2(\dltva\vsign-\vsigb\epsvh)
\end{split}
\end{align}
\begin{align}\label{rpathx1d}
\begin{split}
&\efkol=\parvy\epsvn-\parvz\dltvb+\efkoa\frkte+\efkob\frktu+\efkoc\frktx
  +\parvm(\epsvg-\epsva\dltva)+\parvn(\epsvh-\epsvd\dltva)+\parvo(\Lamrep^2-\dltva^2)\\
  &\qquad+\parvb\parve(\vsigd-\vsigb\dltva)+\parvp(\epsvi-\epsve\dltva)
  +\parvq(\epsva\epsvh-\epsvd\epsvg)+\parvr(\epsva\Lamrep^2-\dltva\epsvg)\\
  &\qquad+\parvs(\epsva\vsigd-\vsigb\epsvg)+\parvt(\epsva\epsvi-\epsve\epsvg)
  -\parve\parvd(\epsvd\vsigd-\vsigb\epsvh)+\parvu(\epsvd\Lamrep^2-\dltva\epsvh)\\
  &\qquad+\parvv(\epsvd\epsvi-\epsve\epsvh)-\ethvu\parve(\epsve\vsigd-\vsigb\epsvi)
  +\parvw(\epsve\Lamrep^2-\dltva\epsvi)+\parve^2(\dltva\vsigd-\vsigb\Lamrep^2)\\
&\efkom=-\parvx\frktu+\parvy\frktc-\parvz\frkta-\efkoa\frktf+\efkoc\frkty
  +\parvm(\vsigc-\epsva\vsigb)+\parvn(\vsign-\epsvd\vsigb)\\
  &\qquad+\parvo(\vsigd-\dltva\vsigb)+\parvb\parve(\vsige-\vsigb^2)+\parvp(\vsiga-\epsve\vsigb)
  +\parvq(\epsva\vsign-\epsvd\vsigc)+\parvr(\epsva\vsigd-\dltva\vsigc)\\
  &\qquad+\parvs(\epsva\vsige-\vsigb\vsigc)+\parvt(\epsva\vsiga-\epsve\vsigc)
  -\parve\parvd(\epsvd\vsige-\vsigb\vsign)+\parvu(\epsvd\vsigd-\dltva\vsign)\\
  &\qquad+\parvv(\epsvd\vsiga-\epsve\vsign)-\ethvu\parve(\epsve\vsige-\vsigb\vsiga)
  +\parvw(\epsve\vsigd-\dltva\vsiga)+\parve^2(\dltva\vsige-\vsigb\vsigd)
\end{split}
\end{align}
\begin{align}\label{rpathx1e}
\begin{split}
&\efkon=\parvx\dltvb-\parvy\frktt+\efkoa\epsvj+\efkob\frkta+\efkoc\frktb
  +\parvm(\epsvc-\epsva\epsve)+\parvn(\epsvf-\epsvd\epsve)+\parvo(\epsvi-\dltva\epsve)\\
  &\qquad+\parvb\parve(\vsiga-\vsigb\epsve)+\parvp(1-\epsve^2)+\parvq(\epsva\epsvf-\epsvd\epsvc)
  +\parvr(\epsva\epsvi-\dltva\epsvc)+\parvs(\epsva\vsiga-\vsigb\epsvc)\\
  &\qquad+\parvt(\epsva-\epsve\epsvc)-\parve\parvd(\epsvd\vsiga-\vsigb\epsvf)
  +\parvu(\epsvd\epsvi-\dltva\epsvf)+\parvv(\epsvd-\epsve\epsvf)\\
  &\qquad-\ethvu\parve(\epsve\vsiga-\vsigb)+\parvw(\epsve\epsvi-\dltva)+\parve^2(\dltva\vsiga-\vsigb\epsvi)\\
&\efkoo=-\parvx\epsvm+\parvz\epsvl+\efkob\frktl+\efkoc\frktm
  +\parvm(\epsva\epsvb-\Omerep^2\epsvd)+\parvn(\epsva\scalr^2-\epsvb\epsvd)\\
  &\qquad+\parvo(\epsva\epsvh-\epsvg\epsvd)+\parvb\parve(\epsva\vsign-\vsigc\epsvd)
  +\parvp(\epsva\epsvf-\epsvc\epsvd)+\parvq(\Omerep^2\scalr^2-\epsvb^2)\\
  &\qquad+\parvr(\Omerep^2\epsvh-\epsvg\epsvb)+\parvs(\Omerep^2\vsign-\vsigc\epsvb)
  +\parvt(\Omerep^2\epsvf-\epsvc\epsvb)-\parve\parvd(\epsvb\vsign-\vsigc\scalr^2)\\
  &\qquad+\parvu(\epsvb\epsvh-\epsvg\scalr^2)+\parvv(\epsvb\epsvf-\epsvc\scalr^2)
  -\ethvu\parve(\epsvc\vsign-\vsigc\epsvf)+\parvw(\epsvc\epsvh-\epsvg\epsvf)\\
  &\qquad+\parve^2(\epsvg\vsign-\vsigc\epsvh)
\end{split}
\end{align}
\begin{align}\label{rpathx1f}
\begin{split}
&\efkop=\parvy\epsvm+\parvz\frkto+\efkob\frktv+\efkoc\frktz
  +\parvm(\epsva\epsvg-\Omerep^2\dltva)+\parvn(\epsva\epsvh-\epsvb\dltva)\\
  &\qquad+\parvo(\epsva\Lamrep^2-\epsvg\dltva)+\parvb\parve(\epsva\vsigd-\vsigc\dltva)
  +\parvp(\epsva\epsvi-\epsvc\dltva)+\parvq(\Omerep^2\epsvh-\epsvb\epsvg)\\
  &\qquad+\parvr(\Omerep^2\Lamrep^2-\epsvg^2)+\parvs(\Omerep^2\vsigd-\vsigc\epsvg)
  +\parvt(\Omerep^2\epsvi-\epsvc\epsvg)-\parve\parvd(\epsvb\vsigd-\vsigc\epsvh)\\
  &\qquad+\parvu(\epsvb\Lamrep^2-\epsvg\epsvh)+\parvv(\epsvb\epsvi-\epsvc\epsvh)
  -\ethvu\parve(\epsvc\vsigd-\vsigc\epsvi)+\parvw(\epsvc\Lamrep^2-\epsvg\epsvi)\\
  &\qquad+\parve^2(\epsvg\vsigd-\vsigc\Lamrep^2)
\end{split}
\end{align}
\begin{align}\label{rpathx1f2}
\begin{split}
&\efkoq=-\parvx\frktv-\parvy\frktl+\parvz\frktp+\efkoc\frkxa
  +\parvm(\epsva\vsigc-\Omerep^2\vsigb)+\parvn(\epsva\vsign-\epsvb\vsigb)+\parvo(\epsva\vsigd-\epsvg\vsigb)\\
  &\qquad+\parvb\parve(\epsva\vsige-\vsigc\vsigb)+\parvp(\epsva\vsiga-\epsvc\vsigb)
  +\parvq(\Omerep^2\vsign-\epsvb\vsigc)+\parvr(\Omerep^2\vsigd-\epsvg\vsigc)\\
  &\qquad+\parvs(\Omerep^2\vsige-\vsigc^2)+\parvt(\Omerep^2\vsiga-\epsvc\vsigc)
  -\parve\parvd(\epsvb\vsige-\vsigc\vsign)+\parvu(\epsvb\vsigd-\epsvg\vsign)\\
  &\qquad+\parvv(\epsvb\vsiga-\epsvc\vsign)-\ethvu\parve(\epsvc\vsige-\vsigc\vsiga)
  +\parvw(\epsvc\vsigd-\epsvg\vsiga)+\parve^2(\epsvg\vsige-\vsigc\vsigd)
\end{split}
\end{align}
\begin{align}\label{rpathx1g}
\begin{split}
&\efkor=-\parvx\frkto-\parvy\epsvl-\efkob\frktp-\efkoc\frktq
  +\parvm(\epsva\epsvc-\Omerep^2\epsve)+\parvn(\epsva\epsvf-\epsvb\epsve)+\parvo(\epsva\epsvi-\epsvg\epsve)\\
  &\qquad+\parvb\parve(\epsva\vsiga-\vsigc\epsve)+\parvp(\epsva-\epsvc\epsve)
  +\parvq(\Omerep^2\epsvf-\epsvb\epsvc)+\parvr(\Omerep^2\epsvi-\epsvg\epsvc)\\
  &\qquad+\parvs(\Omerep^2\vsiga-\vsigc\epsvc)+\parvt(\Omerep^2-\epsvc^2)
  -\parve\parvd(\epsvb\vsiga-\vsigc\epsvf)+\parvu(\epsvb\epsvi-\epsvg\epsvf)\\
  &\qquad+\parvv(\epsvb-\epsvc\epsvf)-\ethvu\parve(\epsvc\vsiga-\vsigc)
  +\parvw(\epsvc\epsvi-\epsvg)+\parve^2(\epsvg\vsiga-\vsigc\epsvi)\\
&\efkos=\parvx\frkth-\parvz\frktn+\efkoa\frktl-\efkoc\frktj
  +\parvm(\epsvd\vsigc-\epsvb\vsigb)+\parvn(\epsvd\vsign-\scalr^2\vsigb)+\parvo(\epsvd\vsigd-\epsvh\vsigb)\\
  &\qquad+\parvb\parve(\epsvd\vsige-\vsign\vsigb)+\parvp(\epsvd\vsiga-\epsvf\vsigb)
  +\parvq(\epsvb\vsign-\scalr^2\vsigc)+\parvr(\epsvb\vsigd-\epsvh\vsigc)\\
  &\qquad+\parvs(\epsvb\vsige-\vsign\vsigc)+\parvt(\epsvb\vsiga-\epsvf\vsigc)
  -\parve\parvd(\scalr^2\vsige-\vsign^2)+\parvu(\scalr^2\vsigd-\epsvh\vsign)\\
  &\qquad+\parvv(\scalr^2\vsiga-\epsvf\vsign)-\ethvu\parve(\epsvf\vsige-\vsign\vsiga)
  +\parvw(\epsvf\vsigd-\epsvh\vsiga)+\parve^2(\epsvh\vsige-\vsign\vsigd)
\end{split}
\end{align}
\begin{align}\label{rpathx1h}
\begin{split}
&\efkot=-\parvz\epsvo-\efkoa\epsvm-\efkob\frkth-\efkoc\frkti
  +\parvm(\epsvd\epsvg-\epsvb\dltva)+\parvn(\epsvd\epsvh-\scalr^2\dltva)\\
  &\qquad+\parvo(\epsvd\Lamrep^2-\epsvh\dltva)+\parvb\parve(\epsvd\vsigd-\vsign\dltva)
  +\parvp(\epsvd\epsvi-\epsvf\dltva)+\parvq(\epsvb\epsvh-\scalr^2\epsvg)\\
  &\qquad+\parvr(\epsvb\Lamrep^2-\epsvh\epsvg)+\parvs(\epsvb\vsigd-\vsign\epsvg)
  +\parvt(\epsvb\epsvi-\epsvf\epsvg)-\parve\parvd(\scalr^2\vsigd-\vsign\epsvh)\\
  &\qquad+\parvu(\scalr^2\Lamrep^2-\epsvh^2)+\parvv(\scalr^2\epsvi-\epsvf\epsvh)
  -\ethvu\parve(\epsvf\vsigd-\vsign\epsvi)+\parvw(\epsvf\Lamrep^2-\epsvh\epsvi)\\
  &\qquad+\parve^2(\epsvh\vsigd-\vsign\Lamrep^2)\\
&\efkou=\parvx\epsvo+\efkoa\epsvl+\efkob\frktn+\efkoc\frkxb
  +\parvm(\epsvd\epsvc-\epsvb\epsve)+\parvn(\epsvd\epsvf-\scalr^2\epsve)+\parvo(\epsvd\epsvi-\epsvh\epsve)\\
  &\qquad+\parvb\parve(\epsvd\vsiga-\vsign\epsve)+\parvp(\epsvd-\epsvf\epsve)
  +\parvq(\epsvb\epsvf-\scalr^2\epsvc)+\parvr(\epsvb\epsvi-\epsvh\epsvc)\\
  &\qquad+\parvs(\epsvb\vsiga-\vsign\epsvc)+\parvt(\epsvb-\epsvf\epsvc)
  -\parve\parvd(\scalr^2\vsiga-\vsign\epsvf)+\parvu(\scalr^2\epsvi-\epsvh\epsvf)\\
  &\qquad+\parvv(\scalr^2-\epsvf^2)-\ethvu\parve(\epsvf\vsiga-\vsign)
  +\parvw(\epsvf\epsvi-\epsvh)+\parve^2(\epsvh\vsiga-\vsign\epsvi)
\end{split}
\end{align}
\begin{align}\label{rpathx1i}
\begin{split}
&\efkov=-\parvx\frktr+\parvy\frktn-\efkoa\frktp+\efkoc\frkxc
  +\parvm(\epsve\vsigc-\epsvc\vsigb)+\parvn(\epsve\vsign-\epsvf\vsigb)+\parvo(\epsve\vsigd-\epsvi\vsigb)\\
  &\qquad+\parvb\parve(\epsve\vsige-\vsiga\vsigb)+\parvp(\epsve\vsiga-\vsigb)
  +\parvq(\epsvc\vsign-\epsvf\vsigc)+\parvr(\epsvc\vsigd-\epsvi\vsigc)\\
  &\qquad+\parvs(\epsvc\vsige-\vsiga\vsigc)+\parvt(\epsvc\vsiga-\vsigc)
  -\parve\parvd(\epsvf\vsige-\vsiga\vsign)+\parvu(\epsvf\vsigd-\epsvi\vsign)\\
  &\qquad+\parvv(\epsvf\vsiga-\vsign)-\ethvu\parve(\vsige-\vsiga^2)
  +\parvw(\vsigd-\epsvi\vsiga)+\parve^2(\epsvi\vsige-\vsiga\vsigd)\\
&\efkow=\parvy\epsvo-\efkoa\frkto+\efkob\frktr+\efkoc\frkts
  +\parvm(\epsve\epsvg-\epsvc\dltva)+\parvn(\epsve\epsvh-\epsvf\dltva)+\parvo(\epsve\Lamrep^2-\epsvi\dltva)\\
  &\qquad+\parvb\parve(\epsve\vsigd-\vsiga\dltva)+\parvp(\epsve\epsvi-\dltva)
  +\parvq(\epsvc\epsvh-\epsvf\epsvg)+\parvr(\epsvc\Lamrep^2-\epsvi\epsvg)\\
  &\qquad+\parvs(\epsvc\vsigd-\vsiga\epsvg)+\parvt(\epsvc\epsvi-\epsvg)
  -\parve\parvd(\epsvf\vsigd-\vsiga\epsvh)+\parvu(\epsvf\Lamrep^2-\epsvi\epsvh)\\
  &\qquad+\parvv(\epsvf\epsvi-\epsvh)-\ethvu\parve(\vsigd-\vsiga\epsvi)
  +\parvw(\Lamrep^2-\epsvi^2)+\parve^2(\epsvi\vsigd-\vsiga\Lamrep^2)
\end{split}
\end{align}
\begin{align}\label{rpathx1j}
\begin{split}
&\efkox=-\parvy\frkth+\parvz\frktr+\efkoa\frktv+\efkoc\frkxd
  +\parvm(\dltva\vsigc-\epsvg\vsigb)+\parvn(\dltva\vsign-\epsvh\vsigb)+\parvo(\dltva\vsigd-\Lamrep^2\vsigb)\\
  &\qquad+\parvb\parve(\dltva\vsige-\vsigd\vsigb)+\parvp(\dltva\vsiga-\epsvi\vsigb)
  +\parvq(\epsvg\vsign-\epsvh\vsigc)+\parvr(\epsvg\vsigd-\Lamrep^2\vsigc)\\
  &\qquad+\parvs(\epsvg\vsige-\vsigd\vsigc)+\parvt(\epsvg\vsiga-\epsvi\vsigc)
  -\parve\parvd(\epsvh\vsige-\vsigd\vsign)+\parvu(\epsvh\vsigd-\Lamrep^2\vsign)\\
  &\qquad+\parvv(\epsvh\vsiga-\epsvi\vsign)-\ethvu\parve(\epsvi\vsige-\vsigd\vsiga)
  +\parvw(\epsvi\vsigd-\Lamrep^2\vsiga)+\parve^2(\Lamrep^2\vsige-\vsigd^2)\\
&\efkoy=\parvx\dltva+\parvy\epsvd+\parvz\epsve+\efkoa\epsva+\efkob\vsigb+\efkoc\vsigg+\parvq\epsvk
  -\parvr\frkte+\parvs\frktf-\parvt\epsvj+\parve\parvd\frktc\\
  &\qquad-\parvu\epsvn+\parvv\frktt-\ethvu\parve\frkta+\parvw\dltvb+\parve^2\frktu
\end{split}
\end{align}
\begin{align}\label{rpathx1j2}
\begin{split}
&\efkoz=\parvx\frkts-\parvy\frkxb+\efkoa\frktq+\efkob\frkxc
  +\parvm(\vsigg\epsvc-\vsigh\epsve)+\parvn(\vsigg\epsvf-\vsigo\epsve)+\parvo(\vsigg\epsvi-\vsigi\epsve)\\
  &\qquad+\parvb\parve(\vsigg\vsiga-\vsigj\epsve)+\parvp(\vsigg-\vsigf\epsve)
  +\parvq(\vsigh\epsvf-\vsigo\epsvc)+\parvr(\vsigh\epsvi-\vsigi\epsvc)\\
  &\qquad+\parvs(\vsigh\vsiga-\vsigj\epsvc)+\parvt(\vsigh-\vsigf\epsvc)
  -\parve\parvd(\vsigo\vsiga-\vsigj\epsvf)+\parvu(\vsigo\epsvi-\vsigi\epsvf)\\
  &\qquad+\parvv(\vsigo-\vsigf\epsvf)-\ethvu\parve(\vsigf\vsiga-\vsigj)
  +\parvw(\vsigf\epsvi-\vsigi)+\parve^2(\vsigi\vsiga-\vsigj\epsvi)
\end{split}
\end{align}
\begin{align}\label{rpathx1k}
\begin{split}
&\efkca=-\parvx\frkti+\parvz\frkxb-\efkoa\frktm-\efkob\frktj
  +\parvm(\vsigg\epsvb-\vsigh\epsvd)+\parvn(\vsigg\scalr^2-\vsigo\epsvd)+\parvo(\vsigg\epsvh-\vsigi\epsvd)\\
  &\qquad+\parvb\parve(\vsigg\vsign-\vsigj\epsvd)+\parvp(\vsigg\epsvf-\vsigf\epsvd)
  +\parvq(\vsigh\scalr^2-\vsigo\epsvb)+\parvr(\vsigh\epsvh-\vsigi\epsvb)\\
  &\qquad+\parvs(\vsigh\vsign-\vsigj\epsvb)+\parvt(\vsigh\epsvf-\vsigf\epsvb)
  -\parve\parvd(\vsigo\vsign-\vsigj\scalr^2)+\parvu(\vsigo\epsvh-\vsigi\scalr^2)\\
  &\qquad+\parvv(\vsigo\epsvf-\vsigf\scalr^2)-\ethvu\parve(\vsigf\vsign-\vsigj\epsvf)
  +\parvw(\vsigf\epsvh-\vsigi\epsvf)+\parve^2(\vsigi\vsign-\vsigj\epsvh)\\
&\efkcb=\parvx\frkxe+\parvz\frkxf+\efkoa\frkxg+\efkob\frkxh+\efkoc\frkxi
  +\parvm(\vsigp\epsvb-\vsigm\epsvd)+\parvn(\vsigp\scalr^2-\vsigq\epsvd)\\
  &\qquad+\parvo(\vsigp\epsvh-\vsigr\epsvd)+\parvb\parve(\vsigp\vsign-\vsigs\epsvd)
  +\parvp(\vsigp\epsvf-\vsigl\epsvd)+\parvq(\vsigm\scalr^2-\vsigq\epsvb)\\
  &\qquad+\parvr(\vsigm\epsvh-\vsigr\epsvb)+\parvs(\vsigm\vsign-\vsigs\epsvb)
  +\parvt(\vsigm\epsvf-\vsigl\epsvb)-\parve\parvd(\vsigq\vsign-\vsigs\scalr^2)\\
  &\qquad+\parvu(\vsigq\epsvh-\vsigr\scalr^2)+\parvv(\vsigq\epsvf-\vsigl\scalr^2)
  -\ethvu\parve(\vsigl\vsign-\vsigs\epsvf)+\parvw(\vsigl\epsvh-\vsigr\epsvf)\\
  &\qquad+\parve^2(\vsigr\vsign-\vsigs\epsvh)
\end{split}
\end{align}
\begin{align}\label{rpathx1l}
\begin{split}
&\efkcc=[\parvx\efkod+\parvy\efkoe+\parvz\efkof+\efkoa\efkog+\efkob\efkoh+\efkoc\efkoi+\parvm\efkoj
  +\parvn\efkok+\parvo\efkol+\parvb\parve\efkom+\parvp\efkon\\
  &\qquad+\parvq\efkoo+\parvr\efkop+\parvs\efkoq+\parvt\efkor-\parve\parvd\efkos+\parvu\efkot
  +\parvv\efkou-\ethvu\parve\efkov+\parvw\efkow+\parve^2\efkox]^{1/2}
\end{split}
\end{align}
\begin{align}\label{rpathx1m}
\begin{split}
&\efkcd=\parvb^2+2\parvc\parvb\epsva-2\parvd\parvb\epsvd+2\parve\parvb\dltva-2\ethvu\parvb\epsve
  +2\parva\parvb\epsvn+2\rhorep\parvb\frktc-2\rhorep\Omerep^2\parvb\epsvk\\
  &\qquad+2\ethvo\parvb\epsvj+2\vphig\vphih\parvb\dltvb+\parvc^2\Omerep^2-2\parvd\parvc\epsvb+2\parve\parvc\epsvg
  -2\ethvu\parvc\epsvc+2\parva\parvc\epsvm\\
  &\qquad-2\rhorep\parvc\frktl-2\vphig\vphih\parvc\frkto+\parvd^2\scalr^2-2\parve\parvd\epsvh+2\ethvu\parvd\epsvf
  -2\ethvo\parvd\epsvl-2\parvd\vphig\vphih\epsvo\\
  &\qquad+\parve^2\Lamrep^2-2\ethvu\parve\epsvi-2\rhorep\parve\frkth+2\rhorep\Omerep^2\parve\epsvm
  +2\ethvo\parve\frkto+\ethvu^2\\
  &\qquad-2\parva\ethvu\epsvo-2\rhorep\ethvu\frktn+2\rhorep\Omerep^2\ethvu\epsvl\\
&\efkce=[\efkcd+\parva^2(\Lamrep^2\scalr^2-\epsvh^2)+2\rhorep\parva(\vsigd\scalr^2-\epsvh\vsign)
  -2\rhorep\Omerep^2\parva(\epsvg\scalr^2-\epsvh\epsvb)\\
  &\qquad+2\ethvo\parva(\epsvi\epsvb-\epsvg\epsvf)+2\vphig\vphih\parva(\epsvi\epsvh-\Lamrep^2\epsvf)
  -2\rhorep^2\Omerep^2(\vsigc\scalr^2-\epsvb\vsign)\\
  &\qquad+\rhorep^2\Omerep^4(\Omerep^2\scalr^2-\epsvb^2)-2\ethvo\rhorep\Omerep^2(\epsvc\epsvb-\Omerep^2\epsvf)
  -2\vphig\vphih\rhorep\Omerep^2(\epsvc\epsvh-\epsvg\epsvf)\\
  &\qquad+\rhorep^2(\vsige\scalr^2-\vsign^2)+2\ethvo\rhorep(\vsiga\epsvb-\vsigc\epsvf)
  +2\vphig\vphih\rhorep(\vsiga\epsvh-\vsigd\epsvf)\\
  &\qquad+\ethvo^2(\Omerep^2-\epsvc^2)+2\vphig\vphih\ethvo(\epsvg-\epsvi\epsvc)
  +\vphig^2\vphih^2(\Lamrep^2-\epsvi^2)]^{1/2}
\end{split}
\end{align}
\begin{align}\label{rpathx1n}
\begin{split}
\efkcf&=\parvg\efkoy+\ethvk(\epsvb\efkog-\Omerep^2\efkoe-\efkot)
  +3\ethvj(2\epsvh\efkog-3\epsvg\efkoe-\efkos-\Omerep^2\efkoo+\epsvb\efkod)\\
  &\quad+\parvi(3\epsvh\efkod+\epsvb\efkoh+\vrhot\efkog+\vrhou\efkoe-2\epsvb\efkop+3\Omerep^2\efkot-3\epsvg\efkoo+\efkca)\\
  &\quad-\ethvq\efkor+3\ethvp\efkow+3\ethvo\efkov-\vphig\vphih\efkoz
  +\ethvt\efkog+3\ethvs\efkod+3\ethvr\efkoh+\vphii\efkoi-\ethvw\efkof\\
  &\quad+\rhorep[(\vrhot+3\vsign)\efkod+4\epsvh\efkoh+\epsvb\efkoi+\vrhov\efkog+5\vrhow\efkoe
    -5\epsvb\efkoq+6\Omerep^2\efkos-5\epsvh\efkop\\
  &\qquad+12\epsvg\efkot+\vrhou\efkoo+\efkcb].
\end{split}
\end{align}
\end{subequations}

\subart{Development of equation \eqnref{main4}}
Bearing the foregoing quantities in mind, we derive
\begin{subequations}\label{rpath2}
\begin{align}\label{rpath2a}
\fdot{\xirep}
&=\dif{[2(\dprod{\vectOme}{\unitplz})(\dprod{\vectOme}{\vectkap})
  -(\dprod{\vectOme}{\vectOme})(\dprod{\vectkap}{\unitplz})]}\beqref{main4c}\nonumber\\
&=2(\dprod{\vectLam}{\unitplz})(\dprod{\vectOme}{\vectkap})
  +2(\dprod{\vectOme}{\unitplz})(\dprod{\vectLam}{\vectkap})
  -2(\dprod{\vectOme}{\vectLam})(\dprod{\vectkap}{\unitplz})
  \nonumber\\
&=2\kaprep(\epsvi\epsva+\epsvc\dltva-\epsvg\epsve)
  \beqref{rot1a}\text{ \& }\eqnref{rxpeed1a}\nonumber\\
&=2\kaprep\vrhoa\beqref{rpath1c}
\end{align}
\begin{align}\label{rpath2b}
\ffdot{\xirep}
&=\dif{[2(\dprod{\vectLam}{\unitplz})(\dprod{\vectOme}{\vectkap})
  +2(\dprod{\vectOme}{\unitplz})(\dprod{\vectLam}{\vectkap})
  -2(\dprod{\vectOme}{\vectLam})(\dprod{\vectkap}{\unitplz})]}
  \beqref{rpath2a}\nonumber\\
\begin{split}
&=2(\dprod{\fdot{\vectLam}}{\unitplz})(\dprod{\vectOme}{\vectkap})
  +2(\dprod{\vectLam}{\unitplz})(\dprod{\vectLam}{\vectkap})
  +2(\dprod{\vectLam}{\unitplz})(\dprod{\vectLam}{\vectkap})
  +2(\dprod{\vectOme}{\unitplz})(\dprod{\fdot{\vectLam}}{\vectkap})\\
  &\qquad-2(\dprod{\vectkap}{\unitplz})[(\dprod{\vectLam}{\vectLam})+(\dprod{\vectOme}{\fdot{\vectLam}})]
\end{split}
\nonumber\\
&=2(\dprod{\fdot{\vectLam}}{\unitplz})(\dprod{\vectOme}{\vectkap})
  +4(\dprod{\vectLam}{\unitplz})(\dprod{\vectLam}{\vectkap})
  +2(\dprod{\vectOme}{\unitplz})(\dprod{\fdot{\vectLam}}{\vectkap})
  -2(\dprod{\vectkap}{\unitplz})[(\dprod{\vectLam}{\vectLam})+(\dprod{\vectOme}{\fdot{\vectLam}})]
\nonumber\\
&=2\kaprep[\vsiga\epsva+2\epsvi\dltva+\epsvc\vsigb-\epsve(\Lamrep^2+\vsigc)]
  \beqref{rot1a}, \eqnref{rxpeed1a}\text{ \& }\eqnref{rpath1a}\nonumber\\
&=2\kaprep\vrhob\beqref{rpath1c}
\end{align}
\begin{align}\label{rpath2c}
\begin{split}
\fffdot{\xirep}
&=[2(\dprod{\fdot{\vectLam}}{\unitplz})(\dprod{\vectOme}{\vectkap})
  +4(\dprod{\vectLam}{\unitplz})(\dprod{\vectLam}{\vectkap})
  +2(\dprod{\vectOme}{\unitplz})(\dprod{\fdot{\vectLam}}{\vectkap})\\
  &\qquad-2(\dprod{\vectkap}{\unitplz})(\dprod{\vectLam}{\vectLam})
  -2(\dprod{\vectkap}{\unitplz})(\dprod{\vectOme}{\fdot{\vectLam}})\dif{]}
  \beqref{rpath2b}
\end{split}
\nonumber\\
\begin{split}
&=2(\dprod{\ffdot{\vectLam}}{\unitplz})(\dprod{\vectOme}{\vectkap})
  +2(\dprod{\fdot{\vectLam}}{\unitplz})(\dprod{\vectLam}{\vectkap})
  +4(\dprod{\fdot{\vectLam}}{\unitplz})(\dprod{\vectLam}{\vectkap})
  +4(\dprod{\vectLam}{\unitplz})(\dprod{\fdot{\vectLam}}{\vectkap})\\
  &\quad+2(\dprod{\vectLam}{\unitplz})(\dprod{\fdot{\vectLam}}{\vectkap})
  +2(\dprod{\vectOme}{\unitplz})(\dprod{\ffdot{\vectLam}}{\vectkap})
  -4(\dprod{\vectkap}{\unitplz})(\dprod{\vectLam}{\fdot{\vectLam}})
  -2(\dprod{\vectkap}{\unitplz})[(\dprod{\vectLam}{\fdot{\vectLam}})+(\dprod{\vectOme}{\ffdot{\vectLam}})]
\end{split}
\nonumber\\
\begin{split}
&=2(\dprod{\ffdot{\vectLam}}{\unitplz})(\dprod{\vectOme}{\vectkap})
  +6(\dprod{\fdot{\vectLam}}{\unitplz})(\dprod{\vectLam}{\vectkap})
  +6(\dprod{\vectLam}{\unitplz})(\dprod{\fdot{\vectLam}}{\vectkap})\\
  &\quad+2(\dprod{\vectOme}{\unitplz})(\dprod{\ffdot{\vectLam}}{\vectkap})
  -6(\dprod{\vectkap}{\unitplz})(\dprod{\vectLam}{\fdot{\vectLam}})
  -2(\dprod{\vectkap}{\unitplz})(\dprod{\vectOme}{\ffdot{\vectLam}})
\end{split}
\nonumber\\
&=2\kaprep[\vsigf\epsva+3\vsiga\dltva+3\epsvi\vsigb+\epsvc\vsigg-\epsve(3\vsigd+\vsigh)]
  \beqref{rot1a}, \eqnref{rxpeed1a}\text{ \& }\eqnref{rpath1a}\nonumber\\
&=2\kaprep(\vsigf\epsva+3\vsiga\dltva+3\epsvi\vsigb+\epsvc\vsigg-\vrhoc)
=2\kaprep\vrhod\beqref{rpath1c}
\end{align}
\end{subequations}
\begin{subequations}\label{rpath3}
\begin{align}\label{rpath3a}
\fdot{\zetarep}
&=\dif{[(\dprod{\vectkap}{\unitplz})(\dprod{\vectOme}{\vectLam})
  -(\dprod{\vectLam}{\unitplz})(\dprod{\vectOme}{\vectkap})]}
  \beqref{main4c}\nonumber\\
\begin{split}
&=(\dprod{\vectkap}{\unitplz})[(\dprod{\vectLam}{\vectLam})+(\dprod{\vectOme}{\fdot{\vectLam}})]
  -(\dprod{\fdot{\vectLam}}{\unitplz})(\dprod{\vectOme}{\vectkap})
  -(\dprod{\vectLam}{\unitplz})(\dprod{\vectLam}{\vectkap})
\end{split}
\nonumber\\
&=\kaprep[\epsve(\Lamrep^2+\vsigc)-\vsiga\epsva-\epsvi\dltva]
  \beqref{rot1a}, \eqnref{rxpeed1a}\text{ \& }\eqnref{rpath1a}\nonumber\\
&=\kaprep\vrhoe\beqref{rpath1c}
\end{align}
\begin{align}\label{rpath3b}
\ffdot{\zetarep}
&=\dif{[(\dprod{\vectkap}{\unitplz})(\dprod{\vectLam}{\vectLam})
  +(\dprod{\vectkap}{\unitplz})(\dprod{\vectOme}{\fdot{\vectLam}})
  -(\dprod{\fdot{\vectLam}}{\unitplz})(\dprod{\vectOme}{\vectkap})
  -(\dprod{\vectLam}{\unitplz})(\dprod{\vectLam}{\vectkap})]}
  \beqref{rpath3a}\nonumber\\
\begin{split}
&=2(\dprod{\vectkap}{\unitplz})(\dprod{\vectLam}{\fdot{\vectLam}})
  +(\dprod{\vectkap}{\unitplz})[(\dprod{\vectLam}{\fdot{\vectLam}})+(\dprod{\vectOme}{\ffdot{\vectLam}})]
  -(\dprod{\ffdot{\vectLam}}{\unitplz})(\dprod{\vectOme}{\vectkap})\\
  &\qquad-(\dprod{\fdot{\vectLam}}{\unitplz})(\dprod{\vectLam}{\vectkap})
  -(\dprod{\fdot{\vectLam}}{\unitplz})(\dprod{\vectLam}{\vectkap})
  -(\dprod{\vectLam}{\unitplz})(\dprod{\fdot{\vectLam}}{\vectkap})
\end{split}
\nonumber\\
\begin{split}
&=(\dprod{\vectkap}{\unitplz})[3(\dprod{\vectLam}{\fdot{\vectLam}})+(\dprod{\vectOme}{\ffdot{\vectLam}})]
  -(\dprod{\ffdot{\vectLam}}{\unitplz})(\dprod{\vectOme}{\vectkap})
  -2(\dprod{\fdot{\vectLam}}{\unitplz})(\dprod{\vectLam}{\vectkap})
  -(\dprod{\vectLam}{\unitplz})(\dprod{\fdot{\vectLam}}{\vectkap})
\end{split}
\nonumber\\
&=\kaprep[\epsve(3\vsigd+\vsigh)-\vsigf\epsva-2\vsiga\dltva-\epsvi\vsigb]
  \beqref{rot1a}, \eqnref{rxpeed1a}\text{ \& }\eqnref{rpath1a}\nonumber\\
&=\kaprep(\vrhoc-\vsigf\epsva-2\vsiga\dltva-\epsvi\vsigb)
=\kaprep\vrhof\beqref{rpath1c}
\end{align}
\begin{align*}
\begin{split}
\fffdot{\zetarep}
&=[3(\dprod{\vectkap}{\unitplz})(\dprod{\vectLam}{\fdot{\vectLam}})
  +(\dprod{\vectkap}{\unitplz})(\dprod{\vectOme}{\ffdot{\vectLam}})
  -(\dprod{\ffdot{\vectLam}}{\unitplz})(\dprod{\vectOme}{\vectkap})\\
  &\qquad-2(\dprod{\fdot{\vectLam}}{\unitplz})(\dprod{\vectLam}{\vectkap})
  -(\dprod{\vectLam}{\unitplz})(\dprod{\fdot{\vectLam}}{\vectkap})\dif{]}
  \beqref{rpath3b}
\end{split}
\nonumber\\
\begin{split}
&=3(\dprod{\vectkap}{\unitplz})[(\dprod{\fdot{\vectLam}}{\fdot{\vectLam}})+(\dprod{\vectLam}{\ffdot{\vectLam}})]
  +(\dprod{\vectkap}{\unitplz})[(\dprod{\vectLam}{\ffdot{\vectLam}})+(\dprod{\vectOme}{\fffdot{\vectLam}})]
  -(\dprod{\fffdot{\vectLam}}{\unitplz})(\dprod{\vectOme}{\vectkap})
  -(\dprod{\ffdot{\vectLam}}{\unitplz})(\dprod{\vectLam}{\vectkap})\\
  &\quad-2(\dprod{\ffdot{\vectLam}}{\unitplz})(\dprod{\vectLam}{\vectkap})
  -2(\dprod{\fdot{\vectLam}}{\unitplz})(\dprod{\fdot{\vectLam}}{\vectkap})
  -(\dprod{\fdot{\vectLam}}{\unitplz})(\dprod{\fdot{\vectLam}}{\vectkap})
  -(\dprod{\vectLam}{\unitplz})(\dprod{\ffdot{\vectLam}}{\vectkap})
\end{split}
\end{align*}
\begin{align}\label{rpath3c}
\begin{split}
&=(\dprod{\vectkap}{\unitplz})[3(\dprod{\fdot{\vectLam}}{\fdot{\vectLam}})
    +4(\dprod{\vectLam}{\ffdot{\vectLam}})+(\dprod{\vectOme}{\fffdot{\vectLam}})]
  -(\dprod{\fffdot{\vectLam}}{\unitplz})(\dprod{\vectOme}{\vectkap})\\
  &\quad-3(\dprod{\ffdot{\vectLam}}{\unitplz})(\dprod{\vectLam}{\vectkap})
  -3(\dprod{\fdot{\vectLam}}{\unitplz})(\dprod{\fdot{\vectLam}}{\vectkap})
  -(\dprod{\vectLam}{\unitplz})(\dprod{\ffdot{\vectLam}}{\vectkap})
\end{split}
\nonumber\\
&=\kaprep[\epsve(3\vsige+4\vsigi+\vsigm)-\vsigl\epsva-3\vsigf\dltva-3\vsiga\vsigb-\epsvi\vsigg]
  \beqref{rot1a}, \eqnref{rxpeed1a}\text{ \& }\eqnref{rpath1a}\nonumber\\
&=\kaprep\vrhog\beqref{rpath1c}.
\end{align}
\end{subequations}
From \eqnref{main4c}, \eqnref{rxpeed1a} and \eqnref{rpath1b}, we have
\begin{equation}\label{rpath4}
\fdot{\etarep}=\dprod{\vectkap}{(\cprod{\unitplz}{\vectLam})}=\kaprep\dltvb,\quad
\ffdot{\etarep}=\dprod{\vectkap}{(\cprod{\unitplz}{\fdot{\vectLam}})}=\kaprep\frkta,\quad
\fffdot{\etarep}=\dprod{\vectkap}{(\cprod{\unitplz}{\ffdot{\vectLam}})}=\kaprep\frktb
\end{equation}
whilst from \eqnref{main4d}, we derive
\begin{subequations}\label{rpath5}
\begin{align}\label{rpath5a}
\fdot{\szer}
&=\dif{\biggl[\zetarep-\frac{\xirep^2}{\etarep}\biggr]}
=\fdot{\zetarep}-\frac{2\xirep\fdot{\xirep}}{\etarep}
  +\frac{\xirep^2\fdot{\etarep}}{\etarep^2}
=\fdot{\zetarep}+\frac{\xirep}{\etarep}\biggl[-2\fdot{\xirep}
  +\frac{\xirep\fdot{\etarep}}{\etarep}\biggr]\nonumber\\
&=\kaprep\vrhoe+\frac{2\kaprep\vphie}{\kaprep\epsvj}\biggl[
    -4\kaprep\vrhoa+\frac{2\kaprep^2\vphie\dltvb}{\kaprep\epsvj}\biggr]
  \beqref{rot4}, \eqnref{rpath2a}, \eqnref{rpath3a}\text{ \& }\eqnref{rpath4}\nonumber\\
&=\kaprep\biggl[\vrhoe-\frac{4\vphie}{\epsvj}\biggl\{
    2\vrhoa-\frac{\dltvb\vphie}{\epsvj}\biggr\}\biggr]
=\kaprep\biggl\{\vrhoe-\frac{4\vrhoh\vphie}{\epsvj}\biggr\}
=\kaprep\vrhoi\beqref{rpath1c}
\end{align}
\begin{align*}
\ffdot{\szer}
&=\dif{\biggl[\fdot{\zetarep}-\frac{2\xirep\fdot{\xirep}}{\etarep}
  +\frac{\xirep^2\fdot{\etarep}}{\etarep^2}\biggr]}\beqref{rpath5a}\nonumber\\
&=\ffdot{\zetarep}-2\biggl[\frac{\fdot{\xirep}^2}{\etarep}+\frac{\xirep\ffdot{\xirep}}{\etarep}
  -\frac{\xirep\fdot{\xirep}\fdot{\etarep}}{\etarep^2}\biggr]
  +\frac{2\xirep\fdot{\xirep}\fdot{\etarep}}{\etarep^2}
  +\frac{\xirep^2\ffdot{\etarep}}{\etarep^2}
  -\frac{2\xirep^2\fdot{\etarep}^2}{\etarep^3}
  \nonumber\\
&=\ffdot{\zetarep}
  -\frac{2\fdot{\xirep}^2}{\etarep}
  -\frac{2\xirep\ffdot{\xirep}}{\etarep}
  +\frac{\xirep^2\ffdot{\etarep}}{\etarep^2}
  +\frac{4\xirep\fdot{\xirep}\fdot{\etarep}}{\etarep^2}
  -\frac{2\xirep^2\fdot{\etarep}^2}{\etarep^3}
=\ffdot{\zetarep}
  +\frac{1}{\etarep}\biggl[-2\fdot{\xirep}^2-2\xirep\ffdot{\xirep}
  +\frac{\xirep}{\etarep}\biggl\{\xirep\ffdot{\etarep}+4\fdot{\xirep}\fdot{\etarep}
  -\frac{2\xirep\fdot{\etarep}^2}{\etarep}\biggr\}\biggr]
\end{align*}
\begin{align}\label{rpath5b}
\begin{split}
&=\kaprep\vrhof+\frac{1}{\kaprep\epsvj}\biggl[-8\kaprep^2\vrhoa^2-8\kaprep^2\vphie\vrhob
  +\frac{2\kaprep\vphie}{\kaprep\epsvj}\biggl\{2\kaprep^2\vphie\frkta+8\kaprep^2\vrhoa\dltvb
  -\frac{4\kaprep^3\vphie\dltvb^2}{\kaprep\epsvj}\biggr\}\biggr]\\
  &\qquad\beqref{rot4}, \eqnref{rpath2}, \eqnref{rpath3b}\text{ \& }\eqnref{rpath4}
\end{split}
\nonumber\\
&=\kaprep\vrhof-\frac{8\kaprep}{\epsvj}\biggl[\vrhoa^2+\vphie\vrhob
  -\frac{\vphie}{2\epsvj}\biggl\{\vphie\frkta+2\dltvb\biggl(2\vrhoa
  -\frac{\vphie\dltvb}{\epsvj}\biggr)\biggr\}\biggr]\nonumber\\
&=\kaprep\vrhof-\frac{8\kaprep}{\epsvj}\biggl[\vrhoj
  -\frac{\vphie}{2\epsvj}\biggl\{\vrhok+\dltvb\vrhoh\biggr\}\biggr]
=\kaprep\vrhon\beqref{rpath1c}\text{ \& }\eqnref{rpath1d}
\end{align}
\begin{align*}
\fffdot{\szer}
&=\dif{\biggl[\ffdot{\zetarep}
  -\frac{2\fdot{\xirep}^2}{\etarep}
  -\frac{2\xirep\ffdot{\xirep}}{\etarep}
  +\frac{4\xirep\fdot{\xirep}\fdot{\etarep}}{\etarep^2}
  +\frac{\xirep^2\ffdot{\etarep}}{\etarep^2}
  -\frac{2\xirep^2\fdot{\etarep}^2}{\etarep^3}\biggr]}
  \beqref{rpath5b}\nonumber\\
\begin{split}
&=\fffdot{\zetarep}
  -2\biggl[\frac{2\fdot{\xirep}\ffdot{\xirep}}{\etarep}
    -\frac{\fdot{\xirep}^2\fdot{\etarep}}{\etarep^2}\biggr]
  -2\biggl[\frac{\fdot{\xirep}\ffdot{\xirep}}{\etarep}
    +\frac{\xirep\fffdot{\xirep}}{\etarep}
    -\frac{\xirep\ffdot{\xirep}\fdot{\etarep}}{\etarep^2}\biggr]
  +4\biggl[\frac{\fdot{\xirep}^2\fdot{\etarep}}{\etarep^2}
    +\frac{\xirep\ffdot{\xirep}\fdot{\etarep}}{\etarep^2}
    +\frac{\xirep\fdot{\xirep}\ffdot{\etarep}}{\etarep^2}
    -\frac{2\xirep\fdot{\xirep}\fdot{\etarep}^2}{\etarep^3}\biggr]\\
  &\quad+\biggl[\frac{2\xirep\fdot{\xirep}\ffdot{\etarep}}{\etarep^2}
    +\frac{\xirep^2\fffdot{\etarep}}{\etarep^2}
    -\frac{2\xirep^2\ffdot{\etarep}\fdot{\etarep}}{\etarep^3}\biggr]
  -2\biggl[\frac{2\xirep\fdot{\xirep}\fdot{\etarep}^2}{\etarep^3}
    +\frac{2\xirep^2\fdot{\etarep}\ffdot{\etarep}}{\etarep^3}
    -\frac{3\xirep^2\fdot{\etarep}^3}{\etarep^4}\biggr]
\end{split}
\nonumber\\
&=\fffdot{\zetarep}
  -\frac{2\xirep\fffdot{\xirep}}{\etarep}
  -\frac{6\fdot{\xirep}\ffdot{\xirep}}{\etarep}
  +\frac{\xirep^2\fffdot{\etarep}}{\etarep^2}
  +\frac{6\fdot{\xirep}^2\fdot{\etarep}}{\etarep^2}
  +\frac{6\xirep\ffdot{\xirep}\fdot{\etarep}}{\etarep^2}
  +\frac{6\xirep\fdot{\xirep}\ffdot{\etarep}}{\etarep^2}
  -\frac{12\xirep\fdot{\xirep}\fdot{\etarep}^2}{\etarep^3}
  -\frac{6\xirep^2\fdot{\etarep}\ffdot{\etarep}}{\etarep^3}
  +\frac{6\xirep^2\fdot{\etarep}^3}{\etarep^4}
\nonumber\\
&=\fffdot{\zetarep}-\frac{1}{\etarep}\biggl[
  2\xirep\fffdot{\xirep}+6\fdot{\xirep}\ffdot{\xirep}
  -\frac{1}{\etarep}\biggl\{\xirep^2\fffdot{\etarep}
    +6\fdot{\xirep}^2\fdot{\etarep}
    +6\xirep\ffdot{\xirep}\fdot{\etarep}
    +6\xirep\fdot{\xirep}\ffdot{\etarep}\biggr\}
  +\frac{\xirep}{\etarep^2}\biggl\{
    12\fdot{\xirep}\fdot{\etarep}^2
    +6\xirep\fdot{\etarep}\ffdot{\etarep}
    -\frac{6\xirep\fdot{\etarep}^3}{\etarep}\biggr\}\biggr]
\end{align*}
\begin{align}\label{rpath5c}
\begin{split}
&=\kaprep\vrhog-\frac{1}{\kaprep\epsvj}\biggl[8\kaprep^2\vphie\vrhod+24\kaprep^2\vrhoa\vrhob
  -\frac{1}{\kaprep\epsvj}\biggl\{4\kaprep^3\vphie^2\frktb+24\kaprep^3\vrhoa^2\dltvb
     +24\kaprep^3\vphie\vrhob\dltvb+24\kaprep^3\vphie\vrhoa\frkta\biggr\}\\
  &\qquad+\frac{2\kaprep\vphie}{\kaprep^2\epsvj^2}\biggl\{24\kaprep^3\vrhoa\dltvb^2
   +12\kaprep^3\vphie\dltvb\frkta-\frac{12\kaprep^4\vphie\dltvb^3}{\kaprep\epsvj}\biggr\}\biggr]
  \beqref{rot4}, \eqnref{rpath2}, \eqnref{rpath3c}\text{ \& }\eqnref{rpath4}
\end{split}
\nonumber\\
\begin{split}
&=\kaprep\vrhog-\frac{8\kaprep}{\epsvj}\biggl[\vphie\vrhod+3\vrhoa\vrhob
  -\frac{1}{2\epsvj}\biggl\{\vphie^2\frktb+6\dltvb(\vrhoa^2+\vphie\vrhob)+6\vphie\vrhoa\frkta\biggr\}\\
  &\qquad+\frac{3\vphie\dltvb}{\epsvj^2}\biggl\{\vphie\frkta
    +\dltvb\biggl(2\vrhoa-\frac{\vphie\dltvb}{\epsvj}\biggr)\biggr\}\biggr]
\end{split}
\nonumber\\
&=\kaprep\vrhog-\frac{8\kaprep}{\epsvj}\biggl[\vphie\vrhod+3\vrhoa\vrhob
  -\frac{1}{2\epsvj}\biggl\{\vphie^2\frktb+6\dltvb\vrhoj+6\vphie\vrhoa\frkta\biggr\}
  +\frac{3\vrhok\vphie\dltvb}{\epsvj^2}\biggr]\beqref{rpath1c}\nonumber\\
&=\kaprep\vrhoo\beqref{rpath1d}.
\end{align}
\end{subequations}

\subart{Derivatives of $\vecta$ and $\gamrep$}
From \eqnref{main4a}, we obtain\footnote{In the expressions for the derivatives of
$\gamrep$, we are to take the positive sign when $\szer>-\etarep\pfreq^2$ and the negative sign
when $\szer<-\etarep\pfreq^2$. The case $\szer=-\etarep\pfreq^2$ is forbidden since $\gamrep$ is
nonzero.}
\begin{subequations}\label{rpath6}
\begin{align}\label{rpath6a}
\fdotg
&=\dif{\biggl[\left|1+\frac{\szer}{\etarep\pfreq^2}\right|^{1/2}\biggr]}
=\absign\frac{1}{2}\left|1+\frac{\szer}{\etarep\pfreq^2}\right|^{-1/2}\dif{\biggl[\frac{\szer}{\etarep\pfreq^2}\biggr]}
=\absign\frac{1}{2\gamrep}\dif{\biggl[\frac{\szer}{\etarep\pfreq^2}\biggr]}\nonumber\\
&=\absign\frac{1}{2\gamrep\pfreq^2}\biggl[\frac{\fdot{\szer}}{\etarep}-\frac{\szer\fdot{\etarep}}{\etarep^2}\biggr]
=\absign\frac{1}{2\gamrep\etarep\pfreq^2}\biggl[\fdot{\szer}-\frac{\szer\fdot{\etarep}}{\etarep}\biggr]\nonumber\\
&=\absign\frac{1}{2\gamrep\kaprep\epsvj\pfreq^2}\biggl[\kaprep\vrhoi-\frac{\kaprep^2\vphig\dltvb}{\kaprep\epsvj}\biggr]
  \beqref{rot4}, \eqnref{rot5a}, \eqnref{rpath5a}\text{ \& }\eqnref{rpath4}\nonumber\\
&=\absign\frac{1}{2\gamrep\epsvj\pfreq^2}\biggl[\vrhoi-\frac{\vphig\dltvb}{\epsvj}\biggr]
=\absign(\vrhom/\vrhol)=\vrhop\beqref{rpath1d}
\end{align}
\begin{align*}
\ffdotg
&=\absign\frac{1}{2\pfreq^2}\dif{\biggl[\frac{\fdot{\szer}}{\gamrep\etarep}-\frac{\szer\fdot{\etarep}}{\gamrep\etarep^2}
   \biggr]}\beqref{rpath6a}\nonumber\\
&=\absign\frac{1}{2\pfreq^2}\biggl[\frac{\ffdot{\szer}}{\gamrep\etarep}
  -\frac{\fdot{\szer}\fdotg}{\gamrep^2\etarep}-\frac{\fdot{\szer}\fdot{\etarep}}{\gamrep\etarep^2}
  -\frac{\fdot{\szer}\fdot{\etarep}}{\gamrep\etarep^2}-\frac{\szer\ffdot{\etarep}}{\gamrep\etarep^2}
  +\frac{\szer\fdot{\etarep}\fdotg}{\gamrep^2\etarep^2}+\frac{2\szer\fdot{\etarep}^2}{\gamrep\etarep^3}\biggr]
\nonumber\\
&=\absign\frac{1}{2\gamrep\etarep\pfreq^2}\biggl[\ffdot{\szer}
  -\fdot{\szer}\biggl\{\frac{\fdotg}{\gamrep}
    +\frac{2\fdot{\etarep}}{\etarep}\biggr\}
  -\frac{\szer}{\etarep}\biggl\{\ffdot{\etarep}
    -\frac{\fdot{\etarep}\fdotg}{\gamrep}
    -\frac{2\fdot{\etarep}^2}{\etarep}\biggr\}\biggr]
\nonumber\\
\begin{split}
&=\absign\frac{1}{2\gamrep\kaprep\epsvj\pfreq^2}\biggl[\kaprep\vrhon
  -\kaprep\vrhoi\biggl\{\frac{\vrhop}{\gamrep}+\frac{2\kaprep\dltvb}{\kaprep\epsvj}\biggr\}
  -\frac{\kaprep\vphig}{\kaprep\epsvj}\biggl\{\kaprep\frkta-\frac{\kaprep\dltvb\vrhop}{\gamrep}
     -\frac{2\kaprep^2\dltvb^2}{\kaprep\epsvj}\biggr\}\biggr]\\
  &\qquad\beqref{rot4}, \eqnref{rot5a}, \eqnref{rpath5}, \eqnref{rpath6a}\text{ \& }\eqnref{rpath4}
\end{split}
\end{align*}
\begin{align}\label{rpath6b}
&=\absign\frac{1}{2\gamrep\epsvj\pfreq^2}\biggl[\vrhon-\frac{\vphig\frkta}{\epsvj}
  -\vrhoi\biggl\{\frac{\vrhop}{\gamrep}+\frac{2\dltvb}{\epsvj}\biggr\}
  +\frac{\dltvb\vphig}{\epsvj}\biggl\{\frac{\vrhop}{\gamrep}+\frac{2\dltvb}{\epsvj}\biggr\}\biggr]\nonumber\\
&=\absign\frac{1}{\vrhol}\biggl[\vrhon-\frac{\vphig\frkta}{\epsvj}
  -\biggl\{\vrhoi-\frac{\dltvb\vphig}{\epsvj}\biggr\}
  \biggl\{\frac{\vrhop}{\gamrep}+\frac{2\dltvb}{\epsvj}\biggr\}\biggr]\nonumber\\
&=\frac{1}{\vrhol}\biggl\{\vrhon-\vrhom\vrhoq-\frac{\vphig\frkta}{\epsvj}\biggr\}
=\vrhor\beqref{rpath1d}
\end{align}
\begin{align*}
\fffdotg
&=\absign\frac{1}{2\pfreq^2}\dif{\biggl[\frac{\ffdot{\szer}}{\gamrep\etarep}
  -\frac{\fdot{\szer}\fdotg}{\gamrep^2\etarep}
  -\frac{2\fdot{\szer}\fdot{\etarep}}{\gamrep\etarep^2}
  -\frac{\szer\ffdot{\etarep}}{\gamrep\etarep^2}
  +\frac{\szer\fdot{\etarep}\fdotg}{\gamrep^2\etarep^2}
  +\frac{2\szer\fdot{\etarep}^2}{\gamrep\etarep^3}\biggr]}\beqref{rpath6b}
\nonumber\\
\begin{split}
&=\absign\frac{1}{2\pfreq^2}\biggl[
  \biggl\{\frac{\fffdot{\szer}}{\gamrep\etarep}
    -\frac{\ffdot{\szer}\fdotg}{\gamrep^2\etarep}
    -\frac{\ffdot{\szer}\fdot{\etarep}}{\gamrep\etarep^2}\biggr\}
  -\biggl\{\frac{\ffdot{\szer}\fdotg}{\gamrep^2\etarep}
    +\frac{\fdot{\szer}\ffdotg}{\gamrep^2\etarep}
    -\frac{2\fdot{\szer}\fdotg^2}{\gamrep^3\etarep}
    -\frac{\fdot{\szer}\fdotg\fdot{\etarep}}{\gamrep^2\etarep^2}\biggr\}\\
  &\quad-2\biggl\{\frac{\ffdot{\szer}\fdot{\etarep}}{\gamrep\etarep^2}
    +\frac{\fdot{\szer}\ffdot{\etarep}}{\gamrep\etarep^2}
    -\frac{\fdot{\szer}\fdot{\etarep}\fdotg}{\gamrep^2\etarep^2}
    -\frac{2\fdot{\szer}\fdot{\etarep}^2}{\gamrep\etarep^3}\biggr\}
  -\biggl\{\frac{\fdot{\szer}\ffdot{\etarep}}{\gamrep\etarep^2}
    +\frac{\szer\fffdot{\etarep}}{\gamrep\etarep^2}
    -\frac{\szer\ffdot{\etarep}\fdotg}{\gamrep^2\etarep^2}
    -\frac{2\szer\ffdot{\etarep}\fdot{\etarep}}{\gamrep\etarep^3}\biggr\}\\
  &\quad+\biggl\{\frac{\fdot{\szer}\fdot{\etarep}\fdotg}{\gamrep^2\etarep^2}
    +\frac{\szer\ffdot{\etarep}\fdotg}{\gamrep^2\etarep^2}
    +\frac{\szer\fdot{\etarep}\ffdotg}{\gamrep^2\etarep^2}
    -\frac{2\szer\fdot{\etarep}\fdotg^2}{\gamrep^3\etarep^2}
    -\frac{2\szer\fdot{\etarep}^2\fdotg}{\gamrep^2\etarep^3}\biggr\}
  +2\biggl\{\frac{\fdot{\szer}\fdot{\etarep}^2}{\gamrep\etarep^3}
    +\frac{2\szer\fdot{\etarep}\ffdot{\etarep}}{\gamrep\etarep^3}
    -\frac{\szer\fdot{\etarep}^2\fdotg}{\gamrep^2\etarep^3}
    -\frac{3\szer\fdot{\etarep}^3}{\gamrep\etarep^4}\biggr\}\biggr]
\end{split}
\end{align*}
\begin{align*}
\begin{split}
&=\absign\frac{1}{2\gamrep\etarep\pfreq^2}\biggl[\fffdot{\szer}
  -\frac{\szer\fffdot{\etarep}}{\etarep}
  -\frac{3\ffdot{\szer}\fdot{\etarep}}{\etarep}
  -\frac{3\fdot{\szer}\ffdot{\etarep}}{\etarep}
  +\frac{6\fdot{\szer}\fdot{\etarep}^2}{\etarep^2}
  +\frac{6\szer\ffdot{\etarep}\fdot{\etarep}}{\etarep^2}
  -\frac{6\szer\fdot{\etarep}^3}{\etarep^3}\\
  &\qquad-\frac{\fdot{\szer}\ffdotg}{\gamrep}
  -\frac{2\ffdot{\szer}\fdotg}{\gamrep}
  +\frac{2\szer\ffdot{\etarep}\fdotg}{\gamrep\etarep}
  +\frac{\szer\fdot{\etarep}\ffdotg}{\gamrep\etarep}
  +\frac{4\fdot{\szer}\fdotg\fdot{\etarep}}{\gamrep\etarep}
  -\frac{4\szer\fdot{\etarep}^2\fdotg}{\gamrep\etarep^2}
  +\frac{2\fdot{\szer}\fdotg^2}{\gamrep^2}
  -\frac{2\szer\fdot{\etarep}\fdotg^2}{\gamrep^2\etarep}\biggr]
\end{split}
\nonumber\\
\begin{split}
&=\absign\frac{1}{2\gamrep\etarep\pfreq^2}\biggl[\fffdot{\szer}
  -\frac{1}{\etarep}\biggl\{\szer\fffdot{\etarep}
    +3\ffdot{\szer}\fdot{\etarep}+3\fdot{\szer}\ffdot{\etarep}\biggr\}
  +\frac{6\fdot{\etarep}}{\etarep^2}\biggl\{\fdot{\szer}\fdot{\etarep}
    +\szer\biggl(\ffdot{\etarep}-\frac{\fdot{\etarep}^2}{\etarep}\biggr)\biggr\}\\
  &\qquad-\frac{1}{\gamrep}\biggl\{\fdot{\szer}\ffdotg+2\ffdot{\szer}\fdotg\biggr\}
  +\frac{\szer}{\gamrep\etarep}\biggl\{2\ffdot{\etarep}\fdotg+\fdot{\etarep}\ffdotg\biggr\}
    +\frac{4\fdotg\fdot{\etarep}}{\gamrep\etarep}\biggl\{\fdot{\szer}-\frac{\szer\fdot{\etarep}}{\etarep}\biggr\}
  +\frac{2\fdotg^2}{\gamrep^2}\biggl\{\fdot{\szer}-\frac{\szer\fdot{\etarep}}{\etarep}\biggr\}\biggr]
\end{split}
\end{align*}
\begin{align*}
\begin{split}
&=\absign\frac{1}{2\gamrep\etarep\pfreq^2}\biggl[\fffdot{\szer}
  -\frac{1}{\etarep}\biggl\{\szer\fffdot{\etarep}
    +3\ffdot{\szer}\fdot{\etarep}+3\fdot{\szer}\ffdot{\etarep}\biggr\}
  +\frac{6\fdot{\etarep}}{\etarep^2}\biggl\{\fdot{\szer}\fdot{\etarep}
    +\szer\biggl(\ffdot{\etarep}-\frac{\fdot{\etarep}^2}{\etarep}\biggr)\biggr\}\\
  &\qquad-\frac{1}{\gamrep}\biggl\{\fdot{\szer}\ffdotg+2\ffdot{\szer}\fdotg\biggr\}
  +\frac{\szer}{\gamrep\etarep}\biggl\{2\ffdot{\etarep}\fdotg+\fdot{\etarep}\ffdotg\biggr\}
  +\frac{2\fdotg}{\gamrep}\biggl\{\frac{2\fdot{\etarep}}{\etarep}+\frac{\fdotg}{\gamrep}\biggr\}
    \biggl\{\fdot{\szer}-\frac{\szer\fdot{\etarep}}{\etarep}\biggr\}\biggr]
\end{split}
\nonumber\\
\begin{split}
&=\absign\frac{1}{2\gamrep\kaprep\epsvj\pfreq^2}\biggl[\kaprep\vrhoo
  -\frac{1}{\kaprep\epsvj}\biggl\{\kaprep^2\vphig\frktb
    +3\kaprep^2\vrhon\dltvb+3\kaprep^2\vrhoi\frkta\biggr\}
  +\frac{6\kaprep\dltvb}{\kaprep^2\epsvj^2}\biggl\{\kaprep^2\vrhoi\dltvb
    +\kaprep\vphig\biggl(\kaprep\frkta-\frac{\kaprep^2\dltvb^2}{\kaprep\epsvj}\biggr)\biggr\}\\
  &\qquad-\frac{1}{\gamrep}\biggl\{\kaprep\vrhoi\vrhor+2\kaprep\vrhon\vrhop\biggr\}
  +\frac{\kaprep\vphig}{\gamrep\kaprep\epsvj}\biggl\{2\kaprep\frkta\vrhop+\kaprep\dltvb\vrhor\biggr\}
  +\frac{2\vrhop}{\gamrep}\biggl\{\frac{2\kaprep\dltvb}{\kaprep\epsvj}+\frac{\vrhop}{\gamrep}\biggr\}
    \biggl\{\kaprep\vrhoi-\frac{\kaprep^2\vphig\dltvb}{\kaprep\epsvj}\biggr\}\biggr]\\
  &\qquad\beqref{rot4b}, \eqnref{rot5a}, \eqnref{rpath5}, \eqnref{rpath4},
    \eqnref{rpath6a}\text{ \& }\eqnref{rpath6b}
\end{split}
\end{align*}
\begin{align}\label{rpath6c}
\begin{split}
&=\absign\frac{1}{\vrhol}\biggl[\vrhoo
  -\frac{1}{\epsvj}\biggl\{\vphig\frktb+3\vrhon\dltvb+3\vrhoi\frkta\biggr\}
  +\frac{6\dltvb}{\epsvj^2}\biggl\{\vrhoi\dltvb+\vphig\biggl(\frkta-\frac{\dltvb^2}{\epsvj}\biggr)\biggr\}\\
  &\qquad-\frac{1}{\gamrep}\biggl\{\vrhoi\vrhor+2\vrhon\vrhop\biggr\}
  +\frac{\vphig}{\gamrep\epsvj}\biggl\{2\frkta\vrhop+\dltvb\vrhor\biggr\}
  +\frac{2\vrhom\vrhop\vrhoq}{\gamrep}\biggr]
=\vrhos\beqref{rpath1d}.
\end{split}
\end{align}
\end{subequations}
Moreover, from \eqnref{main4a}, we derive
\begin{subequations}\label{rpath7}
\begin{align}\label{rpath7a}
\fdota
&=\dif{(\cprod{\vectOme}{\vectu}+\cprod{\vectLam}{\vectr})}
=\cprod{\vectLam}{\vectu}+\cprod{\vectOme}{\vecta}+\cprod{\fdot{\vectLam}}{\vectr}
  +\cprod{\vectLam}{\vectu}
=2\cprod{\vectLam}{\vectu}+\cprod{\fdot{\vectLam}}{\vectr}+\cprod{\vectOme}{\vecta}\nonumber\\
&=2\cprod{\vectLam}{(\cprod{\vectOme}{\vectr})}+\cprod{\fdot{\vectLam}}{\vectr}
  +\cprod{\vectOme}{[(\dprod{\vectOme}{\vectr})\vectOme-\Omerep^2\vectr+\cprod{\vectLam}{\vectr}]}
  \beqref{main4}\nonumber\\
&=2[\vectOme(\dprod{\vectLam}{\vectr})-\vectr(\dprod{\vectOme}{\vectLam})]
  +\cprod{\fdot{\vectLam}}{\vectr}-\Omerep^2(\cprod{\vectOme}{\vectr})
  +\vectLam(\dprod{\vectOme}{\vectr})-\vectr(\dprod{\vectOme}{\vectLam})
  \beqref{alg1}\nonumber\\
&=2\epsvh\vectOme-3\epsvg\vectr+\cprod{\fdot{\vectLam}}{\vectr}-\Omerep^2(\cprod{\vectOme}{\vectr})+\epsvb\vectLam
  \beqref{rot1a}
\end{align}
\begin{align*}
\ffdota
&=\dif{(2\cprod{\vectLam}{\vectu}+\cprod{\fdot{\vectLam}}{\vectr}+\cprod{\vectOme}{\vecta})}
  \beqref{rpath7a}\nonumber\\
&=2\cprod{\fdot{\vectLam}}{\vectu}+2\cprod{\vectLam}{\vecta}+\cprod{\ffdot{\vectLam}}{\vectr}
  +\cprod{\fdot{\vectLam}}{\vectu}+\cprod{\vectLam}{\vecta}+\cprod{\vectOme}{\fdota}
  \nonumber\\
&=3\cprod{\fdot{\vectLam}}{\vectu}+3\cprod{\vectLam}{\vecta}+\cprod{\ffdot{\vectLam}}{\vectr}
  +\cprod{\vectOme}{\fdota}\nonumber\\
\begin{split}
&=3\cprod{\fdot{\vectLam}}{(\cprod{\vectOme}{\vectr})}
  +3\cprod{\vectLam}{[(\dprod{\vectOme}{\vectr})\vectOme-\Omerep^2\vectr+\cprod{\vectLam}{\vectr}]}
  +\cprod{\ffdot{\vectLam}}{\vectr}\\
  &\quad+\cprod{\vectOme}{[2\epsvh\vectOme-3\epsvg\vectr+\cprod{\fdot{\vectLam}}{\vectr}
    -\Omerep^2(\cprod{\vectOme}{\vectr})+\epsvb\vectLam]}\beqref{main4}\text{ \& }\eqnref{rpath7a}
\end{split}
\nonumber\\
\begin{split}
&=3\cprod{\fdot{\vectLam}}{(\cprod{\vectOme}{\vectr})}
  +3\epsvb(\cprod{\vectLam}{\vectOme})
  -3\Omerep^2(\cprod{\vectLam}{\vectr})
  +3[\cprod{\vectLam}{(\cprod{\vectLam}{\vectr})}]
  +\cprod{\ffdot{\vectLam}}{\vectr}\\
  &\quad-3\epsvg(\cprod{\vectOme}{\vectr})
  +\cprod{\vectOme}{(\cprod{\fdot{\vectLam}}{\vectr})}
  -\Omerep^2[\cprod{\vectOme}{(\cprod{\vectOme}{\vectr})}]
  +\epsvb(\cprod{\vectOme}{\vectLam})\beqref{rot1a}
\end{split}
\end{align*}
\begin{align}\label{rpath7b}
\begin{split}
&=3[\vectOme(\dprod{\fdot{\vectLam}}{\vectr})-\vectr(\dprod{\vectOme}{\fdot{\vectLam}})]
  +3\epsvb(\cprod{\vectLam}{\vectOme})
  -3\Omerep^2(\cprod{\vectLam}{\vectr})
  +3[\vectLam(\dprod{\vectLam}{\vectr})-\Lamrep^2\vectr]
  +\cprod{\ffdot{\vectLam}}{\vectr}\\
  &\quad-3\epsvg(\cprod{\vectOme}{\vectr})
  +[\fdot{\vectLam}(\dprod{\vectOme}{\vectr})-\vectr(\dprod{\fdot{\vectLam}}{\vectOme})]
  -\Omerep^2[\vectOme(\dprod{\vectOme}{\vectr})-\Omerep^2\vectr]
  +\epsvb(\cprod{\vectOme}{\vectLam})\beqref{alg1}
\end{split}
\nonumber\\
\begin{split}
&=3\vsign\vectOme-3\vsigc\vectr+3\epsvb(\cprod{\vectLam}{\vectOme})-3\Omerep^2(\cprod{\vectLam}{\vectr})
  +3\epsvh\vectLam-3\Lamrep^2\vectr+\cprod{\ffdot{\vectLam}}{\vectr}\\
  &\quad-3\epsvg(\cprod{\vectOme}{\vectr})+\epsvb\fdot{\vectLam}-\vsigc\vectr
  -\Omerep^2\epsvb\vectOme+\Omerep^4\vectr+\epsvb(\cprod{\vectOme}{\vectLam})
  \beqref{rot1a}\text{ \& }\eqnref{rpath1a}
\end{split}
\nonumber\\
\begin{split}
&=3\epsvh\vectLam+\epsvb\fdot{\vectLam} +(3\vsign-\Omerep^2\epsvb)\vectOme
  +(\Omerep^4-4\vsigc-3\Lamrep^2)\vectr+2\epsvb(\cprod{\vectLam}{\vectOme})\\
  &\quad-3\Omerep^2(\cprod{\vectLam}{\vectr})-3\epsvg(\cprod{\vectOme}{\vectr})
  +(\cprod{\ffdot{\vectLam}}{\vectr})
\end{split}
\nonumber\\
\begin{split}
&=3\epsvh\vectLam+\epsvb\fdot{\vectLam}+\vrhot\vectOme
  +\vrhou\vectr+2\epsvb(\cprod{\vectLam}{\vectOme})
  -3\Omerep^2(\cprod{\vectLam}{\vectr})-3\epsvg(\cprod{\vectOme}{\vectr})\\
  &\qquad+(\cprod{\ffdot{\vectLam}}{\vectr})\beqref{rpath1e}
\end{split}
\end{align}
\begin{align*}
\fffdota
&=\dif{(3\cprod{\fdot{\vectLam}}{\vectu}+3\cprod{\vectLam}{\vecta}+\cprod{\ffdot{\vectLam}}{\vectr}
  +\cprod{\vectOme}{\fdota})}\beqref{rpath7b}\nonumber\\
&=3\cprod{\ffdot{\vectLam}}{\vectu}+3\cprod{\fdot{\vectLam}}{\vecta}+3\cprod{\fdot{\vectLam}}{\vecta}
  +3\cprod{\vectLam}{\fdota}+\cprod{\fffdot{\vectLam}}{\vectr}+\cprod{\ffdot{\vectLam}}{\vectu}
  +\cprod{\vectLam}{\fdota}+\cprod{\vectOme}{\ffdota}\nonumber\\
&=4\cprod{\ffdot{\vectLam}}{\vectu}+6\cprod{\fdot{\vectLam}}{\vecta}
  +4\cprod{\vectLam}{\fdota}+\cprod{\fffdot{\vectLam}}{\vectr}+\cprod{\vectOme}{\ffdota}
\end{align*}
\begin{align*}
\begin{split}
&=\cprod{\fffdot{\vectLam}}{\vectr}
  +4\cprod{\ffdot{\vectLam}}{(\cprod{\vectOme}{\vectr})}
  +6\cprod{\fdot{\vectLam}}{[(\dprod{\vectOme}{\vectr})\vectOme-\Omerep^2\vectr+\cprod{\vectLam}{\vectr}]}\\
  &\quad+4\cprod{\vectLam}{[2\epsvh\vectOme-3\epsvg\vectr+\cprod{\fdot{\vectLam}}{\vectr}
    -\Omerep^2(\cprod{\vectOme}{\vectr})+\epsvb\vectLam]}\\
  &\quad+\cprod{\vectOme}{[3\epsvh\vectLam+\epsvb\fdot{\vectLam}
    +\vrhot\vectOme+\vrhou\vectr+2\epsvb(\cprod{\vectLam}{\vectOme})]}\\
  &\quad+\cprod{\vectOme}{[-3\Omerep^2(\cprod{\vectLam}{\vectr})-3\epsvg(\cprod{\vectOme}{\vectr})
    +(\cprod{\ffdot{\vectLam}}{\vectr})]}
  \beqref{main4}, \eqnref{rpath7a}\text{ \& }\eqnref{rpath7b}
\end{split}
\nonumber\\
\begin{split}
&=\cprod{\fffdot{\vectLam}}{\vectr}
  +4\cprod{\ffdot{\vectLam}}{(\cprod{\vectOme}{\vectr})}
  +6\epsvb(\cprod{\fdot{\vectLam}}{\vectOme})
  -6\Omerep^2(\cprod{\fdot{\vectLam}}{\vectr})
  +6[\cprod{\fdot{\vectLam}}{(\cprod{\vectLam}{\vectr})}]\\
  &\quad+8\epsvh(\cprod{\vectLam}{\vectOme})
  -12\epsvg(\cprod{\vectLam}{\vectr})
  +4[\cprod{\vectLam}{(\cprod{\fdot{\vectLam}}{\vectr})}]
  -4\Omerep^2[\cprod{\vectLam}{(\cprod{\vectOme}{\vectr})}]
  +4\epsvb(\cprod{\vectLam}{\vectLam})\\
  &\quad+3\epsvh(\cprod{\vectOme}{\vectLam})
  +\epsvb(\cprod{\vectOme}{\fdot{\vectLam}})
  +\vrhot(\cprod{\vectOme}{\vectOme})
  +\vrhou(\cprod{\vectOme}{\vectr})
  +2\epsvb[\cprod{\vectOme}{(\cprod{\vectLam}{\vectOme})}]\\
  &\quad-3\Omerep^2[\cprod{\vectOme}{(\cprod{\vectLam}{\vectr})}]
  -3\epsvg[\cprod{\vectOme}{(\cprod{\vectOme}{\vectr})}]
  +\cprod{\vectOme}{(\cprod{\ffdot{\vectLam}}{\vectr})}\beqref{rot1a}
\end{split}
\end{align*}
\begin{align*}
\begin{split}
&=\cprod{\fffdot{\vectLam}}{\vectr}
  +4[\vectOme(\dprod{\ffdot{\vectLam}}{\vectr})-\vectr(\dprod{\ffdot{\vectLam}}{\vectOme})]
  +6\epsvb(\cprod{\fdot{\vectLam}}{\vectOme})
  -6\Omerep^2(\cprod{\fdot{\vectLam}}{\vectr})
  +6[\vectLam(\dprod{\fdot{\vectLam}}{\vectr})-\vectr(\dprod{\fdot{\vectLam}}{\vectLam})]\\
  &\quad+5\epsvh(\cprod{\vectLam}{\vectOme})
  -12\epsvg(\cprod{\vectLam}{\vectr})
  +4[\fdot{\vectLam}(\dprod{\vectLam}{\vectr})-\vectr(\dprod{\vectLam}{\fdot{\vectLam}})]
  -4\Omerep^2[\vectOme(\dprod{\vectLam}{\vectr})-\vectr(\dprod{\vectLam}{\vectOme})]\\
  &\quad+\epsvb(\cprod{\vectOme}{\fdot{\vectLam}})
  +\vrhou(\cprod{\vectOme}{\vectr})
  +2\epsvb[\Omerep^2\vectLam-\vectOme(\dprod{\vectOme}{\vectLam})]
  -3\Omerep^2[\vectLam(\dprod{\vectOme}{\vectr})-\vectr(\dprod{\vectOme}{\vectLam})]\\
  &\quad-3\epsvg[\vectOme(\dprod{\vectOme}{\vectr})-\Omerep^2\vectr]
  +[\ffdot{\vectLam}(\dprod{\vectOme}{\vectr})-\vectr(\dprod{\ffdot{\vectLam}}{\vectOme})]
  \beqref{alg1}
\end{split}
\nonumber\\
\begin{split}
&=\cprod{\fffdot{\vectLam}}{\vectr}+4(\vsigo\vectOme-\vsigh\vectr)
  +6\epsvb(\cprod{\fdot{\vectLam}}{\vectOme})-6\Omerep^2(\cprod{\fdot{\vectLam}}{\vectr})
  +6(\vsign\vectLam-\vsigd\vectr)+5\epsvh(\cprod{\vectLam}{\vectOme})\\
  &\quad-12\epsvg(\cprod{\vectLam}{\vectr})
  +4(\epsvh\fdot{\vectLam}-\vsigd\vectr)-4\Omerep^2(\epsvh\vectOme-\epsvg\vectr)
  +\epsvb(\cprod{\vectOme}{\fdot{\vectLam}})
  +\vrhou(\cprod{\vectOme}{\vectr})\\
  &\quad+2\epsvb(\Omerep^2\vectLam-\epsvg\vectOme)
  -3\Omerep^2(\epsvb\vectLam-\epsvg\vectr)
  -3\epsvg(\epsvb\vectOme-\Omerep^2\vectr)
  +(\epsvb\ffdot{\vectLam}-\vsigh\vectr)
  \beqref{rot1a}\text{ \& }\eqnref{rpath1a}
\end{split}
\end{align*}
\begin{align}\label{rpath7c}
\begin{split}
&=(6\vsign-\Omerep^2\epsvb)\vectLam+4\epsvh\fdot{\vectLam}+\epsvb\ffdot{\vectLam}
  +(4\vsigo-4\Omerep^2\epsvh-5\epsvb\epsvg)\vectOme+5(2\Omerep^2\epsvg-\vsigh-2\vsigd)\vectr\\
  &\quad-5\epsvb(\cprod{\vectOme}{\fdot{\vectLam}})
  -6\Omerep^2(\cprod{\fdot{\vectLam}}{\vectr})+5\epsvh(\cprod{\vectLam}{\vectOme})
  -12\epsvg(\cprod{\vectLam}{\vectr})+\vrhou(\cprod{\vectOme}{\vectr})
  +\cprod{\fffdot{\vectLam}}{\vectr}
\end{split}
\nonumber\\
\begin{split}
&=(\vrhot+3\vsign)\vectLam+4\epsvh\fdot{\vectLam}+\epsvb\ffdot{\vectLam}
  +\vrhov\vectOme+5\vrhow\vectr-5\epsvb(\cprod{\vectOme}{\fdot{\vectLam}})
  -6\Omerep^2(\cprod{\fdot{\vectLam}}{\vectr})\\
  &\quad+5\epsvh(\cprod{\vectLam}{\vectOme})
  -12\epsvg(\cprod{\vectLam}{\vectr})+\vrhou(\cprod{\vectOme}{\vectr})
  +\cprod{\fffdot{\vectLam}}{\vectr}\beqref{rpath1e}.
\end{split}
\end{align}
\end{subequations}

\subart{Development of equations \eqnref{kpath5} through \eqnref{kpath9}}
Equations \eqnref{kpath5} through \eqnref{kpath9} evaluate as
\begin{subequations}\label{rpath8}
\begin{align}\label{rpath8a}
\fdot{\alprep}
&=\dprod{\vectkap}{\fdota}\beqref{kpath5}\nonumber\\
&=\dprod{\vectkap}{[2\epsvh\vectOme-3\epsvg\vectr+\cprod{\fdot{\vectLam}}{\vectr}-\Omerep^2(\cprod{\vectOme}{\vectr})
  +\epsvb\vectLam]}\beqref{rpath7a}\nonumber\\
&=\kaprep(2\epsva\epsvh-3\epsvd\epsvg+\frktc-\Omerep^2\epsvk+\dltva\epsvb)
  \beqref{rot1a}, \eqnref{rxpeed1a}\text{ \& }\eqnref{rpath1b}\nonumber\\
&=\kaprep(\frktc+\epsvb\dltva+2\epsva\epsvh-3\epsvd\epsvg-\Omerep^2\epsvk)
=\kaprep\vrhox\beqref{rpath1e}
\end{align}
\begin{align}\label{rpath8b}
\ffdot{\alprep}
&=\dprod{\vectkap}{\ffdota}\beqref{kpath5}\nonumber\\
\begin{split}
&=\dprod{\vectkap}{}[3\epsvh\vectLam+\epsvb\fdot{\vectLam}
  +\vrhot\vectOme+\vrhou\vectr
  +2\epsvb(\cprod{\vectLam}{\vectOme})
  -3\Omerep^2(\cprod{\vectLam}{\vectr})
  -3\epsvg(\cprod{\vectOme}{\vectr})\\
  &\qquad+(\cprod{\ffdot{\vectLam}}{\vectr})]\beqref{rpath7b}
\end{split}
\nonumber\\
\begin{split}
&=\kaprep[3\epsvh\dltva+\epsvb\vsigb+\vrhot\epsva+\vrhou\epsvd
  +2\epsvb\frkte-3\Omerep^2\epsvn-3\epsvg\epsvk+\frktd]\\
  &\qquad\beqref{rot1a}, \eqnref{rxpeed1a}, \eqnref{rpath1a}\text{ \& }\eqnref{rpath1b}
\end{split}
\nonumber\\
&=\kaprep[\frktd+\epsva\vrhot+\epsvd\vrhou
  +\epsvb(\vsigb+2\frkte)+3(\epsvh\dltva-\Omerep^2\epsvn-\epsvg\epsvk)]
=\kaprep\vrhoy\beqref{rpath1e}
\end{align}
\begin{align*}
\fffdot{\alprep}
&=\dprod{\vectkap}{\fffdota}\beqref{kpath5}\nonumber\\
\begin{split}
&=\dprod{\vectkap}{}[
  (\vrhot+3\vsign)\vectLam+4\epsvh\fdot{\vectLam}+\epsvb\ffdot{\vectLam}
  +\vrhov\vectOme+5\vrhow\vectr-5\epsvb(\cprod{\vectOme}{\fdot{\vectLam}})\\
  &\quad-6\Omerep^2(\cprod{\fdot{\vectLam}}{\vectr})+5\epsvh(\cprod{\vectLam}{\vectOme})
  -12\epsvg(\cprod{\vectLam}{\vectr})+\vrhou(\cprod{\vectOme}{\vectr})
  +\cprod{\fffdot{\vectLam}}{\vectr}]\beqref{rpath7c}
\end{split}
\end{align*}
\begin{align}\label{rpath8c}
\begin{split}
&=\kaprep[\dltva(\vrhot+3\vsign)+4\epsvh\vsigb+\epsvb\vsigg
  +\epsva\vrhov+5\epsvd\vrhow-5\epsvb\frktf-6\Omerep^2\frktc+5\epsvh\frkte\\
  &\quad-12\epsvg\epsvn+\vrhou\epsvk+\frktg]
  \beqref{rot1a}, \eqnref{rxpeed1a}, \eqnref{rpath1a}\text{ \& }\eqnref{rpath1b}
\end{split}
\nonumber\\
\begin{split}
&=\kaprep[\frktg+\epsva\vrhov+\epsvk\vrhou+5\epsvd\vrhow+\epsvb(\vsigg-5\frktf)+\dltva(\vrhot+3\vsign)\\
  &\qquad+\epsvh(4\vsigb+5\frkte)-6(\Omerep^2\frktc+2\epsvg\epsvn)]
=\kaprep\vrhoz\beqref{rpath1e}
\end{split}
\end{align}
\end{subequations}
\begin{subequations}\label{rpath9}
\begin{align}\label{rpath9a}
\fdot{\vthtrep}
&=\frac{1}{\gamrep^2\pfreq^2}\left[\fdot{\alprep}-\frac{2\alprep\fdot{\gamrep}}{\gamrep}\right]
  \beqref{kpath6a}\nonumber\\
&=\frac{1}{\gamrep^2\pfreq^2}\left[\kaprep\vrhox-\frac{2\kaprep\vphid\vrhop}{\gamrep}\right]
  \beqref{rot4a}, \eqnref{rpath6a}, \text{ \& }\eqnref{rpath8a}\nonumber\\
&=\ethva\beqref{rpath1f}
\end{align}
\begin{align}\label{rpath9b}
\ffdot{\vthtrep}
&=\frac{1}{\gamrep^2\pfreq^2}\left[\ffdot{\alprep}-\frac{4\fdot{\alprep}\fdotg}{\gamrep}
   -\frac{2\alprep\ffdotg}{\gamrep}+\frac{6\alprep\fdotg^2}{\gamrep^2}\right]
  \beqref{kpath6b}\nonumber\\
&=\frac{1}{\gamrep^2\pfreq^2}\left[\kaprep\vrhoy-\frac{4\kaprep\vrhox\vrhop}{\gamrep}
   -\frac{2\kaprep\vphid\vrhor}{\gamrep}+\frac{6\kaprep\vphid\vrhop^2}{\gamrep^2}\right]
  \beqref{rot4a}, \eqnref{rpath6}\text{ \& }\eqnref{rpath8}\nonumber\\
&=\ethvb\beqref{rpath1f}
\end{align}
\begin{align}\label{rpath9c}
\fffdot{\vthtrep}
&=\frac{1}{\gamrep^2\pfreq^2}\left[\fffdot{\alprep}-\frac{6\ffdot{\alprep}\fdotg}{\gamrep}
   -\frac{6\fdot{\alprep}\ffdotg}{\gamrep}+\frac{18\fdot{\alprep}\fdotg^2}{\gamrep^2}
   -\frac{2\alprep\fffdotg}{\gamrep}+\frac{18\alprep\fdotg\ffdotg}{\gamrep^2}
   -\frac{24\alprep\fdotg^3}{\gamrep^3}\right]\beqref{kpath6c}\nonumber\\
\begin{split}
&=\frac{1}{\gamrep^2\pfreq^2}\left[\kaprep\vrhoz-\frac{6\kaprep\vrhoy\vrhop}{\gamrep}
   -\frac{6\kaprep\vrhox\vrhor}{\gamrep}+\frac{18\kaprep\vrhox\vrhop^2}{\gamrep^2}
   -\frac{2\kaprep\vphid\vrhos}{\gamrep}+\frac{18\kaprep\vphid\vrhop\vrhor}{\gamrep^2}
   -\frac{24\kaprep\vphid\vrhop^3}{\gamrep^3}\right]\\
  &\qquad\beqref{rot4a}, \eqnref{rpath6}\text{ \& }\eqnref{rpath8}
\end{split}
\nonumber\\
&=\ethvc\beqref{rpath1f}
\end{align}
\end{subequations}
\begin{subequations}\label{rpath10}
\begin{align}\label{rpath10a}
\fdot{\xcons}
&=\fdot{\vthtrep}(1+\vthtrep^2)^{-3/2}\beqref{kpath7a}\nonumber\\
&=\ethva\dltvg^{-3}\beqref{rpath9a}\text{ \& }\eqnref{rxpeed1a}\nonumber\\
\end{align}
\begin{align}\label{rpath10b}
\ffdot{\xcons}
&=(1+\vthtrep^2)^{-5/2}[\ffdot{\vthtrep}(1+\vthtrep^2)-3\vthtrep\fdot{\vthtrep}^2]\beqref{kpath7b}\nonumber\\
&=\dltvg^{-5}(\ethvb\dltvg^2-3\vthtrep\ethva^2)
  \beqref{rpath9}\text{ \& }\eqnref{rxpeed1a}\nonumber\\
&=\ethvd\beqref{rpath1f}
\end{align}
\begin{align}\label{rpath10c}
\fffdot{\xcons}
&=(1+\vthtrep^2)^{-7/2}[\fffdot{\vthtrep}(1+\vthtrep^2)^2-9\vthtrep\fdot{\vthtrep}\ffdot{\vthtrep}(1+\vthtrep^2)
  -3\fdot{\vthtrep}^3(1-4\vthtrep^2)]
  \beqref{kpath7c}\nonumber\\
&=\dltvg^{-7}[\ethvc\dltvg^4-9\vthtrep\ethva\ethvb\dltvg^2-3\ethva^3(1-4\vthtrep^2)]
  \beqref{rpath9}\text{ \& }\eqnref{rxpeed1a}\nonumber\\
&=\ethve\beqref{rpath1f}
\end{align}
\end{subequations}
\begin{subequations}\label{rpath11}
\begin{align}\label{rpath11a}
\fdot{\dragf}
&=\frac{\fdotg\dragf}{\gamrep}+\frac{\xcons\gamrep^2\fdot{\vthtrep}}{4\dragf}
  \beqref{kpath8a}\nonumber\\
&=\frac{\vrhop\dragf}{\gamrep}+\frac{\xcons\gamrep^2\ethva}{4\dragf}
  \beqref{rpath6a}\text{ \& }\eqnref{rpath9a}\nonumber\\
&=\ethvf\beqref{rpath1g}
\end{align}
\begin{align}\label{rpath11b}
\ffdot{\dragf}
&=\dragf\biggl[\frac{\ffdotg}{\gamrep}-\frac{\fdotg^2}{\gamrep^2}\biggr]
   +\fdot{\dragf}\biggl[\frac{\fdotg}{\gamrep}-\frac{\xcons\gamrep^2\fdot{\vthtrep}}{4\dragf^2}\biggr]
   +\frac{\gamrep}{4\dragf}\biggl[\fdot{\xcons}\gamrep\fdot{\vthtrep}
    +2\xcons\fdotg\fdot{\vthtrep}+\xcons\gamrep\ffdot{\vthtrep}\biggr]\beqref{kpath8b}\nonumber\\
\begin{split}
&=\dragf\biggl[\frac{\vrhor}{\gamrep}-\frac{\vrhop^2}{\gamrep^2}\biggr]
   +\ethvf\biggl[\frac{\vrhop}{\gamrep}-\frac{\xcons\gamrep^2\ethva}{4\dragf^2}\biggr]
   +\frac{\gamrep}{4\dragf}\biggl[\frac{\gamrep\ethva^2}{\dltvg^3}
    +2\xcons\vrhop\ethva+\xcons\gamrep\ethvb\biggr]\\
  &\qquad\beqref{rpath6}, \eqnref{rpath9}, \eqnref{rpath10a}\text{ \& }\eqnref{rpath11a}
\end{split}
\nonumber\\
&=\ethvg\beqref{rpath1g}
\end{align}
\begin{align}\label{rpath11c}
\begin{split}
\fffdot{\dragf}
&=\dragf\biggl[\frac{\fffdotg}{\gamrep}-\frac{3\fdotg\ffdotg}{\gamrep^2}+\frac{2\fdotg^3}{\gamrep^3}\biggr]
  +2\fdot{\dragf}\biggl[\frac{\ffdotg}{\gamrep}-\frac{\fdotg^2}{\gamrep^2}
    -\frac{\fdot{\xcons}\gamrep^2\fdot{\vthtrep}}{4\dragf^2}-\frac{\xcons\gamrep\fdotg\fdot{\vthtrep}}{2\dragf^2}
    -\frac{\xcons\gamrep^2\ffdot{\vthtrep}}{4\dragf^2}
    +\frac{\xcons\gamrep^2\fdot{\dragf}\fdot{\vthtrep}}{4\dragf^3}\biggr]
   +\ffdot{\dragf}\biggl[\frac{\fdotg}{\gamrep}-\frac{\xcons\gamrep^2\fdot{\vthtrep}}{4\dragf^2}\biggr]\\
   &\quad+\frac{\fdotg}{4\dragf}\biggl[\fdot{\xcons}\gamrep\fdot{\vthtrep}+2\xcons\fdotg\fdot{\vthtrep}
      +\xcons\gamrep\ffdot{\vthtrep}\biggr]
  +\frac{\gamrep}{4\dragf}\biggl[\ffdot{\xcons}\gamrep\fdot{\vthtrep}+3\fdot{\xcons}\fdotg\fdot{\vthtrep}
    +2\fdot{\xcons}\gamrep\ffdot{\vthtrep}+2\xcons\ffdotg\fdot{\vthtrep}
    +3\xcons\fdotg\ffdot{\vthtrep}+\xcons\gamrep\fffdot{\vthtrep}\biggr]\beqref{kpath8c}
\end{split}
\nonumber\\
\begin{split}
&=\dragf\biggl[\frac{\vrhos}{\gamrep}-\frac{3\vrhop\vrhor}{\gamrep^2}+\frac{2\vrhop^3}{\gamrep^3}\biggr]
  +2\ethvf\biggl[\frac{\vrhor}{\gamrep}-\frac{\vrhop^2}{\gamrep^2}
    -\frac{\gamrep^2\ethva^2}{4\dragf^2\dltvg^3}-\frac{\xcons\gamrep\vrhop\ethva}{2\dragf^2}
    -\frac{\xcons\gamrep^2\ethvb}{4\dragf^2}
    +\frac{\xcons\gamrep^2\ethvf\ethva}{4\dragf^3}\biggr]\\
  &\quad+\ethvg\biggl[\frac{\vrhop}{\gamrep}-\frac{\xcons\gamrep^2\ethva}{4\dragf^2}\biggr]
   +\frac{\vrhop}{4\dragf}\biggl[\frac{\gamrep\ethva^2}{\dltvg^3}+2\xcons\vrhop\ethva
      +\xcons\gamrep\ethvb\biggr]
   +\frac{\gamrep}{4\dragf}\biggl[\gamrep\ethva\ethvd+\frac{3\ethva^2\vrhop}{\dltvg^3}
      +\frac{2\gamrep\ethva\ethvb}{\dltvg^3}\\
    &\quad+2\xcons\vrhor\ethva+3\xcons\vrhop\ethvb+\xcons\gamrep\ethvc\biggr]
  \beqref{rpath6}, \eqnref{rpath9}, \eqnref{rpath10},
    \eqnref{rpath11a}\text{ \& }\eqnref{rpath11b}
\end{split}
\nonumber\\
&=\ethvh\beqref{rpath1g}
\end{align}
\end{subequations}
\begin{subequations}\label{rpath12}
\begin{align}\label{rpath12a}
\fdot{\rhorep}
&=\frac{\dragf\fdot{\xcons}-\xcons\fdot{\dragf}}{4\pfreq\dragf^2}
  \beqref{kpath9a}\nonumber\\
&=\frac{1}{4\pfreq\dragf^2}\biggl[\frac{\dragf\ethva}{\dltvg^3}-\xcons\ethvf\biggr]
  \beqref{rpath10a}\text{ \& }\eqnref{rpath11a}\nonumber\\
&=\ethvi\beqref{rpath1h}
\end{align}
\begin{align}\label{rpath12b}
\ffdot{\rhorep}
&=-\frac{\fdot{\dragf}(\dragf\fdot{\xcons}-\xcons\fdot{\dragf})}{2\pfreq\dragf^3}
   +\frac{\dragf\ffdot{\xcons}-\xcons\ffdot{\dragf}}{4\pfreq\dragf^2}
  \beqref{kpath9b}\nonumber\\
&=-\frac{\ethvf}{2\pfreq\dragf^3}\biggl[\frac{\dragf\ethva}{\dltvg^3}-\xcons\ethvf\biggr]
   +\frac{1}{4\pfreq\dragf^2}\biggl[\dragf\ethvd-\xcons\ethvg\biggr]
   \beqref{rpath10}\text{ \& }\eqnref{rpath11}\nonumber\\
&=\frac{1}{4\pfreq\dragf^2}\biggl[\dragf\ethvd-\xcons\ethvg
   -\frac{2\ethvf}{\dragf}\biggl\{\frac{\dragf\ethva}{\dltvg^3}-\xcons\ethvf\biggr\}\biggr]
   \nonumber\\
&=\ethvj\beqref{rpath1h}
\end{align}
\begin{align}\label{rpath12c}
\fffdot{\rhorep}
&=\frac{(3\fdot{\dragf}^2-\dragf\ffdot{\dragf})(\dragf\fdot{\xcons}-\xcons\fdot{\dragf})}{2\pfreq\dragf^4}
   -\frac{\fdot{\dragf}(\dragf\ffdot{\xcons}-\xcons\ffdot{\dragf})}{\pfreq\dragf^3}
  +\frac{\fdot{\dragf}\ffdot{\xcons}+\dragf\fffdot{\xcons}
     -\fdot{\xcons}\ffdot{\dragf}-\xcons\fffdot{\dragf}}{4\pfreq\dragf^2}
  \beqref{kpath9c}\nonumber\\
\begin{split}
&=\frac{3\ethvf^2-\dragf\ethvg}{2\pfreq\dragf^4}\biggl[\frac{\dragf\ethva}{\dltvg^3}-\xcons\ethvf\biggr]
   -\frac{\ethvf}{\pfreq\dragf^3}\biggl[\dragf\ethvd-\xcons\ethvg\biggr]\\
  &\quad+\frac{1}{4\pfreq\dragf^2}\biggl[\ethvf\ethvd+\dragf\ethve-\frac{\ethva\ethvg}{\dltvg^3}-\xcons\ethvh\biggr]
   \beqref{rpath10}\text{ \& }\eqnref{rpath11}
\end{split}
\nonumber\\
&=\frac{1}{4\pfreq\dragf^2}\biggl[\ethvf\ethvd+\dragf\ethve-\xcons\ethvh-\frac{\ethva\ethvg}{\dltvg^3}
  +\frac{2(3\ethvf^2-\dragf\ethvg)}{\dragf^2}\biggl\{\frac{\dragf\ethva}{\dltvg^3}-\xcons\ethvf\biggr\}
  -\frac{4\ethvf}{\dragf}\biggl\{\dragf\ethvd-\xcons\ethvg\biggr\}\biggr]\nonumber\\
&=\ethvk\beqref{rpath1h}.
\end{align}
\end{subequations}

\subart{Derivatives of $\taurep$ and $\vecte$}
To compute the derivatives of $\taurep$ and $\vecte$, we first derive
\begin{subequations}\label{rpath13}
\begin{align}\label{rpath13a}
\fdot{\sone}
&=\dif{\biggl[\frac{2\rhorep\alprep-\dragf\pfreq}{2\etarep(\szer+\etarep\pfreq^2)}\biggr]}
  \beqref{main4d}\nonumber\\
&=\frac{2\etarep(\szer+\etarep\pfreq^2)\dif{(2\rhorep\alprep-\dragf\pfreq)}
  -(2\rhorep\alprep-\dragf\pfreq)\dif{[2\etarep(\szer+\etarep\pfreq^2)]}}
  {4\etarep^2(\szer+\etarep\pfreq^2)^2}\nonumber\\
&=\frac{2\fdot{\rhorep}\alprep+2\rhorep\fdot{\alprep}-\fdot{\dragf}\pfreq
  -\sone[2\fdot{\etarep}(\szer+\etarep\pfreq^2)+2\etarep(\fdot{\szer}+\fdot{\etarep}\pfreq^2)]}
  {2\etarep(\szer+\etarep\pfreq^2)}\beqref{main4d}\nonumber\\
&=\frac{2\fdot{\rhorep}\alprep+2\rhorep\fdot{\alprep}-\fdot{\dragf}\pfreq
  -2\sone(\fdot{\etarep}\szer+\etarep\fdot{\szer}+2\etarep\fdot{\etarep}\pfreq^2)}
  {2\etarep(\szer+\etarep\pfreq^2)}
\nonumber\\
\begin{split}
&=\frac{2\kaprep\vphid\ethvi+2\rhorep\kaprep\vrhox-\ethvf\pfreq
  -2(\vrhoh/\kaprep)(\kaprep^2\dltvb\vphig+\kaprep^2\epsvj\vrhoi+2\kaprep^2\epsvj\dltvb\pfreq^2)}
  {2\kaprep\epsvj(\kaprep\vphig+\kaprep\epsvj\pfreq^2)}\\
  &\qquad\beqref{rot4}, \eqnref{rot5}, \eqnref{rpath4}, \eqnref{rpath5a}, \eqnref{rpath8a},
  \eqnref{rpath11a}\text{ \& }\eqnref{rpath12a}
\end{split}
\nonumber\\
&=\frac{2\vphid\ethvi+2\rhorep\vrhox-\scalc\ethvf
  -2\vrhoh(\dltvb\vphig+\epsvj\vrhoi+2\epsvj\dltvb\pfreq^2)}
  {2\kaprep\epsvj(\vphig+\epsvj\pfreq^2)}
=\ethvl/\kaprep\beqref{rpath1i}
\end{align}
\begin{align*}
\ffdot{\sone}
&=\dif{\biggl[\frac{2\fdot{\rhorep}\alprep+2\rhorep\fdot{\alprep}-\fdot{\dragf}\pfreq
  -2\sone(\fdot{\etarep}\szer+\etarep\fdot{\szer}+2\etarep\fdot{\etarep}\pfreq^2)}
  {2\etarep(\szer+\etarep\pfreq^2)}\biggr]}\beqref{rpath13a}\nonumber\\
\begin{split}
&=\frac{2\etarep(\szer+\etarep\pfreq^2)\dif{[2\fdot{\rhorep}\alprep+2\rhorep\fdot{\alprep}-\fdot{\dragf}\pfreq
    -2\sone(\fdot{\etarep}\szer+\etarep\fdot{\szer}+2\etarep\fdot{\etarep}\pfreq^2)]}}
    {4\etarep^2(\szer+\etarep\pfreq^2)^2}\\
  &\qquad-\frac{[2\fdot{\rhorep}\alprep+2\rhorep\fdot{\alprep}-\fdot{\dragf}\pfreq-2\sone(\fdot{\etarep}\szer
     +\etarep\fdot{\szer}+2\etarep\fdot{\etarep}\pfreq^2)]\dif{[2\etarep(\szer+\etarep\pfreq^2)]}}
    {4\etarep^2(\szer+\etarep\pfreq^2)^2}
\end{split}
\nonumber\\
&=\frac{\dif{[2\fdot{\rhorep}\alprep+2\rhorep\fdot{\alprep}-\fdot{\dragf}\pfreq]}}
    {2\etarep(\szer+\etarep\pfreq^2)}
  -\frac{\dif{[2\sone(\fdot{\etarep}\szer+\etarep\fdot{\szer}+2\etarep\fdot{\etarep}\pfreq^2)]}}
    {2\etarep(\szer+\etarep\pfreq^2)}
  -\frac{\fdot{\sone}\dif{[2\etarep(\szer+\etarep\pfreq^2)]}}
    {2\etarep(\szer+\etarep\pfreq^2)}\beqref{rpath13a}
\nonumber\\
\begin{split}
&=\frac{2\ffdot{\rhorep}\alprep+2\fdot{\rhorep}\fdot{\alprep}+2\fdot{\rhorep}\fdot{\alprep}
    +2\rhorep\ffdot{\alprep}-\ffdot{\dragf}\pfreq}{2\etarep(\szer+\etarep\pfreq^2)}
  -\frac{2\fdot{\sone}(\fdot{\etarep}\szer+\etarep\fdot{\szer}+2\etarep\fdot{\etarep}\pfreq^2)}
    {2\etarep(\szer+\etarep\pfreq^2)}\\
  &\qquad-\frac{2\sone(\ffdot{\etarep}\szer+\fdot{\etarep}\fdot{\szer}+\fdot{\etarep}\fdot{\szer}
    +\etarep\ffdot{\szer}+2\fdot{\etarep}^2\pfreq^2+2\etarep\ffdot{\etarep}\pfreq^2)}
    {2\etarep(\szer+\etarep\pfreq^2)}
  -\frac{2\fdot{\sone}(\fdot{\etarep}\szer+\etarep\fdot{\szer}+2\etarep\fdot{\etarep}\pfreq^2)}
    {2\etarep(\szer+\etarep\pfreq^2)}
\end{split}
\end{align*}
\begin{align}\label{rpath13b}
\begin{split}
&=\frac{2\ffdot{\rhorep}\alprep+4\fdot{\rhorep}\fdot{\alprep}
    +2\rhorep\ffdot{\alprep}-\ffdot{\dragf}\pfreq}{2\etarep(\szer+\etarep\pfreq^2)}
  -\frac{4\fdot{\sone}(\fdot{\etarep}\szer+\etarep\fdot{\szer}+2\etarep\fdot{\etarep}\pfreq^2)}
    {2\etarep(\szer+\etarep\pfreq^2)}\\
  &\qquad-\frac{2\sone(\ffdot{\etarep}\szer+\etarep\ffdot{\szer}+2\fdot{\etarep}\fdot{\szer}
    +2\fdot{\etarep}^2\pfreq^2+2\etarep\ffdot{\etarep}\pfreq^2)}
    {2\etarep(\szer+\etarep\pfreq^2)}
\end{split}
\nonumber\\
\begin{split}
&=\frac{2\kaprep\vphid\ethvj+4\kaprep\vrhox\ethvi+2\rhorep\kaprep\vrhoy-\ethvg\pfreq}
   {2\kaprep\epsvj(\kaprep\vphig+\kaprep\epsvj\pfreq^2)}
  -\frac{4(\ethvl/\kaprep)(\kaprep^2\dltvb\vphig+\kaprep^2\epsvj\vrhoi+2\kaprep^2\epsvj\dltvb\pfreq^2)}
   {2\kaprep\epsvj(\kaprep\vphig+\kaprep\epsvj\pfreq^2)}\\
  &\qquad-\frac{2(\vphih/\kaprep)(\kaprep^2\vphig\frkta+\kaprep^2\epsvj\vrhon+2\kaprep^2\dltvb\vrhoi
    +2\kaprep^2\dltvb^2\pfreq^2+2\kaprep^2\epsvj\frkta\pfreq^2)}
   {2\kaprep\epsvj(\kaprep\vphig+\kaprep\epsvj\pfreq^2)}\\
  &\qquad\beqref{rot4}, \eqnref{rot5}, \eqnref{rpath4}, \eqnref{rpath5}, \eqnref{rpath8},
  \eqnref{rpath11b}, \eqnref{rpath12}\text{ \& }\eqnref{rpath13a}
\end{split}
\nonumber\\
\begin{split}
&=\frac{2\vphid\ethvj+4\vrhox\ethvi+2\rhorep\vrhoy-\scalc\ethvg}
   {2\kaprep\epsvj(\vphig+\epsvj\pfreq^2)}
  -\frac{4\ethvl(\dltvb\vphig+\epsvj\vrhoi+2\epsvj\dltvb\pfreq^2)}
   {2\kaprep\epsvj(\vphig+\epsvj\pfreq^2)}\\
  &\qquad-\frac{2\vphih(\vphig\frkta+\epsvj\vrhon+2\dltvb\vrhoi
    +2\dltvb^2\pfreq^2+2\epsvj\frkta\pfreq^2)}
   {2\kaprep\epsvj(\vphig+\epsvj\pfreq^2)}
=\ethvm/\kaprep\beqref{rpath1i}
\end{split}
\end{align}
\begin{align*}
\begin{split}
\fffdot{\sone}
&=\dif{\biggl[\frac{2\ffdot{\rhorep}\alprep+4\fdot{\rhorep}\fdot{\alprep}
    +2\rhorep\ffdot{\alprep}-\ffdot{\dragf}\pfreq}{2\etarep(\szer+\etarep\pfreq^2)}\biggr]}
  -\dif{\biggl[\frac{4\fdot{\sone}(\fdot{\etarep}\szer+\etarep\fdot{\szer}+2\etarep\fdot{\etarep}\pfreq^2)}
    {2\etarep(\szer+\etarep\pfreq^2)}\biggr]}\\
  &\qquad-\dif{\biggl[\frac{2\sone(\ffdot{\etarep}\szer+\etarep\ffdot{\szer}+2\fdot{\etarep}\fdot{\szer}
    +2\fdot{\etarep}^2\pfreq^2+2\etarep\ffdot{\etarep}\pfreq^2)}
    {2\etarep(\szer+\etarep\pfreq^2)}\biggr]}\beqref{rpath13b}
\end{split}
\nonumber\\
\begin{split}
&=\frac{2\etarep(\szer+\etarep\pfreq^2)\dif{[2\ffdot{\rhorep}\alprep+4\fdot{\rhorep}\fdot{\alprep}
    +2\rhorep\ffdot{\alprep}-\ffdot{\dragf}\pfreq]}
    -[2\ffdot{\rhorep}\alprep+4\fdot{\rhorep}\fdot{\alprep}+2\rhorep\ffdot{\alprep}-\ffdot{\dragf}\pfreq]
      \dif{[2\etarep(\szer+\etarep\pfreq^2)]}}{4\etarep^2(\szer+\etarep\pfreq^2)^2}\\
  &\quad-\frac{2\etarep(\szer+\etarep\pfreq^2)\dif{[4\fdot{\sone}(\fdot{\etarep}\szer+\etarep\fdot{\szer}
    +2\etarep\fdot{\etarep}\pfreq^2)]}-[4\fdot{\sone}(\fdot{\etarep}\szer+\etarep\fdot{\szer}
    +2\etarep\fdot{\etarep}\pfreq^2)]\dif{[2\etarep(\szer+\etarep\pfreq^2)]}}
    {4\etarep^2(\szer+\etarep\pfreq^2)^2}\\
  &\quad-\frac{2\etarep(\szer+\etarep\pfreq^2)\dif{[2\sone(\ffdot{\etarep}\szer+\etarep\ffdot{\szer}
    +2\fdot{\etarep}\fdot{\szer}+2\fdot{\etarep}^2\pfreq^2+2\etarep\ffdot{\etarep}\pfreq^2)]}}
    {4\etarep^2(\szer+\etarep\pfreq^2)^2}\\
  &\quad+\frac{[2\sone(\ffdot{\etarep}\szer+\etarep\ffdot{\szer}+2\fdot{\etarep}\fdot{\szer}
    +2\fdot{\etarep}^2\pfreq^2+2\etarep\ffdot{\etarep}\pfreq^2)]\dif{[2\etarep(\szer+\etarep\pfreq^2)]}}
    {4\etarep^2(\szer+\etarep\pfreq^2)^2}
\end{split}
\end{align*}
\begin{align*}
\begin{split}
&=\frac{\dif{[2\ffdot{\rhorep}\alprep+4\fdot{\rhorep}\fdot{\alprep}
    +2\rhorep\ffdot{\alprep}-\ffdot{\dragf}\pfreq]}}{2\etarep(\szer+\etarep\pfreq^2)}
    -\frac{\ffdot{\sone}\dif{[2\etarep(\szer+\etarep\pfreq^2)]}}{2\etarep(\szer+\etarep\pfreq^2)}
  -\frac{\dif{[4\fdot{\sone}(\fdot{\etarep}\szer+\etarep\fdot{\szer}
    +2\etarep\fdot{\etarep}\pfreq^2)]}}{2\etarep(\szer+\etarep\pfreq^2)}\\
  &\quad-\frac{\dif{[2\sone(\ffdot{\etarep}\szer+\etarep\ffdot{\szer}
    +2\fdot{\etarep}\fdot{\szer}+2\fdot{\etarep}^2\pfreq^2+2\etarep\ffdot{\etarep}\pfreq^2)]}}
    {2\etarep(\szer+\etarep\pfreq^2)}\beqref{rpath13b}
\end{split}
\nonumber\\
\begin{split}
&=\frac{2\fffdot{\rhorep}\alprep+2\ffdot{\rhorep}\fdot{\alprep}
    +4\ffdot{\rhorep}\fdot{\alprep}+4\fdot{\rhorep}\ffdot{\alprep}
    +2\fdot{\rhorep}\ffdot{\alprep}+2\rhorep\fffdot{\alprep}
    -\fffdot{\dragf}\pfreq}
    {2\etarep(\szer+\etarep\pfreq^2)}
  -\frac{2\ffdot{\sone}(\fdot{\etarep}\szer+\etarep\fdot{\szer}
    +2\etarep\fdot{\etarep}\pfreq^2)}
    {2\etarep(\szer+\etarep\pfreq^2)}\\
  &\quad-\frac{4\ffdot{\sone}(\fdot{\etarep}\szer+\etarep\fdot{\szer}
    +2\etarep\fdot{\etarep}\pfreq^2)+4\fdot{\sone}(\ffdot{\etarep}\szer+\fdot{\etarep}\fdot{\szer}
    +\fdot{\etarep}\fdot{\szer}+\etarep\ffdot{\szer}
    +2\fdot{\etarep}^2\pfreq^2+2\etarep\ffdot{\etarep}\pfreq^2)}
    {2\etarep(\szer+\etarep\pfreq^2)}\\
  &\quad-\frac{2\sone(\fffdot{\etarep}\szer+\ffdot{\etarep}\fdot{\szer}
    +\fdot{\etarep}\ffdot{\szer}+\etarep\fffdot{\szer}
    +2\ffdot{\etarep}\fdot{\szer}+2\fdot{\etarep}\ffdot{\szer}
    +4\fdot{\etarep}\ffdot{\etarep}\pfreq^2
    +2\fdot{\etarep}\ffdot{\etarep}\pfreq^2+2\etarep\fffdot{\etarep}\pfreq^2)}
    {2\etarep(\szer+\etarep\pfreq^2)}\\
  &\quad-\frac{2\fdot{\sone}(\ffdot{\etarep}\szer+\etarep\ffdot{\szer}
    +2\fdot{\etarep}\fdot{\szer}+2\fdot{\etarep}^2\pfreq^2+2\etarep\ffdot{\etarep}\pfreq^2)}
    {2\etarep(\szer+\etarep\pfreq^2)}
\end{split}
\end{align*}
\begin{align}\label{rpath13c}
\begin{split}
&=\frac{2\fffdot{\rhorep}\alprep+2\rhorep\fffdot{\alprep}
    +6\ffdot{\rhorep}\fdot{\alprep}+6\fdot{\rhorep}\ffdot{\alprep}
    -\fffdot{\dragf}\pfreq}
    {2\etarep(\szer+\etarep\pfreq^2)}
  -\frac{6\ffdot{\sone}(\fdot{\etarep}\szer+\etarep\fdot{\szer}+2\etarep\fdot{\etarep}\pfreq^2)}
    {2\etarep(\szer+\etarep\pfreq^2)}\\
  &\quad-\frac{2\sone(\fffdot{\etarep}\szer+\etarep\fffdot{\szer}
    +3\ffdot{\etarep}\fdot{\szer}+3\fdot{\etarep}\ffdot{\szer}
    +6\fdot{\etarep}\ffdot{\etarep}\pfreq^2+2\etarep\fffdot{\etarep}\pfreq^2)}
    {2\etarep(\szer+\etarep\pfreq^2)}
  -\frac{6\fdot{\sone}(\ffdot{\etarep}\szer
    +\etarep\ffdot{\szer}+2\fdot{\etarep}\fdot{\szer}
    +2\fdot{\etarep}^2\pfreq^2+2\etarep\ffdot{\etarep}\pfreq^2)}
    {2\etarep(\szer+\etarep\pfreq^2)}
\end{split}
\nonumber\\
\begin{split}
&=\frac{2\kaprep\vphid\ethvk+2\rhorep\kaprep\vrhoz
    +6\kaprep\vrhox\ethvj+6\kaprep\vrhoy\ethvi-\ethvh\pfreq}
    {2\kaprep\epsvj(\kaprep\vphig+\kaprep\epsvj\pfreq^2)}
  -\frac{6(\ethvm/\kaprep)(\kaprep^2\dltvb\vphig+\kaprep^2\epsvj\vrhoi+2\kaprep^2\dltvb\epsvj\pfreq^2)}
    {2\kaprep\epsvj(\kaprep\vphig+\kaprep\epsvj\pfreq^2)}\\
  &\quad-\frac{2(\vphih/\kaprep)(\kaprep^2\vphig\frktb+\kaprep^2\epsvj\vrhoo
    +3\kaprep^2\vrhoi\frkta+3\kaprep^2\dltvb\vrhon
    +6\kaprep^2\dltvb\frkta\pfreq^2+2\kaprep^2\epsvj\frktb\pfreq^2)}
    {2\kaprep\epsvj(\kaprep\vphig+\kaprep\epsvj\pfreq^2)}\\
  &\quad-\frac{6(\ethvl/\kaprep)(\kaprep^2\vphig\frkta
    +\kaprep^2\epsvj\vrhon+2\kaprep^2\dltvb\vrhoi
    +2\kaprep^2\dltvb^2\pfreq^2+2\kaprep^2\epsvj\frkta\pfreq^2)}
    {2\kaprep\epsvj(\kaprep\vphig+\kaprep\epsvj\pfreq^2)}\\
  &\qquad\beqref{rot4}, \eqnref{rot5}, \eqnref{rpath4}, \eqnref{rpath5}, \eqnref{rpath8},
  \eqnref{rpath11c}, \eqnref{rpath12}, \eqnref{rpath13a}\text{ \& }\eqnref{rpath13b}
\end{split}
\nonumber\\
\begin{split}
&=\frac{2\vphid\ethvk+2\rhorep\vrhoz+6\vrhox\ethvj+6\vrhoy\ethvi-\scalc\ethvh}
    {2\kaprep\epsvj(\vphig+\epsvj\pfreq^2)}
  -\frac{6\ethvm(\dltvb\vphig+\epsvj\vrhoi+2\dltvb\epsvj\pfreq^2)}
    {2\kaprep\epsvj(\vphig+\epsvj\pfreq^2)}\\
  &\quad-\frac{2\vphih(\vphig\frktb+\epsvj\vrhoo+3\vrhoi\frkta+3\dltvb\vrhon
    +6\dltvb\frkta\pfreq^2+2\epsvj\frktb\pfreq^2)}
    {2\kaprep\epsvj(\vphig+\epsvj\pfreq^2)}\\
  &\quad-\frac{6\ethvl(\vphig\frkta+\epsvj\vrhon+2\dltvb\vrhoi
    +2\dltvb^2\pfreq^2+2\epsvj\frkta\pfreq^2)}
    {2\kaprep\epsvj(\vphig+\epsvj\pfreq^2)}
=\ethvn/\kaprep\beqref{rpath1i}
\end{split}
\end{align}
\end{subequations}
\begin{subequations}\label{rpath14}
\begin{align}\label{rpath14a}
\fdot{\stwo}
&=\dif{[\szer\sone]}\beqref{main4d}\nonumber\\
&=\szer\fdot{\sone}+\fdot{\szer}\sone
=(\kaprep\vphig)(\ethvl/\kaprep)+(\kaprep\vrhoi)(\vphih/\kaprep)
  \beqref{rot5}, \eqnref{rpath5a}\text{ \& }\eqnref{rpath13a}\nonumber\\
&=\vphig\ethvl+\vrhoi\vphih=\ethvo\beqref{rpath1j}
\end{align}
\begin{align}\label{rpath14b}
\ffdot{\stwo}
&=\dif{[\szer\fdot{\sone}+\fdot{\szer}\sone]}\beqref{rpath14a}\nonumber\\
&=\fdot{\szer}\fdot{\sone}+\szer\ffdot{\sone}+\ffdot{\szer}\sone+\fdot{\szer}\fdot{\sone}
=2\fdot{\szer}\fdot{\sone}+\szer\ffdot{\sone}+\ffdot{\szer}\sone\nonumber\\
&=2(\kaprep\vrhoi)(\ethvl/\kaprep)+(\kaprep\vphig)(\ethvm/\kaprep)+(\kaprep\vrhon)(\vphih/\kaprep)
  \beqref{rot5}, \eqnref{rpath5}\text{ \& }\eqnref{rpath13}\nonumber\\
&=2\vrhoi\ethvl+\vphig\ethvm+\vrhon\vphih=\ethvp\beqref{rpath1j}
\end{align}
\begin{align}\label{rpath14c}
\fffdot{\stwo}
&=\dif{[2\fdot{\szer}\fdot{\sone}+\szer\ffdot{\sone}+\ffdot{\szer}\sone]}\beqref{rpath14b}\nonumber\\
&=2\ffdot{\szer}\fdot{\sone}+2\fdot{\szer}\ffdot{\sone}+\fdot{\szer}\ffdot{\sone}+\szer\fffdot{\sone}
  +\fffdot{\szer}\sone+\ffdot{\szer}\fdot{\sone}
=3\ffdot{\szer}\fdot{\sone}+3\fdot{\szer}\ffdot{\sone}+\szer\fffdot{\sone}+\fffdot{\szer}\sone\nonumber\\
&=3(\kaprep\vrhon)(\ethvl/\kaprep)+3(\kaprep\vrhoi)(\ethvm/\kaprep)+(\kaprep\vphig)(\ethvn/\kaprep)
  +(\kaprep\vrhoo)(\vphih/\kaprep)\beqref{rot5}, \eqnref{rpath5}\text{ \& }\eqnref{rpath13}\nonumber\\
&=3(\vrhon\ethvl+\vrhoi\ethvm)+\vphig\ethvn+\vrhoo\vphih=\ethvq\beqref{rpath1j}
\end{align}
\end{subequations}
\begin{subequations}\label{rpath15}
\begin{align}\label{rpath15a}
\fdot{\sthr}
&=\dif{[\{\etarep(\dprod{\vectLam}{\unitplz})+4\xirep(\dprod{\vectOme}{\unitplz})\}\sone]}
  \beqref{main4d}\nonumber\\
&=\fdot{\sone}[\etarep(\dprod{\vectLam}{\unitplz})+4\xirep(\dprod{\vectOme}{\unitplz})]
  +\sone[\fdot{\etarep}(\dprod{\vectLam}{\unitplz})+\etarep(\dprod{\fdot{\vectLam}}{\unitplz})
    +4\fdot{\xirep}(\dprod{\vectOme}{\unitplz})+4\xirep(\dprod{\vectLam}{\unitplz})]\nonumber\\
&=\fdot{\sone}(\etarep\epsvi+4\xirep\epsvc)+\sone(\fdot{\etarep}\epsvi+\etarep\vsiga
    +4\fdot{\xirep}\epsvc+4\xirep\epsvi)\beqref{rot1a}\text{ \& }\eqnref{rpath1a}\nonumber\\
&=(\etarep\fdot{\sone}+\fdot{\etarep}\sone+4\xirep\sone)\epsvi
  +(4\xirep\fdot{\sone}+4\fdot{\xirep}\sone)\epsvc
  +\etarep\sone\vsiga\nonumber\\
\begin{split}
&=[\kaprep\epsvj(\ethvl/\kaprep)+\kaprep\dltvb(\vphih/\kaprep)+8\kaprep\vphie(\vphih/\kaprep)]\epsvi
  +[8\kaprep\vphie(\ethvl/\kaprep)+8\kaprep\vrhoa(\vphih/\kaprep)]\epsvc\\
  &\qquad+\kaprep\epsvj(\vphih/\kaprep)\vsiga
  \beqref{rot4}, \eqnref{rot5b}, \eqnref{rpath2a}, \eqnref{rpath4}\text{ \& }\eqnref{rpath13a}
\end{split}
\nonumber\\
&=\epsvj(\epsvi\ethvl+\vphih\vsiga)+\epsvi\vphih(\dltvb+8\vphie)
  +8\epsvc(\vphie\ethvl+\vrhoa\vphih)\nonumber\\
&=\ethvr\beqref{rpath1j}
\end{align}
\begin{align}\label{rpath15b}
\ffdot{\sthr}
&=\dif{[(\etarep\fdot{\sone}+\fdot{\etarep}\sone+4\xirep\sone)(\dprod{\vectLam}{\unitplz})
  +(4\xirep\fdot{\sone}+4\fdot{\xirep}\sone)(\dprod{\vectOme}{\unitplz})
  +\etarep\sone(\dprod{\fdot{\vectLam}}{\unitplz})]}\beqref{rpath15a}\nonumber\\
\begin{split}
&=(\fdot{\etarep}\fdot{\sone}+\etarep\ffdot{\sone}+\ffdot{\etarep}\sone+\fdot{\etarep}\fdot{\sone}
  +4\fdot{\xirep}\sone+4\xirep\fdot{\sone})(\dprod{\vectLam}{\unitplz})
  +(\etarep\fdot{\sone}+\fdot{\etarep}\sone+4\xirep\sone)(\dprod{\fdot{\vectLam}}{\unitplz})\\
  &\quad+(4\fdot{\xirep}\fdot{\sone}+4\xirep\ffdot{\sone}+4\ffdot{\xirep}\sone+4\fdot{\xirep}\fdot{\sone})
    (\dprod{\vectOme}{\unitplz})
  +(4\xirep\fdot{\sone}+4\fdot{\xirep}\sone)(\dprod{\vectLam}{\unitplz})\\
  &\quad+(\fdot{\etarep}\sone+\etarep\fdot{\sone})(\dprod{\fdot{\vectLam}}{\unitplz})
  +\etarep\sone(\dprod{\ffdot{\vectLam}}{\unitplz})
\end{split}
\nonumber\\
\begin{split}
&=(\etarep\ffdot{\sone}+\ffdot{\etarep}\sone+2\fdot{\etarep}\fdot{\sone}
  +8\fdot{\xirep}\sone+8\xirep\fdot{\sone})(\dprod{\vectLam}{\unitplz})
  +(2\etarep\fdot{\sone}+2\fdot{\etarep}\sone+4\xirep\sone)(\dprod{\fdot{\vectLam}}{\unitplz})\\
  &\quad+(4\xirep\ffdot{\sone}+4\ffdot{\xirep}\sone+8\fdot{\xirep}\fdot{\sone})(\dprod{\vectOme}{\unitplz})
  +\etarep\sone(\dprod{\ffdot{\vectLam}}{\unitplz})
\end{split}
\nonumber\\
\begin{split}
&=[\kaprep\epsvj(\ethvm/\kaprep)+\kaprep\frkta(\vphih/\kaprep)+2\kaprep\dltvb(\ethvl/\kaprep)
  +16\kaprep\vrhoa(\vphih/\kaprep)+16\kaprep\vphie(\ethvl/\kaprep)]\epsvi\\
  &\quad+[2\kaprep\epsvj(\ethvl/\kaprep)+2\kaprep\dltvb(\vphih/\kaprep)+8\kaprep\vphie(\vphih/\kaprep)]\vsiga
  +[8\kaprep\vphie(\ethvm/\kaprep)+8\kaprep\vrhob(\vphih/\kaprep)+16\kaprep\vrhoa(\ethvl/\kaprep)]\epsvc\\
  &\quad+\kaprep\epsvj(\vphih/\kaprep)\vsigf
  \beqref{rot1a}, \eqnref{rot4}, \eqnref{rot5b}, \eqnref{rpath1a}, \eqnref{rpath2},
  \eqnref{rpath4}\text{ \& }\eqnref{rpath13}
\end{split}
\nonumber\\
\begin{split}
&=\epsvj(\vphih\vsigf+\epsvi\ethvm)+\epsvi(\frkta\vphih+2\dltvb\ethvl+16\vrhoa\vphih+16\vphie\ethvl)\\
  &\quad+2\vsiga(\epsvj\ethvl+\dltvb\vphih+4\vphie\vphih)+8\epsvc(\vphie\ethvm+\vrhob\vphih+2\vrhoa\ethvl)
\end{split}
\nonumber\\
&=\ethvs\beqref{rpath1j}
\end{align}
\begin{align*}
\begin{split}
\fffdot{\sthr}
&=[(\etarep\ffdot{\sone}+\ffdot{\etarep}\sone+2\fdot{\etarep}\fdot{\sone}
  +8\fdot{\xirep}\sone+8\xirep\fdot{\sone})(\dprod{\vectLam}{\unitplz})
  +(2\etarep\fdot{\sone}+2\fdot{\etarep}\sone+4\xirep\sone)(\dprod{\fdot{\vectLam}}{\unitplz})\\
  &\quad+(4\xirep\ffdot{\sone}+4\ffdot{\xirep}\sone+8\fdot{\xirep}\fdot{\sone})(\dprod{\vectOme}{\unitplz})
  +\etarep\sone(\dprod{\ffdot{\vectLam}}{\unitplz})\dif{]}\beqref{rpath15b}
\end{split}
\nonumber\\
\begin{split}
&=(\fdot{\etarep}\ffdot{\sone}+\etarep\fffdot{\sone}
    +\fffdot{\etarep}\sone+\ffdot{\etarep}\fdot{\sone}
    +2\ffdot{\etarep}\fdot{\sone}+2\fdot{\etarep}\ffdot{\sone}
    +8\ffdot{\xirep}\sone+8\fdot{\xirep}\fdot{\sone}
    +8\fdot{\xirep}\fdot{\sone}+8\xirep\ffdot{\sone})(\dprod{\vectLam}{\unitplz})\\
  &\quad+(\etarep\ffdot{\sone}+\ffdot{\etarep}\sone+2\fdot{\etarep}\fdot{\sone}
    +8\fdot{\xirep}\sone+8\xirep\fdot{\sone})(\dprod{\fdot{\vectLam}}{\unitplz})
  +(2\fdot{\etarep}\fdot{\sone}+2\etarep\ffdot{\sone}
     +2\ffdot{\etarep}\sone+2\fdot{\etarep}\fdot{\sone}
     +4\fdot{\xirep}\sone+4\xirep\fdot{\sone})(\dprod{\fdot{\vectLam}}{\unitplz})\\
  &\quad+(2\etarep\fdot{\sone}+2\fdot{\etarep}\sone+4\xirep\sone)(\dprod{\ffdot{\vectLam}}{\unitplz})
  +(4\fdot{\xirep}\ffdot{\sone}+4\xirep\fffdot{\sone}
     +4\fffdot{\xirep}\sone+4\ffdot{\xirep}\fdot{\sone}
     +8\ffdot{\xirep}\fdot{\sone}+8\fdot{\xirep}\ffdot{\sone})(\dprod{\vectOme}{\unitplz})\\
  &\quad+(4\xirep\ffdot{\sone}+4\ffdot{\xirep}\sone+8\fdot{\xirep}\fdot{\sone})(\dprod{\vectLam}{\unitplz})
  +(\fdot{\etarep}\sone+\etarep\fdot{\sone})(\dprod{\ffdot{\vectLam}}{\unitplz})
  +\etarep\sone(\dprod{\fffdot{\vectLam}}{\unitplz})
\end{split}
\nonumber\\
\begin{split}
&=(\etarep\fffdot{\sone}+\fffdot{\etarep}\sone
    +3\ffdot{\etarep}\fdot{\sone}+3\fdot{\etarep}\ffdot{\sone}
    +12\ffdot{\xirep}\sone+24\fdot{\xirep}\fdot{\sone}
    +12\xirep\ffdot{\sone})(\dprod{\vectLam}{\unitplz})\\
  &\quad+(3\etarep\ffdot{\sone}+3\ffdot{\etarep}\sone+6\fdot{\etarep}\fdot{\sone}
     +12\fdot{\xirep}\sone+12\xirep\fdot{\sone})(\dprod{\fdot{\vectLam}}{\unitplz})
  +(3\etarep\fdot{\sone}+3\fdot{\etarep}\sone+4\xirep\sone)(\dprod{\ffdot{\vectLam}}{\unitplz})\\
  &\quad+(4\xirep\fffdot{\sone}+4\fffdot{\xirep}\sone
     +12\ffdot{\xirep}\fdot{\sone}+12\fdot{\xirep}\ffdot{\sone})(\dprod{\vectOme}{\unitplz})
  +\etarep\sone(\dprod{\fffdot{\vectLam}}{\unitplz})
\end{split}
\end{align*}
\begin{align}\label{rpath15c}
\begin{split}
&=[\kaprep\epsvj(\ethvn/\kaprep)+\kaprep\frktb(\vphih/\kaprep)
    +3\kaprep\frkta(\ethvl/\kaprep)+3\kaprep\dltvb(\ethvm/\kaprep)
    +24\kaprep\vrhob(\vphih/\kaprep)+48\kaprep\vrhoa(\ethvl/\kaprep)\\
    &\quad+24\kaprep\vphie(\ethvm/\kaprep)]\epsvi
  +[3\kaprep\epsvj(\ethvm/\kaprep)+3\kaprep\frkta(\vphih/\kaprep)+6\kaprep\dltvb(\ethvl/\kaprep)
     +24\kaprep\vrhoa(\vphih/\kaprep)\\
     &\quad+24\kaprep\vphie(\ethvl/\kaprep)]\vsiga
  +[3\kaprep\epsvj(\ethvl/\kaprep)+3\kaprep\dltvb(\vphih/\kaprep)+8\kaprep\vphie(\vphih/\kaprep)]
     \vsigf\\
  &\quad+[8\kaprep\vphie(\ethvn/\kaprep)+8\kaprep\vrhod(\vphih/\kaprep)
     +24\kaprep\vrhob(\ethvl/\kaprep)+24\kaprep\vrhoa(\ethvm/\kaprep)]\epsvc
  +\kaprep\epsvj(\vphih/\kaprep)\vsigl\\
  &\quad\beqref{rot1a}, \eqnref{rot4}, \eqnref{rot5b}, \eqnref{rpath1a}, \eqnref{rpath2},
  \eqnref{rpath4}\text{ \& }\eqnref{rpath13}
\end{split}
\nonumber\\
\begin{split}
&=\epsvj(\vphih\vsigl+\epsvi\ethvn)+\epsvi(\frktb\vphih+3\frkta\ethvl+3\dltvb\ethvm+24\vrhob\vphih
    +48\vrhoa\ethvl+24\vphie\ethvm)\\
  &\quad+3\vsiga(\epsvj\ethvm+\frkta\vphih+2\dltvb\ethvl+8\vrhoa\vphih+8\vphie\ethvl)
  +\vsigf(3\epsvj\ethvl+3\dltvb\vphih+8\vphie\vphih)\\
  &\quad+8\epsvc(\vphie\ethvn+\vrhod\vphih+3\vrhob\ethvl+3\vrhoa\ethvm)
\end{split}
\nonumber\\
&=\ethvt\beqref{rpath1j}
\end{align}
\end{subequations}
\begin{subequations}\label{rpath16}
\begin{align}\label{rpath16a}
\fdot{\sfou}
&=\dif{[\{2\xirep\Omerep^2+\etarep(\dprod{\vectOme}{\vectLam})\}\sone]}\beqref{main4d}\nonumber\\
&=[2\fdot{\xirep}\Omerep^2+4\xirep\Omerep\fdot{\Omerep}+\fdot{\etarep}(\dprod{\vectOme}{\vectLam})
    +\etarep(\dprod{\vectLam}{\vectLam})+\etarep(\dprod{\vectOme}{\fdot{\vectLam}})]\sone
  +[2\xirep\Omerep^2+\etarep(\dprod{\vectOme}{\vectLam})]\fdot{\sone}\nonumber\\
&=2\Omerep^2(\fdot{\xirep}\sone+\xirep\fdot{\sone})
  +(\etarep\fdot{\sone}+\fdot{\etarep}\sone+4\xirep\sone)(\dprod{\vectOme}{\vectLam})
    +\etarep\sone[(\dprod{\vectLam}{\vectLam})+(\dprod{\vectOme}{\fdot{\vectLam}})]
  \nonumber\\
\begin{split}
&=2\Omerep^2[2\kaprep\vrhoa(\vphih/\kaprep)+2\kaprep\vphie(\ethvl/\kaprep)]
  +[\kaprep\epsvj(\ethvl/\kaprep)+\kaprep\dltvb(\vphih/\kaprep)+8\kaprep\vphie(\vphih/\kaprep)]\epsvg\\
    &\quad+\kaprep\epsvj(\vphih/\kaprep)(\Lamrep^2+\vsigc)
  \beqref{rot1a}, \eqnref{rot4}, \eqnref{rot5b}, \eqnref{rpath1a}, \eqnref{rpath2a},
  \eqnref{rpath4}\text{ \& }\eqnref{rpath13a}
\end{split}
\nonumber\\
&=4\Omerep^2(\vrhoa\vphih+\vphie\ethvl)+\epsvg(\epsvj\ethvl+\dltvb\vphih+8\vphie\vphih)
    +\epsvj\vphih(\Lamrep^2+\vsigc)\nonumber\\
&=\ethvu\beqref{rpath1j2}
\end{align}
\begin{align}\label{rpath16b}
\ffdot{\sfou}
&=\dif{[2\Omerep^2(\fdot{\xirep}\sone+\xirep\fdot{\sone})
  +(\etarep\fdot{\sone}+\fdot{\etarep}\sone+4\xirep\sone)(\dprod{\vectOme}{\vectLam})
    +\etarep\sone(\dprod{\vectLam}{\vectLam})+\etarep\sone(\dprod{\vectOme}{\fdot{\vectLam}})]}
  \beqref{rpath16a}\nonumber\\
\begin{split}
&=4\Omerep\fdot{\Omerep}(\fdot{\xirep}\sone+\xirep\fdot{\sone})
  +2\Omerep^2(\ffdot{\xirep}\sone+\fdot{\xirep}\fdot{\sone}+\fdot{\xirep}\fdot{\sone}+\xirep\ffdot{\sone})\\
  &\quad+(\fdot{\etarep}\fdot{\sone}+\etarep\ffdot{\sone}+\ffdot{\etarep}\sone+\fdot{\etarep}\fdot{\sone}
    +4\fdot{\xirep}\sone+4\xirep\fdot{\sone})(\dprod{\vectOme}{\vectLam})
  +(\etarep\fdot{\sone}+\fdot{\etarep}\sone+4\xirep\sone)
     [(\dprod{\vectLam}{\vectLam})+(\dprod{\vectOme}{\fdot{\vectLam}})]\\
  &\quad+(\fdot{\etarep}\sone+\etarep\fdot{\sone})(\dprod{\vectLam}{\vectLam})
  +2\etarep\sone(\dprod{\fdot{\vectLam}}{\vectLam})
  +(\fdot{\etarep}\sone+\etarep\fdot{\sone})(\dprod{\vectOme}{\fdot{\vectLam}})
  +\etarep\sone[(\dprod{\vectLam}{\fdot{\vectLam}})+(\dprod{\vectOme}{\ffdot{\vectLam}})]
\end{split}
\nonumber\\
\begin{split}
&=2\Omerep^2(\ffdot{\xirep}\sone+\xirep\ffdot{\sone}+2\fdot{\xirep}\fdot{\sone})
  +(\etarep\ffdot{\sone}+\ffdot{\etarep}\sone+2\fdot{\etarep}\fdot{\sone}
    +8\fdot{\xirep}\sone+8\xirep\fdot{\sone})(\dprod{\vectOme}{\vectLam})\\
  &\quad+(2\etarep\fdot{\sone}+2\fdot{\etarep}\sone+4\xirep\sone)
     [(\dprod{\vectLam}{\vectLam})+(\dprod{\vectOme}{\fdot{\vectLam}})]
  +\etarep\sone[3(\dprod{\fdot{\vectLam}}{\vectLam})+(\dprod{\vectOme}{\ffdot{\vectLam}})]
\end{split}
\nonumber\\
\begin{split}
&=2\Omerep^2[2\kaprep\vrhob(\vphih/\kaprep)+2\kaprep\vphie(\ethvm/\kaprep)+4\kaprep\vrhoa(\ethvl/\kaprep)]
  +[\kaprep\epsvj(\ethvm/\kaprep)+\kaprep\frkta(\vphih/\kaprep)+2\kaprep\dltvb(\ethvl/\kaprep)\\
    &\quad+16\kaprep\vrhoa(\vphih/\kaprep)+16\kaprep\vphie(\ethvl/\kaprep)]\epsvg
  +[2\kaprep\epsvj(\ethvl/\kaprep)+2\kaprep\dltvb(\vphih/\kaprep)+8\kaprep\vphie(\vphih/\kaprep)]
     (\Lamrep^2+\vsigc)\\
  &\quad+\kaprep\epsvj(\vphih/\kaprep)(3\vsigd+\vsigh)
  \beqref{rot1a}, \eqnref{rot4}, \eqnref{rot5b}, \eqnref{rpath1a}, \eqnref{rpath2},
  \eqnref{rpath4}\text{ \& }\eqnref{rpath13}
\end{split}
\nonumber\\
\begin{split}
&=4\Omerep^2(\vrhob\vphih+\vphie\ethvm+2\vrhoa\ethvl)+\epsvg(\epsvj\ethvm+\frkta\vphih+2\dltvb\ethvl
    +16\vrhoa\vphih+16\vphie\ethvl)\\
  &\quad+2(\Lamrep^2+\vsigc)(\epsvj\ethvl+\dltvb\vphih+4\vphie\vphih)+\epsvj\vphih(3\vsigd+\vsigh)
\end{split}
\nonumber\\
&=\ethvv\beqref{rpath1j2}
\end{align}
\begin{align*}
\begin{split}
\fffdot{\sfou}
&=[2\Omerep^2(\ffdot{\xirep}\sone+\xirep\ffdot{\sone}+2\fdot{\xirep}\fdot{\sone})
  +(\etarep\ffdot{\sone}+\ffdot{\etarep}\sone+2\fdot{\etarep}\fdot{\sone}
    +8\fdot{\xirep}\sone+8\xirep\fdot{\sone})(\dprod{\vectOme}{\vectLam})\\
  &\quad+(2\etarep\fdot{\sone}+2\fdot{\etarep}\sone+4\xirep\sone)(\dprod{\vectLam}{\vectLam})
  +(2\etarep\fdot{\sone}+2\fdot{\etarep}\sone+4\xirep\sone)(\dprod{\vectOme}{\fdot{\vectLam}})\\
  &\quad+3\etarep\sone(\dprod{\fdot{\vectLam}}{\vectLam})
  +\etarep\sone(\dprod{\vectOme}{\ffdot{\vectLam}})\dif{]}\beqref{rpath16b}
\end{split}
\nonumber\\
\begin{split}
&=4\Omerep\fdot{\Omerep}(\ffdot{\xirep}\sone+\xirep\ffdot{\sone}+2\fdot{\xirep}\fdot{\sone})
  +2\Omerep^2(\fffdot{\xirep}\sone+\ffdot{\xirep}\fdot{\sone}+\fdot{\xirep}\ffdot{\sone}
     +\xirep\fffdot{\sone}+2\ffdot{\xirep}\fdot{\sone}+2\fdot{\xirep}\ffdot{\sone})\\
  &\quad+(\fdot{\etarep}\ffdot{\sone}+\etarep\fffdot{\sone}
     +\fffdot{\etarep}\sone+\ffdot{\etarep}\fdot{\sone}
     +2\ffdot{\etarep}\fdot{\sone}+2\fdot{\etarep}\ffdot{\sone}
     +8\ffdot{\xirep}\sone+8\fdot{\xirep}\fdot{\sone}
     +8\fdot{\xirep}\fdot{\sone}+8\xirep\ffdot{\sone})(\dprod{\vectOme}{\vectLam})\\
  &\quad+(\etarep\ffdot{\sone}+\ffdot{\etarep}\sone+2\fdot{\etarep}\fdot{\sone}
     +8\fdot{\xirep}\sone+8\xirep\fdot{\sone})[(\dprod{\vectLam}{\vectLam})+(\dprod{\vectOme}{\fdot{\vectLam}})]\\
  &\quad+(2\fdot{\etarep}\fdot{\sone}+2\etarep\ffdot{\sone}
     +2\ffdot{\etarep}\sone+2\fdot{\etarep}\fdot{\sone}
     +4\fdot{\xirep}\sone+4\xirep\fdot{\sone})(\dprod{\vectLam}{\vectLam})
  +2(2\etarep\fdot{\sone}+2\fdot{\etarep}\sone+4\xirep\sone)(\dprod{\fdot{\vectLam}}{\vectLam})\\
  &\quad+(2\fdot{\etarep}\fdot{\sone}+2\etarep\ffdot{\sone}
     +2\ffdot{\etarep}\sone+2\fdot{\etarep}\fdot{\sone}
     +4\fdot{\xirep}\sone+4\xirep\fdot{\sone})(\dprod{\vectOme}{\fdot{\vectLam}})\\
  &\quad+(2\etarep\fdot{\sone}+2\fdot{\etarep}\sone+4\xirep\sone)[(\dprod{\vectLam}{\fdot{\vectLam}})
    +(\dprod{\vectOme}{\ffdot{\vectLam}})]
  +3(\fdot{\etarep}\sone+\etarep\fdot{\sone})(\dprod{\fdot{\vectLam}}{\vectLam})\\
  &\quad+3\etarep\sone[(\dprod{\ffdot{\vectLam}}{\vectLam})+(\dprod{\fdot{\vectLam}}{\fdot{\vectLam}})]
  +(\fdot{\etarep}\sone+\etarep\fdot{\sone})(\dprod{\vectOme}{\ffdot{\vectLam}})
  +\etarep\sone[(\dprod{\vectLam}{\ffdot{\vectLam}})+(\dprod{\vectOme}{\fffdot{\vectLam}})]
\end{split}
\end{align*}
\begin{align}\label{rpath16c}
\begin{split}
&=2\Omerep^2(\fffdot{\xirep}\sone+\xirep\fffdot{\sone}+3\ffdot{\xirep}\fdot{\sone}
    +3\fdot{\xirep}\ffdot{\sone})
  +\etarep\sone[4(\dprod{\ffdot{\vectLam}}{\vectLam})
    +3(\dprod{\fdot{\vectLam}}{\fdot{\vectLam}})+(\dprod{\vectOme}{\fffdot{\vectLam}})]\\
  &\quad+(\etarep\fffdot{\sone}+\fffdot{\etarep}\sone+3\ffdot{\etarep}\fdot{\sone}+3\fdot{\etarep}\ffdot{\sone}
     +12\ffdot{\xirep}\sone+24\fdot{\xirep}\fdot{\sone}+12\xirep\ffdot{\sone})
     (\dprod{\vectOme}{\vectLam})\\
  &\quad+(3\etarep\ffdot{\sone}+3\ffdot{\etarep}\sone+6\fdot{\etarep}\fdot{\sone}
     +12\fdot{\xirep}\sone+12\xirep\fdot{\sone})[(\dprod{\vectOme}{\fdot{\vectLam}})+(\dprod{\vectLam}{\vectLam})]\\
  &\quad+(3\etarep\fdot{\sone}+3\fdot{\etarep}\sone+4\xirep\sone)
    [(\dprod{\vectOme}{\ffdot{\vectLam}})+3(\dprod{\fdot{\vectLam}}{\vectLam})]
\end{split}
\nonumber\\
\begin{split}
&=2\Omerep^2[2\kaprep\vrhod(\vphih/\kaprep)+2\kaprep\vphie(\ethvn/\kaprep)
     +6\kaprep\vrhob(\ethvl/\kaprep)+6\kaprep\vrhoa(\ethvm/\kaprep)]
  +\kaprep\epsvj(\vphih/\kaprep)(4\vsigi+3\vsige+\vsigm)\\
  &\quad+[\kaprep\epsvj(\ethvn/\kaprep)+\kaprep\frktb(\vphih/\kaprep)+3\kaprep\frkta(\ethvl/\kaprep)
     +3\kaprep\dltvb(\ethvm/\kaprep)+24\kaprep\vrhob(\vphih/\kaprep)+48\kaprep\vrhoa(\ethvl/\kaprep)\\
     &\quad+24\kaprep\vphie(\ethvm/\kaprep)]\epsvg
  +[3\kaprep\epsvj(\ethvm/\kaprep)+3\kaprep\frkta(\vphih/\kaprep)+6\kaprep\dltvb(\ethvl/\kaprep)
     +24\kaprep\vrhoa(\vphih/\kaprep)\\
     &\quad+24\kaprep\vphie(\ethvl/\kaprep)](\Lamrep^2+\vsigc)
  +[3\kaprep\epsvj(\ethvl/\kaprep)+3\kaprep\dltvb(\vphih/\kaprep)
     +8\kaprep\vphie(\vphih/\kaprep)](\vsigh+3\vsigd)\\
  &\quad\beqref{rot1a}, \eqnref{rot4}, \eqnref{rot5b}, \eqnref{rpath1a}, \eqnref{rpath2},
  \eqnref{rpath4}\text{ \& }\eqnref{rpath13}
\end{split}
\nonumber\\
\begin{split}
&=4\Omerep^2(\vrhod\vphih+\vphie\ethvn+3\vrhob\ethvl+3\vrhoa\ethvm)+\epsvj\vphih(4\vsigi+3\vsige+\vsigm)\\
  &\quad+\epsvg(\epsvj\ethvn+\frktb\vphih+3\frkta\ethvl+3\dltvb\ethvm+24\vrhob\vphih+48\vrhoa\ethvl
     +24\vphie\ethvm)\\
  &\quad+3(\Lamrep^2+\vsigc)(\epsvj\ethvm+\frkta\vphih+2\dltvb\ethvl+8\vrhoa\vphih+8\vphie\ethvl)\\
  &\quad+(\vsigh+3\vsigd)(3\epsvj\ethvl+3\dltvb\vphih+8\vphie\vphih)
\end{split}
\nonumber\\
&=\ethvw\beqref{rpath1j2}
\end{align}
\end{subequations}
in consequence of which the derivatives of $\taurep$ and $\vecte$ are computed as
\begin{subequations}\label{rpath17}
\begin{align}\label{rpath17a}
\fdote
&=\dif{[\stwo(\cprod{\unitplz}{\vectOme})+\sthr\vectOme-\sfou\unitplz]}
  \beqref{main4b}\nonumber\\
&=\fdot{\stwo}(\cprod{\unitplz}{\vectOme})+\stwo(\cprod{\unitplz}{\vectLam})
  +\fdot{\sthr}\vectOme+\sthr\vectLam-\fdot{\sfou}\unitplz\nonumber\\
\begin{split}
&=\ethvo(\cprod{\unitplz}{\vectOme})+\vphig\vphih(\cprod{\unitplz}{\vectLam})
  +\ethvr\vectOme+\vphii\vectLam-\ethvu\unitplz\\
  &\quad\beqref{rot5}, \eqnref{rpath14a}, \eqnref{rpath15a}\text{ \& }\eqnref{rpath16a}
\end{split}
\end{align}
\begin{align}\label{rpath17b}
\ffdote
&=\dif{[\fdot{\stwo}(\cprod{\unitplz}{\vectOme})+\stwo(\cprod{\unitplz}{\vectLam})
  +\fdot{\sthr}\vectOme+\sthr\vectLam-\fdot{\sfou}\unitplz]}\beqref{rpath17a}\nonumber\\
&=\ffdot{\stwo}(\cprod{\unitplz}{\vectOme})+\fdot{\stwo}(\cprod{\unitplz}{\vectLam})
  +\fdot{\stwo}(\cprod{\unitplz}{\vectLam})+\stwo(\cprod{\unitplz}{\fdot{\vectLam}})
  +\ffdot{\sthr}\vectOme+\fdot{\sthr}\vectLam+\fdot{\sthr}\vectLam+\sthr\fdot{\vectLam}
  -\ffdot{\sfou}\unitplz
\nonumber\\
&=\ffdot{\stwo}(\cprod{\unitplz}{\vectOme})+2\fdot{\stwo}(\cprod{\unitplz}{\vectLam})
  +\stwo(\cprod{\unitplz}{\fdot{\vectLam}})
  +\ffdot{\sthr}\vectOme+2\fdot{\sthr}\vectLam+\sthr\fdot{\vectLam}-\ffdot{\sfou}\unitplz
\nonumber\\
\begin{split}
&=\ethvp(\cprod{\unitplz}{\vectOme})+2\ethvo(\cprod{\unitplz}{\vectLam})
  +\vphig\vphih(\cprod{\unitplz}{\fdot{\vectLam}})
  +\ethvs\vectOme+2\ethvr\vectLam+\vphii\fdot{\vectLam}-\ethvv\unitplz\\
  &\quad\beqref{rot5}, \eqnref{rpath14}, \eqnref{rpath15}\text{ \& }\eqnref{rpath16b}
\end{split}
\end{align}
\begin{align*}
\fffdote
&=\dif{[\ffdot{\stwo}(\cprod{\unitplz}{\vectOme})+2\fdot{\stwo}(\cprod{\unitplz}{\vectLam})
  +\stwo(\cprod{\unitplz}{\fdot{\vectLam}})
  +\ffdot{\sthr}\vectOme+2\fdot{\sthr}\vectLam+\sthr\fdot{\vectLam}-\ffdot{\sfou}\unitplz]}
  \beqref{rpath17b}\nonumber\\
\begin{split}
&=\fffdot{\stwo}(\cprod{\unitplz}{\vectOme})+\ffdot{\stwo}(\cprod{\unitplz}{\vectLam})
  +2\ffdot{\stwo}(\cprod{\unitplz}{\vectLam})+2\fdot{\stwo}(\cprod{\unitplz}{\fdot{\vectLam}})
  +\fdot{\stwo}(\cprod{\unitplz}{\fdot{\vectLam}})+\stwo(\cprod{\unitplz}{\ffdot{\vectLam}})\\
  &\quad+\fffdot{\sthr}\vectOme+\ffdot{\sthr}\vectLam
  +2\ffdot{\sthr}\vectLam+2\fdot{\sthr}\fdot{\vectLam}
  +\fdot{\sthr}\fdot{\vectLam}+\sthr\ffdot{\vectLam}
  -\fffdot{\sfou}\unitplz
\end{split}
\end{align*}
\begin{align}\label{rpath17c}
\begin{split}
&=\fffdot{\stwo}(\cprod{\unitplz}{\vectOme})+3\ffdot{\stwo}(\cprod{\unitplz}{\vectLam})
  +3\fdot{\stwo}(\cprod{\unitplz}{\fdot{\vectLam}})+\stwo(\cprod{\unitplz}{\ffdot{\vectLam}})
  +\fffdot{\sthr}\vectOme+3\ffdot{\sthr}\vectLam+3\fdot{\sthr}\fdot{\vectLam}\\
  &\quad+\sthr\ffdot{\vectLam}-\fffdot{\sfou}\unitplz
\end{split}
\nonumber\\
\begin{split}
&=\ethvq(\cprod{\unitplz}{\vectOme})+3\ethvp(\cprod{\unitplz}{\vectLam})
  +3\ethvo(\cprod{\unitplz}{\fdot{\vectLam}})+\vphig\vphih(\cprod{\unitplz}{\ffdot{\vectLam}})
  +\ethvt\vectOme+3\ethvs\vectLam+3\ethvr\fdot{\vectLam}\\
  &\quad+\vphii\ffdot{\vectLam}-\ethvw\unitplz
  \beqref{rot5}, \eqnref{rpath14}, \eqnref{rpath15}\text{ \& }\eqnref{rpath16c}
\end{split}
\end{align}
\end{subequations}
\begin{subequations}\label{rpath18}
\begin{align}\label{rpath18a}
\fdott
&=2\kaprep^{-2}\dif{[\rhorep\alprep-\etarep\stwo]}\beqref{main4b}\nonumber\\
&=2\kaprep^{-2}(\fdot{\rhorep}\alprep+\rhorep\fdot{\alprep}-\fdot{\etarep}\stwo-\etarep\fdot{\stwo})\nonumber\\
\begin{split}
&=2\kaprep^{-2}(\kaprep\vrhod\ethvi+\rhorep\kaprep\vrhox-\kaprep\dltvb\vphig\vphih-\kaprep\epsvj\ethvo)\\
  &\qquad\beqref{rot4}, \eqnref{rot5c}, \eqnref{rpath4}, \eqnref{rpath8a},
  \eqnref{rpath12a}\text{ \& }\eqnref{rpath14a}
\end{split}
\nonumber\\
&=2\kaprep^{-1}(\vrhod\ethvi+\rhorep\vrhox-\dltvb\vphig\vphih-\epsvj\ethvo)\nonumber\\
&=2(\ethvx/\kaprep)\beqref{rpath1k}
\end{align}
\begin{align}\label{rpath18b}
\ffdott
&=2\kaprep^{-2}\dif{[\fdot{\rhorep}\alprep+\rhorep\fdot{\alprep}-\fdot{\etarep}\stwo-\etarep\fdot{\stwo}]}
  \beqref{rpath18a}\nonumber\\
&=2\kaprep^{-2}(\ffdot{\rhorep}\alprep+\fdot{\rhorep}\fdot{\alprep}
  +\fdot{\rhorep}\fdot{\alprep}+\rhorep\ffdot{\alprep}
  -\ffdot{\etarep}\stwo-\fdot{\etarep}\fdot{\stwo}
  -\fdot{\etarep}\fdot{\stwo}-\etarep\ffdot{\stwo})
  \nonumber\\
&=2\kaprep^{-2}(\ffdot{\rhorep}\alprep+2\fdot{\rhorep}\fdot{\alprep}
  +\rhorep\ffdot{\alprep}-\ffdot{\etarep}\stwo-2\fdot{\etarep}\fdot{\stwo}
  -\etarep\ffdot{\stwo})\nonumber\\
\begin{split}
&=2\kaprep^{-2}(\ethvj\kaprep\vrhod+2\ethvi\kaprep\vrhox
  +\rhorep\kaprep\vrhoy-\kaprep\frkta\vphig\vphih-2\kaprep\dltvb\ethvo-\kaprep\epsvj\ethvp)\\
  &\qquad\beqref{rot4}, \eqnref{rot5c}, \eqnref{rpath4}, \eqnref{rpath8},
  \eqnref{rpath12}\text{ \& }\eqnref{rpath14}
\end{split}
\nonumber\\
&=2\kaprep^{-1}(\ethvj\vrhod+2\vrhox\ethvi+\rhorep\vrhoy-\frkta\vphig\vphih-2\dltvb\ethvo-\epsvj\ethvp)\nonumber\\
&=2(\ethvy/\kaprep)\beqref{rpath1k}
\end{align}
\begin{align}\label{rpath18c}
\fffdott
&=2\kaprep^{-2}\dif{[\ffdot{\rhorep}\alprep+2\fdot{\rhorep}\fdot{\alprep}
  +\rhorep\ffdot{\alprep}-\ffdot{\etarep}\stwo-2\fdot{\etarep}\fdot{\stwo}
  -\etarep\ffdot{\stwo}]}\beqref{rpath18b}\nonumber\\
\begin{split}
&=2\kaprep^{-2}(\fffdot{\rhorep}\alprep+\ffdot{\rhorep}\fdot{\alprep}
  +2\ffdot{\rhorep}\fdot{\alprep}+2\fdot{\rhorep}\ffdot{\alprep}
  +\fdot{\rhorep}\ffdot{\alprep}+\rhorep\fffdot{\alprep}
  -\fffdot{\etarep}\stwo-\ffdot{\etarep}\fdot{\stwo}
  -2\ffdot{\etarep}\fdot{\stwo}-2\fdot{\etarep}\ffdot{\stwo}
  -\fdot{\etarep}\ffdot{\stwo}-\etarep\fffdot{\stwo})
\end{split}
\nonumber\\
&=2\kaprep^{-2}(\fffdot{\rhorep}\alprep+3\ffdot{\rhorep}\fdot{\alprep}
  +3\fdot{\rhorep}\ffdot{\alprep}+\rhorep\fffdot{\alprep}
  -\fffdot{\etarep}\stwo-3\ffdot{\etarep}\fdot{\stwo}
  -3\fdot{\etarep}\ffdot{\stwo}-\etarep\fffdot{\stwo})
\nonumber\\
\begin{split}
&=2\kaprep^{-2}(\ethvk\kaprep\vphid+3\ethvj\kaprep\vrhox
  +3\ethvi\kaprep\vrhoy+\rhorep\kaprep\vrhoz
  -\kaprep\frktb\vphig\vphih-3\kaprep\frkta\ethvo
  -3\kaprep\dltvb\ethvp-\kaprep\epsvj\ethvq)\\
  &\qquad\beqref{rot4}, \eqnref{rot5c}, \eqnref{rpath4}, \eqnref{rpath8},
  \eqnref{rpath12}\text{ \& }\eqnref{rpath14}
\end{split}
\nonumber\\
&=2\kaprep^{-1}(\ethvk\vphid+3\ethvj\vrhox+3\ethvi\vrhoy+\rhorep\vrhoz
  -\frktb\vphig\vphih-3\frkta\ethvo-3\dltvb\ethvp-\epsvj\ethvq)\nonumber\\
&=2(\ethvz/\kaprep)\beqref{rpath1k}.
\end{align}
\end{subequations}

\subart{Development of equation \eqnref{kpath2a}}
The quantities defined by \eqnref{kpath2a} evaluate as
\begin{subequations}\label{rpath19}
\begin{align}\label{rpath19a}
\fdot{\cdkt}
&=\scalc\fdot{\dragf}-\kaprep\fdott\beqref{kpath2a}\nonumber\\
&=\scalc\ethvf-2\ethvx\beqref{rpath11a}\text{ \& }\eqnref{rpath18a}
\end{align}
\begin{align}\label{rpath19b}
\ffdot{\cdkt}
&=\scalc\ffdot{\dragf}-\kaprep\ffdott\beqref{kpath2a}\nonumber\\
&=\scalc\ethvg-2\ethvy\beqref{rpath11b}\text{ \& }\eqnref{rpath18b}
\end{align}
\begin{align}\label{rpath19c}
\fffdot{\cdkt}
&=\scalc\fffdot{\dragf}-\kaprep\fffdott\beqref{kpath2a}\nonumber\\
&=\scalc\ethvh-2\ethvz\beqref{rpath11c}\text{ \& }\eqnref{rpath18c}
\end{align}
\end{subequations}
\begin{subequations}\label{rpath20}
\begin{align}\label{rpath20a}
\vbba
&=\fdot{\rhorep}-1\beqref{kpath2a}\nonumber\\
&=\ethvi-1\beqref{rpath12a}
\end{align}
\begin{align}\label{rpath20b}
\vbbb
&=2\fdot{\rhorep}-1\beqref{kpath2a}\nonumber\\
&=2\ethvi-1\beqref{rpath12a}
\end{align}
\begin{align}\label{rpath20c}
\vbbc
&=3\fdot{\rhorep}-1\beqref{kpath2a}\nonumber\\
&=3\ethvi-1\beqref{rpath12a}
\end{align}
\begin{align}\label{rpath20d}
\vbbd
&=\ffdot{\rhorep}\fdot{\cdkt}-\ffdot{\cdkt}\vbba\beqref{kpath2a}
\nonumber\\
&=\ethvj(\scalc\ethvf-2\ethvx)-(\scalc\ethvg-2\ethvy)(\ethvi-1)
  \beqref{rpath12b}, \eqnref{rpath19}\text{ \& }\eqnref{rpath20a}
\end{align}
\begin{align}\label{rpath20e}
\vbbe
&=\fdot{\cdkt}\vbbb-\rhorep\ffdot{\cdkt}\beqref{kpath2a}
\nonumber\\
&=(\scalc\ethvf-2\ethvx)(2\ethvi-1)-\rhorep(\scalc\ethvg-2\ethvy)
  \beqref{rpath19}\text{ \& }\eqnref{rpath20b}
\end{align}
\begin{align}\label{rpath20f}
\vbbf
&=\vbba\vbbb-\rhorep\ffdot{\rhorep}\beqref{kpath2a}
\nonumber\\
&=(\ethvi-1)(2\ethvi-1)-\rhorep\ethvj
  \beqref{rpath20a}, \eqnref{rpath20b}\text{ \& }\eqnref{rpath12b}.
\end{align}
\end{subequations}

\subart{Development of equation \eqnref{kpath2b}}
The quantities defined by \eqnref{kpath2b} become
\begin{subequations}\label{rpath21}
\begin{align}\label{rpath21a}
\vscra
&=\cprod{\unitkap}{\vecta}\beqref{kpath2b}\nonumber\\
&=\cprod{\unitkap}{[(\dprod{\vectOme}{\vectr})\vectOme-\Omerep^2\vectr+\cprod{\vectLam}{\vectr}]}
  \beqref{main4a}\nonumber\\
&=\epsvb(\cprod{\unitkap}{\vectOme})-\Omerep^2(\cprod{\unitkap}{\vectr})
  +\vectLam(\dprod{\unitkap}{\vectr})-\vectr(\dprod{\unitkap}{\vectLam})
  \beqref{rot1a}\text{ \& }\eqnref{alg1}\nonumber\\
&=\epsvb(\cprod{\unitkap}{\vectOme})-\Omerep^2(\cprod{\unitkap}{\vectr})
  +\epsvd\vectLam-\dltva\vectr\beqref{rot1a}\text{ \& }\eqnref{rxpeed1a}
\end{align}
\begin{align*}
\vscrb
&=\cprod{\unitkap}{\fdota}\beqref{kpath2b}\nonumber\\
&=\cprod{\unitkap}{[2\epsvh\vectOme-3\epsvg\vectr+\cprod{\fdot{\vectLam}}{\vectr}
  -\Omerep^2(\cprod{\vectOme}{\vectr})+\epsvb\vectLam]}\beqref{rpath7a}\nonumber\\
&=\cprod{\unitkap}{[2\epsvh\vectOme-3\epsvg\vectr+\epsvb\vectLam+\cprod{\fdot{\vectLam}}{\vectr}
  -\Omerep^2(\cprod{\vectOme}{\vectr})]}\nonumber\\
\begin{split}
&=2\epsvh(\cprod{\unitkap}{\vectOme})-3\epsvg(\cprod{\unitkap}{\vectr})+\epsvb(\cprod{\unitkap}{\vectLam})
  +\fdot{\vectLam}(\dprod{\unitkap}{\vectr})-\vectr(\dprod{\unitkap}{\fdot{\vectLam}})\\
  &\quad-\Omerep^2[\vectOme(\dprod{\unitkap}{\vectr})-\vectr(\dprod{\unitkap}{\vectOme})]
  \beqref{alg1}
\end{split}
\end{align*}
\begin{align}\label{rpath21b}
\begin{split}
&=2\epsvh(\cprod{\unitkap}{\vectOme})-3\epsvg(\cprod{\unitkap}{\vectr})+\epsvb(\cprod{\unitkap}{\vectLam})
  +\epsvd\fdot{\vectLam}-\vsigb\vectr-\Omerep^2(\epsvd\vectOme-\epsva\vectr)\\
  &\quad\beqref{rot1a}\text{ \& }\eqnref{rpath1a}
\end{split}
\nonumber\\
&=2\epsvh(\cprod{\unitkap}{\vectOme})-3\epsvg(\cprod{\unitkap}{\vectr})+\epsvb(\cprod{\unitkap}{\vectLam})
  +\epsvd\fdot{\vectLam}-\epsvd\Omerep^2\vectOme+(\Omerep^2\epsva-\vsigb)\vectr
\nonumber\\
&=2\epsvh(\cprod{\unitkap}{\vectOme})-3\epsvg(\cprod{\unitkap}{\vectr})+\epsvb(\cprod{\unitkap}{\vectLam})
  +\epsvd\fdot{\vectLam}-\epsvd\Omerep^2\vectOme+\frkya\vectr\beqref{rpath1l}
\end{align}
\begin{align*}
\vscrc
&=\cprod{\unitkap}{\ffdota}\beqref{kpath2b}\nonumber\\
\begin{split}
&=\cprod{\unitkap}{}[3\epsvh\vectLam+\epsvb\fdot{\vectLam}
  +\vrhot\vectOme+\vrhou\vectr+2\epsvb(\cprod{\vectLam}{\vectOme})
  -3\Omerep^2(\cprod{\vectLam}{\vectr})\\
  &\quad-3\epsvg(\cprod{\vectOme}{\vectr})+(\cprod{\ffdot{\vectLam}}{\vectr})]\beqref{rpath7b}
\end{split}
\nonumber\\
\begin{split}
&=3\epsvh(\cprod{\unitkap}{\vectLam})+\epsvb(\cprod{\unitkap}{\fdot{\vectLam}})
  +\vrhot(\cprod{\unitkap}{\vectOme})+\vrhou(\cprod{\unitkap}{\vectr})
  +2\epsvb[\cprod{\unitkap}{(\cprod{\vectLam}{\vectOme})}]\\
  &\quad-3\Omerep^2[\cprod{\unitkap}{(\cprod{\vectLam}{\vectr})}]
  -3\epsvg[\cprod{\unitkap}{(\cprod{\vectOme}{\vectr})}]
  +\cprod{\unitkap}{(\cprod{\ffdot{\vectLam}}{\vectr})}
\end{split}
\nonumber\\
\begin{split}
&=3\epsvh(\cprod{\unitkap}{\vectLam})+\epsvb(\cprod{\unitkap}{\fdot{\vectLam}})
  +\vrhot(\cprod{\unitkap}{\vectOme})+\vrhou(\cprod{\unitkap}{\vectr})
  +2\epsvb[\vectLam(\dprod{\unitkap}{\vectOme})-\vectOme(\dprod{\unitkap}{\vectLam})]\\
  &\quad-3\Omerep^2[\vectLam(\dprod{\unitkap}{\vectr})-\vectr(\dprod{\unitkap}{\vectLam})]
  -3\epsvg[\vectOme(\dprod{\unitkap}{\vectr})-\vectr(\dprod{\unitkap}{\vectOme})]
  +\ffdot{\vectLam}(\dprod{\unitkap}{\vectr})-\vectr(\dprod{\unitkap}{\ffdot{\vectLam}})
  \beqref{alg1}
\end{split}
\end{align*}
\begin{align}\label{rpath21c}
\begin{split}
&=3\epsvh(\cprod{\unitkap}{\vectLam})+\epsvb(\cprod{\unitkap}{\fdot{\vectLam}})
  +\vrhot(\cprod{\unitkap}{\vectOme})+\vrhou(\cprod{\unitkap}{\vectr})
  +2\epsvb(\epsva\vectLam-\dltva\vectOme)\\
  &\quad-3\Omerep^2(\epsvd\vectLam-\dltva\vectr)
  -3\epsvg(\epsvd\vectOme-\epsva\vectr)+\epsvd\ffdot{\vectLam}-\vsigg\vectr
  \beqref{rot1a}, \eqnref{rxpeed1a}\text{ \& }\eqnref{rpath1a}
\end{split}
\nonumber\\
\begin{split}
&=3\epsvh(\cprod{\unitkap}{\vectLam})+\epsvb(\cprod{\unitkap}{\fdot{\vectLam}})
  +\vrhot(\cprod{\unitkap}{\vectOme})+\vrhou(\cprod{\unitkap}{\vectr})
  +2\epsva\epsvb\vectLam-2\epsvb\dltva\vectOme\\
  &\quad-3\Omerep^2\epsvd\vectLam+3\Omerep^2\dltva\vectr-3\epsvd\epsvg\vectOme
  +3\epsva\epsvg\vectr+\epsvd\ffdot{\vectLam}-\vsigg\vectr
\end{split}
\nonumber\\
\begin{split}
&=3\epsvh(\cprod{\unitkap}{\vectLam})+\epsvb(\cprod{\unitkap}{\fdot{\vectLam}})
  +\vrhot(\cprod{\unitkap}{\vectOme})+\vrhou(\cprod{\unitkap}{\vectr})
  +(2\epsva\epsvb-3\Omerep^2\epsvd)\vectLam\\
  &\quad-(2\epsvb\dltva+3\epsvd\epsvg)\vectOme
  +(3\Omerep^2\dltva+3\epsva\epsvg-\vsigg)\vectr+\epsvd\ffdot{\vectLam}
\end{split}
\nonumber\\
\begin{split}
&=3\epsvh(\cprod{\unitkap}{\vectLam})+\epsvb(\cprod{\unitkap}{\fdot{\vectLam}})
  +\vrhot(\cprod{\unitkap}{\vectOme})+\vrhou(\cprod{\unitkap}{\vectr})\\
  &\qquad+\frkyb\vectLam-\frkyc\vectOme+\frkyd\vectr+\epsvd\ffdot{\vectLam}
  \beqref{rpath1l}
\end{split}
\end{align}
\begin{align*}
\vscrd
&=\cprod{\unitkap}{\fdote}\beqref{kpath2b}\nonumber\\
&=\cprod{\unitkap}{[\ethvo(\cprod{\unitplz}{\vectOme})+\vphig\vphih(\cprod{\unitplz}{\vectLam})
  +\ethvr\vectOme+\vphii\vectLam-\ethvu\unitplz]}\beqref{rpath17a}\nonumber\\
&=\cprod{\unitkap}{[\ethvr\vectOme+\vphii\vectLam-\ethvu\unitplz+\ethvo(\cprod{\unitplz}{\vectOme})
  +\vphig\vphih(\cprod{\unitplz}{\vectLam})]}\nonumber\\
&=\ethvr(\cprod{\unitkap}{\vectOme})+\vphii(\cprod{\unitkap}{\vectLam})-\ethvu(\cprod{\unitkap}{\unitplz})
  +\ethvo[\cprod{\unitkap}{(\cprod{\unitplz}{\vectOme})}]+\vphig\vphih[\cprod{\unitkap}{(\cprod{\unitplz}{\vectLam})}]
  \nonumber\\
\begin{split}
&=\ethvr(\cprod{\unitkap}{\vectOme})+\vphii(\cprod{\unitkap}{\vectLam})-\ethvu(\cprod{\unitkap}{\unitplz})
  +\ethvo[\unitplz(\dprod{\unitkap}{\vectOme})-\vectOme(\dprod{\unitkap}{\unitplz})]\\
  &\qquad+\vphig\vphih[\unitplz(\dprod{\unitkap}{\vectLam})-\vectLam(\dprod{\unitkap}{\unitplz})]
  \beqref{alg1}
\end{split}
\end{align*}
\begin{align}\label{rpath21d}
\begin{split}
&=\ethvr(\cprod{\unitkap}{\vectOme})+\vphii(\cprod{\unitkap}{\vectLam})-\ethvu(\cprod{\unitkap}{\unitplz})
  +\ethvo(\epsva\unitplz-\epsve\vectOme)+\vphig\vphih(\dltva\unitplz-\epsve\vectLam)\\
  &\qquad\beqref{rot1a}\text{ \& }\eqnref{rxpeed1a}
\end{split}
\nonumber\\
&=\ethvr(\cprod{\unitkap}{\vectOme})+\vphii(\cprod{\unitkap}{\vectLam})-\ethvu(\cprod{\unitkap}{\unitplz})
  +\ethvo\epsva\unitplz-\ethvo\epsve\vectOme+\vphig\vphih\dltva\unitplz-\vphig\vphih\epsve\vectLam\nonumber\\
&=\ethvr(\cprod{\unitkap}{\vectOme})+\vphii(\cprod{\unitkap}{\vectLam})-\ethvu(\cprod{\unitkap}{\unitplz})
  +(\epsva\ethvo+\vphig\vphih\dltva)\unitplz-\epsve\ethvo\vectOme-\epsve\vphig\vphih\vectLam\nonumber\\
&=\ethvr(\cprod{\unitkap}{\vectOme})+\vphii(\cprod{\unitkap}{\vectLam})-\ethvu(\cprod{\unitkap}{\unitplz})
  +\frkye\unitplz-\epsve\ethvo\vectOme-\epsve\vphig\vphih\vectLam\beqref{rpath1l}
\end{align}
\begin{align*}
\vscre
&=\cprod{\unitkap}{\ffdote}\beqref{kpath2b}\nonumber\\
&=\cprod{\unitkap}{}[\ethvp(\cprod{\unitplz}{\vectOme})+2\ethvo(\cprod{\unitplz}{\vectLam})
  +\vphig\vphih(\cprod{\unitplz}{\fdot{\vectLam}})
  +\ethvs\vectOme+2\ethvr\vectLam+\vphii\fdot{\vectLam}-\ethvv\unitplz]
  \beqref{rpath17b}\nonumber\\
&=\cprod{\unitkap}{}[\ethvs\vectOme+2\ethvr\vectLam+\vphii\fdot{\vectLam}-\ethvv\unitplz
  +\ethvp(\cprod{\unitplz}{\vectOme})+2\ethvo(\cprod{\unitplz}{\vectLam})
  +\vphig\vphih(\cprod{\unitplz}{\fdot{\vectLam}})]\nonumber\\
\begin{split}
&=\ethvs(\cprod{\unitkap}{\vectOme})+2\ethvr(\cprod{\unitkap}{\vectLam})
  +\vphii(\cprod{\unitkap}{\fdot{\vectLam}})-\ethvv(\cprod{\unitkap}{\unitplz})
  +\ethvp[\cprod{\unitkap}{(\cprod{\unitplz}{\vectOme})}]\\
  &\quad+2\ethvo[\cprod{\unitkap}{(\cprod{\unitplz}{\vectLam})}]
  +\vphig\vphih[\cprod{\unitkap}{(\cprod{\unitplz}{\fdot{\vectLam}})}]
\end{split}
\nonumber\\
\begin{split}
&=\ethvs(\cprod{\unitkap}{\vectOme})+2\ethvr(\cprod{\unitkap}{\vectLam})
  +\vphii(\cprod{\unitkap}{\fdot{\vectLam}})-\ethvv(\cprod{\unitkap}{\unitplz})
  +\ethvp[\unitplz(\dprod{\unitkap}{\vectOme})-\vectOme(\dprod{\unitkap}{\unitplz})]\\
  &\quad+2\ethvo[\unitplz(\dprod{\unitkap}{\vectLam})-\vectLam(\dprod{\unitkap}{\unitplz})]
  +\vphig\vphih[\unitplz(\dprod{\unitkap}{\fdot{\vectLam}})-\fdot{\vectLam}(\dprod{\unitkap}{\unitplz})]
  \beqref{alg1}
\end{split}
\end{align*}
\begin{align}\label{rpath21e}
\begin{split}
&=\ethvs(\cprod{\unitkap}{\vectOme})+2\ethvr(\cprod{\unitkap}{\vectLam})
  +\vphii(\cprod{\unitkap}{\fdot{\vectLam}})-\ethvv(\cprod{\unitkap}{\unitplz})
  +\ethvp(\epsva\unitplz-\epsve\vectOme)\\
  &\quad+2\ethvo(\dltva\unitplz-\epsve\vectLam)
  +\vphig\vphih(\vsigb\unitplz-\epsve\fdot{\vectLam})
  \beqref{rot1a}, \eqnref{rxpeed1a}\text{ \& }\eqnref{rpath1a}
\end{split}
\nonumber\\
\begin{split}
&=\ethvs(\cprod{\unitkap}{\vectOme})+2\ethvr(\cprod{\unitkap}{\vectLam})
  +\vphii(\cprod{\unitkap}{\fdot{\vectLam}})-\ethvv(\cprod{\unitkap}{\unitplz})
  +\ethvp\epsva\unitplz-\ethvp\epsve\vectOme\\
  &\quad+2\ethvo\dltva\unitplz-2\ethvo\epsve\vectLam
  +\vphig\vphih\vsigb\unitplz-\vphig\vphih\epsve\fdot{\vectLam}
\end{split}
\nonumber\\
\begin{split}
&=\ethvs(\cprod{\unitkap}{\vectOme})+2\ethvr(\cprod{\unitkap}{\vectLam})
  +\vphii(\cprod{\unitkap}{\fdot{\vectLam}})-\ethvv(\cprod{\unitkap}{\unitplz})
  +(\ethvp\epsva+2\ethvo\dltva+\vphig\vphih\vsigb)\unitplz\\
  &\quad-\ethvp\epsve\vectOme-2\ethvo\epsve\vectLam-\vphig\vphih\epsve\fdot{\vectLam}
\end{split}
\nonumber\\
\begin{split}
&=\ethvs(\cprod{\unitkap}{\vectOme})+2\ethvr(\cprod{\unitkap}{\vectLam})
  +\vphii(\cprod{\unitkap}{\fdot{\vectLam}})-\ethvv(\cprod{\unitkap}{\unitplz})\\
  &\qquad+\frkyf\unitplz-\ethvp\epsve\vectOme-2\ethvo\epsve\vectLam-\vphig\vphih\epsve\fdot{\vectLam}
  \beqref{rpath1l}
\end{split}
\end{align}
\begin{align*}
\vscrf
&=\cprod{\vecta}{\fdota}\beqref{kpath2b}\nonumber\\
&=\cprod{[(\dprod{\vectOme}{\vectr})\vectOme-\Omerep^2\vectr+\cprod{\vectLam}{\vectr}]}
  {[2\epsvh\vectOme-3\epsvg\vectr+\cprod{\fdot{\vectLam}}{\vectr}-\Omerep^2(\cprod{\vectOme}{\vectr})
  +\epsvb\vectLam]}\beqref{main4a}\text{ \& }\eqnref{rpath7a}\nonumber\\
&=\cprod{[\epsvb\vectOme-\Omerep^2\vectr+\cprod{\vectLam}{\vectr}]}
  {[2\epsvh\vectOme-3\epsvg\vectr+\cprod{\fdot{\vectLam}}{\vectr}-\Omerep^2(\cprod{\vectOme}{\vectr})
  +\epsvb\vectLam]}\beqref{rot1a}\nonumber\\
\begin{split}
&=-3\epsvg\epsvb(\cprod{\vectOme}{\vectr})
  +\epsvb[\cprod{\vectOme}{(\cprod{\fdot{\vectLam}}{\vectr})}]
  -\Omerep^2\epsvb[\cprod{\vectOme}{(\cprod{\vectOme}{\vectr})}]
  +\epsvb^2(\cprod{\vectOme}{\vectLam})
  -2\epsvh\Omerep^2(\cprod{\vectr}{\vectOme})\\
  &\quad-\Omerep^2[\cprod{\vectr}{(\cprod{\fdot{\vectLam}}{\vectr})}]
  +\Omerep^4[\cprod{\vectr}{(\cprod{\vectOme}{\vectr})}]
  -\epsvb\Omerep^2(\cprod{\vectr}{\vectLam})
  -2\epsvh[\cprod{\vectOme}{(\cprod{\vectLam}{\vectr})}]\\
  &\quad+3\epsvg[\cprod{\vectr}{(\cprod{\vectLam}{\vectr})}]
  +\cprod{(\cprod{\vectLam}{\vectr})}{(\cprod{\fdot{\vectLam}}{\vectr})}
  -\Omerep^2[\cprod{(\cprod{\vectLam}{\vectr})}{(\cprod{\vectOme}{\vectr})}]
  -\epsvb[\cprod{\vectLam}{(\cprod{\vectLam}{\vectr})}]
\end{split}
\nonumber\\
\begin{split}
&=(2\epsvh\Omerep^2-3\epsvg\epsvb)(\cprod{\vectOme}{\vectr})
  +\epsvb^2(\cprod{\vectOme}{\vectLam})-\epsvb\Omerep^2(\cprod{\vectr}{\vectLam})
  +\epsvb[\cprod{\vectOme}{(\cprod{\fdot{\vectLam}}{\vectr})}]\\
  &\quad-\Omerep^2\epsvb[\cprod{\vectOme}{(\cprod{\vectOme}{\vectr})}]
  -\Omerep^2[\cprod{\vectr}{(\cprod{\fdot{\vectLam}}{\vectr})}]
  +\Omerep^4[\cprod{\vectr}{(\cprod{\vectOme}{\vectr})}]
  -2\epsvh[\cprod{\vectOme}{(\cprod{\vectLam}{\vectr})}]\\
  &\quad+3\epsvg[\cprod{\vectr}{(\cprod{\vectLam}{\vectr})}]
  +\cprod{(\cprod{\vectLam}{\vectr})}{(\cprod{\fdot{\vectLam}}{\vectr})}
  -\Omerep^2[\cprod{(\cprod{\vectLam}{\vectr})}{(\cprod{\vectOme}{\vectr})}]
  -\epsvb[\cprod{\vectLam}{(\cprod{\vectLam}{\vectr})}]
\end{split}
\nonumber\\
\begin{split}
&=(2\epsvh\Omerep^2-3\epsvg\epsvb)(\cprod{\vectOme}{\vectr})
  +\epsvb^2(\cprod{\vectOme}{\vectLam})-\epsvb\Omerep^2(\cprod{\vectr}{\vectLam})
  +\epsvb[\fdot{\vectLam}(\dprod{\vectOme}{\vectr})-\vectr(\dprod{\vectOme}{\fdot{\vectLam}})]\\
  &\quad-\Omerep^2\epsvb[\vectOme(\dprod{\vectOme}{\vectr})-\Omerep^2\vectr]
  -\Omerep^2[\scalr^2\fdot{\vectLam}-\vectr(\dprod{\vectr}{\fdot{\vectLam}})]
  +\Omerep^4[\scalr^2\vectOme-\vectr(\dprod{\vectr}{\vectOme})]\\
  &\quad-2\epsvh[\vectLam(\dprod{\vectOme}{\vectr})-\vectr(\dprod{\vectOme}{\vectLam})]
  +3\epsvg[\scalr^2\vectLam-\vectr(\dprod{\vectr}{\vectLam})]
  +\fdot{\vectLam}[\dprod{\vectr}{(\cprod{\vectLam}{\vectr})}]
  -\vectr[\dprod{\fdot{\vectLam}}{(\cprod{\vectLam}{\vectr})}]\\
  &\quad-\Omerep^2\vectOme[\dprod{\vectr}{(\cprod{\vectLam}{\vectr})}]
  +\Omerep^2\vectr[\dprod{\vectOme}{(\cprod{\vectLam}{\vectr})}]
  -\epsvb[\vectLam(\dprod{\vectLam}{\vectr})-\Lamrep^2\vectr]
  \beqref{alg1}\text{ \& }\eqnref{alg5}
\end{split}
\end{align*}
\begin{align}\label{rpath21f}
\begin{split}
&=(2\epsvh\Omerep^2-3\epsvg\epsvb)(\cprod{\vectOme}{\vectr})
  +\epsvb^2(\cprod{\vectOme}{\vectLam})-\epsvb\Omerep^2(\cprod{\vectr}{\vectLam})
  +\epsvb(\epsvb\fdot{\vectLam}-\vsigc\vectr)
  -\Omerep^2\epsvb(\epsvb\vectOme-\Omerep^2\vectr)\\
  &\quad-\Omerep^2(\scalr^2\fdot{\vectLam}-\vsign\vectr)
  +\Omerep^4(\scalr^2\vectOme-\epsvb\vectr)
  -2\epsvh(\epsvb\vectLam-\epsvg\vectr)
  +3\epsvg(\scalr^2\vectLam-\epsvh\vectr)\\
  &\quad-\frkth\vectr+\Omerep^2\epsvm\vectr-\epsvb(\epsvh\vectLam-\Lamrep^2\vectr)
  \beqref{rot1a}, \eqnref{rpath1a}, \eqnref{rpath1b}\text{ \& }\eqnref{alg4}
\end{split}
\nonumber\\
\begin{split}
&=(2\epsvh\Omerep^2-3\epsvg\epsvb)(\cprod{\vectOme}{\vectr})
  +\epsvb^2(\cprod{\vectOme}{\vectLam})-\epsvb\Omerep^2(\cprod{\vectr}{\vectLam})
  +\epsvb^2\fdot{\vectLam}-\epsvb\vsigc\vectr
  -\Omerep^2\epsvb^2\vectOme+\epsvb\Omerep^4\vectr\\
  &\quad-\Omerep^2\scalr^2\fdot{\vectLam}+\Omerep^2\vsign\vectr
  +\Omerep^4\scalr^2\vectOme-\Omerep^4\epsvb\vectr
  -2\epsvh\epsvb\vectLam+2\epsvh\epsvg\vectr
  +3\epsvg\scalr^2\vectLam-3\epsvg\epsvh\vectr\\
  &\quad-\frkth\vectr+\Omerep^2\epsvm\vectr-\epsvb\epsvh\vectLam+\epsvb\Lamrep^2\vectr
\end{split}
\nonumber\\
\begin{split}
&=(2\epsvh\Omerep^2-3\epsvg\epsvb)(\cprod{\vectOme}{\vectr})
  +\epsvb^2(\cprod{\vectOme}{\vectLam})-\epsvb\Omerep^2(\cprod{\vectr}{\vectLam})
  +\epsvb^2\fdot{\vectLam}-\Omerep^2\scalr^2\fdot{\vectLam}
  -\epsvb\vsigc\vectr+\epsvb\Omerep^4\vectr\\
    &\quad+\Omerep^2\vsign\vectr-\Omerep^4\epsvb\vectr+2\epsvh\epsvg\vectr
    -3\epsvg\epsvh\vectr-\frkth\vectr+\Omerep^2\epsvm\vectr+\epsvb\Lamrep^2\vectr\\
  &\quad-\Omerep^2\epsvb^2\vectOme+\Omerep^4\scalr^2\vectOme
  -2\epsvh\epsvb\vectLam+3\epsvg\scalr^2\vectLam-\epsvb\epsvh\vectLam
\end{split}
\nonumber\\
\begin{split}
&=(2\epsvh\Omerep^2-3\epsvg\epsvb)(\cprod{\vectOme}{\vectr})
  +\epsvb^2(\cprod{\vectOme}{\vectLam})-\epsvb\Omerep^2(\cprod{\vectr}{\vectLam})
  -(\Omerep^2\scalr^2-\epsvb^2)\fdot{\vectLam}\\
  &\quad+[\Omerep^2(\vsign+\epsvm)+\epsvb(\Lamrep^2-\vsigc)-\epsvg\epsvh-\frkth]\vectr
  +\Omerep^2(\Omerep^2\scalr^2-\epsvb^2)\vectOme+3(\epsvg\scalr^2-\epsvh\epsvb)\vectLam
\end{split}
\nonumber\\
\begin{split}
&=\frkyg(\cprod{\vectOme}{\vectr})
  +\epsvb^2(\cprod{\vectOme}{\vectLam})-\epsvb\Omerep^2(\cprod{\vectr}{\vectLam})
  -\vphia^2\fdot{\vectLam}+\frkyh\vectr
  +\Omerep^2\vphia^2\vectOme+3\vphic\vectLam\\
  &\qquad\beqref{rot1b}\text{ \& }\eqnref{rpath1l}
\end{split}
\end{align}
\begin{align*}
\vscrg
&=\cprod{\vecta}{\ffdota}\beqref{kpath2b}\nonumber\\
\begin{split}
&=\cprod{[(\dprod{\vectOme}{\vectr})\vectOme-\Omerep^2\vectr+\cprod{\vectLam}{\vectr}]}{}
  [3\epsvh\vectLam+\epsvb\fdot{\vectLam}
  +\vrhot\vectOme+\vrhou\vectr\\
  &\quad+2\epsvb(\cprod{\vectLam}{\vectOme})
  -3\Omerep^2(\cprod{\vectLam}{\vectr})
  -3\epsvg(\cprod{\vectOme}{\vectr})
  +(\cprod{\ffdot{\vectLam}}{\vectr})]
  \beqref{rpath7b}\text{ \& }\eqnref{main4a}
\end{split}
\nonumber\\
\begin{split}
&=3\epsvh\epsvb(\cprod{\vectOme}{\vectLam})
  +\epsvb^2(\cprod{\vectOme}{\fdot{\vectLam}})
  +\epsvb\vrhou(\cprod{\vectOme}{\vectr})
  +2\epsvb^2[\cprod{\vectOme}{(\cprod{\vectLam}{\vectOme})}]\\
  &\quad-3\Omerep^2\epsvb[\cprod{\vectOme}{(\cprod{\vectLam}{\vectr})}]
  -3\epsvg\epsvb[\cprod{\vectOme}{(\cprod{\vectOme}{\vectr})}]
  +\epsvb[\cprod{\vectOme}{(\cprod{\ffdot{\vectLam}}{\vectr})}]
  -3\epsvh\Omerep^2(\cprod{\vectr}{\vectLam})\\
  &\quad-\epsvb\Omerep^2(\cprod{\vectr}{\fdot{\vectLam}})
  -\Omerep^2\vrhot(\cprod{\vectr}{\vectOme})
  -2\epsvb\Omerep^2[\cprod{\vectr}{(\cprod{\vectLam}{\vectOme})}]
  +3\Omerep^4[\cprod{\vectr}{(\cprod{\vectLam}{\vectr})}]\\
  &\quad+3\epsvg\Omerep^2[\cprod{\vectr}{(\cprod{\vectOme}{\vectr})}]
  -\Omerep^2[\cprod{\vectr}{(\cprod{\ffdot{\vectLam}}{\vectr})}]
  -3\epsvh[\cprod{\vectLam}{(\cprod{\vectLam}{\vectr})}]
  -\epsvb[\cprod{\fdot{\vectLam}}{(\cprod{\vectLam}{\vectr})}]\\
  &\quad-\vrhot[\cprod{\vectOme}{(\cprod{\vectLam}{\vectr})}]
  -\vrhou[\cprod{\vectr}{(\cprod{\vectLam}{\vectr})}]
  +2\epsvb[\cprod{(\cprod{\vectLam}{\vectr})}{(\cprod{\vectLam}{\vectOme})}]\\
  &\quad-3\epsvg[\cprod{(\cprod{\vectLam}{\vectr})}{(\cprod{\vectOme}{\vectr})}]
  +[\cprod{(\cprod{\vectLam}{\vectr})}{(\cprod{\ffdot{\vectLam}}{\vectr})}]
  \beqref{rot1a}
\end{split}
\end{align*}
\begin{align*}
\begin{split}
&=3\epsvh\epsvb(\cprod{\vectOme}{\vectLam})
  +\epsvb^2(\cprod{\vectOme}{\fdot{\vectLam}})
  +(\epsvb\vrhou+\Omerep^2\vrhot)(\cprod{\vectOme}{\vectr})
  -3\epsvh\Omerep^2(\cprod{\vectr}{\vectLam})
  -\epsvb\Omerep^2(\cprod{\vectr}{\fdot{\vectLam}})\\
  &\quad+2\epsvb^2[\Omerep^2\vectLam-\vectOme(\dprod{\vectOme}{\vectLam})]
  -3\Omerep^2\epsvb[\vectLam(\dprod{\vectOme}{\vectr})-\vectr(\dprod{\vectOme}{\vectLam})]
  -3\epsvg\epsvb[\vectOme(\dprod{\vectOme}{\vectr})-\Omerep^2\vectr]\\
  &\quad+\epsvb[\ffdot{\vectLam}(\dprod{\vectOme}{\vectr})-\vectr(\dprod{\vectOme}{\ffdot{\vectLam}})]
  -2\epsvb\Omerep^2[\vectLam(\dprod{\vectr}{\vectOme})-\vectOme(\dprod{\vectLam}{\vectr})]
  +3\Omerep^4[\scalr^2\vectLam-\vectr(\dprod{\vectr}{\vectLam})]\\
  &\quad+3\epsvg\Omerep^2[\scalr^2\vectOme-\vectr(\dprod{\vectOme}{\vectr})]
  -\Omerep^2[\scalr^2\ffdot{\vectLam}-\vectr(\dprod{\vectr}{\ffdot{\vectLam}})]
  -3\epsvh[\vectLam(\dprod{\vectLam}{\vectr})-\Lamrep^2\vectr]\\
  &\quad-\epsvb[\vectLam(\dprod{\fdot{\vectLam}}{\vectr})-\vectr(\dprod{\fdot{\vectLam}}{\vectLam})]
  -\vrhot[\vectLam(\dprod{\vectOme}{\vectr})-\vectr(\dprod{\vectOme}{\vectLam})]
  -\vrhou[\scalr^2\vectLam-\vectr(\dprod{\vectr}{\vectLam})]\\
  &\quad+2\epsvb[\vectLam(\dprod{\vectOme}{(\cprod{\vectLam}{\vectr})})
     -\vectOme(\dprod{\vectLam}{(\cprod{\vectLam}{\vectr})})]
  -3\epsvg[\vectOme(\dprod{\vectr}{(\cprod{\vectLam}{\vectr})})
     -\vectr(\dprod{\vectOme}{(\cprod{\vectLam}{\vectr})})]\\
  &\quad+[\ffdot{\vectLam}(\dprod{\vectr}{(\cprod{\vectLam}{\vectr})})
     -\vectr(\dprod{\ffdot{\vectLam}}{(\cprod{\vectLam}{\vectr})})]
  \beqref{alg1}\text{ \& }\eqnref{alg5}
\end{split}
\nonumber\\
\begin{split}
&=3\epsvh\epsvb(\cprod{\vectOme}{\vectLam})
  +\epsvb^2(\cprod{\vectOme}{\fdot{\vectLam}})
  +(\epsvb\vrhou+\Omerep^2\vrhot)(\cprod{\vectOme}{\vectr})
  -3\epsvh\Omerep^2(\cprod{\vectr}{\vectLam})\\
  &\quad-\epsvb\Omerep^2(\cprod{\vectr}{\fdot{\vectLam}})
  +2\epsvb^2(\Omerep^2\vectLam-\epsvg\vectOme)
  -3\Omerep^2\epsvb(\epsvb\vectLam-\epsvg\vectr)
  -3\epsvg\epsvb(\epsvb\vectOme-\Omerep^2\vectr)\\
  &\quad+\epsvb(\epsvb\ffdot{\vectLam}-\vsigh\vectr)
  -2\epsvb\Omerep^2(\epsvb\vectLam-\epsvh\vectOme)
  +3\Omerep^4(\scalr^2\vectLam-\epsvh\vectr)
  +3\epsvg\Omerep^2(\scalr^2\vectOme-\epsvb\vectr)\\
  &\quad-\Omerep^2(\scalr^2\ffdot{\vectLam}-\vsigo\vectr)
  -3\epsvh(\epsvh\vectLam-\Lamrep^2\vectr)
  -\epsvb(\vsign\vectLam-\vsigd\vectr)
  -\vrhot(\epsvb\vectLam-\epsvg\vectr)\\
  &\quad-\vrhou(\scalr^2\vectLam-\epsvh\vectr)
  +2\epsvb\epsvm\vectLam+3\epsvg\epsvm\vectr-\frkti\vectr
  \beqref{rot1a}, \eqnref{rpath1a}\text{ \& }\eqnref{rpath1b}
\end{split}
\end{align*}
\begin{align*}
\begin{split}
&=3\epsvh\epsvb(\cprod{\vectOme}{\vectLam})
  +\epsvb^2(\cprod{\vectOme}{\fdot{\vectLam}})
  +(\epsvb\vrhou+\Omerep^2\vrhot)(\cprod{\vectOme}{\vectr})
  -3\epsvh\Omerep^2(\cprod{\vectr}{\vectLam})\\
  &\quad-\epsvb\Omerep^2(\cprod{\vectr}{\fdot{\vectLam}})
  +2\epsvb^2\Omerep^2\vectLam-2\epsvb^2\epsvg\vectOme
  -3\Omerep^2\epsvb^2\vectLam+3\Omerep^2\epsvb\epsvg\vectr
  -3\epsvg\epsvb^2\vectOme+3\epsvg\epsvb\Omerep^2\vectr\\
  &\quad+\epsvb^2\ffdot{\vectLam}-\epsvb\vsigh\vectr
  -2\epsvb^2\Omerep^2\vectLam+2\epsvb\epsvh\Omerep^2\vectOme
  +3\Omerep^4\scalr^2\vectLam-3\Omerep^4\epsvh\vectr
  +3\epsvg\Omerep^2\scalr^2\vectOme-3\epsvg\epsvb\Omerep^2\vectr\\
  &\quad-\Omerep^2\scalr^2\ffdot{\vectLam}+\Omerep^2\vsigo\vectr
  -3\epsvh^2\vectLam+3\epsvh\Lamrep^2\vectr
  -\epsvb\vsign\vectLam+\epsvb\vsigd\vectr
  -\vrhot\epsvb\vectLam+\vrhot\epsvg\vectr\\
  &\quad-\vrhou\scalr^2\vectLam+\vrhou\epsvh\vectr
  +2\epsvb\epsvm\vectLam+3\epsvg\epsvm\vectr-\frkti\vectr
\end{split}
\nonumber\\
\begin{split}
&=3\epsvh\epsvb(\cprod{\vectOme}{\vectLam})
  +\epsvb^2(\cprod{\vectOme}{\fdot{\vectLam}})
  +(\epsvb\vrhou+\Omerep^2\vrhot)(\cprod{\vectOme}{\vectr})
  -3\epsvh\Omerep^2(\cprod{\vectr}{\vectLam})
  -\epsvb\Omerep^2(\cprod{\vectr}{\fdot{\vectLam}})\\
  &\quad+(2\epsvb^2\Omerep^2-3\Omerep^2\epsvb^2-2\epsvb^2\Omerep^2
  +3\Omerep^4\scalr^2-3\epsvh^2-\epsvb\vsign-\vrhot\epsvb
  -\vrhou\scalr^2+2\epsvb\epsvm)\vectLam\\
  &\quad+(3\Omerep^2\epsvb\epsvg+3\epsvg\epsvb\Omerep^2-\epsvb\vsigh
  -3\Omerep^4\epsvh-3\epsvg\epsvb\Omerep^2+\Omerep^2\vsigo+3\epsvh\Lamrep^2
  +\epsvb\vsigd+\vrhot\epsvg\\
  &\quad+\vrhou\epsvh+3\epsvg\epsvm-\frkti)\vectr
  +(-2\epsvb^2\epsvg-3\epsvg\epsvb^2+2\epsvb\epsvh\Omerep^2+3\epsvg\Omerep^2\scalr^2)\vectOme
  +(\epsvb^2-\Omerep^2\scalr^2)\ffdot{\vectLam}
\end{split}
\end{align*}
\begin{align*}
\begin{split}
&=3\epsvh\epsvb(\cprod{\vectOme}{\vectLam})
  +\epsvb^2(\cprod{\vectOme}{\fdot{\vectLam}})
  +(\epsvb\vrhou+\Omerep^2\vrhot)(\cprod{\vectOme}{\vectr})
  -3\epsvh\Omerep^2(\cprod{\vectr}{\vectLam})\\
  &\quad-\epsvb\Omerep^2(\cprod{\vectr}{\fdot{\vectLam}})
  +[3\Omerep^2(\Omerep^2\scalr^2-\epsvb^2)-3\epsvh^2-\epsvb\vsign-\vrhot\epsvb
    -\vrhou\scalr^2+2\epsvb\epsvm]\vectLam\\
  &\quad+(3\Omerep^2\epsvb\epsvg-\epsvb\vsigh-3\Omerep^4\epsvh+\Omerep^2\vsigo+3\epsvh\Lamrep^2+\epsvb\vsigd
     +\vrhot\epsvg+\vrhou\epsvh+3\epsvg\epsvm-\frkti)\vectr\\
  &\quad+(3\scalr^2\Omerep^2\epsvg+2\Omerep^2\epsvb\epsvh-5\epsvg\epsvb^2)\vectOme
  -(\Omerep^2\scalr^2-\epsvb^2)\ffdot{\vectLam}
\end{split}
\end{align*}
\begin{align}\label{rpath21g}
\begin{split}
&=3\epsvh\epsvb(\cprod{\vectOme}{\vectLam})
  +\epsvb^2(\cprod{\vectOme}{\fdot{\vectLam}})
  +(\epsvb\vrhou+\Omerep^2\vrhot)(\cprod{\vectOme}{\vectr})
  -3\epsvh\Omerep^2(\cprod{\vectr}{\vectLam})\\
  &\quad-\epsvb\Omerep^2(\cprod{\vectr}{\fdot{\vectLam}})
  +(3\Omerep^2\vphia^2-3\epsvh^2-\epsvb\vsign-\vrhot\epsvb
    -\vrhou\scalr^2+2\epsvb\epsvm)\vectLam\\
  &\quad+(3\Omerep^2\epsvb\epsvg-\epsvb\vsigh-3\Omerep^4\epsvh+\Omerep^2\vsigo+3\epsvh\Lamrep^2+\epsvb\vsigd
     +\vrhot\epsvg+\vrhou\epsvh+3\epsvg\epsvm-\frkti)\vectr\\
  &\quad+(3\scalr^2\Omerep^2\epsvg+2\Omerep^2\epsvb\epsvh-5\epsvg\epsvb^2)\vectOme
  -\vphia^2\ffdot{\vectLam}\beqref{rot1b}
\end{split}
\nonumber\\
\begin{split}
&=3\epsvh\epsvb(\cprod{\vectOme}{\vectLam})
  +\epsvb^2(\cprod{\vectOme}{\fdot{\vectLam}})
  +\frkyi(\cprod{\vectOme}{\vectr})
  -3\epsvh\Omerep^2(\cprod{\vectr}{\vectLam})\\
  &\quad-\epsvb\Omerep^2(\cprod{\vectr}{\fdot{\vectLam}})
  +\frkyj\vectLam+\frkyl\vectr+\frkyk\vectOme-\vphia^2\ffdot{\vectLam}\beqref{rpath1l}
\end{split}
\end{align}
\begin{align*}
\vscrh
&=\cprod{\vecta}{\fdote}\beqref{kpath2b}\nonumber\\
\begin{split}
&=\cprod{[(\dprod{\vectOme}{\vectr})\vectOme-\Omerep^2\vectr+\cprod{\vectLam}{\vectr}]}
  {[\ethvo(\cprod{\unitplz}{\vectOme})+\vphig\vphih(\cprod{\unitplz}{\vectLam})
  +\ethvr\vectOme+\vphii\vectLam-\ethvu\unitplz]}\\
  &\qquad\beqref{main4a}\text{ \& }\eqnref{rpath17a}
\end{split}
\end{align*}
\begin{align*}
\begin{split}
&=\ethvo\epsvb[\cprod{\vectOme}{(\cprod{\unitplz}{\vectOme})}]
  +\vphig\vphih\epsvb[\cprod{\vectOme}{(\cprod{\unitplz}{\vectLam})}]
  +\vphii\epsvb(\cprod{\vectOme}{\vectLam})
  -\ethvu\epsvb(\cprod{\vectOme}{\unitplz})\\
  &\quad-\ethvo\Omerep^2[\cprod{\vectr}{(\cprod{\unitplz}{\vectOme})}]
  -\vphig\vphih\Omerep^2[\cprod{\vectr}{(\cprod{\unitplz}{\vectLam})}]
  -\ethvr\Omerep^2(\cprod{\vectr}{\vectOme})
  -\vphii\Omerep^2(\cprod{\vectr}{\vectLam})\\
  &\quad+\ethvu\Omerep^2(\cprod{\vectr}{\unitplz})
  +\ethvo[\cprod{(\cprod{\vectLam}{\vectr})}{(\cprod{\unitplz}{\vectOme})}]
  +\vphig\vphih[\cprod{(\cprod{\vectLam}{\vectr})}{(\cprod{\unitplz}{\vectLam})}]\\
  &\quad-\ethvr[\cprod{\vectOme}{(\cprod{\vectLam}{\vectr})}]
  -\vphii[\cprod{\vectLam}{(\cprod{\vectLam}{\vectr})}]
  +\ethvu[\cprod{\unitplz}{(\cprod{\vectLam}{\vectr})}]
  \beqref{rot1a}
\end{split}
\nonumber\\
\begin{split}
&=\vphii\epsvb(\cprod{\vectOme}{\vectLam})
  -\ethvu\epsvb(\cprod{\vectOme}{\unitplz})
  -\ethvr\Omerep^2(\cprod{\vectr}{\vectOme})
  -\vphii\Omerep^2(\cprod{\vectr}{\vectLam})
  +\ethvu\Omerep^2(\cprod{\vectr}{\unitplz})\\
  &\quad+\ethvo\epsvb[\Omerep^2\unitplz-\vectOme(\dprod{\unitplz}{\vectOme})]
  +\vphig\vphih\epsvb[\unitplz(\dprod{\vectOme}{\vectLam})-\vectLam(\dprod{\vectOme}{\unitplz})]
  -\ethvo\Omerep^2[\unitplz(\dprod{\vectr}{\vectOme})-\vectOme(\dprod{\unitplz}{\vectr})]\\
  &\quad-\vphig\vphih\Omerep^2[\unitplz(\dprod{\vectr}{\vectLam})-\vectLam(\dprod{\vectr}{\unitplz})]
  +\ethvo[\unitplz(\dprod{\vectOme}{(\cprod{\vectLam}{\vectr})})
     -\vectOme(\dprod{\unitplz}{(\cprod{\vectLam}{\vectr})})]\\
  &\quad+\vphig\vphih[\unitplz(\dprod{\vectLam}{(\cprod{\vectLam}{\vectr})})
     -\vectLam(\dprod{\unitplz}{(\cprod{\vectLam}{\vectr})})]
  -\ethvr[\vectLam(\dprod{\vectOme}{\vectr})-\vectr(\dprod{\vectOme}{\vectLam})]\\
  &\quad-\vphii[\vectLam(\dprod{\vectLam}{\vectr})-\Lamrep^2\vectr]
  +\ethvu[\vectLam(\dprod{\unitplz}{\vectr})-\vectr(\dprod{\unitplz}{\vectLam})]
  \beqref{alg1}\text{ \& }\eqnref{alg5}
\end{split}
\end{align*}
\begin{align*}
\begin{split}
&=\vphii\epsvb(\cprod{\vectOme}{\vectLam})
  -\ethvu\epsvb(\cprod{\vectOme}{\unitplz})
  -\ethvr\Omerep^2(\cprod{\vectr}{\vectOme})
  -\vphii\Omerep^2(\cprod{\vectr}{\vectLam})
  +\ethvu\Omerep^2(\cprod{\vectr}{\unitplz})\\
  &\quad+\ethvo\epsvb(\Omerep^2\unitplz-\epsvc\vectOme)
  +\vphig\vphih\epsvb(\epsvg\unitplz-\epsvc\vectLam)
  -\ethvo\Omerep^2(\epsvb\unitplz-\epsvf\vectOme)\\
  &\quad-\vphig\vphih\Omerep^2(\epsvh\unitplz-\epsvf\vectLam)
  +\ethvo(\epsvm\unitplz-\epsvo\vectOme)
  -\vphig\vphih\epsvo\vectLam
  -\ethvr(\epsvb\vectLam-\epsvg\vectr)\\
  &\quad-\vphii(\epsvh\vectLam-\Lamrep^2\vectr)
  +\ethvu(\epsvf\vectLam-\epsvi\vectr)
  \beqref{rot1a}
\end{split}
\nonumber\\
\begin{split}
&=\vphii\epsvb(\cprod{\vectOme}{\vectLam})
  -\ethvu\epsvb(\cprod{\vectOme}{\unitplz})
  -\ethvr\Omerep^2(\cprod{\vectr}{\vectOme})
  -\vphii\Omerep^2(\cprod{\vectr}{\vectLam})
  +\ethvu\Omerep^2(\cprod{\vectr}{\unitplz})\\
  &\quad+\ethvo\epsvb\Omerep^2\unitplz-\ethvo\epsvb\epsvc\vectOme
  +\vphig\vphih\epsvb\epsvg\unitplz-\vphig\vphih\epsvb\epsvc\vectLam
  -\ethvo\Omerep^2\epsvb\unitplz+\ethvo\Omerep^2\epsvf\vectOme\\
  &\quad-\vphig\vphih\Omerep^2\epsvh\unitplz+\vphig\vphih\Omerep^2\epsvf\vectLam
  +\ethvo\epsvm\unitplz-\ethvo\epsvo\vectOme
  -\vphig\vphih\epsvo\vectLam
  -\ethvr\epsvb\vectLam+\ethvr\epsvg\vectr\\
  &\quad-\vphii\epsvh\vectLam+\vphii\Lamrep^2\vectr
  +\ethvu\epsvf\vectLam-\ethvu\epsvi\vectr
\end{split}
\end{align*}
\begin{align}\label{rpath21h}
\begin{split}
&=\vphii\epsvb(\cprod{\vectOme}{\vectLam})
  -\ethvu\epsvb(\cprod{\vectOme}{\unitplz})
  -\ethvr\Omerep^2(\cprod{\vectr}{\vectOme})
  -\vphii\Omerep^2(\cprod{\vectr}{\vectLam})
  +\ethvu\Omerep^2(\cprod{\vectr}{\unitplz})\\
  &\quad+(\ethvo\epsvb\Omerep^2+\vphig\vphih\epsvb\epsvg
  -\ethvo\Omerep^2\epsvb-\vphig\vphih\Omerep^2\epsvh+\ethvo\epsvm)\unitplz
  +(-\ethvo\epsvb\epsvc+\ethvo\Omerep^2\epsvf-\ethvo\epsvo)\vectOme\\
  &\quad+(-\vphig\vphih\epsvb\epsvc+\vphig\vphih\Omerep^2\epsvf
  -\vphig\vphih\epsvo-\ethvr\epsvb-\vphii\epsvh+\ethvu\epsvf)\vectLam
  +(\ethvr\epsvg+\vphii\Lamrep^2-\ethvu\epsvi)\vectr
\end{split}
\nonumber\\
\begin{split}
&=\vphii\epsvb(\cprod{\vectOme}{\vectLam})
  -\ethvu\epsvb(\cprod{\vectOme}{\unitplz})
  -\ethvr\Omerep^2(\cprod{\vectr}{\vectOme})
  -\vphii\Omerep^2(\cprod{\vectr}{\vectLam})
  +\ethvu\Omerep^2(\cprod{\vectr}{\unitplz})\\
  &\quad+\ethvo(\Omerep^2\epsvf-\epsvb\epsvc-\epsvo)\vectOme
  +(\ethvr\epsvg+\vphii\Lamrep^2-\ethvu\epsvi)\vectr
  +[\vphig\vphih(\epsvb\epsvg-\epsvh\Omerep^2)+\ethvo\epsvm]\unitplz\\
  &\quad+[\ethvu\epsvf-\ethvr\epsvb-\vphii\epsvh+\vphig\vphih(\Omerep^2\epsvf-\epsvb\epsvc-\epsvo)]\vectLam
\end{split}
\nonumber\\
\begin{split}
&=\vphii\epsvb(\cprod{\vectOme}{\vectLam})
  -\ethvu\epsvb(\cprod{\vectOme}{\unitplz})
  -\ethvr\Omerep^2(\cprod{\vectr}{\vectOme})
  -\vphii\Omerep^2(\cprod{\vectr}{\vectLam})
  +\ethvu\Omerep^2(\cprod{\vectr}{\unitplz})\\
  &\quad+\ethvo\frkym\vectOme+\frkyn\vectr+\frkyo\unitplz+\frkyp\vectLam
  \beqref{rpath1l}
\end{split}
\end{align}
\begin{align*}
\vscri
&=\cprod{\vecta}{\ffdote}\beqref{kpath2b}\nonumber\\
\begin{split}
&=\cprod{[(\dprod{\vectOme}{\vectr})\vectOme-\Omerep^2\vectr+\cprod{\vectLam}{\vectr}]}
  {[\ethvp(\cprod{\unitplz}{\vectOme})+2\ethvo(\cprod{\unitplz}{\vectLam})
  +\vphig\vphih(\cprod{\unitplz}{\fdot{\vectLam}})+\ethvs\vectOme}\\
  &\qquad+2\ethvr\vectLam+\vphii\fdot{\vectLam}-\ethvv\unitplz]
  \beqref{main4a}\text{ \& }\eqnref{rpath17b}
\end{split}
\nonumber\\
\begin{split}
&=\ethvp\epsvb[\cprod{\vectOme}{(\cprod{\unitplz}{\vectOme})}]
  +2\ethvo\epsvb[\cprod{\vectOme}{(\cprod{\unitplz}{\vectLam})}]
  +\vphig\vphih\epsvb[\cprod{\vectOme}{(\cprod{\unitplz}{\fdot{\vectLam}})}]
  +2\ethvr\epsvb(\cprod{\vectOme}{\vectLam})\\
  &\quad+\vphii\epsvb(\cprod{\vectOme}{\fdot{\vectLam}})
  -\ethvv\epsvb(\cprod{\vectOme}{\unitplz})
  -\ethvp\Omerep^2[\cprod{\vectr}{(\cprod{\unitplz}{\vectOme})}]
  -2\ethvo\Omerep^2[\cprod{\vectr}{(\cprod{\unitplz}{\vectLam})}]\\
  &\quad-\vphig\vphih\Omerep^2[\cprod{\vectr}{(\cprod{\unitplz}{\fdot{\vectLam}})}]
  -\ethvs\Omerep^2(\cprod{\vectr}{\vectOme})
  -2\ethvr\Omerep^2(\cprod{\vectr}{\vectLam})
  -\vphii\Omerep^2(\cprod{\vectr}{\fdot{\vectLam}})
  +\ethvv\Omerep^2(\cprod{\vectr}{\unitplz})\\
  &\quad+\ethvp[\cprod{(\cprod{\vectLam}{\vectr})}{(\cprod{\unitplz}{\vectOme})}]
  +2\ethvo[\cprod{(\cprod{\vectLam}{\vectr})}{(\cprod{\unitplz}{\vectLam})}]
  +\vphig\vphih[\cprod{(\cprod{\vectLam}{\vectr})}{(\cprod{\unitplz}{\fdot{\vectLam}})}]\\
  &\quad-\ethvs[\cprod{\vectOme}{(\cprod{\vectLam}{\vectr})}]
  -2\ethvr[\cprod{\vectLam}{(\cprod{\vectLam}{\vectr})}]
  -\vphii[\cprod{\fdot{\vectLam}}{(\cprod{\vectLam}{\vectr})}]
  +\ethvv[\cprod{\unitplz}{(\cprod{\vectLam}{\vectr})}]
  \beqref{rot1a}
\end{split}
\end{align*}
\begin{align*}
\begin{split}
&=2\ethvr\epsvb(\cprod{\vectOme}{\vectLam})
  +\vphii\epsvb(\cprod{\vectOme}{\fdot{\vectLam}})
  -\ethvv\epsvb(\cprod{\vectOme}{\unitplz})
  -\ethvs\Omerep^2(\cprod{\vectr}{\vectOme})
  -2\ethvr\Omerep^2(\cprod{\vectr}{\vectLam})\\
  &\quad-\vphii\Omerep^2(\cprod{\vectr}{\fdot{\vectLam}})
  +\ethvv\Omerep^2(\cprod{\vectr}{\unitplz})
  +\ethvp\epsvb[\Omerep^2\unitplz-\vectOme(\dprod{\unitplz}{\vectOme})]
  +2\ethvo\epsvb[\unitplz(\dprod{\vectOme}{\vectLam})-\vectLam(\dprod{\vectOme}{\unitplz})]\\
  &\quad+\vphig\vphih\epsvb[\unitplz(\dprod{\vectOme}{\fdot{\vectLam}})-\fdot{\vectLam}(\dprod{\unitplz}{\vectOme})]
  -\ethvp\Omerep^2[\unitplz(\dprod{\vectr}{\vectOme})-\vectOme(\dprod{\unitplz}{\vectr})]
  -2\ethvo\Omerep^2[\unitplz(\dprod{\vectr}{\vectLam})-\vectLam(\dprod{\vectr}{\unitplz})]\\
  &\quad-\vphig\vphih\Omerep^2[\unitplz(\dprod{\vectr}{\fdot{\vectLam}})-\fdot{\vectLam}(\dprod{\unitplz}{\vectr})]
  +\ethvp[\unitplz(\dprod{\vectOme}{(\cprod{\vectLam}{\vectr})})
    -\vectOme(\dprod{\unitplz}{(\cprod{\vectLam}{\vectr})})]\\
  &\quad+2\ethvo[\unitplz(\dprod{\vectLam}{(\cprod{\vectLam}{\vectr})})
    -\vectLam(\dprod{\unitplz}{(\cprod{\vectLam}{\vectr})})]
  +\vphig\vphih[\unitplz(\dprod{\fdot{\vectLam}}{(\cprod{\vectLam}{\vectr})})
    -\fdot{\vectLam}(\dprod{\unitplz}{(\cprod{\vectLam}{\vectr})})]\\
  &\quad-\ethvs[\vectLam(\dprod{\vectOme}{\vectr})-\vectr(\dprod{\vectOme}{\vectLam})]
  -2\ethvr[\vectLam(\dprod{\vectLam}{\vectr})-\Lamrep^2\vectr]
  -\vphii[\vectLam(\dprod{\fdot{\vectLam}}{\vectr})-\vectr(\dprod{\vectLam}{\fdot{\vectLam}})]\\
  &\quad+\ethvv[\vectLam(\dprod{\unitplz}{\vectr})-\vectr(\dprod{\unitplz}{\vectLam})]
  \beqref{alg1}\text{ \& }\eqnref{alg5}
\end{split}
\end{align*}
\begin{align*}
\begin{split}
&=2\ethvr\epsvb(\cprod{\vectOme}{\vectLam})
  +\vphii\epsvb(\cprod{\vectOme}{\fdot{\vectLam}})
  -\ethvv\epsvb(\cprod{\vectOme}{\unitplz})
  -\ethvs\Omerep^2(\cprod{\vectr}{\vectOme})
  -2\ethvr\Omerep^2(\cprod{\vectr}{\vectLam})\\
  &\quad-\vphii\Omerep^2(\cprod{\vectr}{\fdot{\vectLam}})
  +\ethvv\Omerep^2(\cprod{\vectr}{\unitplz})
  +\ethvp\epsvb(\Omerep^2\unitplz-\epsvc\vectOme)
  +2\ethvo\epsvb(\epsvg\unitplz-\epsvc\vectLam)\\
  &\quad+\vphig\vphih\epsvb(\vsigc\unitplz-\epsvc\fdot{\vectLam})
  -\ethvp\Omerep^2(\epsvb\unitplz-\epsvf\vectOme)
  -2\ethvo\Omerep^2(\epsvh\unitplz-\epsvf\vectLam)\\
  &\quad-\vphig\vphih\Omerep^2(\vsign\unitplz-\epsvf\fdot{\vectLam})
  +\ethvp(\epsvm\unitplz-\epsvo\vectOme)
  -2\ethvo\epsvo\vectLam
  +\vphig\vphih(\frkth\unitplz-\epsvo\fdot{\vectLam})\\
  &\quad-\ethvs(\epsvb\vectLam-\epsvg\vectr)
  -2\ethvr(\epsvh\vectLam-\Lamrep^2\vectr)
  -\vphii(\vsign\vectLam-\vsigd\vectr)\\
  &\quad+\ethvv(\epsvf\vectLam-\epsvi\vectr)
  \beqref{rot1a}, \eqnref{rpath1a}\text{ \& }\eqnref{rpath1b}
\end{split}
\end{align*}
\begin{align*}
\begin{split}
&=2\ethvr\epsvb(\cprod{\vectOme}{\vectLam})
  +\vphii\epsvb(\cprod{\vectOme}{\fdot{\vectLam}})
  -\ethvv\epsvb(\cprod{\vectOme}{\unitplz})
  -\ethvs\Omerep^2(\cprod{\vectr}{\vectOme})
  -2\ethvr\Omerep^2(\cprod{\vectr}{\vectLam})\\
  &\quad-\vphii\Omerep^2(\cprod{\vectr}{\fdot{\vectLam}})
  +\ethvv\Omerep^2(\cprod{\vectr}{\unitplz})
  +\ethvp\epsvb\Omerep^2\unitplz-\ethvp\epsvb\epsvc\vectOme
  +2\ethvo\epsvb\epsvg\unitplz-2\ethvo\epsvb\epsvc\vectLam\\
  &\quad+\vphig\vphih\epsvb\vsigc\unitplz-\vphig\vphih\epsvb\epsvc\fdot{\vectLam}
  -\ethvp\Omerep^2\epsvb\unitplz+\ethvp\Omerep^2\epsvf\vectOme
  -2\ethvo\Omerep^2\epsvh\unitplz+2\ethvo\Omerep^2\epsvf\vectLam\\
  &\quad-\vphig\vphih\Omerep^2\vsign\unitplz+\vphig\vphih\Omerep^2\epsvf\fdot{\vectLam}
  +\ethvp\epsvm\unitplz-\ethvp\epsvo\vectOme
  -2\ethvo\epsvo\vectLam
  +\vphig\vphih\frkth\unitplz-\vphig\vphih\epsvo\fdot{\vectLam}\\
  &\quad-\ethvs\epsvb\vectLam+\ethvs\epsvg\vectr
  -2\ethvr\epsvh\vectLam+2\ethvr\Lamrep^2\vectr
  -\vphii\vsign\vectLam+\vphii\vsigd\vectr
  +\ethvv\epsvf\vectLam-\ethvv\epsvi\vectr
\end{split}
\end{align*}
\begin{align*}
\begin{split}
&=2\ethvr\epsvb(\cprod{\vectOme}{\vectLam})
  +\vphii\epsvb(\cprod{\vectOme}{\fdot{\vectLam}})
  -\ethvv\epsvb(\cprod{\vectOme}{\unitplz})
  -\ethvs\Omerep^2(\cprod{\vectr}{\vectOme})
  -2\ethvr\Omerep^2(\cprod{\vectr}{\vectLam})
  -\vphii\Omerep^2(\cprod{\vectr}{\fdot{\vectLam}})\\
  &\quad+\ethvv\Omerep^2(\cprod{\vectr}{\unitplz})
  +(-\ethvp\epsvb\epsvc+\ethvp\Omerep^2\epsvf-\ethvp\epsvo)\vectOme
  +(-\vphig\vphih\epsvb\epsvc+\vphig\vphih\Omerep^2\epsvf-\vphig\vphih\epsvo)\fdot{\vectLam}\\
  &\quad+(\ethvp\epsvb\Omerep^2+2\ethvo\epsvb\epsvg+\vphig\vphih\epsvb\vsigc
  -\ethvp\Omerep^2\epsvb-2\ethvo\Omerep^2\epsvh-\vphig\vphih\Omerep^2\vsign
  +\ethvp\epsvm+\vphig\vphih\frkth)\unitplz\\
  &\quad+(\ethvs\epsvg+2\ethvr\Lamrep^2+\vphii\vsigd-\ethvv\epsvi)\vectr
  +(-2\ethvo\epsvb\epsvc+2\ethvo\Omerep^2\epsvf-2\ethvo\epsvo
  -\ethvs\epsvb-2\ethvr\epsvh-\vphii\vsign+\ethvv\epsvf)\vectLam
\end{split}
\end{align*}
\begin{align*}
\begin{split}
&=2\ethvr\epsvb(\cprod{\vectOme}{\vectLam})
  +\vphii\epsvb(\cprod{\vectOme}{\fdot{\vectLam}})
  -\ethvv\epsvb(\cprod{\vectOme}{\unitplz})
  -\ethvs\Omerep^2(\cprod{\vectr}{\vectOme})
  -2\ethvr\Omerep^2(\cprod{\vectr}{\vectLam})\\
  &\quad-\vphii\Omerep^2(\cprod{\vectr}{\fdot{\vectLam}})
  +\ethvv\Omerep^2(\cprod{\vectr}{\unitplz})
  +\ethvp(\Omerep^2\epsvf-\epsvb\epsvc-\epsvo)\vectOme
  +\vphig\vphih(\Omerep^2\epsvf-\epsvb\epsvc-\epsvo)\fdot{\vectLam}\\
  &\quad+(\ethvs\epsvg+2\ethvr\Lamrep^2+\vphii\vsigd-\ethvv\epsvi)\vectr
  +[\ethvp\epsvm+2\ethvo(\epsvb\epsvg-\Omerep^2\epsvh)
     +\vphig\vphih(\epsvb\vsigc-\Omerep^2\vsign+\frkth)]\unitplz\\
  &\quad+[\ethvv\epsvf-\ethvs\epsvb-2\ethvr\epsvh-\vphii\vsign
     +2\ethvo(\Omerep^2\epsvf-\epsvb\epsvc-\epsvo)]\vectLam
\end{split}
\end{align*}
\begin{align}\label{rpath21i}
\begin{split}
&=2\ethvr\epsvb(\cprod{\vectOme}{\vectLam})
  +\vphii\epsvb(\cprod{\vectOme}{\fdot{\vectLam}})
  -\ethvv\epsvb(\cprod{\vectOme}{\unitplz})
  -\ethvs\Omerep^2(\cprod{\vectr}{\vectOme})
  -2\ethvr\Omerep^2(\cprod{\vectr}{\vectLam})\\
  &\quad-\vphii\Omerep^2(\cprod{\vectr}{\fdot{\vectLam}})
  +\ethvv\Omerep^2(\cprod{\vectr}{\unitplz})+\ethvp\frkym\vectOme
  +\vphig\vphih\frkym\fdot{\vectLam}+\frkyq\vectr+\frkyr\unitplz+\frkys\vectLam\\
  &\quad\beqref{rpath1l}\text{ \& }\eqnref{rpath1m}
\end{split}
\end{align}
\begin{align*}
\vscrj
&=\cprod{\fdota}{\ffdota}\beqref{kpath2b}\nonumber\\
\begin{split}
&=\cprod{[2\epsvh\vectOme-3\epsvg\vectr+\cprod{\fdot{\vectLam}}{\vectr}
    -\Omerep^2(\cprod{\vectOme}{\vectr})+\epsvb\vectLam]}{}
  [3\epsvh\vectLam+\epsvb\fdot{\vectLam}+\vrhot\vectOme+\vrhou\vectr\\
  &\quad+2\epsvb(\cprod{\vectLam}{\vectOme})
  -3\Omerep^2(\cprod{\vectLam}{\vectr})-3\epsvg(\cprod{\vectOme}{\vectr})
  +(\cprod{\ffdot{\vectLam}}{\vectr})]\beqref{rpath7a}\text{ \& }\eqnref{rpath7b}
\end{split}
\end{align*}
\begin{align*}
\begin{split}
&=6\epsvh^2(\cprod{\vectOme}{\vectLam})
  +2\epsvb\epsvh(\cprod{\vectOme}{\fdot{\vectLam}})
  +2\vrhou\epsvh(\cprod{\vectOme}{\vectr})
  +4\epsvb\epsvh[\cprod{\vectOme}{(\cprod{\vectLam}{\vectOme})}]
  -6\Omerep^2\epsvh[\cprod{\vectOme}{(\cprod{\vectLam}{\vectr})}]\\
  &\quad-6\epsvg\epsvh[\cprod{\vectOme}{(\cprod{\vectOme}{\vectr})}]
  +2\epsvh[\cprod{\vectOme}{(\cprod{\ffdot{\vectLam}}{\vectr})}]
  -9\epsvh\epsvg(\cprod{\vectr}{\vectLam})
  -3\epsvb\epsvg(\cprod{\vectr}{\fdot{\vectLam}})
  -3\vrhot\epsvg(\cprod{\vectr}{\vectOme})\\
  &\quad-6\epsvb\epsvg[\cprod{\vectr}{(\cprod{\vectLam}{\vectOme})}]
  +9\Omerep^2\epsvg[\cprod{\vectr}{(\cprod{\vectLam}{\vectr})}]
  +9\epsvg^2[\cprod{\vectr}{(\cprod{\vectOme}{\vectr})}]
  -3\epsvg[\cprod{\vectr}{(\cprod{\ffdot{\vectLam}}{\vectr})}]\\
  &\quad-3\epsvh[\cprod{\vectLam}{(\cprod{\fdot{\vectLam}}{\vectr})}]
  -\epsvb[\cprod{\fdot{\vectLam}}{(\cprod{\fdot{\vectLam}}{\vectr})}]
  -\vrhot[\cprod{\vectOme}{(\cprod{\fdot{\vectLam}}{\vectr})}]
  -\vrhou[\cprod{\vectr}{(\cprod{\fdot{\vectLam}}{\vectr})}]\\
  &\quad+2\epsvb[\cprod{(\cprod{\fdot{\vectLam}}{\vectr})}{(\cprod{\vectLam}{\vectOme})}]
  -3\Omerep^2[\cprod{(\cprod{\fdot{\vectLam}}{\vectr})}{(\cprod{\vectLam}{\vectr})}]
  -3\epsvg[\cprod{(\cprod{\fdot{\vectLam}}{\vectr})}{(\cprod{\vectOme}{\vectr})}]\\
  &\quad+[\cprod{(\cprod{\fdot{\vectLam}}{\vectr})}{(\cprod{\ffdot{\vectLam}}{\vectr})}]
  +3\epsvh\Omerep^2[\cprod{\vectLam}{(\cprod{\vectOme}{\vectr})}]
  +\epsvb\Omerep^2[\cprod{\fdot{\vectLam}}{(\cprod{\vectOme}{\vectr})}]
  +\vrhot\Omerep^2[\cprod{\vectOme}{(\cprod{\vectOme}{\vectr})}]\\
  &\quad+\vrhou\Omerep^2[\cprod{\vectr}{(\cprod{\vectOme}{\vectr})}]
  -2\epsvb\Omerep^2[\cprod{(\cprod{\vectOme}{\vectr})}{(\cprod{\vectLam}{\vectOme})}]
  +3\Omerep^4[\cprod{(\cprod{\vectOme}{\vectr})}{(\cprod{\vectLam}{\vectr})}]\\
  &\quad-\Omerep^2[\cprod{(\cprod{\vectOme}{\vectr})}{(\cprod{\ffdot{\vectLam}}{\vectr})}]
  +\epsvb^2(\cprod{\vectLam}{\fdot{\vectLam}})
  +\vrhot\epsvb(\cprod{\vectLam}{\vectOme})
  +\vrhou\epsvb(\cprod{\vectLam}{\vectr})
  +2\epsvb^2[\cprod{\vectLam}{(\cprod{\vectLam}{\vectOme})}]\\
  &\quad-3\Omerep^2\epsvb[\cprod{\vectLam}{(\cprod{\vectLam}{\vectr})}]
  -3\epsvg\epsvb[\cprod{\vectLam}{(\cprod{\vectOme}{\vectr})}]
  +\epsvb[\cprod{\vectLam}{(\cprod{\ffdot{\vectLam}}{\vectr})}]
\end{split}
\end{align*}
\begin{align*}
\begin{split}
&=(6\epsvh^2-\vrhot\epsvb)(\cprod{\vectOme}{\vectLam})
  +2\epsvb\epsvh(\cprod{\vectOme}{\fdot{\vectLam}})
  +(2\vrhou\epsvh+3\vrhot\epsvg)(\cprod{\vectOme}{\vectr})
  +(9\epsvh\epsvg+\vrhou\epsvb)(\cprod{\vectLam}{\vectr})\\
  &\quad-3\epsvb\epsvg(\cprod{\vectr}{\fdot{\vectLam}})
  +\epsvb^2(\cprod{\vectLam}{\fdot{\vectLam}})
  +4\epsvb\epsvh[\Omerep^2\vectLam-\vectOme(\dprod{\vectOme}{\vectLam})]
  -6\Omerep^2\epsvh[\vectLam(\dprod{\vectOme}{\vectr})-\vectr(\dprod{\vectOme}{\vectLam})]\\
  &\quad+(\vrhot\Omerep^2-6\epsvg\epsvh)[\vectOme(\dprod{\vectOme}{\vectr})-\Omerep^2\vectr]
  +2\epsvh[\ffdot{\vectLam}(\dprod{\vectOme}{\vectr})-\vectr(\dprod{\vectOme}{\ffdot{\vectLam}})]
  -6\epsvb\epsvg[\vectLam(\dprod{\vectr}{\vectOme})-\vectOme(\dprod{\vectr}{\vectLam})]\\
  &\quad+9\Omerep^2\epsvg[\scalr^2\vectLam-\vectr(\dprod{\vectr}{\vectLam})]
  +(9\epsvg^2+\vrhou\Omerep^2)[\scalr^2\vectOme-\vectr(\dprod{\vectr}{\vectOme})]
  -3\epsvg[\scalr^2\ffdot{\vectLam}-\vectr(\dprod{\ffdot{\vectLam}}{\vectr})]\\
  &\quad-3\epsvh[\fdot{\vectLam}(\dprod{\vectLam}{\vectr})-\vectr(\dprod{\vectLam}{\fdot{\vectLam}})]
  -\epsvb[\fdot{\vectLam}(\dprod{\fdot{\vectLam}}{\vectr})-\fdot{\vectLam}^2\vectr]
  -\vrhot[\fdot{\vectLam}(\dprod{\vectOme}{\vectr})-\vectr(\dprod{\vectOme}{\fdot{\vectLam}})]
  -\vrhou[\scalr^2\fdot{\vectLam}-\vectr(\dprod{\vectr}{\fdot{\vectLam}})]\\
  &\quad+3(\epsvh\Omerep^2-\epsvg\epsvb)[\vectOme(\dprod{\vectLam}{\vectr})-\vectr(\dprod{\vectLam}{\vectOme})]
  +\epsvb\Omerep^2[\vectOme(\dprod{\fdot{\vectLam}}{\vectr})-\vectr(\dprod{\fdot{\vectLam}}{\vectOme})]
  +2\epsvb^2[\vectLam(\dprod{\vectLam}{\vectOme})-\vectLam^2\vectOme]\\
  &\quad-3\Omerep^2\epsvb[\vectLam(\dprod{\vectLam}{\vectr})-\vectLam^2\vectr]
  +\epsvb[\ffdot{\vectLam}(\dprod{\vectLam}{\vectr})-\vectr(\dprod{\vectLam}{\ffdot{\vectLam}})]
  +2\epsvb[\vectLam(\dprod{\vectOme}{(\cprod{\fdot{\vectLam}}{\vectr})})
       -\vectOme(\dprod{\vectLam}{(\cprod{\fdot{\vectLam}}{\vectr})})]\\
  &\quad-3\Omerep^2[\vectLam(\dprod{\vectr}{(\cprod{\fdot{\vectLam}}{\vectr})})
       -\vectr(\dprod{\vectLam}{(\cprod{\fdot{\vectLam}}{\vectr})})]
  -3\epsvg[\vectOme(\dprod{\vectr}{(\cprod{\fdot{\vectLam}}{\vectr})})
       -\vectr(\dprod{\vectOme}{(\cprod{\fdot{\vectLam}}{\vectr})})]\\
  &\quad+[\ffdot{\vectLam}(\dprod{\vectr}{(\cprod{\fdot{\vectLam}}{\vectr})})
      -\vectr(\dprod{\ffdot{\vectLam}}{(\cprod{\fdot{\vectLam}}{\vectr})})]
  -2\epsvb\Omerep^2[\vectLam(\dprod{\vectOme}{(\cprod{\vectOme}{\vectr})})
      -\vectOme(\dprod{\vectLam}{(\cprod{\vectOme}{\vectr})})]\\
  &\quad+3\Omerep^4[\vectLam(\dprod{\vectr}{(\cprod{\vectOme}{\vectr})})
      -\vectr(\dprod{\vectLam}{(\cprod{\vectOme}{\vectr})})]
  -\Omerep^2[\ffdot{\vectLam}(\dprod{\vectr}{(\cprod{\vectOme}{\vectr})})
      -\vectr(\dprod{\ffdot{\vectLam}}{(\cprod{\vectOme}{\vectr})})]
  \beqref{alg1}\text{ \& }\eqnref{alg5}
\end{split}
\end{align*}
\begin{align*}
\begin{split}
&=(6\epsvh^2-\vrhot\epsvb)(\cprod{\vectOme}{\vectLam})
  +2\epsvb\epsvh(\cprod{\vectOme}{\fdot{\vectLam}})
  +(2\vrhou\epsvh+3\vrhot\epsvg)(\cprod{\vectOme}{\vectr})
  +(9\epsvh\epsvg+\vrhou\epsvb)(\cprod{\vectLam}{\vectr})\\
  &\quad-3\epsvb\epsvg(\cprod{\vectr}{\fdot{\vectLam}})
  +\epsvb^2(\cprod{\vectLam}{\fdot{\vectLam}})
  +4\epsvb\epsvh(\Omerep^2\vectLam-\epsvg\vectOme)
  -6\Omerep^2\epsvh(\epsvb\vectLam-\epsvg\vectr)\\
  &\quad+(\vrhot\Omerep^2-6\epsvg\epsvh)(\epsvb\vectOme-\Omerep^2\vectr)
  +2\epsvh(\epsvb\ffdot{\vectLam}-\vsigh\vectr)
  -6\epsvb\epsvg(\epsvb\vectLam-\epsvh\vectOme)
  +9\Omerep^2\epsvg(\scalr^2\vectLam-\epsvh\vectr)\\
  &\quad+(9\epsvg^2+\vrhou\Omerep^2)(\scalr^2\vectOme-\epsvb\vectr)
  -3\epsvg(\scalr^2\ffdot{\vectLam}-\vsigo\vectr)
  -3\epsvh(\epsvh\fdot{\vectLam}-\vsigd\vectr)
  -\epsvb(\vsign\fdot{\vectLam}-\fdot{\vectLam}^2\vectr)\\
  &\quad-\vrhot(\epsvb\fdot{\vectLam}-\vsigc\vectr)
  -\vrhou(\scalr^2\fdot{\vectLam}-\vsign\vectr)
  +3(\epsvh\Omerep^2-\epsvg\epsvb)(\epsvh\vectOme-\epsvg\vectr)
  +\epsvb\Omerep^2(\vsign\vectOme-\vsigc\vectr)\\
  &\quad+2\epsvb^2(\epsvg\vectLam-\vectLam^2\vectOme)
  -3\Omerep^2\epsvb(\epsvh\vectLam-\vectLam^2\vectr)
  +\epsvb(\epsvh\ffdot{\vectLam}-\vsigi\vectr)
  +2\epsvb(-\frktl\vectLam+\frkth\vectOme)
  -3\Omerep^2\frkth\vectr\\
  &\quad-3\epsvg\frktl\vectr-\frktj\vectr
  +2\epsvb\Omerep^2\frktk\vectOme+3\Omerep^4\epsvm\vectr+\Omerep^2\frktm\vectr
  \beqref{rot1a}, \eqnref{rpath1a}, \eqnref{rpath1b}\text{ \& }\eqnref{alg4}
\end{split}
\end{align*}
\begin{align*}
\begin{split}
&=(6\epsvh^2-\vrhot\epsvb)(\cprod{\vectOme}{\vectLam})
  +2\epsvb\epsvh(\cprod{\vectOme}{\fdot{\vectLam}})
  +(2\vrhou\epsvh+3\vrhot\epsvg)(\cprod{\vectOme}{\vectr})
  +(9\epsvh\epsvg+\vrhou\epsvb)(\cprod{\vectLam}{\vectr})\\
  &\quad-3\epsvb\epsvg(\cprod{\vectr}{\fdot{\vectLam}})
  +\epsvb^2(\cprod{\vectLam}{\fdot{\vectLam}})
  +4\epsvb\epsvh\Omerep^2\vectLam-4\epsvb\epsvh\epsvg\vectOme
  -6\Omerep^2\epsvh\epsvb\vectLam+6\Omerep^2\epsvh\epsvg\vectr\\
  &\quad+\epsvb(\vrhot\Omerep^2-6\epsvg\epsvh)\vectOme-\Omerep^2(\vrhot\Omerep^2-6\epsvg\epsvh)\vectr
  +2\epsvh\epsvb\ffdot{\vectLam}-2\epsvh\vsigh\vectr
  -6\epsvb^2\epsvg\vectLam+6\epsvb\epsvg\epsvh\vectOme\\
  &\quad+9\Omerep^2\epsvg\scalr^2\vectLam-9\Omerep^2\epsvg\epsvh\vectr
  +\scalr^2(9\epsvg^2+\vrhou\Omerep^2)\vectOme-\epsvb(9\epsvg^2+\vrhou\Omerep^2)\vectr
  -3\epsvg\scalr^2\ffdot{\vectLam}+3\epsvg\vsigo\vectr-3\epsvh^2\fdot{\vectLam}\\
  &\quad+3\epsvh\vsigd\vectr-\epsvb\vsign\fdot{\vectLam}+\epsvb\fdot{\vectLam}^2\vectr
  -\vrhot\epsvb\fdot{\vectLam}+\vrhot\vsigc\vectr
  -\vrhou\scalr^2\fdot{\vectLam}+\vrhou\vsign\vectr
  +3\epsvh(\epsvh\Omerep^2-\epsvg\epsvb)\vectOme\\
  &\quad-3\epsvg(\epsvh\Omerep^2-\epsvg\epsvb)\vectr
  +\epsvb\Omerep^2\vsign\vectOme-\epsvb\Omerep^2\vsigc\vectr
  +2\epsvb^2\epsvg\vectLam-2\epsvb^2\vectLam^2\vectOme
  -3\Omerep^2\epsvb\epsvh\vectLam+3\Omerep^2\epsvb\vectLam^2\vectr\\
  &\quad+\epsvb\epsvh\ffdot{\vectLam}
  -\epsvb\vsigi\vectr-2\epsvb\frktl\vectLam+2\epsvb\frkth\vectOme
  -3\Omerep^2\frkth\vectr-3\epsvg\frktl\vectr-\frktj\vectr
  +2\epsvb\Omerep^2\frktk\vectOme+3\Omerep^4\epsvm\vectr+\Omerep^2\frktm\vectr
\end{split}
\end{align*}
\begin{align*}
\begin{split}
&=(6\epsvh^2-\vrhot\epsvb)(\cprod{\vectOme}{\vectLam})
  +2\epsvb\epsvh(\cprod{\vectOme}{\fdot{\vectLam}})
  +(2\vrhou\epsvh+3\vrhot\epsvg)(\cprod{\vectOme}{\vectr})
  +(9\epsvh\epsvg+\vrhou\epsvb)(\cprod{\vectLam}{\vectr})\\
  &\quad-3\epsvb\epsvg(\cprod{\vectr}{\fdot{\vectLam}})
  +\epsvb^2(\cprod{\vectLam}{\fdot{\vectLam}})
  +(-3\epsvh^2-\epsvb\vsign-\vrhot\epsvb-\vrhou\scalr^2)\fdot{\vectLam}
  +(2\epsvh\epsvb-3\epsvg\scalr^2+\epsvb\epsvh)\ffdot{\vectLam}\\
  &\quad+(4\epsvb\epsvh\Omerep^2-6\Omerep^2\epsvh\epsvb-6\epsvb^2\epsvg
  +9\Omerep^2\epsvg\scalr^2+2\epsvb^2\epsvg-3\Omerep^2\epsvb\epsvh-2\epsvb\frktl)\vectLam\\
  &\quad+[-4\epsvb\epsvh\epsvg+\epsvb(\vrhot\Omerep^2-6\epsvg\epsvh)
  +6\epsvb\epsvg\epsvh+\scalr^2(9\epsvg^2+\vrhou\Omerep^2)
  +3\epsvh(\epsvh\Omerep^2-\epsvg\epsvb)\\
  &\qquad+\epsvb\Omerep^2\vsign-2\epsvb^2\vectLam^2+2\epsvb\frkth+2\epsvb\Omerep^2\frktk]\vectOme\\
  &\quad+[6\Omerep^2\epsvh\epsvg-\Omerep^2(\vrhot\Omerep^2-6\epsvg\epsvh)
  -2\epsvh\vsigh-9\Omerep^2\epsvg\epsvh-\epsvb(9\epsvg^2+\vrhou\Omerep^2)
  +3\epsvg\vsigo+\epsvb\fdot{\vectLam}^2+\vrhot\vsigc\\
  &\qquad+\vrhou\vsign
  -3\epsvg(\epsvh\Omerep^2-\epsvg\epsvb)-\epsvb\Omerep^2\vsigc
  +3\Omerep^2\epsvb\vectLam^2-\epsvb\vsigi-3\Omerep^2\frkth-3\epsvg\frktl-\frktj\\
  &\qquad+3\Omerep^4\epsvm+\Omerep^2\frktm+3\epsvh\vsigd]\vectr
\end{split}
\end{align*}
\begin{align*}
\begin{split}
&=(6\epsvh^2-\vrhot\epsvb)(\cprod{\vectOme}{\vectLam})
  +2\epsvb\epsvh(\cprod{\vectOme}{\fdot{\vectLam}})
  +(2\vrhou\epsvh+3\vrhot\epsvg)(\cprod{\vectOme}{\vectr})
  +(9\epsvh\epsvg+\vrhou\epsvb)(\cprod{\vectLam}{\vectr})\\
  &\quad-3\epsvb\epsvg(\cprod{\vectr}{\fdot{\vectLam}})
  +\epsvb^2(\cprod{\vectLam}{\fdot{\vectLam}})
  -(3\epsvh^2+\epsvb\vsign+\vrhot\epsvb+\vrhou\scalr^2)\fdot{\vectLam}
  +3(\epsvh\epsvb-\epsvg\scalr^2)\ffdot{\vectLam}\\
  &\quad+[5\Omerep^2(\scalr^2\epsvg-\epsvb\epsvh)+4\epsvg(\scalr^2\Omerep^2-\epsvb^2)-2\epsvb\frktl]\vectLam\\
  &\quad+[7\epsvg(\scalr^2\epsvg-\epsvb\epsvh)+\Omerep^2\epsvb(\vrhot+2\frktk+\vsign)
    +\Omerep^2(\scalr^2\vrhou+3\epsvh^2)+2\epsvb(\frkth-\Lamrep^2\epsvb)+2\scalr^2\epsvg^2]\vectOme\\
  &\quad+[\Omerep^4(3\epsvm-\vrhot)+\Omerep^2(\frktm-3\frkth)+\Omerep^2\epsvb(3\Lamrep^2-\vrhou-\vsigc)
    +\epsvb(\fdot{\vectLam}^2-6\epsvg^2-\vsigi)\\
    &\qquad+\epsvh(3\vsigd-2\vsigh)+3\epsvg(\vsigo-\frktl)+\vrhot\vsigc+\vrhou\vsign-\frktj]\vectr
\end{split}
\end{align*}
\begin{align*}
\begin{split}
&=(6\epsvh^2-\vrhot\epsvb)(\cprod{\vectOme}{\vectLam})
  +2\epsvb\epsvh(\cprod{\vectOme}{\fdot{\vectLam}})
  +(2\vrhou\epsvh+3\vrhot\epsvg)(\cprod{\vectOme}{\vectr})
  +(9\epsvh\epsvg+\vrhou\epsvb)(\cprod{\vectLam}{\vectr})\\
  &\quad-3\epsvb\epsvg(\cprod{\vectr}{\fdot{\vectLam}})
  +\epsvb^2(\cprod{\vectLam}{\fdot{\vectLam}})
  -(3\epsvh^2+\epsvb\vsign+\vrhot\epsvb+\vrhou\scalr^2)\fdot{\vectLam}
  -3(\epsvg\scalr^2-\epsvh\epsvb)\ffdot{\vectLam}\\
  &\quad+[5\Omerep^2(\scalr^2\epsvg-\epsvb\epsvh)+4\epsvg(\scalr^2\Omerep^2-\epsvb^2)-2\epsvb\frktl]\vectLam\\
  &\quad+[7\epsvg(\scalr^2\epsvg-\epsvb\epsvh)+\Omerep^2\epsvb(\vrhot+2\frktk+\vsign)
    +\Omerep^2(\scalr^2\vrhou+3\epsvh^2)+2\epsvb(\frkth-\Lamrep^2\epsvb)+2\scalr^2\epsvg^2]\vectOme\\
  &\quad+[\Omerep^4(3\epsvm-\vrhot)+\Omerep^2(\frktm-3\frkth)+\Omerep^2\epsvb(3\Lamrep^2-\vrhou-\vsigc)
    +\epsvb(\fdot{\vectLam}^2-6\epsvg^2-\vsigi)\\
    &\qquad+\epsvh(3\vsigd-2\vsigh)+3\epsvg(\vsigo-\frktl)+\vrhot\vsigc+\vrhou\vsign-\frktj]\vectr
\end{split}
\end{align*}
\begin{align}\label{rpath21j}
\begin{split}
&=(6\epsvh^2-\vrhot\epsvb)(\cprod{\vectOme}{\vectLam})
  +2\epsvb\epsvh(\cprod{\vectOme}{\fdot{\vectLam}})
  +(2\vrhou\epsvh+3\vrhot\epsvg)(\cprod{\vectOme}{\vectr})
  +(9\epsvh\epsvg+\vrhou\epsvb)(\cprod{\vectLam}{\vectr})\\
  &\quad-3\epsvb\epsvg(\cprod{\vectr}{\fdot{\vectLam}})
  +\epsvb^2(\cprod{\vectLam}{\fdot{\vectLam}})
  +(5\Omerep^2\vphic+4\epsvg\vphia^2-2\epsvb\frktl)\vectLam
  -(3\epsvh^2+\epsvb\vsign+\vrhot\epsvb+\vrhou\scalr^2)\fdot{\vectLam}\\
  &\quad-3\vphic\ffdot{\vectLam}
  +[7\epsvg\vphic+\Omerep^2\epsvb(\vrhot+2\frktk+\vsign)
    +\Omerep^2(\scalr^2\vrhou+3\epsvh^2)+2\epsvb(\frkth-\Lamrep^2\epsvb)+2\scalr^2\epsvg^2]\vectOme\\
  &\quad+[\Omerep^4(3\epsvm-\vrhot)+\Omerep^2(\frktm-3\frkth)+\Omerep^2\epsvb(3\Lamrep^2-\vrhou-\vsigc)
    +\epsvb(\fdot{\vectLam}^2-6\epsvg^2-\vsigi)\\
    &\qquad+\epsvh(3\vsigd-2\vsigh)+3\epsvg(\vsigo-\frktl)+\vrhot\vsigc+\vrhou\vsign-\frktj]\vectr
    \beqref{rot1b}
\end{split}
\nonumber\\
\begin{split}
&=\frkyt(\cprod{\vectOme}{\vectLam})
  +2\epsvb\epsvh(\cprod{\vectOme}{\fdot{\vectLam}})
  +\frkyu(\cprod{\vectOme}{\vectr})
  +\frkyv(\cprod{\vectLam}{\vectr})
  -3\epsvb\epsvg(\cprod{\vectr}{\fdot{\vectLam}})
  +\epsvb^2(\cprod{\vectLam}{\fdot{\vectLam}})\\
  &\qquad+\frkyw\vectLam-\frkyx\fdot{\vectLam}
  -3\vphic\ffdot{\vectLam}+\frkyy\vectOme+\frkyz\vectr
  \beqref{rpath1m}
\end{split}
\end{align}
\begin{align*}
\vscrk
&=\cprod{\fdota}{\fdote}\beqref{kpath2b}\nonumber\\
\begin{split}
&=\cprod{[2\epsvh\vectOme-3\epsvg\vectr+\cprod{\fdot{\vectLam}}{\vectr}
   -\Omerep^2(\cprod{\vectOme}{\vectr})+\epsvb\vectLam]}{[\ethvo(\cprod{\unitplz}{\vectOme})
   +\vphig\vphih(\cprod{\unitplz}{\vectLam})+\ethvr\vectOme+\vphii\vectLam-\ethvu\unitplz]}\\
   &\quad\beqref{rpath7a}\text{ \& }\eqnref{rpath17a}
\end{split}
\nonumber\\
\begin{split}
&=2\ethvo\epsvh[\cprod{\vectOme}{(\cprod{\unitplz}{\vectOme})}]
   +2\vphig\vphih\epsvh[\cprod{\vectOme}{(\cprod{\unitplz}{\vectLam})}]
   +2\vphii\epsvh(\cprod{\vectOme}{\vectLam})
   -2\ethvu\epsvh(\cprod{\vectOme}{\unitplz})\\
   &\quad-3\ethvo\epsvg[\cprod{\vectr}{(\cprod{\unitplz}{\vectOme})}]
   -3\vphig\vphih\epsvg[\cprod{\vectr}{(\cprod{\unitplz}{\vectLam})}]
   -3\ethvr\epsvg(\cprod{\vectr}{\vectOme})
   -3\vphii\epsvg(\cprod{\vectr}{\vectLam})\\
   &\quad+3\ethvu\epsvg(\cprod{\vectr}{\unitplz})
   +\ethvo[\cprod{(\cprod{\fdot{\vectLam}}{\vectr})}{(\cprod{\unitplz}{\vectOme})}]
   +\vphig\vphih[\cprod{(\cprod{\fdot{\vectLam}}{\vectr})}{(\cprod{\unitplz}{\vectLam})}]
   -\ethvr[\cprod{\vectOme}{(\cprod{\fdot{\vectLam}}{\vectr})}]\\
   &\quad-\vphii[\cprod{\vectLam}{(\cprod{\fdot{\vectLam}}{\vectr})}]
   +\ethvu[\cprod{\unitplz}{(\cprod{\fdot{\vectLam}}{\vectr})}]
   -\ethvo\Omerep^2[\cprod{(\cprod{\vectOme}{\vectr})}{(\cprod{\unitplz}{\vectOme})}]
   -\vphig\vphih\Omerep^2[\cprod{(\cprod{\vectOme}{\vectr})}{(\cprod{\unitplz}{\vectLam})}]\\
   &\quad+\ethvr\Omerep^2[\cprod{\vectOme}{(\cprod{\vectOme}{\vectr})}]
   +\vphii\Omerep^2[\cprod{\vectLam}{(\cprod{\vectOme}{\vectr})}]
   -\ethvu\Omerep^2[\cprod{\unitplz}{(\cprod{\vectOme}{\vectr})}]
   +\ethvo\epsvb[\cprod{\vectLam}{(\cprod{\unitplz}{\vectOme})}]\\
   &\quad+\vphig\vphih\epsvb[\cprod{\vectLam}{(\cprod{\unitplz}{\vectLam})}]
   +\ethvr\epsvb(\cprod{\vectLam}{\vectOme})
   -\ethvu\epsvb(\cprod{\vectLam}{\unitplz})
\end{split}
\end{align*}
\begin{align*}
\begin{split}
&=(2\vphii\epsvh-\ethvr\epsvb)(\cprod{\vectOme}{\vectLam})-2\ethvu\epsvh(\cprod{\vectOme}{\unitplz})
  -3\ethvr\epsvg(\cprod{\vectr}{\vectOme})-3\vphii\epsvg(\cprod{\vectr}{\vectLam})\\
  &\quad+3\ethvu\epsvg(\cprod{\vectr}{\unitplz})-\ethvu\epsvb(\cprod{\vectLam}{\unitplz})
  +2\ethvo\epsvh[\Omerep^2\unitplz-\vectOme(\dprod{\unitplz}{\vectOme})]
  +2\vphig\vphih\epsvh[\unitplz(\dprod{\vectOme}{\vectLam})-\vectLam(\dprod{\vectOme}{\unitplz})]\\
  &\quad-3\ethvo\epsvg[\unitplz(\dprod{\vectOme}{\vectr})-\vectOme(\dprod{\unitplz}{\vectr})]
  -3\vphig\vphih\epsvg[\unitplz(\dprod{\vectr}{\vectLam})-\vectLam(\dprod{\vectr}{\unitplz})]
  -\ethvr[\fdot{\vectLam}(\dprod{\vectOme}{\vectr})-\vectr(\dprod{\vectOme}{\fdot{\vectLam}})]\\
  &\quad-\vphii[\fdot{\vectLam}(\dprod{\vectLam}{\vectr})-\vectr(\dprod{\vectLam}{\fdot{\vectLam}})]
  +\ethvu[\fdot{\vectLam}(\dprod{\unitplz}{\vectr})-\vectr(\dprod{\unitplz}{\fdot{\vectLam}})]
  +\ethvr\Omerep^2[\vectOme(\dprod{\vectOme}{\vectr})-\Omerep^2\vectr]\\
  &\quad+\vphii\Omerep^2[\vectOme(\dprod{\vectLam}{\vectr})-\vectr(\dprod{\vectOme}{\vectLam})]
  -\ethvu\Omerep^2[\vectOme(\dprod{\unitplz}{\vectr})-\vectr(\dprod{\unitplz}{\vectOme})]
  +\ethvo\epsvb[\unitplz(\dprod{\vectLam}{\vectOme})-\vectOme(\dprod{\vectLam}{\unitplz})]\\
  &\quad+\vphig\vphih\epsvb[\Lamrep^2\unitplz-\vectLam(\dprod{\vectLam}{\unitplz})]
  +\ethvo[\unitplz(\dprod{\vectOme}{(\cprod{\fdot{\vectLam}}{\vectr})})
     -\vectOme(\dprod{\unitplz}{(\cprod{\fdot{\vectLam}}{\vectr})})]\\
  &\quad+\vphig\vphih[\unitplz(\dprod{\vectLam}{(\cprod{\fdot{\vectLam}}{\vectr})})
     -\vectLam(\dprod{\unitplz}{(\cprod{\fdot{\vectLam}}{\vectr})})]
  -\ethvo\Omerep^2[\unitplz(\dprod{\vectOme}{(\cprod{\vectOme}{\vectr})})
     -\vectOme(\dprod{\unitplz}{(\cprod{\vectOme}{\vectr})})]\\
  &\quad-\vphig\vphih\Omerep^2[\unitplz(\dprod{\vectLam}{(\cprod{\vectOme}{\vectr})})
     -\vectLam(\dprod{\unitplz}{(\cprod{\vectOme}{\vectr})})]
  \beqref{alg1}\text{ \& }\eqnref{alg5}
\end{split}
\end{align*}
\begin{align*}
\begin{split}
&=(2\vphii\epsvh-\ethvr\epsvb)(\cprod{\vectOme}{\vectLam})-2\ethvu\epsvh(\cprod{\vectOme}{\unitplz})
  -3\ethvr\epsvg(\cprod{\vectr}{\vectOme})-3\vphii\epsvg(\cprod{\vectr}{\vectLam})\\
  &\quad+3\ethvu\epsvg(\cprod{\vectr}{\unitplz})-\ethvu\epsvb(\cprod{\vectLam}{\unitplz})
  +2\ethvo\epsvh(\Omerep^2\unitplz-\epsvc\vectOme)
  +2\vphig\vphih\epsvh(\epsvg\unitplz-\epsvc\vectLam)\\
  &\quad-3\ethvo\epsvg(\epsvb\unitplz-\epsvf\vectOme)
  -3\vphig\vphih\epsvg(\epsvh\unitplz-\epsvf\vectLam)
  -\ethvr(\epsvb\fdot{\vectLam}-\vsigc\vectr)
  -\vphii(\epsvh\fdot{\vectLam}-\vsigd\vectr)\\
  &\quad+\ethvu(\epsvf\fdot{\vectLam}-\vsiga\vectr)
  +\ethvr\Omerep^2(\epsvb\vectOme-\Omerep^2\vectr)
  +\vphii\Omerep^2(\epsvh\vectOme-\epsvg\vectr)
  -\ethvu\Omerep^2(\epsvf\vectOme-\epsvc\vectr)\\
  &\quad+\ethvo\epsvb(\epsvg\unitplz-\epsvi\vectOme)
  +\vphig\vphih\epsvb(\Lamrep^2\unitplz-\epsvi\vectLam)
  +\ethvo(-\frktl\unitplz-\frktn\vectOme)
  +\vphig\vphih(-\frkth\unitplz-\frktn\vectLam)\\
  &\quad+\epsvl\ethvo\Omerep^2\vectOme
  -\vphig\vphih\Omerep^2(-\epsvm\unitplz-\epsvl\vectLam)
  \beqref{rot1a}, \eqnref{rpath1a}\text{ \& }\eqnref{rpath1b}
\end{split}
\end{align*}
\begin{align*}
\begin{split}
&=(2\vphii\epsvh-\ethvr\epsvb)(\cprod{\vectOme}{\vectLam})-2\ethvu\epsvh(\cprod{\vectOme}{\unitplz})
  -3\ethvr\epsvg(\cprod{\vectr}{\vectOme})-3\vphii\epsvg(\cprod{\vectr}{\vectLam})\\
  &\quad+3\ethvu\epsvg(\cprod{\vectr}{\unitplz})-\ethvu\epsvb(\cprod{\vectLam}{\unitplz})
  +2\ethvo\epsvh\Omerep^2\unitplz-2\ethvo\epsvh\epsvc\vectOme
  +2\vphig\vphih\epsvh\epsvg\unitplz-2\vphig\vphih\epsvh\epsvc\vectLam\\
  &\quad-3\ethvo\epsvg\epsvb\unitplz+3\ethvo\epsvg\epsvf\vectOme
  -3\vphig\vphih\epsvg\epsvh\unitplz+3\vphig\vphih\epsvg\epsvf\vectLam
  -\ethvr\epsvb\fdot{\vectLam}+\ethvr\vsigc\vectr
  -\vphii\epsvh\fdot{\vectLam}+\vphii\vsigd\vectr\\
  &\quad+\ethvu\epsvf\fdot{\vectLam}-\ethvu\vsiga\vectr
  +\ethvr\Omerep^2\epsvb\vectOme-\ethvr\Omerep^4\vectr
  +\vphii\Omerep^2\epsvh\vectOme-\vphii\Omerep^2\epsvg\vectr
  -\ethvu\Omerep^2\epsvf\vectOme+\ethvu\Omerep^2\epsvc\vectr\\
  &\quad+\ethvo\epsvb\epsvg\unitplz-\ethvo\epsvb\epsvi\vectOme
  +\vphig\vphih\epsvb\Lamrep^2\unitplz-\vphig\vphih\epsvb\epsvi\vectLam
  -\ethvo\frktl\unitplz-\ethvo\frktn\vectOme
  -\vphig\vphih\frkth\unitplz-\vphig\vphih\frktn\vectLam\\
  &\quad+\epsvl\ethvo\Omerep^2\vectOme
  +\vphig\vphih\Omerep^2\epsvm\unitplz+\vphig\vphih\Omerep^2\epsvl\vectLam
\end{split}
\end{align*}
\begin{align*}
\begin{split}
&=(2\vphii\epsvh-\ethvr\epsvb)(\cprod{\vectOme}{\vectLam})-2\ethvu\epsvh(\cprod{\vectOme}{\unitplz})
  -3\ethvr\epsvg(\cprod{\vectr}{\vectOme})-3\vphii\epsvg(\cprod{\vectr}{\vectLam})
  +3\ethvu\epsvg(\cprod{\vectr}{\unitplz})\\
  &\quad-\ethvu\epsvb(\cprod{\vectLam}{\unitplz})
  +(-\ethvr\epsvb-\vphii\epsvh+\ethvu\epsvf)\fdot{\vectLam}
  +(\ethvr\vsigc+\vphii\vsigd-\ethvu\vsiga-\ethvr\Omerep^4-\vphii\Omerep^2\epsvg+\ethvu\Omerep^2\epsvc)\vectr\\
  &\quad+(2\ethvo\epsvh\Omerep^2+2\vphig\vphih\epsvh\epsvg-3\ethvo\epsvg\epsvb
  -3\vphig\vphih\epsvg\epsvh+\ethvo\epsvb\epsvg+\vphig\vphih\epsvb\Lamrep^2-\ethvo\frktl-\vphig\vphih\frkth\\
  &\qquad+\vphig\vphih\Omerep^2\epsvm)\unitplz
  +(-2\ethvo\epsvh\epsvc+3\ethvo\epsvg\epsvf+\ethvr\Omerep^2\epsvb
  +\vphii\Omerep^2\epsvh-\ethvu\Omerep^2\epsvf-\ethvo\epsvb\epsvi
  -\ethvo\frktn+\epsvl\ethvo\Omerep^2)\vectOme\\
  &\quad+(-2\ethvo\epsvh\epsvc+3\ethvo\epsvg\epsvf+\ethvr\Omerep^2\epsvb
  +\vphii\Omerep^2\epsvh-\ethvu\Omerep^2\epsvf-\ethvo\epsvb\epsvi-\ethvo\frktn+\epsvl\ethvo\Omerep^2\\
  &\qquad-2\vphig\vphih\epsvh\epsvc+3\vphig\vphih\epsvg\epsvf-\vphig\vphih\epsvb\epsvi
  -\vphig\vphih\frktn+\vphig\vphih\Omerep^2\epsvl)\vectLam
\end{split}
\end{align*}
\begin{align}\label{rpath21k}
\begin{split}
&=(2\vphii\epsvh-\ethvr\epsvb)(\cprod{\vectOme}{\vectLam})-2\ethvu\epsvh(\cprod{\vectOme}{\unitplz})
  -3\ethvr\epsvg(\cprod{\vectr}{\vectOme})-3\vphii\epsvg(\cprod{\vectr}{\vectLam})\\
  &\quad+3\ethvu\epsvg(\cprod{\vectr}{\unitplz})-\ethvu\epsvb(\cprod{\vectLam}{\unitplz})
  +[\ethvr(\vsigc-\Omerep^4)+\vphii(\vsigd-\Omerep^2\epsvg)+\ethvu(\Omerep^2\epsvc-\vsiga)]\vectr\\
  &\quad+[\Omerep^2(2\epsvh\ethvo+\epsvm\vphig\vphih)-\ethvo(\frktl+2\epsvb\epsvg)
    +\vphig\vphih(\Lamrep^2\epsvb-\frkth-\epsvg\epsvh)]\unitplz\\
  &\quad+[\Omerep^2(\epsvl\ethvo-\epsvf\ethvu+\epsvb\ethvr+\epsvh\vphii)
    +\ethvo(3\epsvf\epsvg-2\epsvc\epsvh-\epsvb\epsvi-\frktn)]\vectOme\\
  &\quad+\vphig\vphih(-2\epsvc\epsvh+3\epsvf\epsvg-\epsvb\epsvi-\frktn+\Omerep^2\epsvl)\vectLam
  +(\ethvu\epsvf-\ethvr\epsvb-\vphii\epsvh)\fdot{\vectLam}
\end{split}
\nonumber\\
\begin{split}
&=\vakpa(\cprod{\vectOme}{\vectLam})-2\ethvu\epsvh(\cprod{\vectOme}{\unitplz})
  -3\ethvr\epsvg(\cprod{\vectr}{\vectOme})-3\vphii\epsvg(\cprod{\vectr}{\vectLam})
  +3\ethvu\epsvg(\cprod{\vectr}{\unitplz})\\
  &\quad-\ethvu\epsvb(\cprod{\vectLam}{\unitplz})
  +\vakpb\vectr+\vakpe\unitplz+\vakpf\vectOme+\vphig\vphih\vakpd\vectLam+\vakpc\fdot{\vectLam}
  \beqref{rpath1n}
\end{split}
\end{align}
\begin{align*}
\vscrl
&=\cprod{\fdota}{\ffdote}\beqref{kpath2b}\nonumber\\
\begin{split}
&=\cprod{[2\epsvh\vectOme-3\epsvg\vectr+\cprod{\fdot{\vectLam}}{\vectr}
  -\Omerep^2(\cprod{\vectOme}{\vectr})+\epsvb\vectLam]}{}[\ethvp(\cprod{\unitplz}{\vectOme})
   +2\ethvo(\cprod{\unitplz}{\vectLam})+\vphig\vphih(\cprod{\unitplz}{\fdot{\vectLam}})\\
  &\qquad+\ethvs\vectOme+2\ethvr\vectLam+\vphii\fdot{\vectLam}
    -\ethvv\unitplz]\beqref{rpath7a}\text{ \& }\eqnref{rpath17b}
\end{split}
\nonumber\\
\begin{split}
&=2\ethvp\epsvh[\cprod{\vectOme}{(\cprod{\unitplz}{\vectOme})}]
  +4\ethvo\epsvh[\cprod{\vectOme}{(\cprod{\unitplz}{\vectLam})}]
  +2\vphig\vphih\epsvh[\cprod{\vectOme}{(\cprod{\unitplz}{\fdot{\vectLam}})}]\\
  &\quad+2\ethvs\epsvh(\cprod{\vectOme}{\vectOme})
  +4\ethvr\epsvh(\cprod{\vectOme}{\vectLam})
  +2\vphii\epsvh(\cprod{\vectOme}{\fdot{\vectLam}})
  -2\ethvv\epsvh(\cprod{\vectOme}{\unitplz})\\
  &\quad-3\ethvp\epsvg[\cprod{\vectr}{(\cprod{\unitplz}{\vectOme})}]
  -6\ethvo\epsvg[\cprod{\vectr}{(\cprod{\unitplz}{\vectLam})}]
  -3\vphig\vphih\epsvg[\cprod{\vectr}{(\cprod{\unitplz}{\fdot{\vectLam}})}]\\
  &\quad-3\ethvs\epsvg(\cprod{\vectr}{\vectOme})
  -6\ethvr\epsvg(\cprod{\vectr}{\vectLam})
  -3\vphii\epsvg(\cprod{\vectr}{\fdot{\vectLam}})
  +3\ethvv\epsvg(\cprod{\vectr}{\unitplz})\\
  &\quad+\ethvp[\cprod{(\cprod{\fdot{\vectLam}}{\vectr})}{(\cprod{\unitplz}{\vectOme})}]
  +2\ethvo[\cprod{(\cprod{\fdot{\vectLam}}{\vectr})}{(\cprod{\unitplz}{\vectLam})}]
  +\vphig\vphih[\cprod{(\cprod{\fdot{\vectLam}}{\vectr})}{(\cprod{\unitplz}{\fdot{\vectLam}})}]\\
  &\quad-\ethvs[\cprod{\vectOme}{(\cprod{\fdot{\vectLam}}{\vectr})}]
  -2\ethvr[\cprod{\vectLam}{(\cprod{\fdot{\vectLam}}{\vectr})}]
  -\vphii[\cprod{\fdot{\vectLam}}{(\cprod{\fdot{\vectLam}}{\vectr})}]
  +\ethvv[\cprod{\unitplz}{(\cprod{\fdot{\vectLam}}{\vectr})}]\\
  &\quad-\ethvp\Omerep^2[\cprod{(\cprod{\vectOme}{\vectr})}{(\cprod{\unitplz}{\vectOme})}]
  -2\ethvo\Omerep^2[\cprod{(\cprod{\vectOme}{\vectr})}{(\cprod{\unitplz}{\vectLam})}]
  -\vphig\vphih\Omerep^2[\cprod{(\cprod{\vectOme}{\vectr})}{(\cprod{\unitplz}{\fdot{\vectLam}})}]\\
  &\quad+\ethvs\Omerep^2[\cprod{\vectOme}{(\cprod{\vectOme}{\vectr})}]
  +2\ethvr\Omerep^2[\cprod{\vectLam}{(\cprod{\vectOme}{\vectr})}]
  +\vphii\Omerep^2[\cprod{\fdot{\vectLam}}{(\cprod{\vectOme}{\vectr})}]
  -\ethvv\Omerep^2[\cprod{\unitplz}{(\cprod{\vectOme}{\vectr})}]\\
  &\quad+\ethvp\epsvb[\cprod{\vectLam}{(\cprod{\unitplz}{\vectOme})}]
  +2\ethvo\epsvb[\cprod{\vectLam}{(\cprod{\unitplz}{\vectLam})}]
  +\vphig\vphih\epsvb[\cprod{\vectLam}{(\cprod{\unitplz}{\fdot{\vectLam}})}]
  +\ethvs\epsvb(\cprod{\vectLam}{\vectOme})\\
  &\quad+2\ethvr\epsvb(\cprod{\vectLam}{\vectLam})
  +\vphii\epsvb(\cprod{\vectLam}{\fdot{\vectLam}})
  -\ethvv\epsvb(\cprod{\vectLam}{\unitplz})
\end{split}
\end{align*}
\begin{align*}
\begin{split}
&=(4\ethvr\epsvh-\ethvs\epsvb)(\cprod{\vectOme}{\vectLam})
  -2\ethvv\epsvh(\cprod{\vectOme}{\unitplz})
  -3\ethvs\epsvg(\cprod{\vectr}{\vectOme})
  -6\ethvr\epsvg(\cprod{\vectr}{\vectLam})\\
  &\quad+3\ethvv\epsvg(\cprod{\vectr}{\unitplz})
  -\ethvv\epsvb(\cprod{\vectLam}{\unitplz})
  +2\vphii\epsvh(\cprod{\vectOme}{\fdot{\vectLam}})
  -3\vphii\epsvg(\cprod{\vectr}{\fdot{\vectLam}})
  +\vphii\epsvb(\cprod{\vectLam}{\fdot{\vectLam}})\\
  &\quad+2\ethvp\epsvh[\cprod{\vectOme}{(\cprod{\unitplz}{\vectOme})}]
  +4\ethvo\epsvh[\cprod{\vectOme}{(\cprod{\unitplz}{\vectLam})}]
  +2\vphig\vphih\epsvh[\cprod{\vectOme}{(\cprod{\unitplz}{\fdot{\vectLam}})}]\\
  &\quad-3\ethvp\epsvg[\cprod{\vectr}{(\cprod{\unitplz}{\vectOme})}]
  -6\ethvo\epsvg[\cprod{\vectr}{(\cprod{\unitplz}{\vectLam})}]
  -3\vphig\vphih\epsvg[\cprod{\vectr}{(\cprod{\unitplz}{\fdot{\vectLam}})}]\\
  &\quad-\ethvs[\cprod{\vectOme}{(\cprod{\fdot{\vectLam}}{\vectr})}]
  -2\ethvr[\cprod{\vectLam}{(\cprod{\fdot{\vectLam}}{\vectr})}]
  -\vphii[\cprod{\fdot{\vectLam}}{(\cprod{\fdot{\vectLam}}{\vectr})}]
  +\ethvv[\cprod{\unitplz}{(\cprod{\fdot{\vectLam}}{\vectr})}]\\
  &\quad+\ethvs\Omerep^2[\cprod{\vectOme}{(\cprod{\vectOme}{\vectr})}]
  +2\ethvr\Omerep^2[\cprod{\vectLam}{(\cprod{\vectOme}{\vectr})}]
  +\vphii\Omerep^2[\cprod{\fdot{\vectLam}}{(\cprod{\vectOme}{\vectr})}]
  -\ethvv\Omerep^2[\cprod{\unitplz}{(\cprod{\vectOme}{\vectr})}]\\
  &\quad+\ethvp\epsvb[\cprod{\vectLam}{(\cprod{\unitplz}{\vectOme})}]
  +2\ethvo\epsvb[\cprod{\vectLam}{(\cprod{\unitplz}{\vectLam})}]
  +\vphig\vphih\epsvb[\cprod{\vectLam}{(\cprod{\unitplz}{\fdot{\vectLam}})}]\\
  &\quad+\ethvp[\cprod{(\cprod{\fdot{\vectLam}}{\vectr})}{(\cprod{\unitplz}{\vectOme})}]
  +2\ethvo[\cprod{(\cprod{\fdot{\vectLam}}{\vectr})}{(\cprod{\unitplz}{\vectLam})}]
  +\vphig\vphih[\cprod{(\cprod{\fdot{\vectLam}}{\vectr})}{(\cprod{\unitplz}{\fdot{\vectLam}})}]\\
  &\quad-\ethvp\Omerep^2[\cprod{(\cprod{\vectOme}{\vectr})}{(\cprod{\unitplz}{\vectOme})}]
  -2\ethvo\Omerep^2[\cprod{(\cprod{\vectOme}{\vectr})}{(\cprod{\unitplz}{\vectLam})}]
  -\vphig\vphih\Omerep^2[\cprod{(\cprod{\vectOme}{\vectr})}{(\cprod{\unitplz}{\fdot{\vectLam}})}]
\end{split}
\end{align*}
\begin{align*}
\begin{split}
&=(4\ethvr\epsvh-\ethvs\epsvb)(\cprod{\vectOme}{\vectLam})
  -2\ethvv\epsvh(\cprod{\vectOme}{\unitplz})
  -3\ethvs\epsvg(\cprod{\vectr}{\vectOme})
  -6\ethvr\epsvg(\cprod{\vectr}{\vectLam})\\
  &\quad+3\ethvv\epsvg(\cprod{\vectr}{\unitplz})
  -\ethvv\epsvb(\cprod{\vectLam}{\unitplz})
  +2\vphii\epsvh(\cprod{\vectOme}{\fdot{\vectLam}})
  -3\vphii\epsvg(\cprod{\vectr}{\fdot{\vectLam}})
  +\vphii\epsvb(\cprod{\vectLam}{\fdot{\vectLam}})\\
  &\quad+2\ethvp\epsvh[\Omerep^2\unitplz-\vectOme(\dprod{\vectOme}{\unitplz})]
  +4\ethvo\epsvh[\unitplz(\dprod{\vectOme}{\vectLam})-\vectLam(\dprod{\vectOme}{\unitplz})]
  +2\vphig\vphih\epsvh[\unitplz(\dprod{\vectOme}{\fdot{\vectLam}})-\fdot{\vectLam}(\dprod{\vectOme}{\unitplz})]\\
  &\quad-3\ethvp\epsvg[\unitplz(\dprod{\vectr}{\vectOme})-\vectOme(\dprod{\vectr}{\unitplz})]
  -6\ethvo\epsvg[\unitplz(\dprod{\vectr}{\vectLam})-\vectLam(\dprod{\vectr}{\unitplz})]
  -3\vphig\vphih\epsvg[\unitplz(\dprod{\vectr}{\fdot{\vectLam}})-\fdot{\vectLam}(\dprod{\vectr}{\unitplz})]\\
  &\quad-\ethvs[\fdot{\vectLam}(\dprod{\vectOme}{\vectr})-\vectr(\dprod{\vectOme}{\fdot{\vectLam}})]
  -2\ethvr[\fdot{\vectLam}(\dprod{\vectLam}{\vectr})-\vectr(\dprod{\vectLam}{\fdot{\vectLam}})]
  -\vphii[\fdot{\vectLam}(\dprod{\fdot{\vectLam}}{\vectr})-\fdot{\vectLam}^2\vectr]\\
  &\quad+\ethvv[\fdot{\vectLam}(\dprod{\unitplz}{\vectr})-\vectr(\dprod{\unitplz}{\fdot{\vectLam}})]
  +\ethvs\Omerep^2[\vectOme(\dprod{\vectOme}{\vectr})-\Omerep^2\vectr]
  +2\ethvr\Omerep^2[\vectOme(\dprod{\vectLam}{\vectr})-\vectr(\dprod{\vectLam}{\vectOme})]\\
  &\quad+\vphii\Omerep^2[\vectOme(\dprod{\fdot{\vectLam}}{\vectr})-\vectr(\dprod{\fdot{\vectLam}}{\vectOme})]
  -\ethvv\Omerep^2[\vectOme(\dprod{\unitplz}{\vectr})-\vectr(\dprod{\unitplz}{\vectOme})]
  +\ethvp\epsvb[\unitplz(\dprod{\vectLam}{\vectOme})-\vectOme(\dprod{\vectLam}{\unitplz})]\\
  &\quad+2\ethvo\epsvb[\Lamrep^2\unitplz-\vectLam(\dprod{\vectLam}{\unitplz})]
  +\vphig\vphih\epsvb[\unitplz(\dprod{\vectLam}{\fdot{\vectLam}})-\fdot{\vectLam}(\dprod{\vectLam}{\unitplz})]
  +\ethvp[\unitplz(\dprod{\vectOme}{(\cprod{\fdot{\vectLam}}{\vectr})})
     -\vectOme(\dprod{\unitplz}{(\cprod{\fdot{\vectLam}}{\vectr})})]\\
  &\quad+2\ethvo[\unitplz(\dprod{\vectLam}{(\cprod{\fdot{\vectLam}}{\vectr})})
     -\vectLam(\dprod{\unitplz}{(\cprod{\fdot{\vectLam}}{\vectr})})]
  +\vphig\vphih[\unitplz(\dprod{\fdot{\vectLam}}{(\cprod{\fdot{\vectLam}}{\vectr})})
     -\fdot{\vectLam}(\dprod{\unitplz}{(\cprod{\fdot{\vectLam}}{\vectr})})]\\
  &\quad-\ethvp\Omerep^2[\unitplz(\dprod{\vectOme}{(\cprod{\vectOme}{\vectr})})
     -\vectOme(\dprod{\unitplz}{(\cprod{\vectOme}{\vectr})})]
  -2\ethvo\Omerep^2[\unitplz(\dprod{\vectLam}{(\cprod{\vectOme}{\vectr})})
     -\vectLam(\dprod{\unitplz}{(\cprod{\vectOme}{\vectr})})]\\
  &\quad-\vphig\vphih\Omerep^2[\unitplz(\dprod{\fdot{\vectLam}}{(\cprod{\vectOme}{\vectr})})
     -\fdot{\vectLam}(\dprod{\unitplz}{(\cprod{\vectOme}{\vectr})})]
  \beqref{alg1}\text{ \& }\eqnref{alg5}
\end{split}
\end{align*}
\begin{align*}
\begin{split}
&=(4\ethvr\epsvh-\ethvs\epsvb)(\cprod{\vectOme}{\vectLam})
  -2\ethvv\epsvh(\cprod{\vectOme}{\unitplz})
  -3\ethvs\epsvg(\cprod{\vectr}{\vectOme})
  -6\ethvr\epsvg(\cprod{\vectr}{\vectLam})\\
  &\quad+3\ethvv\epsvg(\cprod{\vectr}{\unitplz})
  -\ethvv\epsvb(\cprod{\vectLam}{\unitplz})
  +2\vphii\epsvh(\cprod{\vectOme}{\fdot{\vectLam}})
  -3\vphii\epsvg(\cprod{\vectr}{\fdot{\vectLam}})
  +\vphii\epsvb(\cprod{\vectLam}{\fdot{\vectLam}})\\
  &\quad+2\ethvp\epsvh(\Omerep^2\unitplz-\epsvc\vectOme)
  +4\ethvo\epsvh(\epsvg\unitplz-\epsvc\vectLam)
  +2\vphig\vphih\epsvh(\vsigc\unitplz-\epsvc\fdot{\vectLam})
  -3\ethvp\epsvg(\epsvb\unitplz-\epsvf\vectOme)\\
  &\quad-6\ethvo\epsvg(\epsvh\unitplz-\epsvf\vectLam)
  -3\vphig\vphih\epsvg(\vsign\unitplz-\epsvf\fdot{\vectLam})
  -\ethvs(\epsvb\fdot{\vectLam}-\vsigc\vectr)
  -2\ethvr(\epsvh\fdot{\vectLam}-\vsigd\vectr)
  -\vphii(\vsign\fdot{\vectLam}-\fdot{\vectLam}^2\vectr)\\
  &\quad+\ethvv(\epsvf\fdot{\vectLam}-\vsiga\vectr)
  +\ethvs\Omerep^2(\epsvb\vectOme-\Omerep^2\vectr)
  +2\ethvr\Omerep^2(\epsvh\vectOme-\epsvg\vectr)
  +\vphii\Omerep^2(\vsign\vectOme-\vsigc\vectr)
  -\ethvv\Omerep^2(\epsvf\vectOme-\epsvc\vectr)\\
  &\quad+\ethvp\epsvb(\epsvg\unitplz-\epsvi\vectOme)
  +2\ethvo\epsvb(\Lamrep^2\unitplz-\epsvi\vectLam)
  +\vphig\vphih\epsvb(\vsigd\unitplz-\epsvi\fdot{\vectLam})
  +\ethvp(-\frktl\unitplz-\frktn\vectOme)\\
  &\quad+2\ethvo(-\frkth\unitplz-\frktn\vectLam)
  +\vphig\vphih(-\frktn\fdot{\vectLam})
  -\ethvp\Omerep^2(-\epsvl\vectOme)
  -2\ethvo\Omerep^2(-\epsvm\unitplz-\epsvl\vectLam)\\
  &\quad-\vphig\vphih\Omerep^2(\frktl\unitplz-\epsvl\fdot{\vectLam})
  \beqref{rot1a}, \eqnref{rpath1a}\text{ \& }\eqnref{rpath1b}
\end{split}
\end{align*}
\begin{align*}
\begin{split}
&=(4\ethvr\epsvh-\ethvs\epsvb)(\cprod{\vectOme}{\vectLam})
  -2\ethvv\epsvh(\cprod{\vectOme}{\unitplz})
  -3\ethvs\epsvg(\cprod{\vectr}{\vectOme})
  -6\ethvr\epsvg(\cprod{\vectr}{\vectLam})\\
  &\quad+3\ethvv\epsvg(\cprod{\vectr}{\unitplz})
  -\ethvv\epsvb(\cprod{\vectLam}{\unitplz})
  +2\vphii\epsvh(\cprod{\vectOme}{\fdot{\vectLam}})
  -3\vphii\epsvg(\cprod{\vectr}{\fdot{\vectLam}})
  +\vphii\epsvb(\cprod{\vectLam}{\fdot{\vectLam}})\\
  &\quad+2\ethvp\epsvh\Omerep^2\unitplz-2\ethvp\epsvh\epsvc\vectOme
  +4\ethvo\epsvh\epsvg\unitplz-4\ethvo\epsvh\epsvc\vectLam
  +2\vphig\vphih\epsvh\vsigc\unitplz-2\vphig\vphih\epsvh\epsvc\fdot{\vectLam}\\
  &\quad-3\ethvp\epsvg\epsvb\unitplz+3\ethvp\epsvg\epsvf\vectOme
  -6\ethvo\epsvg\epsvh\unitplz+6\ethvo\epsvg\epsvf\vectLam
  -3\vphig\vphih\epsvg\vsign\unitplz+3\vphig\vphih\epsvg\epsvf\fdot{\vectLam}\\
  &\quad-\ethvs\epsvb\fdot{\vectLam}+\ethvs\vsigc\vectr
  -2\ethvr\epsvh\fdot{\vectLam}+2\ethvr\vsigd\vectr
  -\vphii\vsign\fdot{\vectLam}+\vphii\fdot{\vectLam}^2\vectr
  +\ethvv\epsvf\fdot{\vectLam}-\ethvv\vsiga\vectr
  +\ethvs\Omerep^2\epsvb\vectOme-\ethvs\Omerep^4\vectr\\
  &\quad+2\ethvr\Omerep^2\epsvh\vectOme-2\ethvr\Omerep^2\epsvg\vectr
  +\vphii\Omerep^2\vsign\vectOme-\vphii\Omerep^2\vsigc\vectr
  -\ethvv\Omerep^2\epsvf\vectOme+\ethvv\Omerep^2\epsvc\vectr
  +\ethvp\epsvb\epsvg\unitplz-\ethvp\epsvb\epsvi\vectOme\\
  &\quad+2\ethvo\epsvb\Lamrep^2\unitplz-2\ethvo\epsvb\epsvi\vectLam
  +\vphig\vphih\epsvb\vsigd\unitplz-\vphig\vphih\epsvb\epsvi\fdot{\vectLam}
  -\ethvp\frktl\unitplz-\ethvp\frktn\vectOme
  -2\ethvo\frkth\unitplz-2\ethvo\frktn\vectLam\\
  &\quad-\vphig\vphih\frktn\fdot{\vectLam}
  +\ethvp\Omerep^2\epsvl\vectOme
  +2\ethvo\Omerep^2\epsvm\unitplz+2\ethvo\Omerep^2\epsvl\vectLam
  -\vphig\vphih\Omerep^2\frktl\unitplz+\vphig\vphih\Omerep^2\epsvl\fdot{\vectLam}
\end{split}
\end{align*}
\begin{align*}
\begin{split}
&=(4\ethvr\epsvh-\ethvs\epsvb)(\cprod{\vectOme}{\vectLam})
  -2\ethvv\epsvh(\cprod{\vectOme}{\unitplz})
  -3\ethvs\epsvg(\cprod{\vectr}{\vectOme})
  -6\ethvr\epsvg(\cprod{\vectr}{\vectLam})\\
  &\quad+3\ethvv\epsvg(\cprod{\vectr}{\unitplz})
  -\ethvv\epsvb(\cprod{\vectLam}{\unitplz})
  +2\vphii\epsvh(\cprod{\vectOme}{\fdot{\vectLam}})
  -3\vphii\epsvg(\cprod{\vectr}{\fdot{\vectLam}})
  +\vphii\epsvb(\cprod{\vectLam}{\fdot{\vectLam}})\\
  &\quad+[2\ethvp\epsvh\Omerep^2+4\ethvo\epsvh\epsvg+2\vphig\vphih\epsvh\vsigc
    -3\ethvp\epsvg\epsvb-6\ethvo\epsvg\epsvh-3\vphig\vphih\epsvg\vsign
    +\ethvp\epsvb\epsvg+2\ethvo\epsvb\Lamrep^2\\
    &\qquad+\vphig\vphih\epsvb\vsigd-2\ethvo\frkth-\ethvp\frktl+2\ethvo\Omerep^2\epsvm
    -\vphig\vphih\Omerep^2\frktl]\unitplz\\
  &\quad+[-4\ethvo\epsvh\epsvc+6\ethvo\epsvg\epsvf-2\ethvo\epsvb\epsvi
    -2\ethvo\frktn+2\ethvo\Omerep^2\epsvl]\vectLam\\
  &\quad+[\ethvs\vsigc+2\ethvr\vsigd+\vphii\fdot{\vectLam}^2-\ethvv\vsiga-\ethvs\Omerep^4
    -2\ethvr\Omerep^2\epsvg-\vphii\Omerep^2\vsigc+\ethvv\Omerep^2\epsvc]\vectr\\
  &\quad+[-2\ethvp\epsvh\epsvc+3\ethvp\epsvg\epsvf+\ethvs\Omerep^2\epsvb+2\ethvr\Omerep^2\epsvh
    +\vphii\Omerep^2\vsign-\ethvv\Omerep^2\epsvf-\ethvp\epsvb\epsvi
    -\ethvp\frktn+\ethvp\Omerep^2\epsvl]\vectOme\\
  &\quad+[-2\vphig\vphih\epsvh\epsvc+3\vphig\vphih\epsvg\epsvf-\ethvs\epsvb-2\ethvr\epsvh-\vphii\vsign
    +\ethvv\epsvf-\vphig\vphih\epsvb\epsvi-\vphig\vphih\frktn+\vphig\vphih\Omerep^2\epsvl]\fdot{\vectLam}
\end{split}
\end{align*}
\begin{align}\label{rpath21l}
\begin{split}
&=(4\ethvr\epsvh-\ethvs\epsvb)(\cprod{\vectOme}{\vectLam})
  -2\ethvv\epsvh(\cprod{\vectOme}{\unitplz})
  -3\ethvs\epsvg(\cprod{\vectr}{\vectOme})
  -6\ethvr\epsvg(\cprod{\vectr}{\vectLam})\\
  &\quad+3\ethvv\epsvg(\cprod{\vectr}{\unitplz})
  -\ethvv\epsvb(\cprod{\vectLam}{\unitplz})
  +2\vphii\epsvh(\cprod{\vectOme}{\fdot{\vectLam}})
  -3\vphii\epsvg(\cprod{\vectr}{\fdot{\vectLam}})
  +\vphii\epsvb(\cprod{\vectLam}{\fdot{\vectLam}})\\
  &\quad+[2\Omerep^2(\ethvp\epsvh+\ethvo\epsvm)+2\ethvo(\epsvb\Lamrep^2-\epsvg\epsvh-\frkth)
    +\vphig\vphih(-\Omerep^2\frktl+2\epsvh\vsigc-3\epsvg\vsign+\epsvb\vsigd)\\
    &\qquad-\ethvp(2\epsvg\epsvb+\frktl)]\unitplz
  +2\ethvo[\Omerep^2\epsvl-2\epsvh\epsvc+3\epsvg\epsvf-\epsvb\epsvi-\frktn]\vectLam\\
  &\quad+[\Omerep^2(\ethvs\epsvb+2\ethvr\epsvh+\vphii\vsign-\ethvv\epsvf+\ethvp\epsvl)
    +\ethvp(3\epsvg\epsvf-2\epsvh\epsvc-\epsvb\epsvi-\frktn)]\vectOme\\
  &\quad+[\vphig\vphih(3\epsvg\epsvf-2\epsvh\epsvc-\epsvb\epsvi-\frktn+\Omerep^2\epsvl)
    -\ethvs\epsvb-2\ethvr\epsvh-\vphii\vsign+\ethvv\epsvf]\fdot{\vectLam}\\
  &\quad+[\ethvs(\vsigc-\Omerep^4)+2\ethvr(\vsigd-\Omerep^2\epsvg)
    +\vphii(\fdot{\vectLam}^2-\Omerep^2\vsigc)+\ethvv(\Omerep^2\epsvc-\vsiga)]\vectr
\end{split}
\nonumber\\
\begin{split}
&=\vakpg(\cprod{\vectOme}{\vectLam})
  -2\ethvv\epsvh(\cprod{\vectOme}{\unitplz})
  -3\ethvs\epsvg(\cprod{\vectr}{\vectOme})
  -6\ethvr\epsvg(\cprod{\vectr}{\vectLam})\\
  &\quad+3\ethvv\epsvg(\cprod{\vectr}{\unitplz})
  -\ethvv\epsvb(\cprod{\vectLam}{\unitplz})
  +2\vphii\epsvh(\cprod{\vectOme}{\fdot{\vectLam}})
  -3\vphii\epsvg(\cprod{\vectr}{\fdot{\vectLam}})
  +\vphii\epsvb(\cprod{\vectLam}{\fdot{\vectLam}})\\
  &\quad+\vakpi\unitplz+2\ethvo\vakph\vectLam+\vakpj\vectOme
  +\vakpk\fdot{\vectLam}+\vakpl\vectr\beqref{rpath1n}
\end{split}
\end{align}
\begin{align*}
\vscrn
&=\cprod{\fdote}{\ffdota}\beqref{kpath2b}\nonumber\\
\begin{split}
&=\cprod{[\ethvo(\cprod{\unitplz}{\vectOme})+\vphig\vphih(\cprod{\unitplz}{\vectLam})
  +\ethvr\vectOme+\vphii\vectLam-\ethvu\unitplz]}{}[3\epsvh\vectLam+\epsvb\fdot{\vectLam}
  +\vrhot\vectOme+\vrhou\vectr\\
  &\quad+2\epsvb(\cprod{\vectLam}{\vectOme})
  -3\Omerep^2(\cprod{\vectLam}{\vectr})-3\epsvg(\cprod{\vectOme}{\vectr})
  +(\cprod{\ffdot{\vectLam}}{\vectr})]\beqref{rpath7b}\text{ \& }\eqnref{rpath17a}
\end{split}
\nonumber\\
\begin{split}
&=-3\epsvh\ethvo[\cprod{\vectLam}{(\cprod{\unitplz}{\vectOme})}]
  -\epsvb\ethvo[\cprod{\fdot{\vectLam}}{(\cprod{\unitplz}{\vectOme})}]
  -\vrhot\ethvo[\cprod{\vectOme}{(\cprod{\unitplz}{\vectOme})}]
  -\vrhou\ethvo[\cprod{\vectr}{(\cprod{\unitplz}{\vectOme})}]\\
  &\quad+2\epsvb\ethvo[\cprod{(\cprod{\unitplz}{\vectOme})}{(\cprod{\vectLam}{\vectOme})}]
  -3\Omerep^2\ethvo[\cprod{(\cprod{\unitplz}{\vectOme})}{(\cprod{\vectLam}{\vectr})}]
  -3\epsvg\ethvo[\cprod{(\cprod{\unitplz}{\vectOme})}{(\cprod{\vectOme}{\vectr})}]\\
  &\quad+\ethvo[\cprod{(\cprod{\unitplz}{\vectOme})}{(\cprod{\ffdot{\vectLam}}{\vectr})}]
  -3\epsvh\vphig\vphih[\cprod{\vectLam}{(\cprod{\unitplz}{\vectLam})}]
  -\epsvb\vphig\vphih[\cprod{\fdot{\vectLam}}{(\cprod{\unitplz}{\vectLam})}]\\
  &\quad-\vrhot\vphig\vphih[\cprod{\vectOme}{(\cprod{\unitplz}{\vectLam})}]
  -\vrhou\vphig\vphih[\cprod{\vectr}{(\cprod{\unitplz}{\vectLam})}]
  +2\epsvb\vphig\vphih[\cprod{(\cprod{\unitplz}{\vectLam})}{(\cprod{\vectLam}{\vectOme})}]\\
  &\quad-3\Omerep^2\vphig\vphih[\cprod{(\cprod{\unitplz}{\vectLam})}{(\cprod{\vectLam}{\vectr})}]
  -3\epsvg\vphig\vphih[\cprod{(\cprod{\unitplz}{\vectLam})}{(\cprod{\vectOme}{\vectr})}]
  +\vphig\vphih[\cprod{(\cprod{\unitplz}{\vectLam})}{(\cprod{\ffdot{\vectLam}}{\vectr})}]\\
  &\quad+3\epsvh\ethvr(\cprod{\vectOme}{\vectLam})
  +\epsvb\ethvr(\cprod{\vectOme}{\fdot{\vectLam}})
  +\vrhou\ethvr(\cprod{\vectOme}{\vectr})
  +2\epsvb\ethvr[\cprod{\vectOme}{(\cprod{\vectLam}{\vectOme})}]\\
  &\quad-3\Omerep^2\ethvr[\cprod{\vectOme}{(\cprod{\vectLam}{\vectr})}]
  -3\epsvg\ethvr[\cprod{\vectOme}{(\cprod{\vectOme}{\vectr})}]
  +\ethvr[\cprod{\vectOme}{(\cprod{\ffdot{\vectLam}}{\vectr})}]
  +\epsvb\vphii(\cprod{\vectLam}{\fdot{\vectLam}})
  +\vrhot\vphii(\cprod{\vectLam}{\vectOme})\\
  &\quad+\vrhou\vphii(\cprod{\vectLam}{\vectr})
  +2\epsvb\vphii[\cprod{\vectLam}{(\cprod{\vectLam}{\vectOme})}]
  -3\Omerep^2\vphii[\cprod{\vectLam}{(\cprod{\vectLam}{\vectr})}]
  -3\epsvg\vphii[\cprod{\vectLam}{(\cprod{\vectOme}{\vectr})}]\\
  &\quad+\vphii[\cprod{\vectLam}{(\cprod{\ffdot{\vectLam}}{\vectr})}]
  -3\epsvh\ethvu(\cprod{\unitplz}{\vectLam})
  -\epsvb\ethvu(\cprod{\unitplz}{\fdot{\vectLam}})
  -\vrhot\ethvu(\cprod{\unitplz}{\vectOme})
  -\vrhou\ethvu(\cprod{\unitplz}{\vectr})\\
  &\quad-2\epsvb\ethvu[\cprod{\unitplz}{(\cprod{\vectLam}{\vectOme})}]
  +3\Omerep^2\ethvu[\cprod{\unitplz}{(\cprod{\vectLam}{\vectr})}]
  +3\epsvg\ethvu[\cprod{\unitplz}{(\cprod{\vectOme}{\vectr})}]
  -\ethvu[\cprod{\unitplz}{(\cprod{\ffdot{\vectLam}}{\vectr})}]
\end{split}
\end{align*}
\begin{align*}
\begin{split}
&=(3\epsvh\ethvr-\vrhot\vphii)(\cprod{\vectOme}{\vectLam})
  +\vrhou\ethvr(\cprod{\vectOme}{\vectr})
  +\vrhou\vphii(\cprod{\vectLam}{\vectr})
  -\vrhot\ethvu(\cprod{\unitplz}{\vectOme})\\
  &\quad-\vrhou\ethvu(\cprod{\unitplz}{\vectr})
  -3\epsvh\ethvu(\cprod{\unitplz}{\vectLam})
  -\epsvb\ethvu(\cprod{\unitplz}{\fdot{\vectLam}})
  +\epsvb\ethvr(\cprod{\vectOme}{\fdot{\vectLam}})
  +\epsvb\vphii(\cprod{\vectLam}{\fdot{\vectLam}})\\
  &\quad-3\epsvh\ethvo[\cprod{\vectLam}{(\cprod{\unitplz}{\vectOme})}]
  -\epsvb\ethvo[\cprod{\fdot{\vectLam}}{(\cprod{\unitplz}{\vectOme})}]
  -\vrhot\ethvo[\cprod{\vectOme}{(\cprod{\unitplz}{\vectOme})}]
  -\vrhou\ethvo[\cprod{\vectr}{(\cprod{\unitplz}{\vectOme})}]\\
  &\quad-3\epsvh\vphig\vphih[\cprod{\vectLam}{(\cprod{\unitplz}{\vectLam})}]
  -\epsvb\vphig\vphih[\cprod{\fdot{\vectLam}}{(\cprod{\unitplz}{\vectLam})}]
  -\vrhot\vphig\vphih[\cprod{\vectOme}{(\cprod{\unitplz}{\vectLam})}]\\
  &\quad-\vrhou\vphig\vphih[\cprod{\vectr}{(\cprod{\unitplz}{\vectLam})}]
  +2\epsvb\ethvr[\cprod{\vectOme}{(\cprod{\vectLam}{\vectOme})}]
  -3\Omerep^2\ethvr[\cprod{\vectOme}{(\cprod{\vectLam}{\vectr})}]
  -3\epsvg\ethvr[\cprod{\vectOme}{(\cprod{\vectOme}{\vectr})}]\\
  &\quad+\ethvr[\cprod{\vectOme}{(\cprod{\ffdot{\vectLam}}{\vectr})}]
  +2\epsvb\vphii[\cprod{\vectLam}{(\cprod{\vectLam}{\vectOme})}]
  -3\Omerep^2\vphii[\cprod{\vectLam}{(\cprod{\vectLam}{\vectr})}]
  -3\epsvg\vphii[\cprod{\vectLam}{(\cprod{\vectOme}{\vectr})}]\\
  &\quad+\vphii[\cprod{\vectLam}{(\cprod{\ffdot{\vectLam}}{\vectr})}]
  -2\epsvb\ethvu[\cprod{\unitplz}{(\cprod{\vectLam}{\vectOme})}]
  +3\Omerep^2\ethvu[\cprod{\unitplz}{(\cprod{\vectLam}{\vectr})}]
  +3\epsvg\ethvu[\cprod{\unitplz}{(\cprod{\vectOme}{\vectr})}]\\
  &\quad-\ethvu[\cprod{\unitplz}{(\cprod{\ffdot{\vectLam}}{\vectr})}]
  +2\epsvb\ethvo[\cprod{(\cprod{\unitplz}{\vectOme})}{(\cprod{\vectLam}{\vectOme})}]
  -3\Omerep^2\ethvo[\cprod{(\cprod{\unitplz}{\vectOme})}{(\cprod{\vectLam}{\vectr})}]\\
  &\quad-3\epsvg\ethvo[\cprod{(\cprod{\unitplz}{\vectOme})}{(\cprod{\vectOme}{\vectr})}]
  +\ethvo[\cprod{(\cprod{\unitplz}{\vectOme})}{(\cprod{\ffdot{\vectLam}}{\vectr})}]
  +2\epsvb\vphig\vphih[\cprod{(\cprod{\unitplz}{\vectLam})}{(\cprod{\vectLam}{\vectOme})}]\\
  &\quad-3\Omerep^2\vphig\vphih[\cprod{(\cprod{\unitplz}{\vectLam})}{(\cprod{\vectLam}{\vectr})}]
  -3\epsvg\vphig\vphih[\cprod{(\cprod{\unitplz}{\vectLam})}{(\cprod{\vectOme}{\vectr})}]
  +\vphig\vphih[\cprod{(\cprod{\unitplz}{\vectLam})}{(\cprod{\ffdot{\vectLam}}{\vectr})}]
\end{split}
\end{align*}
\begin{align*}
\begin{split}
&=(3\epsvh\ethvr-\vrhot\vphii)(\cprod{\vectOme}{\vectLam})
  +\vrhou\ethvr(\cprod{\vectOme}{\vectr})
  +\vrhou\vphii(\cprod{\vectLam}{\vectr})
  -\vrhot\ethvu(\cprod{\unitplz}{\vectOme})\\
  &\quad-\vrhou\ethvu(\cprod{\unitplz}{\vectr})
  -3\epsvh\ethvu(\cprod{\unitplz}{\vectLam})
  -\epsvb\ethvu(\cprod{\unitplz}{\fdot{\vectLam}})
  +\epsvb\ethvr(\cprod{\vectOme}{\fdot{\vectLam}})
  +\epsvb\vphii(\cprod{\vectLam}{\fdot{\vectLam}})\\
  &\quad-3\epsvh\ethvo[\unitplz(\dprod{\vectLam}{\vectOme})-\vectOme(\dprod{\vectLam}{\unitplz})]
  -\epsvb\ethvo[\unitplz(\dprod{\fdot{\vectLam}}{\vectOme})-\vectOme(\dprod{\unitplz}{\fdot{\vectLam}})]
  -\vrhot\ethvo[\Omerep^2\unitplz-\vectOme(\dprod{\vectOme}{\unitplz})]\\
  &\quad-\vrhou\ethvo[\unitplz(\dprod{\vectr}{\vectOme})-\vectOme(\dprod{\vectr}{\unitplz})]
  -3\epsvh\vphig\vphih[\Lamrep^2\unitplz-\vectLam(\dprod{\vectLam}{\unitplz})]
  -\epsvb\vphig\vphih[\unitplz(\dprod{\fdot{\vectLam}}{\vectLam})-\vectLam(\dprod{\unitplz}{\fdot{\vectLam}})]\\
  &\quad-\vrhot\vphig\vphih[\unitplz(\dprod{\vectOme}{\vectLam})-\vectLam(\dprod{\vectOme}{\unitplz})]
  -\vrhou\vphig\vphih[\unitplz(\dprod{\vectr}{\vectLam})-\vectLam(\dprod{\vectr}{\unitplz})]
  +2\epsvb\ethvr[\Omerep^2\vectLam-\vectOme(\dprod{\vectOme}{\vectLam})]\\
  &\quad-3\Omerep^2\ethvr[\vectLam(\dprod{\vectOme}{\vectr})-\vectr(\dprod{\vectOme}{\vectLam})]
  -3\epsvg\ethvr[\vectOme(\dprod{\vectOme}{\vectr})-\Omerep^2\vectr]
  +\ethvr[\ffdot{\vectLam}(\dprod{\vectOme}{\vectr})-\vectr(\dprod{\vectOme}{\ffdot{\vectLam}})]\\
  &\quad+2\epsvb\vphii[\vectLam(\dprod{\vectLam}{\vectOme})-\Lamrep^2\vectOme]
  -3\Omerep^2\vphii[\vectLam(\dprod{\vectLam}{\vectr})-\Lamrep^2\vectr]
  -3\epsvg\vphii[\vectOme(\dprod{\vectLam}{\vectr})-\vectr(\dprod{\vectLam}{\vectOme})]\\
  &\quad+\vphii[\ffdot{\vectLam}(\dprod{\vectLam}{\vectr})-\vectr(\dprod{\vectLam}{\ffdot{\vectLam}})]
  -2\epsvb\ethvu[\vectLam(\dprod{\unitplz}{\vectOme})-\vectOme(\dprod{\unitplz}{\vectLam})]
  +3\Omerep^2\ethvu[\vectLam(\dprod{\unitplz}{\vectr})-\vectr(\dprod{\unitplz}{\vectLam})]\\
  &\quad+3\epsvg\ethvu[\vectOme(\dprod{\unitplz}{\vectr})-\vectr(\dprod{\unitplz}{\vectOme})]
  -\ethvu[\ffdot{\vectLam}(\dprod{\unitplz}{\vectr})-\vectr(\dprod{\unitplz}{\ffdot{\vectLam}})]
  +2\epsvb\ethvo[\vectLam(\dprod{\vectOme}{(\cprod{\unitplz}{\vectOme})})
     -\vectOme(\dprod{\vectLam}{(\cprod{\unitplz}{\vectOme})})]\\
  &\quad-3\Omerep^2\ethvo[\vectLam(\dprod{\vectr}{(\cprod{\unitplz}{\vectOme})})
     -\vectr(\dprod{\vectLam}{(\cprod{\unitplz}{\vectOme})})]
  -3\epsvg\ethvo[\vectOme(\dprod{\vectr}{(\cprod{\unitplz}{\vectOme})})
     -\vectr(\dprod{\vectOme}{(\cprod{\unitplz}{\vectOme})})]\\
  &\quad+\ethvo[\ffdot{\vectLam}(\dprod{\vectr}{(\cprod{\unitplz}{\vectOme})})
     -\vectr(\dprod{\ffdot{\vectLam}}{(\cprod{\unitplz}{\vectOme})})]
  +2\epsvb\vphig\vphih[\vectLam(\dprod{\vectOme}{(\cprod{\unitplz}{\vectLam})})
     -\vectOme(\dprod{\vectLam}{(\cprod{\unitplz}{\vectLam})})]\\
  &\quad-3\Omerep^2\vphig\vphih[\vectLam(\dprod{\vectr}{(\cprod{\unitplz}{\vectLam})})
     -\vectr(\dprod{\vectLam}{(\cprod{\unitplz}{\vectLam})})]
  -3\epsvg\vphig\vphih[\vectOme(\dprod{\vectr}{(\cprod{\unitplz}{\vectLam})})
     -\vectr(\dprod{\vectOme}{(\cprod{\unitplz}{\vectLam})})]\\
  &\quad+\vphig\vphih[\ffdot{\vectLam}(\dprod{\vectr}{(\cprod{\unitplz}{\vectLam})})
    -\vectr(\dprod{\ffdot{\vectLam}}{(\cprod{\unitplz}{\vectLam})})]
   \beqref{alg1}\text{ \& }\eqnref{alg5}
\end{split}
\end{align*}
\begin{align*}
\begin{split}
&=(3\epsvh\ethvr-\vrhot\vphii)(\cprod{\vectOme}{\vectLam})
  +\vrhou\ethvr(\cprod{\vectOme}{\vectr})
  +\vrhou\vphii(\cprod{\vectLam}{\vectr})
  -\vrhot\ethvu(\cprod{\unitplz}{\vectOme})\\
  &\quad-\vrhou\ethvu(\cprod{\unitplz}{\vectr})
  -3\epsvh\ethvu(\cprod{\unitplz}{\vectLam})
  -\epsvb\ethvu(\cprod{\unitplz}{\fdot{\vectLam}})
  +\epsvb\ethvr(\cprod{\vectOme}{\fdot{\vectLam}})
  +\epsvb\vphii(\cprod{\vectLam}{\fdot{\vectLam}})\\
  &\quad-3\epsvh\ethvo(\epsvg\unitplz-\epsvi\vectOme)
  -\epsvb\ethvo(\vsigc\unitplz-\vsiga\vectOme)
  -\vrhot\ethvo(\Omerep^2\unitplz-\epsvc\vectOme)
  -\vrhou\ethvo(\epsvb\unitplz-\epsvf\vectOme)\\
  &\quad-3\epsvh\vphig\vphih(\Lamrep^2\unitplz-\epsvi\vectLam)
  -\epsvb\vphig\vphih(\vsigd\unitplz-\vsiga\vectLam)
  -\vrhot\vphig\vphih(\epsvg\unitplz-\epsvc\vectLam)
  -\vrhou\vphig\vphih(\epsvh\unitplz-\epsvf\vectLam)\\
  &\quad+2\epsvb\ethvr(\Omerep^2\vectLam-\epsvg\vectOme)
  -3\Omerep^2\ethvr(\epsvb\vectLam-\epsvg\vectr)
  -3\epsvg\ethvr(\epsvb\vectOme-\Omerep^2\vectr)
  +\ethvr(\epsvb\ffdot{\vectLam}-\vsigh\vectr)\\
  &\quad+2\epsvb\vphii(\epsvg\vectLam-\Lamrep^2\vectOme)
  -3\Omerep^2\vphii(\epsvh\vectLam-\Lamrep^2\vectr)
  -3\epsvg\vphii(\epsvh\vectOme-\epsvg\vectr)
  +\vphii(\epsvh\ffdot{\vectLam}-\vsigi\vectr)\\
  &\quad-2\epsvb\ethvu(\epsvc\vectLam-\epsvi\vectOme)
  +3\Omerep^2\ethvu(\epsvf\vectLam-\epsvi\vectr)
  +3\epsvg\ethvu(\epsvf\vectOme-\epsvc\vectr)
  -\ethvu(\epsvf\ffdot{\vectLam}-\vsigf\vectr)\\
  &\quad-2\epsvb\ethvo\frkto\vectOme
  -3\Omerep^2\ethvo(\epsvl\vectLam-\frkto\vectr)
  -3\epsvg\ethvo\epsvl\vectOme
  +\ethvo(\epsvl\ffdot{\vectLam}-\frktq\vectr)
  -2\epsvb\vphig\vphih\frkto\vectLam\\
  &\quad-3\Omerep^2\vphig\vphih\epsvo\vectLam
  -3\epsvg\vphig\vphih(\epsvo\vectOme+\frkto\vectr)
  +\vphig\vphih(\epsvo\ffdot{\vectLam}-\frkts\vectr)
   \beqref{rot1a}, \eqnref{rpath1a}, \eqnref{rpath1b}\text{ \& }\eqnref{alg4}
\end{split}
\end{align*}
\begin{align*}
\begin{split}
&=(3\epsvh\ethvr-\vrhot\vphii)(\cprod{\vectOme}{\vectLam})
  +\vrhou\ethvr(\cprod{\vectOme}{\vectr})
  +\vrhou\vphii(\cprod{\vectLam}{\vectr})
  -\vrhot\ethvu(\cprod{\unitplz}{\vectOme})\\
  &\quad-\vrhou\ethvu(\cprod{\unitplz}{\vectr})
  -3\epsvh\ethvu(\cprod{\unitplz}{\vectLam})
  -\epsvb\ethvu(\cprod{\unitplz}{\fdot{\vectLam}})
  +\epsvb\ethvr(\cprod{\vectOme}{\fdot{\vectLam}})
  +\epsvb\vphii(\cprod{\vectLam}{\fdot{\vectLam}})\\
  &\quad-3\epsvh\ethvo\epsvg\unitplz+3\epsvh\ethvo\epsvi\vectOme
  -\epsvb\ethvo\vsigc\unitplz+\epsvb\ethvo\vsiga\vectOme
  -\vrhot\ethvo\Omerep^2\unitplz+\vrhot\ethvo\epsvc\vectOme
  -\vrhou\ethvo\epsvb\unitplz+\vrhou\ethvo\epsvf\vectOme\\
  &\quad-3\epsvh\vphig\vphih\Lamrep^2\unitplz+3\epsvh\vphig\vphih\epsvi\vectLam
  -\epsvb\vphig\vphih\vsigd\unitplz+\epsvb\vphig\vphih\vsiga\vectLam
  -\vrhot\vphig\vphih\epsvg\unitplz+\vrhot\vphig\vphih\epsvc\vectLam\\
  &\quad-\vrhou\vphig\vphih\epsvh\unitplz+\vrhou\vphig\vphih\epsvf\vectLam
  +2\epsvb\ethvr\Omerep^2\vectLam-2\epsvb\ethvr\epsvg\vectOme
  -3\Omerep^2\ethvr\epsvb\vectLam+3\Omerep^2\ethvr\epsvg\vectr\\
  &\quad-3\epsvg\ethvr\epsvb\vectOme+3\epsvg\ethvr\Omerep^2\vectr
  +\ethvr\epsvb\ffdot{\vectLam}-\ethvr\vsigh\vectr
  +2\epsvb\vphii\epsvg\vectLam-2\epsvb\vphii\Lamrep^2\vectOme
  -3\Omerep^2\vphii\epsvh\vectLam+3\Omerep^2\vphii\Lamrep^2\vectr\\
  &\quad-3\epsvg\vphii\epsvh\vectOme+3\epsvg\vphii\epsvg\vectr
  +\vphii\epsvh\ffdot{\vectLam}-\vphii\vsigi\vectr
  -2\epsvb\ethvu\epsvc\vectLam+2\epsvb\ethvu\epsvi\vectOme
  +3\Omerep^2\ethvu\epsvf\vectLam-3\Omerep^2\ethvu\epsvi\vectr\\
  &\quad+3\epsvg\ethvu\epsvf\vectOme-3\epsvg\ethvu\epsvc\vectr
  -\ethvu\epsvf\ffdot{\vectLam}+\ethvu\vsigf\vectr
  -2\epsvb\ethvo\frkto\vectOme
  -3\Omerep^2\ethvo\epsvl\vectLam+3\Omerep^2\ethvo\frkto\vectr\\
  &\quad-3\epsvg\ethvo\epsvl\vectOme
  +\ethvo\epsvl\ffdot{\vectLam}-\ethvo\frktq\vectr
  -2\epsvb\vphig\vphih\frkto\vectLam
  -3\Omerep^2\vphig\vphih\epsvo\vectLam\\
  &\quad-3\epsvg\vphig\vphih\epsvo\vectOme-3\epsvg\vphig\vphih\frkto\vectr
  +\vphig\vphih\epsvo\ffdot{\vectLam}-\vphig\vphih\frkts\vectr
\end{split}
\end{align*}
\begin{align*}
\begin{split}
&=(3\epsvh\ethvr-\vrhot\vphii)(\cprod{\vectOme}{\vectLam})
  +\vrhou\ethvr(\cprod{\vectOme}{\vectr})
  +\vrhou\vphii(\cprod{\vectLam}{\vectr})
  -\vrhot\ethvu(\cprod{\unitplz}{\vectOme})\\
  &\quad-\vrhou\ethvu(\cprod{\unitplz}{\vectr})
  -3\epsvh\ethvu(\cprod{\unitplz}{\vectLam})
  -\epsvb\ethvu(\cprod{\unitplz}{\fdot{\vectLam}})
  +\epsvb\ethvr(\cprod{\vectOme}{\fdot{\vectLam}})
  +\epsvb\vphii(\cprod{\vectLam}{\fdot{\vectLam}})\\
  &\quad+[-3\epsvh\ethvo\epsvg-\epsvb\ethvo\vsigc-\vrhot\ethvo\Omerep^2
    -\vrhou\ethvo\epsvb-3\epsvh\vphig\vphih\Lamrep^2-\epsvb\vphig\vphih\vsigd
  -\vrhot\vphig\vphih\epsvg-\vrhou\vphig\vphih\epsvh]\unitplz\\
  &\quad+[3\Omerep^2\ethvr\epsvg+3\epsvg\ethvr\Omerep^2-\ethvr\vsigh
    +3\Omerep^2\vphii\Lamrep^2+3\epsvg\vphii\epsvg-\vphii\vsigi
    -3\Omerep^2\ethvu\epsvi-3\epsvg\ethvu\epsvc\\
    &\qquad+\ethvu\vsigf+3\Omerep^2\ethvo\frkto-\ethvo\frktq-3\epsvg\vphig\vphih\frkto
    -\vphig\vphih\frkts]\vectr
  +[\ethvr\epsvb+\vphii\epsvh-\ethvu\epsvf+\ethvo\epsvl+\vphig\vphih\epsvo]\ffdot{\vectLam}\\
  &\quad+[3\epsvh\ethvo\epsvi+\epsvb\ethvo\vsiga+\vrhot\ethvo\epsvc
    +\vrhou\ethvo\epsvf-2\epsvb\ethvr\epsvg-3\epsvg\ethvr\epsvb
    -2\epsvb\vphii\Lamrep^2-3\epsvg\vphii\epsvh\\
    &\qquad+2\epsvb\ethvu\epsvi+3\epsvg\ethvu\epsvf-2\epsvb\ethvo\frkto-3\epsvg\ethvo\epsvl
    -3\epsvg\vphig\vphih\epsvo]\vectOme\\
  &\quad+[3\epsvh\vphig\vphih\epsvi+\epsvb\vphig\vphih\vsiga+\vrhot\vphig\vphih\epsvc
    +\vrhou\vphig\vphih\epsvf+2\epsvb\ethvr\Omerep^2-3\Omerep^2\ethvr\epsvb+2\epsvb\vphii\epsvg\\
    &\qquad-3\Omerep^2\vphii\epsvh-2\epsvb\ethvu\epsvc
    +3\Omerep^2\ethvu\epsvf-3\Omerep^2\ethvo\epsvl-2\epsvb\vphig\vphih\frkto
    -3\Omerep^2\vphig\vphih\epsvo]\vectLam
\end{split}
\end{align*}
\begin{align*}
\begin{split}
&=(3\epsvh\ethvr-\vrhot\vphii)(\cprod{\vectOme}{\vectLam})
  +\vrhou\ethvr(\cprod{\vectOme}{\vectr})
  +\vrhou\vphii(\cprod{\vectLam}{\vectr})
  -\vrhot\ethvu(\cprod{\unitplz}{\vectOme})\\
  &\quad-\vrhou\ethvu(\cprod{\unitplz}{\vectr})
  -3\epsvh\ethvu(\cprod{\unitplz}{\vectLam})
  -\epsvb\ethvu(\cprod{\unitplz}{\fdot{\vectLam}})
  +\epsvb\ethvr(\cprod{\vectOme}{\fdot{\vectLam}})
  +\epsvb\vphii(\cprod{\vectLam}{\fdot{\vectLam}})\\
  &\quad+[-\ethvo(3\epsvh\epsvg+\epsvb\vsigc+\vrhot\Omerep^2+\vrhou\epsvb)
    +\vphig\vphih(-3\epsvh\Lamrep^2-\epsvb\vsigd-\vrhot\epsvg-\vrhou\epsvh)]\unitplz\\
  &\quad+[3\Omerep^2(2\ethvr\epsvg+\vphii\Lamrep^2-\ethvu\epsvi+\ethvo\frkto)
    +\vphii(3\epsvg^2-\vsigi)+\ethvu(\vsigf-3\epsvg\epsvc)\\
    &\qquad-\vphig\vphih(3\epsvg\frkto+\frkts)-\ethvo\frktq-\ethvr\vsigh]\vectr
  +[\ethvr\epsvb+\vphii\epsvh-\ethvu\epsvf+\ethvo\epsvl+\vphig\vphih\epsvo]\ffdot{\vectLam}\\
  &\quad+[\ethvo(3\epsvh\epsvi+\epsvb\vsiga+\vrhot\epsvc+\vrhou\epsvf)
    +2\epsvb(\ethvu\epsvi-\ethvo\frkto-\ethvr\epsvg-\vphii\Lamrep^2)\\
    &\qquad+3\epsvg(\ethvu\epsvf-\ethvo\epsvl-\vphig\vphih\epsvo-\epsvb\ethvr-\vphii\epsvh)]\vectOme\\
  &\quad+[\vphig\vphih(3\epsvh\epsvi+\epsvb\vsiga+\vrhot\epsvc+\vrhou\epsvf)
    -3\Omerep^2(\ethvr\epsvb+\vphii\epsvh-\ethvu\epsvf+\ethvo\epsvl+\vphig\vphih\epsvo)\\
    &\qquad+2\epsvb(\vphii\epsvg+\ethvr\Omerep^2-\ethvu\epsvc-\vphig\vphih\frkto)]\vectLam
\end{split}
\end{align*}
\begin{align}\label{rpath21n}
\begin{split}
&=\vakpm(\cprod{\vectOme}{\vectLam})
  +\vrhou\ethvr(\cprod{\vectOme}{\vectr})
  +\vrhou\vphii(\cprod{\vectLam}{\vectr})
  -\vrhot\ethvu(\cprod{\unitplz}{\vectOme})\\
  &\quad-\vrhou\ethvu(\cprod{\unitplz}{\vectr})
  -3\epsvh\ethvu(\cprod{\unitplz}{\vectLam})
  -\epsvb\ethvu(\cprod{\unitplz}{\fdot{\vectLam}})
  +\epsvb\ethvr(\cprod{\vectOme}{\fdot{\vectLam}})
  +\epsvb\vphii(\cprod{\vectLam}{\fdot{\vectLam}})\\
  &\quad+\vakpo\unitplz+\vakpp\vectr+\vakpq\vectOme
  +\vakpr\vectLam+\vakpn\ffdot{\vectLam}\beqref{rpath1o}
\end{split}
\end{align}
\begin{align*}
\vscro
&=\cprod{\fdote}{\ffdote}\beqref{kpath2b}\nonumber\\
\begin{split}
&=\cprod{[\ethvo(\cprod{\unitplz}{\vectOme})+\vphig\vphih(\cprod{\unitplz}{\vectLam})
  +\ethvr\vectOme+\vphii\vectLam-\ethvu\unitplz]}{[\ethvp(\cprod{\unitplz}{\vectOme})+2\ethvo(\cprod{\unitplz}{\vectLam})}\\
  &\qquad+\vphig\vphih(\cprod{\unitplz}{\fdot{\vectLam}})+\ethvs\vectOme+2\ethvr\vectLam
  +\vphii\fdot{\vectLam}-\ethvv\unitplz]\beqref{rpath17}
\end{split}
\end{align*}
\begin{align*}
\begin{split}
&=2\ethvo^2[\cprod{(\cprod{\unitplz}{\vectOme})}{(\cprod{\unitplz}{\vectLam})}]
  +\vphig\vphih\ethvo[\cprod{(\cprod{\unitplz}{\vectOme})}{(\cprod{\unitplz}{\fdot{\vectLam}})}]
  -\ethvs\ethvo[\cprod{\vectOme}{(\cprod{\unitplz}{\vectOme})}]
  -2\ethvr\ethvo[\cprod{\vectLam}{(\cprod{\unitplz}{\vectOme})}]\\
  &\quad-\vphii\ethvo[\cprod{\fdot{\vectLam}}{(\cprod{\unitplz}{\vectOme})}]
  +\ethvv\ethvo[\cprod{\unitplz}{(\cprod{\unitplz}{\vectOme})}]
  +\ethvp\vphig\vphih[\cprod{(\cprod{\unitplz}{\vectLam})}{(\cprod{\unitplz}{\vectOme})}]
  +\vphig^2\vphih^2[\cprod{(\cprod{\unitplz}{\vectLam})}{(\cprod{\unitplz}{\fdot{\vectLam}})}]\\
  &\quad-\ethvs\vphig\vphih[\cprod{\vectOme}{(\cprod{\unitplz}{\vectLam})}]
  -2\ethvr\vphig\vphih[\cprod{\vectLam}{(\cprod{\unitplz}{\vectLam})}]
  -\vphii\vphig\vphih[\cprod{\fdot{\vectLam}}{(\cprod{\unitplz}{\vectLam})}]
  +\ethvv\vphig\vphih[\cprod{\unitplz}{(\cprod{\unitplz}{\vectLam})}]\\
  &\quad+\ethvp\ethvr[\cprod{\vectOme}{(\cprod{\unitplz}{\vectOme})}]
  +2\ethvo\ethvr[\cprod{\vectOme}{(\cprod{\unitplz}{\vectLam})}]
  +\vphig\vphih\ethvr[\cprod{\vectOme}{(\cprod{\unitplz}{\fdot{\vectLam}})}]
  +2\ethvr^2(\cprod{\vectOme}{\vectLam})
  +\vphii\ethvr(\cprod{\vectOme}{\fdot{\vectLam}})\\
  &\quad-\ethvv\ethvr(\cprod{\vectOme}{\unitplz})
  +\ethvp\vphii[\cprod{\vectLam}{(\cprod{\unitplz}{\vectOme})}]
  +2\ethvo\vphii[\cprod{\vectLam}{(\cprod{\unitplz}{\vectLam})}]
  +\vphig\vphih\vphii[\cprod{\vectLam}{(\cprod{\unitplz}{\fdot{\vectLam}})}]\\
  &\quad+\ethvs\vphii(\cprod{\vectLam}{\vectOme})
  +\vphii^2(\cprod{\vectLam}{\fdot{\vectLam}})
  -\ethvv\vphii(\cprod{\vectLam}{\unitplz})
  -\ethvp\ethvu[\cprod{\unitplz}{(\cprod{\unitplz}{\vectOme})}]
  -2\ethvo\ethvu[\cprod{\unitplz}{(\cprod{\unitplz}{\vectLam})}]\\
  &\quad-\vphig\vphih\ethvu[\cprod{\unitplz}{(\cprod{\unitplz}{\fdot{\vectLam}})}]
  -\ethvs\ethvu(\cprod{\unitplz}{\vectOme})
  -2\ethvr\ethvu(\cprod{\unitplz}{\vectLam})
  -\vphii\ethvu(\cprod{\unitplz}{\fdot{\vectLam}})
\end{split}
\end{align*}
\begin{align*}
\begin{split}
&=(2\ethvr^2-\ethvs\vphii)(\cprod{\vectOme}{\vectLam})
  +\vphii\ethvr(\cprod{\vectOme}{\fdot{\vectLam}})
  +(\ethvs\ethvu-\ethvv\ethvr)(\cprod{\vectOme}{\unitplz})
  +(\ethvv\vphii-2\ethvr\ethvu)(\cprod{\unitplz}{\vectLam})\\
  &\quad-\vphii\ethvu(\cprod{\unitplz}{\fdot{\vectLam}})
  +\vphii^2(\cprod{\vectLam}{\fdot{\vectLam}})
  +(\ethvp\ethvr-\ethvs\ethvo)[\cprod{\vectOme}{(\cprod{\unitplz}{\vectOme})}]
  +(\ethvp\vphii-2\ethvr\ethvo)[\cprod{\vectLam}{(\cprod{\unitplz}{\vectOme})}]\\
  &\quad-\vphii\ethvo[\cprod{\fdot{\vectLam}}{(\cprod{\unitplz}{\vectOme})}]
  +(\ethvv\ethvo-\ethvp\ethvu)[\cprod{\unitplz}{(\cprod{\unitplz}{\vectOme})}]
  +(2\ethvo\ethvr-\ethvs\vphig\vphih)[\cprod{\vectOme}{(\cprod{\unitplz}{\vectLam})}]\\
  &\quad+2(\ethvo\vphii-\ethvr\vphig\vphih)[\cprod{\vectLam}{(\cprod{\unitplz}{\vectLam})}]
  -\vphii\vphig\vphih[\cprod{\fdot{\vectLam}}{(\cprod{\unitplz}{\vectLam})}]
  +(\ethvv\vphig\vphih-2\ethvo\ethvu)[\cprod{\unitplz}{(\cprod{\unitplz}{\vectLam})}]\\
  &\quad+\vphig\vphih\ethvr[\cprod{\vectOme}{(\cprod{\unitplz}{\fdot{\vectLam}})}]
  +\vphig\vphih\vphii[\cprod{\vectLam}{(\cprod{\unitplz}{\fdot{\vectLam}})}]
  -\vphig\vphih\ethvu[\cprod{\unitplz}{(\cprod{\unitplz}{\fdot{\vectLam}})}]\\
  &\quad+2\ethvo^2[\cprod{(\cprod{\unitplz}{\vectOme})}{(\cprod{\unitplz}{\vectLam})}]
  +\vphig\vphih\ethvo[\cprod{(\cprod{\unitplz}{\vectOme})}{(\cprod{\unitplz}{\fdot{\vectLam}})}]
  +\ethvp\vphig\vphih[\cprod{(\cprod{\unitplz}{\vectLam})}{(\cprod{\unitplz}{\vectOme})}]\\
  &\quad+\vphig^2\vphih^2[\cprod{(\cprod{\unitplz}{\vectLam})}{(\cprod{\unitplz}{\fdot{\vectLam}})}]
\end{split}
\end{align*}
\begin{align*}
\begin{split}
&=(2\ethvr^2-\ethvs\vphii)(\cprod{\vectOme}{\vectLam})
  +\vphii\ethvr(\cprod{\vectOme}{\fdot{\vectLam}})
  +(\ethvs\ethvu-\ethvv\ethvr)(\cprod{\vectOme}{\unitplz})
  +(\ethvv\vphii-2\ethvr\ethvu)(\cprod{\unitplz}{\vectLam})
  -\vphii\ethvu(\cprod{\unitplz}{\fdot{\vectLam}})\\
  &\quad+\vphii^2(\cprod{\vectLam}{\fdot{\vectLam}})
  +(\ethvp\ethvr-\ethvs\ethvo)[\Omerep^2\unitplz-\vectOme(\dprod{\vectOme}{\unitplz})]
  +(\ethvp\vphii-2\ethvr\ethvo)[\unitplz(\dprod{\vectLam}{\vectOme})-\vectOme(\dprod{\unitplz}{\vectLam})]\\
  &\quad-\vphii\ethvo[\unitplz(\dprod{\fdot{\vectLam}}{\vectOme})-\vectOme(\dprod{\fdot{\vectLam}}{\unitplz})]
  +(\ethvv\ethvo-\ethvp\ethvu)[\unitplz(\dprod{\unitplz}{\vectOme})-\vectOme]
  +(2\ethvo\ethvr-\ethvs\vphig\vphih)[\unitplz(\dprod{\vectOme}{\vectLam})-\vectLam(\dprod{\vectOme}{\unitplz})]\\
  &\quad+2(\ethvo\vphii-\ethvr\vphig\vphih)[\Lamrep^2\unitplz-\vectLam(\dprod{\vectLam}{\unitplz})]
  -\vphii\vphig\vphih[\unitplz(\dprod{\fdot{\vectLam}}{\vectLam})-\vectLam(\dprod{\fdot{\vectLam}}{\unitplz})]\\
  &\quad+(\ethvv\vphig\vphih-2\ethvo\ethvu)[\unitplz(\dprod{\unitplz}{\vectLam})-\vectLam]
  +\vphig\vphih\ethvr[\unitplz(\dprod{\vectOme}{\fdot{\vectLam}})-\fdot{\vectLam}(\dprod{\vectOme}{\unitplz})]
  +\vphig\vphih\vphii[\unitplz(\dprod{\vectLam}{\fdot{\vectLam}})-\fdot{\vectLam}(\dprod{\vectLam}{\unitplz})]\\
  &\quad-\vphig\vphih\ethvu[\unitplz(\dprod{\unitplz}{\fdot{\vectLam}})-\fdot{\vectLam}]
  +2\ethvo^2[\unitplz(\dprod{\vectLam}{(\cprod{\unitplz}{\vectOme})})
    -\vectLam(\dprod{\unitplz}{(\cprod{\unitplz}{\vectOme})})]\\
  &\quad+\vphig\vphih\ethvo[\unitplz(\dprod{\fdot{\vectLam}}{(\cprod{\unitplz}{\vectOme})})
    -\fdot{\vectLam}(\dprod{\unitplz}{(\cprod{\unitplz}{\vectOme})})]
  +\ethvp\vphig\vphih[\unitplz(\dprod{\vectOme}{(\cprod{\unitplz}{\vectLam})})
    -\vectOme(\dprod{\unitplz}{(\cprod{\unitplz}{\vectLam})})]\\
  &\quad+\vphig^2\vphih^2[\unitplz(\dprod{\fdot{\vectLam}}{(\cprod{\unitplz}{\vectLam})})
    -\fdot{\vectLam}(\dprod{\unitplz}{(\cprod{\unitplz}{\vectLam})})]
   \beqref{alg1}\text{ \& }\eqnref{alg5}
\end{split}
\end{align*}
\begin{align*}
\begin{split}
&=(2\ethvr^2-\ethvs\vphii)(\cprod{\vectOme}{\vectLam})
  +\vphii\ethvr(\cprod{\vectOme}{\fdot{\vectLam}})
  +(\ethvs\ethvu-\ethvv\ethvr)(\cprod{\vectOme}{\unitplz})
  +(\ethvv\vphii-2\ethvr\ethvu)(\cprod{\unitplz}{\vectLam})\\
  &\quad-\vphii\ethvu(\cprod{\unitplz}{\fdot{\vectLam}})
  +\vphii^2(\cprod{\vectLam}{\fdot{\vectLam}})
  +(\ethvp\ethvr-\ethvs\ethvo)(\Omerep^2\unitplz-\epsvc\vectOme)
  +(\ethvp\vphii-2\ethvr\ethvo)(\epsvg\unitplz-\epsvi\vectOme)\\
  &\quad-\vphii\ethvo(\vsigc\unitplz-\vsiga\vectOme)
  +(\ethvv\ethvo-\ethvp\ethvu)(\epsvc\unitplz-\vectOme)
  +(2\ethvo\ethvr-\ethvs\vphig\vphih)(\epsvg\unitplz-\epsvc\vectLam)\\
  &\quad+2(\ethvo\vphii-\ethvr\vphig\vphih)(\Lamrep^2\unitplz-\epsvi\vectLam)
  -\vphii\vphig\vphih(\vsigd\unitplz-\vsiga\vectLam)
  +(\ethvv\vphig\vphih-2\ethvo\ethvu)(\epsvi\unitplz-\vectLam)\\
  &\quad+\vphig\vphih\ethvr(\vsigc\unitplz-\epsvc\fdot{\vectLam})
  +\vphig\vphih\vphii(\vsigd\unitplz-\epsvi\fdot{\vectLam})
  -\vphig\vphih\ethvu(\vsiga\unitplz-\fdot{\vectLam})
  +2\ethvo^2\frkto\unitplz
  +\vphig\vphih\ethvo\frktp\unitplz\\
  &\quad-\ethvp\vphig\vphih\frkto\unitplz+\vphig^2\vphih^2\frktr\unitplz
  \beqref{rot1a}, \eqnref{rpath1a}\text{ \& }\eqnref{rpath1b}
\end{split}
\end{align*}
\begin{align*}
\begin{split}
&=(2\ethvr^2-\ethvs\vphii)(\cprod{\vectOme}{\vectLam})
  +\vphii\ethvr(\cprod{\vectOme}{\fdot{\vectLam}})
  +(\ethvs\ethvu-\ethvv\ethvr)(\cprod{\vectOme}{\unitplz})
  +(\ethvv\vphii-2\ethvr\ethvu)(\cprod{\unitplz}{\vectLam})\\
  &\quad-\vphii\ethvu(\cprod{\unitplz}{\fdot{\vectLam}})
  +\vphii^2(\cprod{\vectLam}{\fdot{\vectLam}})
  +\Omerep^2(\ethvp\ethvr-\ethvs\ethvo)\unitplz-\epsvc(\ethvp\ethvr-\ethvs\ethvo)\vectOme
  +\epsvg(\ethvp\vphii-2\ethvr\ethvo)\unitplz\\
  &\quad-\epsvi(\ethvp\vphii-2\ethvr\ethvo)\vectOme
  -\vsigc\vphii\ethvo\unitplz+\vsiga\vphii\ethvo\vectOme
  +\epsvc(\ethvv\ethvo-\ethvp\ethvu)\unitplz-(\ethvv\ethvo-\ethvp\ethvu)\vectOme\\
  &\quad+\epsvg(2\ethvo\ethvr-\ethvs\vphig\vphih)\unitplz-\epsvc(2\ethvo\ethvr-\ethvs\vphig\vphih)\vectLam
  +2\Lamrep^2(\ethvo\vphii-\ethvr\vphig\vphih)\unitplz-2\epsvi(\ethvo\vphii-\ethvr\vphig\vphih)\vectLam\\
  &\quad-\vsigd\vphii\vphig\vphih\unitplz+\vsiga\vphii\vphig\vphih\vectLam
  +\epsvi(\ethvv\vphig\vphih-2\ethvo\ethvu)\unitplz-(\ethvv\vphig\vphih-2\ethvo\ethvu)\vectLam\\
  &\quad+\vphig\vphih\ethvr(\vsigc\unitplz-\epsvc\fdot{\vectLam})
  +\vsigd\vphig\vphih\vphii\unitplz-\epsvi\vphig\vphih\vphii\fdot{\vectLam}
  -\vsiga\vphig\vphih\ethvu\unitplz+\vphig\vphih\ethvu\fdot{\vectLam}\\
  &\quad+2\ethvo^2\frkto\unitplz+\vphig\vphih\ethvo\frktp\unitplz
  -\ethvp\vphig\vphih\frkto\unitplz+\vphig^2\vphih^2\frktr\unitplz
\end{split}
\end{align*}
\begin{align}\label{rpath21o}
\begin{split}
&=(2\ethvr^2-\ethvs\vphii)(\cprod{\vectOme}{\vectLam})
  +\vphii\ethvr(\cprod{\vectOme}{\fdot{\vectLam}})
  +(\ethvs\ethvu-\ethvv\ethvr)(\cprod{\vectOme}{\unitplz})
  +(\ethvv\vphii-2\ethvr\ethvu)(\cprod{\unitplz}{\vectLam})\\
  &\quad-\vphii\ethvu(\cprod{\unitplz}{\fdot{\vectLam}})
  +\vphii^2(\cprod{\vectLam}{\fdot{\vectLam}})
  +[\Omerep^2(\ethvp\ethvr-\ethvs\ethvo)+\epsvg(\ethvp\vphii-2\ethvr\ethvo)-\vsigc\vphii\ethvo\\
    &\quad+\epsvc(\ethvv\ethvo-\ethvp\ethvu)+\epsvg(2\ethvo\ethvr-\ethvs\vphig\vphih)
    +2\Lamrep^2(\ethvo\vphii-\ethvr\vphig\vphih)-\vsigd\vphii\vphig\vphih\\
    &\qquad+\epsvi(\ethvv\vphig\vphih-2\ethvo\ethvu)
    +\vsigd\vphig\vphih\vphii-\vsiga\vphig\vphih\ethvu+\vsigc\vphig\vphih\ethvr
    +2\ethvo^2\frkto+\vphig\vphih\ethvo\frktp\\
    &\qquad-\ethvp\vphig\vphih\frkto+\vphig^2\vphih^2\frktr]\unitplz
  +[\vphig\vphih\ethvu-\vphig\vphih\epsvc\ethvr-\vphig\vphih\epsvi\vphii]\fdot{\vectLam}\\
  &\quad+[\vsiga\vphii\vphig\vphih-\epsvc(2\ethvo\ethvr-\ethvs\vphig\vphih)
    -2\epsvi(\ethvo\vphii-\ethvr\vphig\vphih)-(\ethvv\vphig\vphih-2\ethvo\ethvu)]\vectLam\\
  &\quad+[\vsiga\vphii\ethvo-\epsvc(\ethvp\ethvr-\ethvs\ethvo)-\epsvi(\ethvp\vphii-2\ethvr\ethvo)
    -(\ethvv\ethvo-\ethvp\ethvu)]\vectOme
\end{split}
\nonumber\\
\begin{split}
&=\vakps(\cprod{\vectOme}{\vectLam})
  +\vphii\ethvr(\cprod{\vectOme}{\fdot{\vectLam}})
  +\vakpt(\cprod{\vectOme}{\unitplz})
  +\vakpu(\cprod{\unitplz}{\vectLam})
  -\vphii\ethvu(\cprod{\unitplz}{\fdot{\vectLam}})
  +\vphii^2(\cprod{\vectLam}{\fdot{\vectLam}})\\
  &\quad+\vakpw\unitplz+\vphig\vphih\vakpv\fdot{\vectLam}+\vakpx\vectLam
  +\vakpy\vectOme\beqref{rpath1o}.
\end{split}
\end{align}
\end{subequations}

\subart{Development of equation \eqnref{kpath2c}}
Having the foregoing derivations in view, we have that
\begin{subequations}\label{rpath22}
\begin{align*}
\vscrp
&=\vbbd\vscra+\vbbe\vscrb+\rhorep\fdot{\cdkt}\vscrc-\ffdot{\cdkt}\vscrd+\fdot{\cdkt}\vscre
  +\vbbf\vscrf+\rhorep\vbba\vscrg\beqref{kpath2c}
\nonumber\\
\begin{split}
&=\vbbd[\epsvb(\cprod{\unitkap}{\vectOme})-\Omerep^2(\cprod{\unitkap}{\vectr})
  +\epsvd\vectLam-\dltva\vectr]\\
  &\quad+\vbbe[2\epsvh(\cprod{\unitkap}{\vectOme})-3\epsvg(\cprod{\unitkap}{\vectr})+\epsvb(\cprod{\unitkap}{\vectLam})
  +\epsvd\fdot{\vectLam}-\epsvd\Omerep^2\vectOme+\frkya\vectr]\\
  &\quad+\rhorep\fdot{\cdkt}[3\epsvh(\cprod{\unitkap}{\vectLam})+\epsvb(\cprod{\unitkap}{\fdot{\vectLam}})
  +\vrhot(\cprod{\unitkap}{\vectOme})+\vrhou(\cprod{\unitkap}{\vectr})+\frkyb\vectLam-\frkyc\vectOme
     +\frkyd\vectr+\epsvd\ffdot{\vectLam}]\\
  &\quad-\ffdot{\cdkt}[\ethvr(\cprod{\unitkap}{\vectOme})+\vphii(\cprod{\unitkap}{\vectLam})-\ethvu(\cprod{\unitkap}{\unitplz})
  +\frkye\unitplz-\epsve\ethvo\vectOme-\epsve\vphig\vphih\vectLam]\\
  &\quad+\fdot{\cdkt}[\ethvs(\cprod{\unitkap}{\vectOme})+2\ethvr(\cprod{\unitkap}{\vectLam})
  +\vphii(\cprod{\unitkap}{\fdot{\vectLam}})-\ethvv(\cprod{\unitkap}{\unitplz})
    +\frkyf\unitplz-\ethvp\epsve\vectOme-2\ethvo\epsve\vectLam-\vphig\vphih\epsve\fdot{\vectLam}]\\
  &\quad+\vbbf[\frkyg(\cprod{\vectOme}{\vectr})+\epsvb^2(\cprod{\vectOme}{\vectLam})-\epsvb\Omerep^2(\cprod{\vectr}{\vectLam})
  -\vphia^2\fdot{\vectLam}+\frkyh\vectr+\Omerep^2\vphia^2\vectOme+3\vphic\vectLam]\\
  &\quad+\rhorep\vbba[3\epsvh\epsvb(\cprod{\vectOme}{\vectLam})+\epsvb^2(\cprod{\vectOme}{\fdot{\vectLam}})
  +\frkyi(\cprod{\vectOme}{\vectr})-3\epsvh\Omerep^2(\cprod{\vectr}{\vectLam})
    -\epsvb\Omerep^2(\cprod{\vectr}{\fdot{\vectLam}})+\frkyj\vectLam+\frkyl\vectr\\
    &\qquad+\frkyk\vectOme-\vphia^2\ffdot{\vectLam}]\beqref{rpath21}
\end{split}
\end{align*}
\begin{align}\label{rpath22a}
\begin{split}
&=(\vbbd\epsvb+2\vbbe\epsvh+\rhorep\fdot{\cdkt}\vrhot-\ffdot{\cdkt}\ethvr+\fdot{\cdkt}\ethvs)(\cprod{\unitkap}{\vectOme})
  +(-\vbbd\Omerep^2-3\vbbe\epsvg+\rhorep\fdot{\cdkt}\vrhou)(\cprod{\unitkap}{\vectr})\\
  &\quad+(\vbbe\epsvb+3\rhorep\fdot{\cdkt}\epsvh-\ffdot{\cdkt}\vphii+2\fdot{\cdkt}\ethvr)(\cprod{\unitkap}{\vectLam})
  +(\rhorep\fdot{\cdkt}\epsvb+\fdot{\cdkt}\vphii)(\cprod{\unitkap}{\fdot{\vectLam}})
  +(\ffdot{\cdkt}\ethvu-\fdot{\cdkt}\ethvv)(\cprod{\unitkap}{\unitplz})\\
  &\quad+(\vbbf\frkyg+\rhorep\vbba\frkyi)(\cprod{\vectOme}{\vectr})
  +(\vbbf\epsvb^2+3\rhorep\vbba\epsvh\epsvb)(\cprod{\vectOme}{\vectLam})
  +\rhorep\vbba\epsvb^2(\cprod{\vectOme}{\fdot{\vectLam}})
  -\rhorep\vbba\epsvb\Omerep^2(\cprod{\vectr}{\fdot{\vectLam}})\\
  &\quad-(\vbbf\epsvb\Omerep^2+3\rhorep\vbba\epsvh\Omerep^2)(\cprod{\vectr}{\vectLam})
  +(\vbbd\epsvd+\rhorep\fdot{\cdkt}\frkyb+\ffdot{\cdkt}\epsve\vphig\vphih-2\fdot{\cdkt}\ethvo\epsve
     +3\vbbf\vphic+\rhorep\vbba\frkyj)\vectLam\\
  &\quad+(-\vbbd\dltva+\vbbe\frkya+\rhorep\fdot{\cdkt}\frkyd+\vbbf\frkyh+\rhorep\vbba\frkyl)\vectr
  +(\vbbe\epsvd-\fdot{\cdkt}\vphig\vphih\epsve-\vbbf\vphia^2)\fdot{\vectLam}
  +(\fdot{\cdkt}\frkyf-\ffdot{\cdkt}\frkye)\unitplz\\
  &\quad+(-\vbbe\epsvd\Omerep^2-\rhorep\fdot{\cdkt}\frkyc+\ffdot{\cdkt}\epsve\ethvo-\fdot{\cdkt}\ethvp\epsve
     +\vbbf\Omerep^2\vphia^2+\rhorep\vbba\frkyk)\vectOme
  +(\rhorep\fdot{\cdkt}\epsvd-\rhorep\vbba\vphia^2)\ffdot{\vectLam}
\end{split}
\end{align}
\begin{align*}
\vscrq
&=-\ffdot{\rhorep}\vscrh+\vbba\vscri+\rhorep^2\vscrj-\vbbb\vscrk+\rhorep\vscrl
  +\rhorep\vscrn+\vscro\beqref{kpath2c}\nonumber\\
\begin{split}
&=-\ffdot{\rhorep}[\vphii\epsvb(\cprod{\vectOme}{\vectLam})-\ethvu\epsvb(\cprod{\vectOme}{\unitplz})
    -\ethvr\Omerep^2(\cprod{\vectr}{\vectOme})-\vphii\Omerep^2(\cprod{\vectr}{\vectLam})\\
    &\qquad+\ethvu\Omerep^2(\cprod{\vectr}{\unitplz})+\ethvo\frkym\vectOme+\frkyn\vectr+\frkyo\unitplz+\frkyp\vectLam]\\
  &\quad+\vbba[2\ethvr\epsvb(\cprod{\vectOme}{\vectLam})+\vphii\epsvb(\cprod{\vectOme}{\fdot{\vectLam}})
    -\ethvv\epsvb(\cprod{\vectOme}{\unitplz})-\ethvs\Omerep^2(\cprod{\vectr}{\vectOme})-2\ethvr\Omerep^2(\cprod{\vectr}{\vectLam})\\
    &\qquad-\vphii\Omerep^2(\cprod{\vectr}{\fdot{\vectLam}})+\ethvv\Omerep^2(\cprod{\vectr}{\unitplz})+\ethvp\frkym\vectOme
    +\vphig\vphih\frkym\fdot{\vectLam}+\frkyq\vectr+\frkyr\unitplz+\frkys\vectLam]\\
  &\quad+\rhorep^2[\frkyt(\cprod{\vectOme}{\vectLam})+2\epsvb\epsvh(\cprod{\vectOme}{\fdot{\vectLam}})
    +\frkyu(\cprod{\vectOme}{\vectr})+\frkyv(\cprod{\vectLam}{\vectr})-3\epsvb\epsvg(\cprod{\vectr}{\fdot{\vectLam}})\\
    &\qquad+\epsvb^2(\cprod{\vectLam}{\fdot{\vectLam}})+\frkyw\vectLam-\frkyx\fdot{\vectLam}
    -3\vphic\ffdot{\vectLam}+\frkyy\vectOme+\frkyz\vectr]\\
  &\quad-\vbbb[\vakpa(\cprod{\vectOme}{\vectLam})-2\ethvu\epsvh(\cprod{\vectOme}{\unitplz})-3\ethvr\epsvg(\cprod{\vectr}{\vectOme})
    -3\vphii\epsvg(\cprod{\vectr}{\vectLam})+3\ethvu\epsvg(\cprod{\vectr}{\unitplz})\\
    &\qquad-\ethvu\epsvb(\cprod{\vectLam}{\unitplz})+\vakpb\vectr+\vakpe\unitplz+\vakpf\vectOme
    +\vphig\vphih\vakpd\vectLam+\vakpc\fdot{\vectLam}]\\
  &\quad+\rhorep[\vakpg(\cprod{\vectOme}{\vectLam})-2\ethvv\epsvh(\cprod{\vectOme}{\unitplz})-3\ethvs\epsvg(\cprod{\vectr}{\vectOme})
    -6\ethvr\epsvg(\cprod{\vectr}{\vectLam})+3\ethvv\epsvg(\cprod{\vectr}{\unitplz})-\ethvv\epsvb(\cprod{\vectLam}{\unitplz})\\
    &\qquad+2\vphii\epsvh(\cprod{\vectOme}{\fdot{\vectLam}})-3\vphii\epsvg(\cprod{\vectr}{\fdot{\vectLam}})
    +\vphii\epsvb(\cprod{\vectLam}{\fdot{\vectLam}})+\vakpi\unitplz+2\ethvo\vakph\vectLam+\vakpj\vectOme
    +\vakpk\fdot{\vectLam}+\vakpl\vectr]\\
  &\quad+\rhorep[\vakpm(\cprod{\vectOme}{\vectLam})+\vrhou\ethvr(\cprod{\vectOme}{\vectr})+\vrhou\vphii(\cprod{\vectLam}{\vectr})
    -\vrhot\ethvu(\cprod{\unitplz}{\vectOme})-\vrhou\ethvu(\cprod{\unitplz}{\vectr})-3\epsvh\ethvu(\cprod{\unitplz}{\vectLam})\\
    &\qquad-\epsvb\ethvu(\cprod{\unitplz}{\fdot{\vectLam}})
    +\epsvb\ethvr(\cprod{\vectOme}{\fdot{\vectLam}})+\epsvb\vphii(\cprod{\vectLam}{\fdot{\vectLam}})
    +\vakpo\unitplz+\vakpp\vectr+\vakpq\vectOme+\vakpr\vectLam+\vakpn\ffdot{\vectLam}]\\
  &\quad+[\vakps(\cprod{\vectOme}{\vectLam})+\vphii\ethvr(\cprod{\vectOme}{\fdot{\vectLam}})
    +\vakpt(\cprod{\vectOme}{\unitplz})+\vakpu(\cprod{\unitplz}{\vectLam})-\vphii\ethvu(\cprod{\unitplz}{\fdot{\vectLam}})
    +\vphii^2(\cprod{\vectLam}{\fdot{\vectLam}})\\
    &\qquad+\vakpw\unitplz+\vphig\vphih\vakpv\fdot{\vectLam}+\vakpx\vectLam+\vakpy\vectOme]
\end{split}
\end{align*}
\begin{align}\label{rpath22b}
\begin{split}
&=(-\ffdot{\rhorep}\ethvr\Omerep^2+\vbba\ethvs\Omerep^2+\rhorep^2\frkyu-3\vbbb\ethvr\epsvg
      +3\rhorep\ethvs\epsvg+\rhorep\vrhou\ethvr)(\cprod{\vectOme}{\vectr})\\
  &\quad+(-\ffdot{\rhorep}\vphii\epsvb+2\vbba\ethvr\epsvb+\rhorep^2\frkyt-\vbbb\vakpa
      +\rhorep\vakpg+\rhorep\vakpm+\vakps)(\cprod{\vectOme}{\vectLam})\\
  &\quad+(\vbba\vphii\epsvb+2\rhorep^2\epsvb\epsvh+2\rhorep\vphii\epsvh+\rhorep\epsvb\ethvr
      +\vphii\ethvr)(\cprod{\vectOme}{\fdot{\vectLam}})\\
  &\quad+(\ffdot{\rhorep}\ethvu\epsvb-\vbba\ethvv\epsvb+2\vbbb\ethvu\epsvh-2\rhorep\ethvv\epsvh
      +\rhorep\vrhot\ethvu+\vakpt)(\cprod{\vectOme}{\unitplz})\\
  &\quad+(-\vbba\vphii\Omerep^2-3\rhorep^2\epsvb\epsvg-3\rhorep\vphii\epsvg)(\cprod{\vectr}{\fdot{\vectLam}})\\
  &\quad+(\ffdot{\rhorep}\vphii\Omerep^2-2\vbba\ethvr\Omerep^2-\rhorep^2\frkyv+3\vbbb\vphii\epsvg
      -6\rhorep\ethvr\epsvg-\rhorep\vrhou\vphii)(\cprod{\vectr}{\vectLam})\\
  &\quad+(-\ffdot{\rhorep}\ethvu\Omerep^2+\vbba\ethvv\Omerep^2-3\vbbb\ethvu\epsvg+3\rhorep\ethvv\epsvg
      +\rhorep\vrhou\ethvu)(\cprod{\vectr}{\unitplz})
  -(\rhorep\epsvb\ethvu+\vphii\ethvu)(\cprod{\unitplz}{\fdot{\vectLam}})\\
  &\quad+(-\vbbb\ethvu\epsvb+\rhorep\ethvv\epsvb-3\rhorep\epsvh\ethvu+\vakpu)(\cprod{\unitplz}{\vectLam})
  +(\rhorep^2\epsvb^2+\rhorep\vphii\epsvb+\rhorep\epsvb\vphii+\vphii^2)(\cprod{\vectLam}{\fdot{\vectLam}})\\
  &\quad+(-\ffdot{\rhorep}\frkyp+\vbba\frkys+\rhorep^2\frkyw-\vbbb\vphig\vphih\vakpd+2\rhorep\ethvo\vakph
      +\rhorep\vakpr+\vakpx)\vectLam\\
  &\quad+(-\ffdot{\rhorep}\frkyn+\vbba\frkyq+\rhorep^2\frkyz-\vbbb\vakpb+\rhorep\vakpl+\rhorep\vakpp)\vectr
  +(-\ffdot{\rhorep}\frkyo+\vbba\frkyr-\vbbb\vakpe+\rhorep\vakpi+\rhorep\vakpo+\vakpw)\unitplz\\
  &\quad+(-\ffdot{\rhorep}\ethvo\frkym+\vbba\ethvp\frkym+\rhorep^2\frkyy-\vbbb\vakpf+\rhorep\vakpj+\rhorep\vakpq+\vakpy)\vectOme\\
  &\quad+(\vbba\vphig\vphih\frkym-\rhorep^2\frkyx-\vbbb\vakpc+\rhorep\vakpk+\vphig\vphih\vakpv)\fdot{\vectLam}
  +(-3\rhorep^2\vphic+\rhorep\vakpn)\ffdot{\vectLam}
\end{split}
\end{align}
\begin{align*}
&\vscrp+\vscrq\nonumber\\
\begin{split}
&=(\vbbd\epsvb+2\vbbe\epsvh+\rhorep\fdot{\cdkt}\vrhot-\ffdot{\cdkt}\ethvr+\fdot{\cdkt}\ethvs)(\cprod{\unitkap}{\vectOme})
  +(-\vbbd\Omerep^2-3\vbbe\epsvg+\rhorep\fdot{\cdkt}\vrhou)(\cprod{\unitkap}{\vectr})\\
  &\quad+(\vbbe\epsvb+3\rhorep\fdot{\cdkt}\epsvh-\ffdot{\cdkt}\vphii+2\fdot{\cdkt}\ethvr)(\cprod{\unitkap}{\vectLam})
  +(\rhorep\fdot{\cdkt}\epsvb+\fdot{\cdkt}\vphii)(\cprod{\unitkap}{\fdot{\vectLam}})
  +(\ffdot{\cdkt}\ethvu-\fdot{\cdkt}\ethvv)(\cprod{\unitkap}{\unitplz})\\
  &\quad+(\vbbf\frkyg+\rhorep\vbba\frkyi)(\cprod{\vectOme}{\vectr})
  +(\vbbf\epsvb^2+3\rhorep\vbba\epsvh\epsvb)(\cprod{\vectOme}{\vectLam})
  +\rhorep\vbba\epsvb^2(\cprod{\vectOme}{\fdot{\vectLam}})
  -\rhorep\vbba\epsvb\Omerep^2(\cprod{\vectr}{\fdot{\vectLam}})\\
  &\quad-(\vbbf\epsvb\Omerep^2+3\rhorep\vbba\epsvh\Omerep^2)(\cprod{\vectr}{\vectLam})
  +(\vbbd\epsvd+\rhorep\fdot{\cdkt}\frkyb+\ffdot{\cdkt}\epsve\vphig\vphih-2\fdot{\cdkt}\ethvo\epsve
     +3\vbbf\vphic+\rhorep\vbba\frkyj)\vectLam\\
  &\quad+(-\vbbd\dltva+\vbbe\frkya+\rhorep\fdot{\cdkt}\frkyd+\vbbf\frkyh+\rhorep\vbba\frkyl)\vectr
  +(\vbbe\epsvd-\fdot{\cdkt}\vphig\vphih\epsve-\vbbf\vphia^2)\fdot{\vectLam}
  +(\fdot{\cdkt}\frkyf-\ffdot{\cdkt}\frkye)\unitplz\\
  &\quad+(-\vbbe\epsvd\Omerep^2-\rhorep\fdot{\cdkt}\frkyc+\ffdot{\cdkt}\epsve\ethvo-\fdot{\cdkt}\ethvp\epsve
     +\vbbf\Omerep^2\vphia^2+\rhorep\vbba\frkyk)\vectOme
  +(\rhorep\fdot{\cdkt}\epsvd-\rhorep\vbba\vphia^2)\ffdot{\vectLam}\\
  &\quad+(-\ffdot{\rhorep}\ethvr\Omerep^2+\vbba\ethvs\Omerep^2+\rhorep^2\frkyu-3\vbbb\ethvr\epsvg
  +3\rhorep\ethvs\epsvg+\rhorep\vrhou\ethvr)(\cprod{\vectOme}{\vectr})\\
  &\quad+(-\ffdot{\rhorep}\vphii\epsvb+2\vbba\ethvr\epsvb+\rhorep^2\frkyt-\vbbb\vakpa
      +\rhorep\vakpg+\rhorep\vakpm+\vakps)(\cprod{\vectOme}{\vectLam})+\cdots
\end{split}
\end{align*}
\begin{align*}
\begin{split}
  \cdots&\quad+(\vbba\vphii\epsvb+2\rhorep^2\epsvb\epsvh+2\rhorep\vphii\epsvh+\rhorep\epsvb\ethvr
      +\vphii\ethvr)(\cprod{\vectOme}{\fdot{\vectLam}})\\
  &\quad+(\ffdot{\rhorep}\ethvu\epsvb-\vbba\ethvv\epsvb+2\vbbb\ethvu\epsvh-2\rhorep\ethvv\epsvh
      +\rhorep\vrhot\ethvu+\vakpt)(\cprod{\vectOme}{\unitplz})\\
  &\quad+(-\vbba\vphii\Omerep^2-3\rhorep^2\epsvb\epsvg-3\rhorep\vphii\epsvg)(\cprod{\vectr}{\fdot{\vectLam}})\\
  &\quad+(\ffdot{\rhorep}\vphii\Omerep^2-2\vbba\ethvr\Omerep^2-\rhorep^2\frkyv+3\vbbb\vphii\epsvg
      -6\rhorep\ethvr\epsvg-\rhorep\vrhou\vphii)(\cprod{\vectr}{\vectLam})\\
  &\quad+(-\ffdot{\rhorep}\ethvu\Omerep^2+\vbba\ethvv\Omerep^2-3\vbbb\ethvu\epsvg+3\rhorep\ethvv\epsvg
      +\rhorep\vrhou\ethvu)(\cprod{\vectr}{\unitplz})
  -(\rhorep\epsvb\ethvu+\vphii\ethvu)(\cprod{\unitplz}{\fdot{\vectLam}})\\
  &\quad+(-\vbbb\ethvu\epsvb+\rhorep\ethvv\epsvb-3\rhorep\epsvh\ethvu+\vakpu)(\cprod{\unitplz}{\vectLam})
  +(\rhorep^2\epsvb^2+\rhorep\vphii\epsvb+\rhorep\epsvb\vphii+\vphii^2)(\cprod{\vectLam}{\fdot{\vectLam}})\\
  &\quad+(-\ffdot{\rhorep}\frkyp+\vbba\frkys+\rhorep^2\frkyw-\vbbb\vphig\vphih\vakpd+2\rhorep\ethvo\vakph
      +\rhorep\vakpr+\vakpx)\vectLam\\
  &\quad+(-\ffdot{\rhorep}\frkyn+\vbba\frkyq+\rhorep^2\frkyz-\vbbb\vakpb+\rhorep\vakpl+\rhorep\vakpp)\vectr
  +(-\ffdot{\rhorep}\frkyo+\vbba\frkyr-\vbbb\vakpe+\rhorep\vakpi+\rhorep\vakpo+\vakpw)\unitplz\\
  &\quad+(-\ffdot{\rhorep}\ethvo\frkym+\vbba\ethvp\frkym+\rhorep^2\frkyy-\vbbb\vakpf+\rhorep\vakpj+\rhorep\vakpq+\vakpy)\vectOme\\
  &\quad+(\vbba\vphig\vphih\frkym-\rhorep^2\frkyx-\vbbb\vakpc+\rhorep\vakpk+\vphig\vphih\vakpv)\fdot{\vectLam}
  +(-3\rhorep^2\vphic+\rhorep\vakpn)\ffdot{\vectLam}
  \beqref{rpath22a}\text{ \& }\eqnref{rpath22b}
\end{split}
\end{align*}
\begin{align*}
\begin{split}
&=(\vbbd\epsvb+2\vbbe\epsvh+\rhorep\fdot{\cdkt}\vrhot-\ffdot{\cdkt}\ethvr+\fdot{\cdkt}\ethvs)(\cprod{\unitkap}{\vectOme})
  +(-\vbbd\Omerep^2-3\vbbe\epsvg+\rhorep\fdot{\cdkt}\vrhou)(\cprod{\unitkap}{\vectr})\\
  &\quad+(\vbbe\epsvb+3\rhorep\fdot{\cdkt}\epsvh-\ffdot{\cdkt}\vphii+2\fdot{\cdkt}\ethvr)(\cprod{\unitkap}{\vectLam})
  +(\rhorep\fdot{\cdkt}\epsvb+\fdot{\cdkt}\vphii)(\cprod{\unitkap}{\fdot{\vectLam}})
  +(\ffdot{\cdkt}\ethvu-\fdot{\cdkt}\ethvv)(\cprod{\unitkap}{\unitplz})\\
  &\quad+(\vbbf\frkyg+\rhorep\vbba\frkyi-\ffdot{\rhorep}\ethvr\Omerep^2+\vbba\ethvs\Omerep^2+\rhorep^2\frkyu
      -3\vbbb\ethvr\epsvg+3\rhorep\ethvs\epsvg+\rhorep\vrhou\ethvr)(\cprod{\vectOme}{\vectr})\\
  &\quad+(\vbbf\epsvb^2+3\rhorep\vbba\epsvh\epsvb-\ffdot{\rhorep}\vphii\epsvb+2\vbba\ethvr\epsvb+\rhorep^2\frkyt
      -\vbbb\vakpa+\rhorep\vakpg+\rhorep\vakpm+\vakps)(\cprod{\vectOme}{\vectLam})\\
  &\quad+(\rhorep\vbba\epsvb^2+\vbba\vphii\epsvb+2\rhorep^2\epsvb\epsvh+2\rhorep\vphii\epsvh+\rhorep\epsvb\ethvr
      +\vphii\ethvr)(\cprod{\vectOme}{\fdot{\vectLam}})\\
  &\quad+(\ffdot{\rhorep}\ethvu\epsvb-\vbba\ethvv\epsvb+2\vbbb\ethvu\epsvh-2\rhorep\ethvv\epsvh
      +\rhorep\vrhot\ethvu+\vakpt)(\cprod{\vectOme}{\unitplz})\\
  &\quad+(-\rhorep\vbba\epsvb\Omerep^2-\vbba\vphii\Omerep^2-3\rhorep^2\epsvb\epsvg
      -3\rhorep\vphii\epsvg)(\cprod{\vectr}{\fdot{\vectLam}})\\
  &\quad+(-\vbbf\epsvb\Omerep^2-3\rhorep\vbba\epsvh\Omerep^2+\ffdot{\rhorep}\vphii\Omerep^2-2\vbba\ethvr\Omerep^2
      -\rhorep^2\frkyv+3\vbbb\vphii\epsvg-6\rhorep\ethvr\epsvg-\rhorep\vrhou\vphii)(\cprod{\vectr}{\vectLam})\\
  &\quad+(-\ffdot{\rhorep}\ethvu\Omerep^2+\vbba\ethvv\Omerep^2-3\vbbb\ethvu\epsvg+3\rhorep\ethvv\epsvg
      +\rhorep\vrhou\ethvu)(\cprod{\vectr}{\unitplz})
  -(\rhorep\epsvb\ethvu+\vphii\ethvu)(\cprod{\unitplz}{\fdot{\vectLam}})\\
  &\quad+(-\vbbb\ethvu\epsvb+\rhorep\ethvv\epsvb-3\rhorep\epsvh\ethvu+\vakpu)(\cprod{\unitplz}{\vectLam})
  +(\rhorep^2\epsvb^2+\rhorep\vphii\epsvb+\rhorep\epsvb\vphii+\vphii^2)(\cprod{\vectLam}{\fdot{\vectLam}})\\
  &\quad+(\vbbd\epsvd+\rhorep\fdot{\cdkt}\frkyb+\ffdot{\cdkt}\epsve\vphig\vphih-2\fdot{\cdkt}\ethvo\epsve
     +3\vbbf\vphic+\rhorep\vbba\frkyj-\ffdot{\rhorep}\frkyp+\vbba\frkys+\rhorep^2\frkyw\\
     &\qquad-\vbbb\vphig\vphih\vakpd+2\rhorep\ethvo\vakph+\rhorep\vakpr+\vakpx)\vectLam+\cdots
\end{split}
\end{align*}
\begin{align}\label{rpath22c}
\begin{split}
  \cdots&\quad+(-\vbbd\dltva+\vbbe\frkya+\rhorep\fdot{\cdkt}\frkyd+\vbbf\frkyh+\rhorep\vbba\frkyl
     -\ffdot{\rhorep}\frkyn+\vbba\frkyq+\rhorep^2\frkyz-\vbbb\vakpb+\rhorep\vakpl+\rhorep\vakpp)\vectr\\
  &\quad+(\fdot{\cdkt}\frkyf-\ffdot{\cdkt}\frkye-\ffdot{\rhorep}\frkyo+\vbba\frkyr-\vbbb\vakpe+\rhorep\vakpi
     +\rhorep\vakpo+\vakpw)\unitplz\\
  &\quad+(-\vbbe\epsvd\Omerep^2-\rhorep\fdot{\cdkt}\frkyc+\ffdot{\cdkt}\epsve\ethvo-\fdot{\cdkt}\ethvp\epsve
     +\vbbf\Omerep^2\vphia^2+\rhorep\vbba\frkyk-\ffdot{\rhorep}\ethvo\frkym+\vbba\ethvp\frkym\\
     &\qquad+\rhorep^2\frkyy-\vbbb\vakpf+\rhorep\vakpj+\rhorep\vakpq+\vakpy)\vectOme\\
  &\quad+(\vbbe\epsvd-\fdot{\cdkt}\vphig\vphih\epsve-\vbbf\vphia^2+\vbba\vphig\vphih\frkym-\rhorep^2\frkyx
     -\vbbb\vakpc+\rhorep\vakpk+\vphig\vphih\vakpv)\fdot{\vectLam}\\
  &\quad+(\rhorep\fdot{\cdkt}\epsvd-\rhorep\vbba\vphia^2-3\rhorep^2\vphic+\rhorep\vakpn)\ffdot{\vectLam}
\end{split}
\end{align}
\begin{align}\label{rpath22d}
\vscrr
&=\fdot{\cdkt}\unitkap+\vbba\vecta+\rhorep\fdota+\fdote\beqref{kpath2c}\nonumber\\
&=(\scalc\ethvf-2\ethvx)\unitkap+(\ethvi-1)\vecta+\rhorep\fdota+\fdote
  \beqref{rpath19a}\text{ \& }\eqnref{rpath20a}
\nonumber\\
\begin{split}
&=(\scalc\ethvf-2\ethvx)\unitkap
  +(\ethvi-1)[(\dprod{\vectOme}{\vectr})\vectOme-\Omerep^2\vectr+\cprod{\vectLam}{\vectr}]
  +\rhorep[2\epsvh\vectOme-3\epsvg\vectr+\cprod{\fdot{\vectLam}}{\vectr}-\Omerep^2(\cprod{\vectOme}{\vectr})\\
    &\quad+\epsvb\vectLam]
  +\ethvo(\cprod{\unitplz}{\vectOme})+\vphig\vphih(\cprod{\unitplz}{\vectLam})+\ethvr\vectOme+\vphii\vectLam-\ethvu\unitplz
  \beqref{main4}, \eqnref{rpath7a}\text{ \& }\eqnref{rpath17a}
\end{split}
\nonumber\\
\begin{split}
&=(\scalc\ethvf-2\ethvx)\unitkap
  +\epsvb(\ethvi-1)\vectOme-\Omerep^2(\ethvi-1)\vectr+(\ethvi-1)(\cprod{\vectLam}{\vectr})
  +2\rhorep\epsvh\vectOme-3\rhorep\epsvg\vectr+\rhorep(\cprod{\fdot{\vectLam}}{\vectr})\\
  &\quad-\rhorep\Omerep^2(\cprod{\vectOme}{\vectr})+\rhorep\epsvb\vectLam
  +\ethvo(\cprod{\unitplz}{\vectOme})+\vphig\vphih(\cprod{\unitplz}{\vectLam})+\ethvr\vectOme
  +\vphii\vectLam-\ethvu\unitplz\beqref{rot1a}
\end{split}
\nonumber\\
\begin{split}
&=(\ethvi-1)(\cprod{\vectLam}{\vectr})
  +\rhorep(\cprod{\fdot{\vectLam}}{\vectr})
  -\rhorep\Omerep^2(\cprod{\vectOme}{\vectr})
  +\ethvo(\cprod{\unitplz}{\vectOme})
  +\vphig\vphih(\cprod{\unitplz}{\vectLam})
  +(\scalc\ethvf-2\ethvx)\unitkap\\
  &\quad+[\epsvb(\ethvi-1)+2\rhorep\epsvh+\ethvr]\vectOme
  -[\Omerep^2(\ethvi-1)+3\rhorep\epsvg]\vectr
  +(\rhorep\epsvb+\vphii)\vectLam-\ethvu\unitplz
\end{split}
\nonumber\\
\begin{split}
&=\parvb\unitkap+\parvc\vectOme-\parvd\vectr
  +\parve\vectLam-\ethvu\unitplz+\parva(\cprod{\vectLam}{\vectr})
  +\rhorep(\cprod{\fdot{\vectLam}}{\vectr})\\
  &\quad-\rhorep\Omerep^2(\cprod{\vectOme}{\vectr})
  +\ethvo(\cprod{\unitplz}{\vectOme})
  +\vphig\vphih(\cprod{\unitplz}{\vectLam})
  \beqref{rpath1p}.
\end{split}
\end{align}
\end{subequations}
Equations \eqnref{rpath19a}, \eqnref{rpath20a} and \eqnref{rpath1p} together give
\begin{align}\label{rpath23}
\begin{split}
&\fdot{\cdkt}=\parvb,\quad\ffdot{\cdkt}=\parvf,\quad\fffdot{\cdkt}=\parvg\\
\vbba=\parva,\quad\vbbb=\parvh,&\quad\vbbc=\parvi,\quad\vbbd=\parvj,\quad\vbbe=\parvk,\quad\vbbf=\parvl
\end{split}
\end{align}
on account of which \eqnref{rpath22c} becomes
\begin{align*}
&\vscrp+\vscrq\nonumber\\
\begin{split}
&=(\parvj\epsvb+2\parvk\epsvh+\rhorep\parvb\vrhot-\parvf\ethvr+\parvb\ethvs)(\cprod{\unitkap}{\vectOme})
  +(-\parvj\Omerep^2-3\parvk\epsvg+\rhorep\parvb\vrhou)(\cprod{\unitkap}{\vectr})\\
  &\quad+(\parvk\epsvb+3\rhorep\parvb\epsvh-\parvf\vphii+2\parvb\ethvr)(\cprod{\unitkap}{\vectLam})
  +(\rhorep\parvb\epsvb+\parvb\vphii)(\cprod{\unitkap}{\fdot{\vectLam}})
  +(\parvf\ethvu-\parvb\ethvv)(\cprod{\unitkap}{\unitplz})\\
  &\quad+(\parvl\frkyg+\rhorep\parva\frkyi-\ethvj\ethvr\Omerep^2+\parva\ethvs\Omerep^2+\rhorep^2\frkyu
      -3\parvh\ethvr\epsvg+3\rhorep\ethvs\epsvg+\rhorep\vrhou\ethvr)(\cprod{\vectOme}{\vectr})\\
  &\quad+(\parvl\epsvb^2+3\rhorep\parva\epsvh\epsvb-\parvf\vphii\epsvb+2\parva\ethvr\epsvb+\rhorep^2\frkyt
      -\parvh\vakpa+\rhorep\vakpg+\rhorep\vakpm+\vakps)(\cprod{\vectOme}{\vectLam})\\
  &\quad+(\rhorep\parva\epsvb^2+\parva\vphii\epsvb+2\rhorep^2\epsvb\epsvh+2\rhorep\vphii\epsvh+\rhorep\epsvb\ethvr
      +\vphii\ethvr)(\cprod{\vectOme}{\fdot{\vectLam}})\\
  &\quad+(\parvf\ethvu\epsvb-\parva\ethvv\epsvb+2\parvh\ethvu\epsvh-2\rhorep\ethvv\epsvh
      +\rhorep\vrhot\ethvu+\vakpt)(\cprod{\vectOme}{\unitplz})\\
  &\quad+(-\rhorep\parva\epsvb\Omerep^2-\parva\vphii\Omerep^2-3\rhorep^2\epsvb\epsvg
      -3\rhorep\vphii\epsvg)(\cprod{\vectr}{\fdot{\vectLam}})\\
  &\quad+(-\parvl\epsvb\Omerep^2-3\rhorep\parva\epsvh\Omerep^2+\parvf\vphii\Omerep^2-2\parva\ethvr\Omerep^2
      -\rhorep^2\frkyv+3\parvh\vphii\epsvg-6\rhorep\ethvr\epsvg-\rhorep\vrhou\vphii)(\cprod{\vectr}{\vectLam})\\
  &\quad+(-\ethvj\ethvu\Omerep^2+\parva\ethvv\Omerep^2-3\parvh\ethvu\epsvg+3\rhorep\ethvv\epsvg
      +\rhorep\vrhou\ethvu)(\cprod{\vectr}{\unitplz})
  -(\rhorep\epsvb\ethvu+\vphii\ethvu)(\cprod{\unitplz}{\fdot{\vectLam}})\\
  &\quad+(-\parvh\ethvu\epsvb+\rhorep\ethvv\epsvb-3\rhorep\epsvh\ethvu+\vakpu)(\cprod{\unitplz}{\vectLam})
  +(\rhorep^2\epsvb^2+\rhorep\vphii\epsvb+\rhorep\epsvb\vphii+\vphii^2)(\cprod{\vectLam}{\fdot{\vectLam}})\\
  &\quad+(\parvj\epsvd+\rhorep\parvb\frkyb+\parvf\epsve\vphig\vphih-2\parvb\ethvo\epsve
     +3\parvl\vphic+\rhorep\parva\frkyj-\parvf\frkyp+\parva\frkys+\rhorep^2\frkyw\\
     &\qquad-\parvh\vphig\vphih\vakpd+2\rhorep\ethvo\vakph+\rhorep\vakpr+\vakpx)\vectLam\\
  &\quad+(-\parvj\dltva+\parvk\frkya+\rhorep\parvb\frkyd+\parvl\frkyh+\rhorep\parva\frkyl
     -\parvf\frkyn+\parva\frkyq+\rhorep^2\frkyz-\parvh\vakpb+\rhorep\vakpl+\rhorep\vakpp)\vectr\\
  &\quad+(\parvb\frkyf-\parvf\frkye-\parvf\frkyo+\parva\frkyr-\parvh\vakpe+\rhorep\vakpi
     +\rhorep\vakpo+\vakpw)\unitplz\\
  &\quad+(-\parvk\epsvd\Omerep^2-\rhorep\parvb\frkyc+\parvf\epsve\ethvo-\parvb\ethvp\epsve
     +\parvl\Omerep^2\vphia^2+\rhorep\parva\frkyk-\parvf\ethvo\frkym+\parva\ethvp\frkym\\
     &\qquad+\rhorep^2\frkyy-\parvh\vakpf+\rhorep\vakpj+\rhorep\vakpq+\vakpy)\vectOme\\
  &\quad+(\parvk\epsvd-\parvb\vphig\vphih\epsve-\parvl\vphia^2+\parva\vphig\vphih\frkym-\rhorep^2\frkyx
     -\parvh\vakpc+\rhorep\vakpk+\vphig\vphih\vakpv)\fdot{\vectLam}\\
  &\quad+(\rhorep\parvb\epsvd-\rhorep\parva\vphia^2-3\rhorep^2\vphic+\rhorep\vakpn)\ffdot{\vectLam}
\end{split}
\end{align*}
\begin{align*}
\begin{split}
&=[\parvj\epsvb+2\parvk\epsvh\vrhot-\parvf\ethvr+\parvb(\rhorep+\ethvs)](\cprod{\unitkap}{\vectOme})
  +(\rhorep\parvb\vrhou-\parvj\Omerep^2-3\parvk\epsvg)(\cprod{\unitkap}{\vectr})\\
  &\quad+[\parvk\epsvb-\parvf\vphii+\parvb(3\rhorep\epsvh+2\ethvr)](\cprod{\unitkap}{\vectLam})
  +\parvb(\rhorep\epsvb+\vphii)(\cprod{\unitkap}{\fdot{\vectLam}})
  +(\parvf\ethvu-\parvb\ethvv)(\cprod{\unitkap}{\unitplz})\\
  &\quad+[\parvl\frkyg+\parva(\rhorep\frkyi+\ethvs\Omerep^2)+\rhorep(\rhorep\frkyu+3\ethvs\epsvg)
      +\ethvr(\rhorep\vrhou-\ethvj\Omerep^2-3\parvh\epsvg)](\cprod{\vectOme}{\vectr})\\
  &\quad+[\vakps-\parvh\vakpa+\epsvb(\parvl\epsvb-\ethvj\vphii+\parva(3\rhorep\epsvh+2\ethvr))
      +\rhorep(\vakpg+\vakpm+\rhorep\frkyt)](\cprod{\vectOme}{\vectLam})\\
  &\quad+[\rhorep\epsvb(\ethvr+\parva\epsvb+2\rhorep\epsvh)
      +\vphii(\ethvr+\parva\epsvb+2\rhorep\epsvh)](\cprod{\vectOme}{\fdot{\vectLam}})\\
  &\quad+[\vakpt+\rhorep\vrhot\ethvu+\epsvb(\ethvj\ethvu-\parva\ethvv)
      +2\epsvh(\parvh\ethvu-\rhorep\ethvv)](\cprod{\vectOme}{\unitplz})\\
  &\quad-[\parva\Omerep^2(\rhorep\epsvb+\vphii)+3\rhorep\epsvg(\rhorep\epsvb
      +\vphii)](\cprod{\vectr}{\fdot{\vectLam}})\\
  &\quad+[\vphii(3\parvh\epsvg+\ethvj\Omerep^2)-\Omerep^2(\parvl\epsvb+\parva(3\rhorep\epsvh+2\ethvr))
      -\rhorep(\rhorep\frkyv+6\ethvr\epsvg+\vrhou\vphii)](\cprod{\vectr}{\vectLam})\\
  &\quad+[\rhorep\vrhou\ethvu+\Omerep^2(\parva\ethvv-\ethvj\ethvu)
      +3\epsvg(\rhorep\ethvv-\parvh\ethvu)](\cprod{\vectr}{\unitplz})
  -\ethvu(\rhorep\epsvb+\vphii)(\cprod{\unitplz}{\fdot{\vectLam}})\\
  &\quad+[\vakpu-\parvh\ethvu\epsvb+\rhorep(\ethvv\epsvb-3\epsvh\ethvu)](\cprod{\unitplz}{\vectLam})
  +[\rhorep\epsvb(\rhorep\epsvb+\vphii)+\vphii(\rhorep\epsvb+\vphii)](\cprod{\vectLam}{\fdot{\vectLam}})\\
  &\quad+[\vakpx-\ethvj\frkyp+3\parvl\vphic+\parvj\epsvd+\parvb(\rhorep\frkyb-2\ethvo\epsve)+\parva(\frkys+\rhorep\frkyj)\\
     &\qquad+\rhorep(\vakpr+\rhorep\frkyw+2\ethvo\vakph)+\vphig\vphih(\parvf\epsve-\parvh\vakpd)]\vectLam+\cdots
\end{split}
\end{align*}
\begin{align*}
\begin{split}
  \cdots&\quad+[\parvk\frkya-\parvj\dltva+\parvl\frkyh-\ethvj\frkyn+\parva\frkyq-\parvh\vakpb
     +\rhorep(\vakpl+\vakpp+\rhorep\frkyz+\parvb\frkyd+\parva\frkyl)]\vectr\\
  &\quad+[\vakpw-\ethvj\frkyo+\parva\frkyr+\parvb\frkyf-\parvf\frkye-\parvh\vakpe
     +\rhorep(\vakpi+\vakpo)]\unitplz\\
  &\quad+[\vakpy-\parvh\vakpf+\Omerep^2(\parvl\vphia^2-\parvk\epsvd)+\ethvo(\parvf\epsve-\ethvj\frkym)\\
     &\qquad+\rhorep(\parva\frkyk-\parvb\frkyc+\vakpj+\vakpq+\rhorep\frkyy)+\ethvp(\parva\frkym-\parvb\epsve)]\vectOme\\
  &\quad+[\parvk\epsvd-\parvl\vphia^2-\parvh\vakpc+\vphig\vphih(\vakpv-\parvb\epsve+\parva\frkym)
     +\rhorep(\vakpk-\rhorep\frkyx)]\fdot{\vectLam}\\
  &\quad+\rhorep(\parvb\epsvd-\parva\vphia^2-3\rhorep\vphic+\vakpn)\ffdot{\vectLam}
\end{split}
\end{align*}
\begin{align}\label{rpath24}
\begin{split}
&=\parvx\vectLam+\parvy\vectr+\parvz\unitplz+\efkoa\vectOme+\efkob\fdot{\vectLam}+\efkoc\ffdot{\vectLam}
  +\parvm(\cprod{\unitkap}{\vectOme})+\parvn(\cprod{\unitkap}{\vectr})+\parvo(\cprod{\unitkap}{\vectLam})\\
  &\quad+\parvb\parve(\cprod{\unitkap}{\fdot{\vectLam}})+\parvp(\cprod{\unitkap}{\unitplz})
  +\parvq(\cprod{\vectOme}{\vectr})+\parvr(\cprod{\vectOme}{\vectLam})
  +\parvs(\cprod{\vectOme}{\fdot{\vectLam}})+\parvt(\cprod{\vectOme}{\unitplz})\\
  &\quad-\parve\parvd(\cprod{\vectr}{\fdot{\vectLam}})
  +\parvu(\cprod{\vectr}{\vectLam})+\parvv(\cprod{\vectr}{\unitplz})
  -\ethvu\parve(\cprod{\unitplz}{\fdot{\vectLam}})
  +\parvw(\cprod{\unitplz}{\vectLam})+\parve^2(\cprod{\vectLam}{\fdot{\vectLam}})\\
  &\quad\beqref{rpath1q}, \eqnref{rpath1r}\text{ \& }\eqnref{rpathx1a}.
\end{split}
\end{align}
Consequently, we derive
\begin{subequations}\label{rpath25}
\begin{align*}
&\dprod{\vectLam}{(\vscrp+\vscrq)}\nonumber\\
\begin{split}
&=\dprod{\vectLam}{[}\parvx\vectLam+\parvy\vectr+\parvz\unitplz+\efkoa\vectOme+\efkob\fdot{\vectLam}+\efkoc\ffdot{\vectLam}
  +\parvm(\cprod{\unitkap}{\vectOme})+\parvn(\cprod{\unitkap}{\vectr})+\parvo(\cprod{\unitkap}{\vectLam})\\
  &\quad+\parvb\parve(\cprod{\unitkap}{\fdot{\vectLam}})+\parvp(\cprod{\unitkap}{\unitplz})
  +\parvq(\cprod{\vectOme}{\vectr})+\parvr(\cprod{\vectOme}{\vectLam})
  +\parvs(\cprod{\vectOme}{\fdot{\vectLam}})+\parvt(\cprod{\vectOme}{\unitplz})\\
  &\quad-\parve\parvd(\cprod{\vectr}{\fdot{\vectLam}})
  +\parvu(\cprod{\vectr}{\vectLam})+\parvv(\cprod{\vectr}{\unitplz})
  -\ethvu\parve(\cprod{\unitplz}{\fdot{\vectLam}})
  +\parvw(\cprod{\unitplz}{\vectLam})+\parve^2(\cprod{\vectLam}{\fdot{\vectLam}})]
  \beqref{rpath24}
\end{split}
\end{align*}
\begin{align*}
\begin{split}
&=\parvx(\dprod{\vectLam}{\vectLam})
  +\parvy(\dprod{\vectLam}{\vectr})
  +\parvz(\dprod{\vectLam}{\unitplz})
  +\efkoa(\dprod{\vectLam}{\vectOme})
  +\efkob(\dprod{\vectLam}{\fdot{\vectLam}})
  +\efkoc(\dprod{\vectLam}{\ffdot{\vectLam}})
  +\parvm[\dprod{\vectLam}{(\cprod{\unitkap}{\vectOme})}]\\
  &\quad+\parvn[\dprod{\vectLam}{(\cprod{\unitkap}{\vectr})}]
  +\parvo[\dprod{\vectLam}{(\cprod{\unitkap}{\vectLam})}]
  +\parvb\parve[\dprod{\vectLam}{(\cprod{\unitkap}{\fdot{\vectLam}})}]
  +\parvp[\dprod{\vectLam}{(\cprod{\unitkap}{\unitplz})}]
  +\parvq[\dprod{\vectLam}{(\cprod{\vectOme}{\vectr})}]\\
  &\quad+\parvr[\dprod{\vectLam}{(\cprod{\vectOme}{\vectLam})}]
  +\parvs[\dprod{\vectLam}{(\cprod{\vectOme}{\fdot{\vectLam}})}]
  +\parvt[\dprod{\vectLam}{(\cprod{\vectOme}{\unitplz})}]
  -\parve\parvd[\dprod{\vectLam}{(\cprod{\vectr}{\fdot{\vectLam}})}]
  +\parvu[\dprod{\vectLam}{(\cprod{\vectr}{\vectLam})}]\\
  &\quad+\parvv[\dprod{\vectLam}{(\cprod{\vectr}{\unitplz})}]
  -\ethvu\parve[\dprod{\vectLam}{(\cprod{\unitplz}{\fdot{\vectLam}})}]
  +\parvw[\dprod{\vectLam}{(\cprod{\unitplz}{\vectLam})}]
  +\parve^2[\dprod{\vectLam}{(\cprod{\vectLam}{\fdot{\vectLam}})}]
\end{split}
\end{align*}
\begin{align}\label{rpath25a}
\begin{split}
&=\parvx\Lamrep^2+\parvy\epsvh+\parvz\epsvi+\efkoa\epsvg+\efkob\vsigd+\efkoc\vsigi-\parvm\frkte
  -\parvn\epsvn-\parvb\parve\frktu+\parvp\dltvb-\parvq\epsvm\\
  &\quad-\parvs\frktv-\parvt\frkto-\parve\parvd\frkth+\parvv\epsvo+\ethvu\parve\frktr
  \beqref{rot1a}, \eqnref{rxpeed1a}, \eqnref{rpath1a}, \eqnref{rpath1b}\text{ \& }\eqnref{rpath1b2}
\end{split}
\nonumber\\
&=\efkod\beqref{rpathx1a}
\end{align}
\begin{align*}
&\dprod{\vectr}{(\vscrp+\vscrq)}\nonumber\\
\begin{split}
&=\dprod{\vectr}{[}\parvx\vectLam+\parvy\vectr+\parvz\unitplz+\efkoa\vectOme+\efkob\fdot{\vectLam}+\efkoc\ffdot{\vectLam}
  +\parvm(\cprod{\unitkap}{\vectOme})+\parvn(\cprod{\unitkap}{\vectr})+\parvo(\cprod{\unitkap}{\vectLam})\\
  &\quad+\parvb\parve(\cprod{\unitkap}{\fdot{\vectLam}})+\parvp(\cprod{\unitkap}{\unitplz})
  +\parvq(\cprod{\vectOme}{\vectr})+\parvr(\cprod{\vectOme}{\vectLam})
  +\parvs(\cprod{\vectOme}{\fdot{\vectLam}})+\parvt(\cprod{\vectOme}{\unitplz})\\
  &\quad-\parve\parvd(\cprod{\vectr}{\fdot{\vectLam}})
  +\parvu(\cprod{\vectr}{\vectLam})+\parvv(\cprod{\vectr}{\unitplz})
  -\ethvu\parve(\cprod{\unitplz}{\fdot{\vectLam}})
  +\parvw(\cprod{\unitplz}{\vectLam})+\parve^2(\cprod{\vectLam}{\fdot{\vectLam}})]
  \beqref{rpath24}
\end{split}
\end{align*}
\begin{align*}
\begin{split}
&=\parvx(\dprod{\vectr}{\vectLam})
  +\parvy(\dprod{\vectr}{\vectr})
  +\parvz(\dprod{\vectr}{\unitplz})
  +\efkoa(\dprod{\vectr}{\vectOme})
  +\efkob(\dprod{\vectr}{\fdot{\vectLam}})
  +\efkoc(\dprod{\vectr}{\ffdot{\vectLam}})
  +\parvm[\dprod{\vectr}{(\cprod{\unitkap}{\vectOme})}]\\
  &\quad+\parvn[\dprod{\vectr}{(\cprod{\unitkap}{\vectr})}]
  +\parvo[\dprod{\vectr}{(\cprod{\unitkap}{\vectLam})}]
  +\parvb\parve[\dprod{\vectr}{(\cprod{\unitkap}{\fdot{\vectLam}})}]
  +\parvp[\dprod{\vectr}{(\cprod{\unitkap}{\unitplz})}]
  +\parvq[\dprod{\vectr}{(\cprod{\vectOme}{\vectr})}]\\
  &\quad+\parvr[\dprod{\vectr}{(\cprod{\vectOme}{\vectLam})}]
  +\parvs[\dprod{\vectr}{(\cprod{\vectOme}{\fdot{\vectLam}})}]
  +\parvt[\dprod{\vectr}{(\cprod{\vectOme}{\unitplz})}]
  -\parve\parvd[\dprod{\vectr}{(\cprod{\vectr}{\fdot{\vectLam}})}]
  +\parvu[\dprod{\vectr}{(\cprod{\vectr}{\vectLam})}]\\
  &\quad+\parvv[\dprod{\vectr}{(\cprod{\vectr}{\unitplz})}]
  -\ethvu\parve[\dprod{\vectr}{(\cprod{\unitplz}{\fdot{\vectLam}})}]
  +\parvw[\dprod{\vectr}{(\cprod{\unitplz}{\vectLam})}]
  +\parve^2[\dprod{\vectr}{(\cprod{\vectLam}{\fdot{\vectLam}})}]
\end{split}
\end{align*}
\begin{align}\label{rpath25b}
\begin{split}
&=\parvx\epsvh+\parvy\scalr^2+\parvz\epsvf+\efkoa\epsvb+\efkob\vsign+\efkoc\vsigo+\parvm\epsvk
  +\parvo\epsvn+\parvb\parve\frktc-\parvp\frktt+\parvr\epsvm\\
  &\quad-\parvs\frktl-\parvt\epsvl-\ethvu\parve\frktn+\parvw\epsvo-\parve^2\frkth
  \beqref{rot1a}, \eqnref{rpath1a}, \eqnref{rpath1b}\text{ \& }\eqnref{rpath1b2}
\end{split}
\nonumber\\
&=\efkoe\beqref{rpathx1a}
\end{align}
\begin{align*}
&\dprod{\unitplz}{(\vscrp+\vscrq)}\nonumber\\
\begin{split}
&=\dprod{\unitplz}{[}\parvx\vectLam+\parvy\vectr+\parvz\unitplz+\efkoa\vectOme+\efkob\fdot{\vectLam}+\efkoc\ffdot{\vectLam}
  +\parvm(\cprod{\unitkap}{\vectOme})+\parvn(\cprod{\unitkap}{\vectr})+\parvo(\cprod{\unitkap}{\vectLam})\\
  &\quad+\parvb\parve(\cprod{\unitkap}{\fdot{\vectLam}})+\parvp(\cprod{\unitkap}{\unitplz})
  +\parvq(\cprod{\vectOme}{\vectr})+\parvr(\cprod{\vectOme}{\vectLam})
  +\parvs(\cprod{\vectOme}{\fdot{\vectLam}})+\parvt(\cprod{\vectOme}{\unitplz})\\
  &\quad-\parve\parvd(\cprod{\vectr}{\fdot{\vectLam}})
  +\parvu(\cprod{\vectr}{\vectLam})+\parvv(\cprod{\vectr}{\unitplz})
  -\ethvu\parve(\cprod{\unitplz}{\fdot{\vectLam}})
  +\parvw(\cprod{\unitplz}{\vectLam})+\parve^2(\cprod{\vectLam}{\fdot{\vectLam}})]
  \beqref{rpath24}
\end{split}
\end{align*}
\begin{align*}
\begin{split}
&=\parvx(\dprod{\unitplz}{\vectLam})
  +\parvy(\dprod{\unitplz}{\vectr})
  +\parvz(\dprod{\unitplz}{\unitplz})
  +\efkoa(\dprod{\unitplz}{\vectOme})
  +\efkob(\dprod{\unitplz}{\fdot{\vectLam}})
  +\efkoc(\dprod{\unitplz}{\ffdot{\vectLam}})
  +\parvm[\dprod{\unitplz}{(\cprod{\unitkap}{\vectOme})}]\\
  &\quad+\parvn[\dprod{\unitplz}{(\cprod{\unitkap}{\vectr})}]
  +\parvo[\dprod{\unitplz}{(\cprod{\unitkap}{\vectLam})}]
  +\parvb\parve[\dprod{\unitplz}{(\cprod{\unitkap}{\fdot{\vectLam}})}]
  +\parvp[\dprod{\unitplz}{(\cprod{\unitkap}{\unitplz})}]
  +\parvq[\dprod{\unitplz}{(\cprod{\vectOme}{\vectr})}]\\
  &\quad+\parvr[\dprod{\unitplz}{(\cprod{\vectOme}{\vectLam})}]
  +\parvs[\dprod{\unitplz}{(\cprod{\vectOme}{\fdot{\vectLam}})}]
  +\parvt[\dprod{\unitplz}{(\cprod{\vectOme}{\unitplz})}]
  -\parve\parvd[\dprod{\unitplz}{(\cprod{\vectr}{\fdot{\vectLam}})}]
  +\parvu[\dprod{\unitplz}{(\cprod{\vectr}{\vectLam})}]\\
  &\quad+\parvv[\dprod{\unitplz}{(\cprod{\vectr}{\unitplz})}]
  -\ethvu\parve[\dprod{\unitplz}{(\cprod{\unitplz}{\fdot{\vectLam}})}]
  +\parvw[\dprod{\unitplz}{(\cprod{\unitplz}{\vectLam})}]
  +\parve^2[\dprod{\unitplz}{(\cprod{\vectLam}{\fdot{\vectLam}})}]
\end{split}
\end{align*}
\begin{align}\label{rpath25c}
\begin{split}
&=\parvx\epsvi+\parvy\epsvf+\parvz+\efkoa\epsvc+\efkob\vsiga+\efkoc\vsigf-\parvm\epsvj
  +\parvn\frktt-\parvo\dltvb-\parvb\parve\frkta+\parvq\epsvl\\
  &\quad+\parvr\frkto+\parvs\frktp+\parve\parvd\frktn-\parvu\epsvo+\parve^2\frktr
  \beqref{rot1a}, \eqnref{rxpeed1a}, \eqnref{rpath1a}, \eqnref{rpath1b}\text{ \& }\eqnref{rpath1b2}
\end{split}
\nonumber\\
&=\efkof\beqref{rpathx1b}
\end{align}
\begin{align*}
&\dprod{\vectOme}{(\vscrp+\vscrq)}\nonumber\\
\begin{split}
&=\dprod{\vectOme}{[}\parvx\vectLam+\parvy\vectr+\parvz\unitplz+\efkoa\vectOme+\efkob\fdot{\vectLam}+\efkoc\ffdot{\vectLam}
  +\parvm(\cprod{\unitkap}{\vectOme})+\parvn(\cprod{\unitkap}{\vectr})+\parvo(\cprod{\unitkap}{\vectLam})\\
  &\quad+\parvb\parve(\cprod{\unitkap}{\fdot{\vectLam}})+\parvp(\cprod{\unitkap}{\unitplz})
  +\parvq(\cprod{\vectOme}{\vectr})+\parvr(\cprod{\vectOme}{\vectLam})
  +\parvs(\cprod{\vectOme}{\fdot{\vectLam}})+\parvt(\cprod{\vectOme}{\unitplz})\\
  &\quad-\parve\parvd(\cprod{\vectr}{\fdot{\vectLam}})
  +\parvu(\cprod{\vectr}{\vectLam})+\parvv(\cprod{\vectr}{\unitplz})
  -\ethvu\parve(\cprod{\unitplz}{\fdot{\vectLam}})
  +\parvw(\cprod{\unitplz}{\vectLam})+\parve^2(\cprod{\vectLam}{\fdot{\vectLam}})]
  \beqref{rpath24}
\end{split}
\end{align*}
\begin{align*}
\begin{split}
&=\parvx(\dprod{\vectOme}{\vectLam})
  +\parvy(\dprod{\vectOme}{\vectr})
  +\parvz(\dprod{\vectOme}{\unitplz})
  +\efkoa(\dprod{\vectOme}{\vectOme})
  +\efkob(\dprod{\vectOme}{\fdot{\vectLam}})
  +\efkoc(\dprod{\vectOme}{\ffdot{\vectLam}})
  +\parvm[\dprod{\vectOme}{(\cprod{\unitkap}{\vectOme})}]\\
  &\quad+\parvn[\dprod{\vectOme}{(\cprod{\unitkap}{\vectr})}]
  +\parvo[\dprod{\vectOme}{(\cprod{\unitkap}{\vectLam})}]
  +\parvb\parve[\dprod{\vectOme}{(\cprod{\unitkap}{\fdot{\vectLam}})}]
  +\parvp[\dprod{\vectOme}{(\cprod{\unitkap}{\unitplz})}]
  +\parvq[\dprod{\vectOme}{(\cprod{\vectOme}{\vectr})}]\\
  &\quad+\parvr[\dprod{\vectOme}{(\cprod{\vectOme}{\vectLam})}]
  +\parvs[\dprod{\vectOme}{(\cprod{\vectOme}{\fdot{\vectLam}})}]
  +\parvt[\dprod{\vectOme}{(\cprod{\vectOme}{\unitplz})}]
  -\parve\parvd[\dprod{\vectOme}{(\cprod{\vectr}{\fdot{\vectLam}})}]
  +\parvu[\dprod{\vectOme}{(\cprod{\vectr}{\vectLam})}]\\
  &\quad+\parvv[\dprod{\vectOme}{(\cprod{\vectr}{\unitplz})}]
  -\ethvu\parve[\dprod{\vectOme}{(\cprod{\unitplz}{\fdot{\vectLam}})}]
  +\parvw[\dprod{\vectOme}{(\cprod{\unitplz}{\vectLam})}]
  +\parve^2[\dprod{\vectOme}{(\cprod{\vectLam}{\fdot{\vectLam}})}]
\end{split}
\end{align*}
\begin{align}\label{rpath25d}
\begin{split}
&=\parvx\epsvg+\parvy\epsvb+\parvz\epsvc+\efkoa\Omerep^2+\efkob\vsigc+\efkoc\vsigh-\parvn\epsvk
  +\parvo\frkte-\parvb\parve\frktf+\parvp\epsvj-\parve\parvd\frktl\\
  &\quad-\parvu\epsvm+\parvv\epsvl+\ethvu\parve\frktp-\parvw\frkto+\parve^2\frktv
  \beqref{rot1a}, \eqnref{rpath1a}, \eqnref{rpath1b}\text{ \& }\eqnref{rpath1b2}
\end{split}
\nonumber\\
&=\efkog\beqref{rpathx1b}
\end{align}
\begin{align*}
&\dprod{\fdot{\vectLam}}{(\vscrp+\vscrq)}\nonumber\\
\begin{split}
&=\dprod{\fdot{\vectLam}}{[}\parvx\vectLam+\parvy\vectr+\parvz\unitplz+\efkoa\vectOme+\efkob\fdot{\vectLam}+\efkoc\ffdot{\vectLam}
  +\parvm(\cprod{\unitkap}{\vectOme})+\parvn(\cprod{\unitkap}{\vectr})+\parvo(\cprod{\unitkap}{\vectLam})\\
  &\quad+\parvb\parve(\cprod{\unitkap}{\fdot{\vectLam}})+\parvp(\cprod{\unitkap}{\unitplz})
  +\parvq(\cprod{\vectOme}{\vectr})+\parvr(\cprod{\vectOme}{\vectLam})
  +\parvs(\cprod{\vectOme}{\fdot{\vectLam}})+\parvt(\cprod{\vectOme}{\unitplz})\\
  &\quad-\parve\parvd(\cprod{\vectr}{\fdot{\vectLam}})
  +\parvu(\cprod{\vectr}{\vectLam})+\parvv(\cprod{\vectr}{\unitplz})
  -\ethvu\parve(\cprod{\unitplz}{\fdot{\vectLam}})
  +\parvw(\cprod{\unitplz}{\vectLam})+\parve^2(\cprod{\vectLam}{\fdot{\vectLam}})]
  \beqref{rpath24}
\end{split}
\end{align*}
\begin{align*}
\begin{split}
&=\parvx(\dprod{\fdot{\vectLam}}{\vectLam})
  +\parvy(\dprod{\fdot{\vectLam}}{\vectr})
  +\parvz(\dprod{\fdot{\vectLam}}{\unitplz})
  +\efkoa(\dprod{\fdot{\vectLam}}{\vectOme})
  +\efkob(\dprod{\fdot{\vectLam}}{\fdot{\vectLam}})
  +\efkoc(\dprod{\fdot{\vectLam}}{\ffdot{\vectLam}})
  +\parvm[\dprod{\fdot{\vectLam}}{(\cprod{\unitkap}{\vectOme})}]\\
  &\quad+\parvn[\dprod{\fdot{\vectLam}}{(\cprod{\unitkap}{\vectr})}]
  +\parvo[\dprod{\fdot{\vectLam}}{(\cprod{\unitkap}{\vectLam})}]
  +\parvb\parve[\dprod{\fdot{\vectLam}}{(\cprod{\unitkap}{\fdot{\vectLam}})}]
  +\parvp[\dprod{\fdot{\vectLam}}{(\cprod{\unitkap}{\unitplz})}]
  +\parvq[\dprod{\fdot{\vectLam}}{(\cprod{\vectOme}{\vectr})}]\\
  &\quad+\parvr[\dprod{\fdot{\vectLam}}{(\cprod{\vectOme}{\vectLam})}]
  +\parvs[\dprod{\fdot{\vectLam}}{(\cprod{\vectOme}{\fdot{\vectLam}})}]
  +\parvt[\dprod{\fdot{\vectLam}}{(\cprod{\vectOme}{\unitplz})}]
  -\parve\parvd[\dprod{\fdot{\vectLam}}{(\cprod{\vectr}{\fdot{\vectLam}})}]
  +\parvu[\dprod{\fdot{\vectLam}}{(\cprod{\vectr}{\vectLam})}]\\
  &\quad+\parvv[\dprod{\fdot{\vectLam}}{(\cprod{\vectr}{\unitplz})}]
  -\ethvu\parve[\dprod{\fdot{\vectLam}}{(\cprod{\unitplz}{\fdot{\vectLam}})}]
  +\parvw[\dprod{\fdot{\vectLam}}{(\cprod{\unitplz}{\vectLam})}]
  +\parve^2[\dprod{\fdot{\vectLam}}{(\cprod{\vectLam}{\fdot{\vectLam}})}]
\end{split}
\end{align*}
\begin{align}\label{rpath25e}
\begin{split}
&=\parvx\vsigd+\parvy\vsign+\parvz\vsiga+\efkoa\vsigc+\efkob\vsige+\efkoc\vsigj+\parvm\frktf
  -\parvn\frktc+\parvo\frktu+\parvp\frkta+\parvq\frktl\\
  &\quad+\parvr\frktv-\parvt\frktp-\parvu\frkth+\parvv\frktn+\parvw\frktr
  \beqref{rpath1a}, \eqnref{rpath1b}\text{ \& }\eqnref{rpath1b2}
\end{split}
\nonumber\\
&=\efkoh\beqref{rpathx1b}
\end{align}
\begin{align*}
&\dprod{\ffdot{\vectLam}}{(\vscrp+\vscrq)}\nonumber\\
\begin{split}
&=\dprod{\ffdot{\vectLam}}{[}\parvx\vectLam+\parvy\vectr+\parvz\unitplz+\efkoa\vectOme+\efkob\fdot{\vectLam}+\efkoc\ffdot{\vectLam}
  +\parvm(\cprod{\unitkap}{\vectOme})+\parvn(\cprod{\unitkap}{\vectr})+\parvo(\cprod{\unitkap}{\vectLam})\\
  &\quad+\parvb\parve(\cprod{\unitkap}{\fdot{\vectLam}})+\parvp(\cprod{\unitkap}{\unitplz})
  +\parvq(\cprod{\vectOme}{\vectr})+\parvr(\cprod{\vectOme}{\vectLam})
  +\parvs(\cprod{\vectOme}{\fdot{\vectLam}})+\parvt(\cprod{\vectOme}{\unitplz})\\
  &\quad-\parve\parvd(\cprod{\vectr}{\fdot{\vectLam}})
  +\parvu(\cprod{\vectr}{\vectLam})+\parvv(\cprod{\vectr}{\unitplz})
  -\ethvu\parve(\cprod{\unitplz}{\fdot{\vectLam}})
  +\parvw(\cprod{\unitplz}{\vectLam})+\parve^2(\cprod{\vectLam}{\fdot{\vectLam}})]
  \beqref{rpath24}
\end{split}
\end{align*}
\begin{align*}
\begin{split}
&=\parvx(\dprod{\ffdot{\vectLam}}{\vectLam})
  +\parvy(\dprod{\ffdot{\vectLam}}{\vectr})
  +\parvz(\dprod{\ffdot{\vectLam}}{\unitplz})
  +\efkoa(\dprod{\ffdot{\vectLam}}{\vectOme})
  +\efkob(\dprod{\ffdot{\vectLam}}{\fdot{\vectLam}})
  +\efkoc(\dprod{\ffdot{\vectLam}}{\ffdot{\vectLam}})
  +\parvm[\dprod{\ffdot{\vectLam}}{(\cprod{\unitkap}{\vectOme})}]\\
  &\quad+\parvn[\dprod{\ffdot{\vectLam}}{(\cprod{\unitkap}{\vectr})}]
  +\parvo[\dprod{\ffdot{\vectLam}}{(\cprod{\unitkap}{\vectLam})}]
  +\parvb\parve[\dprod{\ffdot{\vectLam}}{(\cprod{\unitkap}{\fdot{\vectLam}})}]
  +\parvp[\dprod{\ffdot{\vectLam}}{(\cprod{\unitkap}{\unitplz})}]
  +\parvq[\dprod{\ffdot{\vectLam}}{(\cprod{\vectOme}{\vectr})}]\\
  &\quad+\parvr[\dprod{\ffdot{\vectLam}}{(\cprod{\vectOme}{\vectLam})}]
  +\parvs[\dprod{\ffdot{\vectLam}}{(\cprod{\vectOme}{\fdot{\vectLam}})}]
  +\parvt[\dprod{\ffdot{\vectLam}}{(\cprod{\vectOme}{\unitplz})}]
  -\parve\parvd[\dprod{\ffdot{\vectLam}}{(\cprod{\vectr}{\fdot{\vectLam}})}]
  +\parvu[\dprod{\ffdot{\vectLam}}{(\cprod{\vectr}{\vectLam})}]\\
  &\quad+\parvv[\dprod{\ffdot{\vectLam}}{(\cprod{\vectr}{\unitplz})}]
  -\ethvu\parve[\dprod{\ffdot{\vectLam}}{(\cprod{\unitplz}{\fdot{\vectLam}})}]
  +\parvw[\dprod{\ffdot{\vectLam}}{(\cprod{\unitplz}{\vectLam})}]
  +\parve^2[\dprod{\ffdot{\vectLam}}{(\cprod{\vectLam}{\fdot{\vectLam}})}]
\end{split}
\end{align*}
\begin{align}\label{rpath25f}
\begin{split}
&=\parvx\vsigi+\parvy\vsigo+\parvz\vsigf+\efkoa\vsigh+\efkob\vsigj+\efkoc\vsigk+\parvm\frktw
  -\parvn\frktd+\parvo\frktx+\parvb\parve\frkty+\parvp\frktb\\
  &\quad+\parvq\frktm+\parvr\frktz+\parvs\frkxa-\parvt\frktq+\parve\parvd\frktj-\parvu\frkti
  +\parvv\frkxb-\ethvu\parve\frkxc+\parvw\frkts+\parve^2\frkxd\\
  &\quad\beqref{rpath1a}, \eqnref{rpath1b}\text{ \& }\eqnref{rpath1b2}
\end{split}
\nonumber\\
&=\efkoi\beqref{rpathx1b}
\end{align}
\begin{align*}
&\dprod{(\cprod{\unitkap}{\vectOme})}{(\vscrp+\vscrq)}\nonumber\\
\begin{split}
&=\dprod{(\cprod{\unitkap}{\vectOme})}{[}\parvx\vectLam+\parvy\vectr+\parvz\unitplz+\efkoa\vectOme
  +\efkob\fdot{\vectLam}+\efkoc\ffdot{\vectLam}
  +\parvm(\cprod{\unitkap}{\vectOme})+\parvn(\cprod{\unitkap}{\vectr})+\parvo(\cprod{\unitkap}{\vectLam})\\
  &\quad+\parvb\parve(\cprod{\unitkap}{\fdot{\vectLam}})+\parvp(\cprod{\unitkap}{\unitplz})
  +\parvq(\cprod{\vectOme}{\vectr})+\parvr(\cprod{\vectOme}{\vectLam})
  +\parvs(\cprod{\vectOme}{\fdot{\vectLam}})+\parvt(\cprod{\vectOme}{\unitplz})\\
  &\quad-\parve\parvd(\cprod{\vectr}{\fdot{\vectLam}})
  +\parvu(\cprod{\vectr}{\vectLam})+\parvv(\cprod{\vectr}{\unitplz})
  -\ethvu\parve(\cprod{\unitplz}{\fdot{\vectLam}})
  +\parvw(\cprod{\unitplz}{\vectLam})+\parve^2(\cprod{\vectLam}{\fdot{\vectLam}})]
  \beqref{rpath24}
\end{split}
\end{align*}
\begin{align*}
\begin{split}
&=\parvx[\dprod{\vectLam}{(\cprod{\unitkap}{\vectOme})}]
  +\parvy[\dprod{\vectr}{(\cprod{\unitkap}{\vectOme})}]
  +\parvz[\dprod{\unitplz}{(\cprod{\unitkap}{\vectOme})}]
  +\efkoa[\dprod{\vectOme}{(\cprod{\unitkap}{\vectOme})}]
  +\efkob[\dprod{\fdot{\vectLam}}{(\cprod{\unitkap}{\vectOme})}]\\
  &\quad+\efkoc[\dprod{\ffdot{\vectLam}}{(\cprod{\unitkap}{\vectOme})}]
  +\parvm[\dprod{(\cprod{\unitkap}{\vectOme})}{(\cprod{\unitkap}{\vectOme})}]
  +\parvn[\dprod{(\cprod{\unitkap}{\vectOme})}{(\cprod{\unitkap}{\vectr})}]
  +\parvo[\dprod{(\cprod{\unitkap}{\vectOme})}{(\cprod{\unitkap}{\vectLam})}]\\
  &\quad+\parvb\parve[\dprod{(\cprod{\unitkap}{\vectOme})}{(\cprod{\unitkap}{\fdot{\vectLam}})}]
  +\parvp[\dprod{(\cprod{\unitkap}{\vectOme})}{(\cprod{\unitkap}{\unitplz})}]
  +\parvq[\dprod{(\cprod{\unitkap}{\vectOme})}{(\cprod{\vectOme}{\vectr})}]\\
  &\quad+\parvr[\dprod{(\cprod{\unitkap}{\vectOme})}{(\cprod{\vectOme}{\vectLam})}]
  +\parvs[\dprod{(\cprod{\unitkap}{\vectOme})}{(\cprod{\vectOme}{\fdot{\vectLam}})}]
  +\parvt[\dprod{(\cprod{\unitkap}{\vectOme})}{(\cprod{\vectOme}{\unitplz})}]\\
  &\quad-\parve\parvd[\dprod{(\cprod{\unitkap}{\vectOme})}{(\cprod{\vectr}{\fdot{\vectLam}})}]
  +\parvu[\dprod{(\cprod{\unitkap}{\vectOme})}{(\cprod{\vectr}{\vectLam})}]
  +\parvv[\dprod{(\cprod{\unitkap}{\vectOme})}{(\cprod{\vectr}{\unitplz})}]\\
  &\quad-\ethvu\parve[\dprod{(\cprod{\unitkap}{\vectOme})}{(\cprod{\unitplz}{\fdot{\vectLam}})}]
  +\parvw[\dprod{(\cprod{\unitkap}{\vectOme})}{(\cprod{\unitplz}{\vectLam})}]
  +\parve^2[\dprod{(\cprod{\unitkap}{\vectOme})}{(\cprod{\vectLam}{\fdot{\vectLam}})}]
\end{split}
\end{align*}
\begin{align*}
\begin{split}
&=-\parvx\frkte+\parvy\epsvk-\parvz\epsvj+\efkob\frktf+\efkoc\frktw\\
  &\quad+\parvm[\Omerep^2-(\dprod{\unitkap}{\vectOme})^2]
  +\parvn[(\dprod{\vectOme}{\vectr})-(\dprod{\unitkap}{\vectr})(\dprod{\vectOme}{\unitkap})]
  +\parvo[(\dprod{\vectOme}{\vectLam})-(\dprod{\unitkap}{\vectLam})(\dprod{\vectOme}{\unitkap})]\\
  &\quad+\parvb\parve[(\dprod{\vectOme}{\fdot{\vectLam}})-(\dprod{\unitkap}{\fdot{\vectLam}})(\dprod{\vectOme}{\unitkap})]
  +\parvp[(\dprod{\vectOme}{\unitplz})-(\dprod{\unitkap}{\unitplz})(\dprod{\unitkap}{\vectOme})]
  +\parvq[(\dprod{\unitkap}{\vectOme})(\dprod{\vectOme}{\vectr})-(\dprod{\unitkap}{\vectr})\Omerep^2]\\
  &\quad+\parvr[(\dprod{\unitkap}{\vectOme})(\dprod{\vectOme}{\vectLam})-(\dprod{\unitkap}{\vectLam})\Omerep^2]
  +\parvs[(\dprod{\unitkap}{\vectOme})(\dprod{\vectOme}{\fdot{\vectLam}})-(\dprod{\unitkap}{\fdot{\vectLam}})\Omerep^2]
  +\parvt[(\dprod{\unitkap}{\vectOme})(\dprod{\vectOme}{\unitplz})-(\dprod{\unitkap}{\unitplz})\Omerep^2]\\
  &\quad-\parve\parvd[(\dprod{\unitkap}{\vectr})(\dprod{\vectOme}{\fdot{\vectLam}})-(\dprod{\unitkap}{\fdot{\vectLam}})(\dprod{\vectr}{\vectOme})]
  +\parvu[(\dprod{\unitkap}{\vectr})(\dprod{\vectOme}{\vectLam})-(\dprod{\unitkap}{\vectLam})(\dprod{\vectOme}{\vectr})]\\
  &\quad+\parvv[(\dprod{\unitkap}{\vectr})(\dprod{\vectOme}{\unitplz})-(\dprod{\unitkap}{\unitplz})(\dprod{\vectOme}{\vectr})]
  -\ethvu\parve[(\dprod{\unitkap}{\unitplz})(\dprod{\vectOme}{\fdot{\vectLam}})-(\dprod{\unitkap}{\fdot{\vectLam}})(\dprod{\vectOme}{\unitplz})]\\
  &\quad+\parvw[(\dprod{\unitkap}{\unitplz})(\dprod{\vectOme}{\vectLam})-(\dprod{\unitkap}{\vectLam})(\dprod{\vectOme}{\unitplz})]
  +\parve^2[(\dprod{\unitkap}{\vectLam})(\dprod{\vectOme}{\fdot{\vectLam}})-(\dprod{\unitkap}{\fdot{\vectLam}})(\dprod{\vectOme}{\vectLam})]\\
  &\quad\beqref{rot1a}, \eqnref{rpath1b}, \eqnref{rpath1b2}\text{ \& }\eqnref{alg2}
\end{split}
\end{align*}
\begin{align}\label{rpath25g}
\begin{split}
&=-\parvx\frkte+\parvy\epsvk-\parvz\epsvj+\efkob\frktf+\efkoc\frktw
  +\parvm(\Omerep^2-\epsva^2)+\parvn(\epsvb-\epsvd\epsva)+\parvo(\epsvg-\dltva\epsva)\\
  &\quad+\parvb\parve(\vsigc-\vsigb\epsva)+\parvp(\epsvc-\epsve\epsva)
  +\parvq(\epsva\epsvb-\epsvd\Omerep^2)+\parvr(\epsva\epsvg-\dltva\Omerep^2)\\
  &\quad+\parvs(\epsva\vsigc-\vsigb\Omerep^2)+\parvt(\epsva\epsvc-\epsve\Omerep^2)
  -\parve\parvd(\epsvd\vsigc-\vsigb\epsvb)+\parvu(\epsvd\epsvg-\dltva\epsvb)\\
  &\quad+\parvv(\epsvd\epsvc-\epsve\epsvb)-\ethvu\parve(\epsve\vsigc-\vsigb\epsvc)
  +\parvw(\epsve\epsvg-\dltva\epsvc)+\parve^2(\dltva\vsigc-\vsigb\epsvg)\\
  &\quad\beqref{rot1a}, \eqnref{rxpeed1a}\text{ \& }\eqnref{rpath1a}
\end{split}
\nonumber\\
&=\efkoj\beqref{rpathx1c}
\end{align}
\begin{align*}
&\dprod{(\cprod{\unitkap}{\vectr})}{(\vscrp+\vscrq)}\nonumber\\
\begin{split}
&=\dprod{(\cprod{\unitkap}{\vectr})}{[}\parvx\vectLam+\parvy\vectr+\parvz\unitplz+\efkoa\vectOme
  +\efkob\fdot{\vectLam}+\efkoc\ffdot{\vectLam}
  +\parvm(\cprod{\unitkap}{\vectOme})+\parvn(\cprod{\unitkap}{\vectr})+\parvo(\cprod{\unitkap}{\vectLam})\\
  &\quad+\parvb\parve(\cprod{\unitkap}{\fdot{\vectLam}})+\parvp(\cprod{\unitkap}{\unitplz})
  +\parvq(\cprod{\vectOme}{\vectr})+\parvr(\cprod{\vectOme}{\vectLam})
  +\parvs(\cprod{\vectOme}{\fdot{\vectLam}})+\parvt(\cprod{\vectOme}{\unitplz})\\
  &\quad-\parve\parvd(\cprod{\vectr}{\fdot{\vectLam}})
  +\parvu(\cprod{\vectr}{\vectLam})+\parvv(\cprod{\vectr}{\unitplz})
  -\ethvu\parve(\cprod{\unitplz}{\fdot{\vectLam}})
  +\parvw(\cprod{\unitplz}{\vectLam})+\parve^2(\cprod{\vectLam}{\fdot{\vectLam}})]
  \beqref{rpath24}
\end{split}
\end{align*}
\begin{align*}
\begin{split}
&=\parvx[\dprod{\vectLam}{(\cprod{\unitkap}{\vectr})}]
  +\parvy[\dprod{\vectr}{(\cprod{\unitkap}{\vectr})}]
  +\parvz[\dprod{\unitplz}{(\cprod{\unitkap}{\vectr})}]
  +\efkoa[\dprod{\vectOme}{(\cprod{\unitkap}{\vectr})}]
  +\efkob[\dprod{\fdot{\vectLam}}{(\cprod{\unitkap}{\vectr})}]\\
  &\quad+\efkoc[\dprod{\ffdot{\vectLam}}{(\cprod{\unitkap}{\vectr})}]
  +\parvm[\dprod{(\cprod{\unitkap}{\vectr})}{(\cprod{\unitkap}{\vectOme})}]
  +\parvn[\dprod{(\cprod{\unitkap}{\vectr})}{(\cprod{\unitkap}{\vectr})}]
  +\parvo[\dprod{(\cprod{\unitkap}{\vectr})}{(\cprod{\unitkap}{\vectLam})}]\\
  &\quad+\parvb\parve[\dprod{(\cprod{\unitkap}{\vectr})}{(\cprod{\unitkap}{\fdot{\vectLam}})}]
  +\parvp[\dprod{(\cprod{\unitkap}{\vectr})}{(\cprod{\unitkap}{\unitplz})}]
  +\parvq[\dprod{(\cprod{\unitkap}{\vectr})}{(\cprod{\vectOme}{\vectr})}]\\
  &\quad+\parvr[\dprod{(\cprod{\unitkap}{\vectr})}{(\cprod{\vectOme}{\vectLam})}]
  +\parvs[\dprod{(\cprod{\unitkap}{\vectr})}{(\cprod{\vectOme}{\fdot{\vectLam}})}]
  +\parvt[\dprod{(\cprod{\unitkap}{\vectr})}{(\cprod{\vectOme}{\unitplz})}]\\
  &\quad-\parve\parvd[\dprod{(\cprod{\unitkap}{\vectr})}{(\cprod{\vectr}{\fdot{\vectLam}})}]
  +\parvu[\dprod{(\cprod{\unitkap}{\vectr})}{(\cprod{\vectr}{\vectLam})}]
  +\parvv[\dprod{(\cprod{\unitkap}{\vectr})}{(\cprod{\vectr}{\unitplz})}]\\
  &\quad-\ethvu\parve[\dprod{(\cprod{\unitkap}{\vectr})}{(\cprod{\unitplz}{\fdot{\vectLam}})}]
  +\parvw[\dprod{(\cprod{\unitkap}{\vectr})}{(\cprod{\unitplz}{\vectLam})}]
  +\parve^2[\dprod{(\cprod{\unitkap}{\vectr})}{(\cprod{\vectLam}{\fdot{\vectLam}})}]
\end{split}
\end{align*}
\begin{align*}
\begin{split}
&=-\parvx\epsvn+\parvz\frktt-\efkoa\epsvk-\efkob\frktc-\efkoc\frktd
  +\parvm[(\dprod{\vectr}{\vectOme})-(\dprod{\unitkap}{\vectOme})(\dprod{\unitkap}{\vectr})]
  +\parvn[\scalr^2-(\dprod{\unitkap}{\vectr})^2]\\
  &\quad+\parvo[(\dprod{\vectr}{\vectLam})-(\dprod{\unitkap}{\vectLam})(\dprod{\vectr}{\unitkap})]
  +\parvb\parve[(\dprod{\vectr}{\fdot{\vectLam}})-(\dprod{\unitkap}{\fdot{\vectLam}})(\dprod{\unitkap}{\vectr})]\\
  &\quad+\parvp[(\dprod{\vectr}{\unitplz})-(\dprod{\unitkap}{\unitplz})(\dprod{\unitkap}{\vectr})]
  +\parvq[(\dprod{\unitkap}{\vectOme})\scalr^2-(\dprod{\unitkap}{\vectr})(\dprod{\vectr}{\vectOme})]\\
  &\quad+\parvr[(\dprod{\unitkap}{\vectOme})(\dprod{\vectr}{\vectLam})-(\dprod{\unitkap}{\vectLam})(\dprod{\vectr}{\vectOme})]
  +\parvs[(\dprod{\unitkap}{\vectOme})(\dprod{\vectr}{\fdot{\vectLam}})-(\dprod{\unitkap}{\fdot{\vectLam}})(\dprod{\vectr}{\vectOme})]\\
  &\quad+\parvt[(\dprod{\unitkap}{\vectOme})(\dprod{\vectr}{\unitplz})-(\dprod{\unitkap}{\unitplz})(\dprod{\vectr}{\vectOme})]
  -\parve\parvd[(\dprod{\unitkap}{\vectr})(\dprod{\vectr}{\fdot{\vectLam}})-(\dprod{\unitkap}{\fdot{\vectLam}})\scalr^2]\\
  &\quad+\parvu[(\dprod{\unitkap}{\vectr})(\dprod{\vectr}{\vectLam})-(\dprod{\unitkap}{\vectLam})\scalr^2]
  +\parvv[(\dprod{\unitkap}{\vectr})(\dprod{\vectr}{\unitplz})-(\dprod{\unitkap}{\unitplz})\scalr^2]\\
  &\quad-\ethvu\parve[(\dprod{\unitkap}{\unitplz})(\dprod{\vectr}{\fdot{\vectLam}})-(\dprod{\unitkap}{\fdot{\vectLam}})(\dprod{\vectr}{\unitplz})]
  +\parvw[(\dprod{\unitkap}{\unitplz})(\dprod{\vectr}{\vectLam})-(\dprod{\unitkap}{\vectLam})(\dprod{\vectr}{\unitplz})]\\
  &\quad+\parve^2[(\dprod{\unitkap}{\vectLam})(\dprod{\vectr}{\fdot{\vectLam}})-(\dprod{\unitkap}{\fdot{\vectLam}})(\dprod{\vectr}{\vectLam})]
  \beqref{rot1a}, \eqnref{rpath1b}\text{ \& }\eqnref{alg2}
\end{split}
\end{align*}
\begin{align}\label{rpath25h}
\begin{split}
&=-\parvx\epsvn+\parvz\frktt-\efkoa\epsvk-\efkob\frktc-\efkoc\frktd
  +\parvm(\epsvb-\epsva\epsvd)+\parvn(\scalr^2-\epsvd^2)\\
  &\quad+\parvo(\epsvh-\dltva\epsvd)+\parvb\parve(\vsign-\vsigb\epsvd)
  +\parvp(\epsvf-\epsve\epsvd)+\parvq(\epsva\scalr^2-\epsvd\epsvb)\\
  &\quad+\parvr(\epsva\epsvh-\dltva\epsvb)+\parvs(\epsva\vsign-\vsigb\epsvb)
  +\parvt(\epsva\epsvf-\epsve\epsvb)-\parve\parvd(\epsvd\vsign-\vsigb\scalr^2)\\
  &\quad+\parvu(\epsvd\epsvh-\dltva\scalr^2)+\parvv(\epsvd\epsvf-\epsve\scalr^2)
  -\ethvu\parve(\epsve\vsign-\vsigb\epsvf)+\parvw(\epsve\epsvh-\dltva\epsvf)\\
  &\quad+\parve^2(\dltva\vsign-\vsigb\epsvh)
  \beqref{rot1a}, \eqnref{rxpeed1a}\text{ \& }\eqnref{rpath1a}
\end{split}
\nonumber\\
&=\efkok\beqref{rpathx1c}
\end{align}
\begin{align*}
&\dprod{(\cprod{\unitkap}{\vectLam})}{(\vscrp+\vscrq)}\nonumber\\
\begin{split}
&=\dprod{(\cprod{\unitkap}{\vectLam})}{[}\parvx\vectLam+\parvy\vectr+\parvz\unitplz+\efkoa\vectOme
  +\efkob\fdot{\vectLam}+\efkoc\ffdot{\vectLam}
  +\parvm(\cprod{\unitkap}{\vectOme})+\parvn(\cprod{\unitkap}{\vectr})+\parvo(\cprod{\unitkap}{\vectLam})\\
  &\quad+\parvb\parve(\cprod{\unitkap}{\fdot{\vectLam}})+\parvp(\cprod{\unitkap}{\unitplz})
  +\parvq(\cprod{\vectOme}{\vectr})+\parvr(\cprod{\vectOme}{\vectLam})
  +\parvs(\cprod{\vectOme}{\fdot{\vectLam}})+\parvt(\cprod{\vectOme}{\unitplz})\\
  &\quad-\parve\parvd(\cprod{\vectr}{\fdot{\vectLam}})
  +\parvu(\cprod{\vectr}{\vectLam})+\parvv(\cprod{\vectr}{\unitplz})
  -\ethvu\parve(\cprod{\unitplz}{\fdot{\vectLam}})
  +\parvw(\cprod{\unitplz}{\vectLam})+\parve^2(\cprod{\vectLam}{\fdot{\vectLam}})]
  \beqref{rpath24}
\end{split}
\end{align*}
\begin{align*}
\begin{split}
&=\parvx[\dprod{\vectLam}{(\cprod{\unitkap}{\vectLam})}]
  +\parvy[\dprod{\vectr}{(\cprod{\unitkap}{\vectLam})}]
  +\parvz[\dprod{\unitplz}{(\cprod{\unitkap}{\vectLam})}]
  +\efkoa[\dprod{\vectOme}{(\cprod{\unitkap}{\vectLam})}]
  +\efkob[\dprod{\fdot{\vectLam}}{(\cprod{\unitkap}{\vectLam})}]\\
  &\quad+\efkoc[\dprod{\ffdot{\vectLam}}{(\cprod{\unitkap}{\vectLam})}]
  +\parvm[\dprod{(\cprod{\unitkap}{\vectLam})}{(\cprod{\unitkap}{\vectOme})}]
  +\parvn[\dprod{(\cprod{\unitkap}{\vectLam})}{(\cprod{\unitkap}{\vectr})}]
  +\parvo[\dprod{(\cprod{\unitkap}{\vectLam})}{(\cprod{\unitkap}{\vectLam})}]\\
  &\quad+\parvb\parve[\dprod{(\cprod{\unitkap}{\vectLam})}{(\cprod{\unitkap}{\fdot{\vectLam}})}]
  +\parvp[\dprod{(\cprod{\unitkap}{\vectLam})}{(\cprod{\unitkap}{\unitplz})}]
  +\parvq[\dprod{(\cprod{\unitkap}{\vectLam})}{(\cprod{\vectOme}{\vectr})}]\\
  &\quad+\parvr[\dprod{(\cprod{\unitkap}{\vectLam})}{(\cprod{\vectOme}{\vectLam})}]
  +\parvs[\dprod{(\cprod{\unitkap}{\vectLam})}{(\cprod{\vectOme}{\fdot{\vectLam}})}]
  +\parvt[\dprod{(\cprod{\unitkap}{\vectLam})}{(\cprod{\vectOme}{\unitplz})}]\\
  &\quad-\parve\parvd[\dprod{(\cprod{\unitkap}{\vectLam})}{(\cprod{\vectr}{\fdot{\vectLam}})}]
  +\parvu[\dprod{(\cprod{\unitkap}{\vectLam})}{(\cprod{\vectr}{\vectLam})}]
  +\parvv[\dprod{(\cprod{\unitkap}{\vectLam})}{(\cprod{\vectr}{\unitplz})}]\\
  &\quad-\ethvu\parve[\dprod{(\cprod{\unitkap}{\vectLam})}{(\cprod{\unitplz}{\fdot{\vectLam}})}]
  +\parvw[\dprod{(\cprod{\unitkap}{\vectLam})}{(\cprod{\unitplz}{\vectLam})}]
  +\parve^2[\dprod{(\cprod{\unitkap}{\vectLam})}{(\cprod{\vectLam}{\fdot{\vectLam}})}]
\end{split}
\end{align*}
\begin{align*}
\begin{split}
&=\parvy\epsvn-\parvz\dltvb+\efkoa\frkte+\efkob\frktu+\efkoc\frktx
  +\parvm[(\dprod{\vectLam}{\vectOme})-(\dprod{\unitkap}{\vectOme})(\dprod{\vectLam}{\unitkap})]\\
  &\quad+\parvn[(\dprod{\vectLam}{\vectr})-(\dprod{\unitkap}{\vectr})(\dprod{\vectLam}{\unitkap})]
  +\parvo[\Lamrep^2-(\dprod{\unitkap}{\vectLam})^2]
  +\parvb\parve[(\dprod{\vectLam}{\fdot{\vectLam}})-(\dprod{\unitkap}{\fdot{\vectLam}})(\dprod{\unitkap}{\vectLam})]\\
  &\quad+\parvp[(\dprod{\vectLam}{\unitplz})-(\dprod{\unitkap}{\unitplz})(\dprod{\vectLam}{\unitkap})]
  +\parvq[(\dprod{\unitkap}{\vectOme})(\dprod{\vectLam}{\vectr})-(\dprod{\unitkap}{\vectr})(\dprod{\vectLam}{\vectOme})]\\
  &\quad+\parvr[(\dprod{\unitkap}{\vectOme})\Lamrep^2-(\dprod{\unitkap}{\vectLam})(\dprod{\vectLam}{\vectOme})]
  +\parvs[(\dprod{\unitkap}{\vectOme})(\dprod{\vectLam}{\fdot{\vectLam}})-(\dprod{\unitkap}{\fdot{\vectLam}})(\dprod{\vectLam}{\vectOme})]\\
  &\quad+\parvt[(\dprod{\unitkap}{\vectOme})(\dprod{\vectLam}{\unitplz})-(\dprod{\unitkap}{\unitplz})(\dprod{\vectLam}{\vectOme})]
  -\parve\parvd[(\dprod{\unitkap}{\vectr})(\dprod{\vectLam}{\fdot{\vectLam}})-(\dprod{\unitkap}{\fdot{\vectLam}})(\dprod{\vectLam}{\vectr})]\\
  &\quad+\parvu[(\dprod{\unitkap}{\vectr})\Lamrep^2-(\dprod{\unitkap}{\vectLam})(\dprod{\vectLam}{\vectr})]
  +\parvv[(\dprod{\unitkap}{\vectr})(\dprod{\vectLam}{\unitplz})-(\dprod{\unitkap}{\unitplz})(\dprod{\vectLam}{\vectr})]\\
  &\quad-\ethvu\parve[(\dprod{\unitkap}{\unitplz})(\dprod{\vectLam}{\fdot{\vectLam}})-(\dprod{\unitkap}{\fdot{\vectLam}})(\dprod{\vectLam}{\unitplz})]
  +\parvw[(\dprod{\unitkap}{\unitplz})\Lamrep^2-(\dprod{\unitkap}{\vectLam})(\dprod{\vectLam}{\unitplz})]\\
  &\quad+\parve^2[(\dprod{\unitkap}{\vectLam})(\dprod{\vectLam}{\fdot{\vectLam}})-(\dprod{\unitkap}{\fdot{\vectLam}})\Lamrep^2]
  \beqref{rot1a}, \eqnref{rxpeed1a}, \eqnref{rpath1b}, \eqnref{rpath1b2}\text{ \& }\eqnref{alg2}
\end{split}
\end{align*}
\begin{align}\label{rpath25i}
\begin{split}
&=\parvy\epsvn-\parvz\dltvb+\efkoa\frkte+\efkob\frktu+\efkoc\frktx
  +\parvm(\epsvg-\epsva\dltva)+\parvn(\epsvh-\epsvd\dltva)\\
  &\quad+\parvo(\Lamrep^2-\dltva^2)+\parvb\parve(\vsigd-\vsigb\dltva)
  +\parvp(\epsvi-\epsve\dltva)+\parvq(\epsva\epsvh-\epsvd\epsvg)\\
  &\quad+\parvr(\epsva\Lamrep^2-\dltva\epsvg)+\parvs(\epsva\vsigd-\vsigb\epsvg)
  +\parvt(\epsva\epsvi-\epsve\epsvg)-\parve\parvd(\epsvd\vsigd-\vsigb\epsvh)\\
  &\quad+\parvu(\epsvd\Lamrep^2-\dltva\epsvh)+\parvv(\epsvd\epsvi-\epsve\epsvh)
  -\ethvu\parve(\epsve\vsigd-\vsigb\epsvi)+\parvw(\epsve\Lamrep^2-\dltva\epsvi)\\
  &\quad+\parve^2(\dltva\vsigd-\vsigb\Lamrep^2)
  \beqref{rot1a}, \eqnref{rxpeed1a}\text{ \& }\eqnref{rpath1a}
\end{split}
\nonumber\\
&=\efkol\beqref{rpathx1d}
\end{align}
\begin{align*}
&\dprod{(\cprod{\unitkap}{\fdot{\vectLam}})}{(\vscrp+\vscrq)}\nonumber\\
\begin{split}
&=\dprod{(\cprod{\unitkap}{\fdot{\vectLam}})}{[}\parvx\vectLam+\parvy\vectr+\parvz\unitplz+\efkoa\vectOme
  +\efkob\fdot{\vectLam}+\efkoc\ffdot{\vectLam}
  +\parvm(\cprod{\unitkap}{\vectOme})+\parvn(\cprod{\unitkap}{\vectr})+\parvo(\cprod{\unitkap}{\vectLam})\\
  &\quad+\parvb\parve(\cprod{\unitkap}{\fdot{\vectLam}})+\parvp(\cprod{\unitkap}{\unitplz})
  +\parvq(\cprod{\vectOme}{\vectr})+\parvr(\cprod{\vectOme}{\vectLam})
  +\parvs(\cprod{\vectOme}{\fdot{\vectLam}})+\parvt(\cprod{\vectOme}{\unitplz})\\
  &\quad-\parve\parvd(\cprod{\vectr}{\fdot{\vectLam}})
  +\parvu(\cprod{\vectr}{\vectLam})+\parvv(\cprod{\vectr}{\unitplz})
  -\ethvu\parve(\cprod{\unitplz}{\fdot{\vectLam}})
  +\parvw(\cprod{\unitplz}{\vectLam})+\parve^2(\cprod{\vectLam}{\fdot{\vectLam}})]
  \beqref{rpath24}
\end{split}
\end{align*}
\begin{align*}
\begin{split}
&=\parvx[\dprod{\vectLam}{(\cprod{\unitkap}{\fdot{\vectLam}})}]
  +\parvy[\dprod{\vectr}{(\cprod{\unitkap}{\fdot{\vectLam}})}]
  +\parvz[\dprod{\unitplz}{(\cprod{\unitkap}{\fdot{\vectLam}})}]
  +\efkoa[\dprod{\vectOme}{(\cprod{\unitkap}{\fdot{\vectLam}})}]
  +\efkob[\dprod{\fdot{\vectLam}}{(\cprod{\unitkap}{\fdot{\vectLam}})}]\\
  &\quad+\efkoc[\dprod{\ffdot{\vectLam}}{(\cprod{\unitkap}{\fdot{\vectLam}})}]
  +\parvm[\dprod{(\cprod{\unitkap}{\fdot{\vectLam}})}{(\cprod{\unitkap}{\vectOme})}]
  +\parvn[\dprod{(\cprod{\unitkap}{\fdot{\vectLam}})}{(\cprod{\unitkap}{\vectr})}]
  +\parvo[\dprod{(\cprod{\unitkap}{\fdot{\vectLam}})}{(\cprod{\unitkap}{\vectLam})}]\\
  &\quad+\parvb\parve[\dprod{(\cprod{\unitkap}{\fdot{\vectLam}})}{(\cprod{\unitkap}{\fdot{\vectLam}})}]
  +\parvp[\dprod{(\cprod{\unitkap}{\fdot{\vectLam}})}{(\cprod{\unitkap}{\unitplz})}]
  +\parvq[\dprod{(\cprod{\unitkap}{\fdot{\vectLam}})}{(\cprod{\vectOme}{\vectr})}]\\
  &\quad+\parvr[\dprod{(\cprod{\unitkap}{\fdot{\vectLam}})}{(\cprod{\vectOme}{\vectLam})}]
  +\parvs[\dprod{(\cprod{\unitkap}{\fdot{\vectLam}})}{(\cprod{\vectOme}{\fdot{\vectLam}})}]
  +\parvt[\dprod{(\cprod{\unitkap}{\fdot{\vectLam}})}{(\cprod{\vectOme}{\unitplz})}]\\
  &\quad-\parve\parvd[\dprod{(\cprod{\unitkap}{\fdot{\vectLam}})}{(\cprod{\vectr}{\fdot{\vectLam}})}]
  +\parvu[\dprod{(\cprod{\unitkap}{\fdot{\vectLam}})}{(\cprod{\vectr}{\vectLam})}]
  +\parvv[\dprod{(\cprod{\unitkap}{\fdot{\vectLam}})}{(\cprod{\vectr}{\unitplz})}]\\
  &\quad-\ethvu\parve[\dprod{(\cprod{\unitkap}{\fdot{\vectLam}})}{(\cprod{\unitplz}{\fdot{\vectLam}})}]
  +\parvw[\dprod{(\cprod{\unitkap}{\fdot{\vectLam}})}{(\cprod{\unitplz}{\vectLam})}]
  +\parve^2[\dprod{(\cprod{\unitkap}{\fdot{\vectLam}})}{(\cprod{\vectLam}{\fdot{\vectLam}})}]
\end{split}
\end{align*}
\begin{align*}
\begin{split}
&=-\parvx\frktu+\parvy\frktc-\parvz\frkta-\efkoa\frktf+\efkoc\frkty
  +\parvm[(\dprod{\fdot{\vectLam}}{\vectOme})-(\dprod{\unitkap}{\vectOme})(\dprod{\fdot{\vectLam}}{\unitkap})]\\
  &\quad+\parvn[(\dprod{\fdot{\vectLam}}{\vectr})-(\dprod{\unitkap}{\vectr})(\dprod{\fdot{\vectLam}}{\unitkap})]
  +\parvo[(\dprod{\fdot{\vectLam}}{\vectLam})-(\dprod{\unitkap}{\vectLam})(\dprod{\fdot{\vectLam}}{\unitkap})]
  +\parvb\parve[(\dprod{\fdot{\vectLam}}{\fdot{\vectLam}})-(\dprod{\unitkap}{\fdot{\vectLam}})^2]\\
  &\quad+\parvp[(\dprod{\fdot{\vectLam}}{\unitplz})-(\dprod{\unitkap}{\unitplz})(\dprod{\fdot{\vectLam}}{\unitkap})]
  +\parvq[(\dprod{\unitkap}{\vectOme})(\dprod{\fdot{\vectLam}}{\vectr})-(\dprod{\unitkap}{\vectr})(\dprod{\fdot{\vectLam}}{\vectOme})]\\
  &\quad+\parvr[(\dprod{\unitkap}{\vectOme})(\dprod{\fdot{\vectLam}}{\vectLam})-(\dprod{\unitkap}{\vectLam})(\dprod{\fdot{\vectLam}}{\vectOme})]
  +\parvs[(\dprod{\unitkap}{\vectOme})\fdot{\vectLam}^2-(\dprod{\unitkap}{\fdot{\vectLam}})(\dprod{\fdot{\vectLam}}{\vectOme})]\\
  &\quad+\parvt[(\dprod{\unitkap}{\vectOme})(\dprod{\fdot{\vectLam}}{\unitplz})-(\dprod{\unitkap}{\unitplz})(\dprod{\fdot{\vectLam}}{\vectOme})]
  -\parve\parvd[(\dprod{\unitkap}{\vectr})\fdot{\vectLam}^2-(\dprod{\unitkap}{\fdot{\vectLam}})(\dprod{\fdot{\vectLam}}{\vectr})]\\
  &\quad+\parvu[(\dprod{\unitkap}{\vectr})(\dprod{\fdot{\vectLam}}{\vectLam})-(\dprod{\unitkap}{\vectLam})(\dprod{\fdot{\vectLam}}{\vectr})]
  +\parvv[(\dprod{\unitkap}{\vectr})(\dprod{\fdot{\vectLam}}{\unitplz})-(\dprod{\unitkap}{\unitplz})(\dprod{\fdot{\vectLam}}{\vectr})]\\
  &\quad-\ethvu\parve[(\dprod{\unitkap}{\unitplz})\fdot{\vectLam}^2-(\dprod{\unitkap}{\fdot{\vectLam}})(\dprod{\fdot{\vectLam}}{\unitplz})]
  +\parvw[(\dprod{\unitkap}{\unitplz})(\dprod{\fdot{\vectLam}}{\vectLam})-(\dprod{\unitkap}{\vectLam})(\dprod{\fdot{\vectLam}}{\unitplz})]\\
  &\quad+\parve^2[(\dprod{\unitkap}{\vectLam})\fdot{\vectLam}^2-(\dprod{\unitkap}{\fdot{\vectLam}})(\dprod{\fdot{\vectLam}}{\vectLam})]
  \beqref{rpath1b}, \eqnref{rpath1b2}\text{ \& }\eqnref{alg2}
\end{split}
\end{align*}
\begin{align}\label{rpath25j}
\begin{split}
&=-\parvx\frktu+\parvy\frktc-\parvz\frkta-\efkoa\frktf+\efkoc\frkty
  +\parvm(\vsigc-\epsva\vsigb)+\parvn(\vsign-\epsvd\vsigb)\\
  &\quad+\parvo(\vsigd-\dltva\vsigb)+\parvb\parve(\vsige-\vsigb^2)
  +\parvp(\vsiga-\epsve\vsigb)+\parvq(\epsva\vsign-\epsvd\vsigc)\\
  &\quad+\parvr(\epsva\vsigd-\dltva\vsigc)+\parvs(\epsva\vsige-\vsigb\vsigc)
  +\parvt(\epsva\vsiga-\epsve\vsigc)-\parve\parvd(\epsvd\vsige-\vsigb\vsign)\\
  &\quad+\parvu(\epsvd\vsigd-\dltva\vsign)+\parvv(\epsvd\vsiga-\epsve\vsign)
  -\ethvu\parve(\epsve\vsige-\vsigb\vsiga)+\parvw(\epsve\vsigd-\dltva\vsiga)\\
  &\quad+\parve^2(\dltva\vsige-\vsigb\vsigd)
  \beqref{rot1a}, \eqnref{rxpeed1a}\text{ \& }\eqnref{rpath1a}
\end{split}
\nonumber\\
&=\efkom\beqref{rpathx1d}
\end{align}
\begin{align*}
&\dprod{(\cprod{\unitkap}{\unitplz})}{(\vscrp+\vscrq)}\nonumber\\
\begin{split}
&=\dprod{(\cprod{\unitkap}{\unitplz})}{[}\parvx\vectLam+\parvy\vectr+\parvz\unitplz+\efkoa\vectOme
  +\efkob\fdot{\vectLam}+\efkoc\ffdot{\vectLam}
  +\parvm(\cprod{\unitkap}{\vectOme})+\parvn(\cprod{\unitkap}{\vectr})+\parvo(\cprod{\unitkap}{\vectLam})\\
  &\quad+\parvb\parve(\cprod{\unitkap}{\fdot{\vectLam}})+\parvp(\cprod{\unitkap}{\unitplz})
  +\parvq(\cprod{\vectOme}{\vectr})+\parvr(\cprod{\vectOme}{\vectLam})
  +\parvs(\cprod{\vectOme}{\fdot{\vectLam}})+\parvt(\cprod{\vectOme}{\unitplz})\\
  &\quad-\parve\parvd(\cprod{\vectr}{\fdot{\vectLam}})
  +\parvu(\cprod{\vectr}{\vectLam})+\parvv(\cprod{\vectr}{\unitplz})
  -\ethvu\parve(\cprod{\unitplz}{\fdot{\vectLam}})
  +\parvw(\cprod{\unitplz}{\vectLam})+\parve^2(\cprod{\vectLam}{\fdot{\vectLam}})]
  \beqref{rpath24}
\end{split}
\end{align*}
\begin{align*}
\begin{split}
&=\parvx[\dprod{\vectLam}{(\cprod{\unitkap}{\unitplz})}]
  +\parvy[\dprod{\vectr}{(\cprod{\unitkap}{\unitplz})}]
  +\parvz[\dprod{\unitplz}{(\cprod{\unitkap}{\unitplz})}]
  +\efkoa[\dprod{\vectOme}{(\cprod{\unitkap}{\unitplz})}]
  +\efkob[\dprod{\fdot{\vectLam}}{(\cprod{\unitkap}{\unitplz})}]\\
  &\quad+\efkoc[\dprod{\ffdot{\vectLam}}{(\cprod{\unitkap}{\unitplz})}]
  +\parvm[\dprod{(\cprod{\unitkap}{\unitplz})}{(\cprod{\unitkap}{\vectOme})}]
  +\parvn[\dprod{(\cprod{\unitkap}{\unitplz})}{(\cprod{\unitkap}{\vectr})}]
  +\parvo[\dprod{(\cprod{\unitkap}{\unitplz})}{(\cprod{\unitkap}{\vectLam})}]\\
  &\quad+\parvb\parve[\dprod{(\cprod{\unitkap}{\unitplz})}{(\cprod{\unitkap}{\fdot{\vectLam}})}]
  +\parvp[\dprod{(\cprod{\unitkap}{\unitplz})}{(\cprod{\unitkap}{\unitplz})}]
  +\parvq[\dprod{(\cprod{\unitkap}{\unitplz})}{(\cprod{\vectOme}{\vectr})}]\\
  &\quad+\parvr[\dprod{(\cprod{\unitkap}{\unitplz})}{(\cprod{\vectOme}{\vectLam})}]
  +\parvs[\dprod{(\cprod{\unitkap}{\unitplz})}{(\cprod{\vectOme}{\fdot{\vectLam}})}]
  +\parvt[\dprod{(\cprod{\unitkap}{\unitplz})}{(\cprod{\vectOme}{\unitplz})}]\\
  &\quad-\parve\parvd[\dprod{(\cprod{\unitkap}{\unitplz})}{(\cprod{\vectr}{\fdot{\vectLam}})}]
  +\parvu[\dprod{(\cprod{\unitkap}{\unitplz})}{(\cprod{\vectr}{\vectLam})}]
  +\parvv[\dprod{(\cprod{\unitkap}{\unitplz})}{(\cprod{\vectr}{\unitplz})}]\\
  &\quad-\ethvu\parve[\dprod{(\cprod{\unitkap}{\unitplz})}{(\cprod{\unitplz}{\fdot{\vectLam}})}]
  +\parvw[\dprod{(\cprod{\unitkap}{\unitplz})}{(\cprod{\unitplz}{\vectLam})}]
  +\parve^2[\dprod{(\cprod{\unitkap}{\unitplz})}{(\cprod{\vectLam}{\fdot{\vectLam}})}]
\end{split}
\end{align*}
\begin{align*}
\begin{split}
&=\parvx\dltvb-\parvy\frktt+\efkoa\epsvj+\efkob\frkta+\efkoc\frktb
  +\parvm[(\dprod{\unitplz}{\vectOme})-(\dprod{\unitkap}{\vectOme})(\dprod{\unitplz}{\unitkap})]\\
  &\quad+\parvn[(\dprod{\unitplz}{\vectr})-(\dprod{\unitkap}{\vectr})(\dprod{\unitplz}{\unitkap})]
  +\parvo[(\dprod{\unitplz}{\vectLam})-(\dprod{\unitkap}{\vectLam})(\dprod{\unitkap}{\unitplz})]
  +\parvb\parve[(\dprod{\unitplz}{\fdot{\vectLam}})-(\dprod{\unitkap}{\fdot{\vectLam}})(\dprod{\unitplz}{\unitkap})]\\
  &\quad+\parvp[1-(\dprod{\unitkap}{\unitplz})^2]
  +\parvq[(\dprod{\unitkap}{\vectOme})(\dprod{\unitplz}{\vectr})-(\dprod{\unitkap}{\vectr})(\dprod{\unitplz}{\vectOme})]
  +\parvr[(\dprod{\unitkap}{\vectOme})(\dprod{\unitplz}{\vectLam})-(\dprod{\unitkap}{\vectLam})(\dprod{\unitplz}{\vectOme})]\\
  &\quad+\parvs[(\dprod{\unitkap}{\vectOme})(\dprod{\unitplz}{\fdot{\vectLam}})-(\dprod{\unitkap}{\fdot{\vectLam}})(\dprod{\unitplz}{\vectOme})]
  +\parvt[(\dprod{\unitkap}{\vectOme})-(\dprod{\unitkap}{\unitplz})(\dprod{\unitplz}{\vectOme})]\\
  &\quad-\parve\parvd[(\dprod{\unitkap}{\vectr})(\dprod{\unitplz}{\fdot{\vectLam}})-(\dprod{\unitkap}{\fdot{\vectLam}})(\dprod{\unitplz}{\vectr})]
  +\parvu[(\dprod{\unitkap}{\vectr})(\dprod{\unitplz}{\vectLam})-(\dprod{\unitkap}{\vectLam})(\dprod{\unitplz}{\vectr})]\\
  &\quad+\parvv[(\dprod{\unitkap}{\vectr})-(\dprod{\unitkap}{\unitplz})(\dprod{\unitplz}{\vectr})]
  -\ethvu\parve[(\dprod{\unitkap}{\unitplz})(\dprod{\unitplz}{\fdot{\vectLam}})-(\dprod{\unitkap}{\fdot{\vectLam}})]
  +\parvw[(\dprod{\unitkap}{\unitplz})(\dprod{\unitplz}{\vectLam})-(\dprod{\unitkap}{\vectLam})]\\
  &\quad+\parve^2[(\dprod{\unitkap}{\vectLam})(\dprod{\unitplz}{\fdot{\vectLam}})-(\dprod{\unitkap}{\fdot{\vectLam}})(\dprod{\unitplz}{\vectLam})]
  \beqref{rot1a}, \eqnref{rxpeed1a}, \eqnref{rpath1b}, \eqnref{rpath1b2}\text{ \& }\eqnref{alg2}
\end{split}
\end{align*}
\begin{align}\label{rpath25k}
\begin{split}
&=\parvx\dltvb-\parvy\frktt+\efkoa\epsvj+\efkob\frkta+\efkoc\frktb
  +\parvm(\epsvc-\epsva\epsve)+\parvn(\epsvf-\epsvd\epsve)+\parvo(\epsvi-\dltva\epsve)\\
  &\quad+\parvb\parve(\vsiga-\vsigb\epsve)+\parvp(1-\epsve^2)+\parvq(\epsva\epsvf-\epsvd\epsvc)
  +\parvr(\epsva\epsvi-\dltva\epsvc)+\parvs(\epsva\vsiga-\vsigb\epsvc)\\
  &\quad+\parvt(\epsva-\epsve\epsvc)-\parve\parvd(\epsvd\vsiga-\vsigb\epsvf)
  +\parvu(\epsvd\epsvi-\dltva\epsvf)+\parvv(\epsvd-\epsve\epsvf)\\
  &\quad-\ethvu\parve(\epsve\vsiga-\vsigb)+\parvw(\epsve\epsvi-\dltva)+\parve^2(\dltva\vsiga-\vsigb\epsvi)
  \beqref{rot1a}, \eqnref{rxpeed1a}\text{ \& }\eqnref{rpath1a}
\end{split}
\nonumber\\
&=\efkon\beqref{rpathx1e}
\end{align}
\begin{align*}
&\dprod{(\cprod{\vectOme}{\vectr})}{(\vscrp+\vscrq)}\nonumber\\
\begin{split}
&=\dprod{(\cprod{\vectOme}{\vectr})}{[}\parvx\vectLam+\parvy\vectr+\parvz\unitplz+\efkoa\vectOme
  +\efkob\fdot{\vectLam}+\efkoc\ffdot{\vectLam}
  +\parvm(\cprod{\unitkap}{\vectOme})+\parvn(\cprod{\unitkap}{\vectr})+\parvo(\cprod{\unitkap}{\vectLam})\\
  &\quad+\parvb\parve(\cprod{\unitkap}{\fdot{\vectLam}})+\parvp(\cprod{\unitkap}{\unitplz})
  +\parvq(\cprod{\vectOme}{\vectr})+\parvr(\cprod{\vectOme}{\vectLam})
  +\parvs(\cprod{\vectOme}{\fdot{\vectLam}})+\parvt(\cprod{\vectOme}{\unitplz})\\
  &\quad-\parve\parvd(\cprod{\vectr}{\fdot{\vectLam}})
  +\parvu(\cprod{\vectr}{\vectLam})+\parvv(\cprod{\vectr}{\unitplz})
  -\ethvu\parve(\cprod{\unitplz}{\fdot{\vectLam}})
  +\parvw(\cprod{\unitplz}{\vectLam})+\parve^2(\cprod{\vectLam}{\fdot{\vectLam}})]
  \beqref{rpath24}
\end{split}
\end{align*}
\begin{align*}
\begin{split}
&=\parvx[\dprod{\vectLam}{(\cprod{\vectOme}{\vectr})}]
  +\parvy[\dprod{\vectr}{(\cprod{\vectOme}{\vectr})}]
  +\parvz[\dprod{\unitplz}{(\cprod{\vectOme}{\vectr})}]
  +\efkoa[\dprod{\vectOme}{(\cprod{\vectOme}{\vectr})}]
  +\efkob[\dprod{\fdot{\vectLam}}{(\cprod{\vectOme}{\vectr})}]\\
  &\quad+\efkoc[\dprod{\ffdot{\vectLam}}{(\cprod{\vectOme}{\vectr})}]
  +\parvm[\dprod{(\cprod{\vectOme}{\vectr})}{(\cprod{\unitkap}{\vectOme})}]
  +\parvn[\dprod{(\cprod{\vectOme}{\vectr})}{(\cprod{\unitkap}{\vectr})}]
  +\parvo[\dprod{(\cprod{\vectOme}{\vectr})}{(\cprod{\unitkap}{\vectLam})}]\\
  &\quad+\parvb\parve[\dprod{(\cprod{\vectOme}{\vectr})}{(\cprod{\unitkap}{\fdot{\vectLam}})}]
  +\parvp[\dprod{(\cprod{\vectOme}{\vectr})}{(\cprod{\unitkap}{\unitplz})}]
  +\parvq[\dprod{(\cprod{\vectOme}{\vectr})}{(\cprod{\vectOme}{\vectr})}]\\
  &\quad+\parvr[\dprod{(\cprod{\vectOme}{\vectr})}{(\cprod{\vectOme}{\vectLam})}]
  +\parvs[\dprod{(\cprod{\vectOme}{\vectr})}{(\cprod{\vectOme}{\fdot{\vectLam}})}]
  +\parvt[\dprod{(\cprod{\vectOme}{\vectr})}{(\cprod{\vectOme}{\unitplz})}]\\
  &\quad-\parve\parvd[\dprod{(\cprod{\vectOme}{\vectr})}{(\cprod{\vectr}{\fdot{\vectLam}})}]
  +\parvu[\dprod{(\cprod{\vectOme}{\vectr})}{(\cprod{\vectr}{\vectLam})}]
  +\parvv[\dprod{(\cprod{\vectOme}{\vectr})}{(\cprod{\vectr}{\unitplz})}]\\
  &\quad-\ethvu\parve[\dprod{(\cprod{\vectOme}{\vectr})}{(\cprod{\unitplz}{\fdot{\vectLam}})}]
  +\parvw[\dprod{(\cprod{\vectOme}{\vectr})}{(\cprod{\unitplz}{\vectLam})}]
  +\parve^2[\dprod{(\cprod{\vectOme}{\vectr})}{(\cprod{\vectLam}{\fdot{\vectLam}})}]
\end{split}
\end{align*}
\begin{align*}
\begin{split}
&=-\parvx\epsvm+\parvz\epsvl+\efkob\frktl+\efkoc\frktm
  +\parvm[(\dprod{\vectOme}{\unitkap})(\dprod{\vectr}{\vectOme})-\Omerep^2(\dprod{\vectr}{\unitkap})]
  +\parvn[(\dprod{\vectOme}{\unitkap})\scalr^2-(\dprod{\vectOme}{\vectr})(\dprod{\vectr}{\unitkap})]\\
  &\quad+\parvo[(\dprod{\vectOme}{\unitkap})(\dprod{\vectr}{\vectLam})-(\dprod{\vectOme}{\vectLam})(\dprod{\vectr}{\unitkap})]
  +\parvb\parve[(\dprod{\vectOme}{\unitkap})(\dprod{\vectr}{\fdot{\vectLam}})-(\dprod{\vectOme}{\fdot{\vectLam}})(\dprod{\unitkap}{\vectr})]\\
  &\quad+\parvp[(\dprod{\vectOme}{\unitkap})(\dprod{\vectr}{\unitplz})-(\dprod{\vectOme}{\unitplz})(\dprod{\vectr}{\unitkap})]
  +\parvq[\Omerep^2\scalr^2-(\dprod{\vectOme}{\vectr})^2]\\
  &\quad+\parvr[\Omerep^2(\dprod{\vectr}{\vectLam})-(\dprod{\vectOme}{\vectLam})(\dprod{\vectr}{\vectOme})]
  +\parvs[\Omerep^2(\dprod{\vectr}{\fdot{\vectLam}})-(\dprod{\vectOme}{\fdot{\vectLam}})(\dprod{\vectr}{\vectOme})]\\
  &\quad+\parvt[\Omerep^2(\dprod{\vectr}{\unitplz})-(\dprod{\vectOme}{\unitplz})(\dprod{\vectr}{\vectOme})]
  -\parve\parvd[(\dprod{\vectOme}{\vectr})(\dprod{\vectr}{\fdot{\vectLam}})-(\dprod{\vectOme}{\fdot{\vectLam}})\scalr^2]\\
  &\quad+\parvu[(\dprod{\vectOme}{\vectr})(\dprod{\vectr}{\vectLam})-(\dprod{\vectOme}{\vectLam})\scalr^2]
  +\parvv[(\dprod{\vectOme}{\vectr})(\dprod{\vectr}{\unitplz})-(\dprod{\vectOme}{\unitplz})\scalr^2]\\
  &\quad-\ethvu\parve[(\dprod{\vectOme}{\unitplz})(\dprod{\vectr}{\fdot{\vectLam}})-(\dprod{\vectOme}{\fdot{\vectLam}})(\dprod{\vectr}{\unitplz})]
  +\parvw[(\dprod{\vectOme}{\unitplz})(\dprod{\vectr}{\vectLam})-(\dprod{\vectOme}{\vectLam})(\dprod{\vectr}{\unitplz})]\\
  &\quad+\parve^2[(\dprod{\vectOme}{\vectLam})(\dprod{\vectr}{\fdot{\vectLam}})-(\dprod{\vectOme}{\fdot{\vectLam}})(\dprod{\vectr}{\vectLam})]
  \beqref{rot1a}, \eqnref{rpath1b}\text{ \& }\eqnref{alg2}
\end{split}
\end{align*}
\begin{align}\label{rpath25l}
\begin{split}
&=-\parvx\epsvm+\parvz\epsvl+\efkob\frktl+\efkoc\frktm
  +\parvm(\epsva\epsvb-\Omerep^2\epsvd)+\parvn(\epsva\scalr^2-\epsvb\epsvd)\\
  &\quad+\parvo(\epsva\epsvh-\epsvg\epsvd)+\parvb\parve(\epsva\vsign-\vsigc\epsvd)
  +\parvp(\epsva\epsvf-\epsvc\epsvd)+\parvq(\Omerep^2\scalr^2-\epsvb^2)\\
  &\quad+\parvr(\Omerep^2\epsvh-\epsvg\epsvb)+\parvs(\Omerep^2\vsign-\vsigc\epsvb)
  +\parvt(\Omerep^2\epsvf-\epsvc\epsvb)-\parve\parvd(\epsvb\vsign-\vsigc\scalr^2)\\
  &\quad+\parvu(\epsvb\epsvh-\epsvg\scalr^2)+\parvv(\epsvb\epsvf-\epsvc\scalr^2)
  -\ethvu\parve(\epsvc\vsign-\vsigc\epsvf)+\parvw(\epsvc\epsvh-\epsvg\epsvf)\\
  &\quad+\parve^2(\epsvg\vsign-\vsigc\epsvh)
  \beqref{rot1a}\text{ \& }\eqnref{rpath1a}
\end{split}
\nonumber\\
&=\efkoo\beqref{rpathx1e}
\end{align}
\begin{align*}
&\dprod{(\cprod{\vectOme}{\vectLam})}{(\vscrp+\vscrq)}\nonumber\\
\begin{split}
&=\dprod{(\cprod{\vectOme}{\vectLam})}{[}\parvx\vectLam+\parvy\vectr+\parvz\unitplz+\efkoa\vectOme
  +\efkob\fdot{\vectLam}+\efkoc\ffdot{\vectLam}
  +\parvm(\cprod{\unitkap}{\vectOme})+\parvn(\cprod{\unitkap}{\vectr})+\parvo(\cprod{\unitkap}{\vectLam})\\
  &\quad+\parvb\parve(\cprod{\unitkap}{\fdot{\vectLam}})+\parvp(\cprod{\unitkap}{\unitplz})
  +\parvq(\cprod{\vectOme}{\vectr})+\parvr(\cprod{\vectOme}{\vectLam})
  +\parvs(\cprod{\vectOme}{\fdot{\vectLam}})+\parvt(\cprod{\vectOme}{\unitplz})\\
  &\quad-\parve\parvd(\cprod{\vectr}{\fdot{\vectLam}})
  +\parvu(\cprod{\vectr}{\vectLam})+\parvv(\cprod{\vectr}{\unitplz})
  -\ethvu\parve(\cprod{\unitplz}{\fdot{\vectLam}})
  +\parvw(\cprod{\unitplz}{\vectLam})+\parve^2(\cprod{\vectLam}{\fdot{\vectLam}})]
  \beqref{rpath24}
\end{split}
\end{align*}
\begin{align*}
\begin{split}
&=\parvx[\dprod{\vectLam}{(\cprod{\vectOme}{\vectLam})}]
  +\parvy[\dprod{\vectr}{(\cprod{\vectOme}{\vectLam})}]
  +\parvz[\dprod{\unitplz}{(\cprod{\vectOme}{\vectLam})}]
  +\efkoa[\dprod{\vectOme}{(\cprod{\vectOme}{\vectLam})}]
  +\efkob[\dprod{\fdot{\vectLam}}{(\cprod{\vectOme}{\vectLam})}]\\
  &\quad+\efkoc[\dprod{\ffdot{\vectLam}}{(\cprod{\vectOme}{\vectLam})}]
  +\parvm[\dprod{(\cprod{\vectOme}{\vectLam})}{(\cprod{\unitkap}{\vectOme})}]
  +\parvn[\dprod{(\cprod{\vectOme}{\vectLam})}{(\cprod{\unitkap}{\vectr})}]
  +\parvo[\dprod{(\cprod{\vectOme}{\vectLam})}{(\cprod{\unitkap}{\vectLam})}]\\
  &\quad+\parvb\parve[\dprod{(\cprod{\vectOme}{\vectLam})}{(\cprod{\unitkap}{\fdot{\vectLam}})}]
  +\parvp[\dprod{(\cprod{\vectOme}{\vectLam})}{(\cprod{\unitkap}{\unitplz})}]
  +\parvq[\dprod{(\cprod{\vectOme}{\vectLam})}{(\cprod{\vectOme}{\vectr})}]\\
  &\quad+\parvr[\dprod{(\cprod{\vectOme}{\vectLam})}{(\cprod{\vectOme}{\vectLam})}]
  +\parvs[\dprod{(\cprod{\vectOme}{\vectLam})}{(\cprod{\vectOme}{\fdot{\vectLam}})}]
  +\parvt[\dprod{(\cprod{\vectOme}{\vectLam})}{(\cprod{\vectOme}{\unitplz})}]\\
  &\quad-\parve\parvd[\dprod{(\cprod{\vectOme}{\vectLam})}{(\cprod{\vectr}{\fdot{\vectLam}})}]
  +\parvu[\dprod{(\cprod{\vectOme}{\vectLam})}{(\cprod{\vectr}{\vectLam})}]
  +\parvv[\dprod{(\cprod{\vectOme}{\vectLam})}{(\cprod{\vectr}{\unitplz})}]\\
  &\quad-\ethvu\parve[\dprod{(\cprod{\vectOme}{\vectLam})}{(\cprod{\unitplz}{\fdot{\vectLam}})}]
  +\parvw[\dprod{(\cprod{\vectOme}{\vectLam})}{(\cprod{\unitplz}{\vectLam})}]
  +\parve^2[\dprod{(\cprod{\vectOme}{\vectLam})}{(\cprod{\vectLam}{\fdot{\vectLam}})}]
\end{split}
\end{align*}
\begin{align*}
\begin{split}
&=\parvy\epsvm+\parvz\frkto+\efkob\frktv+\efkoc\frktz
  +\parvm[(\dprod{\vectOme}{\unitkap})(\dprod{\vectLam}{\vectOme})-\Omerep^2(\dprod{\vectLam}{\unitkap})]\\
  &\quad+\parvn[(\dprod{\vectOme}{\unitkap})(\dprod{\vectLam}{\vectr})-(\dprod{\vectOme}{\vectr})(\dprod{\vectLam}{\unitkap})]
  +\parvo[(\dprod{\vectOme}{\unitkap})\Lamrep^2-(\dprod{\vectOme}{\vectLam})(\dprod{\vectLam}{\unitkap})]\\
  &\quad+\parvb\parve[(\dprod{\vectOme}{\unitkap})(\dprod{\vectLam}{\fdot{\vectLam}})-(\dprod{\vectOme}{\fdot{\vectLam}})(\dprod{\vectLam}{\unitkap})]
  +\parvp[(\dprod{\vectOme}{\unitkap})(\dprod{\vectLam}{\unitplz})-(\dprod{\vectOme}{\unitplz})(\dprod{\vectLam}{\unitkap})]\\
  &\quad+\parvq[\Omerep^2(\dprod{\vectLam}{\vectr})-(\dprod{\vectOme}{\vectr})(\dprod{\vectLam}{\vectOme})]
  +\parvr[\Omerep^2\Lamrep^2-(\dprod{\vectOme}{\vectLam})^2]
  +\parvs[\Omerep^2(\dprod{\vectLam}{\fdot{\vectLam}})-(\dprod{\vectOme}{\fdot{\vectLam}})(\dprod{\vectLam}{\vectOme})]\\
  &\quad+\parvt[\Omerep^2(\dprod{\vectLam}{\unitplz})-(\dprod{\vectOme}{\unitplz})(\dprod{\vectLam}{\vectOme})]
  -\parve\parvd[(\dprod{\vectOme}{\vectr})(\dprod{\vectLam}{\fdot{\vectLam}})-(\dprod{\vectOme}{\fdot{\vectLam}})(\dprod{\vectLam}{\vectr})]\\
  &\quad+\parvu[(\dprod{\vectOme}{\vectr})\Lamrep^2-(\dprod{\vectOme}{\vectLam})(\dprod{\vectLam}{\vectr})]
  +\parvv[(\dprod{\vectOme}{\vectr})(\dprod{\vectLam}{\unitplz})-(\dprod{\vectOme}{\unitplz})(\dprod{\vectLam}{\vectr})]\\
  &\quad-\ethvu\parve[(\dprod{\vectOme}{\unitplz})(\dprod{\vectLam}{\fdot{\vectLam}})-(\dprod{\vectOme}{\fdot{\vectLam}})(\dprod{\vectLam}{\unitplz})]
  +\parvw[(\dprod{\vectOme}{\unitplz})\Lamrep^2-(\dprod{\vectOme}{\vectLam})(\dprod{\vectLam}{\unitplz})]\\
  &\quad+\parve^2[(\dprod{\vectOme}{\vectLam})(\dprod{\vectLam}{\fdot{\vectLam}})-(\dprod{\vectOme}{\fdot{\vectLam}})\Lamrep^2]
  \beqref{rot1a}, \eqnref{rpath1b}, \eqnref{rpath1b2}\text{ \& }\eqnref{alg2}
\end{split}
\end{align*}
\begin{align}\label{rpath25m}
\begin{split}
&=\parvy\epsvm+\parvz\frkto+\efkob\frktv+\efkoc\frktz
  +\parvm(\epsva\epsvg-\Omerep^2\dltva)+\parvn(\epsva\epsvh-\epsvb\dltva)\\
  &\quad+\parvo(\epsva\Lamrep^2-\epsvg\dltva)+\parvb\parve(\epsva\vsigd-\vsigc\dltva)
  +\parvp(\epsva\epsvi-\epsvc\dltva)+\parvq(\Omerep^2\epsvh-\epsvb\epsvg)\\
  &\quad+\parvr(\Omerep^2\Lamrep^2-\epsvg^2)+\parvs(\Omerep^2\vsigd-\vsigc\epsvg)
  +\parvt(\Omerep^2\epsvi-\epsvc\epsvg)-\parve\parvd(\epsvb\vsigd-\vsigc\epsvh)\\
  &\quad+\parvu(\epsvb\Lamrep^2-\epsvg\epsvh)+\parvv(\epsvb\epsvi-\epsvc\epsvh)
  -\ethvu\parve(\epsvc\vsigd-\vsigc\epsvi)+\parvw(\epsvc\Lamrep^2-\epsvg\epsvi)\\
  &\quad+\parve^2(\epsvg\vsigd-\vsigc\Lamrep^2)
  \beqref{rot1a}, \eqnref{rxpeed1a}\text{ \& }\eqnref{rpath1a}
\end{split}
\nonumber\\
&=\efkop\beqref{rpathx1f}
\end{align}
\begin{align*}
&\dprod{(\cprod{\vectOme}{\fdot{\vectLam}})}{(\vscrp+\vscrq)}\nonumber\\
\begin{split}
&=\dprod{(\cprod{\vectOme}{\fdot{\vectLam}})}{[}\parvx\vectLam+\parvy\vectr+\parvz\unitplz+\efkoa\vectOme
  +\efkob\fdot{\vectLam}+\efkoc\ffdot{\vectLam}
  +\parvm(\cprod{\unitkap}{\vectOme})+\parvn(\cprod{\unitkap}{\vectr})+\parvo(\cprod{\unitkap}{\vectLam})\\
  &\quad+\parvb\parve(\cprod{\unitkap}{\fdot{\vectLam}})+\parvp(\cprod{\unitkap}{\unitplz})
  +\parvq(\cprod{\vectOme}{\vectr})+\parvr(\cprod{\vectOme}{\vectLam})
  +\parvs(\cprod{\vectOme}{\fdot{\vectLam}})+\parvt(\cprod{\vectOme}{\unitplz})\\
  &\quad-\parve\parvd(\cprod{\vectr}{\fdot{\vectLam}})
  +\parvu(\cprod{\vectr}{\vectLam})+\parvv(\cprod{\vectr}{\unitplz})
  -\ethvu\parve(\cprod{\unitplz}{\fdot{\vectLam}})
  +\parvw(\cprod{\unitplz}{\vectLam})+\parve^2(\cprod{\vectLam}{\fdot{\vectLam}})]
  \beqref{rpath24}
\end{split}
\end{align*}
\begin{align*}
\begin{split}
&=\parvx[\dprod{\vectLam}{(\cprod{\vectOme}{\fdot{\vectLam}})}]
  +\parvy[\dprod{\vectr}{(\cprod{\vectOme}{\fdot{\vectLam}})}]
  +\parvz[\dprod{\unitplz}{(\cprod{\vectOme}{\fdot{\vectLam}})}]
  +\efkoa[\dprod{\vectOme}{(\cprod{\vectOme}{\fdot{\vectLam}})}]
  +\efkob[\dprod{\fdot{\vectLam}}{(\cprod{\vectOme}{\fdot{\vectLam}})}]\\
  &\quad+\efkoc[\dprod{\ffdot{\vectLam}}{(\cprod{\vectOme}{\fdot{\vectLam}})}]
  +\parvm[\dprod{(\cprod{\vectOme}{\fdot{\vectLam}})}{(\cprod{\unitkap}{\vectOme})}]
  +\parvn[\dprod{(\cprod{\vectOme}{\fdot{\vectLam}})}{(\cprod{\unitkap}{\vectr})}]
  +\parvo[\dprod{(\cprod{\vectOme}{\fdot{\vectLam}})}{(\cprod{\unitkap}{\vectLam})}]\\
  &\quad+\parvb\parve[\dprod{(\cprod{\vectOme}{\fdot{\vectLam}})}{(\cprod{\unitkap}{\fdot{\vectLam}})}]
  +\parvp[\dprod{(\cprod{\vectOme}{\fdot{\vectLam}})}{(\cprod{\unitkap}{\unitplz})}]
  +\parvq[\dprod{(\cprod{\vectOme}{\fdot{\vectLam}})}{(\cprod{\vectOme}{\vectr})}]\\
  &\quad+\parvr[\dprod{(\cprod{\vectOme}{\fdot{\vectLam}})}{(\cprod{\vectOme}{\vectLam})}]
  +\parvs[\dprod{(\cprod{\vectOme}{\fdot{\vectLam}})}{(\cprod{\vectOme}{\fdot{\vectLam}})}]
  +\parvt[\dprod{(\cprod{\vectOme}{\fdot{\vectLam}})}{(\cprod{\vectOme}{\unitplz})}]\\
  &\quad-\parve\parvd[\dprod{(\cprod{\vectOme}{\fdot{\vectLam}})}{(\cprod{\vectr}{\fdot{\vectLam}})}]
  +\parvu[\dprod{(\cprod{\vectOme}{\fdot{\vectLam}})}{(\cprod{\vectr}{\vectLam})}]
  +\parvv[\dprod{(\cprod{\vectOme}{\fdot{\vectLam}})}{(\cprod{\vectr}{\unitplz})}]\\
  &\quad-\ethvu\parve[\dprod{(\cprod{\vectOme}{\fdot{\vectLam}})}{(\cprod{\unitplz}{\fdot{\vectLam}})}]
  +\parvw[\dprod{(\cprod{\vectOme}{\fdot{\vectLam}})}{(\cprod{\unitplz}{\vectLam})}]
  +\parve^2[\dprod{(\cprod{\vectOme}{\fdot{\vectLam}})}{(\cprod{\vectLam}{\fdot{\vectLam}})}]
\end{split}
\end{align*}
\begin{align*}
\begin{split}
&=-\parvx\frktv-\parvy\frktl+\parvz\frktp+\efkoc\frkxa
  +\parvm[(\dprod{\vectOme}{\unitkap})(\dprod{\fdot{\vectLam}}{\vectOme})-\Omerep^2(\dprod{\fdot{\vectLam}}{\unitkap})]\\
  &\quad+\parvn[(\dprod{\vectOme}{\unitkap})(\dprod{\fdot{\vectLam}}{\vectr})-(\dprod{\vectOme}{\vectr})(\dprod{\fdot{\vectLam}}{\unitkap})]
  +\parvo[(\dprod{\vectOme}{\unitkap})(\dprod{\fdot{\vectLam}}{\vectLam})-(\dprod{\vectOme}{\vectLam})(\dprod{\fdot{\vectLam}}{\unitkap})]\\
  &\quad+\parvb\parve[(\dprod{\vectOme}{\unitkap})\fdot{\vectLam}^2-(\dprod{\vectOme}{\fdot{\vectLam}})(\dprod{\fdot{\vectLam}}{\unitkap})]
  +\parvp[(\dprod{\vectOme}{\unitkap})(\dprod{\fdot{\vectLam}}{\unitplz})-(\dprod{\vectOme}{\unitplz})(\dprod{\fdot{\vectLam}}{\unitkap})]\\
  &\quad+\parvq[\Omerep^2(\dprod{\fdot{\vectLam}}{\vectr})-(\dprod{\vectOme}{\vectr})(\dprod{\fdot{\vectLam}}{\vectOme})]
  +\parvr[\Omerep^2(\dprod{\fdot{\vectLam}}{\vectLam})-(\dprod{\vectOme}{\vectLam})(\dprod{\fdot{\vectLam}}{\vectOme})]
  +\parvs[\Omerep^2\fdot{\vectLam}^2-(\dprod{\vectOme}{\fdot{\vectLam}})^2]\\
  &\quad+\parvt[\Omerep^2(\dprod{\fdot{\vectLam}}{\unitplz})-(\dprod{\vectOme}{\unitplz})(\dprod{\fdot{\vectLam}}{\vectOme})]
  -\parve\parvd[(\dprod{\vectOme}{\vectr})\fdot{\vectLam}^2-(\dprod{\vectOme}{\fdot{\vectLam}})(\dprod{\fdot{\vectLam}}{\vectr})]\\
  &\quad+\parvu[(\dprod{\vectOme}{\vectr})(\dprod{\fdot{\vectLam}}{\vectLam})-(\dprod{\vectOme}{\vectLam})(\dprod{\fdot{\vectLam}}{\vectr})]
  +\parvv[(\dprod{\vectOme}{\vectr})(\dprod{\fdot{\vectLam}}{\unitplz})-(\dprod{\vectOme}{\unitplz})(\dprod{\fdot{\vectLam}}{\vectr})]\\
  &\quad-\ethvu\parve[(\dprod{\vectOme}{\unitplz})\fdot{\vectLam}^2-(\dprod{\vectOme}{\fdot{\vectLam}})(\dprod{\fdot{\vectLam}}{\unitplz})]
  +\parvw[(\dprod{\vectOme}{\unitplz})(\dprod{\fdot{\vectLam}}{\vectLam})-(\dprod{\vectOme}{\vectLam})(\dprod{\fdot{\vectLam}}{\unitplz})]\\
  &\quad+\parve^2[(\dprod{\vectOme}{\vectLam})\fdot{\vectLam}^2-(\dprod{\vectOme}{\fdot{\vectLam}})(\dprod{\fdot{\vectLam}}{\vectLam})]
  \beqref{rpath1b}, \eqnref{rpath1b2}\text{ \& }\eqnref{alg2}
\end{split}
\end{align*}
\begin{align}\label{rpath25n}
\begin{split}
&=-\parvx\frktv-\parvy\frktl+\parvz\frktp+\efkoc\frkxa
  +\parvm(\epsva\vsigc-\Omerep^2\vsigb)+\parvn(\epsva\vsign-\epsvb\vsigb)+\parvo(\epsva\vsigd-\epsvg\vsigb)\\
  &\quad+\parvb\parve(\epsva\vsige-\vsigc\vsigb)+\parvp(\epsva\vsiga-\epsvc\vsigb)
  +\parvq(\Omerep^2\vsign-\epsvb\vsigc)+\parvr(\Omerep^2\vsigd-\epsvg\vsigc)\\
  &\quad+\parvs(\Omerep^2\vsige-\vsigc^2)+\parvt(\Omerep^2\vsiga-\epsvc\vsigc)
  -\parve\parvd(\epsvb\vsige-\vsigc\vsign)+\parvu(\epsvb\vsigd-\epsvg\vsign)
  +\parvv(\epsvb\vsiga-\epsvc\vsign)\\
  &\quad-\ethvu\parve(\epsvc\vsige-\vsigc\vsiga)
  +\parvw(\epsvc\vsigd-\epsvg\vsiga)+\parve^2(\epsvg\vsige-\vsigc\vsigd)
  \beqref{rot1a}\text{ \& }\eqnref{rpath1a}
\end{split}
\nonumber\\
&=\efkoq\beqref{rpathx1f2}
\end{align}
\begin{align*}
&\dprod{(\cprod{\vectOme}{\unitplz})}{(\vscrp+\vscrq)}\nonumber\\
\begin{split}
&=\dprod{(\cprod{\vectOme}{\unitplz})}{[}\parvx\vectLam+\parvy\vectr+\parvz\unitplz+\efkoa\vectOme
  +\efkob\fdot{\vectLam}+\efkoc\ffdot{\vectLam}
  +\parvm(\cprod{\unitkap}{\vectOme})+\parvn(\cprod{\unitkap}{\vectr})+\parvo(\cprod{\unitkap}{\vectLam})\\
  &\quad+\parvb\parve(\cprod{\unitkap}{\fdot{\vectLam}})+\parvp(\cprod{\unitkap}{\unitplz})
  +\parvq(\cprod{\vectOme}{\vectr})+\parvr(\cprod{\vectOme}{\vectLam})
  +\parvs(\cprod{\vectOme}{\fdot{\vectLam}})+\parvt(\cprod{\vectOme}{\unitplz})\\
  &\quad-\parve\parvd(\cprod{\vectr}{\fdot{\vectLam}})
  +\parvu(\cprod{\vectr}{\vectLam})+\parvv(\cprod{\vectr}{\unitplz})
  -\ethvu\parve(\cprod{\unitplz}{\fdot{\vectLam}})
  +\parvw(\cprod{\unitplz}{\vectLam})+\parve^2(\cprod{\vectLam}{\fdot{\vectLam}})]
  \beqref{rpath24}
\end{split}
\end{align*}
\begin{align*}
\begin{split}
&=\parvx[\dprod{\vectLam}{(\cprod{\vectOme}{\unitplz})}]
  +\parvy[\dprod{\vectr}{(\cprod{\vectOme}{\unitplz})}]
  +\parvz[\dprod{\unitplz}{(\cprod{\vectOme}{\unitplz})}]
  +\efkoa[\dprod{\vectOme}{(\cprod{\vectOme}{\unitplz})}]
  +\efkob[\dprod{\fdot{\vectLam}}{(\cprod{\vectOme}{\unitplz})}]\\
  &\quad+\efkoc[\dprod{\ffdot{\vectLam}}{(\cprod{\vectOme}{\unitplz})}]
  +\parvm[\dprod{(\cprod{\vectOme}{\unitplz})}{(\cprod{\unitkap}{\vectOme})}]
  +\parvn[\dprod{(\cprod{\vectOme}{\unitplz})}{(\cprod{\unitkap}{\vectr})}]
  +\parvo[\dprod{(\cprod{\vectOme}{\unitplz})}{(\cprod{\unitkap}{\vectLam})}]\\
  &\quad+\parvb\parve[\dprod{(\cprod{\vectOme}{\unitplz})}{(\cprod{\unitkap}{\fdot{\vectLam}})}]
  +\parvp[\dprod{(\cprod{\vectOme}{\unitplz})}{(\cprod{\unitkap}{\unitplz})}]
  +\parvq[\dprod{(\cprod{\vectOme}{\unitplz})}{(\cprod{\vectOme}{\vectr})}]\\
  &\quad+\parvr[\dprod{(\cprod{\vectOme}{\unitplz})}{(\cprod{\vectOme}{\vectLam})}]
  +\parvs[\dprod{(\cprod{\vectOme}{\unitplz})}{(\cprod{\vectOme}{\fdot{\vectLam}})}]
  +\parvt[\dprod{(\cprod{\vectOme}{\unitplz})}{(\cprod{\vectOme}{\unitplz})}]\\
  &\quad-\parve\parvd[\dprod{(\cprod{\vectOme}{\unitplz})}{(\cprod{\vectr}{\fdot{\vectLam}})}]
  +\parvu[\dprod{(\cprod{\vectOme}{\unitplz})}{(\cprod{\vectr}{\vectLam})}]
  +\parvv[\dprod{(\cprod{\vectOme}{\unitplz})}{(\cprod{\vectr}{\unitplz})}]\\
  &\quad-\ethvu\parve[\dprod{(\cprod{\vectOme}{\unitplz})}{(\cprod{\unitplz}{\fdot{\vectLam}})}]
  +\parvw[\dprod{(\cprod{\vectOme}{\unitplz})}{(\cprod{\unitplz}{\vectLam})}]
  +\parve^2[\dprod{(\cprod{\vectOme}{\unitplz})}{(\cprod{\vectLam}{\fdot{\vectLam}})}]
\end{split}
\end{align*}
\begin{align*}
\begin{split}
&=-\parvx\frkto-\parvy\epsvl-\efkob\frktp-\efkoc\frktq
  +\parvm[(\dprod{\vectOme}{\unitkap})(\dprod{\unitplz}{\vectOme})-\Omerep^2(\dprod{\unitplz}{\unitkap})]\\
  &\quad+\parvn[(\dprod{\vectOme}{\unitkap})(\dprod{\unitplz}{\vectr})-(\dprod{\vectOme}{\vectr})(\dprod{\unitplz}{\unitkap})]
  +\parvo[(\dprod{\vectOme}{\unitkap})(\dprod{\unitplz}{\vectLam})-(\dprod{\vectOme}{\vectLam})(\dprod{\unitplz}{\unitkap})]\\
  &\quad+\parvb\parve[(\dprod{\vectOme}{\unitkap})(\dprod{\unitplz}{\fdot{\vectLam}})-(\dprod{\vectOme}{\fdot{\vectLam}})(\dprod{\unitplz}{\unitkap})]
  +\parvp[(\dprod{\vectOme}{\unitkap})-(\dprod{\vectOme}{\unitplz})(\dprod{\unitplz}{\unitkap})]\\
  &\quad+\parvq[\Omerep^2(\dprod{\unitplz}{\vectr})-(\dprod{\vectOme}{\vectr})(\dprod{\unitplz}{\vectOme})]
  +\parvr[\Omerep^2(\dprod{\unitplz}{\vectLam})-(\dprod{\vectOme}{\vectLam})(\dprod{\unitplz}{\vectOme})]\\
  &\quad+\parvs[\Omerep^2(\dprod{\unitplz}{\fdot{\vectLam}})-(\dprod{\vectOme}{\fdot{\vectLam}})(\dprod{\unitplz}{\vectOme})]
  +\parvt[\Omerep^2-(\dprod{\vectOme}{\unitplz})^2]
  -\parve\parvd[(\dprod{\vectOme}{\vectr})(\dprod{\unitplz}{\fdot{\vectLam}})-(\dprod{\vectOme}{\fdot{\vectLam}})(\dprod{\unitplz}{\vectr})]\\
  &\quad+\parvu[(\dprod{\vectOme}{\vectr})(\dprod{\unitplz}{\vectLam})-(\dprod{\vectOme}{\vectLam})(\dprod{\unitplz}{\vectr})]
  +\parvv[(\dprod{\vectOme}{\vectr})-(\dprod{\vectOme}{\unitplz})(\dprod{\unitplz}{\vectr})]\\
  &\quad-\ethvu\parve[(\dprod{\vectOme}{\unitplz})(\dprod{\unitplz}{\fdot{\vectLam}})-(\dprod{\vectOme}{\fdot{\vectLam}})]
  +\parvw[(\dprod{\vectOme}{\unitplz})(\dprod{\unitplz}{\vectLam})-(\dprod{\vectOme}{\vectLam})]\\
  &\quad+\parve^2[(\dprod{\vectOme}{\vectLam})(\dprod{\unitplz}{\fdot{\vectLam}})-(\dprod{\vectOme}{\fdot{\vectLam}})(\dprod{\unitplz}{\vectLam})]
  \beqref{rot1a}, \eqnref{rpath1b}\text{ \& }\eqnref{alg2}
\end{split}
\end{align*}
\begin{align}\label{rpath25o}
\begin{split}
&=-\parvx\frkto-\parvy\epsvl-\efkob\frktp-\efkoc\frktq
  +\parvm(\epsva\epsvc-\Omerep^2\epsve)+\parvn(\epsva\epsvf-\epsvb\epsve)+\parvo(\epsva\epsvi-\epsvg\epsve)\\
  &\quad+\parvb\parve(\epsva\vsiga-\vsigc\epsve)+\parvp(\epsva-\epsvc\epsve)
  +\parvq(\Omerep^2\epsvf-\epsvb\epsvc)+\parvr(\Omerep^2\epsvi-\epsvg\epsvc)\\
  &\quad+\parvs(\Omerep^2\vsiga-\vsigc\epsvc)+\parvt(\Omerep^2-\epsvc^2)
  -\parve\parvd(\epsvb\vsiga-\vsigc\epsvf)+\parvu(\epsvb\epsvi-\epsvg\epsvf)+\parvv(\epsvb-\epsvc\epsvf)\\
  &\quad-\ethvu\parve(\epsvc\vsiga-\vsigc)+\parvw(\epsvc\epsvi-\epsvg)+\parve^2(\epsvg\vsiga-\vsigc\epsvi)
  \beqref{rot1a}\text{ \& }\eqnref{rpath1a}
\end{split}
\nonumber\\
&=\efkor\beqref{rpathx1g}
\end{align}
\begin{align*}
&\dprod{(\cprod{\vectr}{\fdot{\vectLam}})}{(\vscrp+\vscrq)}\nonumber\\
\begin{split}
&=\dprod{(\cprod{\vectr}{\fdot{\vectLam}})}{[}\parvx\vectLam+\parvy\vectr+\parvz\unitplz+\efkoa\vectOme
  +\efkob\fdot{\vectLam}+\efkoc\ffdot{\vectLam}
  +\parvm(\cprod{\unitkap}{\vectOme})+\parvn(\cprod{\unitkap}{\vectr})+\parvo(\cprod{\unitkap}{\vectLam})\\
  &\quad+\parvb\parve(\cprod{\unitkap}{\fdot{\vectLam}})+\parvp(\cprod{\unitkap}{\unitplz})
  +\parvq(\cprod{\vectOme}{\vectr})+\parvr(\cprod{\vectOme}{\vectLam})
  +\parvs(\cprod{\vectOme}{\fdot{\vectLam}})+\parvt(\cprod{\vectOme}{\unitplz})\\
  &\quad-\parve\parvd(\cprod{\vectr}{\fdot{\vectLam}})
  +\parvu(\cprod{\vectr}{\vectLam})+\parvv(\cprod{\vectr}{\unitplz})
  -\ethvu\parve(\cprod{\unitplz}{\fdot{\vectLam}})
  +\parvw(\cprod{\unitplz}{\vectLam})+\parve^2(\cprod{\vectLam}{\fdot{\vectLam}})]
  \beqref{rpath24}
\end{split}
\end{align*}
\begin{align*}
\begin{split}
&=\parvx[\dprod{\vectLam}{(\cprod{\vectr}{\fdot{\vectLam}})}]
  +\parvy[\dprod{\vectr}{(\cprod{\vectr}{\fdot{\vectLam}})}]
  +\parvz[\dprod{\unitplz}{(\cprod{\vectr}{\fdot{\vectLam}})}]
  +\efkoa[\dprod{\vectOme}{(\cprod{\vectr}{\fdot{\vectLam}})}]
  +\efkob[\dprod{\fdot{\vectLam}}{(\cprod{\vectr}{\fdot{\vectLam}})}]\\
  &\quad+\efkoc[\dprod{\ffdot{\vectLam}}{(\cprod{\vectr}{\fdot{\vectLam}})}]
  +\parvm[\dprod{(\cprod{\vectr}{\fdot{\vectLam}})}{(\cprod{\unitkap}{\vectOme})}]
  +\parvn[\dprod{(\cprod{\vectr}{\fdot{\vectLam}})}{(\cprod{\unitkap}{\vectr})}]
  +\parvo[\dprod{(\cprod{\vectr}{\fdot{\vectLam}})}{(\cprod{\unitkap}{\vectLam})}]\\
  &\quad+\parvb\parve[\dprod{(\cprod{\vectr}{\fdot{\vectLam}})}{(\cprod{\unitkap}{\fdot{\vectLam}})}]
  +\parvp[\dprod{(\cprod{\vectr}{\fdot{\vectLam}})}{(\cprod{\unitkap}{\unitplz})}]
  +\parvq[\dprod{(\cprod{\vectr}{\fdot{\vectLam}})}{(\cprod{\vectOme}{\vectr})}]\\
  &\quad+\parvr[\dprod{(\cprod{\vectr}{\fdot{\vectLam}})}{(\cprod{\vectOme}{\vectLam})}]
  +\parvs[\dprod{(\cprod{\vectr}{\fdot{\vectLam}})}{(\cprod{\vectOme}{\fdot{\vectLam}})}]
  +\parvt[\dprod{(\cprod{\vectr}{\fdot{\vectLam}})}{(\cprod{\vectOme}{\unitplz})}]\\
  &\quad-\parve\parvd[\dprod{(\cprod{\vectr}{\fdot{\vectLam}})}{(\cprod{\vectr}{\fdot{\vectLam}})}]
  +\parvu[\dprod{(\cprod{\vectr}{\fdot{\vectLam}})}{(\cprod{\vectr}{\vectLam})}]
  +\parvv[\dprod{(\cprod{\vectr}{\fdot{\vectLam}})}{(\cprod{\vectr}{\unitplz})}]\\
  &\quad-\ethvu\parve[\dprod{(\cprod{\vectr}{\fdot{\vectLam}})}{(\cprod{\unitplz}{\fdot{\vectLam}})}]
  +\parvw[\dprod{(\cprod{\vectr}{\fdot{\vectLam}})}{(\cprod{\unitplz}{\vectLam})}]
  +\parve^2[\dprod{(\cprod{\vectr}{\fdot{\vectLam}})}{(\cprod{\vectLam}{\fdot{\vectLam}})}]
\end{split}
\end{align*}
\begin{align*}
\begin{split}
&=\parvx\frkth-\parvz\frktn+\efkoa\frktl-\efkoc\frktj
  +\parvm[(\dprod{\vectr}{\unitkap})(\dprod{\fdot{\vectLam}}{\vectOme})-(\dprod{\vectr}{\vectOme})(\dprod{\fdot{\vectLam}}{\unitkap})]\\
  &\quad+\parvn[(\dprod{\vectr}{\unitkap})(\dprod{\fdot{\vectLam}}{\vectr})-\scalr^2(\dprod{\fdot{\vectLam}}{\unitkap})]
  +\parvo[(\dprod{\vectr}{\unitkap})(\dprod{\fdot{\vectLam}}{\vectLam})-(\dprod{\vectr}{\vectLam})(\dprod{\fdot{\vectLam}}{\unitkap})]\\
  &\quad+\parvb\parve[(\dprod{\vectr}{\unitkap})\fdot{\vectLam}^2-(\dprod{\vectr}{\fdot{\vectLam}})(\dprod{\fdot{\vectLam}}{\unitkap})]
  +\parvp[(\dprod{\vectr}{\unitkap})(\dprod{\fdot{\vectLam}}{\unitplz})-(\dprod{\vectr}{\unitplz})(\dprod{\fdot{\vectLam}}{\unitkap})]\\
  &\quad+\parvq[(\dprod{\vectr}{\vectOme})(\dprod{\fdot{\vectLam}}{\vectr})-\scalr^2(\dprod{\fdot{\vectLam}}{\vectOme})]
  +\parvr[(\dprod{\vectr}{\vectOme})(\dprod{\fdot{\vectLam}}{\vectLam})-(\dprod{\vectr}{\vectLam})(\dprod{\fdot{\vectLam}}{\vectOme})]\\
  &\quad+\parvs[(\dprod{\vectr}{\vectOme})\fdot{\vectLam}^2-(\dprod{\vectr}{\fdot{\vectLam}})(\dprod{\fdot{\vectLam}}{\vectOme})]
  +\parvt[(\dprod{\vectr}{\vectOme})(\dprod{\fdot{\vectLam}}{\unitplz})-(\dprod{\vectr}{\unitplz})(\dprod{\fdot{\vectLam}}{\vectOme})]\\
  &\quad-\parve\parvd[\scalr^2\fdot{\vectLam}^2-(\dprod{\vectr}{\fdot{\vectLam}})^2]
  +\parvu[\scalr^2(\dprod{\fdot{\vectLam}}{\vectLam})-(\dprod{\vectr}{\vectLam})(\dprod{\fdot{\vectLam}}{\vectr})]\\
  &\quad+\parvv[\scalr^2(\dprod{\fdot{\vectLam}}{\unitplz})-(\dprod{\vectr}{\unitplz})(\dprod{\fdot{\vectLam}}{\vectr})]
  -\ethvu\parve[(\dprod{\vectr}{\unitplz})\fdot{\vectLam}^2-(\dprod{\vectr}{\fdot{\vectLam}})(\dprod{\fdot{\vectLam}}{\unitplz})]\\
  &\quad+\parvw[(\dprod{\vectr}{\unitplz})(\dprod{\fdot{\vectLam}}{\vectLam})-(\dprod{\vectr}{\vectLam})(\dprod{\fdot{\vectLam}}{\unitplz})]
  +\parve^2[(\dprod{\vectr}{\vectLam})\fdot{\vectLam}^2-(\dprod{\vectr}{\fdot{\vectLam}})(\dprod{\fdot{\vectLam}}{\vectLam})]\\
  &\quad\beqref{rpath1b}\text{ \& }\eqnref{alg2}
\end{split}
\end{align*}
\begin{align}\label{rpath25p}
\begin{split}
&=\parvx\frkth-\parvz\frktn+\efkoa\frktl-\efkoc\frktj
  +\parvm(\epsvd\vsigc-\epsvb\vsigb)+\parvn(\epsvd\vsign-\scalr^2\vsigb)+\parvo(\epsvd\vsigd-\epsvh\vsigb)\\
  &\quad+\parvb\parve(\epsvd\vsige-\vsign\vsigb)+\parvp(\epsvd\vsiga-\epsvf\vsigb)
  +\parvq(\epsvb\vsign-\scalr^2\vsigc)+\parvr(\epsvb\vsigd-\epsvh\vsigc)\\
  &\quad+\parvs(\epsvb\vsige-\vsign\vsigc)+\parvt(\epsvb\vsiga-\epsvf\vsigc)
  -\parve\parvd(\scalr^2\vsige-\vsign^2)+\parvu(\scalr^2\vsigd-\epsvh\vsign)\\
  &\quad+\parvv(\scalr^2\vsiga-\epsvf\vsign)-\ethvu\parve(\epsvf\vsige-\vsign\vsiga)
  +\parvw(\epsvf\vsigd-\epsvh\vsiga)+\parve^2(\epsvh\vsige-\vsign\vsigd)\\
  &\quad\beqref{rot1a}\text{ \& }\eqnref{rpath1a}
\end{split}
\nonumber\\
&=\efkos\beqref{rpathx1g}
\end{align}
\begin{align*}
&\dprod{(\cprod{\vectr}{\vectLam})}{(\vscrp+\vscrq)}\nonumber\\
\begin{split}
&=\dprod{(\cprod{\vectr}{\vectLam})}{[}\parvx\vectLam+\parvy\vectr+\parvz\unitplz+\efkoa\vectOme
  +\efkob\fdot{\vectLam}+\efkoc\ffdot{\vectLam}
  +\parvm(\cprod{\unitkap}{\vectOme})+\parvn(\cprod{\unitkap}{\vectr})+\parvo(\cprod{\unitkap}{\vectLam})\\
  &\quad+\parvb\parve(\cprod{\unitkap}{\fdot{\vectLam}})+\parvp(\cprod{\unitkap}{\unitplz})
  +\parvq(\cprod{\vectOme}{\vectr})+\parvr(\cprod{\vectOme}{\vectLam})
  +\parvs(\cprod{\vectOme}{\fdot{\vectLam}})+\parvt(\cprod{\vectOme}{\unitplz})\\
  &\quad-\parve\parvd(\cprod{\vectr}{\fdot{\vectLam}})
  +\parvu(\cprod{\vectr}{\vectLam})+\parvv(\cprod{\vectr}{\unitplz})
  -\ethvu\parve(\cprod{\unitplz}{\fdot{\vectLam}})
  +\parvw(\cprod{\unitplz}{\vectLam})+\parve^2(\cprod{\vectLam}{\fdot{\vectLam}})]
  \beqref{rpath24}
\end{split}
\end{align*}
\begin{align*}
\begin{split}
&=\parvx[\dprod{\vectLam}{(\cprod{\vectr}{\vectLam})}]
  +\parvy[\dprod{\vectr}{(\cprod{\vectr}{\vectLam})}]
  +\parvz[\dprod{\unitplz}{(\cprod{\vectr}{\vectLam})}]
  +\efkoa[\dprod{\vectOme}{(\cprod{\vectr}{\vectLam})}]
  +\efkob[\dprod{\fdot{\vectLam}}{(\cprod{\vectr}{\vectLam})}]\\
  &\quad+\efkoc[\dprod{\ffdot{\vectLam}}{(\cprod{\vectr}{\vectLam})}]
  +\parvm[\dprod{(\cprod{\vectr}{\vectLam})}{(\cprod{\unitkap}{\vectOme})}]
  +\parvn[\dprod{(\cprod{\vectr}{\vectLam})}{(\cprod{\unitkap}{\vectr})}]
  +\parvo[\dprod{(\cprod{\vectr}{\vectLam})}{(\cprod{\unitkap}{\vectLam})}]\\
  &\quad+\parvb\parve[\dprod{(\cprod{\vectr}{\vectLam})}{(\cprod{\unitkap}{\fdot{\vectLam}})}]
  +\parvp[\dprod{(\cprod{\vectr}{\vectLam})}{(\cprod{\unitkap}{\unitplz})}]
  +\parvq[\dprod{(\cprod{\vectr}{\vectLam})}{(\cprod{\vectOme}{\vectr})}]\\
  &\quad+\parvr[\dprod{(\cprod{\vectr}{\vectLam})}{(\cprod{\vectOme}{\vectLam})}]
  +\parvs[\dprod{(\cprod{\vectr}{\vectLam})}{(\cprod{\vectOme}{\fdot{\vectLam}})}]
  +\parvt[\dprod{(\cprod{\vectr}{\vectLam})}{(\cprod{\vectOme}{\unitplz})}]\\
  &\quad-\parve\parvd[\dprod{(\cprod{\vectr}{\vectLam})}{(\cprod{\vectr}{\fdot{\vectLam}})}]
  +\parvu[\dprod{(\cprod{\vectr}{\vectLam})}{(\cprod{\vectr}{\vectLam})}]
  +\parvv[\dprod{(\cprod{\vectr}{\vectLam})}{(\cprod{\vectr}{\unitplz})}]\\
  &\quad-\ethvu\parve[\dprod{(\cprod{\vectr}{\vectLam})}{(\cprod{\unitplz}{\fdot{\vectLam}})}]
  +\parvw[\dprod{(\cprod{\vectr}{\vectLam})}{(\cprod{\unitplz}{\vectLam})}]
  +\parve^2[\dprod{(\cprod{\vectr}{\vectLam})}{(\cprod{\vectLam}{\fdot{\vectLam}})}]
\end{split}
\end{align*}
\begin{align*}
\begin{split}
&=-\parvz\epsvo-\efkoa\epsvm-\efkob\frkth-\efkoc\frkti
  +\parvm[(\dprod{\vectr}{\unitkap})(\dprod{\vectLam}{\vectOme})-(\dprod{\vectr}{\vectOme})(\dprod{\vectLam}{\unitkap})]\\
  &\quad+\parvn[(\dprod{\vectr}{\unitkap})(\dprod{\vectLam}{\vectr})-\scalr^2(\dprod{\vectLam}{\unitkap})]
  +\parvo[(\dprod{\vectr}{\unitkap})\Lamrep^2-(\dprod{\vectr}{\vectLam})(\dprod{\vectLam}{\unitkap})]\\
  &\quad+\parvb\parve[(\dprod{\vectr}{\unitkap})(\dprod{\vectLam}{\fdot{\vectLam}})-(\dprod{\vectr}{\fdot{\vectLam}})(\dprod{\vectLam}{\unitkap})]
  +\parvp[(\dprod{\vectr}{\unitkap})(\dprod{\vectLam}{\unitplz})-(\dprod{\vectr}{\unitplz})(\dprod{\vectLam}{\unitkap})]\\
  &\quad+\parvq[(\dprod{\vectr}{\vectOme})(\dprod{\vectLam}{\vectr})-\scalr^2(\dprod{\vectLam}{\vectOme})]
  +\parvr[(\dprod{\vectr}{\vectOme})\Lamrep^2-(\dprod{\vectr}{\vectLam})(\dprod{\vectLam}{\vectOme})]\\
  &\quad+\parvs[(\dprod{\vectr}{\vectOme})(\dprod{\vectLam}{\fdot{\vectLam}})-(\dprod{\vectr}{\fdot{\vectLam}})(\dprod{\vectLam}{\vectOme})]
  +\parvt[(\dprod{\vectr}{\vectOme})(\dprod{\vectLam}{\unitplz})-(\dprod{\vectr}{\unitplz})(\dprod{\vectLam}{\vectOme})]\\
  &\quad-\parve\parvd[\scalr^2(\dprod{\vectLam}{\fdot{\vectLam}})-(\dprod{\vectr}{\fdot{\vectLam}})(\dprod{\vectLam}{\vectr})]
  +\parvu[\scalr^2\Lamrep^2-(\dprod{\vectr}{\vectLam})^2]
  +\parvv[\scalr^2(\dprod{\vectLam}{\unitplz})-(\dprod{\vectr}{\unitplz})(\dprod{\vectLam}{\vectr})]\\
  &\quad-\ethvu\parve[(\dprod{\vectr}{\unitplz})(\dprod{\vectLam}{\fdot{\vectLam}})-(\dprod{\vectr}{\fdot{\vectLam}})(\dprod{\vectLam}{\unitplz})]
  +\parvw[(\dprod{\vectr}{\unitplz})\Lamrep^2-(\dprod{\vectr}{\vectLam})(\dprod{\vectLam}{\unitplz})]\\
  &\quad+\parve^2[(\dprod{\vectr}{\vectLam})(\dprod{\vectLam}{\fdot{\vectLam}})-(\dprod{\vectr}{\fdot{\vectLam}})\Lamrep^2]
  \beqref{rot1a}, \eqnref{rpath1b}\text{ \& }\eqnref{alg2}
\end{split}
\end{align*}
\begin{align}\label{rpath25q}
\begin{split}
&=-\parvz\epsvo-\efkoa\epsvm-\efkob\frkth-\efkoc\frkti
  +\parvm(\epsvd\epsvg-\epsvb\dltva)+\parvn(\epsvd\epsvh-\scalr^2\dltva)+\parvo(\epsvd\Lamrep^2-\epsvh\dltva)\\
  &\quad+\parvb\parve(\epsvd\vsigd-\vsign\dltva)+\parvp(\epsvd\epsvi-\epsvf\dltva)
  +\parvq(\epsvb\epsvh-\scalr^2\epsvg)+\parvr(\epsvb\Lamrep^2-\epsvh\epsvg)\\
  &\quad+\parvs(\epsvb\vsigd-\vsign\epsvg)+\parvt(\epsvb\epsvi-\epsvf\epsvg)
  -\parve\parvd(\scalr^2\vsigd-\vsign\epsvh)+\parvu(\scalr^2\Lamrep^2-\epsvh^2)\\
  &\quad+\parvv(\scalr^2\epsvi-\epsvf\epsvh)-\ethvu\parve(\epsvf\vsigd-\vsign\epsvi)
  +\parvw(\epsvf\Lamrep^2-\epsvh\epsvi)+\parve^2(\epsvh\vsigd-\vsign\Lamrep^2)\\
  &\quad\beqref{rot1a}, \eqnref{rxpeed1a}\text{ \& }\eqnref{rpath1a}
\end{split}
\nonumber\\
&=\efkot\beqref{rpathx1h}
\end{align}
\begin{align*}
&\dprod{(\cprod{\vectr}{\unitplz})}{(\vscrp+\vscrq)}\nonumber\\
\begin{split}
&=\dprod{(\cprod{\vectr}{\unitplz})}{[}\parvx\vectLam+\parvy\vectr+\parvz\unitplz+\efkoa\vectOme
  +\efkob\fdot{\vectLam}+\efkoc\ffdot{\vectLam}
  +\parvm(\cprod{\unitkap}{\vectOme})+\parvn(\cprod{\unitkap}{\vectr})+\parvo(\cprod{\unitkap}{\vectLam})\\
  &\quad+\parvb\parve(\cprod{\unitkap}{\fdot{\vectLam}})+\parvp(\cprod{\unitkap}{\unitplz})
  +\parvq(\cprod{\vectOme}{\vectr})+\parvr(\cprod{\vectOme}{\vectLam})
  +\parvs(\cprod{\vectOme}{\fdot{\vectLam}})+\parvt(\cprod{\vectOme}{\unitplz})\\
  &\quad-\parve\parvd(\cprod{\vectr}{\fdot{\vectLam}})
  +\parvu(\cprod{\vectr}{\vectLam})+\parvv(\cprod{\vectr}{\unitplz})
  -\ethvu\parve(\cprod{\unitplz}{\fdot{\vectLam}})
  +\parvw(\cprod{\unitplz}{\vectLam})+\parve^2(\cprod{\vectLam}{\fdot{\vectLam}})]
  \beqref{rpath24}
\end{split}
\end{align*}
\begin{align*}
\begin{split}
&=\parvx[\dprod{\vectLam}{(\cprod{\vectr}{\unitplz})}]
  +\parvy[\dprod{\vectr}{(\cprod{\vectr}{\unitplz})}]
  +\parvz[\dprod{\unitplz}{(\cprod{\vectr}{\unitplz})}]
  +\efkoa[\dprod{\vectOme}{(\cprod{\vectr}{\unitplz})}]
  +\efkob[\dprod{\fdot{\vectLam}}{(\cprod{\vectr}{\unitplz})}]\\
  &\quad+\efkoc[\dprod{\ffdot{\vectLam}}{(\cprod{\vectr}{\unitplz})}]
  +\parvm[\dprod{(\cprod{\vectr}{\unitplz})}{(\cprod{\unitkap}{\vectOme})}]
  +\parvn[\dprod{(\cprod{\vectr}{\unitplz})}{(\cprod{\unitkap}{\vectr})}]
  +\parvo[\dprod{(\cprod{\vectr}{\unitplz})}{(\cprod{\unitkap}{\vectLam})}]\\
  &\quad+\parvb\parve[\dprod{(\cprod{\vectr}{\unitplz})}{(\cprod{\unitkap}{\fdot{\vectLam}})}]
  +\parvp[\dprod{(\cprod{\vectr}{\unitplz})}{(\cprod{\unitkap}{\unitplz})}]
  +\parvq[\dprod{(\cprod{\vectr}{\unitplz})}{(\cprod{\vectOme}{\vectr})}]\\
  &\quad+\parvr[\dprod{(\cprod{\vectr}{\unitplz})}{(\cprod{\vectOme}{\vectLam})}]
  +\parvs[\dprod{(\cprod{\vectr}{\unitplz})}{(\cprod{\vectOme}{\fdot{\vectLam}})}]
  +\parvt[\dprod{(\cprod{\vectr}{\unitplz})}{(\cprod{\vectOme}{\unitplz})}]\\
  &\quad-\parve\parvd[\dprod{(\cprod{\vectr}{\unitplz})}{(\cprod{\vectr}{\fdot{\vectLam}})}]
  +\parvu[\dprod{(\cprod{\vectr}{\unitplz})}{(\cprod{\vectr}{\vectLam})}]
  +\parvv[\dprod{(\cprod{\vectr}{\unitplz})}{(\cprod{\vectr}{\unitplz})}]\\
  &\quad-\ethvu\parve[\dprod{(\cprod{\vectr}{\unitplz})}{(\cprod{\unitplz}{\fdot{\vectLam}})}]
  +\parvw[\dprod{(\cprod{\vectr}{\unitplz})}{(\cprod{\unitplz}{\vectLam})}]
  +\parve^2[\dprod{(\cprod{\vectr}{\unitplz})}{(\cprod{\vectLam}{\fdot{\vectLam}})}]
\end{split}
\end{align*}
\begin{align*}
\begin{split}
&=\parvx\epsvo+\efkoa\epsvl+\efkob\frktn+\efkoc\frkxb
  +\parvm[(\dprod{\vectr}{\unitkap})(\dprod{\unitplz}{\vectOme})-(\dprod{\vectr}{\vectOme})(\dprod{\unitplz}{\unitkap})]\\
  &\quad+\parvn[(\dprod{\vectr}{\unitkap})(\dprod{\unitplz}{\vectr})-\scalr^2(\dprod{\unitplz}{\unitkap})]
  +\parvo[(\dprod{\vectr}{\unitkap})(\dprod{\unitplz}{\vectLam})-(\dprod{\vectr}{\vectLam})(\dprod{\unitplz}{\unitkap})]\\
  &\quad+\parvb\parve[(\dprod{\vectr}{\unitkap})(\dprod{\unitplz}{\fdot{\vectLam}})-(\dprod{\vectr}{\fdot{\vectLam}})(\dprod{\unitplz}{\unitkap})]
  +\parvp[(\dprod{\vectr}{\unitkap})-(\dprod{\vectr}{\unitplz})(\dprod{\unitplz}{\unitkap})]\\
  &\quad+\parvq[(\dprod{\vectr}{\vectOme})(\dprod{\unitplz}{\vectr})-\scalr^2(\dprod{\unitplz}{\vectOme})]
  +\parvr[(\dprod{\vectr}{\vectOme})(\dprod{\unitplz}{\vectLam})-(\dprod{\vectr}{\vectLam})(\dprod{\unitplz}{\vectOme})]\\
  &\quad+\parvs[(\dprod{\vectr}{\vectOme})(\dprod{\unitplz}{\fdot{\vectLam}})-(\dprod{\vectr}{\fdot{\vectLam}})(\dprod{\unitplz}{\vectOme})]
  +\parvt[(\dprod{\vectr}{\vectOme})-(\dprod{\vectr}{\unitplz})(\dprod{\unitplz}{\vectOme})]\\
  &\quad-\parve\parvd[\scalr^2(\dprod{\unitplz}{\fdot{\vectLam}})-(\dprod{\vectr}{\fdot{\vectLam}})(\dprod{\unitplz}{\vectr})]
  +\parvu[\scalr^2(\dprod{\unitplz}{\vectLam})-(\dprod{\vectr}{\vectLam})(\dprod{\unitplz}{\vectr})]
  +\parvv[\scalr^2-(\dprod{\vectr}{\unitplz})^2]\\
  &\quad-\ethvu\parve[(\dprod{\vectr}{\unitplz})(\dprod{\unitplz}{\fdot{\vectLam}})-(\dprod{\vectr}{\fdot{\vectLam}})]
  +\parvw[(\dprod{\vectr}{\unitplz})(\dprod{\unitplz}{\vectLam})-(\dprod{\vectr}{\vectLam})]\\
  &\quad+\parve^2[(\dprod{\vectr}{\vectLam})(\dprod{\unitplz}{\fdot{\vectLam}})-(\dprod{\vectr}{\fdot{\vectLam}})(\dprod{\unitplz}{\vectLam})]
  \beqref{rot1a}, \eqnref{rpath1b}, \eqnref{rpath1b2}\text{ \& }\eqnref{alg2}
\end{split}
\end{align*}
\begin{align}\label{rpath25r}
\begin{split}
&=\parvx\epsvo+\efkoa\epsvl+\efkob\frktn+\efkoc\frkxb
  +\parvm(\epsvd\epsvc-\epsvb\epsve)+\parvn(\epsvd\epsvf-\scalr^2\epsve)+\parvo(\epsvd\epsvi-\epsvh\epsve)\\
  &\quad+\parvb\parve(\epsvd\vsiga-\vsign\epsve)+\parvp(\epsvd-\epsvf\epsve)
  +\parvq(\epsvb\epsvf-\scalr^2\epsvc)+\parvr(\epsvb\epsvi-\epsvh\epsvc)\\
  &\quad+\parvs(\epsvb\vsiga-\vsign\epsvc)+\parvt(\epsvb-\epsvf\epsvc)
  -\parve\parvd(\scalr^2\vsiga-\vsign\epsvf)+\parvu(\scalr^2\epsvi-\epsvh\epsvf)\\
  &\quad+\parvv(\scalr^2-\epsvf^2)-\ethvu\parve(\epsvf\vsiga-\vsign)
  +\parvw(\epsvf\epsvi-\epsvh)+\parve^2(\epsvh\vsiga-\vsign\epsvi)\\
  &\quad\beqref{rot1a}\text{ \& }\eqnref{rpath1a}
\end{split}
\nonumber\\
&=\efkou\beqref{rpathx1h}
\end{align}
\begin{align*}
&\dprod{(\cprod{\unitplz}{\fdot{\vectLam}})}{(\vscrp+\vscrq)}\nonumber\\
\begin{split}
&=\dprod{(\cprod{\unitplz}{\fdot{\vectLam}})}{[}\parvx\vectLam+\parvy\vectr+\parvz\unitplz+\efkoa\vectOme
  +\efkob\fdot{\vectLam}+\efkoc\ffdot{\vectLam}
  +\parvm(\cprod{\unitkap}{\vectOme})+\parvn(\cprod{\unitkap}{\vectr})+\parvo(\cprod{\unitkap}{\vectLam})\\
  &\quad+\parvb\parve(\cprod{\unitkap}{\fdot{\vectLam}})+\parvp(\cprod{\unitkap}{\unitplz})
  +\parvq(\cprod{\vectOme}{\vectr})+\parvr(\cprod{\vectOme}{\vectLam})
  +\parvs(\cprod{\vectOme}{\fdot{\vectLam}})+\parvt(\cprod{\vectOme}{\unitplz})\\
  &\quad-\parve\parvd(\cprod{\vectr}{\fdot{\vectLam}})
  +\parvu(\cprod{\vectr}{\vectLam})+\parvv(\cprod{\vectr}{\unitplz})
  -\ethvu\parve(\cprod{\unitplz}{\fdot{\vectLam}})
  +\parvw(\cprod{\unitplz}{\vectLam})+\parve^2(\cprod{\vectLam}{\fdot{\vectLam}})]
  \beqref{rpath24}
\end{split}
\end{align*}
\begin{align*}
\begin{split}
&=\parvx[\dprod{\vectLam}{(\cprod{\unitplz}{\fdot{\vectLam}})}]
  +\parvy[\dprod{\vectr}{(\cprod{\unitplz}{\fdot{\vectLam}})}]
  +\parvz[\dprod{\unitplz}{(\cprod{\unitplz}{\fdot{\vectLam}})}]
  +\efkoa[\dprod{\vectOme}{(\cprod{\unitplz}{\fdot{\vectLam}})}]
  +\efkob[\dprod{\fdot{\vectLam}}{(\cprod{\unitplz}{\fdot{\vectLam}})}]\\
  &\quad+\efkoc[\dprod{\ffdot{\vectLam}}{(\cprod{\unitplz}{\fdot{\vectLam}})}]
  +\parvm[\dprod{(\cprod{\unitplz}{\fdot{\vectLam}})}{(\cprod{\unitkap}{\vectOme})}]
  +\parvn[\dprod{(\cprod{\unitplz}{\fdot{\vectLam}})}{(\cprod{\unitkap}{\vectr})}]
  +\parvo[\dprod{(\cprod{\unitplz}{\fdot{\vectLam}})}{(\cprod{\unitkap}{\vectLam})}]\\
  &\quad+\parvb\parve[\dprod{(\cprod{\unitplz}{\fdot{\vectLam}})}{(\cprod{\unitkap}{\fdot{\vectLam}})}]
  +\parvp[\dprod{(\cprod{\unitplz}{\fdot{\vectLam}})}{(\cprod{\unitkap}{\unitplz})}]
  +\parvq[\dprod{(\cprod{\unitplz}{\fdot{\vectLam}})}{(\cprod{\vectOme}{\vectr})}]\\
  &\quad+\parvr[\dprod{(\cprod{\unitplz}{\fdot{\vectLam}})}{(\cprod{\vectOme}{\vectLam})}]
  +\parvs[\dprod{(\cprod{\unitplz}{\fdot{\vectLam}})}{(\cprod{\vectOme}{\fdot{\vectLam}})}]
  +\parvt[\dprod{(\cprod{\unitplz}{\fdot{\vectLam}})}{(\cprod{\vectOme}{\unitplz})}]\\
  &\quad-\parve\parvd[\dprod{(\cprod{\unitplz}{\fdot{\vectLam}})}{(\cprod{\vectr}{\fdot{\vectLam}})}]
  +\parvu[\dprod{(\cprod{\unitplz}{\fdot{\vectLam}})}{(\cprod{\vectr}{\vectLam})}]
  +\parvv[\dprod{(\cprod{\unitplz}{\fdot{\vectLam}})}{(\cprod{\vectr}{\unitplz})}]\\
  &\quad-\ethvu\parve[\dprod{(\cprod{\unitplz}{\fdot{\vectLam}})}{(\cprod{\unitplz}{\fdot{\vectLam}})}]
  +\parvw[\dprod{(\cprod{\unitplz}{\fdot{\vectLam}})}{(\cprod{\unitplz}{\vectLam})}]
  +\parve^2[\dprod{(\cprod{\unitplz}{\fdot{\vectLam}})}{(\cprod{\vectLam}{\fdot{\vectLam}})}]
\end{split}
\end{align*}
\begin{align*}
\begin{split}
&=-\parvx\frktr+\parvy\frktn-\efkoa\frktp+\efkoc\frkxc
  +\parvm[(\dprod{\unitplz}{\unitkap})(\dprod{\fdot{\vectLam}}{\vectOme})-(\dprod{\unitplz}{\vectOme})(\dprod{\fdot{\vectLam}}{\unitkap})]\\
  &\quad+\parvn[(\dprod{\unitplz}{\unitkap})(\dprod{\fdot{\vectLam}}{\vectr})-(\dprod{\unitplz}{\vectr})(\dprod{\fdot{\vectLam}}{\unitkap})]
  +\parvo[(\dprod{\unitplz}{\unitkap})(\dprod{\fdot{\vectLam}}{\vectLam})-(\dprod{\unitplz}{\vectLam})(\dprod{\fdot{\vectLam}}{\unitkap})]\\
  &\quad+\parvb\parve[(\dprod{\unitplz}{\unitkap})\fdot{\vectLam}^2-(\dprod{\unitplz}{\fdot{\vectLam}})(\dprod{\fdot{\vectLam}}{\unitkap})]
  +\parvp[(\dprod{\unitplz}{\unitkap})(\dprod{\fdot{\vectLam}}{\unitplz})-(\dprod{\fdot{\vectLam}}{\unitkap})]\\
  &\quad+\parvq[(\dprod{\unitplz}{\vectOme})(\dprod{\fdot{\vectLam}}{\vectr})-(\dprod{\unitplz}{\vectr})(\dprod{\fdot{\vectLam}}{\vectOme})]
  +\parvr[(\dprod{\unitplz}{\vectOme})(\dprod{\fdot{\vectLam}}{\vectLam})-(\dprod{\unitplz}{\vectLam})(\dprod{\fdot{\vectLam}}{\vectOme})]\\
  &\quad+\parvs[(\dprod{\unitplz}{\vectOme})\fdot{\vectLam}^2-(\dprod{\unitplz}{\fdot{\vectLam}})(\dprod{\fdot{\vectLam}}{\vectOme})]
  +\parvt[(\dprod{\unitplz}{\vectOme})(\dprod{\fdot{\vectLam}}{\unitplz})-(\dprod{\fdot{\vectLam}}{\vectOme})]\\
  &\quad-\parve\parvd[(\dprod{\unitplz}{\vectr})\fdot{\vectLam}^2-(\dprod{\unitplz}{\fdot{\vectLam}})(\dprod{\fdot{\vectLam}}{\vectr})]
  +\parvu[(\dprod{\unitplz}{\vectr})(\dprod{\fdot{\vectLam}}{\vectLam})-(\dprod{\unitplz}{\vectLam})(\dprod{\fdot{\vectLam}}{\vectr})]\\
  &\quad+\parvv[(\dprod{\unitplz}{\vectr})(\dprod{\fdot{\vectLam}}{\unitplz})-(\dprod{\fdot{\vectLam}}{\vectr})]
  -\ethvu\parve[\fdot{\vectLam}^2-(\dprod{\unitplz}{\fdot{\vectLam}})^2]
  +\parvw[(\dprod{\fdot{\vectLam}}{\vectLam})-(\dprod{\unitplz}{\vectLam})(\dprod{\fdot{\vectLam}}{\unitplz})]\\
  &\quad+\parve^2[(\dprod{\unitplz}{\vectLam})\fdot{\vectLam}^2-(\dprod{\unitplz}{\fdot{\vectLam}})(\dprod{\fdot{\vectLam}}{\vectLam})]
  \beqref{rpath1b}, \eqnref{rpath1b2}\text{ \& }\eqnref{alg2}
\end{split}
\end{align*}
\begin{align}\label{rpath25s}
\begin{split}
&=-\parvx\frktr+\parvy\frktn-\efkoa\frktp+\efkoc\frkxc
  +\parvm(\epsve\vsigc-\epsvc\vsigb)+\parvn(\epsve\vsign-\epsvf\vsigb)+\parvo(\epsve\vsigd-\epsvi\vsigb)\\
  &\quad+\parvb\parve(\epsve\vsige-\vsiga\vsigb)+\parvp(\epsve\vsiga-\vsigb)
  +\parvq(\epsvc\vsign-\epsvf\vsigc)+\parvr(\epsvc\vsigd-\epsvi\vsigc)\\
  &\quad+\parvs(\epsvc\vsige-\vsiga\vsigc)+\parvt(\epsvc\vsiga-\vsigc)
  -\parve\parvd(\epsvf\vsige-\vsiga\vsign)+\parvu(\epsvf\vsigd-\epsvi\vsign)\\
  &\quad+\parvv(\epsvf\vsiga-\vsign)-\ethvu\parve(\vsige-\vsiga^2)
  +\parvw(\vsigd-\epsvi\vsiga)+\parve^2(\epsvi\vsige-\vsiga\vsigd)\\
  &\quad\beqref{rot1a}\text{ \& }\eqnref{rpath1a}
\end{split}
\nonumber\\
&=\efkov\beqref{rpathx1i}
\end{align}
\begin{align*}
&\dprod{(\cprod{\unitplz}{\vectLam})}{(\vscrp+\vscrq)}\nonumber\\
\begin{split}
&=\dprod{(\cprod{\unitplz}{\vectLam})}{[}\parvx\vectLam+\parvy\vectr+\parvz\unitplz+\efkoa\vectOme
  +\efkob\fdot{\vectLam}+\efkoc\ffdot{\vectLam}
  +\parvm(\cprod{\unitkap}{\vectOme})+\parvn(\cprod{\unitkap}{\vectr})+\parvo(\cprod{\unitkap}{\vectLam})\\
  &\quad+\parvb\parve(\cprod{\unitkap}{\fdot{\vectLam}})+\parvp(\cprod{\unitkap}{\unitplz})
  +\parvq(\cprod{\vectOme}{\vectr})+\parvr(\cprod{\vectOme}{\vectLam})
  +\parvs(\cprod{\vectOme}{\fdot{\vectLam}})+\parvt(\cprod{\vectOme}{\unitplz})\\
  &\quad-\parve\parvd(\cprod{\vectr}{\fdot{\vectLam}})
  +\parvu(\cprod{\vectr}{\vectLam})+\parvv(\cprod{\vectr}{\unitplz})
  -\ethvu\parve(\cprod{\unitplz}{\fdot{\vectLam}})
  +\parvw(\cprod{\unitplz}{\vectLam})+\parve^2(\cprod{\vectLam}{\fdot{\vectLam}})]
  \beqref{rpath24}
\end{split}
\end{align*}
\begin{align*}
\begin{split}
&=\parvx[\dprod{\vectLam}{(\cprod{\unitplz}{\vectLam})}]
  +\parvy[\dprod{\vectr}{(\cprod{\unitplz}{\vectLam})}]
  +\parvz[\dprod{\unitplz}{(\cprod{\unitplz}{\vectLam})}]
  +\efkoa[\dprod{\vectOme}{(\cprod{\unitplz}{\vectLam})}]
  +\efkob[\dprod{\fdot{\vectLam}}{(\cprod{\unitplz}{\vectLam})}]\\
  &\quad+\efkoc[\dprod{\ffdot{\vectLam}}{(\cprod{\unitplz}{\vectLam})}]
  +\parvm[\dprod{(\cprod{\unitplz}{\vectLam})}{(\cprod{\unitkap}{\vectOme})}]
  +\parvn[\dprod{(\cprod{\unitplz}{\vectLam})}{(\cprod{\unitkap}{\vectr})}]
  +\parvo[\dprod{(\cprod{\unitplz}{\vectLam})}{(\cprod{\unitkap}{\vectLam})}]\\
  &\quad+\parvb\parve[\dprod{(\cprod{\unitplz}{\vectLam})}{(\cprod{\unitkap}{\fdot{\vectLam}})}]
  +\parvp[\dprod{(\cprod{\unitplz}{\vectLam})}{(\cprod{\unitkap}{\unitplz})}]
  +\parvq[\dprod{(\cprod{\unitplz}{\vectLam})}{(\cprod{\vectOme}{\vectr})}]\\
  &\quad+\parvr[\dprod{(\cprod{\unitplz}{\vectLam})}{(\cprod{\vectOme}{\vectLam})}]
  +\parvs[\dprod{(\cprod{\unitplz}{\vectLam})}{(\cprod{\vectOme}{\fdot{\vectLam}})}]
  +\parvt[\dprod{(\cprod{\unitplz}{\vectLam})}{(\cprod{\vectOme}{\unitplz})}]\\
  &\quad-\parve\parvd[\dprod{(\cprod{\unitplz}{\vectLam})}{(\cprod{\vectr}{\fdot{\vectLam}})}]
  +\parvu[\dprod{(\cprod{\unitplz}{\vectLam})}{(\cprod{\vectr}{\vectLam})}]
  +\parvv[\dprod{(\cprod{\unitplz}{\vectLam})}{(\cprod{\vectr}{\unitplz})}]\\
  &\quad-\ethvu\parve[\dprod{(\cprod{\unitplz}{\vectLam})}{(\cprod{\unitplz}{\fdot{\vectLam}})}]
  +\parvw[\dprod{(\cprod{\unitplz}{\vectLam})}{(\cprod{\unitplz}{\vectLam})}]
  +\parve^2[\dprod{(\cprod{\unitplz}{\vectLam})}{(\cprod{\vectLam}{\fdot{\vectLam}})}]
\end{split}
\end{align*}
\begin{align*}
\begin{split}
&=\parvy\epsvo-\efkoa\frkto+\efkob\frktr+\efkoc\frkts
  +\parvm[(\dprod{\unitplz}{\unitkap})(\dprod{\vectLam}{\vectOme})-(\dprod{\unitplz}{\vectOme})(\dprod{\vectLam}{\unitkap})]\\
  &\quad+\parvn[(\dprod{\unitplz}{\unitkap})(\dprod{\vectLam}{\vectr})-(\dprod{\unitplz}{\vectr})(\dprod{\vectLam}{\unitkap})]
  +\parvo[(\dprod{\unitplz}{\unitkap})\Lamrep^2-(\dprod{\unitplz}{\vectLam})(\dprod{\vectLam}{\unitkap})]\\
  &\quad+\parvb\parve[(\dprod{\unitplz}{\unitkap})(\dprod{\vectLam}{\fdot{\vectLam}})-(\dprod{\unitplz}{\fdot{\vectLam}})(\dprod{\vectLam}{\unitkap})]
  +\parvp[(\dprod{\unitplz}{\unitkap})(\dprod{\vectLam}{\unitplz})-(\dprod{\vectLam}{\unitkap})]\\
  &\quad+\parvq[(\dprod{\unitplz}{\vectOme})(\dprod{\vectLam}{\vectr})-(\dprod{\unitplz}{\vectr})(\dprod{\vectLam}{\vectOme})]
  +\parvr[(\dprod{\unitplz}{\vectOme})\Lamrep^2-(\dprod{\unitplz}{\vectLam})(\dprod{\vectLam}{\vectOme})]\\
  &\quad+\parvs[(\dprod{\unitplz}{\vectOme})(\dprod{\vectLam}{\fdot{\vectLam}})-(\dprod{\unitplz}{\fdot{\vectLam}})(\dprod{\vectLam}{\vectOme})]
  +\parvt[(\dprod{\unitplz}{\vectOme})(\dprod{\vectLam}{\unitplz})-(\dprod{\vectLam}{\vectOme})]\\
  &\quad-\parve\parvd[(\dprod{\unitplz}{\vectr})(\dprod{\vectLam}{\fdot{\vectLam}})-(\dprod{\unitplz}{\fdot{\vectLam}})(\dprod{\vectLam}{\vectr})]
  +\parvu[(\dprod{\unitplz}{\vectr})\Lamrep^2-(\dprod{\unitplz}{\vectLam})(\dprod{\vectLam}{\vectr})]\\
  &\quad+\parvv[(\dprod{\unitplz}{\vectr})(\dprod{\vectLam}{\unitplz})-(\dprod{\vectLam}{\vectr})]
  -\ethvu\parve[(\dprod{\vectLam}{\fdot{\vectLam}})-(\dprod{\unitplz}{\fdot{\vectLam}})(\dprod{\vectLam}{\unitplz})]
  +\parvw[\Lamrep^2-(\dprod{\unitplz}{\vectLam})^2]\\
  &\quad+\parve^2[(\dprod{\unitplz}{\vectLam})(\dprod{\vectLam}{\fdot{\vectLam}})-(\dprod{\unitplz}{\fdot{\vectLam}})\Lamrep^2]
  \beqref{rot1a}, \eqnref{rpath1b}\text{ \& }\eqnref{alg2}
\end{split}
\end{align*}
\begin{align}\label{rpath25t}
\begin{split}
&=\parvy\epsvo-\efkoa\frkto+\efkob\frktr+\efkoc\frkts
  +\parvm(\epsve\epsvg-\epsvc\dltva)+\parvn(\epsve\epsvh-\epsvf\dltva)+\parvo(\epsve\Lamrep^2-\epsvi\dltva)\\
  &\quad+\parvb\parve(\epsve\vsigd-\vsiga\dltva)+\parvp(\epsve\epsvi-\dltva)
  +\parvq(\epsvc\epsvh-\epsvf\epsvg)+\parvr(\epsvc\Lamrep^2-\epsvi\epsvg)\\
  &\quad+\parvs(\epsvc\vsigd-\vsiga\epsvg)+\parvt(\epsvc\epsvi-\epsvg)
  -\parve\parvd(\epsvf\vsigd-\vsiga\epsvh)+\parvu(\epsvf\Lamrep^2-\epsvi\epsvh)\\
  &\quad+\parvv(\epsvf\epsvi-\epsvh)-\ethvu\parve(\vsigd-\vsiga\epsvi)
  +\parvw(\Lamrep^2-\epsvi^2)+\parve^2(\epsvi\vsigd-\vsiga\Lamrep^2)\\
  &\quad\beqref{rot1a}, \eqnref{rxpeed1a}\text{ \& }\eqnref{rpath1a}
\end{split}
\nonumber\\
&=\efkow\beqref{rpathx1i}
\end{align}
\begin{align*}
&\dprod{(\cprod{\vectLam}{\fdot{\vectLam}})}{(\vscrp+\vscrq)}\nonumber\\
\begin{split}
&=\dprod{(\cprod{\vectLam}{\fdot{\vectLam}})}{[}\parvx\vectLam+\parvy\vectr+\parvz\unitplz+\efkoa\vectOme
  +\efkob\fdot{\vectLam}+\efkoc\ffdot{\vectLam}
  +\parvm(\cprod{\unitkap}{\vectOme})+\parvn(\cprod{\unitkap}{\vectr})+\parvo(\cprod{\unitkap}{\vectLam})\\
  &\quad+\parvb\parve(\cprod{\unitkap}{\fdot{\vectLam}})+\parvp(\cprod{\unitkap}{\unitplz})
  +\parvq(\cprod{\vectOme}{\vectr})+\parvr(\cprod{\vectOme}{\vectLam})
  +\parvs(\cprod{\vectOme}{\fdot{\vectLam}})+\parvt(\cprod{\vectOme}{\unitplz})\\
  &\quad-\parve\parvd(\cprod{\vectr}{\fdot{\vectLam}})
  +\parvu(\cprod{\vectr}{\vectLam})+\parvv(\cprod{\vectr}{\unitplz})
  -\ethvu\parve(\cprod{\unitplz}{\fdot{\vectLam}})
  +\parvw(\cprod{\unitplz}{\vectLam})+\parve^2(\cprod{\vectLam}{\fdot{\vectLam}})]
  \beqref{rpath24}
\end{split}
\end{align*}
\begin{align*}
\begin{split}
&=\parvx[\dprod{\vectLam}{(\cprod{\vectLam}{\fdot{\vectLam}})}]
  +\parvy[\dprod{\vectr}{(\cprod{\vectLam}{\fdot{\vectLam}})}]
  +\parvz[\dprod{\unitplz}{(\cprod{\vectLam}{\fdot{\vectLam}})}]
  +\efkoa[\dprod{\vectOme}{(\cprod{\vectLam}{\fdot{\vectLam}})}]
  +\efkob[\dprod{\fdot{\vectLam}}{(\cprod{\vectLam}{\fdot{\vectLam}})}]\\
  &\quad+\efkoc[\dprod{\ffdot{\vectLam}}{(\cprod{\vectLam}{\fdot{\vectLam}})}]
  +\parvm[\dprod{(\cprod{\vectLam}{\fdot{\vectLam}})}{(\cprod{\unitkap}{\vectOme})}]
  +\parvn[\dprod{(\cprod{\vectLam}{\fdot{\vectLam}})}{(\cprod{\unitkap}{\vectr})}]
  +\parvo[\dprod{(\cprod{\vectLam}{\fdot{\vectLam}})}{(\cprod{\unitkap}{\vectLam})}]\\
  &\quad+\parvb\parve[\dprod{(\cprod{\vectLam}{\fdot{\vectLam}})}{(\cprod{\unitkap}{\fdot{\vectLam}})}]
  +\parvp[\dprod{(\cprod{\vectLam}{\fdot{\vectLam}})}{(\cprod{\unitkap}{\unitplz})}]
  +\parvq[\dprod{(\cprod{\vectLam}{\fdot{\vectLam}})}{(\cprod{\vectOme}{\vectr})}]\\
  &\quad+\parvr[\dprod{(\cprod{\vectLam}{\fdot{\vectLam}})}{(\cprod{\vectOme}{\vectLam})}]
  +\parvs[\dprod{(\cprod{\vectLam}{\fdot{\vectLam}})}{(\cprod{\vectOme}{\fdot{\vectLam}})}]
  +\parvt[\dprod{(\cprod{\vectLam}{\fdot{\vectLam}})}{(\cprod{\vectOme}{\unitplz})}]\\
  &\quad-\parve\parvd[\dprod{(\cprod{\vectLam}{\fdot{\vectLam}})}{(\cprod{\vectr}{\fdot{\vectLam}})}]
  +\parvu[\dprod{(\cprod{\vectLam}{\fdot{\vectLam}})}{(\cprod{\vectr}{\vectLam})}]
  +\parvv[\dprod{(\cprod{\vectLam}{\fdot{\vectLam}})}{(\cprod{\vectr}{\unitplz})}]\\
  &\quad-\ethvu\parve[\dprod{(\cprod{\vectLam}{\fdot{\vectLam}})}{(\cprod{\unitplz}{\fdot{\vectLam}})}]
  +\parvw[\dprod{(\cprod{\vectLam}{\fdot{\vectLam}})}{(\cprod{\unitplz}{\vectLam})}]
  +\parve^2[\dprod{(\cprod{\vectLam}{\fdot{\vectLam}})}{(\cprod{\vectLam}{\fdot{\vectLam}})}]
\end{split}
\end{align*}
\begin{align*}
\begin{split}
&=-\parvy\frkth+\parvz\frktr+\efkoa\frktv+\efkoc\frkxd
  +\parvm[(\dprod{\vectLam}{\unitkap})(\dprod{\fdot{\vectLam}}{\vectOme})-(\dprod{\vectLam}{\vectOme})(\dprod{\fdot{\vectLam}}{\unitkap})]\\
  &\quad+\parvn[(\dprod{\vectLam}{\unitkap})(\dprod{\fdot{\vectLam}}{\vectr})-(\dprod{\vectLam}{\vectr})(\dprod{\fdot{\vectLam}}{\unitkap})]
  +\parvo[(\dprod{\vectLam}{\unitkap})(\dprod{\fdot{\vectLam}}{\vectLam})-\Lamrep^2(\dprod{\fdot{\vectLam}}{\unitkap})]\\
  &\quad+\parvb\parve[(\dprod{\vectLam}{\unitkap})\fdot{\vectLam}^2-(\dprod{\vectLam}{\fdot{\vectLam}})(\dprod{\fdot{\vectLam}}{\unitkap})]
  +\parvp[(\dprod{\vectLam}{\unitkap})(\dprod{\fdot{\vectLam}}{\unitplz})-(\dprod{\vectLam}{\unitplz})(\dprod{\fdot{\vectLam}}{\unitkap})]\\
  &\quad+\parvq[(\dprod{\vectLam}{\vectOme})(\dprod{\fdot{\vectLam}}{\vectr})-(\dprod{\vectLam}{\vectr})(\dprod{\fdot{\vectLam}}{\vectOme})]
  +\parvr[(\dprod{\vectLam}{\vectOme})(\dprod{\fdot{\vectLam}}{\vectLam})-\Lamrep^2(\dprod{\fdot{\vectLam}}{\vectOme})]\\
  &\quad+\parvs[(\dprod{\vectLam}{\vectOme})\fdot{\vectLam}^2-(\dprod{\vectLam}{\fdot{\vectLam}})(\dprod{\fdot{\vectLam}}{\vectOme})]
  +\parvt[(\dprod{\vectLam}{\vectOme})(\dprod{\fdot{\vectLam}}{\unitplz})-(\dprod{\vectLam}{\unitplz})(\dprod{\fdot{\vectLam}}{\vectOme})]\\
  &\quad-\parve\parvd[(\dprod{\vectLam}{\vectr})\fdot{\vectLam}^2-(\dprod{\vectLam}{\fdot{\vectLam}})(\dprod{\fdot{\vectLam}}{\vectr})]
  +\parvu[(\dprod{\vectLam}{\vectr})(\dprod{\fdot{\vectLam}}{\vectLam})-\Lamrep^2(\dprod{\fdot{\vectLam}}{\vectr})]\\
  &\quad+\parvv[(\dprod{\vectLam}{\vectr})(\dprod{\fdot{\vectLam}}{\unitplz})-(\dprod{\vectLam}{\unitplz})(\dprod{\fdot{\vectLam}}{\vectr})]
  -\ethvu\parve[(\dprod{\vectLam}{\unitplz})\fdot{\vectLam}^2-(\dprod{\vectLam}{\fdot{\vectLam}})(\dprod{\fdot{\vectLam}}{\unitplz})]\\
  &\quad+\parvw[(\dprod{\vectLam}{\unitplz})(\dprod{\fdot{\vectLam}}{\vectLam})-\Lamrep^2(\dprod{\fdot{\vectLam}}{\unitplz})]
  +\parve^2[\Lamrep^2\fdot{\vectLam}^2-(\dprod{\vectLam}{\fdot{\vectLam}})^2]\\
  &\quad\beqref{rpath1b}, \eqnref{rpath1b2}\text{ \& }\eqnref{alg2}
\end{split}
\end{align*}
\begin{align}\label{rpath25u}
\begin{split}
&=-\parvy\frkth+\parvz\frktr+\efkoa\frktv+\efkoc\frkxd
  +\parvm(\dltva\vsigc-\epsvg\vsigb)+\parvn(\dltva\vsign-\epsvh\vsigb)+\parvo(\dltva\vsigd-\Lamrep^2\vsigb)\\
  &\quad+\parvb\parve(\dltva\vsige-\vsigd\vsigb)+\parvp(\dltva\vsiga-\epsvi\vsigb)
  +\parvq(\epsvg\vsign-\epsvh\vsigc)+\parvr(\epsvg\vsigd-\Lamrep^2\vsigc)\\
  &\quad+\parvs(\epsvg\vsige-\vsigd\vsigc)+\parvt(\epsvg\vsiga-\epsvi\vsigc)
  -\parve\parvd(\epsvh\vsige-\vsigd\vsign)+\parvu(\epsvh\vsigd-\Lamrep^2\vsign)\\
  &\quad+\parvv(\epsvh\vsiga-\epsvi\vsign)-\ethvu\parve(\epsvi\vsige-\vsigd\vsiga)
  +\parvw(\epsvi\vsigd-\Lamrep^2\vsiga)+\parve^2(\Lamrep^2\vsige-\vsigd^2)\\
  &\quad\beqref{rot1a}, \eqnref{rxpeed1a}\text{ \& }\eqnref{rpath1a}
\end{split}
\nonumber\\
&=\efkox\beqref{rpathx1j}.
\end{align}
\end{subequations}
Similarly, we derive
\begin{subequations}\label{rpath26}
\begin{align*}
&\dprod{\unitkap}{(\vscrp+\vscrq)}\nonumber\\
\begin{split}
&=\dprod{\unitkap}{[}\parvx\vectLam+\parvy\vectr+\parvz\unitplz+\efkoa\vectOme+\efkob\fdot{\vectLam}+\efkoc\ffdot{\vectLam}
  +\parvm(\cprod{\unitkap}{\vectOme})+\parvn(\cprod{\unitkap}{\vectr})+\parvo(\cprod{\unitkap}{\vectLam})\\
  &\quad+\parvb\parve(\cprod{\unitkap}{\fdot{\vectLam}})+\parvp(\cprod{\unitkap}{\unitplz})
  +\parvq(\cprod{\vectOme}{\vectr})+\parvr(\cprod{\vectOme}{\vectLam})
  +\parvs(\cprod{\vectOme}{\fdot{\vectLam}})+\parvt(\cprod{\vectOme}{\unitplz})\\
  &\quad-\parve\parvd(\cprod{\vectr}{\fdot{\vectLam}})
  +\parvu(\cprod{\vectr}{\vectLam})+\parvv(\cprod{\vectr}{\unitplz})
  -\ethvu\parve(\cprod{\unitplz}{\fdot{\vectLam}})
  +\parvw(\cprod{\unitplz}{\vectLam})+\parve^2(\cprod{\vectLam}{\fdot{\vectLam}})]
  \beqref{rpath24}
\end{split}
\end{align*}
\begin{align*}
\begin{split}
&=\parvx(\dprod{\unitkap}{\vectLam})+\parvy(\dprod{\unitkap}{\vectr})+\parvz(\dprod{\unitkap}{\unitplz})
  +\efkoa(\dprod{\unitkap}{\vectOme})+\efkob(\dprod{\unitkap}{\fdot{\vectLam}})+\efkoc(\dprod{\unitkap}{\ffdot{\vectLam}})
  +\parvq[\dprod{\unitkap}{(\cprod{\vectOme}{\vectr})}]\\
  &\quad+\parvr[\dprod{\unitkap}{(\cprod{\vectOme}{\vectLam})}]
  +\parvs[\dprod{\unitkap}{(\cprod{\vectOme}{\fdot{\vectLam}})}]
  +\parvt[\dprod{\unitkap}{(\cprod{\vectOme}{\unitplz})}]
  -\parve\parvd[\dprod{\unitkap}{(\cprod{\vectr}{\fdot{\vectLam}})}]
  +\parvu[\dprod{\unitkap}{(\cprod{\vectr}{\vectLam})}]\\
  &\quad+\parvv[\dprod{\unitkap}{(\cprod{\vectr}{\unitplz})}]
  -\ethvu\parve[\dprod{\unitkap}{(\cprod{\unitplz}{\fdot{\vectLam}})}]
  +\parvw[\dprod{\unitkap}{(\cprod{\unitplz}{\vectLam})}]
  +\parve^2[\dprod{\unitkap}{(\cprod{\vectLam}{\fdot{\vectLam}})}]
\end{split}
\end{align*}
\begin{align}\label{rpath26a}
\begin{split}
&=\parvx\dltva+\parvy\epsvd+\parvz\epsve+\efkoa\epsva+\efkob\vsigb+\efkoc\vsigg+\parvq\epsvk
  -\parvr\frkte+\parvs\frktf-\parvt\epsvj+\parve\parvd\frktc\\
  &\quad-\parvu\epsvn+\parvv\frktt-\ethvu\parve\frkta+\parvw\dltvb+\parve^2\frktu
  \beqref{rot1a}, \eqnref{rxpeed1a}, \eqnref{rpath1a}, \eqnref{rpath1b}\text{ \& }\eqnref{rpath1b2}
\end{split}
\nonumber\\
&=\efkoy\beqref{rpathx1j}
\end{align}
\begin{align*}
&\dprod{(\cprod{\ffdot{\vectLam}}{\unitplz})}{(\vscrp+\vscrq)}\nonumber\\
\begin{split}
&=\dprod{(\cprod{\ffdot{\vectLam}}{\unitplz})}{[}\parvx\vectLam+\parvy\vectr+\parvz\unitplz+\efkoa\vectOme
  +\efkob\fdot{\vectLam}+\efkoc\ffdot{\vectLam}
  +\parvm(\cprod{\unitkap}{\vectOme})+\parvn(\cprod{\unitkap}{\vectr})+\parvo(\cprod{\unitkap}{\vectLam})\\
  &\quad+\parvb\parve(\cprod{\unitkap}{\fdot{\vectLam}})+\parvp(\cprod{\unitkap}{\unitplz})
  +\parvq(\cprod{\vectOme}{\vectr})+\parvr(\cprod{\vectOme}{\vectLam})
  +\parvs(\cprod{\vectOme}{\fdot{\vectLam}})+\parvt(\cprod{\vectOme}{\unitplz})\\
  &\quad-\parve\parvd(\cprod{\vectr}{\fdot{\vectLam}})
  +\parvu(\cprod{\vectr}{\vectLam})+\parvv(\cprod{\vectr}{\unitplz})
  -\ethvu\parve(\cprod{\unitplz}{\fdot{\vectLam}})
  +\parvw(\cprod{\unitplz}{\vectLam})+\parve^2(\cprod{\vectLam}{\fdot{\vectLam}})]
  \beqref{rpath24}
\end{split}
\end{align*}
\begin{align*}
\begin{split}
&=\parvx[\dprod{\vectLam}{(\cprod{\ffdot{\vectLam}}{\unitplz})}]
  +\parvy[\dprod{\vectr}{(\cprod{\ffdot{\vectLam}}{\unitplz})}]
  +\parvz[\dprod{\unitplz}{(\cprod{\ffdot{\vectLam}}{\unitplz})}]
  +\efkoa[\dprod{\vectOme}{(\cprod{\ffdot{\vectLam}}{\unitplz})}]
  +\efkob[\dprod{\fdot{\vectLam}}{(\cprod{\ffdot{\vectLam}}{\unitplz})}]\\
  &\quad+\efkoc[\dprod{\ffdot{\vectLam}}{(\cprod{\ffdot{\vectLam}}{\unitplz})}]
  +\parvm[\dprod{(\cprod{\ffdot{\vectLam}}{\unitplz})}{(\cprod{\unitkap}{\vectOme})}]
  +\parvn[\dprod{(\cprod{\ffdot{\vectLam}}{\unitplz})}{(\cprod{\unitkap}{\vectr})}]
  +\parvo[\dprod{(\cprod{\ffdot{\vectLam}}{\unitplz})}{(\cprod{\unitkap}{\vectLam})}]\\
  &\quad+\parvb\parve[\dprod{(\cprod{\ffdot{\vectLam}}{\unitplz})}{(\cprod{\unitkap}{\fdot{\vectLam}})}]
  +\parvp[\dprod{(\cprod{\ffdot{\vectLam}}{\unitplz})}{(\cprod{\unitkap}{\unitplz})}]
  +\parvq[\dprod{(\cprod{\ffdot{\vectLam}}{\unitplz})}{(\cprod{\vectOme}{\vectr})}]\\
  &\quad+\parvr[\dprod{(\cprod{\ffdot{\vectLam}}{\unitplz})}{(\cprod{\vectOme}{\vectLam})}]
  +\parvs[\dprod{(\cprod{\ffdot{\vectLam}}{\unitplz})}{(\cprod{\vectOme}{\fdot{\vectLam}})}]
  +\parvt[\dprod{(\cprod{\ffdot{\vectLam}}{\unitplz})}{(\cprod{\vectOme}{\unitplz})}]\\
  &\quad-\parve\parvd[\dprod{(\cprod{\ffdot{\vectLam}}{\unitplz})}{(\cprod{\vectr}{\fdot{\vectLam}})}]
  +\parvu[\dprod{(\cprod{\ffdot{\vectLam}}{\unitplz})}{(\cprod{\vectr}{\vectLam})}]
  +\parvv[\dprod{(\cprod{\ffdot{\vectLam}}{\unitplz})}{(\cprod{\vectr}{\unitplz})}]\\
  &\quad-\ethvu\parve[\dprod{(\cprod{\ffdot{\vectLam}}{\unitplz})}{(\cprod{\unitplz}{\fdot{\vectLam}})}]
  +\parvw[\dprod{(\cprod{\ffdot{\vectLam}}{\unitplz})}{(\cprod{\unitplz}{\vectLam})}]
  +\parve^2[\dprod{(\cprod{\ffdot{\vectLam}}{\unitplz})}{(\cprod{\vectLam}{\fdot{\vectLam}})}]
\end{split}
\end{align*}
\begin{align*}
\begin{split}
&=\parvx\frkts-\parvy\frkxb+\efkoa\frktq+\efkob\frkxc
  +\parvm[(\dprod{\ffdot{\vectLam}}{\unitkap})(\dprod{\unitplz}{\vectOme})-(\dprod{\ffdot{\vectLam}}{\vectOme})(\dprod{\unitplz}{\unitkap})]\\
  &\quad+\parvn[(\dprod{\ffdot{\vectLam}}{\unitkap})(\dprod{\unitplz}{\vectr})-(\dprod{\ffdot{\vectLam}}{\vectr})(\dprod{\unitplz}{\unitkap})]
  +\parvo[(\dprod{\ffdot{\vectLam}}{\unitkap})(\dprod{\unitplz}{\vectLam})-(\dprod{\ffdot{\vectLam}}{\vectLam})(\dprod{\unitplz}{\unitkap})]\\
  &\quad+\parvb\parve[(\dprod{\ffdot{\vectLam}}{\unitkap})(\dprod{\unitplz}{\fdot{\vectLam}})-(\dprod{\ffdot{\vectLam}}{\fdot{\vectLam}})(\dprod{\unitplz}{\unitkap})]
  +\parvp[(\dprod{\ffdot{\vectLam}}{\unitkap})-(\dprod{\ffdot{\vectLam}}{\unitplz})(\dprod{\unitplz}{\unitkap})]\\
  &\quad+\parvq[(\dprod{\ffdot{\vectLam}}{\vectOme})(\dprod{\unitplz}{\vectr})-(\dprod{\ffdot{\vectLam}}{\vectr})(\dprod{\unitplz}{\vectOme})]
  +\parvr[(\dprod{\ffdot{\vectLam}}{\vectOme})(\dprod{\unitplz}{\vectLam})-(\dprod{\ffdot{\vectLam}}{\vectLam})(\dprod{\unitplz}{\vectOme})]\\
  &\quad+\parvs[(\dprod{\ffdot{\vectLam}}{\vectOme})(\dprod{\unitplz}{\fdot{\vectLam}})-(\dprod{\ffdot{\vectLam}}{\fdot{\vectLam}})(\dprod{\unitplz}{\vectOme})]
  +\parvt[(\dprod{\ffdot{\vectLam}}{\vectOme})-(\dprod{\ffdot{\vectLam}}{\unitplz})(\dprod{\unitplz}{\vectOme})]\\
  &\quad-\parve\parvd[(\dprod{\ffdot{\vectLam}}{\vectr})(\dprod{\unitplz}{\fdot{\vectLam}})-(\dprod{\ffdot{\vectLam}}{\fdot{\vectLam}})(\dprod{\unitplz}{\vectr})]
  +\parvu[(\dprod{\ffdot{\vectLam}}{\vectr})(\dprod{\unitplz}{\vectLam})-(\dprod{\ffdot{\vectLam}}{\vectLam})(\dprod{\unitplz}{\vectr})]\\
  &\quad+\parvv[(\dprod{\ffdot{\vectLam}}{\vectr})-(\dprod{\ffdot{\vectLam}}{\unitplz})(\dprod{\unitplz}{\vectr})]
  -\ethvu\parve[(\dprod{\ffdot{\vectLam}}{\unitplz})(\dprod{\unitplz}{\fdot{\vectLam}})-(\dprod{\ffdot{\vectLam}}{\fdot{\vectLam}})]\\
  &\quad+\parvw[(\dprod{\ffdot{\vectLam}}{\unitplz})(\dprod{\unitplz}{\vectLam})-(\dprod{\ffdot{\vectLam}}{\vectLam})]
  +\parve^2[(\dprod{\ffdot{\vectLam}}{\vectLam})(\dprod{\unitplz}{\fdot{\vectLam}})-(\dprod{\ffdot{\vectLam}}{\fdot{\vectLam}})(\dprod{\unitplz}{\vectLam})]\\
  &\quad\beqref{rpath1b}, \eqnref{rpath1b2}\text{ \& }\eqnref{alg2}
\end{split}
\end{align*}
\begin{align}\label{rpath26b}
\begin{split}
&=\parvx\frkts-\parvy\frkxb+\efkoa\frktq+\efkob\frkxc
  +\parvm(\vsigg\epsvc-\vsigh\epsve)+\parvn(\vsigg\epsvf-\vsigo\epsve)+\parvo(\vsigg\epsvi-\vsigi\epsve)\\
  &\quad+\parvb\parve(\vsigg\vsiga-\vsigj\epsve)+\parvp(\vsigg-\vsigf\epsve)
  +\parvq(\vsigh\epsvf-\vsigo\epsvc)+\parvr(\vsigh\epsvi-\vsigi\epsvc)\\
  &\quad+\parvs(\vsigh\vsiga-\vsigj\epsvc)+\parvt(\vsigh-\vsigf\epsvc)
  -\parve\parvd(\vsigo\vsiga-\vsigj\epsvf)+\parvu(\vsigo\epsvi-\vsigi\epsvf)\\
  &\quad+\parvv(\vsigo-\vsigf\epsvf)-\ethvu\parve(\vsigf\vsiga-\vsigj)
  +\parvw(\vsigf\epsvi-\vsigi)+\parve^2(\vsigi\vsiga-\vsigj\epsvi)\\
  &\quad\beqref{rot1a}\text{ \& }\eqnref{rpath1a}
\end{split}
\nonumber\\
&=\efkoz\beqref{rpathx1j2}
\end{align}
\begin{align*}
&\dprod{(\cprod{\ffdot{\vectLam}}{\vectr})}{(\vscrp+\vscrq)}\nonumber\\
\begin{split}
&=\dprod{(\cprod{\ffdot{\vectLam}}{\vectr})}{[}\parvx\vectLam+\parvy\vectr+\parvz\unitplz+\efkoa\vectOme
  +\efkob\fdot{\vectLam}+\efkoc\ffdot{\vectLam}
  +\parvm(\cprod{\unitkap}{\vectOme})+\parvn(\cprod{\unitkap}{\vectr})+\parvo(\cprod{\unitkap}{\vectLam})\\
  &\quad+\parvb\parve(\cprod{\unitkap}{\fdot{\vectLam}})+\parvp(\cprod{\unitkap}{\unitplz})
  +\parvq(\cprod{\vectOme}{\vectr})+\parvr(\cprod{\vectOme}{\vectLam})
  +\parvs(\cprod{\vectOme}{\fdot{\vectLam}})+\parvt(\cprod{\vectOme}{\unitplz})\\
  &\quad-\parve\parvd(\cprod{\vectr}{\fdot{\vectLam}})
  +\parvu(\cprod{\vectr}{\vectLam})+\parvv(\cprod{\vectr}{\unitplz})
  -\ethvu\parve(\cprod{\unitplz}{\fdot{\vectLam}})
  +\parvw(\cprod{\unitplz}{\vectLam})+\parve^2(\cprod{\vectLam}{\fdot{\vectLam}})]
  \beqref{rpath24}
\end{split}
\end{align*}
\begin{align*}
\begin{split}
&=\parvx[\dprod{\vectLam}{(\cprod{\ffdot{\vectLam}}{\vectr})}]
  +\parvy[\dprod{\vectr}{(\cprod{\ffdot{\vectLam}}{\vectr})}]
  +\parvz[\dprod{\unitplz}{(\cprod{\ffdot{\vectLam}}{\vectr})}]
  +\efkoa[\dprod{\vectOme}{(\cprod{\ffdot{\vectLam}}{\vectr})}]
  +\efkob[\dprod{\fdot{\vectLam}}{(\cprod{\ffdot{\vectLam}}{\vectr})}]\\
  &\quad+\efkoc[\dprod{\ffdot{\vectLam}}{(\cprod{\ffdot{\vectLam}}{\vectr})}]
  +\parvm[\dprod{(\cprod{\ffdot{\vectLam}}{\vectr})}{(\cprod{\unitkap}{\vectOme})}]
  +\parvn[\dprod{(\cprod{\ffdot{\vectLam}}{\vectr})}{(\cprod{\unitkap}{\vectr})}]
  +\parvo[\dprod{(\cprod{\ffdot{\vectLam}}{\vectr})}{(\cprod{\unitkap}{\vectLam})}]\\
  &\quad+\parvb\parve[\dprod{(\cprod{\ffdot{\vectLam}}{\vectr})}{(\cprod{\unitkap}{\fdot{\vectLam}})}]
  +\parvp[\dprod{(\cprod{\ffdot{\vectLam}}{\vectr})}{(\cprod{\unitkap}{\unitplz})}]
  +\parvq[\dprod{(\cprod{\ffdot{\vectLam}}{\vectr})}{(\cprod{\vectOme}{\vectr})}]\\
  &\quad+\parvr[\dprod{(\cprod{\ffdot{\vectLam}}{\vectr})}{(\cprod{\vectOme}{\vectLam})}]
  +\parvs[\dprod{(\cprod{\ffdot{\vectLam}}{\vectr})}{(\cprod{\vectOme}{\fdot{\vectLam}})}]
  +\parvt[\dprod{(\cprod{\ffdot{\vectLam}}{\vectr})}{(\cprod{\vectOme}{\unitplz})}]\\
  &\quad-\parve\parvd[\dprod{(\cprod{\ffdot{\vectLam}}{\vectr})}{(\cprod{\vectr}{\fdot{\vectLam}})}]
  +\parvu[\dprod{(\cprod{\ffdot{\vectLam}}{\vectr})}{(\cprod{\vectr}{\vectLam})}]
  +\parvv[\dprod{(\cprod{\ffdot{\vectLam}}{\vectr})}{(\cprod{\vectr}{\unitplz})}]\\
  &\quad-\ethvu\parve[\dprod{(\cprod{\ffdot{\vectLam}}{\vectr})}{(\cprod{\unitplz}{\fdot{\vectLam}})}]
  +\parvw[\dprod{(\cprod{\ffdot{\vectLam}}{\vectr})}{(\cprod{\unitplz}{\vectLam})}]
  +\parve^2[\dprod{(\cprod{\ffdot{\vectLam}}{\vectr})}{(\cprod{\vectLam}{\fdot{\vectLam}})}]
\end{split}
\end{align*}
\begin{align*}
\begin{split}
&=-\parvx\frkti+\parvz\frkxb-\efkoa\frktm-\efkob\frktj
  +\parvm[(\dprod{\ffdot{\vectLam}}{\unitkap})(\dprod{\vectr}{\vectOme})-(\dprod{\ffdot{\vectLam}}{\vectOme})(\dprod{\vectr}{\unitkap})]\\
  &\quad+\parvn[(\dprod{\ffdot{\vectLam}}{\unitkap})\scalr^2-(\dprod{\ffdot{\vectLam}}{\vectr})(\dprod{\vectr}{\unitkap})]
  +\parvo[(\dprod{\ffdot{\vectLam}}{\unitkap})(\dprod{\vectr}{\vectLam})-(\dprod{\ffdot{\vectLam}}{\vectLam})(\dprod{\vectr}{\unitkap})]\\
  &\quad+\parvb\parve[(\dprod{\ffdot{\vectLam}}{\unitkap})(\dprod{\vectr}{\fdot{\vectLam}})-(\dprod{\ffdot{\vectLam}}{\fdot{\vectLam}})(\dprod{\vectr}{\unitkap})]
  +\parvp[(\dprod{\ffdot{\vectLam}}{\unitkap})(\dprod{\vectr}{\unitplz})-(\dprod{\ffdot{\vectLam}}{\unitplz})(\dprod{\vectr}{\unitkap})]\\
  &\quad+\parvq[(\dprod{\ffdot{\vectLam}}{\vectOme})\scalr^2-(\dprod{\ffdot{\vectLam}}{\vectr})(\dprod{\vectr}{\vectOme})]
  +\parvr[(\dprod{\ffdot{\vectLam}}{\vectOme})(\dprod{\vectr}{\vectLam})-(\dprod{\ffdot{\vectLam}}{\vectLam})(\dprod{\vectr}{\vectOme})]\\
  &\quad+\parvs[(\dprod{\ffdot{\vectLam}}{\vectOme})(\dprod{\vectr}{\fdot{\vectLam}})-(\dprod{\ffdot{\vectLam}}{\fdot{\vectLam}})(\dprod{\vectr}{\vectOme})]
  +\parvt[(\dprod{\ffdot{\vectLam}}{\vectOme})(\dprod{\vectr}{\unitplz})-(\dprod{\ffdot{\vectLam}}{\unitplz})(\dprod{\vectr}{\vectOme})]\\
  &\quad-\parve\parvd[(\dprod{\ffdot{\vectLam}}{\vectr})(\dprod{\vectr}{\fdot{\vectLam}})-(\dprod{\ffdot{\vectLam}}{\fdot{\vectLam}})\scalr^2]
  +\parvu[(\dprod{\ffdot{\vectLam}}{\vectr})(\dprod{\vectr}{\vectLam})-(\dprod{\ffdot{\vectLam}}{\vectLam})\scalr^2]\\
  &\quad+\parvv[(\dprod{\ffdot{\vectLam}}{\vectr})(\dprod{\vectr}{\unitplz})-(\dprod{\ffdot{\vectLam}}{\unitplz})\scalr^2]
  -\ethvu\parve[(\dprod{\ffdot{\vectLam}}{\unitplz})(\dprod{\vectr}{\fdot{\vectLam}})-(\dprod{\ffdot{\vectLam}}{\fdot{\vectLam}})(\dprod{\vectr}{\unitplz})]\\
  &\quad+\parvw[(\dprod{\ffdot{\vectLam}}{\unitplz})(\dprod{\vectr}{\vectLam})-(\dprod{\ffdot{\vectLam}}{\vectLam})(\dprod{\vectr}{\unitplz})]
  +\parve^2[(\dprod{\ffdot{\vectLam}}{\vectLam})(\dprod{\vectr}{\fdot{\vectLam}})-(\dprod{\ffdot{\vectLam}}{\fdot{\vectLam}})(\dprod{\vectr}{\vectLam})]\\
  &\quad\beqref{rpath1b}, \eqnref{rpath1b2}\text{ \& }\eqnref{alg2}
\end{split}
\end{align*}
\begin{align}\label{rpath26c}
\begin{split}
&=-\parvx\frkti+\parvz\frkxb-\efkoa\frktm-\efkob\frktj
  +\parvm(\vsigg\epsvb-\vsigh\epsvd)+\parvn(\vsigg\scalr^2-\vsigo\epsvd)+\parvo(\vsigg\epsvh-\vsigi\epsvd)\\
  &\quad+\parvb\parve(\vsigg\vsign-\vsigj\epsvd)+\parvp(\vsigg\epsvf-\vsigf\epsvd)
  +\parvq(\vsigh\scalr^2-\vsigo\epsvb)+\parvr(\vsigh\epsvh-\vsigi\epsvb)\\
  &\quad+\parvs(\vsigh\vsign-\vsigj\epsvb)+\parvt(\vsigh\epsvf-\vsigf\epsvb)
  -\parve\parvd(\vsigo\vsign-\vsigj\scalr^2)+\parvu(\vsigo\epsvh-\vsigi\scalr^2)\\
  &\quad+\parvv(\vsigo\epsvf-\vsigf\scalr^2)-\ethvu\parve(\vsigf\vsign-\vsigj\epsvf)
  +\parvw(\vsigf\epsvh-\vsigi\epsvf)+\parve^2(\vsigi\vsign-\vsigj\epsvh)\\
  &\quad\beqref{rot1a}\text{ \& }\eqnref{rpath1a}
\end{split}
\nonumber\\
&=\efkca\beqref{rpathx1k}
\end{align}
\begin{align*}
&\dprod{(\cprod{\fffdot{\vectLam}}{\vectr})}{(\vscrp+\vscrq)}\nonumber\\
\begin{split}
&=\dprod{(\cprod{\fffdot{\vectLam}}{\vectr})}{[}\parvx\vectLam+\parvy\vectr+\parvz\unitplz+\efkoa\vectOme
  +\efkob\fdot{\vectLam}+\efkoc\ffdot{\vectLam}
  +\parvm(\cprod{\unitkap}{\vectOme})+\parvn(\cprod{\unitkap}{\vectr})+\parvo(\cprod{\unitkap}{\vectLam})\\
  &\quad+\parvb\parve(\cprod{\unitkap}{\fdot{\vectLam}})+\parvp(\cprod{\unitkap}{\unitplz})
  +\parvq(\cprod{\vectOme}{\vectr})+\parvr(\cprod{\vectOme}{\vectLam})
  +\parvs(\cprod{\vectOme}{\fdot{\vectLam}})+\parvt(\cprod{\vectOme}{\unitplz})\\
  &\quad-\parve\parvd(\cprod{\vectr}{\fdot{\vectLam}})
  +\parvu(\cprod{\vectr}{\vectLam})+\parvv(\cprod{\vectr}{\unitplz})
  -\ethvu\parve(\cprod{\unitplz}{\fdot{\vectLam}})
  +\parvw(\cprod{\unitplz}{\vectLam})+\parve^2(\cprod{\vectLam}{\fdot{\vectLam}})]
  \beqref{rpath24}
\end{split}
\end{align*}
\begin{align*}
\begin{split}
&=\parvx[\dprod{\vectLam}{(\cprod{\fffdot{\vectLam}}{\vectr})}]
  +\parvy[\dprod{\vectr}{(\cprod{\fffdot{\vectLam}}{\vectr})}]
  +\parvz[\dprod{\unitplz}{(\cprod{\fffdot{\vectLam}}{\vectr})}]
  +\efkoa[\dprod{\vectOme}{(\cprod{\fffdot{\vectLam}}{\vectr})}]
  +\efkob[\dprod{\fdot{\vectLam}}{(\cprod{\fffdot{\vectLam}}{\vectr})}]\\
  &\quad+\efkoc[\dprod{\ffdot{\vectLam}}{(\cprod{\fffdot{\vectLam}}{\vectr})}]
  +\parvm[\dprod{(\cprod{\fffdot{\vectLam}}{\vectr})}{(\cprod{\unitkap}{\vectOme})}]
  +\parvn[\dprod{(\cprod{\fffdot{\vectLam}}{\vectr})}{(\cprod{\unitkap}{\vectr})}]
  +\parvo[\dprod{(\cprod{\fffdot{\vectLam}}{\vectr})}{(\cprod{\unitkap}{\vectLam})}]\\
  &\quad+\parvb\parve[\dprod{(\cprod{\fffdot{\vectLam}}{\vectr})}{(\cprod{\unitkap}{\fdot{\vectLam}})}]
  +\parvp[\dprod{(\cprod{\fffdot{\vectLam}}{\vectr})}{(\cprod{\unitkap}{\unitplz})}]
  +\parvq[\dprod{(\cprod{\fffdot{\vectLam}}{\vectr})}{(\cprod{\vectOme}{\vectr})}]\\
  &\quad+\parvr[\dprod{(\cprod{\fffdot{\vectLam}}{\vectr})}{(\cprod{\vectOme}{\vectLam})}]
  +\parvs[\dprod{(\cprod{\fffdot{\vectLam}}{\vectr})}{(\cprod{\vectOme}{\fdot{\vectLam}})}]
  +\parvt[\dprod{(\cprod{\fffdot{\vectLam}}{\vectr})}{(\cprod{\vectOme}{\unitplz})}]\\
  &\quad-\parve\parvd[\dprod{(\cprod{\fffdot{\vectLam}}{\vectr})}{(\cprod{\vectr}{\fdot{\vectLam}})}]
  +\parvu[\dprod{(\cprod{\fffdot{\vectLam}}{\vectr})}{(\cprod{\vectr}{\vectLam})}]
  +\parvv[\dprod{(\cprod{\fffdot{\vectLam}}{\vectr})}{(\cprod{\vectr}{\unitplz})}]\\
  &\quad-\ethvu\parve[\dprod{(\cprod{\fffdot{\vectLam}}{\vectr})}{(\cprod{\unitplz}{\fdot{\vectLam}})}]
  +\parvw[\dprod{(\cprod{\fffdot{\vectLam}}{\vectr})}{(\cprod{\unitplz}{\vectLam})}]
  +\parve^2[\dprod{(\cprod{\fffdot{\vectLam}}{\vectr})}{(\cprod{\vectLam}{\fdot{\vectLam}})}]
\end{split}
\end{align*}
\begin{align*}
\begin{split}
&=\parvx\frkxe+\parvz\frkxf+\efkoa\frkxg+\efkob\frkxh+\efkoc\frkxi
  +\parvm[(\dprod{\fffdot{\vectLam}}{\unitkap})(\dprod{\vectr}{\vectOme})-(\dprod{\fffdot{\vectLam}}{\vectOme})(\dprod{\vectr}{\unitkap})]\\
  &\quad+\parvn[(\dprod{\fffdot{\vectLam}}{\unitkap})\scalr^2-(\dprod{\fffdot{\vectLam}}{\vectr})(\dprod{\vectr}{\unitkap})]
  +\parvo[(\dprod{\fffdot{\vectLam}}{\unitkap})(\dprod{\vectr}{\vectLam})-(\dprod{\fffdot{\vectLam}}{\vectLam})(\dprod{\vectr}{\unitkap})]\\
  &\quad+\parvb\parve[(\dprod{\fffdot{\vectLam}}{\unitkap})(\dprod{\vectr}{\fdot{\vectLam}})-(\dprod{\fffdot{\vectLam}}{\fdot{\vectLam}})(\dprod{\vectr}{\unitkap})]
  +\parvp[(\dprod{\fffdot{\vectLam}}{\unitkap})(\dprod{\vectr}{\unitplz})-(\dprod{\fffdot{\vectLam}}{\unitplz})(\dprod{\vectr}{\unitkap})]\\
  &\quad+\parvq[(\dprod{\fffdot{\vectLam}}{\vectOme})\scalr^2-(\dprod{\fffdot{\vectLam}}{\vectr})(\dprod{\vectr}{\vectOme})]
  +\parvr[(\dprod{\fffdot{\vectLam}}{\vectOme})(\dprod{\vectr}{\vectLam})-(\dprod{\fffdot{\vectLam}}{\vectLam})(\dprod{\vectr}{\vectOme})]\\
  &\quad+\parvs[(\dprod{\fffdot{\vectLam}}{\vectOme})(\dprod{\vectr}{\fdot{\vectLam}})-(\dprod{\fffdot{\vectLam}}{\fdot{\vectLam}})(\dprod{\vectr}{\vectOme})]
  +\parvt[(\dprod{\fffdot{\vectLam}}{\vectOme})(\dprod{\vectr}{\unitplz})-(\dprod{\fffdot{\vectLam}}{\unitplz})(\dprod{\vectr}{\vectOme})]\\
  &\quad-\parve\parvd[(\dprod{\fffdot{\vectLam}}{\vectr})(\dprod{\vectr}{\fdot{\vectLam}})-(\dprod{\fffdot{\vectLam}}{\fdot{\vectLam}})\scalr^2]
  +\parvu[(\dprod{\fffdot{\vectLam}}{\vectr})(\dprod{\vectr}{\vectLam})-(\dprod{\fffdot{\vectLam}}{\vectLam})\scalr^2]\\
  &\quad+\parvv[(\dprod{\fffdot{\vectLam}}{\vectr})(\dprod{\vectr}{\unitplz})-(\dprod{\fffdot{\vectLam}}{\unitplz})\scalr^2]
  -\ethvu\parve[(\dprod{\fffdot{\vectLam}}{\unitplz})(\dprod{\vectr}{\fdot{\vectLam}})-(\dprod{\fffdot{\vectLam}}{\fdot{\vectLam}})(\dprod{\vectr}{\unitplz})]\\
  &\quad+\parvw[(\dprod{\fffdot{\vectLam}}{\unitplz})(\dprod{\vectr}{\vectLam})-(\dprod{\fffdot{\vectLam}}{\vectLam})(\dprod{\vectr}{\unitplz})]
  +\parve^2[(\dprod{\fffdot{\vectLam}}{\vectLam})(\dprod{\vectr}{\fdot{\vectLam}})-(\dprod{\fffdot{\vectLam}}{\fdot{\vectLam}})(\dprod{\vectr}{\vectLam})]\\
  &\quad\beqref{rpath1b2}\text{ \& }\eqnref{alg2}
\end{split}
\end{align*}
\begin{align}\label{rpath26d}
\begin{split}
&=\parvx\frkxe+\parvz\frkxf+\efkoa\frkxg+\efkob\frkxh+\efkoc\frkxi
  +\parvm(\vsigp\epsvb-\vsigm\epsvd)+\parvn(\vsigp\scalr^2-\vsigq\epsvd)\\
  &\quad+\parvo(\vsigp\epsvh-\vsigr\epsvd)+\parvb\parve(\vsigp\vsign-\vsigs\epsvd)
  +\parvp(\vsigp\epsvf-\vsigl\epsvd)+\parvq(\vsigm\scalr^2-\vsigq\epsvb)\\
  &\quad+\parvr(\vsigm\epsvh-\vsigr\epsvb)+\parvs(\vsigm\vsign-\vsigs\epsvb)
  +\parvt(\vsigm\epsvf-\vsigl\epsvb)-\parve\parvd(\vsigq\vsign-\vsigs\scalr^2)\\
  &\quad+\parvu(\vsigq\epsvh-\vsigr\scalr^2)+\parvv(\vsigq\epsvf-\vsigl\scalr^2)
  -\ethvu\parve(\vsigl\vsign-\vsigs\epsvf)+\parvw(\vsigl\epsvh-\vsigr\epsvf)\\
  &\quad+\parve^2(\vsigr\vsign-\vsigs\epsvh)
  \beqref{rot1a}\text{ \& }\eqnref{rpath1a}
\end{split}
\nonumber\\
&=\efkcb\beqref{rpathx1k}.
\end{align}
\end{subequations}
As a consequence of the foregoing derivations, we get
\begin{subequations}\label{rpath27}
\begin{align*}
\begin{split}
&|\vscrp+\vscrq|^2\\
&=\dprod{(\vscrp+\vscrq)}{[}\parvx\vectLam+\parvy\vectr+\parvz\unitplz+\efkoa\vectOme+\efkob\fdot{\vectLam}+\efkoc\ffdot{\vectLam}
  +\parvm(\cprod{\unitkap}{\vectOme})+\parvn(\cprod{\unitkap}{\vectr})+\parvo(\cprod{\unitkap}{\vectLam})\\
  &\quad+\parvb\parve(\cprod{\unitkap}{\fdot{\vectLam}})+\parvp(\cprod{\unitkap}{\unitplz})
  +\parvq(\cprod{\vectOme}{\vectr})+\parvr(\cprod{\vectOme}{\vectLam})
  +\parvs(\cprod{\vectOme}{\fdot{\vectLam}})+\parvt(\cprod{\vectOme}{\unitplz})\\
  &\quad-\parve\parvd(\cprod{\vectr}{\fdot{\vectLam}})
  +\parvu(\cprod{\vectr}{\vectLam})+\parvv(\cprod{\vectr}{\unitplz})
  -\ethvu\parve(\cprod{\unitplz}{\fdot{\vectLam}})
  +\parvw(\cprod{\unitplz}{\vectLam})+\parve^2(\cprod{\vectLam}{\fdot{\vectLam}})]
  \beqref{rpath24}
\end{split}
\end{align*}
\begin{align*}
\begin{split}
&=\parvx[\dprod{\vectLam}{(\vscrp+\vscrq)}]
  +\parvy[\dprod{\vectr}{(\vscrp+\vscrq)}]
  +\parvz[\dprod{\unitplz}{(\vscrp+\vscrq)}]
  +\efkoa[\dprod{\vectOme}{(\vscrp+\vscrq)}]\\
  &\quad+\efkob[\dprod{\fdot{\vectLam}}{(\vscrp+\vscrq)}]
  +\efkoc[\dprod{\ffdot{\vectLam}}{(\vscrp+\vscrq)}]
  +\parvm[\dprod{(\vscrp+\vscrq)}{(\cprod{\unitkap}{\vectOme})}]\\
  &\quad+\parvn[\dprod{(\vscrp+\vscrq)}{(\cprod{\unitkap}{\vectr})}]
  +\parvo[\dprod{(\vscrp+\vscrq)}{(\cprod{\unitkap}{\vectLam})}]
  +\parvb\parve[\dprod{(\vscrp+\vscrq)}{(\cprod{\unitkap}{\fdot{\vectLam}})}]\\
  &\quad+\parvp[\dprod{(\vscrp+\vscrq)}{(\cprod{\unitkap}{\unitplz})}]
  +\parvq[\dprod{(\vscrp+\vscrq)}{(\cprod{\vectOme}{\vectr})}]
  +\parvr[\dprod{(\vscrp+\vscrq)}{(\cprod{\vectOme}{\vectLam})}]\\
  &\quad+\parvs[\dprod{(\vscrp+\vscrq)}{(\cprod{\vectOme}{\fdot{\vectLam}})}]
  +\parvt[\dprod{(\vscrp+\vscrq)}{(\cprod{\vectOme}{\unitplz})}]
  -\parve\parvd[\dprod{(\vscrp+\vscrq)}{(\cprod{\vectr}{\fdot{\vectLam}})}]\\
  &\quad+\parvu[\dprod{(\vscrp+\vscrq)}{(\cprod{\vectr}{\vectLam})}]
  +\parvv[\dprod{(\vscrp+\vscrq)}{(\cprod{\vectr}{\unitplz})}]
  -\ethvu\parve[\dprod{(\vscrp+\vscrq)}{(\cprod{\unitplz}{\fdot{\vectLam}})}]\\
  &\quad+\parvw[\dprod{(\vscrp+\vscrq)}{(\cprod{\unitplz}{\vectLam})}]
  +\parve^2[\dprod{(\vscrp+\vscrq)}{(\cprod{\vectLam}{\fdot{\vectLam}})}]
\end{split}
\end{align*}
\begin{align}\label{rpath27a}
\begin{split}
&=\parvx\efkod+\parvy\efkoe+\parvz\efkof+\efkoa\efkog+\efkob\efkoh+\efkoc\efkoi+\parvm\efkoj
  +\parvn\efkok+\parvo\efkol+\parvb\parve\efkom+\parvp\efkon\\
  &\quad+\parvq\efkoo+\parvr\efkop+\parvs\efkoq+\parvt\efkor-\parve\parvd\efkos+\parvu\efkot
  +\parvv\efkou-\ethvu\parve\efkov+\parvw\efkow+\parve^2\efkox\\
  &\quad\beqref{rpath26}
\end{split}
\nonumber\\
\therefore|\vscrp+\vscrq|&=\efkcc\beqref{rpathx1l}
\end{align}
\begin{align*}
\begin{split}
|\vscrr|^2
&=[\parvb\unitkap+\parvc\vectOme-\parvd\vectr+\parve\vectLam-\ethvu\unitplz+\parva(\cprod{\vectLam}{\vectr})
  +\rhorep(\cprod{\fdot{\vectLam}}{\vectr})-\rhorep\Omerep^2(\cprod{\vectOme}{\vectr})
  +\ethvo(\cprod{\unitplz}{\vectOme})\\
  &\quad+\vphig\vphih(\cprod{\unitplz}{\vectLam})]^2\beqref{rpath22d}
\end{split}
\end{align*}
\begin{align*}
\begin{split}
&=\parvb^2
  +2\parvc\parvb(\dprod{\unitkap}{\vectOme})
  -2\parvd\parvb(\dprod{\unitkap}{\vectr})
  +2\parve\parvb(\dprod{\unitkap}{\vectLam})
  -2\ethvu\parvb(\dprod{\unitkap}{\unitplz})
  +2\parva\parvb[\dprod{\unitkap}{(\cprod{\vectLam}{\vectr})}]\\
  &\quad+2\rhorep\parvb[\dprod{\unitkap}{(\cprod{\fdot{\vectLam}}{\vectr})}]
  -2\rhorep\Omerep^2\parvb[\dprod{\unitkap}{(\cprod{\vectOme}{\vectr})}]
  +2\ethvo\parvb[\dprod{\unitkap}{(\cprod{\unitplz}{\vectOme})}]
  +2\vphig\vphih\parvb[\dprod{\unitkap}{(\cprod{\unitplz}{\vectLam})}]\\%
  &\quad+\parvc^2\Omerep^2
  -2\parvd\parvc(\dprod{\vectOme}{\vectr})
  +2\parve\parvc(\dprod{\vectOme}{\vectLam})
  -2\ethvu\parvc(\dprod{\vectOme}{\unitplz})
  +2\parva\parvc[\dprod{\vectOme}{(\cprod{\vectLam}{\vectr})}]\\
  &\quad+2\rhorep\parvc[\dprod{\vectOme}{(\cprod{\fdot{\vectLam}}{\vectr})}]
  +2\vphig\vphih\parvc[\dprod{\vectOme}{(\cprod{\unitplz}{\vectLam})}]%
  +\parvd^2\scalr^2
  -2\parve\parvd(\dprod{\vectr}{\vectLam})
  +2\ethvu\parvd(\dprod{\vectr}{\unitplz})\\
  &\quad-2\ethvo\parvd[\dprod{\vectr}{(\cprod{\unitplz}{\vectOme})}]
  -2\parvd\vphig\vphih[\dprod{\vectr}{(\cprod{\unitplz}{\vectLam})}]%
  +\parve^2\Lamrep^2
  -2\ethvu\parve(\dprod{\vectLam}{\unitplz})
  +2\rhorep\parve[\dprod{\vectLam}{(\cprod{\fdot{\vectLam}}{\vectr})}]\\
  &\quad-2\rhorep\Omerep^2\parve[\dprod{\vectLam}{(\cprod{\vectOme}{\vectr})}]
  +2\ethvo\parve[\dprod{\vectLam}{(\cprod{\unitplz}{\vectOme})}]%
  +\ethvu^2
  -2\parva\ethvu[\dprod{\unitplz}{(\cprod{\vectLam}{\vectr})}]
  -2\rhorep\ethvu[\dprod{\unitplz}{(\cprod{\fdot{\vectLam}}{\vectr})}]\\
  &\quad+2\rhorep\Omerep^2\ethvu[\dprod{\unitplz}{(\cprod{\vectOme}{\vectr})}]%
  +\parva^2[\dprod{(\cprod{\vectLam}{\vectr})}{(\cprod{\vectLam}{\vectr})}]
  +2\rhorep\parva[\dprod{(\cprod{\vectLam}{\vectr})}{(\cprod{\fdot{\vectLam}}{\vectr})}] \\
  &\quad-2\rhorep\Omerep^2\parva[\dprod{(\cprod{\vectLam}{\vectr})}{(\cprod{\vectOme}{\vectr})}]
  +2\ethvo\parva[\dprod{(\cprod{\vectLam}{\vectr})}{(\cprod{\unitplz}{\vectOme})}]
  +2\vphig\vphih\parva[\dprod{(\cprod{\vectLam}{\vectr})}{(\cprod{\unitplz}{\vectLam})}]\\%
  &\quad+\rhorep^2[\dprod{(\cprod{\fdot{\vectLam}}{\vectr})}{(\cprod{\fdot{\vectLam}}{\vectr})}]
  -2\rhorep^2\Omerep^2[\dprod{(\cprod{\vectOme}{\vectr})}{(\cprod{\fdot{\vectLam}}{\vectr})}]
  +2\ethvo\rhorep[\dprod{(\cprod{\fdot{\vectLam}}{\vectr})}{(\cprod{\unitplz}{\vectOme})}]\\
  &\quad+2\vphig\vphih\rhorep[\dprod{(\cprod{\fdot{\vectLam}}{\vectr})}{(\cprod{\unitplz}{\vectLam})}]%
  +\rhorep^2\Omerep^4[\dprod{(\cprod{\vectOme}{\vectr})}{(\cprod{\vectOme}{\vectr})}]
  -2\ethvo\rhorep\Omerep^2[\dprod{(\cprod{\vectOme}{\vectr})}{(\cprod{\unitplz}{\vectOme})}]\\
  &\quad-2\vphig\vphih\rhorep\Omerep^2[\dprod{(\cprod{\vectOme}{\vectr})}{(\cprod{\unitplz}{\vectLam})}]%
  +\ethvo^2[\dprod{(\cprod{\unitplz}{\vectOme})}{(\cprod{\unitplz}{\vectOme})}]
  +2\vphig\vphih\ethvo[\dprod{(\cprod{\unitplz}{\vectOme})}{(\cprod{\unitplz}{\vectLam})}]\\%
  &\quad+\vphig^2\vphih^2[\dprod{(\cprod{\unitplz}{\vectLam})}{(\cprod{\unitplz}{\vectLam})}]
\end{split}
\end{align*}
\begin{align*}
\begin{split}
&=\parvb^2+2\parvc\parvb\epsva-2\parvd\parvb\epsvd+2\parve\parvb\dltva-2\ethvu\parvb\epsve
  +2\parva\parvb\epsvn+2\rhorep\parvb\frktc-2\rhorep\Omerep^2\parvb\epsvk+2\ethvo\parvb\epsvj\\
  &\quad+2\vphig\vphih\parvb\dltvb+\parvc^2\Omerep^2-2\parvd\parvc\epsvb+2\parve\parvc\epsvg
  -2\ethvu\parvc\epsvc+2\parva\parvc\epsvm-2\rhorep\parvc\frktl-2\vphig\vphih\parvc\frkto\\
  &\quad+\parvd^2\scalr^2-2\parve\parvd\epsvh+2\ethvu\parvd\epsvf-2\ethvo\parvd\epsvl
  -2\parvd\vphig\vphih\epsvo+\parve^2\Lamrep^2-2\ethvu\parve\epsvi-2\rhorep\parve\frkth\\
  &\quad+2\rhorep\Omerep^2\parve\epsvm+2\ethvo\parve\frkto+\ethvu^2-2\parva\ethvu\epsvo
  -2\rhorep\ethvu\frktn+2\rhorep\Omerep^2\ethvu\epsvl
  +\parva^2[\dprod{(\cprod{\vectLam}{\vectr})}{(\cprod{\vectLam}{\vectr})}]\\
  &\quad+2\rhorep\parva[\dprod{(\cprod{\vectLam}{\vectr})}{(\cprod{\fdot{\vectLam}}{\vectr})}]
  -2\rhorep\Omerep^2\parva[\dprod{(\cprod{\vectLam}{\vectr})}{(\cprod{\vectOme}{\vectr})}]
  +2\ethvo\parva[\dprod{(\cprod{\vectLam}{\vectr})}{(\cprod{\unitplz}{\vectOme})}]\\
  &\quad+2\vphig\vphih\parva[\dprod{(\cprod{\vectLam}{\vectr})}{(\cprod{\unitplz}{\vectLam})}]
  -2\rhorep^2\Omerep^2[\dprod{(\cprod{\vectOme}{\vectr})}{(\cprod{\fdot{\vectLam}}{\vectr})}]
  +\rhorep^2\Omerep^4[\dprod{(\cprod{\vectOme}{\vectr})}{(\cprod{\vectOme}{\vectr})}]\\
  &\quad-2\ethvo\rhorep\Omerep^2[\dprod{(\cprod{\vectOme}{\vectr})}{(\cprod{\unitplz}{\vectOme})}]
  -2\vphig\vphih\rhorep\Omerep^2[\dprod{(\cprod{\vectOme}{\vectr})}{(\cprod{\unitplz}{\vectLam})}]
  +\rhorep^2[\dprod{(\cprod{\fdot{\vectLam}}{\vectr})}{(\cprod{\fdot{\vectLam}}{\vectr})}]\\
  &\quad+2\ethvo\rhorep[\dprod{(\cprod{\fdot{\vectLam}}{\vectr})}{(\cprod{\unitplz}{\vectOme})}]
  +2\vphig\vphih\rhorep[\dprod{(\cprod{\fdot{\vectLam}}{\vectr})}{(\cprod{\unitplz}{\vectLam})}]
  +\ethvo^2[\dprod{(\cprod{\unitplz}{\vectOme})}{(\cprod{\unitplz}{\vectOme})}]\\
  &\quad+2\vphig\vphih\ethvo[\dprod{(\cprod{\unitplz}{\vectOme})}{(\cprod{\unitplz}{\vectLam})}]
  +\vphig^2\vphih^2[\dprod{(\cprod{\unitplz}{\vectLam})}{(\cprod{\unitplz}{\vectLam})}]
  \beqref{rot1a}, \eqnref{rxpeed1a}, \eqnref{rpath1a}\text{ \& }\eqnref{rpath1b}
\end{split}
\end{align*}
\begin{align*}
\begin{split}
&=\efkcd+\parva^2[\Lamrep^2\scalr^2-(\dprod{\vectLam}{\vectr})^2]
  +2\rhorep\parva[(\dprod{\vectLam}{\fdot{\vectLam}})\scalr^2-(\dprod{\vectLam}{\vectr})(\dprod{\vectr}{\fdot{\vectLam}})]\\
  &\quad-2\rhorep\Omerep^2\parva[(\dprod{\vectLam}{\vectOme})\scalr^2-(\dprod{\vectLam}{\vectr})(\dprod{\vectr}{\vectOme})]
  +2\ethvo\parva[(\dprod{\vectLam}{\unitplz})(\dprod{\vectr}{\vectOme})-(\dprod{\vectLam}{\vectOme})(\dprod{\vectr}{\unitplz})]\\
  &\quad+2\vphig\vphih\parva[(\dprod{\vectLam}{\unitplz})(\dprod{\vectr}{\vectLam})-\Lamrep^2(\dprod{\vectr}{\unitplz})]
  -2\rhorep^2\Omerep^2[(\dprod{\vectOme}{\fdot{\vectLam}})\scalr^2-(\dprod{\vectOme}{\vectr})(\dprod{\vectr}{\fdot{\vectLam}})]\\
  &\quad+\rhorep^2\Omerep^4[\Omerep^2\scalr^2-(\dprod{\vectOme}{\vectr})^2]
  -2\ethvo\rhorep\Omerep^2[(\dprod{\vectOme}{\unitplz})(\dprod{\vectr}{\vectOme})-\Omerep^2(\dprod{\vectr}{\unitplz})]\\
  &\quad-2\vphig\vphih\rhorep\Omerep^2[(\dprod{\vectOme}{\unitplz})(\dprod{\vectr}{\vectLam})-(\dprod{\vectOme}{\vectLam})(\dprod{\vectr}{\unitplz})]
  +\rhorep^2[\fdot{\vectLam}^2\scalr^2-(\dprod{\fdot{\vectLam}}{\vectr})^2]\\
  &\quad+2\ethvo\rhorep[(\dprod{\fdot{\vectLam}}{\unitplz})(\dprod{\vectr}{\vectOme})-(\dprod{\fdot{\vectLam}}{\vectOme})(\dprod{\vectr}{\unitplz})]
  +2\vphig\vphih\rhorep[(\dprod{\fdot{\vectLam}}{\unitplz})(\dprod{\vectr}{\vectLam})-(\dprod{\fdot{\vectLam}}{\vectLam})(\dprod{\vectr}{\unitplz})]\\
  &\quad+\ethvo^2[\Omerep^2-(\dprod{\unitplz}{\Omerep})^2]
  +2\vphig\vphih\ethvo[(\dprod{\vectOme}{\vectLam})-(\dprod{\unitplz}{\vectLam})(\dprod{\vectOme}{\unitplz})]
  +\vphig^2\vphih^2[\Lamrep^2-(\dprod{\unitplz}{\vectLam})^2]\\
  &\quad\beqref{rpathx1m}\text{ \& }\eqnref{alg2}
\end{split}
\end{align*}
\begin{align}\label{rpath27b}
\begin{split}
&=\efkcd+\parva^2(\Lamrep^2\scalr^2-\epsvh^2)
  +2\rhorep\parva(\vsigd\scalr^2-\epsvh\vsign)
  -2\rhorep\Omerep^2\parva(\epsvg\scalr^2-\epsvh\epsvb)
  +2\ethvo\parva(\epsvi\epsvb-\epsvg\epsvf)\\
  &\quad+2\vphig\vphih\parva(\epsvi\epsvh-\Lamrep^2\epsvf)
  -2\rhorep^2\Omerep^2(\vsigc\scalr^2-\epsvb\vsign)
  +\rhorep^2\Omerep^4(\Omerep^2\scalr^2-\epsvb^2)
  -2\ethvo\rhorep\Omerep^2(\epsvc\epsvb-\Omerep^2\epsvf)\\
  &\quad-2\vphig\vphih\rhorep\Omerep^2(\epsvc\epsvh-\epsvg\epsvf)
  +\rhorep^2(\vsige\scalr^2-\vsign^2)
  +2\ethvo\rhorep(\vsiga\epsvb-\vsigc\epsvf)
  +2\vphig\vphih\rhorep(\vsiga\epsvh-\vsigd\epsvf)\\
  &\quad+\ethvo^2(\Omerep^2-\epsvc^2)
  +2\vphig\vphih\ethvo(\epsvg-\epsvi\epsvc)
  +\vphig^2\vphih^2(\Lamrep^2-\epsvi^2)
  \beqref{rot1a}\text{ \& }\eqnref{rpath1a}
\end{split}
\nonumber\\
\therefore|\vscrr|
&=\efkce\beqref{rpathx1m}
\end{align}
\end{subequations}

\subart{Development of equation \eqnref{kpath2d}}
The quantities defined by \eqnref{kpath2d} evaluate as
\begin{subequations}\label{rpath28}
\begin{align}\label{rpath28a}
\alepha
&=\dprod{\unitkap}{(\vscrp+\vscrq)}\beqref{kpath2d}\nonumber\\
&=\efkoy\beqref{rpath26a}
\end{align}
\begin{align}\label{rpath28b}
\alephb
&=\dprod{\vecta}{(\vscrp+\vscrq)}\beqref{kpath2d}\nonumber\\
&=\dprod{(\vscrp+\vscrq)}{[\epsvb\vectOme-\Omerep^2\vectr+\cprod{\vectLam}{\vectr}]}
  \beqref{main4a}\text{ \& }\eqnref{rot1a}\nonumber\\
&=\epsvb[\dprod{\vectOme}{(\vscrp+\vscrq)}]
  -\Omerep^2[\dprod{\vectr}{(\vscrp+\vscrq)}]
  +\dprod{(\cprod{\vectLam}{\vectr})}{(\vscrp+\vscrq)}\nonumber\\
&=\epsvb\efkog-\Omerep^2\efkoe-\efkot
  \beqref{rpath25}
\end{align}
\begin{align}\label{rpath28c}
\alephc
&=\dprod{\fdota}{(\vscrp+\vscrq)}\beqref{kpath2d}\nonumber\\
&=\dprod{(\vscrp+\vscrq)}{[2\epsvh\vectOme-3\epsvg\vectr+\cprod{\fdot{\vectLam}}{\vectr}
  -\Omerep^2(\cprod{\vectOme}{\vectr})+\epsvb\vectLam]}\beqref{rpath7a}\nonumber\\
\begin{split}
&=2\epsvh[\dprod{\vectOme}{(\vscrp+\vscrq)}]-3\epsvg[\dprod{\vectr}{(\vscrp+\vscrq)}]
  +[\dprod{\cprod{(\fdot{\vectLam}}{\vectr})}{(\vscrp+\vscrq)}]\\
  &\qquad-\Omerep^2[\dprod{(\cprod{\vectOme}{\vectr})}{(\vscrp+\vscrq)}]
  +\epsvb[\dprod{\vectLam}{(\vscrp+\vscrq)}]
\end{split}
\nonumber\\
&=2\epsvh\efkog-3\epsvg\efkoe-\efkos-\Omerep^2\efkoo+\epsvb\efkod
  \beqref{rpath25}
\end{align}
\begin{align}\label{rpath28d}
\alephd
&=\dprod{\ffdota}{(\vscrp+\vscrq)}
\beqref{kpath2d}\nonumber\\
\begin{split}
&=\dprod{(\vscrp+\vscrq)}{[}3\epsvh\vectLam+\epsvb\fdot{\vectLam}+\vrhot\vectOme
  +\vrhou\vectr+2\epsvb(\cprod{\vectLam}{\vectOme})
  -3\Omerep^2(\cprod{\vectLam}{\vectr})-3\epsvg(\cprod{\vectOme}{\vectr})\\
  &\qquad+(\cprod{\ffdot{\vectLam}}{\vectr})]\beqref{rpath7b}
\end{split}
\nonumber\\
\begin{split}
&=3\epsvh[\dprod{(\vscrp+\vscrq)}{\vectLam}]
  +\epsvb[\dprod{(\vscrp+\vscrq)}{\fdot{\vectLam}}]
  +\vrhot[\dprod{(\vscrp+\vscrq)}{\vectOme}]
  +\vrhou[\dprod{(\vscrp+\vscrq)}{\vectr}]\\
  &\quad+2\epsvb[\dprod{(\vscrp+\vscrq)}{(\cprod{\vectLam}{\vectOme})}]
  -3\Omerep^2[\dprod{(\vscrp+\vscrq)}{(\cprod{\vectLam}{\vectr})}]
  -3\epsvg[\dprod{(\vscrp+\vscrq)}{(\cprod{\vectOme}{\vectr})}]\\
  &\quad+[\dprod{(\vscrp+\vscrq)}{(\cprod{\ffdot{\vectLam}}{\vectr})}]
\end{split}
\nonumber\\
\begin{split}
&=3\epsvh\efkod+\epsvb\efkoh+\vrhot\efkog+\vrhou\efkoe-2\epsvb\efkop+3\Omerep^2\efkot-3\epsvg\efkoo+\efkca
  \beqref{rpath25}\text{ \& }\eqnref{rpath26c}
\end{split}
\end{align}
\begin{align}\label{rpath28e}
\alephe
&=\dprod{\fffdota}{(\vscrp+\vscrq)}
\beqref{kpath2d}\nonumber\\
\begin{split}
&=\dprod{(\vscrp+\vscrq)}{[}(\vrhot+3\vsign)\vectLam+4\epsvh\fdot{\vectLam}+\epsvb\ffdot{\vectLam}
  +\vrhov\vectOme+5\vrhow\vectr-5\epsvb(\cprod{\vectOme}{\fdot{\vectLam}})
  -6\Omerep^2(\cprod{\fdot{\vectLam}}{\vectr})\\
  &\quad+5\epsvh(\cprod{\vectLam}{\vectOme})
  -12\epsvg(\cprod{\vectLam}{\vectr})+\vrhou(\cprod{\vectOme}{\vectr})
  +\cprod{\fffdot{\vectLam}}{\vectr}]\beqref{rpath7c}
\end{split}
\nonumber\\
\begin{split}
&=(\vrhot+3\vsign)[\dprod{(\vscrp+\vscrq)}{\vectLam}]
  +4\epsvh[\dprod{(\vscrp+\vscrq)}{\fdot{\vectLam}}]
  +\epsvb[\dprod{(\vscrp+\vscrq)}{\ffdot{\vectLam}}]
  +\vrhov[\dprod{(\vscrp+\vscrq)}{\vectOme}]\\
  &\quad+5\vrhow[\dprod{(\vscrp+\vscrq)}{\vectr}]
  -5\epsvb[\dprod{(\vscrp+\vscrq)}{(\cprod{\vectOme}{\fdot{\vectLam}})}]
  -6\Omerep^2[\dprod{(\vscrp+\vscrq)}{(\cprod{\fdot{\vectLam}}{\vectr})}]\\
  &\quad+5\epsvh[\dprod{(\vscrp+\vscrq)}{(\cprod{\vectLam}{\vectOme})}]
  -12\epsvg[\dprod{(\vscrp+\vscrq)}{(\cprod{\vectLam}{\vectr})}]
  +\vrhou[\dprod{(\vscrp+\vscrq)}{(\cprod{\vectOme}{\vectr})}]\\
  &\quad+[\dprod{(\vscrp+\vscrq)}{(\cprod{\fffdot{\vectLam}}{\vectr}})]
\end{split}
\nonumber\\
\begin{split}
&=(\vrhot+3\vsign)\efkod+4\epsvh\efkoh+\epsvb\efkoi+\vrhov\efkog+5\vrhow\efkoe
  -5\epsvb\efkoq+6\Omerep^2\efkos-5\epsvh\efkop\\
  &\quad+12\epsvg\efkot+\vrhou\efkoo+\efkcb
  \beqref{rpath25}\text{ \& }\eqnref{rpath26d}
\end{split}
\end{align}
\begin{align}\label{rpath28f}
\alephf
&=\dprod{\fffdote}{(\vscrp+\vscrq)}
\beqref{kpath2d}\nonumber\\
\begin{split}
&=\dprod{(\vscrp+\vscrq)}{[}\ethvq(\cprod{\unitplz}{\vectOme})+3\ethvp(\cprod{\unitplz}{\vectLam})
  +3\ethvo(\cprod{\unitplz}{\fdot{\vectLam}})+\vphig\vphih(\cprod{\unitplz}{\ffdot{\vectLam}})\\
  &\quad+\ethvt\vectOme+3\ethvs\vectLam+3\ethvr\fdot{\vectLam}
  +\vphii\ffdot{\vectLam}-\ethvw\unitplz]\beqref{rpath17c}
\end{split}
\nonumber\\
\begin{split}
&=\ethvq[\dprod{(\vscrp+\vscrq)}{(\cprod{\unitplz}{\vectOme})}]
  +3\ethvp[\dprod{(\vscrp+\vscrq)}{(\cprod{\unitplz}{\vectLam})}]
  +3\ethvo[\dprod{(\vscrp+\vscrq)}{(\cprod{\unitplz}{\fdot{\vectLam}})}]\\
  &\quad+\vphig\vphih[\dprod{(\vscrp+\vscrq)}{(\cprod{\unitplz}{\ffdot{\vectLam}})}]
  +\ethvt[\dprod{(\vscrp+\vscrq)}{\vectOme}]
  +3\ethvs[\dprod{(\vscrp+\vscrq)}{\vectLam}]\\
  &\quad+3\ethvr[\dprod{(\vscrp+\vscrq)}{\fdot{\vectLam}}]
  +\vphii[\dprod{(\vscrp+\vscrq)}{\ffdot{\vectLam}}]
  -\ethvw[\dprod{(\vscrp+\vscrq)}{\unitplz}]
\end{split}
\nonumber\\
\begin{split}
&=-\ethvq\efkor+3\ethvp\efkow+3\ethvo\efkov-\vphig\vphih\efkoz
  +\ethvt\efkog+3\ethvs\efkod+3\ethvr\efkoh\\
  &\quad+\vphii\efkoi-\ethvw\efkof\beqref{rpath25}\text{ \& }\eqnref{rpath26b}
\end{split}
\end{align}
\end{subequations}
\begin{align}\label{rpath29}
&\fffdot{\cdkt}\alepha+\fffdot{\rhorep}\alephb+3\ffdot{\rhorep}\alephc
  +\vbbc\alephd+\rhorep\alephe+\alephf\nonumber\\
&=\parvg\alepha+\ethvk\alephb+3\ethvj\alephc+\parvi\alephd+\rhorep\alephe+\alephf
  \beqref{rpath23}\text{ \& }\eqnref{rpath12}\nonumber\\
\begin{split}
&=\parvg\efkoy+\ethvk(\epsvb\efkog-\Omerep^2\efkoe-\efkot)
  +3\ethvj(2\epsvh\efkog-3\epsvg\efkoe-\efkos-\Omerep^2\efkoo+\epsvb\efkod)\\
  &\quad+\parvi(3\epsvh\efkod+\epsvb\efkoh+\vrhot\efkog+\vrhou\efkoe-2\epsvb\efkop+3\Omerep^2\efkot-3\epsvg\efkoo+\efkca)\\
  &\quad-\ethvq\efkor+3\ethvp\efkow+3\ethvo\efkov-\vphig\vphih\efkoz
  +\ethvt\efkog+3\ethvs\efkod+3\ethvr\efkoh+\vphii\efkoi-\ethvw\efkof\\
  &\quad+\rhorep[(\vrhot+3\vsign)\efkod+4\epsvh\efkoh+\epsvb\efkoi+\vrhov\efkog+5\vrhow\efkoe
    -5\epsvb\efkoq+6\Omerep^2\efkos-5\epsvh\efkop\\
  &\qquad+12\epsvg\efkot+\vrhou\efkoo+\efkcb]
  \beqref{rpath28}
\end{split}
\nonumber\\
&=\efkcf\beqref{rpathx1n}.
\end{align}

\subart{Results of the computations}
Substituting \eqnref{rpath27}, \eqnref{rpath29}, \eqnref{rpath22d} and \eqnref{rpath24} into
\eqnref{kpath11} leads to
\begin{subequations}\label{rpath30}
\begin{gather}
\bbk=\plusmin\frac{\efkcc}{(\efkce)^3},\quad
\bbt=\frac{\efkcf}{(\efkcc)^2}
\label{rpath30a}\\
\begin{split}
&\frt=\frac{1}{\efkce}\biggl[\parvb\unitkap+\parvc\vectOme-\parvd\vectr+\parve\vectLam-\ethvu\unitplz
  +\parva(\cprod{\vectLam}{\vectr})+\rhorep(\cprod{\fdot{\vectLam}}{\vectr})
  -\rhorep\Omerep^2(\cprod{\vectOme}{\vectr})\\
  &\qquad+\ethvo(\cprod{\unitplz}{\vectOme})+\vphig\vphih(\cprod{\unitplz}{\vectLam})\biggr]
\end{split}
\label{rpath30b}\\
\begin{split}
&\frb=\frac{1}{\efkcc}\biggl[\parvx\vectLam+\parvy\vectr+\parvz\unitplz+\efkoa\vectOme+\efkob\fdot{\vectLam}+\efkoc\ffdot{\vectLam}
  +\parvm(\cprod{\unitkap}{\vectOme})+\parvn(\cprod{\unitkap}{\vectr})+\parvo(\cprod{\unitkap}{\vectLam})\\
  &\qquad+\parvb\parve(\cprod{\unitkap}{\fdot{\vectLam}})+\parvp(\cprod{\unitkap}{\unitplz})
  +\parvq(\cprod{\vectOme}{\vectr})+\parvr(\cprod{\vectOme}{\vectLam})
  +\parvs(\cprod{\vectOme}{\fdot{\vectLam}})+\parvt(\cprod{\vectOme}{\unitplz})\\
  &\qquad-\parve\parvd(\cprod{\vectr}{\fdot{\vectLam}})
  +\parvu(\cprod{\vectr}{\vectLam})+\parvv(\cprod{\vectr}{\unitplz})
  -\ethvu\parve(\cprod{\unitplz}{\fdot{\vectLam}})
  +\parvw(\cprod{\unitplz}{\vectLam})+\parve^2(\cprod{\vectLam}{\fdot{\vectLam}})\biggr]
\end{split}
\label{rpath30c}
\end{gather}
\end{subequations}
as the complete set of equations describing the apparent path of the light source for a rotating observer.

\art{Apparent geometry of obliquated rays}
To evaluate \eqnref{kray4} for a rotating observer, we introduce, in addition to
\eqnref{rot1}, \eqnref{rxpeed1}, \eqnref{rpath1} and \eqnref{rpathx1}, the quantities
\begin{subequations}\label{rray1}
\begin{align}\label{rray1a}
\begin{split}
&\veka=\cdkt\parva-\rhorep\parvb,\quad
\vekb=-\vphig\vphih\epsvl-\parva(\vphij\epsvf-\vphii\epsvb)+\vphig\vphih(\vphij-\vphii\epsvc)\\
&\vekc=\vphii\epsvb-\vphij\epsvf+\vphig\vphih\epsvl,\quad
\vekd=\ethvu\vphii-\vphij(\parva\epsvb+2\rhorep\epsvh+\ethvr)\\
&\veke=-\parvb\epsva-\parva\epsvm-\parva(\vphii\epsvg-\vphij\epsvi)+\rhorep\vphii(\Omerep^4-\vsigc)
     +\rhorep\vphij(\vsiga-\Omerep^2\epsvc)-\rhorep\vphig\vphih\frktp\\
     &\qquad-\rhorep(2\epsvh\Omerep^2-2\epsvg\epsvb-\frktl)-(\ethvr\Omerep^2+\vphii\epsvg-\ethvu\epsvc)
\end{split}
\end{align}
\begin{align}\label{rray1b}
\begin{split}
&\vekf=-\parvb\vphig\vphih\epsva-\parva\vphig\vphih\epsvm+2\rhorep\vphig\vphih(\epsvg\epsvb-\epsvh\Omerep^2)+\vphig\vphih\epsvm\\
     &\qquad+\vphig\vphih(\ethvu\epsvc-\ethvr\Omerep^2-\vphii\epsvg+\vphig\vphih\frkto)
     +\vphii(\ethvo\Omerep^2+\vphig\vphih\epsvg)-\vphij(\ethvo\epsvc+\vphig\vphih\epsvi)\\
&\vekg=\parvb\epsvd+\parvb\vphig\vphih\epsve-\parva\vphia^2-\parva\vphig\vphih(\Omerep^2\epsvf-\epsvb\epsvc-\epsvo)\\
     &\qquad+\rhorep\vphig\vphih(2\epsvh\epsvc-3\epsvg\epsvf+\epsvb\epsvi-\Omerep^2\epsvl)
     +\rhorep\Omerep^2(\vphij\epsvf-\vphii\epsvb)-3\rhorep\vphic\\
     &\qquad+\vphig\vphih(\ethvr\epsvc+\vphii\epsvi-\ethvu)+\ethvo(\vphij-\vphii\epsvc)
     -(\ethvu\epsvf-\ethvr\epsvb-\vphii\epsvh-\ethvo\epsvl)
\end{split}
\end{align}
\begin{align}\label{rray1c}
\begin{split}
&\vekh=\vekf+\rhorep\frkyo+\cdkt(\ethvo\epsva+\vphig\vphih\dltva),\quad
\veki=\veke+\rhorep(\frkyn+\rhorep\frkyh)-\rhorep\cdkt(\vsigb-\Omerep^2\epsva)-\veka\dltva\\
&\vekj=\vekg+\ethvo(\rhorep\frkym-\cdkt\epsve)+\rhorep\Omerep^2(\rhorep\vphia^2-\cdkt\epsvd),\quad
\vekk=\veka\epsvb-\parvb\vphii+\cdkt(\ethvr+2\rhorep\epsvh)\\
&\vekl=\rhorep(\vekc+\cdkt\epsvd-\rhorep\vphia^2),\quad
\vekm=\vekb+\veka\epsvd-\cdkt\vphig\vphih\epsve+\rhorep(\frkyp+3\rhorep\vphic)\\
&\vekn=\veka\Omerep^2+3\rhorep\cdkt\epsvg,\quad
\veko=-\vphii\parvd+\rhorep(\ethvr\Omerep^2+\rhorep\frkyg),\quad
\vekp=\parvb\vphij-\cdkt\ethvu\\
&\vekq=\vekd+\rhorep\ethvu\epsvb,\quad
\vekr=\vphij\parvd-\rhorep\ethvu\Omerep^2
\end{split}
\end{align}
\begin{align}\label{rray1d}
\begin{split}
&\vekr=\vekh+\veki\epsvf+\vekj\epsvc+\vekm\epsvi+\vekl\vsiga-\vekk\epsvj-\vekn\frktt-\cdkt\parve\dltvb
  +\parve^2\frkto+\veko\epsvl+\rhorep\Omerep^2\parve\epsvo\\
&\veks=\vekh\epsvf+\veki\scalr^2+\vekj\epsvb+\vekm\epsvh+\vekl\vsign+\vekk\epsvk+\cdkt\parve\epsvn
  -\vekp\frktt+\parve^2\epsvm+\vekq\epsvl\\
  &\qquad-\vphij\parve\epsvo\\
&\vekt=\vekh\epsvc+\veki\epsvb+\vekj\Omerep^2+\vekm\epsvg+\vekl\vsigc+\vekn\epsvk+\cdkt\parve\frkte
  +\vekp\epsvj+\rhorep\Omerep^2\parve\epsvm\\
  &\qquad-\vekr\epsvl+\vphij\parve\frkto\\
&\veku=\vekh\epsvi+\veki\epsvh+\vekj\epsvg+\vekm\Lamrep^2+\vekl\vsigd
  -\vekk\frkte+\vekn\epsvn+\vekp\dltvb+\vekq\frkto-\veko\epsvm-\vekr\epsvo\\
&\vekv=\vekh\vsiga+\veki\vsign+\vekj\vsigc+\vekm\vsigd+\vekl\vsige+\vekk\frktf+\vekn\frktc+\cdkt\parve\frktu
  +\vekp\frkta+\parve^2\frktv+\vekq\frktp\\
  &\qquad+\veko\frktl+\rhorep\Omerep^2\parve\frkth-\vekr\frktn-\vphij\parve\frktr
\end{split}
\end{align}
\begin{align}\label{rray1e}
\begin{split}
&\vekw=-\vekh\epsvj+\veki\epsvk-\vekm\frkte+\vekl\frktf+\vekk(\Omerep^2-\epsva^2)
  -\vekn(\epsvb-\epsvd\epsva)+\cdkt\parve(\epsvg-\dltva\epsva)\\
  &\qquad+\vekp(\epsvc-\epsve\epsva)+\parve^2(\epsva\epsvg-\dltva\Omerep^2)
  +\vekq(\epsve\Omerep^2-\epsva\epsvc)+\veko(\epsva\epsvb-\epsvd\Omerep^2)\\
  &\qquad-\rhorep\Omerep^2\parve(\epsvd\epsvg-\dltva\epsvb)+\vekr(\epsve\epsvb-\epsvd\epsvc)
  -\vphij\parve(\epsve\epsvg-\dltva\epsvc)\\
&\vekx=\vekh\frktt-\vekj\epsvk-\vekm\epsvn-\vekl\frktc
  +\vekk(\epsvb-\epsva\epsvd)-\vekn(\scalr^2-\epsvd^2)+\cdkt\parve(\epsvh-\dltva\epsvd)\\
  &\qquad+\vekp(\epsvf-\epsve\epsvd)+\parve^2(\epsva\epsvh-\dltva\epsvb)
  +\vekq(\epsve\epsvb-\epsva\epsvf)+\veko(\epsva\scalr^2-\epsvd\epsvb)\\
  &\qquad-\rhorep\Omerep^2\parve(\epsvd\epsvh-\dltva\scalr^2)+\vekr(\epsve\scalr^2-\epsvd\epsvf)
  -\vphij\parve(\epsve\epsvh-\dltva\epsvf)
\end{split}
\end{align}
\begin{align}\label{rray1f}
\begin{split}
&\veky=-\vekh\dltvb+\veki\epsvn+\vekj\frkte+\vekl\frktu
  +\vekk(\epsvg-\epsva\dltva)-\vekn(\epsvh-\epsvd\dltva)+\cdkt\parve(\Lamrep^2-\dltva^2)\\
  &\qquad+\vekp(\epsvi-\epsve\dltva)+\parve^2(\epsva\Lamrep^2-\dltva\epsvg)
  +\vekq(\epsve\epsvg-\epsva\epsvi)+\veko(\epsva\epsvh-\epsvd\epsvg)\\
  &\qquad-\rhorep\Omerep^2\parve(\epsvd\Lamrep^2-\dltva\epsvh)+\vekr(\epsve\epsvh-\epsvd\epsvi)
  -\vphij\parve(\epsve\Lamrep^2-\dltva\epsvi)\\
&\vekz=-\veki\frktt+\vekj\epsvj+\vekm\dltvb+\vekl\frkta
  +\vekk(\epsvc-\epsva\epsve)-\vekn(\epsvf-\epsvd\epsve)+\cdkt\parve(\epsvi-\dltva\epsve)\\
  &\qquad+\vekp(1-\epsve^2)+\parve^2(\epsva\epsvi-\dltva\epsvc)
  +\vekq(\epsve\epsvc-\epsva)+\veko(\epsva\epsvf-\epsvd\epsvc)\\
  &\qquad-\rhorep\Omerep^2\parve(\epsvd\epsvi-\dltva\epsvf)+\vekr(\epsve\epsvf-\epsvd)
  -\vphij\parve(\epsve\epsvi-\dltva)
\end{split}
\end{align}
\begin{align}\label{rray1g}
\begin{split}
&\vpsa=\vekh\frkto+\veki\epsvm+\vekl\frktv
  +\vekk(\epsva\epsvg-\Omerep^2\dltva)-\vekn(\epsva\epsvh-\epsvb\dltva)+\cdkt\parve(\epsva\Lamrep^2-\epsvg\dltva)\\
  &\qquad+\vekp(\epsva\epsvi-\epsvc\dltva)+\parve^2(\Omerep^2\Lamrep^2-\epsvg^2)
  +\vekq(\epsvc\epsvg-\Omerep^2\epsvi)+\veko(\Omerep^2\epsvh-\epsvb\epsvg)\\
  &\qquad-\rhorep\Omerep^2\parve(\epsvb\Lamrep^2-\epsvg\epsvh)+\vekr(\epsvc\epsvh-\epsvb\epsvi)
  -\vphij\parve(\epsvc\Lamrep^2-\epsvg\epsvi)\\
&\vpsb=\veki\epsvl+\vekm\frkto+\vekl\frktp
  +\vekk(\epsve\Omerep^2-\epsvc\epsva)-\vekn(\epsve\epsvb-\epsvf\epsva)+\cdkt\parve(\epsve\epsvg-\epsvi\epsva)\\
  &\qquad+\vekp(\epsve\epsvc-\epsva)+\parve^2(\epsvc\epsvg-\epsvi\Omerep^2)
  +\vekq(\Omerep^2-\epsvc^2)+\veko(\epsvc\epsvb-\epsvf\Omerep^2)\\
  &\qquad-\rhorep\Omerep^2\parve(\epsvf\epsvg-\epsvi\epsvb)
  +\vekr(\epsvb-\epsvf\epsvc)-\vphij\parve(\epsvg-\epsvi\epsv)
\end{split}
\end{align}
\begin{align}\label{rray1h}
\begin{split}
&\vpsc=\vekh\epsvl-\vekm\epsvm+\vekl\frktl
  +\vekk(\epsva\epsvb-\Omerep^2\epsvd)-\vekn(\epsva\scalr^2-\epsvb\epsvd)+\cdkt\parve(\epsva\epsvh-\epsvg\epsvd)\\
  &\qquad+\vekp(\epsva\epsvf-\epsvc\epsvd)+\parve^2(\Omerep^2\epsvh-\epsvg\epsvb)
  +\vekq(\epsvc\epsvb-\Omerep^2\epsvf)+\veko(\Omerep^2\scalr^2-\epsvb^2)\\
  &\qquad-\rhorep\Omerep^2\parve(\epsvb\epsvh-\epsvg\scalr^2)+\vekr(\epsvc\scalr^2-\epsvb\epsvf)
  -\vphij\parve(\epsvc\epsvh-\epsvg\epsvf)\\
&\vpsd=-\vekh\epsvo-\vekj\epsvm-\vekl\frkth
  +\vekk(\epsvd\epsvg-\epsvb\dltva)-\vekn(\epsvd\epsvh-\scalr^2\dltva)+\cdkt\parve(\epsvd\Lamrep^2-\epsvh\dltva)\\
  &\qquad+\vekp(\epsvd\epsvi-\epsvf\dltva)+\parve^2(\epsvb\Lamrep^2-\epsvh\epsvg)
  +\vekq(\epsvf\epsvg-\epsvb\epsvi)+\veko(\epsvb\epsvh-\scalr^2\epsvg)\\
  &\qquad-\rhorep\Omerep^2\parve(\scalr^2\Lamrep^2-\epsvh^2)+\vekr(\epsvf\epsvh-\scalr^2\epsvi)
  -\vphij\parve(\epsvf\Lamrep^2-\epsvh\epsvi)
\end{split}
\end{align}
\begin{align}\label{rray1i}
\begin{split}
&\vpse=-\vekj\epsvl-\vekm\epsvo-\vekl\frktn
  +\vekk(\epsve\epsvb-\epsvc\epsvd)-\vekn(\epsve\scalr^2-\epsvf\epsvd)+\cdkt\parve(\epsve\epsvh-\epsvi\epsvd)\\
  &\qquad+\vekp(\epsve\epsvf-\epsvd)+\parve^2(\epsvc\epsvh-\epsvi\epsvb)
  +\vekq(\epsvb-\epsvc\epsvf)+\veko(\epsvc\scalr^2-\epsvf\epsvb)\\
  &\qquad-\rhorep\Omerep^2\parve(\epsvf\epsvh-\epsvi\scalr^2)+\vekr(\scalr^2-\epsvf^2)
  -\vphij\parve(\epsvh-\epsvi\epsvf)\\
&\vpsf=\veki\epsvo-\vekj\frkto+\vekl\frktr+\vekk(\epsve\epsvg-\epsvc\dltva)
  -\vekn(\epsve\epsvh-\epsvf\dltva)+\cdkt\parve(\epsve\Lamrep^2-\epsvi\dltva)\\
  &\qquad+\vekp(\epsve\epsvi-\dltva)+\parve^2(\epsvc\Lamrep^2-\epsvi\epsvg)
  +\vekq(\epsvg-\epsvc\epsvi)+\veko(\epsvc\epsvh-\epsvf\epsvg)\\
  &\qquad-\rhorep\Omerep^2\parve(\epsvf\Lamrep^2-\epsvi\epsvh)+\vekr(\epsvh-\epsvf\epsvi)
  -\vphij\parve(\Lamrep^2-\epsvi^2)
\end{split}
\end{align}
\begin{align}\label{rray1j}
\begin{split}
&\vpsg=-\veki\frktn+\vekj\frktp+\vekm\frktr+\vekk(\vsigb\epsvc-\vsigc\epsve)
  -\vekn(\vsigb\epsvf-\vsign\epsve)+\cdkt\parve(\vsigb\epsvi-\vsigd\epsve)\\
  &\qquad+\vekp(\vsigb-\vsiga\epsve)+\parve^2(\vsigc\epsvi-\vsigd\epsvc)
  +\vekq(\vsiga\epsvc-\vsigc)+\veko(\vsigc\epsvf-\vsign\epsvc)\\
  &\qquad-\rhorep\Omerep^2\parve(\vsign\epsvi-\vsigd\epsvf)
  +\vekr(\vsiga\epsvf-\vsign)-\vphij\parve(\vsiga\epsvi-\vsigd)\\
&\vpsh=\vekh\frktn-\vekj\frktl-\vekm\frkth+\vekk(\vsigb\epsvb-\vsigc\epsvd)
  -\vekn(\vsigb\scalr^2-\vsign\epsvd)+\cdkt\parve(\vsigb\epsvh-\vsigd\epsvd)\\
  &\qquad+\vekp(\vsigb\epsvf-\vsiga\epsvd)+\parve^2(\vsigc\epsvh-\vsigd\epsvb)
  +\vekq(\vsiga\epsvb-\vsigc\epsvf)+\veko(\vsigc\scalr^2-\vsign\epsvb)\\
  &\qquad-\rhorep\Omerep^2\parve(\vsign\epsvh-\vsigd\scalr^2)+\vekr(\vsiga\scalr^2-\vsign\epsvf)
  -\vphij\parve(\vsiga\epsvh-\vsigd\epsvf)
\end{split}
\end{align}
\begin{align}\label{rray1k}
\begin{split}
&\vpsi=\vekh\frkxb-\vekj\frktm-\vekm\frkti-\vekl\frktj+\vekk(\vsigg\epsvb-\vsigh\epsvd)-\vekn(\vsigg\scalr^2-\vsigo\epsvd)\\
  &\qquad+\cdkt\parve(\vsigg\epsvh-\vsigi\epsvd)+\vekp(\vsigg\epsvf-\vsigf\epsvd)+\parve^2(\vsigh\epsvh-\vsigi\epsvb)
  +\vekq(\vsigf\epsvb-\vsigh\epsvf)\\
  &\qquad+\veko(\vsigh\scalr^2-\vsigo\epsvb)-\rhorep\Omerep^2\parve(\vsigo\epsvh-\vsigi\scalr^2)
  +\vekr(\vsigf\scalr^2-\vsigo\epsvf)-\vphij\parve(\vsigf\epsvh-\vsigi\epsvf)\\
&\vpsj=\vekh\epsve+\veki\epsvd+\vekj\epsva+\vekm\dltva+\vekl\vsigb-\parve^2\frkte+\vekq\epsvj+\veko\epsvk\\
  &\qquad+\rhorep\Omerep^2\parve\epsvn-\vekr\frktt-\vphij\parve\dltvb
\end{split}
\end{align}
\begin{align}\label{rray1l}
\begin{split}
&\vpsk=[\vekh\vekr+\veki\veks+\vekj\vekt+\vekm\veku+\vekl\vekv+\vekk\vekw-\vekn\vekx+\cdkt\parve\veky\\
  &\qquad+\vekp\vekz+\parve^2\vpsa+\vekq\vpsb+\veko\vpsc-\rhorep\Omerep^2\parve\vpsd+\vekr\vpse-\vphij\parve\vpsf]^{1/2}\\
&\vpsl=\parvf\vpsj+\ethvj(\epsvb\vekt-\Omerep^2\veks-\vpsd)
  +\parvh(2\epsvh\vekt-3\epsvg\veks+\vpsh-\Omerep^2\vpsc+\epsvb\veku)\\
  &\qquad+\rhorep(3\epsvh\veku+\epsvb\vekv+\vrhot\vekt+\vrhou\veks-2\epsvb\vpsa
    +3\Omerep^2\vpsd-3\epsvg\vpsc+\vpsi)\\
  &\qquad+\ethvp\vpsb+2\ethvo\vpsf-\vphig\vphih\vpsg+\ethvs\vekt+2\ethvr\veku
  +\vphii\vekv-\ethvv\vekr.
\end{split}
\end{align}
\end{subequations}

\subart{Development of equation \eqnref{kray2a}}
With the above quantities in view, we derive
\begin{subequations}\label{rray2}
\begin{align}\label{rray2a}
\veusa
&=\cprod{\unitkap}{\vectu}\beqref{kray2a}\nonumber\\
&=\cprod{\unitkap}{(\cprod{\vectOme}{\vectr})}\beqref{main4c}\nonumber\\
&=\vectOme(\dprod{\unitkap}{\vectr})-\vectr(\dprod{\unitkap}{\vectOme})\beqref{alg1}\nonumber\\
&=\epsvd\vectOme-\epsva\vectr\beqref{rot1a}
\end{align}
\begin{align}\label{rray2b}
\veusb
&=\cprod{\unitkap}{\vecta}\beqref{kray2a}\nonumber\\
&=\cprod{\unitkap}{(\epsvb\vectOme-\Omerep^2\vectr+\cprod{\vectLam}{\vectr})}
  \beqref{main4a}\text{ \& }\eqnref{rot1a}\nonumber\\
&=\epsvb(\cprod{\unitkap}{\vectOme})
  -\Omerep^2(\cprod{\unitkap}{\vectr})
  +\cprod{\unitkap}{(\cprod{\vectLam}{\vectr})}\nonumber\\
&=\epsvb(\cprod{\unitkap}{\vectOme})
  -\Omerep^2(\cprod{\unitkap}{\vectr})
  +\vectLam(\dprod{\unitkap}{\vectr})-\vectr(\dprod{\unitkap}{\vectLam})
  \beqref{alg1}\nonumber\\
&=\epsvb(\cprod{\unitkap}{\vectOme})-\Omerep^2(\cprod{\unitkap}{\vectr})+\epsvd\vectLam-\dltva\vectr
  \beqref{rot1a}\text{ \& }\eqnref{rxpeed1a}
\end{align}
\begin{align}\label{rray2c}
\veusc
&=\cprod{\unitkap}{\vecte}\beqref{kray2a}\nonumber\\
&=\cprod{\unitkap}{[\stwo(\cprod{\unitplz}{\vectOme})+\sthr\vectOme-\sfou\unitplz]}
  \beqref{main4b}\nonumber\\
&=\stwo[\cprod{\unitkap}{(\cprod{\unitplz}{\vectOme})}]
  +\sthr(\cprod{\unitkap}{\vectOme})
  -\sfou(\cprod{\unitkap}{\unitplz})\nonumber\\
&=\stwo[\unitplz(\dprod{\unitkap}{\vectOme})-\vectOme(\dprod{\unitkap}{\unitplz})]
  +\sthr(\cprod{\unitkap}{\vectOme})
  -\sfou(\cprod{\unitkap}{\unitplz})
  \beqref{alg1}\nonumber\\
&=\stwo(\epsva\unitplz-\epsve\vectOme)+\sthr(\cprod{\unitkap}{\vectOme})
  -\sfou(\cprod{\unitkap}{\unitplz})\beqref{rot1a}\nonumber\\
&=\vphig\vphih(\epsva\unitplz-\epsve\vectOme)+\vphii(\cprod{\unitkap}{\vectOme})
  -\vphij(\cprod{\unitkap}{\unitplz})\beqref{rot5}
\end{align}
\begin{align}\label{rray2d}
\veusd
&=\cprod{\unitkap}{\fdota}\beqref{kray2a}\nonumber\\
&=\cprod{\unitkap}{[2\epsvh\vectOme-3\epsvg\vectr+\cprod{\fdot{\vectLam}}{\vectr}
   -\Omerep^2(\cprod{\vectOme}{\vectr})+\epsvb\vectLam]}\beqref{rpath7a}\nonumber\\
&=2\epsvh(\cprod{\unitkap}{\vectOme})
  -3\epsvg(\cprod{\unitkap}{\vectr})
  +[\cprod{\unitkap}{(\cprod{\fdot{\vectLam}}{\vectr})}]
  -\Omerep^2[\cprod{\unitkap}{(\cprod{\vectOme}{\vectr})}]
  +\epsvb(\cprod{\unitkap}{\vectLam})\nonumber\\
\begin{split}
&=2\epsvh(\cprod{\unitkap}{\vectOme})
  -3\epsvg(\cprod{\unitkap}{\vectr})
  +\epsvb(\cprod{\unitkap}{\vectLam})
  +[\fdot{\vectLam}(\dprod{\unitkap}{\vectr})-\vectr(\dprod{\unitkap}{\fdot{\vectLam}})]\\
  &\quad-\Omerep^2[\vectOme(\dprod{\unitkap}{\vectr})-\vectr(\dprod{\unitkap}{\vectOme})]
  \beqref{alg1}
\end{split}
\nonumber\\
\begin{split}
&=2\epsvh(\cprod{\unitkap}{\vectOme})
  -3\epsvg(\cprod{\unitkap}{\vectr})
  +\epsvb(\cprod{\unitkap}{\vectLam})
  +\epsvd\fdot{\vectLam}-\vsigb\vectr\\
  &\quad-\Omerep^2(\epsvd\vectOme-\epsva\vectr)
  \beqref{rot1a}\text{ \& }\eqnref{rpath1a}
\end{split}
\end{align}
\begin{align}\label{rray2e}
\veuse
&=\cprod{\unitkap}{\fdote}\beqref{kray2a}\nonumber\\
&=\cprod{\unitkap}{[\ethvo(\cprod{\unitplz}{\vectOme})+\vphig\vphih(\cprod{\unitplz}{\vectLam})
  +\ethvr\vectOme+\vphii\vectLam-\ethvu\unitplz]}\beqref{rpath17a}\nonumber\\
&=\ethvo[\cprod{\unitkap}{(\cprod{\unitplz}{\vectOme})}]
  +\vphig\vphih[\cprod{\unitkap}{(\cprod{\unitplz}{\vectLam})}]
  +\ethvr(\cprod{\unitkap}{\vectOme})
  +\vphii(\cprod{\unitkap}{\vectLam})
  -\ethvu(\cprod{\unitkap}{\unitplz})\nonumber\\
\begin{split}
&=\ethvo[\unitplz(\dprod{\unitkap}{\vectOme})-\vectOme(\dprod{\unitkap}{\unitplz})]
  +\vphig\vphih[\unitplz(\dprod{\unitkap}{\vectLam})-\vectLam(\dprod{\unitkap}{\unitplz})]
  +\ethvr(\cprod{\unitkap}{\vectOme})\\
  &\quad+\vphii(\cprod{\unitkap}{\vectLam})
  -\ethvu(\cprod{\unitkap}{\unitplz})
  \beqref{alg1}
\end{split}
\nonumber\\
\begin{split}
&=\ethvo(\epsva\unitplz-\epsve\vectOme)
  +\vphig\vphih(\dltva\unitplz-\epsve\vectLam)
  +\ethvr(\cprod{\unitkap}{\vectOme})
  +\vphii(\cprod{\unitkap}{\vectLam})
  -\ethvu(\cprod{\unitkap}{\unitplz})\\
  &\quad\beqref{rot1a}\text{ \& }\eqnref{rxpeed1a}
\end{split}
\end{align}
\begin{align}\label{rray2f}
\veusf
&=\cprod{\vecta}{\vectu}\beqref{kray2a}\nonumber\\
&=\cprod{(\epsvb\vectOme-\Omerep^2\vectr+\cprod{\vectLam}{\vectr})}{(\cprod{\vectOme}{\vectr})}
  \beqref{main4a}, \eqnref{main4c}\text{ \& }\eqnref{rot1a}\nonumber\\
&=\epsvb[\cprod{\vectOme}{(\cprod{\vectOme}{\vectr})}]
  -\Omerep^2[\cprod{\vectr}{(\cprod{\vectOme}{\vectr})}]
  +\cprod{(\cprod{\vectLam}{\vectr})}{(\cprod{\vectOme}{\vectr})}\nonumber\\
\begin{split}
&=\epsvb[\vectOme(\dprod{\vectOme}{\vectr})-\Omerep^2\vectr]
  -\Omerep^2[\scalr^2\vectOme-\vectr(\dprod{\vectr}{\vectOme})]
  +[\vectOme(\dprod{\vectr}{(\cprod{\vectLam}{\vectr})})-\vectr((\dprod{\vectOme}{(\cprod{\vectLam}{\vectr})}))]\\
  &\quad\beqref{alg1}\text{ \& }\eqnref{alg5}
\end{split}
\nonumber\\
&=\epsvb(\epsvb\vectOme-\Omerep^2\vectr)-\Omerep^2(\scalr^2\vectOme-\epsvb\vectr)
  -\epsvm\vectr\beqref{rot1a}\nonumber\\
&=(\epsvb^2-\Omerep^2\scalr^2)\vectOme-\epsvm\vectr
=-\vphia^2\vectOme-\epsvm\vectr\beqref{rot1b}
\end{align}
\begin{align*}
\veusg
&=\cprod{\vecta}{\vecte}\beqref{kray2a}\nonumber\\
&=\cprod{(\epsvb\vectOme-\Omerep^2\vectr+\cprod{\vectLam}{\vectr})}
  {[\stwo(\cprod{\unitplz}{\vectOme})+\sthr\vectOme-\sfou\unitplz]}
  \beqref{main4a}, \eqnref{main4b}\text{ \& }\eqnref{rot1a}\nonumber\\
\begin{split}
&=\stwo\epsvb[(\cprod{\vectOme}{(\cprod{\unitplz}{\vectOme})}]
  +\sthr\epsvb(\cprod{\vectOme}{\vectOme})
  -\sfou\epsvb(\cprod{\vectOme}{\unitplz})
  -\stwo\Omerep^2[\cprod{\vectr}{(\cprod{\unitplz}{\vectOme})}]\\
  &\quad-\sthr\Omerep^2(\cprod{\vectr}{\vectOme})
  +\sfou\Omerep^2(\cprod{\vectr}{\unitplz})
  +\stwo[\cprod{(\cprod{\vectLam}{\vectr})}{(\cprod{\unitplz}{\vectOme})}]\\
  &\quad-\sthr[\cprod{\vectOme}{(\cprod{\vectLam}{\vectr})}]
  +\sfou[\cprod{\unitplz}{(\cprod{\vectLam}{\vectr})}]
\end{split}
\nonumber\\
\begin{split}
&=\sfou\Omerep^2(\cprod{\vectr}{\unitplz})
  -\sfou\epsvb(\cprod{\vectOme}{\unitplz})
  -\sthr\Omerep^2(\cprod{\vectr}{\vectOme})
  +\stwo\epsvb[\Omerep^2\unitplz-\vectOme(\dprod{\vectOme}{\unitplz})]\\
  &\quad-\stwo\Omerep^2[\unitplz(\dprod{\vectr}{\vectOme})-\vectOme(\dprod{\vectr}{\unitplz})]
  -\sthr[\vectLam(\dprod{\vectOme}{\vectr})-\vectr(\dprod{\vectOme}{\vectLam})]
  +\sfou[\vectLam(\dprod{\unitplz}{\vectr})-\vectr(\dprod{\unitplz}{\vectLam})]\\
  &\quad+\stwo[\unitplz(\dprod{\vectOme}{(\cprod{\vectLam}{\vectr})})-\vectOme(\dprod{\unitplz}{(\cprod{\vectLam}{\vectr})})]
  \beqref{alg1}\text{ \& }\eqnref{alg5}
\end{split}
\end{align*}
\begin{align*}
\begin{split}
&=\sfou\Omerep^2(\cprod{\vectr}{\unitplz})
  -\sfou\epsvb(\cprod{\vectOme}{\unitplz})
  -\sthr\Omerep^2(\cprod{\vectr}{\vectOme})
  +\stwo\epsvb(\Omerep^2\unitplz-\epsvc\vectOme)\\
  &\quad-\stwo\Omerep^2(\epsvb\unitplz-\epsvf\vectOme)
  -\sthr(\epsvb\vectLam-\epsvg\vectr)
  +\sfou(\epsvf\vectLam-\epsvi\vectr)
  +\stwo(\epsvm\unitplz-\epsvo\vectOme)\beqref{rot1a}
\end{split}
\nonumber\\
\begin{split}
&=\sfou\Omerep^2(\cprod{\vectr}{\unitplz})
  -\sfou\epsvb(\cprod{\vectOme}{\unitplz})
  -\sthr\Omerep^2(\cprod{\vectr}{\vectOme})
  +\stwo\epsvb\Omerep^2\unitplz-\stwo\epsvb\epsvc\vectOme\\
  &\quad-\stwo\Omerep^2\epsvb\unitplz+\stwo\Omerep^2\epsvf\vectOme
  -\sthr\epsvb\vectLam+\sthr\epsvg\vectr
  +\sfou\epsvf\vectLam-\sfou\epsvi\vectr
  +\stwo\epsvm\unitplz-\stwo\epsvo\vectOme
\end{split}
\nonumber\\
\begin{split}
&=\sfou\Omerep^2(\cprod{\vectr}{\unitplz})
  -\sfou\epsvb(\cprod{\vectOme}{\unitplz})
  -\sthr\Omerep^2(\cprod{\vectr}{\vectOme})
  +\stwo\epsvb\Omerep^2\unitplz-\stwo\Omerep^2\epsvb\unitplz+\stwo\epsvm\unitplz\\
  &\quad-\stwo\epsvb\epsvc\vectOme+\stwo\Omerep^2\epsvf\vectOme-\stwo\epsvo\vectOme
  -\sthr\epsvb\vectLam+\sfou\epsvf\vectLam
  +\sthr\epsvg\vectr-\sfou\epsvi\vectr
\end{split}
\end{align*}
\begin{align}\label{rray2g}
\begin{split}
&=\sfou\Omerep^2(\cprod{\vectr}{\unitplz})
  -\sfou\epsvb(\cprod{\vectOme}{\unitplz})
  -\sthr\Omerep^2(\cprod{\vectr}{\vectOme})
  +\stwo(\epsvb\Omerep^2-\Omerep^2\epsvb+\epsvm)\unitplz\\
  &\quad+\stwo(\Omerep^2\epsvf-\epsvb\epsvc-\epsvo)\vectOme
  +(\sfou\epsvf-\sthr\epsvb)\vectLam
  +(\sthr\epsvg-\sfou\epsvi)\vectr
\end{split}
\nonumber\\
\begin{split}
&=\vphij\Omerep^2(\cprod{\vectr}{\unitplz})
  -\vphij\epsvb(\cprod{\vectOme}{\unitplz})
  -\vphii\Omerep^2(\cprod{\vectr}{\vectOme})
  +\vphig\vphih\epsvm\unitplz\\
  &\quad+\vphig\vphih(\Omerep^2\epsvf-\epsvb\epsvc-\epsvo)\vectOme
  +(\vphij\epsvf-\vphii\epsvb)\vectLam
  +(\vphii\epsvg-\vphij\epsvi)\vectr
  \beqref{rot5}
\end{split}
\end{align}
\begin{align*}
\veush
&=\cprod{\vecta}{\fdota}\beqref{kray2a}\nonumber\\
&=\cprod{(\epsvb\vectOme-\Omerep^2\vectr+\cprod{\vectLam}{\vectr})}
  {[2\epsvh\vectOme-3\epsvg\vectr+\cprod{\fdot{\vectLam}}{\vectr}-\Omerep^2(\cprod{\vectOme}{\vectr})+\epsvb\vectLam]}\\
   &\quad\beqref{main4a}, \eqnref{rot1a}\text{ \& }\eqnref{rpath7a}\nonumber\\
\begin{split}
&=2\epsvh\epsvb(\cprod{\vectOme}{\vectOme})
  -3\epsvg\epsvb(\cprod{\vectOme}{\vectr})
  +\epsvb[\cprod{\vectOme}{(\cprod{\fdot{\vectLam}}{\vectr})}]
  -\Omerep^2\epsvb[\cprod{\vectOme}{(\cprod{\vectOme}{\vectr})}]\\
  &\quad+\epsvb^2(\cprod{\vectOme}{\vectLam})
  -2\epsvh\Omerep^2(\cprod{\vectr}{\vectOme})
  +3\epsvg\Omerep^2(\cprod{\vectr}{\vectr})
  -\Omerep^2[\cprod{\vectr}{(\cprod{\fdot{\vectLam}}{\vectr})}]
  +\Omerep^4[\cprod{\vectr}{(\cprod{\vectOme}{\vectr})}]\\
  &\quad-\epsvb\Omerep^2(\cprod{\vectr}{\vectLam})
  +2\epsvh[\cprod{(\cprod{\vectLam}{\vectr})}{\vectOme}]
  -3\epsvg[\cprod{(\cprod{\vectLam}{\vectr})}{\vectr}]
  +[\cprod{(\cprod{\vectLam}{\vectr})}{(\cprod{\fdot{\vectLam}}{\vectr})}]\\
  &\quad-\Omerep^2[\cprod{(\cprod{\vectLam}{\vectr})}{(\cprod{\vectOme}{\vectr})}]
  +\epsvb[\cprod{(\cprod{\vectLam}{\vectr})}{\vectLam}]
\end{split}
\nonumber\\
\begin{split}
&=-3\epsvg\epsvb(\cprod{\vectOme}{\vectr})
  +\epsvb^2(\cprod{\vectOme}{\vectLam})
  -2\epsvh\Omerep^2(\cprod{\vectr}{\vectOme})
  -\epsvb\Omerep^2(\cprod{\vectr}{\vectLam})
  +\epsvb[\cprod{\vectOme}{(\cprod{\fdot{\vectLam}}{\vectr})}]\\
  &\quad-\Omerep^2\epsvb[\cprod{\vectOme}{(\cprod{\vectOme}{\vectr})}]
  -\Omerep^2[\cprod{\vectr}{(\cprod{\fdot{\vectLam}}{\vectr})}]
  +\Omerep^4[\cprod{\vectr}{(\cprod{\vectOme}{\vectr})}]
  -2\epsvh[\cprod{\vectOme}{(\cprod{\vectLam}{\vectr})}]\\
  &\quad+3\epsvg[\cprod{\vectr}{(\cprod{\vectLam}{\vectr})}]
  -\epsvb[\cprod{\vectLam}{(\cprod{\vectLam}{\vectr})}]
  +[\cprod{(\cprod{\vectLam}{\vectr})}{(\cprod{\fdot{\vectLam}}{\vectr})}]
  -\Omerep^2[\cprod{(\cprod{\vectLam}{\vectr})}{(\cprod{\vectOme}{\vectr})}]
\end{split}
\end{align*}
\begin{align*}
\begin{split}
&=-3\epsvg\epsvb(\cprod{\vectOme}{\vectr})
  +\epsvb^2(\cprod{\vectOme}{\vectLam})
  -2\epsvh\Omerep^2(\cprod{\vectr}{\vectOme})
  -\epsvb\Omerep^2(\cprod{\vectr}{\vectLam})
  +\epsvb[\fdot{\vectLam}(\dprod{\vectOme}{\vectr})-\vectr(\dprod{\vectOme}{\fdot{\vectLam}})]\\
  &\quad-\Omerep^2\epsvb[\vectOme(\dprod{\vectOme}{\vectr})-\Omerep^2\vectr]
  -\Omerep^2[\scalr^2\fdot{\vectLam}-\vectr(\dprod{\vectr}{\fdot{\vectLam}})]
  +\Omerep^4[\scalr^2\vectOme-\vectr(\dprod{\vectr}{\vectOme})]\\
  &\quad-2\epsvh[\vectLam(\dprod{\vectOme}{\vectr})-\vectr(\dprod{\vectOme}{\vectLam})]
  +3\epsvg[\scalr^2\vectLam-\vectr(\dprod{\vectr}{\vectLam})]
  -\epsvb[\vectLam(\dprod{\vectLam}{\vectr})-\Lamrep^2\vectr]\\
  &\quad+[\fdot{\vectLam}(\dprod{\vectr}{(\cprod{\vectLam}{\vectr})})-\vectr(\dprod{\fdot{\vectLam}}{(\cprod{\vectLam}{\vectr})})]
  -\Omerep^2[\vectOme(\dprod{\vectr}{(\cprod{\vectLam}{\vectr})})-\vectr(\dprod{\vectOme}{(\cprod{\vectLam}{\vectr})})]
\end{split}
\nonumber\\
\begin{split}
&=-3\epsvg\epsvb(\cprod{\vectOme}{\vectr})
  +\epsvb^2(\cprod{\vectOme}{\vectLam})
  -2\epsvh\Omerep^2(\cprod{\vectr}{\vectOme})
  -\epsvb\Omerep^2(\cprod{\vectr}{\vectLam})
  +\epsvb(\epsvb\fdot{\vectLam}-\vsigc\vectr)\\
  &\quad-\Omerep^2\epsvb(\epsvb\vectOme-\Omerep^2\vectr)
  -\Omerep^2(\scalr^2\fdot{\vectLam}-\vsign\vectr)
  +\Omerep^4(\scalr^2\vectOme-\epsvb\vectr)
  -2\epsvh(\epsvb\vectLam-\epsvg\vectr)\\
  &\quad+3\epsvg(\scalr^2\vectLam-\epsvh\vectr)
  -\epsvb(\epsvh\vectLam-\Lamrep^2\vectr)
  -\frkth\vectr+\Omerep^2\epsvm\vectr
  \beqref{rot1a}, \eqnref{rpath1a}\text{ \& }\eqnref{rpath1b}
\end{split}
\end{align*}
\begin{align}\label{rray2h}
\begin{split}
&=(2\epsvh\Omerep^2-3\epsvg\epsvb)(\cprod{\vectOme}{\vectr})
  +\epsvb^2(\cprod{\vectOme}{\vectLam})
  -\epsvb\Omerep^2(\cprod{\vectr}{\vectLam})
  +\epsvb^2\fdot{\vectLam}-\epsvb\vsigc\vectr\\
  &\quad-\Omerep^2\epsvb^2\vectOme+\epsvb\Omerep^4\vectr
  -\Omerep^2\scalr^2\fdot{\vectLam}+\Omerep^2\vsign\vectr
  +\Omerep^4\scalr^2\vectOme-\Omerep^4\epsvb\vectr
  -2\epsvh\epsvb\vectLam+2\epsvh\epsvg\vectr\\
  &\quad+3\epsvg\scalr^2\vectLam-3\epsvg\epsvh\vectr
  -\epsvb\epsvh\vectLam+\epsvb\Lamrep^2\vectr
  +(\Omerep^2\epsvm-\frkth)\vectr
\end{split}
\nonumber\\
\begin{split}
&=(2\epsvh\Omerep^2-3\epsvg\epsvb)(\cprod{\vectOme}{\vectr})
  +\epsvb^2(\cprod{\vectOme}{\vectLam})
  -\epsvb\Omerep^2(\cprod{\vectr}{\vectLam})
  +\epsvb^2\fdot{\vectLam}-\Omerep^2\scalr^2\fdot{\vectLam}\\
  &\quad-\epsvb\vsigc\vectr+\epsvb\Omerep^4\vectr+\Omerep^2\vsign\vectr
  -\Omerep^4\epsvb\vectr+2\epsvh\epsvg\vectr-3\epsvg\epsvh\vectr+\epsvb\Lamrep^2\vectr
  +(\Omerep^2\epsvm-\frkth)\vectr\\
  &\quad-\Omerep^2\epsvb^2\vectOme+\Omerep^4\scalr^2\vectOme
  -2\epsvh\epsvb\vectLam+3\epsvg\scalr^2\vectLam-\epsvb\epsvh\vectLam
\end{split}
\nonumber\\
\begin{split}
&=(2\epsvh\Omerep^2-3\epsvg\epsvb)(\cprod{\vectOme}{\vectr})
  +\epsvb^2(\cprod{\vectOme}{\vectLam})
  -\epsvb\Omerep^2(\cprod{\vectr}{\vectLam})
  +(\epsvb^2-\Omerep^2\scalr^2)\fdot{\vectLam}\\
  &\quad+(-\epsvb\vsigc+\Omerep^2\vsign-\epsvg\epsvh+\epsvb\Lamrep^2+\Omerep^2\epsvm-\frkth)\vectr
  -\Omerep^2(\epsvb^2-\Omerep^2\scalr^2)\vectOme
  +3(\epsvg\scalr^2-\epsvh\epsvb)\vectLam
\end{split}
\nonumber\\
\begin{split}
&=(2\epsvh\Omerep^2-3\epsvg\epsvb)(\cprod{\vectOme}{\vectr})
  +\epsvb^2(\cprod{\vectOme}{\vectLam})
  -\epsvb\Omerep^2(\cprod{\vectr}{\vectLam})
  -\vphia^2\fdot{\vectLam}+\Omerep^2\vphia^2\vectOme+3\vphic\vectLam\\
  &\quad+[\epsvb(\Lamrep^2-\vsigc)+\Omerep^2(\epsvm+\vsign)-\epsvg\epsvh-\frkth]\vectr
  \beqref{rot1b}
\end{split}
\end{align}
\begin{align*}
\veusi
&=\cprod{\vecta}{\fdote}\beqref{kray2a}\nonumber\\
&=\cprod{(\epsvb\vectOme-\Omerep^2\vectr+\cprod{\vectLam}{\vectr})}
  {[\ethvo(\cprod{\unitplz}{\vectOme})+\vphig\vphih(\cprod{\unitplz}{\vectLam})+\ethvr\vectOme+\vphii\vectLam-\ethvu\unitplz]}\\
   &\quad\beqref{main4a}, \eqnref{rot1a}\text{ \& }\eqnref{rpath17a}\nonumber\\
\begin{split}
&=\ethvo\epsvb[\cprod{\vectOme}{(\cprod{\unitplz}{\vectOme})}]
  +\vphig\vphih\epsvb[\cprod{\vectOme}{(\cprod{\unitplz}{\vectLam})}]
  +\ethvr\epsvb(\cprod{\vectOme}{\vectOme})
  +\vphii\epsvb(\cprod{\vectOme}{\vectLam})\\
  &\quad-\ethvu\epsvb(\cprod{\vectOme}{\unitplz})
  -\ethvo\Omerep^2[\cprod{\vectr}{(\cprod{\unitplz}{\vectOme})}]
  -\vphig\vphih\Omerep^2[\cprod{\vectr}{(\cprod{\unitplz}{\vectLam})}]
  -\ethvr\Omerep^2(\cprod{\vectr}{\vectOme})\\
  &\quad-\vphii\Omerep^2(\cprod{\vectr}{\vectLam})
  +\ethvu\Omerep^2(\cprod{\vectr}{\unitplz})
  +\ethvo[\cprod{(\cprod{\vectLam}{\vectr})}{(\cprod{\unitplz}{\vectOme})}]
  +\vphig\vphih[\cprod{(\cprod{\vectLam}{\vectr})}{(\cprod{\unitplz}{\vectLam})}]\\
  &\quad+\ethvr[\cprod{(\cprod{\vectLam}{\vectr})}{\vectOme}]
  +\vphii[\cprod{(\cprod{\vectLam}{\vectr})}{\vectLam}]
  -\ethvu[\cprod{(\cprod{\vectLam}{\vectr})}{\unitplz}]
\end{split}
\nonumber\\
\begin{split}
&=\vphii\epsvb(\cprod{\vectOme}{\vectLam})
  -\ethvu\epsvb(\cprod{\vectOme}{\unitplz})
  -\ethvr\Omerep^2(\cprod{\vectr}{\vectOme})
  -\vphii\Omerep^2(\cprod{\vectr}{\vectLam})
  +\ethvu\Omerep^2(\cprod{\vectr}{\unitplz})\\
  &\quad+\ethvo\epsvb[\cprod{\vectOme}{(\cprod{\unitplz}{\vectOme})}]
  +\vphig\vphih\epsvb[\cprod{\vectOme}{(\cprod{\unitplz}{\vectLam})}]
  -\ethvo\Omerep^2[\cprod{\vectr}{(\cprod{\unitplz}{\vectOme})}]\\
  &\quad-\vphig\vphih\Omerep^2[\cprod{\vectr}{(\cprod{\unitplz}{\vectLam})}]
  -\ethvr[\cprod{\vectOme}{(\cprod{\vectLam}{\vectr})}]
  -\vphii[\cprod{\vectLam}{(\cprod{\vectLam}{\vectr})}]
  +\ethvu[\cprod{\unitplz}{(\cprod{\vectLam}{\vectr})}]\\
  &\quad+\ethvo[\cprod{(\cprod{\vectLam}{\vectr})}{(\cprod{\unitplz}{\vectOme})}]
  +\vphig\vphih[\cprod{(\cprod{\vectLam}{\vectr})}{(\cprod{\unitplz}{\vectLam})}]
\end{split}
\end{align*}
\begin{align*}
\begin{split}
&=\vphii\epsvb(\cprod{\vectOme}{\vectLam})
  -\ethvu\epsvb(\cprod{\vectOme}{\unitplz})
  -\ethvr\Omerep^2(\cprod{\vectr}{\vectOme})
  -\vphii\Omerep^2(\cprod{\vectr}{\vectLam})
  +\ethvu\Omerep^2(\cprod{\vectr}{\unitplz})\\
  &\quad+\ethvo\epsvb[\Omerep^2\unitplz-\vectOme(\dprod{\vectOme}{\unitplz})]
  +\vphig\vphih\epsvb[\unitplz(\dprod{\vectOme}{\vectLam})-\vectLam(\dprod{\vectOme}{\unitplz})]
  -\ethvo\Omerep^2[\unitplz(\dprod{\vectr}{\vectOme})-\vectOme(\dprod{\vectr}{\unitplz})]\\
  &\quad-\vphig\vphih\Omerep^2[\unitplz(\dprod{\vectr}{\vectLam})-\vectLam(\dprod{\vectr}{\unitplz})]
  -\ethvr[\vectLam(\dprod{\vectOme}{\vectr})-\vectr(\dprod{\vectOme}{\vectLam})]
  -\vphii[\vectLam(\dprod{\vectLam}{\vectr})-\Lamrep^2\vectr]\\
  &\quad+\ethvu[\vectLam(\dprod{\unitplz}{\vectr})-\vectr(\dprod{\unitplz}{\vectLam})]
  +\ethvo[\unitplz(\dprod{\vectOme}{(\cprod{\vectLam}{\vectr})})-\vectOme(\dprod{\unitplz}{(\cprod{\vectLam}{\vectr})})]\\
  &\quad+\vphig\vphih[\unitplz(\dprod{\vectLam}{(\cprod{\vectLam}{\vectr})})-\vectLam(\dprod{\unitplz}{(\cprod{\vectLam}{\vectr})})]
  \beqref{alg1}\text{ \& }\eqnref{alg5}
\end{split}
\nonumber\\
\begin{split}
&=\vphii\epsvb(\cprod{\vectOme}{\vectLam})
  -\ethvu\epsvb(\cprod{\vectOme}{\unitplz})
  -\ethvr\Omerep^2(\cprod{\vectr}{\vectOme})
  -\vphii\Omerep^2(\cprod{\vectr}{\vectLam})
  +\ethvu\Omerep^2(\cprod{\vectr}{\unitplz})\\
  &\quad+\ethvo\epsvb(\Omerep^2\unitplz-\epsvc\vectOme)
  +\vphig\vphih\epsvb(\epsvg\unitplz-\epsvc\vectLam)
  -\ethvo\Omerep^2(\epsvb\unitplz-\epsvf\vectOme)\\
  &\quad-\vphig\vphih\Omerep^2(\epsvh\unitplz-\epsvf\vectLam)
  -\ethvr(\epsvb\vectLam-\epsvg\vectr)
  -\vphii(\epsvh\vectLam-\Lamrep^2\vectr)\\
  &\quad+\ethvu(\epsvf\vectLam-\epsvi\vectr)
  +\ethvo(\epsvm\unitplz-\epsvo\vectOme)
  -\vphig\vphih\epsvo\vectLam
  \beqref{rot1a}
\end{split}
\end{align*}
\begin{align}\label{rray2i}
\begin{split}
&=\vphii\epsvb(\cprod{\vectOme}{\vectLam})
  -\ethvu\epsvb(\cprod{\vectOme}{\unitplz})
  -\ethvr\Omerep^2(\cprod{\vectr}{\vectOme})
  -\vphii\Omerep^2(\cprod{\vectr}{\vectLam})
  +\ethvu\Omerep^2(\cprod{\vectr}{\unitplz})\\
  &\quad+\ethvo\epsvb\Omerep^2\unitplz-\ethvo\epsvb\epsvc\vectOme
  +\vphig\vphih\epsvb\epsvg\unitplz-\vphig\vphih\epsvb\epsvc\vectLam
  -\ethvo\Omerep^2\epsvb\unitplz+\ethvo\Omerep^2\epsvf\vectOme\\
  &\quad-\vphig\vphih\Omerep^2\epsvh\unitplz+\vphig\vphih\Omerep^2\epsvf\vectLam
  -\ethvr\epsvb\vectLam+\ethvr\epsvg\vectr
  -\vphii\epsvh\vectLam+\vphii\Lamrep^2\vectr\\
  &\quad+\ethvu\epsvf\vectLam-\ethvu\epsvi\vectr
  +\ethvo\epsvm\unitplz-\ethvo\epsvo\vectOme
  -\vphig\vphih\epsvo\vectLam
\end{split}
\nonumber\\
\begin{split}
&=\vphii\epsvb(\cprod{\vectOme}{\vectLam})
  -\ethvu\epsvb(\cprod{\vectOme}{\unitplz})
  -\ethvr\Omerep^2(\cprod{\vectr}{\vectOme})
  -\vphii\Omerep^2(\cprod{\vectr}{\vectLam})
  +\ethvu\Omerep^2(\cprod{\vectr}{\unitplz})\\
  &\quad+\ethvo\epsvb\Omerep^2\unitplz+\vphig\vphih\epsvb\epsvg\unitplz-\ethvo\Omerep^2\epsvb\unitplz
  -\vphig\vphih\Omerep^2\epsvh\unitplz+\ethvo\epsvm\unitplz
  -\ethvo\epsvb\epsvc\vectOme\\
    &\quad+\ethvo\Omerep^2\epsvf\vectOme-\ethvo\epsvo\vectOme
  -\vphig\vphih\epsvb\epsvc\vectLam+\vphig\vphih\Omerep^2\epsvf\vectLam
  -\ethvr\epsvb\vectLam-\vphii\epsvh\vectLam+\ethvu\epsvf\vectLam\\
    &\quad-\vphig\vphih\epsvo\vectLam
  +\ethvr\epsvg\vectr+\vphii\Lamrep^2\vectr-\ethvu\epsvi\vectr
\end{split}
\nonumber\\
\begin{split}
&=\vphii\epsvb(\cprod{\vectOme}{\vectLam})
  -\ethvu\epsvb(\cprod{\vectOme}{\unitplz})
  -\ethvr\Omerep^2(\cprod{\vectr}{\vectOme})
  -\vphii\Omerep^2(\cprod{\vectr}{\vectLam})
  +\ethvu\Omerep^2(\cprod{\vectr}{\unitplz})\\
  &\quad+[\ethvo\epsvm+\vphig\vphih(\epsvb\epsvg-\Omerep^2\epsvh)]\unitplz
  +\ethvo(\Omerep^2\epsvf-\epsvb\epsvc-\epsvo)\vectOme
  +(\ethvr\epsvg+\vphii\Lamrep^2-\ethvu\epsvi)\vectr\\
  &\quad+[\ethvu\epsvf-\ethvr\epsvb-\vphii\epsvh+\vphig\vphih(\Omerep^2\epsvf-\epsvb\epsvc-\epsvo)]\vectLam
\end{split}
\end{align}
\begin{align}\label{rray2j}
\veusj
&=\cprod{\vectu}{\fdota}\beqref{kray2a}\nonumber\\
&=\cprod{(\cprod{\vectOme}{\vectr})}{[2\epsvh\vectOme-3\epsvg\vectr+\cprod{\fdot{\vectLam}}{\vectr}
   -\Omerep^2(\cprod{\vectOme}{\vectr})+\epsvb\vectLam]}
   \beqref{main4c}\text{ \& }\eqnref{rpath7a}\nonumber\\
\begin{split}
&=-2\epsvh[\cprod{\vectOme}{(\cprod{\vectOme}{\vectr})}]
  +3\epsvg[\cprod{\vectr}{(\cprod{\vectOme}{\vectr})}]
  -\epsvb[\cprod{\vectLam}{(\cprod{\vectOme}{\vectr})}]\\
  &\quad+[\cprod{(\cprod{\vectOme}{\vectr})}{(\cprod{\fdot{\vectLam}}{\vectr})}]
  -\Omerep^2[\cprod{(\cprod{\vectOme}{\vectr})}{(\cprod{\vectOme}{\vectr})}]
\end{split}
\nonumber\\
\begin{split}
&=-2\epsvh[\vectOme(\dprod{\vectOme}{\vectr})-\Omerep^2\vectr]
  +3\epsvg[\scalr^2\vectOme-\vectr(\dprod{\vectr}{\vectOme})]
  -\epsvb[\vectOme(\dprod{\vectLam}{\vectr})-\vectr(\dprod{\vectLam}{\vectOme})]\\
  &\quad+[\fdot{\vectLam}(\dprod{\vectr}{(\cprod{\vectOme}{\vectr})})-\vectr(\dprod{\fdot{\vectLam}}{(\cprod{\vectOme}{\vectr})})]
  \beqref{alg1}\text{ \& }\eqnref{alg5}
\end{split}
\nonumber\\
\begin{split}
&=-2\epsvh(\epsvb\vectOme-\Omerep^2\vectr)
  +3\epsvg(\scalr^2\vectOme-\epsvb\vectr)
  -\epsvb(\epsvh\vectOme-\epsvg\vectr)
  -\frktl\vectr\beqref{rot1a}\text{ \& }\eqnref{rpath1b}
\end{split}
\nonumber\\
\begin{split}
&=-2\epsvh\epsvb\vectOme+2\epsvh\Omerep^2\vectr
  +3\epsvg\scalr^2\vectOme-3\epsvg\epsvb\vectr
  -\epsvb\epsvh\vectOme+\epsvb\epsvg\vectr-\frktl\vectr\\
&=-2\epsvh\epsvb\vectOme+3\epsvg\scalr^2\vectOme-\epsvb\epsvh\vectOme
  +2\epsvh\Omerep^2\vectr-3\epsvg\epsvb\vectr+\epsvb\epsvg\vectr-\frktl\vectr
\end{split}
\nonumber\\
&=3(\epsvg\scalr^2-\epsvh\epsvb)\vectOme+(2\epsvh\Omerep^2-2\epsvg\epsvb-\frktl)\vectr\nonumber\\
&=3\vphic\vectOme+(2\epsvh\Omerep^2-2\epsvg\epsvb-\frktl)\vectr\beqref{rot1b}
\end{align}
\begin{align}\label{rray2k}
\veusk
&=\cprod{\vectu}{\fdote}\beqref{kray2a}\nonumber\\
&=\cprod{(\cprod{\vectOme}{\vectr})}{[\ethvo(\cprod{\unitplz}{\vectOme})+\vphig\vphih(\cprod{\unitplz}{\vectLam})
  +\ethvr\vectOme+\vphii\vectLam-\ethvu\unitplz]}\beqref{rpath17a}\nonumber\\
\begin{split}
&=-\ethvr[\cprod{\vectOme}{(\cprod{\vectOme}{\vectr})}]
  -\vphii[\cprod{\vectLam}{(\cprod{\vectOme}{\vectr})}]
  +\ethvu[\cprod{\unitplz}{(\cprod{\vectOme}{\vectr})}]\\
  &\quad+\ethvo[\cprod{(\cprod{\vectOme}{\vectr})}{(\cprod{\unitplz}{\vectOme})}]
  +\vphig\vphih[\cprod{(\cprod{\vectOme}{\vectr})}{(\cprod{\unitplz}{\vectLam})}]
\end{split}
\nonumber\\
\begin{split}
&=-\ethvr[\vectOme(\dprod{\vectOme}{\vectr})-\Omerep^2\vectr]
  -\vphii[\vectOme(\dprod{\vectLam}{\vectr})-\vectr(\dprod{\vectLam}{\vectOme})]
  +\ethvu[\vectOme(\dprod{\unitplz}{\vectr})-\vectr(\dprod{\unitplz}{\vectOme})]\\
  &\quad+\ethvo[\unitplz(\dprod{\vectOme}{(\cprod{\vectOme}{\vectr})})-\vectOme(\dprod{\unitplz}{(\cprod{\vectOme}{\vectr})})]
  +\vphig\vphih[\unitplz(\dprod{\vectLam}{(\cprod{\vectOme}{\vectr})})-\vectLam(\dprod{\unitplz}{(\cprod{\vectOme}{\vectr})})]
\end{split}
\nonumber\\
\begin{split}
&=-\ethvr(\epsvb\vectOme-\Omerep^2\vectr)
  -\vphii(\epsvh\vectOme-\epsvg\vectr)
  +\ethvu(\epsvf\vectOme-\epsvc\vectr)\\
  &\quad-\ethvo\epsvl\vectOme+\vphig\vphih(-\epsvm\unitplz-\epsvl\vectLam)
  \beqref{rot1a}
\end{split}
\nonumber\\
\begin{split}
&=-\ethvr\epsvb\vectOme+\ethvr\Omerep^2\vectr
  -\vphii\epsvh\vectOme+\vphii\epsvg\vectr
  +\ethvu\epsvf\vectOme-\ethvu\epsvc\vectr
  -\ethvo\epsvl\vectOme-\vphig\vphih\epsvm\unitplz+\vphig\vphih\epsvl\vectLam
\end{split}
\nonumber\\
\begin{split}
&=\vphig\vphih\epsvl\vectLam-\vphig\vphih\epsvm\unitplz
  +\ethvr\Omerep^2\vectr+\vphii\epsvg\vectr-\ethvu\epsvc\vectr
  -\ethvr\epsvb\vectOme-\vphii\epsvh\vectOme+\ethvu\epsvf\vectOme-\ethvo\epsvl\vectOme
\end{split}
\nonumber\\
\begin{split}
&=\vphig\vphih(\epsvl\vectLam-\epsvm\unitplz)
  +(\ethvr\Omerep^2+\vphii\epsvg-\ethvu\epsvc)\vectr
  +(\ethvu\epsvf-\ethvr\epsvb-\vphii\epsvh-\ethvo\epsvl)\vectOme
\end{split}
\end{align}
\begin{align*}
\veusl
&=\cprod{\vecte}{\fdota}\beqref{kray2a}\nonumber\\
&=\cprod{[\stwo(\cprod{\unitplz}{\vectOme})+\sthr\vectOme-\sfou\unitplz]}
   {[2\epsvh\vectOme-3\epsvg\vectr+\cprod{\fdot{\vectLam}}{\vectr}-\Omerep^2(\cprod{\vectOme}{\vectr})+\epsvb\vectLam]}\\
  &\quad\beqref{main4b}\text{ \& }\eqnref{rpath7a}\nonumber\\
\begin{split}
&=-2\epsvh\stwo[\cprod{\vectOme}{(\cprod{\unitplz}{\vectOme})}]
   +3\epsvg\stwo[\cprod{\vectr}{(\cprod{\unitplz}{\vectOme})}]
   +\stwo[\cprod{(\cprod{\unitplz}{\vectOme})}{(\cprod{\fdot{\vectLam}}{\vectr})}]\\
   &\quad-\Omerep^2\stwo[\cprod{(\cprod{\unitplz}{\vectOme})}{(\cprod{\vectOme}{\vectr})}]
   -\epsvb\stwo[\cprod{\vectLam}{(\cprod{\unitplz}{\vectOme})}]
   +2\epsvh\sthr(\cprod{\vectOme}{\vectOme})
   -3\epsvg\sthr(\cprod{\vectOme}{\vectr})\\
   &\quad+\sthr[\cprod{\vectOme}{(\cprod{\fdot{\vectLam}}{\vectr})}]
   -\Omerep^2\sthr[\cprod{\vectOme}{(\cprod{\vectOme}{\vectr})}]
   +\epsvb\sthr(\cprod{\vectOme}{\vectLam})
   -2\epsvh\sfou(\cprod{\unitplz}{\vectOme})\\
   &\quad+3\epsvg\sfou(\cprod{\unitplz}{\vectr})
   -\sfou[\cprod{\unitplz}{(\cprod{\fdot{\vectLam}}{\vectr})}]
   +\Omerep^2\sfou[\cprod{\unitplz}{(\cprod{\vectOme}{\vectr})}]
   -\epsvb\sfou(\cprod{\unitplz}{\vectLam})
\end{split}
\nonumber\\
\begin{split}
&=-3\epsvg\sthr(\cprod{\vectOme}{\vectr})
   +\epsvb\sthr(\cprod{\vectOme}{\vectLam})
   -2\epsvh\sfou(\cprod{\unitplz}{\vectOme})
   +3\epsvg\sfou(\cprod{\unitplz}{\vectr})
   -\epsvb\sfou(\cprod{\unitplz}{\vectLam})\\
   &\quad-2\epsvh\stwo[\cprod{\vectOme}{(\cprod{\unitplz}{\vectOme})}]
   +3\epsvg\stwo[\cprod{\vectr}{(\cprod{\unitplz}{\vectOme})}]
   -\epsvb\stwo[\cprod{\vectLam}{(\cprod{\unitplz}{\vectOme})}]
   +\sthr[\cprod{\vectOme}{(\cprod{\fdot{\vectLam}}{\vectr})}]\\
   &\quad-\Omerep^2\sthr[\cprod{\vectOme}{(\cprod{\vectOme}{\vectr})}]
   -\sfou[\cprod{\unitplz}{(\cprod{\fdot{\vectLam}}{\vectr})}]
   +\Omerep^2\sfou[\cprod{\unitplz}{(\cprod{\vectOme}{\vectr})}]
   +\stwo[\cprod{(\cprod{\unitplz}{\vectOme})}{(\cprod{\fdot{\vectLam}}{\vectr})}]\\
   &\quad-\Omerep^2\stwo[\cprod{(\cprod{\unitplz}{\vectOme})}{(\cprod{\vectOme}{\vectr})}]
\end{split}
\end{align*}
\begin{align*}
\begin{split}
&=-3\epsvg\sthr(\cprod{\vectOme}{\vectr})
   +\epsvb\sthr(\cprod{\vectOme}{\vectLam})
   -2\epsvh\sfou(\cprod{\unitplz}{\vectOme})
   +3\epsvg\sfou(\cprod{\unitplz}{\vectr})
   -\epsvb\sfou(\cprod{\unitplz}{\vectLam})\\
   &\quad-2\epsvh\stwo[\Omerep^2\unitplz-\vectOme(\dprod{\vectOme}{\unitplz})]
   +3\epsvg\stwo[\unitplz(\dprod{\vectr}{\vectOme})-\vectOme(\dprod{\vectr}{\unitplz})]
   -\epsvb\stwo[\unitplz(\dprod{\vectLam}{\vectOme})-\vectOme(\dprod{\vectLam}{\unitplz})]\\
   &\quad+\sthr[\fdot{\vectLam}(\dprod{\vectOme}{\vectr})-\vectr(\dprod{\vectOme}{\fdot{\vectLam}})]
   -\Omerep^2\sthr[\vectOme(\dprod{\vectOme}{\vectr})-\Omerep^2\vectr]
   -\sfou[\fdot{\vectLam}(\dprod{\unitplz}{\vectr})-\vectr(\dprod{\unitplz}{\fdot{\vectLam}})]\\
   &\quad+\Omerep^2\sfou[\vectOme(\dprod{\unitplz}{\vectr})-\vectr(\dprod{\unitplz}{\vectOme})]
   +\stwo[\fdot{\vectLam}(\dprod{\vectr}{(\cprod{\unitplz}{\vectOme})})-\vectr(\dprod{\fdot{\vectLam}}{(\cprod{\unitplz}{\vectOme})})]\\
   &\quad-\Omerep^2\stwo[\vectOme(\dprod{\vectr}{(\cprod{\unitplz}{\vectOme})})-\vectr(\dprod{\vectOme}{(\cprod{\unitplz}{\vectOme})})]
   \beqref{alg1}\text{ \& }\eqnref{alg5}
\end{split}
\nonumber\\
\begin{split}
&=-3\epsvg\sthr(\cprod{\vectOme}{\vectr})
   +\epsvb\sthr(\cprod{\vectOme}{\vectLam})
   -2\epsvh\sfou(\cprod{\unitplz}{\vectOme})
   +3\epsvg\sfou(\cprod{\unitplz}{\vectr})
   -\epsvb\sfou(\cprod{\unitplz}{\vectLam})\\
   &\quad-2\epsvh\stwo(\Omerep^2\unitplz-\epsvc\vectOme)
   +3\epsvg\stwo(\epsvb\unitplz-\epsvf\vectOme)
   -\epsvb\stwo(\epsvg\unitplz-\epsvi\vectOme)
   +\sthr(\epsvb\fdot{\vectLam}-\vsigc\vectr)\\
   &\quad-\Omerep^2\sthr(\epsvb\vectOme-\Omerep^2\vectr)
   -\sfou(\epsvf\fdot{\vectLam}-\vsiga\vectr)
   +\Omerep^2\sfou(\epsvf\vectOme-\epsvc\vectr)
   +\stwo(\epsvl\fdot{\vectLam}-\frktp\vectr)\\
   &\quad-\Omerep^2\stwo\epsvl\vectOme
   \beqref{rot1a}, \eqnref{rpath1a}\text{ \& }\eqnref{rpath1b}
\end{split}
\end{align*}
\begin{align*}
\begin{split}
&=-3\epsvg\sthr(\cprod{\vectOme}{\vectr})
   +\epsvb\sthr(\cprod{\vectOme}{\vectLam})
   -2\epsvh\sfou(\cprod{\unitplz}{\vectOme})
   +3\epsvg\sfou(\cprod{\unitplz}{\vectr})
   -\epsvb\sfou(\cprod{\unitplz}{\vectLam})\\
   &\quad-2\epsvh\stwo\Omerep^2\unitplz+2\epsvh\stwo\epsvc\vectOme
   +3\epsvg\stwo\epsvb\unitplz-3\epsvg\stwo\epsvf\vectOme
   -\epsvb\stwo\epsvg\unitplz+\epsvb\stwo\epsvi\vectOme\\
   &\quad+\sthr\epsvb\fdot{\vectLam}-\sthr\vsigc\vectr
   -\Omerep^2\sthr\epsvb\vectOme+\Omerep^4\sthr\vectr
   -\sfou\epsvf\fdot{\vectLam}+\sfou\vsiga\vectr
   +\Omerep^2\sfou\epsvf\vectOme\\
   &\quad-\Omerep^2\sfou\epsvc\vectr+\stwo\epsvl\fdot{\vectLam}-\stwo\frktp\vectr
   -\Omerep^2\stwo\epsvl\vectOme
\end{split}
\nonumber\\
\begin{split}
&=-3\epsvg\sthr(\cprod{\vectOme}{\vectr})
   +\epsvb\sthr(\cprod{\vectOme}{\vectLam})
   -2\epsvh\sfou(\cprod{\unitplz}{\vectOme})
   +3\epsvg\sfou(\cprod{\unitplz}{\vectr})
   -\epsvb\sfou(\cprod{\unitplz}{\vectLam})\\
   &\quad-2\epsvh\stwo\Omerep^2\unitplz+3\epsvg\stwo\epsvb\unitplz-\epsvb\stwo\epsvg\unitplz
   +2\epsvh\stwo\epsvc\vectOme-3\epsvg\stwo\epsvf\vectOme+\epsvb\stwo\epsvi\vectOme\\
     &\quad-\Omerep^2\sthr\epsvb\vectOme+\Omerep^2\sfou\epsvf\vectOme-\Omerep^2\stwo\epsvl\vectOme
   -\sthr\vsigc\vectr+\Omerep^4\sthr\vectr+\sfou\vsiga\vectr
   -\stwo\frktp\vectr-\Omerep^2\sfou\epsvc\vectr\\
   &\quad+\sthr\epsvb\fdot{\vectLam}-\sfou\epsvf\fdot{\vectLam}
   +\stwo\epsvl\fdot{\vectLam}
\end{split}
\end{align*}
\begin{align}\label{rray2l}
\begin{split}
&=-3\epsvg\sthr(\cprod{\vectOme}{\vectr})
   +\epsvb\sthr(\cprod{\vectOme}{\vectLam})
   -2\epsvh\sfou(\cprod{\unitplz}{\vectOme})
   +3\epsvg\sfou(\cprod{\unitplz}{\vectr})
   -\epsvb\sfou(\cprod{\unitplz}{\vectLam})\\
   &\quad+2\stwo(\epsvg\epsvb-\epsvh\Omerep^2)\unitplz
   +[\stwo(2\epsvh\epsvc-3\epsvg\epsvf+\epsvb\epsvi-\Omerep^2\epsvl)
     +\Omerep^2(\sfou\epsvf-\sthr\epsvb)]\vectOme\\
   &\quad+[\sthr(\Omerep^4-\vsigc)+\sfou(\vsiga-\Omerep^2\epsvc)-\stwo\frktp]\vectr
   +(\sthr\epsvb-\sfou\epsvf+\stwo\epsvl)\fdot{\vectLam}
\end{split}
\nonumber\\
\begin{split}
&=-3\epsvg\vphii(\cprod{\vectOme}{\vectr})
   +\epsvb\vphii(\cprod{\vectOme}{\vectLam})
   -2\epsvh\vphij(\cprod{\unitplz}{\vectOme})
   +3\epsvg\vphij(\cprod{\unitplz}{\vectr})
   -\epsvb\vphij(\cprod{\unitplz}{\vectLam})\\
   &\quad+2\vphig\vphih(\epsvg\epsvb-\epsvh\Omerep^2)\unitplz
   +[\vphig\vphih(2\epsvh\epsvc-3\epsvg\epsvf+\epsvb\epsvi-\Omerep^2\epsvl)
     +\Omerep^2(\vphij\epsvf-\vphii\epsvb)]\vectOme\\
   &\quad+[\vphii(\Omerep^4-\vsigc)+\vphij(\vsiga-\Omerep^2\epsvc)-\vphig\vphih\frktp]\vectr
   +(\vphii\epsvb-\vphij\epsvf+\vphig\vphih\epsvl)\fdot{\vectLam}
   \beqref{rot5}
\end{split}
\end{align}
\begin{align*}
\veusm
&=\cprod{\vecte}{\fdote}\beqref{kray2a}\nonumber\\
&=\cprod{[\stwo(\cprod{\unitplz}{\vectOme})+\sthr\vectOme-\sfou\unitplz]}
  {[\ethvo(\cprod{\unitplz}{\vectOme})+\vphig\vphih(\cprod{\unitplz}{\vectLam})
     +\ethvr\vectOme+\vphii\vectLam-\ethvu\unitplz]}\\
  &\quad\beqref{main4b}\text{ \& }\eqnref{rpath17a}\nonumber\\
\begin{split}
&=\vphig\vphih\stwo[\cprod{(\cprod{\unitplz}{\vectOme})}{(\cprod{\unitplz}{\vectLam})}]
  -\ethvr\stwo[\cprod{\vectOme}{(\cprod{\unitplz}{\vectOme})}]
  -\vphii\stwo[\cprod{\vectLam}{(\cprod{\unitplz}{\vectOme})}]
  +\ethvu\stwo[\cprod{\unitplz}{(\cprod{\unitplz}{\vectOme})}]\\
  &\quad+\ethvo\sthr[\cprod{\vectOme}{(\cprod{\unitplz}{\vectOme})}]
  +\vphig\vphih\sthr[\cprod{\vectOme}{(\cprod{\unitplz}{\vectLam})}]
  +\vphii\sthr(\cprod{\vectOme}{\vectLam})
  -\ethvu\sthr(\cprod{\vectOme}{\unitplz})\\
  &\quad-\ethvo\sfou[\cprod{\unitplz}{(\cprod{\unitplz}{\vectOme})}]
  -\vphig\vphih\sfou[\cprod{\unitplz}{(\cprod{\unitplz}{\vectLam})}]
  -\ethvr\sfou(\cprod{\unitplz}{\vectOme})
  -\vphii\sfou(\cprod{\unitplz}{\vectLam})
\end{split}
\nonumber\\
\begin{split}
&=\vphii\sthr(\cprod{\vectOme}{\vectLam})
  -\ethvu\sthr(\cprod{\vectOme}{\unitplz})
  -\ethvr\sfou(\cprod{\unitplz}{\vectOme})
  -\vphii\sfou(\cprod{\unitplz}{\vectLam})\\
  &\quad-\ethvr\stwo[\cprod{\vectOme}{(\cprod{\unitplz}{\vectOme})}]
  -\vphii\stwo[\cprod{\vectLam}{(\cprod{\unitplz}{\vectOme})}]
  +\ethvu\stwo[\cprod{\unitplz}{(\cprod{\unitplz}{\vectOme})}]\\
  &\quad+\ethvo\sthr[\cprod{\vectOme}{(\cprod{\unitplz}{\vectOme})}]
  +\vphig\vphih\sthr[\cprod{\vectOme}{(\cprod{\unitplz}{\vectLam})}]
  -\ethvo\sfou[\cprod{\unitplz}{(\cprod{\unitplz}{\vectOme})}]\\
  &\quad-\vphig\vphih\sfou[\cprod{\unitplz}{(\cprod{\unitplz}{\vectLam})}]
  +\vphig\vphih\stwo[\cprod{(\cprod{\unitplz}{\vectOme})}{(\cprod{\unitplz}{\vectLam})}]
\end{split}
\end{align*}
\begin{align*}
\begin{split}
&=\vphii\sthr(\cprod{\vectOme}{\vectLam})
  -\ethvu\sthr(\cprod{\vectOme}{\unitplz})
  -\ethvr\sfou(\cprod{\unitplz}{\vectOme})
  -\vphii\sfou(\cprod{\unitplz}{\vectLam})\\
  &\quad-\ethvr\stwo[\Omerep^2\unitplz-\vectOme(\dprod{\vectOme}{\unitplz})]
  -\vphii\stwo[\unitplz(\dprod{\vectLam}{\vectOme})-\vectOme(\dprod{\vectLam}{\unitplz})]
  +\ethvu\stwo[\unitplz(\dprod{\unitplz}{\vectOme})-\vectOme]\\
  &\quad+\ethvo\sthr[\Omerep^2\unitplz-\vectOme(\dprod{\unitplz}{\vectOme})]
  +\vphig\vphih\sthr[\unitplz(\dprod{\vectLam}{\vectOme})-\vectLam(\dprod{\vectOme}{\unitplz})]
  -\ethvo\sfou[\unitplz(\dprod{\unitplz}{\vectOme})-\vectOme]\\
  &\quad-\vphig\vphih\sfou[\unitplz(\dprod{\unitplz}{\vectLam})-\vectLam]
  +\vphig\vphih\stwo[\unitplz(\dprod{\vectLam}{(\cprod{\unitplz}{\vectOme})})-\vectLam(\dprod{\unitplz}{(\cprod{\unitplz}{\vectOme})})]
  \beqref{alg1}\text{ \& }\eqnref{alg5}
\end{split}
\nonumber\\
\begin{split}
&=\vphii\sthr(\cprod{\vectOme}{\vectLam})
  -\ethvu\sthr(\cprod{\vectOme}{\unitplz})
  -\ethvr\sfou(\cprod{\unitplz}{\vectOme})
  -\vphii\sfou(\cprod{\unitplz}{\vectLam})\\
  &\quad-\ethvr\stwo(\Omerep^2\unitplz-\epsvc\vectOme)
  -\vphii\stwo(\epsvg\unitplz-\epsvi\vectOme)
  +\ethvu\stwo(\epsvc\unitplz-\vectOme)\\
  &\quad+\ethvo\sthr(\Omerep^2\unitplz-\epsvc\vectOme)
  +\vphig\vphih\sthr(\epsvg\unitplz-\epsvc\vectLam)
  -\ethvo\sfou(\epsvc\unitplz-\vectOme)\\
  &\quad-\vphig\vphih\sfou(\epsvi\unitplz-\vectLam)
  +\vphig\vphih\stwo\frkto\unitplz
  \beqref{rot1a}\text{ \& }\eqnref{rpath1b}
\end{split}
\end{align*}
\begin{align*}
\begin{split}
&=\vphii\sthr(\cprod{\vectOme}{\vectLam})
  -\ethvu\sthr(\cprod{\vectOme}{\unitplz})
  -\ethvr\sfou(\cprod{\unitplz}{\vectOme})
  -\vphii\sfou(\cprod{\unitplz}{\vectLam})\\
  &\quad-\ethvr\stwo\Omerep^2\unitplz+\ethvr\stwo\epsvc\vectOme
  -\vphii\stwo\epsvg\unitplz+\vphii\stwo\epsvi\vectOme
  +\ethvu\stwo\epsvc\unitplz-\ethvu\stwo\vectOme\\
  &\quad+\ethvo\sthr\Omerep^2\unitplz-\ethvo\sthr\epsvc\vectOme
  +\vphig\vphih\sthr\epsvg\unitplz-\vphig\vphih\sthr\epsvc\vectLam
  -\ethvo\sfou\epsvc\unitplz+\ethvo\sfou\vectOme\\
  &\quad-\vphig\vphih\sfou\epsvi\unitplz+\vphig\vphih\sfou\vectLam
  +\vphig\vphih\stwo\frkto\unitplz
\end{split}
\nonumber\\
\begin{split}
&=\vphii\sthr(\cprod{\vectOme}{\vectLam})
  -\ethvu\sthr(\cprod{\vectOme}{\unitplz})
  -\ethvr\sfou(\cprod{\unitplz}{\vectOme})
  -\vphii\sfou(\cprod{\unitplz}{\vectLam})
  -\ethvr\stwo\Omerep^2\unitplz-\vphii\stwo\epsvg\unitplz\\
    &\quad+\ethvu\stwo\epsvc\unitplz
  +\ethvo\sthr\Omerep^2\unitplz+\vphig\vphih\sthr\epsvg\unitplz-\ethvo\sfou\epsvc\unitplz
  -\vphig\vphih\sfou\epsvi\unitplz+\vphig\vphih\stwo\frkto\unitplz\\
  &\quad+\ethvr\stwo\epsvc\vectOme+\vphii\stwo\epsvi\vectOme-\ethvu\stwo\vectOme
  -\ethvo\sthr\epsvc\vectOme+\ethvo\sfou\vectOme
  -\vphig\vphih\sthr\epsvc\vectLam
  +\vphig\vphih\sfou\vectLam
\end{split}
\end{align*}
\begin{align}\label{rray2m}
\begin{split}
&=\vphii\sthr(\cprod{\vectOme}{\vectLam})
  -\ethvu\sthr(\cprod{\vectOme}{\unitplz})
  -\ethvr\sfou(\cprod{\unitplz}{\vectOme})
  -\vphii\sfou(\cprod{\unitplz}{\vectLam})\\
  &\quad+[\stwo(\ethvu\epsvc-\ethvr\Omerep^2-\vphii\epsvg+\vphig\vphih\frkto)
  +\sthr(\ethvo\Omerep^2+\vphig\vphih\epsvg)-\sfou(\ethvo\epsvc+\vphig\vphih\epsvi)]\unitplz\\
  &\quad+[\stwo(\ethvr\epsvc+\vphii\epsvi-\ethvu)+\ethvo(\sfou-\sthr\epsvc)]\vectOme
  +\vphig\vphih(\sfou-\sthr\epsvc)\vectLam
\end{split}
\nonumber\\
\begin{split}
&=\vphii^2(\cprod{\vectOme}{\vectLam})
  -\ethvu\vphii(\cprod{\vectOme}{\unitplz})
  -\ethvr\vphij(\cprod{\unitplz}{\vectOme})
  -\vphii\vphij(\cprod{\unitplz}{\vectLam})
  +\vphig\vphih(\vphij-\vphii\epsvc)\vectLam\\
  &\quad+[\vphig\vphih(\ethvu\epsvc-\ethvr\Omerep^2-\vphii\epsvg+\vphig\vphih\frkto)
  +\vphii(\ethvo\Omerep^2+\vphig\vphih\epsvg)-\vphij(\ethvo\epsvc+\vphig\vphih\epsvi)]\unitplz\\
  &\quad+[\vphig\vphih(\ethvr\epsvc+\vphii\epsvi-\ethvu)+\ethvo(\vphij-\vphii\epsvc)]\vectOme
  \beqref{rot5}.
\end{split}
\end{align}
\end{subequations}

\subart{Development of equation \eqnref{kray2b}}
\begin{subequations}\label{rray3}
\begin{align*}
\veust
&=(\cdkt\vbba-\rhorep\fdot{\cdkt})\veusb+\cdkt\veusn+\rhorep\veuso
  \beqref{kray2b}\nonumber\\
&=(\cdkt\parva-\rhorep\parvb)\veusb+\cdkt(\rhorep\veusd+\veuse)+\rhorep(\rhorep\veush+\veusi)
  \beqref{kray2b}\text{ \& }\eqnref{rpath23}\nonumber\\
&=\veka\veusb+\cdkt\rhorep\veusd+\cdkt\veuse+\rhorep^2\veush+\rhorep\veusi
  \beqref{rray1a}\nonumber\\
\end{align*}
\begin{align*}
\begin{split}
&=\veka\{\epsvb(\cprod{\unitkap}{\vectOme})-\Omerep^2(\cprod{\unitkap}{\vectr})
     +\epsvd\vectLam-\dltva\vectr\}\\
  &\quad+\cdkt\rhorep\{2\epsvh(\cprod{\unitkap}{\vectOme})-3\epsvg(\cprod{\unitkap}{\vectr})
     +\epsvb(\cprod{\unitkap}{\vectLam})+\epsvd\fdot{\vectLam}-\vsigb\vectr
     -\Omerep^2(\epsvd\vectOme-\epsva\vectr)\}\\
  &\quad+\cdkt\{\ethvo(\epsva\unitplz-\epsve\vectOme)+\vphig\vphih(\dltva\unitplz-\epsve\vectLam)
     +\ethvr(\cprod{\unitkap}{\vectOme})+\vphii(\cprod{\unitkap}{\vectLam})-\ethvu(\cprod{\unitkap}{\unitplz})\}\\
  &\quad+\rhorep\{\vphii\epsvb(\cprod{\vectOme}{\vectLam})-\ethvu\epsvb(\cprod{\vectOme}{\unitplz})
     -\ethvr\Omerep^2(\cprod{\vectr}{\vectOme})-\vphii\Omerep^2(\cprod{\vectr}{\vectLam})+\ethvu\Omerep^2(\cprod{\vectr}{\unitplz})\\
     &\qquad+[\ethvo\epsvm+\vphig\vphih(\epsvb\epsvg-\Omerep^2\epsvh)]\unitplz+\ethvo(\Omerep^2\epsvf-\epsvb\epsvc-\epsvo)\vectOme
     +(\ethvr\epsvg+\vphii\Lamrep^2-\ethvu\epsvi)\vectr\\
     &\qquad+[\ethvu\epsvf-\ethvr\epsvb-\vphii\epsvh+\vphig\vphih(\Omerep^2\epsvf-\epsvb\epsvc-\epsvo)]\vectLam\}\\
  &\quad+\rhorep^2\{(2\epsvh\Omerep^2-3\epsvg\epsvb)(\cprod{\vectOme}{\vectr})+\epsvb^2(\cprod{\vectOme}{\vectLam})
     -\epsvb\Omerep^2(\cprod{\vectr}{\vectLam})-\vphia^2\fdot{\vectLam}+\Omerep^2\vphia^2\vectOme+3\vphic\vectLam\\
     &\qquad+[\epsvb(\Lamrep^2-\vsigc)+\Omerep^2(\epsvm+\vsign)-\epsvg\epsvh-\frkth]\vectr\}
  \beqref{rray2}
\end{split}
\end{align*}
\begin{align*}
\begin{split}
&=\veka\{\epsvb(\cprod{\unitkap}{\vectOme})-\Omerep^2(\cprod{\unitkap}{\vectr})
     +\epsvd\vectLam-\dltva\vectr\}\\
  &\quad+\cdkt\rhorep\{2\epsvh(\cprod{\unitkap}{\vectOme})-3\epsvg(\cprod{\unitkap}{\vectr})
     +\epsvb(\cprod{\unitkap}{\vectLam})+\epsvd\fdot{\vectLam}-\vsigb\vectr
     -\Omerep^2(\epsvd\vectOme-\epsva\vectr)\}\\
  &\quad+\cdkt\{\ethvo(\epsva\unitplz-\epsve\vectOme)+\vphig\vphih(\dltva\unitplz-\epsve\vectLam)
     +\ethvr(\cprod{\unitkap}{\vectOme})+\vphii(\cprod{\unitkap}{\vectLam})-\ethvu(\cprod{\unitkap}{\unitplz})\}\\
  &\quad+\rhorep\{\vphii\epsvb(\cprod{\vectOme}{\vectLam})-\ethvu\epsvb(\cprod{\vectOme}{\unitplz})
     -\ethvr\Omerep^2(\cprod{\vectr}{\vectOme})-\vphii\Omerep^2(\cprod{\vectr}{\vectLam})+\ethvu\Omerep^2(\cprod{\vectr}{\unitplz})\\
     &\qquad+\frkyo\unitplz+\ethvo\frkym\vectOme+\frkyn\vectr+\frkyp\vectLam\}\\
  &\quad+\rhorep^2\{\frkyg(\cprod{\vectOme}{\vectr})+\epsvb^2(\cprod{\vectOme}{\vectLam})
     -\epsvb\Omerep^2(\cprod{\vectr}{\vectLam})-\vphia^2\fdot{\vectLam}+\Omerep^2\vphia^2\vectOme+3\vphic\vectLam
     +\frkyh\vectr\}\beqref{rpath1l}
\end{split}
\nonumber\\
\begin{split}
&=\veka\epsvb(\cprod{\unitkap}{\vectOme})
  -\veka\Omerep^2(\cprod{\unitkap}{\vectr})
  +\veka\epsvd\vectLam
  -\veka\dltva\vectr
  +2\rhorep\cdkt\epsvh(\cprod{\unitkap}{\vectOme})
  -3\rhorep\cdkt\epsvg(\cprod{\unitkap}{\vectr})\\
  &\quad+\rhorep\cdkt\epsvb(\cprod{\unitkap}{\vectLam})
  +\rhorep\cdkt\epsvd\fdot{\vectLam}
  -\rhorep\cdkt\vsigb\vectr
  -\rhorep\cdkt\Omerep^2\epsvd\vectOme
  +\rhorep\cdkt\Omerep^2\epsva\vectr
  +\cdkt\ethvo\epsva\unitplz
  -\cdkt\ethvo\epsve\vectOme\\
  &\quad+\cdkt\vphig\vphih\dltva\unitplz
  -\cdkt\vphig\vphih\epsve\vectLam
  +\cdkt\ethvr(\cprod{\unitkap}{\vectOme})
  +\cdkt\vphii(\cprod{\unitkap}{\vectLam})
  -\cdkt\ethvu(\cprod{\unitkap}{\unitplz})\\
  &\quad+\rhorep\vphii\epsvb(\cprod{\vectOme}{\vectLam})
  -\rhorep\ethvu\epsvb(\cprod{\vectOme}{\unitplz})
  -\rhorep\ethvr\Omerep^2(\cprod{\vectr}{\vectOme})
  -\rhorep\vphii\Omerep^2(\cprod{\vectr}{\vectLam})
  +\rhorep\ethvu\Omerep^2(\cprod{\vectr}{\unitplz})\\
  &\quad+\rhorep\frkyo\unitplz
  +\rhorep\ethvo\frkym\vectOme
  +\rhorep\frkyn\vectr
  +\rhorep\frkyp\vectLam
  +\rhorep^2\frkyg(\cprod{\vectOme}{\vectr})
  +\rhorep^2\epsvb^2(\cprod{\vectOme}{\vectLam})
  -\rhorep^2\epsvb\Omerep^2(\cprod{\vectr}{\vectLam})\\
  &\quad-\rhorep^2\vphia^2\fdot{\vectLam}
  +\rhorep^2\Omerep^2\vphia^2\vectOme
  +3\rhorep^2\vphic\vectLam
  +\rhorep^2\frkyh\vectr
\end{split}
\end{align*}
\begin{align*}
\begin{split}
&=\rhorep\frkyo\unitplz
  +\cdkt\ethvo\epsva\unitplz
  +\cdkt\vphig\vphih\dltva\unitplz
  +\rhorep\frkyn\vectr
  -\veka\dltva\vectr
  -\rhorep\cdkt\vsigb\vectr
  +\rhorep\cdkt\Omerep^2\epsva\vectr
  +\rhorep^2\frkyh\vectr
  +\rhorep\ethvo\frkym\vectOme\\
  &\quad-\rhorep\cdkt\Omerep^2\epsvd\vectOme
  -\cdkt\ethvo\epsve\vectOme
  +\rhorep^2\Omerep^2\vphia^2\vectOme
  +\rhorep\frkyp\vectLam
  +\veka\epsvd\vectLam
  -\cdkt\vphig\vphih\epsve\vectLam
  +3\rhorep^2\vphic\vectLam\\
  &\quad+\rhorep\cdkt\epsvd\fdot{\vectLam}
  -\rhorep^2\vphia^2\fdot{\vectLam}
  +\veka\epsvb(\cprod{\unitkap}{\vectOme})
  +2\rhorep\cdkt\epsvh(\cprod{\unitkap}{\vectOme})
  +\cdkt\ethvr(\cprod{\unitkap}{\vectOme})\\
  &\quad-\veka\Omerep^2(\cprod{\unitkap}{\vectr})
  -3\rhorep\cdkt\epsvg(\cprod{\unitkap}{\vectr})
  +\rhorep\cdkt\epsvb(\cprod{\unitkap}{\vectLam})
  +\cdkt\vphii(\cprod{\unitkap}{\vectLam})\\
  &\quad-\cdkt\ethvu(\cprod{\unitkap}{\unitplz})
  +\rhorep\vphii\epsvb(\cprod{\vectOme}{\vectLam})
  +\rhorep^2\epsvb^2(\cprod{\vectOme}{\vectLam})
  -\rhorep\ethvu\epsvb(\cprod{\vectOme}{\unitplz})
  -\rhorep\ethvr\Omerep^2(\cprod{\vectr}{\vectOme})\\
  &\quad+\rhorep^2\frkyg(\cprod{\vectOme}{\vectr})
  -\rhorep\vphii\Omerep^2(\cprod{\vectr}{\vectLam})
  -\rhorep^2\epsvb\Omerep^2(\cprod{\vectr}{\vectLam})
  +\rhorep\ethvu\Omerep^2(\cprod{\vectr}{\unitplz})
\end{split}
\end{align*}
\begin{align}\label{rray3a}
\begin{split}
&=[\rhorep\frkyo+\cdkt(\ethvo\epsva+\vphig\vphih\dltva)]\unitplz
  +[\rhorep(\frkyn+\rhorep\frkyh)-\rhorep\cdkt(\vsigb-\Omerep^2\epsva)-\veka\dltva]\vectr\\
  &\quad+[\ethvo(\rhorep\frkym-\cdkt\epsve)+\rhorep\Omerep^2(\rhorep\vphia^2-\cdkt\epsvd)]\vectOme
  +[\veka\epsvd-\cdkt\vphig\vphih\epsve+\rhorep(\frkyp+3\rhorep\vphic)]\vectLam\\
  &\quad+\rhorep(\cdkt\epsvd-\rhorep\vphia^2)\fdot{\vectLam}
  +[\veka\epsvb+\cdkt(\ethvr+2\rhorep\epsvh)](\cprod{\unitkap}{\vectOme})
  -(\veka\Omerep^2+3\rhorep\cdkt\epsvg)(\cprod{\unitkap}{\vectr})\\
  &\quad+\cdkt(\rhorep\epsvb+\vphii)(\cprod{\unitkap}{\vectLam})
  -\cdkt\ethvu(\cprod{\unitkap}{\unitplz})
  +\rhorep\epsvb(\vphii+\rhorep\epsvb)(\cprod{\vectOme}{\vectLam})
  -\rhorep\ethvu\epsvb(\cprod{\vectOme}{\unitplz})\\
  &\quad+\rhorep(\ethvr\Omerep^2+\rhorep\frkyg)(\cprod{\vectOme}{\vectr})
  -\rhorep\Omerep^2(\vphii+\rhorep\epsvb)(\cprod{\vectr}{\vectLam})
  +\rhorep\ethvu\Omerep^2(\cprod{\vectr}{\unitplz})
\end{split}
\end{align}
\begin{align*}
\veusu
&=\fdot{\cdkt}\veusp+\vbba\veusq+\rhorep\veusr+\veuss
  \beqref{kray2b}\nonumber\\
&=\parvb(\veusa-\veusc)+\parva(\veusf-\veusg)+\rhorep(\veusl-\veusj)+(\veusm-\veusk)
  \beqref{kray2b}\text{ \& }\eqnref{rpath23}\nonumber\\
\begin{split}
&=\parvb\{\epsvd\vectOme-\epsva\vectr\}
  -\parvb\{\vphig\vphih(\epsva\unitplz-\epsve\vectOme)+\vphii(\cprod{\unitkap}{\vectOme})
      -\vphij(\cprod{\unitkap}{\unitplz})\}
  +\parva\{-\vphia^2\vectOme-\epsvm\vectr\}\\
  &\quad-\parva\{\vphij\Omerep^2(\cprod{\vectr}{\unitplz})-\vphij\epsvb(\cprod{\vectOme}{\unitplz})
      -\vphii\Omerep^2(\cprod{\vectr}{\vectOme})+\vphig\vphih\epsvm\unitplz\\
      &\qquad+\vphig\vphih(\Omerep^2\epsvf-\epsvb\epsvc-\epsvo)\vectOme+(\vphij\epsvf-\vphii\epsvb)\vectLam
      +(\vphii\epsvg-\vphij\epsvi)\vectr\}\\
  &\quad+\rhorep\{-3\epsvg\vphii(\cprod{\vectOme}{\vectr})+\epsvb\vphii(\cprod{\vectOme}{\vectLam})
      -2\epsvh\vphij(\cprod{\unitplz}{\vectOme})+3\epsvg\vphij(\cprod{\unitplz}{\vectr})-\epsvb\vphij(\cprod{\unitplz}{\vectLam})\\
      &\qquad+2\vphig\vphih(\epsvg\epsvb-\epsvh\Omerep^2)\unitplz+[\vphig\vphih(2\epsvh\epsvc-3\epsvg\epsvf
      +\epsvb\epsvi-\Omerep^2\epsvl)+\Omerep^2(\vphij\epsvf-\vphii\epsvb)]\vectOme\\
      &\qquad+[\vphii(\Omerep^4-\vsigc)+\vphij(\vsiga-\Omerep^2\epsvc)-\vphig\vphih\frktp]\vectr
      +(\vphii\epsvb-\vphij\epsvf+\vphig\vphih\epsvl)\fdot{\vectLam}\}\\
  &\quad-\rhorep\{3\vphic\vectOme+(2\epsvh\Omerep^2-2\epsvg\epsvb-\frktl)\vectr\}\\
  &\quad+\{\vphii^2(\cprod{\vectOme}{\vectLam})-\ethvu\vphii(\cprod{\vectOme}{\unitplz})-\ethvr\vphij(\cprod{\unitplz}{\vectOme})
      -\vphii\vphij(\cprod{\unitplz}{\vectLam})+\vphig\vphih(\vphij-\vphii\epsvc)\vectLam\\
      &\qquad+[\vphig\vphih(\ethvu\epsvc-\ethvr\Omerep^2-\vphii\epsvg+\vphig\vphih\frkto)
      +\vphii(\ethvo\Omerep^2+\vphig\vphih\epsvg)-\vphij(\ethvo\epsvc+\vphig\vphih\epsvi)]\unitplz\\
      &\qquad+[\vphig\vphih(\ethvr\epsvc+\vphii\epsvi-\ethvu)+\ethvo(\vphij-\vphii\epsvc)]\vectOme\}\\
  &\quad-\{\vphig\vphih(\epsvl\vectLam-\epsvm\unitplz)+(\ethvr\Omerep^2+\vphii\epsvg-\ethvu\epsvc)\vectr
      +(\ethvu\epsvf-\ethvr\epsvb-\vphii\epsvh-\ethvo\epsvl)\vectOme\}
\end{split}
\end{align*}
\begin{align*}
\begin{split}
&=\parvb\epsvd\vectOme
  -\parvb\epsva\vectr
  -\parvb\vphig\vphih\epsva\unitplz
  +\parvb\vphig\vphih\epsve\vectOme
  -\parvb\vphii(\cprod{\unitkap}{\vectOme})
  +\parvb\vphij(\cprod{\unitkap}{\unitplz})
  -\parva\vphia^2\vectOme\\
  &\quad-\parva\epsvm\vectr
  -\parva\vphij\Omerep^2(\cprod{\vectr}{\unitplz})
  +\parva\vphij\epsvb(\cprod{\vectOme}{\unitplz})
  +\parva\vphii\Omerep^2(\cprod{\vectr}{\vectOme})
  -\parva\vphig\vphih\epsvm\unitplz\\
  &\quad-\parva\vphig\vphih(\Omerep^2\epsvf-\epsvb\epsvc-\epsvo)\vectOme
  -\parva(\vphij\epsvf-\vphii\epsvb)\vectLam
  -\parva(\vphii\epsvg-\vphij\epsvi)\vectr\\
  &\quad-3\rhorep\epsvg\vphii(\cprod{\vectOme}{\vectr})
  +\rhorep\epsvb\vphii(\cprod{\vectOme}{\vectLam})
  -2\rhorep\epsvh\vphij(\cprod{\unitplz}{\vectOme})
  +3\rhorep\epsvg\vphij(\cprod{\unitplz}{\vectr})
  -\rhorep\epsvb\vphij(\cprod{\unitplz}{\vectLam})\\
  &\quad+2\rhorep\vphig\vphih(\epsvg\epsvb-\epsvh\Omerep^2)\unitplz
  +\rhorep[\vphig\vphih(2\epsvh\epsvc-3\epsvg\epsvf+\epsvb\epsvi-\Omerep^2\epsvl)+\Omerep^2(\vphij\epsvf-\vphii\epsvb)]\vectOme\\
  &\quad+\rhorep[\vphii(\Omerep^4-\vsigc)+\vphij(\vsiga-\Omerep^2\epsvc)-\vphig\vphih\frktp]\vectr
  +\rhorep(\vphii\epsvb-\vphij\epsvf+\vphig\vphih\epsvl)\fdot{\vectLam}
  -3\rhorep\vphic\vectOme\\
  &\quad-\rhorep(2\epsvh\Omerep^2-2\epsvg\epsvb-\frktl)\vectr
  +\vphii^2(\cprod{\vectOme}{\vectLam})
  -\ethvu\vphii(\cprod{\vectOme}{\unitplz})
  -\ethvr\vphij(\cprod{\unitplz}{\vectOme})
  -\vphii\vphij(\cprod{\unitplz}{\vectLam})\\
  &\quad+\vphig\vphih(\vphij-\vphii\epsvc)\vectLam
  +[\vphig\vphih(\ethvr\epsvc+\vphii\epsvi-\ethvu)+\ethvo(\vphij-\vphii\epsvc)]\vectOme\\
  &\quad-\vphig\vphih\epsvl\vectLam+\vphig\vphih\epsvm\unitplz
  -(\ethvr\Omerep^2+\vphii\epsvg-\ethvu\epsvc)\vectr
  -(\ethvu\epsvf-\ethvr\epsvb-\vphii\epsvh-\ethvo\epsvl)\vectOme\\
  &\quad+[\vphig\vphih(\ethvu\epsvc-\ethvr\Omerep^2-\vphii\epsvg+\vphig\vphih\frkto)
      +\vphii(\ethvo\Omerep^2+\vphig\vphih\epsvg)-\vphij(\ethvo\epsvc+\vphig\vphih\epsvi)]\unitplz
\end{split}
\end{align*}
\begin{align*}
\begin{split}
&=-\parvb\epsva\vectr
  -\parva\epsvm\vectr
  -\parva(\vphii\epsvg-\vphij\epsvi)\vectr
  +\rhorep[\vphii(\Omerep^4-\vsigc)+\vphij(\vsiga-\Omerep^2\epsvc)-\vphig\vphih\frktp]\vectr\\
  &\quad-\rhorep(2\epsvh\Omerep^2-2\epsvg\epsvb-\frktl)\vectr
  -(\ethvr\Omerep^2+\vphii\epsvg-\ethvu\epsvc)\vectr\\
  &\quad-\parvb\vphig\vphih\epsva\unitplz
  -\parva\vphig\vphih\epsvm\unitplz
  +2\rhorep\vphig\vphih(\epsvg\epsvb-\epsvh\Omerep^2)\unitplz+\vphig\vphih\epsvm\unitplz\\
  &\quad+[\vphig\vphih(\ethvu\epsvc-\ethvr\Omerep^2-\vphii\epsvg+\vphig\vphih\frkto)
      +\vphii(\ethvo\Omerep^2+\vphig\vphih\epsvg)-\vphij(\ethvo\epsvc+\vphig\vphih\epsvi)]\unitplz\\
  &\quad+\parvb\epsvd\vectOme
  +\parvb\vphig\vphih\epsve\vectOme
  -\parva\vphia^2\vectOme
  -\parva\vphig\vphih(\Omerep^2\epsvf-\epsvb\epsvc-\epsvo)\vectOme\\
  &\quad+\rhorep[\vphig\vphih(2\epsvh\epsvc-3\epsvg\epsvf+\epsvb\epsvi-\Omerep^2\epsvl)+\Omerep^2(\vphij\epsvf-\vphii\epsvb)]\vectOme
  -3\rhorep\vphic\vectOme\\
  &\quad+[\vphig\vphih(\ethvr\epsvc+\vphii\epsvi-\ethvu)+\ethvo(\vphij-\vphii\epsvc)]\vectOme
  -(\ethvu\epsvf-\ethvr\epsvb-\vphii\epsvh-\ethvo\epsvl)\vectOme
  -\vphig\vphih\epsvl\vectLam\\
  &\quad-\parva(\vphij\epsvf-\vphii\epsvb)\vectLam
  +\vphig\vphih(\vphij-\vphii\epsvc)\vectLam
  +\rhorep(\vphii\epsvb-\vphij\epsvf+\vphig\vphih\epsvl)\fdot{\vectLam}
  -\parvb\vphii(\cprod{\unitkap}{\vectOme})\\
  &\quad+\parvb\vphij(\cprod{\unitkap}{\unitplz})
  -\parva\vphij\Omerep^2(\cprod{\vectr}{\unitplz})
  +3\rhorep\epsvg\vphij(\cprod{\unitplz}{\vectr})
  +\parva\vphij\epsvb(\cprod{\vectOme}{\unitplz})
  -\ethvu\vphii(\cprod{\vectOme}{\unitplz})\\
  &\quad-2\rhorep\epsvh\vphij(\cprod{\unitplz}{\vectOme})
  -\ethvr\vphij(\cprod{\unitplz}{\vectOme})
  +\parva\vphii\Omerep^2(\cprod{\vectr}{\vectOme})
  -3\rhorep\epsvg\vphii(\cprod{\vectOme}{\vectr})\\
  &\quad+\rhorep\epsvb\vphii(\cprod{\vectOme}{\vectLam})
  +\vphii^2(\cprod{\vectOme}{\vectLam})
  -\rhorep\epsvb\vphij(\cprod{\unitplz}{\vectLam})
  -\vphii\vphij(\cprod{\unitplz}{\vectLam})
\end{split}
\end{align*}
\begin{align*}
\begin{split}
&=[-\parvb\epsva-\parva\epsvm-\parva(\vphii\epsvg-\vphij\epsvi)+\rhorep\vphii(\Omerep^4-\vsigc)
     +\rhorep\vphij(\vsiga-\Omerep^2\epsvc)-\rhorep\vphig\vphih\frktp\\
     &\qquad-\rhorep(2\epsvh\Omerep^2-2\epsvg\epsvb-\frktl)-(\ethvr\Omerep^2+\vphii\epsvg-\ethvu\epsvc)]\vectr\\
  &\quad+[-\parvb\vphig\vphih\epsva-\parva\vphig\vphih\epsvm+2\rhorep\vphig\vphih(\epsvg\epsvb-\epsvh\Omerep^2)+\vphig\vphih\epsvm\\
     &\qquad+\vphig\vphih(\ethvu\epsvc-\ethvr\Omerep^2-\vphii\epsvg+\vphig\vphih\frkto)
     +\vphii(\ethvo\Omerep^2+\vphig\vphih\epsvg)-\vphij(\ethvo\epsvc+\vphig\vphih\epsvi)]\unitplz\\
  &\quad+[\parvb\epsvd+\parvb\vphig\vphih\epsve-\parva\vphia^2-\parva\vphig\vphih(\Omerep^2\epsvf-\epsvb\epsvc-\epsvo)\\
     &\qquad+\rhorep\vphig\vphih(2\epsvh\epsvc-3\epsvg\epsvf+\epsvb\epsvi-\Omerep^2\epsvl)
     +\rhorep\Omerep^2(\vphij\epsvf-\vphii\epsvb)-3\rhorep\vphic\\
     &\qquad+\vphig\vphih(\ethvr\epsvc+\vphii\epsvi-\ethvu)+\ethvo(\vphij-\vphii\epsvc)
     -(\ethvu\epsvf-\ethvr\epsvb-\vphii\epsvh-\ethvo\epsvl)]\vectOme\\
  &\quad+[-\vphig\vphih\epsvl-\parva(\vphij\epsvf-\vphii\epsvb)+\vphig\vphih(\vphij-\vphii\epsvc)]\vectLam
  +\rhorep(\vphii\epsvb-\vphij\epsvf+\vphig\vphih\epsvl)\fdot{\vectLam}\\
  &\quad-\parvb\vphii(\cprod{\unitkap}{\vectOme})
  +\parvb\vphij(\cprod{\unitkap}{\unitplz})
  +\vphij(\parva\Omerep^2+3\rhorep\epsvg)(\cprod{\unitplz}{\vectr})\\
  &\quad+[\ethvu\vphii-\vphij(\parva\epsvb+2\rhorep\epsvh+\ethvr)](\cprod{\unitplz}{\vectOme})
  -\vphii(\parva\Omerep^2+3\rhorep\epsvg)(\cprod{\vectOme}{\vectr})\\
  &\quad+\vphii(\rhorep\epsvb+\vphii)(\cprod{\vectOme}{\vectLam})
  -\vphij(\rhorep\epsvb+\vphii)(\cprod{\unitplz}{\vectLam})
\end{split}
\end{align*}
\begin{align}\label{rray3b}
\begin{split}
&=\veke\vectr+\vekf\unitplz+\vekg\vectOme
  +\vekb\vectLam+\rhorep\vekc\fdot{\vectLam}
  -\parvb\vphii(\cprod{\unitkap}{\vectOme})
  +\parvb\vphij(\cprod{\unitkap}{\unitplz})
  +\vphij\parvd(\cprod{\unitplz}{\vectr})\\
  &\quad+\vekd(\cprod{\unitplz}{\vectOme})
  -\vphii\parvd(\cprod{\vectOme}{\vectr})
  +\vphii\parve(\cprod{\vectOme}{\vectLam})
  -\vphij\parve(\cprod{\unitplz}{\vectLam})\\
  &\quad\beqref{rpath1p}, \eqnref{rray1a}\text{ \& }\eqnref{rray1b}.
\end{split}
\end{align}
\end{subequations}
Accordingly, we obtain
\begin{align*}
\begin{split}
\veust+\veusu
&=[\rhorep\frkyo+\cdkt(\ethvo\epsva+\vphig\vphih\dltva)]\unitplz
  +[\rhorep(\frkyn+\rhorep\frkyh)-\rhorep\cdkt(\vsigb-\Omerep^2\epsva)-\veka\dltva]\vectr\\
  &\quad+[\ethvo(\rhorep\frkym-\cdkt\epsve)+\rhorep\Omerep^2(\rhorep\vphia^2-\cdkt\epsvd)]\vectOme
  +[\veka\epsvd-\cdkt\vphig\vphih\epsve+\rhorep(\frkyp+3\rhorep\vphic)]\vectLam\\
  &\quad+\rhorep(\cdkt\epsvd-\rhorep\vphia^2)\fdot{\vectLam}
  +[\veka\epsvb+\cdkt(\ethvr+2\rhorep\epsvh)](\cprod{\unitkap}{\vectOme})
  -(\veka\Omerep^2+3\rhorep\cdkt\epsvg)(\cprod{\unitkap}{\vectr})\\
  &\quad+\cdkt(\rhorep\epsvb+\vphii)(\cprod{\unitkap}{\vectLam})
  -\cdkt\ethvu(\cprod{\unitkap}{\unitplz})
  +\rhorep\epsvb(\vphii+\rhorep\epsvb)(\cprod{\vectOme}{\vectLam})
  -\rhorep\ethvu\epsvb(\cprod{\vectOme}{\unitplz})\\
  &\quad+\rhorep(\ethvr\Omerep^2+\rhorep\frkyg)(\cprod{\vectOme}{\vectr})
  -\rhorep\Omerep^2(\vphii+\rhorep\epsvb)(\cprod{\vectr}{\vectLam})
  +\rhorep\ethvu\Omerep^2(\cprod{\vectr}{\unitplz})\\
  &\quad+\veke\vectr+\vekf\unitplz+\vekg\vectOme
  +\vekb\vectLam+\rhorep\vekc\fdot{\vectLam}
  -\parvb\vphii(\cprod{\unitkap}{\vectOme})
  +\parvb\vphij(\cprod{\unitkap}{\unitplz})
  +\vphij\parvd(\cprod{\unitplz}{\vectr})\\
  &\quad+\vekd(\cprod{\unitplz}{\vectOme})
  -\vphii\parvd(\cprod{\vectOme}{\vectr})
  +\vphii\parve(\cprod{\vectOme}{\vectLam})
  -\vphij\parve(\cprod{\unitplz}{\vectLam})\beqref{rray3}
\end{split}
\end{align*}
\begin{align*}
\begin{split}
&=\vekf\unitplz+[\rhorep\frkyo+\cdkt(\ethvo\epsva+\vphig\vphih\dltva)]\unitplz
  +\veke\vectr+[\rhorep(\frkyn+\rhorep\frkyh)-\rhorep\cdkt(\vsigb-\Omerep^2\epsva)-\veka\dltva]\vectr\\
  &\quad+\vekg\vectOme+[\ethvo(\rhorep\frkym-\cdkt\epsve)+\rhorep\Omerep^2(\rhorep\vphia^2-\cdkt\epsvd)]\vectOme
  +\vekb\vectLam+[\veka\epsvd-\cdkt\vphig\vphih\epsve+\rhorep(\frkyp+3\rhorep\vphic)]\vectLam\\
  &\quad+\rhorep\vekc\fdot{\vectLam}+\rhorep(\cdkt\epsvd-\rhorep\vphia^2)\fdot{\vectLam}
  -\parvb\vphii(\cprod{\unitkap}{\vectOme})+[\veka\epsvb+\cdkt(\ethvr+2\rhorep\epsvh)](\cprod{\unitkap}{\vectOme})\\
  &\quad-(\veka\Omerep^2+3\rhorep\cdkt\epsvg)(\cprod{\unitkap}{\vectr})
  +\cdkt(\rhorep\epsvb+\vphii)(\cprod{\unitkap}{\vectLam})
  +\parvb\vphij(\cprod{\unitkap}{\unitplz})-\cdkt\ethvu(\cprod{\unitkap}{\unitplz})\\
  &\quad+\vphii\parve(\cprod{\vectOme}{\vectLam})
  +\rhorep\epsvb(\vphii+\rhorep\epsvb)(\cprod{\vectOme}{\vectLam})
  +\vekd(\cprod{\unitplz}{\vectOme})
  -\rhorep\ethvu\epsvb(\cprod{\vectOme}{\unitplz})
  -\vphii\parvd(\cprod{\vectOme}{\vectr})\\
  &\quad+\rhorep(\ethvr\Omerep^2+\rhorep\frkyg)(\cprod{\vectOme}{\vectr})
  -\rhorep\Omerep^2(\vphii+\rhorep\epsvb)(\cprod{\vectr}{\vectLam})
  +\vphij\parvd(\cprod{\unitplz}{\vectr})
  +\rhorep\ethvu\Omerep^2(\cprod{\vectr}{\unitplz})\\
  &\quad-\vphij\parve(\cprod{\unitplz}{\vectLam})
\end{split}
\end{align*}
\begin{align*}
\begin{split}
&=[\vekf+\rhorep\frkyo+\cdkt(\ethvo\epsva+\vphig\vphih\dltva)]\unitplz
  +[\veke+\rhorep(\frkyn+\rhorep\frkyh)-\rhorep\cdkt(\vsigb-\Omerep^2\epsva)-\veka\dltva]\vectr\\
  &\quad+[\vekg+\ethvo(\rhorep\frkym-\cdkt\epsve)+\rhorep\Omerep^2(\rhorep\vphia^2-\cdkt\epsvd)]\vectOme
  +[\vekb+\veka\epsvd-\cdkt\vphig\vphih\epsve+\rhorep(\frkyp+3\rhorep\vphic)]\vectLam\\
  &\quad+\rhorep(\vekc+\cdkt\epsvd-\rhorep\vphia^2)\fdot{\vectLam}
  +[\veka\epsvb-\parvb\vphii+\cdkt(\ethvr+2\rhorep\epsvh)](\cprod{\unitkap}{\vectOme})\\
  &\quad-(\veka\Omerep^2+3\rhorep\cdkt\epsvg)(\cprod{\unitkap}{\vectr})
  +\cdkt(\rhorep\epsvb+\vphii)(\cprod{\unitkap}{\vectLam})
  +(\parvb\vphij-\cdkt\ethvu)(\cprod{\unitkap}{\unitplz})\\
  &\quad+[\vphii\parve+\rhorep\epsvb(\vphii+\rhorep\epsvb)](\cprod{\vectOme}{\vectLam})
  +(\vekd+\rhorep\ethvu\epsvb)(\cprod{\unitplz}{\vectOme})
  +[-\vphii\parvd+\rhorep(\ethvr\Omerep^2+\rhorep\frkyg)](\cprod{\vectOme}{\vectr})\\
  &\quad-\rhorep\Omerep^2(\vphii+\rhorep\epsvb)(\cprod{\vectr}{\vectLam})
  +(\vphij\parvd-\rhorep\ethvu\Omerep^2)(\cprod{\unitplz}{\vectr})
  -\vphij\parve(\cprod{\unitplz}{\vectLam})
\end{split}
\end{align*}
\begin{align}\label{rray4}
\begin{split}
&=\vekh\unitplz+\veki\vectr+\vekj\vectOme+\vekm\vectLam+\vekl\fdot{\vectLam}+\vekk(\cprod{\unitkap}{\vectOme})
  -\vekn(\cprod{\unitkap}{\vectr})+\cdkt\parve(\cprod{\unitkap}{\vectLam})+\vekp(\cprod{\unitkap}{\unitplz})\\
  &\quad+\parve^2(\cprod{\vectOme}{\vectLam})+\vekq(\cprod{\unitplz}{\vectOme})+\veko(\cprod{\vectOme}{\vectr})
  -\rhorep\Omerep^2\parve(\cprod{\vectr}{\vectLam})+\vekr(\cprod{\unitplz}{\vectr})-\vphij\parve(\cprod{\unitplz}{\vectLam})\\
  &\quad\beqref{rpath1p}\text{ \& }\eqnref{rray1c}.
\end{split}
\end{align}

\subart{Computation of the magnitude of $\veust+\veusu$}
To calculate the magnitude of the vector $\veust+\veusu$, we first derive
\begin{subequations}\label{rray5}
\begin{align}\label{rray5a}
\begin{split}
&\dprod{\unitplz}{(\veust+\veusu)}\\
&=\dprod{\unitplz}{[}\vekh\unitplz+\veki\vectr+\vekj\vectOme+\vekm\vectLam
  +\vekl\fdot{\vectLam}+\vekk(\cprod{\unitkap}{\vectOme})-\vekn(\cprod{\unitkap}{\vectr})\\
  &\quad+\cdkt\parve(\cprod{\unitkap}{\vectLam})+\vekp(\cprod{\unitkap}{\unitplz})
  +\parve^2(\cprod{\vectOme}{\vectLam})+\vekq(\cprod{\unitplz}{\vectOme})+\veko(\cprod{\vectOme}{\vectr})
  -\rhorep\Omerep^2\parve(\cprod{\vectr}{\vectLam})\\
  &\quad+\vekr(\cprod{\unitplz}{\vectr})-\vphij\parve(\cprod{\unitplz}{\vectLam})]\beqref{rray4}\\
\end{split}
\nonumber\\
\begin{split}
&=\vekh(\dprod{\unitplz}{\unitplz})
  +\veki(\dprod{\unitplz}{\vectr})
  +\vekj(\dprod{\unitplz}{\vectOme})
  +\vekm(\dprod{\unitplz}{\vectLam})
  +\vekl(\dprod{\unitplz}{\fdot{\vectLam}})
  +\vekk[\dprod{\unitplz}{(\cprod{\unitkap}{\vectOme})}]\\
  &\quad-\vekn[\dprod{\unitplz}{(\cprod{\unitkap}{\vectr})}]
  +\cdkt\parve[\dprod{\unitplz}{(\cprod{\unitkap}{\vectLam})}]
  +\vekp[\dprod{\unitplz}{(\cprod{\unitkap}{\unitplz})}]
  +\parve^2[\dprod{\unitplz}{(\cprod{\vectOme}{\vectLam})}]
  +\vekq[\dprod{\unitplz}{(\cprod{\unitplz}{\vectOme})}]\\
  &\quad+\veko[\dprod{\unitplz}{(\cprod{\vectOme}{\vectr})}]
  -\rhorep\Omerep^2\parve[\dprod{\unitplz}{(\cprod{\vectr}{\vectLam})}]
  +\vekr[\dprod{\unitplz}{(\cprod{\unitplz}{\vectr})}]
  -\vphij\parve[\dprod{\unitplz}{(\cprod{\unitplz}{\vectLam})}]
\end{split}
\nonumber\\
\begin{split}
&=\vekh+\veki\epsvf+\vekj\epsvc+\vekm\epsvi+\vekl\vsiga-\vekk\epsvj-\vekn\frktt-\cdkt\parve\dltvb
  +\parve^2\frkto+\veko\epsvl+\rhorep\Omerep^2\parve\epsvo\\
  &\qquad\beqref{rot1a}, \eqnref{rxpeed1a}, \eqnref{rpath1a}\text{ \& }\eqnref{rpath1b}
\end{split}
\nonumber\\
&=\vekr\beqref{rray1d}
\end{align}
\begin{align}\label{rray5b}
\begin{split}
&\dprod{\vectr}{(\veust+\veusu)}\\
&=\dprod{\vectr}{[}\vekh\unitplz+\veki\vectr+\vekj\vectOme+\vekm\vectLam
  +\vekl\fdot{\vectLam}+\vekk(\cprod{\unitkap}{\vectOme})-\vekn(\cprod{\unitkap}{\vectr})\\
  &\quad+\cdkt\parve(\cprod{\unitkap}{\vectLam})+\vekp(\cprod{\unitkap}{\unitplz})
  +\parve^2(\cprod{\vectOme}{\vectLam})+\vekq(\cprod{\unitplz}{\vectOme})+\veko(\cprod{\vectOme}{\vectr})
  -\rhorep\Omerep^2\parve(\cprod{\vectr}{\vectLam})\\
  &\quad+\vekr(\cprod{\unitplz}{\vectr})-\vphij\parve(\cprod{\unitplz}{\vectLam})]\beqref{rray4}
\end{split}
\nonumber\\
\begin{split}
&=\vekh(\dprod{\vectr}{\unitplz})
  +\veki(\dprod{\vectr}{\vectr})
  +\vekj(\dprod{\vectr}{\vectOme})
  +\vekm(\dprod{\vectr}{\vectLam})
  +\vekl(\dprod{\vectr}{\fdot{\vectLam}})
  +\vekk[\dprod{\vectr}{(\cprod{\unitkap}{\vectOme})}]\\
  &\quad-\vekn[\dprod{\vectr}{(\cprod{\unitkap}{\vectr})}]
  +\cdkt\parve[\dprod{\vectr}{(\cprod{\unitkap}{\vectLam})}]
  +\vekp[\dprod{\vectr}{(\cprod{\unitkap}{\unitplz})}]
  +\parve^2[\dprod{\vectr}{(\cprod{\vectOme}{\vectLam})}]
  +\vekq[\dprod{\vectr}{(\cprod{\unitplz}{\vectOme})}]\\
  &\quad+\veko[\dprod{\vectr}{(\cprod{\vectOme}{\vectr})}]
  -\rhorep\Omerep^2\parve[\dprod{\vectr}{(\cprod{\vectr}{\vectLam})}]
  +\vekr[\dprod{\vectr}{(\cprod{\unitplz}{\vectr})}]
  -\vphij\parve[\dprod{\vectr}{(\cprod{\unitplz}{\vectLam})}]
\end{split}
\nonumber\\
\begin{split}
&=\vekh\epsvf+\veki\scalr^2+\vekj\epsvb+\vekm\epsvh+\vekl\vsign+\vekk\epsvk+\cdkt\parve\epsvn-\vekp\frktt
  +\parve^2\epsvm+\vekq\epsvl-\vphij\parve\epsvo\\
  &\qquad\beqref{rot1a}, \eqnref{rpath1a}\text{ \& }\eqnref{rpath1b}
\end{split}
\nonumber\\
&=\veks\beqref{rray1d}
\end{align}
\begin{align}\label{rray5c}
\begin{split}
&\dprod{\vectOme}{(\veust+\veusu)}\\
&=\dprod{\vectOme}{[}\vekh\unitplz+\veki\vectr+\vekj\vectOme+\vekm\vectLam
  +\vekl\fdot{\vectLam}+\vekk(\cprod{\unitkap}{\vectOme})-\vekn(\cprod{\unitkap}{\vectr})\\
  &\quad+\cdkt\parve(\cprod{\unitkap}{\vectLam})+\vekp(\cprod{\unitkap}{\unitplz})
  +\parve^2(\cprod{\vectOme}{\vectLam})+\vekq(\cprod{\unitplz}{\vectOme})+\veko(\cprod{\vectOme}{\vectr})
  -\rhorep\Omerep^2\parve(\cprod{\vectr}{\vectLam})\\
  &\quad+\vekr(\cprod{\unitplz}{\vectr})-\vphij\parve(\cprod{\unitplz}{\vectLam})]\beqref{rray4}
\end{split}
\nonumber\\
\begin{split}
&=\vekh(\dprod{\vectOme}{\unitplz})
  +\veki(\dprod{\vectOme}{\vectr})
  +\vekj(\dprod{\vectOme}{\vectOme})
  +\vekm(\dprod{\vectOme}{\vectLam})
  +\vekl(\dprod{\vectOme}{\fdot{\vectLam}})
  +\vekk[\dprod{\vectOme}{(\cprod{\unitkap}{\vectOme})}]\\
  &\quad-\vekn[\dprod{\vectOme}{(\cprod{\unitkap}{\vectr})}]
  +\cdkt\parve[\dprod{\vectOme}{(\cprod{\unitkap}{\vectLam})}]
  +\vekp[\dprod{\vectOme}{(\cprod{\unitkap}{\unitplz})}]
  +\parve^2[\dprod{\vectOme}{(\cprod{\vectOme}{\vectLam})}]
  +\vekq[\dprod{\vectOme}{(\cprod{\unitplz}{\vectOme})}]\\
  &\quad+\veko[\dprod{\vectOme}{(\cprod{\vectOme}{\vectr})}]
  -\rhorep\Omerep^2\parve[\dprod{\vectOme}{(\cprod{\vectr}{\vectLam})}]
  +\vekr[\dprod{\vectOme}{(\cprod{\unitplz}{\vectr})}]
  -\vphij\parve[\dprod{\vectOme}{(\cprod{\unitplz}{\vectLam})}]
\end{split}
\nonumber\\
\begin{split}
&=\vekh\epsvc+\veki\epsvb+\vekj\Omerep^2+\vekm\epsvg+\vekl\vsigc+\vekn\epsvk+\cdkt\parve\frkte
  +\vekp\epsvj+\rhorep\Omerep^2\parve\epsvm-\vekr\epsvl\\
  &\qquad+\vphij\parve\frkto
  \beqref{rot1a}, \eqnref{rpath1a}\text{ \& }\eqnref{rpath1b}
\end{split}
\nonumber\\
&=\vekt\beqref{rray1d}
\end{align}
\begin{align}\label{rray5d}
\begin{split}
&\dprod{\vectLam}{(\veust+\veusu)}\\
&=\dprod{\vectLam}{[}\vekh\unitplz+\veki\vectr+\vekj\vectOme+\vekm\vectLam
  +\vekl\fdot{\vectLam}+\vekk(\cprod{\unitkap}{\vectOme})-\vekn(\cprod{\unitkap}{\vectr})\\
  &\quad+\cdkt\parve(\cprod{\unitkap}{\vectLam})+\vekp(\cprod{\unitkap}{\unitplz})
  +\parve^2(\cprod{\vectOme}{\vectLam})+\vekq(\cprod{\unitplz}{\vectOme})+\veko(\cprod{\vectOme}{\vectr})
  -\rhorep\Omerep^2\parve(\cprod{\vectr}{\vectLam})\\
  &\quad+\vekr(\cprod{\unitplz}{\vectr})-\vphij\parve(\cprod{\unitplz}{\vectLam})]\beqref{rray4}
\end{split}
\nonumber\\
\begin{split}
&=\vekh(\dprod{\vectLam}{\unitplz})
  +\veki(\dprod{\vectLam}{\vectr})
  +\vekj(\dprod{\vectLam}{\vectOme})
  +\vekm(\dprod{\vectLam}{\vectLam})
  +\vekl(\dprod{\vectLam}{\fdot{\vectLam}})
  +\vekk[\dprod{\vectLam}{(\cprod{\unitkap}{\vectOme})}]\\
  &\quad-\vekn[\dprod{\vectLam}{(\cprod{\unitkap}{\vectr})}]
  +\cdkt\parve[\dprod{\vectLam}{(\cprod{\unitkap}{\vectLam})}]
  +\vekp[\dprod{\vectLam}{(\cprod{\unitkap}{\unitplz})}]
  +\parve^2[\dprod{\vectLam}{(\cprod{\vectOme}{\vectLam})}]
  +\vekq[\dprod{\vectLam}{(\cprod{\unitplz}{\vectOme})}]\\
  &\quad+\veko[\dprod{\vectLam}{(\cprod{\vectOme}{\vectr})}]
  -\rhorep\Omerep^2\parve[\dprod{\vectLam}{(\cprod{\vectr}{\vectLam})}]
  +\vekr[\dprod{\vectLam}{(\cprod{\unitplz}{\vectr})}]
  -\vphij\parve[\dprod{\vectLam}{(\cprod{\unitplz}{\vectLam})}]
\end{split}
\nonumber\\
\begin{split}
&=\vekh\epsvi+\veki\epsvh+\vekj\epsvg+\vekm\Lamrep^2+\vekl\vsigd
  -\vekk\frkte+\vekn\epsvn+\vekp\dltvb+\vekq\frkto-\veko\epsvm-\vekr\epsvo\\
  &\qquad\beqref{rot1a}, \eqnref{rxpeed1a}, \eqnref{rpath1a}\text{ \& }\eqnref{rpath1b}
\end{split}
\nonumber\\
&=\veku\beqref{rray1d}
\end{align}
\begin{align}\label{rray5e}
\begin{split}
&\dprod{\fdot{\vectLam}}{(\veust+\veusu)}\\
&=\dprod{\fdot{\vectLam}}{[}\vekh\unitplz+\veki\vectr+\vekj\vectOme+\vekm\vectLam
  +\vekl\fdot{\vectLam}+\vekk(\cprod{\unitkap}{\vectOme})-\vekn(\cprod{\unitkap}{\vectr})\\
  &\quad+\cdkt\parve(\cprod{\unitkap}{\vectLam})+\vekp(\cprod{\unitkap}{\unitplz})
  +\parve^2(\cprod{\vectOme}{\vectLam})+\vekq(\cprod{\unitplz}{\vectOme})+\veko(\cprod{\vectOme}{\vectr})
  -\rhorep\Omerep^2\parve(\cprod{\vectr}{\vectLam})\\
  &\quad+\vekr(\cprod{\unitplz}{\vectr})-\vphij\parve(\cprod{\unitplz}{\vectLam})]\beqref{rray4}
\end{split}
\nonumber\\
\begin{split}
&=\vekh(\dprod{\fdot{\vectLam}}{\unitplz})
  +\veki(\dprod{\fdot{\vectLam}}{\vectr})
  +\vekj(\dprod{\fdot{\vectLam}}{\vectOme})
  +\vekm(\dprod{\fdot{\vectLam}}{\vectLam})
  +\vekl(\dprod{\fdot{\vectLam}}{\fdot{\vectLam}})
  +\vekk[\dprod{\fdot{\vectLam}}{(\cprod{\unitkap}{\vectOme})}]\\
  &\quad-\vekn[\dprod{\fdot{\vectLam}}{(\cprod{\unitkap}{\vectr})}]
  +\cdkt\parve[\dprod{\fdot{\vectLam}}{(\cprod{\unitkap}{\vectLam})}]
  +\vekp[\dprod{\fdot{\vectLam}}{(\cprod{\unitkap}{\unitplz})}]
  +\parve^2[\dprod{\fdot{\vectLam}}{(\cprod{\vectOme}{\vectLam})}]
  +\vekq[\dprod{\fdot{\vectLam}}{(\cprod{\unitplz}{\vectOme})}]\\
  &\quad+\veko[\dprod{\fdot{\vectLam}}{(\cprod{\vectOme}{\vectr})}]
  -\rhorep\Omerep^2\parve[\dprod{\fdot{\vectLam}}{(\cprod{\vectr}{\vectLam})}]
  +\vekr[\dprod{\fdot{\vectLam}}{(\cprod{\unitplz}{\vectr})}
  -\vphij\parve[\dprod{\fdot{\vectLam}}{(\cprod{\unitplz}{\vectLam})}]
\end{split}
\nonumber\\
\begin{split}
&=\vekh\vsiga+\veki\vsign+\vekj\vsigc+\vekm\vsigd+\vekl\vsige+\vekk\frktf+\vekn\frktc+\cdkt\parve\frktu
  +\vekp\frkta+\parve^2\frktv+\vekq\frktp\\
  &\quad+\veko\frktl+\rhorep\Omerep^2\parve\frkth-\vekr\frktn-\vphij\parve\frktr
  \beqref{rpath1a}, \eqnref{rpath1b}\text{ \& }\eqnref{rpath1b2}
\end{split}
\nonumber\\
&=\vekv\beqref{rray1e}
\end{align}
\begin{align*}
\begin{split}
&\dprod{(\cprod{\unitkap}{\vectOme})}{(\veust+\veusu)}\\
&=\dprod{(\cprod{\unitkap}{\vectOme})}{[}\vekh\unitplz+\veki\vectr+\vekj\vectOme+\vekm\vectLam
  +\vekl\fdot{\vectLam}+\vekk(\cprod{\unitkap}{\vectOme})-\vekn(\cprod{\unitkap}{\vectr})\\
  &\quad+\cdkt\parve(\cprod{\unitkap}{\vectLam})+\vekp(\cprod{\unitkap}{\unitplz})
  +\parve^2(\cprod{\vectOme}{\vectLam})+\vekq(\cprod{\unitplz}{\vectOme})+\veko(\cprod{\vectOme}{\vectr})
  -\rhorep\Omerep^2\parve(\cprod{\vectr}{\vectLam})\\
  &\quad+\vekr(\cprod{\unitplz}{\vectr})-\vphij\parve(\cprod{\unitplz}{\vectLam})]\beqref{rray4}
\end{split}
\nonumber\\
\begin{split}
&=\vekh[\dprod{\unitplz}{(\cprod{\unitkap}{\vectOme})}]
  +\veki[\dprod{\vectr}{(\cprod{\unitkap}{\vectOme})}]
  +\vekj[\dprod{\vectOme}{(\cprod{\unitkap}{\vectOme})}]
  +\vekm[\dprod{\vectLam}{(\cprod{\unitkap}{\vectOme})}]\\
  &\quad+\vekl[\dprod{\fdot{\vectLam}}{(\cprod{\unitkap}{\vectOme})}]
  +\vekk[\dprod{(\cprod{\unitkap}{\vectOme})}{(\cprod{\unitkap}{\vectOme})}]
  -\vekn[\dprod{(\cprod{\unitkap}{\vectOme})}{(\cprod{\unitkap}{\vectr})}]\\
  &\quad+\cdkt\parve[\dprod{(\cprod{\unitkap}{\vectOme})}{(\cprod{\unitkap}{\vectLam})}]
  +\vekp[\dprod{(\cprod{\unitkap}{\vectOme})}{(\cprod{\unitkap}{\unitplz})}]
  +\parve^2[\dprod{(\cprod{\unitkap}{\vectOme})}{(\cprod{\vectOme}{\vectLam})}]\\
  &\quad+\vekq[\dprod{(\cprod{\unitkap}{\vectOme})}{(\cprod{\unitplz}{\vectOme})}]
  +\veko[\dprod{(\cprod{\unitkap}{\vectOme})}{(\cprod{\vectOme}{\vectr})}]
  -\rhorep\Omerep^2\parve[\dprod{(\cprod{\unitkap}{\vectOme})}{(\cprod{\vectr}{\vectLam})}]\\
  &\quad+\vekr[\dprod{(\cprod{\unitkap}{\vectOme})}{(\cprod{\unitplz}{\vectr})}]
  -\vphij\parve[\dprod{(\cprod{\unitkap}{\vectOme})}{(\cprod{\unitplz}{\vectLam})}]
\end{split}
\end{align*}
\begin{align}\label{rray5f}
\begin{split}
&=-\vekh\epsvj+\veki\epsvk-\vekm\frkte+\vekl\frktf
  +\vekk[\Omerep^2-(\dprod{\unitkap}{\vectOme})^2]\\
  &\quad-\vekn[(\dprod{\vectOme}{\vectr})-(\dprod{\unitkap}{\vectr})(\dprod{\vectOme}{\unitkap})]
  +\cdkt\parve[(\dprod{\vectOme}{\vectLam})-(\dprod{\unitkap}{\vectLam})(\dprod{\vectOme}{\unitkap})]\\
  &\quad+\vekp[(\dprod{\vectOme}{\unitplz})-(\dprod{\unitkap}{\unitplz})(\dprod{\vectOme}{\unitkap})]
  +\parve^2[(\dprod{\unitkap}{\vectOme})(\dprod{\vectOme}{\vectLam})-(\dprod{\unitkap}{\vectLam})\Omerep^2]\\
  &\quad+\vekq[(\dprod{\unitkap}{\unitplz})\Omerep^2-(\dprod{\unitkap}{\vectOme})(\dprod{\vectOme}{\unitplz})]
  +\veko[(\dprod{\unitkap}{\vectOme})(\dprod{\vectOme}{\vectr})-(\dprod{\unitkap}{\vectr})\Omerep^2]\\
  &\quad-\rhorep\Omerep^2\parve[(\dprod{\unitkap}{\vectr})(\dprod{\vectOme}{\vectLam})-(\dprod{\unitkap}{\vectLam})(\dprod{\vectOme}{\vectr})]
  +\vekr[(\dprod{\unitkap}{\unitplz})(\dprod{\vectOme}{\vectr})-(\dprod{\unitkap}{\vectr})(\dprod{\vectOme}{\unitplz})]\\
  &\quad-\vphij\parve[(\dprod{\unitkap}{\unitplz})(\dprod{\vectOme}{\vectLam})-(\dprod{\unitkap}{\vectLam})(\dprod{\vectOme}{\unitplz})]
  \beqref{rot1a}, \eqnref{rpath1b}\text{ \& }\eqnref{alg2}
\end{split}
\nonumber\\
\begin{split}
&=-\vekh\epsvj+\veki\epsvk-\vekm\frkte+\vekl\frktf+\vekk(\Omerep^2-\epsva^2)
  -\vekn(\epsvb-\epsvd\epsva)+\cdkt\parve(\epsvg-\dltva\epsva)\\
  &\quad+\vekp(\epsvc-\epsve\epsva)+\parve^2(\epsva\epsvg-\dltva\Omerep^2)
  +\vekq(\epsve\Omerep^2-\epsva\epsvc)+\veko(\epsva\epsvb-\epsvd\Omerep^2)\\
  &\quad-\rhorep\Omerep^2\parve(\epsvd\epsvg-\dltva\epsvb)+\vekr(\epsve\epsvb-\epsvd\epsvc)
  -\vphij\parve(\epsve\epsvg-\dltva\epsvc)
  \beqref{rot1a}\text{ \& }\eqnref{rxpeed1a}
\end{split}
\nonumber\\
&=\vekw\beqref{rray1e}
\end{align}
\begin{align*}
\begin{split}
&\dprod{(\cprod{\unitkap}{\vectr})}{(\veust+\veusu)}\\
&=\dprod{(\cprod{\unitkap}{\vectr})}{[}\vekh\unitplz+\veki\vectr+\vekj\vectOme+\vekm\vectLam
  +\vekl\fdot{\vectLam}+\vekk(\cprod{\unitkap}{\vectOme})-\vekn(\cprod{\unitkap}{\vectr})\\
  &\quad+\cdkt\parve(\cprod{\unitkap}{\vectLam})+\vekp(\cprod{\unitkap}{\unitplz})
  +\parve^2(\cprod{\vectOme}{\vectLam})+\vekq(\cprod{\unitplz}{\vectOme})+\veko(\cprod{\vectOme}{\vectr})
  -\rhorep\Omerep^2\parve(\cprod{\vectr}{\vectLam})\\
  &\quad+\vekr(\cprod{\unitplz}{\vectr})-\vphij\parve(\cprod{\unitplz}{\vectLam})]\beqref{rray4}
\end{split}
\nonumber\\
\begin{split}
&=\vekh[\dprod{\unitplz}{(\cprod{\unitkap}{\vectr})}]
  +\veki[\dprod{\vectr}{(\cprod{\unitkap}{\vectr})}]
  +\vekj[\dprod{\vectOme}{(\cprod{\unitkap}{\vectr})}]
  +\vekm[\dprod{\vectLam}{(\cprod{\unitkap}{\vectr})}]\\
  &\quad+\vekl[\dprod{\fdot{\vectLam}}{(\cprod{\unitkap}{\vectr})}]
  +\vekk[\dprod{(\cprod{\unitkap}{\vectr})}{(\cprod{\unitkap}{\vectOme})}]
  -\vekn[\dprod{(\cprod{\unitkap}{\vectr})}{(\cprod{\unitkap}{\vectr})}]\\
  &\quad+\cdkt\parve[\dprod{(\cprod{\unitkap}{\vectr})}{(\cprod{\unitkap}{\vectLam})}]
  +\vekp[\dprod{(\cprod{\unitkap}{\vectr})}{(\cprod{\unitkap}{\unitplz})}]
  +\parve^2[\dprod{(\cprod{\unitkap}{\vectr})}{(\cprod{\vectOme}{\vectLam})}]\\
  &\quad+\vekq[\dprod{(\cprod{\unitkap}{\vectr})}{(\cprod{\unitplz}{\vectOme})}]
  +\veko[\dprod{(\cprod{\unitkap}{\vectr})}{(\cprod{\vectOme}{\vectr})}]
  -\rhorep\Omerep^2\parve[\dprod{(\cprod{\unitkap}{\vectr})}{(\cprod{\vectr}{\vectLam})}]\\
  &\quad+\vekr[\dprod{(\cprod{\unitkap}{\vectr})}{(\cprod{\unitplz}{\vectr})}]
  -\vphij\parve[\dprod{(\cprod{\unitkap}{\vectr})}{(\cprod{\unitplz}{\vectLam})}]
\end{split}
\end{align*}
\begin{align}\label{rray5g}
\begin{split}
&=\vekh\frktt-\vekj\epsvk-\vekm\epsvn-\vekl\frktc
  +\vekk[(\dprod{\vectr}{\vectOme})-(\dprod{\unitkap}{\vectOme})(\dprod{\vectr}{\unitkap})]\\
  &\quad-\vekn[\scalr^2-(\dprod{\unitkap}{\vectr})^2]
  +\cdkt\parve[(\dprod{\vectr}{\vectLam})-(\dprod{\unitkap}{\vectLam})(\dprod{\vectr}{\unitkap})]\\
  &\quad+\vekp[(\dprod{\vectr}{\unitplz})-(\dprod{\unitkap}{\unitplz})(\dprod{\vectr}{\unitkap})]
  +\parve^2[(\dprod{\unitkap}{\vectOme})(\dprod{\vectr}{\vectLam})-(\dprod{\unitkap}{\vectLam})(\dprod{\vectr}{\vectOme})]\\
  &\quad+\vekq[(\dprod{\unitkap}{\unitplz})(\dprod{\vectr}{\vectOme})-(\dprod{\unitkap}{\vectOme})(\dprod{\vectr}{\unitplz})]
  +\veko[(\dprod{\unitkap}{\vectOme})\scalr^2-(\dprod{\unitkap}{\vectr})(\dprod{\vectr}{\vectOme})]\\
  &\quad-\rhorep\Omerep^2\parve[(\dprod{\unitkap}{\vectr})(\dprod{\vectr}{\vectLam})-(\dprod{\unitkap}{\vectLam})\scalr^2]
  +\vekr[(\dprod{\unitkap}{\unitplz})\scalr^2-(\dprod{\unitkap}{\vectr})(\dprod{\vectr}{\unitplz})]\\
  &\quad-\vphij\parve[(\dprod{\unitkap}{\unitplz})(\dprod{\vectr}{\vectLam})-(\dprod{\unitkap}{\vectLam})(\dprod{\vectr}{\unitplz})]
  \beqref{rot1a}, \eqnref{rpath1b}\text{ \& }\eqnref{alg2}
\end{split}
\nonumber\\
\begin{split}
&=\vekh\frktt-\vekj\epsvk-\vekm\epsvn-\vekl\frktc
  +\vekk(\epsvb-\epsva\epsvd)-\vekn(\scalr^2-\epsvd^2)+\cdkt\parve(\epsvh-\dltva\epsvd)\\
  &\quad+\vekp(\epsvf-\epsve\epsvd)+\parve^2(\epsva\epsvh-\dltva\epsvb)
  +\vekq(\epsve\epsvb-\epsva\epsvf)+\veko(\epsva\scalr^2-\epsvd\epsvb)\\
  &\quad-\rhorep\Omerep^2\parve(\epsvd\epsvh-\dltva\scalr^2)+\vekr(\epsve\scalr^2-\epsvd\epsvf)
  -\vphij\parve(\epsve\epsvh-\dltva\epsvf)
  \beqref{rot1a}\text{ \& }\eqnref{rxpeed1a}
\end{split}
\nonumber\\
&=\vekx\beqref{rray1e}
\end{align}
\begin{align*}
\begin{split}
&\dprod{(\cprod{\unitkap}{\vectLam})}{(\veust+\veusu)}\\
&=\dprod{(\cprod{\unitkap}{\vectLam})}{[}\vekh\unitplz+\veki\vectr+\vekj\vectOme+\vekm\vectLam
  +\vekl\fdot{\vectLam}+\vekk(\cprod{\unitkap}{\vectOme})-\vekn(\cprod{\unitkap}{\vectr})\\
  &\quad+\cdkt\parve(\cprod{\unitkap}{\vectLam})+\vekp(\cprod{\unitkap}{\unitplz})
  +\parve^2(\cprod{\vectOme}{\vectLam})+\vekq(\cprod{\unitplz}{\vectOme})+\veko(\cprod{\vectOme}{\vectr})
  -\rhorep\Omerep^2\parve(\cprod{\vectr}{\vectLam})\\
  &\quad+\vekr(\cprod{\unitplz}{\vectr})-\vphij\parve(\cprod{\unitplz}{\vectLam})]\beqref{rray4}
\end{split}
\nonumber\\
\begin{split}
&=\vekh[\dprod{\unitplz}{(\cprod{\unitkap}{\vectLam})}]
  +\veki[\dprod{\vectr}{(\cprod{\unitkap}{\vectLam})}]
  +\vekj[\dprod{\vectOme}{(\cprod{\unitkap}{\vectLam})}]
  +\vekm[\dprod{\vectLam}{(\cprod{\unitkap}{\vectLam})}]\\
  &\quad+\vekl[\dprod{\fdot{\vectLam}}{(\cprod{\unitkap}{\vectLam})}]
  +\vekk[\dprod{(\cprod{\unitkap}{\vectLam})}{(\cprod{\unitkap}{\vectOme})}]
  -\vekn[\dprod{(\cprod{\unitkap}{\vectLam})}{(\cprod{\unitkap}{\vectr})}]\\
  &\quad+\cdkt\parve[\dprod{(\cprod{\unitkap}{\vectLam})}{(\cprod{\unitkap}{\vectLam})}]
  +\vekp[\dprod{(\cprod{\unitkap}{\vectLam})}{(\cprod{\unitkap}{\unitplz})}]
  +\parve^2[\dprod{(\cprod{\unitkap}{\vectLam})}{(\cprod{\vectOme}{\vectLam})}]\\
  &\quad+\vekq[\dprod{(\cprod{\unitkap}{\vectLam})}{(\cprod{\unitplz}{\vectOme})}]
  +\veko[\dprod{(\cprod{\unitkap}{\vectLam})}{(\cprod{\vectOme}{\vectr})}]
  -\rhorep\Omerep^2\parve[\dprod{(\cprod{\unitkap}{\vectLam})}{(\cprod{\vectr}{\vectLam})}]\\
  &\quad+\vekr[\dprod{(\cprod{\unitkap}{\vectLam})}{(\cprod{\unitplz}{\vectr})}]
  -\vphij\parve[\dprod{(\cprod{\unitkap}{\vectLam})}{(\cprod{\unitplz}{\vectLam})}]
\end{split}
\end{align*}
\begin{align}\label{rray5h}
\begin{split}
&=-\vekh\dltvb+\veki\epsvn+\vekj\frkte+\vekl\frktu
  +\vekk[(\dprod{\vectLam}{\vectOme})-(\dprod{\unitkap}{\vectOme})(\dprod{\vectLam}{\unitkap})]\\
  &\quad-\vekn[(\dprod{\vectLam}{\vectr})-(\dprod{\unitkap}{\vectr})(\dprod{\vectLam}{\unitkap})]
  +\cdkt\parve[\Lamrep^2-(\dprod{\unitkap}{\vectLam})^2]\\
  &\quad+\vekp[(\dprod{\vectLam}{\unitplz})-(\dprod{\unitkap}{\unitplz})(\dprod{\vectLam}{\unitkap})]
  +\parve^2[(\dprod{\unitkap}{\vectOme})\Lamrep^2-(\dprod{\unitkap}{\vectLam})(\dprod{\vectLam}{\vectOme})]\\
  &\quad+\vekq[(\dprod{\unitkap}{\unitplz})(\dprod{\vectLam}{\vectOme})-(\dprod{\unitkap}{\vectOme})(\dprod{\vectLam}{\unitplz})]
  +\veko[(\dprod{\unitkap}{\vectOme})(\dprod{\vectLam}{\vectr})-(\dprod{\unitkap}{\vectr})(\dprod{\vectLam}{\vectOme})]\\
  &\quad-\rhorep\Omerep^2\parve[(\dprod{\unitkap}{\vectr})\Lamrep^2-(\dprod{\unitkap}{\vectLam})(\dprod{\vectLam}{\vectr})]
  +\vekr[(\dprod{\unitkap}{\unitplz})(\dprod{\vectLam}{\vectr})-(\dprod{\unitkap}{\vectr})(\dprod{\vectLam}{\unitplz})]\\
  &\quad-\vphij\parve[(\dprod{\unitkap}{\unitplz})\Lamrep^2-(\dprod{\unitkap}{\vectLam})(\dprod{\vectLam}{\unitplz})]
  \beqref{rot1a}, \eqnref{rxpeed1a}, \eqnref{rpath1b}, \eqnref{rpath1b2}\text{ \& }\eqnref{alg2}
\end{split}
\nonumber\\
\begin{split}
&=-\vekh\dltvb+\veki\epsvn+\vekj\frkte+\vekl\frktu
  +\vekk(\epsvg-\epsva\dltva)-\vekn(\epsvh-\epsvd\dltva)+\cdkt\parve(\Lamrep^2-\dltva^2)\\
  &\quad+\vekp(\epsvi-\epsve\dltva)+\parve^2(\epsva\Lamrep^2-\dltva\epsvg)
  +\vekq(\epsve\epsvg-\epsva\epsvi)+\veko(\epsva\epsvh-\epsvd\epsvg)\\
  &\quad-\rhorep\Omerep^2\parve(\epsvd\Lamrep^2-\dltva\epsvh)+\vekr(\epsve\epsvh-\epsvd\epsvi)
  -\vphij\parve(\epsve\Lamrep^2-\dltva\epsvi)
  \beqref{rot1a}\text{ \& }\eqnref{rxpeed1a}
\end{split}
\nonumber\\
&=\veky\beqref{rray1f}
\end{align}
\begin{align*}
\begin{split}
&\dprod{(\cprod{\unitkap}{\unitplz})}{(\veust+\veusu)}\\
&=\dprod{(\cprod{\unitkap}{\unitplz})}{[}\vekh\unitplz+\veki\vectr+\vekj\vectOme+\vekm\vectLam
  +\vekl\fdot{\vectLam}+\vekk(\cprod{\unitkap}{\vectOme})-\vekn(\cprod{\unitkap}{\vectr})\\
  &\quad+\cdkt\parve(\cprod{\unitkap}{\vectLam})+\vekp(\cprod{\unitkap}{\unitplz})
  +\parve^2(\cprod{\vectOme}{\vectLam})+\vekq(\cprod{\unitplz}{\vectOme})+\veko(\cprod{\vectOme}{\vectr})
  -\rhorep\Omerep^2\parve(\cprod{\vectr}{\vectLam})\\
  &\quad+\vekr(\cprod{\unitplz}{\vectr})-\vphij\parve(\cprod{\unitplz}{\vectLam})]\beqref{rray4}
\end{split}
\nonumber\\
\begin{split}
&=\vekh[\dprod{\unitplz}{(\cprod{\unitkap}{\unitplz})}]
  +\veki[\dprod{\vectr}{(\cprod{\unitkap}{\unitplz})}]
  +\vekj[\dprod{\vectOme}{(\cprod{\unitkap}{\unitplz})}]
  +\vekm[\dprod{\vectLam}{(\cprod{\unitkap}{\unitplz})}]\\
  &\quad+\vekl[\dprod{\fdot{\vectLam}}{(\cprod{\unitkap}{\unitplz})}]
  +\vekk[\dprod{(\cprod{\unitkap}{\unitplz})}{(\cprod{\unitkap}{\vectOme})}]
  -\vekn[\dprod{(\cprod{\unitkap}{\unitplz})}{(\cprod{\unitkap}{\vectr})}]\\
  &\quad+\cdkt\parve[\dprod{(\cprod{\unitkap}{\unitplz})}{(\cprod{\unitkap}{\vectLam})}]
  +\vekp[\dprod{(\cprod{\unitkap}{\unitplz})}{(\cprod{\unitkap}{\unitplz})}]
  +\parve^2[\dprod{(\cprod{\unitkap}{\unitplz})}{(\cprod{\vectOme}{\vectLam})}]\\
  &\quad+\vekq[\dprod{(\cprod{\unitkap}{\unitplz})}{(\cprod{\unitplz}{\vectOme})}]
  +\veko[\dprod{(\cprod{\unitkap}{\unitplz})}{(\cprod{\vectOme}{\vectr})}]
  -\rhorep\Omerep^2\parve[\dprod{(\cprod{\unitkap}{\unitplz})}{(\cprod{\vectr}{\vectLam})}]\\
  &\quad+\vekr[\dprod{(\cprod{\unitkap}{\unitplz})}{(\cprod{\unitplz}{\vectr})}]
  -\vphij\parve[\dprod{(\cprod{\unitkap}{\unitplz})}{(\cprod{\unitplz}{\vectLam})}]
\end{split}
\end{align*}
\begin{align}\label{rray5i}
\begin{split}
&=-\veki\frktt+\vekj\epsvj+\vekm\dltvb+\vekl\frkta
  +\vekk[(\dprod{\unitplz}{\vectOme})-(\dprod{\unitkap}{\vectOme})(\dprod{\unitplz}{\unitkap})]\\
  &\quad-\vekn[(\dprod{\unitplz}{\vectr})-(\dprod{\unitkap}{\vectr})(\dprod{\unitplz}{\unitkap})]
  +\cdkt\parve[(\dprod{\unitplz}{\vectLam})-(\dprod{\unitkap}{\vectLam})(\dprod{\unitplz}{\unitkap})]
  +\vekp[1-(\dprod{\unitkap}{\unitplz})^2]\\
  &\quad+\parve^2[(\dprod{\unitkap}{\vectOme})(\dprod{\unitplz}{\vectLam})-(\dprod{\unitkap}{\vectLam})(\dprod{\unitplz}{\vectOme})]
  +\vekq[(\dprod{\unitkap}{\unitplz})(\dprod{\unitplz}{\vectOme})-(\dprod{\unitkap}{\vectOme})]\\
  &\quad+\veko[(\dprod{\unitkap}{\vectOme})(\dprod{\unitplz}{\vectr})-(\dprod{\unitkap}{\vectr})(\dprod{\unitplz}{\vectOme})]
  -\rhorep\Omerep^2\parve[(\dprod{\unitkap}{\vectr})(\dprod{\unitplz}{\vectLam})-(\dprod{\unitkap}{\vectLam})(\dprod{\unitplz}{\vectr})]\\
  &\quad+\vekr[(\dprod{\unitkap}{\unitplz})(\dprod{\unitplz}{\vectr})-(\dprod{\unitkap}{\vectr})]
  -\vphij\parve[(\dprod{\unitkap}{\unitplz})(\dprod{\unitplz}{\vectLam})-(\dprod{\unitkap}{\vectLam})]\\
  &\quad\beqref{rot1a}, \eqnref{rxpeed1a}, \eqnref{rpath1b}\text{ \& }\eqnref{alg2}
\end{split}
\nonumber\\
\begin{split}
&=-\veki\frktt+\vekj\epsvj+\vekm\dltvb+\vekl\frkta
  +\vekk(\epsvc-\epsva\epsve)-\vekn(\epsvf-\epsvd\epsve)+\cdkt\parve(\epsvi-\dltva\epsve)\\
  &\quad+\vekp(1-\epsve^2)+\parve^2(\epsva\epsvi-\dltva\epsvc)
  +\vekq(\epsve\epsvc-\epsva)+\veko(\epsva\epsvf-\epsvd\epsvc)\\
  &\quad-\rhorep\Omerep^2\parve(\epsvd\epsvi-\dltva\epsvf)+\vekr(\epsve\epsvf-\epsvd)
  -\vphij\parve(\epsve\epsvi-\dltva)
  \beqref{rot1a}\text{ \& }\eqnref{rxpeed1a}
\end{split}
\nonumber\\
&=\vekz\beqref{rray1f}
\end{align}
\begin{align*}
\begin{split}
&\dprod{(\cprod{\vectOme}{\vectLam})}{(\veust+\veusu)}\\
&=\dprod{(\cprod{\vectOme}{\vectLam})}{[}\vekh\unitplz+\veki\vectr+\vekj\vectOme+\vekm\vectLam
  +\vekl\fdot{\vectLam}+\vekk(\cprod{\unitkap}{\vectOme})-\vekn(\cprod{\unitkap}{\vectr})\\
  &\quad+\cdkt\parve(\cprod{\unitkap}{\vectLam})+\vekp(\cprod{\unitkap}{\unitplz})
  +\parve^2(\cprod{\vectOme}{\vectLam})+\vekq(\cprod{\unitplz}{\vectOme})+\veko(\cprod{\vectOme}{\vectr})
  -\rhorep\Omerep^2\parve(\cprod{\vectr}{\vectLam})\\
  &\quad+\vekr(\cprod{\unitplz}{\vectr})-\vphij\parve(\cprod{\unitplz}{\vectLam})]\beqref{rray4}
\end{split}
\nonumber\\
\begin{split}
&=\vekh[\dprod{\unitplz}{(\cprod{\vectOme}{\vectLam})}]
  +\veki[\dprod{\vectr}{(\cprod{\vectOme}{\vectLam})}]
  +\vekj[\dprod{\vectOme}{(\cprod{\vectOme}{\vectLam})}]
  +\vekm[\dprod{\vectLam}{(\cprod{\vectOme}{\vectLam})}]\\
  &\quad+\vekl[\dprod{\fdot{\vectLam}}{(\cprod{\vectOme}{\vectLam})}]
  +\vekk[\dprod{(\cprod{\vectOme}{\vectLam})}{(\cprod{\unitkap}{\vectOme})}]
  -\vekn[\dprod{(\cprod{\vectOme}{\vectLam})}{(\cprod{\unitkap}{\vectr})}]\\
  &\quad+\cdkt\parve[\dprod{(\cprod{\vectOme}{\vectLam})}{(\cprod{\unitkap}{\vectLam})}]
  +\vekp[\dprod{(\cprod{\vectOme}{\vectLam})}{(\cprod{\unitkap}{\unitplz})}]
  +\parve^2[\dprod{(\cprod{\vectOme}{\vectLam})}{(\cprod{\vectOme}{\vectLam})}]\\
  &\quad+\vekq[\dprod{(\cprod{\vectOme}{\vectLam})}{(\cprod{\unitplz}{\vectOme})}]
  +\veko[\dprod{(\cprod{\vectOme}{\vectLam})}{(\cprod{\vectOme}{\vectr})}]
  -\rhorep\Omerep^2\parve[\dprod{(\cprod{\vectOme}{\vectLam})}{(\cprod{\vectr}{\vectLam})}]\\
  &\quad+\vekr[\dprod{(\cprod{\vectOme}{\vectLam})}{(\cprod{\unitplz}{\vectr})}]
  -\vphij\parve[\dprod{(\cprod{\vectOme}{\vectLam})}{(\cprod{\unitplz}{\vectLam})}]
\end{split}
\end{align*}
\begin{align}\label{rray5j}
\begin{split}
&=\vekh\frkto+\veki\epsvm+\vekl\frktv
  +\vekk[(\dprod{\vectOme}{\unitkap})(\dprod{\vectLam}{\vectOme})-\Omerep^2(\dprod{\vectLam}{\unitkap})]\\
  &\quad-\vekn[(\dprod{\vectOme}{\unitkap})(\dprod{\vectLam}{\vectr})-(\dprod{\vectOme}{\vectr})(\dprod{\vectLam}{\unitkap})]
  +\cdkt\parve[(\dprod{\vectOme}{\unitkap})\Lamrep^2-(\dprod{\vectOme}{\vectLam})(\dprod{\vectLam}{\unitkap})]\\
  &\quad+\vekp[(\dprod{\vectOme}{\unitkap})(\dprod{\vectLam}{\unitplz})-(\dprod{\vectOme}{\unitplz})(\dprod{\vectLam}{\unitkap})]
  +\parve^2[\Omerep^2\Lamrep^2-(\dprod{\vectOme}{\vectLam})^2]\\
  &\quad+\vekq[(\dprod{\vectOme}{\unitplz})(\dprod{\vectLam}{\vectOme})-\Omerep^2(\dprod{\vectLam}{\unitplz})]
  +\veko[\Omerep^2(\dprod{\vectLam}{\vectr})-(\dprod{\vectOme}{\vectr})(\dprod{\vectLam}{\vectOme})]\\
  &\quad-\rhorep\Omerep^2\parve[(\dprod{\vectOme}{\vectr})\Lamrep^2-(\dprod{\vectOme}{\vectLam})(\dprod{\vectLam}{\vectr})]
  +\vekr[(\dprod{\vectOme}{\unitplz})(\dprod{\vectLam}{\vectr})-(\dprod{\vectOme}{\vectr})(\dprod{\vectLam}{\unitplz})]\\
  &\quad-\vphij\parve[(\dprod{\vectOme}{\unitplz})\Lamrep^2-(\dprod{\vectOme}{\vectLam})(\dprod{\vectLam}{\unitplz})]
  \beqref{rot1a}, \eqnref{rpath1b}, \eqnref{rpath1b2}\text{ \& }\eqnref{alg2}
\end{split}
\nonumber\\
\begin{split}
&=\vekh\frkto+\veki\epsvm+\vekl\frktv
  +\vekk(\epsva\epsvg-\Omerep^2\dltva)-\vekn(\epsva\epsvh-\epsvb\dltva)+\cdkt\parve(\epsva\Lamrep^2-\epsvg\dltva)\\
  &\quad+\vekp(\epsva\epsvi-\epsvc\dltva)+\parve^2(\Omerep^2\Lamrep^2-\epsvg^2)
  +\vekq(\epsvc\epsvg-\Omerep^2\epsvi)+\veko(\Omerep^2\epsvh-\epsvb\epsvg)\\
  &\quad-\rhorep\Omerep^2\parve(\epsvb\Lamrep^2-\epsvg\epsvh)+\vekr(\epsvc\epsvh-\epsvb\epsvi)
  -\vphij\parve(\epsvc\Lamrep^2-\epsvg\epsvi)
  \beqref{rot1a}\text{ \& }\eqnref{rxpeed1a}
\end{split}
\nonumber\\
&=\vpsa\beqref{rray1g}
\end{align}
\begin{align*}
\begin{split}
&\dprod{(\cprod{\unitplz}{\vectOme})}{(\veust+\veusu)}\\
&=\dprod{(\cprod{\unitplz}{\vectOme})}{[}\vekh\unitplz+\veki\vectr+\vekj\vectOme+\vekm\vectLam
  +\vekl\fdot{\vectLam}+\vekk(\cprod{\unitkap}{\vectOme})-\vekn(\cprod{\unitkap}{\vectr})\\
  &\quad+\cdkt\parve(\cprod{\unitkap}{\vectLam})+\vekp(\cprod{\unitkap}{\unitplz})
  +\parve^2(\cprod{\vectOme}{\vectLam})+\vekq(\cprod{\unitplz}{\vectOme})+\veko(\cprod{\vectOme}{\vectr})
  -\rhorep\Omerep^2\parve(\cprod{\vectr}{\vectLam})\\
  &\quad+\vekr(\cprod{\unitplz}{\vectr})-\vphij\parve(\cprod{\unitplz}{\vectLam})]\beqref{rray4}
\end{split}
\nonumber\\
\begin{split}
&=\vekh[\dprod{\unitplz}{(\cprod{\unitplz}{\vectOme})}]
  +\veki[\dprod{\vectr}{(\cprod{\unitplz}{\vectOme})}]
  +\vekj[\dprod{\vectOme}{(\cprod{\unitplz}{\vectOme})}]
  +\vekm[\dprod{\vectLam}{(\cprod{\unitplz}{\vectOme})}]\\
  &\quad+\vekl[\dprod{\fdot{\vectLam}}{(\cprod{\unitplz}{\vectOme})}]
  +\vekk[\dprod{(\cprod{\unitplz}{\vectOme})}{(\cprod{\unitkap}{\vectOme})}]
  -\vekn[\dprod{(\cprod{\unitplz}{\vectOme})}{(\cprod{\unitkap}{\vectr})}]\\
  &\quad+\cdkt\parve[\dprod{(\cprod{\unitplz}{\vectOme})}{(\cprod{\unitkap}{\vectLam})}]
  +\vekp[\dprod{(\cprod{\unitplz}{\vectOme})}{(\cprod{\unitkap}{\unitplz})}]
  +\parve^2[\dprod{(\cprod{\unitplz}{\vectOme})}{(\cprod{\vectOme}{\vectLam})}]\\
  &\quad+\vekq[\dprod{(\cprod{\unitplz}{\vectOme})}{(\cprod{\unitplz}{\vectOme})}]
  +\veko[\dprod{(\cprod{\unitplz}{\vectOme})}{(\cprod{\vectOme}{\vectr})}]
  -\rhorep\Omerep^2\parve[\dprod{(\cprod{\unitplz}{\vectOme})}{(\cprod{\vectr}{\vectLam})}]\\
  &\quad+\vekr[\dprod{(\cprod{\unitplz}{\vectOme})}{(\cprod{\unitplz}{\vectr})}]
  -\vphij\parve[\dprod{(\cprod{\unitplz}{\vectOme})}{(\cprod{\unitplz}{\vectLam})}]
\end{split}
\end{align*}
\begin{align}\label{rray5k}
\begin{split}
&=\veki\epsvl+\vekm\frkto+\vekl\frktp
  +\vekk[(\dprod{\unitplz}{\unitkap})\Omerep^2-(\dprod{\unitplz}{\vectOme})(\dprod{\vectOme}{\unitkap})]\\
  &\quad-\vekn[(\dprod{\unitplz}{\unitkap})(\dprod{\vectOme}{\vectr})-(\dprod{\unitplz}{\vectr})(\dprod{\vectOme}{\unitkap})]
  +\cdkt\parve[(\dprod{\unitplz}{\unitkap})(\dprod{\vectOme}{\vectLam})-(\dprod{\unitplz}{\vectLam})(\dprod{\vectOme}{\unitkap})]\\
  &\quad+\vekp[(\dprod{\unitplz}{\unitkap})(\dprod{\vectOme}{\unitplz})-(\dprod{\vectOme}{\unitkap})]
  +\parve^2[(\dprod{\unitplz}{\vectOme})(\dprod{\vectOme}{\vectLam})-(\dprod{\unitplz}{\vectLam})\Omerep^2]
  +\vekq[\Omerep^2-(\dprod{\unitplz}{\vectOme})^2]\\
  &\quad+\veko[(\dprod{\unitplz}{\vectOme})(\dprod{\vectOme}{\vectr})-(\dprod{\unitplz}{\vectr})\Omerep^2]
  -\rhorep\Omerep^2\parve[(\dprod{\unitplz}{\vectr})(\dprod{\vectOme}{\vectLam})-(\dprod{\unitplz}{\vectLam})(\dprod{\vectOme}{\vectr})]\\
  &\quad+\vekr[(\dprod{\vectOme}{\vectr})-(\dprod{\unitplz}{\vectr})(\dprod{\vectOme}{\unitplz})]
  -\vphij\parve[(\dprod{\vectOme}{\vectLam})-(\dprod{\unitplz}{\vectLam})(\dprod{\vectOme}{\unitplz})]\\
  &\quad\beqref{rot1a}, \eqnref{rpath1b}\text{ \& }\eqnref{alg2}
\end{split}
\nonumber\\
\begin{split}
&=\veki\epsvl+\vekm\frkto+\vekl\frktp
  +\vekk(\epsve\Omerep^2-\epsvc\epsva)-\vekn(\epsve\epsvb-\epsvf\epsva)+\cdkt\parve(\epsve\epsvg-\epsvi\epsva)\\
  &\quad+\vekp(\epsve\epsvc-\epsva)+\parve^2(\epsvc\epsvg-\epsvi\Omerep^2)
  +\vekq(\Omerep^2-\epsvc^2)+\veko(\epsvc\epsvb-\epsvf\Omerep^2)\\
  &\quad-\rhorep\Omerep^2\parve(\epsvf\epsvg-\epsvi\epsvb)
  +\vekr(\epsvb-\epsvf\epsvc)-\vphij\parve(\epsvg-\epsvi\epsvc)
  \beqref{rot1a}
\end{split}
\nonumber\\
&=\vpsb\beqref{rray1g}
\end{align}
\begin{align*}
\begin{split}
&\dprod{(\cprod{\vectOme}{\vectr})}{(\veust+\veusu)}\\
&=\dprod{(\cprod{\vectOme}{\vectr})}{[}\vekh\unitplz+\veki\vectr+\vekj\vectOme+\vekm\vectLam
  +\vekl\fdot{\vectLam}+\vekk(\cprod{\unitkap}{\vectOme})-\vekn(\cprod{\unitkap}{\vectr})\\
  &\quad+\cdkt\parve(\cprod{\unitkap}{\vectLam})+\vekp(\cprod{\unitkap}{\unitplz})
  +\parve^2(\cprod{\vectOme}{\vectLam})+\vekq(\cprod{\unitplz}{\vectOme})+\veko(\cprod{\vectOme}{\vectr})
  -\rhorep\Omerep^2\parve(\cprod{\vectr}{\vectLam})\\
  &\quad+\vekr(\cprod{\unitplz}{\vectr})-\vphij\parve(\cprod{\unitplz}{\vectLam})]\beqref{rray4}
\end{split}
\nonumber\\
\begin{split}
&=\vekh[\dprod{\unitplz}{(\cprod{\vectOme}{\vectr})}]
  +\veki[\dprod{\vectr}{(\cprod{\vectOme}{\vectr})}]
  +\vekj[\dprod{\vectOme}{(\cprod{\vectOme}{\vectr})}]
  +\vekm[\dprod{\vectLam}{(\cprod{\vectOme}{\vectr})}]\\
  &\quad+\vekl[\dprod{\fdot{\vectLam}}{(\cprod{\vectOme}{\vectr})}]
  +\vekk[\dprod{(\cprod{\vectOme}{\vectr})}{(\cprod{\unitkap}{\vectOme})}]
  -\vekn[\dprod{(\cprod{\vectOme}{\vectr})}{(\cprod{\unitkap}{\vectr})}]\\
  &\quad+\cdkt\parve[\dprod{(\cprod{\vectOme}{\vectr})}{(\cprod{\unitkap}{\vectLam})}]
  +\vekp[\dprod{(\cprod{\vectOme}{\vectr})}{(\cprod{\unitkap}{\unitplz})}]
  +\parve^2[\dprod{(\cprod{\vectOme}{\vectr})}{(\cprod{\vectOme}{\vectLam})}]\\
  &\quad+\vekq[\dprod{(\cprod{\vectOme}{\vectr})}{(\cprod{\unitplz}{\vectOme})}]
  +\veko[\dprod{(\cprod{\vectOme}{\vectr})}{(\cprod{\vectOme}{\vectr})}]
  -\rhorep\Omerep^2\parve[\dprod{(\cprod{\vectOme}{\vectr})}{(\cprod{\vectr}{\vectLam})}]\\
  &\quad+\vekr[\dprod{(\cprod{\vectOme}{\vectr})}{(\cprod{\unitplz}{\vectr})}]
  -\vphij\parve[\dprod{(\cprod{\vectOme}{\vectr})}{(\cprod{\unitplz}{\vectLam})}]
\end{split}
\end{align*}
\begin{align}\label{rray5l}
\begin{split}
&=\vekh\epsvl-\vekm\epsvm+\vekl\frktl
  +\vekk[(\dprod{\vectOme}{\unitkap})(\dprod{\vectr}{\vectOme})-\Omerep^2(\dprod{\vectr}{\unitkap})]\\
  &\quad-\vekn[(\dprod{\vectOme}{\unitkap})\scalr^2-(\dprod{\vectOme}{\vectr})(\dprod{\vectr}{\unitkap})]
  +\cdkt\parve[(\dprod{\vectOme}{\unitkap})(\dprod{\vectr}{\vectLam})-(\dprod{\vectOme}{\vectLam})(\dprod{\vectr}{\unitkap})]\\
  &\quad+\vekp[(\dprod{\vectOme}{\unitkap})(\dprod{\vectr}{\unitplz})-(\dprod{\vectOme}{\unitplz})(\dprod{\vectr}{\unitkap})]
  +\parve^2[\Omerep^2(\dprod{\vectr}{\vectLam})-(\dprod{\vectOme}{\vectLam})(\dprod{\vectr}{\vectOme})]\\
  &\quad+\vekq[(\dprod{\vectOme}{\unitplz})(\dprod{\vectr}{\vectOme})-\Omerep^2(\dprod{\vectr}{\unitplz})]
  +\veko[\Omerep^2\scalr^2-(\dprod{\vectOme}{\vectr})^2]\\
  &\quad-\rhorep\Omerep^2\parve[(\dprod{\vectOme}{\vectr})(\dprod{\vectr}{\vectLam})-(\dprod{\vectOme}{\vectLam})\scalr^2]
  +\vekr[(\dprod{\vectOme}{\unitplz})\scalr^2-(\dprod{\vectOme}{\vectr})(\dprod{\vectr}{\unitplz})]\\
  &\quad-\vphij\parve[(\dprod{\vectOme}{\unitplz})(\dprod{\vectr}{\vectLam})-(\dprod{\vectOme}{\vectLam})(\dprod{\vectr}{\unitplz})]
  \beqref{rot1a}, \eqnref{rpath1b}\text{ \& }\eqnref{alg2}
\end{split}
\nonumber\\
\begin{split}
&=\vekh\epsvl-\vekm\epsvm+\vekl\frktl
  +\vekk(\epsva\epsvb-\Omerep^2\epsvd)-\vekn(\epsva\scalr^2-\epsvb\epsvd)+\cdkt\parve(\epsva\epsvh-\epsvg\epsvd)\\
  &\quad+\vekp(\epsva\epsvf-\epsvc\epsvd)+\parve^2(\Omerep^2\epsvh-\epsvg\epsvb)
  +\vekq(\epsvc\epsvb-\Omerep^2\epsvf)+\veko(\Omerep^2\scalr^2-\epsvb^2)\\
  &\quad-\rhorep\Omerep^2\parve(\epsvb\epsvh-\epsvg\scalr^2)+\vekr(\epsvc\scalr^2-\epsvb\epsvf)
  -\vphij\parve(\epsvc\epsvh-\epsvg\epsvf)
  \beqref{rot1a}
\end{split}
\nonumber\\
&=\vpsc\beqref{rray1h}
\end{align}
\begin{align*}
\begin{split}
&\dprod{(\cprod{\vectr}{\vectLam})}{(\veust+\veusu)}\\
&=\dprod{(\cprod{\vectr}{\vectLam})}{[}\vekh\unitplz+\veki\vectr+\vekj\vectOme+\vekm\vectLam
  +\vekl\fdot{\vectLam}+\vekk(\cprod{\unitkap}{\vectOme})-\vekn(\cprod{\unitkap}{\vectr})\\
  &\quad+\cdkt\parve(\cprod{\unitkap}{\vectLam})+\vekp(\cprod{\unitkap}{\unitplz})
  +\parve^2(\cprod{\vectOme}{\vectLam})+\vekq(\cprod{\unitplz}{\vectOme})+\veko(\cprod{\vectOme}{\vectr})
  -\rhorep\Omerep^2\parve(\cprod{\vectr}{\vectLam})\\
  &\quad+\vekr(\cprod{\unitplz}{\vectr})-\vphij\parve(\cprod{\unitplz}{\vectLam})]\beqref{rray4}
\end{split}
\nonumber\\
\begin{split}
&=\vekh[\dprod{\unitplz}{(\cprod{\vectr}{\vectLam})}]
  +\veki[\dprod{\vectr}{(\cprod{\vectr}{\vectLam})}]
  +\vekj[\dprod{\vectOme}{(\cprod{\vectr}{\vectLam})}]
  +\vekm[\dprod{\vectLam}{(\cprod{\vectr}{\vectLam})}]\\
  &\quad+\vekl[\dprod{\fdot{\vectLam}}{(\cprod{\vectr}{\vectLam})}]
  +\vekk[\dprod{(\cprod{\vectr}{\vectLam})}{(\cprod{\unitkap}{\vectOme})}]
  -\vekn[\dprod{(\cprod{\vectr}{\vectLam})}{(\cprod{\unitkap}{\vectr})}]\\
  &\quad+\cdkt\parve[\dprod{(\cprod{\vectr}{\vectLam})}{(\cprod{\unitkap}{\vectLam})}]
  +\vekp[\dprod{(\cprod{\vectr}{\vectLam})}{(\cprod{\unitkap}{\unitplz})}]
  +\parve^2[\dprod{(\cprod{\vectr}{\vectLam})}{(\cprod{\vectOme}{\vectLam})}]\\
  &\quad+\vekq[\dprod{(\cprod{\vectr}{\vectLam})}{(\cprod{\unitplz}{\vectOme})}]
  +\veko[\dprod{(\cprod{\vectr}{\vectLam})}{(\cprod{\vectOme}{\vectr})}]
  -\rhorep\Omerep^2\parve[\dprod{(\cprod{\vectr}{\vectLam})}{(\cprod{\vectr}{\vectLam})}]\\
  &\quad+\vekr[\dprod{(\cprod{\vectr}{\vectLam})}{(\cprod{\unitplz}{\vectr})}]
  -\vphij\parve[\dprod{(\cprod{\vectr}{\vectLam})}{(\cprod{\unitplz}{\vectLam})}]
\end{split}
\end{align*}
\begin{align}\label{rray5m}
\begin{split}
&=-\vekh\epsvo-\vekj\epsvm-\vekl\frkth
  +\vekk[(\dprod{\vectr}{\unitkap})(\dprod{\vectLam}{\vectOme})-(\dprod{\vectr}{\vectOme})(\dprod{\vectLam}{\unitkap})]\\
  &\quad-\vekn[(\dprod{\vectr}{\unitkap})(\dprod{\vectLam}{\vectr})-\scalr^2(\dprod{\vectLam}{\unitkap})]
  +\cdkt\parve[(\dprod{\vectr}{\unitkap})\Lamrep^2-(\dprod{\vectr}{\vectLam})(\dprod{\vectLam}{\unitkap})]\\
  &\quad+\vekp[(\dprod{\vectr}{\unitkap})(\dprod{\vectLam}{\unitplz})-(\dprod{\vectr}{\unitplz})(\dprod{\vectLam}{\unitkap})]
  +\parve^2[(\dprod{\vectr}{\vectOme})\Lamrep^2-(\dprod{\vectr}{\vectLam})(\dprod{\vectLam}{\vectOme})]\\
  &\quad+\vekq[(\dprod{\vectr}{\unitplz})(\dprod{\vectLam}{\vectOme})-(\dprod{\vectr}{\vectOme})(\dprod{\vectLam}{\unitplz})]
  +\veko[(\dprod{\vectr}{\vectOme})(\dprod{\vectLam}{\vectr})-\scalr^2(\dprod{\vectLam}{\vectOme})]\\
  &\quad-\rhorep\Omerep^2\parve[\scalr^2\Lamrep^2-(\dprod{\vectr}{\vectLam})^2]
  +\vekr[(\dprod{\vectr}{\unitplz})(\dprod{\vectLam}{\vectr})-\scalr^2(\dprod{\vectLam}{\unitplz})]\\
  &\quad-\vphij\parve[(\dprod{\vectr}{\unitplz})\Lamrep^2-(\dprod{\vectr}{\vectLam})(\dprod{\vectLam}{\unitplz})]
  \beqref{rot1a}, \eqnref{rpath1b}\text{ \& }\eqnref{alg2}
\end{split}
\nonumber\\
\begin{split}
&=-\vekh\epsvo-\vekj\epsvm-\vekl\frkth
  +\vekk(\epsvd\epsvg-\epsvb\dltva)-\vekn(\epsvd\epsvh-\scalr^2\dltva)+\cdkt\parve(\epsvd\Lamrep^2-\epsvh\dltva)\\
  &\quad+\vekp(\epsvd\epsvi-\epsvf\dltva)+\parve^2(\epsvb\Lamrep^2-\epsvh\epsvg)
  +\vekq(\epsvf\epsvg-\epsvb\epsvi)+\veko(\epsvb\epsvh-\scalr^2\epsvg)\\
  &\quad-\rhorep\Omerep^2\parve(\scalr^2\Lamrep^2-\epsvh^2)+\vekr(\epsvf\epsvh-\scalr^2\epsvi)
  -\vphij\parve(\epsvf\Lamrep^2-\epsvh\epsvi)
  \beqref{rot1a}\text{ \& }\eqnref{rxpeed1a}
\end{split}
\nonumber\\
&=\vpsd\beqref{rray1h}
\end{align}
\begin{align*}
\begin{split}
&\dprod{(\cprod{\unitplz}{\vectr})}{(\veust+\veusu)}\\
&=\dprod{(\cprod{\unitplz}{\vectr})}{[}\vekh\unitplz+\veki\vectr+\vekj\vectOme+\vekm\vectLam
  +\vekl\fdot{\vectLam}+\vekk(\cprod{\unitkap}{\vectOme})-\vekn(\cprod{\unitkap}{\vectr})\\
  &\quad+\cdkt\parve(\cprod{\unitkap}{\vectLam})+\vekp(\cprod{\unitkap}{\unitplz})
  +\parve^2(\cprod{\vectOme}{\vectLam})+\vekq(\cprod{\unitplz}{\vectOme})+\veko(\cprod{\vectOme}{\vectr})
  -\rhorep\Omerep^2\parve(\cprod{\vectr}{\vectLam})\\
  &\quad+\vekr(\cprod{\unitplz}{\vectr})-\vphij\parve(\cprod{\unitplz}{\vectLam})]\beqref{rray4}
\end{split}
\nonumber\\
\begin{split}
&=\vekh[\dprod{\unitplz}{(\cprod{\unitplz}{\vectr})}]
  +\veki[\dprod{\vectr}{(\cprod{\unitplz}{\vectr})}]
  +\vekj[\dprod{\vectOme}{(\cprod{\unitplz}{\vectr})}]
  +\vekm[\dprod{\vectLam}{(\cprod{\unitplz}{\vectr})}]\\
  &\quad+\vekl[\dprod{\fdot{\vectLam}}{(\cprod{\unitplz}{\vectr})}]
  +\vekk[\dprod{(\cprod{\unitplz}{\vectr})}{(\cprod{\unitkap}{\vectOme})}]
  -\vekn[\dprod{(\cprod{\unitplz}{\vectr})}{(\cprod{\unitkap}{\vectr})}]\\
  &\quad+\cdkt\parve[\dprod{(\cprod{\unitplz}{\vectr})}{(\cprod{\unitkap}{\vectLam})}]
  +\vekp[\dprod{(\cprod{\unitplz}{\vectr})}{(\cprod{\unitkap}{\unitplz})}]
  +\parve^2[\dprod{(\cprod{\unitplz}{\vectr})}{(\cprod{\vectOme}{\vectLam})}]\\
  &\quad+\vekq[\dprod{(\cprod{\unitplz}{\vectr})}{(\cprod{\unitplz}{\vectOme})}]
  +\veko[\dprod{(\cprod{\unitplz}{\vectr})}{(\cprod{\vectOme}{\vectr})}]
  -\rhorep\Omerep^2\parve[\dprod{(\cprod{\unitplz}{\vectr})}{(\cprod{\vectr}{\vectLam})}]\\
  &\quad+\vekr[\dprod{(\cprod{\unitplz}{\vectr})}{(\cprod{\unitplz}{\vectr})}]
  -\vphij\parve[\dprod{(\cprod{\unitplz}{\vectr})}{(\cprod{\unitplz}{\vectLam})}]
\end{split}
\end{align*}
\begin{align}\label{rray5n}
\begin{split}
&=-\vekj\epsvl-\vekm\epsvo-\vekl\frktn
  +\vekk[(\dprod{\unitplz}{\unitkap})(\dprod{\vectr}{\vectOme})-(\dprod{\unitplz}{\vectOme})(\dprod{\vectr}{\unitkap})]\\
  &\quad-\vekn[(\dprod{\unitplz}{\unitkap})\scalr^2-(\dprod{\unitplz}{\vectr})(\dprod{\vectr}{\unitkap})]
  +\cdkt\parve[(\dprod{\unitplz}{\unitkap})(\dprod{\vectr}{\vectLam})-(\dprod{\unitplz}{\vectLam})(\dprod{\vectr}{\unitkap})]\\
  &\quad+\vekp[(\dprod{\unitplz}{\unitkap})(\dprod{\vectr}{\unitplz})-(\dprod{\vectr}{\unitkap})]
  +\parve^2[(\dprod{\unitplz}{\vectOme})(\dprod{\vectr}{\vectLam})-(\dprod{\unitplz}{\vectLam})(\dprod{\vectr}{\vectOme})]\\
  &\quad+\vekq[(\dprod{\vectr}{\vectOme})-(\dprod{\unitplz}{\vectOme})(\dprod{\vectr}{\unitplz})]
  +\veko[(\dprod{\unitplz}{\vectOme})\scalr^2-(\dprod{\unitplz}{\vectr})(\dprod{\vectr}{\vectOme})]\\
  &\quad-\rhorep\Omerep^2\parve[(\dprod{\unitplz}{\vectr})(\dprod{\vectr}{\vectLam})-(\dprod{\unitplz}{\vectLam})\scalr^2]
  +\vekr[\scalr^2-(\dprod{\unitplz}{\vectr})^2]\\
  &\quad-\vphij\parve[(\dprod{\vectr}{\vectLam})-(\dprod{\unitplz}{\vectLam})(\dprod{\vectr}{\unitplz})]
  \beqref{rot1a}, \eqnref{rpath1b}\text{ \& }\eqnref{alg2}
\end{split}
\nonumber\\
\begin{split}
&=-\vekj\epsvl-\vekm\epsvo-\vekl\frktn
  +\vekk(\epsve\epsvb-\epsvc\epsvd)-\vekn(\epsve\scalr^2-\epsvf\epsvd)+\cdkt\parve(\epsve\epsvh-\epsvi\epsvd)\\
  &\quad+\vekp(\epsve\epsvf-\epsvd)+\parve^2(\epsvc\epsvh-\epsvi\epsvb)
  +\vekq(\epsvb-\epsvc\epsvf)+\veko(\epsvc\scalr^2-\epsvf\epsvb)\\
  &\quad-\rhorep\Omerep^2\parve(\epsvf\epsvh-\epsvi\scalr^2)+\vekr(\scalr^2-\epsvf^2)
  -\vphij\parve(\epsvh-\epsvi\epsvf)
  \beqref{rot1a}
\end{split}
\nonumber\\
&=\vpse\beqref{rray1i}
\end{align}
\begin{align*}
\begin{split}
&\dprod{(\cprod{\unitplz}{\vectLam})}{(\veust+\veusu)}\\
&=\dprod{(\cprod{\unitplz}{\vectLam})}{[}\vekh\unitplz+\veki\vectr+\vekj\vectOme+\vekm\vectLam
  +\vekl\fdot{\vectLam}+\vekk(\cprod{\unitkap}{\vectOme})-\vekn(\cprod{\unitkap}{\vectr})\\
  &\quad+\cdkt\parve(\cprod{\unitkap}{\vectLam})+\vekp(\cprod{\unitkap}{\unitplz})
  +\parve^2(\cprod{\vectOme}{\vectLam})+\vekq(\cprod{\unitplz}{\vectOme})+\veko(\cprod{\vectOme}{\vectr})
  -\rhorep\Omerep^2\parve(\cprod{\vectr}{\vectLam})\\
  &\quad+\vekr(\cprod{\unitplz}{\vectr})-\vphij\parve(\cprod{\unitplz}{\vectLam})]\beqref{rray4}
\end{split}
\nonumber\\
\begin{split}
&=\vekh[\dprod{\unitplz}{(\cprod{\unitplz}{\vectLam})}]
  +\veki[\dprod{\vectr}{(\cprod{\unitplz}{\vectLam})}]
  +\vekj[\dprod{\vectOme}{(\cprod{\unitplz}{\vectLam})}]
  +\vekm[\dprod{\vectLam}{(\cprod{\unitplz}{\vectLam})}]\\
  &\quad+\vekl[\dprod{\fdot{\vectLam}}{(\cprod{\unitplz}{\vectLam})}]
  +\vekk[\dprod{(\cprod{\unitplz}{\vectLam})}{(\cprod{\unitkap}{\vectOme})}]
  -\vekn[\dprod{(\cprod{\unitplz}{\vectLam})}{(\cprod{\unitkap}{\vectr})}]\\
  &\quad+\cdkt\parve[\dprod{(\cprod{\unitplz}{\vectLam})}{(\cprod{\unitkap}{\vectLam})}]
  +\vekp[\dprod{(\cprod{\unitplz}{\vectLam})}{(\cprod{\unitkap}{\unitplz})}]
  +\parve^2[\dprod{(\cprod{\unitplz}{\vectLam})}{(\cprod{\vectOme}{\vectLam})}]\\
  &\quad+\vekq[\dprod{(\cprod{\unitplz}{\vectLam})}{(\cprod{\unitplz}{\vectOme})}]
  +\veko[\dprod{(\cprod{\unitplz}{\vectLam})}{(\cprod{\vectOme}{\vectr})}]
  -\rhorep\Omerep^2\parve[\dprod{(\cprod{\unitplz}{\vectLam})}{(\cprod{\vectr}{\vectLam})}]\\
  &\quad+\vekr[\dprod{(\cprod{\unitplz}{\vectLam})}{(\cprod{\unitplz}{\vectr})}]
  -\vphij\parve[\dprod{(\cprod{\unitplz}{\vectLam})}{(\cprod{\unitplz}{\vectLam})}]
\end{split}
\end{align*}
\begin{align}\label{rray5o}
\begin{split}
&=\veki\epsvo-\vekj\frkto+\vekl\frktr
  +\vekk[(\dprod{\unitplz}{\unitkap})(\dprod{\vectLam}{\vectOme})-(\dprod{\unitplz}{\vectOme})(\dprod{\vectLam}{\unitkap})]\\
  &\quad-\vekn[(\dprod{\unitplz}{\unitkap})(\dprod{\vectLam}{\vectr})-(\dprod{\unitplz}{\vectr})(\dprod{\vectLam}{\unitkap})]
  +\cdkt\parve[(\dprod{\unitplz}{\unitkap})\Lamrep^2-(\dprod{\unitplz}{\vectLam})(\dprod{\vectLam}{\unitkap})]\\
  &\quad+\vekp[(\dprod{\unitplz}{\unitkap})(\dprod{\vectLam}{\unitplz})-(\dprod{\vectLam}{\unitkap})]
  +\parve^2[(\dprod{\unitplz}{\vectOme})\Lamrep^2-(\dprod{\unitplz}{\vectLam})(\dprod{\vectLam}{\vectOme})]\\
  &\quad+\vekq[(\dprod{\vectLam}{\vectOme})-(\dprod{\unitplz}{\vectOme})(\dprod{\vectLam}{\unitplz})]
  +\veko[(\dprod{\unitplz}{\vectOme})(\dprod{\vectLam}{\vectr})-(\dprod{\unitplz}{\vectr})(\dprod{\vectLam}{\vectOme})]\\
  &\quad-\rhorep\Omerep^2\parve[(\dprod{\unitplz}{\vectr})\Lamrep^2-(\dprod{\unitplz}{\vectLam})(\dprod{\vectLam}{\vectr})]
  +\vekr[(\dprod{\vectLam}{\vectr})-(\dprod{\unitplz}{\vectr})(\dprod{\vectLam}{\unitplz})]\\
  &\quad-\vphij\parve[\Lamrep^2-(\dprod{\unitplz}{\vectLam})^2]
  \beqref{rot1a}, \eqnref{rpath1b}\text{ \& }\eqnref{alg2}
\end{split}
\nonumber\\
\begin{split}
&=\veki\epsvo-\vekj\frkto+\vekl\frktr+\vekk(\epsve\epsvg-\epsvc\dltva)
  -\vekn(\epsve\epsvh-\epsvf\dltva)+\cdkt\parve(\epsve\Lamrep^2-\epsvi\dltva)\\
  &\quad+\vekp(\epsve\epsvi-\dltva)+\parve^2(\epsvc\Lamrep^2-\epsvi\epsvg)
  +\vekq(\epsvg-\epsvc\epsvi)+\veko(\epsvc\epsvh-\epsvf\epsvg)\\
  &\quad-\rhorep\Omerep^2\parve(\epsvf\Lamrep^2-\epsvi\epsvh)+\vekr(\epsvh-\epsvf\epsvi)
  -\vphij\parve(\Lamrep^2-\epsvi^2)
  \beqref{rot1a}\text{ \& }\eqnref{rxpeed1a}
\end{split}
\nonumber\\
&=\vpsf\beqref{rray1i}
\end{align}
\begin{align*}
\begin{split}
&\dprod{(\cprod{\fdot{\vectLam}}{\unitplz})}{(\veust+\veusu)}\\
&=\dprod{(\cprod{\fdot{\vectLam}}{\unitplz})}{[}\vekh\unitplz+\veki\vectr+\vekj\vectOme+\vekm\vectLam
  +\vekl\fdot{\vectLam}+\vekk(\cprod{\unitkap}{\vectOme})-\vekn(\cprod{\unitkap}{\vectr})\\
  &\quad+\cdkt\parve(\cprod{\unitkap}{\vectLam})+\vekp(\cprod{\unitkap}{\unitplz})
  +\parve^2(\cprod{\vectOme}{\vectLam})+\vekq(\cprod{\unitplz}{\vectOme})+\veko(\cprod{\vectOme}{\vectr})
  -\rhorep\Omerep^2\parve(\cprod{\vectr}{\vectLam})\\
  &\quad+\vekr(\cprod{\unitplz}{\vectr})-\vphij\parve(\cprod{\unitplz}{\vectLam})]\beqref{rray4}
\end{split}
\nonumber\\
\begin{split}
&=\vekh[\dprod{\unitplz}{(\cprod{\fdot{\vectLam}}{\unitplz})}]
  +\veki[\dprod{\vectr}{(\cprod{\fdot{\vectLam}}{\unitplz})}]
  +\vekj[\dprod{\vectOme}{(\cprod{\fdot{\vectLam}}{\unitplz})}]
  +\vekm[\dprod{\vectLam}{(\cprod{\fdot{\vectLam}}{\unitplz})}]\\
  &\quad+\vekl[\dprod{\fdot{\vectLam}}{(\cprod{\fdot{\vectLam}}{\unitplz})}]
  +\vekk[\dprod{(\cprod{\fdot{\vectLam}}{\unitplz})}{(\cprod{\unitkap}{\vectOme})}]
  -\vekn[\dprod{(\cprod{\fdot{\vectLam}}{\unitplz})}{(\cprod{\unitkap}{\vectr})}]\\
  &\quad+\cdkt\parve[\dprod{(\cprod{\fdot{\vectLam}}{\unitplz})}{(\cprod{\unitkap}{\vectLam})}]
  +\vekp[\dprod{(\cprod{\fdot{\vectLam}}{\unitplz})}{(\cprod{\unitkap}{\unitplz})}]
  +\parve^2[\dprod{(\cprod{\fdot{\vectLam}}{\unitplz})}{(\cprod{\vectOme}{\vectLam})}]\\
  &\quad+\vekq[\dprod{(\cprod{\fdot{\vectLam}}{\unitplz})}{(\cprod{\unitplz}{\vectOme})}]
  +\veko[\dprod{(\cprod{\fdot{\vectLam}}{\unitplz})}{(\cprod{\vectOme}{\vectr})}]
  -\rhorep\Omerep^2\parve[\dprod{(\cprod{\fdot{\vectLam}}{\unitplz})}{(\cprod{\vectr}{\vectLam})}]\\
  &\quad+\vekr[\dprod{(\cprod{\fdot{\vectLam}}{\unitplz})}{(\cprod{\unitplz}{\vectr})}]
  -\vphij\parve[\dprod{(\cprod{\fdot{\vectLam}}{\unitplz})}{(\cprod{\unitplz}{\vectLam})}]
\end{split}
\end{align*}
\begin{align}\label{rray5p}
\begin{split}
&=-\veki\frktn+\vekj\frktp+\vekm\frktr
  +\vekk[(\dprod{\fdot{\vectLam}}{\unitkap})(\dprod{\unitplz}{\vectOme})-(\dprod{\fdot{\vectLam}}{\vectOme})(\dprod{\unitplz}{\unitkap})]\\
  &\quad-\vekn[(\dprod{\fdot{\vectLam}}{\unitkap})(\dprod{\unitplz}{\vectr})-(\dprod{\fdot{\vectLam}}{\vectr})(\dprod{\unitplz}{\unitkap})]
  +\cdkt\parve[(\dprod{\fdot{\vectLam}}{\unitkap})(\dprod{\unitplz}{\vectLam})-(\dprod{\fdot{\vectLam}}{\vectLam})(\dprod{\unitplz}{\unitkap})]\\
  &\quad+\vekp[(\dprod{\fdot{\vectLam}}{\unitkap})-(\dprod{\fdot{\vectLam}}{\unitplz})(\dprod{\unitplz}{\unitkap})]
  +\parve^2[(\dprod{\fdot{\vectLam}}{\vectOme})(\dprod{\unitplz}{\vectLam})-(\dprod{\fdot{\vectLam}}{\vectLam})(\dprod{\unitplz}{\vectOme})]\\
  &\quad+\vekq[(\dprod{\fdot{\vectLam}}{\unitplz})(\dprod{\unitplz}{\vectOme})-(\dprod{\fdot{\vectLam}}{\vectOme})]
  +\veko[(\dprod{\fdot{\vectLam}}{\vectOme})(\dprod{\unitplz}{\vectr})-(\dprod{\fdot{\vectLam}}{\vectr})(\dprod{\unitplz}{\vectOme})]\\
  &\quad-\rhorep\Omerep^2\parve[(\dprod{\fdot{\vectLam}}{\vectr})(\dprod{\unitplz}{\vectLam})-(\dprod{\fdot{\vectLam}}{\vectLam})(\dprod{\unitplz}{\vectr})]
  +\vekr[(\dprod{\fdot{\vectLam}}{\unitplz})(\dprod{\unitplz}{\vectr})-(\dprod{\fdot{\vectLam}}{\vectr})]\\
  &\quad-\vphij\parve[(\dprod{\fdot{\vectLam}}{\unitplz})(\dprod{\unitplz}{\vectLam})-(\dprod{\fdot{\vectLam}}{\vectLam})]
  \beqref{rpath1b}\text{ \& }\eqnref{alg2}
\end{split}
\nonumber\\
\begin{split}
&=-\veki\frktn+\vekj\frktp+\vekm\frktr+\vekk(\vsigb\epsvc-\vsigc\epsve)
  -\vekn(\vsigb\epsvf-\vsign\epsve)+\cdkt\parve(\vsigb\epsvi-\vsigd\epsve)\\
  &\quad+\vekp(\vsigb-\vsiga\epsve)+\parve^2(\vsigc\epsvi-\vsigd\epsvc)
  +\vekq(\vsiga\epsvc-\vsigc)+\veko(\vsigc\epsvf-\vsign\epsvc)\\
  &\quad-\rhorep\Omerep^2\parve(\vsign\epsvi-\vsigd\epsvf)
  +\vekr(\vsiga\epsvf-\vsign)-\vphij\parve(\vsiga\epsvi-\vsigd)
  \beqref{rot1a}\text{ \& }\eqnref{rpath1a}
\end{split}
\nonumber\\
&=\vpsg\beqref{rray1j}
\end{align}
\begin{align*}
\begin{split}
&\dprod{(\cprod{\fdot{\vectLam}}{\vectr})}{(\veust+\veusu)}\\
&=\dprod{(\cprod{\fdot{\vectLam}}{\vectr})}{[}\vekh\unitplz+\veki\vectr+\vekj\vectOme+\vekm\vectLam
  +\vekl\fdot{\vectLam}+\vekk(\cprod{\unitkap}{\vectOme})-\vekn(\cprod{\unitkap}{\vectr})\\
  &\quad+\cdkt\parve(\cprod{\unitkap}{\vectLam})+\vekp(\cprod{\unitkap}{\unitplz})
  +\parve^2(\cprod{\vectOme}{\vectLam})+\vekq(\cprod{\unitplz}{\vectOme})+\veko(\cprod{\vectOme}{\vectr})
  -\rhorep\Omerep^2\parve(\cprod{\vectr}{\vectLam})\\
  &\quad+\vekr(\cprod{\unitplz}{\vectr})-\vphij\parve(\cprod{\unitplz}{\vectLam})]\beqref{rray4}
\end{split}
\nonumber\\
\begin{split}
&=\vekh[\dprod{\unitplz}{(\cprod{\fdot{\vectLam}}{\vectr})}]
  +\veki[\dprod{\vectr}{(\cprod{\fdot{\vectLam}}{\vectr})}]
  +\vekj[\dprod{\vectOme}{(\cprod{\fdot{\vectLam}}{\vectr})}]
  +\vekm[\dprod{\vectLam}{(\cprod{\fdot{\vectLam}}{\vectr})}]\\
  &\quad+\vekl[\dprod{\fdot{\vectLam}}{(\cprod{\fdot{\vectLam}}{\vectr})}]
  +\vekk[\dprod{(\cprod{\fdot{\vectLam}}{\vectr})}{(\cprod{\unitkap}{\vectOme})}]
  -\vekn[\dprod{(\cprod{\fdot{\vectLam}}{\vectr})}{(\cprod{\unitkap}{\vectr})}]\\
  &\quad+\cdkt\parve[\dprod{(\cprod{\fdot{\vectLam}}{\vectr})}{(\cprod{\unitkap}{\vectLam})}]
  +\vekp[\dprod{(\cprod{\fdot{\vectLam}}{\vectr})}{(\cprod{\unitkap}{\unitplz})}]
  +\parve^2[\dprod{(\cprod{\fdot{\vectLam}}{\vectr})}{(\cprod{\vectOme}{\vectLam})}]\\
  &\quad+\vekq[\dprod{(\cprod{\fdot{\vectLam}}{\vectr})}{(\cprod{\unitplz}{\vectOme})}]
  +\veko[\dprod{(\cprod{\fdot{\vectLam}}{\vectr})}{(\cprod{\vectOme}{\vectr})}]
  -\rhorep\Omerep^2\parve[\dprod{(\cprod{\fdot{\vectLam}}{\vectr})}{(\cprod{\vectr}{\vectLam})}]\\
  &\quad+\vekr[\dprod{(\cprod{\fdot{\vectLam}}{\vectr})}{(\cprod{\unitplz}{\vectr})}]
  -\vphij\parve[\dprod{(\cprod{\fdot{\vectLam}}{\vectr})}{(\cprod{\unitplz}{\vectLam})}]
\end{split}
\end{align*}
\begin{align}\label{rray5q}
\begin{split}
&=\vekh\frktn-\vekj\frktl-\vekm\frkth
  +\vekk[(\dprod{\fdot{\vectLam}}{\unitkap})(\dprod{\vectr}{\vectOme})-(\dprod{\fdot{\vectLam}}{\vectOme})(\dprod{\vectr}{\unitkap})]\\
  &\quad-\vekn[(\dprod{\fdot{\vectLam}}{\unitkap})\scalr^2-(\dprod{\fdot{\vectLam}}{\vectr})(\dprod{\vectr}{\unitkap})]
  +\cdkt\parve[(\dprod{\fdot{\vectLam}}{\unitkap})(\dprod{\vectr}{\vectLam})-(\dprod{\fdot{\vectLam}}{\vectLam})(\dprod{\vectr}{\unitkap})]\\
  &\quad+\vekp[(\dprod{\fdot{\vectLam}}{\unitkap})(\dprod{\vectr}{\unitplz})-(\dprod{\fdot{\vectLam}}{\unitplz})(\dprod{\vectr}{\unitkap})]
  +\parve^2[(\dprod{\fdot{\vectLam}}{\vectOme})(\dprod{\vectr}{\vectLam})-(\dprod{\fdot{\vectLam}}{\vectLam})(\dprod{\vectr}{\vectOme})]\\
  &\quad+\vekq[(\dprod{\fdot{\vectLam}}{\unitplz})(\dprod{\vectr}{\vectOme})-(\dprod{\fdot{\vectLam}}{\vectOme})(\dprod{\vectr}{\unitplz})]
  +\veko[(\dprod{\fdot{\vectLam}}{\vectOme})\scalr^2-(\dprod{\fdot{\vectLam}}{\vectr})(\dprod{\vectr}{\vectOme})]\\
  &\quad-\rhorep\Omerep^2\parve[(\dprod{\fdot{\vectLam}}{\vectr})(\dprod{\vectr}{\vectLam})-(\dprod{\fdot{\vectLam}}{\vectLam})\scalr^2]
  +\vekr[(\dprod{\fdot{\vectLam}}{\unitplz})\scalr^2-(\dprod{\fdot{\vectLam}}{\vectr})(\dprod{\vectr}{\unitplz})]\\
  &\quad-\vphij\parve[(\dprod{\fdot{\vectLam}}{\unitplz})(\dprod{\vectr}{\vectLam})-(\dprod{\fdot{\vectLam}}{\vectLam})(\dprod{\vectr}{\unitplz})]
  \beqref{rpath1b}\text{ \& }\eqnref{alg2}
\end{split}
\nonumber\\
\begin{split}
&=\vekh\frktn-\vekj\frktl-\vekm\frkth+\vekk(\vsigb\epsvb-\vsigc\epsvd)
  -\vekn(\vsigb\scalr^2-\vsign\epsvd)+\cdkt\parve(\vsigb\epsvh-\vsigd\epsvd)\\
  &\quad+\vekp(\vsigb\epsvf-\vsiga\epsvd)+\parve^2(\vsigc\epsvh-\vsigd\epsvb)
  +\vekq(\vsiga\epsvb-\vsigc\epsvf)+\veko(\vsigc\scalr^2-\vsign\epsvb)\\
  &\quad-\rhorep\Omerep^2\parve(\vsign\epsvh-\vsigd\scalr^2)+\vekr(\vsiga\scalr^2-\vsign\epsvf)
  -\vphij\parve(\vsiga\epsvh-\vsigd\epsvf)
  \beqref{rot1a}\text{ \& }\eqnref{rpath1a}
\end{split}
\nonumber\\
&=\vpsh\beqref{rray1j}
\end{align}
\begin{align*}
\begin{split}
&\dprod{(\cprod{\ffdot{\vectLam}}{\vectr})}{(\veust+\veusu)}\\
&=\dprod{(\cprod{\ffdot{\vectLam}}{\vectr})}{[}\vekh\unitplz+\veki\vectr+\vekj\vectOme+\vekm\vectLam
  +\vekl\fdot{\vectLam}+\vekk(\cprod{\unitkap}{\vectOme})-\vekn(\cprod{\unitkap}{\vectr})\\
  &\quad+\cdkt\parve(\cprod{\unitkap}{\vectLam})+\vekp(\cprod{\unitkap}{\unitplz})
  +\parve^2(\cprod{\vectOme}{\vectLam})+\vekq(\cprod{\unitplz}{\vectOme})+\veko(\cprod{\vectOme}{\vectr})
  -\rhorep\Omerep^2\parve(\cprod{\vectr}{\vectLam})\\
  &\quad+\vekr(\cprod{\unitplz}{\vectr})-\vphij\parve(\cprod{\unitplz}{\vectLam})]\beqref{rray4}
\end{split}
\nonumber\\
\begin{split}
&=\vekh[\dprod{\unitplz}{(\cprod{\ffdot{\vectLam}}{\vectr})}]
  +\veki[\dprod{\vectr}{(\cprod{\ffdot{\vectLam}}{\vectr})}]
  +\vekj[\dprod{\vectOme}{(\cprod{\ffdot{\vectLam}}{\vectr})}]
  +\vekm[\dprod{\vectLam}{(\cprod{\ffdot{\vectLam}}{\vectr})}]\\
  &\quad+\vekl[\dprod{\fdot{\vectLam}}{(\cprod{\ffdot{\vectLam}}{\vectr})}]
  +\vekk[\dprod{(\cprod{\ffdot{\vectLam}}{\vectr})}{(\cprod{\unitkap}{\vectOme})}]
  -\vekn[\dprod{(\cprod{\ffdot{\vectLam}}{\vectr})}{(\cprod{\unitkap}{\vectr})}]\\
  &\quad+\cdkt\parve[\dprod{(\cprod{\ffdot{\vectLam}}{\vectr})}{(\cprod{\unitkap}{\vectLam})}]
  +\vekp[\dprod{(\cprod{\ffdot{\vectLam}}{\vectr})}{(\cprod{\unitkap}{\unitplz})}]
  +\parve^2[\dprod{(\cprod{\ffdot{\vectLam}}{\vectr})}{(\cprod{\vectOme}{\vectLam})}]\\
  &\quad+\vekq[\dprod{(\cprod{\ffdot{\vectLam}}{\vectr})}{(\cprod{\unitplz}{\vectOme})}]
  +\veko[\dprod{(\cprod{\ffdot{\vectLam}}{\vectr})}{(\cprod{\vectOme}{\vectr})}]
  -\rhorep\Omerep^2\parve[\dprod{(\cprod{\ffdot{\vectLam}}{\vectr})}{(\cprod{\vectr}{\vectLam})}]\\
  &\quad+\vekr[\dprod{(\cprod{\ffdot{\vectLam}}{\vectr})}{(\cprod{\unitplz}{\vectr})}]
  -\vphij\parve[\dprod{(\cprod{\ffdot{\vectLam}}{\vectr})}{(\cprod{\unitplz}{\vectLam})}]
\end{split}
\end{align*}
\begin{align*}
\begin{split}
&=\vekh\frkxb-\vekj\frktm-\vekm\frkti-\vekl\frktj
  +\vekk[(\dprod{\ffdot{\vectLam}}{\unitkap})(\dprod{\vectr}{\vectOme})-(\dprod{\ffdot{\vectLam}}{\vectOme})(\dprod{\vectr}{\unitkap})]\\
  &\quad-\vekn[(\dprod{\ffdot{\vectLam}}{\unitkap})\scalr^2-(\dprod{\ffdot{\vectLam}}{\vectr})(\dprod{\vectr}{\unitkap})]
  +\cdkt\parve[(\dprod{\ffdot{\vectLam}}{\unitkap})(\dprod{\vectr}{\vectLam})-(\dprod{\ffdot{\vectLam}}{\vectLam})(\dprod{\vectr}{\unitkap})]\\
  &\quad+\vekp[(\dprod{\ffdot{\vectLam}}{\unitkap})(\dprod{\vectr}{\unitplz})-(\dprod{\ffdot{\vectLam}}{\unitplz})(\dprod{\vectr}{\unitkap})]
  +\parve^2[(\dprod{\ffdot{\vectLam}}{\vectOme})(\dprod{\vectr}{\vectLam})-(\dprod{\ffdot{\vectLam}}{\vectLam})(\dprod{\vectr}{\vectOme})]\\
  &\quad+\vekq[(\dprod{\ffdot{\vectLam}}{\unitplz})(\dprod{\vectr}{\vectOme})-(\dprod{\ffdot{\vectLam}}{\vectOme})(\dprod{\vectr}{\unitplz})]
  +\veko[(\dprod{\ffdot{\vectLam}}{\vectOme})\scalr^2-(\dprod{\ffdot{\vectLam}}{\vectr})(\dprod{\vectr}{\vectOme})]\\
  &\quad-\rhorep\Omerep^2\parve[(\dprod{\ffdot{\vectLam}}{\vectr})(\dprod{\vectr}{\vectLam})-(\dprod{\ffdot{\vectLam}}{\vectLam})\scalr^2]
  +\vekr[(\dprod{\ffdot{\vectLam}}{\unitplz})\scalr^2-(\dprod{\ffdot{\vectLam}}{\vectr})(\dprod{\vectr}{\unitplz})]\\
  &\quad-\vphij\parve[(\dprod{\ffdot{\vectLam}}{\unitplz})(\dprod{\vectr}{\vectLam})-(\dprod{\ffdot{\vectLam}}{\vectLam})(\dprod{\vectr}{\unitplz})]
  \beqref{rpath1b},\eqnref{rpath1b2}\text{ \& }\eqnref{alg2}
\end{split}
\end{align*}
\begin{align}\label{rray5r}
\begin{split}
&=\vekh\frkxb-\vekj\frktm-\vekm\frkti-\vekl\frktj+\vekk(\vsigg\epsvb-\vsigh\epsvd)
  -\vekn(\vsigg\scalr^2-\vsigo\epsvd)+\cdkt\parve(\vsigg\epsvh-\vsigi\epsvd)\\
  &\quad+\vekp(\vsigg\epsvf-\vsigf\epsvd)+\parve^2(\vsigh\epsvh-\vsigi\epsvb)
  +\vekq(\vsigf\epsvb-\vsigh\epsvf)+\veko(\vsigh\scalr^2-\vsigo\epsvb)\\
  &\quad-\rhorep\Omerep^2\parve(\vsigo\epsvh-\vsigi\scalr^2)
  +\vekr(\vsigf\scalr^2-\vsigo\epsvf)-\vphij\parve(\vsigf\epsvh-\vsigi\epsvf)
  \beqref{rot1a}\text{ \& }\eqnref{rpath1a}
\end{split}
\nonumber\\
&=\vpsi\beqref{rray1k}
\end{align}
\begin{align}\label{rray5s}
\begin{split}
&\dprod{\unitkap}{(\veust+\veusu)}\\
&=\dprod{\unitkap}{[}\vekh\unitplz+\veki\vectr+\vekj\vectOme+\vekm\vectLam
  +\vekl\fdot{\vectLam}+\vekk(\cprod{\unitkap}{\vectOme})-\vekn(\cprod{\unitkap}{\vectr})\\
  &\quad+\cdkt\parve(\cprod{\unitkap}{\vectLam})+\vekp(\cprod{\unitkap}{\unitplz})
  +\parve^2(\cprod{\vectOme}{\vectLam})+\vekq(\cprod{\unitplz}{\vectOme})+\veko(\cprod{\vectOme}{\vectr})
  -\rhorep\Omerep^2\parve(\cprod{\vectr}{\vectLam})\\
  &\quad+\vekr(\cprod{\unitplz}{\vectr})-\vphij\parve(\cprod{\unitplz}{\vectLam})]\beqref{rray4}
\end{split}
\nonumber\\
\begin{split}
&=\vekh(\dprod{\unitkap}{\unitplz})
  +\veki(\dprod{\unitkap}{\vectr})
  +\vekj(\dprod{\unitkap}{\vectOme})
  +\vekm(\dprod{\unitkap}{\vectLam})
  +\vekl(\dprod{\unitkap}{\fdot{\vectLam}})
  +\vekk[\dprod{\unitkap}{(\cprod{\unitkap}{\vectOme})}]\\
  &\quad-\vekn[\dprod{\unitkap}{(\cprod{\unitkap}{\vectr})}]
  +\cdkt\parve[\dprod{\unitkap}{(\cprod{\unitkap}{\vectLam})}]
  +\vekp[\dprod{\unitkap}{(\cprod{\unitkap}{\unitplz})}]
  +\parve^2[\dprod{\unitkap}{(\cprod{\vectOme}{\vectLam})}]\\
  &\quad+\vekq[\dprod{\unitkap}{(\cprod{\unitplz}{\vectOme})}]
  +\veko[\dprod{\unitkap}{(\cprod{\vectOme}{\vectr})}]
  -\rhorep\Omerep^2\parve[\dprod{\unitkap}{(\cprod{\vectr}{\vectLam})}]
  +\vekr[\dprod{\unitkap}{(\cprod{\unitplz}{\vectr})}]\\
  &\quad-\vphij\parve[\dprod{\unitkap}{(\cprod{\unitplz}{\vectLam})}]
\end{split}
\nonumber\\
\begin{split}
&=\vekh\epsve+\veki\epsvd+\vekj\epsva+\vekm\dltva+\vekl\vsigb-\parve^2\frkte+\vekq\epsvj+\veko\epsvk\\
  &\quad+\rhorep\Omerep^2\parve\epsvn-\vekr\frktt-\vphij\parve\dltvb
  \beqref{rot1a}, \eqnref{rxpeed1a}, \eqnref{rpath1a}\text{ \& }\eqnref{rpath1b}
\end{split}
\nonumber\\
&=\vpsj\beqref{rray1k}.
\end{align}
\end{subequations}
As a result of the foregoing derivations, we obtain
\begin{align}\label{rray6}
\begin{split}
|\veust+\veusu|^2
&=\dprod{(\veust+\veusu)}{[}\vekh\unitplz+\veki\vectr+\vekj\vectOme+\vekm\vectLam
  +\vekl\fdot{\vectLam}+\vekk(\cprod{\unitkap}{\vectOme})-\vekn(\cprod{\unitkap}{\vectr})\\
  &\quad+\cdkt\parve(\cprod{\unitkap}{\vectLam})+\vekp(\cprod{\unitkap}{\unitplz})
  +\parve^2(\cprod{\vectOme}{\vectLam})+\vekq(\cprod{\unitplz}{\vectOme})+\veko(\cprod{\vectOme}{\vectr})\\
  &\quad-\rhorep\Omerep^2\parve(\cprod{\vectr}{\vectLam})
  +\vekr(\cprod{\unitplz}{\vectr})-\vphij\parve(\cprod{\unitplz}{\vectLam})]\beqref{rray4}\\
&=\vekh[\dprod{\unitplz}{(\veust+\veusu)}]
  +\veki[\dprod{\vectr}{(\veust+\veusu)}]
  +\vekj[\dprod{\vectOme}{(\veust+\veusu)}]
  +\vekm[\dprod{\vectLam}{(\veust+\veusu)}]\\
  &\quad+\vekl[\dprod{\fdot{\vectLam}}{(\veust+\veusu)}]
  +\vekk[\dprod{(\veust+\veusu)}{(\cprod{\unitkap}{\vectOme})}]
  -\vekn[\dprod{(\veust+\veusu)}{(\cprod{\unitkap}{\vectr})}]\\
  &\quad+\cdkt\parve[\dprod{(\veust+\veusu)}{(\cprod{\unitkap}{\vectLam})}]
  +\vekp[\dprod{(\veust+\veusu)}{(\cprod{\unitkap}{\unitplz})}]
  +\parve^2[\dprod{(\veust+\veusu)}{(\cprod{\vectOme}{\vectLam})}]\\
  &\quad+\vekq[\dprod{(\veust+\veusu)}{(\cprod{\unitplz}{\vectOme})}]
  +\veko[\dprod{(\veust+\veusu)}{(\cprod{\vectOme}{\vectr})}]
  -\rhorep\Omerep^2\parve[\dprod{(\veust+\veusu)}{(\cprod{\vectr}{\vectLam})}]\\
  &\quad+\vekr[\dprod{(\veust+\veusu)}{(\cprod{\unitplz}{\vectr})}]
  -\vphij\parve[\dprod{(\veust+\veusu)}{(\cprod{\unitplz}{\vectLam})}]
\end{split}
\nonumber\\
\begin{split}
&=\vekh\vekr+\veki\veks+\vekj\vekt+\vekm\veku+\vekl\vekv+\vekk\vekw-\vekn\vekx+\cdkt\parve\veky+\vekp\vekz\\
  &\quad+\parve^2\vpsa+\vekq\vpsb+\veko\vpsc-\rhorep\Omerep^2\parve\vpsd+\vekr\vpse-\vphij\parve\vpsf
  \beqref{rray5}
\end{split}
\nonumber\\
\therefore|\veust+\veusu|
&=\vpsk\beqref{rray1l}.
\end{align}

\subart{Development of equation \eqnref{kray2c}}
Furthermore, we derive
\begin{subequations}\label{rray7}
\begin{align}\label{rray7a}
\imaa
&=\dprod{\unitkap}{(\veust+\veusu)}\beqref{kray2c}\nonumber\\
&=\vpsj\beqref{rray5s}
\end{align}
\begin{align}\label{rray7b}
\imab
&=\dprod{\vecta}{(\veust+\veusu)}\beqref{kray2c}\nonumber\\
&=\dprod{(\epsvb\vectOme-\Omerep^2\vectr+\cprod{\vectLam}{\vectr})}{(\veust+\veusu)}
  \beqref{main4a}\text{ \& }\eqnref{rot1a}\nonumber\\
&=\epsvb[\dprod{\vectOme}{(\veust+\veusu)}]-\Omerep^2[\dprod{\vectr}{(\veust+\veusu)}]
  +[\dprod{(\veust+\veusu)}{(\cprod{\vectLam}{\vectr})}]\nonumber\\
&=\epsvb\vekt-\Omerep^2\veks-\vpsd\beqref{rray5}
\end{align}
\begin{align}\label{rray7c}
\imac
&=\dprod{\fdota}{(\veust+\veusu)}\beqref{kray2c}\nonumber\\
&=\dprod{(\veust+\veusu)}{[2\epsvh\vectOme-3\epsvg\vectr+\cprod{\fdot{\vectLam}}{\vectr}
   -\Omerep^2(\cprod{\vectOme}{\vectr})+\epsvb\vectLam]}\beqref{rpath7a}\nonumber\\
\begin{split}
&=2\epsvh[\dprod{\vectOme}{(\veust+\veusu)}]-3\epsvg[\dprod{\vectr}{(\veust+\veusu)}]
  +[\dprod{(\veust+\veusu)}{(\cprod{\fdot{\vectLam}}{\vectr}})]\\
  &\quad-\Omerep^2[\dprod{(\veust+\veusu)}{(\cprod{\vectOme}{\vectr})}]
  +\epsvb[\dprod{\vectLam}{(\veust+\veusu)}]
\end{split}
\nonumber\\
&=2\epsvh\vekt-3\epsvg\veks+\vpsh-\Omerep^2\vpsc+\epsvb\veku
  \beqref{rray5}
\end{align}
\begin{align}\label{rray7d}
\imad
&=\dprod{\ffdota}{(\veust+\veusu)}\beqref{kray2c}\nonumber\\
\begin{split}
&=\dprod{(\veust+\veusu)}{[}3\epsvh\vectLam+\epsvb\fdot{\vectLam}+\vrhot\vectOme
  +\vrhou\vectr+2\epsvb(\cprod{\vectLam}{\vectOme})\\
  &\quad-3\Omerep^2(\cprod{\vectLam}{\vectr})-3\epsvg(\cprod{\vectOme}{\vectr})
  +(\cprod{\ffdot{\vectLam}}{\vectr})]\beqref{rpath7b}
\end{split}
\nonumber\\
\begin{split}
&=3\epsvh[\dprod{\vectLam}{(\veust+\veusu)}]
  +\epsvb[\dprod{\fdot{\vectLam}}{(\veust+\veusu)}]
  +\vrhot[\dprod{\vectOme}{(\veust+\veusu)}]
  +\vrhou[\dprod{\vectr}{(\veust+\veusu)}]\\
  &\quad+2\epsvb[\dprod{(\veust+\veusu)}{(\cprod{\vectLam}{\vectOme})}]
  -3\Omerep^2[\dprod{(\veust+\veusu)}{(\cprod{\vectLam}{\vectr})}]\\
  &\quad-3\epsvg[\dprod{(\veust+\veusu)}{(\cprod{\vectOme}{\vectr})}]
  +[\dprod{(\veust+\veusu)}{(\cprod{\ffdot{\vectLam}}{\vectr})}]
\end{split}
\nonumber\\
&=3\epsvh\veku+\epsvb\vekv+\vrhot\vekt+\vrhou\veks-2\epsvb\vpsa
  +3\Omerep^2\vpsd-3\epsvg\vpsc+\vpsi\beqref{rray5}
\end{align}
\begin{align}\label{rray7e}
\imae
&=\dprod{\ffdote}{(\veust+\veusu)}\beqref{kray2c}\nonumber\\
&=\dprod{(\veust+\veusu)}{[}\ethvp(\cprod{\unitplz}{\vectOme})+2\ethvo(\cprod{\unitplz}{\vectLam})
  +\vphig\vphih(\cprod{\unitplz}{\fdot{\vectLam}})+\ethvs\vectOme
  +2\ethvr\vectLam+\vphii\fdot{\vectLam}-\ethvv\unitplz]
\nonumber\\
\begin{split}
&=\ethvp[\dprod{(\veust+\veusu)}{(\cprod{\unitplz}{\vectOme})}]
  +2\ethvo[\dprod{(\veust+\veusu)}{(\cprod{\unitplz}{\vectLam})}]
  +\vphig\vphih[\dprod{(\veust+\veusu)}{(\cprod{\unitplz}{\fdot{\vectLam}})}]\\
  &\quad+\ethvs[\dprod{\vectOme}{(\veust+\veusu)}]
  +2\ethvr[\dprod{\vectLam}{(\veust+\veusu)}]
  +\vphii[\dprod{\fdot{\vectLam}}{(\veust+\veusu)}]
  -\ethvv[\dprod{\unitplz}{(\veust+\veusu)}]
\end{split}
\nonumber\\
&=\ethvp\vpsb+2\ethvo\vpsf-\vphig\vphih\vpsg+\ethvs\vekt+2\ethvr\veku
  +\vphii\vekv-\ethvv\vekr\beqref{rray5}
\end{align}
\end{subequations}
from which we obtain
\begin{align}\label{rray8}
&\ffdot{\cdkt}\imaa+\ffdot{\rhorep}\imab+\vbbb\imac+\rhorep\imad+\imae\nonumber\\
&=\parvf\imaa+\ethvj\imab+\parvh\imac+\rhorep\imad+\imae
  \beqref{rpath12b}\text{ \& }\eqnref{rpath23}
\nonumber\\
\begin{split}
&=\parvf\vpsj+\ethvj(\epsvb\vekt-\Omerep^2\veks-\vpsd)
  +\parvh(2\epsvh\vekt-3\epsvg\veks+\vpsh-\Omerep^2\vpsc+\epsvb\veku)\\
  &\quad+\rhorep(3\epsvh\veku+\epsvb\vekv+\vrhot\vekt+\vrhou\veks-2\epsvb\vpsa
    +3\Omerep^2\vpsd-3\epsvg\vpsc+\vpsi)\\
  &\quad+\ethvp\vpsb+2\ethvo\vpsf-\vphig\vphih\vpsg+\ethvs\vekt+2\ethvr\veku
  +\vphii\vekv-\ethvv\vekr\beqref{rray7}
\end{split}
\nonumber\\
&=\vpsl\beqref{rray1l}.
\end{align}
We note also that
\begin{align}\label{rray9}
\vectups
&=\cdkt\unitkap+\rhorep\vecta-\vectu+\vecte\beqref{kpath1c}\nonumber\\
\begin{split}
&=\cdkt\unitkap+\rhorep(\epsvb\vectOme-\Omerep^2\vectr+\cprod{\vectLam}{\vectr})-(\cprod{\vectOme}{\vectr})
  +\stwo(\cprod{\unitplz}{\vectOme})+\sthr\vectOme-\sfou\unitplz\\
  &\qquad\beqref{main4}\text{ \& }\eqnref{rot1a}
\end{split}
\nonumber\\
&=\cdkt\unitkap+\rhorep(\epsvb\vectOme-\Omerep^2\vectr+\cprod{\vectLam}{\vectr})-(\cprod{\vectOme}{\vectr})
  +\vphig\vphih(\cprod{\unitplz}{\vectOme})+\vphii\vectOme-\vphij\unitplz\beqref{rpath5}\nonumber\\
&=\cdkt\unitkap+(\vphii+\rhorep\epsvb)\vectOme-\vphij\unitplz-\rhorep\Omerep^2\vectr+\rhorep(\cprod{\vectLam}{\vectr})
  -(\cprod{\vectOme}{\vectr})+\vphig\vphih(\cprod{\unitplz}{\vectOme})
\nonumber\\
&=\cdkt\unitkap+\parve\vectOme-\vphij\unitplz-\rhorep\Omerep^2\vectr+\rhorep(\cprod{\vectLam}{\vectr})
  -(\cprod{\vectOme}{\vectr})+\vphig\vphih(\cprod{\unitplz}{\vectOme})\beqref{rpath1p}.
\end{align}

\subart{Results of the computations}
It follows by substituting \eqnref{rray9}, \eqnref{rray8}, \eqnref{rray6} and \eqnref{rray4}
into \eqnref{kray4} that
\begin{subequations}\label{rray10}
\begin{align}\label{rray10a}
\bbkbar=\plusmin\frac{\vpsk}{\scalc^3\rcal^3},\qquad
\bbtbar=\frac{\vpsl}{(\vpsk)^2}
\end{align}
\begin{align}\label{rray10b}
\begin{split}
&\frtbar=\frac{1}{\scalc\rcal}\biggl[\cdkt\unitkap-\vphij\unitplz-\rhorep\Omerep^2\vectr+\parve\vectOme
  -(\cprod{\vectOme}{\vectr})+\rhorep(\cprod{\vectLam}{\vectr})+\vphig\vphih(\cprod{\unitplz}{\vectOme})\biggr]\\
&\frbbar=\frac{1}{\vpsk}\biggl[\vekh\unitplz+\veki\vectr+\vekj\vectOme+\vekm\vectLam+\vekl\fdot{\vectLam}
  +\vekk(\cprod{\unitkap}{\vectOme})-\vekn(\cprod{\unitkap}{\vectr})+\cdkt\parve(\cprod{\unitkap}{\vectLam})\\
  &\qquad+\vekp(\cprod{\unitkap}{\unitplz})+\parve^2(\cprod{\vectOme}{\vectLam})+\vekq(\cprod{\unitplz}{\vectOme})
  +\veko(\cprod{\vectOme}{\vectr})-\rhorep\Omerep^2\parve(\cprod{\vectr}{\vectLam})\\
  &\qquad+\vekr(\cprod{\unitplz}{\vectr})-\vphij\parve(\cprod{\unitplz}{\vectLam})\biggr]
\end{split}
\end{align}
\end{subequations}
is the complete set of equations describing the apparent geometry of obliquated rays for
a rotating observer.

\section{Gravitational obliquation}\label{S_GRAOB}
\art{Apparent direction to a light source}
To evaluate \eqnref{grad6} for a gravitating observer, we introduce the quantities~\footnote{We note
here that the angle between $\vectz$ and $\vectr$ is the true anomaly of the observer's orbit while the
eccentricity (or Laplace-Runge-Lenz) vector $\vectz$ is perpendicular to the latus rectum of the orbit and the
Hamilton vector $\cprod{\vectz}{\vecth}$ is parallel to the latus rectum.}
\begin{subequations}\label{ogrv1}
\begin{align}\label{ogrv1a}
\begin{split}
&\epsva=\dprod{\unitkap}{\unitpos},\quad
\epsvb=\dprod{\unitkap}{\unitplz},\quad
\epsvc=\dprod{\unitpos}{\unitplz},\quad
\epsvd=\dprod{\unitpos}{\vectz},\quad
\epsve=\dprod{\unitkap}{(\cprod{\vectr}{\vecth})}\\
&\epsvf=\dprod{\unitplz}{(\cprod{\vectr}{\vecth})}\ne0,\quad
\epsvg=\dprod{\unitkap}{(\cprod{\vectz}{\vecth})},\quad
\epsvh=\dprod{\unitpos}{(\cprod{\vectz}{\vecth})},\quad
\epsvi=\dprod{\unitplz}{(\cprod{\vectz}{\vecth})}\\
\end{split}
\end{align}
\begin{align}\label{ogrv1b}
\begin{split}
&\qquad\vphia=\scalq/\scalr^2,\quad
\vphib=\scalq/\scalr,\quad
\vphic=[(\epsvb\epsve)/\epsvf]^{1/2},\quad
\vphid=\vphic^2-1\ne0,\quad
\vphie=\left|\vphid\right|^{1/2}\\
&\vphif=(2\rhorep\vphia\epsva+\dragf\scalc)/(2\epsvf\vphid),\quad
\vphig=-2(\rhorep\vphia\epsva+\epsvf\vphif\vphic^2),\quad
\vphih=(\scalz^2+2\scalq\epsvd+\scalq^2)^{1/2}/\scalh\\
&\vphii=(\epsvg+\vphih\epsve)/(\scalh^2\vphih),\quad
\vphij=\epsvh/(\scalh^2\vphih),\quad
\vphik=\vphif[\scalr\epsvb(\scalq+\epsvd)+\scalh^{-2}\epsve(\epsvi+\vphib\epsvf)]\\
&\qquad\qquad\vphil=(\scalr^2\scalh^2\epsvb^2+2\epsvb\epsve\epsvf+\epsve^2)^{1/2},\quad
\vphim=\vphih^2[\vphik+\vphih(\rhorep\vphia\vphij-\vphig\vphii)]
\end{split}
\end{align}
\begin{align}\label{ogrv1c}
\begin{split}
&\vphin=2\dragf\scalc\vphih^2(2\vphif\epsve\epsvb-\vphig)
  +\vphih^2\vphif(\vphif\vphil^2-2\rhorep\vphia\epsve\epsvc-4\vphig\epsvb\epsve)
  +\vphih^2\vphig(\vphig-2\rhorep\vphia\epsva)\\
&\vphio=2\dragf\scalc\vphih\vphii(\vphik-\vphih\vphig\vphii)
  -\vphih^2\vphig\vphii(2\rhorep\vphia\vphij-\vphig\vphii)
  +\vphik^2+2\vphik\vphih(\rhorep\vphia\vphij-\vphig\vphii).
\end{split}
\end{align}
\end{subequations}

\subart{Preliminary calculations}
In view of the above quantities, we derive
\begin{subequations}\label{ogrv2}
\begin{align}\label{ogrv2a}
\alprep
&=\dprod{\vectkap}{\vecta}\beqref{main2b}\nonumber\\
&=\dprod{\vectkap}{(-\scalq\vectr/\scalr^3)}\beqref{main5a}\nonumber\\
&=-(\scalq/\scalr^3)(\dprod{\vectkap}{\vectr})=-\kaprep(\scalq/\scalr^2)\epsva\beqref{ogrv1a}\nonumber\\
&=-\kaprep\vphia\epsva\beqref{ogrv1b}
\end{align}
\begin{align}
\scalm
&=\dprod{\vectkap}{(\cprod{\vectr}{\vecth})}\beqref{main5c}\nonumber\\
&=\kaprep[\dprod{\unitkap}{(\cprod{\vectr}{\vecth})}]\nonumber\\
&=\kaprep\epsve\beqref{ogrv1a}
\label{ogrv2b}\\
\scaln
&=\dprod{\unitplz}{(\cprod{\vectr}{\vecth})}\beqref{main5c}\nonumber\\
&=\epsvf\beqref{ogrv1a}
\label{ogrv2c}
\end{align}
\begin{align}\label{ogrv2d}
\kcons^2
&=(\scalm/\scaln)(\dprod{\vectkap}{\unitplz})\beqref{main5c}\nonumber\\
&=(\scalm/\scaln)(\kaprep\epsvb)\beqref{ogrv1a}\nonumber\\
&=\kaprep^2(\epsve/\epsvf)\epsvb\beqref{ogrv2b}\text{ \& }\eqnref{ogrv2c}\nonumber\\
\therefore\kcons
&=\kaprep\vphic\beqref{ogrv1b}
\end{align}
\begin{align}\label{ogrv2e}
\gamrep^2
&=\left|1-\frac{\kcons^2}{\kaprep^2}\right|\beqref{main5a}\nonumber\\
&=\left|1-\vphic^2\right|\beqref{ogrv2d}\nonumber\\
&=\left|\vphic^2-1\right|=\left|\vphid\right|\beqref{ogrv1b}\nonumber\\
\therefore\gamrep
&=\vphie\beqref{ogrv1b}
\end{align}
\begin{align}\label{ogrv2f}
\fzer
&=\frac{2\rhorep\alprep-\dragf\pfreq}{\scaln(\kaprep^2-\kcons^2)}
  \beqref{main5d}\nonumber\\
&=\frac{-2\rhorep\kaprep\vphia\epsva-\dragf\pfreq}{\epsvf(\kaprep^2-\kaprep^2\vphic^2)}
  \beqref{ogrv2a}, \eqnref{ogrv2c}\text{ \& }\eqnref{ogrv2d}\nonumber\\
&=\frac{-2\rhorep\kaprep\vphia\epsva-\dragf\pfreq}{\kaprep^2\epsvf(1-\vphic^2)}
=\frac{2\rhorep\vphia\epsva+\dragf\scalc}{\kaprep\epsvf(\vphic^2-1)}\beqref{main2b}\nonumber\\
&=2\vphif/\kaprep\beqref{ogrv1b}
\end{align}
\begin{align}
\fone
&=\frac{(\dprod{\vectkap}{\unitplz})\fzer}{2}\beqref{main5d}\nonumber\\
&=\frac{(\kaprep\epsvb)(2\vphif/\kaprep)}{2}\beqref{ogrv1a}\text{ \& }\eqnref{ogrv2f}\nonumber\\
&=\epsvb\vphif
\label{ogrv2g}\\
\ftwo
&=\frac{\scalm\fzer}{2}\beqref{main5d}\nonumber\\
&=\frac{(\kaprep\epsve)(2\vphif/\kaprep)}{2}\beqref{ogrv2b}\text{ \& }\eqnref{ogrv2f}\nonumber\\
&=\epsve\vphif.
\label{ogrv2h}
\end{align}
\end{subequations}
In addition, we have
\begin{subequations}\label{ogrv3}
\begin{align}\label{ogrv3a}
\taurep
&=(2\rhorep\alprep-\scaln\kcons^2\fzer)\kaprep^{-2}\beqref{main5b}\nonumber\\
&=(-2\rhorep\kaprep\vphia\epsva-\scaln\kcons^2\fzer)\kaprep^{-2}\beqref{ogrv2a}\nonumber\\
&=[-2\rhorep\kaprep\vphia\epsva-\epsvf(\kaprep^2\vphic^2)(2\vphif/\kaprep)]\kaprep^{-2}
  \beqref{ogrv2c}, \eqnref{ogrv2d}\text{ \& }\eqnref{ogrv2f}\nonumber\\
&=-2(\rhorep\vphia\epsva+\epsvf\vphif\vphic^2)/\kaprep\nonumber\\
&=\vphig/\kaprep\beqref{ogrv1b}
\end{align}
\begin{align}\label{ogrv3b}
\scalu^2
&=\scalh^{-4}(\cprod{\vectz}{\vecth}+\cprod{\scalq\unitpos}{\vecth})^2\beqref{main5c}\nonumber\\
&=\scalh^{-4}[\dprod{(\cprod{\vectz}{\vecth})}{(\cprod{\vectz}{\vecth})}
  +2\dprod{(\cprod{\vectz}{\vecth})}{(\cprod{\scalq\unitpos}{\vecth})}
  +\dprod{(\cprod{\scalq\unitpos}{\vecth})}{(\cprod{\scalq\unitpos}{\vecth})}]
\nonumber\\
&=\scalh^{-4}[\scalz^2\scalh^2-(\dprod{\vectz}{\vecth})^2
  +2(\dprod{\vectz}{\scalq\unitpos})\scalh^2-2(\dprod{\vectz}{\vecth})(\dprod{\vecth}{\scalq\unitpos})
  +\scalq^2\scalh^2-(\dprod{\scalq\unitpos}{\vecth})^2]\beqref{alg2}
\nonumber\\
&=\scalh^{-4}[\scalz^2\scalh^2+2\scalq\scalh^2(\dprod{\vectz}{\unitpos})+\scalq^2\scalh^2]\beqref{main5c}\nonumber\\
&=\scalh^{-2}(\scalz^2+2\scalq\epsvd+\scalq^2)\beqref{ogrv1a}\nonumber\\
\therefore\scalu
&=\vphih\beqref{ogrv1b}
\end{align}
\end{subequations}
\begin{subequations}\label{ogrv4}
\begin{align}\label{ogrv4a}
\cos\lamrep
&=\dprod{(\vectkap/\kaprep)}{(\vecta/\scala)}\text{ by definition of $\lamrep$}\nonumber\\
&=\dprod{\unitkap}{\unitpos}\beqref{main5a}\nonumber\\
&=\epsva\beqref{ogrv1a}
\end{align}
\begin{align}\label{ogrv4b}
\cos\phirep
&=\dprod{(\vectkap/\kaprep)}{(\vectu/\scalu)}\text{ by definition of $\phirep$}\nonumber\\
&=\dprod{(\unitkap/\scalu)}{[\scalh^{-2}(\cprod{\vectz}{\vecth}+\cprod{\scalq\unitpos}{\vecth})]}
  \beqref{main5c}\nonumber\\
&=\scalu^{-1}\scalh^{-2}[\dprod{\unitkap}{(\cprod{\vectz}{\vecth})}+\scalq\dprod{\unitkap}{(\cprod{\unitpos}{\vecth})}]\nonumber\\
&=\vphih^{-1}\scalh^{-2}[\dprod{\unitkap}{(\cprod{\vectz}{\vecth})}+(\scalq/\scalr)\dprod{\unitkap}{(\cprod{\vectr}{\vecth})}]
  \beqref{ogrv3b}\nonumber\\
&=\vphih^{-1}\scalh^{-2}[\epsvg+(\scalq/\scalr)\epsve]\beqref{ogrv1a}\nonumber\\
&=\vphih^{-1}\scalh^{-2}(\epsvg+\vphib\epsve)
=\vphii\beqref{ogrv1b}
\end{align}
\begin{align}\label{ogrv4c}
\cos\thtrep
&=\dprod{(\vecta/\scala)}{(\vectu/\scalu)}\text{ by definition of $\thtrep$}\nonumber\\
&=\dprod{\unitpos}{(\vectu/\scalu)}\beqref{main5a}\nonumber\\
&=\dprod{(\unitpos/\scalu)}{[\scalh^{-2}(\cprod{\vectz}{\vecth}+\cprod{\scalq\unitpos}{\vecth})]}
  \beqref{main5c}\nonumber\\
&=\vphih^{-1}\scalh^{-2}[\dprod{\unitpos}{(\cprod{\vectz}{\vecth})}
  +\scalq\dprod{\unitpos}{(\cprod{\unitpos}{\vecth})}]\beqref{ogrv3b}\nonumber\\
&=\vphih^{-1}\scalh^{-2}\epsvh\beqref{ogrv1a}\nonumber\\
&=\vphij\beqref{ogrv1b}.
\end{align}
\end{subequations}

\subart{Development of equation \eqnref{grad2a}}
The various quantities defined in \eqnref{grad2a} evaluate as
\begin{subequations}\label{ogrv5}
\begin{align}\label{ogrv5a}
\bcal
&=\dprod{\vecte}{\vectc}\beqref{grad2a}\nonumber\\
&=\dprod{\vectc}{[\fone(\cprod{\vectr}{\vecth})+\ftwo\unitplz]}\beqref{main5b}\nonumber\\
&=\fone\scalc[\dprod{\unitkap}{(\cprod{\vectr}{\vecth})}]
  +\ftwo\scalc(\dprod{\unitkap}{\unitplz})\beqref{main1}\nonumber\\
&=\fone\scalc\epsve+\ftwo\scalc\epsvb\beqref{ogrv1a}\nonumber\\
&=\epsvb\vphif\scalc\epsve+\epsve\vphif\scalc\epsvb
=2\scalc\vphif\epsve\epsvb\beqref{ogrv2g}\text{ \& }\eqnref{ogrv2h}
\end{align}
\begin{align*}
\dcal
&=\dprod{\vecte}{\vectu}\beqref{grad2a}\nonumber\\
&=\scalh^{-2}\dprod{[\fone(\cprod{\vectr}{\vecth})+\ftwo\unitplz]}
  {(\cprod{\vectz}{\vecth}+\cprod{\scalq\unitpos}{\vecth})}
   \beqref{main5b}\text{ \& }\eqnref{main5c}\nonumber\\
\begin{split}
&=\scalh^{-2}\fone[\dprod{(\cprod{\vectr}{\vecth})}{(\cprod{\vectz}{\vecth})}]
  +\scalq\scalh^{-2}\fone[\dprod{(\cprod{\vectr}{\vecth})}{(\cprod{\unitpos}{\vecth})}]
  +\scalh^{-2}\ftwo[\dprod{\unitplz}{(\cprod{\vectz}{\vecth})}]\\
  &\quad+\scalq\scalh^{-2}\ftwo[\dprod{\unitplz}{(\cprod{\unitpos}{\vecth})}]
\end{split}
\nonumber\\
\begin{split}
&=\scalh^{-2}\fone[(\dprod{\vectr}{\vectz})\scalh^2-(\dprod{\vectr}{\vecth})(\dprod{\vecth}{\vectz})]
  +\scalq\scalh^{-2}\fone[(\dprod{\vectr}{\unitpos})\scalh^2-(\dprod{\vectr}{\vecth})(\dprod{\vecth}{\unitpos})]\\
  &\quad+\scalh^{-2}\ftwo[\dprod{\unitplz}{(\cprod{\vectz}{\vecth})}]
  +(\scalq/\scalr)\scalh^{-2}\ftwo[\dprod{\unitplz}{(\cprod{\vectr}{\vecth})}]\beqref{alg2}
\end{split}
\end{align*}
\begin{align}\label{ogrv5b}
&=\scalr\fone(\dprod{\unitpos}{\vectz})+\scalq\fone\scalr
  +\scalh^{-2}\ftwo[\dprod{\unitplz}{(\cprod{\vectz}{\vecth})}]
  +(\scalq/\scalr)\scalh^{-2}\ftwo[\dprod{\unitplz}{(\cprod{\vectr}{\vecth})}]\beqref{main5c}\nonumber\\
&=\scalr\fone\epsvd+\scalq\fone\scalr+\scalh^{-2}\ftwo\epsvi
  +(\scalq/\scalr)\scalh^{-2}\ftwo\epsvf\beqref{ogrv1a}\nonumber\\
&=\scalr\fone(\scalq+\epsvd)+\scalh^{-2}\ftwo[\epsvi+(\scalq/\scalr)\epsvf]\nonumber\\
&=\scalr\epsvb\vphif(\scalq+\epsvd)+\scalh^{-2}\epsve\vphif(\epsvi+\vphib\epsvf)
  \beqref{ogrv1b}, \eqnref{ogrv2g}\text{ \& }\eqnref{ogrv2h}\nonumber\\
&=\vphik\beqref{ogrv1b}
\end{align}
\begin{align}\label{ogrv5c}
\acal
&=\dprod{\vecte}{\vecta}\beqref{grad2a}\nonumber\\
&=\dprod{[\fone(\cprod{\vectr}{\vecth})+\ftwo\unitplz]}{(-\vphia\unitpos)}
  \beqref{main5a}, \eqnref{main5b}\text{ \& }\eqnref{ogrv1b}\nonumber\\
&=-\vphia\fone[\dprod{\unitpos}{(\cprod{\vectr}{\vecth})}]
  -\vphia\ftwo(\dprod{\unitplz}{\unitpos})\nonumber\\
&=-\vphia\epsve\vphif\epsvc\beqref{ogrv1a}\text{ \& }\eqnref{ogrv2h}
\end{align}
\begin{align}\label{ogrv5d}
\hcal
&=\dprod{\vecte}{\vectkap}\beqref{grad2a}\nonumber\\
&=\dprod{[\fone(\cprod{\vectr}{\vecth})+\ftwo\unitplz]}{\vectkap}\beqref{main5b}\nonumber\\
&=\kaprep\fone[\dprod{\unitkap}{(\cprod{\vectr}{\vecth})}]+\kaprep\ftwo(\dprod{\unitplz}{\unitkap})
=\kaprep\fone\epsve+\kaprep\ftwo\epsvb\beqref{ogrv1a}\nonumber\\
&=\kaprep\epsvb\vphif\epsve+\kaprep\epsve\vphif\epsvb
=2\kaprep\epsvb\vphif\epsve\beqref{ogrv2g}\text{ \& }\eqnref{ogrv2h}
\end{align}
\begin{align}\label{ogrv5e}
\ecal
&=\dprod{\vecte}{\vecte}\beqref{grad2a}\nonumber\\
&=\dprod{[\fone(\cprod{\vectr}{\vecth})+\ftwo\unitplz]}
  {[\fone(\cprod{\vectr}{\vecth})+\ftwo\unitplz]}\beqref{main5b}\nonumber\\
&=\fone^2[\dprod{(\cprod{\vectr}{\vecth})}{(\cprod{\vectr}{\vecth})}]
  +2\fone\ftwo[\dprod{\unitplz}{(\cprod{\vectr}{\vecth})}]
  +\ftwo^2(\dprod{\unitplz}{\unitplz})\nonumber\\
&=\fone^2[\scalr^2\scalh^2-(\dprod{\vectr}{\vecth})^2]+2\fone\ftwo\epsvf+\ftwo^2
  \beqref{ogrv1a}\text{ \& }\eqnref{alg2}\nonumber\\
&=\fone^2\scalr^2\scalh^2+2\fone\ftwo\epsvf+\ftwo^2\beqref{main5c}\nonumber\\
&=(\epsvb\vphif)^2\scalr^2\scalh^2+2(\epsvb\vphif)(\epsve\vphif)\epsvf+(\epsve\vphif)^2
  \beqref{ogrv2g}\text{ \& }\eqnref{ogrv2h}\nonumber\\
&=\vphif^2(\scalr^2\scalh^2\epsvb^2+2\epsvb\epsve\epsvf+\epsve^2)
=\vphif^2\vphil^2\beqref{ogrv1b}.
\end{align}
\end{subequations}

\subart{Development of equations \eqnref{grad2b} through \eqnref{grad2d}}
Consequently, from the above derivations, we obtain
\begin{subequations}\label{ogrv6}
\begin{align}
\lcal_0
&=\dcal-\scalu\kaprep\taurep\cos\phirep\beqref{grad2b}\nonumber\\
&=\vphik-\vphih\vphig\vphii\beqref{ogrv5b}, \eqnref{ogrv3b}, \eqnref{ogrv3a}\text{ \& }\eqnref{ogrv4b}
\label{ogrv6a}\\
\lcal_1
&=\rhorep\scala\cos\thtrep-\kaprep\taurep\cos\phirep\beqref{grad2b}\nonumber\\
&=\rhorep\vphia\vphij-\vphig\vphii
  \beqref{main5a}, \eqnref{ogrv1b}, \eqnref{ogrv4c}, \eqnref{ogrv3a}\text{ \& }\eqnref{ogrv4b}
\label{ogrv6b}\\
\lcal_2
&=\lcal_1+\rhorep\scala\cos\thtrep\beqref{grad2b}\nonumber\\
&=\rhorep\vphia\vphij-\vphig\vphii+\rhorep\vphia\vphij
  \beqref{ogrv6b}, \eqnref{main5a}, \eqnref{ogrv1b}\text{ \& }\eqnref{ogrv4c}\nonumber\\
&=2\rhorep\vphia\vphij-\vphig\vphii
\label{ogrv6c}
\end{align}
\end{subequations}
\begin{subequations}\label{ogrv7}
\begin{align}\label{ogrv7a}
\ncal_1
&=2\dragf\scalu^2(\bcal-\taurep\pfreq)\beqref{grad2c}\nonumber\\
&=2\dragf\scalu^2(2\scalc\vphif\epsve\epsvb-\scalc\kaprep\taurep)
  \beqref{ogrv5a}\text{ \& }\eqnref{main2b}\nonumber\\
&=2\dragf\vphih^2(2\scalc\vphif\epsve\epsvb-\scalc\vphig)
  \beqref{ogrv3b}\text{ \& }\eqnref{ogrv3a}\nonumber\\
&=2\dragf\scalc\vphih^2(2\vphif\epsve\epsvb-\vphig)
\end{align}
\begin{align}\label{ogrv7b}
\ncal_2
&=2\dragf\scalu\scalc\lcal_0\cos\phirep\beqref{grad2c}\nonumber\\
&=2\dragf(\vphih)\scalc(\vphik-\vphih\vphig\vphii)(\vphii)
  \beqref{ogrv3b}, \eqnref{ogrv6a}\text{ \& }\eqnref{ogrv4b}\nonumber\\
&=2\dragf\scalc\vphih\vphii(\vphik-\vphih\vphig\vphii)
\end{align}
\begin{align}\label{ogrv7c}
\ncal_3
&=\scalu^2[\ecal+2(\rhorep\acal-\taurep\hcal)]\beqref{grad2c}\nonumber\\
&=\vphih^2[\vphif^2\vphil^2+2\rhorep(-\vphia\epsve\vphif\epsvc)-2\taurep(2\kaprep\epsvb\vphif\epsve)]
  \beqref{ogrv3b}, \eqnref{ogrv5e}, \eqnref{ogrv5c}\text{ \& }\eqnref{ogrv5d}\nonumber\\
&=\vphih^2(\vphif^2\vphil^2-2\rhorep\vphia\epsve\vphif\epsvc-4\taurep\kaprep\epsvb\vphif\epsve)\nonumber\\
&=\vphih^2(\vphif^2\vphil^2-2\rhorep\vphia\epsve\vphif\epsvc-4\vphig\epsvb\vphif\epsve)
  \beqref{ogrv3a}\nonumber\\
&=\vphih^2\vphif(\vphif\vphil^2-2\rhorep\vphia\epsve\epsvc-4\vphig\epsvb\epsve)
\end{align}
\begin{align}\label{ogrv7d}
\ncal_4
&=\scalu^2\kaprep\taurep(\kaprep\taurep-2\rhorep\scala\cos\lamrep)
  \beqref{grad2d}\nonumber\\
&=\vphih^2\vphig(\vphig-2\rhorep\vphia\epsva)
  \beqref{ogrv3b}, \eqnref{ogrv3a}, \eqnref{main5a}, \eqnref{ogrv1b}\text{ \& }\eqnref{ogrv4a}
\end{align}
\begin{align}\label{ogrv7e}
\ncal_5
&=\scalu^2\kaprep\taurep\lcal_2\cos\phirep\beqref{grad2d}\nonumber\\
&=\vphih^2\vphig(2\rhorep\vphia\vphij-\vphig\vphii)\vphii
  \beqref{ogrv3b}, \eqnref{ogrv3a}, \eqnref{ogrv6c}\text{ \& }\eqnref{ogrv4b}\nonumber\\
&=\vphih^2\vphig\vphii(2\rhorep\vphia\vphij-\vphig\vphii)
\end{align}
\begin{align}\label{ogrv7f}
\ncal_6
&=\dcal(\dcal+2\scalu\lcal_1)\beqref{grad2d}\nonumber\\
&=\vphik[\vphik+2\vphih(\rhorep\vphia\vphij-\vphig\vphii)]
  \beqref{ogrv5b}, \eqnref{ogrv3b}\text{ \& }\eqnref{ogrv6b}\nonumber\\
&=\vphik^2+2\vphik\vphih(\rhorep\vphia\vphij-\vphig\vphii)
\end{align}
\end{subequations}
\begin{subequations}\label{ogrv8}
\begin{align}
2\scalu^2(\dcal+\scalu\lcal_1)
&=\vphih^2[\vphik+\vphih(\rhorep\vphia\vphij-\vphig\vphii)]
  \beqref{ogrv3b}, \eqnref{ogrv5b}\text{ \& }\eqnref{ogrv6b}\nonumber\\
&=\vphim\beqref{ogrv1b}
\label{ogrv8a}\\
\begin{split}
\ncal_1+\ncal_3+\ncal_4
&=2\dragf\scalc\vphih^2(2\vphif\epsve\epsvb-\vphig)
  +\vphih^2\vphif(\vphif\vphil^2-2\rhorep\vphia\epsve\epsvc-4\vphig\epsvb\epsve)\\
  &\quad+\vphih^2\vphig(\vphig-2\rhorep\vphia\epsva)\beqref{ogrv7}
\end{split}
\nonumber\\
&=\vphin\beqref{ogrv1c}
\label{ogrv8b}\\
\begin{split}
\ncal_2-\ncal_5+\ncal_6
&=2\dragf\scalc\vphih\vphii(\vphik-\vphih\vphig\vphii)
  -\vphih^2\vphig\vphii(2\rhorep\vphia\vphij-\vphig\vphii)\\
  &\quad+\vphik^2+2\vphik\vphih(\rhorep\vphia\vphij-\vphig\vphii)\beqref{ogrv7}
\end{split}
\nonumber\\
&=\vphio\beqref{ogrv1c}.
\label{ogrv8c}
\end{align}
\end{subequations}

\subart{Results of the computations}
By substituting \eqnref{ogrv8}, \eqnref{ogrv6a} and \eqnref{ogrv4} into \eqnref{grad2w}, we get
\begin{subequations}\label{ogrv9}
\begin{align}\label{ogrv9a}
\begin{split}
&\lcal=(\vphik-\vphih\vphig\vphii)/(\betrep\scalc^2),\quad
\pcal=(\vphin-\vphim)/(\btcf),\quad
\ncal=(\vphin-\vphio)/(\btcf)\\
&\gcal=\lcal-\betrep+\dragf\vphii+\rhorep\sigrep\vphij,\quad
\rcal=[\pcal+\dragf^2+\betrep^2+\rhorep^2\sigrep^2+2\dragf(\rhorep\sigrep\epsva-\betrep\vphii)]^{1/2}\\
&\qquad\fcal=[\ncal+\dragf^2(1-\vphii^2)+\rhorep^2\sigrep^2(1-\vphij^2)+2\dragf\rhorep\sigrep(\epsva-\vphii\vphij)]^{1/2}
\end{split}
\end{align}
while from \eqnref{grad6} and \eqnref{ogrv4}, we get
\begin{align}\label{ogrv9b}
\tan\psirep=\frac{[\ncal+\dragf^2(1-\vphii^2)+\rhorep^2\sigrep^2(1-\vphij^2)
  +2\dragf\rhorep\sigrep(\epsva-\vphii\vphij)]^{1/2}}
  {\lcal-\betrep+\dragf\vphii+\rhorep\sigrep\vphij}
\end{align}
and by \eqnref{main2a}, \eqnref{main2b}, \eqnref{main5a}, \eqnref{main6}, \eqnref{ogrv2a}, \eqnref{ogrv3} and
\eqnref{ogrv2e}, we have
\begin{align}\label{ogrv9c}
\begin{split}
&\qquad\qquad\qquad\rhorep=\frac{\xcons}{4\dragf\pfreq},\quad
\xcons=\frac{\vthtrep}{\sqrt{1+\vthtrep^2}},\quad
\dragf=\gamrep\left\{\frac{1+\sqrt{1+\vthtrep^2}}{2}\right\}^{1/2}\\
&\vthtrep=\alprep/(\gamrep\pfreq)^2,\quad
\alprep=-\kaprep\vphia\epsva,\quad
\gamrep=\vphie,\quad
\betrep=\vphih/\scalc,\quad
\sigrep=\vphia/\scalc,\quad
\cdkt=\scalc\dragf-\vphig.
\end{split}
\end{align}
\end{subequations}
Equations \eqnref{ogrv1} and \eqnref{ogrv9} give a complete prescription for
calculating $\psirep$ for a gravitating observer.

\art{Apparent drift of a light source}
To evaluate \eqnref{kas7} for a gravitating observer, it is convenient to introduce the
following quantities in addition to those given by \eqnref{ogrv1} and \eqnref{ogrv9}
\begin{subequations}\label{gxpeed1}
\begin{align}\label{gxpeed1a}
\begin{split}
&\dltva=\dprod{\unitkap}{\vecth},\quad
\dltvb=\dprod{\unitkap}{\vectz},\quad
\dltvc=\dprod{\unitplz}{\vecth},\quad
\dltvd=\dprod{\unitplz}{\vectz},\quad
\dltve=\dprod{\unitkap}{(\cprod{\vecth}{\unitplz})},\quad
\dltvf=\dprod{\unitpos}{(\cprod{\unitkap}{\vectz})}
\end{split}
\end{align}
\begin{align}\label{gxpeed1b}
\begin{split}
&\efkta=\dltvb+\scalq\epsva,\quad
\efktb=\epsvd+\scalq,\quad
\efktc=\dltvd+\scalq\epsvc,\quad
\efktd=\scalh^{-2}(\cdkt\dltva+\epsve\vphif\dltvc)\\
&\efkte=\scalh^{-2}(\cdkt\efkta-\rhorep\vphia\efktb-\scalr\epsvb\vphif\epsvh+\epsve\vphif\efktc),\quad
\efktf=\epsve[\vphif\dltve+(\rhorep\vphia/\scalr)]\\
&\efktg=\scalr\epsvb\vphif\scalh^2-\scalq,\quad
\efkth=\efktf-\dltvb+\epsva\efktg,\quad
\efkti=(1+\vthtrep^2)^{1/2},\quad
\efktj=\dragf/(\gamrep^2\pfreq)\\
&\efktk=(\efktj-\rhorep\vthtrep\efkti^2)/(4\dragf^2\pfreq^2\efkti^3),\quad
\efktl=(2\kaprep\vphia\epsva\efktk-\rhorep)/(2\scalr^2\kaprep^2\epsvf\vphid),\quad
\efktm=\rhorep/\scalr^2\\
\end{split}
\end{align}
\begin{align}\label{gxpeed1c}
\begin{split}
&\efktn=\efktl\epsve\epsvb+(\efktm/\kaprep^2),\quad
\efkto=2(\efktm/\kaprep)-2\vphia\epsva(\efktk/\scalr^2)-2\kaprep\epsvf\efktl\vphic^2,\quad
\efktp=(\betrep\scalc^2\fcal)^{-1}\\
&\efktq=\fcal/\scalh^2,\quad
\efktr=\gcal\efktp(\efkte\efkta+\efktd\vphih^2\dltva)-\efktq(\epsvg+\vphib\epsve),\quad
\efkts=-\vphia(\gcal\efkte\efktb\efktp-\epsvh\efktq)\\
&\efktt=\kaprep^2\efktq(\efktl\epsve\dltvc\dltva-2\efktn\scalh^2),\quad
\efktu=\kaprep^2\efktq\efktl\dltvc\epsve\efkta,\quad
\efktv=2\kaprep^2\gcal\efktp\efkte\efktn\\
&\efktw=\kaprep^2\gcal\efktp\efkte\efktl\epsve\dltvc,\quad
\efktx=\kaprep^2\gcal\efktp\efktd\vphih^2\efktl\epsve\dltvc.
\end{split}
\end{align}
\end{subequations}

\subart{Preliminary calculations}
We derive, in view of the above quantities,
\begin{subequations}\label{gxpeed2}
\begin{align}\label{gxpeed2a}
\cprod{\vectu}{\unitkap}
&=-\cprod{\unitkap}{\vectu}
=-\scalh^{-2}\cprod{\unitkap}{(\cprod{\vectz}{\vecth}+\cprod{\scalq\unitpos}{\vecth})}
  \beqref{main5c}\nonumber\\
&=-\scalh^{-2}[\cprod{\unitkap}{(\cprod{\vectz}{\vecth})}
  +\scalq\cprod{\unitkap}{(\cprod{\unitpos}{\vecth})}]\nonumber\\
&=-\scalh^{-2}[\vectz(\dprod{\unitkap}{\vecth})-\vecth(\dprod{\unitkap}{\vectz})]
  -\scalq\scalh^{-2}[\unitpos(\dprod{\unitkap}{\vecth})-\vecth(\dprod{\unitkap}{\unitpos})]
  \beqref{alg1}\nonumber\\
&=-\scalh^{-2}(\dltva\vectz-\dltvb\vecth)
  -\scalq\scalh^{-2}(\dltva\unitpos-\epsva\vecth)
  \beqref{gxpeed1a}\text{ \& }\eqnref{ogrv1a}\nonumber\\
&=-\scalh^{-2}[\dltva\vectz-\dltvb\vecth+\scalq(\dltva\unitpos-\epsva\vecth)]\nonumber\\
&=\scalh^{-2}[(\dltvb+\scalq\epsva)\vecth-\dltva\vectz-\scalq\dltva\unitpos]
=\scalh^{-2}(\efkta\vecth-\dltva\vectz-\scalq\dltva\unitpos)\beqref{gxpeed1b}
\end{align}
\begin{align}\label{gxpeed2b}
\cprod{\vectu}{\vecta}&=-\cprod{\vecta}{\vectu}
=-\scalh^{-2}\cprod{(-\vphia\unitpos)}{(\cprod{\vectz}{\vecth}+\cprod{\scalq\unitpos}{\vecth})}
  \beqref{main5a}, \eqnref{ogrv1b}\text{ \& }\eqnref{main5c}\nonumber\\
&=\vphia\scalh^{-2}[\cprod{\unitpos}{(\cprod{\vectz}{\vecth})}
  +\scalq\cprod{\unitpos}{(\cprod{\unitpos}{\vecth})}]\nonumber\\
&=\vphia\scalh^{-2}[\vectz(\dprod{\unitpos}{\vecth})-\vecth(\dprod{\unitpos}{\vectz})]
  +\scalq\vphia\scalh^{-2}[\unitpos(\dprod{\unitpos}{\vecth})-\vecth(\dprod{\unitpos}{\unitpos})]
  \beqref{alg1}\nonumber\\
&=-\vphia\scalh^{-2}\epsvd\vecth-\scalq\vphia\scalh^{-2}\vecth
  \beqref{main5c}\text{ \& }\eqnref{ogrv1a}\nonumber\\
&=-\vphia\scalh^{-2}(\epsvd+\scalq)\vecth
=-\scalh^{-2}\vphia\efktb\vecth\beqref{gxpeed1b}
\end{align}
\begin{align}\label{gxpeed2c}
\cprod{\vectu}{(\cprod{\vectr}{\vecth})}
&=\vectr(\dprod{\vectu}{\vecth})-\vecth(\dprod{\vectu}{\vectr})
=-\vecth(\dprod{\vectu}{\vectr})\beqref{alg1}\text{ \& }\eqnref{main5c}\nonumber\\
&=-\scalh^{-2}[\dprod{\vectr}{(\cprod{\vectz}{\vecth}+\cprod{\scalq\unitpos}{\vecth})}]\vecth
  \beqref{main5c}\nonumber\\
&=-\scalr\scalh^{-2}[\dprod{\unitpos}{(\cprod{\vectz}{\vecth})}]\vecth
=-\scalh^{-2}\scalr\epsvh\vecth\beqref{ogrv1a}
\end{align}
\begin{align}\label{gxpeed2d}
\cprod{\vectu}{\unitplz}&=-\cprod{\unitplz}{\vectu}
=-\scalh^{-2}\cprod{\unitplz}{(\cprod{\vectz}{\vecth}+\cprod{\scalq\unitpos}{\vecth})}
  \beqref{main5c}\nonumber\\
&=-\scalh^{-2}[\cprod{\unitplz}{(\cprod{\vectz}{\vecth})}]
  -\scalq\scalh^{-2}[\cprod{\unitplz}{(\cprod{\unitpos}{\vecth})}]\nonumber\\
&=-\scalh^{-2}[\vectz(\dprod{\unitplz}{\vecth})-\vecth(\dprod{\unitplz}{\vectz})]
  -\scalq\scalh^{-2}[\unitpos(\dprod{\unitplz}{\vecth})-\vecth(\dprod{\unitplz}{\unitpos})]
  \beqref{alg1}\nonumber\\
&=-\scalh^{-2}(\dltvc\vectz-\dltvd\vecth)-\scalq\scalh^{-2}(\dltvc\unitpos-\epsvc\vecth)
  \beqref{ogrv1a}\text{ \& }\eqnref{gxpeed1a}\nonumber\\
&=-\scalh^{-2}[\dltvc\vectz-\dltvd\vecth+\scalq(\dltvc\unitpos-\epsvc\vecth)]\nonumber\\
&=\scalh^{-2}[(\dltvd+\scalq\epsvc)\vecth-\dltvc\vectz-\scalq\dltvc\unitpos]
=\scalh^{-2}(\efktc\vecth-\dltvc\vectz-\scalq\dltvc\unitpos)\beqref{gxpeed1b}.
\end{align}
\end{subequations}
\begin{subequations}\label{gxpeed3}
\begin{align*}
\cprod{\vectu}{\vectups}
&=\cprod{\vectu}{(\cdkt\unitkap+\rhorep\vecta-\vectu+\vecte)}
  \beqref{kpath1c}\nonumber\\
&=\cdkt(\cprod{\vectu}{\unitkap})+\rhorep(\cprod{\vectu}{\vecta})
  +(\cprod{\vectu}{\vecte})\nonumber\\
\begin{split}
&=\cdkt\scalh^{-2}(\efkta\vecth-\dltva\vectz-\scalq\dltva\unitpos)
  +\rhorep(-\vphia\scalh^{-2}\efktb\vecth)
  +\cprod{\vectu}{[\fone(\cprod{\vectr}{\vecth})+\ftwo\unitplz]}\\
  &\quad\beqref{gxpeed2a}, \eqnref{gxpeed2b}\text{ \& }\eqnref{main5b}
\end{split}
\end{align*}
\begin{align}\label{gxpeed3a}
&=\scalh^{-2}(\cdkt\efkta-\rhorep\vphia\efktb)\vecth
  -\cdkt\scalh^{-2}(\dltva\vectz+\scalq\dltva\unitpos)
  +\fone[\cprod{\vectu}{(\cprod{\vectr}{\vecth})}]
  +\ftwo(\cprod{\vectu}{\unitplz})
\nonumber\\
\begin{split}
&=\scalh^{-2}(\cdkt\efkta-\rhorep\vphia\efktb)\vecth
  -\cdkt\scalh^{-2}(\dltva\vectz+\scalq\dltva\unitpos)
  +\fone(-\scalr\scalh^{-2}\epsvh\vecth)\\
  &\quad+\ftwo\scalh^{-2}(\efktc\vecth-\dltvc\vectz-\scalq\dltvc\unitpos)
  \beqref{gxpeed2c}\text{ \& }\eqnref{gxpeed2d}
\end{split}
\nonumber\\
&=\scalh^{-2}(\cdkt\efkta-\rhorep\vphia\efktb-\fone\scalr\epsvh+\ftwo\efktc)\vecth
  -\scalh^{-2}(\cdkt\dltva+\ftwo\dltvc)(\vectz+\scalq\unitpos)
\nonumber\\
\begin{split}
&=\scalh^{-2}(\cdkt\efkta-\rhorep\vphia\efktb-\scalr\epsvb\vphif\epsvh+\epsve\vphif\efktc)\vecth
  -\scalh^{-2}(\cdkt\dltva+\epsve\vphif\dltvc)(\vectz+\scalq\unitpos)\\
  &\quad\beqref{ogrv2g}\text{ \& }\eqnref{ogrv2h}
\end{split}
\nonumber\\
&=\efkte\vecth-\efktd(\vectz+\scalq\unitpos)\beqref{gxpeed1b}
\end{align}
\begin{align}\label{gxpeed3b}
\cprod{\vecth}{\vectups}
&=\cprod{\vecth}{(\cdkt\unitkap+\rhorep\vecta-\vectu+\vecte)}
  \beqref{kpath1c}\nonumber\\
&=\cdkt(\cprod{\vecth}{\unitkap})+\rhorep(\cprod{\vecth}{\vecta})-(\cprod{\vecth}{\vectu})+(\cprod{\vecth}{\vecte})
  \nonumber\\
\begin{split}
&=\cdkt(\cprod{\vecth}{\unitkap})+\rhorep[\cprod{\vecth}{(-\vphia\unitpos)}]
  -\scalh^{-2}[\cprod{\vecth}{(\cprod{\vectz}{\vecth}+\cprod{\scalq\unitpos}{\vecth})}]\\
  &\quad+\cprod{\vecth}{[\fone(\cprod{\vectr}{\vecth})+\ftwo\unitplz]}
  \beqref{main5a}, \eqnref{ogrv1b}, \eqnref{main5b}\text{ \& }\eqnref{main5c}
\end{split}
\nonumber\\
\begin{split}
&=\cdkt(\cprod{\vecth}{\unitkap})-\rhorep\vphia(\cprod{\vecth}{\unitpos})
  -\scalh^{-2}[\cprod{\vecth}{(\cprod{\vectz}{\vecth})}]
  -\scalq\scalh^{-2}[\cprod{\vecth}{(\cprod{\unitpos}{\vecth})}]\\
  &\quad+\fone[\cprod{\vecth}{(\cprod{\vectr}{\vecth})}]
  +\ftwo(\cprod{\vecth}{\unitplz})
\end{split}
\nonumber\\
\begin{split}
&=\cdkt(\cprod{\vecth}{\unitkap})-\rhorep\vphia(\cprod{\vecth}{\unitpos})
  -\scalh^{-2}[\vectz\scalh^2-\vecth(\dprod{\vecth}{\vectz})]
  -\scalh^{-2}\scalq[\unitpos\scalh^2-\vecth(\dprod{\vecth}{\unitpos})]\\
  &\quad+\fone[\vectr\scalh^2-\vecth(\dprod{\vecth}{\vectr})]
  +\ftwo(\cprod{\vecth}{\unitplz})\beqref{alg1}
\end{split}
\nonumber\\
&=\cdkt(\cprod{\vecth}{\unitkap})-\rhorep\vphia(\cprod{\vecth}{\unitpos})-\vectz-\scalq\unitpos
  +\fone\scalh^2\vectr+\ftwo(\cprod{\vecth}{\unitplz})\beqref{main5c}\nonumber\\
\begin{split}
&=-\vectz+(\scalr\epsvb\vphif\scalh^2-\scalq)\unitpos+\cdkt(\cprod{\vecth}{\unitkap})
  -\rhorep\vphia(\cprod{\vecth}{\unitpos})+\epsve\vphif(\cprod{\vecth}{\unitplz})\\
  &\quad\beqref{ogrv2g}\text{ \& }\eqnref{ogrv2h}
\end{split}
\nonumber\\
&=-\vectz+\efktg\unitpos+\cdkt(\cprod{\vecth}{\unitkap})-\rhorep\vphia(\cprod{\vecth}{\unitpos})
  +\epsve\vphif(\cprod{\vecth}{\unitplz})\beqref{gxpeed1b}
\end{align}
\begin{align}\label{gxpeed3c}
\begin{split}
\dprod{\unitkap}{(\cprod{\vecth}{\vectups})}
&=-\dprod{\unitkap}{\vectz}
  +\efktg(\dprod{\unitkap}{\unitpos})
  +\cdkt[\dprod{\unitkap}{(\cprod{\vecth}{\unitkap})}]
  -\rhorep\vphia[\dprod{\unitkap}{(\cprod{\vecth}{\unitpos})}]\\
  &\quad+\epsve\vphif[\dprod{\unitkap}{(\cprod{\vecth}{\unitplz})}]
  \beqref{gxpeed3b}
\end{split}
\nonumber\\
&=-\dltvb+\efktg\epsva+(\rhorep\vphia/\scalr)\epsve+\epsve\vphif\dltve
  \beqref{ogrv1a}\text{ \& }\eqnref{gxpeed1a}\nonumber\\
&=-\dltvb+\efktg\epsva+\epsve[\vphif\dltve+(\rhorep\vphia/\scalr)]\nonumber\\
&=\efktf-\dltvb+\epsva\efktg\beqref{gxpeed1b}\nonumber\\
&=\efkth\beqref{gxpeed1b}
\end{align}
\end{subequations}
\begin{align}\label{gxpeed4}
\scalq\unitpos
&=\cprod{\vecth}{\vectu}-\vectz\beqref{main5c}\nonumber\\
\therefore\vecta
&=-(\scalq/\scalr^2)\unitpos=\scalr^{-2}(\vectz-\cprod{\vecth}{\vectu})\beqref{main5a}.
\end{align}

\subart{Development of equation \eqnref{kas2b}}
The various quantities defined by \eqnref{kas2b} evaluate as
\begin{subequations}\label{gxpeed5}
\begin{align}\label{gxpeed5a}
\vga
&=\ugdiv{\vectkap}{\vecta}\beqref{kas2b}\nonumber\\
&=\ugdiv{\vectkap}{[\scalr^{-2}(\vectz-\cprod{\vecth}{\vectu})]}
  \beqref{gxpeed4}\nonumber\\
&=\scalr^{-2}\ugdiv{\vectkap}{(\vectz-\cprod{\vecth}{\vectu})}
=-\scalr^{-2}\ugdiv{\vectkap}{(\cprod{\vecth}{\vectu})}\beqref{clc42}\nonumber\\
&=-\scalr^{-2}\{\cprod{\vecth}{[\ugdiv{\vectkap}{\vectu}]}-\cprod{\vectu}{[\ugdiv{\vectkap}{\vecth}]}\}
  \beqref{clc43}\nonumber\\
&=-\scalr^{-2}\{\cprod{\vecth}{[\ugdiv{\vectkap}{\vectu}]}\}
=\scalr^{-2}(\cprod{\vectkap}{\vecth})\beqref{clc41}
\end{align}
\begin{align}\label{gxpeed5b}
\vgb
&=\ucurl{\vecta}\beqref{kas2b}\nonumber\\
&=\ucurl{[\scalr^{-2}(\vectz+\cprod{\vectu}{\vecth})]}\beqref{gxpeed4}\nonumber\\
&=\scalr^{-2}[\ucurl{(\vectz+\cprod{\vectu}{\vecth})}]
=\scalr^{-2}[\ucurl{(\cprod{\vectu}{\vecth})}]\beqref{clc42}\nonumber\\
&=\scalr^{-2}[\vectu(\udivg{\vecth})-\ugdiv{\vectu}{\vecth}+\ugdiv{\vecth}{\vectu}-\vecth(\udivg{\vectu})]
  \beqref{clc25}\nonumber\\
&=\scalr^{-2}[\ugdiv{\vecth}{\vectu}-\vecth(\udivg{\vectu})]
=\scalr^{-2}(\vecth-3\vecth)\beqref{clc41}\text{ \& }\eqnref{clc11}\nonumber\\
&=-2\scalr^{-2}\vecth
\end{align}
\begin{align}\label{gxpeed5c}
\vgc
&=\ugdiv{\vectu}{\vecta}\beqref{kas2b}\nonumber\\
&=\ugdiv{\vectu}{[\scalr^{-2}(\vectz+\cprod{\vectu}{\vecth})]}\beqref{gxpeed4}\nonumber\\
&=\scalr^{-2}[\ugdiv{\vectu}{(\vectz+\cprod{\vectu}{\vecth})}]
=\scalr^{-2}[\ugdiv{\vectu}{(\cprod{\vectu}{\vecth})}]\beqref{clc42}\nonumber\\
&=\scalr^{-2}\{\cprod{\vectu}{[\ugdiv{\vectu}{\vecth}]}-\cprod{\vecth}{[\ugdiv{\vectu}{\vectu}]}\}
  \beqref{clc43}\nonumber\\
&=-\scalr^{-2}\{\cprod{\vecth}{[\ugdiv{\vectu}{\vectu}]}\}
=-\scalr^{-2}(\cprod{\vecth}{\vectu})\beqref{clc41}\nonumber\\
&=-\scalr^{-2}(\vectz+\scalq\unitpos)\beqref{main5c}
\end{align}
\begin{align}\label{gxpeed5d}
\vgd
&=\ugdiv{\vectups}{\vecta}\beqref{kas2b}\nonumber\\
&=\ugdiv{\vectups}{[\scalr^{-2}(\vectz+\cprod{\vectu}{\vecth})]}
  \beqref{gxpeed4}\nonumber\\
&=\scalr^{-2}[\ugdiv{\vectups}{(\vectz+\cprod{\vectu}{\vecth})}]
=\scalr^{-2}[\ugdiv{\vectups}{(\cprod{\vectu}{\vecth})}]
  \beqref{clc42}\nonumber\\
&=\scalr^{-2}\{\cprod{\vectu}{[\ugdiv{\vectups}{\vecth}]}
  -\cprod{\vecth}{[\ugdiv{\vectups}{\vectu}]}\}\beqref{clc43}\nonumber\\
&=-\scalr^{-2}\{\cprod{\vecth}{[\ugdiv{\vectups}{\vectu}]}\}
=-\scalr^{-2}(\cprod{\vecth}{\vectups})\beqref{clc41}
\end{align}
\begin{align}\label{gxpeed5e}
\vge
&=\ugrad{\gamrep}\beqref{kas2b}\nonumber\\
&=\ugrad{\left|1-\frac{\kcons^2}{\kaprep^2}\right|^{1/2}}
  \beqref{main5a}\nonumber\\
&=0\beqref{main5c}
\end{align}
\begin{align}
\vgf
&=\vga+\cprod{\vectkap}{\vgb}\beqref{kas2b}\nonumber\\
&=\scalr^{-2}(\cprod{\vectkap}{\vecth})-2\scalr^{-2}(\cprod{\vectkap}{\vecth})
  \beqref{gxpeed5a}\text{ \& }\eqnref{gxpeed5b}\nonumber\\
&=\scalr^{-2}(\cprod{\vecth}{\vectkap})
\label{gxpeed5f}\\
\vgg
&=\vchb\vgf-\vchc\vge\beqref{kas2b}\nonumber\\
&=\vchb\scalr^{-2}(\cprod{\vecth}{\vectkap})
  \beqref{gxpeed5f}\text{ \& }\eqnref{gxpeed5e}
\label{gxpeed5g}
\end{align}
\begin{align}
\vgh
&=\vcha\vgf-\vchf\vge\beqref{kas2b}\nonumber\\
&=\vcha\scalr^{-2}(\cprod{\vecth}{\vectkap})
  \beqref{gxpeed5f}\text{ \& }\eqnref{gxpeed5e}
\label{gxpeed5h}\\
\vgi
&=\vchg\vgf-\vchh\vge\beqref{kas2b}\nonumber\\
&=\vchg\scalr^{-2}(\cprod{\vecth}{\vectkap})
  \beqref{gxpeed5f}\text{ \& }\eqnref{gxpeed5e}
\label{gxpeed5i}
\end{align}
\end{subequations}
\begin{align}\label{gxpeed6}
\vchg
&=\vche\left[\frac{\vchb}{\vchd^2}-\frac{\vthtrep\vcha}{\dragf}\right]
=\frac{(1+\vthtrep^2)^{-1/2}}{4\dragf\pfreq}\left[
  \frac{1}{(1+\vthtrep^2)\gamrep^2\pfreq^2}-\frac{\vthtrep\rhorep}{\dragf\pfreq}\right]
  \beqref{kas2a}\nonumber\\
&=\frac{(1+\vthtrep^2)^{-3/2}}{4\dragf\pfreq}\left[
  \frac{1}{\gamrep^2\pfreq^2}-\frac{\rhorep\vthtrep(1+\vthtrep^2)}{\dragf\pfreq}\right]
=\frac{1}{4\dragf^2\pfreq^2\efkti^3}\left[\frac{\dragf}{\gamrep^2\pfreq}-\rhorep\vthtrep\efkti^2\right]
  \beqref{gxpeed1b}\nonumber\\
&=(\efktj-\rhorep\vthtrep\efkti^2)/(4\dragf^2\pfreq^2\efkti^3)
=\efktk\beqref{gxpeed1b}
\end{align}
\begin{subequations}\label{gxpeed7}
\begin{align}\label{gxpeed7a}
\ugrad{\fzer}
&=\ugrad{\left[\frac{2\rhorep\alprep-\dragf\pfreq}{\scaln(\kaprep^2-\kcons^2)}
  \right]}\beqref{main5d}\nonumber\\
&=[\scaln(\kaprep^2-\kcons^2)]^{-1}\ugrad{(2\rhorep\alprep-\dragf\pfreq)}
  \beqref{clc33}\text{ \& }\eqnref{main5c}\nonumber\\
&=[\scaln(\kaprep^2-\kcons^2)]^{-1}(2\rhorep\ugrad{\alprep}+2\alprep\ugrad{\rhorep}
  -\pfreq\ugrad{\dragf})\beqref{clc33}\nonumber\\
&=[\scaln(\kaprep^2-\kcons^2)]^{-1}(2\rhorep\vgf+2\alprep\vgi
  -\pfreq\vgh)\beqref{kas3}\nonumber\\
&=\scalr^{-2}[\scaln(\kaprep^2-\kcons^2)]^{-1}(2\rhorep+2\alprep\vchg
  -\pfreq\vcha)(\cprod{\vecth}{\vectkap})\beqref{gxpeed5}\nonumber\\
&=\scalr^{-2}[\scaln(\kaprep^2-\kaprep^2\vphic^2)]^{-1}(\rhorep+2\alprep\vchg)
  (\cprod{\vecth}{\vectkap})\beqref{kas2a}\text{ \& }\eqnref{ogrv2d}\nonumber\\
&=[\scalr^2\kaprep^2\epsvf(1-\vphic^2)]^{-1}(\rhorep-2\kaprep\vphia\epsva\vchg)
  (\cprod{\vecth}{\vectkap})\beqref{ogrv2a}\text{ \& }\eqnref{ogrv2c}\nonumber\\
&=[(\rhorep-2\kaprep\vphia\epsva\efktk)/(-\scalr^2\kaprep^2\epsvf\vphid)]
  (\cprod{\vecth}{\vectkap})\beqref{ogrv1b}\text{ \& }\eqnref{gxpeed6}\nonumber\\
&=2\efktl(\cprod{\vecth}{\vectkap})\beqref{gxpeed1b}
\end{align}
\begin{align}
\ugrad{\fone}
&=\ugrad{[(1/2)(\dprod{\vectkap}{\unitplz})\fzer]}\beqref{main5d}\nonumber\\
&=(1/2)(\dprod{\vectkap}{\unitplz})\ugrad{\fzer}\beqref{clc32}\nonumber\\
&=(1/2)(\kaprep\epsvb)[2\efktl(\cprod{\vecth}{\vectkap})]\beqref{ogrv1a}\text{ \& }\eqnref{gxpeed7a}\nonumber\\
&=\kaprep\epsvb\efktl(\cprod{\vecth}{\vectkap})
\label{gxpeed7b}\\
\ugrad{\ftwo}
&=\ugrad{[(1/2)\scalm\fzer]}\beqref{main5d}\nonumber\\
&=(1/2)\scalm\ugrad{\fzer}\beqref{clc32}\nonumber\\
&=(1/2)(\kaprep\epsve)[2\efktl(\cprod{\vecth}{\vectkap})]
  \beqref{ogrv2b}\text{ \& }\eqnref{gxpeed7a}\nonumber\\
&=\kaprep\epsve\efktl(\cprod{\vecth}{\vectkap})
\label{gxpeed7c}.
\end{align}
\end{subequations}

\subart{Development of equation \eqnref{kas2c}}
The quantities defined by \eqnref{kas2c} become
\begin{subequations}\label{gxpeed8}
\begin{align}\label{gxpeed8a}
\vja
&=\ucurl{\vecte}\beqref{kas2c}\nonumber\\
&=\curl{[\fone(\cprod{\vectr}{\vecth})+\ftwo\unitplz]}
  \beqref{main5b}\nonumber\\
&=\fone[\ucurl{(\cprod{\vectr}{\vecth})}]-\cprod{(\cprod{\vectr}{\vecth})}{(\ugrad{\fone})}
  +\ftwo(\ucurl{\unitplz})-\cprod{\unitplz}{(\ugrad{\ftwo})}
  \beqref{clc24}\nonumber\\
&=-\cprod{(\cprod{\vectr}{\vecth})}{(\ugrad{\fone})}-\cprod{\unitplz}{(\ugrad{\ftwo})}
  \nonumber\\
&=-\cprod{(\cprod{\vectr}{\vecth})}{[\kaprep\epsvb\efktl(\cprod{\vecth}{\vectkap})]}
  -\cprod{\unitplz}{[\kaprep\epsve\efktl(\cprod{\vecth}{\vectkap})]}
  \beqref{gxpeed7b}\text{ \& }\eqnref{gxpeed7c}\nonumber\\
&=-\kaprep\epsvb\efktl[\vecth(\dprod{\vectkap}{(\cprod{\vectr}{\vecth})})-\vectkap(\dprod{\vecth}{(\cprod{\vectr}{\vecth})})]
  -\kaprep\epsve\efktl[\vecth(\dprod{\unitplz}{\vectkap})-\vectkap(\dprod{\unitplz}{\vecth})]
  \beqref{alg1}\text{ \& }\eqnref{alg5}\nonumber\\
&=-\kaprep^2\epsvb\efktl[\vecth(\dprod{\unitkap}{(\cprod{\vectr}{\vecth})})]
  -\kaprep^2\epsve\efktl[\vecth(\dprod{\unitplz}{\unitkap})-\unitkap(\dprod{\unitplz}{\vecth})]\nonumber\\
&=-\kaprep^2\epsvb\efktl\epsve\vecth-\kaprep^2\epsve\efktl(\epsvb\vecth-\dltvc\unitkap)
  \beqref{ogrv1a}\text{ \& }\eqnref{gxpeed1a}\nonumber\\
&=-2\kaprep^2\efktl\epsve\epsvb\vecth+\kaprep^2\efktl\epsve\dltvc\unitkap
\end{align}
\begin{align}\label{gxpeed8b}
\vjb
&=\ugdiv{\vectu}{\vecte}\beqref{kas2c}\nonumber\\
&=\ugdiv{\vectu}{[\fone(\cprod{\vectr}{\vecth})+\ftwo\unitplz]}
  \beqref{main5b}\nonumber\\
&=(\cprod{\vectr}{\vecth})[\dprod{\vectu}{(\ugrad{\fone})}]+\fone[\ugdiv{\vectu}{(\cprod{\vectr}{\vecth})}]
  +\unitplz[\dprod{\vectu}{(\ugrad{\ftwo})}]+\ftwo[\ugdiv{\vectu}{\unitplz}]
  \beqref{clc42}\nonumber\\
&=(\cprod{\vectr}{\vecth})[\dprod{\vectu}{(\ugrad{\fone})}]
  +\unitplz[\dprod{\vectu}{(\ugrad{\ftwo})}]\nonumber\\
&=\kaprep\epsvb\efktl(\cprod{\vectr}{\vecth})[\dprod{\vectu}{(\cprod{\vecth}{\vectkap})}]
  +\kaprep\epsve\efktl\unitplz[\dprod{\vectu}{(\cprod{\vecth}{\vectkap})}]
  \beqref{gxpeed7}\nonumber\\
&=\kaprep\efktl[\dprod{\vectkap}{(\cprod{\vectu}{\vecth})}][\epsvb(\cprod{\vectr}{\vecth})+\epsve\unitplz]
  \beqref{alg4}\nonumber\\
&=\kaprep\efktl[\dprod{\vectkap}{(-\vectz-\scalq\unitpos)}][\epsvb(\cprod{\vectr}{\vecth})+\epsve\unitplz]
  \beqref{main5c}\nonumber\\
&=-\kaprep^2\efktl(\dltvb +\scalq\epsva)[\epsvb(\cprod{\vectr}{\vecth})+\epsve\unitplz]
  \beqref{ogrv1a}\text{ \& }\eqnref{gxpeed1a}\nonumber\\
&=-\kaprep^2\efktl\efkta[\epsve\unitplz+\epsvb(\cprod{\vectr}{\vecth})]\beqref{gxpeed1b}
\end{align}
\begin{align}\label{gxpeed8c}
\vjc
&=\ugdiv{\vectups}{\vecte}\beqref{kas2c}\nonumber\\
&=\ugdiv{\vectups}{[\fone(\cprod{\vectr}{\vecth})+\ftwo\unitplz]}
  \beqref{main5b}\nonumber\\
\begin{split}
&=(\cprod{\vectr}{\vecth})[\dprod{\vectups}{(\ugrad{\fone})}]+\fone[\ugdiv{\vectups}{(\cprod{\vectr}{\vecth})}]
  +\unitplz[\dprod{\vectups}{(\ugrad{\ftwo})}]\\
  &\quad+\ftwo[\ugdiv{\vectups}{\unitplz}]\beqref{clc42}
\end{split}
\nonumber\\
&=(\cprod{\vectr}{\vecth})[\dprod{\vectups}{(\ugrad{\fone})}]
  +\unitplz[\dprod{\vectups}{(\ugrad{\ftwo})}]\nonumber\\
&=\kaprep\epsvb\efktl(\cprod{\vectr}{\vecth})[\dprod{\vectups}{(\cprod{\vecth}{\vectkap})}]
  +\kaprep\epsve\efktl\unitplz[\dprod{\vectups}{(\cprod{\vecth}{\vectkap})}]
  \beqref{gxpeed7}\nonumber\\
&=-\kaprep\efktl[\epsve\unitplz+\epsvb(\cprod{\vectr}{\vecth})][\dprod{\vectkap}{(\cprod{\vecth}{\vectups})}]
  \beqref{alg4}\nonumber\\
&=-\kaprep^2\efktl\efkth[\epsve\unitplz+\epsvb(\cprod{\vectr}{\vecth})]\beqref{gxpeed3c}
\end{align}
\begin{align}\label{gxpeed8d}
\vjd
&=\vja+\rhorep\vgb\beqref{kas2c}\nonumber\\
&=-2\kaprep^2\efktl\epsve\epsvb\vecth+\kaprep^2\efktl\epsve\dltvc\unitkap
  +\rhorep(-2\scalr^{-2}\vecth)\beqref{gxpeed5b}\text{ \& }\eqnref{gxpeed8a}\nonumber\\
&=-2(\kaprep^2\efktl\epsve\epsvb+\efktm)\vecth
  +\kaprep^2\efktl\epsve\dltvc\unitkap\beqref{gxpeed1b}\nonumber\\
&=-2\kaprep^2\efktn\vecth+\kaprep^2\efktl\epsve\dltvc\unitkap\beqref{gxpeed1c}
\end{align}
\begin{align}\label{gxpeed8e}
\vje
&=\vjb+\rhorep\vgc\beqref{kas2c}\nonumber\\
&=-\kaprep^2\efktl\efkta[\epsve\unitplz+\epsvb(\cprod{\vectr}{\vecth})]
  +\rhorep[-\scalr^{-2}(\vectz+\scalq\unitpos)]
  \beqref{gxpeed8b}\text{ \& }\eqnref{gxpeed5c}\nonumber\\
&=-\kaprep^2\efktl\efkta\epsve\unitplz-\efktm\vectz-\scalq\efktm\unitpos
  -\kaprep^2\efktl\efkta\epsvb(\cprod{\vectr}{\vecth})
\end{align}
\begin{align}\label{gxpeed8f}
\vjf
&=\vjc+\rhorep\vgd\beqref{kas2c}\nonumber\\
&=-\kaprep^2\efktl\efkth[\epsve\unitplz+\epsvb(\cprod{\vectr}{\vecth})]
  +\rhorep[-\scalr^{-2}(\cprod{\vecth}{\vectups})]
  \beqref{gxpeed8c}\text{ \& }\eqnref{gxpeed5d}\nonumber\\
&=-\kaprep^2\efktl\efkth\epsve\unitplz-\kaprep^2\efktl\efkth\epsvb(\cprod{\vectr}{\vecth})
  -\efktm(\cprod{\vecth}{\vectups})\beqref{gxpeed1c}
\end{align}
\begin{align}\label{gxpeed8g}
\vjg
&=\ugrad{\taurep}\beqref{kas2c}\nonumber\\
&=\ugrad{[\kaprep^{-2}(2\rhorep\alprep-\scaln\kcons^2\fzer)]}
  \beqref{main5b}\nonumber\\
&=\kaprep^{-2}(2\rhorep\ugrad{\alprep}+2\alprep\ugrad{\rhorep}
   -\scaln\kcons^2\ugrad{\fzer})\beqref{clc33}\nonumber\\
&=\kaprep^{-2}[2\rhorep\vgf+2\alprep\vgi-2\scaln\kcons^2\efktl(\cprod{\vecth}{\vectkap})]
  \beqref{kas3}\text{ \& }\eqnref{gxpeed7a}\nonumber\\
&=\kaprep^{-2}[2\rhorep\scalr^{-2}(\cprod{\vecth}{\vectkap})
  +2\alprep\vchg\scalr^{-2}(\cprod{\vecth}{\vectkap})
  -2\scaln\kcons^2\efktl(\cprod{\vecth}{\vectkap})]
  \beqref{gxpeed5f}\text{ \& }\eqnref{gxpeed5i}\nonumber\\
&=\kaprep^{-2}(2\efktm-2\kaprep\vphia\epsva\vchg\scalr^{-2}-2\epsvf\kcons^2\efktl)(\cprod{\vecth}{\vectkap})
  \beqref{gxpeed1b}, \eqnref{ogrv2a}\text{ \& }\eqnref{ogrv2c}\nonumber\\
&=[2(\efktm/\kaprep)-2\vphia\epsva(\efktk/\scalr^2)-2\kaprep\epsvf\efktl\vphic^2](\cprod{\vecth}{\unitkap})
  \beqref{gxpeed6}\text{ \& }\eqnref{ogrv2d}\nonumber\\
&=\efkto(\cprod{\vecth}{\unitkap})\beqref{gxpeed1c}
\end{align}
\begin{align}\label{gxpeed8h}
\vjh
&=(\cprod{\vectu}{\vectups})/|\cprod{\vectu}{\vectups}|\beqref{kas2c}\nonumber\\
&=(\betrep\scalc^2\fcal)^{-1}(\cprod{\vectu}{\vectups})\beqref{grad5c}\nonumber\\
&=(\betrep\scalc^2\fcal)^{-1}(\efkte\vecth-\efktd\vectz-\scalq\efktd\unitpos)
  \beqref{gxpeed3a}\nonumber\\
&=\efktp(\efkte\vecth-\efktd\vectz-\scalq\efktd\unitpos)\beqref{gxpeed1c}
\end{align}
\begin{align*}
\vji
&=\cprod{\vjh}{\vectu}\beqref{kas2c}\nonumber\\
&=\cprod{[\efktp(\efkte\vecth-\efktd\vectz-\scalq\efktd\unitpos)]}{\vectu}
  \beqref{gxpeed8h}\nonumber\\
&=\efktp[\efkte(\cprod{\vecth}{\vectu})-\efktd(\cprod{\vectz}{\vectu})-\scalq\efktd(\cprod{\unitpos}{\vectu})]
  \nonumber\\
&=\efktp[\efkte(\vectz+\scalq\unitpos)-\efktd(\cprod{\vectz}{\vectu})+\scalq\efktd(\vecth/\scalr)]
  \beqref{main5c}\nonumber\\
&=\efktp[\efkte(\vectz+\scalq\unitpos)+\vphib\efktd\vecth+\efktd(\cprod{\vectu}{\vectz})]
  \beqref{ogrv1b}
\end{align*}
\begin{align}\label{gxpeed8i}
&=\efktp[\efkte(\vectz+\scalq\unitpos)+\vphib\efktd\vecth+\efktd(\cprod{\vectu}{\{\cprod{\vecth}{\vectu}-\scalq\unitpos\}})]
  \beqref{main5c}\nonumber\\
&=\efktp[\efkte(\vectz+\scalq\unitpos)+\vphib\efktd\vecth+\efktd\cprod{\vectu}{(\cprod{\vecth}{\vectu})}
  -\scalq\efktd(\cprod{\vectu}{\unitpos})]\nonumber\\
&=\efktp[\efkte(\vectz+\scalq\unitpos)+\vphib\efktd\vecth+\efktd\cprod{\vectu}{(\cprod{\vecth}{\vectu})}
  -(\scalq/\scalr)\efktd\vecth]\beqref{main5c}\nonumber\\
&=\efktp[\efkte(\vectz+\scalq\unitpos)+\vphib\efktd\vecth+\efktd\cprod{\vectu}{(\cprod{\vecth}{\vectu})}
  -\vphib\efktd\vecth]\beqref{ogrv1b}\nonumber\\
&=\efktp[\efkte(\vectz+\scalq\unitpos)+\efktd(\vecth\scalu^2-\vectu(\dprod{\vectu}{\vecth}))]
  \beqref{alg1}\nonumber\\
&=\efktp\efkte\vectz+\scalq\efktp\efkte\unitpos+\efktp\efktd\vphih^2\vecth
  \beqref{main5c}\text{ \& }\eqnref{ogrv3b}
\end{align}
\begin{align}\label{gxpeed8j}
\vjj
&=\gcal\vji-\fcal\vectu\beqref{kas2c}\nonumber\\
&=\gcal[\efktp\efkte\vectz+\scalq\efktp\efkte\unitpos+\efktp\efktd\vphih^2\vecth]-\fcal\scalh^{-2}
  [(\cprod{\vectz}{\vecth})+\scalq(\cprod{\unitpos}{\vecth})]
  \beqref{gxpeed8i}\text{ \& }\eqnref{main5c}\nonumber\\
&=\gcal\efktp\efkte\vectz+\scalq\gcal\efktp\efkte\unitpos+\gcal\efktp\efktd\vphih^2\vecth
  -\efktq(\cprod{\vectz}{\vecth})-\scalq\efktq(\cprod{\unitpos}{\vecth})\beqref{gxpeed1c}.
\end{align}
\end{subequations}
From the foregoing derivations, we obtain
\begin{subequations}\label{gxpeed9}
\begin{align}\label{gxpeed9a}
\dprod{\unitkap}{\vjj}
&=\dprod{\unitkap}{[\gcal\efktp\efkte\vectz+\scalq\gcal\efktp\efkte\unitpos+\gcal\efktp\efktd\vphih^2\vecth
  -\efktq(\cprod{\vectz}{\vecth})-\scalq\efktq(\cprod{\unitpos}{\vecth})]}
  \beqref{gxpeed8j}\nonumber\\
&=\gcal\efktp\efkte(\dprod{\unitkap}{\vectz})
  +\scalq\gcal\efktp\efkte(\dprod{\unitkap}{\unitpos})
  +\gcal\efktp\efktd\vphih^2(\dprod{\unitkap}{\vecth})
  -\efktq[\dprod{\unitkap}{(\cprod{\vectz}{\vecth})}]
  -\scalq\efktq[\dprod{\unitkap}{(\cprod{\unitpos}{\vecth})}]
\nonumber\\
&=\gcal\efktp\efkte\dltvb+\scalq\gcal\efktp\efkte\epsva+\gcal\efktp\efktd\vphih^2\dltva
  -\efktq\epsvg-(\scalq/\scalr)\efktq\epsve\beqref{gxpeed1a}\text{ \& }\eqnref{ogrv1a}\nonumber\\
&=\gcal\efktp[\efkte(\dltvb+\scalq\epsva)+\efktd\vphih^2\dltva]-\efktq(\epsvg+\vphib\epsve)
  \beqref{ogrv1b}\nonumber\\
&=\gcal\efktp(\efkte\efkta+\efktd\vphih^2\dltva)-\efktq(\epsvg+\vphib\epsve)
  \beqref{gxpeed1b}\nonumber\\
&=\efktr\beqref{gxpeed1c}
\end{align}
\begin{align}\label{gxpeed9b}
\begin{split}
&\dprod{\vectc}{\vjj}
=\scalc(\dprod{\unitkap}{\vjj})
=\scalc\efktr\beqref{gxpeed9a}\\
&\dprod{\vectkap}{\vjj}
=\kaprep(\dprod{\unitkap}{\vjj})
=\kaprep\efktr\beqref{gxpeed9a}
\end{split}
\end{align}
\begin{align}\label{gxpeed9c}
\dprod{\vecta}{\vjj}
&=\dprod{(-\vphia\unitpos)}{\vjj}\beqref{main5a}\text{ \& }\eqnref{ogrv1b}\nonumber\\
&=-\vphia\dprod{\unitpos}{[\gcal\efktp\efkte\vectz+\scalq\gcal\efktp\efkte\unitpos+\gcal\efktp\efktd\vphih^2\vecth
  -\efktq(\cprod{\vectz}{\vecth})-\scalq\efktq(\cprod{\unitpos}{\vecth})]}
  \beqref{gxpeed8j}\nonumber\\
\begin{split}
&=-\vphia[\gcal\efktp\efkte(\dprod{\unitpos}{\vectz})
  +\scalq\gcal\efktp\efkte(\dprod{\unitpos}{\unitpos})
  +\gcal\efktp\efktd\vphih^2(\dprod{\unitpos}{\vecth})
  -\efktq(\dprod{\unitpos}{(\cprod{\vectz}{\vecth})})
  -\scalq\efktq(\dprod{\unitpos}{(\cprod{\unitpos}{\vecth})})]
\end{split}
\nonumber\\
&=-\vphia(\gcal\efktp\efkte\epsvd+\scalq\gcal\efktp\efkte-\efktq\epsvh)
  \beqref{ogrv1a}\text{ \& }\eqnref{main5c}\nonumber\\
&=-\gcal\efktp\efkte\vphia(\epsvd+\scalq)+\vphia\efktq\epsvh
=-\gcal\efktp\efkte\vphia\efktb+\vphia\efktq\epsvh\beqref{gxpeed1b}\nonumber\\
&=-\vphia(\gcal\efkte\efktb\efktp-\epsvh\efktq)
=\efkts\beqref{gxpeed1c}
\end{align}
\begin{align*}
\begin{split}
\cprod{\vjj}{\vjd}
&=-\cprod{\vjd}{\vjj}
=-\cprod{(-2\kaprep^2\efktn\vecth+\kaprep^2\efktl\epsve\dltvc\unitkap)}{[}
  \gcal\efktp\efkte\vectz+\scalq\gcal\efktp\efkte\unitpos+\gcal\efktp\efktd\vphih^2\vecth\\
  &\quad-\efktq(\cprod{\vectz}{\vecth})-\scalq\efktq(\cprod{\unitpos}{\vecth})]
  \beqref{gxpeed8d}\text{ \& }\eqnref{gxpeed8j}
\end{split}
\nonumber\\
\begin{split}
&=2\gcal\efktp\efkte\kaprep^2\efktn(\cprod{\vecth}{\vectz})
  +2\scalq\gcal\efktp\efkte\kaprep^2\efktn(\cprod{\vecth}{\unitpos})
  +2\gcal\efktp\efktd\vphih^2\kaprep^2\efktn(\cprod{\vecth}{\vecth})\\
  &\quad-2\efktq\kaprep^2\efktn[\cprod{\vecth}{(\cprod{\vectz}{\vecth})}]
  -2\scalq\efktq\kaprep^2\efktn[\cprod{\vecth}{(\cprod{\unitpos}{\vecth})}]
  -\gcal\efktp\efkte\kaprep^2\efktl\epsve\dltvc(\cprod{\unitkap}{\vectz})\\
  &\quad-\scalq\gcal\efktp\efkte\kaprep^2\efktl\epsve\dltvc(\cprod{\unitkap}{\unitpos})
  -\gcal\efktp\efktd\vphih^2\kaprep^2\efktl\epsve\dltvc(\cprod{\unitkap}{\vecth})
  +\efktq\kaprep^2\efktl\epsve\dltvc[\cprod{\unitkap}{(\cprod{\vectz}{\vecth})}]\\
  &\quad+\scalq\efktq\kaprep^2\efktl\epsve\dltvc[\cprod{\unitkap}{(\cprod{\unitpos}{\vecth})}]
\end{split}
\end{align*}
\begin{align*}
\begin{split}
&=2\gcal\efktp\efkte\kaprep^2\efktn(\cprod{\vecth}{\vectz})
  +2\scalq\gcal\efktp\efkte\kaprep^2\efktn(\cprod{\vecth}{\unitpos})
  -\gcal\efktp\efkte\kaprep^2\efktl\epsve\dltvc(\cprod{\unitkap}{\vectz})\\
  &\quad-\scalq\gcal\efktp\efkte\kaprep^2\efktl\epsve\dltvc(\cprod{\unitkap}{\unitpos})
  -\gcal\efktp\efktd\vphih^2\kaprep^2\efktl\epsve\dltvc(\cprod{\unitkap}{\vecth})
  -2\efktq\kaprep^2\efktn[\cprod{\vecth}{(\cprod{\vectz}{\vecth})}]\\
  &\quad-2\scalq\efktq\kaprep^2\efktn[\cprod{\vecth}{(\cprod{\unitpos}{\vecth})}]
  +\efktq\kaprep^2\efktl\epsve\dltvc[\cprod{\unitkap}{(\cprod{\vectz}{\vecth})}]
  +\scalq\efktq\kaprep^2\efktl\epsve\dltvc[\cprod{\unitkap}{(\cprod{\unitpos}{\vecth})}]
\end{split}
\nonumber\\
\begin{split}
&=2\gcal\efktp\efkte\kaprep^2\efktn(\cprod{\vecth}{\vectz})
  +2\scalq\gcal\efktp\efkte\kaprep^2\efktn(\cprod{\vecth}{\unitpos})
  -\gcal\efktp\efkte\kaprep^2\efktl\epsve\dltvc(\cprod{\unitkap}{\vectz})
  -\scalq\gcal\efktp\efkte\kaprep^2\efktl\epsve\dltvc(\cprod{\unitkap}{\unitpos})\\
  &\quad-\gcal\efktp\efktd\vphih^2\kaprep^2\efktl\epsve\dltvc(\cprod{\unitkap}{\vecth})
  -2\efktq\kaprep^2\efktn[\vectz\scalh^2-\vecth(\dprod{\vecth}{\vectz})]
  -2\scalq\efktq\kaprep^2\efktn[\unitpos\scalh^2-\vecth(\dprod{\vecth}{\unitpos})]\\
  &\quad+\efktq\kaprep^2\efktl\epsve\dltvc[\vectz(\dprod{\unitkap}{\vecth})-\vecth(\dprod{\unitkap}{\vectz})]
  +\scalq\efktq\kaprep^2\efktl\epsve\dltvc[\unitpos(\dprod{\unitkap}{\vecth})-\vecth(\dprod{\unitkap}{\unitpos})]
  \beqref{alg1}
\end{split}
\end{align*}
\begin{align*}
\begin{split}
&=2\gcal\efktp\efkte\kaprep^2\efktn(\cprod{\vecth}{\vectz})
  +2\scalq\gcal\efktp\efkte\kaprep^2\efktn(\cprod{\vecth}{\unitpos})
  -\gcal\efktp\efkte\kaprep^2\efktl\epsve\dltvc(\cprod{\unitkap}{\vectz})\\
  &\quad-\scalq\gcal\efktp\efkte\kaprep^2\efktl\epsve\dltvc(\cprod{\unitkap}{\unitpos})
  -\gcal\efktp\efktd\vphih^2\kaprep^2\efktl\epsve\dltvc(\cprod{\unitkap}{\vecth})
  -2\efktq\kaprep^2\efktn\scalh^2\vectz
  -2\scalq\efktq\kaprep^2\efktn\scalh^2\unitpos\\
  &\quad+\efktq\kaprep^2\efktl\epsve\dltvc[(\dprod{\unitkap}{\vecth})\vectz-(\dprod{\unitkap}{\vectz})\vecth]
  +\scalq\efktq\kaprep^2\efktl\epsve\dltvc[(\dprod{\unitkap}{\vecth})\unitpos-(\dprod{\unitkap}{\unitpos})\vecth]
  \beqref{main5c}
\end{split}
\nonumber\\
\begin{split}
&=2\gcal\efktp\efkte\kaprep^2\efktn(\cprod{\vecth}{\vectz})
  +2\scalq\gcal\efktp\efkte\kaprep^2\efktn(\cprod{\vecth}{\unitpos})
  -\gcal\efktp\efkte\kaprep^2\efktl\epsve\dltvc(\cprod{\unitkap}{\vectz})\\
  &\quad-\scalq\gcal\efktp\efkte\kaprep^2\efktl\epsve\dltvc(\cprod{\unitkap}{\unitpos})
  -\gcal\efktp\efktd\vphih^2\kaprep^2\efktl\epsve\dltvc(\cprod{\unitkap}{\vecth})
  -2\efktq\kaprep^2\efktn\scalh^2\vectz
  -2\scalq\efktq\kaprep^2\efktn\scalh^2\unitpos\\
  &\quad+\efktq\kaprep^2\efktl\epsve\dltvc(\dltva\vectz-\dltvb\vecth)
  +\scalq\efktq\kaprep^2\efktl\epsve\dltvc(\dltva\unitpos-\epsva\vecth)
  \beqref{ogrv1a}\text{ \& }\eqnref{gxpeed1a}
\end{split}
\end{align*}
\begin{align*}
\begin{split}
&=2\gcal\efktp\efkte\kaprep^2\efktn(\cprod{\vecth}{\vectz})
  +2\scalq\gcal\efktp\efkte\kaprep^2\efktn(\cprod{\vecth}{\unitpos})
  -\gcal\efktp\efkte\kaprep^2\efktl\epsve\dltvc(\cprod{\unitkap}{\vectz})\\
  &\quad-\scalq\gcal\efktp\efkte\kaprep^2\efktl\epsve\dltvc(\cprod{\unitkap}{\unitpos})
  -\gcal\efktp\efktd\vphih^2\kaprep^2\efktl\epsve\dltvc(\cprod{\unitkap}{\vecth})
  -2\efktq\kaprep^2\efktn\scalh^2\vectz
  +\efktq\kaprep^2\efktl\epsve\dltvc\dltva\vectz\\
  &\quad-2\scalq\efktq\kaprep^2\efktn\scalh^2\unitpos
  +\scalq\efktq\kaprep^2\efktl\epsve\dltvc\dltva\unitpos
  -\efktq\kaprep^2\efktl\epsve\dltvc\dltvb\vecth
  -\scalq\efktq\kaprep^2\efktl\epsve\dltvc\epsva\vecth
\end{split}
\nonumber\\
\begin{split}
&=2\kaprep^2\gcal\efktp\efkte\efktn(\cprod{\vecth}{\vectz})
  +2\kaprep^2\scalq\gcal\efktp\efkte\efktn(\cprod{\vecth}{\unitpos})
  -\kaprep^2\gcal\efktp\efkte\efktl\epsve\dltvc(\cprod{\unitkap}{\vectz})\\
  &\quad-\kaprep^2\scalq\gcal\efktp\efkte\efktl\epsve\dltvc(\cprod{\unitkap}{\unitpos})
  -\kaprep^2\gcal\efktp\efktd\vphih^2\efktl\epsve\dltvc(\cprod{\unitkap}{\vecth})
  +\kaprep^2\efktq(\efktl\epsve\dltvc\dltva-2\efktn\scalh^2)\vectz\\
  &\quad+\kaprep^2\scalq\efktq(\efktl\epsve\dltvc\dltva-2\efktn\scalh^2)\unitpos
  -\kaprep^2\efktq\efktl\dltvc\epsve(\dltvb+\scalq\epsva)\vecth
\end{split}
\end{align*}
\begin{align}\label{gxpeed9d}
\begin{split}
&=\efktt\vectz+\scalq\efktt\unitpos-\kaprep^2\efktq\efktl\dltvc\epsve\efkta\vecth
  +2\kaprep^2\gcal\efktp\efkte\efktn(\cprod{\vecth}{\vectz})
  +2\kaprep^2\scalq\gcal\efktp\efkte\efktn(\cprod{\vecth}{\unitpos})\\
  &\quad-\kaprep^2\gcal\efktp\efkte\efktl\epsve\dltvc(\cprod{\unitkap}{\vectz})
  -\kaprep^2\scalq\gcal\efktp\efkte\efktl\epsve\dltvc(\cprod{\unitkap}{\unitpos})
  -\kaprep^2\gcal\efktp\efktd\vphih^2\efktl\epsve\dltvc(\cprod{\unitkap}{\vecth})\\
  &\quad\beqref{gxpeed1b}\text{ \& }\eqnref{gxpeed1c}
\end{split}
\nonumber\\
\begin{split}
&=\efktt\vectz+\scalq\efktt\unitpos-\efktu\vecth
  +\efktv(\cprod{\vecth}{\vectz})+\scalq\efktv(\cprod{\vecth}{\unitpos})
  -\efktw(\cprod{\unitkap}{\vectz})-\scalq\efktw(\cprod{\unitkap}{\unitpos})\\
  &\quad-\efktx(\cprod{\unitkap}{\vecth})\beqref{gxpeed1c}
\end{split}
\end{align}
\end{subequations}
\begin{align*}
&\xcala\vectu-\xcalb\vectups+\xcalc\vjf-\xcald\vje+(\dprod{\vectc}{\vjj})\vgh
 -(\dprod{\vectkap}{\vjj})\vjg+(\dprod{\vecta}{\vjj})\vgi+\cprod{\vjj}{\vjd}
\nonumber\\
&=\xcala\vectu-\xcalb\vectups+\xcalc\vjf-\xcald\vje+\scalc\efktr\vgh
  -\kaprep\efktr\vjg+\efkts\vgi+\cprod{\vjj}{\vjd}\beqref{gxpeed9}\nonumber\\
\begin{split}
&=\xcala\vectu-\xcalb(\cdkt\unitkap+\rhorep\vecta-\vectu+\vecte)+\xcalc\vjf-\xcald\vje
  +\scalc\efktr[\vcha\scalr^{-2}(\cprod{\vecth}{\vectkap})]-\kaprep\efktr\vjg\\
  &\quad+\efkts[\vchg\scalr^{-2}(\cprod{\vecth}{\vectkap})]+\cprod{\vjj}{\vjd}
  \beqref{kpath1c}, \eqnref{gxpeed5h}\text{ \& }\eqnref{gxpeed5i}
\end{split}
\nonumber\\
\begin{split}
&=(\xcala+\xcalb)\vectu-\rhorep\xcalb\vecta-\xcalb\vecte+\xcalc\vjf-\xcald\vje
  -\kaprep\efktr\vjg-\cdkt\xcalb\unitkap+\scalc\kaprep\efktr\vcha\scalr^{-2}(\cprod{\vecth}{\unitkap})\\
  &\quad+\kaprep\efkts\vchg\scalr^{-2}(\cprod{\vecth}{\unitkap})+\cprod{\vjj}{\vjd}
\end{split}
\nonumber\\
\begin{split}
&=(\xcala+\xcalb)\vectu-\rhorep\xcalb\vecta-\xcalb\vecte+\xcalc\vjf-\xcald\vje
  -\kaprep\efktr\vjg-\cdkt\xcalb\unitkap\\
  &\quad+\scalr^{-2}(\rhorep\efktr+\kaprep\efkts\efktk)(\cprod{\vecth}{\unitkap})
  +\cprod{\vjj}{\vjd}
  \beqref{main2b}, \eqnref{kas2a}\text{ \& }\eqnref{gxpeed6}
\end{split}
\end{align*}
\begin{align*}
\begin{split}
&=(\xcala+\xcalb)\vectu-\rhorep\xcalb\vecta-\xcalb\vecte
  +\xcalc[-\kaprep^2\efktl\efkth\epsve\unitplz-\kaprep^2\efktl\efkth\epsvb(\cprod{\vectr}{\vecth})-\efktm(\cprod{\vecth}{\vectups})]\\
  &\quad-\xcald[-\kaprep^2\efktl\efkta\epsve\unitplz-\efktm\vectz-\scalq\efktm\unitpos-\kaprep^2\efktl\efkta\epsvb(\cprod{\vectr}{\vecth})]
  -\kaprep\efktr[\efkto(\cprod{\vecth}{\unitkap})]\\
  &\quad-\cdkt\xcalb\unitkap+\scalr^{-2}(\rhorep\efktr+\kaprep\efkts\efktk)(\cprod{\vecth}{\unitkap})
  +\cprod{\vjj}{\vjd}
  \beqref{gxpeed8f}, \eqnref{gxpeed8e}\text{ \& }\eqnref{gxpeed8g}
\end{split}
\nonumber\\
\begin{split}
&=(\xcala+\xcalb)\vectu-\rhorep\xcalb\vecta-\xcalb\vecte
  -\xcalc\kaprep^2\efktl\efkth\epsve\unitplz
  -\xcalc\kaprep^2\efktl\efkth\epsvb(\cprod{\vectr}{\vecth})\\
  &\quad-\xcalc\efktm[-\vectz+\efktg\unitpos+\cdkt(\cprod{\vecth}{\unitkap})-\rhorep\vphia(\cprod{\vecth}{\unitpos})
     +\epsve\vphif(\cprod{\vecth}{\unitplz})]\\
  &\quad+\xcald\kaprep^2\efktl\efkta\epsve\unitplz
  +\xcald\efktm\vectz
  +\xcald\scalq\efktm\unitpos
  +\xcald\kaprep^2\efktl\efkta\epsvb(\cprod{\vectr}{\vecth})
  -\kaprep\efktr\efkto(\cprod{\vecth}{\unitkap})\\
  &\quad-\cdkt\xcalb\unitkap+\scalr^{-2}(\rhorep\efktr+\kaprep\efkts\efktk)(\cprod{\vecth}{\unitkap})
  +\cprod{\vjj}{\vjd}
  \beqref{gxpeed3b}
\end{split}
\end{align*}
\begin{align*}
\begin{split}
&=(\xcala+\xcalb)\vectu-\rhorep\xcalb\vecta-\xcalb\vecte
  -\cdkt\xcalb\unitkap
  +\xcald\kaprep^2\efktl\efkta\epsve\unitplz-\xcalc\kaprep^2\efktl\efkth\epsve\unitplz\\
  &\quad+\xcald\efktm\vectz+\xcalc\efktm\vectz
  -\xcalc\efktm\efktg\unitpos+\xcald\scalq\efktm\unitpos
  -\xcalc\efktm\epsve\vphif(\cprod{\vecth}{\unitplz})\\
  &\quad-\xcalc\kaprep^2\efktl\efkth\epsvb(\cprod{\vectr}{\vecth})
  +\xcald\kaprep^2\efktl\efkta\epsvb(\cprod{\vectr}{\vecth})
  +\xcalc\efktm\rhorep\vphia(\cprod{\vecth}{\unitpos})\\
  &\quad-\xcalc\efktm\cdkt(\cprod{\vecth}{\unitkap})
  -\kaprep\efktr\efkto(\cprod{\vecth}{\unitkap})
  +\scalr^{-2}(\rhorep\efktr+\kaprep\efkts\efktk)(\cprod{\vecth}{\unitkap})
  +\cprod{\vjj}{\vjd}
\end{split}
\nonumber\\
\begin{split}
&=(\xcala+\xcalb)\vectu-\rhorep\xcalb\vecta-\xcalb\vecte
  -\cdkt\xcalb\unitkap
  +\kaprep^2\epsve\efktl(\xcald\efkta-\xcalc\efkth)\unitplz\\
  &\quad+\efktm(\xcald+\xcalc)\vectz
  +\efktm(\scalq\xcald-\xcalc\efktg)\unitpos
  -\xcalc\efktm\epsve\vphif(\cprod{\vecth}{\unitplz})\\
  &\quad+[\scalr\kaprep^2\epsvb\efktl(\xcald\efkta-\xcalc\efkth)-\rhorep\xcalc\efktm\vphia](\cprod{\unitpos}{\vecth})\\
  &\quad+[\scalr^{-2}(\rhorep\efktr+\kaprep\efkts\efktk)-\cdkt\xcalc\efktm-\kaprep\efktr\efkto](\cprod{\vecth}{\unitkap})
  +\cprod{\vjj}{\vjd}
\end{split}
\end{align*}
\begin{align*}
\begin{split}
&=\scalh^{-2}(\xcala+\xcalb)(\cprod{\vectz}{\vecth}+\cprod{\scalq\unitpos}{\vecth})
  -\rhorep\xcalb(-\vphia\unitpos)
  -\xcalb[\fone(\cprod{\vectr}{\vecth})+\ftwo\unitplz]\\
  &\quad-\cdkt\xcalb\unitkap
  +\kaprep^2\epsve\efktl(\xcald\efkta-\xcalc\efkth)\unitplz
  +\efktm(\xcald+\xcalc)\vectz
  +\efktm(\scalq\xcald-\xcalc\efktg)\unitpos\\
  &\quad-\xcalc\efktm\epsve\vphif(\cprod{\vecth}{\unitplz})
  +[\scalr\kaprep^2\epsvb\efktl(\xcald\efkta-\xcalc\efkth)-\rhorep\xcalc\efktm\vphia](\cprod{\unitpos}{\vecth})\\
  &\quad+[\scalr^{-2}(\rhorep\efktr+\kaprep\efkts\efktk)-\cdkt\xcalc\efktm-\kaprep\efktr\efkto](\cprod{\vecth}{\unitkap})
  +\cprod{\vjj}{\vjd}
  \beqref{main5}\text{ \& }\eqnref{ogrv1b}
\end{split}
\nonumber\\
\begin{split}
&=\scalh^{-2}(\xcala+\xcalb)(\cprod{\vectz}{\vecth})
  +\scalq\scalh^{-2}(\xcala+\xcalb)(\cprod{\unitpos}{\vecth})
  +\rhorep\xcalb\vphia\unitpos
  -\xcalb\epsvb\vphif(\cprod{\vectr}{\vecth})\\
  &\quad-\xcalb\epsve\vphif\unitplz
  -\cdkt\xcalb\unitkap
  +\kaprep^2\epsve\efktl(\xcald\efkta-\xcalc\efkth)\unitplz
  +\efktm(\xcald+\xcalc)\vectz
  +\efktm(\scalq\xcald-\xcalc\efktg)\unitpos\\
  &\quad-\xcalc\efktm\epsve\vphif(\cprod{\vecth}{\unitplz})
  +[\scalr\kaprep^2\epsvb\efktl(\xcald\efkta-\xcalc\efkth)-\rhorep\xcalc\efktm\vphia](\cprod{\unitpos}{\vecth})\\
  &\quad+[\scalr^{-2}(\rhorep\efktr+\kaprep\efkts\efktk)-\cdkt\xcalc\efktm-\kaprep\efktr\efkto](\cprod{\vecth}{\unitkap})
  +\cprod{\vjj}{\vjd}
  \beqref{ogrv2g}\text{ \& }\eqnref{ogrv2h}
\end{split}
\end{align*}
\begin{align*}
\begin{split}
&=-\cdkt\xcalb\unitkap
  -\xcalb\epsve\vphif\unitplz
  +\kaprep^2\epsve\efktl(\xcald\efkta-\xcalc\efkth)\unitplz
  +\efktm(\xcald+\xcalc)\vectz\\
  &\quad+\rhorep\xcalb\vphia\unitpos
  +\efktm(\scalq\xcald-\xcalc\efktg)\unitpos
  +\scalh^{-2}(\xcala+\xcalb)(\cprod{\vectz}{\vecth})
  -\xcalc\efktm\epsve\vphif(\cprod{\vecth}{\unitplz})\\
  &\quad-\scalr\xcalb\epsvb\vphif(\cprod{\unitpos}{\vecth})
  +\scalq\scalh^{-2}(\xcala+\xcalb)(\cprod{\unitpos}{\vecth})
  +[\scalr\kaprep^2\epsvb\efktl(\xcald\efkta-\xcalc\efkth)-\rhorep\xcalc\efktm\vphia](\cprod{\unitpos}{\vecth})\\
  &\quad+[\scalr^{-2}(\rhorep\efktr+\kaprep\efkts\efktk)-\cdkt\xcalc\efktm-\kaprep\efktr\efkto](\cprod{\vecth}{\unitkap})
  +\cprod{\vjj}{\vjd}
\end{split}
\nonumber\\
\begin{split}
&=\efktm(\xcald+\xcalc)\vectz-\cdkt\xcalb\unitkap
  +[\kaprep^2\epsve\efktl(\xcald\efkta-\xcalc\efkth)-\xcalb\epsve\vphif]\unitplz\\
  &\quad+[\rhorep\xcalb\vphia+\efktm(\scalq\xcald-\xcalc\efktg)]\unitpos
  +\scalh^{-2}(\xcala+\xcalb)(\cprod{\vectz}{\vecth})
  -\xcalc\efktm\epsve\vphif(\cprod{\vecth}{\unitplz})\\
  &\quad+[\scalq\scalh^{-2}(\xcala+\xcalb)+\scalr\kaprep^2\epsvb\efktl(\xcald\efkta-\xcalc\efkth)
  -\scalr\xcalb\epsvb\vphif-\rhorep\xcalc\efktm\vphia](\cprod{\unitpos}{\vecth})\\
  &\quad+[\scalr^{-2}(\rhorep\efktr+\kaprep\efkts\efktk)-\cdkt\xcalc\efktm-\kaprep\efktr\efkto](\cprod{\vecth}{\unitkap})
  +\cprod{\vjj}{\vjd}
\end{split}
\end{align*}
\begin{align*}
\begin{split}
&=\efktm(\xcald+\xcalc)\vectz-\cdkt\xcalb\unitkap
  +[\kaprep^2\epsve\efktl(\xcald\efkta-\xcalc\efkth)-\xcalb\epsve\vphif]\unitplz\\
  &\quad+[\rhorep\xcalb\vphia+\efktm(\scalq\xcald-\xcalc\efktg)]\unitpos
  +\scalh^{-2}(\xcala+\xcalb)(\cprod{\vectz}{\vecth})
  -\xcalc\efktm\epsve\vphif(\cprod{\vecth}{\unitplz})\\
  &\quad+[\scalq\scalh^{-2}(\xcala+\xcalb)+\scalr\kaprep^2\epsvb\efktl(\xcald\efkta-\xcalc\efkth)
  -\scalr\xcalb\epsvb\vphif-\rhorep\xcalc\efktm\vphia](\cprod{\unitpos}{\vecth})\\
  &\quad+[\scalr^{-2}(\rhorep\efktr+\kaprep\efkts\efktk)-\cdkt\xcalc\efktm-\kaprep\efktr\efkto](\cprod{\vecth}{\unitkap})
  +\efktt\vectz+\scalq\efktt\unitpos-\efktu\vecth+\efktv(\cprod{\vecth}{\vectz})\\
  &\quad+\scalq\efktv(\cprod{\vecth}{\unitpos})
  -\efktw(\cprod{\unitkap}{\vectz})-\scalq\efktw(\cprod{\unitkap}{\unitpos})
  -\efktx(\cprod{\unitkap}{\vecth})\beqref{gxpeed9d}
\end{split}
\nonumber\\
\begin{split}
&=\efktt\vectz+\efktm(\xcald+\xcalc)\vectz-\efktu\vecth-\cdkt\xcalb\unitkap
  +[\kaprep^2\epsve\efktl(\xcald\efkta-\xcalc\efkth)-\xcalb\epsve\vphif]\unitplz\\
  &\quad+\scalq\efktt\unitpos+[\rhorep\xcalb\vphia+\efktm(\scalq\xcald-\xcalc\efktg)]\unitpos
  -\efktw(\cprod{\unitkap}{\vectz})\\
  &\quad-\scalq\efktw(\cprod{\unitkap}{\unitpos})
  -\efktv(\cprod{\vectz}{\vecth})+\scalh^{-2}(\xcala+\xcalb)(\cprod{\vectz}{\vecth})
  -\xcalc\efktm\epsve\vphif(\cprod{\vecth}{\unitplz})\\
  &\quad+[\scalq\scalh^{-2}(\xcala+\xcalb)+\scalr\kaprep^2\epsvb\efktl(\xcald\efkta-\xcalc\efkth)
  -\scalr\xcalb\epsvb\vphif-\rhorep\xcalc\efktm\vphia](\cprod{\unitpos}{\vecth})
  -\scalq\efktv(\cprod{\unitpos}{\vecth})\\
  &\quad+\efktx(\cprod{\vecth}{\unitkap})
  +[\scalr^{-2}(\rhorep\efktr+\kaprep\efkts\efktk)-\cdkt\xcalc\efktm-\kaprep\efktr\efkto](\cprod{\vecth}{\unitkap})
\end{split}
\end{align*}
\begin{align}\label{gxpeed10}
\begin{split}
&=-\efktu\vecth-\cdkt\xcalb\unitkap+[\efktt+\efktm(\xcald+\xcalc)]\vectz
  +[-\xcalb\epsve\vphif+\kaprep^2\epsve\efktl(\xcald\efkta-\xcalc\efkth)]\unitplz\\
  &\quad+[\scalq\efktt+\rhorep\xcalb\vphia+\efktm(\scalq\xcald-\xcalc\efktg)]\unitpos
  -\efktw(\cprod{\unitkap}{\vectz})
  -\scalq\efktw(\cprod{\unitkap}{\unitpos})
  -\xcalc\efktm\epsve\vphif(\cprod{\vecth}{\unitplz})\\
  &\quad+[-\scalr\xcalb\epsvb\vphif-\rhorep\xcalc\efktm\vphia-\scalq\efktv
  +\scalq\scalh^{-2}(\xcala+\xcalb)+\scalr\kaprep^2\epsvb\efktl(\xcald\efkta-\xcalc\efkth)](\cprod{\unitpos}{\vecth})\\
  &\quad+[\efktx-\cdkt\xcalc\efktm-\kaprep\efktr\efkto+\scalr^{-2}(\rhorep\efktr+\kaprep\efkts\efktk)](\cprod{\vecth}{\unitkap})
  +[-\efktv+\scalh^{-2}(\xcala+\xcalb)](\cprod{\vectz}{\vecth}).
\end{split}
\end{align}

\subart{Results of the computations}
It follows from \eqnref{kas7} and \eqnref{gxpeed10} that
\begin{subequations}\label{gxpeed11}
\begin{equation}\label{gxpeed11a}
\begin{split}
\angus&=\frac{1}{\betrep\scalc^2\rcal^2}\biggl[-\efktu\vecth-\cdkt\xcalb\unitkap+\vtta\vectz+\vttb\unitplz
  +\vttc\unitpos-\efktw(\cprod{\unitkap}{\vectz})-\scalq\efktw(\cprod{\unitkap}{\unitpos})\\
  &\qquad-\vttd(\cprod{\vecth}{\unitplz})+\vtte(\cprod{\unitpos}{\vecth})
  +\vttf(\cprod{\vecth}{\unitkap})+\vttg(\cprod{\vectz}{\vecth})\biggr]
\end{split}
\end{equation}
where
\begin{align}\label{gxpeed11b}
\begin{split}
&\vtta=\efktt+\efktm(\xcald+\xcalc),\quad
\vttb=-\xcalb\epsve\vphif+\kaprep^2\epsve\efktl(\xcald\efkta-\xcalc\efkth)\\
&\vttc=\scalq\efktt+\rhorep\xcalb\vphia+\efktm(\scalq\xcald-\xcalc\efktg),\quad
\vttd=\xcalc\efktm\epsve\vphif\\
&\vtte=-\scalr\xcalb\epsvb\vphif-\rhorep\xcalc\efktm\vphia-\scalq\efktv+\scalq\scalh^{-2}(\xcala+\xcalb)
   +\scalr\kaprep^2\epsvb\efktl(\xcald\efkta-\xcalc\efkth)\\
&\vttf=\efktx-\cdkt\xcalc\efktm-\kaprep\efktr\efkto+\scalr^{-2}(\rhorep\efktr+\kaprep\efkts\efktk),\quad
\vttg=-\efktv+\scalh^{-2}(\xcala+\xcalb)
\end{split}
\end{align}
and with $\rcal, \gcal, \fcal$ given by \eqnref{ogrv9a},
\begin{align}\label{gxpeed11c}
\begin{split}
\xcala&=\frac{\rcal^2(\betrep+\gcal)}{\betrep\fcal},\quad
\xcalb=\frac{\rcal^2+\betrep\gcal}{\fcal},\quad
\xcalc=\frac{\betrep\gcal}{\fcal},\quad
\xcald=\frac{\rcal^2}{\fcal}.
\end{split}
\end{align}
Furthermore, we have from \eqnref{kas1c}, \eqnref{main5a} and \eqnref{gxpeed11a} that
\begin{align}\label{gxpeed11d}
\aspua
&=\fdot{\psirep}/\scala=(\dprod{\angus}{\vecta})/\scala=-(\dprod{\unitpos}{\angus})\nonumber\\
\begin{split}
&=\frac{1}{\betrep\scalc^2\rcal^2}\biggl[\efktu(\dprod{\unitpos}{\vecth})
  +\cdkt\xcalb(\dprod{\unitpos}{\unitkap})
  -\vtta(\dprod{\unitpos}{\vectz})
  -\vttb(\dprod{\unitpos}{\unitplz})
  -\vttc(\dprod{\unitpos}{\unitpos})\\
  &\quad+\efktw[\dprod{\unitpos}{(\cprod{\unitkap}{\vectz})}]
  +\scalq\efktw[\dprod{\unitpos}{(\cprod{\unitkap}{\unitpos})}]
  +\vttd[\dprod{\unitpos}{(\cprod{\vecth}{\unitplz})}]
  -\vtte[\dprod{\unitpos}{(\cprod{\unitpos}{\vecth})}]\\
  &\quad-\vttf[\dprod{\unitpos}{(\cprod{\vecth}{\unitkap})}]
  -\vttg[\dprod{\unitpos}{(\cprod{\vectz}{\vecth})}]\biggr]
\end{split}
\nonumber\\
\begin{split}
&=\frac{1}{\betrep\scalc^2\rcal^2}\biggl[\cdkt\xcalb\epsva-\vtta\epsvd-\vttb\epsvc-\vttc
  +\efktw\dltvf+\vttd(\epsvf/\scalr)-\vttf(\epsve/\scalr)-\vttg\epsvh\biggr]\\
  &\quad\beqref{ogrv1a}\text{ \& }\eqnref{gxpeed1a}.
\end{split}
\end{align}
\end{subequations}
Equations \eqnref{gxpeed1} and \eqnref{gxpeed11} completely determine the slope and the variation of obliquation
for a gravitating observer.

\art{Apparent path of a light source}
Equation \eqnref{kpath11} can be evaluated for a gravitating observer as follows.
We introduce the following quantities in addition to those defined by \eqnref{ogrv1},
\eqnref{ogrv9} and \eqnref{gxpeed1},
\begin{subequations}\label{gpath1}
\begin{align}\label{gpath1a}
\begin{split}
&\vsiga=\dprod{\vectr}{(\cprod{\unitkap}{\unitplz})},\quad
\vsigb=\dprod{\vectr}{(\cprod{\unitplz}{\vectz})},\quad
\vsigc=\dprod{\unitkap}{(\cprod{\unitplz}{\vectz})}
\end{split}
\end{align}
\begin{align}\label{gpath1b}
\begin{split}
&\vrhoa=\scalq/\scalr^3,\quad
\vrhob=\epsvh/(\scalr\scalh^2),\quad
\vrhoc=\vrhoa(\efkta+3\vrhob\epsve),\quad
\vrhod=\vrhoa(\efktc+3\vrhob\epsvf)\\
&\vrhoe=\epsvb(\epsve\efktc-\epsvf\efkta)/(2\vphic\epsvf^2),\quad
\vrhof=\vrhoe(\vrhoe\epsvf-2\efktc\vphic)/(\vphic\epsvf),\quad
\vrhog=\absign(\vphic\vrhoe)/\vphie\\
&\vrhoh=[\vrhof\efktc\vphic+\vrhoe(3\vrhof\epsvf-\vrhoe\efktc-2\vrhoa\epsvf\vphic)]/(\vphic\epsvf),\quad
\vrhoi=(1/\vphie)[\absign(\vrhoe^2-\vrhof\vphic)-\vrhog^2]\\
&\vrhoj=(1/\vphie)[\absign(\vrhoh\vphic-3\vrhoe\vrhof)-3\vrhog\vrhoi],\quad
\vrhok=\vphih/\scalr,\quad
\vrhol=2\vrhoa+15\vrhob^2-3\vrhok^2\\
&\vrhom=\vrhoa(9\vrhok-8\vrhoa-45\vrhob^2),\quad
\vrhon=15\vrhoa\vrhob(2\vrhoa-3\vrhok+7\vrhob^2),\quad
\vrhoo=\scalh^{-2}(\epsvg+\epsve\vphib)\\
&
\vrhop=\vrhoa(3\scalr\vrhob\epsva-\vrhoo),\quad
\vrhoq=\vrhoa(6\vrhob\vrhoo-\scalr\vrhol\epsva),\quad
\vrhor=\vrhom\vrhoo+\scalr\vrhon\epsva
\end{split}
\end{align}
\begin{align}\label{gpath1c}
\begin{split}
&\vrhos=\frac{1}{\vphie^2\pfreq^2}\biggl[\vrhop+\frac{2\vphia\epsva\vrhog}{\vphie}\biggr],\quad
\vrhot=\frac{1}{\vphie^2\pfreq^2}\biggl[\vrhoq-\frac{4\vrhop\vrhog}{\vphie}
   +\frac{2\vphia\epsva\vrhoi}{\vphie}-\frac{6\vphia\epsva\vrhog^2}{\vphie^2}\biggr]\\
&\vrhou=\frac{1}{\vphie^2\pfreq^2}\biggl[\vrhor-\frac{6\vrhoq\vrhog}{\vphie}
   -\frac{6\vrhop\vrhoi}{\vphie}+\frac{2\vphia\epsva\vrhoj}{\vphie}+\frac{18\vrhop\vrhog^2}{\vphie^2}
   -\frac{18\vphia\epsva\vrhog\vrhoi}{\vphie^2}+\frac{24\vphia\epsva\vrhog^3}{\vphie^3}\biggr]\\
&\vrhov=\vrhos/\efkti^3,\quad
\vrhow=(\vrhot\efkti^2-3\vthtrep\vrhos^2)/\efkti^5,\quad
\vrhox=[\vrhou\efkti^4-9\kaprep\vthtrep\vrhos\vrhot\efkti^2-3\kaprep^2\vrhos^3(1-4\vthtrep^2)]/\efkti^7
\end{split}
\end{align}
\begin{align}\label{gpath1d}
\begin{split}
&\ethva=\dragf\biggl[\frac{\vrhog}{\vphie}+\frac{\kaprep\xcons\vphie^2\vrhos}{4\dragf^2}\biggr],\quad
\ethvb=\frac{\vrhog}{\vphie}-\frac{\kaprep\xcons\vphie^2\vrhos}{4\dragf^2},\quad
\ethvc=\frac{\kaprep}{4\dragf}\biggl[\kaprep\vrhov\vphie\vrhos+2\xcons\vrhog\vrhos+\xcons\vphie\vrhot\biggr]\\
&\ethvd=\frac{\vrhoi}{\vphie}-\frac{\vrhog^2}{\vphie^2},\quad
\ethve=\dragf\ethvd+\ethva\ethvb+\vphie\ethvc,\quad
\ethvf=\dragf\biggl[\frac{\vrhoj}{\vphie}-\frac{3\vrhog\vrhoi}{\vphie^2}+\frac{2\vrhog^3}{\vphie^3}\biggr]\\
&\ethvg=\ethvf+\ethve\ethvb+\vrhog\ethvc
  +2\ethva\biggl[\ethvd-\frac{\kaprep^2\vrhov\vphie^2\vrhos}{4\dragf^2}-\frac{\kaprep\xcons\vphie\vrhog\vrhos}{2\dragf^2}
    -\frac{\kaprep\xcons\vphie^2\vrhot}{4\dragf^2}
    +\frac{\kaprep\xcons\vphie^2\ethva\vrhos}{4\dragf^3}\biggr]\\
  &\qquad+\frac{\kaprep\vphie}{4\dragf}\biggl[\kaprep\biggl\{\vrhow\vphie\vrhos+3\vrhov\vrhog\vrhos
    +2\vrhov\vphie\vrhot\biggr\}+\xcons\biggl\{2\vrhoi\vrhos+3\vrhog\vrhot+\vphie\vrhou\biggr\}\biggr]
\end{split}
\end{align}
\begin{align}\label{gpath1e}
\begin{split}
&\ethvh=\frac{\kaprep\dragf\vrhov-\xcons\ethva}{4\pfreq\dragf^2},\quad
\ethvi=-\frac{\ethva(\dragf\kaprep\vrhov-\xcons\ethva)}{2\pfreq\dragf^3}
   +\frac{\dragf\kaprep\vrhow-\xcons\ethve}{4\pfreq\dragf^2}\\
&\ethvj=\frac{(3\ethva^2-\dragf\ethve)(\dragf\kaprep\vrhov-\xcons\ethva)}{2\pfreq\dragf^4}
   -\frac{\ethva(\dragf\kaprep\vrhow-\xcons\ethve)}{\pfreq\dragf^3}
  +\frac{\kaprep(\ethva\vrhow+\dragf\vrhox-\vrhov\ethve)-\xcons\ethvg}{4\pfreq\dragf^2}\\
&\ethvk=\frac{\rhorep\vrhop-\ethvh\vphia\epsva-\frac{1}{2}\ethva\scalc-\vphif\vphid\efktc+2\epsvf\vphif\vphic\vrhoe}{\kaprep\epsvf\vphid}\\
&\ethvl=\frac{\epsva\vphia\ethvi-\rhorep\vrhoq-2\ethvh\vrhop+\frac{1}{2}\ethve\scalc
    +4\kaprep\ethvk\vrhoe\epsvf\vphic+4\vphif\vphic\efktc\vrhoe
    -2\vrhoe^2\vphif\epsvf+2\vrhof\vphif\epsvf\vphic}{\kaprep\epsvf\vphid}\\
  &\qquad-\frac{2\kaprep\ethvk\efktc-\epsvf\vrhoa\vphif}{\kaprep\epsvf}
\end{split}
\end{align}
\begin{align}\label{gpath1f}
\begin{split}
\ethvm&=-\frac{\rhorep\vrhor-\ethvj\vphia\epsva
     +3\ethvi\vrhop+3\ethvh\vrhoq-\frac{1}{2}\ethvg\scalc}{\kaprep\epsvf\vphid}
  +\frac{\efktc\vphid-2\epsvf\vphic\vrhoe}{\epsvf\vphid}
      \biggl\{\ethvl+\frac{2\kaprep\ethvk\efktc-\vrhoa\epsvf\vphif}{\kaprep\epsvf}\biggr\}\\
  &\quad-\frac{\epsvf[\kaprep(-2\ethvl\efktc+2\ethvk\vrhoa\epsvf+\ethvk\vrhoa\epsvf)+\vrhod\vphif]
    +\efktc(2\kaprep\ethvk\efktc-\vrhoa\epsvf\vphif)}{\kaprep\epsvf^2}\\
  &\quad-\frac{8\kaprep\vrhoe\ethvk\efktc\vphic
    -4\vrhoe\vrhoa\epsvf\vphif\vphic-4\efktc\vrhoe^2\vphif
    +6\efktc\vrhof\vphif\vphic-2\vrhoe^2\efktc\vphif}{\kaprep\epsvf\vphid}\\
  &\quad-\frac{4\kaprep\vrhoe\ethvl\epsvf\vphic-6\kaprep\vrhoe^2\ethvk\epsvf
    -6\vrhoe\vrhof\vphif\epsvf+6\kaprep\vrhof\ethvk\epsvf\vphic+2\vrhoh\vphif\epsvf\vphic}{\kaprep\epsvf\vphid}
\end{split}
\end{align}
\begin{align}\label{gpath1g}
\begin{split}
&\ethvn=-\efkta\vphif-\kaprep\epsve\ethvk,\quad
\ethvo=-\vrhoa\epsve\vphif+\kaprep(2\efkta\ethvk+\epsve\ethvl),\quad
\ethvp=\epsvb(\kaprep\ethvl-\vphif\vrhoa)\\
&\ethvq=\vrhoc\vphif+\kaprep(3\vrhoa\epsve\ethvk-3\efkta\ethvl+\epsve\ethvm),\quad
\ethvr=\epsvb(\kaprep\ethvm+3\kaprep\ethvk\vrhoa+3\vrhoa\vrhob\vphif)\\
&\ethvs=\epsvb(3\kaprep\ethvl-\vrhoa\vphif),\quad
\ethvt=\rhorep\vrhop-\ethvh\vphia\epsva+\efktc\vphic^2\vphif-2\vrhoe\epsvf\vphic\vphif+\kaprep\ethvk\epsvf\vphic^2\\
&\ethvu=-\epsva\vphia\ethvi+2\ethvh\vrhop+\rhorep\vrhoq+\vrhoa\epsvf\vphic^2\vphif
  +4\efktc\vrhoe\vphic\vphif-2\kaprep\efktc\ethvk\vphic^2-2\vrhoe^2\epsvf\vphif\\
  &\qquad+2\vrhof\epsvf\vphic\vphif+4\kaprep\vrhoe\ethvk\epsvf\vphic
  -\kaprep\epsvf\ethvl\vphic^2\\
&\ethvv=-\vphia\epsva\ethvj+3\vrhop\ethvi+3\vrhoq\ethvh+\vrhor\rhorep-\vrhod\vphic^2\vphif
     -\kaprep\epsvf\ethvk\vrhoa\vphic^2+6\vrhoe\vrhoa\epsvf\vphic\vphif\\
  &\qquad+6\vrhoe^2\efktc\vphif -6\vrhof\efktc\vphic\vphif-12\kaprep\vrhoe\ethvk\efktc\vphic
     -2\kaprep\ethvk\vrhoa\epsvf\vphic^2+3\kaprep\efktc\ethvl\vphic^2+4\vrhoe\vrhof\epsvf\vphif\\
  &\qquad+6\kaprep\vrhoe^2\ethvk\epsvf
  +2\vrhoe\vrhof\epsvf\vphif-2\vrhoh\epsvf\vphic\vphif-6\kaprep\vrhof\ethvk\epsvf\vphic
  -6\kaprep\vrhoe\ethvl\epsvf\vphic-\kaprep\epsvf\ethvm\vphic^2
\end{split}
\end{align}
\begin{align}\label{gpath1h}
\begin{split}
&\frkya=\scalc\ethva-2\ethvt,\quad
\frkyb=\scalc\ethve-2\ethvu,\quad
\frkyc=\scalc\ethvg-2\ethvv,\quad
\frkyd=\ethvh-1,\quad
\frkye=2\ethvh-1\\
&\frkyf=3\ethvh-1,\quad
\frkyg=\ethvi\frkya-\frkyb\frkyd,\quad
\frkyh=\frkya\frkye-\rhorep\frkyb,\quad
\frkyi=\frkyd\frkye-\rhorep\ethvi\\
&\frkyj=\vrhoa^2(\vrhol-18\vrhob^2),\quad
\frkyk=\vphib\epsvb(\vrhoa\vphif+2\kaprep\vrhob\ethvk),\quad
\frkyl=2\vphib(\kaprep\vrhoa\epsvb\ethvk+\vrhob\ethvp)\\
&\frkym=\epsvb\vphib[6\vrhoa\vrhob\vphif+\kaprep\ethvk(\vrhol-6\vrhob^2)],\quad
\frkyn=2\kaprep\ethvk\ethvn+\ethvo\vphif,\quad
\frkyo=\ethvp\ethvn+\kaprep\epsvb\ethvk\ethvo\\
&\frkyp=2\kaprep^2\epsvb\ethvk^2-\ethvp\vphif,\quad
\frkyq=\scalr(\epsvc\frkyo+\epsvb\frkyp\efktb),\quad
\frkyr=\vphif\frkyb+2\kaprep\ethvk\frkya\\
&\frkys=\frkyh-6\rhorep\vrhob\frkya,\quad
\frkyt=\frkyi-6\rhorep\vrhob\frkyd,\quad
\frkyu=\ethvo\frkya-\ethvn\frkyb,\quad
\frkyv=\kaprep\epsvb\ethvk\frkyb+\ethvp\frkya\\
&\frkyw=3\vrhob\frkyh-\rhorep\vrhol\frkya-\frkyg,\quad
\frkyx=\vphia\frkyw+\scalq\epsvb\frkyr,\quad
\frkyy=\scalr\epsva\frkyv+\vrhoa^2\frkyt,\quad
\frkyz=\scalz^2+\scalq\epsvd
\end{split}
\end{align}
\begin{align}\label{gpath1i}
\begin{split}
&\vakpa=\vrhoa(\ethvn\ethvi-\ethvo\frkyd)-3\vrhoa\vrhob(\ethvn\frkye-\rhorep\ethvo)
    +\rhorep\vrhoa\vrhol\ethvn-\epsvb\vphib\frkyn\\
&\vakpb=\vrhoa\epsvb[\vphif(3\vrhob\frkye-\ethvi-\rhorep\vrhol)+2\kaprep\ethvk(3\rhorep\vrhob-\frkyd)]\\
&\vakpc=\vrhoa(\ethvn\frkye-\rhorep\ethvo-6\rhorep\vrhob\ethvn),\quad
\vakpd=\vrhoa\epsvb(-\vphif\frkye-2\rhorep\kaprep\ethvk+6\rhorep\vrhob\vphif)\\
&\vakpe=\vphib(\kaprep\epsvb\ethvk\ethvi+\ethvp\frkyd)+\rhorep(\rhorep\frkyj-\frkyl+\frkym)-\frkyk\frkye-\frkyq\\
&\vakpf=\vrhoa\frkyd-3\rhorep\vrhoa\vrhob+\vphib\epsvb\vphif,\quad
\vakpg=\dltva\frkyv+\dltvc\frkyo,\quad
\vakph=\vakpe-\frkyy
\end{split}
\end{align}
\begin{align}\label{gpath1j}
\begin{split}
&\vakpi=[\frkya^2+2\ethvn\frkya\epsvb-2\epsvb\vphif\frkya\dltvb-2\rhorep\vrhoa\frkya\vrhoo-2\scalr\vakpf\frkya\epsva
  -2\kaprep\epsvb\ethvk\frkya\epsve+\ethvn^2\\
  &\qquad-2\epsvb\vphif\ethvn\dltvd-2\vrhoa\ethvn(\rhorep/\scalh^2)(\epsvi+\epsvf\vphib)
  -2\scalr\vakpf\ethvn\epsvc-2\kaprep\epsvb\ethvk\ethvn\epsvf
  +\scalz^2\epsvb^2\vphif^2\\
  &\qquad-2\rhorep\scalr\vrhob\vphib\vrhoa\epsvb\vphif+2\scalr\vakpf\epsvb\vphif\epsvd
  -2\scalr\kaprep\epsvb^2\ethvk\vphif\epsvh+(\rhorep\vrhoa\vphih)^2+2\rhorep\scalr^2\vakpf\vrhoa\vrhob\\
  &\qquad+2\kaprep\scalr\efktb\epsvb\ethvk\rhorep\vrhoa+(\scalr\vakpf)^2+(\kaprep\scalr\scalh\epsvb\ethvk)^2]^{1/2}\\
&\vakpj=\vakpg\scalr^2+\frkyu\vsiga+\scalr\epsvb\frkyr\dltvf+\epsvb\frkyn\vsigb
  -\vrhoa\frkys\dltva-\vakpc\dltvc\\
&\vakpk=\scalh^2\vakph+\epsve(\frkyx/\scalr)-\frkyu\dltve+\epsvb\frkyr\epsvg+\epsvb\frkyn\epsvi
  -\vakpa\epsvf+\scalr\vakpb\epsvh+\vrhoa\frkys\efkta+\vakpc\efktc+\vakpd\frkyz\\
&\vakpl=(\vakph/\scalr)\epsve+\frkyx(1-\epsva^2)+\frkyu(\epsvc-\epsvb\epsva)
  +\epsvb\frkyr(\epsvd-\dltvb\epsva)+\epsvb\frkyn(\epsvb\epsvd-\dltvb\epsvc)\\
  &\qquad+\scalr\vakpa(\epsva\epsvc-\epsvb)+\scalr\vakpb(\epsva\epsvd-\dltvb)
  -\vrhoa\frkys(\scalr\vrhob-\vrhoo\epsva)-\vakpc(\scalr\epsvb\vrhob-\vrhoo\epsvc)\\
  &\qquad-\vakpd(\scalr\dltvb\vrhob-\vrhoo\epsvd)
\end{split}
\end{align}
\begin{align}\label{gpath1k}
\begin{split}
&\vakpm=\vakpg\vsiga-\vakph\dltve+\frkyx(\epsvc-\epsva\epsvb)+\frkyu(1-\epsvb^2)
  +\epsvb\frkyr(\dltvd-\dltvb\epsvb)+\epsvb\frkyn(\epsvb\dltvd-\dltvb)\\
  &\qquad+\scalr\vakpa(\epsva-\epsvb\epsvc)+\scalr\vakpb(\epsva\dltvd-\dltvb\epsvc)
  +\vrhoa\frkys\vrhoo\epsvb+\vakpc\vrhoo+\vakpd\vrhoo\dltvd\\
  &\qquad-\scalh^{-2}(\vrhoa\frkys+\vakpc\epsvb+\vakpd\dltvb)(\epsvi+\epsvf\vphib)\\
&\vakpn=\scalr\vakpg\dltvf+\vakph\epsvg+\frkyx(\epsvd-\epsva\dltvb)+\frkyu(\dltvd-\epsvb\dltvb)
  +\epsvb\frkyr(\scalz^2-\dltvb^2)+\epsvb\frkyn(\epsvb\scalz^2-\dltvb\dltvd)\\
  &\qquad+\scalr\vakpa(\epsva\dltvd-\epsvb\epsvd)+\scalr\vakpb(\epsva\scalz^2-\dltvb\epsvd)
  +\vrhoa\frkys(\scalr\vrhob\vphib+\vrhoo\dltvb)+\vakpc(\scalr\epsvb\vrhob\vphib+\vrhoo\dltvd)\\
  &\qquad+\vakpd(\scalr\vrhob\vphib\dltvb+\vrhoo\scalz^2)\\
&\vakpo=\vakpg\vsigb+\vakph\epsvi+\frkyx(\epsvb\epsvd-\epsvc\dltvb)+\frkyu(\epsvb\dltvd-\dltvb)
  +\epsvb\frkyr(\epsvb\scalz^2-\dltvd\dltvb)+\epsvb\frkyn(\scalz^2-\dltvd^2)\\
  &\qquad+\scalr\vakpa(\epsvc\dltvd-\epsvd)
  +\scalr\vakpb(\epsvc\scalz^2-\dltvd\epsvd)
  +\scalr\frkys\epsvb\vrhoa\vrhob\vphib+\scalr\vakpc\vrhob\vphib+\scalr\vakpd\dltvd\vrhob\vphib\\
  &\qquad+\scalh^{-2}(\vrhoa\frkys\dltvb+\vakpc\dltvd+\vakpd\scalz^2)(\epsvi+\epsvf\vphib)
\end{split}
\end{align}
\begin{align}\label{gpath1l}
\begin{split}
&\vakpp=-\vakph\epsvf+\scalr\frkyx(\epsva\epsvc-\epsvb)+\scalr\frkyu(\epsva-\epsvc\epsvb)
  +\scalr\epsvb\frkyr(\epsva\dltvd-\epsvd\epsvb)+\scalr\epsvb\frkyn(\epsvc\dltvd-\epsvd)\\
  &\qquad+\scalr^2\vakpa(1-\epsvc^2)+\scalr^2\vakpb(\dltvd-\epsvd\epsvc)
  +\scalr^2(\epsvb\frkys\vrhoa\vrhob+\vakpc\vrhob+\vakpd\dltvd\vrhob)\\
  &\qquad-(\scalr/\scalh^2)(\vrhoa\frkys\epsva+\vakpc\epsvc+\vakpd\epsvd)(\epsvi+\epsvf\vphib)\\
&\vakpq=\scalr\vakph\epsvh+\scalr\frkyx(\epsva\epsvd-\dltvb)+\scalr\frkyu(\epsva\dltvd-\epsvc\dltvb)
  +\scalr\epsvb\frkyr(\epsva\scalz^2-\epsvd\dltvb)\\
  &\qquad+\scalr\epsvb\frkyn(\epsvc\scalz^2-\epsvd\dltvd)+\scalr^2\vakpa(\dltvd-\epsvc\epsvd)
  +\scalr^2\vakpb(\scalz^2-\epsvd^2)+\scalr^2\vrhoa\frkys\vrhob(\epsva\vphib+\dltvb)\\
  &\qquad+\scalr^2\vakpc\vrhob(\epsvc\vphib+\dltvd)+\scalr^2\vakpd\vrhob(\epsvd\vphib+\scalz^2)\\
&\vakpr=\vakpg\dltva-\vakph\efkta+\frkyx(\scalr\vrhob-\epsva\vrhoo)
  -\frkyu\epsvb\vrhoo-\epsvb\frkyr(\scalr\vrhob\vphib+\dltvb\vrhoo)\\
  &\qquad-\epsvb^2\frkyn\scalr\vrhob\vphib-\scalr^2\vakpa\epsvb\vrhob-\scalr^2\vakpb\vrhob(\epsva\vphib+\dltvb)
  -\vrhoa\frkys(\vphih^2-\vrhoo^2)-\vakpc\epsvb\vphih^2\\
  &\qquad-\vakpd(\dltvb\vphih^2+\scalr\vrhoo\vrhob\vphib)
  +\scalh^{-2}(\frkyu-\epsvb\frkyn\dltvb+\vakpa\scalr\epsva+\vakpc\vrhoo)(\epsvi+\epsvf\vphib)
\end{split}
\end{align}
\begin{align}\label{gpath1m}
\begin{split}
&\vakps=\vakpg\dltvc-\vakph\efktc+\frkyx(\scalr\epsvb\vrhob-\epsvc\vrhoo)
  -\frkyu\vrhoo-\epsvb\frkyr(\scalr\epsvb\vrhob\vphib+\dltvd\vrhoo)-\scalr\epsvb\frkyn\vrhob\vphib\\
  &\qquad-\scalr^2\vrhob[\vakpa+\vakpb(\epsvc\vphib+\dltvd)]-\vphih^2(\vrhoa\frkys\epsvb+\vakpc+\vakpd\dltvd)
  +\scalh^{-4}\vakpc(\epsvi+\epsvf\vphib)^2\\
  &\qquad+\scalh^{-2}(\frkyu\epsvb-\epsvb\frkyn\dltvd+\scalr\vakpa\epsvc+\vrhoo\vrhoa\frkys
  -\scalr\vakpd\vrhob\vphib)(\epsvi+\epsvf\vphib)\\
&\vakpt=-\vakph\frkyz+\frkyx(\scalr\dltvb\vrhob-\epsvd\vrhoo)
  -\frkyu\dltvd\vrhoo-\epsvb\frkyr(\scalr\dltvb\vrhob\vphib+\scalz^2\vrhoo)
  -\scalr\epsvb\frkyn\dltvd\vrhob\vphib\\
  &\qquad-\scalr^2\vakpa\dltvd\vrhob
  -\scalr^2\vakpb\vrhob(\epsvd\vphib+\scalz^2)
  -\vrhoa\frkys(\dltvb\vphih^2+\scalr\vrhob\vphib\vrhoo)-\vakpc\dltvd\vphih^2\\
  &\qquad-\vakpd(\scalz^2\vphih^2-\scalr^2\vrhob^2\vphib^2)
  +\scalh^{-2}(\frkyu\dltvb-\scalz^2\epsvb\frkyn+\scalr\vakpa\epsvd-\scalr\vakpc\vrhob\vphib)(\epsvi+\epsvf\vphib)
\end{split}
\end{align}
\begin{align}\label{gpath1n}
\begin{split}
&\vakpu=[\vakpg\vakpj+\vakph\vakpk+\frkyx\vakpl+\frkyu\vakpm+\epsvb\frkyr\vakpn
  +\epsvb\frkyn\vakpo+\vakpa\vakpp+\vakpb\vakpq-\vrhoa\frkys\vakpr\\
  &\qquad-\vakpc\vakps-\vakpd\vakpt]^{1/2}\\
&\vakpv=\scalr\vakpg\epsva+\vakph\dltva+\epsvb\frkyn\vsigc-\vakpa\vsiga-\scalr\vakpb\dltvf
  -(\vakpc/\scalh^2)(\dltvc\efkta-\efktc\dltva)+(\vakpd/\scalh^2)\frkyz\dltva\\
&\vakpw=\scalr\vakpg\epsvc+\vakph\dltvc-(\frkyx/\scalr)\vsiga-\epsvb\frkyr\vsigc-\vakpb\vsigb
  +\vrhoa(\frkys/\scalh^2)(\dltvc\efkta-\efktc\dltva)+(\vakpd/\scalh^2)\frkyz\dltvc\\
&\vakpx=\scalr\vakpg\epsvd-\frkyx\dltvf+\frkyu\vsigc+\vakpa\vsigb
  -(\frkyz/\scalh^2)(\frkys\vrhoa\dltva+\vakpc\dltvc)\\
&\vakpy=-\scalr\frkyx\dltva+\scalr\frkyu(\epsva\dltvc-\epsvc\dltva)
  -\scalr\epsvb\epsvd(\frkyr\dltva+\frkyn\dltvc)+\scalr^2(\vakpa\dltvc+\vrhoa\frkys\vrhob\dltva+\vakpc\vrhob\dltvc)\\
&\vakpz=-\frkyx\epsvd\dltva+\frkyu(\dltvb\dltvc-\dltvd\dltva)-\scalz^2\epsvb(\frkyr\dltva+\frkyn\dltvc)
  +\scalr(\vakpg\epsvh+\vakpa\epsvd\dltvc)\\
  &\qquad-\scalr\vrhob\vphib(\vrhoa\frkys\dltva+\vakpc\dltvc)
\end{split}
\end{align}
\begin{align}\label{gpath1o}
\begin{split}
&\parva=\scalh^{-2}(\vakpz+\vphib\vakpy),\quad
\parvb=\ethvq\vakpw-\ethvs\vakpx-\vphib\ethvs\vakpj+\ethvr\vakpy\\
&\parvc=-\vakpj\ethvj+3\ethvi(3\vrhob\vakpj-\parva)+\frkyf(6\vrhob\parva-\vrhol\vakpj),\quad
\parvd=\vrhoa\vphib(\rhorep/\scalh^2)+\kaprep\epsvb\ethvk\\
&\parve=\scalh^{-2}(\vrhoa\frkys\dltva+\vakpc\dltvc),\quad
\parvf=\vakph+\scalh^{-2}(\vrhoa\frkys\efkta+\vakpc\efktc+\vakpd\frkyz),\quad
\parvg=\scalr\vakpg-\scalq\parve.
\end{split}
\end{align}
\end{subequations}

\subart{Development of \eqnref{main5}}
Bearing the foregoing quantities in mind, we derive
\begin{subequations}\label{gpath2}
\begin{align}\label{gpath2a}
\fdot{\scalm}
&=\dif{[\dprod{\vectkap}{(\cprod{\vectr}{\vecth})}]}\beqref{main5c}\nonumber\\
&=\dprod{\vectkap}{(\cprod{\vectu}{\vecth})}
=-\dprod{\vectkap}{(\cprod{\vecth}{\vectu})}\nonumber\\
&=-\dprod{\vectkap}{\vectz}-\scalq(\dprod{\vectkap}{\unitpos})\beqref{main5c}\nonumber\\
&=-\kaprep(\dltvb+\scalq\epsva)\beqref{ogrv1a}\text{ \& }\eqnref{gxpeed1a}\nonumber\\
&=-\kaprep\efkta\beqref{gxpeed1b}
\end{align}
\begin{align}\label{gpath2b}
\ffdot{\scalm}
&=\dif{[-\dprod{\vectkap}{\vectz}-\scalq(\dprod{\vectkap}{\unitpos})]}\beqref{gpath2a}\nonumber\\
&=-\scalq(\dprod{\vectkap}{\fdot{\unitpos}})
=-\scalq\scalr^{-3}[\dprod{\vectkap}{\{\cprod{\vectr}{(\cprod{\vectu}{\vectr})}\}}]\nonumber\\
&=-\scalq\scalr^{-3}[\dprod{\vectkap}{(\cprod{\vectr}{\vecth})}]\beqref{main5c}\nonumber\\
&=-\kaprep\vrhoa\epsve\beqref{gpath1b}\text{ \& }\eqnref{ogrv1a}
\end{align}
\begin{align}\label{gpath2c}
\fffdot{\scalm}
&=\dif{[-\scalq\scalr^{-3}\{\dprod{\vectkap}{(\cprod{\vectr}{\vecth})}\}]}\beqref{gpath2b}\nonumber\\
&=-\scalq\dif{[\scalr^{-3}]}[\dprod{\vectkap}{(\cprod{\vectr}{\vecth})}]
  -\scalq\scalr^{-3}\dif{[\dprod{\vectkap}{(\cprod{\vectr}{\vecth})}]}\nonumber\\
&=3\scalq\scalr^{-4}(\dprod{\unitpos}{\vectu})[\dprod{\vectkap}{(\cprod{\vectr}{\vecth})}]
  -\scalq\scalr^{-3}[\dprod{\vectkap}{(\cprod{\vectu}{\vecth})}]\nonumber\\
&=3\scalq\scalr^{-4}\scalh^{-2}[\dprod{\unitpos}{(\cprod{\vectz}{\vecth})}][\dprod{\vectkap}{(\cprod{\vectr}{\vecth})}]
  -\scalq\scalr^{-3}[\dprod{\vectkap}{(-\vectz-\scalq\unitpos)}]\beqref{main5c}\nonumber\\
&=3\scalq\scalr^{-4}\scalh^{-2}\epsvh\epsve\kaprep-\scalq\scalr^{-3}(-\kaprep\dltvb-\scalq\kaprep\epsva)
  \beqref{ogrv1a}\text{ \& }\eqnref{gxpeed1a}\nonumber\\
&=3\scalq\scalr^{-4}\scalh^{-2}\epsvh\epsve\kaprep+\kaprep\scalq\scalr^{-3}\efkta
  \beqref{gxpeed1b}\nonumber\\
&=\kaprep\scalq\scalr^{-3}(\efkta+3\scalr^{-1}\scalh^{-2}\epsvh\epsve)
=\kaprep\vrhoa(\efkta+3\vrhob\epsve)
=\kaprep\vrhoc\beqref{gpath1b}
\end{align}
\end{subequations}
\begin{subequations}\label{gpath3}
\begin{align}\label{gpath3a}
\fdot{\scaln}
&=\dif{[\dprod{\unitplz}{(\cprod{\vectr}{\vecth})}]}\beqref{main5c}\nonumber\\
&=\dprod{\unitplz}{(\cprod{\vectu}{\vecth})}
=-\dprod{\unitplz}{(\cprod{\vecth}{\vectu})}\nonumber\\
&=-\dprod{\unitplz}{\vectz}-\scalq(\dprod{\unitplz}{\unitpos})\beqref{main5c}\nonumber\\
&=-(\dltvd+\scalq\epsvc)\beqref{ogrv1a}\text{ \& }\eqnref{gxpeed1a}\nonumber\\
&=-\efktc\beqref{gxpeed1b}
\end{align}
\begin{align}\label{gpath3b}
\ffdot{\scaln}
&=\dif{[-\dprod{\unitplz}{\vectz}-\scalq(\dprod{\unitplz}{\unitpos})]}\beqref{gpath3a}\nonumber\\
&=-\scalq(\dprod{\unitplz}{\fdot{\unitpos}})
=-\scalq\scalr^{-3}[\dprod{\unitplz}{\{\cprod{\vectr}{(\cprod{\vectu}{\vectr})}\}}]\nonumber\\
&=-\scalq\scalr^{-3}[\dprod{\unitplz}{(\cprod{\vectr}{\vecth})}]\beqref{main5c}\nonumber\\
&=-\vrhoa\epsvf\beqref{gpath1b}\text{ \& }\eqnref{ogrv1a}
\end{align}
\begin{align}\label{gpath3c}
\fffdot{\scaln}
&=\dif{[-\scalq\scalr^{-3}\{\dprod{\unitplz}{(\cprod{\vectr}{\vecth})}\}]}\beqref{gpath3b}\nonumber\\
&=-\scalq\dif{[\scalr^{-3}]}[\dprod{\unitplz}{(\cprod{\vectr}{\vecth})}]
  -\scalq\scalr^{-3}\dif{[\dprod{\unitplz}{(\cprod{\vectr}{\vecth})}]}\nonumber\\
&=3\scalq\scalr^{-4}(\dprod{\unitpos}{\vectu})[\dprod{\unitplz}{(\cprod{\vectr}{\vecth})}]
  -\scalq\scalr^{-3}[\dprod{\unitplz}{(\cprod{\vectu}{\vecth})}]\nonumber\\
&=3\scalq\scalr^{-4}\scalh^{-2}[\dprod{\unitpos}{(\cprod{\vectz}{\vecth})}][\dprod{\unitplz}{(\cprod{\vectr}{\vecth})}]
  -\scalq\scalr^{-3}[\dprod{\unitplz}{(-\vectz-\scalq\unitpos)}]\beqref{main5c}\nonumber\\
&=3\scalq\scalr^{-4}\scalh^{-2}\epsvh\epsvf-\scalq\scalr^{-3}(-\dltvd-\scalq\epsvc)
  \beqref{ogrv1a}\text{ \& }\eqnref{gxpeed1a}\nonumber\\
&=3\scalq\scalr^{-4}\scalh^{-2}\epsvh\epsvf+\scalq\scalr^{-3}\efktc
  \beqref{gxpeed1b}\nonumber\\
&=\scalq\scalr^{-3}(\efktc+3\scalr^{-1}\scalh^{-2}\epsvh\epsvf)
=\vrhoa(\efktc+3\vrhob\epsvf)
=\vrhod\beqref{gpath1b}
\end{align}
\begin{align}\label{gpath3d}
\scaln\ffdot{\scalm}-\scalm\ffdot{\scaln}
&=\scaln(-\kaprep\vrhoa\epsve)-\scalm(-\vrhoa\epsvf)
  \beqref{gpath2b}\text{ \& }\eqnref{gpath3b}\nonumber\\
&=-\kaprep\vrhoa\epsve\scaln+\vrhoa\epsvf\scalm\nonumber\\
&=-\kaprep\vrhoa\epsve(\epsvf)+\vrhoa\epsvf(\kaprep\epsve)
  \beqref{ogrv2b}\text{ \& }\eqnref{ogrv2c}\nonumber\\
&=-\kaprep\vrhoa\epsve\epsvf+\kaprep\vrhoa\epsvf\epsve = 0
\end{align}
\end{subequations}
\begin{subequations}\label{gpath4}
\begin{align}\label{gpath4a}
\dif{[\kcons^2]}
&=\dif{[(\scalm/\scaln)(\dprod{\vectkap}{\unitplz})]}\beqref{main5c}\nonumber\\
&=(\dprod{\vectkap}{\unitplz})\dif{[(\scalm/\scaln)]}
=(\dprod{\vectkap}{\unitplz})[(\scaln\fdot{\scalm}-\scalm\fdot{\scaln})/\scaln^2]\nonumber\\
\therefore\fdot{\kcons}
&=(\dprod{\vectkap}{\unitplz})[(\scaln\fdot{\scalm}-\scalm\fdot{\scaln})/(2\kcons\scaln^2)]\nonumber\\
&=\kaprep\epsvb(\scaln\fdot{\scalm}-\scalm\fdot{\scaln})/(2\kcons\scaln^2)\beqref{ogrv1a}\nonumber\\
&=\kaprep\epsvb[\scaln(-\kaprep\efkta)-\scalm(-\efktc)]/(2\kcons\scaln^2)
   \beqref{gpath2a}\text{ \& }\eqnref{gpath3a}\nonumber\\
&=\kaprep\epsvb[\epsvf(-\kaprep\efkta)-\kaprep\epsve(-\efktc)]/(2\kaprep\vphic\epsvf^2)
   \beqref{ogrv2b}, \eqnref{ogrv2c}\text{ \& }\eqnref{ogrv2d}\nonumber\\
&=\kaprep\epsvb(-\kaprep\epsvf\efkta+\kaprep\epsve\efktc)/(2\kaprep\vphic\epsvf^2)\nonumber\\
&=\kaprep\epsvb(\epsve\efktc-\epsvf\efkta)/(2\vphic\epsvf^2)
=\kaprep\vrhoe\beqref{gpath1b}
\end{align}
\begin{align*}
\ffdot{\kcons}
&=\dif{[(\dprod{\vectkap}{\unitplz})\{(\scaln\fdot{\scalm}-\scalm\fdot{\scaln})/(2\kcons\scaln^2)\}]}
=(\dprod{\vectkap}{\unitplz})\dif{[(\scaln\fdot{\scalm}-\scalm\fdot{\scaln})/(2\kcons\scaln^2)]}
  \beqref{gpath4a}\nonumber\\
&=(\dprod{\vectkap}{\unitplz})\biggl[\frac{(2\kcons\scaln^2)\dif{(\scaln\fdot{\scalm}-\scalm\fdot{\scaln})}
  -(\scaln\fdot{\scalm}-\scalm\fdot{\scaln})\dif{(2\kcons\scaln^2)}}{(2\kcons\scaln^2)^2}\biggr]
  \nonumber\\
&=(\dprod{\vectkap}{\unitplz})\biggl[\frac{(2\kcons\scaln^2)\dif{(\scaln\fdot{\scalm}-\scalm\fdot{\scaln})}}{(2\kcons\scaln^2)^2}\biggr]
  -(\dprod{\vectkap}{\unitplz})\biggl[\frac{(\scaln\fdot{\scalm}-\scalm\fdot{\scaln})\dif{(2\kcons\scaln^2)}}{(2\kcons\scaln^2)^2}\biggr]
  \nonumber\\
&=(\dprod{\vectkap}{\unitplz})\biggl[\frac{\dif{(\scaln\fdot{\scalm}-\scalm\fdot{\scaln})}}{2\kcons\scaln^2}\biggr]
  -\fdot{\kcons}\biggl[\frac{\dif{(2\kcons\scaln^2)}}{2\kcons\scaln^2}\biggr]\beqref{gpath4a}
  \nonumber\\
&=(\dprod{\vectkap}{\unitplz})\biggl[\frac{\fdot{\scaln}\fdot{\scalm}+\scaln\ffdot{\scalm}
     -\fdot{\scalm}\fdot{\scaln}-\scalm\ffdot{\scaln}}{2\kcons\scaln^2}\biggr]
  -\fdot{\kcons}\biggl[\frac{2\fdot{\kcons}\scaln^2+4\kcons\scaln\fdot{\scaln}}{2\kcons\scaln^2}\biggr]
\end{align*}
\begin{align}\label{gpath4b}
&=\frac{(\dprod{\vectkap}{\unitplz})(\scaln\ffdot{\scalm}-\scalm\ffdot{\scaln})
  -2\fdot{\kcons}(\fdot{\kcons}\scaln^2+2\kcons\scaln\fdot{\scaln})}{2\kcons\scaln^2}
=-\frac{\fdot{\kcons}(\fdot{\kcons}\scaln+2\kcons\fdot{\scaln})}{\kcons\scaln}
  \beqref{gpath3d}
\nonumber\\
&=-\frac{(\kaprep\vrhoe)[(\kaprep\vrhoe)\scaln+2\kcons(-\efktc)]}{\kcons\scaln}
  \beqref{gpath3a}\text{ \& }\eqnref{gpath4a}
\nonumber\\
&=-\frac{\kaprep\vrhoe[\kaprep\vrhoe(\epsvf)-2\efktc(\kaprep\vphic)]}{(\kaprep\vphic)(\epsvf)}
  \beqref{ogrv2c}\text{ \& }\eqnref{ogrv2d}
\nonumber\\
&=-\kaprep\vrhoe(\vrhoe\epsvf-2\efktc\vphic)/(\vphic\epsvf)
=-\kaprep\vrhof\beqref{gpath1b}
\end{align}
\begin{align*}
\fffdot{\kcons}
&=\dif{\biggl[-\frac{\fdot{\kcons}(\fdot{\kcons}\scaln+2\kcons\fdot{\scaln})}{\kcons\scaln}\biggr]}
  \beqref{gpath4b}\nonumber\\
&=\frac{-(\kcons\scaln)\dif{[\fdot{\kcons}(\fdot{\kcons}\scaln-2\kcons\fdot{\scaln})]}
  +[\fdot{\kcons}(\fdot{\kcons}\scaln+2\kcons\fdot{\scaln})]\dif{(\kcons\scaln)}}{(\kcons\scaln)^2}
  \nonumber\\
&=\frac{-\dif{[\fdot{\kcons}(\fdot{\kcons}\scaln-2\kcons\fdot{\scaln})]}
  -\ffdot{\kcons}\dif{(\kcons\scaln)}}{\kcons\scaln}\beqref{gpath4b}
  \nonumber\\
&=\frac{-\ffdot{\kcons}(\fdot{\kcons}\scaln-2\kcons\fdot{\scaln})
  -\fdot{\kcons}(\ffdot{\kcons}\scaln+\fdot{\kcons}\fdot{\scaln}-2\fdot{\kcons}\fdot{\scaln}-2\kcons\ffdot{\scaln})
  -\ffdot{\kcons}(\fdot{\kcons}\scaln+\kcons\fdot{\scaln})}{\kcons\scaln}
  \nonumber\\
&=\frac{-\ffdot{\kcons}\fdot{\kcons}\scaln+2\ffdot{\kcons}\kcons\fdot{\scaln}
  -\fdot{\kcons}\ffdot{\kcons}\scaln+\fdot{\kcons}^2\fdot{\scaln}+2\kcons\fdot{\kcons}\ffdot{\scaln}
  -\ffdot{\kcons}\fdot{\kcons}\scaln-\ffdot{\kcons}\kcons\fdot{\scaln}}{\kcons\scaln}
  \nonumber\\
&=\frac{\ffdot{\kcons}\kcons\fdot{\scaln}
  -3\fdot{\kcons}\ffdot{\kcons}\scaln+\fdot{\kcons}^2\fdot{\scaln}+2\kcons\fdot{\kcons}\ffdot{\scaln}}{\kcons\scaln}
\end{align*}
\begin{align}\label{gpath4c}
\begin{split}
&=\frac{(-\kaprep\vrhof)\kcons(-\efktc)
  -3(\kaprep\vrhoe)(-\kaprep\vrhof)\scaln+(\kaprep\vrhoe)^2(-\efktc)+2\kcons(\kaprep\vrhoe)(-\vrhoa\epsvf)}{\kcons\scaln}\\
  &\quad\beqref{gpath3}, \eqnref{gpath4a}\text{ \& }\eqnref{gpath4b}
\end{split}
\nonumber\\
&=\frac{\kaprep\vrhof\efktc\kcons+3\kaprep^2\vrhoe\vrhof\scaln-\kaprep^2\vrhoe^2\efktc
  -2\kaprep\vrhoe\vrhoa\epsvf\kcons}{\kcons\scaln}
  \nonumber\\
&=\frac{\kaprep\vrhof\efktc(\kaprep\vphic)+3\kaprep^2\vrhoe\vrhof(\epsvf)-\kaprep^2\vrhoe^2\efktc
  -2\kaprep\vrhoe\vrhoa\epsvf(\kaprep\vphic)}{(\kaprep\vphic)(\epsvf)}\beqref{ogrv2c}\text{ \& }\eqnref{ogrv2d}
  \nonumber\\
&=\frac{\kaprep^2\vrhof\efktc\vphic+3\kaprep^2\vrhoe\vrhof\epsvf-\kaprep^2\vrhoe^2\efktc
  -2\kaprep^2\vrhoe\vrhoa\epsvf\vphic}{\kaprep\vphic\epsvf}
  \nonumber\\
&=\kaprep[\vrhof\efktc\vphic+\vrhoe(3\vrhof\epsvf-\vrhoe\efktc-2\vrhoa\epsvf\vphic)]/(\vphic\epsvf)
=\kaprep\vrhoh\beqref{gpath1b}.
\end{align}
\end{subequations}

\subart{Derivatives of $\vecta$ and $\gamrep$}
Thus, from the foregoing derivations, we obtain\footnote{In the expressions for the derivatives of
$\gamrep$, we are to take the positive sign when $\kcons>\kaprep$ and the negative sign
when $\kcons<\kaprep$. The case $\kcons=\kaprep$ is forbidden since $\gamrep$ is nonzero.}
\begin{subequations}\label{gpath5}
\begin{align}\label{gpath5a}
\fdotg
&=\dif{\left[\biggl|1-\frac{\kcons^2}{\kaprep^2}\biggr|^{1/2}\right]}\beqref{main5a}\nonumber\\
&=\absign\frac{1}{2}\biggl|1-\frac{\kcons^2}{\kaprep^2}\biggr|^{-1/2}\dif{\biggl[\frac{\kcons^2}{\kaprep^2}\biggr]}
=\absign\frac{1}{2\gamrep}\biggl[\frac{2\kcons\fdot{\kcons}}{\kaprep^2}\biggr]
  \beqref{main5a}\nonumber\\
&=\absign\frac{\kcons\fdot{\kcons}}{\gamrep\kaprep^2}
=\absign\frac{(\kaprep\vphic)(\kaprep\vrhoe)}{\vphie\kaprep^2}
  \beqref{gpath4a}, \eqnref{ogrv2d}\text{ \& }\eqnref{ogrv2e}\nonumber\\
&=\absign(\vphic\vrhoe)/\vphie
=\vrhog\beqref{gpath1b}
\end{align}
\begin{align}\label{gpath5b}
\ffdotg
&=\absign\dif{\biggl[\frac{\kcons\fdot{\kcons}}{\gamrep\kaprep^2}\biggr]}
  \beqref{gpath5a}\nonumber\\
&=\absign\frac{1}{\kaprep^2}\biggl[\frac{\gamrep\dif{(\kcons\fdot{\kcons})}-(\kcons\fdot{\kcons})\fdotg}{\gamrep^2}\biggr]
=\absign\biggl[\frac{\gamrep\dif{(\kcons\fdot{\kcons})}}{\gamrep^2\kaprep^2}
  -\frac{(\kcons\fdot{\kcons})\fdotg}{\gamrep^2\kaprep^2}\biggr]\nonumber\\
&=\absign\biggl[\frac{\fdot{\kcons}^2+\kcons\ffdot{\kcons}}{\gamrep\kaprep^2}\biggr]
  -\biggl[\frac{\fdotg^2}{\gamrep}\biggr]\beqref{gpath5a}\nonumber\\
&=\absign\biggl[\frac{(\kaprep\vrhoe)^2+\kcons(-\kaprep\vrhof)}{\gamrep\kaprep^2}\biggr]
  -\biggl[\frac{(\vrhog)^2}{\gamrep}\biggr]\beqref{gpath4}\text{ \& }\eqnref{gpath5a}\nonumber\\
&=\absign\biggl[\frac{\kaprep^2\vrhoe^2-\kaprep\vrhof\kcons}{\gamrep\kaprep^2}\biggr]
  -\biggl[\frac{\vrhog^2}{\gamrep}\biggr]
=\absign\biggl[\frac{\kaprep^2\vrhoe^2-\kaprep\vrhof(\kaprep\vphic)}{(\vphie)\kaprep^2}\biggr]
  -\biggl[\frac{\vrhog^2}{(\vphie)}\biggr]\beqref{ogrv2d}\text{ \& }\eqnref{ogrv2e}\nonumber\\
&=(1/\vphie)[\absign(\vrhoe^2-\vrhof\vphic)-\vrhog^2]
=\vrhoi\beqref{gpath1b}
\end{align}
\begin{align*}
\fffdotg
&=\absign\dif{\biggl[\frac{\fdot{\kcons}^2+\kcons\ffdot{\kcons}}{\gamrep\kaprep^2}\biggr]}
  -\dif{\biggl[\frac{\fdotg^2}{\gamrep}\biggr]}\beqref{gpath5b}\nonumber\\
&=\absign\frac{1}{\kaprep^2}\biggl[\frac{\gamrep\dif{(\fdot{\kcons}^2+\kcons\ffdot{\kcons})}
     -(\fdot{\kcons}^2+\kcons\ffdot{\kcons})\fdotg}{\gamrep^2} \biggr]
  -\biggl[\frac{\gamrep\dif{(\fdotg^2)}-\fdotg^3}{\gamrep^2}\biggr]\nonumber\\
&=\absign\biggl[\frac{\gamrep\dif{(\fdot{\kcons}^2+\kcons\ffdot{\kcons})}}{\gamrep^2\kaprep^2}\biggr]
  -\biggl[\absign\frac{(\fdot{\kcons}^2+\kcons\ffdot{\kcons})\fdotg}{\gamrep^2\kaprep^2} \biggr]
  -\biggl[\frac{\gamrep\dif{(\fdotg^2)}}{\gamrep^2}\biggr]
  +\biggl[\frac{\fdotg^3}{\gamrep^2}\biggr]\nonumber\\
&=\absign\biggl[\frac{\gamrep\dif{(\fdot{\kcons}^2+\kcons\ffdot{\kcons})}}{\gamrep^2\kaprep^2}\biggr]
  -\biggl[\frac{\gamrep\dif{(\fdotg^2)}}{\gamrep^2}\biggr]
  -\frac{\fdotg}{\gamrep}\biggl[\absign\frac{(\fdot{\kcons}^2+\kcons\ffdot{\kcons})}{\gamrep\kaprep^2}
     -\frac{\fdotg^2}{\gamrep}\biggr]
\end{align*}
\begin{align}\label{gpath5c}
&=\absign\biggl[\frac{2\fdot{\kcons}\ffdot{\kcons}+\fdot{\kcons}\ffdot{\kcons}+\kcons\fffdot{\kcons}}{\gamrep\kaprep^2}\biggr]
  -\biggl[\frac{2\fdotg\ffdotg}{\gamrep}\biggr]
  -\frac{\fdotg\ffdotg}{\gamrep}\beqref{gpath5b}\nonumber\\
&=\absign\biggl[\frac{3\fdot{\kcons}\ffdot{\kcons}+\kcons\fffdot{\kcons}}{\gamrep\kaprep^2}\biggr]
  -\frac{3\fdotg\ffdotg}{\gamrep}\nonumber\\
&=\absign\biggl[\frac{3(\kaprep\vrhoe)(-\kaprep\vrhof)+\kcons(\kaprep\vrhoh)}{\gamrep\kaprep^2}\biggr]
  -\frac{3(\vrhog)(\vrhoi)}{\gamrep}\beqref{gpath4}, \eqnref{gpath5a}\text{ \& }\eqnref{gpath5b}\nonumber\\
&=\absign\biggl[\frac{-3\kaprep^2\vrhoe\vrhof+\kaprep\vrhoh\kcons}{\gamrep\kaprep^2}\biggr]
  -\frac{3\vrhog\vrhoi}{\gamrep}\nonumber\\
&=\absign\biggl[\frac{-3\kaprep^2\vrhoe\vrhof+\kaprep\vrhoh(\kaprep\vphic)}{(\vphie)\kaprep^2}\biggr]
  -\frac{3\vrhog\vrhoi}{(\vphie)}\beqref{ogrv2d}\text{ \& }\eqnref{ogrv2e}\nonumber\\
&=(1/\vphie)[\absign(\vrhoh\vphic-3\vrhoe\vrhof)-3\vrhog\vrhoi]
=\vrhoj\beqref{gpath1b}.
\end{align}
\end{subequations}
We have also that
\begin{subequations}\label{gpath6}
\begin{align}\label{gpath6a}
\fdota
&=\dif{(-\scalq\vectr/\scalr^3)}\beqref{main5a}\nonumber\\
&=-\scalq\biggl[\frac{\scalr^3\vectu-3\scalr^2\fdot{\scalr}\vectr}{\scalr^6}\biggr]
=-\frac{\scalq\vectu}{\scalr^3}+\frac{3\scalq\fdot{\scalr}\vectr}{\scalr^4}
=-\frac{\scalq\vectu}{\scalr^3}+\frac{3\scalq(\dprod{\unitpos}{\vectu})\unitpos}{\scalr^3}\nonumber\\
&=-\vrhoa\vectu+3\vrhoa(\dprod{\unitpos}{\vectu})\unitpos\beqref{gpath1b}\nonumber\\
&=-\vrhoa\vectu+3\vrhoa[\scalh^{-2}\dprod{\unitpos}{(\cprod{\vectz}{\vecth})}]\unitpos\beqref{main5c}\nonumber\\
&=-\vrhoa\vectu+3\vrhoa[\scalh^{-2}\epsvh]\unitpos\beqref{ogrv1a}\nonumber\\
&=-\vrhoa\vectu+3\scalr\vrhoa\vrhob\unitpos\beqref{gpath1b}\nonumber\\
&=-\vrhoa\vectu+3\vrhoa\vrhob\vectr
\end{align}
\begin{align*}
\ffdota
&=\dif{\biggl[-\frac{\scalq\vectu}{\scalr^3}\biggr]}
  +\dif{\biggl[\frac{3\scalq(\dprod{\vectr}{\vectu})\vectr}{\scalr^5}\biggr]}
  \beqref{gpath6a}\nonumber\\
&=-\scalq\biggl[\frac{\scalr^3\vecta-3\scalr^2\fdot{\scalr}\vectu}{\scalr^6}\biggr]
  +3\scalq\biggl[\frac{\scalr^5\dif{[(\dprod{\vectr}{\vectu})\vectr]}}{\scalr^{10}}
      -\frac{5\scalr^4\fdot{\scalr}[(\dprod{\vectr}{\vectu})\vectr]}{\scalr^{10}}\biggr]
  \nonumber\\
&=-\scalq\biggl[
      \frac{\vecta}{\scalr^3}
      -\frac{3\fdot{\scalr}\vectu}{\scalr^4}
      -\frac{3\dif{(\dprod{\vectr}{\vectu})}\vectr}{\scalr^5}
      -\frac{3(\dprod{\vectr}{\vectu})\vectu}{\scalr^5}
      +\frac{15\fdot{\scalr}(\dprod{\vectr}{\vectu})\vectr}{\scalr^6}
   \biggr]
  \nonumber\\
&=-\scalq\biggl[
      \frac{\vecta}{\scalr^3}
      -\frac{3\vectr\dif{(\dprod{\vectr}{\vectu})}}{\scalr^5}
      -\frac{3(\dprod{\vectr}{\vectu})\vectu}{\scalr^5}
      -\frac{3\fdot{\scalr}\vectu}{\scalr^4}
      +\frac{15\fdot{\scalr}(\dprod{\vectr}{\vectu})\vectr}{\scalr^6}
   \biggr]
  \nonumber\\
&=-\frac{\scalq\vecta}{\scalr^3}
  +\frac{3\scalq\vectr\dif{(\dprod{\vectr}{\vectu})}}{\scalr^5}
  +\frac{3\scalq(\dprod{\vectr}{\vectu})\vectu}{\scalr^5}
  +\frac{3\scalq\fdot{\scalr}\vectu}{\scalr^4}
  -\frac{15\scalq\fdot{\scalr}(\dprod{\vectr}{\vectu})\vectr}{\scalr^6}
\end{align*}
\begin{align*}
&=-\frac{\scalq\vecta}{\scalr^3}
  +\frac{3\scalq(\dprod{\vectu}{\vectu})\vectr}{\scalr^5}
  +\frac{3\scalq(\dprod{\vectr}{\vecta})\vectr}{\scalr^5}
  +\frac{3\scalq(\dprod{\vectr}{\vectu})\vectu}{\scalr^5}
  +\frac{3\scalq\fdot{\scalr}\vectu}{\scalr^4}
  -\frac{15\scalq\fdot{\scalr}(\dprod{\vectr}{\vectu})\vectr}{\scalr^6}
  \nonumber\\
&=-\frac{\scalq\vecta}{\scalr^3}
  +\frac{3\scalq(\dprod{\vectu}{\vectu})\vectr}{\scalr^5}
  +\frac{3\scalq(\dprod{\vectr}{\vecta})\vectr}{\scalr^5}
  +\frac{3\scalq(\dprod{\vectr}{\vectu})\vectu}{\scalr^5}
  +\frac{3\scalq(\dprod{\vectr}{\vectu})\vectu}{\scalr^5}
  -\frac{15\scalq(\dprod{\vectr}{\vectu})^2\vectr}{\scalr^7}
  \nonumber\\
&=\frac{\scalq^2\vectr}{\scalr^6}
  +\frac{3\scalq(\dprod{\vectu}{\vectu})\vectr}{\scalr^5}
  +\frac{3\scalq(\dprod{\vectr}{\vecta})\vectr}{\scalr^5}
  +\frac{6\scalq(\dprod{\vectr}{\vectu})\vectu}{\scalr^5}
  -\frac{15\scalq(\dprod{\vectr}{\vectu})^2\vectr}{\scalr^7}
  \beqref{main5a}\nonumber\\
\begin{split}
&=\frac{\scalq^2\vectr}{\scalr^6}
  +\frac{3\scalq\vphih^2\vectr}{\scalr^5}
  -\frac{3\scalq^2\vectr}{\scalr^6}
  +\frac{6\scalq[\scalh^{-2}\dprod{\vectr}{(\cprod{\vectz}{\vecth})}]\vectu}{\scalr^5}
  -\frac{15\scalq[\scalh^{-2}\dprod{\vectr}{(\cprod{\vectz}{\vecth})}]^2\vectr}{\scalr^7}\\
  &\quad\beqref{main5a}, \eqnref{ogrv3b}\text{ \& }\eqnref{main5c}
\end{split}
\end{align*}
\begin{align}\label{gpath6b}
&=\frac{3\scalq\vphih^2\vectr}{\scalr^5}
  -\frac{2\scalq^2\vectr}{\scalr^6}
  +\frac{6\scalq\epsvh\vectu}{\scalh^2\scalr^4}
  -\frac{15\scalq\epsvh^2\vectr}{\scalh^4\scalr^5}
  \beqref{ogrv1a}
\nonumber\\
&=\frac{3\scalq\vphih^2\vectr}{\scalr^5}
  -\frac{2\scalq^2\vectr}{\scalr^6}
  +\frac{6\scalq\vrhob\vectu}{\scalr^3}
  -\frac{15\scalq\vrhob^2\vectr}{\scalr^3}
  \beqref{gpath1b}\nonumber\\
&=3\vrhoa(\vphih/\scalr)^2\vectr-2\vrhoa^2\vectr+6\vrhoa\vrhob\vectu-15\vrhoa\vrhob^2\vectr
  \beqref{gpath1b}\nonumber\\
&=6\vrhoa\vrhob\vectu-\vrhoa[2\vrhoa+15\vrhob^2-3(\vphih/\scalr)^2]\vectr\nonumber\\
&=6\vrhoa\vrhob\vectu-\vrhoa(2\vrhoa+15\vrhob^2-3\vrhok^2)\vectr
=6\vrhoa\vrhob\vectu-\vrhoa\vrhol\vectr\beqref{gpath1b}
\end{align}
\begin{align*}
\fffdota
&=\dif{\biggl[
  \frac{\scalq^2\vectr}{\scalr^6}
  +\frac{3\scalq(\dprod{\vectu}{\vectu})\vectr}{\scalr^5}
  +\frac{3\scalq(\dprod{\vectr}{\vecta})\vectr}{\scalr^5}
  +\frac{6\scalq(\dprod{\vectr}{\vectu})\vectu}{\scalr^5}
  -\frac{15\scalq(\dprod{\vectr}{\vectu})^2\vectr}{\scalr^7}
  \biggr]}\beqref{gpath6b}\nonumber\\
\begin{split}
&=\scalq^2\biggl[\frac{\scalr^6\vectu-6\scalr^5\fdot{\scalr}\vectr}{\scalr^{12}}\biggr]
  +3\scalq\biggl[\frac{\scalr^5[2(\dprod{\vecta}{\vectu})\vectr+(\dprod{\vectu}{\vectu})\vectu]
     -5\scalr^4\fdot{\scalr}[(\dprod{\vectu}{\vectu})\vectr]}{\scalr^{10}}\biggr]\\
  &\quad+3\scalq\biggl[\frac{\scalr^5[(\dprod{\vectr}{\vecta})\vectu+(\dprod{\vectu}{\vecta}+\dprod{\vectr}{\fdota})\vectr]
     -5\scalr^4\fdot{\scalr}[(\dprod{\vectr}{\vecta})\vectr]}{\scalr^{10}}\biggr]\\
  &\quad+6\scalq\biggl[\frac{\scalr^5[(\dprod{\vectr}{\vectu})\vecta+(\dprod{\vectu}{\vectu}+\dprod{\vectr}{\vecta})\vectu]
     -5\scalr^4\fdot{\scalr}[(\dprod{\vectr}{\vectu})\vectu]}{\scalr^{10}}\biggr]\\
  &\quad-15\scalq\biggl[\frac{\scalr^7[(\dprod{\vectr}{\vectu})^2\vectu+2(\dprod{\vectr}{\vectu})(\dprod{\vectu}{\vectu}+\dprod{\vectr}{\vecta})\vectr]
     -7\scalr^6\fdot{\scalr}[(\dprod{\vectr}{\vectu})^2\vectr]}{\scalr^{14}}\biggr]
\end{split}
\end{align*}
\begin{align*}
\begin{split}
&=\scalq^2\biggl[\frac{\scalr^6\vectu}{\scalr^{12}}\biggr]
  -\scalq^2\biggl[\frac{6\scalr^5\fdot{\scalr}\vectr}{\scalr^{12}}\biggr]
  +3\scalq\biggl[\frac{2\scalr^5(\dprod{\vecta}{\vectu})\vectr}{\scalr^{10}}\biggr]
  +3\scalq\biggl[\frac{\scalr^5(\dprod{\vectu}{\vectu})\vectu}{\scalr^{10}}\biggr]
  +3\scalq\biggl[\frac{-5\scalr^4\fdot{\scalr}(\dprod{\vectu}{\vectu})\vectr}{\scalr^{10}}\biggr]\\
  &\quad+3\scalq\biggl[\frac{\scalr^5(\dprod{\vectr}{\vecta})\vectu}{\scalr^{10}}\biggr]
  +3\scalq\biggl[\frac{\scalr^5(\dprod{\vectu}{\vecta})\vectr}{\scalr^{10}}\biggr]
  +3\scalq\biggl[\frac{\scalr^5(\dprod{\vectr}{\fdota})\vectr}{\scalr^{10}}\biggr]
  +3\scalq\biggl[\frac{-5\scalr^4\fdot{\scalr}(\dprod{\vectr}{\vecta})\vectr}{\scalr^{10}}\biggr]\\
  &\quad+6\scalq\biggl[\frac{\scalr^5(\dprod{\vectr}{\vectu})\vecta}{\scalr^{10}}\biggr]
  +6\scalq\biggl[\frac{\scalr^5(\dprod{\vectu}{\vectu})\vectu}{\scalr^{10}}\biggr]
  +6\scalq\biggl[\frac{\scalr^5(\dprod{\vectr}{\vecta})\vectu}{\scalr^{10}}\biggr]
  +6\scalq\biggl[\frac{-5\scalr^4\fdot{\scalr}(\dprod{\vectr}{\vectu})\vectu}{\scalr^{10}}\biggr]\\
  &\quad-15\scalq\biggl[\frac{\scalr^7(\dprod{\vectr}{\vectu})^2\vectu}{\scalr^{14}}\biggr]
  -15\scalq\biggl[\frac{2\scalr^7(\dprod{\vectr}{\vectu})(\dprod{\vectu}{\vectu})\vectr}{\scalr^{14}}\biggr]
  -15\scalq\biggl[\frac{2\scalr^7(\dprod{\vectr}{\vectu})(\dprod{\vectr}{\vecta})\vectr}{\scalr^{14}}\biggr]
  -15\scalq\biggl[\frac{-7\scalr^6\fdot{\scalr}(\dprod{\vectr}{\vectu})^2\vectr}{\scalr^{14}}\biggr]
\end{split}
\end{align*}
\begin{align*}
\begin{split}
&=\frac{\scalq^2\vectu}{\scalr^6}
  -\frac{6\scalq^2\fdot{\scalr}\vectr}{\scalr^7}
  +\frac{6\scalq(\dprod{\vecta}{\vectu})\vectr}{\scalr^5}
  +\frac{3\scalq\scalu^2\vectu}{\scalr^5}
  -\frac{15\scalq\scalu^2\fdot{\scalr}\vectr}{\scalr^6}
  +\frac{3\scalq(\dprod{\vectr}{\vecta})\vectu}{\scalr^5}
  +\frac{3\scalq(\dprod{\vectu}{\vecta})\vectr}{\scalr^5}\\
  &\quad+\frac{3\scalq(\dprod{\vectr}{\fdota})\vectr}{\scalr^5}
  -\frac{15\scalq\fdot{\scalr}(\dprod{\vectr}{\vecta})\vectr}{\scalr^6}
  +\frac{6\scalq(\dprod{\vectr}{\vectu})\vecta}{\scalr^5}
  +\frac{6\scalq\scalu^2\vectu}{\scalr^5}
  +\frac{6\scalq(\dprod{\vectr}{\vecta})\vectu}{\scalr^5}
  -\frac{30\scalq\fdot{\scalr}(\dprod{\vectr}{\vectu})\vectu}{\scalr^6}\\
  &\quad-\frac{15\scalq(\dprod{\vectr}{\vectu})^2\vectu}{\scalr^7}
  -\frac{30\scalq\scalu^2(\dprod{\vectr}{\vectu})\vectr}{\scalr^7}
  -\frac{30\scalq(\dprod{\vectr}{\vectu})(\dprod{\vectr}{\vecta})\vectr}{\scalr^7}
  +\frac{105\scalq\fdot{\scalr}(\dprod{\vectr}{\vectu})^2\vectr}{\scalr^8}
\end{split}
\nonumber\\
\begin{split}
&=\frac{\scalq^2\vectu}{\scalr^6}
  -\frac{6\scalq^2(\dprod{\unitpos}{\vectu})\vectr}{\scalr^7}
  +\frac{6\scalq(\dprod{\vecta}{\vectu})\vectr}{\scalr^5}
  +\frac{3\scalq\scalu^2\vectu}{\scalr^5}
  -\frac{15\scalq\scalu^2(\dprod{\unitpos}{\vectu})\vectr}{\scalr^6}
  +\frac{3\scalq(\dprod{\vectr}{\vecta})\vectu}{\scalr^5}
  +\frac{3\scalq(\dprod{\vectu}{\vecta})\vectr}{\scalr^5}\\
  &\quad+\frac{3\scalq(\dprod{\vectr}{\fdota})\vectr}{\scalr^5}
  -\frac{15\scalq(\dprod{\unitpos}{\vectu})(\dprod{\vectr}{\vecta})\vectr}{\scalr^6}
  +\frac{6\scalq(\dprod{\vectr}{\vectu})\vecta}{\scalr^5}
  +\frac{6\scalq\scalu^2\vectu}{\scalr^5}
  +\frac{6\scalq(\dprod{\vectr}{\vecta})\vectu}{\scalr^5}
  -\frac{30\scalq(\dprod{\unitpos}{\vectu})(\dprod{\vectr}{\vectu})\vectu}{\scalr^6}\\
  &\quad-\frac{15\scalq(\dprod{\vectr}{\vectu})^2\vectu}{\scalr^7}
  -\frac{30\scalq\scalu^2(\dprod{\vectr}{\vectu})\vectr}{\scalr^7}
  -\frac{30\scalq(\dprod{\vectr}{\vectu})(\dprod{\vectr}{\vecta})\vectr}{\scalr^7}
  +\frac{105\scalq(\dprod{\unitpos}{\vectu})(\dprod{\vectr}{\vectu})^2\vectr}{\scalr^8}
\end{split}
\end{align*}
\begin{align*}
\begin{split}
&=\frac{\scalq^2\vectu}{\scalr^6}
  -\frac{6\scalq^2(\dprod{\unitpos}{\vectu})\unitpos}{\scalr^6}
  +\frac{6\scalq(\dprod{\vecta}{\vectu})\unitpos}{\scalr^4}
  +\frac{3\scalq\scalu^2\vectu}{\scalr^5}
  -\frac{15\scalq\scalu^2(\dprod{\unitpos}{\vectu})\unitpos}{\scalr^5}
  +\frac{3\scalq(\dprod{\unitpos}{\vecta})\vectu}{\scalr^4}
  +\frac{3\scalq(\dprod{\vectu}{\vecta})\unitpos}{\scalr^4}\\
  &\quad+\frac{3\scalq(\dprod{\unitpos}{\fdota})\unitpos}{\scalr^3}
  -\frac{15\scalq(\dprod{\unitpos}{\vectu})(\dprod{\unitpos}{\vecta})\unitpos}{\scalr^4}
  +\frac{6\scalq(\dprod{\unitpos}{\vectu})\vecta}{\scalr^4}
  +\frac{6\scalq\scalu^2\vectu}{\scalr^5}
  +\frac{6\scalq(\dprod{\unitpos}{\vecta})\vectu}{\scalr^4}
  -\frac{30\scalq(\dprod{\unitpos}{\vectu})^2\vectu}{\scalr^5}\\
  &\quad-\frac{15\scalq(\dprod{\unitpos}{\vectu})^2\vectu}{\scalr^5}
  -\frac{30\scalq\scalu^2(\dprod{\unitpos}{\vectu})\unitpos}{\scalr^5}
  -\frac{30\scalq(\dprod{\unitpos}{\vectu})(\dprod{\unitpos}{\vecta})\unitpos}{\scalr^4}
  +\frac{105\scalq(\dprod{\unitpos}{\vectu})^3\unitpos}{\scalr^5}
\end{split}
\nonumber\\
\begin{split}
&=\vrhoa^2\vectu
  -6\vrhoa^2(\epsvh/\scalh^2)\unitpos
  +6(\vrhoa/\scalr)(-\vphia\dprod{\unitpos}{\vectu})\unitpos
  +3\vrhoa(\scalu^2/\scalr^2)\vectu
  -15\vrhoa(\scalu^2/\scalr^2)(\epsvh/\scalh^2)\unitpos\\
  &\quad+3(\vrhoa/\scalr)(-\vphia)\vectu
  +3(\vrhoa/\scalr)(-\vphia\dprod{\unitpos}{\vectu})\unitpos
  +3\vrhoa[\dprod{\unitpos}{(-\vrhoa\vectu+3\vrhoa\vrhob\vectr)}]\unitpos\\
  &\quad-15\vrhoa[\epsvh/(\scalr\scalh^2)](-\vphia)\unitpos
  +6(\vrhoa/\scalr)(\epsvh/\scalh^2)(-\vphia\unitpos)
  +6\vrhoa(\scalu^2/\scalr^2)\vectu
  +6(\vrhoa/\scalr)(-\vphia)\vectu\\
  &\quad-30\vrhoa[\epsvh/(\scalr\scalh^2)]^2\vectu
  -15\vrhoa[\epsvh/(\scalr\scalh^2)]^2\vectu
  -30\vrhoa\scalr(\scalu^2/\scalr^2)[\epsvh/(\scalr\scalh^2)]\unitpos
  -30\vrhoa[\epsvh/(\scalr\scalh^2)](-\vphia)\unitpos\\
  &\quad+105\scalr\vrhoa[\epsvh/(\scalr\scalh^2)]^3\unitpos
  \beqref{main5a}, \eqnref{main5c}, \eqnref{ogrv1a}, \eqnref{ogrv1b}, \eqnref{gpath1b}\text{ \& }\eqnref{gpath6a}
\end{split}
\end{align*}
\begin{align*}
\begin{split}
&=\vrhoa^2\vectu
  -6\scalr\vrhob\vrhoa^2\unitpos
  -6\vphia\vrhoa[\epsvh/(\scalr\scalh^2)]\unitpos
  +9\vrhoa(\scalu^2/\scalr^2)\vectu
  -15\scalr\vrhoa\vrhob(\scalu^2/\scalr^2)\unitpos
  -3\vphia(\vrhoa/\scalr)\vectu\\
  &\quad-3\vphia\vrhoa[\epsvh/(\scalr\scalh^2)]\unitpos
  +3\vrhoa[-\vrhoa(\epsvh/\scalh^2)+3\scalr\vrhoa\vrhob]\unitpos
  +15\vphia\vrhoa\vrhob\unitpos
  -6\vphia\vrhoa\vrhob\unitpos\\
  &\quad-6\vphia(\vrhoa/\scalr)\vectu
  -30\vrhoa\vrhob^2\vectu
  -15\vrhoa\vrhob^2\vectu
  -30\scalr\vrhoa\vrhob(\scalu^2/\scalr^2)\unitpos
  +30\vphia\vrhoa\vrhob\unitpos
  +105\scalr\vrhoa\vrhob^3\unitpos\\
  &\quad\beqref{main5c}, \eqnref{ogrv1a}, \eqnref{ogrv1b}\text{ \& }\eqnref{gpath1b}
\end{split}
\nonumber\\
\begin{split}
&=\vrhoa^2\vectu
  -6\scalr\vrhob\vrhoa^2\unitpos
  -6\vphia\vrhoa\vrhob\unitpos
  +9\vrhoa(\vphih^2/\scalr^2)\vectu
  -15\scalr\vrhoa\vrhob(\vphih^2/\scalr^2)\unitpos
  -3\vrhoa^2\vectu\\
  &\quad-3\vphia\vrhoa\vrhob\unitpos
  +3\vrhoa[-\scalr\vrhoa\vrhob+3\scalr\vrhoa\vrhob]\unitpos
  +15\vphia\vrhoa\vrhob\unitpos
  -6\vphia\vrhoa\vrhob\unitpos\\
  &\quad-6\vrhoa^2\vectu
  -30\vrhoa\vrhob^2\vectu
  -15\vrhoa\vrhob^2\vectu
  -30\scalr\vrhoa\vrhob(\vphih^2/\scalr^2)\unitpos
  +30\vphia\vrhoa\vrhob\unitpos
  +105\scalr\vrhoa\vrhob^3\unitpos\\
  &\quad\beqref{gpath1b}, \eqnref{ogrv3b}\text{ \& }\eqnref{ogrv1b}
\end{split}
\end{align*}
\begin{align}\label{gpath6c}
\begin{split}
&=(\vrhoa^2+9\vrhoa\vrhok-3\vrhoa^2-6\vrhoa^2-30\vrhoa\vrhob^2-15\vrhoa\vrhob^2)\vectu\\
  &\quad+(-6\scalr\vrhob\vrhoa^2-6\vphia\vrhoa\vrhob-15\scalr\vrhoa\vrhob\vrhok-3\vphia\vrhoa\vrhob
    +6\scalr\vrhoa^2\vrhob+15\vphia\vrhoa\vrhob-6\vphia\vrhoa\vrhob\\
    &\qquad-30\scalr\vrhoa\vrhob\vrhok+30\vphia\vrhoa\vrhob+105\scalr\vrhoa\vrhob^3)\unitpos
    \beqref{gpath1b}
\end{split}
\nonumber\\
\begin{split}
&=\vrhoa(9\vrhok-8\vrhoa-45\vrhob^2)\vectu
  +(-6\scalr\vrhob\vrhoa^2-6\vphia\vrhoa\vrhob-3\vphia\vrhoa\vrhob+15\vphia\vrhoa\vrhob-6\vphia\vrhoa\vrhob\\
    &\qquad+30\vphia\vrhoa\vrhob-15\scalr\vrhoa\vrhob\vrhok+6\scalr\vrhoa^2\vrhob
    -30\scalr\vrhoa\vrhob\vrhok+105\scalr\vrhoa\vrhob^3)\unitpos
\end{split}
\nonumber\\
&=\vrhoa(9\vrhok-8\vrhoa-45\vrhob^2)\vectu+\vrhoa\vrhob(30\vphia-45\scalr\vrhok+105\scalr\vrhob^2)\unitpos\nonumber\\
&=\vrhoa(9\vrhok-8\vrhoa-45\vrhob^2)\vectu+\vrhoa\vrhob(30\scalr\vrhoa-45\scalr\vrhok+105\scalr\vrhob^2)\unitpos
  \beqref{ogrv1b}\text{ \& }\eqnref{gpath1b}\nonumber\\
&=\vrhoa(9\vrhok-8\vrhoa-45\vrhob^2)\vectu+15\vrhoa\vrhob(2\vrhoa-3\vrhok+7\vrhob^2)\vectr
=\vrhom\vectu+\vrhon\vectr\beqref{gpath1b}.
\end{align}
\end{subequations}

\subart{Development of equations \eqnref{kpath5} through \eqnref{kpath9}}
We derive also
\begin{subequations}\label{gpath7}
\begin{align}\label{gpath7a}
\dprod{\unitkap}{\vectu}
&=\scalh^{-2}[\dprod{\unitkap}{(\cprod{\vectz}{\vecth}+\cprod{\scalq\unitpos}{\vecth})}]
  \beqref{main5c}\nonumber\\
&=\scalh^{-2}[\dprod{\unitkap}{(\cprod{\vectz}{\vecth})}]
  +\scalq\scalh^{-2}[\dprod{\unitkap}{(\cprod{\unitpos}{\vecth})}]\nonumber\\
&=\scalh^{-2}\epsvg+\scalh^{-2}\epsve(\scalq/\scalr)\beqref{ogrv1a}\nonumber\\
&=\scalh^{-2}(\epsvg+\epsve\vphib)\beqref{ogrv1b}\nonumber\\
&=\vrhoo\beqref{gpath1b}
\end{align}
\begin{align}\label{gpath7b}
\dprod{\unitpos}{\vectu}
&=\scalh^{-2}[\dprod{\unitpos}{(\cprod{\vectz}{\vecth}+\cprod{\scalq\unitpos}{\vecth})}]
  \beqref{main5c}\nonumber\\
&=\scalh^{-2}[\dprod{\unitpos}{(\cprod{\vectz}{\vecth})}]\nonumber\\
&=\epsvh/\scalh^2\beqref{ogrv1a}\nonumber\\
&=\scalr\vrhob\beqref{gpath1b}
\end{align}
\begin{align}\label{gpath7c}
\dprod{\unitplz}{\vectu}
&=\scalh^{-2}[\dprod{\unitplz}{(\cprod{\vectz}{\vecth}+\cprod{\scalq\unitpos}{\vecth})}]
  \beqref{main5c}\nonumber\\
&=\scalh^{-2}[\dprod{\unitplz}{(\cprod{\vectz}{\vecth})}]
  +\scalq\scalh^{-2}[\dprod{\unitplz}{(\cprod{\unitpos}{\vecth})}]\nonumber\\
&=\scalh^{-2}\epsvi+\scalh^{-2}\epsvf(\scalq/\scalr)\beqref{ogrv1a}\nonumber\\
&=\scalh^{-2}(\epsvi+\epsvf\vphib)\beqref{ogrv1b}
\end{align}
\begin{align}\label{gpath7d}
\dprod{\vectz}{\vectu}
&=\scalh^{-2}[\dprod{\vectz}{(\cprod{\vectz}{\vecth}+\cprod{\scalq\unitpos}{\vecth})}]
  \beqref{main5c}\nonumber\\
&=\scalh^{-2}[\dprod{\vectz}{(\cprod{\vectz}{\vecth})}]
  +\scalq\scalh^{-2}[\dprod{\vectz}{(\cprod{\unitpos}{\vecth})}]\nonumber\\
&=-\scalh^{-2}\epsvh(\scalq/\scalr)
=-\scalh^{-2}\epsvh\vphib\beqref{ogrv1a}\text{ \& }\eqnref{ogrv1b}\nonumber\\
&=-\scalr\vrhob\vphib\beqref{gpath1b}
\end{align}
\begin{align}\label{gpath7e}
\dprod{(\cprod{\vectr}{\vecth})}{\vectu}
&=\scalh^{-2}[\dprod{(\cprod{\vectr}{\vecth})}{(\cprod{\vectz}{\vecth}+\cprod{\scalq\unitpos}{\vecth})}]
  \beqref{main5c}\nonumber\\
&=\scalh^{-2}[\dprod{(\cprod{\vectr}{\vecth})}{(\cprod{\vectz}{\vecth})}]
  +\scalq\scalr\scalh^{-2}[\dprod{(\cprod{\unitpos}{\vecth})}{(\cprod{\unitpos}{\vecth})}]\nonumber\\
&=\scalh^{-2}[(\dprod{\vectr}{\vectz})(\dprod{\vecth}{\vecth})-(\dprod{\vectr}{\vecth})(\dprod{\vecth}{\vectz})]
  +\scalq\scalr\scalh^{-2}[\scalh^2-(\dprod{\unitpos}{\vecth})^2]
  \beqref{alg2}\nonumber\\
&=\scalh^{-2}(\scalr\epsvd\scalh^2)+\scalq\scalr\scalh^{-2}(\scalh^2)
  \beqref{ogrv1a}\text{ \& }\eqnref{main5c}\nonumber\\
&=\scalr(\epsvd+\scalq)
=\scalr\efktb\beqref{gxpeed1b}
\end{align}
\begin{align}\label{gpath7f}
\dprod{(\cprod{\vectz}{\vecth})}{\vectu}
&=\scalh^{-2}[\dprod{(\cprod{\vectz}{\vecth})}{(\cprod{\vectz}{\vecth}+\cprod{\scalq\unitpos}{\vecth})}]
  \beqref{main5c}\nonumber\\
&=\scalh^{-2}[\dprod{(\cprod{\vectz}{\vecth})}{(\cprod{\vectz}{\vecth})}]
  +\scalq\scalh^{-2}[\dprod{(\cprod{\vectz}{\vecth})}{(\cprod{\unitpos}{\vecth})}]
  \nonumber\\
&=\scalh^{-2}[\scalz^2\scalh^2-(\dprod{\vectz}{\vecth})^2]
  +\scalq\scalh^{-2}[(\dprod{\vectz}{\unitpos})\scalh^2-(\dprod{\vectz}{\vecth})(\dprod{\unitpos}{\vecth})]
  \beqref{alg2}\nonumber\\
&=\scalh^{-2}(\scalz^2\scalh^2)+\scalq\scalh^{-2}(\epsvd\scalh^2)
  \beqref{ogrv1a}\text{ \& }\eqnref{main5c}\nonumber\\
&=\scalz^2+\scalq\epsvd=\frkyz\beqref{gpath1h}.
\end{align}
\end{subequations}
Equations \eqnref{kpath5} through \eqnref{kpath9} therefore evaluate as
\begin{subequations}\label{gpath8}
\begin{align}\label{gpath8a}
\fdot{\alprep}
&=\dprod{\vectkap}{\fdota}\beqref{kpath5}\nonumber\\
&=\dprod{\vectkap}{(-\vrhoa\vectu+3\vrhoa\vrhob\vectr)}\beqref{gpath6a}\nonumber\\
&=-\kaprep\vrhoa(\dprod{\unitkap}{\vectu})+3\kaprep\vrhoa\vrhob(\dprod{\unitkap}{\vectr})\nonumber\\
&=-\kaprep\vrhoa\vrhoo+3\kaprep\scalr\vrhoa\vrhob\epsva\beqref{gpath7a}\text{ \& }\eqnref{ogrv1a}\nonumber\\
&=\kaprep\vrhoa(3\scalr\vrhob\epsva-\vrhoo)
=\kaprep\vrhop\beqref{gpath1b}
\end{align}
\begin{align}\label{gpath8b}
\ffdot{\alprep}
&=\dprod{\vectkap}{\ffdota}\beqref{kpath5}\nonumber\\
&=\dprod{\vectkap}{(6\vrhoa\vrhob\vectu-\vrhoa\vrhol\vectr)}\beqref{gpath6b}\nonumber\\
&=6\kaprep\vrhoa\vrhob(\dprod{\unitkap}{\vectu})-\kaprep\scalr\vrhoa\vrhol(\dprod{\unitkap}{\unitpos})\nonumber\\
&=6\kaprep\vrhoa\vrhob\vrhoo-\kaprep\scalr\vrhoa\vrhol\epsva\beqref{gpath7a}\text{ \& }\eqnref{ogrv1a}\nonumber\\
&=\kaprep\vrhoa(6\vrhob\vrhoo-\scalr\vrhol\epsva)
=\kaprep\vrhoq\beqref{gpath1b}
\end{align}
\begin{align}\label{gpath8c}
\fffdot{\alprep}
&=\dprod{\vectkap}{\fffdota}\beqref{kpath5}\nonumber\\
&=\dprod{\vectkap}{(\vrhom\vectu+\vrhon\vectr)}\beqref{gpath6c}\nonumber\\
&=\kaprep\vrhom(\dprod{\unitkap}{\vectu})+\kaprep\scalr\vrhon(\dprod{\unitkap}{\unitpos})\nonumber\\
&=\kaprep\vrhom\vrhoo+\kaprep\scalr\vrhon\epsva\beqref{gpath7a}\text{ \& }\eqnref{ogrv1a}\nonumber\\
&=\kaprep(\vrhom\vrhoo+\scalr\vrhon\epsva)
=\kaprep\vrhor\beqref{gpath1b}
\end{align}
\end{subequations}
\begin{subequations}\label{gpath9}
\begin{align}\label{gpath9a}
\fdot{\vthtrep}
&=\frac{1}{\gamrep^2\pfreq^2}\biggl[\fdot{\alprep}-\frac{2\alprep\fdot{\gamrep}}{\gamrep}\biggr]
  \beqref{kpath6a}\nonumber\\
&=\frac{1}{\vphie^2\pfreq^2}\biggl[\fdot{\alprep}-\frac{2(-\kaprep\vphia\epsva)\fdot{\gamrep}}{\vphie}\biggr]
  \beqref{ogrv2a}\text{ \& }\eqnref{ogrv2e}\nonumber\\
&=\frac{1}{\vphie^2\pfreq^2}\biggl[(\kaprep\vrhop)+\frac{2\kaprep\vphia\epsva(\vrhog)}{\vphie}\biggr]
  \beqref{gpath5a}\text{ \& }\eqnref{gpath8a}\nonumber\\
&=\frac{\kaprep}{\vphie^2\pfreq^2}\biggl[\vrhop+\frac{2\vphia\epsva\vrhog}{\vphie}\biggr]
=\kaprep\vrhos\beqref{gpath1c}
\end{align}
\begin{align}\label{gpath9b}
\ffdot{\vthtrep}
&=\frac{1}{\gamrep^2\pfreq^2}\biggl[\ffdot{\alprep}-\frac{4\fdot{\alprep}\fdotg}{\gamrep}
   -\frac{2\alprep\ffdotg}{\gamrep}+\frac{6\alprep\fdotg^2}{\gamrep^2}\biggr]
  \beqref{kpath6b}\nonumber\\
&=\frac{1}{\vphie^2\pfreq^2}\biggl[\ffdot{\alprep}-\frac{4\fdot{\alprep}\fdotg}{\vphie}
   -\frac{2(-\kaprep\vphia\epsva)\ffdotg}{\vphie}+\frac{6(-\kaprep\vphia\epsva)\fdotg^2}{\vphie^2}\biggr]
  \beqref{ogrv2a}\text{ \& }\eqnref{ogrv2e}\nonumber\\
&=\frac{1}{\vphie^2\pfreq^2}\biggl[(\kaprep\vrhoq)-\frac{4(\kaprep\vrhop)(\vrhog)}{\vphie}
   +\frac{2\kaprep\vphia\epsva(\vrhoi)}{\vphie}-\frac{6\kaprep\vphia\epsva(\vrhog)^2}{\vphie^2}\biggr]
  \beqref{gpath5}\text{ \& }\eqnref{gpath8}\nonumber\\
&=\frac{\kaprep}{\vphie^2\pfreq^2}\biggl[\vrhoq-\frac{4\vrhop\vrhog}{\vphie}
   +\frac{2\vphia\epsva\vrhoi}{\vphie}-\frac{6\vphia\epsva\vrhog^2}{\vphie^2}\biggr]
=\kaprep\vrhot\beqref{gpath1c}
\end{align}
\begin{align}\label{gpath9c}
\fffdot{\vthtrep}
&=\frac{1}{\gamrep^2\pfreq^2}\biggl[\fffdot{\alprep}-\frac{6\ffdot{\alprep}\fdotg}{\gamrep}
   -\frac{6\fdot{\alprep}\ffdotg}{\gamrep}+\frac{18\fdot{\alprep}\fdotg^2}{\gamrep^2}
   -\frac{2\alprep\fffdotg}{\gamrep}+\frac{18\alprep\fdotg\ffdotg}{\gamrep^2}
   -\frac{24\alprep\fdotg^3}{\gamrep^3}\biggr]
  \beqref{kpath6c}\nonumber\\
\begin{split}
&=\frac{1}{\vphie^2\pfreq^2}\biggl[\fffdot{\alprep}-\frac{6\ffdot{\alprep}\fdotg}{\vphie}
   -\frac{6\fdot{\alprep}\ffdotg}{\vphie}+\frac{18\fdot{\alprep}\fdotg^2}{\vphie^2}
   -\frac{2(-\kaprep\vphia\epsva)\fffdotg}{\vphie}+\frac{18(-\kaprep\vphia\epsva)\fdotg\ffdotg}{\vphie^2}
   -\frac{24(-\kaprep\vphia\epsva)\fdotg^3}{\vphie^3}\biggr]\\
  &\qquad\beqref{ogrv2a}\text{ \& }\eqnref{ogrv2e}
\end{split}
\nonumber\\
\begin{split}
&=\frac{1}{\vphie^2\pfreq^2}\biggl[(\kaprep\vrhor)-\frac{6(\kaprep\vrhoq)(\vrhog)}{\vphie}
   -\frac{6(\kaprep\vrhop)(\vrhoi)}{\vphie}+\frac{18(\kaprep\vrhop)(\vrhog)^2}{\vphie^2}
   +\frac{2\kaprep\vphia\epsva(\vrhoj)}{\vphie}
   -\frac{18\kaprep\vphia\epsva(\vrhog)(\vrhoi)}{\vphie^2}\\
   &\qquad+\frac{24\kaprep\vphia\epsva(\vrhog)^3}{\vphie^3}\biggr]
  \beqref{gpath5}\text{ \& }\eqnref{gpath8}
\end{split}
\nonumber\\
&=\frac{\kaprep}{\vphie^2\pfreq^2}\biggl[\vrhor-\frac{6\vrhoq\vrhog}{\vphie}
   -\frac{6\vrhop\vrhoi}{\vphie}+\frac{2\vphia\epsva\vrhoj}{\vphie}
   +\frac{18\vrhop\vrhog^2}{\vphie^2}
   -\frac{18\vphia\epsva\vrhog\vrhoi}{\vphie^2}
   +\frac{24\vphia\epsva\vrhog^3}{\vphie^3}\biggr]
\nonumber\\
&=\kaprep\vrhou\beqref{gpath1c}
\end{align}
\end{subequations}
\begin{subequations}\label{gpath10}
\begin{align}\label{gpath10a}
\fdot{\xcons}
&=\fdot{\vthtrep}(1+\vthtrep^2)^{-3/2}\beqref{kpath7a}\nonumber\\
&=\fdot{\vthtrep}/\efkti^3\beqref{gxpeed1b}\nonumber\\
&=(\kaprep\vrhos)/\efkti^3\beqref{gpath9a}\nonumber\\
&=\kaprep\vrhov\beqref{gpath1c}
\end{align}
\begin{align}\label{gpath10b}
\ffdot{\xcons}
&=(1+\vthtrep^2)^{-5/2}[\ffdot{\vthtrep}(1+\vthtrep^2)-3\vthtrep\fdot{\vthtrep}^2]
  \beqref{kpath7b}\nonumber\\
&=(1/\efkti^5)(\ffdot{\vthtrep}\efkti^2-3\vthtrep\fdot{\vthtrep}^2)
  \beqref{gxpeed1b}\nonumber\\
&=(1/\efkti^5)[(\kaprep\vrhot)\efkti^2-3\vthtrep(\kaprep\vrhos)^2]
  \beqref{gpath9}\nonumber\\
&=\kaprep(\vrhot\efkti^2-3\vthtrep\vrhos^2)/\efkti^5
=\kaprep\vrhow\beqref{gpath1c}
\end{align}
\begin{align}\label{gpath10c}
\fffdot{\xcons}
&=(1+\vthtrep^2)^{-7/2}[\fffdot{\vthtrep}(1+\vthtrep^2)^2-9\vthtrep\fdot{\vthtrep}\ffdot{\vthtrep}(1+\vthtrep^2)
  -3\fdot{\vthtrep}^3(1-4\vthtrep^2)]
  \beqref{kpath7c}\nonumber\\
&=[\fffdot{\vthtrep}\efkti^4-9\vthtrep\fdot{\vthtrep}\ffdot{\vthtrep}\efkti^2
  -3\fdot{\vthtrep}^3(1-4\vthtrep^2)]/\efkti^7
  \beqref{gxpeed1b}\nonumber\\
&=[(\kaprep\vrhou)\efkti^4-9\vthtrep(\kaprep\vrhos)(\kaprep\vrhot)\efkti^2-3(\kaprep\vrhos)^3(1-4\vthtrep^2)]/\efkti^7
  \beqref{gpath9}\nonumber\\
&=\kaprep[\vrhou\efkti^4-9\kaprep\vthtrep\vrhos\vrhot\efkti^2-3\kaprep^2\vrhos^3(1-4\vthtrep^2)]/\efkti^7
=\kaprep\vrhox\beqref{gpath1c}
\end{align}
\end{subequations}
\begin{subequations}\label{gpath11}
\begin{align}\label{gpath11a}
\fdot{\dragf}
&=\frac{\fdotg\dragf}{\gamrep}+\frac{\xcons\gamrep^2\fdot{\vthtrep}}{4\dragf}
  \beqref{kpath8a}\nonumber\\
&=\frac{\fdotg\dragf}{\vphie}+\frac{\xcons(\vphie)^2\fdot{\vthtrep}}{4\dragf}
  \beqref{ogrv2e}\nonumber\\
&=\frac{\vrhog\dragf}{\vphie}+\frac{\xcons\vphie^2\kaprep\vrhos}{4\dragf}
  \beqref{gpath5a}\text{ \& }\eqnref{gpath9a}\nonumber\\
&=\dragf\biggl[\frac{\vrhog}{\vphie}+\frac{\kaprep\xcons\vphie^2\vrhos}{4\dragf^2}\biggr]
=\ethva\beqref{gpath1d}
\end{align}
\begin{align}\label{gpath11b}
\ffdot{\dragf}
&=\dragf\biggl[\frac{\ffdotg}{\gamrep}-\frac{\fdotg^2}{\gamrep^2}\biggr]
   +\fdot{\dragf}\biggl[\frac{\fdotg}{\gamrep}-\frac{\xcons\gamrep^2\fdot{\vthtrep}}{4\dragf^2}\biggr]
   +\frac{\gamrep}{4\dragf}\biggl[\fdot{\xcons}\gamrep\fdot{\vthtrep}
    +2\xcons\fdotg\fdot{\vthtrep}+\xcons\gamrep\ffdot{\vthtrep}\biggr]
  \beqref{kpath8b}\nonumber\\
\begin{split}
&=\dragf\biggl[\frac{\ffdotg}{\vphie}-\frac{\fdotg^2}{\vphie^2}\biggr]
   +\fdot{\dragf}\biggl[\frac{\fdotg}{\vphie}-\frac{\xcons(\vphie)^2\fdot{\vthtrep}}{4\dragf^2}\biggr]
   +\frac{\vphie}{4\dragf}\biggl[\fdot{\xcons}(\vphie)\fdot{\vthtrep}
    +2\xcons\fdotg\fdot{\vthtrep}+\xcons(\vphie)\ffdot{\vthtrep}\biggr]
    \beqref{ogrv2e}
\end{split}
\nonumber\\
\begin{split}
&=\dragf\biggl[\frac{\vrhoi}{\vphie}-\frac{\vrhog^2}{\vphie^2}\biggr]
   +\ethva\biggl[\frac{\vrhog}{\vphie}-\frac{\xcons\vphie^2(\kaprep\vrhos)}{4\dragf^2}\biggr]
   +\frac{\vphie}{4\dragf}\biggl[(\kaprep\vrhov)\vphie(\kaprep\vrhos)
    +2\xcons\vrhog(\kaprep\vrhos)+\xcons\vphie(\kaprep\vrhot)\biggr]\\
    &\qquad\beqref{gpath5}, \eqnref{gpath9}, \eqnref{gpath10a}, \text{ \& }\eqnref{gpath11a}
\end{split}
\nonumber\\
&=\dragf\biggl[\frac{\vrhoi}{\vphie}-\frac{\vrhog^2}{\vphie^2}\biggr]
   +\ethva\biggl[\frac{\vrhog}{\vphie}-\frac{\kaprep\xcons\vphie^2\vrhos}{4\dragf^2}\biggr]
   +\frac{\kaprep\vphie}{4\dragf}\biggl[\kaprep\vrhov\vphie\vrhos
    +2\xcons\vrhog\vrhos+\xcons\vphie\vrhot\biggr]\nonumber\\
&=\dragf\ethvd+\ethva\ethvb+\vphie\ethvc
=\ethve\beqref{gpath1d}
\end{align}
\begin{align*}
\begin{split}
\fffdot{\dragf}
&=\dragf\biggl[\frac{\fffdotg}{\gamrep}-\frac{3\fdotg\ffdotg}{\gamrep^2}+\frac{2\fdotg^3}{\gamrep^3}\biggr]
  +2\fdot{\dragf}\biggl[\frac{\ffdotg}{\gamrep}-\frac{\fdotg^2}{\gamrep^2}
    -\frac{\fdot{\xcons}\gamrep^2\fdot{\vthtrep}}{4\dragf^2}-\frac{\xcons\gamrep\fdotg\fdot{\vthtrep}}{2\dragf^2}
    -\frac{\xcons\gamrep^2\ffdot{\vthtrep}}{4\dragf^2}
    +\frac{\xcons\gamrep^2\fdot{\dragf}\fdot{\vthtrep}}{4\dragf^3}\biggr]
   +\ffdot{\dragf}\biggl[\frac{\fdotg}{\gamrep}-\frac{\xcons\gamrep^2\fdot{\vthtrep}}{4\dragf^2}\biggr]\\
   &\quad+\frac{\fdotg}{4\dragf}\biggl[\fdot{\xcons}\gamrep\fdot{\vthtrep}+2\xcons\fdotg\fdot{\vthtrep}
      +\xcons\gamrep\ffdot{\vthtrep}\biggr]
  +\frac{\gamrep}{4\dragf}\biggl[\ffdot{\xcons}\gamrep\fdot{\vthtrep}+3\fdot{\xcons}\fdotg\fdot{\vthtrep}
    +2\fdot{\xcons}\gamrep\ffdot{\vthtrep}+2\xcons\ffdotg\fdot{\vthtrep}
    +3\xcons\fdotg\ffdot{\vthtrep}+\xcons\gamrep\fffdot{\vthtrep}\biggr]
\end{split}
\nonumber\\
\begin{split}
&=\dragf\biggl[\frac{\fffdotg}{\gamrep}-\frac{3\fdotg\ffdotg}{\gamrep^2}+\frac{2\fdotg^3}{\gamrep^3}\biggr]
  +2\fdot{\dragf}\biggl[\frac{\ffdotg}{\gamrep}-\frac{\fdotg^2}{\gamrep^2}
    -\frac{\fdot{\xcons}\gamrep^2\fdot{\vthtrep}}{4\dragf^2}-\frac{\xcons\gamrep\fdotg\fdot{\vthtrep}}{2\dragf^2}
    -\frac{\xcons\gamrep^2\ffdot{\vthtrep}}{4\dragf^2}
    +\frac{\xcons\gamrep^2\fdot{\dragf}\fdot{\vthtrep}}{4\dragf^3}\biggr]
   +\ffdot{\dragf}\ethvb\\
   &\quad+\fdotg\ethvc
  +\frac{\gamrep}{4\dragf}\biggl[\ffdot{\xcons}\gamrep\fdot{\vthtrep}+3\fdot{\xcons}\fdotg\fdot{\vthtrep}
    +2\fdot{\xcons}\gamrep\ffdot{\vthtrep}+2\xcons\ffdotg\fdot{\vthtrep}
    +3\xcons\fdotg\ffdot{\vthtrep}+\xcons\gamrep\fffdot{\vthtrep}\biggr]
    \beqref{gpath11b}\text{ \& }\eqnref{gpath1d}
\end{split}
\end{align*}
\begin{align*}
\begin{split}
&=\dragf\biggl[\frac{\vrhoj}{\vphie}-\frac{3\vrhog\vrhoi}{\vphie^2}+\frac{2\vrhog^3}{\vphie^3}\biggr]
  +2\fdot{\dragf}\biggl[\frac{\vrhoi}{\vphie}-\frac{\vrhog^2}{\vphie^2}
    -\frac{\fdot{\xcons}\vphie^2\fdot{\vthtrep}}{4\dragf^2}-\frac{\xcons\vphie\vrhog\fdot{\vthtrep}}{2\dragf^2}
    -\frac{\xcons\vphie^2\ffdot{\vthtrep}}{4\dragf^2}
    +\frac{\xcons\vphie^2\fdot{\dragf}\fdot{\vthtrep}}{4\dragf^3}\biggr]
   +\ffdot{\dragf}\ethvb\\
   &\quad+\vrhog\ethvc
  +\frac{\vphie}{4\dragf}\biggl[\ffdot{\xcons}\vphie\fdot{\vthtrep}+3\fdot{\xcons}\vrhog\fdot{\vthtrep}
    +2\fdot{\xcons}\vphie\ffdot{\vthtrep}+2\xcons\vrhoi\fdot{\vthtrep}
    +3\xcons\vrhog\ffdot{\vthtrep}+\xcons\vphie\fffdot{\vthtrep}\biggr]
    \beqref{ogrv2e}\text{ \& }\eqnref{gpath5}
\end{split}
\nonumber\\
\begin{split}
&=\ethvf
  +2\ethva\biggl[\ethvd-\frac{(\kaprep\vrhov)\vphie^2(\kaprep\vrhos)}{4\dragf^2}-\frac{\xcons\vphie\vrhog(\kaprep\vrhos)}{2\dragf^2}
    -\frac{\xcons\vphie^2(\kaprep\vrhot)}{4\dragf^2}+\frac{\xcons\vphie^2\ethva(\kaprep\vrhos)}{4\dragf^3}\biggr]
  +\ethve\ethvb+\vrhog\ethvc\\
  &\quad+\frac{\vphie}{4\dragf}\biggl[(\kaprep\vrhow)\vphie(\kaprep\vrhos)+3(\kaprep\vrhov)\vrhog(\kaprep\vrhos)
    +2(\kaprep\vrhov)\vphie(\kaprep\vrhot)+2\xcons\vrhoi(\kaprep\vrhos)
    +3\xcons\vrhog(\kaprep\vrhot)+\xcons\vphie(\kaprep\vrhou)\biggr]\\
  &\quad\beqref{gpath1d}, \eqnref{gpath9}, \eqnref{gpath10}, \eqnref{gpath11a}\text{ \& }\eqnref{gpath11b}
\end{split}
\end{align*}
\begin{align}\label{gpath11c}
\begin{split}
&=\ethvf+\ethve\ethvb+\vrhog\ethvc
  +2\ethva\biggl[\ethvd-\frac{\kaprep^2\vrhov\vphie^2\vrhos}{4\dragf^2}-\frac{\kaprep\xcons\vphie\vrhog\vrhos}{2\dragf^2}
    -\frac{\kaprep\xcons\vphie^2\vrhot}{4\dragf^2}
    +\frac{\kaprep\xcons\vphie^2\ethva\vrhos}{4\dragf^3}\biggr]\\
  &\quad+\frac{\kaprep\vphie}{4\dragf}\biggl[\kaprep\biggl\{\vrhow\vphie\vrhos+3\vrhov\vrhog\vrhos
    +2\vrhov\vphie\vrhot\biggr\}+\xcons\biggl\{2\vrhoi\vrhos+3\vrhog\vrhot+\vphie\vrhou\biggr\}\biggr]
\end{split}
\nonumber\\
&=\ethvg\beqref{gpath1d}
\end{align}
\end{subequations}
\begin{subequations}\label{gpath12}
\begin{align}\label{gpath12a}
\fdot{\rhorep}
&=\frac{\dragf\fdot{\xcons}-\xcons\fdot{\dragf}}{4\pfreq\dragf^2}
  \beqref{kpath9a}\nonumber\\
&=\frac{\dragf(\kaprep\vrhov)-\xcons(\ethva)}{4\pfreq\dragf^2}
  \beqref{gpath10a}\text{ \& }\eqnref{gpath11a}\nonumber\\
&=\frac{\dragf\kaprep\vrhov-\xcons\ethva}{4\pfreq\dragf^2}
=\ethvh\beqref{gpath1e}
\end{align}
\begin{align}\label{gpath12b}
\ffdot{\rhorep}
&=-\frac{\fdot{\dragf}(\dragf\fdot{\xcons}-\xcons\fdot{\dragf})}{2\pfreq\dragf^3}
   +\frac{\dragf\ffdot{\xcons}-\xcons\ffdot{\dragf}}{4\pfreq\dragf^2}
  \beqref{kpath9b}\nonumber\\
&=-\frac{\ethva[\dragf(\kaprep\vrhov)-\xcons\ethva]}{2\pfreq\dragf^3}
   +\frac{\dragf(\kaprep\vrhow)-\xcons\ethve}{4\pfreq\dragf^2}
  \beqref{gpath10}\text{ \& }\eqnref{gpath11}\nonumber\\
&=-\frac{\ethva(\dragf\kaprep\vrhov-\xcons\ethva)}{2\pfreq\dragf^3}
   +\frac{\dragf\kaprep\vrhow-\xcons\ethve}{4\pfreq\dragf^2}
=\ethvi\beqref{gpath1e}
\end{align}
\begin{align}\label{gpath12c}
\fffdot{\rhorep}
&=\frac{(3\fdot{\dragf}^2-\dragf\ffdot{\dragf})(\dragf\fdot{\xcons}-\xcons\fdot{\dragf})}{2\pfreq\dragf^4}
   -\frac{\fdot{\dragf}(\dragf\ffdot{\xcons}-\xcons\ffdot{\dragf})}{\pfreq\dragf^3}
  +\frac{\fdot{\dragf}\ffdot{\xcons}+\dragf\fffdot{\xcons}
     -\fdot{\xcons}\ffdot{\dragf}-\xcons\fffdot{\dragf}}{4\pfreq\dragf^2}
  \beqref{kpath9c}\nonumber\\
\begin{split}
&=\frac{(3\ethva^2-\dragf\ethve)[\dragf(\kaprep\vrhov)-\xcons\ethva]}{2\pfreq\dragf^4}
   -\frac{\ethva[\dragf(\kaprep\vrhow)-\xcons\ethve]}{\pfreq\dragf^3}
  +\frac{\ethva(\kaprep\vrhow)+\dragf(\kaprep\vrhox)
     -(\kaprep\vrhov)\ethve-\xcons\ethvg}{4\pfreq\dragf^2}\\
  &\quad\beqref{gpath10}\text{ \& }\eqnref{gpath11}
\end{split}
\nonumber\\
&=\frac{(3\ethva^2-\dragf\ethve)(\dragf\kaprep\vrhov-\xcons\ethva)}{2\pfreq\dragf^4}
   -\frac{\ethva(\dragf\kaprep\vrhow-\xcons\ethve)}{\pfreq\dragf^3}
  +\frac{\kaprep(\ethva\vrhow+\dragf\vrhox-\vrhov\ethve)-\xcons\ethvg}{4\pfreq\dragf^2}\nonumber\\
&=\ethvj\beqref{gpath1e}.
\end{align}
\end{subequations}

\subart{Derivatives of $\taurep$ and $\vecte$}
To calculate the derivatives of $\taurep$ and $\vecte$ we first derive
\begin{subequations}\label{gpath13}
\begin{align*}
\fdot{\fzer}
&=\dif{\biggl[\frac{2\rhorep\alprep-\dragf\pfreq}{\scaln(\kaprep^2-\kcons^2)}\biggr]}
  \beqref{main5d}\nonumber\\
&=\frac{[\scaln(\kaprep^2-\kcons^2)]\dif{(2\rhorep\alprep-\dragf\pfreq)}-(2\rhorep\alprep-\dragf\pfreq)\dif{[\scaln(\kaprep^2-\kcons^2)]}}
  {[\scaln(\kaprep^2-\kcons^2)]^2}\nonumber\\
&=\frac{\dif{(2\rhorep\alprep-\dragf\pfreq)}}{\scaln(\kaprep^2-\kcons^2)}
  -\frac{\fzer\dif{[\scaln(\kaprep^2-\kcons^2)]}}{\scaln(\kaprep^2-\kcons^2)}
  \beqref{main5d}\nonumber\\
&=\frac{2\rhorep\fdot{\alprep}+2\fdot{\rhorep}\alprep-\fdot{\dragf}\pfreq}{\scaln(\kaprep^2-\kcons^2)}
  -\frac{\fzer[\fdot{\scaln}(\kaprep^2-\kcons^2)+\scaln(-2\kcons\fdot{\kcons})]}{\scaln(\kaprep^2-\kcons^2)}
  \nonumber\\
&=\frac{2\rhorep\fdot{\alprep}+2\fdot{\rhorep}\alprep-\fdot{\dragf}\pfreq
  -\fzer\fdot{\scaln}(\kaprep^2-\kcons^2)+2\fzer\scaln\kcons\fdot{\kcons}}{\scaln(\kaprep^2-\kcons^2)}
\end{align*}
\begin{align}\label{gpath13a}
\begin{split}
&=\frac{2\rhorep(\kaprep\vrhop)+2\ethvh\alprep-\ethva\pfreq
  -\fzer(-\efktc)(\kaprep^2-\kcons^2)+2\fzer\scaln\kcons(\kaprep\vrhoe)}{\scaln(\kaprep^2-\kcons^2)}\\
  &\quad\beqref{gpath8a}, \eqnref{gpath12a}, \eqnref{gpath11a}, \eqnref{gpath3a}\text{ \& }\eqnref{gpath4a}
\end{split}
\nonumber\\
&=\frac{2\kaprep\rhorep\vrhop+2\ethvh(-\kaprep\vphia\epsva)-\ethva\pfreq
  +(2\vphif/\kaprep)\efktc(\kaprep^2-\kaprep^2\vphic^2)
  +2\kaprep(2\vphif/\kaprep)\epsvf(\kaprep\vphic)\vrhoe}{\epsvf(\kaprep^2-\kaprep^2\vphic^2)}
  \beqref{ogrv2}
\nonumber\\
&=\frac{2\kaprep\rhorep\vrhop-2\kaprep\ethvh\vphia\epsva-\ethva\pfreq
  +2\kaprep\vphif\efktc(1-\vphic^2)+4\kaprep\vphif\epsvf\vphic\vrhoe}{\kaprep^2\epsvf(1-\vphic^2)}
\nonumber\\
&=\frac{2\rhorep\vrhop-2\ethvh\vphia\epsva-\ethva\scalc
  -2\vphif\vphid\efktc+4\epsvf\vphif\vphic\vrhoe}{-\kaprep\epsvf\vphid}
  \beqref{main2b}\text{ \& }\eqnref{ogrv1b}\nonumber\\
&=-2\ethvk\beqref{gpath1e}
\end{align}
\begin{align*}
\ffdot{\fzer}
&=\dif{\biggl[\frac{2\rhorep\fdot{\alprep}+2\fdot{\rhorep}\alprep-\fdot{\dragf}\pfreq
  -\fzer\fdot{\scaln}(\kaprep^2-\kcons^2)+2\fzer\scaln\kcons\fdot{\kcons}}{\scaln(\kaprep^2-\kcons^2)}\biggr]}
  \beqref{gpath13a}\nonumber\\
\begin{split}
&=\frac{[\scaln(\kaprep^2-\kcons^2)]
  \dif{[2\rhorep\fdot{\alprep}+2\fdot{\rhorep}\alprep-\fdot{\dragf}\pfreq-\fzer\fdot{\scaln}(\kaprep^2-\kcons^2)
     +2\fzer\scaln\kcons\fdot{\kcons}]}}{[\scaln(\kaprep^2-\kcons^2)]^2}\\
  &\quad-\frac{[2\rhorep\fdot{\alprep}+2\fdot{\rhorep}\alprep-\fdot{\dragf}\pfreq-\fzer\fdot{\scaln}(\kaprep^2-\kcons^2)
     +2\fzer\scaln\kcons\fdot{\kcons}]\dif{[\scaln(\kaprep^2-\kcons^2)]}}{[\scaln(\kaprep^2-\kcons^2)]^2}
\end{split}
\nonumber\\
&=\frac{\dif{[2\rhorep\fdot{\alprep}+2\fdot{\rhorep}\alprep-\fdot{\dragf}\pfreq-\fzer\fdot{\scaln}(\kaprep^2-\kcons^2)
     +2\fzer\scaln\kcons\fdot{\kcons}]}}{\scaln(\kaprep^2-\kcons^2)}
  -\frac{\fdot{\fzer}\dif{[\scaln(\kaprep^2-\kcons^2)]}}{\scaln(\kaprep^2-\kcons^2)}
  \beqref{gpath13a}
\end{align*}
\begin{align*}
\begin{split}
&=\frac{\dif{(2\rhorep\fdot{\alprep}+2\fdot{\rhorep}\alprep)}}{\scaln(\kaprep^2-\kcons^2)}
  -\frac{\dif{(\fdot{\dragf}\pfreq)}}{\scaln(\kaprep^2-\kcons^2)}
  -\frac{\dif{[\fzer\fdot{\scaln}(\kaprep^2-\kcons^2)]}}{\scaln(\kaprep^2-\kcons^2)}
  +\frac{2\dif{(\fzer\scaln\kcons\fdot{\kcons})}}{\scaln(\kaprep^2-\kcons^2)}
  -\frac{\fdot{\fzer}\dif{[\scaln(\kaprep^2-\kcons^2)]}}{\scaln(\kaprep^2-\kcons^2)}
\end{split}
\nonumber\\
\begin{split}
&=\frac{2\fdot{\rhorep}\fdot{\alprep}+2\rhorep\ffdot{\alprep}+2\ffdot{\rhorep}\alprep
      +2\fdot{\rhorep}\fdot{\alprep}}{\scaln(\kaprep^2-\kcons^2)}
  -\frac{\ffdot{\dragf}\pfreq}{\scaln(\kaprep^2-\kcons^2)}
  -\frac{\fdot{\fzer}\fdot{\scaln}(\kaprep^2-\kcons^2)+\fzer\ffdot{\scaln}(\kaprep^2-\kcons^2)
      +\fzer\fdot{\scaln}(-2\kcons\fdot{\kcons})}{\scaln(\kaprep^2-\kcons^2)}\\
  &\quad+\frac{2\fdot{\fzer}\scaln\kcons\fdot{\kcons}+2\fzer\fdot{\scaln}\kcons\fdot{\kcons}+2\fzer\scaln\fdot{\kcons}^2
      +2\fzer\scaln\kcons\ffdot{\kcons}}{\scaln(\kaprep^2-\kcons^2)}
  -\frac{\fdot{\fzer}[\fdot{\scaln}(\kaprep^2-\kcons^2)+\scaln(-2\kcons\fdot{\kcons})]}{\scaln(\kaprep^2-\kcons^2)}
\end{split}
\nonumber\\
\begin{split}
&=\frac{2\rhorep\ffdot{\alprep}+2\ffdot{\rhorep}\alprep+4\fdot{\rhorep}\fdot{\alprep}-\ffdot{\dragf}\pfreq}{\scaln(\kaprep^2-\kcons^2)}
  -\frac{\fdot{\fzer}\fdot{\scaln}(\kaprep^2-\kcons^2)+\fzer\ffdot{\scaln}(\kaprep^2-\kcons^2)
      -2\fzer\fdot{\scaln}\kcons\fdot{\kcons}}{\scaln(\kaprep^2-\kcons^2)}\\
  &\quad+\frac{2\fdot{\fzer}\scaln\kcons\fdot{\kcons}+2\fzer\fdot{\scaln}\kcons\fdot{\kcons}+2\fzer\scaln\fdot{\kcons}^2
      +2\fzer\scaln\kcons\ffdot{\kcons}}{\scaln(\kaprep^2-\kcons^2)}
  -\frac{\fdot{\fzer}\fdot{\scaln}(\kaprep^2-\kcons^2)-2\fdot{\fzer}\scaln\kcons\fdot{\kcons}}{\scaln(\kaprep^2-\kcons^2)}
\end{split}
\end{align*}
\begin{align*}
\begin{split}
&=\frac{2\rhorep\ffdot{\alprep}+2\ffdot{\rhorep}\alprep+4\fdot{\rhorep}\fdot{\alprep}-\ffdot{\dragf}\pfreq}{\scaln(\kaprep^2-\kcons^2)}
  -\frac{2\fdot{\fzer}\fdot{\scaln}+\fzer\ffdot{\scaln}}{\scaln}
  +\frac{4\fdot{\fzer}\scaln\kcons\fdot{\kcons}+4\fzer\fdot{\scaln}\kcons\fdot{\kcons}+2\fzer\scaln\fdot{\kcons}^2
      +2\fzer\scaln\kcons\ffdot{\kcons}}{\scaln(\kaprep^2-\kcons^2)}
\end{split}
\nonumber\\
\begin{split}
&=\frac{2\rhorep(\kaprep\vrhoq)+2\ethvi\alprep+4\ethvh(\kaprep\vrhop)-\ethve\pfreq}{\scaln(\kaprep^2-\kcons^2)}
  -\frac{2(-2\ethvk)(-\efktc)+\fzer(-\vrhoa\epsvf)}{\scaln}\\
  &\quad+\frac{4(-2\ethvk)\scaln\kcons(\kaprep\vrhoe)+4\fzer(-\efktc)\kcons(\kaprep\vrhoe)+2\fzer\scaln(\kaprep\vrhoe)^2
      +2\fzer\scaln\kcons(-\kaprep\vrhof)}{\scaln(\kaprep^2-\kcons^2)}\\
  &\quad\beqref{gpath8}, \eqnref{gpath12}, \eqnref{gpath11}, \eqnref{gpath13a}, \eqnref{gpath3}\text{ \& }\eqnref{gpath4}
\end{split}
\nonumber\\
\begin{split}
&=\frac{2\kaprep\rhorep\vrhoq+2\ethvi\alprep+4\kaprep\ethvh\vrhop-\ethve\pfreq}{\scaln(\kaprep^2-\kcons^2)}
  -\frac{4\ethvk\efktc-\epsvf\vrhoa\fzer}{\scaln}\\
  &\quad+\frac{-8\kaprep\ethvk\vrhoe\scaln\kcons-4\kaprep\efktc\vrhoe\fzer\kcons+2\kaprep^2\vrhoe^2\fzer\scaln
      -2\kaprep\vrhof\fzer\scaln\kcons}{\scaln(\kaprep^2-\kcons^2)}
\end{split}
\end{align*}
\begin{align*}
\begin{split}
&=\frac{2\kaprep\rhorep\vrhoq+2\ethvi(-\kaprep\vphia\epsva)+4\kaprep\ethvh\vrhop-\ethve\pfreq}{\epsvf(\kaprep^2-\kaprep^2\vphic^2)}
  -\frac{4\ethvk\efktc-\epsvf\vrhoa(2\vphif/\kaprep)}{\epsvf}\\
  &\quad+\frac{-8\kaprep\ethvk\vrhoe\epsvf(\kaprep\vphic)-4\kaprep\efktc\vrhoe(2\vphif/\kaprep)(\kaprep\vphic)
      +2\kaprep^2\vrhoe^2(2\vphif/\kaprep)\epsvf
      -2\kaprep\vrhof(2\vphif/\kaprep)\epsvf(\kaprep\vphic)}{\epsvf(\kaprep^2-\kaprep^2\vphic^2)}\\
  &\quad\beqref{ogrv2}
\end{split}
\nonumber\\
\begin{split}
&=\frac{2\rhorep\vrhoq-2\epsva\vphia\ethvi+4\ethvh\vrhop-\ethve\scalc}{-\kaprep\epsvf\vphid}
  -\frac{4\kaprep\ethvk\efktc-2\epsvf\vrhoa\vphif}{\kaprep\epsvf}\\
  &\quad+\frac{-8\kaprep\ethvk\vrhoe\epsvf\vphic-8\vphif\vphic\efktc\vrhoe
      +4\vrhoe^2\vphif\epsvf-4\vrhof\vphif\epsvf\vphic}{-\kaprep\epsvf\vphid}
  \beqref{main2b}\text{ \& }\eqnref{ogrv1b}
\end{split}
\end{align*}
\begin{align}\label{gpath13b}
\begin{split}
&=\frac{2\epsva\vphia\ethvi-2\rhorep\vrhoq-4\ethvh\vrhop+\ethve\scalc
    +8\kaprep\ethvk\vrhoe\epsvf\vphic+8\vphif\vphic\efktc\vrhoe
    -4\vrhoe^2\vphif\epsvf+4\vrhof\vphif\epsvf\vphic}{\kaprep\epsvf\vphid}\\
  &\qquad-\frac{4\kaprep\ethvk\efktc-2\epsvf\vrhoa\vphif}{\kaprep\epsvf}
=2\ethvl\beqref{gpath1e}
\end{split}
\end{align}
\begin{align*}
\begin{split}
\fffdot{\fzer}
&=\dif{\biggl[\frac{2\rhorep\ffdot{\alprep}+2\ffdot{\rhorep}\alprep+4\fdot{\rhorep}\fdot{\alprep}
      -\ffdot{\dragf}\pfreq}{\scaln(\kaprep^2-\kcons^2)}\biggr]}
  -\dif{\biggl[\frac{2\fdot{\fzer}\fdot{\scaln}+\fzer\ffdot{\scaln}}{\scaln}\biggr]}
  +\dif{\biggl[\frac{4\fdot{\fzer}\scaln\kcons\fdot{\kcons}+4\fzer\fdot{\scaln}\kcons\fdot{\kcons}+2\fzer\scaln\fdot{\kcons}^2
      +2\fzer\scaln\kcons\ffdot{\kcons}}{\scaln(\kaprep^2-\kcons^2)}\biggr]} \\
  &\quad\beqref{gpath13b}
\end{split}
\nonumber\\
\begin{split}
&=\frac{[\scaln(\kaprep^2-\kcons^2)]\dif{[2\rhorep\ffdot{\alprep}+2\ffdot{\rhorep}\alprep+4\fdot{\rhorep}\fdot{\alprep}
          -\ffdot{\dragf}\pfreq]}
      -[2\rhorep\ffdot{\alprep}+2\ffdot{\rhorep}\alprep+4\fdot{\rhorep}\fdot{\alprep}
          -\ffdot{\dragf}\pfreq]\dif{[\scaln(\kaprep^2-\kcons^2)]}}{[\scaln(\kaprep^2-\kcons^2)]^2}\\
  &\quad-\frac{\scaln\dif{[2\fdot{\fzer}\fdot{\scaln}+\fzer\ffdot{\scaln}]}
      -[2\fdot{\fzer}\fdot{\scaln}+\fzer\ffdot{\scaln}]\fdot{\scaln}}{\scaln^2}
  +\frac{[\scaln(\kaprep^2-\kcons^2)]\dif{[4\fdot{\fzer}\scaln\kcons\fdot{\kcons}+4\fzer\fdot{\scaln}\kcons\fdot{\kcons}
      +2\fzer\scaln\fdot{\kcons}^2+2\fzer\scaln\kcons\ffdot{\kcons}]}}{[\scaln(\kaprep^2-\kcons^2)]^2}\\
  &\quad-\frac{[4\fdot{\fzer}\scaln\kcons\fdot{\kcons}+4\fzer\fdot{\scaln}\kcons\fdot{\kcons}+2\fzer\scaln\fdot{\kcons}^2
         +2\fzer\scaln\kcons\ffdot{\kcons}]\dif{[\scaln(\kaprep^2-\kcons^2)]}}{[\scaln(\kaprep^2-\kcons^2)]^2}
\end{split}
\end{align*}
\begin{align*}
\begin{split}
&=\frac{\dif{[2\rhorep\ffdot{\alprep}+2\ffdot{\rhorep}\alprep+4\fdot{\rhorep}\fdot{\alprep}
          -\ffdot{\dragf}\pfreq]}}{\scaln(\kaprep^2-\kcons^2)}
  -\frac{[2\rhorep\ffdot{\alprep}+2\ffdot{\rhorep}\alprep+4\fdot{\rhorep}\fdot{\alprep}
          -\ffdot{\dragf}\pfreq]\dif{[\scaln(\kaprep^2-\kcons^2)]}}{[\scaln(\kaprep^2-\kcons^2)]^2}\\
  &\quad-\frac{\scaln\dif{[2\fdot{\fzer}\fdot{\scaln}+\fzer\ffdot{\scaln}]}
      -[2\fdot{\fzer}\fdot{\scaln}+\fzer\ffdot{\scaln}]\fdot{\scaln}}{\scaln^2}
  +\frac{\dif{[4\fdot{\fzer}\scaln\kcons\fdot{\kcons}+4\fzer\fdot{\scaln}\kcons\fdot{\kcons}
      +2\fzer\scaln\fdot{\kcons}^2+2\fzer\scaln\kcons\ffdot{\kcons}]}}{\scaln(\kaprep^2-\kcons^2)}\\
  &\quad-\frac{[4\fdot{\fzer}\scaln\kcons\fdot{\kcons}+4\fzer\fdot{\scaln}\kcons\fdot{\kcons}+2\fzer\scaln\fdot{\kcons}^2
         +2\fzer\scaln\kcons\ffdot{\kcons}]\dif{[\scaln(\kaprep^2-\kcons^2)]}}{[\scaln(\kaprep^2-\kcons^2)]^2}
\end{split}
\nonumber\\
\begin{split}
&=\frac{\dif{[2\rhorep\ffdot{\alprep}+2\ffdot{\rhorep}\alprep+4\fdot{\rhorep}\fdot{\alprep}
          -\ffdot{\dragf}\pfreq]}}{\scaln(\kaprep^2-\kcons^2)}
  -\frac{\dif{[\scaln(\kaprep^2-\kcons^2)]}}{\scaln(\kaprep^2-\kcons^2)}\biggl[\ffdot{\fzer}+
         \frac{2\fdot{\fzer}\fdot{\scaln}+\fzer\ffdot{\scaln}}{\scaln}\biggr]\\
  &\quad-\frac{\scaln\dif{[2\fdot{\fzer}\fdot{\scaln}+\fzer\ffdot{\scaln}]}
      -[2\fdot{\fzer}\fdot{\scaln}+\fzer\ffdot{\scaln}]\fdot{\scaln}}{\scaln^2}
  +\frac{\dif{[4\fdot{\fzer}\scaln\kcons\fdot{\kcons}+4\fzer\fdot{\scaln}\kcons\fdot{\kcons}
      +2\fzer\scaln\fdot{\kcons}^2+2\fzer\scaln\kcons\ffdot{\kcons}]}}{\scaln(\kaprep^2-\kcons^2)}\\
  &\quad\beqref{gpath13b}
\end{split}
\end{align*}
\begin{align*}
\begin{split}
&=\frac{2\fdot{\rhorep}\ffdot{\alprep}+2\rhorep\fffdot{\alprep}+2\fffdot{\rhorep}\alprep+2\ffdot{\rhorep}\fdot{\alprep}
     +4\ffdot{\rhorep}\fdot{\alprep}+4\fdot{\rhorep}\ffdot{\alprep}-\fffdot{\dragf}\pfreq}{\scaln(\kaprep^2-\kcons^2)}
  -\frac{\fdot{\scaln}(\kaprep^2-\kcons^2)+\scaln(-2\kcons\fdot{\kcons})}{\scaln(\kaprep^2-\kcons^2)}\biggl[\ffdot{\fzer}+
         \frac{2\fdot{\fzer}\fdot{\scaln}+\fzer\ffdot{\scaln}}{\scaln}\biggr]\\
  &\quad-\frac{\scaln[2\ffdot{\fzer}\fdot{\scaln}+2\fdot{\fzer}\ffdot{\scaln}+\fdot{\fzer}\ffdot{\scaln}+\fzer\fffdot{\scaln}]}{\scaln^2}
  +\frac{[2\fdot{\fzer}\fdot{\scaln}+\fzer\ffdot{\scaln}]\fdot{\scaln}}{\scaln^2}
  +\frac{4\ffdot{\fzer}\scaln\kcons\fdot{\kcons}+4\fdot{\fzer}\fdot{\scaln}\kcons\fdot{\kcons}
    +4\fdot{\fzer}\scaln\fdot{\kcons}^2+4\fdot{\fzer}\scaln\kcons\ffdot{\kcons}}{\scaln(\kaprep^2-\kcons^2)}\\
  &\quad\frac{+4\fdot{\fzer}\fdot{\scaln}\kcons\fdot{\kcons}+4\fzer\ffdot{\scaln}\kcons\fdot{\kcons}+4\fzer\fdot{\scaln}\fdot{\kcons}^2
    +4\fzer\fdot{\scaln}\kcons\ffdot{\kcons}+2\fdot{\fzer}\scaln\fdot{\kcons}^2
    +2\fzer\fdot{\scaln}\fdot{\kcons}^2+4\fzer\scaln\fdot{\kcons}\ffdot{\kcons}}{\scaln(\kaprep^2-\kcons^2)}\\
  &\quad+\frac{2\fdot{\fzer}\scaln\kcons\ffdot{\kcons}+2\fzer\fdot{\scaln}\kcons\ffdot{\kcons}+2\fzer\scaln\fdot{\kcons}\ffdot{\kcons}
    +2\fzer\scaln\kcons\fffdot{\kcons}}{\scaln(\kaprep^2-\kcons^2)}
\end{split}
\end{align*}
\begin{align*}
\begin{split}
&=\frac{2\rhorep\fffdot{\alprep}+2\fffdot{\rhorep}\alprep
     +6\ffdot{\rhorep}\fdot{\alprep}+6\fdot{\rhorep}\ffdot{\alprep}-\fffdot{\dragf}\pfreq}{\scaln(\kaprep^2-\kcons^2)}
  -\frac{\fdot{\scaln}(\kaprep^2-\kcons^2)-2\scaln\kcons\fdot{\kcons}}{\scaln(\kaprep^2-\kcons^2)}\biggl[\ffdot{\fzer}+
         \frac{2\fdot{\fzer}\fdot{\scaln}+\fzer\ffdot{\scaln}}{\scaln}\biggr]\\
  &\quad-\frac{\scaln[2\ffdot{\fzer}\fdot{\scaln}+2\fdot{\fzer}\ffdot{\scaln}+\fdot{\fzer}\ffdot{\scaln}+\fzer\fffdot{\scaln}]
    -\fdot{\scaln}[2\fdot{\fzer}\fdot{\scaln}+\fzer\ffdot{\scaln}]}{\scaln^2}\\
  &\quad+\frac{8\fdot{\fzer}\fdot{\scaln}\kcons\fdot{\kcons}
    +4\fzer\ffdot{\scaln}\kcons\fdot{\kcons}+4\fzer\fdot{\scaln}\fdot{\kcons}^2
    +6\fzer\fdot{\scaln}\kcons\ffdot{\kcons}+2\fzer\fdot{\scaln}\fdot{\kcons}^2}{\scaln(\kaprep^2-\kcons^2)}\\
  &\quad+\frac{4\ffdot{\fzer}\scaln\kcons\fdot{\kcons}+6\fdot{\fzer}\scaln\fdot{\kcons}^2
    +6\fzer\scaln\fdot{\kcons}\ffdot{\kcons}+6\fdot{\fzer}\scaln\kcons\ffdot{\kcons}+2\fzer\scaln\kcons\fffdot{\kcons}}
     {\scaln(\kaprep^2-\kcons^2)}
\end{split}
\end{align*}
\begin{align*}
\begin{split}
&=\frac{2\rhorep(\kaprep\vrhor)+2\ethvj\alprep
     +6\ethvi(\kaprep\vrhop)+6\ethvh(\kaprep\vrhoq)-\ethvg\pfreq}{\scaln(\kaprep^2-\kcons^2)}\\
  &\quad-\frac{(-\efktc)(\kaprep^2-\kcons^2)-2\scaln\kcons(\kaprep\vrhoe)}{\scaln(\kaprep^2-\kcons^2)}\biggl[(2\ethvl)+
         \frac{2(-2\ethvk)(-\efktc)+\fzer(-\vrhoa\epsvf)}{\scaln}\biggr]\\
  &\quad-\frac{\scaln[2(2\ethvl)(-\efktc)+2(-2\ethvk)(-\vrhoa\epsvf)+(-2\ethvk)(-\vrhoa\epsvf)+\fzer(\vrhod)]
    -(-\efktc)[2(-2\ethvk)(-\efktc)+\fzer(-\vrhoa\epsvf)]}{\scaln^2}\\
  &\quad+\frac{8(-2\ethvk)(-\efktc)\kcons(\kaprep\vrhoe)
    +4\fzer(-\vrhoa\epsvf)\kcons(\kaprep\vrhoe)+4\fzer(-\efktc)(\kaprep\vrhoe)^2
    +6\fzer(-\efktc)\kcons(-\kaprep\vrhof)+2\fzer(-\efktc)(\kaprep\vrhoe)^2}{\scaln(\kaprep^2-\kcons^2)}\\
  &\quad+\frac{4(2\ethvl)\scaln\kcons(\kaprep\vrhoe)+6(-2\ethvk)\scaln(\kaprep\vrhoe)^2
    +6\fzer\scaln(\kaprep\vrhoe)(-\kaprep\vrhof)+6(-2\ethvk)\scaln\kcons(-\kaprep\vrhof)+2\fzer\scaln\kcons(\kaprep\vrhoh)}
     {\scaln(\kaprep^2-\kcons^2)}\\
  &\quad\beqref{gpath8}, \eqnref{gpath12}, \eqnref{gpath11}, \eqnref{gpath3}, \eqnref{gpath4},
    \eqnref{gpath13a}\text{ \& }\eqnref{gpath13b}
\end{split}
\end{align*}
\begin{align*}
\begin{split}
&=\frac{2\kaprep\rhorep\vrhor+2\ethvj\alprep
     +6\kaprep\ethvi\vrhop+6\kaprep\ethvh\vrhoq-\ethvg\pfreq}{\scaln(\kaprep^2-\kcons^2)}
  -\frac{-\efktc(\kaprep^2-\kcons^2)-2\kaprep\vrhoe\scaln\kcons}{\scaln(\kaprep^2-\kcons^2)}\biggl[2\ethvl+
         \frac{4\ethvk\efktc-\vrhoa\epsvf\fzer}{\scaln}\biggr]\\
  &\quad-\frac{\scaln[-4\ethvl\efktc+4\ethvk\vrhoa\epsvf+2\ethvk\vrhoa\epsvf+\vrhod\fzer]
    +\efktc[4\ethvk\efktc-\vrhoa\epsvf\fzer]}{\scaln^2}\\
  &\quad+\frac{16\kaprep\vrhoe\ethvk\efktc\kcons
    -4\kaprep\vrhoe\vrhoa\epsvf\fzer\kcons-4\kaprep^2\efktc\vrhoe^2\fzer
    +6\kaprep\efktc\vrhof\fzer\kcons-2\kaprep^2\vrhoe^2\efktc\fzer}{\scaln(\kaprep^2-\kcons^2)}\\
  &\quad+\frac{8\kaprep\vrhoe\ethvl\scaln\kcons-12\kaprep^2\vrhoe^2\ethvk\scaln
    -6\kaprep^2\vrhoe\vrhof\fzer\scaln+12\kaprep\vrhof\ethvk\scaln\kcons+2\kaprep\vrhoh\fzer\scaln\kcons}
     {\scaln(\kaprep^2-\kcons^2)}
\end{split}
\end{align*}
\begin{align*}
\begin{split}
&=\frac{2\kaprep\rhorep\vrhor+2\ethvj(-\kaprep\vphia\epsva)
     +6\kaprep\ethvi\vrhop+6\kaprep\ethvh\vrhoq-\ethvg\pfreq}{\epsvf(\kaprep^2-\kaprep^2\vphic^2)}\\
  &\quad-\frac{-\efktc(\kaprep^2-\kaprep^2\vphic^2)-2\kaprep\vrhoe\epsvf(\kaprep\vphic)}{\epsvf(\kaprep^2-\kaprep^2\vphic^2)}
       \biggl[2\ethvl+\frac{4\ethvk\efktc-\vrhoa\epsvf(2\vphif/\kaprep)}{\epsvf}\biggr]\\
  &\quad-\frac{\epsvf[-4\ethvl\efktc+4\ethvk\vrhoa\epsvf+2\ethvk\vrhoa\epsvf+\vrhod(2\vphif/\kaprep)]
    +\efktc[4\ethvk\efktc-\vrhoa\epsvf(2\vphif/\kaprep)]}{\epsvf^2}\\
  &\quad+\frac{16\kaprep\vrhoe\ethvk\efktc(\kaprep\vphic)
    -4\kaprep\vrhoe\vrhoa\epsvf(2\vphif/\kaprep)(\kaprep\vphic)-4\kaprep^2\efktc\vrhoe^2(2\vphif/\kaprep)
    +6\kaprep\efktc\vrhof(2\vphif/\kaprep)(\kaprep\vphic)}{\epsvf(\kaprep^2-\kaprep^2\vphic^2)}\\
  &\quad+\frac{-2\kaprep^2\vrhoe^2\efktc(2\vphif/\kaprep)+8\kaprep\vrhoe\ethvl\epsvf(\kaprep\vphic)-12\kaprep^2\vrhoe^2\ethvk\epsvf
      -6\kaprep^2\vrhoe\vrhof(2\vphif/\kaprep)\epsvf}{\epsvf(\kaprep^2-\kaprep^2\vphic^2)}\\
  &\quad+\frac{12\kaprep\vrhof\ethvk\epsvf(\kaprep\vphic)
     +2\kaprep\vrhoh(2\vphif/\kaprep)\epsvf(\kaprep\vphic)}{\epsvf(\kaprep^2-\kaprep^2\vphic^2)}
   \beqref{ogrv2}
\end{split}
\end{align*}
\begin{align*}
\begin{split}
&=\frac{2\rhorep\vrhor-2\ethvj\vphia\epsva
     +6\ethvi\vrhop+6\ethvh\vrhoq-\ethvg\scalc}{-\kaprep\epsvf\vphid}
  -\frac{\efktc\vphid-2\epsvf\vphic\vrhoe}{-\epsvf\vphid}
      \biggl\{2\ethvl+\frac{4\kaprep\ethvk\efktc-2\vrhoa\epsvf\vphif}{\kaprep\epsvf}\biggr\}\\
  &\quad-\frac{\epsvf[\kaprep(-4\ethvl\efktc+4\ethvk\vrhoa\epsvf+2\ethvk\vrhoa\epsvf)+2\vrhod\vphif]
    +\efktc(4\kaprep\ethvk\efktc-2\vrhoa\epsvf\vphif)}{\kaprep\epsvf^2}\\
  &\quad+\frac{16\kaprep\vrhoe\ethvk\efktc\vphic
    -8\vrhoe\vrhoa\epsvf\vphif\vphic-8\efktc\vrhoe^2\vphif
    +12\efktc\vrhof\vphif\vphic-4\vrhoe^2\efktc\vphif}{-\kaprep\epsvf\vphid}\\
  &\quad+\frac{8\kaprep\vrhoe\ethvl\epsvf\vphic-12\kaprep\vrhoe^2\ethvk\epsvf
    -12\vrhoe\vrhof\vphif\epsvf+12\kaprep\vrhof\ethvk\epsvf\vphic+4\vrhoh\vphif\epsvf\vphic}
     {-\kaprep\epsvf\vphid}\beqref{main2b}\text{ \& }\eqnref{ogrv1b}
\end{split}
\end{align*}
\begin{align}\label{gpath13c}
\begin{split}
&=2\biggl[-\frac{\rhorep\vrhor-\ethvj\vphia\epsva
     +3\ethvi\vrhop+3\ethvh\vrhoq-\frac{1}{2}\ethvg\scalc}{\kaprep\epsvf\vphid}
  +\frac{\efktc\vphid-2\epsvf\vphic\vrhoe}{\epsvf\vphid}
      \biggl\{\ethvl+\frac{2\kaprep\ethvk\efktc-\vrhoa\epsvf\vphif}{\kaprep\epsvf}\biggr\}\\
  &\quad-\frac{\epsvf[\kaprep(-2\ethvl\efktc+2\ethvk\vrhoa\epsvf+\ethvk\vrhoa\epsvf)+\vrhod\vphif]
    +\efktc(2\kaprep\ethvk\efktc-\vrhoa\epsvf\vphif)}{\kaprep\epsvf^2}\\
  &\quad-\frac{8\kaprep\vrhoe\ethvk\efktc\vphic
    -4\vrhoe\vrhoa\epsvf\vphif\vphic-4\efktc\vrhoe^2\vphif
    +6\efktc\vrhof\vphif\vphic-2\vrhoe^2\efktc\vphif}{\kaprep\epsvf\vphid}\\
  &\quad-\frac{4\kaprep\vrhoe\ethvl\epsvf\vphic-6\kaprep\vrhoe^2\ethvk\epsvf
    -6\vrhoe\vrhof\vphif\epsvf+6\kaprep\vrhof\ethvk\epsvf\vphic+2\vrhoh\vphif\epsvf\vphic}
     {\kaprep\epsvf\vphid}\biggr]
\end{split}
\nonumber\\
&=2\ethvm\beqref{gpath1f}.
\end{align}
\end{subequations}
In consequence of the foregoing derivations, we obtain
\begin{subequations}\label{gpath14}
\begin{align}\label{gpath14a}
\fdot{\fone}
&=(1/2)\dif{[(\dprod{\vectkap}{\unitplz})\fzer]}\beqref{main5d}\nonumber\\
&=(\kaprep/2)(\dprod{\unitkap}{\unitplz})\fdot{\fzer}\nonumber\\
&=(\kaprep/2)(\epsvb)(-2\ethvk)\beqref{ogrv1a}\text{ \& }\eqnref{gpath13a}\nonumber\\
&=-\kaprep\epsvb\ethvk
\end{align}
\begin{align}\label{gpath14b}
\ffdot{\fone}
&=(\kaprep/2)(\dprod{\unitkap}{\unitplz})\ffdot{\fzer}\beqref{gpath14a}\nonumber\\
&=(\kaprep/2)\epsvb(2\ethvl)\beqref{ogrv1a}\text{ \& }\eqnref{gpath13b}\nonumber\\
&=\kaprep\epsvb\ethvl
\end{align}
\begin{align}\label{gpath14c}
\fffdot{\fone}
&=(\kaprep/2)(\dprod{\unitkap}{\unitplz})\fffdot{\fzer}\beqref{gpath14b}\nonumber\\
&=(\kaprep/2)\epsvb(2\ethvm)\beqref{ogrv1a}\text{ \& }\eqnref{gpath13c}\nonumber\\
&=\kaprep\epsvb\ethvm
\end{align}
\end{subequations}
\begin{subequations}\label{gpath15}
\begin{align}\label{gpath15a}
\fdot{\ftwo}
&=(1/2)\dif{(\scalm\fzer)}\beqref{main5d}\nonumber\\
&=(1/2)(\fdot{\scalm}\fzer+\scalm\fdot{\fzer})\nonumber\\
&=(1/2)[(-\kaprep\efkta)(2\vphif/\kaprep)+(\kaprep\epsve)(-2\ethvk)]
  \beqref{gpath2a}, \eqnref{ogrv2f}, \eqnref{ogrv2b}\text{ \& }\eqnref{gpath13a}\nonumber\\
&=-\efkta\vphif-\kaprep\epsve\ethvk
=\ethvn\beqref{gpath1g}
\end{align}
\begin{align}\label{gpath15b}
\ffdot{\ftwo}
&=(1/2)\dif{(\fdot{\scalm}\fzer+\scalm\fdot{\fzer})}\beqref{gpath15a}\nonumber\\
&=(1/2)(\ffdot{\scalm}\fzer+\fdot{\scalm}\fdot{\fzer}+\fdot{\scalm}\fdot{\fzer}+\scalm\ffdot{\fzer})\nonumber\\
&=(1/2)(\ffdot{\scalm}\fzer+2\fdot{\scalm}\fdot{\fzer}+\scalm\ffdot{\fzer})\nonumber\\
&=(1/2)[(-\kaprep\vrhoa\epsve)(2\vphif/\kaprep)+2(-\kaprep\efkta)(-2\ethvk)+(\kaprep\epsve)(2\ethvl)]
  \beqref{gpath2}, \eqnref{ogrv2}\text{ \& }\eqnref{gpath13}\nonumber\\
&=-\vrhoa\epsve\vphif+\kaprep(2\efkta\ethvk+\epsve\ethvl)
=\ethvo\beqref{gpath1g}
\end{align}
\begin{align}\label{gpath15c}
\fffdot{\ftwo}
&=(1/2)\dif{(\ffdot{\scalm}\fzer+2\fdot{\scalm}\fdot{\fzer}+\scalm\ffdot{\fzer})}
  \beqref{gpath15b}\nonumber\\
&=(1/2)(\fffdot{\scalm}\fzer+\ffdot{\scalm}\fdot{\fzer}+2\ffdot{\scalm}\fdot{\fzer}+2\fdot{\scalm}\ffdot{\fzer}
  +\fdot{\scalm}\ffdot{\fzer}+\scalm\fffdot{\fzer})\nonumber\\
&=(1/2)(\fffdot{\scalm}\fzer+3\ffdot{\scalm}\fdot{\fzer}+3\fdot{\scalm}\ffdot{\fzer}+\scalm\fffdot{\fzer})\nonumber\\
\begin{split}
&=(1/2)[(\kaprep\vrhoc)(2\vphif/\kaprep)+3(-\kaprep\vrhoa\epsve)(-2\ethvk)+3(-\kaprep\efkta)(2\ethvl)+(\kaprep\epsve)(2\ethvm)]\\
  &\quad\beqref{gpath2}, \eqnref{ogrv2}\text{ \& }\eqnref{gpath13}
\end{split}
\nonumber\\
&=\vrhoc\vphif+\kaprep(3\vrhoa\epsve\ethvk-3\efkta\ethvl+\epsve\ethvm)
=\ethvq\beqref{gpath1g}.
\end{align}
\end{subequations}
The derivatives of $\vecte$ are computed as
\begin{subequations}\label{gpath16}
\begin{align}\label{gpath16a}
\fdot{\vecte}
&=\dif{[\fone(\cprod{\vectr}{\vecth})+\ftwo\unitplz]}\beqref{main5b}\nonumber\\
&=\fdot{\fone}(\cprod{\vectr}{\vecth})+\fone(\cprod{\vectu}{\vecth})+\fdot{\ftwo}\unitplz
  \nonumber\\
&=\fdot{\fone}(\cprod{\vectr}{\vecth})-\fone(\vectz+\scalq\unitpos)+\fdot{\ftwo}\unitplz
  \beqref{main5c}\nonumber\\
&=(-\kaprep\epsvb\ethvk)(\cprod{\vectr}{\vecth})-(\epsvb\vphif)(\vectz+\scalq\unitpos)+(\ethvn)\unitplz
  \beqref{gpath14a}, \eqnref{ogrv2g}\text{ \& }\eqnref{gpath15a}\nonumber\\
&=-\kaprep\epsvb\ethvk(\cprod{\vectr}{\vecth})-\epsvb\vphif(\vectz+\scalq\unitpos)+\ethvn\unitplz
  \nonumber\\
&=\ethvn\unitplz-\epsvb\vphif\vectz-\scalq\epsvb\vphif\unitpos-\kaprep\epsvb\ethvk(\cprod{\vectr}{\vecth})
\end{align}
\begin{align}\label{gpath16b}
\ffdot{\vecte}
&=\dif{[\fdot{\fone}(\cprod{\vectr}{\vecth})+\fone(\cprod{\vectu}{\vecth})+\fdot{\ftwo}\unitplz]}
  \beqref{gpath16a}\nonumber\\
&=\ffdot{\fone}(\cprod{\vectr}{\vecth})+\fdot{\fone}(\cprod{\vectu}{\vecth})
  +\fdot{\fone}(\cprod{\vectu}{\vecth})+\fone(\cprod{\vecta}{\vecth})
  +\ffdot{\ftwo}\unitplz\nonumber\\
&=\ffdot{\fone}(\cprod{\vectr}{\vecth})+2\fdot{\fone}(\cprod{\vectu}{\vecth})+\fone(\cprod{\vecta}{\vecth})+\ffdot{\ftwo}\unitplz
  \nonumber\\
&=(\kaprep\epsvb\ethvl)(\cprod{\vectr}{\vecth})+2(-\kaprep\epsvb\ethvk)(\cprod{\vectu}{\vecth})
  +(\epsvb\vphif)(\cprod{\vecta}{\vecth})+\ethvo\unitplz
  \beqref{gpath14}, \eqnref{ogrv2g}\text{ \& }\eqnref{gpath15b}\nonumber\\
&=\kaprep\epsvb\ethvl(\cprod{\vectr}{\vecth})-2\kaprep\epsvb\ethvk(-\vectz-\scalq\unitpos)
  +\epsvb\vphif[\cprod{(-\scalq\vectr/\scalr^3)}{\vecth}]+\ethvo\unitplz
  \beqref{main5a}\text{ \& }\eqnref{main5c}\nonumber\\
&=\kaprep\epsvb\ethvl(\cprod{\vectr}{\vecth})+2\kaprep\epsvb\ethvk(\vectz+\scalq\unitpos)
  -\epsvb\vphif\vrhoa(\cprod{\vectr}{\vecth})+\ethvo\unitplz
  \beqref{gpath1b}\nonumber\\
&=\ethvo\unitplz+2\kaprep\epsvb\ethvk\vectz+2\kaprep\scalq\epsvb\ethvk\unitpos
   +\epsvb(\kaprep\ethvl-\vphif\vrhoa)(\cprod{\vectr}{\vecth})\nonumber\\
&=\ethvo\unitplz+2\kaprep\epsvb\ethvk\vectz+2\kaprep\scalq\epsvb\ethvk\unitpos
   +\ethvp(\cprod{\vectr}{\vecth})\beqref{gpath1g}
\end{align}
\begin{align*}
\fffdot{\vecte}
&=\dif{[\ffdot{\fone}(\cprod{\vectr}{\vecth})+2\fdot{\fone}(\cprod{\vectu}{\vecth})+\fone(\cprod{\vecta}{\vecth})+\ffdot{\ftwo}\unitplz]}
  \beqref{gpath16b}\nonumber\\
&=\fffdot{\fone}(\cprod{\vectr}{\vecth})+\ffdot{\fone}(\cprod{\vectu}{\vecth})
  +2\ffdot{\fone}(\cprod{\vectu}{\vecth})+2\fdot{\fone}(\cprod{\vecta}{\vecth})
  +\fdot{\fone}(\cprod{\vecta}{\vecth})+\fone(\cprod{\fdota}{\vecth})
  +\fffdot{\ftwo}\unitplz
\nonumber\\
&=\fffdot{\fone}(\cprod{\vectr}{\vecth})+3\ffdot{\fone}(\cprod{\vectu}{\vecth})+3\fdot{\fone}(\cprod{\vecta}{\vecth})
  +\fone(\cprod{\fdota}{\vecth})+\fffdot{\ftwo}\unitplz\nonumber\\
\begin{split}
&=\fffdot{\fone}(\cprod{\vectr}{\vecth})+3\ffdot{\fone}(\cprod{\vectu}{\vecth})+3\fdot{\fone}[\cprod{(-\vrhoa\vectr)}{\vecth}]
  +\fone[\cprod{(-\vrhoa\vectu+3\vrhoa\vrhob\vectr)}{\vecth}]+\fffdot{\ftwo}\unitplz\\
  &\quad\beqref{main5a}, \eqnref{gpath1b}\text{ \& }\eqnref{gpath6a}
\end{split}
\end{align*}
\begin{align}\label{gpath16c}
&=\fffdot{\fone}(\cprod{\vectr}{\vecth})+3\ffdot{\fone}(\cprod{\vectu}{\vecth})-3\fdot{\fone}\vrhoa(\cprod{\vectr}{\vecth})
  +\fone[-\vrhoa(\cprod{\vectu}{\vecth})+3\vrhoa\vrhob(\cprod{\vectr}{\vecth})]+\fffdot{\ftwo}\unitplz
  \nonumber\\
&=(\fffdot{\fone}-3\fdot{\fone}\vrhoa+3\vrhoa\vrhob\fone)(\cprod{\vectr}{\vecth})
  +(3\ffdot{\fone}-\vrhoa\fone)(\cprod{\vectu}{\vecth})+\fffdot{\ftwo}\unitplz
  \nonumber\\
\begin{split}
&=[(\kaprep\epsvb\ethvm)-3(-\kaprep\epsvb\ethvk)\vrhoa+3\vrhoa\vrhob(\epsvb\vphif)](\cprod{\vectr}{\vecth})
  +[3(\kaprep\epsvb\ethvl)-\vrhoa(\epsvb\vphif)](\cprod{\vectu}{\vecth})+\ethvq\unitplz\\
  &\quad\beqref{gpath14}, \eqnref{ogrv2g}\text{ \& }\eqnref{gpath15c}
\end{split}
\nonumber\\
&=\epsvb(\kaprep\ethvm+3\kaprep\ethvk\vrhoa+3\vrhoa\vrhob\vphif)(\cprod{\vectr}{\vecth})
  +\epsvb(3\kaprep\ethvl-\vrhoa\vphif)(\cprod{\vectu}{\vecth})+\ethvq\unitplz
  \nonumber\\
&=\ethvr(\cprod{\vectr}{\vecth})+\ethvs(-\vectz-\scalq\unitpos)+\ethvq\unitplz
  \beqref{gpath1g}\text{ \& }\eqnref{main5c}\nonumber\\
&=\ethvq\unitplz-\ethvs\vectz-\scalq\ethvs\unitpos+\ethvr(\cprod{\vectr}{\vecth})
\end{align}
\end{subequations}
\begin{subequations}\label{gpath17}
\begin{align}\label{gpath17a}
\fdot{\taurep}
&=\dif{[(2\rhorep\alprep-\scaln\kcons^2\fzer)\kaprep^{-2}]}\beqref{main5b}\nonumber\\
&=\kaprep^{-2}(2\fdot{\rhorep}\alprep+2\rhorep\fdot{\alprep}-\fdot{\scaln}\kcons^2\fzer
  -2\scaln\kcons\fdot{\kcons}\fzer-\scaln\kcons^2\fdot{\fzer})
  \nonumber\\
\begin{split}
&=\kaprep^{-2}[2\ethvh\alprep+2\rhorep(\kaprep\vrhop)-(-\efktc)\kcons^2\fzer
  -2\scaln\kcons(\kaprep\vrhoe)\fzer-\scaln\kcons^2(-2\ethvk)]\\
  &\quad\beqref{gpath12a}, \eqnref{gpath8a}, \eqnref{gpath3a}, \eqnref{gpath4a}\text{ \& }\eqnref{gpath13a}
\end{split}
\nonumber\\
&=\kaprep^{-2}(2\ethvh\alprep+2\kaprep\rhorep\vrhop+\efktc\kcons^2\fzer
  -2\kaprep\vrhoe\scaln\kcons\fzer+2\ethvk\scaln\kcons^2)\nonumber\\
\begin{split}
&=\kaprep^{-2}[2\ethvh(-\kaprep\vphia\epsva)+2\kaprep\rhorep\vrhop+\efktc(\kaprep\vphic)^2(2\vphif/\kaprep)
  -2\kaprep\vrhoe\epsvf(\kaprep\vphic)(2\vphif/\kaprep)\\
  &\qquad+2\ethvk\epsvf(\kaprep\vphic)^2]\beqref{ogrv2}
\end{split}
\nonumber\\
&=\kaprep^{-2}(-2\kaprep\ethvh\vphia\epsva+2\kaprep\rhorep\vrhop+2\kaprep\efktc\vphic^2\vphif
  -4\kaprep\vrhoe\epsvf\vphic\vphif+2\kaprep^2\ethvk\epsvf\vphic^2)
  \nonumber\\
&=2\kaprep^{-1}(\rhorep\vrhop-\ethvh\vphia\epsva+\efktc\vphic^2\vphif-2\vrhoe\epsvf\vphic\vphif+\kaprep\ethvk\epsvf\vphic^2)
=2\ethvt/\kaprep\beqref{gpath1g}
\end{align}
\begin{align*}
\ffdot{\taurep}
&=\kaprep^{-2}\dif{(2\fdot{\rhorep}\alprep+2\rhorep\fdot{\alprep}-\fdot{\scaln}\kcons^2\fzer
  -2\scaln\kcons\fdot{\kcons}\fzer-\scaln\kcons^2\fdot{\fzer})}
  \beqref{gpath17a}\nonumber\\
\begin{split}
&=\kaprep^{-2}[2\ffdot{\rhorep}\alprep+2\fdot{\rhorep}\fdot{\alprep}
  +2\fdot{\rhorep}\fdot{\alprep}+2\rhorep\ffdot{\alprep}
  -\ffdot{\scaln}\kcons^2\fzer-2\fdot{\scaln}\kcons\fdot{\kcons}\fzer-\fdot{\scaln}\kcons^2\fdot{\fzer}
  -2\fdot{\scaln}\kcons\fdot{\kcons}\fzer-2\scaln\fdot{\kcons}^2\fzer\\
  &\quad-2\scaln\kcons\ffdot{\kcons}\fzer-2\scaln\kcons\fdot{\kcons}\fdot{\fzer}
  -\fdot{\scaln}\kcons^2\fdot{\fzer}-2\scaln\kcons\fdot{\kcons}\fdot{\fzer}-\scaln\kcons^2\ffdot{\fzer}]
\end{split}
\nonumber\\
\begin{split}
&=\kaprep^{-2}(2\ffdot{\rhorep}\alprep+4\fdot{\rhorep}\fdot{\alprep}+2\rhorep\ffdot{\alprep}
  -\ffdot{\scaln}\kcons^2\fzer-4\fdot{\scaln}\kcons\fdot{\kcons}\fzer
  -2\fdot{\scaln}\kcons^2\fdot{\fzer}-2\scaln\fdot{\kcons}^2\fzer\\
  &\quad-2\scaln\kcons\ffdot{\kcons}\fzer-4\scaln\kcons\fdot{\kcons}\fdot{\fzer}
  -\scaln\kcons^2\ffdot{\fzer})
\end{split}
\end{align*}
\begin{align*}
\begin{split}
&=\kaprep^{-2}[2\ethvi\alprep+4\ethvh(\kaprep\vrhop)+2\rhorep(\kaprep\vrhoq)
  -(-\vrhoa\epsvf)\kcons^2\fzer-4(-\efktc)\kcons(\kaprep\vrhoe)\fzer-2(-\efktc)\kcons^2(-2\ethvk)\\
  &\quad-2\scaln(\kaprep\vrhoe)^2\fzer
  -2\scaln\kcons(-\kaprep\vrhof)\fzer-4\scaln\kcons(\kaprep\vrhoe)(-2\ethvk)
  -\scaln\kcons^2(2\ethvl)]\\
  &\quad\beqref{gpath12}, \eqnref{gpath8}, \eqnref{gpath3}, \eqnref{gpath4}\text{ \& }\eqnref{gpath13}
\end{split}
\nonumber\\
\begin{split}
&=\kaprep^{-2}(2\ethvi\alprep+4\kaprep\ethvh\vrhop+2\kaprep\rhorep\vrhoq
  +\vrhoa\epsvf\kcons^2\fzer+4\kaprep\efktc\vrhoe\kcons\fzer-4\efktc\ethvk\kcons^2\\
  &\quad-2\kaprep^2\vrhoe^2\scaln\fzer
  +2\kaprep\vrhof\scaln\kcons\fzer+8\kaprep\vrhoe\ethvk\scaln\kcons
  -2\ethvl\scaln\kcons^2)
\end{split}
\nonumber\\
\begin{split}
&=\kaprep^{-2}[2\ethvi(-\kaprep\vphia\epsva)+4\kaprep\ethvh\vrhop+2\kaprep\rhorep\vrhoq
  +\vrhoa\epsvf(\kaprep\vphic)^2(2\vphif/\kaprep)+4\kaprep\efktc\vrhoe(\kaprep\vphic)(2\vphif/\kaprep)\\
  &\quad-4\efktc\ethvk(\kaprep\vphic)^2-2\kaprep^2\vrhoe^2\epsvf(2\vphif/\kaprep)
  +2\kaprep\vrhof\epsvf(\kaprep\vphic)(2\vphif/\kaprep)+8\kaprep\vrhoe\ethvk\epsvf(\kaprep\vphic)\\
  &\quad-2\ethvl\epsvf(\kaprep\vphic)^2]
  \beqref{ogrv2}
\end{split}
\end{align*}
\begin{align}\label{gpath17b}
\begin{split}
&=2\kaprep^{-1}(-\epsva\vphia\ethvi+2\ethvh\vrhop+\rhorep\vrhoq
  +\vrhoa\epsvf\vphic^2\vphif+4\efktc\vrhoe\vphic\vphif
  -2\kaprep\efktc\ethvk\vphic^2\\
  &\quad-2\vrhoe^2\epsvf\vphif
  +2\vrhof\epsvf\vphic\vphif+4\kaprep\vrhoe\ethvk\epsvf\vphic
  -\kaprep\epsvf\ethvl\vphic^2)
\end{split}
\nonumber\\
&=2\ethvu/\kaprep\beqref{gpath1g}
\end{align}
\begin{align*}
\begin{split}
\fffdot{\taurep}
&=\kaprep^{-2}[\dif{(2\ffdot{\rhorep}\alprep+4\fdot{\rhorep}\fdot{\alprep}+2\rhorep\ffdot{\alprep}
  -\ffdot{\scaln}\kcons^2\fzer-4\fdot{\scaln}\kcons\fdot{\kcons}\fzer
  -2\fdot{\scaln}\kcons^2\fdot{\fzer}-2\scaln\fdot{\kcons}^2\fzer)}\\
  &\quad+\dif{(-2\scaln\kcons\ffdot{\kcons}\fzer-4\scaln\kcons\fdot{\kcons}\fdot{\fzer}
  -\scaln\kcons^2\ffdot{\fzer})}\beqref{gpath17b}
\end{split}
\nonumber\\
\begin{split}
&=\kaprep^{-2}[2\fffdot{\rhorep}\alprep+2\ffdot{\rhorep}\fdot{\alprep}
  +4\ffdot{\rhorep}\fdot{\alprep}+4\fdot{\rhorep}\ffdot{\alprep}
  +2\fdot{\rhorep}\ffdot{\alprep}+2\rhorep\fffdot{\alprep}
  -\fffdot{\scaln}\kcons^2\fzer-2\ffdot{\scaln}\kcons\fdot{\kcons}\fzer-\ffdot{\scaln}\kcons^2\fdot{\fzer}\\
  &\quad-4\ffdot{\scaln}\kcons\fdot{\kcons}\fzer-4\fdot{\scaln}\fdot{\kcons}^2\fzer
     -4\fdot{\scaln}\kcons\ffdot{\kcons}\fzer-4\fdot{\scaln}\kcons\fdot{\kcons}\fdot{\fzer}
  -2\ffdot{\scaln}\kcons^2\fdot{\fzer}-4\fdot{\scaln}\kcons\fdot{\kcons}\fdot{\fzer}-2\fdot{\scaln}\kcons^2\ffdot{\fzer}\\
  &\quad-2\fdot{\scaln}\fdot{\kcons}^2\fzer-4\scaln\fdot{\kcons}\ffdot{\kcons}\fzer-2\scaln\fdot{\kcons}^2\fdot{\fzer}
  -2\fdot{\scaln}\kcons\ffdot{\kcons}\fzer-2\scaln\fdot{\kcons}\ffdot{\kcons}\fzer-2\scaln\kcons\fffdot{\kcons}\fzer
     -2\scaln\kcons\ffdot{\kcons}\fdot{\fzer}\\
  &\quad-4\fdot{\scaln}\kcons\fdot{\kcons}\fdot{\fzer}-4\scaln\fdot{\kcons}^2\fdot{\fzer}
     -4\scaln\kcons\ffdot{\kcons}\fdot{\fzer}-4\scaln\kcons\fdot{\kcons}\ffdot{\fzer}
  -\fdot{\scaln}\kcons^2\ffdot{\fzer}-2\scaln\kcons\fdot{\kcons}\ffdot{\fzer}-\scaln\kcons^2\fffdot{\fzer}]
\end{split}
\nonumber\\
\begin{split}
&=\kaprep^{-2}[2\fffdot{\rhorep}\alprep+6\ffdot{\rhorep}\fdot{\alprep}
  +6\fdot{\rhorep}\ffdot{\alprep}+2\rhorep\fffdot{\alprep}
  -\fffdot{\scaln}\kcons^2\fzer-\ffdot{\scaln}\kcons^2\fdot{\fzer}
  -6\ffdot{\scaln}\kcons\fdot{\kcons}\fzer-6\fdot{\scaln}\fdot{\kcons}^2\fzer\\
  &\quad-6\fdot{\scaln}\kcons\ffdot{\kcons}\fzer-12\fdot{\scaln}\kcons\fdot{\kcons}\fdot{\fzer}
  -2\ffdot{\scaln}\kcons^2\fdot{\fzer}-3\fdot{\scaln}\kcons^2\ffdot{\fzer}
  -4\scaln\fdot{\kcons}\ffdot{\kcons}\fzer-6\scaln\fdot{\kcons}^2\fdot{\fzer}\\
  &\quad-2\scaln\fdot{\kcons}\ffdot{\kcons}\fzer-2\scaln\kcons\fffdot{\kcons}\fzer
     -6\scaln\kcons\ffdot{\kcons}\fdot{\fzer}
  -6\scaln\kcons\fdot{\kcons}\ffdot{\fzer}-\scaln\kcons^2\fffdot{\fzer}]
\end{split}
\end{align*}
\begin{align*}
\begin{split}
&=\kaprep^{-2}[2\ethvj\alprep+6\ethvi(\kaprep\vrhop)
  +6\ethvh(\kaprep\vrhoq)+2\rhorep(\kaprep\vrhor)
  -\vrhod\kcons^2\fzer-(-\vrhoa\epsvf)\kcons^2(-2\ethvk)\\
  &\quad-6(-\vrhoa\epsvf)\kcons(\kaprep\vrhoe)\fzer
  -6(-\efktc)(\kaprep\vrhoe)^2\fzer
  -6(-\efktc)\kcons(-\kaprep\vrhof)\fzer-12(-\efktc)\kcons(\kaprep\vrhoe)(-2\ethvk)\\
  &\quad-2(-\vrhoa\epsvf)\kcons^2(-2\ethvk)-3(-\efktc)\kcons^2(2\ethvl)
  -4\scaln(\kaprep\vrhoe)(-\kaprep\vrhof)\fzer\\
  &\quad-6\scaln(\kaprep\vrhoe)^2(-2\ethvk)
  -2\scaln(\kaprep\vrhoe)(-\kaprep\vrhof)\fzer-2\scaln\kcons(\kaprep\vrhoh)\fzer
     -6\scaln\kcons(-\kaprep\vrhof)(-2\ethvk)\\
  &\quad-6\scaln\kcons(\kaprep\vrhoe)(2\ethvl)-\scaln\kcons^2(2\ethvm)]
  \beqref{gpath12}, \eqnref{gpath8}, \eqnref{gpath3}, \eqnref{gpath4}\text{ \& }\eqnref{gpath13}
\end{split}
\nonumber\\
\begin{split}
&=\kaprep^{-2}[2\ethvj\alprep+6\kaprep\vrhop\ethvi
  +6\kaprep\vrhoq\ethvh+2\kaprep\vrhor\rhorep
  -\vrhod\kcons^2\fzer-2\ethvk\vrhoa\epsvf\kcons^2
  +6\kaprep\vrhoe\vrhoa\epsvf\kcons\fzer\\
  &\quad+6\kaprep^2\vrhoe^2\efktc\fzer
  -6\kaprep\vrhof\efktc\kcons\fzer-24\kaprep\vrhoe\ethvk\efktc\kcons
  -4\ethvk\vrhoa\epsvf\kcons^2+6\ethvl\efktc\kcons^2\\
  &\quad+4\kaprep^2\vrhoe\vrhof\scaln\fzer
  +12\kaprep^2\vrhoe^2\ethvk\scaln
  +2\kaprep^2\vrhoe\vrhof\scaln\fzer-2\kaprep\vrhoh\scaln\kcons\fzer
     -12\kaprep\vrhof\ethvk\scaln\kcons\\
  &\quad-12\kaprep\vrhoe\ethvl\scaln\kcons-2\ethvm\scaln\kcons^2]
\end{split}
\end{align*}
\begin{align*}
\begin{split}
&=\kaprep^{-2}[2\ethvj(-\kaprep\vphia\epsva)+6\kaprep\vrhop\ethvi
  +6\kaprep\vrhoq\ethvh+2\kaprep\vrhor\rhorep
  -\vrhod(\kaprep\vphic)^2(2\vphif/\kaprep)-2\ethvk\vrhoa\epsvf(\kaprep\vphic)^2\\
  &\quad+6\kaprep\vrhoe\vrhoa\epsvf(\kaprep\vphic)(2\vphif/\kaprep)
  +6\kaprep^2\vrhoe^2\efktc(2\vphif/\kaprep)
  -6\kaprep\vrhof\efktc(\kaprep\vphic)(2\vphif/\kaprep)-24\kaprep\vrhoe\ethvk\efktc(\kaprep\vphic)\\
  &\quad-4\ethvk\vrhoa\epsvf(\kaprep\vphic)^2
  +6\ethvl\efktc(\kaprep\vphic)^2
  +4\kaprep^2\vrhoe\vrhof\epsvf(2\vphif/\kaprep)
  +12\kaprep^2\vrhoe^2\ethvk\epsvf\\
  &\quad+2\kaprep^2\vrhoe\vrhof\epsvf(2\vphif/\kaprep)-2\kaprep\vrhoh\epsvf(\kaprep\vphic)(2\vphif/\kaprep)
     -12\kaprep\vrhof\ethvk\epsvf(\kaprep\vphic)\\
  &\quad-12\kaprep\vrhoe\ethvl\epsvf(\kaprep\vphic)-2\ethvm\epsvf(\kaprep\vphic)^2]
  \beqref{ogrv2}
\end{split}
\nonumber\\
\begin{split}
&=\kaprep^{-2}[-2\kaprep\vphia\epsva\ethvj+6\kaprep\vrhop\ethvi
  +6\kaprep\vrhoq\ethvh+2\kaprep\vrhor\rhorep
  -2\vrhod\kaprep\vphic^2\vphif-2\kaprep^2\epsvf\ethvk\vrhoa\vphic^2\\
  &\quad+12\kaprep\vrhoe\vrhoa\epsvf\vphic\vphif
  +12\kaprep\vrhoe^2\efktc\vphif
  -12\kaprep\vrhof\efktc\vphic\vphif-24\kaprep^2\vrhoe\ethvk\efktc\vphic\\
  &\quad-4\kaprep^2\ethvk\vrhoa\epsvf\vphic^2
  +6\kaprep^2\efktc\ethvl\vphic^2+8\kaprep\vrhoe\vrhof\epsvf\vphif+12\kaprep^2\vrhoe^2\ethvk\epsvf\\
  &\quad+4\kaprep\vrhoe\vrhof\epsvf\vphif-4\kaprep\vrhoh\epsvf\vphic\vphif
     -12\kaprep^2\vrhof\ethvk\epsvf\vphic
  -12\kaprep^2\vrhoe\ethvl\epsvf\vphic-2\kaprep^2\epsvf\ethvm\vphic^2]
\end{split}
\end{align*}
\begin{align}\label{gpath17c}
\begin{split}
&=2\kaprep^{-1}[-\vphia\epsva\ethvj+3\vrhop\ethvi+3\vrhoq\ethvh+\vrhor\rhorep-\vrhod\vphic^2\vphif-\kaprep\epsvf\ethvk\vrhoa\vphic^2\\
  &\quad+6\vrhoe\vrhoa\epsvf\vphic\vphif+6\vrhoe^2\efktc\vphif -6\vrhof\efktc\vphic\vphif-12\kaprep\vrhoe\ethvk\efktc\vphic\\
  &\quad-2\kaprep\ethvk\vrhoa\epsvf\vphic^2+3\kaprep\efktc\ethvl\vphic^2+4\vrhoe\vrhof\epsvf\vphif+6\kaprep\vrhoe^2\ethvk\epsvf\\
  &\quad+2\vrhoe\vrhof\epsvf\vphif-2\vrhoh\epsvf\vphic\vphif-6\kaprep\vrhof\ethvk\epsvf\vphic
  -6\kaprep\vrhoe\ethvl\epsvf\vphic-\kaprep\epsvf\ethvm\vphic^2]
\end{split}
\nonumber\\
&=2\ethvv/\kaprep\beqref{gpath1g}.
\end{align}
\end{subequations}

\subart{Development of equation \eqnref{kpath2a}}
We are now ready to evaluate the quantities defined by \eqnref{kpath2}. We start with
\begin{subequations}\label{gpath18}
\begin{align}\label{gpath18a}
\fdot{\cdkt}
&=\scalc\fdot{\dragf}-\kaprep\fdott\beqref{kpath2a}\nonumber\\
&=\scalc\ethva-\kaprep(2\ethvt/\kaprep)\beqref{gpath11a}\text{ \& }\eqnref{gpath17a}\nonumber\\
&=\scalc\ethva-2\ethvt
=\frkya\beqref{gpath1h}
\end{align}
\begin{align}\label{gpath18b}
\ffdot{\cdkt}
&=\scalc\ffdot{\dragf}-\kaprep\ffdott\beqref{kpath2a}\nonumber\\
&=\scalc\ethve-\kaprep(2\ethvu/\kaprep)\beqref{gpath11b}\text{ \& }\eqnref{gpath17b}\nonumber\\
&=\scalc\ethve-2\ethvu
=\frkyb\beqref{gpath1h}
\end{align}
\begin{align}\label{gpath18c}
\fffdot{\cdkt}
&=\scalc\fffdot{\dragf}-\kaprep\fffdott\beqref{kpath2a}\nonumber\\
&=\scalc\ethvg-\kaprep(2\ethvv/\kaprep)\beqref{gpath11c}\text{ \& }\eqnref{gpath17c}\nonumber\\
&=\scalc\ethvg-2\ethvv
=\frkyc\beqref{gpath1h}
\end{align}
\end{subequations}
\begin{subequations}\label{gpath19}
\begin{align}\label{gpath19a}
\vbba
&=\fdot{\rhorep}-1\beqref{kpath2a}\nonumber\\
&=\ethvh-1\beqref{gpath12a}\nonumber\\
&=\frkyd\beqref{gpath1h}
\end{align}
\begin{align}\label{gpath19b}
\vbbb
&=2\fdot{\rhorep}-1\beqref{kpath2a}\nonumber\\
&=2\ethvh-1\beqref{gpath12a}\nonumber\\
&=\frkye\beqref{gpath1h}
\end{align}
\begin{align}\label{gpath19c}
\vbbc
&=3\fdot{\rhorep}-1\beqref{kpath2a}\nonumber\\
&=3\ethvh-1\beqref{gpath12a}\nonumber\\
&=\frkyf\beqref{gpath1h}
\end{align}
\begin{align}\label{gpath19d}
\vbbd
&=\ffdot{\rhorep}\fdot{\cdkt}-\ffdot{\cdkt}\vbba\beqref{kpath2a}\nonumber\\
&=\ethvi\frkya-\frkyb\frkyd\beqref{gpath12b}, \eqnref{gpath18}\text{ \& }\eqnref{gpath19a}\nonumber\\
&=\frkyg\beqref{gpath1h}
\end{align}
\begin{align}\label{gpath19e}
\vbbe
&=\fdot{\cdkt}\vbbb-\rhorep\ffdot{\cdkt}\beqref{kpath2a}\nonumber\\
&=\frkya\frkye-\rhorep\frkyb\beqref{gpath18}\text{ \& }\eqnref{gpath19b}\nonumber\\
&=\frkyh\beqref{gpath1h}
\end{align}
\begin{align}\label{gpath19f}
\vbbf
&=\vbba\vbbb-\rhorep\ffdot{\rhorep}\beqref{kpath2a}\nonumber\\
&=\frkyd\frkye-\rhorep\ethvi\beqref{gpath12b}, \eqnref{gpath19a}\text{ \& }\eqnref{gpath19b}\nonumber\\
&=\frkyi\beqref{gpath1h}.
\end{align}
\end{subequations}

\subart{Development of equation \eqnref{kpath2b}}
We have furthermore that
\begin{subequations}\label{gpath20}
\begin{align}\label{gpath20a}
\vscra
&=\cprod{\unitkap}{\vecta}\beqref{kpath2b}\nonumber\\
&=\cprod{\unitkap}{(-\vrhoa\vectr)}\beqref{main5a}\text{ \& }\eqnref{gpath1b}\nonumber\\
&=-\vrhoa(\cprod{\unitkap}{\vectr})
\end{align}
\begin{align}\label{gpath20b}
\vscrb
&=\cprod{\unitkap}{\fdota}\beqref{kpath2b}\nonumber\\
&=\cprod{\unitkap}{(-\vrhoa\vectu+3\vrhoa\vrhob\vectr)}\beqref{gpath6a}\nonumber\\
&=-\vrhoa(\cprod{\unitkap}{\vectu})+3\vrhoa\vrhob(\cprod{\unitkap}{\vectr})
\end{align}
\begin{align}\label{gpath20c}
\vscrc
&=\cprod{\unitkap}{\ffdota}\beqref{kpath2b}\nonumber\\
&=\cprod{\unitkap}{(6\vrhoa\vrhob\vectu-\vrhoa\vrhol\vectr)}\beqref{gpath6b}\nonumber\\
&=6\vrhoa\vrhob(\cprod{\unitkap}{\vectu})-\vrhoa\vrhol(\cprod{\unitkap}{\vectr})
\end{align}
\begin{align}\label{gpath20d}
\vscrd
&=\cprod{\unitkap}{\fdote}\beqref{kpath2b}\nonumber\\
&=\cprod{\unitkap}{[\ethvn\unitplz-\epsvb\vphif\vectz-\scalq\epsvb\vphif\unitpos-\kaprep\epsvb\ethvk(\cprod{\vectr}{\vecth})]}
  \beqref{gpath16a}\nonumber\\
&=\ethvn(\cprod{\unitkap}{\unitplz})-\epsvb\vphif(\cprod{\unitkap}{\vectz})-\scalq\epsvb\vphif(\cprod{\unitkap}{\unitpos})
  -\kaprep\epsvb\ethvk[\cprod{\unitkap}{(\cprod{\vectr}{\vecth})}]\nonumber\\
&=\ethvn(\cprod{\unitkap}{\unitplz})-\epsvb\vphif(\cprod{\unitkap}{\vectz})-\scalq\epsvb\vphif(\cprod{\unitkap}{\unitpos})
  -\kaprep\epsvb\ethvk[\vectr(\dprod{\unitkap}{\vecth})-\vecth(\dprod{\unitkap}{\vectr})]\beqref{alg1}\nonumber\\
&=\ethvn(\cprod{\unitkap}{\unitplz})-\epsvb\vphif(\cprod{\unitkap}{\vectz})-\scalq\epsvb\vphif(\cprod{\unitkap}{\unitpos})
  -\kaprep\epsvb\ethvk(\dltva\vectr-\scalr\epsva\vecth)\beqref{gxpeed1a}\text{ \& }\eqnref{ogrv1a}\nonumber\\
&=\kaprep\scalr\epsva\epsvb\ethvk\vecth-\kaprep\epsvb\ethvk\dltva\vectr
  +\ethvn(\cprod{\unitkap}{\unitplz})-\epsvb\vphif(\cprod{\unitkap}{\vectz})-\scalq\epsvb\vphif(\cprod{\unitkap}{\unitpos})
\end{align}
\begin{align}\label{gpath20e}
\vscre
&=\cprod{\unitkap}{\ffdote}\beqref{kpath2b}\nonumber\\
&=\cprod{\unitkap}{[\ethvo\unitplz+2\kaprep\epsvb\ethvk\vectz+2\kaprep\scalq\epsvb\ethvk\unitpos
   +\ethvp(\cprod{\vectr}{\vecth})]}\beqref{gpath16b}\nonumber\\
&=\ethvo(\cprod{\unitkap}{\unitplz})+2\kaprep\epsvb\ethvk(\cprod{\unitkap}{\vectz})
  +2\kaprep\scalq\epsvb\ethvk(\cprod{\unitkap}{\unitpos})
   +\ethvp[\cprod{\unitkap}{(\cprod{\vectr}{\vecth})}]\nonumber\\
&=\ethvo(\cprod{\unitkap}{\unitplz})+2\kaprep\epsvb\ethvk(\cprod{\unitkap}{\vectz})
  +2\kaprep\scalq\epsvb\ethvk(\cprod{\unitkap}{\unitpos})
   +\ethvp[\vectr(\dprod{\unitkap}{\vecth})-\vecth(\dprod{\unitkap}{\vectr})]
   \beqref{alg1}\nonumber\\
&=\ethvo(\cprod{\unitkap}{\unitplz})+2\kaprep\epsvb\ethvk(\cprod{\unitkap}{\vectz})
  +2\kaprep\scalq\epsvb\ethvk(\cprod{\unitkap}{\unitpos})
   +\ethvp(\dltva\vectr-\scalr\epsva\vecth)
   \beqref{gxpeed1a}\text{ \& }\eqnref{ogrv1a}\nonumber\\
&=\ethvp\dltva\vectr-\scalr\ethvp\epsva\vecth+\ethvo(\cprod{\unitkap}{\unitplz})+2\kaprep\epsvb\ethvk(\cprod{\unitkap}{\vectz})
  +2\kaprep\scalq\epsvb\ethvk(\cprod{\unitkap}{\unitpos})
\end{align}
\begin{align}\label{gpath20f}
\vscrf
&=\cprod{\vecta}{\fdota}\beqref{kpath2b}\nonumber\\
&=\cprod{(-\vrhoa\vectr)}{(-\vrhoa\vectu+3\vrhoa\vrhob\vectr)}
  \beqref{main5a}, \eqnref{gpath1b}\text{ \& }\eqnref{gpath6a}\nonumber\\
&=-\vrhoa^2(\cprod{\vectu}{\vectr})
=-\vrhoa^2\vecth\beqref{main5c}
\end{align}
\begin{align}\label{gpath20g}
\vscrg
&=\cprod{\vecta}{\ffdota}\beqref{kpath2b}\nonumber\\
&=\cprod{(-\vrhoa\vectr)}{(6\vrhoa\vrhob\vectu-\vrhoa\vrhol\vectr)}
  \beqref{main5a}, \eqnref{gpath1b}\text{ \& }\eqnref{gpath6b}\nonumber\\
&=6\vrhoa^2\vrhob(\cprod{\vectu}{\vectr})
=6\vrhoa^2\vrhob\vecth\beqref{main5c}
\end{align}
\begin{align}\label{gpath20h}
\vscrh
&=\cprod{\vecta}{\fdote}\beqref{kpath2b}\nonumber\\
&=\cprod{(-\vrhoa\vectr)}{[\ethvn\unitplz-\epsvb\vphif\vectz-\scalq\epsvb\vphif\unitpos-\kaprep\epsvb\ethvk(\cprod{\vectr}{\vecth})]}
  \beqref{main5a}, \eqnref{gpath1b}\text{ \& }\eqnref{gpath16a}\nonumber\\
&=-\vrhoa\ethvn(\cprod{\vectr}{\unitplz})+\vrhoa\epsvb\vphif(\cprod{\vectr}{\vectz})
  +\kaprep\vrhoa\epsvb\ethvk[\cprod{\vectr}{(\cprod{\vectr}{\vecth})}]\nonumber\\
&=-\vrhoa\ethvn(\cprod{\vectr}{\unitplz})+\vrhoa\epsvb\vphif(\cprod{\vectr}{\vectz})
  +\kaprep\vrhoa\epsvb\ethvk[(\dprod{\vectr}{\vecth})\vectr-\scalr^2\vecth]\beqref{alg1}\nonumber\\
&=-\vrhoa\ethvn(\cprod{\vectr}{\unitplz})+\vrhoa\epsvb\vphif(\cprod{\vectr}{\vectz})
  -\kaprep\vrhoa\epsvb\scalr^2\ethvk\vecth\beqref{main5c}\nonumber\\
&=-\vrhoa\ethvn(\cprod{\vectr}{\unitplz})+\vrhoa\epsvb\vphif(\cprod{\vectr}{\vectz})
  -\kaprep\vphib\epsvb\ethvk\vecth\beqref{ogrv1b}\text{ \& }\eqnref{gpath1b}
\end{align}
\begin{align}\label{gpath20i}
\vscri
&=\cprod{\vecta}{\ffdote}\beqref{kpath2b}\nonumber\\
&=\cprod{(-\vrhoa\vectr)}{[\ethvo\unitplz+2\kaprep\epsvb\ethvk\vectz+2\kaprep\scalq\epsvb\ethvk\unitpos
   +\ethvp(\cprod{\vectr}{\vecth})]}
  \beqref{main5a}, \eqnref{gpath1b}\text{ \& }\eqnref{gpath16b}\nonumber\\
&=-\vrhoa\ethvo(\cprod{\vectr}{\unitplz})-2\kaprep\vrhoa\epsvb\ethvk(\cprod{\vectr}{\vectz})
   -\vrhoa\ethvp[\cprod{\vectr}{(\cprod{\vectr}{\vecth})}]\nonumber\\
&=-\vrhoa\ethvo(\cprod{\vectr}{\unitplz})-2\kaprep\vrhoa\epsvb\ethvk(\cprod{\vectr}{\vectz})
   -\vrhoa\ethvp[(\dprod{\vectr}{\vecth})\vectr-\scalr^2\vecth]\beqref{alg1}\nonumber\\
&=-\vrhoa\ethvo(\cprod{\vectr}{\unitplz})-2\kaprep\vrhoa\epsvb\ethvk(\cprod{\vectr}{\vectz})
   +\vrhoa\scalr^2\ethvp\vecth\beqref{main5c}\nonumber\\
&=-\vrhoa\ethvo(\cprod{\vectr}{\unitplz})-2\kaprep\vrhoa\epsvb\ethvk(\cprod{\vectr}{\vectz})
   +\vphib\ethvp\vecth\beqref{ogrv1b}\text{ \& }\eqnref{gpath1b}
\end{align}
\begin{align}\label{gpath20j}
\vscrj
&=\cprod{\fdota}{\ffdota}\beqref{kpath2b}\nonumber\\
&=\cprod{(-\vrhoa\vectu+3\vrhoa\vrhob\vectr)}{(6\vrhoa\vrhob\vectu-\vrhoa\vrhol\vectr)}
  \beqref{gpath6}\nonumber\\
&=-6\vrhoa^2\vrhob(\cprod{\vectu}{\vectu})+\vrhoa^2\vrhol(\cprod{\vectu}{\vectr})
  +18\vrhoa^2\vrhob^2(\cprod{\vectr}{\vectu})-3\vrhoa^2\vrhob\vrhol(\cprod{\vectr}{\vectr})\nonumber\\
&=(\vrhoa^2\vrhol-18\vrhoa^2\vrhob^2)(\cprod{\vectu}{\vectr})
=\vrhoa^2(\vrhol-18\vrhob^2)\vecth\beqref{main5c}\nonumber\\
&=\frkyj\vecth\beqref{gpath1h}
\end{align}
\begin{align*}
\vscrk
&=\cprod{\fdota}{\fdote}\beqref{kpath2b}\nonumber\\
&=\cprod{(-\vrhoa\vectu+3\vrhoa\vrhob\vectr)}{[\ethvn\unitplz-\epsvb\vphif\vectz-\scalq\epsvb\vphif\unitpos
  -\kaprep\epsvb\ethvk(\cprod{\vectr}{\vecth})]}\beqref{gpath6a}\text{ \& }\eqnref{gpath16a}\nonumber\\
\begin{split}
&=-\vrhoa\ethvn(\cprod{\vectu}{\unitplz})
  +\vrhoa\epsvb\vphif(\cprod{\vectu}{\vectz})
  +\scalq\vrhoa\epsvb\vphif(\cprod{\vectu}{\unitpos})
  +\kaprep\vrhoa\epsvb\ethvk[\cprod{\vectu}{(\cprod{\vectr}{\vecth})}]\\
  &\quad+3\vrhoa\vrhob\ethvn(\cprod{\vectr}{\unitplz})
  -3\vrhoa\vrhob\epsvb\vphif(\cprod{\vectr}{\vectz})
  -3\scalq\vrhoa\vrhob\epsvb\vphif(\cprod{\vectr}{\unitpos})
  -3\kaprep\vrhoa\vrhob\epsvb\ethvk[\cprod{\vectr}{(\cprod{\vectr}{\vecth})}]
\end{split}
\nonumber\\
\begin{split}
&=-\vrhoa\ethvn(\cprod{\vectu}{\unitplz})
  +\vrhoa\epsvb\vphif(\cprod{\vectu}{\vectz})
  +\scalq\vrhoa\epsvb\vphif(\cprod{\vectu}{\unitpos})
  +\kaprep\vrhoa\epsvb\ethvk[(\dprod{\vectu}{\vecth})\vectr-(\dprod{\vectu}{\vectr})\vecth]\\
  &\quad+3\vrhoa\vrhob\ethvn(\cprod{\vectr}{\unitplz})
  -3\vrhoa\vrhob\epsvb\vphif(\cprod{\vectr}{\vectz})
  -3\kaprep\vrhoa\vrhob\epsvb\ethvk[(\dprod{\vectr}{\vecth})\vectr-\scalr^2\vecth]
  \beqref{alg1}
\end{split}
\end{align*}
\begin{align}\label{gpath20k}
\begin{split}
&=-\vrhoa\ethvn(\cprod{\vectu}{\unitplz})
  +\vrhoa\epsvb\vphif(\cprod{\vectu}{\vectz})
  +(\scalq/\scalr)\vrhoa\epsvb\vphif\vecth
  +\kaprep\vrhoa\epsvb\ethvk(-\scalr^2\vrhob\vecth)\\
  &\quad+3\vrhoa\vrhob\ethvn(\cprod{\vectr}{\unitplz})
  -3\vrhoa\vrhob\epsvb\vphif(\cprod{\vectr}{\vectz})
  -3\kaprep\vrhoa\vrhob\epsvb\ethvk(-\scalr^2\vecth)
  \beqref{main5c}\text{ \& }\eqnref{gpath7b}
\end{split}
\nonumber\\
\begin{split}
&=-\vrhoa\ethvn(\cprod{\vectu}{\unitplz})
  +\vrhoa\epsvb\vphif(\cprod{\vectu}{\vectz})
  +\vphib\vrhoa\epsvb\vphif\vecth
  -\kaprep\vphib\vrhob\epsvb\ethvk\vecth\\
  &\quad+3\vrhoa\vrhob\ethvn(\cprod{\vectr}{\unitplz})
  -3\vrhoa\vrhob\epsvb\vphif(\cprod{\vectr}{\vectz})
  +3\kaprep\vphib\vrhob\epsvb\ethvk\vecth\beqref{ogrv1b}\text{ \& }\eqnref{gpath1b}
\end{split}
\nonumber\\
\begin{split}
&=-\vrhoa\ethvn(\cprod{\vectu}{\unitplz})
  +\vrhoa\epsvb\vphif(\cprod{\vectu}{\vectz})
  +3\vrhoa\vrhob\ethvn(\cprod{\vectr}{\unitplz})
  -3\vrhoa\vrhob\epsvb\vphif(\cprod{\vectr}{\vectz})\\
  &\quad+\vphib\epsvb(\vrhoa\vphif+2\kaprep\vrhob\ethvk)\vecth
\end{split}
\nonumber\\
\begin{split}
&=-\vrhoa\ethvn(\cprod{\vectu}{\unitplz})
  +\vrhoa\epsvb\vphif(\cprod{\vectu}{\vectz})
  +3\vrhoa\vrhob\ethvn(\cprod{\vectr}{\unitplz})\\
  &\quad-3\vrhoa\vrhob\epsvb\vphif(\cprod{\vectr}{\vectz})+\frkyk\vecth
  \beqref{gpath1h}
\end{split}
\end{align}
\begin{align*}
\vscrl
&=\cprod{\fdota}{\ffdote}\beqref{kpath2b}\nonumber\\
&=\cprod{(-\vrhoa\vectu+3\vrhoa\vrhob\vectr)}{[\ethvo\unitplz+2\kaprep\epsvb\ethvk\vectz+2\kaprep\scalq\epsvb\ethvk\unitpos
   +\ethvp(\cprod{\vectr}{\vecth})]}\beqref{gpath6a}\text{ \& }\eqnref{gpath16b}\nonumber\\
\begin{split}
&=-\vrhoa\ethvo(\cprod{\vectu}{\unitplz})
  -2\kaprep\vrhoa\epsvb\ethvk(\cprod{\vectu}{\vectz})
  -2\kaprep\scalq\vrhoa\epsvb\ethvk(\cprod{\vectu}{\unitpos})
  -\vrhoa\ethvp[\cprod{\vectu}{(\cprod{\vectr}{\vecth})}]\\
  &\quad+3\vrhoa\vrhob\ethvo(\cprod{\vectr}{\unitplz})
  +6\kaprep\vrhoa\vrhob\epsvb\ethvk(\cprod{\vectr}{\vectz})
  +6\kaprep\scalq\vrhoa\vrhob\epsvb\ethvk(\cprod{\vectr}{\unitpos})
  +3\vrhoa\vrhob\ethvp[\cprod{\vectr}{(\cprod{\vectr}{\vecth})}]
\end{split}
\nonumber\\
\begin{split}
&=-\vrhoa\ethvo(\cprod{\vectu}{\unitplz})
  -2\kaprep\vrhoa\epsvb\ethvk(\cprod{\vectu}{\vectz})
  -2\kaprep\scalq\vrhoa\epsvb\ethvk(\cprod{\vectu}{\unitpos})
  -\vrhoa\ethvp[(\dprod{\vectu}{\vecth})\vectr-(\dprod{\vectu}{\vectr})\vecth]\\
  &\quad+3\vrhoa\vrhob\ethvo(\cprod{\vectr}{\unitplz})
  +6\kaprep\vrhoa\vrhob\epsvb\ethvk(\cprod{\vectr}{\vectz})
  +3\vrhoa\vrhob\ethvp[(\dprod{\vectr}{\vecth})\vectr-\scalr^2\vecth]
  \beqref{alg1}
\end{split}
\end{align*}
\begin{align}\label{gpath20l}
\begin{split}
&=-\vrhoa\ethvo(\cprod{\vectu}{\unitplz})
  -2\kaprep\vrhoa\epsvb\ethvk(\cprod{\vectu}{\vectz})
  -2\kaprep(\scalq/\scalr)\vrhoa\epsvb\ethvk\vecth
  -\vrhoa\ethvp(-\scalr^2\vrhob\vecth)\\
  &\quad+3\vrhoa\vrhob\ethvo(\cprod{\vectr}{\unitplz})
  +6\kaprep\vrhoa\vrhob\epsvb\ethvk(\cprod{\vectr}{\vectz})
  +3\vrhoa\vrhob\ethvp(-\scalr^2\vecth)
  \beqref{main5c}\text{ \& }\eqnref{gpath7b}
\end{split}
\nonumber\\
\begin{split}
&=-\vrhoa\ethvo(\cprod{\vectu}{\unitplz})
  -2\kaprep\vrhoa\epsvb\ethvk(\cprod{\vectu}{\vectz})
  -2\kaprep\vphib\vrhoa\epsvb\ethvk\vecth
  +\vphib\ethvp\vrhob\vecth\\
  &\quad+3\vrhoa\vrhob\ethvo(\cprod{\vectr}{\unitplz})
  +6\kaprep\vrhoa\vrhob\epsvb\ethvk(\cprod{\vectr}{\vectz})
  -3\vphib\vrhob\ethvp\vecth
  \beqref{ogrv1b}\text{ \& }\eqnref{gpath1b}
\end{split}
\nonumber\\
\begin{split}
&=-\vrhoa\ethvo(\cprod{\vectu}{\unitplz})
  -2\kaprep\vrhoa\epsvb\ethvk(\cprod{\vectu}{\vectz})
  +3\vrhoa\vrhob\ethvo(\cprod{\vectr}{\unitplz})
  +6\kaprep\vrhoa\vrhob\epsvb\ethvk(\cprod{\vectr}{\vectz})\\
  &\quad-2\vphib(\kaprep\vrhoa\epsvb\ethvk+\vrhob\ethvp)\vecth
\end{split}
\nonumber\\
\begin{split}
&=-\vrhoa\ethvo(\cprod{\vectu}{\unitplz})
  -2\kaprep\vrhoa\epsvb\ethvk(\cprod{\vectu}{\vectz})
  +3\vrhoa\vrhob\ethvo(\cprod{\vectr}{\unitplz})\\
  &\quad+6\kaprep\vrhoa\vrhob\epsvb\ethvk(\cprod{\vectr}{\vectz})
  -\frkyl\vecth\beqref{gpath1h}
\end{split}
\end{align}
\begin{align*}
\vscrn
&=\cprod{\fdote}{\ffdota}\beqref{kpath2b}\nonumber\\
&=\cprod{[\ethvn\unitplz-\epsvb\vphif\vectz-\scalq\epsvb\vphif\unitpos
  -\kaprep\epsvb\ethvk(\cprod{\vectr}{\vecth})]}{(6\vrhoa\vrhob\vectu-\vrhoa\vrhol\vectr)}
  \beqref{gpath16a}\text{ \& }\eqnref{gpath6b}\nonumber\\
\begin{split}
&=-6\vrhoa\vrhob\ethvn(\cprod{\vectu}{\unitplz})
  +6\vrhoa\vrhob\epsvb\vphif(\cprod{\vectu}{\vectz})
  +6\scalq\vrhoa\vrhob\epsvb\vphif(\cprod{\vectu}{\unitpos})
  +6\kaprep\vrhoa\vrhob\epsvb\ethvk[\cprod{\vectu}{(\cprod{\vectr}{\vecth})}]\\
  &\quad+\vrhoa\vrhol\ethvn(\cprod{\vectr}{\unitplz})
  -\epsvb\vrhoa\vrhol\vphif(\cprod{\vectr}{\vectz})
  -\scalq\epsvb\vrhoa\vrhol\vphif(\cprod{\vectr}{\unitpos})
  -\kaprep\epsvb\vrhoa\vrhol\ethvk[\cprod{\vectr}{(\cprod{\vectr}{\vecth})}]
\end{split}
\nonumber\\
\begin{split}
&=-6\vrhoa\vrhob\ethvn(\cprod{\vectu}{\unitplz})
  +6\vrhoa\vrhob\epsvb\vphif(\cprod{\vectu}{\vectz})
  +6\scalq\vrhoa\vrhob\epsvb\vphif(\cprod{\vectu}{\unitpos})
  +6\kaprep\vrhoa\vrhob\epsvb\ethvk[(\dprod{\vectu}{\vecth})\vectr-(\dprod{\vectu}{\vectr})\vecth]\\
  &\quad+\vrhoa\vrhol\ethvn(\cprod{\vectr}{\unitplz})
  -\epsvb\vrhoa\vrhol\vphif(\cprod{\vectr}{\vectz})
  -\kaprep\epsvb\vrhoa\vrhol\ethvk[(\dprod{\vectr}{\vecth})\vectr-\scalr^2\vecth]
  \beqref{alg1}
\end{split}
\end{align*}
\begin{align}\label{gpath20m}
\begin{split}
&=-6\vrhoa\vrhob\ethvn(\cprod{\vectu}{\unitplz})
  +6\vrhoa\vrhob\epsvb\vphif(\cprod{\vectu}{\vectz})
  +6(\scalq/\scalr)\vrhoa\vrhob\epsvb\vphif\vecth
  +6\kaprep\vrhoa\vrhob\epsvb\ethvk(-\scalr^2\vrhob\vecth)\\
  &\quad+\vrhoa\vrhol\ethvn(\cprod{\vectr}{\unitplz})
  -\epsvb\vrhoa\vrhol\vphif(\cprod{\vectr}{\vectz})
  -\kaprep\epsvb\vrhoa\vrhol\ethvk(-\scalr^2\vecth)
  \beqref{main5c}\text{ \& }\eqnref{gpath7b}
\end{split}
\nonumber\\
\begin{split}
&=-6\vrhoa\vrhob\ethvn(\cprod{\vectu}{\unitplz})
  +6\vrhoa\vrhob\epsvb\vphif(\cprod{\vectu}{\vectz})
  +6\vphib\vrhoa\vrhob\epsvb\vphif\vecth
  -6\kaprep\vphib\vrhob^2\epsvb\ethvk\vecth\\
  &\quad+\vrhoa\vrhol\ethvn(\cprod{\vectr}{\unitplz})
  -\epsvb\vrhoa\vrhol\vphif(\cprod{\vectr}{\vectz})
  +\kaprep\epsvb\vphib\vrhol\ethvk\vecth
  \beqref{ogrv1b}\text{ \& }\eqnref{gpath1b}
\end{split}
\nonumber\\
\begin{split}
&=-6\vrhoa\vrhob\ethvn(\cprod{\vectu}{\unitplz})
  +6\vrhoa\vrhob\epsvb\vphif(\cprod{\vectu}{\vectz})
  +\vrhoa\vrhol\ethvn(\cprod{\vectr}{\unitplz})
  -\epsvb\vrhoa\vrhol\vphif(\cprod{\vectr}{\vectz})\\
  &\quad+\epsvb\vphib[6\vrhoa\vrhob\vphif+\kaprep\ethvk(\vrhol-6\vrhob^2)]\vecth
\end{split}
\nonumber\\
\begin{split}
&=-6\vrhoa\vrhob\ethvn(\cprod{\vectu}{\unitplz})
  +6\vrhoa\vrhob\epsvb\vphif(\cprod{\vectu}{\vectz})
  +\vrhoa\vrhol\ethvn(\cprod{\vectr}{\unitplz})\\
  &\quad-\epsvb\vrhoa\vrhol\vphif(\cprod{\vectr}{\vectz})
  +\frkym\vecth\beqref{gpath1h}
\end{split}
\end{align}
\begin{align*}
\vscro
&=\cprod{\fdote}{\ffdote}\beqref{kpath2b}\nonumber\\
\begin{split}
&=\cprod{[\ethvn\unitplz-\epsvb\vphif\vectz-\scalq\epsvb\vphif\unitpos
  -\kaprep\epsvb\ethvk(\cprod{\vectr}{\vecth})]}{[\ethvo\unitplz+2\kaprep\epsvb\ethvk\vectz+2\kaprep\scalq\epsvb\ethvk\unitpos
   +\ethvp(\cprod{\vectr}{\vecth})]}\beqref{gpath16}
\end{split}
\nonumber\\
\begin{split}
&=2\kaprep\epsvb\ethvk\ethvn(\cprod{\unitplz}{\vectz})
  +2\kaprep\scalq\epsvb\ethvk\ethvn(\cprod{\unitplz}{\unitpos})
  +\ethvp\ethvn[\cprod{\unitplz}{(\cprod{\vectr}{\vecth})}]
  -\ethvo\epsvb\vphif(\cprod{\vectz}{\unitplz})\\
  &\quad-2\kaprep\scalq\epsvb^2\ethvk\vphif(\cprod{\vectz}{\unitpos})
  -\ethvp\epsvb\vphif[\cprod{\vectz}{(\cprod{\vectr}{\vecth})}]
  -\ethvo\scalq\epsvb\vphif(\cprod{\unitpos}{\unitplz})\\
  &\quad-2\kaprep\epsvb^2\ethvk\scalq\vphif(\cprod{\unitpos}{\vectz})
  -\ethvp\scalq\epsvb\vphif[\cprod{\unitpos}{(\cprod{\vectr}{\vecth})}]
  +\kaprep\epsvb\ethvk\ethvo[\cprod{\unitplz}{(\cprod{\vectr}{\vecth})}]\\
  &\quad+2\kaprep^2\epsvb^2\ethvk^2[\cprod{\vectz}{(\cprod{\vectr}{\vecth})}]
  +2\scalq\kaprep^2\epsvb^2\ethvk^2[\cprod{\unitpos}{(\cprod{\vectr}{\vecth})}]
\end{split}
\nonumber\\
\begin{split}
&=\epsvb(2\kaprep\ethvk\ethvn+\ethvo\vphif)(\cprod{\unitplz}{\vectz})
  +\scalq\epsvb(2\kaprep\ethvk\ethvn+\ethvo\vphif)(\cprod{\unitplz}{\unitpos})
  +(\ethvp\ethvn+\kaprep\epsvb\ethvk\ethvo)[\cprod{\unitplz}{(\cprod{\vectr}{\vecth})}]\\
  &\quad+\epsvb(2\kaprep^2\epsvb\ethvk^2-\ethvp\vphif)[\cprod{\vectz}{(\cprod{\vectr}{\vecth})}]
  +\scalq\epsvb(2\kaprep^2\epsvb\ethvk^2-\ethvp\vphif)[\cprod{\unitpos}{(\cprod{\vectr}{\vecth})}]
\end{split}
\end{align*}
\begin{align}\label{gpath20n}
\begin{split}
&=\epsvb\frkyn(\cprod{\unitplz}{\vectz})+\scalq\epsvb\frkyn(\cprod{\unitplz}{\unitpos})
  +\frkyo[(\dprod{\unitplz}{\vecth})\vectr-(\dprod{\unitplz}{\vectr})\vecth]\\
  &\quad+\epsvb\frkyp[(\dprod{\vectz}{\vecth})\vectr-(\dprod{\vectz}{\vectr})\vecth]
  +\scalq\epsvb\frkyp[(\dprod{\unitpos}{\vecth})\vectr-\scalr\vecth]
  \beqref{gpath1h}\text{ \& }\eqnref{alg1}
\end{split}
\nonumber\\
\begin{split}
&=\epsvb\frkyn(\cprod{\unitplz}{\vectz})+\scalq\epsvb\frkyn(\cprod{\unitplz}{\unitpos})
  +\frkyo(\dltvc\vectr-\scalr\epsvc\vecth)+\epsvb\frkyp(-\scalr\epsvd\vecth)\\
  &\quad+\scalq\epsvb\frkyp(-\scalr\vecth)
  \beqref{main5c}, \eqnref{gxpeed1a}\text{ \& }\eqnref{ogrv1a}
\end{split}
\nonumber\\
&=\epsvb\frkyn(\cprod{\unitplz}{\vectz})+\scalq\epsvb\frkyn(\cprod{\unitplz}{\unitpos})
  +\frkyo\dltvc\vectr-\scalr[\epsvc\frkyo+\epsvb\frkyp(\epsvd+\scalq)]\vecth
\nonumber\\
&=\epsvb\frkyn(\cprod{\unitplz}{\vectz})+\scalq\epsvb\frkyn(\cprod{\unitplz}{\unitpos})
  +\frkyo\dltvc\vectr-\scalr(\epsvc\frkyo+\epsvb\frkyp\efktb)\vecth\beqref{gxpeed1b}
  \nonumber\\
&=\epsvb\frkyn(\cprod{\unitplz}{\vectz})+\scalq\epsvb\frkyn(\cprod{\unitplz}{\unitpos})
  +\frkyo\dltvc\vectr-\frkyq\vecth\beqref{gpath1h}.
\end{align}
\end{subequations}

\subart{Development of equation \eqnref{kpath2c}}
From the foregoing derivations, we obtain
\begin{subequations}\label{gpath21}
\begin{align*}
\vscrp
&=\vbbd\vscra+\vbbe\vscrb+\rhorep\fdot{\cdkt}\vscrc-\ffdot{\cdkt}\vscrd+\fdot{\cdkt}\vscre
  +\vbbf\vscrf+\rhorep\vbba\vscrg\beqref{kpath2c}\nonumber\\
&=\frkyg\vscra+\frkyh\vscrb+\rhorep\frkya\vscrc-\frkyb\vscrd+\frkya\vscre
  +\frkyi\vscrf+\rhorep\frkyd\vscrg\beqref{gpath18}\text{ \& }\eqnref{gpath19}\nonumber\\
\begin{split}
&=\frkyg[-\vrhoa(\cprod{\unitkap}{\vectr})]
  +\frkyh[-\vrhoa(\cprod{\unitkap}{\vectu})+3\vrhoa\vrhob(\cprod{\unitkap}{\vectr})]
  +\rhorep\frkya[6\vrhoa\vrhob(\cprod{\unitkap}{\vectu})-\vrhoa\vrhol(\cprod{\unitkap}{\vectr})]\\
  &\quad-\frkyb[\kaprep\scalr\epsva\epsvb\ethvk\vecth-\kaprep\epsvb\ethvk\dltva\vectr+\ethvn(\cprod{\unitkap}{\unitplz})
     -\epsvb\vphif(\cprod{\unitkap}{\vectz})-\scalq\epsvb\vphif(\cprod{\unitkap}{\unitpos})]\\
  &\quad+\frkya[\ethvp\dltva\vectr-\scalr\ethvp\epsva\vecth+\ethvo(\cprod{\unitkap}{\unitplz})
     +2\kaprep\epsvb\ethvk(\cprod{\unitkap}{\vectz})+2\kaprep\scalq\epsvb\ethvk(\cprod{\unitkap}{\unitpos})]\\
  &\quad+\frkyi[-\vrhoa^2\vecth]+\rhorep\frkyd[6\vrhoa^2\vrhob\vecth]\beqref{gpath20}
\end{split}
\nonumber\\
\begin{split}
&=-\frkyg\vrhoa(\cprod{\unitkap}{\vectr})
  -\frkyh\vrhoa(\cprod{\unitkap}{\vectu})+3\frkyh\vrhoa\vrhob(\cprod{\unitkap}{\vectr})
  +6\rhorep\frkya\vrhoa\vrhob(\cprod{\unitkap}{\vectu})-\rhorep\frkya\vrhoa\vrhol(\cprod{\unitkap}{\vectr})\\
  &\quad-\frkyb\kaprep\scalr\epsva\epsvb\ethvk\vecth+\frkyb\kaprep\epsvb\ethvk\dltva\vectr-\frkyb\ethvn(\cprod{\unitkap}{\unitplz})
     +\frkyb\epsvb\vphif(\cprod{\unitkap}{\vectz})+\frkyb\scalq\epsvb\vphif(\cprod{\unitkap}{\unitpos})\\
  &\quad+\frkya\ethvp\dltva\vectr-\frkya\scalr\ethvp\epsva\vecth+\frkya\ethvo(\cprod{\unitkap}{\unitplz})
     +2\frkya\kaprep\epsvb\ethvk(\cprod{\unitkap}{\vectz})+2\frkya\kaprep\scalq\epsvb\ethvk(\cprod{\unitkap}{\unitpos})\\
  &\quad-\frkyi\vrhoa^2\vecth+6\rhorep\frkyd\vrhoa^2\vrhob\vecth
\end{split}
\end{align*}
\begin{align*}
\begin{split}
&=-\frkyg\vrhoa(\cprod{\unitkap}{\vectr})+3\frkyh\vrhoa\vrhob(\cprod{\unitkap}{\vectr})
  -\rhorep\frkya\vrhoa\vrhol(\cprod{\unitkap}{\vectr})+\frkyb\scalq\epsvb\vphif(\cprod{\unitkap}{\unitpos})
  +2\frkya\kaprep\scalq\epsvb\ethvk(\cprod{\unitkap}{\unitpos})\\
  &\quad-\frkyh\vrhoa(\cprod{\unitkap}{\vectu})+6\rhorep\frkya\vrhoa\vrhob(\cprod{\unitkap}{\vectu})
  -\frkyb\ethvn(\cprod{\unitkap}{\unitplz})+\frkya\ethvo(\cprod{\unitkap}{\unitplz})
  +\frkyb\epsvb\vphif(\cprod{\unitkap}{\vectz})\\
  &\quad+2\frkya\kaprep\epsvb\ethvk(\cprod{\unitkap}{\vectz})
  +\frkyb\kaprep\epsvb\ethvk\dltva\vectr+\frkya\ethvp\dltva\vectr
  -\frkyb\kaprep\scalr\epsva\epsvb\ethvk\vecth-\frkya\scalr\ethvp\epsva\vecth\\
  &\quad-\frkyi\vrhoa^2\vecth+6\rhorep\frkyd\vrhoa^2\vrhob\vecth
\end{split}
\nonumber\\
\begin{split}
&=(-\scalr\frkyg\vrhoa+3\scalr\frkyh\vrhoa\vrhob-\scalr\rhorep\frkya\vrhoa\vrhol+\frkyb\scalq\epsvb\vphif
    +2\frkya\kaprep\scalq\epsvb\ethvk)(\cprod{\unitkap}{\unitpos})\\
  &\quad+(-\frkyh\vrhoa+6\rhorep\frkya\vrhoa\vrhob)(\cprod{\unitkap}{\vectu})
  +(-\frkyb\ethvn+\frkya\ethvo)(\cprod{\unitkap}{\unitplz})
  +(\frkyb\epsvb\vphif+2\frkya\kaprep\epsvb\ethvk)(\cprod{\unitkap}{\vectz})\\
  &\quad+(\frkyb\kaprep\epsvb\ethvk\dltva+\frkya\ethvp\dltva)\vectr
  +(-\frkyb\kaprep\scalr\epsva\epsvb\ethvk-\frkya\scalr\ethvp\epsva-\frkyi\vrhoa^2+6\rhorep\frkyd\vrhoa^2\vrhob)\vecth
\end{split}
\nonumber\\
\begin{split}
&=(-\vphia\frkyg+3\vphia\vrhob\frkyh-\rhorep\vphia\vrhol\frkya+\scalq\epsvb\vphif\frkyb
    +2\kaprep\scalq\epsvb\ethvk\frkya)(\cprod{\unitkap}{\unitpos})\\
  &\quad+\vrhoa(6\rhorep\vrhob\frkya-\frkyh)(\cprod{\unitkap}{\vectu})
  +(\ethvo\frkya-\ethvn\frkyb)(\cprod{\unitkap}{\unitplz})
  +\epsvb(\vphif\frkyb+2\kaprep\ethvk\frkya)(\cprod{\unitkap}{\vectz})\\
  &\quad+\dltva(\kaprep\epsvb\ethvk\frkyb+\ethvp\frkya)\vectr
  +[-\scalr\epsva(\kaprep\epsvb\ethvk\frkyb+\ethvp\frkya)-\vrhoa^2(\frkyi-6\rhorep\vrhob\frkyd)]\vecth
  \beqref{ogrv1b}\text{ \& }\eqnref{gpath1b}
\end{split}
\end{align*}
\begin{align}\label{gpath21a}
\begin{split}
&=[\vphia(3\vrhob\frkyh-\rhorep\vrhol\frkya-\frkyg)+\scalq\epsvb(\vphif\frkyb+2\kaprep\ethvk\frkya)]
   (\cprod{\unitkap}{\unitpos})\\
  &\quad-\vrhoa(\frkyh-6\rhorep\vrhob\frkya)(\cprod{\unitkap}{\vectu})
  +(\ethvo\frkya-\ethvn\frkyb)(\cprod{\unitkap}{\unitplz})
  +\epsvb(\vphif\frkyb+2\kaprep\ethvk\frkya)(\cprod{\unitkap}{\vectz})\\
  &\quad+\dltva(\kaprep\epsvb\ethvk\frkyb+\ethvp\frkya)\vectr
  -[\scalr\epsva(\kaprep\epsvb\ethvk\frkyb+\ethvp\frkya)+\vrhoa^2(\frkyi-6\rhorep\vrhob\frkyd)]\vecth
\end{split}
\nonumber\\
\begin{split}
&=(\vphia\frkyw+\scalq\epsvb\frkyr)(\cprod{\unitkap}{\unitpos})-\vrhoa\frkys(\cprod{\unitkap}{\vectu})
  +\frkyu(\cprod{\unitkap}{\unitplz})+\epsvb\frkyr(\cprod{\unitkap}{\vectz})\\
  &\quad+\dltva\frkyv\vectr-(\scalr\epsva\frkyv+\vrhoa^2\frkyt)\vecth
  \beqref{gpath1h}
\end{split}
\nonumber\\
\begin{split}
&=\frkyx(\cprod{\unitkap}{\unitpos})-\vrhoa\frkys(\cprod{\unitkap}{\vectu})
  +\frkyu(\cprod{\unitkap}{\unitplz})+\epsvb\frkyr(\cprod{\unitkap}{\vectz})
  +\dltva\frkyv\vectr-\frkyy\vecth\beqref{gpath1h}
\end{split}
\end{align}
\begin{align*}
\vscrq
&=-\ffdot{\rhorep}\vscrh+\vbba\vscri+\rhorep^2\vscrj-\vbbb\vscrk+\rhorep\vscrl
  +\rhorep\vscrn+\vscro\beqref{kpath2c}\nonumber\\
&=-\ethvi\vscrh+\frkyd\vscri+\rhorep^2\vscrj-\frkye\vscrk+\rhorep\vscrl
  +\rhorep\vscrn+\vscro\beqref{gpath12b}\text{ \& }\eqnref{gpath19}\nonumber\\
\begin{split}
&=-\ethvi[-\vrhoa\ethvn(\cprod{\vectr}{\unitplz})+\vrhoa\epsvb\vphif(\cprod{\vectr}{\vectz})
    -\kaprep\vphib\epsvb\ethvk\vecth]
  +\frkyd[-\vrhoa\ethvo(\cprod{\vectr}{\unitplz})-2\kaprep\vrhoa\epsvb\ethvk(\cprod{\vectr}{\vectz})
     +\vphib\ethvp\vecth]\\
  &\quad+\rhorep^2[\frkyj\vecth]
  -\frkye[-\vrhoa\ethvn(\cprod{\vectu}{\unitplz})+\vrhoa\epsvb\vphif(\cprod{\vectu}{\vectz})
    +3\vrhoa\vrhob\ethvn(\cprod{\vectr}{\unitplz})-3\vrhoa\vrhob\epsvb\vphif(\cprod{\vectr}{\vectz})+\frkyk\vecth]\\
  &\quad+\rhorep[-\vrhoa\ethvo(\cprod{\vectu}{\unitplz})-2\kaprep\vrhoa\epsvb\ethvk(\cprod{\vectu}{\vectz})
    +3\vrhoa\vrhob\ethvo(\cprod{\vectr}{\unitplz})+6\kaprep\vrhoa\vrhob\epsvb\ethvk(\cprod{\vectr}{\vectz})-\frkyl\vecth]\\
  &\quad+\rhorep[-6\vrhoa\vrhob\ethvn(\cprod{\vectu}{\unitplz})+6\vrhoa\vrhob\epsvb\vphif(\cprod{\vectu}{\vectz})
    +\vrhoa\vrhol\ethvn(\cprod{\vectr}{\unitplz})-\epsvb\vrhoa\vrhol\vphif(\cprod{\vectr}{\vectz})+\frkym\vecth]\\
  &\quad+[\epsvb\frkyn(\cprod{\unitplz}{\vectz})+\scalq\epsvb\frkyn(\cprod{\unitplz}{\unitpos})
    +\frkyo\dltvc\vectr-\frkyq\vecth]
  \beqref{gpath20}
\end{split}
\end{align*}
\begin{align*}
\begin{split}
&=\vrhoa\ethvn\ethvi(\cprod{\vectr}{\unitplz})-\vrhoa\epsvb\vphif\ethvi(\cprod{\vectr}{\vectz})
    +\kaprep\vphib\epsvb\ethvk\ethvi\vecth
  -\vrhoa\ethvo\frkyd(\cprod{\vectr}{\unitplz})-2\kaprep\vrhoa\epsvb\ethvk\frkyd(\cprod{\vectr}{\vectz})
     +\vphib\ethvp\frkyd\vecth\\
  &\quad+\rhorep^2\frkyj\vecth
  +\vrhoa\ethvn\frkye(\cprod{\vectu}{\unitplz})-\vrhoa\epsvb\vphif\frkye(\cprod{\vectu}{\vectz})
    -3\vrhoa\vrhob\ethvn\frkye(\cprod{\vectr}{\unitplz})+3\vrhoa\vrhob\epsvb\vphif\frkye(\cprod{\vectr}{\vectz})-\frkyk\frkye\vecth\\
  &\quad-\rhorep\vrhoa\ethvo(\cprod{\vectu}{\unitplz})-2\rhorep\kaprep\vrhoa\epsvb\ethvk(\cprod{\vectu}{\vectz})
    +3\rhorep\vrhoa\vrhob\ethvo(\cprod{\vectr}{\unitplz})+6\rhorep\kaprep\vrhoa\vrhob\epsvb\ethvk(\cprod{\vectr}{\vectz})
    -\rhorep\frkyl\vecth\\
  &\quad-6\rhorep\vrhoa\vrhob\ethvn(\cprod{\vectu}{\unitplz})+6\rhorep\vrhoa\vrhob\epsvb\vphif(\cprod{\vectu}{\vectz})
    +\rhorep\vrhoa\vrhol\ethvn(\cprod{\vectr}{\unitplz})-\rhorep\epsvb\vrhoa\vrhol\vphif(\cprod{\vectr}{\vectz})+\rhorep\frkym\vecth\\
  &\quad+\epsvb\frkyn(\cprod{\unitplz}{\vectz})+\scalq\epsvb\frkyn(\cprod{\unitplz}{\unitpos})
    +\frkyo\dltvc\vectr-\frkyq\vecth
\end{split}
\nonumber\\
\begin{split}
&=(\vrhoa\ethvn\ethvi-\vrhoa\ethvo\frkyd-3\vrhoa\vrhob\ethvn\frkye+3\rhorep\vrhoa\vrhob\ethvo
    +\rhorep\vrhoa\vrhol\ethvn)(\cprod{\vectr}{\unitplz})
  -(\scalq/\scalr)\epsvb\frkyn(\cprod{\vectr}{\unitplz})\\
  &\quad+(-\vrhoa\epsvb\vphif\ethvi-2\kaprep\vrhoa\epsvb\ethvk\frkyd+3\vrhoa\vrhob\epsvb\vphif\frkye
    +6\rhorep\kaprep\vrhoa\vrhob\epsvb\ethvk-\rhorep\epsvb\vrhoa\vrhol\vphif)(\cprod{\vectr}{\vectz})\\
  &\quad+(\vrhoa\ethvn\frkye-\rhorep\vrhoa\ethvo-6\rhorep\vrhoa\vrhob\ethvn)(\cprod{\vectu}{\unitplz})
  +(-\vrhoa\epsvb\vphif\frkye-2\rhorep\kaprep\vrhoa\epsvb\ethvk+6\rhorep\vrhoa\vrhob\epsvb\vphif)(\cprod{\vectu}{\vectz})\\
  &\quad+\epsvb\frkyn(\cprod{\unitplz}{\vectz})
  +\kaprep\vphib\epsvb\ethvk\ethvi\vecth+\vphib\ethvp\frkyd\vecth+\rhorep^2\frkyj\vecth
  -\frkyk\frkye\vecth-\rhorep\frkyl\vecth+\rhorep\frkym\vecth-\frkyq\vecth
  +\frkyo\dltvc\vectr
\end{split}
\end{align*}
\begin{align}\label{gpath21b}
\begin{split}
&=[\vrhoa(\ethvn\ethvi-\ethvo\frkyd)-3\vrhoa\vrhob(\ethvn\frkye-\rhorep\ethvo)
    +\rhorep\vrhoa\vrhol\ethvn-\epsvb\vphib\frkyn](\cprod{\vectr}{\unitplz})\\
  &\quad+\vrhoa\epsvb[\vphif(3\vrhob\frkye-\ethvi-\rhorep\vrhol)
    +2\kaprep\ethvk(3\rhorep\vrhob-\frkyd)](\cprod{\vectr}{\vectz})\\
  &\quad+\vrhoa(\ethvn\frkye-\rhorep\ethvo-6\rhorep\vrhob\ethvn)(\cprod{\vectu}{\unitplz})
  +\vrhoa\epsvb(-\vphif\frkye-2\rhorep\kaprep\ethvk+6\rhorep\vrhob\vphif)(\cprod{\vectu}{\vectz})\\
  &\quad+\epsvb\frkyn(\cprod{\unitplz}{\vectz})
  +[\vphib(\kaprep\epsvb\ethvk\ethvi+\ethvp\frkyd)+\rhorep(\rhorep\frkyj
    -\frkyl+\frkym)-\frkyk\frkye-\frkyq]\vecth\\
  &\quad+\frkyo\dltvc\vectr
  \beqref{ogrv1b}
\end{split}
\nonumber\\
\begin{split}
&=\vakpa(\cprod{\vectr}{\unitplz})+\vakpb(\cprod{\vectr}{\vectz})
  +\vakpc(\cprod{\vectu}{\unitplz})+\vakpd(\cprod{\vectu}{\vectz})
  +\epsvb\frkyn(\cprod{\unitplz}{\vectz})\\
  &\quad+\vakpe\vecth+\frkyo\dltvc\vectr\beqref{gpath1i}
\end{split}
\end{align}
\begin{align}\label{gpath21c}
\vscrr
&=\fdot{\cdkt}\unitkap+\vbba\vecta+\rhorep\fdota+\fdote\beqref{kpath2c}\nonumber\\
&=\frkya\unitkap+\frkyd\vecta+\rhorep\fdota+\fdote\beqref{gpath18a}\text{ \& }\eqnref{gpath19a}\nonumber\\
\begin{split}
&=\frkya\unitkap+\frkyd(-\vrhoa\vectr)+\rhorep(-\vrhoa\vectu+3\vrhoa\vrhob\vectr)
  +[\ethvn\unitplz-\epsvb\vphif\vectz-\scalq\epsvb\vphif\unitpos-\kaprep\epsvb\ethvk(\cprod{\vectr}{\vecth})]\\
  &\quad\beqref{main5a}, \eqnref{gpath1b}, \eqnref{gpath6a}\text{ \& }\eqnref{gpath16a}
\end{split}
\nonumber\\
&=\frkya\unitkap-\vrhoa\frkyd\vectr-\rhorep\vrhoa\vectu+3\rhorep\vrhoa\vrhob\vectr
  +\ethvn\unitplz-\epsvb\vphif\vectz-\scalq\epsvb\vphif\unitpos-\kaprep\epsvb\ethvk(\cprod{\vectr}{\vecth})
  \nonumber\\
&=\frkya\unitkap+\ethvn\unitplz-\epsvb\vphif\vectz-\rhorep\vrhoa\vectu
  +[-\vrhoa\frkyd+3\rhorep\vrhoa\vrhob-(\scalq/\scalr)\epsvb\vphif]\vectr
  -\kaprep\epsvb\ethvk(\cprod{\vectr}{\vecth})
  \nonumber\\
&=\frkya\unitkap+\ethvn\unitplz-\epsvb\vphif\vectz-\rhorep\vrhoa\vectu
  -(\vrhoa\frkyd-3\rhorep\vrhoa\vrhob+\vphib\epsvb\vphif)\vectr
  -\kaprep\epsvb\ethvk(\cprod{\vectr}{\vecth})
  \beqref{ogrv1b}\nonumber\\
&=\frkya\unitkap+\ethvn\unitplz-\epsvb\vphif\vectz-\rhorep\vrhoa\vectu-\vakpf\vectr
  -\kaprep\epsvb\ethvk(\cprod{\vectr}{\vecth})
  \beqref{gpath1i}
\end{align}
\begin{align}\label{gpath21d}
\begin{split}
\vscrp+\vscrq
&=\frkyx(\cprod{\unitkap}{\unitpos})-\vrhoa\frkys(\cprod{\unitkap}{\vectu})
  +\frkyu(\cprod{\unitkap}{\unitplz})+\epsvb\frkyr(\cprod{\unitkap}{\vectz})
  +\dltva\frkyv\vectr-\frkyy\vecth\\
  &\quad+\vakpa(\cprod{\vectr}{\unitplz})+\vakpb(\cprod{\vectr}{\vectz})
  +\vakpc(\cprod{\vectu}{\unitplz})+\vakpd(\cprod{\vectu}{\vectz})
  +\epsvb\frkyn(\cprod{\unitplz}{\vectz})\\
  &\quad+\vakpe\vecth+\frkyo\dltvc\vectr\beqref{gpath21a}\text{ \& }\eqnref{gpath21b}
\end{split}
\nonumber\\
\begin{split}
&=(\dltva\frkyv+\dltvc\frkyo)\vectr
  +(\vakpe-\frkyy)\vecth
  +\frkyx(\cprod{\unitkap}{\unitpos})
  +\frkyu(\cprod{\unitkap}{\unitplz})
  +\epsvb\frkyr(\cprod{\unitkap}{\vectz})\\
  &\quad+\epsvb\frkyn(\cprod{\unitplz}{\vectz})
  +\vakpa(\cprod{\vectr}{\unitplz})
  +\vakpb(\cprod{\vectr}{\vectz})
  -\vrhoa\frkys(\cprod{\unitkap}{\vectu})
  -\vakpc(\cprod{\unitplz}{\vectu})
  -\vakpd(\cprod{\vectz}{\vectu})
\end{split}
\nonumber\\
\begin{split}
&=\vakpg\vectr+\vakph\vecth
  +\frkyx(\cprod{\unitkap}{\unitpos})
  +\frkyu(\cprod{\unitkap}{\unitplz})
  +\epsvb\frkyr(\cprod{\unitkap}{\vectz})
  +\epsvb\frkyn(\cprod{\unitplz}{\vectz})
  +\vakpa(\cprod{\vectr}{\unitplz})\\
  &\quad+\vakpb(\cprod{\vectr}{\vectz})
  -\vrhoa\frkys(\cprod{\unitkap}{\vectu})
  -\vakpc(\cprod{\unitplz}{\vectu})
  -\vakpd(\cprod{\vectz}{\vectu})
  \beqref{gpath1i}.
\end{split}
\end{align}
\end{subequations}

\subart{Magnitude of the vector $\vscrr$}
We derive
\begin{align*}
\begin{split}
|\vscrr|^2=(\dprod{\vscrr}{\vscrr})
&=\dprod{[\frkya\unitkap+\ethvn\unitplz-\epsvb\vphif\vectz-\rhorep\vrhoa\vectu-\vakpf\vectr
  -\kaprep\epsvb\ethvk(\cprod{\vectr}{\vecth})]}{}[\frkya\unitkap+\ethvn\unitplz\\
  &\quad-\epsvb\vphif\vectz-\rhorep\vrhoa\vectu-\vakpf\vectr-\kaprep\epsvb\ethvk(\cprod{\vectr}{\vecth})]
  \beqref{gpath21c}
\end{split}
\nonumber\\
\begin{split}
&=\frkya^2(\dprod{\unitkap}{\unitkap})
  +2\ethvn\frkya(\dprod{\unitkap}{\unitplz})
  -2\epsvb\vphif\frkya(\dprod{\unitkap}{\vectz})
  -2\rhorep\vrhoa\frkya(\dprod{\unitkap}{\vectu})\\
  &\quad-2\vakpf\frkya(\dprod{\unitkap}{\vectr})
  -2\kaprep\epsvb\ethvk\frkya[\dprod{\unitkap}{(\cprod{\vectr}{\vecth})}]
  +\ethvn^2(\dprod{\unitplz}{\unitplz})
  -2\epsvb\vphif\ethvn(\dprod{\unitplz}{\vectz})\\
  &\quad-2\rhorep\vrhoa\ethvn(\dprod{\unitplz}{\vectu})
  -2\vakpf\ethvn(\dprod{\unitplz}{\vectr})
  -2\kaprep\epsvb\ethvk\ethvn[\dprod{\unitplz}{(\cprod{\vectr}{\vecth})}]
  +\epsvb^2\vphif^2(\dprod{\vectz}{\vectz})\\
  &\quad+2\rhorep\vrhoa\epsvb\vphif(\dprod{\vectz}{\vectu})
  +2\vakpf\epsvb\vphif(\dprod{\vectz}{\vectr})
  +2\kaprep\epsvb^2\ethvk\vphif[\dprod{\vectz}{(\cprod{\vectr}{\vecth})}]
  +\rhorep^2\vrhoa^2(\dprod{\vectu}{\vectu})\\
  &\quad+2\vakpf\rhorep\vrhoa(\dprod{\vectu}{\vectr})
  +2\kaprep\epsvb\ethvk\rhorep\vrhoa[\dprod{\vectu}{(\cprod{\vectr}{\vecth})}]
  +\vakpf^2(\dprod{\vectr}{\vectr})
  +2\kaprep\epsvb\ethvk\vakpf[\dprod{\vectr}{(\cprod{\vectr}{\vecth})}]\\
  &\quad+\kaprep^2\epsvb^2\ethvk^2[\dprod{(\cprod{\vectr}{\vecth})}{(\cprod{\vectr}{\vecth})}]
\end{split}
\end{align*}
\begin{align}\label{gpath22}
\begin{split}
&=\frkya^2
  +2\ethvn\frkya\epsvb-2\epsvb\vphif\frkya\dltvb-2\rhorep\vrhoa\frkya\vrhoo-2\scalr\vakpf\frkya\epsva
  -2\kaprep\epsvb\ethvk\frkya\epsve+\ethvn^2\\
  &\quad-2\epsvb\vphif\ethvn\dltvd-2\rhorep\vrhoa\ethvn[\scalh^{-2}(\epsvi+\epsvf\vphib)]
  -2\scalr\vakpf\ethvn\epsvc-2\kaprep\epsvb\ethvk\ethvn\epsvf
  +\scalz^2\epsvb^2\vphif^2\\
  &\quad+2\rhorep\vrhoa\epsvb\vphif(-\scalr\vrhob\vphib)+2\scalr\vakpf\epsvb\vphif\epsvd
  -2\scalr\kaprep\epsvb^2\ethvk\vphif\epsvh+\rhorep^2\vrhoa^2\vphih^2\\
  &\quad+2\scalr\vakpf\rhorep\vrhoa(\scalr\vrhob)+2\kaprep\epsvb\ethvk\rhorep\vrhoa(\scalr\efktb)
  +\scalr^2\vakpf^2+\kaprep^2\epsvb^2\ethvk^2[\scalr^2\scalh^2-(\dprod{\vectr}{\vecth})^2]\\
  &\quad\beqref{ogrv1a}, \eqnref{ogrv3b}, \eqnref{gxpeed1a}, \eqnref{gpath7}\text{ \& }\eqnref{alg2}
\end{split}
\nonumber\\
\begin{split}
&=\frkya^2+2\ethvn\frkya\epsvb-2\epsvb\vphif\frkya\dltvb-2\rhorep\vrhoa\frkya\vrhoo-2\scalr\vakpf\frkya\epsva
  -2\kaprep\epsvb\ethvk\frkya\epsve+\ethvn^2\\
  &\quad-2\epsvb\vphif\ethvn\dltvd-2\vrhoa\ethvn(\rhorep/\scalh^2)(\epsvi+\epsvf\vphib)
  -2\scalr\vakpf\ethvn\epsvc-2\kaprep\epsvb\ethvk\ethvn\epsvf
  +\scalz^2\epsvb^2\vphif^2\\
  &\quad-2\rhorep\scalr\vrhob\vphib\vrhoa\epsvb\vphif+2\scalr\vakpf\epsvb\vphif\epsvd
  -2\scalr\kaprep\epsvb^2\ethvk\vphif\epsvh+\rhorep^2\vrhoa^2\vphih^2\\
  &\quad+2\rhorep\scalr^2\vakpf\vrhoa\vrhob+2\kaprep\scalr\efktb\epsvb\ethvk\rhorep\vrhoa
  +\scalr^2\vakpf^2+\kaprep^2\scalr^2\scalh^2\epsvb^2\ethvk^2\beqref{main5c}
\end{split}
\nonumber\\
\therefore|\vscrr|
&=\vakpi\beqref{gpath1j}.
\end{align}

\subart{Magnitude of the vector $\vscrp+\vscrq$}
We derive also
\begin{subequations}\label{gpath23}
\begin{align}\label{gpath23a}
&\dprod{\vectr}{(\vscrp+\vscrq)}\nonumber\\
\begin{split}
&=\dprod{\vectr}{[}\vakpg\vectr+\vakph\vecth
  +\frkyx(\cprod{\unitkap}{\unitpos})
  +\frkyu(\cprod{\unitkap}{\unitplz})
  +\epsvb\frkyr(\cprod{\unitkap}{\vectz})
  +\epsvb\frkyn(\cprod{\unitplz}{\vectz})
  +\vakpa(\cprod{\vectr}{\unitplz})\\
  &\quad+\vakpb(\cprod{\vectr}{\vectz})
  -\vrhoa\frkys(\cprod{\unitkap}{\vectu})
  -\vakpc(\cprod{\unitplz}{\vectu})
  -\vakpd(\cprod{\vectz}{\vectu})]\beqref{gpath21d}
\end{split}
\nonumber\\
\begin{split}
&=\vakpg(\dprod{\vectr}{\vectr})
  +\vakph(\dprod{\vectr}{\vecth})
  +\frkyx[\dprod{\vectr}{(\cprod{\unitkap}{\unitpos})}]
  +\frkyu[\dprod{\vectr}{(\cprod{\unitkap}{\unitplz})}]
  +\epsvb\frkyr[\dprod{\vectr}{(\cprod{\unitkap}{\vectz})}]
  +\epsvb\frkyn[\dprod{\vectr}{(\cprod{\unitplz}{\vectz})}]\\
  &\quad+\vakpa[\dprod{\vectr}{(\cprod{\vectr}{\unitplz})}]
  +\vakpb[\dprod{\vectr}{(\cprod{\vectr}{\vectz})}]
  -\vrhoa\frkys[\dprod{\vectr}{(\cprod{\unitkap}{\vectu})}]
  -\vakpc[\dprod{\vectr}{(\cprod{\unitplz}{\vectu})}]
  -\vakpd[\dprod{\vectr}{(\cprod{\vectz}{\vectu})}]
\end{split}
\nonumber\\
\begin{split}
&=\vakpg\scalr^2+\frkyu\vsiga+\scalr\epsvb\frkyr\dltvf+\epsvb\frkyn\vsigb
  -\vrhoa\frkys[\dprod{\unitkap}{(\cprod{\vectu}{\vectr})}]
  -\vakpc[\dprod{\unitplz}{(\cprod{\vectu}{\vectr})}]
  -\vakpd[\dprod{\vectz}{(\cprod{\vectu}{\vectr})}]\\
  &\quad\beqref{main5c}, \eqnref{gpath1a}, \eqnref{gxpeed1a}\text{ \& }\eqnref{alg4}
\end{split}
\nonumber\\
\begin{split}
&=\vakpg\scalr^2+\frkyu\vsiga+\scalr\epsvb\frkyr\dltvf+\epsvb\frkyn\vsigb
  -\vrhoa\frkys(\dprod{\unitkap}{\vecth})-\vakpc(\dprod{\unitplz}{\vecth})-\vakpd(\dprod{\vectz}{\vecth})
  \beqref{main5c}
\end{split}
\nonumber\\
\begin{split}
&=\vakpg\scalr^2+\frkyu\vsiga+\scalr\epsvb\frkyr\dltvf+\epsvb\frkyn\vsigb
  -\vrhoa\frkys\dltva-\vakpc\dltvc\beqref{gxpeed1a}
\end{split}
\nonumber\\
&=\vakpj\beqref{gpath1j}
\end{align}
\begin{align*}
&\dprod{\vecth}{(\vscrp+\vscrq)}\nonumber\\
\begin{split}
&=\dprod{\vecth}{[}\vakpg\vectr+\vakph\vecth
  +\frkyx(\cprod{\unitkap}{\unitpos})
  +\frkyu(\cprod{\unitkap}{\unitplz})
  +\epsvb\frkyr(\cprod{\unitkap}{\vectz})
  +\epsvb\frkyn(\cprod{\unitplz}{\vectz})
  +\vakpa(\cprod{\vectr}{\unitplz})\\
  &\quad+\vakpb(\cprod{\vectr}{\vectz})
  -\vrhoa\frkys(\cprod{\unitkap}{\vectu})
  -\vakpc(\cprod{\unitplz}{\vectu})
  -\vakpd(\cprod{\vectz}{\vectu})]\beqref{gpath21d}
\end{split}
\nonumber\\
\begin{split}
&=\vakpg(\dprod{\vecth}{\vectr})
  +\vakph(\dprod{\vecth}{\vecth})
  +\frkyx[\dprod{\vecth}{(\cprod{\unitkap}{\unitpos})}]
  +\frkyu[\dprod{\vecth}{(\cprod{\unitkap}{\unitplz})}]
  +\epsvb\frkyr[\dprod{\vecth}{(\cprod{\unitkap}{\vectz})}]
  +\epsvb\frkyn[\dprod{\vecth}{(\cprod{\unitplz}{\vectz})}]\\
  &\quad+\vakpa[\dprod{\vecth}{(\cprod{\vectr}{\unitplz})}]
  +\vakpb[\dprod{\vecth}{(\cprod{\vectr}{\vectz})}]
  -\vrhoa\frkys[\dprod{\vecth}{(\cprod{\unitkap}{\vectu})}]
  -\vakpc[\dprod{\vecth}{(\cprod{\unitplz}{\vectu})}]
  -\vakpd[\dprod{\vecth}{(\cprod{\vectz}{\vectu})}]
\end{split}
\nonumber\\
\begin{split}
&=\scalh^2\vakph+(\frkyx/\scalr)\epsve-\frkyu\dltve+\epsvb\frkyr\epsvg
  +\epsvb\frkyn\epsvi-\vakpa\epsvf+\scalr\vakpb\epsvh
  -\vrhoa\frkys[\dprod{\unitkap}{(\cprod{\vectu}{\vecth})}]\\
  &\quad-\vakpc[\dprod{\unitplz}{(\cprod{\vectu}{\vecth})}]
  -\vakpd[\dprod{\vectz}{(\cprod{\vectu}{\vecth})}]
  \beqref{ogrv1a}, \eqnref{gxpeed1a}\text{ \& }\eqnref{alg4}
\end{split}
\end{align*}
\begin{align}\label{gpath23b}
\begin{split}
&=\scalh^2\vakph+(\frkyx/\scalr)\epsve-\frkyu\dltve+\epsvb\frkyr\epsvg
  +\epsvb\frkyn\epsvi-\vakpa\epsvf+\scalr\vakpb\epsvh
  -\vrhoa\frkys[\dprod{\unitkap}{(-\vectz-\scalq\unitpos)}]\\
  &\quad-\vakpc[\dprod{\unitplz}{(-\vectz-\scalq\unitpos)}]
  -\vakpd[\dprod{\vectz}{(-\vectz-\scalq\unitpos)}]
  \beqref{main5c}
\end{split}
\nonumber\\
\begin{split}
&=\scalh^2\vakph+(\frkyx/\scalr)\epsve-\frkyu\dltve+\epsvb\frkyr\epsvg
  +\epsvb\frkyn\epsvi-\vakpa\epsvf+\scalr\vakpb\epsvh
  -\vrhoa\frkys[-(\dprod{\unitkap}{\vectz})-\scalq(\dprod{\unitkap}{\unitpos})]\\
  &\quad-\vakpc[-(\dprod{\unitplz}{\vectz})-\scalq(\dprod{\unitplz}{\unitpos})]
  -\vakpd[-(\dprod{\vectz}{\vectz})-\scalq(\dprod{\vectz}{\unitpos})]
\end{split}
\nonumber\\
\begin{split}
&=\scalh^2\vakph+\epsve(\frkyx/\scalr)-\frkyu\dltve+\epsvb\frkyr\epsvg
  +\epsvb\frkyn\epsvi-\vakpa\epsvf+\scalr\vakpb\epsvh-\vrhoa\frkys(-\dltvb-\scalq\epsva)\\
  &\quad-\vakpc(-\dltvd-\scalq\epsvc)-\vakpd(-\scalz^2-\scalq\epsvd)
  \beqref{ogrv1a}\text{ \& }\eqnref{gxpeed1a}
\end{split}
\nonumber\\
\begin{split}
&=\scalh^2\vakph+\epsve(\frkyx/\scalr)-\frkyu\dltve+\epsvb\frkyr\epsvg+\epsvb\frkyn\epsvi
  -\vakpa\epsvf+\scalr\vakpb\epsvh+\vrhoa\frkys\efkta+\vakpc\efktc+\vakpd\frkyz\\
  &\quad\beqref{gxpeed1b}\text{ \& }\eqnref{gpath1h}
\end{split}
\nonumber\\
&=\vakpk\beqref{gpath1j}
\end{align}
\begin{align*}
&\dprod{(\cprod{\unitkap}{\unitpos})}{(\vscrp+\vscrq)}\nonumber\\
\begin{split}
&=\dprod{(\cprod{\unitkap}{\unitpos})}{[}\vakpg\vectr+\vakph\vecth
  +\frkyx(\cprod{\unitkap}{\unitpos})
  +\frkyu(\cprod{\unitkap}{\unitplz})
  +\epsvb\frkyr(\cprod{\unitkap}{\vectz})
  +\epsvb\frkyn(\cprod{\unitplz}{\vectz})
  +\vakpa(\cprod{\vectr}{\unitplz})\\
  &\quad+\vakpb(\cprod{\vectr}{\vectz})
  -\vrhoa\frkys(\cprod{\unitkap}{\vectu})
  -\vakpc(\cprod{\unitplz}{\vectu})
  -\vakpd(\cprod{\vectz}{\vectu})]\beqref{gpath21d}
\end{split}
\nonumber\\
\begin{split}
&=\vakpg[\dprod{\vectr}{(\cprod{\unitkap}{\unitpos})}]
  +\vakph[\dprod{\vecth}{(\cprod{\unitkap}{\unitpos})}]
  +\frkyx[\dprod{(\cprod{\unitkap}{\unitpos})}{(\cprod{\unitkap}{\unitpos})}]
  +\frkyu[\dprod{(\cprod{\unitkap}{\unitpos})}{(\cprod{\unitkap}{\unitplz})}]\\
  &\quad+\epsvb\frkyr[\dprod{(\cprod{\unitkap}{\unitpos})}{(\cprod{\unitkap}{\vectz})}]
  +\epsvb\frkyn[\dprod{(\cprod{\unitkap}{\unitpos})}{(\cprod{\unitplz}{\vectz})}]
  +\vakpa[\dprod{(\cprod{\unitkap}{\unitpos})}{(\cprod{\vectr}{\unitplz})}]\\
  &\quad+\vakpb[\dprod{(\cprod{\unitkap}{\unitpos})}{(\cprod{\vectr}{\vectz})}]
  -\vrhoa\frkys[\dprod{(\cprod{\unitkap}{\unitpos})}{(\cprod{\unitkap}{\vectu})}]
  -\vakpc[\dprod{(\cprod{\unitkap}{\unitpos})}{(\cprod{\unitplz}{\vectu})}]\\
  &\quad-\vakpd[\dprod{(\cprod{\unitkap}{\unitpos})}{(\cprod{\vectz}{\vectu})}]
\end{split}
\end{align*}
\begin{align}\label{gpath23c}
\begin{split}
&=(\vakph/\scalr)\epsve
  +\frkyx[1-(\dprod{\unitkap}{\unitpos})^2]
  +\frkyu[(\dprod{\unitpos}{\unitplz})-(\dprod{\unitkap}{\unitplz})(\dprod{\unitpos}{\unitkap})]
  +\epsvb\frkyr[(\dprod{\unitpos}{\vectz})-(\dprod{\unitkap}{\vectz})(\dprod{\unitpos}{\unitkap})]\\
  &\quad+\epsvb\frkyn[(\dprod{\unitkap}{\unitplz})(\dprod{\unitpos}{\vectz})-(\dprod{\unitkap}{\vectz})(\dprod{\unitpos}{\unitplz})]
  +\vakpa[(\dprod{\unitkap}{\vectr})(\dprod{\unitpos}{\unitplz})-(\dprod{\unitkap}{\unitplz})(\dprod{\unitpos}{\vectr})]\\
  &\quad+\vakpb[(\dprod{\unitkap}{\vectr})(\dprod{\unitpos}{\vectz})-(\dprod{\unitkap}{\vectz})(\dprod{\unitpos}{\vectr})]
  -\vrhoa\frkys[(\dprod{\unitpos}{\vectu})-(\dprod{\unitkap}{\vectu})(\dprod{\unitpos}{\unitkap})]\\
  &\quad-\vakpc[(\dprod{\unitkap}{\unitplz})(\dprod{\unitpos}{\vectu})-(\dprod{\unitkap}{\vectu})(\dprod{\unitpos}{\unitplz})]
  -\vakpd[(\dprod{\unitkap}{\vectz})(\dprod{\unitpos}{\vectu})-(\dprod{\unitkap}{\vectu})(\dprod{\unitpos}{\vectz})]\\
  &\quad\beqref{ogrv1a}\text{ \& }\eqnref{alg2}
\end{split}
\nonumber\\
\begin{split}
&=(\vakph/\scalr)\epsve+\frkyx(1-\epsva^2)+\frkyu(\epsvc-\epsvb\epsva)
  +\epsvb\frkyr(\epsvd-\dltvb\epsva)+\epsvb\frkyn(\epsvb\epsvd-\dltvb\epsvc)\\
  &\quad+\vakpa(\scalr\epsva\epsvc-\scalr\epsvb)+\vakpb(\scalr\epsva\epsvd-\scalr\dltvb)
  -\vrhoa\frkys(\scalr\vrhob-\vrhoo\epsva)-\vakpc(\scalr\epsvb\vrhob-\vrhoo\epsvc)\\
  &\quad-\vakpd(\scalr\dltvb\vrhob-\vrhoo\epsvd)
  \beqref{ogrv1a}, \eqnref{gxpeed1a}\text{ \& }\eqnref{gpath7}
\end{split}
\nonumber\\
&=\vakpl\beqref{gpath1j}
\end{align}
\begin{align*}
&\dprod{(\cprod{\unitkap}{\unitplz})}{(\vscrp+\vscrq)}\nonumber\\
\begin{split}
&=\dprod{(\cprod{\unitkap}{\unitplz})}{[}\vakpg\vectr+\vakph\vecth
  +\frkyx(\cprod{\unitkap}{\unitpos})
  +\frkyu(\cprod{\unitkap}{\unitplz})
  +\epsvb\frkyr(\cprod{\unitkap}{\vectz})
  +\epsvb\frkyn(\cprod{\unitplz}{\vectz})
  +\vakpa(\cprod{\vectr}{\unitplz})\\
  &\quad+\vakpb(\cprod{\vectr}{\vectz})
  -\vrhoa\frkys(\cprod{\unitkap}{\vectu})
  -\vakpc(\cprod{\unitplz}{\vectu})
  -\vakpd(\cprod{\vectz}{\vectu})]\beqref{gpath21d}
\end{split}
\nonumber\\
\begin{split}
&=\vakpg[\dprod{\vectr}{(\cprod{\unitkap}{\unitplz})}]
  +\vakph[\dprod{\vecth}{(\cprod{\unitkap}{\unitplz})}]
  +\frkyx[\dprod{(\cprod{\unitkap}{\unitplz})}{(\cprod{\unitkap}{\unitpos})}]
  +\frkyu[\dprod{(\cprod{\unitkap}{\unitplz})}{(\cprod{\unitkap}{\unitplz})}]\\
  &\quad+\epsvb\frkyr[\dprod{(\cprod{\unitkap}{\unitplz})}{(\cprod{\unitkap}{\vectz})}]
  +\epsvb\frkyn[\dprod{(\cprod{\unitkap}{\unitplz})}{(\cprod{\unitplz}{\vectz})}]
  +\vakpa[\dprod{(\cprod{\unitkap}{\unitplz})}{(\cprod{\vectr}{\unitplz})}]\\
  &\quad+\vakpb[\dprod{(\cprod{\unitkap}{\unitplz})}{(\cprod{\vectr}{\vectz})}]
  -\vrhoa\frkys[\dprod{(\cprod{\unitkap}{\unitplz})}{(\cprod{\unitkap}{\vectu})}]
  -\vakpc[\dprod{(\cprod{\unitkap}{\unitplz})}{(\cprod{\unitplz}{\vectu})}]\\
  &\quad-\vakpd[\dprod{(\cprod{\unitkap}{\unitplz})}{(\cprod{\vectz}{\vectu})}]
\end{split}
\end{align*}
\begin{align*}
\begin{split}
&=\vakpg\vsiga-\vakph\dltve
  +\frkyx[(\dprod{\unitplz}{\unitpos})-(\dprod{\unitkap}{\unitpos})(\dprod{\unitplz}{\unitkap})]
  +\frkyu[1-(\dprod{\unitkap}{\unitplz})^2]\\
  &\quad+\epsvb\frkyr[(\dprod{\unitplz}{\vectz})-(\dprod{\unitkap}{\vectz})(\dprod{\unitplz}{\unitkap})]
  +\epsvb\frkyn[(\dprod{\unitkap}{\unitplz})(\dprod{\unitplz}{\vectz})-(\dprod{\unitkap}{\vectz})]
  +\vakpa[(\dprod{\unitkap}{\vectr})-(\dprod{\unitkap}{\unitplz})(\dprod{\unitplz}{\vectr})]\\
  &\quad+\vakpb[(\dprod{\unitkap}{\vectr})(\dprod{\unitplz}{\vectz})-(\dprod{\unitkap}{\vectz})(\dprod{\unitplz}{\vectr})]
  -\vrhoa\frkys[(\dprod{\unitplz}{\vectu})-(\dprod{\unitkap}{\vectu})(\dprod{\unitplz}{\unitkap})]
  -\vakpc[(\dprod{\unitkap}{\unitplz})(\dprod{\unitplz}{\vectu})-(\dprod{\unitkap}{\vectu})]\\
  &\quad-\vakpd[(\dprod{\unitkap}{\vectz})(\dprod{\unitplz}{\vectu})-(\dprod{\unitkap}{\vectu})(\dprod{\unitplz}{\vectz})]
  \beqref{gpath1a}, \eqnref{gxpeed1a}\text{ \& }\eqnref{alg2}
\end{split}
\nonumber\\
\begin{split}
&=\vakpg\vsiga-\vakph\dltve+\frkyx(\epsvc-\epsva\epsvb)+\frkyu(1-\epsvb^2)
  +\epsvb\frkyr(\dltvd-\dltvb\epsvb)+\epsvb\frkyn(\epsvb\dltvd-\dltvb)
  +\vakpa(\scalr\epsva-\scalr\epsvb\epsvc)\\
  &\quad+\vakpb(\scalr\epsva\dltvd-\scalr\dltvb\epsvc)
  -\vrhoa\frkys[(\dprod{\unitplz}{\vectu})-\vrhoo\epsvb]
  -\vakpc[\epsvb(\dprod{\unitplz}{\vectu})-\vrhoo]
  -\vakpd[\dltvb(\dprod{\unitplz}{\vectu})-\vrhoo\dltvd]\\
  &\quad\beqref{ogrv1a}, \eqnref{gxpeed1a}\text{ \& }\eqnref{gpath7}
\end{split}
\end{align*}
\begin{align}\label{gpath23d}
\begin{split}
&=\vakpg\vsiga-\vakph\dltve+\frkyx(\epsvc-\epsva\epsvb)+\frkyu(1-\epsvb^2)
  +\epsvb\frkyr(\dltvd-\dltvb\epsvb)+\epsvb\frkyn(\epsvb\dltvd-\dltvb)\\
  &\quad+\vakpa(\scalr\epsva-\scalr\epsvb\epsvc)
  +\vakpb(\scalr\epsva\dltvd-\scalr\dltvb\epsvc)
  -\vrhoa\frkys(\dprod{\unitplz}{\vectu})+\vrhoa\frkys\vrhoo\epsvb\\
  &\quad-\vakpc\epsvb(\dprod{\unitplz}{\vectu})+\vakpc\vrhoo
  -\vakpd\dltvb(\dprod{\unitplz}{\vectu})+\vakpd\vrhoo\dltvd
\end{split}
\nonumber\\
\begin{split}
&=\vakpg\vsiga-\vakph\dltve+\frkyx(\epsvc-\epsva\epsvb)+\frkyu(1-\epsvb^2)
  +\epsvb\frkyr(\dltvd-\dltvb\epsvb)+\epsvb\frkyn(\epsvb\dltvd-\dltvb)\\
  &\quad+\scalr\vakpa(\epsva-\epsvb\epsvc)
  +\scalr\vakpb(\epsva\dltvd-\dltvb\epsvc)
  +\vrhoa\frkys\vrhoo\epsvb+\vakpc\vrhoo+\vakpd\vrhoo\dltvd\\
  &\quad-(\vrhoa\frkys+\vakpc\epsvb+\vakpd\dltvb)(\dprod{\unitplz}{\vectu})
\end{split}
\nonumber\\
\begin{split}
&=\vakpg\vsiga-\vakph\dltve+\frkyx(\epsvc-\epsva\epsvb)+\frkyu(1-\epsvb^2)
  +\epsvb\frkyr(\dltvd-\dltvb\epsvb)+\epsvb\frkyn(\epsvb\dltvd-\dltvb)\\
  &\quad+\scalr\vakpa(\epsva-\epsvb\epsvc)
  +\scalr\vakpb(\epsva\dltvd-\dltvb\epsvc)
  +\vrhoa\frkys\vrhoo\epsvb+\vakpc\vrhoo+\vakpd\vrhoo\dltvd\\
  &\quad-\scalh^{-2}(\vrhoa\frkys+\vakpc\epsvb+\vakpd\dltvb)(\epsvi+\epsvf\vphib)
  \beqref{gpath7c}
\end{split}
\nonumber\\
&=\vakpm\beqref{gpath1k}
\end{align}
\begin{align*}
&\dprod{(\cprod{\unitkap}{\vectz})}{(\vscrp+\vscrq)}\nonumber\\
\begin{split}
&=\dprod{(\cprod{\unitkap}{\vectz})}{[}\vakpg\vectr+\vakph\vecth
  +\frkyx(\cprod{\unitkap}{\unitpos})
  +\frkyu(\cprod{\unitkap}{\unitplz})
  +\epsvb\frkyr(\cprod{\unitkap}{\vectz})
  +\epsvb\frkyn(\cprod{\unitplz}{\vectz})
  +\vakpa(\cprod{\vectr}{\unitplz})\\
  &\quad+\vakpb(\cprod{\vectr}{\vectz})
  -\vrhoa\frkys(\cprod{\unitkap}{\vectu})
  -\vakpc(\cprod{\unitplz}{\vectu})
  -\vakpd(\cprod{\vectz}{\vectu})]\beqref{gpath21d}
\end{split}
\nonumber\\
\begin{split}
&=\vakpg[\dprod{\vectr}{(\cprod{\unitkap}{\vectz})}]
  +\vakph[\dprod{\vecth}{(\cprod{\unitkap}{\vectz})}]
  +\frkyx[\dprod{(\cprod{\unitkap}{\vectz})}{(\cprod{\unitkap}{\unitpos})}]
  +\frkyu[\dprod{(\cprod{\unitkap}{\vectz})}{(\cprod{\unitkap}{\unitplz})}]\\
  &\quad+\epsvb\frkyr[\dprod{(\cprod{\unitkap}{\vectz})}{(\cprod{\unitkap}{\vectz})}]
  +\epsvb\frkyn[\dprod{(\cprod{\unitkap}{\vectz})}{(\cprod{\unitplz}{\vectz})}]
  +\vakpa[\dprod{(\cprod{\unitkap}{\vectz})}{(\cprod{\vectr}{\unitplz})}]\\
  &\quad+\vakpb[\dprod{(\cprod{\unitkap}{\vectz})}{(\cprod{\vectr}{\vectz})}]
  -\vrhoa\frkys[\dprod{(\cprod{\unitkap}{\vectz})}{(\cprod{\unitkap}{\vectu})}]
  -\vakpc[\dprod{(\cprod{\unitkap}{\vectz})}{(\cprod{\unitplz}{\vectu})}]\\
  &\quad-\vakpd[\dprod{(\cprod{\unitkap}{\vectz})}{(\cprod{\vectz}{\vectu})}]
\end{split}
\end{align*}
\begin{align}\label{gpath23e}
\begin{split}
&=\scalr\vakpg\dltvf+\vakph\epsvg
  +\frkyx[(\dprod{\vectz}{\unitpos})-(\dprod{\unitkap}{\unitpos})(\dprod{\vectz}{\unitkap})]
  +\frkyu[(\dprod{\vectz}{\unitplz})-(\dprod{\unitkap}{\unitplz})(\dprod{\vectz}{\unitkap})]\\
  &\quad+\epsvb\frkyr[\scalz^2-(\dprod{\unitkap}{\vectz})^2]
  +\epsvb\frkyn[(\dprod{\unitkap}{\unitplz})\scalz^2-(\dprod{\unitkap}{\vectz})(\dprod{\vectz}{\unitplz})]
  +\vakpa[(\dprod{\unitkap}{\vectr})(\dprod{\vectz}{\unitplz})-(\dprod{\unitkap}{\unitplz})(\dprod{\vectz}{\vectr})]\\
  &\quad+\vakpb[(\dprod{\unitkap}{\vectr})\scalz^2-(\dprod{\unitkap}{\vectz})(\dprod{\vectz}{\vectr})]
  -\vrhoa\frkys[(\dprod{\vectz}{\vectu})-(\dprod{\unitkap}{\vectu})(\dprod{\vectz}{\unitkap})]
  -\vakpc[(\dprod{\unitkap}{\unitplz})(\dprod{\vectz}{\vectu})-(\dprod{\unitkap}{\vectu})(\dprod{\vectz}{\unitplz})]\\
  &\quad-\vakpd[(\dprod{\unitkap}{\vectz})(\dprod{\vectz}{\vectu})-(\dprod{\unitkap}{\vectu})\scalz^2]
  \beqref{ogrv1a}, \eqnref{gxpeed1a}\text{ \& }\eqnref{alg2}
\end{split}
\nonumber\\
\begin{split}
&=\scalr\vakpg\dltvf+\vakph\epsvg+\frkyx(\epsvd-\epsva\dltvb)+\frkyu(\dltvd-\epsvb\dltvb)
  +\epsvb\frkyr(\scalz^2-\dltvb^2)+\epsvb\frkyn(\epsvb\scalz^2-\dltvb\dltvd)\\
  &\quad+\vakpa(\scalr\epsva\dltvd-\scalr\epsvb\epsvd)
  +\vakpb(\scalr\epsva\scalz^2-\scalr\dltvb\epsvd)
  -\vrhoa\frkys(-\scalr\vrhob\vphib-\vrhoo\dltvb)
  -\vakpc[\epsvb(-\scalr\vrhob\vphib)-\vrhoo\dltvd]\\
  &\quad-\vakpd[\dltvb(-\scalr\vrhob\vphib)-\vrhoo\scalz^2]
  \beqref{ogrv1a}, \eqnref{gxpeed1a}\text{ \& }\eqnref{gpath7}
\end{split}
\nonumber\\
\begin{split}
&=\scalr\vakpg\dltvf+\vakph\epsvg+\frkyx(\epsvd-\epsva\dltvb)+\frkyu(\dltvd-\epsvb\dltvb)
  +\epsvb\frkyr(\scalz^2-\dltvb^2)+\epsvb\frkyn(\epsvb\scalz^2-\dltvb\dltvd)\\
  &\quad+\scalr\vakpa(\epsva\dltvd-\epsvb\epsvd)+\scalr\vakpb(\epsva\scalz^2-\dltvb\epsvd)
  +\vrhoa\frkys(\scalr\vrhob\vphib+\vrhoo\dltvb)+\vakpc(\scalr\epsvb\vrhob\vphib+\vrhoo\dltvd)\\
  &\quad+\vakpd(\scalr\vrhob\vphib\dltvb+\vrhoo\scalz^2)
=\vakpn\beqref{gpath1k}
\end{split}
\end{align}
\begin{align*}
&\dprod{(\cprod{\unitplz}{\vectz})}{(\vscrp+\vscrq)}\nonumber\\
\begin{split}
&=\dprod{(\cprod{\unitplz}{\vectz})}{[}\vakpg\vectr+\vakph\vecth
  +\frkyx(\cprod{\unitkap}{\unitpos})
  +\frkyu(\cprod{\unitkap}{\unitplz})
  +\epsvb\frkyr(\cprod{\unitkap}{\vectz})
  +\epsvb\frkyn(\cprod{\unitplz}{\vectz})
  +\vakpa(\cprod{\vectr}{\unitplz})\\
  &\quad+\vakpb(\cprod{\vectr}{\vectz})
  -\vrhoa\frkys(\cprod{\unitkap}{\vectu})
  -\vakpc(\cprod{\unitplz}{\vectu})
  -\vakpd(\cprod{\vectz}{\vectu})]\beqref{gpath21d}
\end{split}
\nonumber\\
\begin{split}
&=\vakpg[\dprod{\vectr}{(\cprod{\unitplz}{\vectz})}]
  +\vakph[\dprod{\vecth}{(\cprod{\unitplz}{\vectz})}]
  +\frkyx[\dprod{(\cprod{\unitplz}{\vectz})}{(\cprod{\unitkap}{\unitpos})}]
  +\frkyu[\dprod{(\cprod{\unitplz}{\vectz})}{(\cprod{\unitkap}{\unitplz})}]\\
  &\quad+\epsvb\frkyr[\dprod{(\cprod{\unitplz}{\vectz})}{(\cprod{\unitkap}{\vectz})}]
  +\epsvb\frkyn[\dprod{(\cprod{\unitplz}{\vectz})}{(\cprod{\unitplz}{\vectz})}]
  +\vakpa[\dprod{(\cprod{\unitplz}{\vectz})}{(\cprod{\vectr}{\unitplz})}]\\
  &\quad+\vakpb[\dprod{(\cprod{\unitplz}{\vectz})}{(\cprod{\vectr}{\vectz})}]
  -\vrhoa\frkys[\dprod{(\cprod{\unitplz}{\vectz})}{(\cprod{\unitkap}{\vectu})}]
  -\vakpc[\dprod{(\cprod{\unitplz}{\vectz})}{(\cprod{\unitplz}{\vectu})}]\\
  &\quad-\vakpd[\dprod{(\cprod{\unitplz}{\vectz})}{(\cprod{\vectz}{\vectu})}]
\end{split}
\end{align*}
\begin{align*}
\begin{split}
&=\vakpg\vsigb+\vakph\epsvi
  +\frkyx[(\dprod{\unitplz}{\unitkap})(\dprod{\vectz}{\unitpos})-(\dprod{\unitplz}{\unitpos})(\dprod{\vectz}{\unitkap})]
  +\frkyu[(\dprod{\unitplz}{\unitkap})(\dprod{\vectz}{\unitplz})-(\dprod{\vectz}{\unitkap})]\\
  &\quad+\epsvb\frkyr[(\dprod{\unitplz}{\unitkap})\scalz^2-(\dprod{\unitplz}{\vectz})(\dprod{\vectz}{\unitkap})]
  +\epsvb\frkyn[\scalz^2-(\dprod{\unitplz}{\vectz})^2]
  +\vakpa[(\dprod{\unitplz}{\vectr})(\dprod{\vectz}{\unitplz})-(\dprod{\vectz}{\vectr})]\\
  &\quad+\vakpb[(\dprod{\unitplz}{\vectr})\scalz^2-(\dprod{\unitplz}{\vectz})(\dprod{\vectz}{\vectr})]
  -\vrhoa\frkys[(\dprod{\unitplz}{\unitkap})(\dprod{\vectz}{\vectu})-(\dprod{\unitplz}{\vectu})(\dprod{\vectz}{\unitkap})]\\
  &\quad-\vakpc[(\dprod{\vectz}{\vectu})-(\dprod{\unitplz}{\vectu})(\dprod{\vectz}{\unitplz})]
  -\vakpd[(\dprod{\unitplz}{\vectz})(\dprod{\vectz}{\vectu})-(\dprod{\unitplz}{\vectu})\scalz^2]
  \beqref{gpath1a}, \eqnref{ogrv1a}\text{ \& }\eqnref{alg2}
\end{split}
\nonumber\\
\begin{split}
&=\vakpg\vsigb+\vakph\epsvi+\frkyx(\epsvb\epsvd-\epsvc\dltvb)+\frkyu(\epsvb\dltvd-\dltvb)\\
  &\quad+\epsvb\frkyr(\epsvb\scalz^2-\dltvd\dltvb)+\epsvb\frkyn(\scalz^2-\dltvd^2)
  +\vakpa(\scalr\epsvc\dltvd-\scalr\epsvd)\\
  &\quad+\vakpb(\scalr\epsvc\scalz^2-\scalr\dltvd\epsvd)
  -\vrhoa\frkys[\epsvb(-\scalr\vrhob\vphib)-(\dprod{\unitplz}{\vectu})\dltvb]
  -\vakpc[(-\scalr\vrhob\vphib)-(\dprod{\unitplz}{\vectu})\dltvd]\\
  &\quad-\vakpd[\dltvd(-\scalr\vrhob\vphib)-(\dprod{\unitplz}{\vectu})\scalz^2]
  \beqref{ogrv1a}, \eqnref{gxpeed1a}\text{ \& }\eqnref{gpath7}
\end{split}
\end{align*}
\begin{align}\label{gpath23f}
\begin{split}
&=\vakpg\vsigb+\vakph\epsvi+\frkyx(\epsvb\epsvd-\epsvc\dltvb)+\frkyu(\epsvb\dltvd-\dltvb)
  +\epsvb\frkyr(\epsvb\scalz^2-\dltvd\dltvb)+\epsvb\frkyn(\scalz^2-\dltvd^2)\\
  &\quad+\scalr\vakpa(\epsvc\dltvd-\epsvd)
  +\scalr\vakpb(\epsvc\scalz^2-\dltvd\epsvd)
  +\vrhoa\frkys\epsvb(\scalr\vrhob\vphib)+\vrhoa\frkys\dltvb(\dprod{\unitplz}{\vectu})
  +\vakpc\scalr\vrhob\vphib\\
  &\quad+\vakpc\dltvd(\dprod{\unitplz}{\vectu})
  +\vakpd\dltvd(\scalr\vrhob\vphib)+\vakpd\scalz^2(\dprod{\unitplz}{\vectu})
\end{split}
\nonumber\\
\begin{split}
&=\vakpg\vsigb+\vakph\epsvi+\frkyx(\epsvb\epsvd-\epsvc\dltvb)+\frkyu(\epsvb\dltvd-\dltvb)
  +\epsvb\frkyr(\epsvb\scalz^2-\dltvd\dltvb)+\epsvb\frkyn(\scalz^2-\dltvd^2)\\
  &\quad+\scalr\vakpa(\epsvc\dltvd-\epsvd)
  +\scalr\vakpb(\epsvc\scalz^2-\dltvd\epsvd)
  +\scalr\frkys\epsvb\vrhoa\vrhob\vphib+\scalr\vakpc\vrhob\vphib+\scalr\vakpd\dltvd\vrhob\vphib\\
  &\quad+(\vrhoa\frkys\dltvb+\vakpc\dltvd+\vakpd\scalz^2)(\dprod{\unitplz}{\vectu})
\end{split}
\nonumber\\
\begin{split}
&=\vakpg\vsigb+\vakph\epsvi+\frkyx(\epsvb\epsvd-\epsvc\dltvb)+\frkyu(\epsvb\dltvd-\dltvb)
  +\epsvb\frkyr(\epsvb\scalz^2-\dltvd\dltvb)+\epsvb\frkyn(\scalz^2-\dltvd^2)\\
  &\quad+\scalr\vakpa(\epsvc\dltvd-\epsvd)
  +\scalr\vakpb(\epsvc\scalz^2-\dltvd\epsvd)
  +\scalr\frkys\epsvb\vrhoa\vrhob\vphib+\scalr\vakpc\vrhob\vphib+\scalr\vakpd\dltvd\vrhob\vphib\\
  &\quad+\scalh^{-2}(\vrhoa\frkys\dltvb+\vakpc\dltvd+\vakpd\scalz^2)(\epsvi+\epsvf\vphib)\beqref{gpath7c}
\end{split}
\nonumber\\
&=\vakpo\beqref{gpath1k}
\end{align}
\begin{align*}
&\dprod{(\cprod{\vectr}{\unitplz})}{(\vscrp+\vscrq)}\nonumber\\
\begin{split}
&=\dprod{(\cprod{\vectr}{\unitplz})}{[}\vakpg\vectr+\vakph\vecth
  +\frkyx(\cprod{\unitkap}{\unitpos})
  +\frkyu(\cprod{\unitkap}{\unitplz})
  +\epsvb\frkyr(\cprod{\unitkap}{\vectz})
  +\epsvb\frkyn(\cprod{\unitplz}{\vectz})
  +\vakpa(\cprod{\vectr}{\unitplz})\\
  &\quad+\vakpb(\cprod{\vectr}{\vectz})
  -\vrhoa\frkys(\cprod{\unitkap}{\vectu})
  -\vakpc(\cprod{\unitplz}{\vectu})
  -\vakpd(\cprod{\vectz}{\vectu})]\beqref{gpath21d}
\end{split}
\nonumber\\
\begin{split}
&=\vakpg[\dprod{\vectr}{(\cprod{\vectr}{\unitplz})}]
  +\vakph[\dprod{\vecth}{(\cprod{\vectr}{\unitplz})}]
  +\frkyx[\dprod{(\cprod{\vectr}{\unitplz})}{(\cprod{\unitkap}{\unitpos})}]
  +\frkyu[\dprod{(\cprod{\vectr}{\unitplz})}{(\cprod{\unitkap}{\unitplz})}]\\
  &\quad+\epsvb\frkyr[\dprod{(\cprod{\vectr}{\unitplz})}{(\cprod{\unitkap}{\vectz})}]
  +\epsvb\frkyn[\dprod{(\cprod{\vectr}{\unitplz})}{(\cprod{\unitplz}{\vectz})}]
  +\vakpa[\dprod{(\cprod{\vectr}{\unitplz})}{(\cprod{\vectr}{\unitplz})}]\\
  &\quad+\vakpb[\dprod{(\cprod{\vectr}{\unitplz})}{(\cprod{\vectr}{\vectz})}]
  -\vrhoa\frkys[\dprod{(\cprod{\vectr}{\unitplz})}{(\cprod{\unitkap}{\vectu})}]
  -\vakpc[\dprod{(\cprod{\vectr}{\unitplz})}{(\cprod{\unitplz}{\vectu})}]\\
  &\quad-\vakpd[\dprod{(\cprod{\vectr}{\unitplz})}{(\cprod{\vectz}{\vectu})}]
\end{split}
\end{align*}
\begin{align*}
\begin{split}
&=-\vakph\epsvf
  +\frkyx[(\dprod{\vectr}{\unitkap})(\dprod{\unitplz}{\unitpos})-(\dprod{\vectr}{\unitpos})(\dprod{\unitplz}{\unitkap})]
  +\frkyu[(\dprod{\vectr}{\unitkap})-(\dprod{\vectr}{\unitplz})(\dprod{\unitplz}{\unitkap})]\\
  &\quad+\epsvb\frkyr[(\dprod{\vectr}{\unitkap})(\dprod{\unitplz}{\vectz})-(\dprod{\vectr}{\vectz})(\dprod{\unitplz}{\unitkap})]
  +\epsvb\frkyn[(\dprod{\vectr}{\unitplz})(\dprod{\unitplz}{\vectz})-(\dprod{\vectr}{\vectz})]
  +\vakpa[\scalr^2-(\dprod{\vectr}{\unitplz})^2]\\
  &\quad+\vakpb[\scalr^2(\dprod{\unitplz}{\vectz})-(\dprod{\vectr}{\vectz})(\dprod{\unitplz}{\vectr})]
  -\vrhoa\frkys[(\dprod{\vectr}{\unitkap})(\dprod{\unitplz}{\vectu})-(\dprod{\vectr}{\vectu})(\dprod{\unitplz}{\unitkap})]\\
  &\quad-\vakpc[(\dprod{\vectr}{\unitplz})(\dprod{\unitplz}{\vectu})-(\dprod{\vectr}{\vectu})]
  -\vakpd[(\dprod{\vectr}{\vectz})(\dprod{\unitplz}{\vectu})-(\dprod{\vectr}{\vectu})(\dprod{\unitplz}{\vectz})]
  \beqref{ogrv1a}\text{ \& }\eqnref{alg2}
\end{split}
\nonumber\\
\begin{split}
&=-\vakph\epsvf+\frkyx(\scalr\epsva\epsvc-\scalr\epsvb)+\frkyu(\scalr\epsva-\scalr\epsvc\epsvb)
  +\epsvb\frkyr(\scalr\epsva\dltvd-\scalr\epsvd\epsvb)
  +\epsvb\frkyn(\scalr\epsvc\dltvd-\scalr\epsvd)\\
  &\quad+\vakpa(\scalr^2-\scalr^2\epsvc^2)
  +\vakpb(\scalr^2\dltvd-\scalr^2\epsvd\epsvc)
  -\vrhoa\frkys[\scalr\epsva(\dprod{\unitplz}{\vectu})-\scalr(\scalr\vrhob)\epsvb]
  -\vakpc[\scalr\epsvc(\dprod{\unitplz}{\vectu})-\scalr(\scalr\vrhob)]\\
  &\quad-\vakpd[\scalr\epsvd(\dprod{\unitplz}{\vectu})-\scalr(\scalr\vrhob)\dltvd]
  \beqref{ogrv1a}, \eqnref{gxpeed1a}\text{ \& }\eqnref{gpath7}
\end{split}
\nonumber\\
\begin{split}
&=-\vakph\epsvf+\scalr\frkyx(\epsva\epsvc-\epsvb)+\scalr\frkyu(\epsva-\epsvc\epsvb)
  +\scalr\epsvb\frkyr(\epsva\dltvd-\epsvd\epsvb)
  +\scalr\epsvb\frkyn(\epsvc\dltvd-\epsvd)\\
  &\quad+\scalr^2\vakpa(1-\epsvc^2)
  +\scalr^2\vakpb(\dltvd-\epsvd\epsvc)
  -\vrhoa\frkys\scalr\epsva(\dprod{\unitplz}{\vectu})+\scalr^2\epsvb\frkys\vrhoa\vrhob
  -\vakpc\scalr\epsvc(\dprod{\unitplz}{\vectu})+\scalr^2\vakpc\vrhob\\
  &\quad-\vakpd\scalr\epsvd(\dprod{\unitplz}{\vectu})+\scalr^2\vakpd\dltvd\vrhob
\end{split}
\end{align*}
\begin{align}\label{gpath23g}
\begin{split}
&=-\vakph\epsvf+\scalr\frkyx(\epsva\epsvc-\epsvb)+\scalr\frkyu(\epsva-\epsvc\epsvb)
  +\scalr\epsvb\frkyr(\epsva\dltvd-\epsvd\epsvb)
  +\scalr\epsvb\frkyn(\epsvc\dltvd-\epsvd)\\
  &\quad+\scalr^2\vakpa(1-\epsvc^2)
  +\scalr^2\vakpb(\dltvd-\epsvd\epsvc)
  +\scalr^2\epsvb\frkys\vrhoa\vrhob+\scalr^2\vakpc\vrhob+\scalr^2\vakpd\dltvd\vrhob\\
  &\quad-\scalr(\vrhoa\frkys\epsva+\vakpc\epsvc+\vakpd\epsvd)(\dprod{\unitplz}{\vectu})
\end{split}
\nonumber\\
\begin{split}
&=-\vakph\epsvf+\scalr\frkyx(\epsva\epsvc-\epsvb)+\scalr\frkyu(\epsva-\epsvc\epsvb)
  +\scalr\epsvb\frkyr(\epsva\dltvd-\epsvd\epsvb)
  +\scalr\epsvb\frkyn(\epsvc\dltvd-\epsvd)\\
  &\quad+\scalr^2\vakpa(1-\epsvc^2)
  +\scalr^2\vakpb(\dltvd-\epsvd\epsvc)
  +\scalr^2\epsvb\frkys\vrhoa\vrhob+\scalr^2\vakpc\vrhob+\scalr^2\vakpd\dltvd\vrhob\\
  &\quad-(\scalr/\scalh^2)(\vrhoa\frkys\epsva+\vakpc\epsvc+\vakpd\epsvd)(\epsvi+\epsvf\vphib)
  \beqref{gpath7c}
\end{split}
\nonumber\\
&=\vakpp\beqref{gpath1l}
\end{align}
\begin{align*}
&\dprod{(\cprod{\vectr}{\vectz})}{(\vscrp+\vscrq)}\nonumber\\
\begin{split}
&=\dprod{(\cprod{\vectr}{\vectz})}{[}\vakpg\vectr+\vakph\vecth
  +\frkyx(\cprod{\unitkap}{\unitpos})
  +\frkyu(\cprod{\unitkap}{\unitplz})
  +\epsvb\frkyr(\cprod{\unitkap}{\vectz})
  +\epsvb\frkyn(\cprod{\unitplz}{\vectz})
  +\vakpa(\cprod{\vectr}{\unitplz})\\
  &\quad+\vakpb(\cprod{\vectr}{\vectz})
  -\vrhoa\frkys(\cprod{\unitkap}{\vectu})
  -\vakpc(\cprod{\unitplz}{\vectu})
  -\vakpd(\cprod{\vectz}{\vectu})]\beqref{gpath21d}
\end{split}
\nonumber\\
\begin{split}
&=\vakpg[\dprod{\vectr}{(\cprod{\vectr}{\vectz})}]
  +\vakph[\dprod{\vecth}{(\cprod{\vectr}{\vectz})}]
  +\frkyx[\dprod{(\cprod{\vectr}{\vectz})}{(\cprod{\unitkap}{\unitpos})}]
  +\frkyu[\dprod{(\cprod{\vectr}{\vectz})}{(\cprod{\unitkap}{\unitplz})}]\\
  &\quad+\epsvb\frkyr[\dprod{(\cprod{\vectr}{\vectz})}{(\cprod{\unitkap}{\vectz})}]
  +\epsvb\frkyn[\dprod{(\cprod{\vectr}{\vectz})}{(\cprod{\unitplz}{\vectz})}]
  +\vakpa[\dprod{(\cprod{\vectr}{\vectz})}{(\cprod{\vectr}{\unitplz})}]\\
  &\quad+\vakpb[\dprod{(\cprod{\vectr}{\vectz})}{(\cprod{\vectr}{\vectz})}]
  -\vrhoa\frkys[\dprod{(\cprod{\vectr}{\vectz})}{(\cprod{\unitkap}{\vectu})}]
  -\vakpc[\dprod{(\cprod{\vectr}{\vectz})}{(\cprod{\unitplz}{\vectu})}]\\
  &\quad-\vakpd[\dprod{(\cprod{\vectr}{\vectz})}{(\cprod{\vectz}{\vectu})}]
\end{split}
\end{align*}
\begin{align}\label{gpath23h}
\begin{split}
&=\scalr\vakph\epsvh
  +\frkyx[(\dprod{\vectr}{\unitkap})(\dprod{\vectz}{\unitpos})-(\dprod{\vectr}{\unitpos})(\dprod{\vectz}{\unitkap})]
  +\frkyu[(\dprod{\vectr}{\unitkap})(\dprod{\vectz}{\unitplz})-(\dprod{\vectr}{\unitplz})(\dprod{\vectz}{\unitkap})]\\
  &\quad+\epsvb\frkyr[(\dprod{\vectr}{\unitkap})\scalz^2-(\dprod{\vectr}{\vectz})(\dprod{\vectz}{\unitkap})]
  +\epsvb\frkyn[(\dprod{\vectr}{\unitplz})\scalz^2-(\dprod{\vectr}{\vectz})(\dprod{\vectz}{\unitplz})]
  +\vakpa[\scalr^2(\dprod{\vectz}{\unitplz})-(\dprod{\vectr}{\unitplz})(\dprod{\vectz}{\vectr})]\\
  &\quad+\vakpb[\scalr^2\scalz^2-(\dprod{\vectr}{\vectz})^2]
  -\vrhoa\frkys[(\dprod{\vectr}{\unitkap})(\dprod{\vectz}{\vectu})-(\dprod{\vectr}{\vectu})(\dprod{\vectz}{\unitkap})]
  -\vakpc[(\dprod{\vectr}{\unitplz})(\dprod{\vectz}{\vectu})-(\dprod{\vectr}{\vectu})(\dprod{\vectz}{\unitplz})]\\
  &\quad-\vakpd[(\dprod{\vectr}{\vectz})(\dprod{\vectz}{\vectu})-(\dprod{\vectr}{\vectu})\scalz^2]
  \beqref{ogrv1a}\text{ \& }\eqnref{alg2}
\end{split}
\nonumber\\
\begin{split}
&=\scalr\vakph\epsvh+\frkyx(\scalr\epsva\epsvd-\scalr\dltvb)+\frkyu(\scalr\epsva\dltvd-\scalr\epsvc\dltvb)
  +\epsvb\frkyr(\scalr\epsva\scalz^2-\scalr\epsvd\dltvb)
  +\epsvb\frkyn(\scalr\epsvc\scalz^2-\scalr\epsvd\dltvd)\\
  &\quad+\vakpa(\scalr^2\dltvd-\scalr^2\epsvc\epsvd)
  +\vakpb(\scalr^2\scalz^2-\scalr^2\epsvd^2)
  -\vrhoa\frkys[\scalr\epsva(-\scalr\vrhob\vphib)-\scalr(\scalr\vrhob)\dltvb]\\
  &\quad-\vakpc[\scalr\epsvc(-\scalr\vrhob\vphib)-\scalr(\scalr\vrhob)\dltvd]
  -\vakpd[\scalr\epsvd(-\scalr\vrhob\vphib)-\scalr(\scalr\vrhob)\scalz^2]
  \beqref{ogrv1a}, \eqnref{gxpeed1a}\text{ \& }\eqnref{gpath7}
\end{split}
\nonumber\\
\begin{split}
&=\scalr\vakph\epsvh+\scalr\frkyx(\epsva\epsvd-\dltvb)
  +\scalr\frkyu(\epsva\dltvd-\epsvc\dltvb)
  +\scalr\epsvb\frkyr(\epsva\scalz^2-\epsvd\dltvb)
  +\scalr\epsvb\frkyn(\epsvc\scalz^2-\epsvd\dltvd)\\
  &\quad+\scalr^2\vakpa(\dltvd-\epsvc\epsvd)
  +\scalr^2\vakpb(\scalz^2-\epsvd^2)
  +\scalr^2\vrhoa\frkys\vrhob(\epsva\vphib+\dltvb)
  +\scalr^2\vakpc\vrhob(\epsvc\vphib+\dltvd)\\
  &\quad+\scalr^2\vakpd\vrhob(\epsvd\vphib+\scalz^2)
\end{split}
\nonumber\\
&=\vakpq\beqref{gpath1l}
\end{align}
\begin{align*}
&\dprod{(\cprod{\unitkap}{\vectu})}{(\vscrp+\vscrq)}\nonumber\\
\begin{split}
&=\dprod{(\cprod{\unitkap}{\vectu})}{[}\vakpg\vectr+\vakph\vecth
  +\frkyx(\cprod{\unitkap}{\unitpos})
  +\frkyu(\cprod{\unitkap}{\unitplz})
  +\epsvb\frkyr(\cprod{\unitkap}{\vectz})
  +\epsvb\frkyn(\cprod{\unitplz}{\vectz})
  +\vakpa(\cprod{\vectr}{\unitplz})\\
  &\quad+\vakpb(\cprod{\vectr}{\vectz})
  -\vrhoa\frkys(\cprod{\unitkap}{\vectu})
  -\vakpc(\cprod{\unitplz}{\vectu})
  -\vakpd(\cprod{\vectz}{\vectu})]\beqref{gpath21d}
\end{split}
\nonumber\\
\begin{split}
&=\vakpg[\dprod{\vectr}{(\cprod{\unitkap}{\vectu})}]
  +\vakph[\dprod{\vecth}{(\cprod{\unitkap}{\vectu})}]
  +\frkyx[\dprod{(\cprod{\unitkap}{\vectu})}{(\cprod{\unitkap}{\unitpos})}]
  +\frkyu[\dprod{(\cprod{\unitkap}{\vectu})}{(\cprod{\unitkap}{\unitplz})}]\\
  &\quad+\epsvb\frkyr[\dprod{(\cprod{\unitkap}{\vectu})}{(\cprod{\unitkap}{\vectz})}]
  +\epsvb\frkyn[\dprod{(\cprod{\unitkap}{\vectu})}{(\cprod{\unitplz}{\vectz})}]
  +\vakpa[\dprod{(\cprod{\unitkap}{\vectu})}{(\cprod{\vectr}{\unitplz})}]\\
  &\quad+\vakpb[\dprod{(\cprod{\unitkap}{\vectu})}{(\cprod{\vectr}{\vectz})}]
  -\vrhoa\frkys[\dprod{(\cprod{\unitkap}{\vectu})}{(\cprod{\unitkap}{\vectu})}]
  -\vakpc[\dprod{(\cprod{\unitkap}{\vectu})}{(\cprod{\unitplz}{\vectu})}]\\
  &\quad-\vakpd[\dprod{(\cprod{\unitkap}{\vectu})}{(\cprod{\vectz}{\vectu})}]
\end{split}
\end{align*}
\begin{align*}
\begin{split}
&=\vakpg[\dprod{\unitkap}{(\cprod{\vectu}{\vectr})}]
  +\vakph[\dprod{\unitkap}{(\cprod{\vectu}{\vecth})}]
  +\frkyx[(\dprod{\vectu}{\unitpos})-(\dprod{\unitkap}{\unitpos})(\dprod{\vectu}{\unitkap})]
  +\frkyu[(\dprod{\vectu}{\unitplz})-(\dprod{\unitkap}{\unitplz})(\dprod{\vectu}{\unitkap})]\\
  &\quad+\epsvb\frkyr[(\dprod{\vectu}{\vectz})-(\dprod{\unitkap}{\vectz})(\dprod{\vectu}{\unitkap})]
  +\epsvb\frkyn[(\dprod{\unitkap}{\unitplz})(\dprod{\vectu}{\vectz})-(\dprod{\unitkap}{\vectz})(\dprod{\vectu}{\unitplz})]\\
  &\quad+\vakpa[(\dprod{\unitkap}{\vectr})(\dprod{\vectu}{\unitplz})-(\dprod{\unitkap}{\unitplz})(\dprod{\vectu}{\vectr})]
  +\vakpb[(\dprod{\unitkap}{\vectr})(\dprod{\vectu}{\vectz})-(\dprod{\unitkap}{\vectz})(\dprod{\vectu}{\vectr})]
  -\vrhoa\frkys[\scalu^2-(\dprod{\unitkap}{\vectu})^2]\\
  &\quad-\vakpc[(\dprod{\unitkap}{\unitplz})\scalu^2-(\dprod{\unitkap}{\vectu})(\dprod{\vectu}{\unitplz})]
  -\vakpd[(\dprod{\unitkap}{\vectz})\scalu^2-(\dprod{\unitkap}{\vectu})(\dprod{\vectu}{\vectz})]
  \beqref{alg2}\text{ \& }\eqnref{alg4}
\end{split}
\nonumber\\
\begin{split}
&=\vakpg(\dprod{\unitkap}{\vecth})+\vakph[\dprod{\unitkap}{(-\vectz-\scalq\unitpos)}]
  +\frkyx(\scalr\vrhob-\epsva\vrhoo)+\frkyu[(\dprod{\vectu}{\unitplz})-\epsvb\vrhoo]\\
  &\quad+\epsvb\frkyr(-\scalr\vrhob\vphib-\dltvb\vrhoo)
  +\epsvb\frkyn[\epsvb(-\scalr\vrhob\vphib)-\dltvb(\dprod{\vectu}{\unitplz})]\\
  &\quad+\vakpa[\scalr\epsva(\dprod{\vectu}{\unitplz})-\scalr\epsvb(\scalr\vrhob)]
  +\vakpb[\scalr\epsva(-\scalr\vrhob\vphib)-\scalr\dltvb(\scalr\vrhob)]
  -\vrhoa\frkys(\vphih^2-\vrhoo^2)\\
  &\quad-\vakpc[\epsvb\vphih^2-\vrhoo(\dprod{\vectu}{\unitplz})]
  -\vakpd[\dltvb\vphih^2-\vrhoo(-\scalr\vrhob\vphib)]
  \beqref{main5c}, \eqnref{gpath7}\text{ \& }\eqnref{ogrv3b}
\end{split}
\end{align*}
\begin{align*}
\begin{split}
&=\vakpg\dltva-\vakph(\dltvb+\scalq\epsva)+\frkyx(\scalr\vrhob-\epsva\vrhoo)+\frkyu(\dprod{\vectu}{\unitplz})
  -\frkyu\epsvb\vrhoo-\epsvb\frkyr(\scalr\vrhob\vphib+\dltvb\vrhoo)\\
  &\quad-\epsvb\frkyn\scalr\epsvb\vrhob\vphib-\epsvb\frkyn\dltvb(\dprod{\vectu}{\unitplz})
  +\vakpa\scalr\epsva(\dprod{\vectu}{\unitplz})
  -\scalr^2\vakpa\epsvb\vrhob-\scalr^2\vakpb\vrhob(\epsva\vphib+\dltvb)\\
  &\quad-\vrhoa\frkys(\vphih^2-\vrhoo^2)
  -\vakpc\epsvb\vphih^2+\vakpc\vrhoo(\dprod{\vectu}{\unitplz})
  -\vakpd(\dltvb\vphih^2+\scalr\vrhoo\vrhob\vphib)
  \beqref{ogrv1a}\text{ \& }\eqnref{gxpeed1a}
\end{split}
\nonumber\\
\begin{split}
&=\vakpg\dltva-\vakph(\dltvb+\scalq\epsva)+\frkyx(\scalr\vrhob-\epsva\vrhoo)
  -\frkyu\epsvb\vrhoo-\epsvb\frkyr(\scalr\vrhob\vphib+\dltvb\vrhoo)\\
  &\quad-\epsvb\frkyn\scalr\epsvb\vrhob\vphib
  -\scalr^2\vakpa\epsvb\vrhob-\scalr^2\vakpb\vrhob(\epsva\vphib+\dltvb)-\vrhoa\frkys(\vphih^2-\vrhoo^2)
  -\vakpc\epsvb\vphih^2\\
  &\quad-\vakpd(\dltvb\vphih^2+\scalr\vrhoo\vrhob\vphib)
  +\frkyu(\dprod{\vectu}{\unitplz})-\epsvb\frkyn\dltvb(\dprod{\vectu}{\unitplz})
  +\vakpa\scalr\epsva(\dprod{\vectu}{\unitplz})+\vakpc\vrhoo(\dprod{\vectu}{\unitplz})
\end{split}
\end{align*}
\begin{align}\label{gpath23i}
\begin{split}
&=\vakpg\dltva-\vakph\efkta+\frkyx(\scalr\vrhob-\epsva\vrhoo)
  -\frkyu\epsvb\vrhoo-\epsvb\frkyr(\scalr\vrhob\vphib+\dltvb\vrhoo)\\
  &\quad-\epsvb\frkyn\scalr\epsvb\vrhob\vphib
  -\scalr^2\vakpa\epsvb\vrhob-\scalr^2\vakpb\vrhob(\epsva\vphib+\dltvb)-\vrhoa\frkys(\vphih^2-\vrhoo^2)
  -\vakpc\epsvb\vphih^2\\
  &\quad-\vakpd(\dltvb\vphih^2+\scalr\vrhoo\vrhob\vphib)
  +(\frkyu-\epsvb\frkyn\dltvb+\vakpa\scalr\epsva+\vakpc\vrhoo)(\dprod{\vectu}{\unitplz})
  \beqref{gxpeed1b}
\end{split}
\nonumber\\
\begin{split}
&=\vakpg\dltva-\vakph\efkta+\frkyx(\scalr\vrhob-\epsva\vrhoo)
  -\frkyu\epsvb\vrhoo-\epsvb\frkyr(\scalr\vrhob\vphib+\dltvb\vrhoo)\\
  &\quad-\epsvb^2\frkyn\scalr\vrhob\vphib
  -\scalr^2\vakpa\epsvb\vrhob-\scalr^2\vakpb\vrhob(\epsva\vphib+\dltvb)-\vrhoa\frkys(\vphih^2-\vrhoo^2)
  -\vakpc\epsvb\vphih^2\\
  &\quad-\vakpd(\dltvb\vphih^2+\scalr\vrhoo\vrhob\vphib)
  +\scalh^{-2}(\frkyu-\epsvb\frkyn\dltvb+\vakpa\scalr\epsva+\vakpc\vrhoo)(\epsvi+\epsvf\vphib)
  \beqref{gpath7c}
\end{split}
\nonumber\\
&=\vakpr\beqref{gpath1l}
\end{align}
\begin{align*}
&\dprod{(\cprod{\unitplz}{\vectu})}{(\vscrp+\vscrq)}\nonumber\\
\begin{split}
&=\dprod{(\cprod{\unitplz}{\vectu})}{[}\vakpg\vectr+\vakph\vecth
  +\frkyx(\cprod{\unitkap}{\unitpos})
  +\frkyu(\cprod{\unitkap}{\unitplz})
  +\epsvb\frkyr(\cprod{\unitkap}{\vectz})
  +\epsvb\frkyn(\cprod{\unitplz}{\vectz})
  +\vakpa(\cprod{\vectr}{\unitplz})\\
  &\quad+\vakpb(\cprod{\vectr}{\vectz})
  -\vrhoa\frkys(\cprod{\unitkap}{\vectu})
  -\vakpc(\cprod{\unitplz}{\vectu})
  -\vakpd(\cprod{\vectz}{\vectu})]\beqref{gpath21d}
\end{split}
\nonumber\\
\begin{split}
&=\vakpg[\dprod{\vectr}{(\cprod{\unitplz}{\vectu})}]
  +\vakph[\dprod{\vecth}{(\cprod{\unitplz}{\vectu})}]
  +\frkyx[\dprod{(\cprod{\unitplz}{\vectu})}{(\cprod{\unitkap}{\unitpos})}]
  +\frkyu[\dprod{(\cprod{\unitplz}{\vectu})}{(\cprod{\unitkap}{\unitplz})}]\\
  &\quad+\epsvb\frkyr[\dprod{(\cprod{\unitplz}{\vectu})}{(\cprod{\unitkap}{\vectz})}]
  +\epsvb\frkyn[\dprod{(\cprod{\unitplz}{\vectu})}{(\cprod{\unitplz}{\vectz})}]
  +\vakpa[\dprod{(\cprod{\unitplz}{\vectu})}{(\cprod{\vectr}{\unitplz})}]\\
  &\quad+\vakpb[\dprod{(\cprod{\unitplz}{\vectu})}{(\cprod{\vectr}{\vectz})}]
  -\vrhoa\frkys[\dprod{(\cprod{\unitplz}{\vectu})}{(\cprod{\unitkap}{\vectu})}]
  -\vakpc[\dprod{(\cprod{\unitplz}{\vectu})}{(\cprod{\unitplz}{\vectu})}]\\
  &\quad-\vakpd[\dprod{(\cprod{\unitplz}{\vectu})}{(\cprod{\vectz}{\vectu})}]
\end{split}
\end{align*}
\begin{align*}
\begin{split}
&=\vakpg[\dprod{\unitplz}{(\cprod{\vectu}{\vectr})}]
  +\vakph[\dprod{\unitplz}{(\cprod{\vectu}{\vecth})}]
  +\frkyx[(\dprod{\unitplz}{\unitkap})(\dprod{\vectu}{\unitpos})-(\dprod{\unitplz}{\unitpos})(\dprod{\vectu}{\unitkap})]
  +\frkyu[(\dprod{\unitplz}{\unitkap})(\dprod{\vectu}{\unitplz})-(\dprod{\vectu}{\unitkap})]\\
  &\quad+\epsvb\frkyr[(\dprod{\unitplz}{\unitkap})(\dprod{\vectu}{\vectz})-(\dprod{\unitplz}{\vectz})(\dprod{\vectu}{\unitkap})]
  +\epsvb\frkyn[(\dprod{\vectu}{\vectz})-(\dprod{\unitplz}{\vectz})(\dprod{\vectu}{\unitplz})]
  +\vakpa[(\dprod{\unitplz}{\vectr})(\dprod{\vectu}{\unitplz})-(\dprod{\vectu}{\vectr})]\\
  &\quad+\vakpb[(\dprod{\unitplz}{\vectr})(\dprod{\vectu}{\vectz})-(\dprod{\unitplz}{\vectz})(\dprod{\vectu}{\vectr})]
  -\vrhoa\frkys[(\dprod{\unitplz}{\unitkap})\scalu^2-(\dprod{\unitplz}{\vectu})(\dprod{\vectu}{\unitkap})]
  -\vakpc[\scalu^2-(\dprod{\unitplz}{\vectu})^2]\\
  &\quad-\vakpd[(\dprod{\unitplz}{\vectz})\scalu^2-(\dprod{\unitplz}{\vectu})(\dprod{\vectu}{\vectz})]
  \beqref{alg4}\text{ \& }\eqnref{alg2}
\end{split}
\nonumber\\
\begin{split}
&=\vakpg(\dprod{\unitplz}{\vecth})
  +\vakph[\dprod{\unitplz}{(-\vectz-\scalq\unitpos)}]
  +\frkyx[\epsvb(\scalr\vrhob)-\epsvc\vrhoo]
  +\frkyu[\epsvb(\dprod{\vectu}{\unitplz})-\vrhoo]\\
  &\quad+\epsvb\frkyr[\epsvb(-\scalr\vrhob\vphib)-\dltvd\vrhoo]
  +\epsvb\frkyn[(-\scalr\vrhob\vphib)-\dltvd(\dprod{\vectu}{\unitplz})]
  +\vakpa[\scalr\epsvc(\dprod{\vectu}{\unitplz})-\scalr(\scalr\vrhob)]\\
  &\quad+\vakpb[\scalr\epsvc(-\scalr\vrhob\vphib)-\scalr\dltvd(\scalr\vrhob)]
  -\vrhoa\frkys[\epsvb\vphih^2-(\dprod{\unitplz}{\vectu})\vrhoo]
  -\vakpc[\vphih^2-(\dprod{\unitplz}{\vectu})^2]\\
  &\quad-\vakpd[\dltvd\vphih^2-(\dprod{\unitplz}{\vectu})(-\scalr\vrhob\vphib)]
  \beqref{main5c}, \eqnref{ogrv1a}, \eqnref{ogrv3b}\text{ \& }\eqnref{gxpeed1a}
\end{split}
\nonumber\\
\begin{split}
&=\vakpg\dltvc+\vakph(-\dltvd-\scalq\epsvc)+\frkyx(\scalr\epsvb\vrhob-\epsvc\vrhoo)
  +\frkyu\epsvb(\dprod{\vectu}{\unitplz})-\frkyu\vrhoo\\
  &\quad-\epsvb\frkyr(\scalr\epsvb\vrhob\vphib+\dltvd\vrhoo)
  -\epsvb\frkyn\scalr\vrhob\vphib-\epsvb\frkyn\dltvd(\dprod{\vectu}{\unitplz})
  +\scalr\vakpa\epsvc(\dprod{\vectu}{\unitplz})-\scalr^2\vakpa\vrhob\\
  &\quad-\scalr^2\vakpb\vrhob(\epsvc\vphib+\dltvd)
  -\vrhoa\frkys\epsvb\vphih^2+\vrhoo\vrhoa\frkys(\dprod{\unitplz}{\vectu})
  -\vakpc\vphih^2+\vakpc(\dprod{\unitplz}{\vectu})^2\\
  &\quad-\vakpd\dltvd\vphih^2-\scalr\vakpd\vrhob\vphib(\dprod{\unitplz}{\vectu})
  \beqref{gxpeed1a}
\end{split}
\end{align*}
\begin{align}\label{gpath23j}
\begin{split}
&=\vakpg\dltvc-\vakph(\dltvd+\scalq\epsvc)+\frkyx(\scalr\epsvb\vrhob-\epsvc\vrhoo)
  -\frkyu\vrhoo-\epsvb\frkyr(\scalr\epsvb\vrhob\vphib+\dltvd\vrhoo)
  -\epsvb\frkyn\scalr\vrhob\vphib\\
  &\quad-\scalr^2\vakpa\vrhob-\scalr^2\vakpb\vrhob(\epsvc\vphib+\dltvd)-\vrhoa\frkys\epsvb\vphih^2-\vakpc\vphih^2
  -\vakpd\dltvd\vphih^2+\frkyu\epsvb(\dprod{\vectu}{\unitplz})-\epsvb\frkyn\dltvd(\dprod{\vectu}{\unitplz})\\
  &\quad+\scalr\vakpa\epsvc(\dprod{\vectu}{\unitplz})+\vrhoo\vrhoa\frkys(\dprod{\unitplz}{\vectu})
  +\vakpc(\dprod{\unitplz}{\vectu})^2-\scalr\vakpd\vrhob\vphib(\dprod{\unitplz}{\vectu})
\end{split}
\nonumber\\
\begin{split}
&=\vakpg\dltvc-\vakph(\dltvd+\scalq\epsvc)+\frkyx(\scalr\epsvb\vrhob-\epsvc\vrhoo)
  -\frkyu\vrhoo-\epsvb\frkyr(\scalr\epsvb\vrhob\vphib+\dltvd\vrhoo)\\
  &\quad-\epsvb\frkyn\scalr\vrhob\vphib
  -\scalr^2\vakpa\vrhob-\scalr^2\vakpb\vrhob(\epsvc\vphib+\dltvd)-\vrhoa\frkys\epsvb\vphih^2-\vakpc\vphih^2
  -\vakpd\dltvd\vphih^2\\
  &\quad+\vakpc(\dprod{\unitplz}{\vectu})^2+(\frkyu\epsvb-\epsvb\frkyn\dltvd+\scalr\vakpa\epsvc+\vrhoo\vrhoa\frkys
  -\scalr\vakpd\vrhob\vphib)(\dprod{\unitplz}{\vectu})
\end{split}
\nonumber\\
\begin{split}
&=\vakpg\dltvc-\vakph\efktc+\frkyx(\scalr\epsvb\vrhob-\epsvc\vrhoo)
  -\frkyu\vrhoo-\epsvb\frkyr(\scalr\epsvb\vrhob\vphib+\dltvd\vrhoo)
  -\scalr\epsvb\frkyn\vrhob\vphib\\
  &\quad-\scalr^2\vrhob[\vakpa+\vakpb(\epsvc\vphib+\dltvd)]
  -\vphih^2(\vrhoa\frkys\epsvb+\vakpc+\vakpd\dltvd)
  +\scalh^{-4}\vakpc(\epsvi+\epsvf\vphib)^2\\
  &\quad+\scalh^{-2}(\frkyu\epsvb-\epsvb\frkyn\dltvd+\scalr\vakpa\epsvc+\vrhoo\vrhoa\frkys
  -\scalr\vakpd\vrhob\vphib)(\epsvi+\epsvf\vphib)\beqref{gpath7c}\text{ \& }\eqnref{gxpeed1b}
\end{split}
\nonumber\\
&=\vakps\beqref{gpath1m}
\end{align}
\begin{align*}
&\dprod{(\cprod{\vectz}{\vectu})}{(\vscrp+\vscrq)}\nonumber\\
\begin{split}
&=\dprod{(\cprod{\vectz}{\vectu})}{[}\vakpg\vectr+\vakph\vecth
  +\frkyx(\cprod{\unitkap}{\unitpos})
  +\frkyu(\cprod{\unitkap}{\unitplz})
  +\epsvb\frkyr(\cprod{\unitkap}{\vectz})
  +\epsvb\frkyn(\cprod{\unitplz}{\vectz})
  +\vakpa(\cprod{\vectr}{\unitplz})\\
  &\quad+\vakpb(\cprod{\vectr}{\vectz})
  -\vrhoa\frkys(\cprod{\unitkap}{\vectu})
  -\vakpc(\cprod{\unitplz}{\vectu})
  -\vakpd(\cprod{\vectz}{\vectu})]\beqref{gpath21d}
\end{split}
\nonumber\\
\begin{split}
&=\vakpg[\dprod{\vectr}{(\cprod{\vectz}{\vectu})}]
  +\vakph[\dprod{\vecth}{(\cprod{\vectz}{\vectu})}]
  +\frkyx[\dprod{(\cprod{\vectz}{\vectu})}{(\cprod{\unitkap}{\unitpos})}]
  +\frkyu[\dprod{(\cprod{\vectz}{\vectu})}{(\cprod{\unitkap}{\unitplz})}]\\
  &\quad+\epsvb\frkyr[\dprod{(\cprod{\vectz}{\vectu})}{(\cprod{\unitkap}{\vectz})}]
  +\epsvb\frkyn[\dprod{(\cprod{\vectz}{\vectu})}{(\cprod{\unitplz}{\vectz})}]
  +\vakpa[\dprod{(\cprod{\vectz}{\vectu})}{(\cprod{\vectr}{\unitplz})}]\\
  &\quad+\vakpb[\dprod{(\cprod{\vectz}{\vectu})}{(\cprod{\vectr}{\vectz})}]
  -\vrhoa\frkys[\dprod{(\cprod{\vectz}{\vectu})}{(\cprod{\unitkap}{\vectu})}]
  -\vakpc[\dprod{(\cprod{\vectz}{\vectu})}{(\cprod{\unitplz}{\vectu})}]\\
  &\quad-\vakpd[\dprod{(\cprod{\vectz}{\vectu})}{(\cprod{\vectz}{\vectu})}]
\end{split}
\end{align*}
\begin{align*}
\begin{split}
&=\vakpg[\dprod{\vectz}{(\cprod{\vectu}{\vectr})}]
  +\vakph[\dprod{\vectz}{(\cprod{\vectu}{\vecth})}]
  +\frkyx[(\dprod{\vectz}{\unitkap})(\dprod{\vectu}{\unitpos})-(\dprod{\vectz}{\unitpos})(\dprod{\vectu}{\unitkap})]\\
  &\quad+\frkyu[(\dprod{\vectz}{\unitkap})(\dprod{\vectu}{\unitplz})-(\dprod{\vectz}{\unitplz})(\dprod{\vectu}{\unitkap})]
  +\epsvb\frkyr[(\dprod{\vectz}{\unitkap})(\dprod{\vectu}{\vectz})-\scalz^2(\dprod{\vectu}{\unitkap})]\\
  &\quad+\epsvb\frkyn[(\dprod{\vectz}{\unitplz})(\dprod{\vectu}{\vectz})-\scalz^2(\dprod{\vectu}{\unitplz})]
  +\vakpa[(\dprod{\vectz}{\vectr})(\dprod{\vectu}{\unitplz})-(\dprod{\vectz}{\unitplz})(\dprod{\vectu}{\vectr})]\\
  &\quad+\vakpb[(\dprod{\vectz}{\vectr})(\dprod{\vectu}{\vectz})-\scalz^2(\dprod{\vectu}{\vectr})]
  -\vrhoa\frkys[(\dprod{\vectz}{\unitkap})\scalu^2-(\dprod{\vectz}{\vectu})(\dprod{\vectu}{\unitkap})]
  -\vakpc[(\dprod{\vectz}{\unitplz})\scalu^2-(\dprod{\vectz}{\vectu})(\dprod{\vectu}{\unitplz})]\\
  &\quad-\vakpd[\scalz^2\scalu^2-(\dprod{\vectz}{\vectu})^2]
  \beqref{alg2}\text{ \& }\eqnref{alg4}
\end{split}
\nonumber\\
\begin{split}
&=\vakpg(\dprod{\vectz}{\vecth})
  +\vakph[\dprod{\vectz}{(-\vectz-\scalq\unitpos)}]
  +\frkyx[\dltvb(\scalr\vrhob)-\epsvd\vrhoo]
  +\frkyu[\dltvb(\dprod{\vectu}{\unitplz})-\dltvd\vrhoo]\\
  &\quad+\epsvb\frkyr[\dltvb(-\scalr\vrhob\vphib)-\scalz^2\vrhoo]
  +\epsvb\frkyn[\dltvd(-\scalr\vrhob\vphib)-\scalz^2(\dprod{\vectu}{\unitplz})]
  +\vakpa[\scalr\epsvd(\dprod{\vectu}{\unitplz})-\scalr\dltvd(\scalr\vrhob)]\\
  &\quad+\vakpb[\scalr\epsvd(-\scalr\vrhob\vphib)-\scalr\scalz^2(\scalr\vrhob)]
  -\vrhoa\frkys[\dltvb\vphih^2-(-\scalr\vrhob\vphib)\vrhoo]
  -\vakpc[\dltvd\vphih^2-(-\scalr\vrhob\vphib)(\dprod{\vectu}{\unitplz})]\\
  &\quad-\vakpd[\scalz^2\vphih^2-(-\scalr\vrhob\vphib)^2]
  \beqref{main5c}, \eqnref{gxpeed1a}, \eqnref{gpath7}\text{ \& }\eqnref{ogrv3b}
\end{split}
\end{align*}
\begin{align*}
\begin{split}
&=\vakph(-\scalz^2-\scalq\epsvd)
  +\frkyx(\scalr\dltvb\vrhob-\epsvd\vrhoo)
  +\frkyu\dltvb(\dprod{\vectu}{\unitplz})-\frkyu\dltvd\vrhoo\\
  &\quad-\epsvb\frkyr(\scalr\dltvb\vrhob\vphib+\scalz^2\vrhoo)
  -\scalr\epsvb\frkyn\dltvd\vrhob\vphib-\scalz^2\epsvb\frkyn(\dprod{\vectu}{\unitplz})
  +\scalr\vakpa\epsvd(\dprod{\vectu}{\unitplz})-\scalr^2\vakpa\dltvd\vrhob\\
  &\quad-\scalr^2\vakpb\vrhob(\epsvd\vphib+\scalz^2)
  -\vrhoa\frkys(\dltvb\vphih^2+\scalr\vrhob\vphib\vrhoo)
  -\vakpc\dltvd\vphih^2-\scalr\vakpc\vrhob\vphib(\dprod{\vectu}{\unitplz})\\
  &\quad-\vakpd(\scalz^2\vphih^2-\scalr^2\vrhob^2\vphib^2)
  \beqref{main5c}, \eqnref{ogrv1a}\text{ \& }\eqnref{gxpeed1a}
\end{split}
\nonumber\\
\begin{split}
&=-\vakph(\scalz^2+\scalq\epsvd)+\frkyx(\scalr\dltvb\vrhob-\epsvd\vrhoo)
  -\frkyu\dltvd\vrhoo-\epsvb\frkyr(\scalr\dltvb\vrhob\vphib+\scalz^2\vrhoo)
  -\scalr\epsvb\frkyn\dltvd\vrhob\vphib-\scalr^2\vakpa\dltvd\vrhob\\
  &\quad-\scalr^2\vakpb\vrhob(\epsvd\vphib+\scalz^2)
  -\vrhoa\frkys(\dltvb\vphih^2+\scalr\vrhob\vphib\vrhoo)-\vakpc\dltvd\vphih^2
  -\vakpd(\scalz^2\vphih^2-\scalr^2\vrhob^2\vphib^2)\\
  &\quad+\frkyu\dltvb(\dprod{\vectu}{\unitplz})
  -\scalz^2\epsvb\frkyn(\dprod{\vectu}{\unitplz})
  +\scalr\vakpa\epsvd(\dprod{\vectu}{\unitplz})
  -\scalr\vakpc\vrhob\vphib(\dprod{\vectu}{\unitplz})
\end{split}
\end{align*}
\begin{align}\label{gpath23k}
\begin{split}
&=-\vakph\frkyz+\frkyx(\scalr\dltvb\vrhob-\epsvd\vrhoo)
  -\frkyu\dltvd\vrhoo-\epsvb\frkyr(\scalr\dltvb\vrhob\vphib+\scalz^2\vrhoo)
  -\scalr\epsvb\frkyn\dltvd\vrhob\vphib-\scalr^2\vakpa\dltvd\vrhob\\
  &\quad-\scalr^2\vakpb\vrhob(\epsvd\vphib+\scalz^2)
  -\vrhoa\frkys(\dltvb\vphih^2+\scalr\vrhob\vphib\vrhoo)-\vakpc\dltvd\vphih^2
  -\vakpd(\scalz^2\vphih^2-\scalr^2\vrhob^2\vphib^2)\\
  &\quad+(\frkyu\dltvb-\scalz^2\epsvb\frkyn+\scalr\vakpa\epsvd-\scalr\vakpc\vrhob\vphib)(\dprod{\vectu}{\unitplz})
  \beqref{gpath1h}
\end{split}
\nonumber\\
\begin{split}
&=-\vakph\frkyz+\frkyx(\scalr\dltvb\vrhob-\epsvd\vrhoo)
  -\frkyu\dltvd\vrhoo-\epsvb\frkyr(\scalr\dltvb\vrhob\vphib+\scalz^2\vrhoo)
  -\scalr\epsvb\frkyn\dltvd\vrhob\vphib-\scalr^2\vakpa\dltvd\vrhob\\
  &\quad-\scalr^2\vakpb\vrhob(\epsvd\vphib+\scalz^2)
  -\vrhoa\frkys(\dltvb\vphih^2+\scalr\vrhob\vphib\vrhoo)-\vakpc\dltvd\vphih^2
  -\vakpd(\scalz^2\vphih^2-\scalr^2\vrhob^2\vphib^2)\\
  &\quad+\scalh^{-2}(\frkyu\dltvb-\scalz^2\epsvb\frkyn+\scalr\vakpa\epsvd-\scalr\vakpc\vrhob\vphib)
  (\epsvi+\epsvf\vphib)\beqref{gpath7c}
\end{split}
\nonumber\\
&=\vakpt\beqref{gpath1m}
\end{align}
\end{subequations}
\begin{align}\label{gpath24}
&|\vscrp+\vscrq|^2
=\dprod{(\vscrp+\vscrq)}{(\vscrp+\vscrq)}\nonumber\\
\begin{split}
&=\dprod{(\vscrp+\vscrq)}{[}\vakpg\vectr+\vakph\vecth
  +\frkyx(\cprod{\unitkap}{\unitpos})
  +\frkyu(\cprod{\unitkap}{\unitplz})
  +\epsvb\frkyr(\cprod{\unitkap}{\vectz})
  +\epsvb\frkyn(\cprod{\unitplz}{\vectz})\\
  &\quad+\vakpa(\cprod{\vectr}{\unitplz})
  +\vakpb(\cprod{\vectr}{\vectz})
  -\vrhoa\frkys(\cprod{\unitkap}{\vectu})
  -\vakpc(\cprod{\unitplz}{\vectu})
  -\vakpd(\cprod{\vectz}{\vectu})]\beqref{gpath21d}
\end{split}
\nonumber\\
\begin{split}
&=\vakpg[\dprod{(\vscrp+\vscrq)}{\vectr}]
  +\vakph[\dprod{(\vscrp+\vscrq)}{\vecth}]
  +\frkyx[\dprod{(\vscrp+\vscrq)}{(\cprod{\unitkap}{\unitpos})}]
  +\frkyu[\dprod{(\vscrp+\vscrq)}{(\cprod{\unitkap}{\unitplz})}]\\
  &\quad+\epsvb\frkyr[\dprod{(\vscrp+\vscrq)}{(\cprod{\unitkap}{\vectz})}]
  +\epsvb\frkyn[\dprod{(\vscrp+\vscrq)}{(\cprod{\unitplz}{\vectz})}]
  +\vakpa[\dprod{(\vscrp+\vscrq)}{(\cprod{\vectr}{\unitplz})}]\\
  &\quad+\vakpb[\dprod{(\vscrp+\vscrq)}{(\cprod{\vectr}{\vectz})}]
  -\vrhoa\frkys[\dprod{(\vscrp+\vscrq)}{(\cprod{\unitkap}{\vectu})}]
  -\vakpc[\dprod{(\vscrp+\vscrq)}{(\cprod{\unitplz}{\vectu})}]\\
  &\quad-\vakpd[\dprod{(\vscrp+\vscrq)}{(\cprod{\vectz}{\vectu})}]
\end{split}
\nonumber\\
\begin{split}
&=\vakpg\vakpj+\vakph\vakpk+\frkyx\vakpl+\frkyu\vakpm+\epsvb\frkyr\vakpn+\epsvb\frkyn\vakpo+\vakpa\vakpp
  +\vakpb\vakpq-\vrhoa\frkys\vakpr-\vakpc\vakps\\
  &\quad-\vakpd\vakpt\beqref{gpath23}
\end{split}
\nonumber\\
&\therefore|\vscrp+\vscrq|=\vakpu\beqref{gpath1n}.
\end{align}

\subart{Development of equation \eqnref{kpath2d}}
To evaluate the quantities defined by eqnref{kpath2d}, we first derive
\begin{subequations}\label{gpath25}
\begin{align}\label{gpath25a}
\cprod{\vectz}{\vectu}
&=\scalh^{-2}[\cprod{\vectz}{(\cprod{\vectz}{\vecth}+\scalq\cprod{\unitpos}{\vecth})}]
  \beqref{main5c}\nonumber\\
&=\scalh^{-2}[\cprod{\vectz}{(\cprod{\vectz}{\vecth})}+\scalq\cprod{\vectz}{(\cprod{\unitpos}{\vecth})}]
  \nonumber\\
&=\scalh^{-2}[\vectz(\dprod{\vectz}{\vecth})-\scalz^2\vecth+\scalq\unitpos(\dprod{\vectz}{\vecth})
  -\scalq\vecth(\dprod{\vectz}{\unitpos})]\beqref{alg1}\nonumber\\
&=\scalh^{-2}(-\scalz^2\vecth-\scalq\epsvd\vecth)\beqref{main5c}\text{ \& }\eqnref{ogrv1a}\nonumber\\
&=-\scalh^{-2}(\scalz^2+\scalq\epsvd)\vecth
=-\scalh^{-2}\frkyz\vecth\beqref{gpath1h}
\end{align}
\end{subequations}
\begin{subequations}\label{gpath26}
\begin{align}\label{gpath26a}
&\dprod{\unitkap}{(\vscrp+\vscrq)}\nonumber\\
\begin{split}
&=\dprod{\unitkap}{[}\vakpg\vectr+\vakph\vecth
  +\frkyx(\cprod{\unitkap}{\unitpos})
  +\frkyu(\cprod{\unitkap}{\unitplz})
  +\epsvb\frkyr(\cprod{\unitkap}{\vectz})
  +\epsvb\frkyn(\cprod{\unitplz}{\vectz})
  +\vakpa(\cprod{\vectr}{\unitplz})\\
  &\quad+\vakpb(\cprod{\vectr}{\vectz})
  -\vrhoa\frkys(\cprod{\unitkap}{\vectu})
  -\vakpc(\cprod{\unitplz}{\vectu})
  -\vakpd(\cprod{\vectz}{\vectu})]\beqref{gpath21d}
\end{split}
\nonumber\\
\begin{split}
&=\vakpg(\dprod{\unitkap}{\vectr})
  +\vakph(\dprod{\unitkap}{\vecth})
  +\frkyx[\dprod{\unitkap}{(\cprod{\unitkap}{\unitpos})}]
  +\frkyu[\dprod{\unitkap}{(\cprod{\unitkap}{\unitplz})}]
  +\epsvb\frkyr[\dprod{\unitkap}{(\cprod{\unitkap}{\vectz})}]
  +\epsvb\frkyn[\dprod{\unitkap}{(\cprod{\unitplz}{\vectz})}]\\
  &\quad+\vakpa[\dprod{\unitkap}{(\cprod{\vectr}{\unitplz})}]
  +\vakpb[\dprod{\unitkap}{(\cprod{\vectr}{\vectz})}]
  -\vrhoa\frkys[\dprod{\unitkap}{(\cprod{\unitkap}{\vectu})}]
  -\vakpc[\dprod{\unitkap}{(\cprod{\unitplz}{\vectu})}]
  -\vakpd[\dprod{\unitkap}{(\cprod{\vectz}{\vectu})}]
\end{split}
\nonumber\\
\begin{split}
&=\scalr\vakpg\epsva+\vakph\dltva+\epsvb\frkyn\vsigc-\vakpa\vsiga-\scalr\vakpb\dltvf
  -\scalh^{-2}\vakpc[\dprod{\unitkap}{(\dltvc\vectz+\scalq\dltvc\unitpos-\efktc\vecth)}]\\
  &\quad-\scalh^{-2}\vakpd[\dprod{\unitkap}{(-\frkyz\vecth)}]
  \beqref{ogrv1a}, \eqnref{gxpeed1a}, \eqnref{gpath1a},
  \eqnref{gxpeed2d}\text{ \& }\eqnref{gpath25a}
\end{split}
\nonumber\\
\begin{split}
&=\scalr\vakpg\epsva+\vakph\dltva+\epsvb\frkyn\vsigc-\vakpa\vsiga-\scalr\vakpb\dltvf
  -\scalh^{-2}\vakpc[\dltvc(\dprod{\unitkap}{\vectz})+\scalq\dltvc(\dprod{\unitkap}{\unitpos})-\efktc(\dprod{\unitkap}{\vecth})]\\
  &\quad+\scalh^{-2}\vakpd\frkyz(\dprod{\unitkap}{\vecth})
\end{split}
\nonumber\\
\begin{split}
&=\scalr\vakpg\epsva+\vakph\dltva+\epsvb\frkyn\vsigc-\vakpa\vsiga-\scalr\vakpb\dltvf
  -\scalh^{-2}\vakpc(\dltvc\dltvb+\scalq\dltvc\epsva-\efktc\dltva)\\
  &\quad+\scalh^{-2}\vakpd\frkyz\dltva\beqref{gxpeed1a}\text{ \& }\eqnref{ogrv1a}
\end{split}
\nonumber\\
&=\scalr\vakpg\epsva+\vakph\dltva+\epsvb\frkyn\vsigc-\vakpa\vsiga-\scalr\vakpb\dltvf
  -\scalh^{-2}\vakpc[\dltvc(\dltvb+\scalq\epsva)-\efktc\dltva]+\scalh^{-2}\vakpd\frkyz\dltva
\nonumber\\
&=\scalr\vakpg\epsva+\vakph\dltva+\epsvb\frkyn\vsigc-\vakpa\vsiga-\scalr\vakpb\dltvf
  -(\vakpc/\scalh^2)(\dltvc\efkta-\efktc\dltva)+(\vakpd/\scalh^2)\frkyz\dltva\beqref{gxpeed1b}
\nonumber\\
&=\vakpv\beqref{gpath1n}
\end{align}
\begin{align}\label{gpath26b}
&\dprod{\unitplz}{(\vscrp+\vscrq)}\nonumber\\
\begin{split}
&=\dprod{\unitplz}{[}\vakpg\vectr+\vakph\vecth
  +\frkyx(\cprod{\unitkap}{\unitpos})
  +\frkyu(\cprod{\unitkap}{\unitplz})
  +\epsvb\frkyr(\cprod{\unitkap}{\vectz})
  +\epsvb\frkyn(\cprod{\unitplz}{\vectz})
  +\vakpa(\cprod{\vectr}{\unitplz})\\
  &\quad+\vakpb(\cprod{\vectr}{\vectz})
  -\vrhoa\frkys(\cprod{\unitkap}{\vectu})
  -\vakpc(\cprod{\unitplz}{\vectu})
  -\vakpd(\cprod{\vectz}{\vectu})]\beqref{gpath21d}
\end{split}
\nonumber\\
\begin{split}
&=\vakpg(\dprod{\unitplz}{\vectr})
  +\vakph(\dprod{\unitplz}{\vecth})
  +\frkyx[\dprod{\unitplz}{(\cprod{\unitkap}{\unitpos})}]
  +\frkyu[\dprod{\unitplz}{(\cprod{\unitkap}{\unitplz})}]
  +\epsvb\frkyr[\dprod{\unitplz}{(\cprod{\unitkap}{\vectz})}]
  +\epsvb\frkyn[\dprod{\unitplz}{(\cprod{\unitplz}{\vectz})}]\\
  &\quad+\vakpa[\dprod{\unitplz}{(\cprod{\vectr}{\unitplz})}]
  +\vakpb[\dprod{\unitplz}{(\cprod{\vectr}{\vectz})}]
  -\vrhoa\frkys[\dprod{\unitplz}{(\cprod{\unitkap}{\vectu})}]
  -\vakpc[\dprod{\unitplz}{(\cprod{\unitplz}{\vectu})}]
  -\vakpd[\dprod{\unitplz}{(\cprod{\vectz}{\vectu})}]
\end{split}
\nonumber\\
\begin{split}
&=\scalr\vakpg\epsvc+\vakph\dltvc-(\frkyx/\scalr)\vsiga-\epsvb\frkyr\vsigc-\vakpb\vsigb
  +\vrhoa\frkys[\dprod{\unitkap}{(\cprod{\unitplz}{\vectu})}]\\
  &\quad-\vakpd[\dprod{\unitplz}{(\cprod{\vectz}{\vectu})}]
  \beqref{ogrv1a}, \eqnref{gxpeed1a}, \eqnref{gpath1a}\text{ \& }\eqnref{alg4}
\end{split}
\nonumber\\
\begin{split}
&=\scalr\vakpg\epsvc+\vakph\dltvc-(\frkyx/\scalr)\vsiga-\epsvb\frkyr\vsigc-\vakpb\vsigb
  +\vrhoa(\frkys/\scalh^2)[\dprod{\unitkap}{(\dltvc\vectz+\scalq\dltvc\unitpos-\efktc\vecth)}]\\
  &\quad-(\vakpd/\scalh^2)[\dprod{\unitplz}{(-\frkyz\vecth)}]
  \beqref{gxpeed2d}\text{ \& }\eqnref{gpath25a}
\end{split}
\nonumber\\
\begin{split}
&=\scalr\vakpg\epsvc+\vakph\dltvc-(\frkyx/\scalr)\vsiga-\epsvb\frkyr\vsigc-\vakpb\vsigb
  +\vrhoa(\frkys/\scalh^2)(\dltvc\dltvb+\scalq\dltvc\epsva-\efktc\dltva)
  +(\vakpd/\scalh^2)\frkyz\dltvc\\
  &\quad\beqref{ogrv1a}\text{ \& }\eqnref{gxpeed1a}
\end{split}
\nonumber\\
&=\scalr\vakpg\epsvc+\vakph\dltvc-(\frkyx/\scalr)\vsiga-\epsvb\frkyr\vsigc-\vakpb\vsigb
  +\vrhoa(\frkys/\scalh^2)[\dltvc(\dltvb+\scalq\epsva)-\efktc\dltva]+(\vakpd/\scalh^2)\frkyz\dltvc
\nonumber\\
&=\scalr\vakpg\epsvc+\vakph\dltvc-(\frkyx/\scalr)\vsiga-\epsvb\frkyr\vsigc-\vakpb\vsigb
  +\vrhoa(\frkys/\scalh^2)(\dltvc\efkta-\efktc\dltva)+(\vakpd/\scalh^2)\frkyz\dltvc\beqref{gxpeed1b}
\nonumber\\
&=\vakpw\beqref{gpath1n}
\end{align}
\begin{align}\label{gpath26c}
&\dprod{\vectz}{(\vscrp+\vscrq)}\nonumber\\
\begin{split}
&=\dprod{\vectz}{[}\vakpg\vectr+\vakph\vecth
  +\frkyx(\cprod{\unitkap}{\unitpos})
  +\frkyu(\cprod{\unitkap}{\unitplz})
  +\epsvb\frkyr(\cprod{\unitkap}{\vectz})
  +\epsvb\frkyn(\cprod{\unitplz}{\vectz})
  +\vakpa(\cprod{\vectr}{\unitplz})\\
  &\quad+\vakpb(\cprod{\vectr}{\vectz})
  -\vrhoa\frkys(\cprod{\unitkap}{\vectu})
  -\vakpc(\cprod{\unitplz}{\vectu})
  -\vakpd(\cprod{\vectz}{\vectu})]\beqref{gpath21d}
\end{split}
\nonumber\\
\begin{split}
&=\vakpg(\dprod{\vectz}{\vectr})
  +\vakph(\dprod{\vectz}{\vecth})
  +\frkyx[\dprod{\vectz}{(\cprod{\unitkap}{\unitpos})}]
  +\frkyu[\dprod{\vectz}{(\cprod{\unitkap}{\unitplz})}]
  +\epsvb\frkyr[\dprod{\vectz}{(\cprod{\unitkap}{\vectz})}]
  +\epsvb\frkyn[\dprod{\vectz}{(\cprod{\unitplz}{\vectz})}]\\
  &\quad+\vakpa[\dprod{\vectz}{(\cprod{\vectr}{\unitplz})}]
  +\vakpb[\dprod{\vectz}{(\cprod{\vectr}{\vectz})}]
  -\vrhoa\frkys[\dprod{\vectz}{(\cprod{\unitkap}{\vectu})}]
  -\vakpc[\dprod{\vectz}{(\cprod{\unitplz}{\vectu})}]
  -\vakpd[\dprod{\vectz}{(\cprod{\vectz}{\vectu})}]
\end{split}
\nonumber\\
\begin{split}
&=\scalr\vakpg\epsvd-\frkyx\dltvf+\frkyu\vsigc+\vakpa\vsigb
  +\vrhoa\frkys[\dprod{\unitkap}{(\cprod{\vectz}{\vectu})}]\\
  &\quad+\vakpc[\dprod{\unitplz}{(\cprod{\vectz}{\vectu})}]
  \beqref{main5c}, \eqnref{ogrv1a}, \eqnref{gxpeed1a}, \eqnref{gpath1a}\text{ \& }\eqnref{alg4}
\end{split}
\nonumber\\
&=\scalr\vakpg\epsvd-\frkyx\dltvf+\frkyu\vsigc+\vakpa\vsigb
  +\vrhoa(\frkys/\scalh^2)[\dprod{\unitkap}{(-\frkyz\vecth)}]
  +(\vakpc/\scalh^2)[\dprod{\unitplz}{(-\frkyz\vecth)}]
  \beqref{gpath25a}
\nonumber\\
&=\scalr\vakpg\epsvd-\frkyx\dltvf+\frkyu\vsigc+\vakpa\vsigb-\vrhoa(\frkys/\scalh^2)(\frkyz\dltva)
  -(\vakpc/\scalh^2)(\frkyz\dltvc)\beqref{gxpeed1a}
\nonumber\\
&=\scalr\vakpg\epsvd-\frkyx\dltvf+\frkyu\vsigc+\vakpa\vsigb-(\frkyz/\scalh^2)(\frkys\vrhoa\dltva+\vakpc\dltvc)
\nonumber\\
&=\vakpx\beqref{gpath1n}
\end{align}
\begin{align*}
&\dprod{(\cprod{\vectr}{\vecth})}{(\vscrp+\vscrq)}\nonumber\\
\begin{split}
&=\dprod{(\cprod{\vectr}{\vecth})}{[}\vakpg\vectr+\vakph\vecth
  +\frkyx(\cprod{\unitkap}{\unitpos})
  +\frkyu(\cprod{\unitkap}{\unitplz})
  +\epsvb\frkyr(\cprod{\unitkap}{\vectz})
  +\epsvb\frkyn(\cprod{\unitplz}{\vectz})
  +\vakpa(\cprod{\vectr}{\unitplz})\\
  &\quad+\vakpb(\cprod{\vectr}{\vectz})
  -\vrhoa\frkys(\cprod{\unitkap}{\vectu})
  -\vakpc(\cprod{\unitplz}{\vectu})
  -\vakpd(\cprod{\vectz}{\vectu})]\beqref{gpath21d}
\end{split}
\nonumber\\
\begin{split}
&=\vakpg[\dprod{\vectr}{(\cprod{\vectr}{\vecth})}]
  +\vakph[\dprod{\vecth}{(\cprod{\vectr}{\vecth})}]
  +\frkyx[\dprod{(\cprod{\vectr}{\vecth})}{(\cprod{\unitkap}{\unitpos})}]
  +\frkyu[\dprod{(\cprod{\vectr}{\vecth})}{(\cprod{\unitkap}{\unitplz})}]\\
  &\quad+\epsvb\frkyr[\dprod{(\cprod{\vectr}{\vecth})}{(\cprod{\unitkap}{\vectz})}]
  +\epsvb\frkyn[\dprod{(\cprod{\vectr}{\vecth})}{(\cprod{\unitplz}{\vectz})}]
  +\vakpa[\dprod{(\cprod{\vectr}{\vecth})}{(\cprod{\vectr}{\unitplz})}]\\
  &\quad+\vakpb[\dprod{(\cprod{\vectr}{\vecth})}{(\cprod{\vectr}{\vectz})}]
  -\vrhoa\frkys[\dprod{(\cprod{\vectr}{\vecth})}{(\cprod{\unitkap}{\vectu})}]
  -\vakpc[\dprod{(\cprod{\vectr}{\vecth})}{(\cprod{\unitplz}{\vectu})}]
  -\vakpd[\dprod{(\cprod{\vectr}{\vecth})}{(\cprod{\vectz}{\vectu})}]
\end{split}
\end{align*}
\begin{align}\label{gpath26d}
\begin{split}
&=\frkyx[(\dprod{\vectr}{\unitkap})(\dprod{\vecth}{\unitpos})-(\dprod{\vectr}{\unitpos})(\dprod{\vecth}{\unitkap})]
  +\frkyu[(\dprod{\vectr}{\unitkap})(\dprod{\vecth}{\unitplz})-(\dprod{\vectr}{\unitplz})(\dprod{\vecth}{\unitkap})]\\
  &\quad+\epsvb\frkyr[(\dprod{\vectr}{\unitkap})(\dprod{\vecth}{\vectz})-(\dprod{\vectr}{\vectz})(\dprod{\vecth}{\unitkap})]
  +\epsvb\frkyn[(\dprod{\vectr}{\unitplz})(\dprod{\vecth}{\vectz})-(\dprod{\vectr}{\vectz})(\dprod{\vecth}{\unitplz})]\\
  &\quad+\vakpa[\scalr^2(\dprod{\vecth}{\unitplz})-(\dprod{\vectr}{\unitplz})(\dprod{\vecth}{\vectr})]
  +\vakpb[\scalr^2(\dprod{\vecth}{\vectz})-(\dprod{\vectr}{\vectz})(\dprod{\vecth}{\vectr})]\\
  &\quad-\vrhoa\frkys[(\dprod{\vectr}{\unitkap})(\dprod{\vecth}{\vectu})-(\dprod{\vectr}{\vectu})(\dprod{\vecth}{\unitkap})]
  -\vakpc[(\dprod{\vectr}{\unitplz})(\dprod{\vecth}{\vectu})-(\dprod{\vectr}{\vectu})(\dprod{\vecth}{\unitplz})]\\
  &\quad-\vakpd[(\dprod{\vectr}{\vectz})(\dprod{\vecth}{\vectu})-(\dprod{\vectr}{\vectu})(\dprod{\vecth}{\vectz})]
  \beqref{alg2}
\end{split}
\nonumber\\
\begin{split}
&=\frkyx(-\scalr\dltva)+\frkyu(\scalr\epsva\dltvc-\scalr\epsvc\dltva)+\epsvb\frkyr(-\scalr\epsvd\dltva)
  +\epsvb\frkyn(-\scalr\epsvd\dltvc)+\vakpa(\scalr^2\dltvc)\\
  &\quad-\vrhoa\frkys[-\scalr(\scalr\vrhob)\dltva]
  -\vakpc[-\scalr(\scalr\vrhob)\dltvc]\beqref{main5c}, \eqnref{gxpeed1a}\text{ \& }\eqnref{gpath7b}
\end{split}
\nonumber\\
&=-\scalr\frkyx\dltva+\scalr\frkyu(\epsva\dltvc-\epsvc\dltva)
  -\scalr\epsvb\epsvd(\frkyr\dltva+\frkyn\dltvc)
  +\scalr^2(\vakpa\dltvc+\vrhoa\frkys\vrhob\dltva+\vakpc\vrhob\dltvc)
\nonumber\\
&=\vakpy\beqref{gpath1n}
\end{align}
\begin{align*}
&\dprod{(\cprod{\vectz}{\vecth})}{(\vscrp+\vscrq)}\nonumber\\
\begin{split}
&=\dprod{(\cprod{\vectz}{\vecth})}{[}\vakpg\vectr+\vakph\vecth
  +\frkyx(\cprod{\unitkap}{\unitpos})
  +\frkyu(\cprod{\unitkap}{\unitplz})
  +\epsvb\frkyr(\cprod{\unitkap}{\vectz})
  +\epsvb\frkyn(\cprod{\unitplz}{\vectz})
  +\vakpa(\cprod{\vectr}{\unitplz})\\
  &\quad+\vakpb(\cprod{\vectr}{\vectz})
  -\vrhoa\frkys(\cprod{\unitkap}{\vectu})
  -\vakpc(\cprod{\unitplz}{\vectu})
  -\vakpd(\cprod{\vectz}{\vectu})]\beqref{gpath21d}
\end{split}
\nonumber\\
\begin{split}
&=\vakpg[\dprod{\vectr}{(\cprod{\vectz}{\vecth})}]
  +\vakph[\dprod{\vecth}{(\cprod{\vectz}{\vecth})}]
  +\frkyx[\dprod{(\cprod{\vectz}{\vecth})}{(\cprod{\unitkap}{\unitpos})}]
  +\frkyu[\dprod{(\cprod{\vectz}{\vecth})}{(\cprod{\unitkap}{\unitplz})}]\\
  &\quad+\epsvb\frkyr[\dprod{(\cprod{\vectz}{\vecth})}{(\cprod{\unitkap}{\vectz})}]
  +\epsvb\frkyn[\dprod{(\cprod{\vectz}{\vecth})}{(\cprod{\unitplz}{\vectz})}]
  +\vakpa[\dprod{(\cprod{\vectz}{\vecth})}{(\cprod{\vectr}{\unitplz})}]\\
  &\quad+\vakpb[\dprod{(\cprod{\vectz}{\vecth})}{(\cprod{\vectr}{\vectz})}]
  -\vrhoa\frkys[\dprod{(\cprod{\vectz}{\vecth})}{(\cprod{\unitkap}{\vectu})}]
  -\vakpc[\dprod{(\cprod{\vectz}{\vecth})}{(\cprod{\unitplz}{\vectu})}]\\
  &\quad-\vakpd[\dprod{(\cprod{\vectz}{\vecth})}{(\cprod{\vectz}{\vectu})}]
\end{split}
\end{align*}
\begin{align}\label{gpath26e}
\begin{split}
&=\scalr\vakpg\epsvh
  +\frkyx[(\dprod{\vectz}{\unitkap})(\dprod{\vecth}{\unitpos})-(\dprod{\vectz}{\unitpos})(\dprod{\vecth}{\unitkap})]
  +\frkyu[(\dprod{\vectz}{\unitkap})(\dprod{\vecth}{\unitplz})-(\dprod{\vectz}{\unitplz})(\dprod{\vecth}{\unitkap})]\\
  &\quad+\epsvb\frkyr[(\dprod{\vectz}{\unitkap})(\dprod{\vecth}{\vectz})-\scalz^2(\dprod{\vecth}{\unitkap})]
  +\epsvb\frkyn[(\dprod{\vectz}{\unitplz})(\dprod{\vecth}{\vectz})-\scalz^2(\dprod{\vecth}{\unitplz})]\\
  &\quad+\vakpa[(\dprod{\vectz}{\vectr})(\dprod{\vecth}{\unitplz})-(\dprod{\vectz}{\unitplz})(\dprod{\vecth}{\vectr})]
  +\vakpb[(\dprod{\vectz}{\vectr})(\dprod{\vecth}{\vectz})-\scalz^2(\dprod{\vecth}{\vectr})]\\
  &\quad-\vrhoa\frkys[(\dprod{\vectz}{\unitkap})(\dprod{\vecth}{\vectu})-(\dprod{\vectz}{\vectu})(\dprod{\vecth}{\unitkap})]
  -\vakpc[(\dprod{\vectz}{\unitplz})(\dprod{\vecth}{\vectu})-(\dprod{\vectz}{\vectu})(\dprod{\vecth}{\unitplz})]\\
  &\quad-\vakpd[\scalz^2(\dprod{\vecth}{\vectu})-(\dprod{\vectz}{\vectu})(\dprod{\vecth}{\vectz})]
  \beqref{ogrv1a}\text{ \& }\eqnref{alg2}
\end{split}
\nonumber\\
\begin{split}
&=\scalr\vakpg\epsvh+\frkyx(-\epsvd\dltva)+\frkyu(\dltvb\dltvc-\dltvd\dltva)
  +\epsvb\frkyr(-\scalz^2\dltva)+\epsvb\frkyn(-\scalz^2\dltvc)+\vakpa(\scalr\epsvd\dltvc)\\
  &\quad-\vrhoa\frkys[-(-\scalr\vrhob\vphib)\dltva]-\vakpc[-(-\scalr\vrhob\vphib)\dltvc]
  \beqref{main5c}, \eqnref{ogrv1a}, \eqnref{gxpeed1a}\text{ \& }\eqnref{gpath7d}
\end{split}
\nonumber\\
&=\scalr\vakpg\epsvh-\frkyx\epsvd\dltva+\frkyu(\dltvb\dltvc-\dltvd\dltva)-\scalz^2\epsvb(\frkyr\dltva+\frkyn\dltvc)
  +\scalr(\vakpa\epsvd\dltvc-\vrhoa\frkys\vrhob\vphib\dltva-\vakpc\vrhob\vphib\dltvc)
\nonumber\\
&=-\frkyx\epsvd\dltva+\frkyu(\dltvb\dltvc-\dltvd\dltva)-\scalz^2\epsvb(\frkyr\dltva+\frkyn\dltvc)
  +\scalr(\vakpg\epsvh+\vakpa\epsvd\dltvc)-\scalr\vrhob\vphib(\vrhoa\frkys\dltva+\vakpc\dltvc)
\nonumber\\
&=\vakpz\beqref{gpath1n}
\end{align}
\begin{align}\label{gpath26f}
\dprod{\vectu}{(\vscrp+\vscrq)}
&=\scalh^{-2}[(\dprod{\cprod{\vectz}{\vecth}+\scalq\cprod{\unitpos}{\vecth})}{(\vscrp+\vscrq)}]
  \beqref{main5c}\nonumber\\
&=\scalh^{-2}[\dprod{(\cprod{\vectz}{\vecth})}{(\vscrp+\vscrq)}+
  (\scalq/\scalr)\dprod{(\cprod{\vectr}{\vecth})}{(\vscrp+\vscrq)}]
  \nonumber\\
&=\scalh^{-2}(\vakpz+\vphib\vakpy)\beqref{ogrv1b}, \eqnref{gpath26d}\text{ \& }\eqnref{gpath26e}\nonumber\\
&=\parva\beqref{gpath1o}.
\end{align}
\end{subequations}
Consequently, we obtain
\begin{subequations}\label{gpath27}
\begin{align}\label{gpath27a}
\alepha
&=\dprod{\unitkap}{(\vscrp+\vscrq)}\beqref{kpath2d}\nonumber\\
&=\vakpv\beqref{gpath26a}
\end{align}
\begin{align}\label{gpath27b}
\alephb
&=\dprod{\vecta}{(\vscrp+\vscrq)}\beqref{kpath2d}\nonumber\\
&=\dprod{(-\vrhoa\vectr)}{(\vscrp+\vscrq)}\beqref{main5a}\text{ \& }\eqnref{gpath1b}\nonumber\\
&=-\vrhoa\vakpj\beqref{gpath23a}
\end{align}
\begin{align}\label{gpath27c}
\alephc
&=\dprod{\fdota}{(\vscrp+\vscrq)}\beqref{kpath2d}\nonumber\\
&=\dprod{(-\vrhoa\vectu+3\vrhoa\vrhob\vectr)}{(\vscrp+\vscrq)}\beqref{gpath6a}\nonumber\\
&=-\vrhoa[\dprod{\vectu}{(\vscrp+\vscrq)}]+3\vrhoa\vrhob[\dprod{\vectr}{(\vscrp+\vscrq)}]\nonumber\\
&=-\vrhoa\parva+3\vrhoa\vrhob\vakpj\beqref{gpath26f}\text{ \& }\eqnref{gpath23a}\nonumber\\
&=\vrhoa(3\vrhob\vakpj-\parva)
\end{align}
\begin{align}\label{gpath27d}
\alephd
&=\dprod{\ffdota}{(\vscrp+\vscrq)}\beqref{kpath2d}\nonumber\\
&=\dprod{(6\vrhoa\vrhob\vectu-\vrhoa\vrhol\vectr)}{(\vscrp+\vscrq)}\beqref{gpath6b}\nonumber\\
&=6\vrhoa\vrhob[\dprod{\vectu}{(\vscrp+\vscrq)}]-\vrhoa\vrhol[\dprod{\vectr}{(\vscrp+\vscrq)}]\nonumber\\
&=6\vrhoa\vrhob\parva-\vrhoa\vrhol\vakpj\beqref{gpath26f}\text{ \& }\eqnref{gpath23a}\nonumber\\
&=\vrhoa(6\vrhob\parva-\vrhol\vakpj)
\end{align}
\begin{align}\label{gpath27e}
\alephe
&=\dprod{\fffdota}{(\vscrp+\vscrq)}\beqref{kpath2d}\nonumber\\
&=\dprod{(\vrhom\vectu+\vrhon\vectr)}{(\vscrp+\vscrq)}\beqref{gpath6c}\nonumber\\
&=\vrhom[\dprod{\vectu}{(\vscrp+\vscrq)}]+\vrhon[\dprod{\vectr}{(\vscrp+\vscrq)}]\nonumber\\
&=\vrhom\parva+\vrhon\vakpj\beqref{gpath26f}\text{ \& }\eqnref{gpath23a}
\end{align}
\begin{align}\label{gpath27f}
\alephf
&=\dprod{\fffdote}{(\vscrp+\vscrq)}\beqref{kpath2d}\nonumber\\
&=\dprod{[\ethvq\unitplz-\ethvs\vectz-\scalq\ethvs\unitpos+\ethvr(\cprod{\vectr}{\vecth})]}{(\vscrp+\vscrq)}
  \beqref{gpath16c}\nonumber\\
&=\ethvq[\dprod{\unitplz}{(\vscrp+\vscrq)}]
  -\ethvs[\dprod{\vectz}{(\vscrp+\vscrq)}]
  -\scalq\ethvs[\dprod{\unitpos}{(\vscrp+\vscrq)}]
  +\ethvr[\dprod{(\cprod{\vectr}{\vecth})}{(\vscrp+\vscrq)}]
  \nonumber\\
&=\ethvq\vakpw-\ethvs\vakpx-(\scalq/\scalr)\ethvs[\dprod{\vectr}{(\vscrp+\vscrq)}]
  +\ethvr\vakpy\beqref{gpath26}\nonumber\\
&=\ethvq\vakpw-\ethvs\vakpx-\vphib\ethvs\vakpj+\ethvr\vakpy
  \beqref{gpath23a}\text{ \& }\eqnref{ogrv1b}\nonumber\\
&=\parvb\beqref{gpath1o}
\end{align}
\end{subequations}
from which we derive
\begin{align}\label{gpath28}
&\fffdot{\cdkt}\alepha+\fffdot{\rhorep}\alephb+3\ffdot{\rhorep}\alephc+\vbbc\alephd+\rhorep\alephe+\alephf
\nonumber\\
\begin{split}
&=\fffdot{\cdkt}(\vakpv)+\fffdot{\rhorep}(-\vrhoa\vakpj)+3\ffdot{\rhorep}[\vrhoa(3\vrhob\vakpj-\parva)]
  +\vbbc[\vrhoa(6\vrhob\parva-\vrhol\vakpj)]+\rhorep(\vrhom\parva+\vrhon\vakpj)+\parvb\\
  &\quad\beqref{gpath27}
\end{split}
\nonumber\\
\begin{split}
&=\vakpv(\fffdot{\cdkt})-\vrhoa\vakpj(\fffdot{\rhorep})+3\vrhoa(3\vrhob\vakpj-\parva)(\ffdot{\rhorep})
  +\vrhoa(6\vrhob\parva-\vrhol\vakpj)(\vbbc)+\rhorep(\vrhom\parva+\vrhon\vakpj)+\parvb
\end{split}
\nonumber\\
\begin{split}
&=\vakpv(\frkyc)-\vrhoa\vakpj(\ethvj)+3\vrhoa(3\vrhob\vakpj-\parva)(\ethvi)
  +\vrhoa(6\vrhob\parva-\vrhol\vakpj)(\frkyf)+\rhorep(\vrhom\parva+\vrhon\vakpj)+\parvb\\
  &\quad\beqref{gpath18}, \eqnref{gpath12}\text{ \& }\eqnref{gpath19c}
\end{split}
\nonumber\\
&=\vakpv\frkyc+\vrhoa[-\vakpj\ethvj+3\ethvi(3\vrhob\vakpj-\parva)+\frkyf(6\vrhob\parva-\vrhol\vakpj)]
  +\rhorep(\vrhom\parva+\vrhon\vakpj)+\parvb\nonumber\\
&=\parvb+\vakpv\frkyc+\vrhoa\parvc+\rhorep(\vrhom\parva+\vrhon\vakpj)
  \beqref{gpath1o}.
\end{align}
Moreover, we derive
\begin{subequations}\label{gpath29}
\begin{align}\label{gpath29a}
\vscrr
&=\frkya\unitkap+\ethvn\unitplz-\epsvb\vphif\vectz-\rhorep\vrhoa\vectu-\vakpf\vectr
  -\kaprep\epsvb\ethvk(\cprod{\vectr}{\vecth})\beqref{gpath21c}\nonumber\\
&=\frkya\unitkap+\ethvn\unitplz-\epsvb\vphif\vectz-\vakpf\vectr
  -\scalh^{-2}\rhorep\vrhoa(\cprod{\vectz}{\vecth}+\scalq\cprod{\unitpos}{\vecth})
  -\kaprep\epsvb\ethvk(\cprod{\vectr}{\vecth})\beqref{main5c}\nonumber\\
&=\frkya\unitkap+\ethvn\unitplz-\epsvb\vphif\vectz-\vakpf\vectr
  -\scalh^{-2}\rhorep\vrhoa(\cprod{\vectz}{\vecth})-\scalh^{-2}\rhorep\vrhoa(\scalq/\scalr)(\cprod{\vectr}{\vecth})
  -\kaprep\epsvb\ethvk(\cprod{\vectr}{\vecth})\nonumber\\
&=\frkya\unitkap+\ethvn\unitplz-\epsvb\vphif\vectz-\vakpf\vectr
  -\scalh^{-2}\rhorep\vrhoa(\cprod{\vectz}{\vecth})
  -(\scalh^{-2}\rhorep\vrhoa\vphib+\kaprep\epsvb\ethvk)(\cprod{\vectr}{\vecth})
  \beqref{ogrv1b}\nonumber\\
&=\frkya\unitkap+\ethvn\unitplz-\epsvb\vphif\vectz-\vakpf\vectr
  -\scalh^{-2}\rhorep\vrhoa(\cprod{\vectz}{\vecth})-\parvd(\cprod{\vectr}{\vecth})
  \beqref{gpath1o}
\end{align}
\begin{align*}
\begin{split}
\vscrp+\vscrq
&=\vakpg\vectr+\vakph\vecth+\frkyx(\cprod{\unitkap}{\unitpos})+\frkyu(\cprod{\unitkap}{\unitplz})
  +\epsvb\frkyr(\cprod{\unitkap}{\vectz})+\epsvb\frkyn(\cprod{\unitplz}{\vectz})+\vakpa(\cprod{\vectr}{\unitplz})\\
  &\quad+\vakpb(\cprod{\vectr}{\vectz})-\vrhoa\frkys(\cprod{\unitkap}{\vectu})
  -\vakpc(\cprod{\unitplz}{\vectu})-\vakpd(\cprod{\vectz}{\vectu})
  \beqref{gpath21d}
\end{split}
\nonumber\\
\begin{split}
&=\vakpg\vectr+\vakph\vecth+\frkyx(\cprod{\unitkap}{\unitpos})+\frkyu(\cprod{\unitkap}{\unitplz})
  +\epsvb\frkyr(\cprod{\unitkap}{\vectz})+\epsvb\frkyn(\cprod{\unitplz}{\vectz})+\vakpa(\cprod{\vectr}{\unitplz})\\
  &\quad+\vakpb(\cprod{\vectr}{\vectz})-\vrhoa\frkys[\scalh^{-2}(\dltva\vectz+\scalq\dltva\unitpos-\efkta\vecth)]
  -\vakpc[\scalh^{-2}(\dltvc\vectz+\scalq\dltvc\unitpos-\efktc\vecth)]\\
  &\quad-\vakpd[-\scalh^{-2}\frkyz\vecth]
  \beqref{gxpeed2}\text{ \& }\eqnref{gpath25a}
\end{split}
\nonumber\\
\begin{split}
&=\vakpg\vectr+\vakph\vecth+\frkyx(\cprod{\unitkap}{\unitpos})+\frkyu(\cprod{\unitkap}{\unitplz})
  +\epsvb\frkyr(\cprod{\unitkap}{\vectz})+\epsvb\frkyn(\cprod{\unitplz}{\vectz})+\vakpa(\cprod{\vectr}{\unitplz})\\
  &\quad+\vakpb(\cprod{\vectr}{\vectz})-\scalh^{-2}\vrhoa\frkys\dltva\vectz-\scalq\scalh^{-2}\vrhoa\frkys\dltva\unitpos
  +\scalh^{-2}\vrhoa\frkys\efkta\vecth-\scalh^{-2}\vakpc\dltvc\vectz-\scalq\scalh^{-2}\vakpc\dltvc\unitpos\\
  &\quad+\scalh^{-2}\vakpc\efktc\vecth+\scalh^{-2}\vakpd\frkyz\vecth
\end{split}
\end{align*}
\begin{align}\label{gpath29b}
\begin{split}
&=\scalr\vakpg\unitpos-\scalq\scalh^{-2}\vrhoa\frkys\dltva\unitpos-\scalq\scalh^{-2}\vakpc\dltvc\unitpos
  +\vakph\vecth+\scalh^{-2}\vrhoa\frkys\efkta\vecth+\scalh^{-2}\vakpc\efktc\vecth+\scalh^{-2}\vakpd\frkyz\vecth\\
  &\quad-\scalh^{-2}\vrhoa\frkys\dltva\vectz-\scalh^{-2}\vakpc\dltvc\vectz
  +\frkyx(\cprod{\unitkap}{\unitpos})+\frkyu(\cprod{\unitkap}{\unitplz})
  +\epsvb\frkyr(\cprod{\unitkap}{\vectz})+\epsvb\frkyn(\cprod{\unitplz}{\vectz})\\
  &\quad+\vakpa(\cprod{\vectr}{\unitplz})+\vakpb(\cprod{\vectr}{\vectz})
\end{split}
\nonumber\\
\begin{split}
&=[\scalr\vakpg-\scalq\scalh^{-2}(\vrhoa\frkys\dltva+\vakpc\dltvc)]\unitpos
  +[\vakph+\scalh^{-2}(\vrhoa\frkys\efkta+\vakpc\efktc+\vakpd\frkyz)]\vecth\\
  &\quad-\scalh^{-2}(\vrhoa\frkys\dltva+\vakpc\dltvc)\vectz
  +\frkyx(\cprod{\unitkap}{\unitpos})+\frkyu(\cprod{\unitkap}{\unitplz})
  +\epsvb\frkyr(\cprod{\unitkap}{\vectz})+\epsvb\frkyn(\cprod{\unitplz}{\vectz})\\
  &\quad+\vakpa(\cprod{\vectr}{\unitplz})+\vakpb(\cprod{\vectr}{\vectz})
\end{split}
\nonumber\\
\begin{split}
&=\parvg\unitpos+\parvf\vecth-\parve\vectz+\frkyx(\cprod{\unitkap}{\unitpos})+\frkyu(\cprod{\unitkap}{\unitplz})
  +\epsvb\frkyr(\cprod{\unitkap}{\vectz})+\epsvb\frkyn(\cprod{\unitplz}{\vectz})\\
  &\quad+\vakpa(\cprod{\vectr}{\unitplz})+\vakpb(\cprod{\vectr}{\vectz}).
\end{split}
\end{align}
\end{subequations}

\subart{Results of the computations}
Substituting \eqnref{gpath29}, \eqnref{gpath28}, \eqnref{gpath24} and \eqnref{gpath22}
into \eqnref{kpath11} gives
\begin{subequations}\label{gpath30}
\begin{align}\label{gpath30a}
\bbk=\plusmin\frac{\vakpu}{(\vakpi)^3},\quad
\bbt=\frac{\parvb+\vakpv\frkyc+\vrhoa\parvc+\rhorep(\vrhom\parva+\vrhon\vakpj)}{(\vakpu)^2}
\end{align}
\begin{align}\label{gpath30b}
\begin{aligned}
\frt&=\frac{1}{\vakpi}\biggl[\frkya\unitkap+\ethvn\unitplz-\epsvb\vphif\vectz-\vakpf\vectr
  -\vrhoa(\rhorep/\scalh^2)(\cprod{\vectz}{\vecth})-\parvd(\cprod{\vectr}{\vecth})\biggr]\\
\frb&=\frac{1}{\vakpu}\biggl[\parvg\unitpos+\parvf\vecth-\parve\vectz+\frkyx(\cprod{\unitkap}{\unitpos})
  +\frkyu(\cprod{\unitkap}{\unitplz})+\epsvb\frkyr(\cprod{\unitkap}{\vectz})+\epsvb\frkyn(\cprod{\unitplz}{\vectz})\\
  &\qquad+\vakpa(\cprod{\vectr}{\unitplz})+\vakpb(\cprod{\vectr}{\vectz})\biggr]
\end{aligned}
\end{align}
\end{subequations}
as the complete set of equations describing the apparent path of the light source for a gravitating observer.

\art{Apparent geometry of obliquated rays}
To evaluate \eqnref{kray4} for a gravitating observer, we introduce, in addition to
\eqnref{ogrv1}, \eqnref{gxpeed1} and \eqnref{gpath1}, the quantities
\begin{subequations}\label{gray1}
\begin{align}\label{gray1a}
\begin{split}
&\veka=\scalh^{-2}[\efktc\ethvn+\scalr\kaprep\epsvb\epsvh\ethvk-\epsvb\vphif(\frkyz+\scalh^2\vphib)],\quad
\vekb=\scalh^{-2}\dltvc\ethvn\\
&\vekc=\ethvn+\kaprep\epsve\ethvk,\quad
\vekd=\scalr\epsvb\vphif(\epsvc\vekc-\epsvb\vphif\efktb),\quad
\veke=\vrhoa/\scalh^2,\quad
\vekf=\epsve\vphif\dltvc\veke\\
&\vekg=\vphif(2\vphib\vrhob\epsvb+\veke\epsve\efktc),\quad
\vekh=\cdkt\frkyd-\rhorep\frkya,\quad
\veki=\cdkt\dltva(\rhorep\veke\vphib+\kaprep\epsvb\ethvk)\\
&\vekj=\rhorep(\cdkt\veke\efkta-\rhorep\vrhoa^2)+\kaprep\epsvb\ethvk(\scalr\cdkt\epsva-\rhorep\vphib),\quad
\vekk=\vrhoa(3\rhorep\cdkt\vrhob-\vekh)-\cdkt\epsvb\vphif\vphib\\
&\vekl=\vekb-\rhorep\vekf+\frkya(\dltva/\scalh^2),\quad
\vekm=\frkya\dltva(\vphib/\scalh^2)-\epsvb\vphif(\dltva\frkya+\dltvc\vekc)+\vphib(\vekb-\rhorep\vekf)\\
&\vekn=\vekd-\veka+\scalh^{-2}(\vphia\efktb\frkyd-\frkya\efkta)+\scalr\epsvb\vphif(\epsva\frkya-\scalr\vrhoa\frkyd)
  +\rhorep(\vekg-3\vrhoa\vrhob)\\
&\veko=\epsve\vphif[\vrhoa(3\rhorep\vrhob-\frkyd)-\epsvb\vphib\vphif],\quad
\vekp=\vekj+\vekn,\quad
\vekq=\vekl-\rhorep\cdkt\veke\dltva\\
&\vekr=\vekm-\veki,\quad
\veks=\cdkt\ethvn-\epsve\vphif\frkya,\quad
\vekt=\veko+\rhorep\vrhoa\ethvn
\end{split}
\end{align}
\begin{align}\label{gray1b}
\begin{split}
&\veku=\vekp\scalh^2+\vekk\epsve-\cdkt\epsvb\vphif\epsvg-\veks\dltve+\vekt\epsvf
  +\scalr\rhorep\vrhoa\epsvb\vphif\epsvh-\epsvb\epsve\vphif^2\epsvi\\
&\vekv=\vekq\scalz^2+\scalr(\vekr\epsvd-\vekk\dltvf)+\veks\vsigc-\vekt\vsigb\\
&\vekw=\scalr(\vekq\epsvd+\scalr\vekr-\cdkt\epsvb\vphif\dltvf)+\veks\vsiga-\epsvb\epsve\vsigb\vphif^2\\
&\vekx=\vekp\epsve-\scalr\vekq\dltvf+\scalr^2\vekk(1-\epsva^2)-\scalr\cdkt\epsvb\vphif(\epsvd-\dltvb\epsva)
  +\scalr\veks(\epsvc-\epsvb\epsva)\\
  &\qquad+\scalr^2\vekt(\epsvb-\epsva\epsvc)+\scalr^2\rhorep\vrhoa\epsvb\vphif(\epsva\epsvd-\dltvb)
  -\scalr\epsvb\epsve\vphif^2(\epsvb\epsvd-\dltvb\epsvc)\\
&\veky=\vekp\epsvg+\scalr\vekr\dltvf+\scalr\vekk(\epsvd-\epsva\dltvb)-\cdkt\epsvb\vphif(\scalz^2-\dltvb^2)
  +\veks(\dltvd-\epsvb\dltvb)\\
  &\qquad+\scalr\vekt(\epsvb\epsvd-\epsva\dltvd)
  +\scalr\rhorep\vrhoa\epsvb\vphif(\scalz^2\epsva-\dltvb\epsvd)
  -\epsvb\epsve\vphif^2(\scalz^2\epsvb-\dltvb\dltvd)\\
&\vekz=-\vekp\dltve+\vekq\vsigc+\vekr\vsiga+\scalr\vekk(\epsvc-\epsva\epsvb)
  -\cdkt\epsvb\vphif(\dltvd-\dltvb\epsvb)+\veks(1-\epsvb^2)\\
  &\qquad+\scalr\vekt(\epsvb\epsvc-\epsva)
  +\scalr\rhorep\vrhoa\epsvb\vphif(\epsva\dltvd-\dltvb\epsvc)
  -\epsvb\epsve\vphif^2(\epsvb\dltvd-\dltvb)
\end{split}
\end{align}
\begin{align}\label{gray1c}
\begin{split}
&\vpsa=\vekp\epsvf-\vekq\vsigb+\scalr^2\vekk(\epsvb-\epsvc\epsva)
  -\scalr\cdkt\epsvb\vphif(\epsvb\epsvd-\dltvd\epsva)+\scalr\veks(\epsvb\epsvc-\epsva)\\
  &\qquad+\scalr^2\vekt(1-\epsvc^2)-\scalr\epsvb\epsve\vphif^2(\epsvd-\dltvd\epsvc)
  +\scalr^2\rhorep\vrhoa\epsvb\vphif(\epsvc\epsvd-\dltvd)\\
&\vpsb=\vekp\epsvi+\vekr\vsigb+\scalr\vekk(\epsvb\epsvd-\epsvc\dltvb)
  -\cdkt\epsvb\vphif(\scalz^2\epsvb-\dltvd\dltvb)+\veks(\epsvb\dltvd-\dltvb)\\
  &\qquad+\scalr\vekt(\epsvd-\epsvc\dltvd)-\epsvb\epsve\vphif^2(\scalz^2-\dltvd^2)
  +\scalr\rhorep\vrhoa\epsvb\vphif(\scalz^2\epsvc-\dltvd\epsvd)\\
&\vpsc=\scalr\vekp\epsvh+\scalr^2\vekk(\epsva\epsvd-\dltvb)
  -\scalr\cdkt\epsvb\vphif(\scalz^2\epsva-\epsvd\dltvb)
  +\scalr\veks(\epsva\dltvd-\epsvc\dltvb)\\
  &\qquad+\scalr^2\vekt(\epsvc\epsvd-\dltvd)
  -\scalr\epsvb\epsve\vphif^2(\scalz^2\epsvc-\epsvd\dltvd)
  +\scalr^2\rhorep\vrhoa\epsvb\vphif(\scalz^2-\epsvd^2)\\
&\vpsd=(\vekp\veku+\vekq\vekv+\vekr\vekw+\vekk\vekx-\cdkt\epsvb\vphif\veky+\veks\vekz+\vekt\vpsa\\
  &\qquad-\epsvb\epsve\vphif^2\vpsb+\rhorep\vrhoa\epsvb\vphif\vpsc)^{1/2}\\
\end{split}
\end{align}
\begin{align}\label{gray1d}
\begin{split}
&\vpse=\vekp\dltva+\vekq\dltvb+\scalr\vekr\epsva+\vekt\vsiga-\scalr\rhorep\vrhoa\epsvb\vphif\dltvf
  -\epsvb\epsve\vphif^2\vsigc\\
&\vpsf=-\scalr\vekq\vrhob\vphib+\scalr^2\vekr\vrhob-\vekk\dltva +\epsvb\vphif\frkyz\dltva(\cdkt/\scalh^2)
  -(\veks/\scalh^2)(\efktc\dltva-\dltvc\efkta)\\
  &\qquad-\vekt\dltvc+\epsvb\epsve\vphif^2\frkyz(\dltvc/\scalh^2)\\
&\vpsg=\vekp\dltvc+\vekq\dltvd+\scalr\vekr\epsvc-\vekk\vsiga+\cdkt\epsvb\vphif\vsigc
  -\rhorep\vrhoa\epsvb\vphif\vsigb\\
&\vpsh=\scalr[-\vekq\epsvh-\scalr(\vekk\dltva+\vekt\dltvc)+\veks(\epsva\dltvc-\epsvc\dltva)
  +\epsvb\epsvd\vphif(\cdkt\dltva+\epsve\vphif\dltvc)]\\
&\vpsi=\frkyb\vpse+\ethvo\vpsg+\ethvp\vpsh-\vrhoa[\ethvi\vekw+\frkye(\vpsf-3\vrhob\vekw)+\rhorep(\vrhol\vekw-6\vrhob\vpsf)]\\
  &\qquad+2\kaprep\epsvb\ethvk(\vekv+\vphib\vekw)\\
&\vpsj=\epsvb\vphif-(\vphib/\scalh^2)
\end{split}
\end{align}
\end{subequations}

\subart{Development of equation \eqnref{kray2a}}
In view of the above quantities, we derive
\begin{subequations}\label{gray2}
\begin{align}
\veusa
&=\cprod{\unitkap}{\vectu}\beqref{kray2a}\nonumber\\
&=\scalh^{-2}(\dltva\vectz+\scalq\dltva\unitpos-\efkta\vecth)
  \beqref{gxpeed2a}
\label{gray2a}\\
\veusb
&=\cprod{\unitkap}{\vecta}\beqref{kray2a}\nonumber\\
&=-\vrhoa(\cprod{\unitkap}{\vectr})\beqref{gpath20a}
\label{gray2b}
\end{align}
\begin{align}\label{gray2c}
\veusc
&=\cprod{\unitkap}{\vecte}\beqref{kray2a}\nonumber\\
&=\cprod{\unitkap}{[\fone(\cprod{\vectr}{\vecth})+\ftwo\unitplz]}\beqref{main5b}\nonumber\\
&=\fone[\cprod{\unitkap}{(\cprod{\vectr}{\vecth})}]
  +\ftwo(\cprod{\unitkap}{\unitplz})\nonumber\\
&=\fone[\vectr(\dprod{\unitkap}{\vecth})-\vecth(\dprod{\unitkap}{\vectr})]
  +\ftwo(\cprod{\unitkap}{\unitplz})\beqref{alg1}\nonumber\\
&=\fone(\dltva\vectr-\scalr\epsva\vecth)
  +\ftwo(\cprod{\unitkap}{\unitplz})\beqref{ogrv1a}\text{ \& }\eqnref{gxpeed1a}\nonumber\\
&=\epsvb\vphif(\dltva\vectr-\scalr\epsva\vecth)
  +\epsve\vphif(\cprod{\unitkap}{\unitplz})\beqref{ogrv2g}\text{ \& }\eqnref{ogrv2h}
\end{align}
\begin{align}
\veusd
&=\cprod{\unitkap}{\fdota}\beqref{kray2a}\nonumber\\
&=-\vrhoa(\cprod{\unitkap}{\vectu})+3\vrhoa\vrhob(\cprod{\unitkap}{\vectr})
  \beqref{gpath20b}\nonumber\\
&=(\vrhoa/\scalh^2)(\efkta\vecth-\dltva\vectz-\scalq\dltva\unitpos)
  +3\vrhoa\vrhob(\cprod{\unitkap}{\vectr})\beqref{gxpeed2a}
\label{gray2d}\\
\veuse
&=\cprod{\unitkap}{\fdote}\beqref{kray2a}\nonumber\\
&=\kaprep\scalr\epsva\epsvb\ethvk\vecth-\kaprep\epsvb\ethvk\dltva\vectr
  +\ethvn(\cprod{\unitkap}{\unitplz})-\epsvb\vphif(\cprod{\unitkap}{\vectz})-\scalq\epsvb\vphif(\cprod{\unitkap}{\unitpos})
  \beqref{gpath20d}
\label{gray2e}
\end{align}
\begin{align}
\veusf
&=\cprod{\vecta}{\vectu}\beqref{kray2a}\nonumber\\
&=\scalh^{-2}\vphia\efktb\vecth\beqref{gxpeed2b}
\label{gray2f}\\
\veusg
&=\cprod{\vecta}{\vecte}\beqref{kray2a}\nonumber\\
&=\cprod{(-\vrhoa\vectr)}{[\fone(\cprod{\vectr}{\vecth})+\ftwo\unitplz]}
  \beqref{main5}\text{ \& }\eqnref{gpath1b}\nonumber\\
&=-\vrhoa\fone[\cprod{\vectr}{(\cprod{\vectr}{\vecth})}]-\vrhoa\ftwo(\cprod{\vectr}{\unitplz})
  \nonumber\\
&=-\vrhoa\fone[\vectr(\dprod{\vectr}{\vecth})-\scalr^2\vecth]-\vrhoa\ftwo(\cprod{\vectr}{\unitplz})
  \beqref{alg1}\nonumber\\
&=\scalr^2\vrhoa\fone\vecth-\vrhoa\ftwo(\cprod{\vectr}{\unitplz})
  \beqref{main5c}\nonumber\\
&=\scalr^2\vrhoa\epsvb\vphif\vecth-\vrhoa\epsve\vphif(\cprod{\vectr}{\unitplz})
  \beqref{ogrv2g}\text{ \& }\eqnref{ogrv2h}
\label{gray2g}
\end{align}
\begin{align}
\veush
&=\cprod{\vecta}{\fdota}\beqref{kray2a}\nonumber\\
&=-\vrhoa^2\vecth\beqref{gpath20f}
\label{gray2h}\\
\veusi
&=\cprod{\vecta}{\fdote}\beqref{kray2a}\nonumber\\
&=-\vrhoa\ethvn(\cprod{\vectr}{\unitplz})+\vrhoa\epsvb\vphif(\cprod{\vectr}{\vectz})
  -\kaprep\vphib\epsvb\ethvk\vecth\beqref{gpath20h}
\label{gray2i}\\
\veusj
&=\cprod{\vectu}{\fdota}\beqref{kray2a}\nonumber\\
&=\cprod{\vectu}{(-\vrhoa\vectu+3\vrhoa\vrhob\vectr)}\beqref{gpath6a}\nonumber\\
&=3\vrhoa\vrhob(\cprod{\vectu}{\vectr})
=3\vrhoa\vrhob\vecth\beqref{main5c}
\label{gray2j}
\end{align}
\begin{align}\label{gray2k}
\veusk
&=\cprod{\vectu}{\fdote}\beqref{kray2a}\nonumber\\
&=\cprod{\vectu}{[\ethvn\unitplz-\epsvb\vphif\vectz-\scalq\epsvb\vphif\unitpos-\kaprep\epsvb\ethvk(\cprod{\vectr}{\vecth})]}
  \beqref{gpath16a}\nonumber\\
&=\ethvn(\cprod{\vectu}{\unitplz})-\epsvb\vphif(\cprod{\vectu}{\vectz})
  -\scalq\epsvb\vphif(\cprod{\vectu}{\unitpos})-\kaprep\epsvb\ethvk[\cprod{\vectu}{(\cprod{\vectr}{\vecth})}]
  \nonumber\\
&=\ethvn(\cprod{\vectu}{\unitplz})-\epsvb\vphif(\cprod{\vectu}{\vectz})
  -\epsvb\vphif(\scalq/\scalr)(\cprod{\vectu}{\vectr})-\kaprep\epsvb\ethvk(-\scalh^{-2}\scalr\epsvh\vecth)
  \beqref{gxpeed2c}\nonumber\\
&=\ethvn(\cprod{\vectu}{\unitplz})-\epsvb\vphif(\cprod{\vectu}{\vectz})
  -\epsvb\vphif\vphib\vecth+\scalh^{-2}\scalr\kaprep\epsvb\epsvh\ethvk\vecth
  \beqref{main5c}\text{ \& }\eqnref{ogrv1b}\nonumber\\
\begin{split}
&=\ethvn[\scalh^{-2}(\efktc\vecth-\dltvc\vectz-\scalq\dltvc\unitpos)]
  +\epsvb\vphif[-\scalh^{-2}\frkyz\vecth]-\epsvb\vphif\vphib\vecth\\
  &\quad+\scalh^{-2}\scalr\kaprep\epsvb\epsvh\ethvk\vecth
  \beqref{gxpeed2d}\text{ \& }\eqnref{gpath25a}
\end{split}
\nonumber\\
&=\scalh^{-2}\efktc\ethvn\vecth-\scalh^{-2}\dltvc\ethvn\vectz-\scalh^{-2}\scalq\dltvc\ethvn\unitpos
  -\scalh^{-2}\epsvb\vphif\frkyz\vecth
  -\epsvb\vphif\vphib\vecth+\scalh^{-2}\scalr\kaprep\epsvb\epsvh\ethvk\vecth
  \nonumber\\
&=\scalh^{-2}[\efktc\ethvn+\scalr\kaprep\epsvb\epsvh\ethvk-\epsvb\vphif(\frkyz+\scalh^2\vphib)]\vecth
  -\scalh^{-2}\dltvc\ethvn(\vectz+\scalq\unitpos)\nonumber\\
&=\veka\vecth-\vekb(\vectz+\scalq\unitpos)\beqref{gray1a}
\end{align}
\begin{align*}
\veusl
&=\cprod{\vecte}{\fdota}\beqref{kray2a}\nonumber\\
&=\cprod{[\fone(\cprod{\vectr}{\vecth})+\ftwo\unitplz]}{(-\vrhoa\vectu+3\vrhoa\vrhob\vectr)}
  \beqref{main5b}\text{ \& }\eqnref{gpath6a}\nonumber\\
&=\vrhoa\fone[\cprod{\vectu}{(\cprod{\vectr}{\vecth})}]
  -3\vrhoa\vrhob\fone[\cprod{\vectr}{(\cprod{\vectr}{\vecth})}]
  -\vrhoa\ftwo(\cprod{\unitplz}{\vectu})
  +3\vrhoa\vrhob\ftwo(\cprod{\unitplz}{\vectr})
\nonumber\\
\begin{split}
&=\vrhoa\fone[\vectr(\dprod{\vectu}{\vecth})-\vecth(\dprod{\vectu}{\vectr})]
  -3\vrhoa\vrhob\fone[\vectr(\dprod{\vectr}{\vecth})-\scalr^2\vecth]
  +\vrhoa\ftwo[\scalh^{-2}(\efktc\vecth-\dltvc\vectz-\scalq\dltvc\unitpos)]\\
  &\quad+3\vrhoa\vrhob\ftwo(\cprod{\unitplz}{\vectr})
  \beqref{alg1}\text{ \& }\eqnref{gxpeed2d}
\end{split}
\nonumber\\
\begin{split}
&=\vrhoa\fone[-\scalr(\scalr\vrhob)\vecth]
  -3\vrhoa\vrhob\fone(-\scalr^2\vecth)
  +\ftwo(\vrhoa/\scalh^2)(\efktc\vecth-\dltvc\vectz-\scalq\dltvc\unitpos)\\
  &\quad+3\vrhoa\vrhob\ftwo(\cprod{\unitplz}{\vectr})
  \beqref{main5c}\text{ \& }\eqnref{gpath7b}
\end{split}
\end{align*}
\begin{align}\label{gray2l}
\begin{split}
&=-\scalr^2\vrhoa\fone\vrhob\vecth+3\scalr^2\vrhoa\vrhob\fone\vecth+\ftwo(\vrhoa/\scalh^2)\efktc\vecth
  -\ftwo\dltvc(\vrhoa/\scalh^2)(\vectz+\scalq\unitpos)+3\vrhoa\vrhob\ftwo(\cprod{\unitplz}{\vectr})
\end{split}
\nonumber\\
\begin{split}
&=[2\vphib\vrhob\fone+(\vrhoa/\scalh^2)\efktc\ftwo]\vecth
  -\ftwo\dltvc(\vrhoa/\scalh^2)(\vectz+\scalq\unitpos)
  +3\vrhoa\vrhob\ftwo(\cprod{\unitplz}{\vectr})\\
  &\quad\beqref{ogrv1b}\text{ \& }\eqnref{gpath1b}
\end{split}
\nonumber\\
\begin{split}
&=\vphif[2\vphib\vrhob\epsvb+(\vrhoa/\scalh^2)\epsve\efktc]\vecth
  -\epsve\vphif\dltvc(\vrhoa/\scalh^2)(\vectz+\scalq\unitpos)
  +3\vrhoa\vrhob\epsve\vphif(\cprod{\unitplz}{\vectr})\\
  &\quad\beqref{ogrv2g}\text{ \& }\eqnref{ogrv2h}
\end{split}
\nonumber\\
&=\vekg\vecth-\vekf(\vectz+\scalq\unitpos)
  +3\vrhoa\vrhob\epsve\vphif(\cprod{\unitplz}{\vectr})
  \beqref{gray1a}
\end{align}
\begin{align*}
\veusm
&=\cprod{\vecte}{\fdote}\beqref{kray2a}\nonumber\\
&=\cprod{[\fone(\cprod{\vectr}{\vecth})+\ftwo\unitplz]}{[\ethvn\unitplz-\epsvb\vphif\vectz
  -\scalq\epsvb\vphif\unitpos-\kaprep\epsvb\ethvk(\cprod{\vectr}{\vecth})]}
  \beqref{main5b}\text{ \& }\eqnref{gpath16a}\nonumber\\
\begin{split}
&=-\ethvn\fone[\cprod{\unitplz}{(\cprod{\vectr}{\vecth})}]
  +\epsvb\vphif\fone[\cprod{\vectz}{(\cprod{\vectr}{\vecth})}]
  +\scalq\epsvb\vphif\fone[\cprod{\unitpos}{(\cprod{\vectr}{\vecth})}]\\
  &\quad-\epsvb\vphif\ftwo(\cprod{\unitplz}{\vectz})
  -\scalq\epsvb\vphif\ftwo(\cprod{\unitplz}{\unitpos})
  -\kaprep\epsvb\ethvk\ftwo[\cprod{\unitplz}{(\cprod{\vectr}{\vecth})}]
\end{split}
\nonumber\\
\begin{split}
&=-\ethvn\fone[\vectr(\dprod{\unitplz}{\vecth})-\vecth(\dprod{\unitplz}{\vectr})]
  +\epsvb\vphif\fone[\vectr(\dprod{\vectz}{\vecth})-\vecth(\dprod{\vectz}{\vectr})]
  +\scalq\epsvb\vphif\fone[\vectr(\dprod{\unitpos}{\vecth})-\vecth(\dprod{\unitpos}{\vectr})]\\
  &\quad-\epsvb\vphif\ftwo(\cprod{\unitplz}{\vectz})
  -\scalq\epsvb\vphif\ftwo(\cprod{\unitplz}{\unitpos})
  -\kaprep\epsvb\ethvk\ftwo[\vectr(\dprod{\unitplz}{\vecth})-\vecth(\dprod{\unitplz}{\vectr})]
  \beqref{alg1}
\end{split}
\nonumber\\
\begin{split}
&=-\ethvn\fone(\dltvc\vectr-\scalr\epsvc\vecth)
  +\epsvb\vphif\fone(-\scalr\epsvd\vecth)
  +\scalq\epsvb\vphif\fone(-\scalr\vecth)
  -\epsvb\vphif\ftwo(\cprod{\unitplz}{\vectz})\\
  &\quad-\scalq\epsvb\vphif\ftwo(\cprod{\unitplz}{\unitpos})
  -\kaprep\epsvb\ethvk\ftwo(\dltvc\vectr-\scalr\epsvc\vecth)
  \beqref{main5c}, \eqnref{ogrv1a}\text{ \& }\eqnref{gxpeed1a}
\end{split}
\end{align*}
\begin{align}\label{gray2m}
\begin{split}
&=-\ethvn\epsvb\vphif(\dltvc\vectr-\scalr\epsvc\vecth)
  +\epsvb^2\vphif^2(-\scalr\epsvd\vecth)
  +\scalq\epsvb^2\vphif^2(-\scalr\vecth)
  -\epsvb\epsve\vphif^2(\cprod{\unitplz}{\vectz})\\
  &\quad-\scalq\epsvb\epsve\vphif^2(\cprod{\unitplz}{\unitpos})
  -\kaprep\epsvb\ethvk\epsve\vphif(\dltvc\vectr-\scalr\epsvc\vecth)
  \beqref{ogrv2g}\text{ \& }\eqnref{ogrv2h}
\end{split}
\nonumber\\
\begin{split}
&=-\ethvn\epsvb\vphif\dltvc\vectr+\scalr\ethvn\epsvb\vphif\epsvc\vecth
  -\scalr\epsvb^2\vphif^2\epsvd\vecth-\scalr\scalq\epsvb^2\vphif^2\vecth
  -\epsvb\epsve\vphif^2(\cprod{\unitplz}{\vectz})\\
  &\quad-\scalq\epsvb\epsve\vphif^2(\cprod{\unitplz}{\unitpos})
  -\kaprep\epsvb\ethvk\epsve\vphif\dltvc\vectr
  +\scalr\kaprep\epsvb\ethvk\epsve\vphif\epsvc\vecth
\end{split}
\nonumber\\
\begin{split}
&=-\epsvb\vphif\dltvc(\ethvn+\kaprep\epsve\ethvk)\vectr
  +\scalr\epsvb\vphif[\epsvc(\ethvn+\kaprep\epsve\ethvk)-\epsvb\vphif(\epsvd+\scalq)]\vecth\\
  &\quad-\epsvb\epsve\vphif^2(\cprod{\unitplz}{\vectz})
  -\scalq\epsvb\epsve\vphif^2(\cprod{\unitplz}{\unitpos})
\end{split}
\nonumber\\
\begin{split}
&=-\epsvb\vphif\dltvc\vekc\vectr+\vekd\vecth
  -\epsvb\epsve\vphif^2(\cprod{\unitplz}{\vectz})
  -\scalq\epsvb\epsve\vphif^2(\cprod{\unitplz}{\unitpos})
  \beqref{gray1a}.
\end{split}
\end{align}
\end{subequations}

\subart{Development of equation \eqnref{kray2b}}
Using the foregoing equations, we derive
\begin{subequations}\label{gray3}
\begin{align*}
\veust
&=(\cdkt\vbba-\rhorep\fdot{\cdkt})\veusb+\cdkt\veusn+\rhorep\veuso
  \beqref{kray2b}\nonumber\\
&=(\cdkt\frkyd-\rhorep\frkya)\veusb+\cdkt(\rhorep\veusd+\veuse)+\rhorep(\rhorep\veush+\veusi)
  \beqref{gpath19a}, \eqnref{gpath18a}\text{ \& }\eqnref{kray2b}\nonumber\\
&=\vekh\veusb+\rhorep\cdkt\veusd+\cdkt\veuse+\rhorep^2\veush+\rhorep\veusi
  \beqref{gray1a}\nonumber\\
\begin{split}
&=\vekh[-\vrhoa(\cprod{\unitkap}{\vectr})]
  +\rhorep\cdkt[\veke(\efkta\vecth-\dltva\vectz-\scalq\dltva\unitpos)+3\vrhoa\vrhob(\cprod{\unitkap}{\vectr})]\\
  &\quad+\cdkt[\kaprep\scalr\epsva\epsvb\ethvk\vecth-\kaprep\epsvb\ethvk\dltva\vectr+\ethvn(\cprod{\unitkap}{\unitplz})
     -\epsvb\vphif(\cprod{\unitkap}{\vectz})-\scalq\epsvb\vphif(\cprod{\unitkap}{\unitpos})]
  +\rhorep^2[-\vrhoa^2\vecth]\\
  &\quad+\rhorep[-\vrhoa\ethvn(\cprod{\vectr}{\unitplz})+\vrhoa\epsvb\vphif(\cprod{\vectr}{\vectz})
    -\kaprep\vphib\epsvb\ethvk\vecth]
  \beqref{gray1a}\text{ \& }\eqnref{gray2}
\end{split}
\nonumber\\
\begin{split}
&=-\vekh\vrhoa(\cprod{\unitkap}{\vectr})
  +\rhorep\cdkt\veke\efkta\vecth-\rhorep\cdkt\veke\dltva\vectz-\scalq\rhorep\cdkt\veke\dltva\unitpos
  +3\rhorep\cdkt\vrhoa\vrhob(\cprod{\unitkap}{\vectr})\\
  &\quad+\kaprep\scalr\cdkt\epsva\epsvb\ethvk\vecth-\kaprep\cdkt\epsvb\ethvk\dltva\vectr+\cdkt\ethvn(\cprod{\unitkap}{\unitplz})
     -\cdkt\epsvb\vphif(\cprod{\unitkap}{\vectz})-\scalq\cdkt\epsvb\vphif(\cprod{\unitkap}{\unitpos})
  -\rhorep^2\vrhoa^2\vecth\\
  &\quad-\rhorep\vrhoa\ethvn(\cprod{\vectr}{\unitplz})+\rhorep\vrhoa\epsvb\vphif(\cprod{\vectr}{\vectz})
    -\kaprep\rhorep\vphib\epsvb\ethvk\vecth
\end{split}
\end{align*}
\begin{align}\label{gray3a}
\begin{split}
&=\rhorep\cdkt\veke\efkta\vecth+\kaprep\scalr\cdkt\epsva\epsvb\ethvk\vecth
  -\rhorep^2\vrhoa^2\vecth-\kaprep\rhorep\vphib\epsvb\ethvk\vecth
  -\rhorep\cdkt\veke\dltva\vectz
  -\rhorep\cdkt\veke\dltva(\scalq/\scalr)\vectr-\kaprep\cdkt\epsvb\ethvk\dltva\vectr\\
  &\quad-\vekh\vrhoa(\cprod{\unitkap}{\vectr})+3\rhorep\cdkt\vrhoa\vrhob(\cprod{\unitkap}{\vectr})
  -\cdkt\epsvb\vphif(\scalq/\scalr)(\cprod{\unitkap}{\vectr})
  -\cdkt\epsvb\vphif(\cprod{\unitkap}{\vectz})
  +\cdkt\ethvn(\cprod{\unitkap}{\unitplz})\\
  &\quad-\rhorep\vrhoa\ethvn(\cprod{\vectr}{\unitplz})
  +\rhorep\vrhoa\epsvb\vphif(\cprod{\vectr}{\vectz})
\end{split}
\nonumber\\
\begin{split}
&=[\rhorep(\cdkt\veke\efkta-\rhorep\vrhoa^2)+\kaprep\epsvb\ethvk(\scalr\cdkt\epsva-\rhorep\vphib)]\vecth
  -\rhorep\cdkt\veke\dltva\vectz
  -\cdkt\dltva(\rhorep\veke\vphib+\kaprep\epsvb\ethvk)\vectr\\
  &\quad+[\vrhoa(3\rhorep\cdkt\vrhob-\vekh)-\cdkt\epsvb\vphif\vphib](\cprod{\unitkap}{\vectr})
  -\cdkt\epsvb\vphif(\cprod{\unitkap}{\vectz})
  +\cdkt\ethvn(\cprod{\unitkap}{\unitplz})\\
  &\quad-\rhorep\vrhoa\ethvn(\cprod{\vectr}{\unitplz})
  +\rhorep\vrhoa\epsvb\vphif(\cprod{\vectr}{\vectz})
  \beqref{ogrv1b}
\end{split}
\nonumber\\
\begin{split}
&=\vekj\vecth-\rhorep\cdkt\veke\dltva\vectz-\veki\vectr
  +\vekk(\cprod{\unitkap}{\vectr})-\cdkt\epsvb\vphif(\cprod{\unitkap}{\vectz})
  +\cdkt\ethvn(\cprod{\unitkap}{\unitplz})\\
  &\quad-\rhorep\vrhoa\ethvn(\cprod{\vectr}{\unitplz})
  +\rhorep\vrhoa\epsvb\vphif(\cprod{\vectr}{\vectz})
  \beqref{gray1a}
\end{split}
\end{align}
\begin{align*}
\veusu
&=\fdot{\cdkt}\veusp+\vbba\veusq+\rhorep\veusr+\veuss
  \beqref{kray2b}\nonumber\\
&=\frkya(\veusa-\veusc)+\frkyd(\veusf-\veusg)+\rhorep(\veusl-\veusj)+\veusm-\veusk
  \beqref{gpath18a}, \eqnref{gpath19a}\text{ \& }\eqnref{kray2b}\nonumber\\
&=\frkya\veusa-\frkya\veusc+\frkyd\veusf-\frkyd\veusg+\rhorep\veusl
  -\rhorep\veusj+\veusm-\veusk\nonumber\\
\begin{split}
&=\frkya[\scalh^{-2}(\dltva\vectz+\scalq\dltva\unitpos-\efkta\vecth)]
  -\frkya[\epsvb\vphif(\dltva\vectr-\scalr\epsva\vecth)+\epsve\vphif(\cprod{\unitkap}{\unitplz})]
  +\frkyd[\scalh^{-2}\vphia\efktb\vecth]\\
  &\quad-\frkyd[\scalr^2\vrhoa\epsvb\vphif\vecth-\vrhoa\epsve\vphif(\cprod{\vectr}{\unitplz})]
  +\rhorep[\vekg\vecth-\vekf(\vectz+\scalq\unitpos)+3\vrhoa\vrhob\epsve\vphif(\cprod{\unitplz}{\vectr})]
  -\rhorep[3\vrhoa\vrhob\vecth]\\
  &\quad-\epsvb\vphif\dltvc\vekc\vectr+\vekd\vecth
  -\epsvb\epsve\vphif^2(\cprod{\unitplz}{\vectz})
  -\scalq\epsvb\epsve\vphif^2(\cprod{\unitplz}{\unitpos})
  -[\veka\vecth-\vekb(\vectz+\scalq\unitpos)]
  \beqref{gray2}
\end{split}
\nonumber\\
\begin{split}
&=\frkya(\dltva/\scalh^2)\vectz+\scalq\frkya(\dltva/\scalh^2)\unitpos-\frkya(\efkta/\scalh^2)\vecth
  -\epsvb\vphif\dltva\frkya\vectr+\scalr\epsva\epsvb\vphif\frkya\vecth-\epsve\vphif\frkya(\cprod{\unitkap}{\unitplz})\\
  &\quad+\vphia\efktb(\frkyd/\scalh^2)\vecth
  -\scalr^2\vrhoa\epsvb\vphif\frkyd\vecth+\vrhoa\epsve\vphif\frkyd(\cprod{\vectr}{\unitplz})
  +\rhorep\vekg\vecth-\rhorep\vekf\vectz-\scalq\rhorep\vekf\unitpos\\
  &\quad+3\rhorep\vrhoa\vrhob\epsve\vphif(\cprod{\unitplz}{\vectr})
  -3\rhorep\vrhoa\vrhob\vecth
  -\epsvb\vphif\dltvc\vekc\vectr+\vekd\vecth
  -\epsvb\epsve\vphif^2(\cprod{\unitplz}{\vectz})
  -\scalq\epsvb\epsve\vphif^2(\cprod{\unitplz}{\unitpos})\\
  &\quad-\veka\vecth+\vekb\vectz+\scalq\vekb\unitpos
\end{split}
\end{align*}
\begin{align}\label{gray3b}
\begin{split}
&=\frkya(\dltva/\scalh^2)\vectz-\rhorep\vekf\vectz+\vekb\vectz
  +\frkya(\dltva/\scalh^2)(\scalq/\scalr)\vectr-\epsvb\vphif\dltva\frkya\vectr\\
  &\quad-\rhorep\vekf(\scalq/\scalr)\vectr-\epsvb\vphif\dltvc\vekc\vectr+\vekb(\scalq/\scalr)\vectr
  -\frkya(\efkta/\scalh^2)\vecth+\scalr\epsva\epsvb\vphif\frkya\vecth+\vphia\efktb(\frkyd/\scalh^2)\vecth\\
  &\quad-\scalr^2\vrhoa\epsvb\vphif\frkyd\vecth+\rhorep\vekg\vecth-3\rhorep\vrhoa\vrhob\vecth
  +\vekd\vecth-\veka\vecth
  -\epsve\vphif\frkya(\cprod{\unitkap}{\unitplz})
  -\epsvb\epsve\vphif^2(\cprod{\unitplz}{\vectz})\\
  &\quad+\vrhoa\epsve\vphif\frkyd(\cprod{\vectr}{\unitplz})
  +3\rhorep\vrhoa\vrhob\epsve\vphif(\cprod{\unitplz}{\vectr})
  -\epsvb\epsve\vphif^2(\scalq/\scalr)(\cprod{\unitplz}{\vectr})
\end{split}
\nonumber\\
\begin{split}
&=[\vekb-\rhorep\vekf+\frkya(\dltva/\scalh^2)]\vectz
  +[\frkya\dltva(\vphib/\scalh^2)-\epsvb\vphif(\dltva\frkya+\dltvc\vekc)+\vphib(\vekb-\rhorep\vekf)]\vectr\\
  &\quad+[\vekd-\veka+\scalh^{-2}(\vphia\efktb\frkyd-\frkya\efkta)
  +\scalr\epsvb\vphif(\epsva\frkya-\scalr\vrhoa\frkyd)+\rhorep(\vekg-3\vrhoa\vrhob)]\vecth\\
  &\quad-\epsve\vphif\frkya(\cprod{\unitkap}{\unitplz})
  -\epsvb\epsve\vphif^2(\cprod{\unitplz}{\vectz})
  +\epsve\vphif[\vrhoa(3\rhorep\vrhob-\frkyd)-\epsvb\vphib\vphif](\cprod{\unitplz}{\vectr})
  \beqref{ogrv1b}
\end{split}
\nonumber\\
&=\vekl\vectz+\vekm\vectr+\vekn\vecth
  -\epsve\vphif\frkya(\cprod{\unitkap}{\unitplz})
  -\epsvb\epsve\vphif^2(\cprod{\unitplz}{\vectz})
  +\veko(\cprod{\unitplz}{\vectr})\beqref{gray1a}
\end{align}
\end{subequations}
\begin{align}\label{gray4}
&\veust+\veusu\nonumber\\
\begin{split}
&=\vekj\vecth-\rhorep\cdkt\veke\dltva\vectz-\veki\vectr
  +\vekk(\cprod{\unitkap}{\vectr})-\cdkt\epsvb\vphif(\cprod{\unitkap}{\vectz})
  +\cdkt\ethvn(\cprod{\unitkap}{\unitplz})\\
  &\quad-\rhorep\vrhoa\ethvn(\cprod{\vectr}{\unitplz})
  +\rhorep\vrhoa\epsvb\vphif(\cprod{\vectr}{\vectz})
  +\vekl\vectz+\vekm\vectr+\vekn\vecth
  -\epsve\vphif\frkya(\cprod{\unitkap}{\unitplz})\\
  &\quad-\epsvb\epsve\vphif^2(\cprod{\unitplz}{\vectz})
  +\veko(\cprod{\unitplz}{\vectr})\beqref{gray3}
\end{split}
\nonumber\\
\begin{split}
&=\vekj\vecth+\vekn\vecth-\rhorep\cdkt\veke\dltva\vectz+\vekl\vectz-\veki\vectr+\vekm\vectr
  +\vekk(\cprod{\unitkap}{\vectr})-\cdkt\epsvb\vphif(\cprod{\unitkap}{\vectz})+\cdkt\ethvn(\cprod{\unitkap}{\unitplz})\\
  &\quad-\epsve\vphif\frkya(\cprod{\unitkap}{\unitplz})
  +\rhorep\vrhoa\ethvn(\cprod{\unitplz}{\vectr})+\veko(\cprod{\unitplz}{\vectr})
  +\rhorep\vrhoa\epsvb\vphif(\cprod{\vectr}{\vectz})
  -\epsvb\epsve\vphif^2(\cprod{\unitplz}{\vectz})
\end{split}
\nonumber\\
\begin{split}
&=(\vekj+\vekn)\vecth+(\vekl-\rhorep\cdkt\veke\dltva)\vectz+(\vekm-\veki)\vectr
  +\vekk(\cprod{\unitkap}{\vectr})-\cdkt\epsvb\vphif(\cprod{\unitkap}{\vectz})\\
  &\quad+(\cdkt\ethvn-\epsve\vphif\frkya)(\cprod{\unitkap}{\unitplz})
  +(\veko+\rhorep\vrhoa\ethvn)(\cprod{\unitplz}{\vectr})
  +\rhorep\vrhoa\epsvb\vphif(\cprod{\vectr}{\vectz})
  -\epsvb\epsve\vphif^2(\cprod{\unitplz}{\vectz})
\end{split}
\nonumber\\
\begin{split}
&=\vekp\vecth+\vekq\vectz+\vekr\vectr+\vekk(\cprod{\unitkap}{\vectr})-\cdkt\epsvb\vphif(\cprod{\unitkap}{\vectz})
  +\veks(\cprod{\unitkap}{\unitplz})+\vekt(\cprod{\unitplz}{\vectr})\\
  &\quad+\rhorep\vrhoa\epsvb\vphif(\cprod{\vectr}{\vectz})
  -\epsvb\epsve\vphif^2(\cprod{\unitplz}{\vectz})\beqref{gray1a}.
\end{split}
\end{align}

\subart{Computation of the magnitude of $\veust+\veusu$}
To compute the magnitude of vector $\veust+\veusu$, we first derive
\begin{subequations}\label{gray5}
\begin{align}\label{gray5a}
&\dprod{\vecth}{(\veust+\veusu)}\nonumber\\
\begin{split}
&=\dprod{\vecth}{[}\vekp\vecth+\vekq\vectz+\vekr\vectr+\vekk(\cprod{\unitkap}{\vectr})-\cdkt\epsvb\vphif(\cprod{\unitkap}{\vectz})
  +\veks(\cprod{\unitkap}{\unitplz})+\vekt(\cprod{\unitplz}{\vectr})\\
  &\quad+\rhorep\vrhoa\epsvb\vphif(\cprod{\vectr}{\vectz})
  -\epsvb\epsve\vphif^2(\cprod{\unitplz}{\vectz})]\beqref{gray4}
\end{split}
\nonumber\\
\begin{split}
&=\vekp(\dprod{\vecth}{\vecth})+\vekq(\dprod{\vecth}{\vectz})+\vekr(\dprod{\vecth}{\vectr})
  +\vekk[\dprod{\vecth}{(\cprod{\unitkap}{\vectr})}]-\cdkt\epsvb\vphif[\dprod{\vecth}{(\cprod{\unitkap}{\vectz})}]
  +\veks[\dprod{\vecth}{(\cprod{\unitkap}{\unitplz})}]\\
  &\quad+\vekt[\dprod{\vecth}{(\cprod{\unitplz}{\vectr})}]
  +\rhorep\vrhoa\epsvb\vphif[\dprod{\vecth}{(\cprod{\vectr}{\vectz})}]
  -\epsvb\epsve\vphif^2[\dprod{\vecth}{(\cprod{\unitplz}{\vectz})}]
\end{split}
\nonumber\\
&=\vekp\scalh^2+\vekk\epsve-\cdkt\epsvb\vphif\epsvg-\veks\dltve+\vekt\epsvf
  +\scalr\rhorep\vrhoa\epsvb\vphif\epsvh-\epsvb\epsve\vphif^2\epsvi
  \beqref{main5c}, \eqnref{ogrv1a}\text{ \& }\eqnref{gxpeed1a}\nonumber\\
&=\veku\beqref{gray1b}
\end{align}
\begin{align}\label{gray5b}
&\dprod{\vectz}{(\veust+\veusu)}\nonumber\\
\begin{split}
&=\dprod{\vectz}{[}\vekp\vecth+\vekq\vectz+\vekr\vectr+\vekk(\cprod{\unitkap}{\vectr})-\cdkt\epsvb\vphif(\cprod{\unitkap}{\vectz})
  +\veks(\cprod{\unitkap}{\unitplz})+\vekt(\cprod{\unitplz}{\vectr})\\
  &\quad+\rhorep\vrhoa\epsvb\vphif(\cprod{\vectr}{\vectz})
  -\epsvb\epsve\vphif^2(\cprod{\unitplz}{\vectz})]\beqref{gray4}
\end{split}
\nonumber\\
\begin{split}
&=\vekp(\dprod{\vectz}{\vecth})+\vekq(\dprod{\vectz}{\vectz})+\vekr(\dprod{\vectz}{\vectr})
  +\vekk[\dprod{\vectz}{(\cprod{\unitkap}{\vectr})}]-\cdkt\epsvb\vphif[\dprod{\vectz}{(\cprod{\unitkap}{\vectz})}]
  +\veks[\dprod{\vectz}{(\cprod{\unitkap}{\unitplz})}]\\
  &\quad+\vekt[\dprod{\vectz}{(\cprod{\unitplz}{\vectr})}]
  +\rhorep\vrhoa\epsvb\vphif[\dprod{\vectz}{(\cprod{\vectr}{\vectz})}]
  -\epsvb\epsve\vphif^2[\dprod{\vectz}{(\cprod{\unitplz}{\vectz})}]
\end{split}
\nonumber\\
&=\vekq\scalz^2+\scalr\vekr\epsvd-\scalr\vekk\dltvf+\veks\vsigc-\vekt\vsigb
  \beqref{main5c}, \eqnref{ogrv1a}, \eqnref{gxpeed1a}\text{ \& }\eqnref{gpath1a}\nonumber\\
&=\vekv\beqref{gray1b}
\end{align}
\begin{align}\label{gray5c}
&\dprod{\vectr}{(\veust+\veusu)}\nonumber\\
\begin{split}
&=\dprod{\vectr}{[}\vekp\vecth+\vekq\vectz+\vekr\vectr+\vekk(\cprod{\unitkap}{\vectr})-\cdkt\epsvb\vphif(\cprod{\unitkap}{\vectz})
  +\veks(\cprod{\unitkap}{\unitplz})+\vekt(\cprod{\unitplz}{\vectr})\\
  &\quad+\rhorep\vrhoa\epsvb\vphif(\cprod{\vectr}{\vectz})
  -\epsvb\epsve\vphif^2(\cprod{\unitplz}{\vectz})]\beqref{gray4}
\end{split}
\nonumber\\
\begin{split}
&=\vekp(\dprod{\vectr}{\vecth})+\vekq(\dprod{\vectr}{\vectz})+\vekr(\dprod{\vectr}{\vectr})
  +\vekk[\dprod{\vectr}{(\cprod{\unitkap}{\vectr})}]-\cdkt\epsvb\vphif[\dprod{\vectr}{(\cprod{\unitkap}{\vectz})}]
  +\veks[\dprod{\vectr}{(\cprod{\unitkap}{\unitplz})}]\\
  &\quad+\vekt[\dprod{\vectr}{(\cprod{\unitplz}{\vectr})}]
  +\rhorep\vrhoa\epsvb\vphif[\dprod{\vectr}{(\cprod{\vectr}{\vectz})}]
  -\epsvb\epsve\vphif^2[\dprod{\vectr}{(\cprod{\unitplz}{\vectz})}]
\end{split}
\nonumber\\
&=\scalr\vekq\epsvd+\scalr^2\vekr-\scalr\cdkt\epsvb\vphif\dltvf+\veks\vsiga-\epsvb\epsve\vphif^2\vsigb
  \beqref{main5c}, \eqnref{ogrv1a}, \eqnref{gxpeed1a}\text{ \& }\eqnref{gpath1a}\nonumber\\
&=\vekw\beqref{gray1b}
\end{align}
\begin{align*}
&\dprod{(\cprod{\unitkap}{\vectr})}{(\veust+\veusu)}\nonumber\\
\begin{split}
&=\dprod{(\cprod{\unitkap}{\vectr})}{[}\vekp\vecth+\vekq\vectz+\vekr\vectr
  +\vekk(\cprod{\unitkap}{\vectr})-\cdkt\epsvb\vphif(\cprod{\unitkap}{\vectz})
  +\veks(\cprod{\unitkap}{\unitplz})+\vekt(\cprod{\unitplz}{\vectr})\\
  &\quad+\rhorep\vrhoa\epsvb\vphif(\cprod{\vectr}{\vectz})
  -\epsvb\epsve\vphif^2(\cprod{\unitplz}{\vectz})]\beqref{gray4}
\end{split}
\nonumber\\
\begin{split}
&=\vekp[\dprod{\vecth}{(\cprod{\unitkap}{\vectr})}]
  +\vekq[\dprod{\vectz}{(\cprod{\unitkap}{\vectr})}]
  +\vekr[\dprod{\vectr}{(\cprod{\unitkap}{\vectr})}]
  +\vekk[\dprod{(\cprod{\unitkap}{\vectr})}{(\cprod{\unitkap}{\vectr})}]\\
  &\quad-\cdkt\epsvb\vphif[\dprod{(\cprod{\unitkap}{\vectr})}{(\cprod{\unitkap}{\vectz})}]
  +\veks[\dprod{(\cprod{\unitkap}{\vectr})}{(\cprod{\unitkap}{\unitplz})}]
  +\vekt[\dprod{(\cprod{\unitkap}{\vectr})}{(\cprod{\unitplz}{\vectr})}]\\
  &\quad+\rhorep\vrhoa\epsvb\vphif[\dprod{(\cprod{\unitkap}{\vectr})}{(\cprod{\vectr}{\vectz})}]
  -\epsvb\epsve\vphif^2[\dprod{(\cprod{\unitkap}{\vectr})}{(\cprod{\unitplz}{\vectz})}]
\end{split}
\nonumber\\
\begin{split}
&=\vekp\epsve-\scalr\vekq\dltvf
  +\vekk[\scalr^2-(\dprod{\unitkap}{\vectr})^2]
  -\cdkt\epsvb\vphif[(\dprod{\vectr}{\vectz})-(\dprod{\unitkap}{\vectz})(\dprod{\vectr}{\unitkap})]\\
  &\quad+\veks[(\dprod{\vectr}{\unitplz})-(\dprod{\unitkap}{\unitplz})(\dprod{\vectr}{\unitkap})]
  +\vekt[(\dprod{\unitkap}{\unitplz})\scalr^2-(\dprod{\unitkap}{\vectr})(\dprod{\vectr}{\unitplz})]
  +\rhorep\vrhoa\epsvb\vphif[(\dprod{\unitkap}{\vectr})(\dprod{\vectr}{\vectz})-(\dprod{\unitkap}{\vectz})\scalr^2]\\
  &\quad-\epsvb\epsve\vphif^2[(\dprod{\unitkap}{\unitplz})(\dprod{\vectr}{\vectz})-(\dprod{\unitkap}{\vectz})(\dprod{\vectr}{\unitplz})]
  \beqref{ogrv1a}, \eqnref{gxpeed1a}\text{ \& }\eqnref{alg2}
\end{split}
\end{align*}
\begin{align}\label{gray5d}
\begin{split}
&=\vekp\epsve-\scalr\vekq\dltvf+\vekk(\scalr^2-\scalr^2\epsva^2)
  -\cdkt\epsvb\vphif(\scalr\epsvd-\scalr\dltvb\epsva)
  +\veks(\scalr\epsvc-\scalr\epsvb\epsva)
  +\vekt(\epsvb\scalr^2-\scalr^2\epsva\epsvc)\\
  &\quad+\rhorep\vrhoa\epsvb\vphif(\scalr^2\epsva\epsvd-\dltvb\scalr^2)
  -\epsvb\epsve\vphif^2(\scalr\epsvb\epsvd-\scalr\dltvb\epsvc)
  \beqref{ogrv1a}\text{ \& }\eqnref{gxpeed1a}
\end{split}
\nonumber\\
\begin{split}
&=\vekp\epsve-\scalr\vekq\dltvf+\scalr^2\vekk(1-\epsva^2)
  -\scalr\cdkt\epsvb\vphif(\epsvd-\dltvb\epsva)
  +\scalr\veks(\epsvc-\epsvb\epsva)
  +\scalr^2\vekt(\epsvb-\epsva\epsvc)\\
  &\quad+\scalr^2\rhorep\vrhoa\epsvb\vphif(\epsva\epsvd-\dltvb)
  -\scalr\epsvb\epsve\vphif^2(\epsvb\epsvd-\dltvb\epsvc)
\end{split}
\nonumber\\
&=\vekx\beqref{gray1b}
\end{align}
\begin{align*}
&\dprod{(\cprod{\unitkap}{\vectz})}{(\veust+\veusu)}\nonumber\\
\begin{split}
&=\dprod{(\cprod{\unitkap}{\vectz})}{[}\vekp\vecth+\vekq\vectz+\vekr\vectr
  +\vekk(\cprod{\unitkap}{\vectr})-\cdkt\epsvb\vphif(\cprod{\unitkap}{\vectz})
  +\veks(\cprod{\unitkap}{\unitplz})+\vekt(\cprod{\unitplz}{\vectr})\\
  &\quad+\rhorep\vrhoa\epsvb\vphif(\cprod{\vectr}{\vectz})
  -\epsvb\epsve\vphif^2(\cprod{\unitplz}{\vectz})]\beqref{gray4}
\end{split}
\nonumber\\
\begin{split}
&=\vekp[\dprod{\vecth}{(\cprod{\unitkap}{\vectz})}]
  +\vekq[\dprod{\vectz}{(\cprod{\unitkap}{\vectz})}]
  +\vekr[\dprod{\vectr}{(\cprod{\unitkap}{\vectz})}]
  +\vekk[\dprod{(\cprod{\unitkap}{\vectz})}{(\cprod{\unitkap}{\vectr})}]\\
  &\quad-\cdkt\epsvb\vphif[\dprod{(\cprod{\unitkap}{\vectz})}{(\cprod{\unitkap}{\vectz})}]
  +\veks[\dprod{(\cprod{\unitkap}{\vectz})}{(\cprod{\unitkap}{\unitplz})}]
  +\vekt[\dprod{(\cprod{\unitkap}{\vectz})}{(\cprod{\unitplz}{\vectr})}]\\
  &\quad+\rhorep\vrhoa\epsvb\vphif[\dprod{(\cprod{\unitkap}{\vectz})}{(\cprod{\vectr}{\vectz})}]
  -\epsvb\epsve\vphif^2[\dprod{(\cprod{\unitkap}{\vectz})}{(\cprod{\unitplz}{\vectz})}]
\end{split}
\nonumber\\
\begin{split}
&=\vekp\epsvg
  +\scalr\vekr\dltvf
  +\vekk[(\dprod{\vectz}{\vectr})-(\dprod{\unitkap}{\vectr})(\dprod{\vectz}{\unitkap})]
  -\cdkt\epsvb\vphif[\scalz^2-(\dprod{\unitkap}{\vectz})^2]
  +\veks[(\dprod{\vectz}{\unitplz})-(\dprod{\unitkap}{\unitplz})(\dprod{\vectz}{\unitkap})]\\
  &\quad+\vekt[(\dprod{\unitkap}{\unitplz})(\dprod{\vectz}{\vectr})-(\dprod{\unitkap}{\vectr})(\dprod{\vectz}{\unitplz})]
  +\rhorep\vrhoa\epsvb\vphif[(\dprod{\unitkap}{\vectr})\scalz^2-(\dprod{\unitkap}{\vectz})(\dprod{\vectz}{\vectr})]\\
  &\quad-\epsvb\epsve\vphif^2[(\dprod{\unitkap}{\unitplz})\scalz^2-(\dprod{\unitkap}{\vectz})(\dprod{\vectz}{\unitplz})]
  \beqref{ogrv1a}, \eqnref{gxpeed1a}\text{ \& }\eqnref{alg2}
\end{split}
\end{align*}
\begin{align}\label{gray5e}
\begin{split}
&=\vekp\epsvg+\scalr\vekr\dltvf+\vekk(\scalr\epsvd-\scalr\epsva\dltvb)-\cdkt\epsvb\vphif(\scalz^2-\dltvb^2)
  +\veks(\dltvd-\epsvb\dltvb)+\vekt(\scalr\epsvb\epsvd-\scalr\epsva\dltvd)\\
  &\quad+\rhorep\vrhoa\epsvb\vphif(\scalr\scalz^2\epsva-\scalr\dltvb\epsvd)
  -\epsvb\epsve\vphif^2(\epsvb\scalz^2-\dltvb\dltvd)
  \beqref{ogrv1a}\text{ \& }\eqnref{gxpeed1a}
\end{split}
\nonumber\\
\begin{split}
&=\vekp\epsvg+\scalr\vekr\dltvf+\scalr\vekk(\epsvd-\epsva\dltvb)-\cdkt\epsvb\vphif(\scalz^2-\dltvb^2)
  +\veks(\dltvd-\epsvb\dltvb)+\scalr\vekt(\epsvb\epsvd-\epsva\dltvd)\\
  &\quad+\scalr\rhorep\vrhoa\epsvb\vphif(\scalz^2\epsva-\dltvb\epsvd)
  -\epsvb\epsve\vphif^2(\scalz^2\epsvb-\dltvb\dltvd)
\end{split}
\nonumber\\
&=\veky\beqref{gray1b}
\end{align}
\begin{align*}
&\dprod{(\cprod{\unitkap}{\unitplz})}{(\veust+\veusu)}\nonumber\\
\begin{split}
&=\dprod{(\cprod{\unitkap}{\unitplz})}{[}\vekp\vecth+\vekq\vectz+\vekr\vectr
  +\vekk(\cprod{\unitkap}{\vectr})-\cdkt\epsvb\vphif(\cprod{\unitkap}{\vectz})
  +\veks(\cprod{\unitkap}{\unitplz})+\vekt(\cprod{\unitplz}{\vectr})\\
  &\quad+\rhorep\vrhoa\epsvb\vphif(\cprod{\vectr}{\vectz})
  -\epsvb\epsve\vphif^2(\cprod{\unitplz}{\vectz})]\beqref{gray4}
\end{split}
\nonumber\\
\begin{split}
&=\vekp[\dprod{\vecth}{(\cprod{\unitkap}{\unitplz})}]
  +\vekq[\dprod{\vectz}{(\cprod{\unitkap}{\unitplz})}]
  +\vekr[\dprod{\vectr}{(\cprod{\unitkap}{\unitplz})}]
  +\vekk[\dprod{(\cprod{\unitkap}{\unitplz})}{(\cprod{\unitkap}{\vectr})}]\\
  &\quad-\cdkt\epsvb\vphif[\dprod{(\cprod{\unitkap}{\unitplz})}{(\cprod{\unitkap}{\vectz})}]
  +\veks[\dprod{(\cprod{\unitkap}{\unitplz})}{(\cprod{\unitkap}{\unitplz})}]
  +\vekt[\dprod{(\cprod{\unitkap}{\unitplz})}{(\cprod{\unitplz}{\vectr})}]\\
  &\quad+\rhorep\vrhoa\epsvb\vphif[\dprod{(\cprod{\unitkap}{\unitplz})}{(\cprod{\vectr}{\vectz})}]
  -\epsvb\epsve\vphif^2[\dprod{(\cprod{\unitkap}{\unitplz})}{(\cprod{\unitplz}{\vectz})}]
\end{split}
\nonumber\\
\begin{split}
&=-\vekp\dltve+\vekq\vsigc+\vekr\vsiga
  +\vekk[(\dprod{\unitplz}{\vectr})-(\dprod{\unitkap}{\vectr})(\dprod{\unitplz}{\unitkap})]
  -\cdkt\epsvb\vphif[(\dprod{\unitplz}{\vectz})-(\dprod{\unitkap}{\vectz})(\dprod{\unitplz}{\unitkap})]\\
  &\quad+\veks[1-(\dprod{\unitkap}{\unitplz})^2]
  +\vekt[(\dprod{\unitkap}{\unitplz})(\dprod{\unitplz}{\vectr})-(\dprod{\unitkap}{\vectr})]
  +\rhorep\vrhoa\epsvb\vphif[(\dprod{\unitkap}{\vectr})(\dprod{\unitplz}{\vectz})-(\dprod{\unitkap}{\vectz})(\dprod{\unitplz}{\vectr})]\\
  &\quad-\epsvb\epsve\vphif^2[(\dprod{\unitkap}{\unitplz})(\dprod{\unitplz}{\vectz})-(\dprod{\unitkap}{\vectz})]
  \beqref{gxpeed1a}, \eqnref{gpath1a}\text{ \& }\eqnref{alg2}
\end{split}
\end{align*}
\begin{align}\label{gray5f}
\begin{split}
&=-\vekp\dltve+\vekq\vsigc+\vekr\vsiga
  +\vekk(\scalr\epsvc-\scalr\epsva\epsvb)
  -\cdkt\epsvb\vphif(\dltvd-\dltvb\epsvb)
  +\veks(1-\epsvb^2)
  +\vekt(\scalr\epsvb\epsvc-\scalr\epsva)\\
  &\quad+\rhorep\vrhoa\epsvb\vphif(\scalr\epsva\dltvd-\scalr\dltvb\epsvc)
  -\epsvb\epsve\vphif^2(\epsvb\dltvd-\dltvb)
  \beqref{ogrv1a}\text{ \& }\eqnref{gxpeed1a}
\end{split}
\nonumber\\
\begin{split}
&=-\vekp\dltve+\vekq\vsigc+\vekr\vsiga+\scalr\vekk(\epsvc-\epsva\epsvb)
  -\cdkt\epsvb\vphif(\dltvd-\dltvb\epsvb)+\veks(1-\epsvb^2)
  +\scalr\vekt(\epsvb\epsvc-\epsva)\\
  &\quad+\scalr\rhorep\vrhoa\epsvb\vphif(\epsva\dltvd-\dltvb\epsvc)
  -\epsvb\epsve\vphif^2(\epsvb\dltvd-\dltvb)
\end{split}
\nonumber\\
&=\vekz\beqref{gray1b}
\end{align}
\begin{align*}
&\dprod{(\cprod{\unitplz}{\vectr})}{(\veust+\veusu)}\nonumber\\
\begin{split}
&=\dprod{(\cprod{\unitplz}{\vectr})}{[}\vekp\vecth+\vekq\vectz+\vekr\vectr
  +\vekk(\cprod{\unitkap}{\vectr})-\cdkt\epsvb\vphif(\cprod{\unitkap}{\vectz})
  +\veks(\cprod{\unitkap}{\unitplz})+\vekt(\cprod{\unitplz}{\vectr})\\
  &\quad-\epsvb\epsve\vphif^2(\cprod{\unitplz}{\vectz})
  +\rhorep\vrhoa\epsvb\vphif(\cprod{\vectr}{\vectz})]\beqref{gray4}
\end{split}
\nonumber\\
\begin{split}
&=\vekp[\dprod{\vecth}{(\cprod{\unitplz}{\vectr})}]
  +\vekq[\dprod{\vectz}{(\cprod{\unitplz}{\vectr})}]
  +\vekr[\dprod{\vectr}{(\cprod{\unitplz}{\vectr})}]
  +\vekk[\dprod{(\cprod{\unitplz}{\vectr})}{(\cprod{\unitkap}{\vectr})}]\\
  &\quad-\cdkt\epsvb\vphif[\dprod{(\cprod{\unitplz}{\vectr})}{(\cprod{\unitkap}{\vectz})}]
  +\veks[\dprod{(\cprod{\unitplz}{\vectr})}{(\cprod{\unitkap}{\unitplz})}]
  +\vekt[\dprod{(\cprod{\unitplz}{\vectr})}{(\cprod{\unitplz}{\vectr})}]\\
  &\quad-\epsvb\epsve\vphif^2[\dprod{(\cprod{\unitplz}{\vectr})}{(\cprod{\unitplz}{\vectz})}]
  +\rhorep\vrhoa\epsvb\vphif[\dprod{(\cprod{\unitplz}{\vectr})}{(\cprod{\vectr}{\vectz})}]
\end{split}
\nonumber\\
\begin{split}
&=\vekp\epsvf-\vekq\vsigb
  +\vekk[(\dprod{\unitplz}{\unitkap})\scalr^2-(\dprod{\unitplz}{\vectr})(\dprod{\vectr}{\unitkap})]
  -\cdkt\epsvb\vphif[(\dprod{\unitplz}{\unitkap})(\dprod{\vectr}{\vectz})-(\dprod{\unitplz}{\vectz})(\dprod{\vectr}{\unitkap})]\\
  &\quad+\veks[(\dprod{\unitplz}{\unitkap})(\dprod{\vectr}{\unitplz})-(\dprod{\vectr}{\unitkap})]
  +\vekt[\scalr^2-(\dprod{\unitplz}{\vectr})^2]
  -\epsvb\epsve\vphif^2[(\dprod{\vectr}{\vectz})-(\dprod{\unitplz}{\vectz})(\dprod{\vectr}{\unitplz})]\\
  &\quad+\rhorep\vrhoa\epsvb\vphif[(\dprod{\unitplz}{\vectr})(\dprod{\vectr}{\vectz})-(\dprod{\unitplz}{\vectz})\scalr^2]
  \beqref{ogrv1a}, \eqnref{gpath1a}\text{ \& }\eqnref{alg2}
\end{split}
\end{align*}
\begin{align}\label{gray5g}
\begin{split}
&=\vekp\epsvf-\vekq\vsigb
  +\vekk(\epsvb\scalr^2-\scalr^2\epsvc\epsva)
  -\cdkt\epsvb\vphif(\scalr\epsvb\epsvd-\scalr\dltvd\epsva)
  +\veks(\scalr\epsvb\epsvc-\scalr\epsva)
  +\vekt(\scalr^2-\scalr^2\epsvc^2)\\
  &\quad-\epsvb\epsve\vphif^2(\scalr\epsvd-\scalr\dltvd\epsvc)
  +\rhorep\vrhoa\epsvb\vphif(\scalr^2\epsvc\epsvd-\scalr^2\dltvd)
  \beqref{ogrv1a}\text{ \& }\eqnref{gxpeed1a}
\end{split}
\nonumber\\
\begin{split}
&=\vekp\epsvf-\vekq\vsigb
  +\scalr^2\vekk(\epsvb-\epsvc\epsva)
  -\scalr\cdkt\epsvb\vphif(\epsvb\epsvd-\dltvd\epsva)
  +\scalr\veks(\epsvb\epsvc-\epsva)
  +\scalr^2\vekt(1-\epsvc^2)\\
  &\quad-\scalr\epsvb\epsve\vphif^2(\epsvd-\dltvd\epsvc)
  +\scalr^2\rhorep\vrhoa\epsvb\vphif(\epsvc\epsvd-\dltvd)
\end{split}
\nonumber\\
&=\vpsa\beqref{gray1c}
\end{align}
\begin{align*}
&\dprod{(\cprod{\unitplz}{\vectz})}{(\veust+\veusu)}\nonumber\\
\begin{split}
&=\dprod{(\cprod{\unitplz}{\vectz})}{[}\vekp\vecth+\vekq\vectz+\vekr\vectr
  +\vekk(\cprod{\unitkap}{\vectr})-\cdkt\epsvb\vphif(\cprod{\unitkap}{\vectz})
  +\veks(\cprod{\unitkap}{\unitplz})+\vekt(\cprod{\unitplz}{\vectr})\\
  &\quad-\epsvb\epsve\vphif^2(\cprod{\unitplz}{\vectz})
  +\rhorep\vrhoa\epsvb\vphif(\cprod{\vectr}{\vectz})]\beqref{gray4}
\end{split}
\nonumber\\
\begin{split}
&=\vekp[\dprod{\vecth}{(\cprod{\unitplz}{\vectz})}]
  +\vekq[\dprod{\vectz}{(\cprod{\unitplz}{\vectz})}]
  +\vekr[\dprod{\vectr}{(\cprod{\unitplz}{\vectz})}]
  +\vekk[\dprod{(\cprod{\unitplz}{\vectz})}{(\cprod{\unitkap}{\vectr})}]\\
  &\quad-\cdkt\epsvb\vphif[\dprod{(\cprod{\unitplz}{\vectz})}{(\cprod{\unitkap}{\vectz})}]
  +\veks[\dprod{(\cprod{\unitplz}{\vectz})}{(\cprod{\unitkap}{\unitplz})}]
  +\vekt[\dprod{(\cprod{\unitplz}{\vectz})}{(\cprod{\unitplz}{\vectr})}]\\
  &\quad-\epsvb\epsve\vphif^2[\dprod{(\cprod{\unitplz}{\vectz})}{(\cprod{\unitplz}{\vectz})}]
  +\rhorep\vrhoa\epsvb\vphif[\dprod{(\cprod{\unitplz}{\vectz})}{(\cprod{\vectr}{\vectz})}]
\end{split}
\nonumber\\
\begin{split}
&=\vekp\epsvi+\vekr\vsigb
  +\vekk[(\dprod{\unitplz}{\unitkap})(\dprod{\vectz}{\vectr})-(\dprod{\unitplz}{\vectr})(\dprod{\vectz}{\unitkap})]
  -\cdkt\epsvb\vphif[(\dprod{\unitplz}{\unitkap})\scalz^2-(\dprod{\unitplz}{\vectz})(\dprod{\vectz}{\unitkap})]\\
  &\quad+\veks[(\dprod{\unitplz}{\unitkap})(\dprod{\vectz}{\unitplz})-(\dprod{\vectz}{\unitkap})]
  +\vekt[(\dprod{\vectz}{\vectr})-(\dprod{\unitplz}{\vectr})(\dprod{\vectz}{\unitplz})]
  -\epsvb\epsve\vphif^2[\scalz^2-(\dprod{\unitplz}{\vectz})^2]\\
  &\quad+\rhorep\vrhoa\epsvb\vphif[(\dprod{\unitplz}{\vectr})\scalz^2-(\dprod{\unitplz}{\vectz})(\dprod{\vectz}{\vectr})]
  \beqref{ogrv1a}, \eqnref{gpath1a}\text{ \& }\eqnref{alg2}
\end{split}
\end{align*}
\begin{align}\label{gray5h}
\begin{split}
&=\vekp\epsvi+\vekr\vsigb+\vekk(\scalr\epsvb\epsvd-\scalr\epsvc\dltvb)
  -\cdkt\epsvb\vphif(\scalz^2\epsvb-\dltvd\dltvb)+\veks(\epsvb\dltvd-\dltvb)
  +\vekt(\scalr\epsvd-\scalr\epsvc\dltvd)\\
  &\quad-\epsvb\epsve\vphif^2(\scalz^2-\dltvd^2)
  +\rhorep\vrhoa\epsvb\vphif(\scalr\scalz^2\epsvc-\scalr\dltvd\epsvd)
  \beqref{ogrv1a}\text{ \& }\eqnref{gxpeed1a}
\end{split}
\nonumber\\
\begin{split}
&=\vekp\epsvi+\vekr\vsigb+\scalr\vekk(\epsvb\epsvd-\epsvc\dltvb)
  -\cdkt\epsvb\vphif(\scalz^2\epsvb-\dltvd\dltvb)+\veks(\epsvb\dltvd-\dltvb)
  +\scalr\vekt(\epsvd-\epsvc\dltvd)\\
  &\quad-\epsvb\epsve\vphif^2(\scalz^2-\dltvd^2)
  +\scalr\rhorep\vrhoa\epsvb\vphif(\scalz^2\epsvc-\dltvd\epsvd)
\end{split}
\nonumber\\
&=\vpsb\beqref{gray1c}
\end{align}
\begin{align*}
&\dprod{(\cprod{\vectr}{\vectz})}{(\veust+\veusu)}\nonumber\\
\begin{split}
&=\dprod{(\cprod{\vectr}{\vectz})}{[}\vekp\vecth+\vekq\vectz+\vekr\vectr
  +\vekk(\cprod{\unitkap}{\vectr})-\cdkt\epsvb\vphif(\cprod{\unitkap}{\vectz})
  +\veks(\cprod{\unitkap}{\unitplz})+\vekt(\cprod{\unitplz}{\vectr})\\
  &\quad-\epsvb\epsve\vphif^2(\cprod{\unitplz}{\vectz})
  +\rhorep\vrhoa\epsvb\vphif(\cprod{\vectr}{\vectz})]\beqref{gray4}
\end{split}
\nonumber\\
\begin{split}
&=\vekp[\dprod{\vecth}{(\cprod{\vectr}{\vectz})}]
  +\vekq[\dprod{\vectz}{(\cprod{\vectr}{\vectz})}]
  +\vekr[\dprod{\vectr}{(\cprod{\vectr}{\vectz})}]
  +\vekk[\dprod{(\cprod{\vectr}{\vectz})}{(\cprod{\unitkap}{\vectr})}]\\
  &\quad-\cdkt\epsvb\vphif[\dprod{(\cprod{\vectr}{\vectz})}{(\cprod{\unitkap}{\vectz})}]
  +\veks[\dprod{(\cprod{\vectr}{\vectz})}{(\cprod{\unitkap}{\unitplz})}]
  +\vekt[\dprod{(\cprod{\vectr}{\vectz})}{(\cprod{\unitplz}{\vectr})}]\\
  &\quad-\epsvb\epsve\vphif^2[\dprod{(\cprod{\vectr}{\vectz})}{(\cprod{\unitplz}{\vectz})}]
  +\rhorep\vrhoa\epsvb\vphif[\dprod{(\cprod{\vectr}{\vectz})}{(\cprod{\vectr}{\vectz})}]
\end{split}
\nonumber\\
\begin{split}
&=\scalr\vekp\epsvh
  +\vekk[(\dprod{\vectr}{\unitkap})(\dprod{\vectz}{\vectr})-\scalr^2(\dprod{\vectz}{\unitkap})]
  -\cdkt\epsvb\vphif[(\dprod{\vectr}{\unitkap})\scalz^2-(\dprod{\vectr}{\vectz})(\dprod{\vectz}{\unitkap})]\\
  &\quad+\veks[(\dprod{\vectr}{\unitkap})(\dprod{\vectz}{\unitplz})-(\dprod{\vectr}{\unitplz})(\dprod{\vectz}{\unitkap})]
  +\vekt[(\dprod{\vectr}{\unitplz})(\dprod{\vectz}{\vectr})-\scalr^2(\dprod{\vectz}{\unitplz})]\\
  &\quad-\epsvb\epsve\vphif^2[(\dprod{\vectr}{\unitplz})\scalz^2-(\dprod{\vectr}{\vectz})(\dprod{\vectz}{\unitplz})]
  +\rhorep\vrhoa\epsvb\vphif[\scalr^2\scalz^2-(\dprod{\vectr}{\vectz})^2]
  \beqref{ogrv1a}\text{ \& }\eqnref{alg2}
\end{split}
\end{align*}
\begin{align}\label{gray5i}
\begin{split}
&=\scalr\vekp\epsvh
  +\vekk(\scalr^2\epsva\epsvd-\scalr^2\dltvb)
  -\cdkt\epsvb\vphif(\scalr\scalz^2\epsva-\scalr\epsvd\dltvb)
  +\veks(\scalr\epsva\dltvd-\scalr\epsvc\dltvb)
  +\vekt(\scalr^2\epsvc\epsvd-\scalr^2\dltvd)\\
  &\quad-\epsvb\epsve\vphif^2(\scalr\scalz^2\epsvc-\scalr\epsvd\dltvd)
  +\rhorep\vrhoa\epsvb\vphif(\scalr^2\scalz^2-\scalr^2\epsvd^2)
  \beqref{ogrv1a}\text{ \& }\eqnref{gxpeed1a}
\end{split}
\nonumber\\
\begin{split}
&=\scalr\vekp\epsvh+\scalr^2\vekk(\epsva\epsvd-\dltvb)-\scalr\cdkt\epsvb\vphif(\scalz^2\epsva-\epsvd\dltvb)
  +\scalr\veks(\epsva\dltvd-\epsvc\dltvb)+\scalr^2\vekt(\epsvc\epsvd-\dltvd)\\
  &\quad-\scalr\epsvb\epsve\vphif^2(\scalz^2\epsvc-\epsvd\dltvd)
  +\scalr^2\rhorep\vrhoa\epsvb\vphif(\scalz^2-\epsvd^2)
\end{split}
\nonumber\\
&=\vpsc\beqref{gray1c}
\end{align}
\end{subequations}
all of which lead to
\begin{align}\label{gray6}
&|\veust+\veusu|^2=\dprod{(\veust+\veusu)}{(\veust+\veusu)}\nonumber\\
\begin{split}
&=\dprod{(\veust+\veusu)}{[}\vekp\vecth+\vekq\vectz+\vekr\vectr
  +\vekk(\cprod{\unitkap}{\vectr})-\cdkt\epsvb\vphif(\cprod{\unitkap}{\vectz})
  +\veks(\cprod{\unitkap}{\unitplz})+\vekt(\cprod{\unitplz}{\vectr})\\
  &\quad-\epsvb\epsve\vphif^2(\cprod{\unitplz}{\vectz})
  +\rhorep\vrhoa\epsvb\vphif(\cprod{\vectr}{\vectz})]\beqref{gray4}
\end{split}
\nonumber\\
\begin{split}
&=\vekp[\dprod{\vecth}{(\veust+\veusu)}]
  +\vekq[\dprod{\vectz}{(\veust+\veusu)}]
  +\vekr[\dprod{\vectr}{(\veust+\veusu)}]
  +\vekk[\dprod{(\veust+\veusu)}{(\cprod{\unitkap}{\vectr})}]\\
  &\quad-\cdkt\epsvb\vphif[\dprod{(\veust+\veusu)}{(\cprod{\unitkap}{\vectz})}]
  +\veks[\dprod{(\veust+\veusu)}{(\cprod{\unitkap}{\unitplz})}]
  +\vekt[\dprod{(\veust+\veusu)}{(\cprod{\unitplz}{\vectr})}]\\
  &\quad-\epsvb\epsve\vphif^2[\dprod{(\veust+\veusu)}{(\cprod{\unitplz}{\vectz})}]
  +\rhorep\vrhoa\epsvb\vphif[\dprod{(\veust+\veusu)}{(\cprod{\vectr}{\vectz})}]
\end{split}
\nonumber\\
\begin{split}
&=\vekp\veku+\vekq\vekv+\vekr\vekw+\vekk\vekx-\cdkt\epsvb\vphif\veky+\veks\vekz+\vekt\vpsa\\
  &\quad-\epsvb\epsve\vphif^2\vpsb+\rhorep\vrhoa\epsvb\vphif\vpsc\beqref{gray5}
\end{split}
\nonumber\\
&\therefore|\veust+\veusu|
=\vpsd\beqref{gray1c}.
\end{align}

\subart{Development of equation \eqnref{kray2c}}
To evaluate the quantities defined by \eqnref{kray2c}, we proceed by first deriving
\begin{subequations}\label{gray7}
\begin{align}\label{gray7a}
&\dprod{\unitkap}{(\veust+\veusu)}\nonumber\\
\begin{split}
&=\dprod{\unitkap}{[}\vekp\vecth+\vekq\vectz+\vekr\vectr+\vekk(\cprod{\unitkap}{\vectr})-\cdkt\epsvb\vphif(\cprod{\unitkap}{\vectz})
  +\veks(\cprod{\unitkap}{\unitplz})+\vekt(\cprod{\unitplz}{\vectr})\\
  &\quad+\rhorep\vrhoa\epsvb\vphif(\cprod{\vectr}{\vectz})
  -\epsvb\epsve\vphif^2(\cprod{\unitplz}{\vectz})]\beqref{gray4}
\end{split}
\nonumber\\
\begin{split}
&=\vekp(\dprod{\unitkap}{\vecth})+\vekq(\dprod{\unitkap}{\vectz})+\vekr(\dprod{\unitkap}{\vectr})
  +\vekk[\dprod{\unitkap}{(\cprod{\unitkap}{\vectr})}]-\cdkt\epsvb\vphif[\dprod{\unitkap}{(\cprod{\unitkap}{\vectz})}]
  +\veks[\dprod{\unitkap}{(\cprod{\unitkap}{\unitplz})}]\\
  &\quad+\vekt[\dprod{\unitkap}{(\cprod{\unitplz}{\vectr})}]
  +\rhorep\vrhoa\epsvb\vphif[\dprod{\unitkap}{(\cprod{\vectr}{\vectz})}]
  -\epsvb\epsve\vphif^2[\dprod{\unitkap}{(\cprod{\unitplz}{\vectz})}]
\end{split}
\nonumber\\
&=\vekp\dltva+\vekq\dltvb+\scalr\vekr\epsva+\vekt\vsiga-\scalr\rhorep\vrhoa\epsvb\vphif\dltvf-\epsvb\epsve\vphif^2\vsigc
  \beqref{ogrv1a}, \eqnref{gxpeed1a}\text{ \& }\eqnref{gpath1a}\nonumber\\
&=\vpse\beqref{gray1d}
\end{align}
\begin{align*}
&\dprod{\vectu}{(\veust+\veusu)}\nonumber\\
\begin{split}
&=\dprod{\vectu}{[}\vekp\vecth+\vekq\vectz+\vekr\vectr+\vekk(\cprod{\unitkap}{\vectr})-\cdkt\epsvb\vphif(\cprod{\unitkap}{\vectz})
  +\veks(\cprod{\unitkap}{\unitplz})+\vekt(\cprod{\unitplz}{\vectr})\\
  &\quad+\rhorep\vrhoa\epsvb\vphif(\cprod{\vectr}{\vectz})
  -\epsvb\epsve\vphif^2(\cprod{\unitplz}{\vectz})]\beqref{gray4}
\end{split}
\nonumber\\
\begin{split}
&=\vekp(\dprod{\vectu}{\vecth})+\vekq(\dprod{\vectu}{\vectz})+\vekr(\dprod{\vectu}{\vectr})
  +\vekk[\dprod{\vectu}{(\cprod{\unitkap}{\vectr})}]-\cdkt\epsvb\vphif[\dprod{\vectu}{(\cprod{\unitkap}{\vectz})}]
  +\veks[\dprod{\vectu}{(\cprod{\unitkap}{\unitplz})}]\\
  &\quad+\vekt[\dprod{\vectu}{(\cprod{\unitplz}{\vectr})}]
  +\rhorep\vrhoa\epsvb\vphif[\dprod{\vectu}{(\cprod{\vectr}{\vectz})}]
  -\epsvb\epsve\vphif^2[\dprod{\vectu}{(\cprod{\unitplz}{\vectz})}]
\end{split}
\nonumber\\
\begin{split}
&=\vekq(-\scalr\vrhob\vphib)+\scalr\vekr(\scalr\vrhob)
  -\vekk[\dprod{\unitkap}{(\cprod{\vectu}{\vectr})}]
  -\cdkt\epsvb\vphif[\dprod{\unitkap}{(\cprod{\vectz}{\vectu})}]
  -\veks[\dprod{\unitkap}{(\cprod{\vectu}{\unitplz})}]\\
  &\quad-\vekt[\dprod{\unitplz}{(\cprod{\vectu}{\vectr})}]
  +\rhorep\vrhoa\epsvb\vphif[\dprod{\vectz}{(\cprod{\vectu}{\vectr})}]
  -\epsvb\epsve\vphif^2[\dprod{\unitplz}{(\cprod{\vectz}{\vectu})}]
  \beqref{main5c}, \eqnref{gpath7}\text{ \& }\eqnref{alg4}
\end{split}
\end{align*}
\begin{align}\label{gray7b}
\begin{split}
&=-\scalr\vekq\vrhob\vphib+\scalr^2\vekr\vrhob-\vekk(\dprod{\unitkap}{\vecth})
  -\cdkt\epsvb\vphif[\dprod{\unitkap}{(-\scalh^{-2}\frkyz\vecth)}]
  -(\veks/\scalh^2)[\dprod{\unitkap}{(\efktc\vecth-\dltvc\vectz-\scalq\dltvc\unitpos)}]\\
  &\quad-\vekt(\dprod{\unitplz}{\vecth})
  +\rhorep\vrhoa\epsvb\vphif(\dprod{\vectz}{\vecth})
  -\epsvb\epsve\vphif^2[\dprod{\unitplz}{(-\scalh^{-2}\frkyz\vecth)}]
  \beqref{main5c}, \eqnref{gxpeed2}\text{ \& }\eqnref{gpath25a}
\end{split}
\nonumber\\
\begin{split}
&=-\scalr\vekq\vrhob\vphib+\scalr^2\vekr\vrhob-\vekk\dltva
  +\epsvb\vphif\frkyz\dltva(\cdkt/\scalh^2)
  -(\veks/\scalh^2)[\efktc\dltva-\dltvc(\dltvb+\scalq\epsva)]\\
  &\quad-\vekt\dltvc+\epsvb\epsve\vphif^2\frkyz(\dltvc/\scalh^2)
  \beqref{gxpeed1a}, \eqnref{ogrv1a}\text{ \& }\eqnref{main5c}
\end{split}
\nonumber\\
\begin{split}
&=-\scalr\vekq\vrhob\vphib+\scalr^2\vekr\vrhob-\vekk\dltva
  +\epsvb\vphif\frkyz\dltva(\cdkt/\scalh^2)
  -(\veks/\scalh^2)(\efktc\dltva-\dltvc\efkta)\\
  &\quad-\vekt\dltvc+\epsvb\epsve\vphif^2\frkyz(\dltvc/\scalh^2)
  \beqref{gxpeed1b}
\end{split}
\nonumber\\
&=\vpsf\beqref{gray1d}
\end{align}
\begin{align}\label{gray7c}
&\dprod{\unitplz}{(\veust+\veusu)}\nonumber\\
\begin{split}
&=\dprod{\unitplz}{[}\vekp\vecth+\vekq\vectz+\vekr\vectr+\vekk(\cprod{\unitkap}{\vectr})-\cdkt\epsvb\vphif(\cprod{\unitkap}{\vectz})
  +\veks(\cprod{\unitkap}{\unitplz})+\vekt(\cprod{\unitplz}{\vectr})\\
  &\quad+\rhorep\vrhoa\epsvb\vphif(\cprod{\vectr}{\vectz})
  -\epsvb\epsve\vphif^2(\cprod{\unitplz}{\vectz})]\beqref{gray4}
\end{split}
\nonumber\\
\begin{split}
&=\vekp(\dprod{\unitplz}{\vecth})+\vekq(\dprod{\unitplz}{\vectz})+\vekr(\dprod{\unitplz}{\vectr})
  +\vekk[\dprod{\unitplz}{(\cprod{\unitkap}{\vectr})}]-\cdkt\epsvb\vphif[\dprod{\unitplz}{(\cprod{\unitkap}{\vectz})}]
  +\veks[\dprod{\unitplz}{(\cprod{\unitkap}{\unitplz})}]\\
  &\quad+\vekt[\dprod{\unitplz}{(\cprod{\unitplz}{\vectr})}]
  +\rhorep\vrhoa\epsvb\vphif[\dprod{\unitplz}{(\cprod{\vectr}{\vectz})}]
  -\epsvb\epsve\vphif^2[\dprod{\unitplz}{(\cprod{\unitplz}{\vectz})}]
\end{split}
\nonumber\\
&=\vekp\dltvc+\vekq\dltvd+\scalr\vekr\epsvc-\vekk\vsiga+\cdkt\epsvb\vphif\vsigc-\rhorep\vrhoa\epsvb\vphif\vsigb
  \beqref{ogrv1a}, \eqnref{gxpeed1a}\text{ \& }\eqnref{gpath1a}\nonumber\\
&=\vpsg\beqref{gray1d}
\end{align}
\begin{align*}
&\dprod{(\cprod{\vectr}{\vecth})}{(\veust+\veusu)}\nonumber\\
\begin{split}
&=\dprod{(\cprod{\vectr}{\vecth})}{[}\vekp\vecth+\vekq\vectz+\vekr\vectr
  +\vekk(\cprod{\unitkap}{\vectr})-\cdkt\epsvb\vphif(\cprod{\unitkap}{\vectz})
  +\veks(\cprod{\unitkap}{\unitplz})+\vekt(\cprod{\unitplz}{\vectr})\\
  &\quad-\epsvb\epsve\vphif^2(\cprod{\unitplz}{\vectz})
  +\rhorep\vrhoa\epsvb\vphif(\cprod{\vectr}{\vectz})]\beqref{gray4}
\end{split}
\nonumber\\
\begin{split}
&=\vekp[\dprod{\vecth}{(\cprod{\vectr}{\vecth})}]
  +\vekq[\dprod{\vectz}{(\cprod{\vectr}{\vecth})}]
  +\vekr[\dprod{\vectr}{(\cprod{\vectr}{\vecth})}]
  +\vekk[\dprod{(\cprod{\vectr}{\vecth})}{(\cprod{\unitkap}{\vectr})}]\\
  &\quad-\cdkt\epsvb\vphif[\dprod{(\cprod{\vectr}{\vecth})}{(\cprod{\unitkap}{\vectz})}]
  +\veks[\dprod{(\cprod{\vectr}{\vecth})}{(\cprod{\unitkap}{\unitplz})}]
  +\vekt[\dprod{(\cprod{\vectr}{\vecth})}{(\cprod{\unitplz}{\vectr})}]\\
  &\quad-\epsvb\epsve\vphif^2[\dprod{(\cprod{\vectr}{\vecth})}{(\cprod{\unitplz}{\vectz})}]
  +\rhorep\vrhoa\epsvb\vphif[\dprod{(\cprod{\vectr}{\vecth})}{(\cprod{\vectr}{\vectz})}]
\end{split}
\end{align*}
\begin{align}\label{gray7d}
\begin{split}
&=-\scalr\vekq\epsvh
  +\vekk[(\dprod{\vectr}{\unitkap})(\dprod{\vecth}{\vectr})-\scalr^2(\dprod{\vecth}{\unitkap})]
  -\cdkt\epsvb\vphif[(\dprod{\vectr}{\unitkap})(\dprod{\vecth}{\vectz})-(\dprod{\vectr}{\vectz})(\dprod{\vecth}{\unitkap})]\\
  &\quad+\veks[(\dprod{\vectr}{\unitkap})(\dprod{\vecth}{\unitplz})-(\dprod{\vectr}{\unitplz})(\dprod{\vecth}{\unitkap})]
  +\vekt[(\dprod{\vectr}{\unitplz})(\dprod{\vecth}{\vectr})-\scalr^2(\dprod{\vecth}{\unitplz})]\\
  &\quad-\epsvb\epsve\vphif^2[(\dprod{\vectr}{\unitplz})(\dprod{\vecth}{\vectz})-(\dprod{\vectr}{\vectz})(\dprod{\vecth}{\unitplz})]
  +\rhorep\vrhoa\epsvb\vphif[\scalr^2(\dprod{\vecth}{\vectz})-(\dprod{\vectr}{\vectz})(\dprod{\vecth}{\vectr})]
  \beqref{ogrv1a}\text{ \& }\eqnref{alg2}
\end{split}
\nonumber\\
\begin{split}
&=-\scalr\vekq\epsvh+\vekk(-\scalr^2\dltva)-\cdkt\epsvb\vphif(-\scalr\epsvd\dltva)
  +\veks(\scalr\epsva\dltvc-\scalr\epsvc\dltva)+\vekt(-\scalr^2\dltvc)\\
  &\quad-\epsvb\epsve\vphif^2(-\scalr\epsvd\dltvc)
  \beqref{main5c}, \eqnref{ogrv1a}\text{ \& }\eqnref{alg2}
\end{split}
\nonumber\\
&=-\scalr\vekq\epsvh-\scalr^2\vekk\dltva+\scalr\cdkt\epsvb\vphif\epsvd\dltva
  +\scalr\veks(\epsva\dltvc-\epsvc\dltva)-\scalr^2\vekt\dltvc+\scalr\epsvb\epsve\vphif^2\epsvd\dltvc
\nonumber\\
&=\scalr[-\vekq\epsvh-\scalr(\vekk\dltva+\vekt\dltvc)+\veks(\epsva\dltvc-\epsvc\dltva)
  +\epsvb\epsvd\vphif(\cdkt\dltva+\epsve\vphif\dltvc)]
\nonumber\\
&=\vpsh\beqref{gray1d}
\end{align}
\end{subequations}
from which we obtain
\begin{subequations}\label{gray8}
\begin{align}
\imaa
&=\dprod{\unitkap}{(\veust+\veusu)}\beqref{kray2c}\nonumber\\
&=\vpse\beqref{gray7a}\label{gray8a}\\
\imab
&=\dprod{\vecta}{(\veust+\veusu)}\beqref{kray2c}\nonumber\\
&=\dprod{(-\vrhoa\vectr)}{(\veust+\veusu)}\beqref{main5a}\text{ \& }\eqnref{gpath1b}\nonumber\\
&=-\vrhoa\vekw\beqref{gray5c}\label{gray8b}
\end{align}
\begin{align}
\imac
&=\dprod{\fdota}{(\veust+\veusu)}\beqref{kray2c}\nonumber\\
&=\dprod{(-\vrhoa\vectu+3\vrhoa\vrhob\vectr)}{(\veust+\veusu)}\beqref{gpath6a}\nonumber\\
&=-\vrhoa\vpsf+3\vrhoa\vrhob\vekw\beqref{gray5c}\text{ \& }\eqnref{gray7b}\label{gray8c}\\
\imad
&=\dprod{\ffdota}{(\veust+\veusu)}\beqref{kray2c}\nonumber\\
&=\dprod{(6\vrhoa\vrhob\vectu-\vrhoa\vrhol\vectr)}{(\veust+\veusu)}\beqref{gpath6b}\nonumber\\
&=6\vrhoa\vrhob\vpsf-\vrhoa\vrhol\vekw\beqref{gray5c}\text{ \& }\eqnref{gray7b}\label{gray8d}
\end{align}
\begin{align}\label{gray8e}
\imae
&=\dprod{\ffdote}{(\veust+\veusu)}\beqref{kray2c}\nonumber\\
&=\dprod{[\ethvo\unitplz+2\kaprep\epsvb\ethvk\vectz+2\kaprep\scalq\epsvb\ethvk\unitpos+\ethvp(\cprod{\vectr}{\vecth})]}
  {(\veust+\veusu)}\beqref{gpath16b}\nonumber\\
&=\ethvo\vpsg+2\kaprep\epsvb\ethvk\vekv+2\kaprep(\scalq/\scalr)\epsvb\ethvk\vekw+\ethvp\vpsh
  \beqref{gray5}\text{ \& }\eqnref{gray7}\nonumber\\
&=\ethvo\vpsg+2\kaprep\epsvb\ethvk\vekv+2\kaprep\vphib\epsvb\ethvk\vekw+\ethvp\vpsh
  \beqref{ogrv1b}.
\end{align}
\end{subequations}
We have also that
\begin{subequations}\label{gray9}
\begin{align}\label{gray9a}
&\ffdot{\cdkt}\imaa+\ffdot{\rhorep}\imab+\vbbb\imac+\rhorep\imad+\imae\nonumber\\
&=\frkyb\imaa+\ethvi\imab+\frkye\imac+\rhorep\imad+\imae
  \beqref{gpath18b}, \eqnref{gpath12b}\text{ \& }\eqnref{gpath19b}\nonumber\\
\begin{split}
&=\frkyb\vpse+\ethvi(-\vrhoa\vekw)+\frkye(-\vrhoa\vpsf+3\vrhoa\vrhob\vekw)+\rhorep(6\vrhoa\vrhob\vpsf-\vrhoa\vrhol\vekw)\\
  &\quad+\ethvo\vpsg+2\kaprep\epsvb\ethvk\vekv+2\kaprep\vphib\epsvb\ethvk\vekw+\ethvp\vpsh\beqref{gray8}
\end{split}
\nonumber\\
\begin{split}
&=\frkyb\vpse+\ethvo\vpsg+\ethvp\vpsh-\vrhoa[\ethvi\vekw+\frkye(\vpsf-3\vrhob\vekw)+\rhorep(\vrhol\vekw-6\vrhob\vpsf)]\\
  &\quad+2\kaprep\epsvb\ethvk(\vekv+\vphib\vekw)
\end{split}
\nonumber\\
&=\vpsi\beqref{gray1d}
\end{align}
\begin{align}\label{gray9b}
&\cdkt\unitkap+\rhorep\vecta-\vectu+\vecte\nonumber\\
\begin{split}
&=\cdkt\unitkap+\rhorep(-\vrhoa\vectr)-[\scalh^{-2}(\cprod{\vectz}{\vecth}+\cprod{\scalq\unitpos}{\vecth})]
  +\fone(\cprod{\vectr}{\vecth})+\ftwo\unitplz\beqref{main5}\text{ \& }\eqnref{gpath1b}
\end{split}
\nonumber\\
&=\cdkt\unitkap-\rhorep\vrhoa\vectr-\scalh^{-2}(\cprod{\vectz}{\vecth})-\scalh^{-2}(\scalq/\scalr)(\cprod{\vectr}{\vecth})
  +\fone(\cprod{\vectr}{\vecth})+\ftwo\unitplz\nonumber\\
&=\cdkt\unitkap+\ftwo\unitplz-\rhorep\vrhoa\vectr-\scalh^{-2}(\cprod{\vectz}{\vecth})
  +(\fone-\scalh^{-2}\vphib)(\cprod{\vectr}{\vecth})\beqref{ogrv1b}\nonumber\\
&=\cdkt\unitkap+\epsve\vphif\unitplz-\rhorep\vrhoa\vectr-\scalh^{-2}(\cprod{\vectz}{\vecth})
  +(\epsvb\vphif-\scalh^{-2}\vphib)(\cprod{\vectr}{\vecth})\beqref{ogrv2g}\text{ \& }\eqnref{ogrv2h}\nonumber\\
&=\cdkt\unitkap+\epsve\vphif\unitplz-\rhorep\vrhoa\vectr-(1/\scalh^2)(\cprod{\vectz}{\vecth})
  +\vpsj(\cprod{\vectr}{\vecth})\beqref{gray1d}.
\end{align}
\end{subequations}

\subart{Results of the computations}
By substituting \eqnref{gray9}, \eqnref{gray6} and \eqnref{gray4} into \eqnref{kray4}, we finally get
\begin{subequations}\label{gray10}
\begin{align}\label{gray10a}
\bbkbar=\plusmin\frac{\vpsd}{\scalc^3\rcal^3},\qquad
\bbtbar=\frac{\vpsi}{(\vpsd)^2}
\end{align}
\begin{align}\label{gray10b}
\begin{split}
&\frtbar=\frac{1}{\scalc\rcal}\biggl[\cdkt\unitkap+\epsve\vphif\unitplz-\rhorep\vrhoa\vectr-(1/\scalh^2)(\cprod{\vectz}{\vecth})
  +\vpsj(\cprod{\vectr}{\vecth})\biggr]\\
&\frbbar=\frac{1}{\vpsd}\biggl[\vekp\vecth+\vekq\vectz
  +\vekr\vectr+\vekk(\cprod{\unitkap}{\vectr})-\cdkt\epsvb\vphif(\cprod{\unitkap}{\vectz})
  +\veks(\cprod{\unitkap}{\unitplz})+\vekt(\cprod{\unitplz}{\vectr})\\
  &\qquad+\rhorep\vrhoa\epsvb\vphif(\cprod{\vectr}{\vectz})
  -\epsvb\epsve\vphif^2(\cprod{\unitplz}{\vectz})\biggr]
\end{split}
\end{align}
\end{subequations}
as the complete set of equations describing the apparent geometry of obliquated rays for a gravitating observer.

\part{Illustrations}\label{partthree}
\begin{wisdom}{Stephen J. Gould (1941 - 2002)}
The beauty of nature lies in detail; the message in generality.
Optimal appreciation demands both and I know of no better tactic than the
illustration of exciting principles by well-chosen particulars.
\end{wisdom}

\section{Effects of constant acceleration}\label{ilconacc}
\art{On the apparent direction to a light source}
When an observer translates with a constant acceleration $\vecta=\fdot{\vectu}$, the velocity $\vectu$
and the position $\vectr$ of the observer at any instant $\scalt$ are given by
\begin{subequations}\label{constd1}
\begin{align}\label{constd1a}
\vectu=\vecta\scalt+\vectuo,\quad
\vectr=\frac{1}{2}\vecta\scalt^2+\vectuo\scalt+\vectro
\end{align}
where $\vectuo$ and $\vectro$ are respectively the observer's velocity and position at instant $\scalt=0$.
By squaring the second of these equations and rearranging its terms, we get
\begin{align}\label{constd1b}
\scala^2\scalt^2+2(\dprod{\vecta}{\vectuo})\scalt+2\dprod{\vecta}{(\vectro-\vectr)}=0
\end{align}
while by squaring the second equation in \eqnref{constd1a} and taking \eqnref{constd1b} into
account, we get
\begin{align}\label{constd1c}
\scalu^2=\scaluo^2+2\scala\scalt\scaluo\cos\thetao+\scala^2\scalt^2,\quad
\scalu^2=\scaluo^2+2\dprod{\vecta}{(\vectr-\vectro)}
\end{align}
where $\thetao$ is the angle between $\vecta$ and $\vectuo$. Consequently, if $\betao=\scaluo/\scalc$,
then by \eqnref{main6} and \eqnref{constd1c},
\begin{align}\label{constd1c2}
\betrep(\scalt)=[\betao^2+2\sigrep\scalt\betao\cos\thetao+\sigrep^2\scalt^2]^{1/2},\quad
\betrep(\vectr)=[\betao^2+2\dprod{(\vectsig/\scalc)}{(\vectr-\vectro)}]^{1/2}.
\end{align}
If we solve the first equation in \eqnref{constd1c} for $\scala\scalt$ and take the second equation into account,
we shall obtain
\begin{align}\label{constd1d}
\scala\scalt
=-\scaluo\cos\thetao\pm(\scalu^2-\scaluo^2\sin^2\thetao)^{1/2}
=-\scaluo\cos\thetao\pm[\scaluo^2\cos^2\thetao+2\dprod{\vecta}{(\vectr-\vectro)}]^{1/2}.
\end{align}
Also, by multiplying the first equation in \eqnref{constd1a} vectorwise by $\vecta$, one may show that
\begin{align}\label{constd1e}
\scalu\sin\thtrep=\scaluo\sin\thetao
\end{align}
where $\thtrep$ is the angle between $\vecta$ and $\vectu$ (cf.~\figref{FIG1}).
These results have been known since Galileo brought them forcefully to the
attention of philosophers; our task will be to study their kineoptical consequences
in some detail.
\end{subequations}

The acceleration $\vecta$ and the wave vector $\vectkap$ being constant, the angle $\lamrep$ between
them at any instant does not change in the course of the observer's motion, whence
\begin{subequations}\label{constd2}
\begin{align}\label{constd2a}
\lamrep(\scalt)=\lamrep(\vectr)=\lambdao
\end{align}
where $\lambdao$ is the angle between these vectors at instant $\scalt=0$. From \eqnref{constd1c} and
\eqnref{constd1e}, we have~\footnote{We remind the reader that throughout this work and in accordance with
\eqnref{main6}, the case $\vectu=\zvect$ is to be excluded from consideration.}
\begin{align}\label{constd2b}
\thtrep(\scalt)=\arcsin\left\{\frac{\scaluo\sin\thetao}
  {\sqrt{\scaluo^2+2\scala\scalt\scaluo\cos\thetao+\scala^2\scalt^2}}\right\},\quad
\thtrep(\vectr)=\arcsin\left\{\frac{\scaluo\sin\thetao}
  {\sqrt{\scaluo^2+2\dprod{\vecta}{(\vectr-\vectro)}}}\right\}.
\end{align}
By multiplying through the first equation in \eqnref{constd1a} scalarwise by $\unitkap$ and taking
\eqnref{constd1d} into consideration, one may show that
\begin{align}\label{constd2c}
\begin{split}
&\qquad\qquad\qquad\phirep(\scalt)=\arccos\left\{\frac{\scaluo\cos\phio+\scala\scalt\cos\lambdao}
  {\sqrt{\scaluo^2+2\scala\scalt\scaluo\cos\thetao+\scala^2\scalt^2}}\right\}\\
&\phirep(\vectr)=\arccos\left\{\frac{\scaluo(\cos\phio-\cos\lambdao\cos\thetao)
  \pm\cos\lambdao\sqrt{\scaluo^2\cos^2\thetao+2\dprod{\vecta}{(\vectr-\vectro)}}}
  {\sqrt{\scaluo^2+2\dprod{\vecta}{(\vectr-\vectro)}}}\right\}
\end{split}
\end{align}
where $\phirep$ is the angle between $\unitkap$ and $\vectu$, and $\phio$ is the angle between
$\unitkap$ and $\vectuo$.
\end{subequations}

\subart{Translation with zero acceleration}
When the observer translates without accelerating ($\vecta=\zvect$), we have $\sigrep=0$ and
$\murep=0$ by \eqnref{main6}. The various quantities defined by \eqnref{trob2c} become
\begin{subequations}\label{constd3}
\begin{align}\label{constd3a}
\rhorep=0,\quad\xcons=0,\quad\dragf=1,\quad\vthtrep=0,\quad\cdkt=\scalc.
\end{align}
Putting $\vecta=\zvect$ into \eqnref{constd2c} and $\sigrep=0$ into \eqnref{constd1c2} leads to
\begin{align}\label{constd3b}
\phirep=\phio,\quad\betrep=\betao
\end{align}
so that, in view of \eqnref{constd3a} and \eqnref{constd3b}, the quantities defined by \eqnref{trob2b}
reduce to
\begin{align}\label{constd3c}
\gcal=-\betao+\cos\phio,\quad
\rcal^2=1+\betao^2-2\betao\cos\phio,\quad
\fcal^2=\sin^2\phio.
\end{align}
\end{subequations}
Substituting the value of $\rcal$ from \eqnref{constd3c} into \eqnref{grad5a} gives the ray speed as
\begin{subequations}\label{constd4}
\begin{align}\label{constd4a}
\upsrep=\scalc(1+\betao^2-2\betao\cos\phio)^{1/2}
\end{align}
while by putting the values of $\fcal$ and $\gcal$ from \eqnref{constd3c} into \eqnref{trob2b2}, we get
the angular displacement of the light source as
\begin{align}\label{constd4b}
\psirep=\arctan\left\{\frac{\sin\phio}{-\betao+\cos\phio}\right\}.
\end{align}
\end{subequations}
This well known result~\cite{Selleri01} is equivalent to \eqnref{abob6a}. We conclude that when
an observer translates without accelerating, polarization and dispersion have no effect on the
apparent displacement of a light source or on the speed of a light ray.

\subart{Rectilinear translation with nonzero acceleration}\label{art28b}
If the initial velocity $\vectuo$ of the observer is directed parallel to the acceleration $\vecta$,
the first equation in \eqnref{constd1a} shows that the instantaneous velocity $\vectu$ will
remain parallel to the acceleration at all times. Indeed, substituting $\thetao=0$ and $\phio=\lambdao$
into \eqnref{constd2b} and \eqnref{constd2c} gives (cf. \figref{FIG1})
\begin{subequations}\label{constd5}
\begin{align}\label{constd5a}
\thtrep=\thetao=0,\quad\phirep=\phio=\lambdao=\lamrep
\end{align}
where we have taken advantage of \eqnref{constd2a}. Substituting \eqnref{constd5a} into
\eqnref{constd1c2} leads to
\begin{align}\label{constd5b}
\betrep(\scalt)=\betao+\sigrep\scalt,\quad
\betrep(\vectr)=[\betao^2+2(\sigrep/\scalc)|\vectr-\vectro|]^{1/2}.
\end{align}
Again in view of \eqnref{constd5a}, we have from \eqnref{trob2c} that
\begin{align}\label{constd5c}
\begin{split}
&\qquad\qquad\vthtrep=\murep\cos\phio,\quad
\dragf=\left\{\frac{1+\sqrt{1+\vthtrep^2}}{2}\right\}^{1/2}\\
&\xcons=\vthtrep(1+\vthtrep^2)^{-1/2},\quad
\rhorep=\xcons/(4\dragf\pfreq),\quad
\cdkt=\scalc(\dragf-2\rhorep\sigrep\cos\phio)
\end{split}
\end{align}
while \eqnref{trob4a} becomes
\begin{align}\label{constd5d}
\begin{split}
&\fcal=\absv{\dragf-2\rhorep\sigrep\cos\phio}\sin\phio,\quad
\gcal=\dragf\cos\phio-\betrep-\rhorep\sigrep\cos2\phio\\
&\rcal^2=\dragf^2+\betrep^2-2\dragf\betrep\cos\phio
    +\rhorep\sigrep(\rhorep\sigrep-2\dragf\cos\phio+2\betrep\cos2\phio).
\end{split}
\end{align}
\end{subequations}
Substituting the value of $\rcal$ from \eqnref{constd5d} into \eqnref{grad5a} gives the ray speed as
\begin{subequations}\label{constd6}
\begin{align}\label{constd6a}
\upsrep=\scalc[\dragf^2+\betrep^2-2\dragf\betrep\cos\phio
    +\rhorep\sigrep(\rhorep\sigrep-2\dragf\cos\phio+2\betrep\cos2\phio)]^{1/2}
\end{align}
while by putting the values of $\fcal$ and $\gcal$ from \eqnref{constd5d} into \eqnref{trob2b2}, we get
the angular displacement of the light source as
\begin{align}\label{constd6b}
\psirep=\arctan\left\{\frac{\absv{\dragf-2\rhorep\sigrep\cos\phio}\sin\phio}
  {\dragf\cos\phio-\betrep-\rhorep\sigrep\cos2\phio}\right\}.
\end{align}
\end{subequations}
We conclude that for an observer in accelerated rectilinear translation, both acceleration and
dispersion will be observed to affect the speed of a light ray as well as the angular displacement
of a light source.

\example{Transverse line of incidence}\label{exa1}
If the light ray is incident~\footnote{We remind the reader of the importance of distinguishing the line of
incidence (loi) of a ray from its line of sight (los) since insufficient attention to this distinction has proved to
be a rich source of errors for careless minds.} at right angles to the observer's
velocity, so that $\phio=\halfpi$, then by \eqnref{constd5c},
\begin{subequations}\label{constd7}
\begin{align}\label{constd7a}
\vthtrep=0,\quad\dragf=1,\quad\xcons=0,\quad\rhorep=0,\quad\cdkt=\scalc.
\end{align}
From \eqnref{constd6a} and \eqnref{constd6b}, we obtain
\begin{align}\label{constd7b}
\upsrep=\scalc(1+\betrep^2)^{1/2},\quad
\psirep=\arctan(-1/\betrep)
\end{align}
\end{subequations}
which shows that the ray speed and the angular displacement of the light source depend
only on the instantaneous velocity of the observer. We conclude that there are indeed situations
in which obliquation is determined by the instantaneous
velocity of the observer and is independent of the observer's acceleration as postulated by the
so-called clock or locality hypothesis of aclassical physics~\protect\cite{Brown01, Mashhoon01}.
Generally speaking, however, obliquation depends explicitly on the observer's acceleration
although its effects are easy to eliminate by an appropriate choice of geometry.

\example{Semi-transverse line of incidence}
If the light ray is incident at $45^\circ$ to the observer's velocity, so that $\phio=45^\circ$ and
$\cos\phio=\sin\phio=1/\sqrt{2}=\sqrt{2}/2$, then by \eqnref{constd5c},
\begin{subequations}\label{constd8}
\begin{align}\label{constd8a}
\vthtrep=\murep/\sqrt{2},\quad
\cdkt=\scalc(\dragf-\rhorep\sigrep\sqrt{2})
\end{align}
while from \eqnref{constd6}, we get
\begin{align}\label{constd8b}
\upsrep=\scalc[\dragf^2+\betrep(\betrep-\dragf\sqrt{2})+\rhorep\sigrep(\rhorep\sigrep-\dragf\sqrt{2})]^{1/2},\quad
\psirep=\arctan\left\{\frac{\absv{\dragf-\rhorep\sigrep\sqrt{2}}}{\dragf-\betrep\sqrt{2}}\right\}
\end{align}
\end{subequations}
which shows the effects of acceleration (via $\sigrep$) and dispersion (via $\rhorep$) on the ray speed $\upsrep$
and on the angular displacement $\psirep$ of the light source. We conclude in view of \eqnref{constd5c} and
\eqnref{constd8a} that since $\rhorep\ne0$ and $\sigrep\ne0$ unless $\scala=0$, the effects of acceleration
and dispersion cannot be strictly eliminated in this case unless the observer ceases to accelerate.

\example{Ultragamma approximation for semi-transverse loi}
When the quantity $\murep$ is so small that expressions containing its third and
higher powers can be neglected ($\murep\ll1$), we have from \eqnref{constd8a} that the third and
higher powers of $\vthtrep$ can also be neglected. Accordingly, by \eqnref{constd5c} and \eqnref{constd8a},
we have the following approximations
\begin{subequations}\label{constd9}
\begin{align}\label{constd9a}
\dragf\approx1+\frac{\murep^2}{16},\quad
\xcons\approx\frac{\murep}{\sqrt{2}},\quad
\rhorep\approx\frac{\murep}{4\pfreq\sqrt{2}},\quad
\cdkt\approx\scalc\left\{1-\frac{3\murep^2}{16}\right\}
\end{align}
where we have used the fact that $\murep=\sigrep/\pfreq$ by \eqnref{main6}.
Bearing this fact in mind in addition to \eqnref{constd8a} and \eqnref{constd9a}, it is easy to show that
for $\dragf-\rhorep\sigrep\sqrt{2}\ge0$, \eqnref{constd8b} takes the form
\begin{align}\label{constd9b}
\upsrep\approx\scalc\left\{1+\betrep(\betrep-\sqrt{2})-\frac{\murep^2}{8}\left(1+\frac{\betrep}{\sqrt{2}}\right)\right\}^{1/2},\quad
\psirep\approx\arctan\left\{1+\frac{\murep^2}{4}-\betrep\sqrt{2}\left(1+\frac{3\murep^2}{16}\right)\right\}^{-1}.
\end{align}
\end{subequations}
We conclude that in this approximation, the effects of acceleration and dispersion
on the ray speed and on the angular displacement of the light source are of second and higher orders in $\murep$.

\example{Infraradio approximation for semi-transverse loi}
When the quantity $\murep$ is so large that expressions containing its first and
higher powers are much greater than $\sqrt{2}$, we have from \eqnref{constd8a} that the first and
higher powers of $\vthtrep$ are much greater than $1$. Accordingly, by \eqnref{constd5c}
and \eqnref{constd8a}, we have the following approximations
\begin{subequations}\label{constd10}
\begin{align}\label{constd10a}
\dragf\approx\left\{\frac{\murep}{\sqrt{8}}\right\}^{1/2},\quad
\xcons\approx1,\quad
\rhorep\approx\frac{1}{4\pfreq}\left\{\frac{\sqrt{8}}{\murep}\right\}^{1/2},\quad
\cdkt\approx0.
\end{align}
Equation \eqnref{constd8b} takes the form
\begin{align}\label{constd10b}
\upsrep\approx\scalc\left\{\betrep^2+\scalb\left(\frac{\scalb}{4}-\betrep\right)\right\}^{1/2},\quad
\scalb=\left\{\frac{\murep}{\sqrt{2}}\right\}^{1/2},\quad
\psirep\approx\arctan0
\end{align}
\end{subequations}
from which we conclude that, in this approximation, the effects of acceleration and dispersion
on the ray speed manifest at first order in $\murep$ while the light source suffers an angular
displacement that places it in a direction parallel or antiparallel to the observer's velocity.

\scholium{Prosaic character of superluminal velocities}
Each expression for the ray speed $\upsrep$ in the foregoing examples imposes a constraint on $\betrep$
for a given $\murep$ since the radicand in the expression must be positive in order for the expression
to hold. These constraints are however geometric and specific to each example. We may therefore not regard
any of them as a fundamental statement applicable to
all obliquation phenomena. This is particularly important when the constraint is such as to require that
$\betrep<1$, because in this case those who have not learnt to give due diligence to the demands of
epistemological completeness may be tempted to suppose (in view of the claim of many proponents of
aclassical physics to this effect) that this constraint is an Act of Nature.
We emphasize therefore that in this work we neither require $\betrep<1$ nor place any apriori
constraint on $\betrep$. This is illustrated by the foregoing examples which admit $\betrep\ge1$
without contradictions.

\subart{Coradial translation with nonzero acceleration}\label{art28c}
When an observer translates with a constant acceleration directed along the line of incidence of
a light ray, we have $\lambdao=0, \thetao=\phio$ which upon substitution into \eqnref{constd2} yields
(cf. \figref{FIG1})
\begin{subequations}\label{constd11}
\begin{align}\label{constd11a}
\thetao=\phio,\quad\thtrep=\phirep,\quad\lambdao=\lamrep=0
\end{align}
on account of which, by \eqnref{constd2b} and \eqnref{constd2c},
\begin{align}\label{constd11b}
\phirep(\scalt)=\arcsin\left\{\frac{\scaluo^2\sin^2\phio}
  {\scaluo^2+2\scala\scalt\scaluo\cos\phio+\scala^2\scalt^2}\right\}^{1/2},\quad
\phirep(\vectr)=\arcsin\left\{\frac{\scaluo^2\sin^2\phio}
  {\scaluo^2+2\dprod{\vecta}{(\vectr-\vectro)}}\right\}^{1/2}.
\end{align}
Substituting \eqnref{constd11a} into \eqnref{constd1c2} leads to
\begin{align}\label{constd11c}
\betrep(\scalt)=[\betao^2+2\sigrep\scalt\betao\cos\phio+\sigrep^2\scalt^2]^{1/2},\quad
\betrep(\vectr)=[\betao^2+2\dprod{(\vectsig/\scalc)}{(\vectr-\vectro)}]^{1/2}.
\end{align}
\end{subequations}
In view of \eqnref{constd11a}, we have from \eqnref{trob2c} that
\begin{subequations}\label{constd12}
\begin{align}\label{constd12a}
\begin{split}
&\qquad\qquad\vthtrep=\murep,\quad
\dragf=\left\{\frac{1+\sqrt{1+\vthtrep^2}}{2}\right\}^{1/2}\\
&\xcons=\vthtrep(1+\vthtrep^2)^{-1/2},\quad
\rhorep=\xcons/(4\dragf\pfreq),\quad
\cdkt=\scalc(\dragf-2\rhorep\sigrep)
\end{split}
\end{align}
while according to \eqnref{trob5a},
\begin{align}\label{constd12b}
\begin{split}
&\fcal=\absv{\dragf-\rhorep\sigrep}\sin\phirep,\quad
\gcal=-\betrep+(\dragf-\rhorep\sigrep)\cos\phirep\\
&\rcal^2=(\dragf-\rhorep\sigrep)^2+\betrep^2-(2\dragf+\rhorep\sigrep)\betrep\cos\phirep.
\end{split}
\end{align}
\end{subequations}
Substituting the value of $\rcal$ from \eqnref{constd12b} into \eqnref{grad5a} gives the ray speed as
\begin{subequations}\label{constd13}
\begin{align}\label{constd13a}
\upsrep=\scalc[\betrep^2-\betrep(2\dragf+\rhorep\sigrep)\cos\phirep+(\dragf-\rhorep\sigrep)^2]^{1/2}.
\end{align}
Furthermore, by putting the values of $\fcal$ and $\gcal$ from \eqnref{constd12b} into \eqnref{trob2b2}, we get
the angular displacement of the light source as
\begin{align}\label{constd13b}
\psirep=\arctan\left\{\frac{\absv{\dragf-\rhorep\sigrep}\sin\phirep}{-\betrep+(\dragf-\rhorep\sigrep)\cos\phirep}\right\}
\end{align}
\end{subequations}
which, while similar in form to \eqnref{trob5b}, differs from \eqnref{trob5b} in that $\phirep$
is required to satisfy \eqnref{constd11b}.
We conclude that for an observer in accelerated coradial translation, both acceleration and
dispersion will be observed to affect the speed of a light ray as well as the angular displacement
of a light source.

\example{Coradial translation with transverse loi}
It may seem that one can obtain an interesting and useful result by putting $\phirep=\halfpi$ into
\eqnref{constd13}. One difficulty with this circumstance is that since $\phirep$ is a function of time
by \eqnref{constd11b}, the condition $\phirep=\halfpi$ can be established only for a brief instant. A more
serious difficulty is that since the observer's acceleration is parallel to the line of incidence of the light
ray, the condition $\phirep=\halfpi$ requires the observer's velocity to be perpendicular to the acceleration.
The acceleration being constant, however, this requirement can be satisfied only at the instant when the observer's
velocity is zero. We conclude that putting $\phirep=\halfpi$ into \eqnref{constd13} does not lead to a
very useful result.

\art{On the apparent drift of a light source}
For an observer translating with constant acceleration, the velocity and acceleration of the observer are
independent quantities, the acceleration being constant while the velocity varies, so that the formulae of
\artref{a_txpeed} are applicable. In view of \eqnref{constd1}, we have from \eqnref{tspeed3}
that
\begin{subequations}\label{consts1}
\begin{align}\label{consts1a}
\angus=\vtta\unitkap+\vttb\vectuo+(\vttb\scalt-\vttc)\vecta
\end{align}
\begin{align}\label{consts1b}
\vtte=\frac{\betrep(\rcal^2+\betrep\gcal)}{\fcal\rcal^2\scalu^2},\quad
\vttd=\frac{\betrep+\gcal}{\fcal\scalu^2},\quad
\vttc=\vtte\rhorep,\quad
\vttb=\vttd+\vtte,\quad
\vtta=-\vtte\cdkt.
\end{align}
Furthermore, by \eqnref{kas1c}, \eqnref{tspeed4} and \eqnref{consts1a}, we get
\begin{align}\label{consts1c}
\begin{split}
&\qquad\qquad\qquad\quad\fdot{\psirep}=\scala[\vtta\cos\lamrep+\vttb\scaluo\cos\thetao+\scala(\vttb\scalt-\vttc)],\quad
\aspua=\fdot{\psirep}/\scala\\
&\angus^2=\vtta^2+2\vtta\vttb\scaluo\cos\phio+\vttb^2\scaluo^2+\scala(\vttb\scalt-\vttc)
  [\scala(\vttb\scalt-\vttc)+2\vtta\cos\lamrep+2\vttb\scaluo\cos\thetao]
\end{split}
\end{align}
which gives the drift $\fdot{\psirep}$ and the magnitude of the obliquation gradient $\angus$ as functions
of time $\scalt$, the variation of obliquation $\aspua$ being defined only when the observer has a
nonzero acceleration.
\end{subequations}

\subart{Translation with zero acceleration}
When the observer translates without accelerating, the obliquation gradient $\angus$ is well defined and can be
calculated by first substituting \eqnref{constd3} into \eqnref{consts1b} to get
\begin{subequations}\label{consts2}
\begin{align}\label{consts2a}
\vtte=\frac{\betao(1-\betao\cos\phio)}{\scaluo^2(1+\betao^2-2\betao\cos\phio)\sin\phio},\quad
\vttd=\frac{\cos\phio}{\scaluo^2\sin\phio},\quad
\vttc=0
\end{align}
\begin{align}\label{consts2b}
\vttb=\frac{\betao+(1-2\betao\cos\phio)\cos\phio}{\scaluo^2(1+\betao^2-2\betao\cos\phio)\sin\phio},\quad
\vtta=\frac{\betao\cos\phio-1}{\scaluo(1+\betao^2-2\betao\cos\phio)\sin\phio}
\end{align}
and then putting $\vecta=\zvect$ into \eqnref{consts1a} and \eqnref{consts1c} to get
\begin{align}\label{consts2c}
\angus=\vtta\unitkap+\vttb\vectuo,\quad
\angus^2=\vtta^2+2\vtta\vttb\scaluo\cos\phio+\vttb^2\scaluo^2,\quad
\fdot{\psirep}=0.
\end{align}
\end{subequations}
We conclude that when an observer translates without accelerating, the angular displacement of a light
source will not change in the course of the observer's motion regardless of the line of incidence of
a light ray from the source to the observer.

\scholium{Interpretation of obliquation gradient}
So long as the obliquation angle $\psirep$ depends on the observer's velocity $\vectu$, the
gradient of $\psirep$ with respect to $\vectu$ is well defined even for a constant $\vectu$ because the
gradient gives the amount by which $\psirep$ changes if and when $\vectu$ changes
without explicitly requiring $\vectu$ to be variable. When $\vectu$ does vary, as it must when an observer
accelerates, the obliquation gradient gives the amount by which the apparent position
of a light source changes for a unit change in the observer's velocity at any given instant. But when $\vectu$
is fixed, as it must be when an observer does not accelerate, the obliquation gradient gives the amount
by which the apparent position of a light source differs for two observers, each moving
with a constant velocity, for every unit difference in the observers' velocities. This case deserves
clarification because those familiar with the prevailing mode of expression in aclassical physics may
be tempted to interprete it by introducing a space filled with observers, all moving with different
but constant velocities, and may therefore be led to suppose that obliquation gradient by its very nature
requires a multiplicity of observers.

\subart{Rectilinear translation with nonzero acceleration}
When an observer translates rectilinearly with a constant acceleration, the results of \artref{art28b} are
applicable. Introducing
\begin{align}\label{consts3}
\delrep=\frac{\dragf(\dragf-\betrep\cos\phio)+\rhorep\sigrep(\rhorep\sigrep-2\dragf\cos\phio+\betrep\cos2\phio)}
  {\dragf^2+\betrep^2-2\dragf\betrep\cos\phio+\rhorep\sigrep(\rhorep\sigrep-2\dragf\cos\phio+2\betrep\cos2\phio)}
\end{align}
we have by substituting \eqnref{constd5d} and the value of $\cdkt$ from \eqnref{constd5c} into \eqnref{consts1b},
\begin{subequations}\label{consts4}
\begin{align}\label{consts4a}
\vtte=\frac{\betrep\delrep}{\scalu^2(\dragf-2\rhorep\sigrep\cos\phio)\sin\phio},\quad
\vttd=\frac{\dragf\cos\phio-\rhorep\sigrep\cos2\phio}{\scalu^2(\dragf-2\rhorep\sigrep\cos\phio)\sin\phio}
\end{align}
\begin{align}\label{consts4b}
\vttc=\vtte\rhorep,\quad
\vttb=\vttd+\vtte,\quad
\vtta=-\delrep/(\scalu\sin\phio).
\end{align}
\end{subequations}
Also, by using \eqnref{constd5a} in \eqnref{consts1c}, we get
\begin{align}\label{consts5}
\begin{split}
&\qquad\qquad\qquad\quad\fdot{\psirep}=\scala[\vtta\cos\phio+\vttb\scaluo+\scala(\vttb\scalt-\vttc)],\quad
\aspua=\fdot{\psirep}/\scala\\
&\angus^2=\vtta^2+2\vtta\vttb\scaluo\cos\phio+\vttb^2\scaluo^2+\scala(\vttb\scalt-\vttc)
  [\scala(\vttb\scalt-\vttc)+2\vtta\cos\phio+2\vttb\scaluo]
\end{split}
\end{align}
from which we conclude that the apparent displacement of a light source will in general
vary with time when an observer translates rectilinearly with a constant nonzero acceleration.

\example{Transverse line of incidence}
If a light ray is incident at right angles to the observer's velocity, so that \eqnref{constd7a} applies,
then \eqnref{consts4} will reduce to
\begin{subequations}\label{consts6}
\begin{align}\label{consts6a}
\begin{split}
&\delrep=\frac{1}{1+\betrep^2},\quad
\vtte=\frac{\betrep}{\scalu^2(1+\betrep^2)},\quad
\vttd=0,\quad
\vttc=0\\
&\qquad\quad\vttb=\frac{\betrep}{\scalu^2(1+\betrep^2)},\quad
\vtta=-\frac{1}{\scalu(1+\betrep^2)}
\end{split}
\end{align}
on account of which \eqnref{consts5} will become
\begin{align}\label{consts6b}
\fdot{\psirep}=\frac{\sigrep}{1+\betrep^2},\quad
\aspua=\frac{1}{\scalc(1+\betrep^2)},\quad
\angus^2=\frac{1}{\scalu^2(1+\betrep^2)}
\end{align}
\end{subequations}
where $\betrep$ satisfies \eqnref{constd5b}. We conclude that in this scenario the apparent drift of
the light source is affected by acceleration but not by dispersion.

\art{On the apparent path of a light source}
To calculate the apparent path of a light source for an observer translating with a constant acceleration,
we note that for this observer, \eqnref{tpath1xa} reduces to
\begin{subequations}\label{constp1}
\begin{align}\label{constp1a}
\begin{split}
&\vsiga=0,\quad\vsigb=0,\quad\vsigc=0,\quad\vsigd=0,\quad\vsige=0,\quad\vsigf=0,\quad\vsigg=0,\quad\vsigh=0\\
&\vsigi=0,\quad\vsigj=0,\quad\vsigk=0,\quad\vsigl=0,\quad\vsigm=0,\quad\vsign=0,\quad\vsigo=0,\quad\vsigp=0\\
&\vsigq=0,\quad\vsigr=0,\quad\vsigs=0,\quad\vsigt=0,\quad\vsigu=0,\quad\vsigv=0.
\end{split}
\end{align}
Ignoring $\vrhoa, \vrhob, \vrhoc$ because they will not be needed in the calculations, we also have
by \eqnref{tpath1},
\begin{align}\label{constp1b}
\begin{split}
&\vrhod=-1,\quad\vrhoe=0,\quad\vrhof=0,\quad\vrhog=0,\quad\vrhoh=0,\quad\vrhoi=0,\quad\vrhoj=0,\quad\vrhok=0\\
&\vrhol=0,\quad\vrhom=0,\quad\vrhon=0,\quad\vrhoo=0,\quad\vrhop=1,\quad\vrhoq=\scala,\quad\vrhor=0,\quad\vrhos=0\\
&\vrhot=0,\quad\vrhou=0,\quad\vrhov=0,\quad\vrhow=0,\quad\vrhox=0,\quad\vrhoy=0.
\end{split}
\end{align}
\end{subequations}
Substituting these values into \eqnref{tpath17} gives~\footnote{We use $\botto$ to denote an indeterminate
or undefined scalar or vector.}
\begin{align}\label{constp2}
\bbk=0,\quad\bbt=0,\quad\frt=-\vecta/\scala,\quad\frb=\botto
\end{align}
from which we conclude that the apparent path of the light source is a straight line in a direction
antiparallel to the observer's acceleration whatever maybe the line of incidence of the light ray from
the source to the observer. We conclude further that if the observer's motion is
not accelerated, the light source will not have an apparent path ($\frt=\botto$) and will therefore
be apparently stationary~\cite{Page25}.

\art{On the apparent geometry of rays}
To study the apparent geometry of obliquated rays for an observer translating with a constant acceleration, we
note that for this observer, \eqnref{tray1a} reduces to
\begin{subequations}\label{constr1}
\begin{align}\label{constr1a}
\begin{split}
&\qquad\vpsa=\dprod{\unitkap}{(\cprod{\vecta}{\vectu})},\quad\vpsb=\zvect,\quad\vpsc=\zvect\\
&\vpsd=\zvect,\quad\vpse=\zvect,\quad\vpsf=\zvect,\quad\vpsg=\zvect,\quad\vpsh=\zvect
\end{split}
\end{align}
so that, with \eqnref{constp1}, the remaining quantities defined in \eqnref{tray1} are given by
\begin{align}\label{constr1b}
\begin{split}
&\qquad\veka=-\cdkt,\quad\vekb=-\vpsa,\quad\vekc=0,\quad\vekd=0,\quad\veke=0\\
&\vekf=\scala^2\cdkt^2\sin^2\lamrep,\quad
\vekg=\scala^2\scalu^2\sin^2\thtrep,\quad
\vekh=\cdkt\scalu\scala^2(\cos\lamrep\cos\thtrep-\cos\phirep).
\end{split}
\end{align}
\end{subequations}
We substitute \eqnref{constr1} into \eqnref{tray6} to get, in view of \eqnref{constp1b} and \eqnref{constd1a},
\begin{align}\label{constr2}
\begin{split}
&\bbkbar=\plusmin\Pirep/(\scalc\rcal^3),\quad\bbtbar=0,\quad
\frtbar=\frac{\cdkt\unitkap-\vectuo+(\rhorep-\scalt)\vecta}{\scalc\rcal},\quad
\frbbar=\frac{\cprod{\vecta}{(\cdkt\unitkap-\vectuo)}}{\scalc^2\Pirep}\\
&\qquad\Pirep=\sigrep[\betrep^2\sin^2\thtrep+(\cdkt/\scalc)^2\sin^2\lamrep
   +2\betrep(\cdkt/\scalc)(\cos\lamrep\cos\thtrep-\cos\phirep)]^{1/2}.
\end{split}
\end{align}
We conclude that for an observer translating with a constant acceleration, the light ray is curved
but lies entirely in a plane whatever the line of incidence of the ray maybe.

\example{Rectilinear translation}
When the observer's motion is rectilinear, the quantities featured in \eqnref{constr2} have the values
calculated in \artref{art28b}. Substituting \eqnref{constd5a} into \eqnref{constr2}
and using the value of $\cdkt$ from \eqnref{constd5c} gives
\begin{subequations}\label{constr3}
\begin{align}\label{constr3a}
\Pirep=\sigrep\absv{\dragf-2\rhorep\sigrep\cos\phio}\sin\phio.
\end{align}
In particular, when the light ray is incident at right angles to the observer's initial velocity $\vectuo$,
we have $\phio=\halfpi$ which reduces \eqnref{constd5d} to $\rcal^2=1+\betrep^2$ and \eqnref{constr3a} to
$\Pirep=\sigrep$. It follows by putting these values into \eqnref{constr2} that
\begin{align}\label{constr3b}
\bbkbar=\plusmin\frac{\sigrep}{\scalc(1+\betrep^2)^{3/2}}
\approx\left\{
\begin{aligned}
&\plusmin\sigrep/\scalc&\text{ if }\betrep\ll1\\
&\plusmin(\sigrep/\scalc)\betrep^{-3}&\text{ if }\betrep\gg1
\end{aligned}
\right.
\end{align}
\end{subequations}
where $\betrep$ satisfies \eqnref{constd5b}. We conclude that dispersion has no effect on the curvature of the ray.
However, unlike the corresponding obliquation angle given by \eqnref{constd7b}, the ray curvature depends explicitly
on the observer's acceleration and not only on the instantaneous velocity of the observer. Hence those who wish to
uphold the socalled locality hypothesis mentioned in \exaref{exa1} ought to bear in mind that while the obliquation
angle may depend only the instantaneous velocity of the observer, the curvature of the ray does in fact depend on
the observer's acceleration, on account of which one may not claim, even in aclassical physics, that acceleration 
has no effect on obliquation.

\example{Contigency of obliquated light rays}\label{xmpl:conti}
\newcommand\tgat[1]{\frtbar^{(#1)}}
\newcommand\rayarcl{\overline{\scals}}
If P and Q are two points on a ray, and if the tangent vectors to the ray at these points are respectively
$\tgat{P}$ and $\tgat{Q}$, the angle between these vectors in the limit as Q approaches P is called
the contigency of the ray at P~(\cite{Struik88}, pg 14). By definition, then,
\begin{subequations}\label{constr4}
\begin{align}\label{constr4a}
\textd\vphirep=\bbkbar\;\textd\rayarcl
\end{align}
where $\textd\rayarcl$ is an element of the ray and $\textd\vphirep$ is an element of the contigency angle.
Observing that $\textd\rayarcl/\textd\scalt=\upsrep$ is the ray speed, the above equation can be rewritten as
\begin{align}\label{constr4b}
\fdot{\vphirep}=\bbkbar\upsrep
\end{align}
\end{subequations}
where $\fdot{\vphirep}$ may be called the variation of contigency. From \eqnref{constd7b},
\eqnref{constr3b} and \eqnref{constr4b} we obtain
\begin{subequations}\label{constr5}
\begin{align}\label{constr5a}
\fdot{\vphirep}=\plusmin\frac{\sigrep}{(1+\betrep^2)^3}
\end{align}
as the variation of contigency for a light ray incident at right angles to the initial velocity of
an observer translating rectilinearly with constant acceleration. Moreover, comparing \eqnref{constr5a}
with \eqnref{consts6b}, we get
\begin{align}\label{constr5b}
\sigrep^2\fdot{\vphirep}=\fdot{\psirep}^3
\end{align}
\end{subequations}
which relates the variation of obliquation at any instant to the corresponding variation of contigency
at that instant.

\scholium{Deviation of obliquated light rays}
Consider a point moving along a light ray with the ray speed $\upsrep$, and let this point be at position P at
time $\scalt_P$ and at position Q at time $\scalt_Q$. If we integrate the variation of contigency $\fdot{\vphirep}$
from $\scalt=\scalt_P$ to $\scalt=\scalt_Q$, the result will be the angle $\Delrep\vphirep$ between the tangent
vectors to the ray at P and Q. This so-called bending angle or deviation constitutes a global measure of the ray
curvature. It is a useful physical quantity when the ray lies entirely in one plane between P and Q.
If the ray does not lie entirely in one plane between these positions, the physical usefulness of the bending
angle is questionable. And even when the ray lies entirely in one plane between these positions, the usefulness
of the bending angle can still be questioned on the grounds that $\Delrep\vphirep=0$ does not necessarily
imply zero curvature everywhere between the two positions, which may be interpreted as showing that the bending
angle is too crude a measure of curvature. Since the problems we are investigating in this work may admit rays
that are both curved and twisted, we shall not in general be concerned with the bending angle in this work.
The variation of contigency will be found to be much more suitable for our purposes because it places more
stringent constraints on conflicting kineoptical theories.

\art{On the apparent frequency of rays}
The apparent frequency of a light ray can be calculated for an observer translating with a constant
acceleration by substituting the appropriate value of $\rcal$ into \eqnref{kfreq4}. In this way we shall find that
the apparent ray frequency $\afreq$ is affected in general by both acceleration and dispersion.

\subart{Apparent ray frequency for nonaccelerated observers}
When an observer translates without accelerating, $\rcal$ has the value given by \eqnref{constd3c}.
Using this value in \eqnref{kfreq4} gives
\begin{align}\label{constf1}
(\afreq/\pfreq)^2=1+\betao^2-2\betao\cos\phio
\end{align}
so that for $\cos\phio=0$, which corresponds to the situation shown in \figref{FIG5A}, we have
\begin{subequations}\label{constf2}
\begin{align}\label{constf2a}
\afreq=\pfreq\sqrt{1+\betao^2}
\end{align}
and for $\cos\phio=\betao$, corresponding to the situation shown in \figref{FIG5B}, we have
\begin{align}\label{constf2b}
\afreq=\pfreq\sqrt{1-\betao^2}.
\end{align}
\end{subequations}
We conclude that in either of the transverse situations shown in \figref{FIG5}, the apparent
ray frequency depends on second and higher order terms in $\betao$ when the observer translates
without accelerating~\cite{Sher68, Tatum85, Baird00, Bonizzoni00}. Also, for a light ray incident in a
direction parallel to the observer's velocity ($\cos\phio=1$), we get
\begin{subequations}\label{constf3}
\begin{align}\label{constf3a}
\afreq=\pfreq\absv{1-\betao}
\end{align}
while for a light ray incident in a direction antiparallel to the observer's velocity ($\cos\phio=-1$), we get
\begin{align}\label{constf3b}
\afreq=\pfreq(1+\betao).
\end{align}
We conclude that in either of these so-called longitudinal or radial situations, the apparent ray frequency satisfies
the usual classical formulae attributed to Doppler.
\end{subequations}

\subart{Apparent ray frequency for accelerated observers}
When an observer translates rectilinearly with a constant acceleration, $\rcal$ has the value given by \eqnref{constd5d}.
Substituting this value into \eqnref{kfreq4} gives
\begin{align}\label{constf4}
(\afreq/\pfreq)^2=\dragf^2+\betrep^2-2\dragf\betrep\cos\phio
    +\rhorep\sigrep(\rhorep\sigrep-2\dragf\cos\phio+2\betrep\cos2\phio)
\end{align}
where the various quantities have the values calculated in \artref{art28b}.

\newcommand{\betat}{\betrep_t}
\example{Transverse line of incidence}
For a light ray incident at right angles to the observer's initial velocity ($\cos\phio=0$), using
\eqnref{constd7a} in \eqnref{constf4} gives
\begin{align}\label{constf5}
\afreq=\pfreq\sqrt{1+\betrep^2}
\end{align}
where $\betrep$ satisfies \eqnref{constd5b}. Comparing this result with \eqnref{constf2a}, we conclude that in this
situation, the apparent frequency of the ray depends only on the instantaneous velocity of the observer and not
explicitly on the observer's acceleration (cf. \exaref{exa1}). We may also consider the situation corresponding
to $\cos\phio=\betrep$. But in this case we must bear in mind that since $\betrep$ varies with time by \eqnref{constd5b},
the condition $\cos\phio=\betrep$ can be established only for a brief instant. Designating the value of $\betrep$ at the
instant when this condition holds good by $\betat$, \eqnref{constd5c} and \eqnref{constf4} give
\begin{subequations}\label{constf6}
\begin{align}\label{constf6a}
(\afreq/\pfreq)^2
=\dragf^2+\betat^2-2\dragf\betat^2+\rhorep\sigrep[\rhorep\sigrep-2\dragf\betat+2\betat(2\betat^2-1)]
\end{align}
\begin{align}\label{constf6b}
\dragf=\left\{\frac{1+\sqrt{1+\murep^2\betat^2}}{2}\right\}^{1/2},\quad
\rhorep=\frac{1}{4\dragf\pfreq}\left\{\frac{\murep\betat}{\sqrt{1+\murep^2\betat^2}}\right\}
\end{align}
\end{subequations}
from which we conclude that, at this particular instant, the apparent ray frequency is affected by both dispersion
and acceleration.

\example{Longitudinal line of incidence}
For a light ray incident in a direction parallel to the initial velocity of the observer ($\cos\phio=1$), we have
by \eqnref{constd5c} and \eqnref{constf4} that
\begin{subequations}\label{constf7}
\begin{align}\label{constf7a}
(\afreq/\pfreq)^2=\dragf^2+\betrep^2-2\dragf\betrep+\rhorep\sigrep(\rhorep\sigrep-2\dragf+2\betrep),\quad
\vthtrep=\murep
\end{align}
while for a light ray incident in a direction antiparallel to the initial velocity of the observer ($\cos\phio=-1$),
we have
\begin{align}\label{constf7b}
(\afreq/\pfreq)^2=\dragf^2+\betrep^2+2\dragf\betrep+\rhorep\sigrep(\rhorep\sigrep+2\dragf+2\betrep),\quad
\vthtrep=-\murep
\end{align}
where $\dragf$ and $\rhorep$ are given in both cases by
\begin{align}\label{constf7c}
\dragf=\left\{\frac{1+\sqrt{1+\vthtrep^2}}{2}\right\}^{1/2},\quad
\rhorep=\frac{1}{4\dragf\pfreq}\left\{\frac{\vthtrep}{\sqrt{1+\vthtrep^2}}\right\}.
\end{align}
We conclude that acceleration and dispersion have effects on the apparent frequency of a ray incident in a
direction parallel or antiparallel to the observer's velocity.
\end{subequations}

\example{Ultragamma approximation for longitudinal loi}
When the quantity $\murep$ is so small that expressions containing its third and higher powers can be neglected
($\murep\ll1$), we have by \eqnref{constf7a} or \eqnref{constf7b} that the same condition holds good for the
quantity $\vthtrep$. Equation \eqnref{constf7c} then becomes
\begin{subequations}\label{constf8}
\begin{align}\label{constf8a}
\dragf\approx1+\frac{\vthtrep^2}{8},\quad
\rhorep\approx\frac{\vthtrep}{4\pfreq}
\end{align}
which upon substitution into \eqnref{constf7a} gives
\begin{align}\label{constf8b}
(\afreq/\pfreq)^2\approx(1-\betrep)^2-\frac{\murep^2(1-\betrep)}{4}
\end{align}
and upon substitution into \eqnref{constf7b} gives
\begin{align}\label{constf8c}
(\afreq/\pfreq)^2\approx(1+\betrep)^2-\frac{\murep^2(1+\betrep)}{4}.
\end{align}
\end{subequations}
We conclude that in this approximation, the effects of acceleration and dispersion on the apparent ray frequency
are of second and higher orders in $\murep$. Moreover, when the observer's motion is such that $\betrep<1$ at
all instants, we see that the effect of acceleration is such as to reduce the ray frequency
regardless of whether the line of incidence of the ray is parallel or antiparallel to the observer's velocity.

\example{Infraradio approximation for longitudinal loi}
When the quantity $\murep$ is so large that expressions containing its first and higher powers are much greater than 1
($\murep\gg1$), we have by \eqnref{constf7a} or \eqnref{constf7b} that the same condition holds good for $\vthtrep$.
In this case we have by \eqnref{constf7c} that
\begin{subequations}\label{constf9}
\begin{align}\label{constf9a}
\dragf\approx\sqrt{\frac{|\vthtrep|}{2}},\quad
\rhorep\approx\frac{\sgn\vthtrep}{4\pfreq}\sqrt{\frac{2}{|\vthtrep|}}
\end{align}
which upon substitution into \eqnref{constf7a} gives
\begin{align}\label{constf9b}
\afreq\approx\pfreq(\scalb-\betrep),\quad\scalb=\sqrt{\murep/8}
\end{align}
and upon substitution into \eqnref{constf7b} gives
\begin{align}\label{constf9c}
\afreq\approx\pfreq(\scalb+\betrep),\quad\scalb=\sqrt{\murep/8}.
\end{align}
\end{subequations}
We conclude that in this approximation, the effects of acceleration and dispersion on the apparent ray frequency
manifest at first order in the square root of $\murep$. We conclude also that when the observer's motion is such that
$\murep>8$, the effect of acceleration is such as to enhance the ray frequency regardless of whether the line of
incidence of the ray is parallel or antiparallel to the observer's velocity.

\section{Effects of centripetal acceleration}\label{CENTRIACC}
\art{On the apparent direction to a light source}
Throughout this section we consider a light ray incident in the plane of motion of an observer moving with constant
linear speed $\scaluo$ and constant angular speed $\Omegao$ in a circle of radius $\scalro$. Then at any instant
$\scalt$, the observer's position $\vectr$, velocity $\vectu$ and acceleration $\vecta$ are given by
\begin{subequations}\label{centrid1}
\begin{align}\label{centrid1a}
\begin{split}
&\vectr=\scalro(\unitkap\cos\azimuth+\unitjap\sin\azimuth),\quad
\vectu=\Omegao\scalro(-\unitkap\sin\azimuth+\unitjap\cos\azimuth)\\
&\qquad\quad\vecta=-\Omegao^2\vectr=-\Omegao(\cprod{\vectu}{\unitiap\,}),\quad
\azimuth(\scalt)=\Omegao\scalt
\end{split}
\end{align}
where $\unitjap$ is a unit vector in the plane of the observer's motion satisfying $\dprod{\unitkap}{\unitjap}=0$,
while $\unitiap=\cprod{\unitkap}{\unitjap}$ is a unit vector perpendicular to the plane of the observer's motion, and
$\azimuth$ is the phase or azimuth of the observer's position with respect to $\unitkap$.
Observing from these equations that
\begin{align}\label{centrid1b}
\begin{split}
&\qquad\quad\dprod{\unitkap}{\vectu}=-\Omegao\scalro\sin\azimuth,\quad
\cprod{\unitkap}{\vectu}=\unitiap\,\Omegao\scalro\cos\azimuth\\
&\dprod{\vectu}{\vecta}=0,\quad
\dprod{\unitkap}{\vecta}=-\Omegao^2\scalro\cos\azimuth,\quad
\cprod{\unitkap}{\vecta}=-\unitiap\,\Omegao^2\scalro\sin\azimuth
\end{split}
\end{align}
we obtain (cf. \figref{FIG1})
\begin{align}\label{centrid1c}
\cos\thtrep=0,\quad
\cos\phirep=\sin\lamrep=-\sin\azimuth,\quad
\sin\phirep=-\cos\lamrep=\cos\azimuth.
\end{align}
\end{subequations}
Consequently, \eqnref{trob2b} becomes
\begin{subequations}\label{centrid2}
\begin{align}\label{centrid2a}
\begin{split}
\fcal&=\absv{\dragf\cos\azimuth+\rhorep\sigrep\cos2\azimuth},\quad
\gcal=-\betrep-\dragf\sin\azimuth-\rhorep\sigrep\sin2\azimuth\\
\rcal^2&=\dragf^2+\betrep^2+2\dragf\betrep\sin\azimuth+\rhorep\sigrep(\rhorep\sigrep
  +2\dragf\cos\azimuth+2\betrep\sin2\azimuth)
\end{split}
\end{align}
while \eqnref{trob2c} becomes
\begin{align}\label{centrid2b}
\begin{split}
&\qquad\qquad\vthtrep=-\murep\cos\azimuth,\quad
\dragf=\left\{\frac{1+\sqrt{1+\vthtrep^2}}{2}\right\}^{1/2}\\
&\xcons=\vthtrep/(1+\vthtrep^2)^{1/2},\quad
\rhorep=\xcons/(4\dragf\pfreq),\quad
\cdkt=\scalc(\dragf+2\rhorep\sigrep\cos\azimuth).
\end{split}
\end{align}
Moreover, substituting the values of $\fcal$ and $\gcal$ from \eqnref{centrid2a} into \eqnref{trob2b2}
gives the instantaneous obliquation angle of the light source as
\begin{align}\label{centrid2c}
\tan\psirep=\frac{\absv{\dragf\cos\azimuth+\rhorep\sigrep\cos2\azimuth}}
  {-\betrep-\dragf\sin\azimuth-\rhorep\sigrep\sin2\azimuth}.
\end{align}
\end{subequations}
We shall find it convenient to say that the observer is at new phase when $\cos\azimuth=1$, at first quarter when
$\sin\azimuth=1$, at full phase when $\cos\azimuth=-1$, and at last quarter when $\sin\azimuth=-1$.

\subart{Apparent direction for observers at new phase}
When an observer is at new phase, \eqnref{centrid2a} reduces to
\begin{subequations}\label{centrid3}
\begin{align}\label{centrid3a}
\fcal=\absv{\rhorep\sigrep+\dragf},\quad
\gcal=-\betrep,\quad
\rcal^2=\dragf^2+\betrep^2+\rhorep\sigrep(\rhorep\sigrep+2\dragf)
\end{align}
while \eqnref{centrid2b} reduces to
\begin{align}\label{centrid3b}
\begin{split}
&\vthtrep=-\murep,\quad
\dragf=\left\{\frac{1+\sqrt{1+\murep^2}}{2}\right\}^{1/2},\quad
\xcons=\frac{-\murep}{\sqrt{1+\murep^2}}\\
&\qquad\rhorep=\frac{1}{4\dragf\pfreq}\left\{\frac{-\murep}{\sqrt{1+\murep^2}}\right\},\quad
\cdkt=\scalc(\dragf+2\rhorep\sigrep)
\end{split}
\end{align}
\end{subequations}
where we have taken advantage of \eqnref{main6}. The ray speed $\upsrep$ and the angle $\psirep$ of obliquation
are obtained by substituting the value of $\rcal$ from \eqnref{centrid3a} into \eqnref{grad5a} and
putting $\cos\azimuth=1$ into \eqnref{centrid2c} to get
\begin{align}\label{centrid4}
\upsrep=\scalc[\betrep^2+(\rhorep\sigrep+\dragf)^2]^{1/2},\quad
\tan\psirep=-\absv{\rhorep\sigrep+\dragf}/\betrep.
\end{align}
We conclude that when the observer is at new phase, acceleration and dispersion affect both the ray speed and the
angular displacement of the light source.

\example{Ultragamma approximation}
If $\murep\ll1$, then to a second order accuracy in $\murep$, we have from \eqnref{centrid3b} that
\begin{subequations}\label{centrid5}
\begin{align}\label{centrid5a}
\dragf\approx1+(\murep^2/8),\quad
\rhorep\approx-\murep/(4\pfreq)
\end{align}
which reduces \eqnref{centrid4} to
\begin{align}\label{centrid5b}
\upsrep\approx\scalc\left\{1+\betrep^2-\frac{\murep^2}{4}\right\}^{1/2},\quad
\tan\psirep\approx-\frac{1}{\betrep}\left\{1-\frac{\murep^2}{8}\right\}.
\end{align}
\end{subequations}
We conclude that in this approximation, the effects of acceleration and dispersion on the ray speed and
on the apparent position of the light source are of second and higher orders in $\murep$.

\example{Infraradio approximation}
If $\murep\gg 1$, then to all orders of accuracy in $\murep$, \eqnref{centrid3b} gives
\begin{subequations}\label{centrid6}
\begin{align}\label{centrid6a}
\dragf\approx\sqrt{\murep/2},\quad
\rhorep\approx-(4\pfreq)^{-1}\sqrt{2/\murep}
\end{align}
which reduces \eqnref{centrid4} to
\begin{align}\label{centrid6b}
\upsrep\approx\scalc\biggl\{\betrep^2+\frac{\murep}{8}\biggr\}^{1/2},\quad
\tan\psirep\approx-\frac{1}{\betrep}\biggl\{\frac{\murep}{8}\biggr\}^{1/2}.
\end{align}
\end{subequations}
We conclude that in this case the effects of acceleration and dispersion on the ray speed and
on the apparent position of the light source manifest at first order in $\murep$.

\example{Relevance to low frequency radio astronomy}
To put some numbers into the above calculations, let $c\approx3\times10^8\text{ms}^{-1}$ and take
$\scala\approx6\times10^{-2}\text{ms}^{-2}$ to represent the earth's acceleration towards the sun.
Then we have $\murep\pfreq\approx200\text{pHz}$. For nanohertz infraradio waves with
$\pfreq\approx1\text{nHz}$, we get $\murep\approx0.2\ll1$ which indicates that \eqnref{centrid5b} gives a
good approximation for $\psirep$. But for picohertz infraradio waves with $\pfreq\approx1\text{pHz}$,
we get $\murep\approx200\gg1$ which indicates that \eqnref{centrid6b} gives a better approximation
for $\psirep$. These numbers suggest that if the earth's orbital motion around the sun can be assumed
to be circular and with constant speed, then the effect of the earth's acceleration on obliquation
can be detected with reasonable certainty if an infraradio survey of the sky is performed
in the picohertz range, for in this range we should find that $\tan\psirep$ is about five times
larger than what would be expected on the basis of other theories. Thus we have reasons to look
forward to technological advances in extremely low frequency radio astronomy which may one day
make such surveys possible.

\subart{Apparent direction for observers at first quarter}
When an observer is at first quarter, \eqnref{centrid2b} reduces to
\begin{subequations}\label{centrid7}
\begin{align}\label{centrid7a}
\vthtrep=0,\quad\dragf=1,\quad\xcons=0,\quad\rhorep=0,\quad\cdkt=\scalc
\end{align}
while \eqnref{centrid2a} reduces to
\begin{align}\label{centrid7b}
\fcal=0,\quad\gcal=-(1+\betrep),\quad\rcal=1+\betrep.
\end{align}
\end{subequations}
Substituting the value of $\rcal$ from \eqnref{centrid7b} into \eqnref{grad5a} and
putting $\sin\azimuth=1$ into \eqnref{centrid2c} gives
\begin{align}\label{centrid8}
\upsrep=\scalc(1+\betrep),\quad
\tan\psirep=0.
\end{align}
We conclude that when the observer is at first quarter, the ray speed is not affected by acceleration and dispersion
while the light source suffers no angular displacement.

\subart{Apparent direction for observers at full phase}
When an observer is at full phase, \eqnref{centrid2b} reduces to
\begin{subequations}\label{centrid9}
\begin{align}\label{centrid9a}
\begin{split}
&\vthtrep=\murep,\quad
\dragf=\left\{\frac{1+\sqrt{1+\murep^2}}{2}\right\}^{1/2},\quad
\xcons=\left\{\frac{\murep}{\sqrt{1+\murep^2}}\right\}\\
&\qquad\rhorep=\frac{1}{4\dragf\pfreq}\left\{\frac{\murep}{\sqrt{1+\murep^2}}\right\},\quad
\cdkt=\scalc(\dragf-2\rhorep\sigrep)
\end{split}
\end{align}
while \eqnref{centrid2a} reduces to
\begin{align}\label{centrid9b}
\fcal=\absv{\rhorep\sigrep-\dragf},\quad
\gcal=-\betrep,\quad
\rcal^2=\dragf^2+\betrep^2+\rhorep\sigrep(\rhorep\sigrep-2\dragf).
\end{align}
\end{subequations}
Substituting the value of $\rcal$ from \eqnref{centrid9b} into \eqnref{grad5a} and
putting $\cos\azimuth=-1$ into \eqnref{centrid2c} gives
\begin{align}\label{centrid10}
\upsrep=\scalc[\betrep^2+(\rhorep\sigrep-\dragf)^2]^{1/2},\quad
\tan\psirep=-\absv{\rhorep\sigrep-\dragf}/\betrep.
\end{align}
We conclude that when the observer is at full phase, acceleration and dispersion affect both the ray speed and the
angular displacement of the light source.

\subart{Apparent direction for observers at last quarter}
When an observer is at last quarter, \eqnref{centrid2b} reduces to
\begin{subequations}\label{centrid11}
\begin{align}\label{centrid11a}
\vthtrep=0,\quad\dragf=1,\quad\xcons=0,\quad\rhorep=0,\quad\cdkt=\scalc
\end{align}
while \eqnref{centrid2a} reduces to
\begin{align}\label{centrid11b}
\fcal=0,\quad\gcal=1-\betrep,\quad\rcal=\absv{1-\betrep}.
\end{align}
\end{subequations}
Substituting the value of $\rcal$ from \eqnref{centrid11b} into \eqnref{grad5a} and
putting $\sin\azimuth=-1$ into \eqnref{centrid2c} gives
\begin{align}\label{centrid12}
\upsrep=\scalc\absv{1-\betrep},\quad
\tan\psirep=0.
\end{align}
We conclude that when the observer is at last quarter, the ray speed is not affected by acceleration and dispersion
while the light source suffers no angular displacement.

\example{Obliquation as a vector transport problem}
The foregoing calculations show that when the observer is at new phase, the light source is displaced at an angle
$\psirep$ to the observer's velocity in accordance with \eqnref{centrid4}. As the observer progresses to first quarter,
the displacement of the light source reduces and finally vanishes when the observer reaches first quarter, at which point
the apparent direction to the light source coincides with the true direction. As the observer progresses further to full
phase, the light source again suffers a displacement which reaches a maximum value when the observer is at full phase. This
displacement in turn diminishes as the observer advances to last quarter, at which point the apparent direction to the
light source coincides again with the true direction. As the observer finally moves from last quarter to new phase, the
light source again suffers a displacement that reaches a maximum value when the observer is at new phase. It appears
therefore that one may treat obliquation as a problem of transporting the ray velocity vector along the path of the
observer. But while this mode of expression may be more convenient for mathematicians, it seems to add nothing
to the physics of the phenomena being investigated, and will therefore not be considered further in this work.

\art{On the apparent drift of a light source}
For an observer translating with centripetal acceleration, the velocity and acceleration of the observer are
dependent quantities, the acceleration being related to the velocity as in \eqnref{centrid1a}, so that the formulae of
\artref{a_txpeed} are not applicable. Since the applicable general formulae for this case were not derived
in that article, we shall not study the apparent drift of a light source for an observer translating with 
centripetal acceleration in the present article. The results which ought to have been derived here can however 
be obtained as a special case of the results obtained in \artref{prec_adls} where we study the same problem for 
a rotating or precessing observer.

\art{On the apparent path of a light source}
To calculate the apparent path of the light source, we first note that by \eqnref{centrid1a},
\begin{subequations}\label{centrip1}
\begin{align}\label{centrip1a}
\fdota=-\Omegao^2\vectu,\quad
\ffdota=\Omegao^4\vectr,\quad
\fffdota=\Omegao^4\vectu
\end{align}
\begin{align}\label{centrip1b}
\begin{split}
&\dprod{\unitkap}{\vectr}=\scalro\cos\azimuth,\quad
\cprod{\unitkap}{\vectr}=\unitiap\,\scalro\sin\azimuth,\quad
\dprod{\vectu}{\vectr}=0,\quad
\cprod{\vectu}{\vectr}=-\unitiap\,\Omegao\scalro^2\\
&\dprod{\unitkap}{\vecta}=-\accparr,\quad
\cprod{\unitkap}{\vecta}=-\unitiap\,\accnorm,\quad
\dprod{\unitkap}{\fdota}=\Omegao\accnorm,\quad
\cprod{\unitkap}{\fdota}=-\unitiap\,\Omegao\accparr\\
&\dprod{\unitkap}{\ffdota}=\Omegao^2\accparr,\quad
\cprod{\unitkap}{\ffdota}=\unitiap\,\Omegao^2\accnorm,\quad
\dprod{\fdota}{\ffdota}=0,\quad
\cprod{\fdota}{\ffdota}=\unitiap\,\scala^2\Omegao^3\\
&\dprod{\vecta}{\fdota}=0,\quad
\cprod{\vecta}{\fdota}=\unitiap\,\scala^2\Omegao,\quad
\dprod{\vecta}{\ffdota}=-\Omegao^2\scala^2,\quad
\cprod{\vecta}{\ffdota}=\zvect
\end{split}
\end{align}
where we have introduced the convenient quantities
\begin{align}\label{centrip1c}
\begin{split}
&\accparr=\scala\cos\azimuth,\quad\accnorm=\scala\sin\azimuth,\quad
\sigparr=\sigrep\cos\azimuth\\
&\signorm=\sigrep\sin\azimuth,\quad
\muparr=\murep\cos\azimuth,\quad\munorm=\murep\sin\azimuth.
\end{split}
\end{align}
\end{subequations}
Substituting \eqnref{centrip1b} into \eqnref{tpath1xa} and taking \eqnref{centrid1} into account, we get
\begin{subequations}\label{centrip2}
\begin{align}\label{centrip2a}
\vsiga=\pfreq\Omegao\signorm,\quad
\vsigb=\pfreq\Omegao^2\sigparr,\quad
\vsigc=-\pfreq\Omegao^3\signorm
\end{align}
\begin{align}\label{centrip2b}
\begin{split}
&\vsigd=0,\quad\vsige=-\scala^2\Omegao^2,\quad\vsigf=0,\quad
\vsigg=\scala^2\Omegao^2,\quad\vsigh=0,\quad\vsigi=-\scala^2\Omegao^4\\
&\quad\vsigj=\scala^2\Omegao^4,\quad\vsigk=0,\quad
\vsigl=\scala^2\Omegao^6,\quad\vsigm=0,\quad\vsign=0,\quad\vsigo=0\\
&\quad\vsigp=0,\quad\vsigq=0,\quad\vsigr=0,\quad\vsigs=0,\quad\vsigt=0,\quad\vsigu=0,\quad\vsigv=0
\end{split}
\end{align}
\end{subequations}
where we have used the fact that $\pfreq=\scalc\kaprep,\;\sigrep=\scala/\scalc,\;\murep=\sigrep/\pfreq$ by \eqnref{main2b}
and \eqnref{main6}. Putting these values into \eqnref{tpath1f} gives $\vrhoy=0$ which upon substitution into \eqnref{tpath17}
yields $\bbt=0$. We conclude that the apparent path of the light source is a plane curve. Moreover, since the various vectors
in the expression for $\frb$ in \eqnref{tpath17} are parallel or antiparallel to $\unitiap$ by \eqnref{centrip1b}, we
conclude also that the apparent path of the light source lies in the plane of the observer's motion.

\subart{Apparent path for observers at new phase}
When an observer is at new phase, we have $\accparr=\scala,\,\accnorm=0$ which reduces \eqnref{centrip2a} to
\begin{subequations}\label{centrip3}
\begin{align}\label{centrip3a}
\vsiga=0,\quad\vsigb=\pfreq\Omegao^2\sigrep,\quad\vsigc=0.
\end{align}
The various quantities defined in \eqnref{tpath1} become, in view of \eqnref{centrid3b}, \eqnref{centrip2b}
and \eqnref{centrip3a},
\begin{align}\label{centrip3b}
\begin{split}
&\vrhoa=(1+\murep^2)^{1/2},\quad
\vrhob=1-4\murep^2,\quad
\vrhoc=(1/\vrhoa^3)-(\xcons/2\dragf)^2,\quad
\vrhod=-1\\
&\vrhoe=-\xcons(\Omegao/\vrhoa)^2,\quad
\vrhof=0,\quad
\vrhog=-\auxvb\sigrep\Omegao,\quad
\vrhoh=0,\quad
\vrhoi=\Omegao\auxva\auxvb,\quad
\vrhoj=0,\quad
\vrhok=0\\
&\vrhol=-\auxvb\scala\Omegao(2\auxva-1),\quad
\vrhom=0,\quad
\vrhon=\vrhol,\quad
\vrhoo=-\scala\auxvb^2(2\auxva-1),\quad
\vrhop=1+\auxva\auxvb^2\\
&\vrhoq=\scala(1+\auxvb^2)^{1/2},\quad
\vrhor=0,\quad
\vrhos=-\scala^3\Omegao^2(1+3\auxva\auxvb^2),\quad
\vrhot=0,\quad
\vrhou=\scala^4\Omegao^2(1+3\auxva\auxvb^2)\\
&\vrhov=0,\quad
\vrhow=\scala^4\Omegao^4(1+3\auxva\auxvb^2),\quad
\vrhox=\scala^2\Omegao|1+3\auxva\auxvb^2|,\quad
\vrhoy=0
\end{split}
\end{align}
\end{subequations}
where we have introduced the quantities
\begin{subequations}\label{centrip4}
\begin{align}\label{centrip4a}
&\quad\auxva=\auxvao\left\{\auxvao-\frac{\murep^2}{4\dragf^2}\right\}
\left\{
\begin{aligned}
&\ge0&\text{if }\murep^2\le3\\
&<0&\text{if }\murep^2>3
\end{aligned}
\right.
,\quad
\auxvb=\auxvao\auxvbo,\quad
\retronp=\frac{1+3\auxva\auxvb^2}{|1+3\auxva\auxvb^2|}
\end{align}
with $\dragf, \auxvao$ and $\auxvbo$ given by
\begin{align}\label{centrip4b}
\dragf=\left\{\frac{1+\sqrt{1+\murep^2}}{2}\right\}^{1/2},\quad
\auxvao=\frac{1}{\sqrt{1+\murep^2}},\quad
\auxvbo=\frac{\murep\Omegao}{4\dragf\pfreq}.
\end{align}
\end{subequations}
In view of \eqnref{centrip3}, \eqnref{centrip1} and \eqnref{centrid1}, \eqnref{tpath17} becomes
\begin{align}\label{centrip5}
\begin{split}
&\bbk=\plusmin\frac{|1+3\auxva\auxvb^2|}{\scalu\bigl(1+\auxvb^2\bigr)^{3/2}},\quad
\bbt=0,\quad
\frt=\frac{\unitkap+\unitjap\,\auxvb}{\sqrt{1+\auxvb^2}},\quad
\frb=\retronp\unitiap\\
&\qquad\qquad\frn=\frac{\retronp(\unitjap-\unitkap\,\auxvb)}{\sqrt{1+\auxvb^2}},\quad
\frc=\plusmin\scalu\left\{\frac{\bigl(1+\auxvb^2\bigr)^{3/2}}{|1+3\auxva\auxvb^2|}\right\}\frn.
\end{split}
\end{align}
We conclude that at the instant when the observer is at new phase, the apparent path of the light source will
be a curved line if $\retronp\ne\botto$ and a straight line if $\retronp=\botto$.

\example{Apparent path for small accelerations}
When the acceleration of the observer is so small as to be negligible ($\murep\approx0$), we have in
\eqnref{centrip4} that
\begin{subequations}\label{centrip6}
\begin{align}\label{centrip6a}
\auxva\approx1,\quad\auxvb\approx0
\end{align}
which upon substitution into \eqnref{centrip5} gives
\begin{align}\label{centrip6b}
\bbk=\plusmin\frac{1}{\scalu},\quad
\frt=\unitkap,\quad
\frn=\unitjap,\quad
\frc=\plusmin\scalu\frn.
\end{align}
\end{subequations}
We conclude that if the true position of the light source is considered to be at the center of curvature of its apparent path,
a vector drawn from its apparent position to its true position will be of the same magnitude and direction as the observer's
velocity, a result that was first established with the greatest authority and originality by Hamilton~\cite{Tait66}.
We conclude further that at the instant in question, the sense of the apparent motion of the light source is parallel to the
line of incidence of a light ray from the source to the observer, and therefore directed towards the observer.

\example{Apparent path in the ultragamma limit}
When the observer's acceleration is not entirely negligible but the quantities $\murep$ and $\Omegao/\pfreq$ are small enough
that their third and higher powers may be neglected, we have that \eqnref{centrid5a} holds. Substituting this equation into
\eqnref{centrip4} gives
\begin{subequations}\label{centrip7}
\begin{align}\label{centrip7a}
\auxva\approx1-\frac{5\murep^2}{4},\quad\auxvb\approx\auxvbo
\end{align}
which reduces \eqnref{centrip5} to
\begin{align}\label{centrip7b}
\bbk\approx\plusmin\frac{1}{\scalu}\left\{1+\frac{3\auxvbo^2}{2}\right\},\quad
\frt\approx\left\{1-\frac{\auxvbo^2}{2}\right\}\unitkap
    +\auxvbo\,\unitjap
\end{align}
\begin{align}\label{centrip7c}
\frn\approx\left\{1-\frac{\auxvbo^2}{2}\right\}\unitjap
   -\auxvbo\,\unitkap,\quad
\frc\approx\plusmin\scalu\left\{1-\frac{3\auxvbo^2}{2}\right\}\frn.
\end{align}
\end{subequations}
We conclude that, strictly speaking, the vector from the apparent position of the light source to its true position
has a magnitude and direction which differ from those of the observer's velocity by small but finitesimal measures.

\example{Apparent path in the infraradio limit}
When the observer's acceleration is not entirely negligible and the quantity $\murep$ is large enough
that its first and higher powers dominate any expression containing them, we have that \eqnref{centrid6a} holds. Substituting
this equation into \eqnref{centrip4} gives
\begin{subequations}\label{centrip8}
\begin{align}\label{centrip8a}
\auxva\approx-\frac{1}{2},\quad\auxvb\approx\frac{\Omegao}{4\pfreq}\sqrt{\frac{2}{\murep}}
\end{align}
on account of which \eqnref{centrip5} becomes, provided $\Omegao/\pfreq$ is not too large,
\begin{align}\label{centrip8b}
\bbk\approx\plusmin\frac{1}{\scalu}\left\{1-\frac{3\Omegao^2}{8\murep\pfreq^2}\right\},\quad
\frt\approx\left\{1-\frac{\Omegao^2}{16\murep\pfreq^2}\right\}\left\{\unitkap+\biggl(\frac{\Omegao}{4\pfreq}
  \sqrt{\frac{2}{\murep}}\biggr)\unitjap\right\}
\end{align}
\begin{align}\label{centrip8c}
\frn\approx\left\{1-\frac{\Omegao^2}{16\murep\pfreq^2}\right\}\left\{\unitjap-\biggl(\frac{\Omegao}{4\pfreq}
  \sqrt{\frac{2}{\murep}}\biggr)\unitkap\right\},\quad
\frc\approx\plusmin\scalu\left\{1+\frac{3\Omegao^2}{8\murep\pfreq^2}\right\}\frn.
\end{align}
\end{subequations}
We conclude that in this approximation also, the magnitude and direction of a vector from the apparent position of
the light source to its true position differ from those of the observer's velocity by small but finitesimal measures.

\subart{Apparent path for observers at first quarter}
When an observer is at first quarter, we have $\accparr=0,\,\accnorm=\scala$ which reduces \eqnref{centrip2a} to
\begin{subequations}\label{centrip9}
\begin{align}\label{centrip9a}
\vsiga=\sigrep\pfreq\Omegao,\quad
\vsigb=0,\quad
\vsigc=-\sigrep\pfreq\Omegao^3.
\end{align}
In view of \eqnref{centrid7a}, \eqnref{centrip2b} and \eqnref{centrip9a},
the various quantities defined in \eqnref{tpath1} become
\begin{align}\label{centrip9b}
\begin{split}
&\vrhoa=1,\quad
\vrhob=1,\quad
\vrhoc=1,\quad
\vrhod=\auxvbo-1,\quad
\vrhoe=0,\quad
\vrhof=-\murep\Omegao^3(1+3\murep^2)\\
&\vrhog=4\dragf\pfreq^2\auxvbo^2,\quad
\vrhoh=0,\quad
\vrhoi=0,\quad
\vrhoj=-\auxvbo\Omegao^2(1+3\murep^2)-12\auxvbo^3\pfreq^2,\quad
\vrhok=0\\
&\vrhol=-3\scala\auxvbo\Omegao,\quad
\vrhom=0,\quad
\vrhon=3\scala\auxvbo\Omegao(\auxvbo-1),\quad
\vrhoo=0,\quad
\vrhop=(\auxvbo-1)(2\auxvbo-1)\\
&\vrhoq=\scala|\auxvbo-1|,\quad
\vrhor=\scala^3\Omegao(\auxvbo^2-1),\quad
\vrhos=0,\quad
\vrhot=-\scala^3\Omegao^3(\auxvbo^2-1)\\
&\vrhou=-\scala^4\Omegao^2(\auxvbo^2-1),\quad
\vrhov=0,\quad
\vrhow=-\scala^4\Omegao^4(\auxvbo^2-1),\quad
\vrhox=\scala^2\Omegao|\auxvbo^2-1|,\quad
\vrhoy=0
\end{split}
\end{align}
where $\auxvbo$ is given by \eqnref{centrip4}, and we shall introduce for convenience~\footnote{
We define the function $\ugn(x)$ to be such that
\begin{equation*}
\ugn(x)=\left\{
\begin{aligned}
&+1&\text{if }x>0\\
&-1&\text{if }x<0\\
&\botto&\text{if }x=0
\end{aligned}
\right.
\end{equation*}
}
\begin{align}\label{centrip9c}
\retrofq=\ugn(\auxvbo-1).
\end{align}
\end{subequations}
In view of \eqnref{centrip9}, \eqnref{centrip1} and \eqnref{centrid1}, \eqnref{tpath17} becomes
\begin{subequations}\label{centrip10}
\begin{align}\label{centrip10a}
\begin{split}
&\bbk=\plusmin\frac{|\auxvbo^2-1|}{\scalu|\auxvbo-1|^3},\quad
\bbt=0,\quad
\frt=-\retrofq\,\unitjap,\quad
\frb=-\retrofq\,\unitiap
\end{split}
\end{align}
\begin{align}\label{centrip10b}
\begin{split}
\frn=-\retrofq^2\unitkap,\quad
\frc=\plusmin\scalu\left\{\frac{|\auxvbo-1|^3}{|\auxvbo^2-1|}\right\}\frn.
\end{split}
\end{align}
\end{subequations}
We conclude that at the instant when the observer is at first quarter, the apparent motion of the light source may
be prograde ($\retrofq=-1$), retrograde ($\retrofq=+1$) or stationary ($\retrofq=\botto$).

\subart{Apparent path for observers at full phase}
When an observer is at full phase, we have $\accparr=-\scala,\,\accnorm=0$ which reduces \eqnref{centrip2a} to
\begin{subequations}\label{centrip11}
\begin{align}\label{centrip11a}
\vsiga=0,\quad\vsigb=-\pfreq\Omegao^2\sigrep,\quad\vsigc=0.
\end{align}
The various quantities defined in \eqnref{tpath1} become, in view of \eqnref{centrid9a}, \eqnref{centrip2b}
and \eqnref{centrip11a},
\begin{align}\label{centrip11b}
\begin{split}
&\vrhoa=(1+\murep^2)^{1/2},\quad
\vrhob=1-4\murep^2,\quad
\vrhoc=(1/\vrhoa^3)-(\xcons/2\dragf)^2,\quad
\vrhod=-1\\
&\vrhoe=-\xcons(\Omegao/\vrhoa)^2,\quad
\vrhof=0,\quad
\vrhog=-\auxvb\sigrep\Omegao,\quad
\vrhoh=0,\quad
\vrhoi=-\auxva\auxvb\Omegao,\quad
\vrhoj=0,\quad
\vrhok=0\\
&\vrhol=\auxvb\scala\Omegao(1+2\auxva),\quad
\vrhom=0,\quad
\vrhon=\vrhol,\quad
\vrhoo=-\scala\auxvb^2(1+2\auxva),\quad
\vrhop=1+\auxva\auxvb^2\\
&\vrhoq=\scala(1+\auxvb^2)^{1/2},\quad
\vrhor=0,\quad
\vrhos=\scala^3\Omegao^2(1-\auxva\auxvb^2),\quad
\vrhot=0,\quad
\vrhou=\scala^4\Omegao^2(1-\auxva\auxvb^2)\\
&\vrhov=0,\quad
\vrhow=\scala^4\Omegao^4(1-\auxva\auxvb^2),\quad
\vrhox=\scala^2\Omegao|1-\auxva\auxvb^2|,\quad
\vrhoy=0
\end{split}
\end{align}
where $\auxva$ and $\auxvb$ are given by \eqnref{centrip4}, and we shall find it convenient to introduce
\begin{align}\label{centrip11c}
\retrofp=\ugn(1-\auxva\auxvb^2).
\end{align}
\end{subequations}
In view of \eqnref{centrip11}, \eqnref{centrip1} and \eqnref{centrid1}, \eqnref{tpath17} becomes
\begin{align}\label{centrip12}
\begin{split}
&\bbk=\plusmin\frac{|1-\auxva\auxvb^2|}{\scalu(1+\auxvb^2)^{3/2}},\quad
\bbt=0,\quad
\frt=\frac{-\unitkap+\unitjap\,\auxvb}{\sqrt{1+\auxvb^2}},\quad
\frb=\retrofp\unitiap\\
&\qquad\frn=\frac{-\retrofp(\unitjap+\unitkap\,\auxvb)}{\sqrt{1+\auxvb^2}},\quad
\frc=\plusmin\scalu\left\{\frac{(1+\auxvb^2)^{3/2}}{|1-\auxva\auxvb^2|}\right\}\frn.
\end{split}
\end{align}
We conclude that at the instant when the observer is at full phase, the apparent path of the light source will
be a curved line if $\retrofp\ne\botto$ and a straight line if $\retrofp=\botto$.

\subart{Apparent path for observers at last quarter}
When an observer is at last quarter, we have $\accparr=0,\,\accnorm=-\scala$ which reduces \eqnref{centrip2a} to
\begin{subequations}\label{centrip13}
\begin{align}\label{centrip13a}
\vsiga=-\sigrep\pfreq\Omegao,\quad\vsigb=0,\quad\vsigc=\sigrep\pfreq\Omegao^3.
\end{align}
The various quantities defined in \eqnref{tpath1} become, in view of \eqnref{centrid11a}, \eqnref{centrip2b}
and \eqnref{centrip13a},
\begin{align}\label{centrip13b}
\begin{split}
&\vrhoa=1,\quad
\vrhob=1,\quad
\vrhoc=1,\quad
\vrhod=-1-\auxvbo,\quad
\vrhoe=0,\quad
\vrhof=\murep\Omegao^3(1+3\murep^2)\\
&\vrhog=4\dragf\pfreq^2\auxvbo^2,\quad
\vrhoh=0,\quad
\vrhoi=0,\quad
\vrhoj=\auxvbo\Omegao^2(1+3\murep^2)+12\pfreq^2\auxvbo^3,\quad
\vrhok=0\\
&\vrhol=-3\scala\auxvbo\Omegao,\quad
\vrhom=0,\quad
\vrhon=-3\scala\auxvbo\Omegao(1+\auxvbo),\quad
\vrhoo=0,\quad
\vrhop=(1+\auxvbo)(1+2\auxvbo)\\
&\vrhoq=\scala(1+\auxvbo),\quad
\vrhor=-\scala^3\Omegao(\auxvbo^2-1),\quad
\vrhos=0,\quad
\vrhot=\scala^3\Omegao^3(\auxvbo^2-1)\\
&\vrhou=-\scala^4\Omegao^2(\auxvbo^2-1),\quad
\vrhov=0,\quad
\vrhow=-\scala^4\Omegao^4(\auxvbo^2-1),\quad
\vrhox=\scala^2\Omegao|1-\auxvbo^2|,\quad
\vrhoy=0
\end{split}
\end{align}
where $\auxvbo$ is given by \eqnref{centrip4}, and we shall find it convenient to introduce
\begin{align}\label{centrip13c}
\retrolq=\ugn(1-\auxvbo).
\end{align}
\end{subequations}
In view of \eqnref{centrip13}, \eqnref{centrip1} and \eqnref{centrid1}, \eqnref{tpath17} becomes
\begin{align}\label{centrip14}
\begin{split}
&\bbk=\plusmin\frac{|\auxvbo-1|}{\scalu(\auxvbo+1)^2},\quad
\bbt=0,\quad
\frt=-\unitjap,\quad
\frb=\retrolq\unitiap\\
&\qquad\frn=\retrolq\unitkap,\quad
\frc=\plusmin\scalu\left\{\frac{(\auxvbo+1)^2}{|\auxvbo-1|}\right\}\frn.
\end{split}
\end{align}
We conclude that at the instant when the observer is at last quarter, the apparent path of the light source will
be a curved line if $\retrolq\ne\botto$ and a straight line if $\retrolq=\botto$.

\example{Elongation in the apparent path of a light source}
Let $\retrofp>0$ and $\retronp>0$. Then when the observer is at full phase or new phase, we have by \eqnref{centrip12}
and \eqnref{centrip5} that
\begin{subequations}\label{centrip15}
\begin{align}\label{centrip15a}
\frn^F=\frac{-\unitjap-\unitkap\,\auxvb}{\sqrt{1+\auxvb^2}},\quad
\frn^N=\frac{\unitjap-\unitkap\,\auxvb}{\sqrt{1+\auxvb^2}}
\end{align}
where $\frn^F$ and $\frn^N$ are unit vectors drawn from the apparent position of the light
source to the center of curvature of its apparent path at full phase and new phase respectively. It follows that,
unless $\auxvb=0$, the angle $\vphirep$ between these vectors will be different from $\fullpi$. To calculate the
amount by which $\vphirep$ differs from $\fullpi$, we take the scalar product of the vectors to get
\begin{align}\label{centrip15b}
\cos\vphirep=\frac{\auxvb^2-1}{\auxvb^2+1}
\end{align}
so that to a third order accuracy in $\murep$, \eqnref{centrip4} permits us to write
\begin{align}\label{centrip15c}
\cos\vphirep\approx\frac{\auxvbo^2-1}{\auxvbo^2+1},\quad
\auxvbo\approx\frac{\murep\Omegao}{4\pfreq}
\end{align}
\end{subequations}
from which the elongation $\fullpi-\vphirep$ can be easily calculated. We conjecture that when the light
source is not stationary but moving, either inherently or as a result of observational errors, elongation may induce
a precession in the apparent path of the source since the center of curvature of the path will no longer be fixed
in space~\footnote{To examine this question more rigorously, it is necessary to first consider the problem of whether
or not the velocity of a light ray may depend on the velocity of its source, a problem that is outside the scope of
this work but will be treated in detail when we study the kineoptics of intrinsic redshifts.}.

\art{On the apparent geometry of rays}
To study the apparent geometry of obliquated rays for an observer in circular motion, we substitute \eqnref{centrid1}
and \eqnref{centrip1} into \eqnref{tray1} to get, in view of \eqnref{centrip2},
\begin{subequations}\label{centrir1}
\begin{align}\label{centrir1a}
\begin{split}
&\vpsa=0,\quad\vpsb=0,\quad\vpsc=0,\quad\vpsd=0,\quad\vpse=0\\
&\qquad\quad\vpsf=0,\quad\vpsg=\Omegao\accnorm,\quad\vpsh=-\Omegao^2\scalu^2
\end{split}
\end{align}
\begin{align}\label{centrir1b}
\begin{split}
&\veka=\cdkt\vrhod,\quad\vekb=0,\quad\vekc=0,\quad\vekd=0,\quad\veke=0\\
&\vekf=[\cdkt\vrhod\accnorm+\rhorep\Omegao(\cdkt\accparr-\rhorep\scala^2)]^2,\quad
\vekg=\scala^2\scalu^2\vrhod^2\\
&\qquad\vekh=\scala\scalu\vrhod[\cdkt\vrhod\accnorm+\rhorep\Omegao(\cdkt\accparr-\rhorep\scala^2)]
\end{split}
\end{align}
\end{subequations}
where we have used the fact that in all the cases we shall be considering, $\vrhok=0$ by
\eqnref{centrip3b}, \eqnref{centrip9b}, \eqnref{centrip11b} and \eqnref{centrip13b}.

\subart{Apparent ray geometry for observers at new phase}
When an observer is at new phase, we have $\accparr=\scala, \accnorm=0$ by definition while $\vrhod=-1$ by
\eqnref{centrip3b}. Equation \eqnref{centrir1b} becomes, in view of \eqnref{centrid3b} and \eqnref{centrip4},
\begin{subequations}\label{centrir2}
\begin{align}\label{centrir2a}
\begin{split}
&\veka=-\scalc(\delrep-\betrep\auxvb),\quad
\vekb=0,\quad\vekc=0,\quad\vekd=0,\quad\veke=0\\
&\qquad\vekf=(\scalc\scala\auxvb\delrep)^2,\quad
\vekg=(\scala\scalu)^2,\quad
\vekh=\scala\scalu(\scalc\scala\auxvb\delrep)
\end{split}
\end{align}
where we have introduced, with $\auxvb$ and $\dragf$ given by \eqnref{centrip4},
\begin{align}\label{centrir2b}
\delrep=\dragf-\betrep\auxvb,\quad
\retronpx=\ugn(\betrep+\delrep\auxvb)
\end{align}
\end{subequations}
so that by using \eqnref{centrid3}, \eqnref{centrir2} and \eqnref{centrip3b} in \eqnref{tray6}, we get
\begin{subequations}\label{centrir3}
\begin{align}\label{centrir3a}
\bbkbar=\plusmin\Pirep/(\scalc\rcal^3),\quad
\bbtbar=0,\quad
\frtbar=(\delrep\,\unitkap-\betrep\,\unitjap\,)/\rcal,\quad
\frbbar=\retronpx\unitiap
\end{align}
\begin{align}\label{centrir3b}
\Pirep=\sigrep|\betrep+\delrep\auxvb|,\quad
\rcal^2=\betrep^2+\delrep^2.
\end{align}
\end{subequations}
We conclude that at the instant when the observer is at new phase, the ray is curved but lies entirely in a plane.

\subart{Apparent ray geometry for observers at first quarter}
When an observer is at first quarter, we have $\accparr=0, \accnorm=\scala$ by definition while $\vrhod=\auxvbo-1$ by
\eqnref{centrip9b}. Equation \eqnref{centrir1b} becomes, in view of \eqnref{centrid7a} and \eqnref{centrip4},
\begin{align}\label{centrir4}
\begin{split}
&\quad\veka=\scalc(\auxvbo-1),\quad\vekb=0,\quad\vekc=0,\quad\vekd=0,\quad\veke=0\\
&\vekf=\scalc^2\scala^2(\auxvbo-1)^2,\quad
\vekg=\scala^2\scalu^2(\auxvbo-1)^2,\quad
\vekh=\scalc\scala^2\scalu(\auxvbo-1)^2.
\end{split}
\end{align}
Using \eqnref{centrid7}, \eqnref{centrir4} and \eqnref{centrip9} in \eqnref{tray6}, we get
\begin{subequations}\label{centrir5}
\begin{align}\label{centrir5a}
\bbkbar=\plusmin\Pirep/(\scalc\rcal^2),\quad
\bbtbar=0,\quad
\frtbar=\unitkap,\quad
\frbbar=-\retrofq\unitiap
\end{align}
\begin{align}\label{centrir5b}
\Pirep=\sigrep|\auxvbo-1|,\quad
\rcal=1+\betrep
\end{align}
\end{subequations}
where $\retrofq$ is given by \eqnref{centrip9c}. We conclude that at the instant when the observer is at first quarter,
the ray is curved even though the observer is moving along the line of incidence of the ray.

\subart{Apparent ray geometry for observers at full phase}
When an observer is at full phase, we have $\accparr=-\scala, \accnorm=0$ by definition while $\vrhod=-1$ by
\eqnref{centrip11b}. Equation \eqnref{centrir1b} becomes, in view of \eqnref{centrid9a} and \eqnref{centrip4},
\begin{align}\label{centrir6}
\begin{split}
&\veka=-\scalc(\delrep-\betrep\auxvb),\quad\vekb=0,\quad\vekc=0,\quad\vekd=0,\quad\veke=0\\
&\qquad\vekf=(\scalc\scala\auxvb\delrep)^2,\quad
\vekg=\scala^2\scalu^2,\quad
\vekh=\scala\scalu(\scalc\scala\auxvb\delrep)
\end{split}
\end{align}
where $\delrep$ is given by \eqnref{centrir2b}, and by using \eqnref{centrid9}, \eqnref{centrir6} and \eqnref{centrip11b}
in \eqnref{tray6}, we get with $\retronpx$ given by \eqnref{centrir2b},
\begin{subequations}\label{centrir7}
\begin{align}\label{centrir7a}
\begin{split}
&\bbkbar=\plusmin\Pirep/(\scalc\rcal^3),\quad
\bbtbar=0,\quad
\frtbar=(\delrep\,\unitkap+\betrep\,\unitjap\,)/\rcal,\quad
\frbbar=\retronpx\unitiap
\end{split}
\end{align}
\begin{align}\label{centrir7b}
\Pirep=\sigrep|\betrep+\delrep\auxvb|,\quad
\rcal^2=\betrep^2+\delrep^2.
\end{align}
\end{subequations}
We conclude that at the instant when the observer is at full phase, the ray is curved to the same extent as
when the observer is at new phase.

\subart{Apparent ray geometry for observers at last quarter}
When an observer is at last quarter, we have $\accparr=0, \accnorm=-\scala$ by definition while $\vrhod=-(\auxvbo+1)$ by
\eqnref{centrip13b}. Equation \eqnref{centrir1b} becomes, in view of \eqnref{centrid11a} and \eqnref{centrip4},
\begin{subequations}\label{centrir8}
\begin{align}\label{centrir8a}
\begin{split}
&\quad\veka=-\scalc(\auxvbo+1),\quad\vekb=0,\quad\vekc=0,\quad\vekd=0,\quad\veke=0\\
&\vekf=\scalc^2\scala^2(\auxvbo+1)^2,\quad
\vekg=\scala^2\scalu^2(\auxvbo+1)^2,\quad
\vekh=-\scala^2\scalu\scalc(\auxvbo+1)^2.
\end{split}
\end{align}
Introducing the quantity
\begin{align}\label{centrir8b}
\retrolqx=\ugn(1-\betrep)
\end{align}
\end{subequations}
and using \eqnref{centrid11}, \eqnref{centrir8} and \eqnref{centrip13} in \eqnref{tray6}, we get
\begin{subequations}\label{centrir9}
\begin{align}\label{centrir9a}
\begin{split}
&\bbkbar=\plusmin\Pirep/(\scalc\rcal^3),\quad
\bbtbar=0,\quad
\frtbar=\retrolqx\unitkap,\quad
\frbbar=-\retrolqx\unitiap
\end{split}
\end{align}
\begin{align}\label{centrir9b}
\Pirep=\sigrep\rcal(\auxvbo+1),\quad
\rcal=|1-\betrep|.
\end{align}
\end{subequations}
We conclude that at the instant when the observer is at last quarter, the ray is curved provided $\betrep\ne1$.
For $\betrep=1$, we see that the ray will not propagate relative to the observer and the light source will therefore
be imperceptible to the observer at last quarter.

\art{On the apparent frequency of rays}\label{art_centrif}
To calculate the apparent frequency of a light ray for an observer in circular motion with constant speed,
we substitute the value of $\rcal$ from \eqnref{centrid2a} into \eqnref{kfreq4}. In this way we shall obtain
\begin{align}\label{centrif1}
(\afreq/\pfreq)^2=\rcal^2,\quad
\rcal^2=\dragf^2+\betrep^2+2\dragf\betrep\sin\azimuth+\rhorep\sigrep(\rhorep\sigrep
  +2\dragf\cos\azimuth+2\betrep\sin2\azimuth)
\end{align}
which shows that the apparent ray frequency $\afreq$ is affected in general by both acceleration and dispersion.

\example{Ray frequency for observers at new phase}
For an observer at new phase, $\rcal$ has the value given by \eqnref{centrir3b}. Putting this value into \eqnref{centrif1}
yields
\begin{align}\label{centrif2}
\afreq=\pfreq\sqrt{\delrep^2+\betrep^2}
\end{align}
where $\delrep$ is given by \eqnref{centrir2b}. We conclude that at the instant when the observer is at new phase, the effect
of the observer's motion is to enhance the ray frequency.

\example{Ray frequency for observers at first quarter}
For an observer at first quarter, $\rcal$ has the value given by \eqnref{centrir5b}, and by substituting this value into
\eqnref{centrif1}, we get
\begin{align}\label{centrif3}
\afreq=\pfreq(1+\betrep)
\end{align}
from which we conclude that at the instant when the observer is at first quarter, the apparent frequency of the ray depends
only on the instantaneous speed of the observer and not on the observer's acceleration.

\example{Ray frequency for observers at full phase}
For an observer at new phase, $\rcal$ has the value given by \eqnref{centrir7b}. Substituting this value into \eqnref{centrif1}
leads to
\begin{align}\label{centrif4}
\afreq=\pfreq\sqrt{\delrep^2+\betrep^2}
\end{align}
where $\delrep$ is given by \eqnref{centrir2b}. We conclude that at the instant when the observer is at full phase, the effect
of the observer's motion is to enhance the ray frequency by the same amount as when the observer is at new phase.

\example{Ray frequency for observers at last quarter}
For an observer at last quarter, $\rcal$ has the value given by \eqnref{centrir9b}. Using this value of $\rcal$ in
\eqnref{centrif1}, we get
\begin{align}\label{centrif5}
\afreq=\pfreq|1-\betrep|
\end{align}
from which we conclude that at the instant when the observer is at last quarter, the effect of the observer's motion is to
reduce the ray frequency if $\betrep<1$ and to enhance the ray frequency if $\betrep>1$. For $\betrep=1$, the ray does not
propagate relative to the observer and the ray frequency therefore vanishes.

\section{Effects of precessional acceleration}
\art{On the apparent direction to a light source}
Throughout this section we consider an observer rotating with a constant angular speed $\Omerep$ about an axis which precesses
at a constant rate $\absv{\precrate}$ about a fixed direction $\precdir$ so that
\begin{align}\label{precd1}
\vectLam=\cprod{\precvect}{\vectOme},\quad
\fdot{\vectLam}=(\dprod{\precvect}{\vectOme})\precvect-\precrate^2\vectOme,\quad
\ffdot{\vectLam}=-\precrate^2(\cprod{\precvect}{\vectOme}),\quad
\fffdot{\vectLam}=-\precrate^2(\dprod{\precvect}{\vectOme})\precvect+\precrate^4\vectOme
\end{align}
where $\precvect$ is a vector with magnitude $\absv{\precrate}$ parallel to $\precdir$ for $\precrate>0$
and antiparallel to $\precdir$ for $\precrate<0$. Considering a light source whose
rays are incident in a direction $\unitkap$ parallel to $\precdir$ and are linearly polarized in a direction $\unitplz$
perpendicular to $\precdir$, we introduce a coordinates system with $\unitkap$, $\unitplz$ and $\unitjap=\cprod{\unitplz}{\unitkap}$
as basis vectors. Then we have
\begin{subequations}\label{precd2}
\begin{align}\label{precd2a}
\begin{split}
&\precdir=\unitkap,\quad
\vectOme=\Omerep(\unitjap\sin\latitm\cos\longtm+\unitplz\sin\latitm\sin\longtm+\unitkap\cos\latitm)\\
&\qquad\vectr=\scalr(\unitjap\sin\latitr\cos\longtr+\unitplz\sin\latitr\sin\longtr+\unitkap\cos\latitr)
\end{split}
\end{align}
\begin{align}\label{precd2b}
\zeropi\le\longtm<\doublepi,\quad
\zeropi\le\longtr<\doublepi,\quad
\zeropi\le\latitm\le\fullpi,\quad
\zeropi\le\latitr\le\fullpi
\end{align}
\end{subequations}
where $\latitm$ is the angle between $\unitkap$ and $\vectOme$, $\latitr$ is the angle
between $\unitkap$ and $\vectr$, $\longtm$ is the angle between $\unitjap$ and the projection of $\vectOme$ on the plane of
$\unitjap$ and $\unitplz$, while $\longtr$ is the angle between $\unitjap$ and the projection of $\vectr$ on the same plane.
In this coordinates system \eqnref{precd1} becomes
\begin{subequations}\label{precd3}
\begin{align}\label{precd3a}
\begin{split}
&\vectLam=\Omerep\precrate(\unitplz\sin\latitm\cos\longtm-\unitjap\sin\latitm\sin\longtm),\quad
\fdot{\vectLam}=-\Omerep\precrate^2(\unitjap\sin\latitm\cos\longtm+\unitplz\sin\latitm\sin\longtm)\\
&\ffdot{\vectLam}=-\Omerep\precrate^3(\unitplz\sin\latitm\cos\longtm-\unitjap\sin\latitm\sin\longtm),\quad
\fffdot{\vectLam}=\Omerep\precrate^4(\unitjap\sin\latitm\cos\longtm+\unitplz\sin\latitm\sin\longtm)
\end{split}
\end{align}
while the angle $\anglea$ between $\vectr$ and $\vectOme$ is given by the cosine law
\begin{align}\label{precd3b}
\cos\anglea=\cos\latitm\cos\latitr+\sin\latitm\sin\latitr\cos\angleb,\quad
\angleb=\longtm-\longtr,\quad
\zeropi\le\anglea\le\fullpi.
\end{align}
These equations are however much too general to be useful for the purposes of illustration. We shall suppose therefore in
the sequel that
\begin{align}\label{precd3c}
\cos\latitm=\cos\anglea=0,\quad\cos\angleb=0,\quad\cos\longtr=\sin\longtm,\quad\sin\longtr=-\cos\longtm
\end{align}
since the loss of generality incurred with this supposition is more than balanced by the instructiveness of its results.
\end{subequations}

\subart{Effects of polarization}
It was noted in \secref{INTRO} that the propagation condition $\dprod{\vectkap}{(\cprod{\unitplz}{\vectOme})}\ne0$ needs to hold
in order for the light rays to propagate relative to the rotating observer as regular rays. When this condition is violated, the
rays degenerate into a mode where they either cease to propagate or propagate with a velocity that is no longer given by \eqnref{main1}.
In terms of the particular situation being considered here, this condition requires that the angular velocity $\vectOme$ be noncollinear
with the polarization direction $\unitplz$. More generally, as the axis of rotation of the observer precesses about the line of
incidence of the rays, it will be carried into the plane of $\unitkap$ and $\unitplz$ at two instants corresponding to
$\longtm=\halfpi$ and $\longtm=\threeqpi$. At these instants the rays will no longer propagate as regular rays. We conclude that
if light rays from the cat's eye nebula are linearly polarized and incident at right angles to the ecliptic, then observational
astronomers will do well to study the photometry of light from this nebula when the earth's axis of rotation lies in a plane
determined by the polarization direction of the rays and the normal to the ecliptic.

\subart{Effects of acceleration}
To study the effects of acceleration on the obliquation angle of the light source, we use \eqnref{precd2} and \eqnref{precd3} to get
\begin{align}\label{precd4}
\begin{split}
\cprod{\unitplz}{\vectOme}&=-\unitkap\Omerep\cos\longtm,\quad
\cprod{\unitplz}{\vectLam}=\unitkap\Omerep\precrate\sin\longtm\\
\cprod{\vectLam}{\vectr}&=\scalr\Omerep\precrate(\unitjap\cos\longtm\cos\latitr+\unitplz\sin\longtm\cos\latitr)\\
\cprod{\vectOme}{\vectr}&=\scalr\Omerep(-\unitkap\sin\latitr-\unitplz\cos\latitr\cos\longtm+\unitjap\cos\latitr\sin\longtm)
\end{split}
\end{align}
which upon substitution into \eqnref{rot1} yields
\begin{subequations}\label{precd5}
\begin{align}\label{precd5a}
\begin{split}
&\epsva=0,\quad
\epsvb=0,\quad
\epsvc=\Omerep\sin\longtm,\quad
\epsvd=\scalr\cos\latitr,\quad
\epsve=0\\
&\epsvf=\scalr\sin\latitr\sin\longtr,\quad
\epsvg=0,\quad
\epsvh=-\scalr\Omerep\precrate\sin\latitr,\quad
\epsvi=\Omerep\precrate\cos\longtm\\
&\epsvj=-\Omerep\cos\longtm\ne0,\quad
\epsvk=-\scalr\Omerep\sin\latitr,\quad
\epsvl=-\scalr\Omerep\cos\latitr\cos\longtm\\
&\epsvm=\scalr\Omerep^2\precrate\cos\latitr,\quad
\epsvn=0,\quad
\epsvo=\scalr\Omerep\precrate\sin\longtm\cos\latitr
\end{split}
\end{align}
\begin{align}\label{precd5b}
\begin{split}
&\vphia=\scalr\Omerep,\quad
\vphib=\scalr\rotacc\Omerep^2,\quad
\vphic=0,\quad
\vphid=-\scalr\Omerep^2\cos\latitr,\quad
\vphie=0,\quad
\vphif=0,\quad
\vphig=0\\
&\vphih=-\cdkt/(2\pfreq^2\Omerep^2\cos^2\longtm),\quad
\vphii=(\cdkt\precrate)/(2\pfreq^2),\quad
\vphij=0,\quad
\vphik=-2\rhorep\scalr\Omerep^2\cos\latitr\\
&\vphil=1,\quad
\vphim=0,\quad
\vphin=0,\quad
\vphio=0,\quad
\vphip=(\scalr\precrate^2\Omerep^2\cdkt\cos\latitr)/(2\pfreq^2)\\
&\vphiq=-(\cdkt\precrate\Omerep)/(2\pfreq^2),\quad
\vphir=-4\rhorep\Omerep\sin\latitr\cos\latitr,\quad
\vphis=2\rhorep\Omerep(\cdkt+\scalc\dragf)\sin\latitr\cos\latitr\\
&\vphit=-\sin\latitr,\quad
\vphiu=-(\cos\latitr)/\rotacc,\quad
\vphiv=0
\end{split}
\end{align}
\end{subequations}
where, by using \eqnref{precd5} in \eqnref{rot12d}, we have
\begin{subequations}\label{precd6}
\begin{align}\label{prec6a}
\begin{split}
&\qquad\redOme=\Omerep/\pfreq,\quad
\redGam=\precrate/(2\pfreq),\quad
\GamOme=\precrate/\Omerep,\quad
\rotacc=(1+\GamOme^2\cos^2\latitr)^{1/2}\\
&\vthtrep=-\betrep\redOme\cos\latitr,\quad
\alprep=-\betrep\Omerep\pfreq\cos\latitr,\quad
\gamrep=1,\quad
\betrep=(\scalr\Omerep)/\scalc,\quad
\sigrep=\rotacc\Omerep\betrep
\end{split}
\end{align}
\begin{align}\label{prec6b}
\begin{split}
&\qquad\xcons=\frac{\vthtrep}{\sqrt{1+\vthtrep^2}},\quad
\dragf=\left\{\frac{1+\sqrt{1+\vthtrep^2}}{2}\right\}^{1/2},\quad
\rhorep=\frac{\xcons}{4\dragf\pfreq}\\
&\precvw=\rhorep\Omerep\betrep\cos\latitr,\quad
\precvu=\dragf+2\precvw,\quad
\cdkt=\scalc\precvu,\quad
\precvv=\redGam(4\precvu\precvw+\redOme^2\precvu^2)^{1/2}.
\end{split}
\end{align}
\end{subequations}
Accordingly, \eqnref{rot12a} and \eqnref{rot12b} become
\begin{subequations}\label{precd7}
\begin{align}\label{precd7a}
\begin{split}
&\lcal=-2\precvw\sin\latitr,\quad
\pcal=\precvv^2+4\precvw(\dragf+\betrep\sin\latitr)\\
&\qquad\ncal=\precvv^2+4\precvw(\dragf\cos^2\latitr-\precvw\sin^2\latitr)
\end{split}
\end{align}
\begin{align}\label{precd7b}
\begin{split}
&\gcal=-\betrep-\precvu\sin\latitr\\
&\rcal^2=\precvv^2+\rhorep^2\sigrep^2+\dragf^2+\betrep^2+4\precvw(\dragf+\betrep\sin\latitr)-2\dragf(\precvw-\betrep\sin\latitr)\\
&\fcal^2=\precvv^2+\rhorep^2\sigrep^2+\dragf^2\cos^2\latitr+4\precvw(\dragf\cos^2\latitr-\precvw\sin^2\latitr)-2\dragf\precvw
\end{split}
\end{align}
\end{subequations}
which upon substitution into \eqnref{rot12c} yields
\begin{align}\label{precd8}
\tan\psirep=
\frac{[\precvv^2+\rhorep^2\sigrep^2+\dragf^2\cos^2\latitr+4\precvw(\dragf\cos^2\latitr-\precvw\sin^2\latitr)-2\dragf\precvw]^{1/2}}
   {-\betrep-\precvu\sin\latitr}.
\end{align}
We conclude that the obliquation angle of the light source depends in general on the precession of the observer and the frequency
of the light rays.

\example{Selfconsistency of classical kineoptics}
When the observer's axis of rotation does not precess ($\precrate=0$), we have by \eqnref{precd6} that $\redGam=0,\,
\GamOme=0,\,\rotacc=1,\,\sigrep=\Omerep\betrep,\text{ and }\precvv=0$. Under these conditions, \eqnref{precd7} reduces --- as
it should --- to the corresponding equation for an observer in centripetal motion that was treated in \secref{CENTRIACC}. Considering
that the formulae which describe the propagation of light rays for observers in translation and rotation differ greatly from
one another as described in \secref{INTRO}, we have here a fine indication of the selfconsistency of our treatment of classical
kineoptics.

\subart{Obliquation at transverse position}
When the observer's position is at right angles to the axis of precession ($\sin\latitr=\pm1$), we have by \eqnref{precd6}
and \eqnref{precd7} that
\begin{subequations}\label{precd9}
\begin{align}\label{precd9a}
\begin{split}
&\rotacc=1,\quad
\vthtrep=0,\quad
\alprep=0,\quad
\sigrep=\betrep\Omerep,\quad
\xcons=0,\quad
\dragf=1\\
&\qquad\rhorep=0,\quad
\precvw=0,\quad
\precvu=1,\quad
\cdkt=\scalc,\quad
\precvv=\redGam\redOme
\end{split}
\end{align}
\begin{align}\label{precd9b}
\gcal=-\betrep\mp1,\quad
\rcal^2=\redGam^2\redOme^2+(1\pm\betrep)^2,\quad
\fcal^2=\redGam^2\redOme^2
\end{align}
\end{subequations}
where the upper signs correspond to $\sin\latitr=+1$ and the lower signs correspond to $\sin\latitr=-1$.
By \eqnref{precd8}, \eqnref{grad5a} and \eqnref{precd9b}, we obtain
\begin{align}\label{precd10}
\upsrep=\scalc\left[\redGam^2\redOme^2+(1\pm\betrep)^2\right]^{1/2},\quad
\tan\psirep=-\left\{\frac{\redOme}{\betrep\pm1}\right\}\absv{\redGam}
\end{align}
from which we conclude that the apparent angular displacement of the light source depends on the precession of the observer
as well as on the frequency of the light rays.

\subart{Obliquation at longitudinal position}
When the observer's position is parallel or antiparallel to the axis of precession ($\cos\latitr=\pm1$), we have by \eqnref{precd6}
and \eqnref{precd7} that
\begin{subequations}\label{precd11}
\begin{align}\label{prec11a}
\rotacc=\sqrt{1+\GamOme^2},\quad
\vthtrep=\mp\betrep\redOme,\quad
\alprep=\mp\betrep\Omerep\pfreq,\quad
\sigrep=\betrep\Omerep\rotacc
\end{align}
\begin{align}\label{prec11b}
\begin{split}
&\qquad\xcons=\mp\betrep(\redOme/\auxva),\quad
\dragf=\sqrt{(1+\auxva)/2},\quad
\rhorep=\xcons/(4\dragf\pfreq)\\
&\precvw=\pm\rhorep\Omerep\betrep,\quad
\precvu=\dragf+2\precvw,\quad
\cdkt=\scalc\precvu,\quad
\precvv=\redGam(4\precvu\precvw+\redOme^2\precvu^2)^{1/2}
\end{split}
\end{align}
\begin{align}\label{precd11c}
\gcal=-\betrep,\quad
\rcal^2=\betrep^2+\auxvb^2+(\dragf+\precvw)^2,\quad
\fcal^2=\auxvb^2+(\dragf+\precvw)^2
\end{align}
where, with the upper signs corresponding to $\cos\latitr=+1$ and the lower signs corresponding to $\cos\latitr=-1$,
we have introduced
\begin{align}\label{prec11d}
\auxva=\sqrt{1+\betrep^2\redOme^2},\quad
\auxvb=\sqrt{\precvv^2+\precvw^2\GamOme^2}.
\end{align}
\end{subequations}
Thus by \eqnref{precd8}, \eqnref{grad5a} and \eqnref{precd11c}, we obtain
\begin{align}\label{precd12}
\upsrep=\scalc\left[\betrep^2+\auxvb^2+(\dragf+\precvw)^2\right]^{1/2},\quad
\tan\psirep=-\frac{\sqrt{\auxvb^2+(\dragf+\precvw)^2}}{\betrep}
\end{align}
from which more convenient approximations may be obtained.

\art{On the apparent drift of a light source}\label{prec_adls}
To calculate the apparent drift of the light source, we assume for simplicity that $\sin\latitr=+1$ so
that \eqnref{precd9} holds. Substituting \eqnref{precd2} through \eqnref{precd7} into \eqnref{rxpeed1} under this
assumption then gives
\begin{subequations}\label{precs1}
\begin{align}\label{precs1a}
\begin{split}
&\dltva=0,\quad
\dltvb=\Omerep\precrate\sin\longtm,\quad
\dltvc=0,\quad
\dltvd=\scalc\Omerep\redOme\redGam,\quad
\dltve=0,\quad
\dltvf=0,\quad
\dltvg=1\\
&\dltvh=1/(4\pfreq^3),\quad
\dltvi=0,\quad
\dltvj=-\Omerep^2\precrate\cos^2\longtm,\quad
\dltvk=0,\quad
\dltvl=0,\quad
\dltvm=0\\
&\dltvn=0,\quad
\dltvo=0,\quad
\dltvp=0,\quad
\dltvq=0,\quad
\dltvr=0,\quad
\dltvs=0,\quad
\dltvt=0,\quad
\dltvu=0
\end{split}
\end{align}
\begin{align}\label{precs1b}
\efkta=\scalr,\quad
\efktb=0,\quad
\efktc=\scalr\Omerep\redGam\redOme,\quad
\efktd=0,\quad
\efkte=0,\quad
\efktf=0.
\end{align}
\end{subequations}
Substituting \eqnref{precs1} into \eqnref{rxpeed10b} in view of \eqnref{kas1c} and \eqnref{kas7b} leads to
\begin{subequations}\label{precs2}
\begin{align}\label{precs2a}
\begin{split}
&\vtta=-\scalc\precsx,\quad
\vttb=0,\quad
\vttc=0,\quad
\vttd=0,\quad
\vtte=-\scalc\precsy\\
&\vttf=0,\quad
\vttg=0,\quad
\vtth=0,\quad
\vtti=0,\quad
\vttj=\precsz
\end{split}
\end{align}
where we have introduced, with $\fcal$ given by \eqnref{precd9b},
\begin{align}\label{precs2b}
\precsx=\fcal+\frac{1+\betrep}{\fcal},\quad
\precsy=\frac{\precrate}{2\pfreq^2}\left\{\fcal+\frac{1+\betrep}{\fcal}\right\},\quad
\precsz=-\frac{(1+\betrep)+\fcal^2(1-\betrep)}{\betrep\fcal}.
\end{align}
\end{subequations}
Taking \eqnref{precd2} and \eqnref{precd4} into account, we substitute \eqnref{precs2}
into \eqnref{rxpeed10a} and \eqnref{rxpeed10c} to get
\begin{align}\label{precs3}
\angus=-\frac{1}{\betrep\scalc\rcal^2}\biggl[(\precsx+\betrep\precsz)\unitkap
  +\precsy\Omerep(\unitjap\cos\longtm+\unitplz\sin\longtm)\biggr],\quad
\aspua=0,\quad
\fdot{\psirep}=0
\end{align}
from which we conclude that at this particular instant, the apparent angular displacement $\psirep$ of the light
source is not varying with time.

\art{On the apparent path of a light source}
The contents of this article have been omitted in so far as I have not been able to put the formulae into a form that is
sufficiently simple, general, and illustrative of some pertinent principle.

\art{On the apparent geometry of rays}
The contents of this article have been omitted in so far as I have not been able to put the formulae into a form that is
sufficiently simple, general, and illustrative of some pertinent principle.

\art{On the apparent frequency of rays}
To study the apparent ray frequency for the precessing observer, we substitute the value of $\rcal$ from \eqnref{precd7b}
into \eqnref{kfreq4} to get
\begin{align}\label{precf1}
(\afreq/\pfreq)^2
=\precvv^2+\rhorep^2\sigrep^2+\dragf^2+\betrep^2+4\precvw(\dragf+\betrep\sin\latitr)-2\dragf(\precvw-\betrep\sin\latitr)
\end{align}
where $\precvv, \rhorep, \sigrep, \dragf, \betrep$ and $\precvw$ are given by \eqnref{precd6}. We conclude that the apparent
ray frequency $\afreq$ is affected in general by both dispersion and precession.

\subart{Apparent ray frequency at transverse position}
When the observer's position is at right angles to the axis of precession ($\sin\latitr=\pm1$), the various quantities defined
in \eqnref{precd6} have the values given by \eqnref{precd9a}. Putting these values into \eqnref{precf1} gives
\begin{align}\label{precf2}
\afreq=\pfreq\sqrt{\redGam^2\redOme^2+(1\pm\betrep)^2}
\end{align}
as the apparent ray frequency for the observer. In particular, if the observer were to rotate without precessing, the problem
will reduce to that of an observer in centripetal motion at first or last quarter, and the apparent ray frequencies given by
\eqnref{precf2} will coincide with those obtained in \artref{art_centrif}.

\subart{Apparent ray frequency at longitudinal position}
When the observer's position is collinear with the axis of precession ($\cos\latitr=\pm1$), the quantities defined
in \eqnref{precd6} have the values given by \eqnref{precd11}. Using these values in \eqnref{precf1} gives
\begin{align}\label{precf3}
\afreq=\pfreq\sqrt{\betrep^2+\auxvb^2+(\dragf+\precvw)^2}
\end{align}
as the apparent ray frequency for the observer. If the observer rotates without precessing, the problem
reduces to that of an observer in centripetal motion at new or full phase, and it is not difficult to show that the apparent
ray frequencies obtained from \eqnref{precf3} agree with those obtained previously in \artref{art_centrif}.

\section{Effects of gravitational acceleration}
\art{On the apparent direction to a light source}
Throughout this section we consider a gravitating observer and a light source whose rays are linearly polarized in a direction
that lies in the orbital plane of the observer's motion and are incident in a direction perpendicular to this plane. Thus
if $\unitvz$ is a unit vector directed from the dynamical focus of the orbit towards perihelion while $\unitvy$
is a unit positive normal to the plane of the orbit and $\unitvx=\cprod{\unitvy}{\unitvz}$, then
\begin{subequations}\label{gravd1}
\begin{align}\label{gravd1a}
\begin{split}
&\vectz=\scalz\unitvz,\quad
\vecth=-\scalh\unitvy,\quad
\vectr=\scalr(\unitvz\cos\tanomaly+\unitvx\sin\tanomaly)\\
&\qquad\vectkap=-\kaprep\unitvy,\quad
\unitplz=(\unitvz\cos\panomaly+\unitvx\sin\panomaly)
\end{split}
\end{align}
\begin{align}\label{gravd1b}
\begin{split}
&\qquad\quad\cprod{\vectr}{\vecth}=\scalr\scalh(\unitvx\cos\tanomaly-\unitvz\sin\tanomaly),\quad
\cprod{\vectz}{\vecth}=\scalz\scalh\unitvx,\quad
\cprod{\unitkap}{\vectz}=-\scalz\unitvx\\
&\cprod{\unitkap}{\unitplz}=\unitvz\sin\panomaly-\unitvx\cos\panomaly,\quad
\cprod{\unitkap}{\vectr}=\scalr(\unitvz\sin\tanomaly-\unitvx\cos\tanomaly),\quad
\cprod{\unitplz}{\vectz}=\unitvy\scalz\sin\panomaly
\end{split}
\end{align}
\end{subequations}
where the true anomaly $\tanomaly$ is the instantaneous angle between $\unitpos$ and $\unitvz$, and the polarization anomaly
$\panomaly$ is the constant angle between $\unitplz$ and $\unitvz$.

\subart{Effects of polarization}
We have remarked earlier in \secref{INTRO} that the propagation condition $\dprod{\unitplz}{(\cprod{\vectr}{\vecth})}\ne0$
must hold in order for the light rays in consideration to propagate relative to a gravitating observer. If the rays are
polarized in a direction that lies in the orbital plane of the observer, then there are two positions in the observer's
orbit at which the vectors $\unitpos$ and $\unitplz$ are parallel or antiparallel. When the observer is at either of
these positions, the propagation condition is violated, the light rays will not propagate relative to the observer
and the light source will be imperceptible to the observer. Assuming therefore that light rays from the cat's eye nebula
are linearly polarized and incident nearly perpendicularly to the earth's orbital plane, their polarization
direction will lie in this plane and there should be two positions on the earth's orbit at which these rays will not propagate
relative to an observer on the earth. Observational astronomers are therefore advised to pay close attention to the
photometry of light from this nebula as a function of the earth's position in its orbit.

\example{Permanently imperceptible light rays}
For light rays that are linearly polarized in a direction normal to the orbital plane of a gravitating observer, the propagation
condition is violated at all instants. Hence at no instant will the light rays propagate relative to the observer, from which
we conclude that the light source will be permanently imperceptible to the observer, all other things being equal.

\example{Transmission of radio signals to artificial satellites}
When a linearly polarized radio signal is transmitted to an artificial satellite, it is essential that the propagation
condition holds in order for the satellite to receive the signal. If this condition is violated, say by making the signal's
polarization direction to be nearly perpendicular to the satellite's orbital plane, then the satellite should find it
quite difficult to receive the signal, all other things being equal.

\subart{Effects of acceleration}
To study the effects of acceleration on obliquation for a gravitating observer, we observe that on account of \eqnref{gravd1},
the various quantities defined in \eqnref{ogrv1} become
\begin{subequations}\label{gravd2}
\begin{align}\label{gravd2a}
\begin{split}
&\quad\epsva=0,\quad
\epsvb=0,\quad
\epsvc=\cos\anodif,\quad
\epsvd=\scalz\cos\tanomaly,\quad
\epsve=0\\
&\epsvf=-\scalr\scalh\sin\anodif\ne0,\quad
\epsvg=0,\quad
\epsvh=\scalz\scalh\sin\tanomaly,\quad
\epsvi=\scalz\scalh\sin\panomaly
\end{split}
\end{align}
\begin{align}\label{gravd2b}
\begin{split}
&\vphia=\potent/\scalr,\quad
\vphib=\potent,\quad
\vphic=0,\quad
\vphid=-1,\quad
\vphie=1,\quad
\vphif=\scalc/(2\scalr\scalh\sin\anodif)\\
&\qquad\vphig=0,\quad
\vphih=\qhratio\orbspd,\quad
\vphii=0,\quad
\vphij=(\eccentty\sin\tanomaly)/\orbspd,\quad
\vphik=0\\
&\qquad\qquad\qquad\vphil=0,\quad
\vphim=0,\quad
\vphin=0,\quad
\vphio=0
\end{split}
\end{align}
\end{subequations}
where we have introduced~\cite{Boulet91, Danby92}
\begin{subequations}\label{gravd3}
\begin{align}\label{gravd3a}
\begin{split}
&\potent=\scalq/\scalr,\quad
\scala=\scalq/\scalr^2,\quad
\anodif=\tanomaly-\panomaly,\quad
\eccentty=\scalz/\scalq,\quad
\qhratio=\scalq/\scalh,\quad
\zhratio=\scalz/\scalh\\
&\qquad\semilat=\scalh^2/\scalq=\scalr(1+\eccentty\cos\tanomaly),\quad
\orbspd=(\eccentty^2+2\eccentty\cos\tanomaly+1)^{1/2}
\end{split}
\end{align}
and used the fact that by \eqnref{ogrv9c},
\begin{align}\label{gravd3b}
\begin{split}
&\quad\alprep=0,\quad
\gamrep=1,\quad
\vthtrep=0,\quad
\betrep=(\qhratio\orbspd)/\scalc\\
&\sigrep=\potent/(\scalr\scalc),\quad
\cdkt=\scalc,\quad
\xcons=0,\quad
\rhorep=0,\quad
\dragf=1.
\end{split}
\end{align}
\end{subequations}
In view of the foregoing equations, the quantities featured in \eqnref{ogrv9a} reduce to
\begin{subequations}\label{gravd4}
\begin{align}\label{gravd4a}
\begin{split}
\lcal=0,\quad
\pcal=0,\quad
\ncal=0,\quad
\gcal=-\betrep,\quad
\rcal=(1+\betrep^2)^{1/2},\quad
\fcal=1
\end{split}
\end{align}
which upon substitution into \eqnref{ogrv9b} and \eqnref{grad5a} yields
\begin{align}\label{gravd4b}
\upsrep=\scalc(1+\betrep^2)^{1/2},\quad
\tan\psirep=-\frac{1}{\betrep}.
\end{align}
\end{subequations}
We conclude that the obliquation angle of the light source depends only on the instantaneous speed of the observer
and not explicitly on the observer's acceleration.

\art{On the apparent drift of a light source}
To calculate the apparent drift of the light source, we substitute \eqnref{gravd1}, \eqnref{gravd2}, \eqnref{gravd3}
and \eqnref{gravd4a} into \eqnref{gxpeed1} to get
\begin{subequations}\label{gravs1}
\begin{align}\label{gravs1a}
\dltva=\scalh,\quad
\dltvb=0,\quad
\dltvc=0,\quad
\dltvd=\scalz\cos\panomaly,\quad
\dltve=0,\quad
\dltvf=-\scalz\sin\tanomaly
\end{align}
\begin{align}\label{gravs1b}
\begin{split}
&\efkta=0,\quad
\efktb=\scalq+\scalz\cos\tanomaly,\quad
\efktc=\scalz\cos\panomaly+\scalq\cos\anodif,\quad
\efktd=\scalc/\scalh,\quad
\efkte=0,\quad
\efktf=0\\
&\quad\efktg=-\scalq,\quad
\efkth=0,\quad
\efkti=1,\quad
\efktj=1/\pfreq,\quad
\efktk=1/(4\pfreq^3),\quad
\efktl=0,\quad
\efktm=0
\end{split}
\end{align}
\begin{align}\label{gravs1c}
\begin{split}
&\efktn=0,\quad
\efkto=0,\quad
\efktp=1/(\betrep\scalc^2),\quad
\efktq=1/\scalh^2,\quad
\efktr=-(\qhratio^2\orbspd^2)/\scalc\\
&\efkts=(\potent\zhratio\sin\tanomaly)/\scalr,\quad
\efktt=0,\quad
\efktu=0,\quad
\efktv=0,\quad
\efktw=0,\quad
\efktx=0.
\end{split}
\end{align}
\end{subequations}
Using \eqnref{gravd4a} in \eqnref{gxpeed11c} leads to
\begin{subequations}\label{gravs2}
\begin{align}\label{gravs2a}
\xcala=0,\quad
\xcalb=1,\quad
\xcalc=-\betrep^2,\quad
\xcald=1+\betrep^2
\end{align}
while using \eqnref{gravd2}, \eqnref{gravd3}, \eqnref{gravs1} and \eqnref{gravs2a} in \eqnref{gxpeed11b} yields
\begin{align}\label{gravs2b}
\begin{split}
&\qquad\vtta=0,\quad
\vttb=0,\quad
\vttc=0,\quad
\vttd=0\\
&\vtte=1/\semilat,\quad
\vttf=(\potent\zhratio\sin\tanomaly)/(4\scalr^3\pfreq^3),\quad
\vttg=\scalh^{-2}.
\end{split}
\end{align}
\end{subequations}
By substituting \eqnref{gravd2a}, \eqnref{gravd3b}, \eqnref{gravd4a}, \eqnref{gravs1} and \eqnref{gravs2}
into \eqnref{gxpeed11a} and \eqnref{gxpeed11d}, we obtain
\begin{align}\label{gravs3}
\angus=\frac{\unitvx\,(\zhratio+\qhratio\cos\tanomaly)+\unitvy\scalc-\unitvz\,\qhratio\sin\tanomaly}{\betrep\scalc^2(1+\betrep^2)},\quad
\aspua=-\frac{\eccentty\sin\tanomaly}{\orbspd\scalc(1+\betrep^2)},\quad
\fdot{\psirep}=-\frac{\sigrep\eccentty\sin\tanomaly}{\orbspd(1+\betrep^2)}
\end{align}
from which we conclude that dispersion and polarization have no effects on the slope and variation of obliquation for the observer.

\art{On the apparent path of a light source}
To study the effects of acceleration, dispersion and polarization on the apparent path of the light source, we substitute
\eqnref{gravd2}, \eqnref{gravd3} and \eqnref{gravs1} into \eqnref{gpath1} to get~\footnote{We use $\toppo$
to denote values that are not needed for the calculations and are therefore not evaluated for convenience.}
\begin{subequations}\label{gravp1}
\begin{align}\label{gravp1a}
\begin{split}
&\vsiga=-\scalr\sin\anodif,\quad
\vsigb=0,\quad
\vsigc=-\scalz\sin\panomaly\\
&\vrhoa=\scalq/\scalr^3,\quad
\vrhob=(\zhratio\sin\tanomaly)/\scalr,\quad
\vrhoc=0,\quad
\vrhod=\scalq^2(\cos\anodif+\eccentty\cos\panomaly-3\eccentty\sin\tanomaly\sin\anodif)/\scalr^3\\
&\vrhoe=0,\quad
\vrhof=0,\quad
\vrhog=0,\quad
\vrhoh=0,\quad
\vrhoi=0,\quad
\vrhoj=0,\quad
\vrhok=(\qhratio\orbspd)/\scalr\\
&\vrhol=(2\potent+15\zhratio^2\sin^2\tanomaly-3\qhratio^2\orbspd^2)/\scalr^2,\quad
\vrhom=\scalq(9\qhratio\orbspd\scalr^2-8\scalq-45\scalr\zhratio^2\sin^2\tanomaly)/\scalr^6\\
&\vrhon=15(\scalq\zhratio\sin\tanomaly)(2\scalq-3\qhratio\orbspd\scalr^2+7\scalr\zhratio^2\sin^2\tanomaly)/\scalr^7,\quad
\vrhoo=0,\quad
\vrhop=0,\quad
\vrhoq=0\\
&\vrhor=0,\quad
\vrhos=0,\quad
\vrhot=0,\quad
\vrhou=0,\quad
\vrhov=0,\quad
\vrhow=0,\quad
\vrhox=0
\end{split}
\end{align}
\begin{align}\label{gravp1b}
\begin{split}
&\ethva=0,\quad
\ethvb=0,\quad
\ethvc=0,\quad
\ethvd=0,\quad
\ethve=0,\quad
\ethvf=0,\quad
\ethvg=0,\quad
\ethvh=0,\quad
\ethvi=0\\
&\ethvj=0,\quad
\ethvk=\toppo,\quad
\ethvl=\toppo,\quad
\ethvm=\toppo,\quad
\ethvn=0,\quad
\ethvo=0,\quad
\ethvp=0,\quad
\ethvq=0\\
&\ethvr=0,\quad
\ethvs=0,\quad
\ethvt=0,\quad
\ethvu=0,\quad
\ethvv=0\\
&\frkya=0,\quad
\frkyb=0,\quad
\frkyc=0,\quad
\frkyd=-1,\quad
\frkye=-1,\quad
\frkyf=-1,\quad
\frkyg=0,\quad
\frkyh=0\\
&\frkyi=1,\quad
\frkyj=\scalq^2(2\potent-3\zhratio^2\sin^2\tanomaly-3\qhratio^2\orbspd^2)/\scalr^8,\quad
\frkyk=0,\quad
\frkyl=0,\quad
\frkym=0\\
&\frkyn=0,\quad
\frkyo=0,\quad
\frkyp=0,\quad
\frkyq=0,\quad
\frkyr=0,\quad
\frkys=0,\quad
\frkyt=1,\quad
\frkyu=0\\
&\frkyv=0,\quad
\frkyw=0,\quad
\frkyx=0,\quad
\frkyy=\vrhoa^2,\quad
\frkyz=\scalq\scalz(\eccentty+\cos\tanomaly)
\end{split}
\end{align}
\begin{align}\label{gravp1c}
\begin{split}
&\vakpa=0,\quad
\vakpb=0,\quad
\vakpc=0,\quad
\vakpd=0,\quad
\vakpe=0,\quad
\vakpf=-\scalq/\scalr^3,\quad
\vakpg=0\\
&\vakph=-\scalq^2/\scalr^6,\quad
\vakpi=\scalq/\scalr^2,\quad
\vakpj=0,\quad
\vakpk=-\scalh^2\scalq^2/\scalr^6,\quad
\vakpl=0,\quad
\vakpm=0\\
&\vakpn=0,\quad
\vakpo=-(\scalq^2\scalz\scalh\sin\panomaly)/\scalr^6,\quad
\vakpp=-(\scalq^2\scalh\sin\anodif)/\scalr^5,\quad
\vakpq=-(\scalq^2\scalz\scalh\sin\tanomaly)/\scalr^5\\
&\vakpr=0,\quad
\vakps=\scalq^3(\cos\anodif+\eccentty\cos\panomaly)/\scalr^6,\quad
\vakpt=\scalz\scalq^3(\eccentty+\cos\tanomaly)/\scalr^6,\quad
\vakpu=(\scalh\scalq^2)/\scalr^6\\
&\vakpv=-(\scalh\scalq^2)/\scalr^6,\quad
\vakpw=0,\quad
\vakpx=0,\quad
\vakpy=0,\quad
\vakpz=0\\
&\parva=0,\quad
\parvb=0,\quad
\parvc=0,\quad
\parvd=0,\quad
\parve=0,\quad
\parvf=-\scalq^2/\scalr^6,\quad
\parvg=0.
\end{split}
\end{align}
\end{subequations}
Substituting \eqnref{gravp1} into \eqnref{gpath30} and taking \eqnref{gravd1}, \eqnref{gravd2}, \eqnref{gravd3}
and \eqnref{gravs1} into account, we obtain
\begin{align}\label{gravp2}
\begin{split}
&\bbk=\plusmin1/\qhratio,\quad
\bbt=0,\quad
\frt=\unitvz\cos\tanomaly+\unitvx\sin\tanomaly,\quad
\frb=\unitvy\\
&\qquad\quad\frn=\unitvx\cos\tanomaly-\unitvz\sin\tanomaly,\quad
\frc=\plusmin\qhratio\frn
\end{split}
\end{align}
from which we conclude that the apparent path of the light source is a circle with a radius equal in magnitude to the quantity
$\qhratio$ and lying in a plane parallel to the orbital plane of the observer.

\scholium{Interpretation of the true position of a light source}
If we consider the center of curvature of the apparent path of a light source to represent the true position of the light source,
it follows from \eqnref{gravp2} that a vector drawn from the apparent position of the light source to its true position does not
have the same magnitude as the speed $\qhratio\orbspd$ of the observer. On the other hand, if we consider the true position of
a light source to be such that a vector drawn from the apparent position of the light source to its true position necessarily
has the same magnitude as the speed of the observer, as Hamilton seemed to have supposed, then it will follow from \eqnref{gravp2}
that although the apparent path of the light is an exact circle, the true position of the light source will not be at the center
of this circle unless the observer's orbit is exactly circular ($\orbspd=1$), a conclusion that was first established by
Hamilton~\cite{Tait66}.

\art{On the apparent geometry of rays}
To study the apparent geometry of the light rays, we substitute \eqnref{gravd2}, \eqnref{gravd3}, \eqnref{gravs1} and
\eqnref{gravp1} into \eqnref{gray1} to get
\begin{subequations}\label{gravr1}
\begin{align}\label{gravr1a}
\begin{split}
&\veka=0,\quad
\vekb=0,\quad
\vekc=0,\quad
\vekd=0,\quad
\veke=\atilde/\scalh^2,\quad
\vekf=0,\quad
\vekg=0\\
&\vekh=-\scalc,\quad
\veki=0,\quad
\vekj=0,\quad
\vekk=\atilde\scalc,\quad
\vekl=0,\quad
\vekm=0,\quad
\vekn=-\atilde\\
&\veko=0,\quad
\vekp=-\atilde,\quad
\vekq=0,\quad
\vekr=0,\quad
\veks=0,\quad
\vekt=0,\quad
\veku=-\atilde\scalh^2\\
&\vekv=\scala\scalc\scalz\sin\tanomaly,\quad
\vekw=0,\quad
\vekx=\potent\scalc,\quad
\veky=\scala\scalc\scalz\cos\tanomaly,\quad
\vekz=\scala\scalc\cos\anodif
\end{split}
\end{align}
\begin{align}\label{gravr1b}
\begin{split}
&\vpsa=\scala\semilat\qhratio\sin\anodif,\quad
\vpsb=-\potent\scala\semilat\zhratio\sin\panomaly,\quad
\vpsc=-\scala\scalz\semilat\qhratio\sin\tanomaly,\quad
\vpsd=\scala\scalc\brquot,\quad
\vpse=-\scalh\atilde\\
&\qquad\vpsf=-\scalh\scalc\atilde,\quad
\vpsg=\scala\scalc\sin\anodif,\quad
\vpsh=-\potent\scalh\scalc,\quad
\vpsi=-\scalh\scalc\atilde^2,\quad
\vpsj=-\potent/\scalh^2
\end{split}
\end{align}
\end{subequations}
where we have introduced the convenient quantities
\begin{align}\label{gravr2}
\atilde=\scala/\scalr,\quad
\hoverc=\scalh/\scalc,\quad
\brquot=\left\{1+\frac{\hoverc^2}{\scalr^2}\right\}^{1/2}.
\end{align}
Substituting \eqnref{gravr1} into \eqnref{gray10} and taking \eqnref{gravd1}, \eqnref{gravd2}, \eqnref{gravd3},
\eqnref{gravs1} and \eqnref{gravp1} into account, we obtain
\begin{subequations}\label{gravr3}
\begin{align}\label{gravr3a}
\bbkbar=\plusmin\frac{\sigrep\brquot}{\scalc\rcal^3},\quad
\bbtbar=-\frac{\hoverc}{\scalr^2\brquot^2}
\end{align}
\begin{align}\label{gravr3b}
\frtbar=\frac{1}{\scalc\rcal}\biggl[\unitvz\,\qhratio\sin\tanomaly-\unitvx\,(\zhratio+\qhratio\cos\tanomaly)-\unitvy\,\scalc\biggr],\quad
\frbbar=\frac{1}{\scalr\brquot}\biggl[\unitvz\,\scalr\sin\tanomaly-\unitvx\,\scalr\cos\tanomaly+\unitvy\,\hoverc\biggr]
\end{align}
\end{subequations}
where $\rcal$ is given by \eqnref{gravd4a}. We conclude that the light ray has both a curvature and a torsion neither of which
is affected by polarization or dispersion.

\example{Apparent ray geometry for observers in circular orbits}
When an observer is in a circular orbit, the quantities $\scalr$ and $\betrep$ have constant values by \eqnref{gravd3}.
The curvature and torsion of the light ray then have constant nonzero values by \eqnref{gravr3a}, and we conclude therefore
that the light ray is a circular helix. Moreover, since $\hoverc>0$ by \eqnref{gravr2}, we have $\bbtbar<0$ by \eqnref{gravr3a}.
We conclude also that the light ray is a left handed helix.

\example{Some clouds may be kineoptical illusions}
When we consider a multitude of light rays each of which is in the form of a circular helix, it would seem that when the light source
is sufficiently distant from the observer, that these rays may give the illusion of a cloud surrounding the light source. If this be so,
then perhaps it may not be entirely without merit to conjecture that such apparent clouds do exist, and that such a cloud can in fact be
perceived in popular images of the cat's eye nebula.

\scholium{Acceleration-induced light bending is kineoptical}
It is often said by those who ought to have reasoned better that when an observer in accelerated motion perceives a light ray to
be curvilinear instead of rectilinear, that the curvature of the ray is indeed due the observer's acceleration, but that since
gravitational and nongravitational accelerations are indistiguishable, that the curvature of the ray can be ascribed to gravity.
These careless minds spare no effort to assure us that gravity and the curvature of the ray are a manifestation of the geometry
of a ``spacetime'' continuum, nor do they show kind regards towards anyone who dares to voice an objection to their immaculate
doctrines, however reasoned such objections may be. It will suffice for us to say that the curvature of the ray is a kineoptical
effect due to the acceleration of the observer, and that it makes no difference whatsoever to the phenomena whether this acceleration
is caused by the action of gravity or by the action of fairies~\footnote{So long as the dependence of the acceleration
on time and position remains the same.}. From the kineoptical viewpoint, therefore, to ascribe the curvature
of the ray in principle to a particular cause of acceleration is to indulge oneself in profound nonsense.

\art{On the apparent frequency of rays}
To study the apparent ray frequency for the observer, we substitute the value of $\rcal$ from \eqnref{gravd4a}
into \eqnref{kfreq4} to get
\begin{align}\label{gravf1}
\afreq=\pfreq\sqrt{1+\betrep^2}
\end{align}
where $\betrep$ is given by \eqnref{gravd3b}. We conclude that the apparent ray frequency $\afreq$ depends only on the
instantaneous speed of the observer and not explicitly on the observer's acceleration.

\example{Effects of eccentricity on apparent ray frequency}
Substituting the value of $\betrep$ from \eqnref{gravd3b} into \eqnref{gravf1}, we obtain
\begin{subequations}\label{gravf2}
\begin{align}\label{gravf2a}
\afreq=\pfreq\sqrt{1+\frac{\dilfact\potent}{\scalc^2}},\quad
\dilfact=\frac{\scalq\scalr\orbspd}{\scalh^2}.
\end{align}
Observing now that by \eqnref{gravd3},
\begin{align}\label{gravf2b}
\orbspd=\frac{\scalh^2}{\scalq}\left\{\frac{2}{\scalr}-\frac{1-\eccentty^2}{\semilat}\right\},
\end{align}
we substitute this value of $\orbspd$ into \eqnref{gravf2a} to get
\begin{align}\label{gravf2c}
\dilfact=2-\left\{\frac{1-\eccentty^2}{1+\eccentty\cos\tanomaly}\right\}
\end{align}
\end{subequations}
which permits the effects of eccentricity on the apparent frequency of the light ray to be determined.

\example{Apparent ray frequency at perihelion}
When the observer is at perihelion, we have $\tanomaly=\zeropi$ so that by \eqnref{gravf2} and \eqnref{gravd3}, the apparent ray
frequency satisfies
\begin{align}\label{gravf3}
\afreq_p=\pfreq\sqrt{1+\frac{\potent_p(1+\eccentty)}{\scalc^2}},\quad
\potent_p=\qhratio^2(1+\eccentty)
\end{align}
where $\afreq_p$ and $\potent_p$ are respectively the values of $\afreq$ and $\potent$ when the observer is at perihelion.

\example{Apparent ray frequency at aphelion}
When the observer is at aphelion, we have $\tanomaly=\fullpi$ so that by \eqnref{gravf2} and \eqnref{gravd3}, the apparent ray
frequency satisfies
\begin{align}\label{gravf4}
\afreq_a=\pfreq\sqrt{1+\frac{\potent_a(1-\eccentty)}{\scalc^2}},\quad
\potent_a=\qhratio^2(1-\eccentty)
\end{align}
where $\afreq_a$ and $\potent_a$ are respectively the values of $\afreq$ and $\potent$ when the observer is at aphelion.

\example{Frequency shift for observers in eccentric orbits}
From \eqnref{gravf3} and \eqnref{gravf4}, we find that the apparent ray frequencies at aphelion and perihelion differ by a
small amount given to a second order accuracy in $\qhratio/\scalc$ by
\begin{align}\label{gravf5}
\Delrep\omerep=\biggl\{\frac{\scalq\eccentty}{\scalh\scalc}\biggr\}^2\pfreq
\end{align}
where $\Delrep\omerep=\afreq_p-\afreq_a$. We conclude therefore that if the frequency of light rays from the cat's eye nebula
is measured when the earth is at perihelion and when the earth is at aphelion, the two measurements should differ by the
amount given by \eqnref{gravf5}, all other things being equal.

\appendix
\newcommand{\xa}{\scal{a}}
\newcommand{\xb}{\scal{b}}
\newcommand{\xc}{\scal{c}}
\newcommand{\xd}{\scal{d}}
\newcommand{\xu}{\scal{u}}
\newcommand{\va}{\vect{a}}
\newcommand{\vb}{\vect{b}}
\newcommand{\vc}{\vect{c}}
\newcommand{\vd}{\vect{d}}
\newcommand{\vu}{\vect{u}}
\newcommand{\php}{\phirep}
\newcommand{\psp}{\psirep}
\section{Appendices}
\subsection{Vector Identities (Algebra)}
Let $\va,\vb,\vc,\vd$ be vectors. Then,
\begin{equation}\label{alg1}
\cprod{\va}{(\cprod{\vb}{\vc})}
=\vb(\dprod{\va}{\vc})-\vc(\dprod{\va}{\vb})
\end{equation}
\begin{equation}\label{alg2}
\dprod{(\cprod{\va}{\vb})}{(\cprod{\vc}{\vd})}
=(\dprod{\va}{\vc})(\dprod{\vb}{\vd})-(\dprod{\va}{\vd})(\dprod{\vb}{\vc})
\end{equation}
\begin{equation}\label{alg3}
(\cprod{\va}{\vb})^2+(\dprod{\va}{\vb})^2=\xa^2\xb^2
\end{equation}
\begin{equation}\label{alg4}
\dprod{\va}{(\cprod{\vb}{\vc})}
=\dprod{\vb}{(\cprod{\vc}{\va})}
=\dprod{\vc}{(\cprod{\va}{\vb})}
\end{equation}
\begin{equation}\label{alg5}
\cprod{(\cprod{\va}{\vb})}{(\cprod{\vc}{\vd})}
=\vc[\dprod{\vd}{(\cprod{\va}{\vb})}]-\vd[\dprod{\vc}{(\cprod{\va}{\vb})}]
\end{equation}

\subsection{Vector Identities (Calculus)}
Let $\va,\vb,\vc,\vd$ be vector fields with respect to a vector $\vu$. Let $\phirep,\psirep$
be scalar fields with respect to $\vu$. Let $n$ be an integer. Then,
\begin{equation}\label{clc11}
\udivg{\vu}=3
\end{equation}
\begin{equation}\label{clc12}
\udivg{(\ucurl{\va})}=0
\end{equation}
\begin{equation}\label{clc13}
\udivg{(\cprod{\va}{\vb})}=\dprod{\vb}{(\ucurl{\va})}-\dprod{\va}{(\ucurl{\vb})}
\end{equation}

\begin{equation}\label{clc21}
\ucurl{\vu}=\zvect
\end{equation}
\begin{equation}\label{clc22}
\ucurl{(\ugrad{\php})}=\zvect
\end{equation}
\begin{equation}\label{clc23}
\ucurl{(\ucurl{\va})}=\ugrad{(\udivg{\va})}-{\ugrad{}}^2\va
\end{equation}
\begin{equation}\label{clc24}
\ucurl{(\php\va)}=\php(\ucurl{\va})-\cprod{\va}{(\ugrad{\php})}
\end{equation}
\begin{equation}\label{clc25}
\ucurl{(\cprod{\va}{\vb})}=\va(\udivg{\vb})-\ugdiv{\va}{\vb}+\ugdiv{\vb}{\va}-\vb(\udivg{\va})
\end{equation}

\begin{equation}\label{clc31}
\ugrad{(\xu^n)}=n\xu^{n-2}\vu
\end{equation}
\begin{equation}\label{clc32}
\ugrad{[\php(\psp)]}=(\textd\php/\textd\psp)(\ugrad{\psp})
\end{equation}
\begin{equation}\label{clc33}
\ugrad{(\php\psp)}=\php(\ugrad{\psp})+\psp(\ugrad{\php})
\end{equation}
\begin{equation}\label{clc34}
\ugrad{(\dprod{\va}{\vu})}=\va+\ugdiv{\vu}{\va}+\cprod{\vu}{(\ucurl{\va})}
\end{equation}
\begin{equation}\label{clc35}
\xa(\ugrad{\xa})=\cprod{\va}{(\ucurl{\va})}+\ugdiv{\va}{\va}
\end{equation}
\begin{equation}\label{clc36}
\ugrad{(\dprod{\va}{\vb})}=\cprod{\va}{(\ucurl{\vb})}+\ugdiv{\va}{\vb}
  +\ugdiv{\vb}{\va}+\cprod{\vb}{(\ucurl{\va})}
\end{equation}
\begin{equation}\label{clc37}
|\cprod{\va}{\vu}|[\ugrad{|\cprod{\va}{\vu}|}]=
  \xu^2\xa(\ugrad{\xa})+\xa^2\vu-(\dprod{\va}{\vu})\ugrad{(\dprod{\va}{\vu})}
\end{equation}

\begin{equation}\label{clc41}
\ugdiv{\va}{\vu}=\va
\end{equation}
\begin{equation}\label{clc42}
\ugdiv{\va}{(\php\vb)}=\vb[\dprod{\va}{(\ugrad{\php})}]+\php[\ugdiv{\va}{\vb}]
\end{equation}
\begin{equation}\label{clc43}
\ugdiv{\va}{(\cprod{\vb}{\vc})}=\cprod{\vb}{[\ugdiv{\va}{\vc}]}-\cprod{\vc}{[\ugdiv{\va}{\vb}]}
\end{equation}

\subsection{Important Riders}
For the convenience of the reader, we state here the conditions under which \eqnref{main1}
was derived~\cite{Adewole01b}. This equation assumes that the strengths or intensities of electric and magnetic
fields are characterized by vector quantities, that magnetic and electric fluxes (or charges) and currents
are defined and measured in accordance with Hertz's theory, and that in the space region of interest,
\begin{itemize}
\item the Gauss-Ostrogradsky divergence theorem holds for the field vectors,
\item there is conservation of magnetic and electric fluxes or charges,
\item there is a stationary, homogeneous and isotropic medium with no boundaries,
\item the light waves are plane, monochromatic, linearly polarized, have a complex frequency with a real wave vector,
\item the observer is in accelerated translational, rotational or gravitational motion.
\end{itemize}
When any or some of these conditions are not satisfied in a space region, the calculations in this monograph
may need to be reworked in the appropriate manners.

\bibliographystyle{unsrt}
\bibliography{xbib}

\begin{thebibliography}{10}

\bibitem{Adewole01b}
A.~I.~A. Adewole.
\newblock Light propagation for accelerated observers.
\newblock Research Paper GEM-2001-01C, Adequest Corporation, 2001.
\newblock See also http://xxx.lanl.gov/abs/physics/0104069.

\bibitem{Phipps91}
T.~E.~Phipps Jr.
\newblock Stellar aberration from the standpoint of the radiation convection
  hypothesis.
\newblock {\em Phys. Essays}, 4:368--372, 1991.

\bibitem{Phipps93}
T.~E.~Phipps Jr.
\newblock On {Hertz's} invariant form of {Maxwell's} equations.
\newblock {\em Phys. Essays}, 6:249--256, 1993.

\bibitem{Liebscher98}
D.-E. Liebscher \&~P. Brosche.
\newblock Aberration and relativity.
\newblock {\em Astron. Nachr.}, 319:309--318, 1998.
\newblock See also http://www.aip.de/\verb ~ lie/publications/plj346.html.

\bibitem{Bohm96}
D.~Bohm.
\newblock {\em The special theory of relativity}.
\newblock Routledge, London, 1996.

\bibitem{Atwater74}
H.~A. Atwater.
\newblock Non-simultaneity in the aberration of starlight.
\newblock {\em Am. J. Physics}, 42:1022--1024, 1974.

\bibitem{Howard95}
A.~Howard \& L. Kitchen \&~S. Dance.
\newblock Relativistic raytracing: simulating the visual appearance of rapidly
  moving objects.
\newblock http://www.cs.mu.oz.au/\verb#~#andrbh/raytrace/raytrace.html, 1995.

\bibitem{Rau98}
R.~Rau \& D. Weiskopf \&~H. Ruder.
\newblock Special relativity in virtual reality.
\newblock In H.~C. Hege \&~K. Polthier, editor, {\em Mathematical
  Visualization}, pages 269--279. Springer, 1998.

\bibitem{Weiskopf99}
D.~Weiskopf \& U. Kraus \&~H. Ruder.
\newblock Searchlight and {Doppler} effects in the visualization of special
  relativity: {A} corrected derivation of the transformation of radiance.
\newblock {\em {ACM} Transactions On Graphics}, 18:278--292, 1999.
\newblock See also
  http://wwwvis.informatik.uni-stuttgart.de/\verb#~#weiskopf/publications.

\bibitem{Savage99}
C.~M. Savage \& A.~C. Searle.
\newblock Visualising special relativity.
\newblock {\em The Physicist (Australian Institute Of Physics)}, 36:*, 1999.
\newblock See also http://www.anu.edu.au/Physics/Searle/.

\bibitem{Sardin01}
G.~Sardin.
\newblock Measure of absolute speed through the {Bradley} aberration of light
  beams on a {three-axis} frame.
\newblock {\em Europhys. Lett.}, 53:310--316, 2001.

\bibitem{Kassner02}
K.~Kassner.
\newblock Why the {Bradley} aberration cannot be used to measure absolute
  speeds. {A} comment.
\newblock arXiv e-print, 2002.
\newblock astro-ph/0203056.

\bibitem{Collins61}
O.~C. Collins \& W.~F. Hiltner.
\newblock The determination of the velocity vector of a space vehicle from
  apparent changes in the angular separation of stars.
\newblock {\em Astron. Journal}, 66:41--41, 1961.
\newblock Presented at the 107th meeting of the American Astronomical Society,
  December 1960.

\bibitem{Tagaki56}
S.~Tagaki.
\newblock Note on the computation of the apparent place of a star. {III}.
\newblock {\em Publications of the astronomical society of {Japan}}, 8:18--26,
  1956.

\bibitem{Klioner01}
S.~A. Klioner.
\newblock Practical relativistic model of microarcsecond astrometry in space.
\newblock arXiv e-print, 2001.
\newblock astro-ph/0107457.

\bibitem{Porter50}
J.~G. Porter \& D.~H. Saddler.
\newblock Stellar aberration.
\newblock {\em M. N. R. A. S.}, 110:467--476, 1950.

\bibitem{Stumpff79}
P.~Stumpff.
\newblock The rigorous treatment of stellar aberration and {Doppler} shift, and
  the barycentric motion of the earth.
\newblock {\em Astron. Astrophys.}, 78:229--238, 1979.

\bibitem{Ron86}
C.~Ron \&~J. Vondr{\'a}k.
\newblock Expansion of annual aberration into trigonometric series.
\newblock {\em Bull. Astron. Inst. Czechosl.}, 37:96--103, 1986.

\bibitem{Scott64}
F.~P. Scott.
\newblock A method for evaluating the elliptic {E} terms of the aberration.
\newblock {\em Astron. Journal}, 69:372--373, 1964.

\bibitem{Bradley29}
J.~Bradley.
\newblock An account of a new discovered motion of fixed stars.
\newblock {\em Phil. Trans. Roy. Soc.}, 35:637--660, 1729.

\bibitem{Fresnel18}
A.~Fresnel.
\newblock Sur l'influence du mouvement de terre dans quelques ph{\'e}nom{\`
  e}nes d'optique.
\newblock {\em Oeuvres Compl{\` e}tes}, 2:627--*, 1818.

\bibitem{Rothman95}
{M. Rothman,} {T. Rothman}~\& {P. Anninos}.
\newblock Reference frames for stellar aberration.
\newblock Technical Report TR033, National Center For Supercomputing
  Applications, 1995.
\newblock See also http://archive.ncsa.uiuc.edu/Pubs/TechReports/TR033.html.

\bibitem{Einstein05}
A.~Einstein.
\newblock Zur elektrodynamik bewegter k{\"o}rper.
\newblock {\em Ann. Physik}, 17:891--921, 1905.
\newblock See also http://www.fourmilab.ch/etexts/einstein/specrel/www/.

\bibitem{Brown01}
K.~S. Brown.
\newblock Reflections on relativity.
\newblock http://www.mathpages.com/rr/rrtoc.htm.
\newblock Sections 2.5 \& 2.6.

\bibitem{Thuring78}
B.~Th{\"u}ring \&~F. Schmeidler.
\newblock Vertikalkreis-beobachtungen der aberration des lichtes von sternen
  verscheidenenen spektraltyps.
\newblock {\em Astron. Nachr.}, 299:55--58, 1978.

\bibitem{Hewitt00}
J.~N. Hewitt.
\newblock A low frequency radio array.
\newblock Meeting Transparencies LIGO-G000247-00-D, LIGO Scientific
  Collaboration, 2000.

\bibitem{Hamilton47}
W.~R. Hamilton.
\newblock The hodograph, or a new method of expressing in symbolical language
  the newtonian law of attraction.
\newblock {\em Proc. Roy. Irish Acad.}, 3:344--353, 1847.

\bibitem{Stumpff80}
P.~Stumpff.
\newblock On the relationship between classical and relativistic theory of
  stellar aberration.
\newblock {\em Astron. Astrophys.}, 84:257--259, 1980.

\bibitem{Selleri01}
F.~Selleri.
\newblock Space and time should be preferred to spacetime {1}.
\newblock In K.~Rudnicki, editor, {\em Gravitation, Electromagnetism And
  Cosmology: Toward A New Synthesis}, pages 57--72. Apeiron, 2001.

\bibitem{Mashhoon01}
B.~Mashhoon.
\newblock Gravitation and nonlocality.
\newblock arXiv e-print, 2001.
\newblock gr-qc/0112058.

\bibitem{Page25}
L.~Page.
\newblock On the aberration of light.
\newblock {\em Astrophysical Journal}, 61:70--72, 1925.

\bibitem{Struik88}
D.~J. Struik.
\newblock {\em Lectures on classical differential geometry}.
\newblock Dover Publications, Inc., New York, second edition, 1988.

\bibitem{Sher68}
D.~Sher.
\newblock The relativistic {Doppler} effect.
\newblock {\em J. Roy. Astron. Soc. Can.}, 62:105--112, 1968.

\bibitem{Tatum85}
J.~B. Tatum.
\newblock The oblique {Doppler} effect.
\newblock {\em J. Roy. Astron. Soc. Can.}, 79:302--311, 1985.

\bibitem{Baird00}
E.~Baird.
\newblock Transverse redshift effects without special relativity.
\newblock arXiv e-print, 2000.
\newblock physics/0010074.

\bibitem{Bonizzoni00}
I.~Bonizzoni \&~G. Giuliani.
\newblock The interpretations by experimenters of experiments on {``time
  dilation''}: 1940--1970 circa.
\newblock arXiv e-print, 2000.
\newblock physics/0008012.

\bibitem{Tait66}
P.~G. Tait.
\newblock {Sir William Rowan Hamilton}.
\newblock {\em North British Review}, 45:37--74, 1866.

\bibitem{Boulet91}
D.~L. Boulet.
\newblock {\em Methods of orbit determination for the microcomputer}.
\newblock Willmann-Bell Inc., Virginia, 1991.

\bibitem{Danby92}
J.~M.~A. Danby.
\newblock {\em Fundamentals of celestial mechanics}.
\newblock Willmann-Bell Inc., Virginia, 1992.

\end{thebibliography}
\printindex
\end{document}